# Proceedings of the 16th Virtual International Meeting on Fully 3D Image Reconstruction in Radiology and Nuclear Medicine

**Editors:** Georg Schramm, Ahmadreza Rezaei, Kris Thielemans and Johan Nuyts

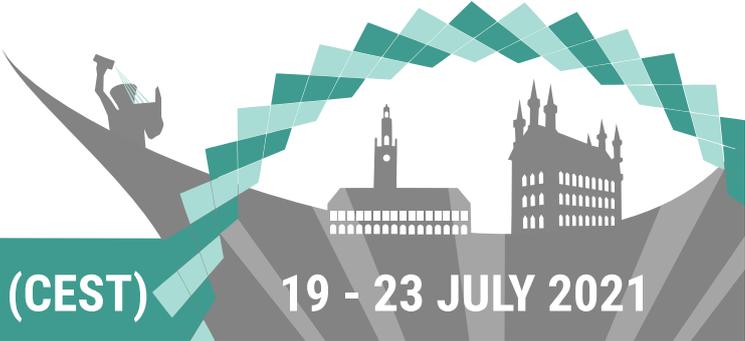

16th **Virtual** International Meeting on
Fully 3D Image Reconstruction in
Radiology and Nuclear Medicine

FULLY3D  Leuven Belgium (CEST)    19 - 23 JULY 2021

# Preface

Welcome to the Proceedings of the 2021 International Meeting on Fully Three-Dimensional Image Reconstruction in Radiology and Nuclear Medicine, which was to be held in Leuven, Belgium, but had to be transformed in a virtual meeting because of the covid-19 pandemic.

This was the 16th in a series of meetings that have served as one of the major forums for presentation of new results in the field of 3D image reconstruction, primarily with applications in x-ray computer tomography, PET and SPECT. The proceedings of the 2021 meeting are deposited in arXiv, and these proceedings and those of all past meetings are archived at http://www.fully3d.org/index.html.

Over the life-time of the meeting the focus has shifted to reflect recent developments in the field. Many of the major developments in fully 3D PET and SPECT imaging were first presented at Fully3D, as were the key results for analytic reconstruction methods in cone beam x-ray CT. Also in this meeting, a broad range of topics has been presented. As expected, deep learning methods are being used increasingly: they are used as post-processing tools or incorporated in the reconstruction, often aiming at improved noise and/or artifact suppression. This year we had three keynote speakers. The talk by *Prof. Bart Preneel*, COSIC KU Leuven and IMEC, Belgium, an expert in cryptography, cybersecurity and privacy, was entitled "The Future of Security and Privacy". *Prof. Koen Van Laere*, Head of Nuclear Medicine at KU Leuven, discussed "Ultrahigh spatial resolution for PET - The holy grail in clinical nuclear medicine?", and *Prof. Marcel Van Herk*, University of Manchester, UK, talked on "History and future of image guided radiotherapy", which he helped and helps shaping.

Fully3D has always been an independent meeting and we have continued the tradition. We are therefore particularly grateful to our sponsors (listed on the next page) for their valuable financial support. In this regard we would like to thank Scott Metzler and Samuel Matej, chairs of Fully3D 2019, for passing on starting funding to us. We would also like to express our appreciation to the Scientific Committee for their prompt reviews of the large number of papers submitted to the meeting and to our team members at KU Leuven and UCL for their valuable help. We would especially like to thank Ann Moerenhout and the KU Leuven Congress Office, for taking care of a powerful conference website and providing excellent administrative support.

Johan Nuyts and Kris Thielemans, *conference chairs*
Georg Schramm and Ahmadreza Rezaei, *organizing committee*

# Remembrance of the virtual poster sessions

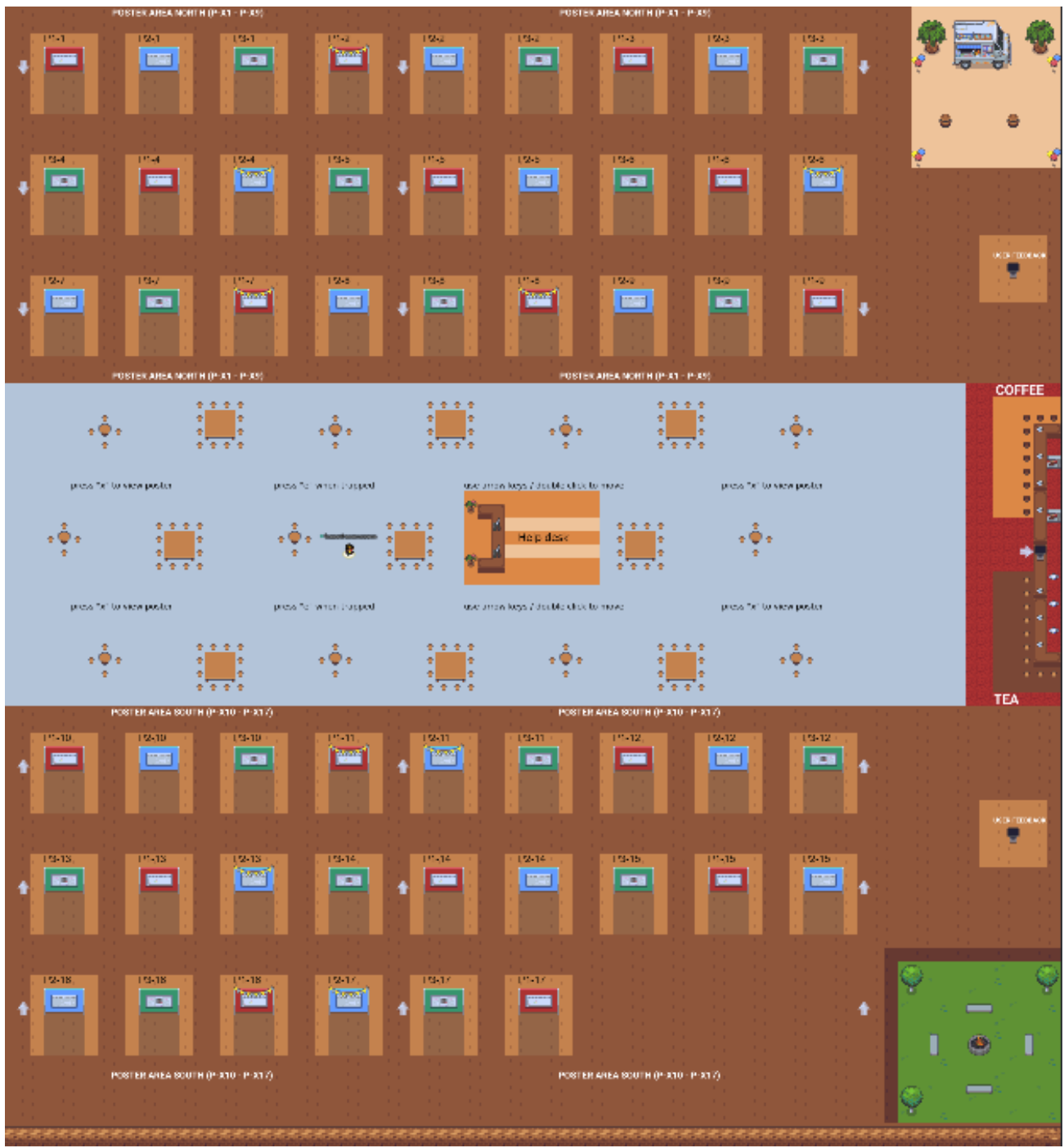

# Many thanks to our sponsors

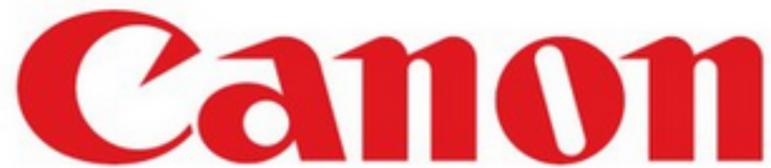

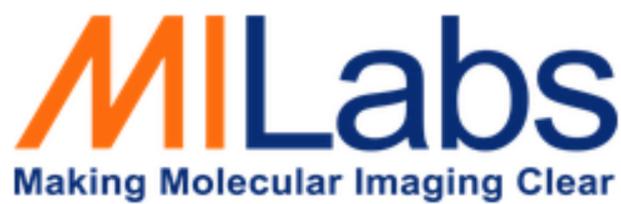

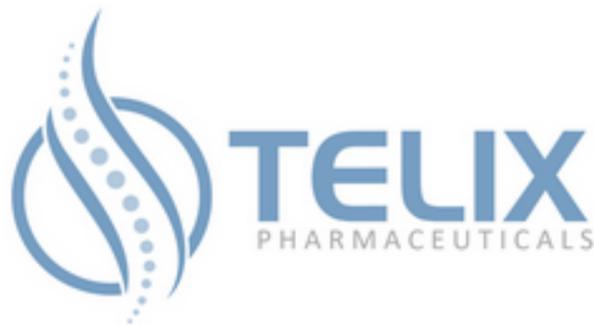

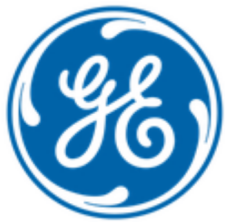

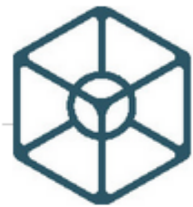

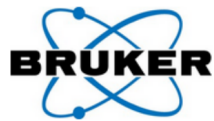

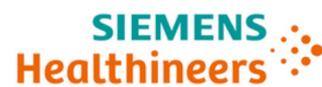

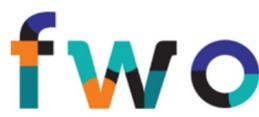

## Conference co-chairs

Johan Nuyts, *KU Leuven, Belgium*
Kris Thielemans, *University College London, United Kingdom*

## Local organizing committee

Johan Nuyts, *KU Leuven, Belgium*
Ahmadreza Rezaei, *KU Leuven, Belgium*
Georg Schramm, *KU Leuven, Belgium*
Alain Seret, *University of Liège, Belgium*
Kris Thielemans, *University College London, United Kingdom*
Stefaan Vandenberghe, *Ghent University, Belgium*
Stephan Walrand, *UC Louvain Hospital, Belgium*

## Award committee

Alexandre Bousse, *French National Institute of Health and Medical Research (INSERM), France*
Se Young Chun, *Seoul National University, Republic of Korea*
Brian Hutton (chair), *University College London, United Kingdom*
William Lionheart, *University of Manchester, United Kingdom*
Jingyan Xu, *Johns Hopkins University, United States*

## Award winners

### Winners of the panel oral award

Pengwei Wu et al., *Johns Hopkins University, United States*
*Using Uncertainty in Deep Learning Reconstruction for Cone-Beam CT of the Brain*

Matthew Tivnan et al., *Johns Hopkins University and Hospital of the Univ. of Pennsylvania, United States*
*High-Sensitivity Iodine Imaging by Combining Spectral CT Technologies*

### Winners of the participant oral choice

Daniel Punzet et al., *Otto-von-Guericke-University Magdeburg, Germany*
*Prior-aided Volume of Interest CBCT Image Reconstruction*

### Winners of the panel poster award

Kaichao Liang et al., *Tsinghua University, Beijing, China*
*Solid angle non-uniformity correction in frequency domain for X-ray diffraction computed tomography*

Ana Radutoiu et al., *Technical University of Munich, Germany*
*Laplace-Beltrami Regularization for Anisotropic X-Ray Dark-field Tomography*

### Winners of the participant poster choice

Buxin Chen et al., *University of Chicago, United States*
*90-plus-90 DECT imaging*

# Scientific committee

Adam Alessio, *Michigan State University, United States*
Simon Robert Arridge, *University College London, United Kingdom*
Evren Asma, *Canon Medical Research, United States*
Freek Beekman, *TU Delft, The Netherlands*
Yannick Berker, *German Cancer Research Center (DKFZ), Germany*
Alexandre Bousse, *LaTIM INSERM, France*
Richard Carson, *Yale University, United States*
Guang-Hong Chen, *University of Wisconsin-Madison, United States*
Rolf Clackdoyle, *Université Grenoble Alpes, France*
Maurizio Conti, *Siemens Medical Solutions, United States*
Margaret Daube-Witherspoon, *University of Pennsylvania, United States*
Bruno De Man, *GE Research, United States*
Michel Defrise, *Vrije Universiteit Brussel, Belgium*
Matthias Ehrhardt, *University of Bath, United Kingdom*
Jeff Fessler, *University of Michigan, United States*
Thomas Flohr, *Siemens Healthcare GmbH, Germany*
Eric Frey, *Johns Hopkins University, United States*
Hewei Gao, *Tsinghua University, China*
Howard Gifford, *University of Houston, United States*
Stephen Glick, *Office of Science and Engineering Labs CDRH FDA MD, United States*
Kuang Gong, *Massachusetts General Hospital, United States*
Paul Gravel, *Yale University, United States*
Jens Gregor, *University of Tennessee, United States*
Grant Theodore Gullberg, *University of California San Francisco, United States*
Jiang Hsieh, *GE Healthcare, United States*
Brian Hutton, *University College London, United Kingdom*
Xun Jia, *UT Southwestern Medical Center, United States*
Jakob Sauer Jorgensen, *Technical University, Denmark*
Marc Kachelrieß, *DKFZ, Germany*
Dan J. Kadrmas, *University of Utah, United States*
Joel Karp, *University of Pennsylvania, United States*
Alexander Katsevich, *University of Central Florida, United States*
Paul Kinahan, *University of Washington, United States*
Michael King, *University of Massachusetts Medical School, United States*
Thomas Koehler, *Philips Research, Germany*
Hiroyuki Kudo, *University of Tsukuba, Japan*
Mathew Kupinski, *University of Arizona, United States*
Patrick La Riviere, *University of Chicago, United States*
Tobias Lasser, *Technical University of Munich, Germany*
Guenter Lauritsch, *Siemens Healthcare GmbH, Germany*
Quanzheng Li, *Massachusetts General Hospital, Harvard Medical School, United States*
Yusheng Li, *University of Pennsylvania, United States*
Jerome Liang, *Stony Brook University, United States*
Nicole Maass, *Siemens Healthcare GmbH, Germany*
Samuel Matej, *University of Pennsylvania, United States*
Scott D. Metzler, *University of Pennsylvania, United States*
Stephen Moore, *University of Pennsylvania, United States*
Xuanqin Mou, *Xi'an Jiaotong University, China*
Klaus Mueller, *Stony Brook University, United States*
Peter Noël, *University of Pennsylvania, United States*
Frederic Noo, *University of Utah, United States*
Johan Nuyts, *KU Leuven, Belgium*
Jed D. Pack, *GE Global Research Center, NY, United States*

Tinsu Pan, *University of Texas M.D. Anderson Cancer Center, United States*
Xiaochuan Pan, *University of Chicago, United States*
Vladimir Panin, *Siemens Medical Solutions, United States*
Amir Pourmorteza, *Emory University School of Medicine, United States*
Jinyi Qi, *University of California, Davis, United States*
Magdalena Rafecas, *Universität zu Lübeck, Germany*
Arman Rahmim, *University of British Columbia, Canada*
Andrew Reader, *King's College London, United Kingdom*
Ahmadreza Rezaei, *KU Leuven, Belgium*
Cyril Riddell, *GE Healthcare, Buc, France*
André Salomon, *Philips, Germany*
Ken Sauer, *University of Notre Dame, United States*
Georg Schramm, *KU Leuven, Belgium*
Alain Seret, *University of Liège, Belgium*
Emil Sidky, *University of Chicago, United States*
Arkadiusz Sitek, *IBM Watson Health Imaging, United States*
Joseph Webster Stayman, *Johns Hopkins University, United States*
Charles Stearns, *GE Healthcare, United States*
Karl Stierstorfer, *Siemens Healthcare GmbH, Germany*
Suleman Surti, *University of Pennsylvania, United States*
Ken Taguchi, *Johns Hopkins University, United States*
Xiangyang Tang, *Emory University School of Medicine, United States*
Kris Thielemans, *University College London, United Kingdom*
Benjamin Tsui, *Johns Hopkins University, United States*
Stefaan Vandenberghe, *Ghent University, Belgium*
Dimitris Visvikis, *LaTIM INSERM, France*
Adam Wang, *Stanford University, United States*
Ge Wang, *RPI, United States*
Guobao Wang, *University of California Davis Medical Center, United States*
Wenli Wang, *Avant Tomography Consulting, LLC, United States*
Charles Watson, *United States*
Glenn Wells, *University of Ottawa Heart Institute, Canada*
Wei Xu, *Brookhaven National Laboratory, United States*
Hengyong Yu, *University of Massachusetts Lowell, United States*
Larry Zeng, *Utah Valley University, United States*

## Technical support

Ander Biguri, *University College London, United Kingdom*
Masoud Elhami Asl, *KU Leuven, Belgium*
Kjell Erlandsson, *University College London, United Kingdom*
Robert Twyman, *University College London, United Kingdom*
Marina Vergara, *KU Leuven, Belgium*

# Contents

# Chapter 1

# Oral Session - Spectral CT

**session chairs**
Ken Sauer, *University of Notre Dame (United States)*
Evangelos Papoutsellis, *Science and Technology Facilities Council, UKRI (United Kingdom)*

# Learned Material Decomposition for Photon Counting CT

Alma Eguizabal[1], Mats U. Persson[2], and Ozan Öktem[1]

[1]Department of Mathematics, KTH Royal Institute of Technology, Stockholm, Sweden
[2]Department of Physics, KTH Royal Institute of Technology, Stockholm, Sweden

**Abstract** Photon-counting detectors are expected to be a great improvement for CT technology. One of its strong potentials is the accuracy in material discrimination in the CT image. Material decomposition is typically solved in a model-based manner, such as a maximum-likelihood estimation of the material concentrations. In this abstract we propose a series of methods to solve successfully the material decomposition that combine model-based and data-driven strategies in a good balance with "physics-aware" deep neural networks. We present three different deep learning approaches and compare their performance with a model-based maximum-likelihood estimate. The deep learning methods Significantly outperformed the maximum-likelihood estimate. The results show that physics-aware deep neural networks is a promising approach for photon-counting CT reconstruction.

## 1 Introduction

Photon Counting Computed Tomography (PCCT) is a cutting-edge technology, that is gaining important prominence in the future of CT [1] [2] [3]. Photon-count detectors provide great improvements. One of these is the enhancement in the material decomposition of CT data, opening new and better possibilities for accurate clinical studies. However, the material decomposition methods are still an on-going research area due to the very late launch of this technology of detectors.

Material decomposition is a non-linear inverse problem that consists in obtaining material concentration signals from the measured PCCT data, and it typically solved in the projection domain, that is, before reconstructing the sinograms. The most accepted solution to this inverse problem is a maximum-likelihood estimation [4] [5] [6]. This is a model-based method that does not take any advantage of available training data and relies on an accurately defined forward model. This forward model depends on the spectral responses of the detector and source, and it is usually obtained by a calibration process that may introduce imprecision. Furthermore, the optimization solvers may become slow or provide noisy results due to the ill-posedness and non-linearity of the problem.

Several solutions have been proposed in the last few years to improve the material decomposition. The authors in [6] introduce regularization to improve noise performance. They prevent certain singularities of the Poisson likelihood with a least-squares approximation and add a regularization term. However, how to choose a good regularization functional and its parameters remains an open question. Another method is to use deep learning, which allows including prior information from training data. In [7] the authors consider a U-Net to solve the inverse problem. This architecture is well-known in medical image processing, nevertheless it is not specific for this material decomposition problem and is agnostic to the physics and statistics of the problem. In order to improve the image quality, it is desirable to incorporate information about the physics of the image acquisition. In the area of CT reconstruction there are interesting solutions that already consider a good balance between model- and data- based approaches. In [8] and [9] the authors propose neural network architectures that are inspired by iterative solutions to the reconstruction problem, where a series of residual convolutional blocks mimic the updates. In this abstract we propose to use this type of solutions in the material decomposition. We aim to design a new a better method to solve the material decomposition in PCCT. For this, we search for a good balance between model-based and data-driven approaches, considering the physics and statistics behind the problem in our solution, as well as making use of state-of-the-art deep learning and available training data.

## 2 Materials and Methods

In this section we overview the technology of PCCT and the models behind it. We also formulate the material decomposition as an inverse problem, and present the typical model-based solution (a maximum-likelihood estimation) as well as our proposed deep learning strategies.

### 2.1 Photon Counting CT and material decomposition

A Photon Counting detector consists in a multi-bin system with $B > 2$ energy bins. Each of the bins, $j = 1, \ldots, B$, registers the projected energy from different sections of the energy spectrum, and therefore has a different energy response. Let us consider a simplified 2-dimensional $(x, y)$ image space, and a detector model that is uniform along its length. The expected value of the number of photon counts in bin $j$ at projection line $\ell$ follows the polychromatic Beer-Lambert law, given as

$$\lambda_j^\ell = \int_0^\infty \omega_j(E) \exp\left(-\int_\ell \mu(x, y; E) ds\right) dE, \quad (1)$$

where $\omega_j(E)$ models an ensemble of effects: an energy dependent X-ray source, the detector efficiency, and the energy response in bin $j$ [1]. In order to perform material decomposition in photon-counting CT, we assume that the X-ray attenuation coefficient, $\mu(x, y; E)$, can be linearly decomposed into $M$ components as $\mu(x, y; E) \approx \sum_{m=1}^{M} \alpha_m(x, y) \tau_m(E)$, where $M$ is the number of potential materials in the image. The decomposition is typically considered in the sinogram domain (before reconstruction). Therefore, the target variable is the line integral of the materials, defined as

$$a_m(\ell) = \int_\ell \alpha_m(x, y) ds = \mathscr{R}(\alpha_m), \qquad (2)$$

where $\mathscr{R}$ is the Radon transformation operator.

## 2.2 The inverse problem

The material decomposition is a non-linear inverse problem that consists in mapping the measured photon counts from the multi-bin detector to the material line integrals as defined in Eq. (2). Let us define the Hilbert spaces $X$ for the material variables (solution space) and $Y$ as the photon counts (measurement space). The solution variable $a \in X$ is a vector containing the components of every material, i.e., $a = [a_1(\ell), \ldots, a_M(\ell)]$ (for simplicity, and without loss of generality, we omit the line $\ell$ from notation).

Given the expected value of bin counts as described in Eq. (1), and line integrals of materials in Eq. (2), the forward operator $\mathscr{F} : X \to Y$ considers the poly-chromatic Beer-Lambert law on each $j$th component as

$$\lambda_j(a) = \int_0^\infty \omega_j(E) \exp\left(-\sum_{m=1}^{M} a_m \tau_m(E)\right) dE, \qquad (3)$$

so that the forward model is given by

$$\mathscr{F}(a) = [\lambda_1(a), \lambda_2(a), \ldots, \lambda_B(a)], \qquad (4)$$

and the measured data $y \in Y$ by

$$y = [y_1, \ldots, y_B]^T, \qquad (5)$$

where a Poisson noise model is considered in the photon counts, so each $j$th energy component in the measured data distributes as $y_j \sim \text{Poisson}(\lambda_j(a))$.

## 2.3 A maximum-likelihood solution

Most accepted methods to solve the material decomposition are a model-based. The solution to this non-linear inverse problem is often interpreted as a Maximum Likelihood (ML) estimation of $a$ [5] [4]. This estimation consists, after applying log and simplifying the Poisson likelihood expression, in solving the following optimization problem:

$$\begin{array}{ll} \underset{a}{\text{minimize}} & \sum_{j=1}^{B} (\lambda_j(a) - y_j \log(\lambda_j(a))) \\ \text{subject to} & a_i \geq 0 \;\; \forall i = 1, \ldots, M \end{array} \qquad (6)$$

where the cost function $\mathscr{L}(a) = \sum_{j=1}^{B} (\lambda_j(a) - y_j \log(\lambda_j(a)))$ is the negative log-likelihood function. Then, considering an iterative algorithm, such

as, for instance, a Log-Barrier method for constrained optimization [10], a solution is found

$$a_{ML} = \arg\min \mathscr{L}(a). \qquad (7)$$

However, this iterative solution may be computationally expensive to obtain. Also, the convexity of the problem is often ignored in the choice of the solver [6].

## 2.4 Deep Learning for the material decomposition

Deep learning together with training data can improve the model-based solutions, such as the ML estimate. In this abstract, we present three different deep learning architectures designed to provide a good balance between the physics and statistics models, and the training data and learning strategies. The proposed architectures contain a set of convolutional residual blocks (ResBlocks in Fig. 1). Each block consists in three convolutional layers followed by a rectified linear unit activation [11].

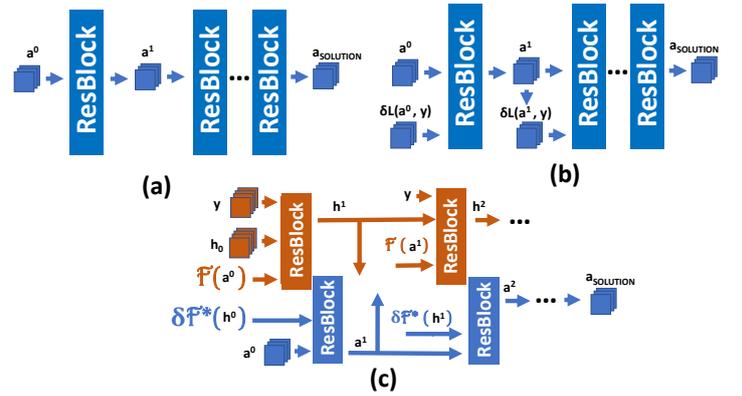

**Figure 1:** Proposed deep learning architectures. (a) A post-processing technique that mimics updates with residual blocks. (b) An unrolled gradient-descent scheme, that also incorporates the gradient on the likelihood in every update (or block). (c) An unrolled primal-dual, that also considers a dual variable, so two different chains of residual blocks are trained in the network (red: the dual chain, blue the primal chain).

### 2.4.1 Learned Denoiser

We first propose a data-driven post-processing ML. This is a residual net that can denoise the output from the ML optimization methods. After each ResBlock, as represented in Fig. 1, the new solution is updated as

$$a^{n+1} = a^n - \Psi_{\theta^n}(a^n), \qquad (8)$$

where $\Psi_{\theta^n}(a^n)$ is the $n$th residual block, parametrized by $\theta^n$, which represents the values of the convolution filters, and with a starting point in the ML estimate, i.e., $a^o = a_{ML}$. These blocks represent a few more learned iterative steps of an ML estimation solver. The physics and statistical model are implicitly contained in the starting point $a_{ML}$.

### 2.4.2 Unrolled gradient-descent iterations

Our next proposed architecture moves one step further, and, following the same philosophy of mimicking iterative updates, we consider a gradient-descent iteration scheme. A gradient-descent solution to the ML estimation would involve iterative updates of the form $a^{n+1} = a^n - \gamma \delta \mathscr{L}(a^n, y)$, that is, the derivative of the likelihood term $\mathscr{L}(a^n, y)$ is used. The unrolled gradient-descent learns the update function on each $n$th iteration, i.e., $\Psi_{\theta^n}$. Therefore, the expression of the residual block operation remains

$$a^{n+1} = a^n - \Psi_{\theta^n}\left(a^n, \delta \mathscr{L}(a^n, y)\right), \qquad (9)$$

with a starting point at $a^o = a_{ML}$, and where the function for $\mathscr{L}(a^n, y)$ is determined as $\frac{\partial \mathscr{L}}{\partial a} = [\frac{\partial \mathscr{L}}{\partial a_1}, \ldots, \frac{\partial \mathscr{L}}{\partial a_M}]$ with elements

$$\frac{\partial \mathscr{L}}{\partial a_m} = \sum_{j=1}^{B} (\frac{y_j}{\lambda_j} - 1) \int_0^\infty \tau_m w_j \exp\left(-\sum_{i=1}^{M} a_i \tau_i(E)\right) dE. \qquad (10)$$

We therefore incorporate information from the physics and the statistics providing structure at the input of every block creating a dependency with the derivative of the likelihood function.

### 2.4.3 Unrolled primal-dual iterations

The next proposed architecture is a learned primal-dual scheme, inspired by the Primal Dual Hybrid Gradient used in regularized optimization problems in tomographic reconstruction [11]. This architecture incorporates a dual variable, $h_n \in Y$ (in data space), so the network alternates between two different types of residual blocks: one in the dual space and one in the primal space (as illustrated in Fig. 1). The network performs then in every pair of blocks

$$h^{n+1} = h^n - \Gamma_{\kappa^n}\left(h^n, \mathscr{F}(a^n), y\right),$$
$$a^{n+1} = a^n - \Psi_{\theta^n}\left(a^n, [\delta \mathscr{F}(a^n)]^* h^{n+1}\right), \qquad (11)$$

with a starting point of $a^o = a_{ML}$. The derivative of the forward operator follows the expression

$$\frac{\partial \lambda_j}{\partial a_m} = -\int_0^\infty \tau_m w_j \exp\left(-\sum_{i=1}^{M} a_i \tau_i(E)\right) dE, \qquad (12)$$

for every combination of $j$ and $m$. Consequently, the physics model is explicitly incorporated to the architecture by means of the forward operator $\mathscr{F}(a^n)$ and the adjoint of its derivative $[\delta \mathscr{F}(a^n)]^*$.

### 2.5 Antropomorphic phantoms and training process

Our antropomorphic phantoms are based on the KiTS19 Challenge kidney dataset [12]. Three materials are considered: bone, soft-tissue and iodine as contrast agent. The simulated values of $a$ are concentration-based. For bone and soft-tissue these are approximated from the Hounsfield units. Iodine is placed on the tumors (masks are available in KITS19) and concentration is random between 0 and 10 mg/ml. One example of a antropomorphic phantom is shown in Fig. 2. Using the ODL library [9], we simulate 600 two-dimensional spectral CT cases. We consider an 8-bin silicon PCCT detector Model Carlo simulated response, and a 120 keV X-ray source. The image size is 512x512 pixels and the current-time product approximately 100 mAs. A flat 9 mm aluminum filter is assumed. Fan-beam geometry with 512 angles and 512 detector elements is used in the Radon transformations, and a filtered back projection (FBP) to calculate the inverses. From these samples, 500 are used as training and 100 as test. For the deep learning calculations we use PyTorch and one NVIDIA GPU GeForce RTX 2080 Ti. We use Adam optimizer with a learning rate of $10^{-4}$.

## 3 Results and discussion

We have conducted a simulation study to compare our proposed deep learning methods to the classical maximum-likelihood estimation. First, we generate a set of simulations using antropomorphic numerical phantoms constructed from real energy integrating CT volumes (not PCCT) with patients of kidney cancer. Then, we discuss the results of solving the material decomposition with the different competing methods.

### 3.1 Test results

In this study investigate the ability of deep-learning sinogram processing for improving the material basis decomposition. The benchmark and starting point of the study is a solution to the ML estimation, $a_{ML}$. As shown in Table 1, the results obtained after applying deep learning have improved considerably the ML estimate: all three proposed architectures enhanced the test MSE of the material decomposition. Also, the three proposed methods have a similar order of training parameters and computational time.

In order to also get a qualitative evaluation of the results, we have computed the virtual monoenergetic image at 70 keV, using the calculated concentrations of bone and soft-tissue in $a$ and the FBP transformation. We also depict the overlay of iodine and bone concentrations, as shown in Fig. 2. From this evaluation, we have observed that, even though the post-processing approach is best is terms of test MSE, the resolution of the reconstructions from the unrolled approaches seems better (see Fig. 2). In particular the unrolled gradient-descent is the best in terms of resolution quality (this is also reflected in the SSIM in Table 1), while the unrolled primal-dual looks more noisy. On the other hand, the material concentrations are more accurate in the unrolled primal-dual than in the other two. We have also realized that, in both unrolled test results there are more outlier cases, that is, particularly bad test samples that decrease the average of the test MSE.

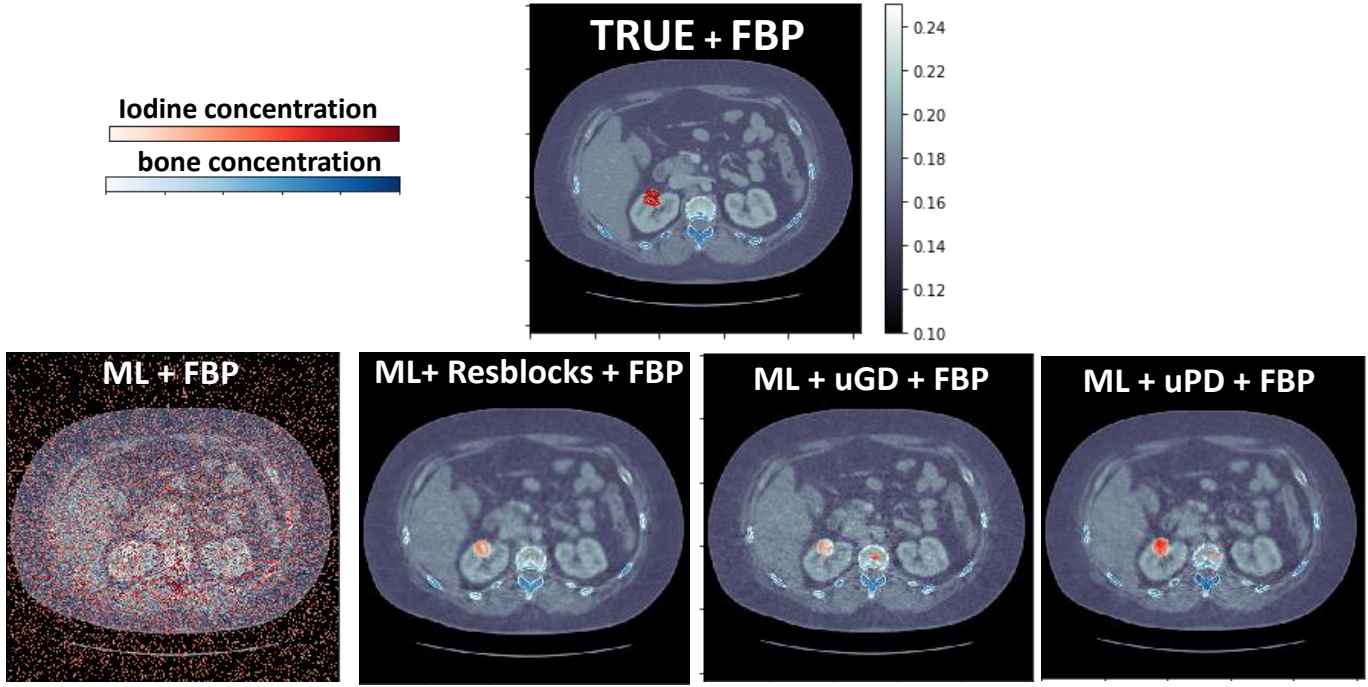

**Figure 2:** Qualitative results. A virtual monoenergy image with material overlay. The antropomorphic phantom is one sample from KiTS19. Filter-back projection (FBP) has been considered as reconstruction strategy. ML corresponds to Maximum Lilehood, Resblocks referes to the Learned Denoise, uGD to the unrolled Gradient Descent, and uPD to the unrolled Primal Dual.

|  | *ML* | *LD* | *uGD* | *uPD* |
|---|---|---|---|---|
| MSE | 0.78 | 0.05 | 0.07 | 0.17 |
| error mean | 0.029 | 0.021 | 0.018 | 0.020 |
| SSIM | 0.27 | 0.45 | 0.83 | 0.49 |

**Table 1:** Quantitative results (MSE: mean squared error; error mean and SSIM: structural similarity index metric) of Maximum Likelihood (ML), Learned Denoiser (LD), unrolled Gradient Descent (uGD) and unrolled Primal Dual (uPD) . Averaged over 100 test samples.

Also, we have observed that the unrolled networks may require a more sophisticated training than the learned denoiser, that is, a good initialization or a dynamic learning rate could significantly boost the performance of the unrolled approaches. Considering a good training strategy will be part of our future work.

## 4   Conclusion

We propose to combine the benefits from new photon counting detectors with the outstanding performance of deep learning in medical image analysis. In order to find a good balance between model-based and data-driven, inside our proposed deep learning solutions we have considered the structure of the underlying inverse problem and the physics from PCCT. We present three different deep learning methods to solve the material decomposition. Our proposed data-driven approaches have improved considerably the accuracy of the typically used model-based ML estimates.

## Acknowledgements

This study was funded by Swedish Foundation of Strategic Research under Grant AM13-0049, by MedtechLabs and by the European Union's Horizon 2020 research and innovation programme under the Marie Sklodowska-Curie grant agreement No 795747. M. U. Persson and A. Eguizabal disclose financial interests in Prismatic Sensors AB and research collaboration with GE Healthcare.

# A constrained dynamical one-step method for spectral CT

Frédéric Jolivet[1], Rob Heylen[1], Georg Schramm[1], and Johan Nuyts[1]

[1]Department of Imaging and Pathology, Division of Nuclear Medicine, KU/UZ Leuven, Leuven, Belgium

**Abstract** Dual energy CT is a promising technique for several medical applications, including dynamic angiography. Recently, a dynamical two-step method has been proposed : first, the water and iodine sinograms are computed from the multi-energy sinograms, then, a dynamic image of the iodine contrast is reconstructed using 4D Total-Variation (TV) constrained reconstruction from the iodine sinogram. In contrast to the 2-step methods, one-step methods use a model relating directly the multi-material images to the multi-energy sinograms. This kind of methods are well-known to reduce the noise correlation between the material images by avoiding the intermediate decomposition step. In this work we propose a dynamical one-step method using an empirical model and based on the Non-Linear Primal–Dual Hybrid Gradient Method (NL-PDHGM). From simulations which consider a CBCT system with dual layer spectral detector and a dynamic software phantom, we compare reconstructions obtained with the dynamical 2-step method and the proposed dynamical one-step method.

## 1 Introduction

Dual-energy CBCT data are of interest to medical applications, notably with the possibility to decompose the object onto some physical (photo-electric/comptom,...) or materials basis (water/bone,water/iodine,...) [1]. The material decomposition problem can be tackled by different strategies. First of all, in 2-step methods the materials sinograms are computed from the multi-energy sinograms, then, a reconstruction method (FDK, iterative methods,...) is used to reconstruct material specific images from the multi-material decomposed sinograms [2, 3]. In general, the material decomposition step greatly amplifies noise due to the ill-conditioning of the inversion step in the basis change. In contrast to the 2-step methods, one-step methods propose to solve the decomposition problem in a constrained one-step inversion, i.e. estimate multi-material reconstructions images from multi-energy sinograms [4–6]. All theses methods consider a static object. On the other hand, several works proposed approaches for dynamic reconstructions based on the 4D TV regularization with different medical applications including in cardiac, thoracic, pulmonary and brain imaging [7–9]. These methods require to solve a non-smooth large-scale optimization problem, therefore it is crucial to use an efficient optimization strategy to have an acceptable computation time. In the last decade, many works proposed computationally efficient implementations based on the primal-dual optimization algorithm of Chambolle and Pock [10] for dynamic reconstructions [11–13] as well as for one-step methods [4]. Recently we have proposed a dynamical iodine reconstruction based on a 2-step method with data acquired with dual-energy (DE) CBCT devices [14] with a motion-correction extension [15]. The main objective of these works is to pro-pose a dynamic iodine reconstruction which can be used to visualize the flow of contrast agent through the brain vasculature, which has a large diagnostic potential in the acute ischemic stroke workflow. In this work we propose a dynamical one-step method to tackle this reconstruction problem. This dynamical one-step method is based on a polynomial empirical imaging model and the Non-Linear Primal–Dual Hybrid Gradient Method (NL-PDHGM) [16] which is a non-linear adaptation of the Chambolle-Pock method [10]. To assess the capabilities of this algorithm we simulate data of a dual-energy dynamic angiography of the brain. For these simulations we consider a CBCT system that obtains dual-energy data by using a stack of two detector layers, where the first layer acts as an energy dependent filter for the second. From these simulated data, we compare reconstructions obtained by the proposed dynamical one-step method with the dynamical 2-step method [14].

## 2 The dynamical one-step method

### 2.1 An empirical model

In this work, we consider the following empirical model which estimates the expectation of the log-converted measured dual-energy sinogram as follows,

$$m_c(l_w, l_i) = a_{5c}l_w^2 + a_{4c}l_i^2 + a_{3c}l_w l_i + a_{2c}l_w + a_{1c}l_i \qquad (1)$$

where $c$ is the index of the detector layer, $l_i$ (respectively $l_w$) is the iodine equivalent thickness (respectively the water equivalent thickness). Polynomial coefficients $a_c$ are estimated by fitting with a set of attenuation values observed by each detector layer, for different combinations of water and iodine thicknesses. These attenuation values can be obtained with calibrated data [2, 6, 17] or calculated using a physical model which requires knowledge of the source spectrum and the detector response. In this work, where the method is evaluated with simulations, we use the latter strategy. For dual-energy CBCT data, we define $\tilde{\mathbf{m}}_c$ the vectorized version of the empirical model (Eq.1). Therefore, each energy layer sinogram $\mathbf{s}_c$ can be expressed as,

$$\mathbf{s}_c = \tilde{\mathbf{m}}_c(\mathbf{s}_w, \mathbf{s}_i) + \mathbf{e}_c \qquad (2)$$

where $\mathbf{s}_i$ (respectively $\mathbf{s}_w$) is the iodine sinogram (respectively the water sinogram) and $\mathbf{e}_c$ is the error vector between measurements and the empirical model (including detection noise, electronic noise and modeling errors).

## 2.2 The proposed dynamical one-step method

Most of the time, one-step methods consider a static object [5]. Assuming errors vectors $\mathbf{e}_{c=1,2}$ as non-correlated gaussian noise with constant variance in (Eq.2), the data fidelity term $F$ of the conventional one-step approach (proportional to the negative log-likelihood), will be expressed as,

$$F(x_w, x_i) = \sum_{c=1}^{2} \left\| \tilde{\mathbf{m}}_c \left( \mathbf{A}\mathbf{x}_w, \mathbf{A}\mathbf{x}_i \right) - \mathbf{s}_c \right\|_2^2 \qquad (3)$$

where $\mathbf{A}$ denotes the forward tomographic projector matrix and $\mathbf{x}_i$ (respectively $\mathbf{x}_w$) is the iodine image (respectively the water image). In this work we propose a dynamical one-step method which considers a static water image $\mathbf{x}_w$ and $T$ different iodine images $\mathbf{x}_{i,1},...,\mathbf{x}_{i,T}$ to have a dynamic reconstruction of the iodine according to $T$ time frames. Therefore, the data fidelity of the proposed dynamical one-step method is defined as,

$$\widetilde{F}(\mathbf{x}_w, \mathbf{x}_{i,.}) = \sum_{c=1}^{2} \sum_{p=1}^{P} \left\| \tilde{\mathbf{m}}_c \left( \mathbf{A}_p \mathbf{x}_w, \tilde{\mathbf{A}}_p \mathbf{x}_{i,.} \right) - \mathbf{s}_{c,p} \right\|_2^2 \qquad (4)$$

where $\mathbf{s}_{c,p}$ the projection with index $p$ measured by the detector layer $c$, $\mathbf{A}_p$ denotes the forward tomographic projector matrix associated to the projection index $p$, and $\tilde{\mathbf{A}}_p$ is the linear operator including the forward tomographic projector matrix associated to the projection index $p$ and the linear time interpolation. Then, we have $\tilde{\mathbf{A}}_p = \mathbf{A}_p \mathbf{L}_p$, where $\mathbf{L}_p$ is a linear interpolator along the time dimension associated to the projection index $p$ [8, 11]. For example, if $\mathbf{x}_{i,.}$ contains ten time frames ($T = 10$) and the dual-energy projections $\mathbf{s}_{c,p}$ has been acquired at the phase $\frac{p-1}{P-1} = 0.47$, then $\mathbf{L}_p \mathbf{x}_{i,.} = 0.3\mathbf{x}_{i,4} + 0.7\mathbf{x}_{i,5}$.

In this work, we aim to reconstruct the dynamic iodine image from a single CBCT acquisition over 360 degrees or even less. In our example with ten time frames (T=10) and a CBCT acquisition over 360 degrees, each time frame $\mathbf{x}_{i,t}$ is linked only with projections over $\frac{360}{T} = 36$ degrees. This problem is severely ill posed, so good 3D+time regularization is mandatory. That is why we propose a regularized dynamical one-step method which can be written as the following constrained optimization problem,

$$\{\hat{\mathbf{x}}_w, \hat{\mathbf{x}}_{i,.}\} \in \underset{x_w, x_{i,.} \geq 0}{\mathrm{argmin}} \; \widetilde{F}(\mathbf{x}_w, \mathbf{x}_{i,.}) + R(\mathbf{x}_w, \mathbf{x}_{i,.}) + \mathbf{I}_\Omega(\mathbf{x}_{i,.}) \quad (5)$$

where $\Omega$ is the set of sequences of volumes in which all voxels included in a static mask remain constant over time, $\mathbf{I}_\Omega$ is the convex indicator function, and the regularization function $R$ which introduces sparsity constraints,

$$R(\mathbf{x}_w, \mathbf{x}_{i,.}) = \beta_1 \|\mathbf{x}_w\|_{TV3D} + \beta_2 \|\mathbf{x}_{i,.}\|_{TV4D} + \beta_3 \|\mathbf{x}_{i,.}\|_1 \quad (6)$$

where $\beta$ are regularization hyper-parameters, $\|.\|_{TV3D}$ is the conventional 3D isotropic Total-Variation and $\|.\|_{TV4D}$ is the 3D+time Total-Variation defined as,

$$\|\mathbf{x}\|_{TV4D} = \sum_{jkvt} |(\nabla_{4D}\mathbf{x})_{jkvt}| \qquad (7)$$

with,

$$|(\nabla_{4D}\mathbf{x})_{jkvt}| = \left( \sum_{d=1}^{3} (\nabla_d \mathbf{x})_{jkvt}^2 + (\gamma \nabla_4 \mathbf{x})_{jkvt}^2 \right)^{1/2} \qquad (8)$$

where the $\nabla_d$ is matrix corresponding to the first order finite difference over the dimension $d$, and $\gamma$ factor gives a different weight in the time direction. Note that in the Algorithm 1, $\nabla_{4D}$ represent the finite difference for a 4D volume, whereas $\nabla_{3D}$ represent the finite difference for a 3D volume.

## 3 Reconstruction algorithm

To solve the non-smooth and non-linear optimization problem (Eq.5), we propose to use the Exact NL-PDHGM framework [16]. In Algorithm 1, we detail the proposed algorithm to solve (Eq.5).

---
**Algorithm 1:** Proposed reconstruction algorithm

---
1  Initialize all variables, choose $\theta \in [0,1]$ and $\tau, \sigma \geq 0$ ;
2  **for** $n = 0$ to niter-1 **do**
3    $\quad \mathbf{y}_{1,c}^{n+1} = \frac{2}{2+\sigma} \left( \mathbf{y}_{1,c}^n + \sigma \left( \tilde{\mathbf{m}}_c \left( \mathbf{A}\bar{\mathbf{x}}_w, \tilde{\mathbf{A}}\bar{\mathbf{x}}_{i,.} \right) - \mathbf{s}_c \right) \right)$
4    $\quad \mathbf{x}_{1,w}^{n+1} = \mathbf{A}^T \left( \sum_{c=1}^{2} \left( 2a_{5c}\mathbf{A}\bar{\mathbf{x}}_w + a_{3c}\tilde{\mathbf{A}}\bar{\mathbf{x}}_{i,.} + a_{2c}\mathbf{1} \right) \cdot \mathbf{y}_{1,c}^{n+1} \right)$
5    $\quad \mathbf{x}_{1,i,.}^{n+1} = \tilde{\mathbf{A}}^T \left( \sum_{c=1}^{2} \left( 2a_{4c}\tilde{\mathbf{A}}\bar{\mathbf{x}}_{i,.} + a_{3c}\mathbf{A}\bar{\mathbf{x}}_w + a_{1c}\mathbf{1} \right) \cdot \mathbf{y}_{1,c}^{n+1} \right)$
6    $\quad \mathbf{y}_{2,w}^{n+1} = proj_{\beta_1 P} \left( \mathbf{y}_{2,w}^n + \sigma \nabla_{3D}\bar{\mathbf{x}}_w \right)$
7    $\quad \mathbf{y}_{2,i,.}^{n+1} = proj_{\beta_2 P} \left( \mathbf{y}_{2,i,.}^n + \sigma \nabla_{4D}\bar{\mathbf{x}}_{i,.} \right)$
8    $\quad \mathbf{x}_{2,w}^{n+1} = -div \left( \mathbf{y}_{2,w}^{n+1} \right)$
9    $\quad \mathbf{x}_{2,i,.}^{n+1} = -div \left( \mathbf{y}_{2,i,.}^{n+1} \right)$
10   $\quad \mathbf{x}_w^{n+1} = proj_{\mathbb{R}^{N+}} \left( \mathbf{x}_w^n - \tau \mathbf{x}_{1,w}^{n+1} - \tau \mathbf{x}_{2,w}^{n+1} \right)$
11   $\quad \mathbf{x}_{i,.}^{n+1} = S_{\tau\beta_3}^+ \left( \mathbf{x}_{i,.}^n - \tau \mathbf{x}_{1,i,.}^{n+1} - \tau \mathbf{x}_{2,i,.}^{n+1} \right)$
12   $\quad \bar{\mathbf{x}}_w^{n+1} = \mathbf{x}_w^{n+1} + \theta \left( \mathbf{x}_w^{n+1} - \mathbf{x}_w^n \right)$
13   $\quad$ **for** t = 1 to T **do** $\; \bar{\mathbf{x}}_{i,t}^{n+1} = \mathbf{x}_{i,t}^{n+1} + \theta \left( \mathbf{x}_{i,t}^{n+1} - \mathbf{x}_{i,t}^n \right)$ **end**
14   $\quad \bar{\mathbf{x}}_{i,.}^{n+1} = prox_{\mathbf{I}_\Omega} \left( \bar{\mathbf{x}}_{i,.}^{n+1} \right)$
15  **end**

---

The operator $\cdot$ in lines 4-5 represents the element-wise product. In lines 6-7 the projection on the set $proj_{\beta_i P}$ projects each gradient voxel-wise onto the $\ell_2$-ball of radius $\beta_i$, while in line 11 the positive soft-thresholding operator $S_\alpha^+$ is applied voxel-wise:

$$S_\alpha^+(\mathbf{u})_j = \begin{cases} \mathbf{u}_j - \dfrac{\alpha}{2} & \text{if} \quad \mathbf{u}_j > \dfrac{\alpha}{2} \\ 0 & \text{if} \quad \mathbf{u}_j \leq \dfrac{\alpha}{2}. \end{cases} \qquad (9)$$

In line 10, $proj_{\mathbb{R}^{N+}}$ enforces each element of a vector in $\mathbb{R}^N$ to be positive, while in the line 14 the proximity operator $prox_{\mathbf{I}_\Omega}$ of the indicator function $\mathbf{I}_\Omega$ is defined by :

$$prox_{\mathbf{I}_\Omega}(\mathbf{u})_{j,t} = \begin{cases} \sum_{t=1}^{T} \mathbf{u}_{j,t}/T & \text{if} \quad \mathbf{u}_{j,t} \in \Omega \\ \mathbf{u}_{j,t} & \text{if} \quad \mathbf{u}_{j,t} \notin \Omega. \end{cases} \qquad (10)$$

In Algorithm 1, the forward projection $\mathbf{A}$ (respectively $\tilde{\mathbf{A}}$) and backward projection $\mathbf{A}^T$ (respectively $\tilde{\mathbf{A}}^T$) are scaled such that the largest eigenvalue of $\mathbf{A}^T\mathbf{A}$ (respectively $\tilde{\mathbf{A}}^T\tilde{\mathbf{A}}$) equals one. These scaling constants are obtained via power iterations. Also the difference finite operators and their adjoint are normalized accordingly.

## 4 Simulations: context and experiment

### 4.1 A dynamic brainweb phantom

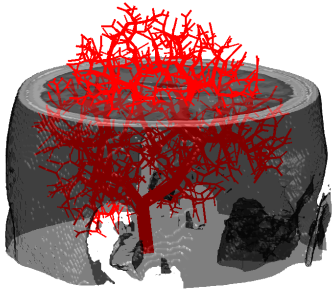

**Figure 1:** Dynamic brain phantom.

To test the capabilities of the prosposed one-step method and compare it with the dynamical 2-step method [14], we simulate dual-energy CBCT scans from a dynamic brain phantom. This dynamic brain phantom (see Fig.1) is similar to the phantom described in [14]. The static part of this brain phantom is based on the brainweb phantom [18], a voxelized head phantom composed by 10 different tissue classes. For the dynamic part, we generated a vascular tree, which has a wide initial artery segment low in the brain, and generation of arterial output terminals was constrained to gray/white matter tissue classes. Each tube segment was considered to possess laminar flow so that dispersion and time delays can be calculated analytically [19], therefore a realistic time behaviour of the flow of contrast is obtained. In addition, no draining venous network was simulated, and the contrast will disappear at the arterial output terminals of the vascular tree. The artificial vascular tree was voxelized, and added to the brainweb phantom as an additional dynamic class. For each projection in the sinogram, an appropriate time point was calculated, and the corresponding iodine contribution from the vascular tree determined. Forward projection was obtained with a dual-energy CBCT simulator. The energy spectrum for each detector layer was simulated with a discretization of 1 KeV from 12 KeV to 150 KeV.

### 4.2 Experiment

Data were simulated to reproduce acquisitions of a dual-energy dynamic angiography of the brain. The simulated CBCT system uses a 2D detector of $198\times256$ pixels with a 1.48 mm pitch and 620 acquisitions angles over 205 degrees. The distance source-detector is 1195 mm whereas the distance object-detector is 390 mm. We simulated a source voltage of 120 kV and the tube current was set to 1.25 mAs. A photonic poissonian noise is added for each dual-energy measurement before the log transform was applied.

## 5 Results

Fig. 2 shows a comparison between the dynamical 2-step method [14] (first row), the proposed dynamical one-step method (second row) and the ground truth (third row). Note that Fig.2 shows a Maximum Intensity Projection (MIP) of results. In both methods, we reconstruct 10 times frames (T=10) of $181\times217\times181$ voxels with a 1 mm$^3$ voxel size. For each method Fig.2 shows only 5 digital subtractions between times frames and the first time frame (one in two). For both methods every time frame was initialized with the same static iodine reconstruction from a static 2-step method, and the one-step method is also initialized with the water reconstruction from this same method. In both methods, we use the same static mask, obtained from a thresholding on a combination between water and iodine static reconstruction, to define $\Omega$ in Eq.5. The comparison shows that one-step method provides better dynamical reconstructions of the iodine than 2-step method.

## 6 Conclusion & Discussion

We have proposed a dynamical one-step method for dual-energy CT including sparsity constraints and based on the optimization strategy NL-PDHGM [16]. This algorithm converges to a critical point which is a local minimum but without guarantee to be the global minimum [16]. Using a dual energy CBCT simulation obtained from a dynamic brain phantom, we compare the proposed dynamical method with the dynamical 2-step method [14]. This promising result should be validated on real patient data. For clinical application, fast computation times are important. For both methods, when we initialize with a static 2-step reconstruction, 200 iterations are enough to reach practical convergence. A good initialization is crucial to reduce drastically numbers of iterations for convergence, while a GPU implementation is another key to reduce the time of calculation. Another critical point for a good dynamical reconstruction is to have a good static mask. A perspective will be to optimize this mask to have an accurate estimation of static voxels (typically the skull is static). To conclude, note that an extension for data from energy-resolved photon counting detectors is straightforward.

## Acknowledgements

This work was done under the NEXIS project, that has received funding from the European Union's Horizon 2020 Research and Innovations Program (Grant Agreement 780026).

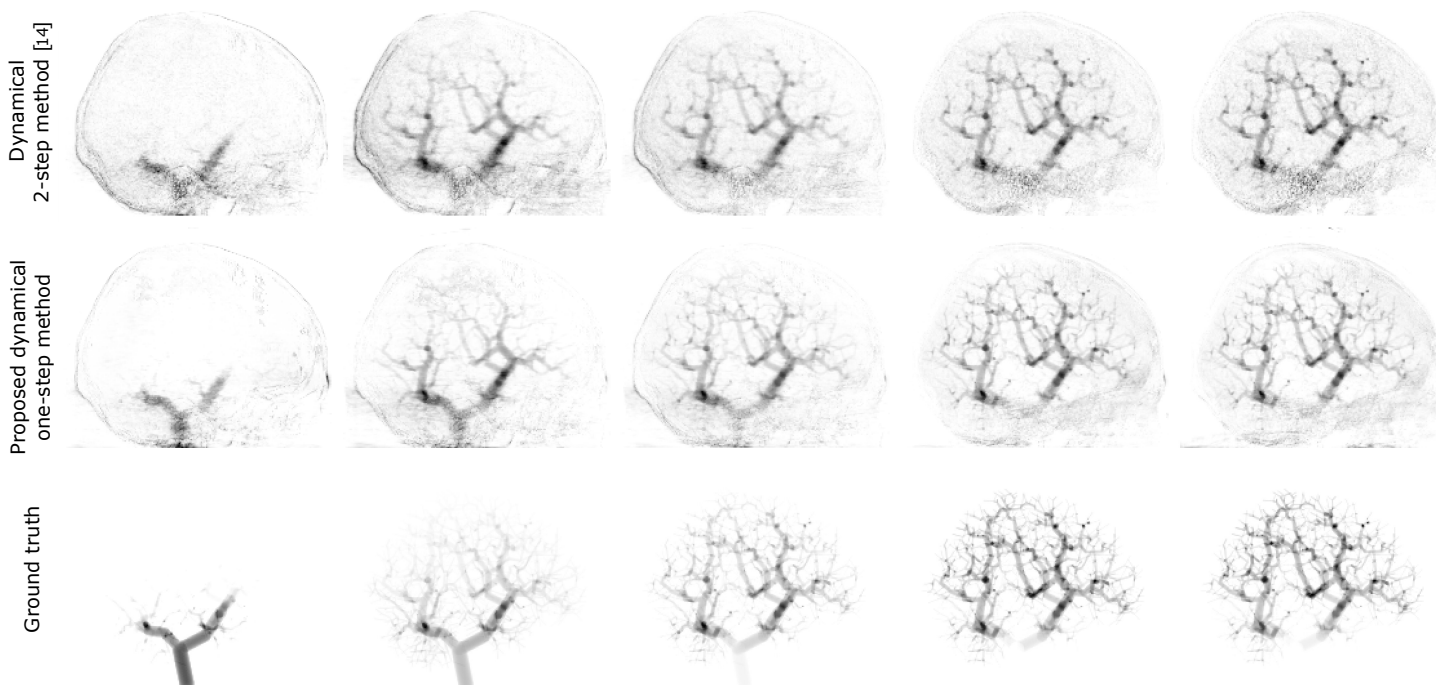

**Figure 2:** MIP visualization of dynamical iodine reconstructions obtained with the dynamical 2-step method [14] (first row) and the proposed dynamical one-step method (second row), and the ground truth (third row).

The authors are grateful to Klaus Jürgen Engel, Bernd Menser and Matthias Simon from Philips Research for providing the simulated data used in this work and for many helpful discussions.

# High-Sensitivity Iodine Imaging by Combining Spectral CT Technologies

Matthew Tivnan[1], Grace Gang[1], Wenchao Cao[2], Nadav Shapira[2], Peter B. Noël[2], and J. Webster Stayman[1]

[1]Department of Biomedical Engineering, Johns Hopkins University, Baltimore, MD, USA
[2]Department of Radiology, Hospital of the University of Pennsylvania, Philadelphia, PA, USA

**Abstract** Spectral CT offers enhanced material discrimination over single-energy systems and enables quantitative estimation of basis material density images. Water/iodine decomposition in contrast-enhanced CT is one of the most widespread applications of this technology in the clinic. However, low concentrations of iodine can be difficult to estimate accurately, limiting potential clinical applications and/or raising injected contrast agent requirements. We seek high-sensitivity spectral CT system designs which minimize noise in water/iodine density estimates. In this work, we present a model-driven framework for spectral CT system design optimization to maximize material separability. We apply this tool to optimize the sensitivity spectra on a spectral CT test bench using a hybrid design which combines source kVp control and k-edge filtration. Following design optimization, we scanned a water/iodine phantom with the hybrid spectral CT system and performed dose-normalized comparisons to two single-technique designs which use only kVp control or only k-edge filtration. The material decomposition results show that the hybrid system reduces both standard deviation and cross-material noise correlations compared to the designs where the constituent technologies are used individually.

## 1 Introduction

Spectral CT systems use data acquisition schemes involving varied spectral sensitivities across photon energies. Thus, compared to single-energy CT systems, spectral CT systems can provide more information about the energy-dependent attenuation of the subject being scanned. In particular, spectral CT enables estimation of basis material densities which has many benefits in quantitative clinical imaging.

Contrast-enhanced imaging of iodine is one of the most widespread clinical applications of spectral CT. Iodine density estimates show contrast-agent concentration, and water density estimates provide virtual non-contrast enhanced images for characterizing patient anatomy [1]. However, the similarity of the attenuation spectra of water and iodine make accurate material decomposition and density estimation challenging. As compared to standard estimation of overall attenuation (as provided by single-energy CT systems), the relative noise is much higher in the individual basis material density estimates provided by spectral CT [2]. For some clinical applications, such as imaging pancreatic cancer, very low levels of differential contrast-enhancement can have a meaningful impact on diagnosis [3]. Therefore, development of spectral CT systems which are capable of high-sensitivity water/iodine decomposition is an important goal with direct clinical implications.

One way to improve sensitivity is with advanced data processing. For example, model-based approaches have been widely adopted in single-energy CT for their improved dose-image quality tradeoffs. In this work, we use a direct one-step model-based material decomposition (MBMD) algorithm rather than a two-step reconstruction-decomposition approach, allowing for incorporation of measurement statistics as well as advanced regularization approaches to help reduce noise.

Another strategy for improving sensitivity, and the focus of this work, is to optimize the spectral CT system design. There are several technologies which can modulate the spectral sensitivity of a CT system: rapid kVp switching [4], multiple x-ray sources [5], source filtration (e.g. k-edge filters) [6], dual-layer/multi-layer detectors [7], photon counting detectors [8]. Each of these spectral modulation technologies have potentially tuneable design parameters such as kVp separation or k-edge filter thickness. There is evidence that combining these methods can improve sensitivity [9]. In previous work, we demonstrated that combining spectral modulation technologies into a hybrid system offers greater control over designed spectral sensitivities [10] [11]. Our previous simulation results showed that joint optimization of these design parameters in a hybrid system can result in higher sensitivity than systems using the constituent spectral modulators individually.

Recently, we have also proposed a mathematical formula for *material separability index* which models the performance of a spectral CT system and its advantage over single-energy CT. The metric takes into account polyenergetic models of x-ray sources, attenuation, and detector sensitivity, as well as noise and correlations [12].

In this work, we apply these theoretical models to the system design optimization of a prototype spectral CT test bench which incorporates both kVp control as well as k-edge filtration. We describe the specific spectral CT model that includes polyenergetic x-ray physics as well as a quantitative metric of water/iodine separability for the modeled system. We describe a spectral CT system design optimization process for a physical prototype including parameters of the design space and optimization methods. Finally, we present the results of a water/iodine imaging study which compares the optimized hybrid kVp control/k-edge filtration design with kVp control individually and k-edge filtration individually.

## 2 Methods

### 2.1 Spectral CT Physical Models

A general forward model for spectral CT is given by:

$$\bar{\mathbf{y}}(\mathbf{x}) = \mathbf{GS} \exp\left(-\mathbf{QAx}\right), \tag{1}$$

where $\mathbf{x}$ is a column vector containing basis material densities, $\mathbf{A}$ is a forward projection operator, $\mathbf{Q}$ contains mass attenuation coefficients of basis materials, $\mathbf{S}$ models the system spectral sensitivity, and $\mathbf{G}$ is a gain. This work will

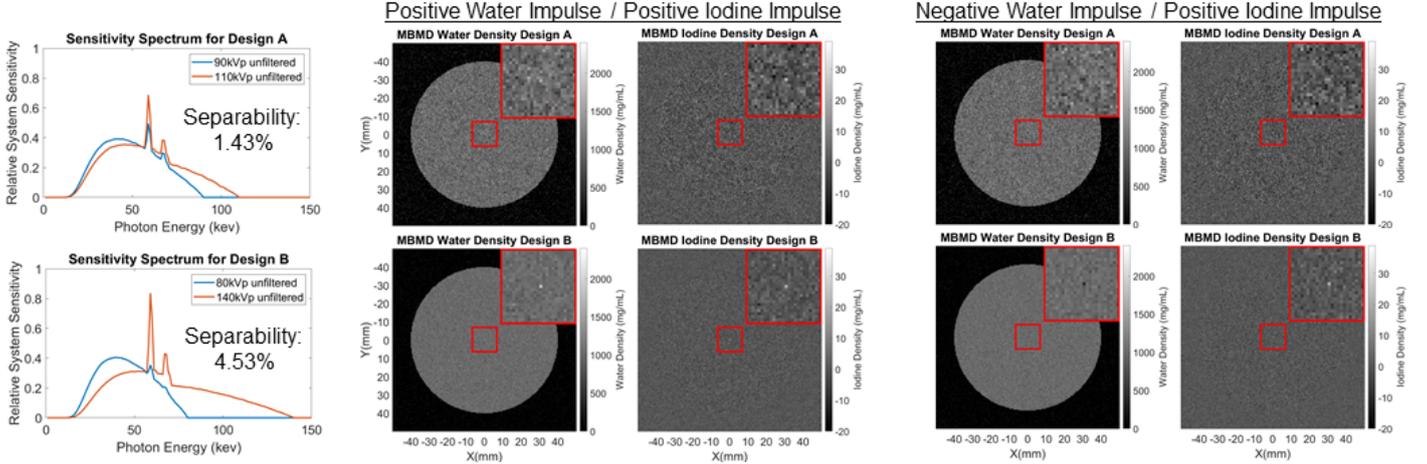

**Figure 1:** Comparison between two spectral CT system designs. Design B results in a greater separability index than design A. As shown by the basis material densiity estimates, the noise correlation is lower in design B and therefore it is easier to detect a positive impulse of idoine plus a negative impulse of water.

focus on modeling flat-panel energy-integrating detectors, for which we can expand **S** as follows:

$$\bar{\mathbf{y}}(\mathbf{x}) = \mathbf{GS_2S_1S_0} \exp\left(-\mathbf{QAx}\right), \quad (2)$$

where $\mathbf{S_0}$ models the spectrum of photons emitted by the source (including filtration), $\mathbf{S_1}$ models the probability of interaction with the scintillator, and $\mathbf{S_2}$ models the generation of secondary quanta in the scintillator, as well as detection and integration by the photodetector. With this spectral CT system model, we can simulate systems with different designs and predict how they will respond to different multi-material objects. For example, we can predict the energy attenuated by an object, $\mathbf{x}$, using the following formula:

$$\text{Dose} = \varepsilon^T \mathbf{S_0}\left(1 - \exp\left(-\mathbf{QAx}\right)\right), \quad (3)$$

where $\varepsilon$ is a vector containing the energy, in mJ.

We can also use this model to estimate material density images from spectral CT measurements via MBMD. Assuming a multivariate gaussian noise model with mean, $\bar{\mathbf{y}}(\mathbf{x})$, and covariance, $\Sigma_y$, a maximum-likelihood estimator of basis material densities, $\hat{\mathbf{x}}(\mathbf{y})$, is given by the following formulae:

$$\Phi(\mathbf{x}, \mathbf{y}) = \left(\mathbf{y} - \bar{\mathbf{y}}(\mathbf{x})\right)^T \Sigma_y^{-1} \left(\mathbf{y} - \bar{\mathbf{y}}(\mathbf{x})\right) \quad (4)$$

$$\hat{\mathbf{x}}(\mathbf{y}) = \underset{\mathbf{x}}{\text{argmin}} \ \Phi(\mathbf{x}, \mathbf{y}). \quad (5)$$

Note that the above formula does not include a regularization term. While we focus on unbiased estimators in this work, further noise reduction is possible using the cross-material regularization strategies described in [13] [14]. To perform the numerical optimization, we use the separable parabolic surrogates algorithm described in [15] and a cross material preconditioner defined in [16] to improve convergence rates. We note that in following sections this MBMD approach will be used for water/iodine decomposition. However, the same estimator can be applied as single-material model-based iterative reconstruction (MBIR) with built-in polyenergetic corrections by using water as the sole basis material.

## 2.2 Material Separability

In this section we present a quantitative metric for material separability originally described in [12]. The metric applies

to maximum-likelihood MBMD and is based on the Fisher information matrix which is defined below:

$$\mathbf{F} = \mathbf{A}^T\mathbf{Q}^T\mathbf{D}^T\mathbf{S}^T\mathbf{G}^T\Sigma_y^{-1}\mathbf{GSDQA} \quad (6)$$

$$\mathbf{D} = D\{\exp\left(-\mathbf{QAx}\right)\}. \quad (7)$$

Note that this expression is object-dependent. That is, different objects, as described by $\mathbf{x}$, will affect the weights in $\mathbf{D}$ and will therefore have an impact on $\mathbf{F}$. This is not surprising since CT image quality is known to be object-dependent. The detectability index of a signal, $\mathbf{w}$, which is in the same multi-material image space as $\mathbf{x}$, is defined below:

$$d^2(\mathbf{w}) = \mathbf{w}^T\Sigma_x^{-1}\mathbf{w} = \mathbf{w}^T\mathbf{Fw}. \quad (8)$$

For two signals, $\mathbf{w_A}$ and $\mathbf{w_B}$ which are normalized such that $d(\mathbf{w_A}) = d(\mathbf{w_B})$, the separability index is defined by the ratio between the detectability of their difference to the detectability of their sum as shown below:

$$s^2(\mathbf{w_A}, \mathbf{w_B}) = \frac{(\mathbf{w_A} - \mathbf{w_B})^T\mathbf{F}(\mathbf{w_A} - \mathbf{w_B})}{(\mathbf{w_A} + \mathbf{w_B})^T\mathbf{F}(\mathbf{w_A} + \mathbf{w_B})}. \quad (9)$$

In theory this formula can characterize the separability of any two signals, but for the purposes of characterizing water/iodine separability, we may assume that $\mathbf{w_A}$ is a voxel impulse of iodine only at a certain position and $\mathbf{w_B}$ is a voxel impulse of water at the same position. Note that a single-energy CT system would be capable of detecting $(\mathbf{w_A} + \mathbf{w_B})$ but it would be effectively impossible to detect $(\mathbf{w_A} - \mathbf{w_B})$ leading to a material separability index near 0.0%. In contrast, a spectral CT system with two or more distinct spectral channels should be capable of detecting the differential signal.

## 2.3 Spectral CT System Design Optimization

Our goal in this section is to apply the theoretical models above to maximize water/iodine separability in a prototype spectral CT test bench using a combination of kVp control and k-edge filtration. The spectral sensitivity design consists of two channels. Each channel was parameterized by three quantities: kVp setting, filtration material, and exposure. We explored six possible kVp settings: 70, 80, 90, 100, 110, and 120 kVp. Eight possible filter materials and thicknesses were 250 $\mu m$ praseodymium, 250 $\mu m$ erbium, 127 $\mu m$ lutetium, 250 $\mu m$ hafnium, 100 $\mu m$ tungsten, 100 $\mu m$ gold, 250 $\mu m$ lead, and no filter. For each spectrum there are 48 unique combinations of filter material and kVp settings. Therefore, there are 1128 unique spectral profiles. Exposure settings were optimized in a nested optimization for each possible shape.

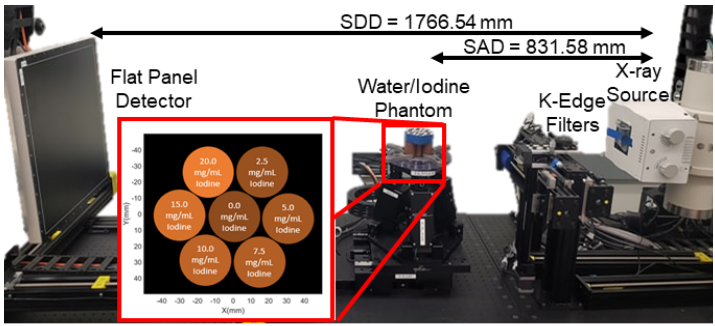

**Figure 2:** Hybrid method spectral CT test bench and water/iodine phantom.

For each possible design, we established a physical model, and computed the separability index for a 0.5 mm voxel impulse of iodine and water at the center of an 80 mm cylinder of water. The final design was chosen as the one which maximized the water/iodine separability metric. The process above was also repeated for the two spectral modulation technologies acting individually. That is, the design space was constrained to cases with no filtration for kVp control design. For the k-edge filtration design, the space was constrained to designs with static kVp settings. For the hybrid design, we used the full parameter space which includes combinations of kVp settings and filtration materials.

The three optimized designs were implemented physically on the x-ray CT test bench as shown in Figure 2. We constructed a water/iodine phantom using cylindrical targets designed for CT calibration. Each cylinder is approximately 27 mm in diameter and the composition of each cylinder has been designed to match the attenuation spectra of water plus some concentration of iodine. The nominal iodine concentrations are 0.0, 2.5, 5.0, 7.5, 10.0, 15.0, and 20.0 mg/mL and they are arranged as shown in Figure 2. We used a 2-dimensional fan-beam system geometry with a source-to-axis distance of 831.58 mm and a source-to-detector distance of 1766.54 mm. The detector pixel size is $0.278 \times 0.278$ mm and the central 60 detector rows were binned to produce the one-dimensional projection measurements for each view.

Exposure settings were calibrated using preliminary scans for each design and approximating photon counts according to the variance in the gain measurements. A voxelized approximation of the water/iodine phantom was used to approximate the dose attenuated by the phantom as defined by (3). The target exposure was then established in such a way that the predicted dose attenuated by the phantom was normalized to 1 mJ and the ratio of exposures was matched to the design op-timization results. The source mAs was scaled in proportion to the ratio between the target exposure and initial exposure estimates. The system spectral sensitivity was calibrated by scanning a phantom containing known concentrations of water and iodine and fitting a parameterized spectral model. After calibration was complete, we scanned the water/iodine phantom using each of the three optimized designs. Two-dimensional material density images with $200 \times 200$ voxels of size 0.5 mm were reconstructed via MBMD using 1000 iterations of the separable quadratic surrogates algorithm described in [15]. We also ran a standard model-based iterative reconstruction to estimate attenuation. This was accomplished using the same polyenergetic model used for MBMD but with a single-material (water) basis. To evaluate image quality, we computed means and cross-material covariances in 7 ROIs centered on each cylinder for the water and iodine basis material density images resulting from MBMD. We also computed variance in the same ROIs for the MBIR reconstructed image results.

## 3 Results

The optimized kVp control design was found to be 48.47% of photons at 70 kVp and 51.53% of photons at 120kVp. This design results in a water/iodine separability index of 2.54%. This result in not particularly surprising, since this is the largest possible kVp separation, matching intuition that spectra which are very different from one another enable greater material separability. The optimized k-edge filtration design was found to be 58.16% of photons at 70 kVp with a 250 $\mu m$ praseodymium filter and 41.84% of photons at 70 kVp with a 250 $\mu m$ lead filter. This design results in a water/iodine separability index of 7.32%. The optimized hybrid design was found to be 34.88% of photons at 70 kVp with a 250$\mu m$ lead filter and 65.12% of photons at 120 kVp with a 250 $\mu m$ praseodymium filter. This design results in a water/iodine separability index of 12.52%.

The sample mean of estimated density for all ROIs was within 10% of the nominal value for both water and iodine for all three designs. As shown in Figure 3 sample variance in the attenuation image estimated via MBIR is comparable for all three designs. This is expected because the total dose absorbed by the phantom was normalized for each design. The sample covariance in the water and iodine density images estimated with MBMD shows that the noise is much lower for the hybrid design than for either of the spectral modulation technologies acting individually. The correlation coefficient is also closer to zero for the hybrid design than either of the individual designs.

## 4 Conclusion

In this work we apply a previously developed quantitative metric of material separability based on the Fisher information matrix. We have shown that this metric can be used to optimize spectral CT system design for higher sensitivity and demonstrated efficacy in a physical system. Furthermore, the results of the imaging study show that spectral CT systems which use a combination of multiple spectral modulation technologies have the potential for improved performance with respect to designs using the constituent individual technologies.

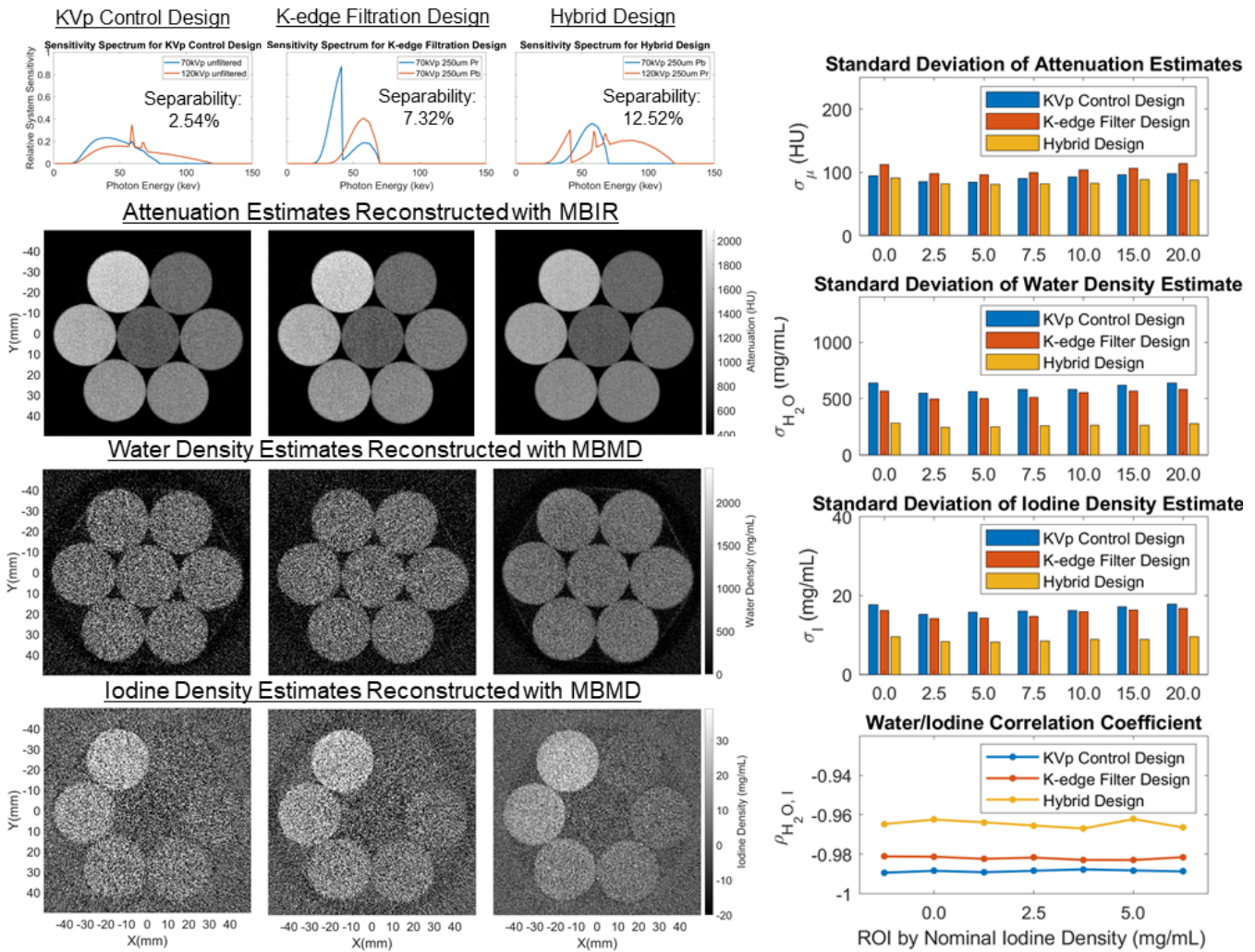

**Figure 3:** Spectral sensitives for each design, MBIR and MBMD reconstruction results, and covariance metrics by ROI.

There are several limitations in this preliminary work which should be acknowledged. First, we used a relatively small phantom compared to the size of a human patient. Material decomposition will be a more poorly conditioned problem for larger objects. Additionally, we did not incorporate a scatter model. For low-contrast applications, scatter and other systemic biases are a concern and must be addressed. These topics are the subject of ongoing studies.

Despite these limitations, the results suggest that hybrid design is a promising strategy for spectral CT with the potential to overcome low-concentration limits of more traditional single-technology spectral methods. While this work has concentrated on two more traditional spectral technologies, other combinations including other source-side modulation schemes or energy-sensitive detectors (such as dual-layer or photon counting detectors) could provide additional advantages. Jointly optimizing such hybrid systems for material separability may potentially have a significant benefit for clin-ical applications involving high sensitivity to low contrast concentrations, and will be investigated in future work.

### Acknowledgements
This work was supported, in part, by NIH grant R21EB026849.

# [1]Single-pixel Neural Network for Charge Sharing Correction and Material Decomposition in Spectral CT

Shengzi Zhao[1,2], Katsuyuki Taguchi[3], Ao Zheng[1,2], Kaichao Liang[1,2], Muge Du[1,2] and Yuxiang Xing[1,2]

[1]Department of Engineering Physics, Tsinghua University, Beijing, China

[2]Key Laboratory of Particle & Radiation Imaging (Tsinghua University), Ministry of Education, Beijing, China

[3]Radiological Physics Division, The Russell H. Morgan Department of Radiology and Radiological Science, Johns Hopkins University School of Medicine, Maryland, America

**Abstract** Photon counting detector based spectral CT imaging has gained a lot of attention in these days. Basis material decomposition is a commonly used step in spectral CT reconstruction. The aim of this work is to take advantage of the power of neural networks to find a method that can get accurate basis material decomposition. To research on the charge sharing correction, we constructed a pixel-wise network architecture as a starting point, named single pixel interweaving network (SPIN). Then we improved it to a patch-wise operation, named patch-wise interweaving network (PIN). To help the network focus on charge sharing better, data simulating random material thickness was used for training. For SPIN, it could deal with charge sharing to some extent, but the bias was unsatisfying. For PIN, it used the information from neighboring pixels and resulted in less bias and total error. Moreover, to reduce the influence of noise, we used the averages of noise realizations to approximate the noiseless input and train the PIN. This method decreased the bias a lot, but was poor with denoising. PIN provided a potential way to get more accurate basis material thickness by using additional information extracted from neighboring detector pixels.

## 1 Introduction

Over the past decades, spectral CT has been widely researched and used. It can take use of attenuation at different energy and decompose the imaged object into two or three basis materials. The decomposition results of material thickness give valuable information of object materials. Spectral CT can decrease the effect of beam hardening and increase the ability of distinguishing different materials.

Photon counting detectors (PCDs) can record the number and energy of incident photons and provide detailed spectral information. Detailed spectral information could improve the quality of CT images and reduce the X-ray dose [1]. However, the performance of PCDs is not ideal because of charge sharing, pile-up and some other problems. They cause spectral distortion and limit the application of PCD. This paper focuses on charge sharing. When a photon injects the PCD, the charges from it are not constrained in one detector pixel and then collected and detected by several neighboring pixels. As Fig. 1 shows, charge sharing includes spill-in sharing and spill-out sharing. Considering a 3×3 region and the yellow pixel is the pixel of interest (POI). In Fig. 1(a), a photon injects a neighbor pixel and some charges diffuse into the POI. Then POI records more photons and this is spill-in charge sharing. In Fig. 1(b), a photon injects the POI and some charges diffuse into a neighbor pixel. Then POI records lower energy and this is spill-out charge sharing. Both of them distort the detected spectrum and generally speaking increase bias and noise of spectral CT images.

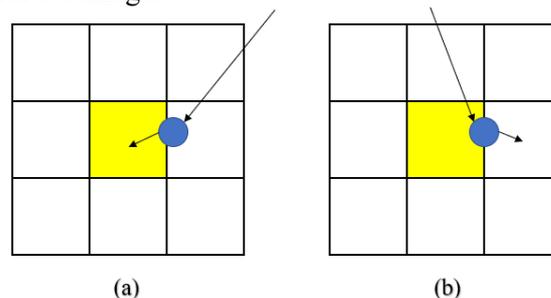

Figure 1. Two types of charge sharing. (a) spill-in sharing to a pixel of interest; (b) spill-out to a pixel of interest

To improve the performance of PCDs, many methods or new detectors have been researched. Lee et al used algorithm based on charge sharing model to correct spectral distortion [2]. Some detectors were invented to correct counts or energy errors according to communications between pixels [3]. Some PCDs were used to get undistorted spectrum by rejecting charge sharing events [4]. Hsieh's detector records extra information about charge sharing and deals with it later [5]. These methods have different problems, including long processing time, insufficient accuracy and difficulty of data processing. Except traditional methods, neural network has been used to solve spectrum distortion problem recently. Holbrook et al proposed Image Domain CNN and Projection Domain CNN to correct spectral distortion separately [6]. Energy integrating detector data were used as labels for the two networks to correct PCD data. Li et al designed WGAN to correct spectral distortion. Two subnetworks in generator dealt with pile up and charge sharing separately and connect by a UNet [7]. It achieved obvious noise suppression and accurate spectral correction in both projection and reconstruction domains. Touch et al used an ANN to reverse the spectral distortion [8]. Zheng et al analyzed the factors of spectrum distortion, then designed WeaveNet, which has advantages on spectrum correction [9].

Following the idea of WeaveNet, we designed a network for single pixel at the start point, which are used to deal with charge sharing problem. Then we improved the architecture to take advantage of the relationship between the pixel of interest and its neighboring pixels to get basis material thickness with smaller relative error. We applied the network to conventional PCD data to get basis material thickness. Furthermore, in order to reduce the influence of noise, we used the means of noise realizations as input to train the network. Simulation results are demonstrated in this paper. Taguchi invented Multi-energy Inter-pixel Coincidence Counters (MEICC) [10][11]. By recording the number and direction of charge sharing events, the CRLB calculated in the light of MEICC data decreases by at least 50% compared to conventional PCDs. In the future, we are to take advantages of the additional MEICC data to improve material decomposition accuracy.

[1] This work is supported by NIH R21 EB029739, U.S.A. and National Natural Science Foundation of China (No.61771279, No.62031020).

## 2 Methods

### 2.1 Basic model

For spectral CT, several energy bins are set for detecting process. Each detector pixel counts the incident photons in different energy bin.

When a photon injects an object, some processes will happen, including photoelectric effect, Compton effect, electron pair effect and so on. Because of them, the energies and numbers of the photons decay. Attenuation coefficient $\mu$ is used to characterize the decaying process. It's the function of material types and energy and can be expressed by basis function composition, and with basis material decomposition, it can be written as:

$$\mu(E) = \sum_{i=1}^{N_{\text{material}}} a_i \mu_i(E) \qquad (1)$$

Where $a_i$ is the equivalent thickness of the $i^{\text{th}}$ basis material and $N_{\text{material}}$ is the number of basis materials.

In this work, we set

$$N_{\text{material}} = 2 \qquad (2)$$

The photon numbers in the two energy bins are as follow.

$$\overline{N_L} = N_0 \int_{\text{Low energy bin}} S(E) e^{-\mu_1(E)T_1 - \mu_2(E)T_2} dE \qquad (3)$$

$$\overline{N_H} = N_0 \int_{\text{High energy bin}} S(E) e^{-\mu_1(E)T_1 - \mu_2(E)T_2} dE \qquad (4)$$

Here, $\overline{N_L}$ and $\overline{N_H}$ are the means of photon numbers in the low/high energy bin, $N_0$ is the initial photon number, $S(E)$ is the spectrum, $T_1$ and $T_2$ are the basis material thickness, $\mu_1(E)$ and $\mu_2(E)$ are the attenuation coefficients of the two basis materials. Two virtual energies, $E_L$ and $E_H$, are chosen to calculate equivalent line integrals of attenuation coefficients, $Y_L$ and $Y_H$.

$$\begin{bmatrix} Y_L \\ Y_H \end{bmatrix} = \begin{bmatrix} \mu_1(E_L) & \mu_2(E_L) \\ \mu_1(E_H) & \mu_2(E_H) \end{bmatrix} \times \begin{bmatrix} T_1 \\ T_2 \end{bmatrix} \qquad (5)$$

$$\begin{bmatrix} T_1 \\ T_2 \end{bmatrix} = \begin{bmatrix} \mu_1(E_L) & \mu_2(E_L) \\ \mu_1(E_H) & \mu_2(E_H) \end{bmatrix}^{-1} \times \begin{bmatrix} Y_L \\ Y_H \end{bmatrix} \qquad (6)$$

According to Eq. (6), we can calculate basis material thickness when the line integrals are obtained.

### 2.2 The architecture of single pixel interweaving network (SPIN)

Charge sharing happens in each detector pixel. Therefore, as a starting point, we focus on a single pixel case. We borrowed the idea from Zheng's Interweaving Network[6].

It has two branches to correct spectrum and deals with the whole CT image. The architecture of SPIN is as Fig. 2.

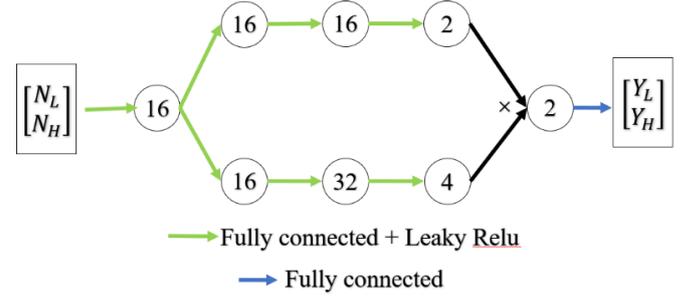

Figure 2. The architecture of single pixel interweaving network. The circled numbers represent numbers of nodes in layers. The inputs are the photon numbers of POI and the outputs are the line integrals.

SPIN is a two branch fully connected network. Each circled number is the number of nodes in each layer. The inputs, $N_L$ and $N_H$, are the photon numbers in low/high energy bin of the pixel of interest. The outputs, $Y_L$ and $Y_H$, are the line integrals of attenuation coefficients at two virtual energies. The green and blue arrows in the firgue represent fully connected layer with and without leaky relu activations. At the end of the two branches, there is matrix multiplication to connect them together. The output of the branch below, a $4 \times 1$ vector, is transformed to a $2 \times 2$ matrix, and then be multiplied with the output of the branch above.

The loss function we used to train SPIN is

$$L_{\text{SPIN}} = \frac{1}{2}\left(\left(\frac{Y_L - Y_L^*}{Y_L^*}\right)^2 + \left(\frac{Y_H - Y_H^*}{Y_H^*}\right)^2\right) \qquad (7)$$

Here, $Y_L^*$ and $Y_H^*$ are the ground truths of line integrals.

### 2.3 The architecture of patch interweaving network (PIN)

Let's analyze charge sharing by a $5 \times 5$ patch, as shown in Fig. 3. The POI is the yellow one. The blue ones are the first neighboring pixels and the green ones are the second neighboring pixels. Focused on the POI, charge sharing happens between blue pixels and the yellow central pixel, as the black arrows shown, which represent the direction of charge sharing. For data of the POI, the extra counts and photon energy changes caused by charge sharing is related to the photon numbers of the blue ones. Therefore, we need these information in order to correct charge sharing. Focusing on the blue pixels, their data are further related to the green ones, the same as the POI. Therefore, the data from green pixels is helpful to correct charge sharing of POI. Hence, charge sharing correction is a cascading problem. Moreover, the pixel is farther away from the POI, the effect mentioned before is smaller. Combined, a $9 \times 9$ patch is chosen in our study.

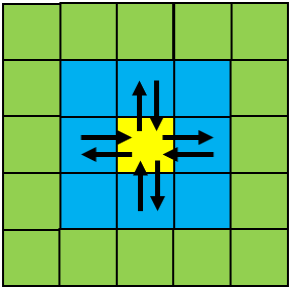

Figure 3. A 5×5 patch to show the relationship of charge sharing. It happens between POI and its blue neighboring pixels.

Based on the above analysis, we are to correct charge sharing patchwise, hence propose PIN. The architecture of PIN is as Fig. 4. PIN is a two-branch convolution network. The meshes represent patches in each layer. The black numbers are the size of patches and the red numbers are the numbers of channels. The green arrows represent convolution. The inputs are the photon numbers in the low/high energy bin of each pixel within the patch. The outputs are the line integrals of attenuation at the two virtual energies of the POI.

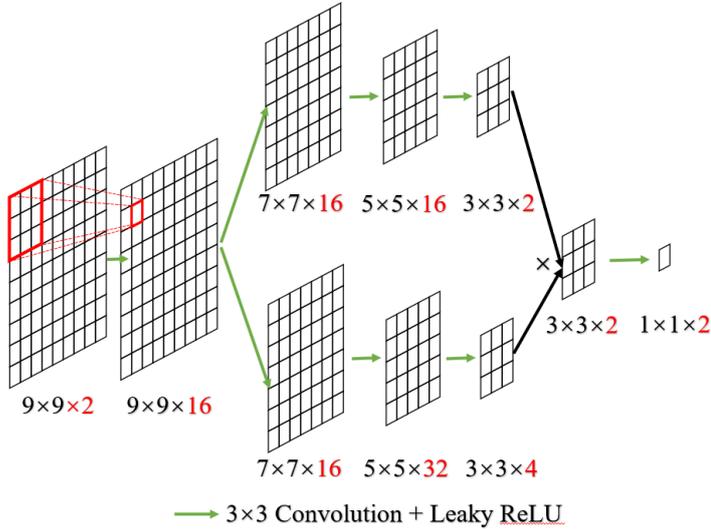

Figure 4. The structure of patch interweaving network. Black numbers are the patch size in each layer. Red numbers are the channel numbers. The inputs are the photon numbers of each pixel within the patch. The outputs are the line integrals.

The loss function we used to train PIN is

$$L_{\text{PIN}} = \frac{1}{2}\left(\left(\frac{Y_{\text{L}} - Y_{\text{L}}^*}{Y_{\text{L}}^*}\right)^2 + \left(\frac{Y_{\text{H}} - Y_{\text{H}}^*}{Y_{\text{H}}^*}\right)^2\right) \qquad (8)$$

Here, $Y_{\text{L}}^*$ and $Y_{\text{H}}^*$ are the ground truths of line integrals. Finally, we calculate the basis material thickness by using the outputs of the networks according to the formular (7).

## 2.4 Configuration for data simulation

We build our dataset by simulating X-ray passing through two materials as shown in Fig. 5. We set the thickness of basis materials for each detector pixel.

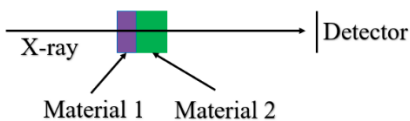

Figure 5. Diagram of data simulation. X-ray passes through material 1 and material 2, then injects the detector.

There could be several ways to configure thickness distribution within a patch, as shown in Fig. 6. Part (a) shows a flat-field and an edge. For a flat-field, all pixels have the same thickness. For an edge, all pixels can be devided into two spatially connected parts. The two parts have different thickness but uniform within each part. The two forms are realistic, but have structural information we should avoid, because the information might mislead the network and we hope the network learn the principle of charge sharing only. Part (b) shows a patch with randomly distributed thickness without any structural information. We use data from patches of random thickness to train and test our networks.

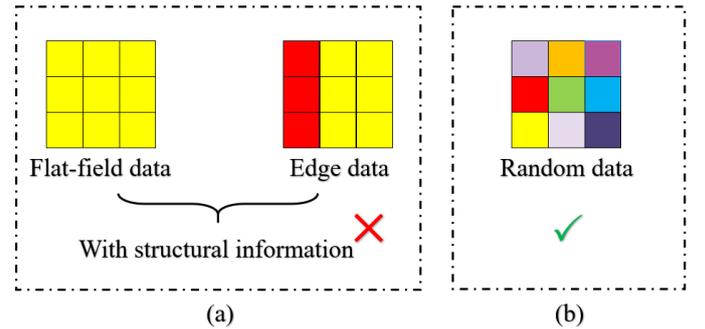

Figure 6. Different data forms. (a) has structure information. (b) doesn't have structure information.

## 3 Experimental studies

To train and test the networks, Monte Carlo simulation was done to generate noisy conventional data. Water and bone were chosen as basis materials. Human's head is our targeted object, so the range of water thickness is from 0cm to 12cm and the range of bone thickness is from 0cm to 2cm. Specificly, in training dataset,

$$T_{\text{w}} = \{1,2,\dots,40\} \times 0.3\text{cm} \qquad (9)$$
$$T_{\text{b}} = \{1,2,\dots,10\} \times 0.2\text{cm} \qquad (10)$$

where $T_{\text{w}}$ is the water thickness and $T_{\text{b}}$ is the bone thickness. The ground truths of line integrals $Y_{\text{L}}$ and $Y_{\text{H}}$ are calculated according to Eq. (5). There are 15 noise realizations for each thickness.

In test dataset,

$$T_{\text{w}} = 0.5, 2.5, 5, 7.5, 10\text{cm} \qquad (11)$$
$$T_{\text{b}} = 0.3, 0.5, 0.9, 1.5, 1.9\text{cm} \qquad (12)$$

which is non-overlap of training setting. There are 90 noise realizations for each thickness.

## 3.1 Results from SPIN and PIN

We analyzed the bias and standard deviation of basis material thickness for the test dataset. Bias of material thickness for difference water and bone composition resulted from SPIN and PIN were shown in Fig. 7. The PIN gets smaller bias and more uniform results. This tells us that the information from neighboring pixels helps the network to correct charge sharing better. We also plotted bias curves at 0.9cm bone for water and at 5cm water for bone in Fig. 7(b) with error bars representing standard deviation. We can see the standard deviations are similar for most points, but the bias of PIN is smaller than SPIN. In Fig. 7(c), we show the distribution of bias and total error for water and bone to further illustrate their performance. While the bias is smaller, the total error has decreased a lot for PIN.

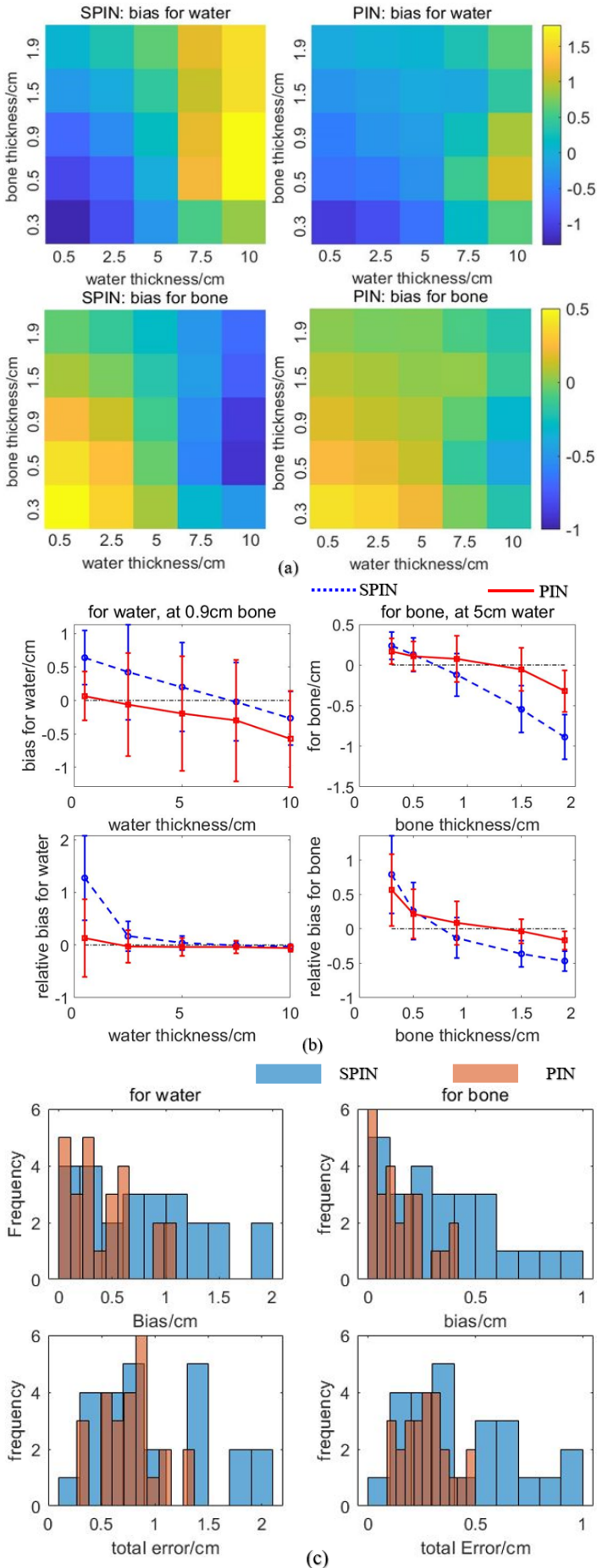

Figure 7. Results for SPIN and PIN. (a) are color maps showing the distribution of bias for bone and water. (b) are profiles at 0.9cm bone and 5cm water with error bars. (c) are histograms showing the distribution of bias and total error.

### 3.2 Improvement for PIN

There is still room for improvement according to the results in the last part. We used noisy data to train PIN and the noise might interfer the learning process. We tried to use the means of noise data for different thickness to train PIN:

$$\widehat{N_L} = \sum_{m=1}^{15} N_{L,m}, m \in \{1,2,...,15\} \qquad (13)$$

$$\widehat{N_H} = \sum_{n=1}^{15} N_{H,n}, n \in \{1,2,...,15\} \qquad (14)$$

Here, $N_{L,m}$ and $N_{H,n}$ are the noise realizations in low/high energy bin. We used these means to approximate the ground truths of photon numbers without noise to eliminate the influence of noise. In this way, we hope PIN can concentrate on charge sharing more and give better results.
The results are in Fig. 8.
In Fig, 8(a), bias distribution for two materials by SPIN and PIN are demonstrated. It is obvious that the bias is much smaller than those in Fig. 7. For all the points, the absolute bias is close to zero. We also plotted the profiles at 0.9cm bone for water and at 5cm water for bone with error bars, in Fig. 8(b).
Although the bias is satisfying, the standard deviations become larger. Because we almost removed all the noise information in the training dataset by using the means of noise realizations, so the network is poor in dealing with noise. Histograms in Fig. 8(c) show the distribution of bias and total error. Because of poor denoising, the total error is slightly larger compared with using noisy data.

## 4 Discussion

In this paper, we propose pixel-wise charge sharing correction by neural network. The network dealt with the data in $9 \times 9$ patch and learned relationship of photon numbers between central pixel and neighboring pixels. By comparing SPIN and PIN trained with noisy data, we verified that using information from neighboring pixels was effective and it could estimate thickness of basis materials with small bias for most conditions. By using mean data to train PIN, we further lower the bias of results and verified that using ground truths without niose during training could be an effective way for charge sharing correction. However, without providing enough information about noise, the performance can be affected by noise significantly.
In the experiment, the relative error is still not satisfying for situations of small material thickness. Charge sharing is different in low energy bin and high energy bin, but our current loss fuction doesn't model this. That leads to different error levels in results between low energy bin and high energy bin, and between water and bone. We will further improve our design to improve the network

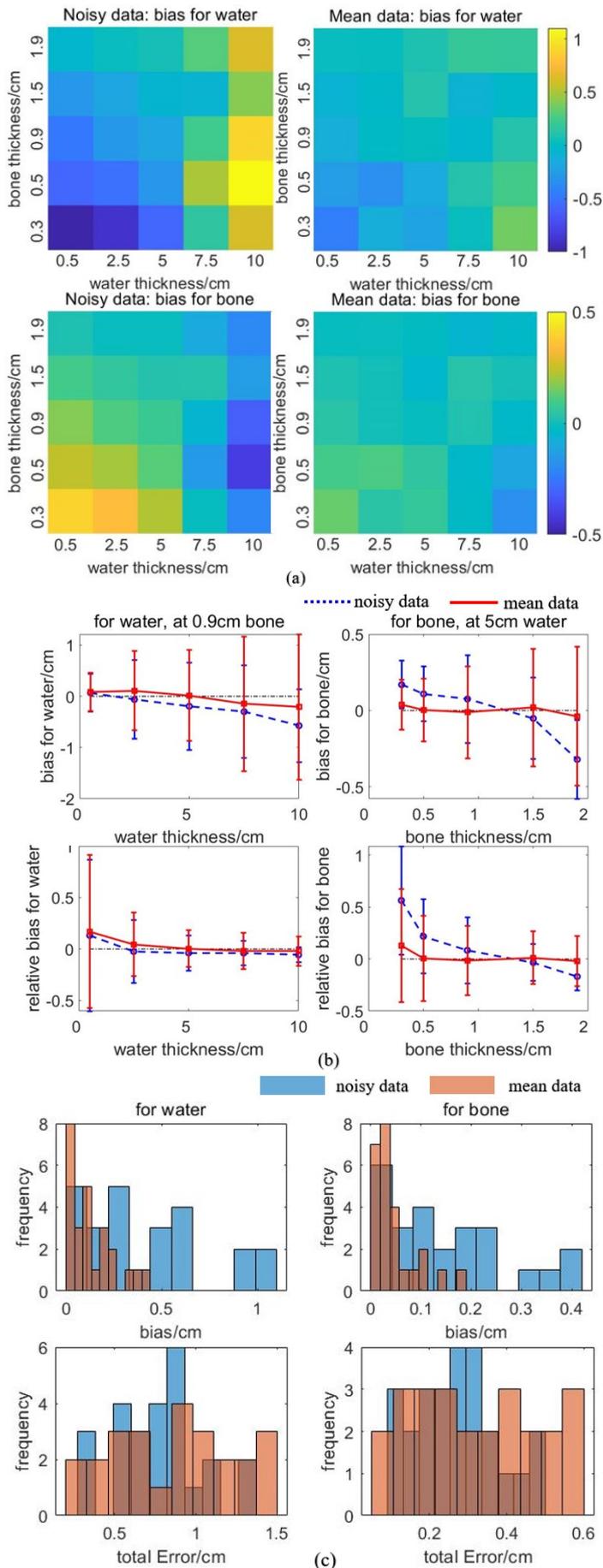

Figure 8. Results for training with noisy data and mean data. (a) are color maps showing the distribution of bias for bone and water. (b) are profiles at 0.9cm bone and 5cm water with error bars. (c) are histograms showing the distribution of bias and total error.

performance.

There are also other shortages. In this paper, we used supervised training method. Therefore, the ground truths of line integrals are necessary. In reality, it is hard to get these ground truths. Using mean is a possible and reasonable choice to as a substitute for ground truths. Another solution is unsupervised training method or semi-supervised training method, which is to be studied in our future work.

## 5 Conclusion

This paper proposes a patch interweaving network to correct charge sharing. Different from conventional methods and neural networks for whole images, our network focuses on the relationship between neighboring pixels and central pixel, and uses photon numbers in neighboring pixels to help correct charge sharing. It has good potential to deal with MEICC data and calculate basis material thickness with very small relative error.

# Non-Contrast Head CT Imaging Using Dual Energy: A Study on Bias Induced by Imperfections in Energy Response and Material Knowledge

Viktor Haase[1,2], Karl Stierstorfer[1], Katharina Hahn[1], Harald Schöndube[1], Andreas Maier[2], and Frédéric Noo[3]

[1]Siemens Healthcare GmbH, Forchheim, Germany
[2]Department of Computer Science, Friedrich-Alexander-Universität Erlangen-Nürnberg, Erlangen, Germany
[3]Department of Radiology and Imaging Sciences, University of Utah, Salt Lake City, USA

**Abstract** Diagnostic of stroke and proton therapy treatment planning require accurate non-contrast head CT imaging. Dual energy CT offers improvements in this field by enabling material decomposition and virtual mono-energetic images. Such a computation in the projection space requires an energy response model for the CT system and a selection of materials for data decomposition. In this work, we use incorrect response models and materials to study the effects of imperfections on the quality of the material decomposition. The experiments are carried out in fanbeam geometry with noise-free computer simulated data of two head phantoms. The results show that small errors in the energy response can quickly lead to small inhomogeneity and to beam hardening errors between close bones. They also show that the difference between soft tissues and water lead to soft tissue leakage in the bone image, but may not have a negative effect on the mono-energetic image.

## 1 Introduction

Accurate non-contrast head computed tomography (CT) imaging is critical for effective management of stroke patients. There are two types of stroke requiring urgent clinical care: hemorrhagic stroke, which corresponds to a vessel rupture; and ischemic stroke, which corresponds to a vessel blockage. Treatment for the second type can only be initiated once hemorrhagic stroke has been ruled out, which cannot be done with contrast agent. Accurate non-contrast head CT imaging is also needed in the context of treatment planning for brain cancer with proton-based therapy. To be most effective, this treatment option requires precise computation of proton-stopping power ratios using CT.

Despite the many advances of CT over the last decades, head CT imaging remains a challenging problem due to X-ray beam hardening. The entire brain is affected, with major variations across patient populations due to differences in bone thicknesses; and the base of the skull is the most difficult region, as the bones are particularly thick and complex in shape in this region. Dual energy CT (DE-CT) may offer further improvements, and for this reason recently been under investigation (see, e.g., [1–3]).

A number of important factors can affect the accuracy of DE-CT for non-contrast head imaging. There are physical effects like quantum noise and scatter; and there are data modeling aspects used for image reconstruction. Here, we are interested in assessing the effect of imperfections in data modeling, under the assumption of no noise and perfect scatter subtraction. From a statistical viewpoint, such an assessment is akin to performing a bias study. Understanding the magnitude and appearance of bias due to various imperfections in system modeling is as important as understanding the effect of noise and scatter, as it provides an upper bound on the best reachable image quality.

For this work, we assume that DE-CT is implemented as two CT scan repetitions, with tube voltage being the only difference between the two scans. Hence, two measurements are obtained for each detector pixel and view angle, and these can be separated into material-length measurements prior to filtered-backprojection reconstruction. Two modeling aspects are studied: assumptions regarding the energy response, and assumptions regarding the two materials used for DE data decomposition. The study relies on computer-simulated data.

## 2 Materials and Methods

The experiments are performed in fanbeam geometry. The basis for the simulations are two phantoms. First, we use the FORBILD head phantom, as described in 2-D in [4], which is a simplified representation of a human head based on (clipped) ellipses. This phantom was specifically designed to challenge image reconstruction algorithms in terms of contrast and shapes. Following [4], the FORBILD head phantom can be defined so as to represent a physical object composed of two materials, chosen here as water and bone. Second, Subject 04 from the BrainWeb family of anthropomorphic head phantoms [5] is used; these were created from slices of patient MRI data labeled pixel-by-pixel for their containing tissue. For our purpose, all 12 materials used in the BrainWeb phantoms are modeled using the X-ray mass attenuation coefficients of tissues suggested in Schneider et al. [6]. The two phantoms are forward projected either analytically, in case of the FORBILD head phantom, or using the distance-driven method for the BrainWeb phantom. The parameters used to model the CT geometry correspond to state-of-the-art scanners and can be found in Table 1.

The energy response of the scanner is modeled analytically as a normalized product of a filtered source spectrum $S$ and a detector response $D$, each depending on the energy $E$. Examples of realistic source spectra and detector response based on Monte Carlo transport of photons can be found in the literature [7], [8]. For the dual energy data simulation, the projections are computed for two different kVp-settings of the X-ray tube, namely the high energy scan at 140 kVp and the low energy scan at 80 kVp. Similar to real CT systems, we model two pre-object filtrations of the X-ray spectrum, using a flat pre-filter followed by a bowtie-shaped filter. The pre-filter is defined as 0.25 mm of copper and the material of the bowtie filter is aluminum. The filters are designed to closely match the filtered source spectrum of a real CT system. The use of a bowtie filter makes the source spectrum depend on the detector pixel location. Mathematically, the normalized energy response for each tube voltage (index $k$) and detector pixel (coordinate $u$) can be written as

$$W_k(E,u) = \frac{S_k(E,u)D(E)}{\int_0^{E_{max}} S_k(E,u)D(E)dE}, \text{ with } k = \{140, 80\} \text{kVp}.$$

**Table 1:** Parameters of scanner geometry

| | |
|---|---|
| Scan trajectory | 360° |
| Number of projections | 2304 |
| Number of detector channels | 736 |
| Source to detector distance | 108.56 cm |
| Source trajectory radius | 59.5 cm |
| Angular detector width | 0.067864° |
| Flying focal spot option | off |

After the high and low energy projection data is created, the material decomposition is performed in the projection domain. Water and bone are used for the two decomposition materials. The decomposition itself proceeds as follows. First, we simulate measurements of different lengths of water and bone using the energy response model. The resulting pairs of material lengths and dual energy measurements are used to model the relationship with two polynomials using least-square fitting. Then, the polynomials are applied to transform the high and low energy measurements into projection data of the two materials.

Finally, the projections of each material are individually reconstructed using fanbeam filtered backprojection with a Hanning apodization window. The reconstructed material images are further used to calculate virtual mono-energetic images, the linear attenuation coefficients of water and bone at the desired energy acting as weight for the linear combination of the water and bone images.

To simulate various imperfections in energy response, we make the following changes to the energy response used for the material decomposition:

- Option 1: Simplification of the detector response by only using the linear attenuation and the thickness of the detector material, $Gd_2O_2S$, also known as Gadox.

- Option 2: Reduction of the pre-filter thickness by 8%.

- Option 3: Removal of the flat pre-filter and application of a narrower bowtie-filter used for cardiac imaging.

Options 1 and 2 represent a rather small error for the assumptions on the detector response or the source spectrum, respectively. By contrast, option 3 is meant to simulate a major mismatch for the applied source spectrum.

For the study on imperfections on the material knowledge, we use two versions of the BrainWeb phantom. The first version is the original one, which involves 11 soft tissues plus bone. The second version is obtained from the original one by replacing each soft tissue with water of a specific mass density, chosen so that the two versions of the phantom are identical at 80 keV.

## 3 Results

### 3.1 FORBILD head phantom

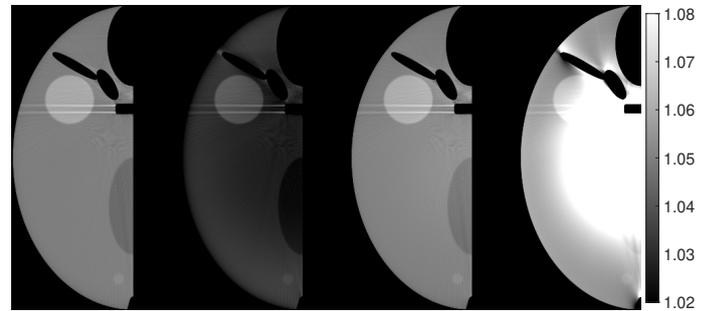

**Figure 1:** Reconstructed water images for the FORBILD head phantom. Left to right: Correct energy response, incorrect options 1-3. To save space, only the left half of the phantom is displayed.

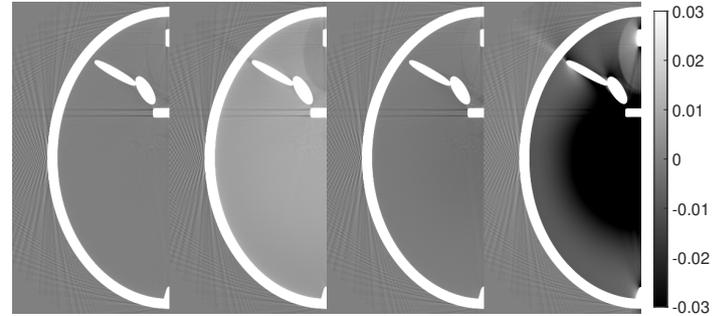

**Figure 2:** Reconstructed bone images for the FORBILD head phantom. Left to right: Correct energy response, incorrect options 1-3.

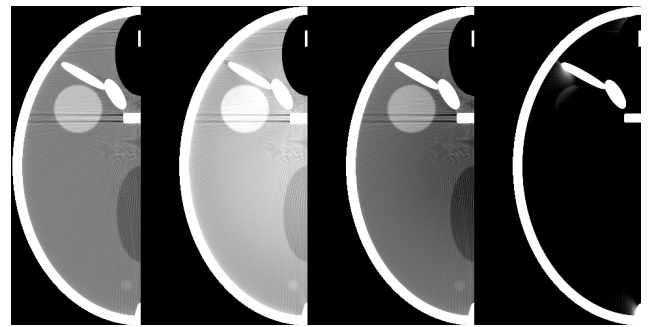

**Figure 3:** Reconstructed virtual mono-energetic images of the FORBILD head phantom corresponding to 50 keV. Left to right: Correct energy response, incorrect options 1-3. C/W=50/40 HU.

| Image | Corr. | Opt. 1 | Opt. 2 | Opt. 3 |
|---|---|---|---|---|
| Water, mean | 1.050 | 1.031 | 1.051 | 1.083 |
| Water, std | 0.001 | 0.003 | 0.002 | 0.034 |
| Bone, mean | 0.000 | 0.008 | $-0.002$ | $-0.027$ |
| Bone, std | 0.001 | 0.001 | 0.001 | 0.023 |
| Mono [HU], mean | 50.2 | 59.5 | 46.0 | $-13.0$ |
| Mono [HU], std | 1.7 | 2.5 | 2.5 | 50.4 |

**Table 2:** Quantitative measurements for the FORBILD head phantom for the correct energy response and incorrect options 1-3. The mean value and the standard deviation over pixels covering the brain matter (relative density of 1.05 and 50 HU at 50 keV) are reported for the water image, the bone image, and the mono image.

Fig. 1 shows the reconstructed water images with a gray scale centered around the relative density of brain matter to water, namely 1.05. The left image acts as ground truth as the correct energy response is used in the decomposition. Incorrect option 1 leads to a clearly visible underestimation of water with a cupping shape, whereas option 2 leads to a slight, difficult to notice overestimation. Incorrect option 3 leads to a major error.

The reconstructed bone images are presented in Fig. 2. Here, the colorbar is centered around 0, to highlight leakage of water into the bone image. As expected, the observations are very similar and consistent with those made for the water images.

When the two material images are combined into a virtual mono-energetic image at 50 keV, as in Fig. 3, a light cupping artifact is now clearly visible for incorrect option 2. In addition to the capping artifact in option 1, there is a beam hardening error where the sinus bone gets closest to the skull. Option 3 leads to strong artifacts making the grayscale window too narrow for close inspection.

The visual impressions are supported by quantitative measurements given in Table 2, where mean values and standard deviations over the brain matter region (i.e., the pixels of groundtruth value equal to 1.05) are reported.

### 3.2  BrainWeb phantom

We start with the results of the experiments on the imperfect material knowledge. As presented in Fig. 4, the decomposition into water and bone is not exact when the object contains other soft-tissue materials. Some brain tissue becomes visible in the bone image, showing that the soft tissues require a small component along the bone material when expressed as a mixture of water and bone.

However, when looking at the 50 keV mono-energetic images, the perfect (Fig. 5) and imperfect material knowledge (Fig. 6) yield similar accuracy relative to the ground truth. Quantitatively, the RMSD in a large region covering the brain is equal to 6.26 HU in the perfect case and to 6.46 HU in the imperfect case. In both cases, the deviation from the ground truth appears dominated by sampling errors; also, resolution losses and the narrow grayscale window cause bone blooming. Note that the contrast between gray and white matter is not expected to be the same in both cases, as the soft tissues are represented by different materials with identical attenuation value occurring only at 80 keV.

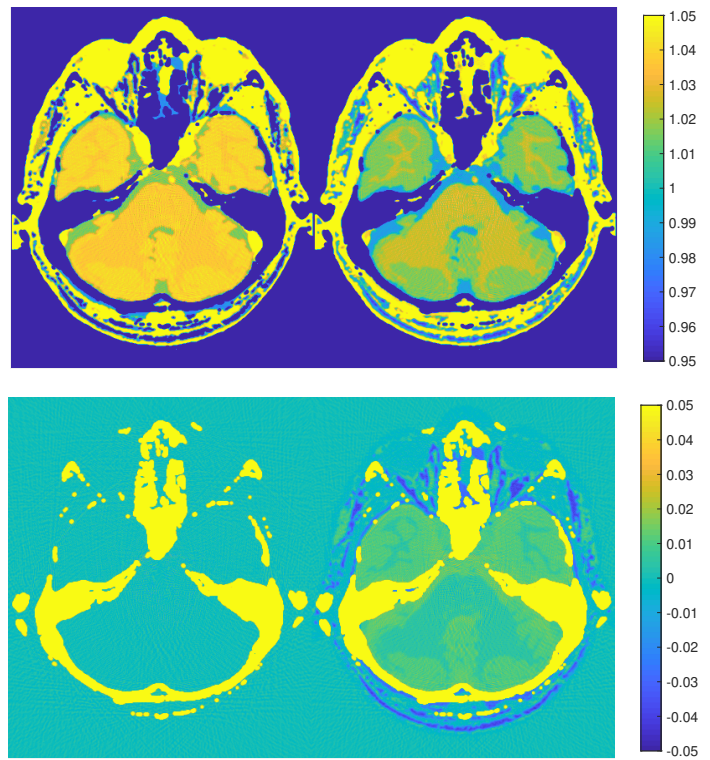

**Figure 4:** Reconstructed images of materials for the BrainWeb data. Top row: water images, bottom row: bone images. In both cases, the modified phantom containing only water and bone (left) is compared to the original one with 12 materials (right).

When repeating the experiments on the imperfect energy response with the BrainWeb phantom, similar observations as for the FORBILD phantom can be made. For the sake of space, we only show images for incorrect option 1 applied to the modified version of the phantom, which only contains water and bone. The errors induced by the imperfect response are visible in all images, bone, water, and mono, especially in the posterior fossa region. Over the same brain region as previously used, a mean of 0.010 and a standard deviation of 0.003 can be measured over the pixels in the bone image; and the mono image has an RMSD of 13.0 HU relative to the ground truth.

### 4  Discussion and Conclusion

In this 2-D simulation study, we investigated the bias induced by imperfections in energy response or material knowledge for non-contrast head DE-CT. As expected, the results firstly show that major imperfections lead to major error in the reconstructed images. Secondly, the results show that small errors in energy response can quickly lead to non-uniformity errors (capping/cupping) and unattractive beam hardening errors between bones that are close to each other. These errors are not ideal for detection of small bleeding or computation of proton stopping power ratio. Thirdly, an imperfect material knowledge reduces the quality of the material decomposition but may not have much impact on the virtual mono-energetic image. Note that our observations are limited to one specific spectra separation between the high and low energy scans. We plan to further complement this study with alternative spectra separations, as well as imperfections in the definition of the bone material and inspections of other slices within the BrainWeb phantom.

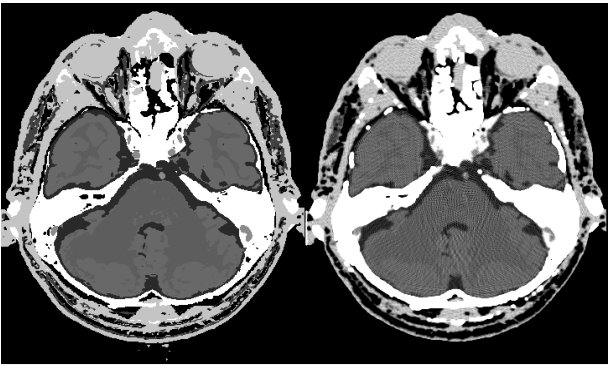

**Figure 5:** Reconstructed virtual mono-energetic image of the BrainWeb phantom for 50 keV next to the ground truth (left). Modified phantom containing only water and bone. C/W=50/100 HU.

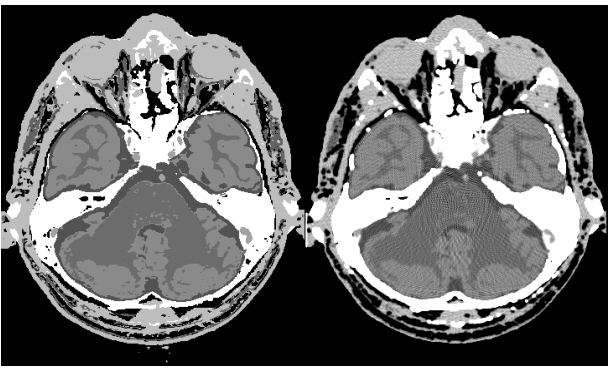

**Figure 6:** Reconstructed virtual mono-energetic image of the BrainWeb phantom for 50 keV next to the ground truth (left). Original phantom made of 12 different materials. C/W=50/100 HU.

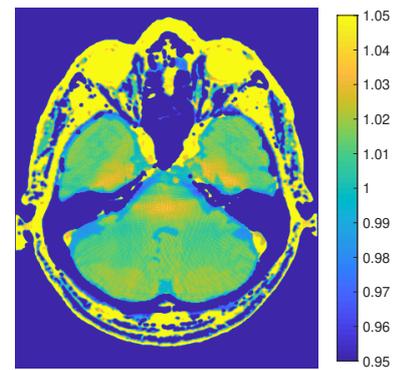

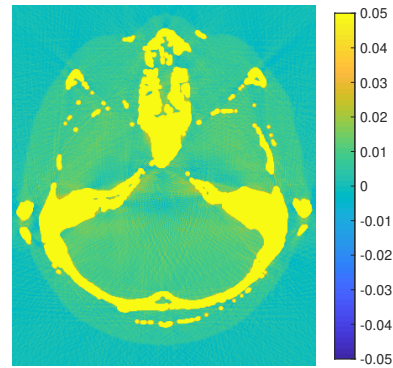

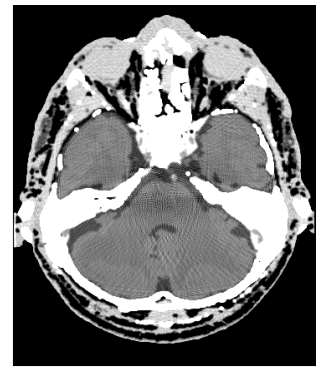

**Figure 7:** Modified BrainWeb phantom consisting of water and bone only. Results based on the incorrect response option 1. From top to bottom: water image, bone image, and 50 keV mono-energetic image (C/W=50/100 HU).

# Chapter 2

# Oral Session - Advanced image reconstruction methods 1

**session chairs**
Grant Theodore Gullberg, *University of California San Francisco (United States)*
Alessandro Perelli, *French National Institute of Health and Medical Research (INSERM) (France)*

# Do additional data improve region-of-interest reconstruction ?

Michel Defrise[1], Rolf Clackdoyle[2], Laurent Desbat[2], Frédéric Noo[3], and Johan Nuyts[4]

[1]Nuclear Medicine, Vrije Universiteit Brussel, Brussels, Belgium.
[2]TIMC-IMAG Laboratory (CNRS UMR 5525), Université Grenoble Alpes, Grenoble, France.
[3]UCAIR, Dept of Radiology, University of Utah, Salt Lake City, Utah.
[4]Nuclear Medicine, KU Leuven, Leuven, Belgium.

**Abstract** We consider reconstruction problems in CT where an exact reconstruction of a region-of-interest (ROI) of the object is possible from a limited data set. Does the measurement of additional line integrals, which do not cross the ROI, improve the reconstruction within the ROI? We consider a general discrete model of this question and prove a theorem, which defines a condition under which the ROI estimation cannot be improved by the additional data. The theorem is illustrated by a small dimension toy problem and by simulated data of the thorax phantom.

## 1 Introduction

This work deals with region-of-interest (ROI) reconstruction in CT. The two-dimensional (2D) filtered backprojection algorithm requires measuring all line integrals crossing the object, but several papers in the last 20 years have shown that unique and stable reconstruction of ROIs of the object is possible from specific subsets of line integrals. Examples include the long object problem in 3D cone-beam CT [1], the super-short scan 2D algorithm [2], and several 2D and 3D configurations based on the relation between the Hilbert transform and the backprojection of the data derivative [3, 4].

Consider for instance the 2D CT problem in Figure 1 and the limited data subset consisting of all line integrals crossing the ROI $A$. In the absence of noise these limited data are sufficient to reconstruct $A$ [3]; in this case measuring additional line integrals would not improve the already perfect estimation within the ROI. The situation is different when the data are noisy, because the measured line integrals crossing $A$ are contaminated by the regions of the object outside $A$ (e.g. by the "spine"). Measuring additional line integrals, which do not cross $A$ but convey information on these contaminating regions, might therefore improve the image estimation inside $A$.

The goal of this work is to determine conditions for when additional measurements will, or will not, improve ROI reconstruction. We consider a general discrete model of the problem, introduced in section 2. Section 3 defines a condition under which the ROI estimation cannot be improved by the additional data. An illustration with a toy problem is given in section 4, and a preliminary numerical example is proposed in section 5.

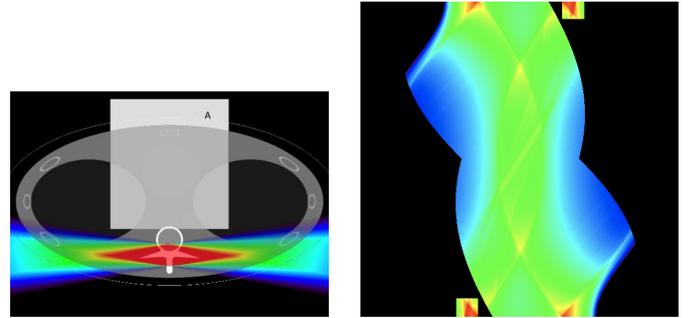

**Figure 1:** Left: the thorax phantom with superimposed the ROI $A$ (white rectangle). The thin white line is the boundary of the assumed object support. Right: The limited sinogram data $y$ (lines crossing the ROI $A$), and the added data $z$ (two small rectangles). The colored band in the left plot is the backprojection of unit added data $z \equiv 1$ (i.e. the backprojection of a sinogram equal to 1 in the two small rectangles and equal to 0 elsewhere).

## 2 A discrete model

The data model is $<y> = Sx$, with the data $y \in \mathbf{R}^m$, the object $x \in \mathbf{R}^n$, and a known $m \times n$ system matrix $S$. The solution space $\mathbf{R}^n = \mathbf{R}^{n_A} \oplus \mathbf{R}^{n_B}$ is the direct sum of two subspaces $A = \{x \in \mathbf{R}^n \,|\, x_j = 0, j = n_A + 1, \ldots, n\}$ and $B = \{x \in \mathbf{R}^n \,|\, x_j = 0, j = 1, \ldots n_A\}$. In the CT case, $S$ is the system matrix that maps the image vector $x$ onto the discretized sinogram subset $y$, and $A$ is the ROI. The voxels are indexed such that the first $n_A$ components of $x$ are the ROI voxels. In this discrete setting all prior information on the support of the object is introduced in the choice of the set of $n$ "voxels".

- **Assumption A1.** The system matrix $S$ is singular but the data $y$, if noise free, determine $x$ within the ROI $A$:

$$Sx = 0 \Rightarrow x_j = 0 \text{ for } j = 1, \ldots, n_A. \qquad (1)$$

- **Assumption A2.** The covariance matrix of the data, $W = <(y - <y>)(y - <y>)^T>$, is diagonal with uniform noise variance $\sigma^2$: $W = \sigma^2 Id$.

Because $S$ is singular, the least-squares (LS) estimate is not unique, and the usual approach is to select the LS solution of minimum norm (generalized solution),

$$x^\dagger = S^\dagger y = \arg\min_{x \in N(S)^\perp} L(x) \qquad (2)$$

with $S^\dagger$ the Moore-Penrose (generalized inverse) of $S$, $N(S) = \{x \in \mathbf{R}^n \,|\, Sx = 0\}$ the null-space of $S$, and $L(x) =$ $\|y - Sx\|^2$. The orthogonal projector onto the null-space of the system matrix is denoted $P_{N(S)}$, and $P_{N(S)^\perp} = Id - P_{N(S)} = S^\dagger S$.

Using the same notations the matrix of the projection onto the ROI subspace $A$ is:

$$(P_A x)_j = \begin{cases} x_j & j = 1, \ldots, n_A \\ 0 & j = n_A + 1, \ldots, n \end{cases} \quad (3)$$

and $P_B = Id - P_A$. It can be shown that assumption A1 is equivalent to the identity $P_A P_{N(S)} = 0$ and also to the identity $P_{N(S)} P_A = 0$. Using these projectors we express any $x \in \mathbf{R}^n$ as $x = x_A + x_B + x_N$, where

$$
\begin{aligned}
x_A &= P_A x = P_{N(S)^\perp} P_A x \text{ is the ROI component,} \\
x_B &= P_{N(S)^\perp} P_B x \text{ is the visible component of the region } B, \\
x_N &= P_{N(S)} x = P_{N(S)} P_B x \text{ is the null-space component,} \\
\|x\|^2 &= \|x_A\|^2 + \|x_B\|^2 + \|x_N\|^2. \quad (4)
\end{aligned}
$$

The following properties are straightforward extensions of well-known properties of minimum norm LS solutions [6] when A1 and A2 hold:

- *Lemma 1.* $x^\dagger$ is an unbiased estimate in the ROI A.

- *Lemma 2.* The covariance of $x^\dagger$ is $\sigma^2 F^\dagger$, with $F^\dagger$ the Moore-Penrose inverse of the Fisher matrix $F = S^T S$.

- *Lemma 3.* Let $H$ be any $n \times m$ matrix such that $P_A H S = P_A P_{N(S)^\perp}$. Then the estimator $\hat{x} = Hy$ has the same expectation as $x^\dagger$ within $A$ and $P_A(Cov(\hat{x}) - Cov(x^\dagger))P_A$ is a non-negative matrix. This means that $x^\dagger$ has the lowest variance among all unbiased linear estimates of $x$ within the ROI $A$.

## 3 Do additional data improve the estimation in the ROI ?

Consider an additional data vector $z \in \mathbf{R}^q$ with mean value $<z> = Cx$, for some known $q \times n$ matrix $C$. We assume that A2 also holds for the data $z$. The $k$-th added data $z_k \in \mathbf{R}$ is defined by the LOR vector $c_k = C_{k,.} \in \mathbf{R}^n$, for $k = 1, \ldots, q$.

- **Assumption A3**. The additional LORs do not intersect the ROI $A$, i.e. $C_{k,j} = 0$ for $j = 1, \ldots, n_A$ and $k = 1, \ldots, q$. This can also be written $P_A C^T = 0$.

- **Assumption A4**. The rows of $C$ are linearly independent and are linearly independent of the rows of $S$ : $Range(C) = \mathbf{R}^q$ and $Range(C^T) \cap Range(S^T) = \{0\}$.

In the context of tomography, the matrices $S^T$ and $C^T$ correspond to the backprojection operation. The second condition in A4 then means that it is impossible to find a non-zero pair of vectors $y \in \mathbf{R}^m$ and $z \in \mathbf{R}^q$ which yield the same image when backprojected onto the image space $\mathbf{R}^n$. In the example of Figure 1a, the backprojection of any added data vector $z$ is zero outside the colored region. Therefore, assumption A4 is satisfied unless one can find a data vector $y$ (set of values for the LORs crossing the square ROI A), the backprojection of which, $S^T y$, is zero everywhere except within the colored region, where it must be equal to $C^T z$.

We will also use the following equivalent formulation of assumption A4 (proof in Appendix):

*Lemma 4.* $Range(C) = \mathbf{R}^q$ and $Range(C^T) \cap Range(S^T) = \{0\} \Leftrightarrow Range(CP_{N(S)}) = \mathbf{R}^q$.

The main result of this work is the following theorem:

*Theorem 1.* If the assumptions A1-A4 are verified, the added data $z$ do not improve the estimation of $x$ in the ROI $A$.

Some remarks are in order.

- Theorem 1 provides a *necessary condition*: if A1, A2 and A3 are satisfied, the additional data are susceptible to improving the estimation in the ROI $A$ only if assumption A4 does not hold, i.e. if there exists a non-zero pair $y \in \mathbf{R}^m$ and $z \in \mathbf{R}^q$ such that $C^T z = S^T y$.

- Theorem 1 is easily generalized if the data covariance $W$ is any positive diagonal matrix, by substituting $S \to W^{-1/2} S$ and $y \to W^{-1/2} y$, and similarly for $C$ and $z$.

- Theorem 1 can be generalized to Poisson data, replacing the least-squares solution $x^\dagger$ by the minimizer of the Kullback-Leibler divergence. This generalization, however, only holds for the unconstrained estimator, i.e. if one does not enforce the non-negativity of $x^\dagger$.

- A generalization to a continuous setting is not straightforward because the generalized inverse is not a continuous operator for ill-posed problems such as the Radon transform.

### Proof of Theorem 1

Let $x^\dagger$ be the generalized solution (2) of the original problem $y = Sx$. Decompose that solution according to (4), thus $x^\dagger = x_A^\dagger + x_B^\dagger + x_N^\dagger$, with $x_N^\dagger = 0$ because it is the minimum norm LS solution. For any $x = x_A + x_B + x_N$,

$$L(x^\dagger) = \|y - Sx_A^\dagger - Sx_B^\dagger\|^2 \le L(x) = \|y - Sx_A - Sx_B\|^2. \quad (5)$$

Quantities with the added data $z$ will be denoted by a tilde. The generalized solution with the added data is found by minimizing the combined data fit

$$
\begin{aligned}
\tilde{L}(x) &= \|y - Sx\|^2 + \|z - Cx\|^2 \\
&= \|y - Sx_A - Sx_B\|^2 + \|z - Cx_B - Cx_N\|^2, \quad (6)
\end{aligned}
$$

(the second line uses assumptions A1 and A3). Consider the following system of $q$ equations:

$$CP_{N(S)} w = z - Cx_B^\dagger. \quad (7)$$

This equation has at least one exact solution for $w$ because by Lemma 4, assumption A4 implies that $Range(CP_{N(S)}) = \mathbf{R}^n$. Therefore (7) is a consistent system of equations for any right hand side vector. Let $w^*$ be the minimum norm solution of (7). One easily verifies that $w^* \in N(S)$. We will verify that $\tilde{x}^\dagger = \tilde{x}_A^\dagger + \tilde{x}_B^\dagger + \tilde{x}_N^\dagger$ with

$$\tilde{x}_A^\dagger = x_A^\dagger \; ; \; \tilde{x}_B^\dagger = x_B^\dagger \; ; \; \tilde{x}_N^\dagger = w^*, \tag{8}$$

is the generalized solution with added data, i.e. $\tilde{x}^\dagger$ is the minimum norm minimizer of $\tilde{L}(x)$. To see this, note first from (8) and (5) that for any $x = x_A + x_B + x_N \in \mathbf{R}^n$,

$$\|y - S\tilde{x}_A^\dagger - S\tilde{x}_B^\dagger\|^2 = \|y - Sx_A^\dagger - Sx_B^\dagger\|^2 \le \|y - Sx_A - Sx_B\|^2 \tag{9}$$

because $x^\dagger$ is a minimizer of $L(x)$. In addition, with $\tilde{x}_N^\dagger = w^*$ chosen as a solution of (7),

$$\begin{aligned} \|z - C\tilde{x}^\dagger\|^2 &= \|z - C\tilde{x}_B^\dagger - C\tilde{x}_N^\dagger\|^2 = \|z - Cx_B^\dagger - Cw^*\|^2 = 0 \\ &\le \|z - Cx_B - Cx_N\|^2 \end{aligned} \tag{10}$$

for all $x = x_A + x_B + x_N \in \mathbf{R}^n$. Adding this inequality to (9),

$$\begin{aligned} &\|y - S\tilde{x}_A^\dagger - S\tilde{x}_B^\dagger\|^2 + \|z - C\tilde{x}_B^\dagger - C\tilde{x}_N^\dagger\|^2 \\ \le\; &\|y - Sx_A - Sx_B\|^2 + \|z - Cx_B - Cx_N\|^2 \end{aligned} \tag{11}$$

and therefore $\tilde{L}(\tilde{x}^\dagger) \le \tilde{L}(x)$ for all $x \in \mathbf{R}^n$.

Having shown that $\tilde{x}^\dagger$ defined by (8) minimizes $\tilde{L}$, it remains to verify that it is the minimum norm minimizer. Suppose $\hat{x} = \hat{x}_A + \hat{x}_B + \hat{x}_N$ also minimizes $\tilde{L}$, so that $\tilde{L}(\hat{x}) = \tilde{L}(\tilde{x}^\dagger)$, i.e.

$$L(\hat{x}) + \|z - C\hat{x}\|^2 = L(\tilde{x}^\dagger) + \|z - C\tilde{x}^\dagger\|^2 = L(x^\dagger) \tag{12}$$

where we used $\|z - C\tilde{x}^\dagger\| = 0$ (eq. (10)) and $L(\tilde{x}^\dagger) = L(x^\dagger)$ (first equality in (9)). Since $x^\dagger$ minimizes $L(x)$ and $\|z - C\hat{x}\| \ge 0$, equation (12) implies $L(\hat{x}) = L(x^\dagger)$. Therefore $\hat{x} \in x^\dagger + N(S)$, and in particular $\hat{x}_B = x_B^\dagger = \tilde{x}_B^\dagger$. Inserting this again in (12), $\hat{x}$ must be solution of (7),

$$\|z - C\hat{x}_B - C\hat{x}_N\|^2 = \|z - C\tilde{x}_B^\dagger - C\hat{x}_N\|^2 = 0. \tag{13}$$

But $\tilde{x}_N^\dagger = w^*$ is the minimum norm solution of that equation so that by (4) $\tilde{x}^\dagger$ is the minimum norm minimizer of $\tilde{L}$.

Recalling the optimal property of the generalized inverse (Lemma 3), and noting from (8) that $\tilde{x}_A^\dagger = x_A^\dagger$, completes the proof that the added measurement $z$ does not improve the estimation of $x$ in the ROI A.

## 4　Illustration with a toy problem

We illustrate the previous result with the toy problem sketched in Figure 2, with $n = 4$ "voxels" and the $3 \times 4$ system matrix

$$S = \begin{pmatrix} 1 & 1 & 0 & 0 \\ 1 & 0 & 1 & 1 \\ 0 & 1 & 1 & 1 \end{pmatrix} \tag{14}$$

The null-space of $S$ is spanned by the vector $(0\ 0\ 1/\sqrt{2}\ -1/\sqrt{2})^T$ hence the two first components of $x$ can be reconstructed exactly from $y = Sx$: the ROI is $A = \{x \in \mathbf{R}^4 \,|\, x_3 = x_4 = 0\}$. Assuming uniform noise ($\sigma = 1$), the covariance of $x^\dagger$ is the Moore-Penrose inverse of the Fisher matrix $F = S^T S$:

$$Cov(x^\dagger) = F^\dagger = \frac{1}{8} \begin{pmatrix} 6 & -2 & -1 & -1 \\ -2 & 6 & -1 & -1 \\ -1 & -1 & 3/2 & 3/2 \\ -1 & -1 & 3/2 & 3/2 \end{pmatrix} \tag{15}$$

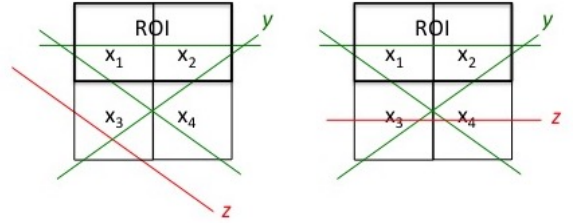

**Figure 2:** Sketch of the toy problem. The limited data $y$ correspond to the green LORs. The additional data $z$ corresponds to the red LOR. The estimation within the ROI is improved only in the case shown in the right plot.

Add a single additional measurement ($q = 1$) with $1 \times 4$ matrix $C = (0\ 0\ 1\ 0)$, which satisfies $P_A C^T = 0$ (A3). The combined system matrix with the added data is

$$\tilde{S} = \begin{pmatrix} S \\ C \end{pmatrix} = \begin{pmatrix} 1 & 1 & 0 & 0 \\ 1 & 0 & 1 & 1 \\ 0 & 1 & 1 & 1 \\ 0 & 0 & 1 & 0 \end{pmatrix} \tag{16}$$

The covariance of the estimator $\tilde{x}^\dagger$ derived from the combined data $(y, z)$ is

$$Cov(\tilde{x}^\dagger) = (\tilde{S}^T \tilde{S})^\dagger = \frac{1}{4} \begin{pmatrix} 3 & -1 & 0 & -1 \\ -1 & 3 & 0 & -1 \\ 0 & 0 & 4 & -4 \\ -1 & -1 & -4 & 7 \end{pmatrix} \tag{17}$$

Comparing (15) with (17) shows that $Cov(\tilde{x}^\dagger)_{i,j \in \{1,2\}} = Cov(x^\dagger)_{i,j \in \{1,2\}}$: the added data $z = Cx$ does not improve the estimation in the ROI. This result is expected from Theorem 1, because $\big(Range(C^T) = \{x \in \mathbf{R}^4 \,|\, x_1 = x_2 = x_4 = 0\}\big) \cap \big(Range(S^T) = \{x \in \mathbf{R}^4 \,|\, x_3 = x_4\}\big) = \{0\}$, hence A4 is satisfied.

If the added measurement is replaced by $C = (0\ 0\ 1\ 1)$, as in the right plot in Figure 2, the combined system matrix becomes

$$\tilde{S} = \begin{pmatrix} S \\ C \end{pmatrix} = \begin{pmatrix} 1 & 1 & 0 & 0 \\ 1 & 0 & 1 & 1 \\ 0 & 1 & 1 & 1 \\ 0 & 0 & 1 & 1 \end{pmatrix} \tag{18}$$

In this case $Range(C^T) = \{x \in \mathbf{R}^4 \,|\, x_1 = x_2 = 0, x_3 = x_4\}$ has a non-trivial intersection with $Range(S^T)$ hence A4 is not satisfied. The covariance matrix is:

$$Cov(\tilde{x}^\dagger) = \tilde{F}^\dagger = \frac{1}{14} \begin{pmatrix} 10 & -4 & -1 & -1 \\ -4 & 10 & -1 & -1 \\ -1 & -1 & 3/2 & 3/2 \\ -1 & -1 & 3/2 & 3/2 \end{pmatrix} \quad (19)$$

Comparing (15) with (19) shows that $Cov(x^\dagger)_{i,j \in \{1,2\}} - Cov(\tilde{x}^\dagger)_{i,j \in \{1,2\}}$ is a non-negative matrix, with eigenvalues 0 and 1/14. We conclude that the added data improves the covariance in the ROI $A$ (in this particular example, only the estimation of the sum $x_1 + x_2$ is improved).

*Remark.* When a single data item is added ($q = 1$), as in this toy problem, explicit expressions of $\tilde{F}^\dagger$ can be found. Denote $c \in \mathbf{R}^n$ the unique row of the $1 \times n$ matrix $C$. When $c \in Range(S^T)$ the condition A4 is not satisfied and the covariance with the added data is given by (Theorem 3 in [7])

$$\tilde{F}^\dagger = (F + c c^T)^\dagger = F^\dagger - \frac{1}{\beta} F^\dagger c c^T F^\dagger \quad (20)$$

where $\beta = 1 + c^T F^\dagger c$.

## 5   The thorax phantom

This section presents results with one slice of the thorax phantom (Figure 1), a rectangular ROI $A$, and the object support in a slightly larger ellipse. The phantom intensity values are rescaled such that the value in the background is 1. The data $y = Sx$ contain all rays intersecting $A$. In a continuous setting these data would allow an exact reconstruction of the ROI [3, 4]. Here we consider a discrete approximation of this problem, with the image matrix and the (full) parallel beam sinogram discretized in a $600 \times 600$ pixel matrix (so that $n = 360000$ and $m = 177052$).

Even in discretized setting, the inversion of the Radon transform is an ill-conditioned problem. The Moore-Penrose solution $x^\dagger$ considered in Theorem 1 must be regularized, and the relevance of that theorem is therefore questionable. An added difficulty is to define regularization parameters, which act in a similar way for the full data and for the limited data. Here we apply the Landweber algorithm with a fixed relaxation parameter and a fixed number of iteration, this choice was motivated by the excellent stability of this algorithm and its property to smoothly and progressively introducing increasing spatial frequencies.

We consider additional data $z = Cx$ corresponding to a set of roughly horizontal lines through the phantom (Figure 1); these lines do not intersect $A$. The number of additional LORs is $q = 2946$. A heuristic argument suggesting that A4 is valid for this geometry is based on the central section theorem. Consider in Figure 3a the point $P$ located in the spine. We argue that the backprojection of the additional data in a neighborhood of point $P$ cannot be the same as the backprojection of the original data. The data $y(\phi, s)$ correspond to all LORs crossing the ROI $A$; therefore, at point $P$ the backprojection $(S^T y)(P)$ only involves the angular range shown by the red lines that connect $P$ to the lower corners of the ROI. Similarly, the backprojection $(C^T z)(P)$ of the added data $z(\phi, s)$ (the two small rectangles in Figure 1b) only involves the angular range shown by the blue lines. Therefore the limited angular ranges covered by the data $y$ and by the additional data $z$ at point $P$ do not overlap. Assuming that the central section theorem can be applied locally at $P$, the support of the Fourier transform of $S^T y$ and $C^T z$ near $P$ are shown in Figure 3b, and because these supports do not overlap, one conjectures that A4 holds true because it is impossible to find functions $y(\phi, s)$ and $z(\phi, s)$ yielding identical backprojections. We have no proof of this conjecture.

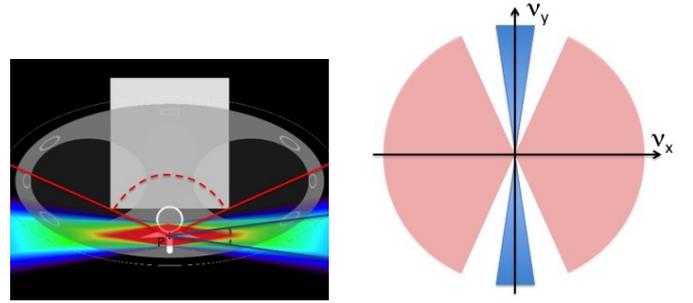

**Figure 3:** Left: the geometry of Figure 1, with the thorax phantom and the rectangular ROI $A$. The red and blue lines are the limits of the angular ranges covered when backprojecting the data and added data at a point $P$ located in the spine. The two angular ranges do not overlap. Right: assuming local shift invariance, the supports of the 2D Fourier transform of the backprojection of the data (red sector) and added data (blue sector) near $P$.

Data were simulated analytically. Uniform gaussian noise was added (post-log, $\sigma = 0.02 \times$ the maximum in the complete sinogram). We use 1000 iterations of the Landweber algorithm, with a fixed relaxation parameter ($0.0116 = 0.95/\|S\|^2$), and we apply a post-reconstruction gaussian smoothing with FWHM of 2 pixels (truncated to a $11 \times 11$ discrete kernel). There was no positivity constraint. Figure 4 shows mean images and standard deviation images (40 reconstructions from independent noise realizations).

The differences between the reconstructions from full sinogram, ROI data $y$, and combined ROI and added data $(y, z)$ were calculated within the ROI $A$. Results are given in Table 1 for the mean and for the standard deviation images:

$$e_{Mean} = \frac{\|P_A(<\hat{x}_1> - <\hat{x}_2>)\|}{\|P_A <\hat{x}_1>\|}$$

$$e_{Stdev} = \frac{\|P_A(stdev(\hat{x}_1) - stdev(\hat{x}_2))\|}{\|P_A stdev(\hat{x}_1)\|} \quad (21)$$

where $\hat{x}_1$ and $\hat{x}_2$ are the two reconstructions being compared, and $<>$ and $stdev()$ are the sample mean and standard deviation images over the 40 noise realizations.

Even though the data are noisy, the three reconstructions are very similar within the ROI, with differences smaller than the differences between the reconstructions from noise-free and from noisy data. Even adding the whole missing data to the limited ROI data has only a small impact on the variance of the Landweber reconstructions interrupted at 1000 iterations. The same simulated data were reconstructed using 200 iterations of the conjugate gradient algorithm (restarted when $L(x^{k+1}) > L(x^k)$), again with post-smoothing (2 pixels FWHM). Although different converging algorithms may follow different paths to the pseudo-solution, a similar observation holds as with the Landweber reconstructions (Figure 5 and Table 2): the added data has a small impact on the image and its variance within the ROI. Outside the ROI however the standard deviation is much larger with the conjugate gradient reconstruction than with the Landweber reconstruction (Table 3). We tentatively attribute this difference to the introduction by the conjugate gradient algorithm of singular components with smaller singular values. Refering to known properties of the finite Hilbert transform [8], these components are known to have significant values only outside the ROI.

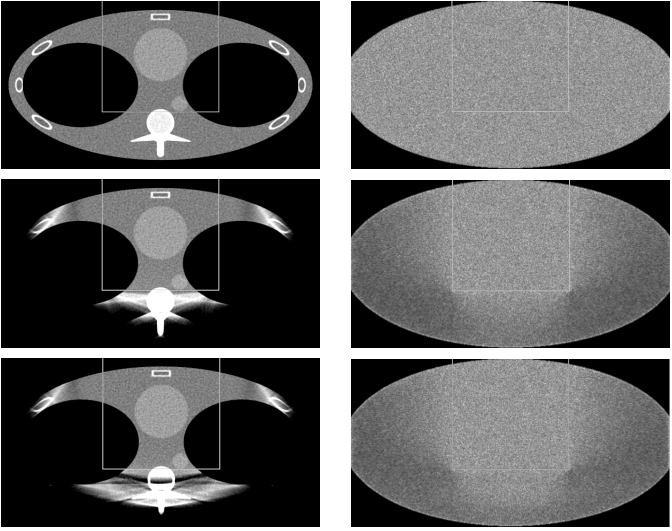

**Figure 4:** Reconstructions of the thorax phantom. Landweber algorithm, 1000 iterations, post-smoothed 2 pixels FWHM. Left column: mean reconstruction (scale 0,8,1.2). Right column: standard deviation images (scale 0, 0.15). Top row: reconstruction from the full sinogram. Middle row: reconstruction using the ROI data only. Bottom row: reconstruction using the ROI+ additional data. Thin white line : boundary of the ROI.

**Table 1:** Relative mean square root differences (RMS) between pairs of mean reconstructions and pairs of standard deviation images. See eq. (21). The RMS includes only the pixels in ROI $A$. Reconstructions with 1000 Landweber iterations from complete data ("full"), ROI data (lines crossing $A$),"ROI", and combined data ("ROI+").

|  | image $\hat{x}_1$ | image $\hat{x}_2$ | RMS difference |
|---|---|---|---|
| $e_{Mean}$ | full | full, no noise | 0.0180 |
|  | full | ROI | 0.0122 |
|  | full | ROI+ | 0.0080 |
|  | ROI | ROI+ | 0.0055 |
| $e_{Stdev}$ | full | ROI | 0.0044 |
|  | full | ROI+ | 0.0044 |
|  | ROI | ROI+ | 0.0007 |

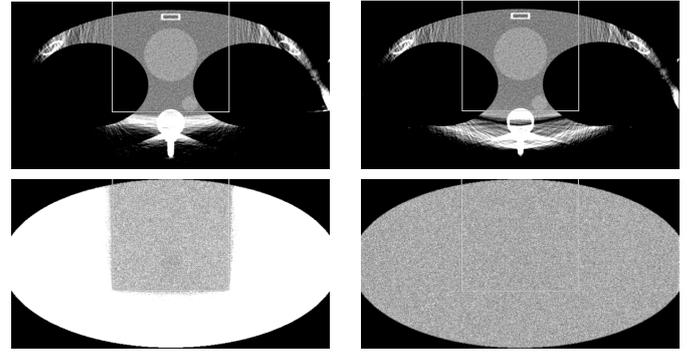

**Figure 5:** Top row: Mean reconstructions (scale 0.8, 1.2). Conjugate gradient algorithm, 200 iterations, post-smoothed 2 pixels FWHM. Left: ROI data only. Right: ROI+ added data. Bottom row: Standard deviation images (scale 0, 0.15). Left: ROI data only. Right: full data. Thin white line : boundary of the ROI.

**Table 2:** Same as Table 1 for the conjugate gradient reconstructions.

|  | image $\hat{x}_1$ | image $\hat{x}_2$ | RMS difference |
|---|---|---|---|
| $e_{Mean}$ | full | full, no noise | 0.0211 |
|  | full | ROI | 0.0101 |
|  | full | ROI+ | 0.0097 |
|  | ROI | ROI+ | 0.0040 |
| $e_{Stdev}$ | full | ROI | 0.1130 |
|  | full | ROI+ | 0.1156 |
|  | ROI | ROI+ | 0.0087 |

## 6 Conclusion

Even when a region of interest can be exactly reconstructed from a limited tomographic data set in the absence of noise, it appears reasonable to expect that the measurement of additional line integrals should improve the ROI in the presence of noise. We show that this is not necessarily the case, and introduce in Theorem 1 a condition, which, when satisfied, implies that the additional data do not improve the estimation in the ROI. The theorem only applies to the unbiased minimum norm least-squares solution. Evaluating the relevance of this result requires further numerical tests and a generalization of the theorem to regularized solutions.

**Table 3:** Mean standard deviation in the images reconstructed from the full data (two first rows) and from the ROI data only (two last rows). Recall that the background value in the thorax phantom is equal to 1.

|  |  | Mean st. dev. | Landweber | conj. gradient |
|---|---|---|---|---|
| full data | within ROI | | 0.087 | 0.098 |
|  | outside ROI | | 0.084 | 0.095 |
| ROI data | within ROI | | 0.087 | 0.107 |
|  | outside ROI | | 0.066 | 0.261 |

## 7 Appendix: proof of Lemma 4.

*Lemma 4.* $Range(C) = \mathbf{R}^q$ and $Range(C^T) \cap Range(S^T) = \{0\} \leftrightarrow Range(CP_{N(S)}) = \mathbf{R}^q$.

*Proof.*

- $\rightarrow$. Suppose that $Range(C) = \mathbf{R}^q$ and $Range(C^T) \cap Range(S^T) = \{0\}$. Suppose by contradiction that $Range(CP_{N(S)}) \subset \neq \mathbf{R}^q$. Because $\mathbf{R}^q = Range(CP_{N(S)}) \bigoplus N(P_{N(S)}C^T)$ this implies that $N(P_{N(S)}C^T) \neq \{0\}$ so there is a $z \in \mathbf{R}^q$ such that $z \neq 0$ and $P_{N(S)}C^Tz = 0$, i.e. $C^Tz \in N(S)^{\perp} = Range(S^T)$. Note also that $N(C^T) = \{0\}$ because $Range(C) = \mathbf{R}^q$ and therefore $C^Tz \neq 0$. There is therefore a vector $0 \neq C^Tz \in Range(C^T) \cap Range(S^T)$, a contradiction.

- $\leftarrow$. Suppose that $Range(CP_{N(S)}) = \mathbf{R}^q$. This immediately implies $Range(C) = \mathbf{R}^q$. Next, suppose by contradiction that $Range(C^T) \cap Range(S^T) \neq \{0\}$, so there is a pair $y \in \mathbf{R}^m$ and $z \in \mathbf{R}^q$ such that $C^Tz = S^Ty = x \neq 0$. But $S^Ty \in Range(S^T) = N(S)^{\perp}$ therefore $x \in N(S)^{\perp}$, and $P_{N(S)}x = P_{N(S)}C^Tz = 0$. Therefore $N(P_{N(S)}C^T) \neq \{0\} \rightarrow Range(CP_{N(S)}) \subset \mathbf{R}^q$, a contradiction.

**Acknowledgment**: This work is partly supported by the ANR (ROIdoré, ANR-17-CE19-0006-01)

# Implementation of the virtual fan-beam method for 2D region-of-interest reconstruction from truncated data

Mathurin Charles[1], Rolf Clackdoyle[1], and Simon Rit[2]

[1]Université Grenoble Alpes, CNRS, TIMC UMR 5525, Grenoble, France

[2]Université de Lyon, INSA-Lyon, Université Claude Bernard Lyon 1, UJM-Saint Etienne, CNRS, INSERM, CREATIS UMR 5220, U1206, F-69373, Lyon, France

**Abstract** In the context of two-dimensional (2D) image reconstruction from truncated projections, we describe five implementations, each based on a different formula derived from the virtual fan-beam (VFB) method. Three formulae are already known: (a) and (b) perform the back-projection in the parallel-beam geometry and (c) performs the back-projection in the virtual fan-beam geometry. Two new formulae, (d) and (e), perform the back-projection in the acquisition geometry. Our simulation results using the Shepp-Logan phantom suggest that the best accuracy is obtained from the implementations of formulae (b), (d) and (e).

## 1 Introduction

The reconstruction of a region-of-interest (ROI) in two-dimensional (2D) tomography from truncated data is possible with both iterative and analytical methods. Iterative methods are more flexible but analytical methods, based on exact inversion formulae, are significantly faster. Many iterative methods have been used to solve this problem, e.g. maximum-likelihood expectation-maximization (ML-EM) [1, 2]. The analytical solutions follow two different approaches: the virtual fan-beam (VFB) method, which is the focus of this paper, and the differentiated back-projection (DBP) method. The VFB method was mainly introduced in [3, 4]. The DBP method was developed simultaneously by several groups [5–7]. Both methods are relevant as they can each solve particular ROI reconstruction problems that the other cannot [5].

The principle of the VFB method is that, since many exact analytical reconstruction formulae require *non-truncated* projections, one identifies virtual source points for which the corresponding virtual projections are non-truncated. The real truncated projections are then rebinned into these virtual non-truncated projections. To do so, we define the field-of-view (FOV), which is the region viewed by every source position. Considering a full scan acquisition trajectory, it follows that every line passing through the FOV is measured so any FOV point which is also outside the convex hull of the object is a valid virtual source point. We then use super-short-scan formulae, which enable exact reconstruction inside the convex hull of the super-short-scan trajectory, in case of non-truncated projections.

In this work, we use super-short-scan formulae from [8] with the VFB method, but many other super-short-scan formulae have been proposed for 2D ROI reconstruction, either for a circular trajectory [9–11] or a free-form trajectory [12].

Previous contributions in the VFB area were applied to truncated parallel-beam projections. The filtered sinogram was either computed after explicitly rebinning to the virtual fan-beam geometry [4], or with a shift-variant "convolution" [3]. In both cases, parallel back-projection was used.

In this work, we apply these two approaches to truncated fan-beam projections along a circular source trajectory instead of a parallel-beam sinogram. For the first approach, we rebin the acquired fan-beam projections into virtual non-truncated fan-beam projections. These virtual projections are suitably filtered and then rebinned into filtered projections corresponding either to a parallel-beam geometry (formulae (a) and (b) below) or to the fan-beam acquisition geometry (formula (d)), also called the real geometry (as opposed to the virtual geometry). The back-projection is computed in the corresponding geometry. We note that the reconstruction is also possible with a back-projection directly in the virtual fan-beam geometry (formula (c)), thus avoiding the second rebinning step. These four formulae assume that both virtual and real geometries have a circular source trajectory. Following the same approach as [3] we also derive in proposition 2 a direct formula of the filtered projections in the geometry of the acquired truncated projections (formula (e)).

## 2 Theory

### 2.1 Notation

Let $f$ denote the 2D object density to be reconstructed. The parallel-beam projections of $f$ are defined by $p(\phi, s) = \int_{\mathbb{R}} f(l\vec{\theta}_\phi + s\vec{\eta}_\phi) \, dl$ where $\vec{\theta}_\phi = (\cos\phi, \sin\phi)$ and $\vec{\eta}_\phi = (-\sin\phi, \cos\phi)$. Let $h_F(s) = \int_{\mathbb{R}} |\sigma| e^{2i\pi\sigma s} \, d\sigma$ denote the ramp filter. The parallel-beam ramp filtered projections are defined by

$$p_F(\phi, s) = \int_{\mathbb{R}} h_F(s - s') p(\phi, s') \, ds'. \tag{1}$$

Let $h_H(s) = \int_{\mathbb{R}} -i \, \text{sign}(\sigma) e^{2i\pi\sigma s} \, d\sigma$ denote the Hilbert filter. The parallel-beam Hilbert filtered projections are defined by

$$p_H(\phi, s) = \int_{\mathbb{R}} h_H(s - s') p(\phi, s') \, ds'. \tag{2}$$

The fan-beam projections of $f$ for a circular source trajectory of radius $R$ are defined by $g^R(\lambda, \gamma) = \int_0^\infty f(R\vec{\theta}_\lambda + t\vec{\theta}_{\lambda+\pi+\gamma}) \, dt$ where $\lambda \in \Lambda \subset [0, 2\pi)$ and $R\vec{\theta}_\Lambda$ is the set of vertices (fan-beam source locations) of the trajectory. The fan-beam Hilbert filtered projections are defined by

$$g_H^R(\lambda, \gamma) = \int_{-\pi}^{\pi} h_H(\sin(\gamma - \gamma')) \, g^R(\lambda, \gamma') \, \mathrm{d}\gamma'. \quad (3)$$

The fan-beam 'differentiated and Hilbert filtered' projections are defined by

$$g_F^R(\lambda, \gamma) = \int_{-\pi}^{\pi} h_H(\sin(\gamma - \gamma')) \left( \frac{\partial}{\partial \lambda} - \frac{\partial}{\partial \gamma'} \right) g^R(\lambda, \gamma') \, \mathrm{d}\gamma'. \quad (4)$$

## 2.2 Use of the VFB method

In our formulae, the key step is the computation of $g_H^R(\lambda, \gamma)$ or $g_F^R(\lambda, \gamma)$. Equations (3) and (4) require non-truncated fan-beam projections. Therefore, we determine a virtual source trajectory for which the associated fan-beam projections are non-truncated, we rebin the initial data into this virtual geometry and we compute $g_H^R$ or $g_F^R$ for this trajectory.

The fan-beam ray parameters $(\lambda_1, \gamma_1)$ and $(\lambda_2, \gamma_2)$ (for circular source trajectories of radius $R_1$ and $R_2$, respectively) are linked by $G_1 : \{R_1 \sin \gamma_1 = R_2 \sin \gamma_2$ and $\lambda_1 + \gamma_1 = \lambda_2 + \gamma_2\}$ (see figure 1). To apply this, we assume that the virtual source trajectory is an arc of circle of radius $R_1$. We use $g^{R_1}(\lambda_1, \gamma_1) = g^{R_2}(\lambda_2, \gamma_2)$ where $G_1$ is satisfied to rebin the truncated projections acquired with a source trajectory of radius $R_2$ (with $R_2 \geq R_1$) into the virtual source trajectory.

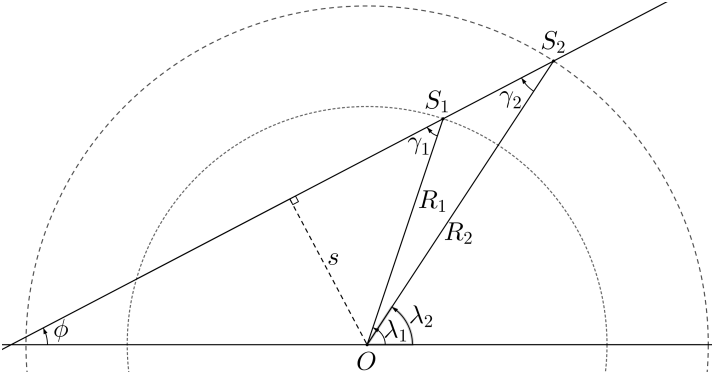

**Figure 1:** A ray of parameters $(\phi, s)$ in parallel geometry and the ray parameters $(\lambda_i, \gamma_i)$ at the point $S_i$ for a circular trajectory of radius $R_i$ with $i \in 1, 2$. The angles $(\lambda_i, \gamma_i)$, measured counterclockwise, verify $s = -R_i \sin \gamma_i$ and $\phi = \lambda_i + \gamma_i$.

## 2.3 Formulae with parallel-beam back-projection

The parallel-beam parameters $(\phi, s)$ and the fan-beam parameters for a virtual source trajectory of radius $R_1$ $(\lambda_1, \gamma_1)$ of a ray are linked by either $P_1 : \{s = -R \sin \gamma$ and $\phi = \lambda + \gamma\}$ or $P_2 : \{s = R \sin \gamma$ and $\phi = \lambda + \gamma + \pi\}$..

A first reconstruction **formula (a)**, derived from [8, eq. (8), (10)], is

$$f(\vec{x}) = \frac{1}{4\pi} \int_0^{2\pi} \left[ \frac{\partial}{\partial s} p_H(\phi, s) \right] \Big|_{s = \vec{x} \cdot \vec{\eta}_\phi} \mathrm{d}\phi \quad (5)$$

where the available values of $p_H$ are obtained through the virtual filtered projections $g_H^{R_1}$ using:

$$\begin{cases} P_1 \implies p_H(\phi, s) = -g_H^{R_1}(\lambda_1, \gamma_1), \\ P_2 \implies p_H(\phi, s) = g_H^{R_1}(\lambda_1, \gamma_1). \end{cases} \quad (6)$$

A second reconstruction **formula (b)**, derived from [8, eq. (2), (14)], is

$$f(\vec{x}) = \frac{1}{2} \int_0^{2\pi} \left[ p_F(\phi, s) \right] \Big|_{s = \vec{x} \cdot \vec{\eta}_\phi} \mathrm{d}\phi \quad (7)$$

where the available values of $p_F$ are obtained through the virtual filtered projections $g_F^{R_1}$ using:

$$P_1 \text{ or } P2 \implies p_F(\phi, s) = -\frac{1}{2\pi R_1 \cos(\gamma_1)} g_F^{R_1}(\lambda_1, \gamma_1). \quad (8)$$

In both formulae, we replace the half rotation over $[0, \pi)$ by a full rotation over $[0, 2\pi)$ since it reduces numerical artefacts.

## 2.4 Formulae with fan-beam back-projection

A third reconstruction **formula (c)**, derived from [8, eq. (33), (34)], is

$$f(\vec{x}) = -\frac{1}{2\pi} \int_{\Lambda_{R_1}} \frac{1}{||R_1 \vec{\theta}_{\lambda_1} - \vec{x}||} w^{R_1}(\lambda_1, \gamma_{\vec{x}, \lambda_1}) g_F^{R_1}(\lambda_1, \gamma_{\vec{x}, \lambda_1}) \, \mathrm{d}\lambda_1 \quad (9)$$

where $\Lambda_{R_1} \subset [0, 2\pi)$ is the angle extent of the virtual source trajectory, $\gamma_{\vec{x}, \lambda_1} = \arctan(-\vec{x} \cdot \vec{\eta}_{\lambda_1}/(R_1 - \vec{x} \cdot \vec{\theta}_{\lambda_1}))$, $w^R(\lambda, \gamma) = c^R(\lambda)/(c^R(\lambda) + c^R(\lambda + \pi + 2\gamma))$ and $c^R$ is a smooth $2\pi$-periodic function such that $\lambda \notin \Lambda_R \implies c^R(\lambda) = 0$ (see [8, eq. (46)] for more details).

In formula (c), the back-projection is done along the virtual source trajectory. To apply the same formula along the real source trajectory, we need to determine $g_F^{R_2}$. Since the real projections are truncated, we cannot compute it directly with (4). However, one can show the following:

**Proposition 1.**

$$G_1 \implies \frac{g_F^{R_1}(\lambda_1, \gamma_1)}{R_1 \cos(\gamma_1)} = \frac{g_F^{R_2}(\lambda_2, \gamma_2)}{R_2 \cos(\gamma_2)} \quad (10)$$

This proposition yields the fourth reconstruction **formula (d)**:

$$f(\vec{x}) = -\frac{1}{2\pi} \int_0^{2\pi} \frac{1}{||R_2 \vec{\theta}_{\lambda_2} - \vec{x}||} w^{R_2}(\lambda_2, \gamma_{\vec{x}, \lambda_2}) g_F^{R_2}(\lambda_2, \gamma_{\vec{x}, \lambda_2}) \, \mathrm{d}\lambda_2 \quad (11)$$

where $g_F^{R_2}(\lambda_2, \gamma_2) = R_2 \cos(\gamma_2) g_F^{R_1}(\lambda_1, \gamma_1)/(R_1 \cos(\gamma_1))$, $\gamma_{\vec{x}, \lambda_2} = \arctan(-\vec{x} \cdot \vec{\eta}_{\lambda_2}/(R_2 - \vec{x} \cdot \vec{\theta}_{\lambda_2}))$ and $w^{R_1}(\lambda_1, \gamma_1) = w^{R_2}(\lambda_2, \gamma_2)$ (to preserve the redundancy weight $w$ associated to each ray).

Proposition 1 enables computation of the values of $g_F^{R_2}$ by rebinning the real truncated projections $g^{R_2}$ into virtual non-truncated projections $g^{R_1}$, computing the virtual filtered projections $g_F^{R_1}$, and then rebinning back these virtual filtered projections into the real filtered projections $g_F^{R_2}$. However, it is also possible to directly compute $g_F^{R_2}$ from its truncated projections:

**Proposition 2.**

$$g_F^{R_2}(\lambda_2, \gamma_2) = \frac{R_2 \cos(\gamma_2)}{\sqrt{R_1^2 - R_2^2 \sin^2(\gamma_2)}} \int_{-\gamma_m}^{\gamma_m} h_H(\sin(\Delta_{R_1}^{R_2}(\gamma_2, \gamma_2')))$$

$$(\partial_1 - \partial_2) g^{R_2}(\lambda_2 + \gamma_2 - \gamma_2' - \Delta_{R_1}^{R_2}(\gamma_2, \gamma_2'), \gamma_2') \, d\gamma_2' \tag{12}$$

provided that $R_1 \vec{\theta}_{\lambda_2 + \gamma_2 - \arcsin\left((R_2/R_1)\sin\gamma_2\right)}$ (which corresponds to $R_1 \vec{\theta}_{\lambda_1}$ when $G_1$ is satisfied) is outside the convex hull of the object and that the tangent to the circle of radius $R_1$ at this point does not intersect the object, where $R_2 \geq R_1$, $\gamma_m = \arcsin(R_1/R_2)$, $|\gamma_2| \leq \gamma_m$, $\Delta_{R_1}^{R_2}(\gamma_2, \gamma_2') = \arcsin\left((R_2/R_1)\sin\gamma_2\right) - \arcsin\left((R_2/R_1)\sin\gamma_2'\right)$ and $\partial_i$ is the derivative according to the $i$-th variable.

So this yields the fifth reconstruction **formula (e)**, using the same back-projection as (d) but with $g_F^{R_2}$ computed by proposition 2.

## 2.5 Simulations

For numerical experiments, we used the geometry defined in figure 2. All values are in arbitrary units (a.u.). The object to reconstruct is the classical 2D Shepp-Logan phantom. The center of the phantom was at $(0, 0.15)$ with $(0,0)$ the center of rotation. The reconstructed images were computed on a square grid of dimensions $[-1,1]^2$ with $\Delta x = 1/200$ (i.e. a square grid of 401 x 401 pixels). The real source trajectory along $[0, 2\pi)$ was sampled with an angular spacing $\Delta\alpha_2$ of 0.5 degree (i.e. with 720 vertices) and each fan-beam truncated projection was sampled with an angular spacing $\Delta\gamma_2 = \Delta x/R_2$ for $\gamma_2$ in $[-\gamma_m, \gamma_m]$ (i.e. with 325 rays). Similarly, the virtual source trajectory along $[0°, 36°)\cup[144°, 360°)$ was sampled with an angular spacing $\Delta\alpha_1$ of 0.5 degree (i.e. with 505 vertices) and each fan-beam projection was sampled with an angular spacing $\Delta\gamma_1 = \Delta x/R_1$ ($\Delta\gamma_2$ and $\Delta\gamma_1$ were chosen so that the spacing between two rays at the center of rotation is equal to $\Delta x$ for both real and virtual rays) for $\gamma_1$ in $[-\pi, \pi]$ (i.e. with 1005 rays). For formulae (a) and (b), the parallel projections along $[0, 2\pi)$ were sampled with an angular spacing $\Delta\phi$ of 0.5 degree (i.e. with 720 projections) and each parallel projection was sampled over $[-1, 1]$ with $\Delta s = \Delta x$ (so each projection consists of 401 parallel rays).

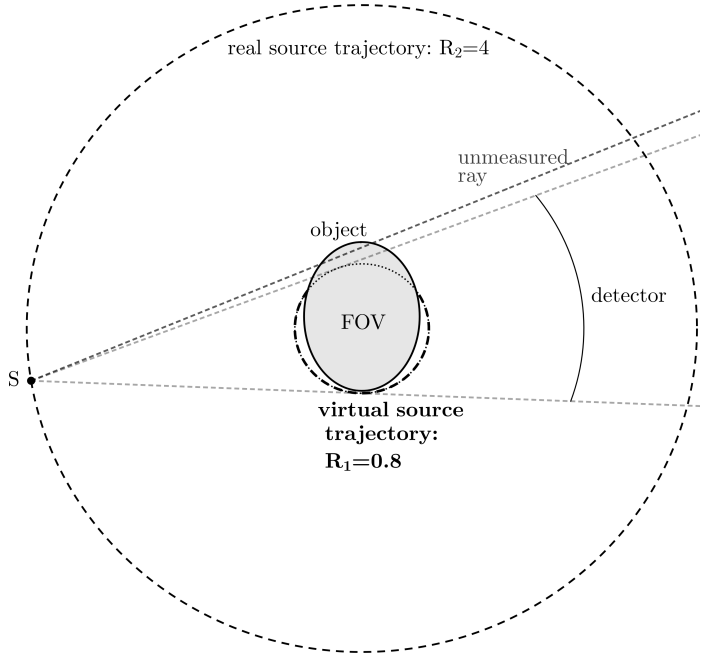

**Figure 2:** The real source trajectory is a circle of radius $R_2 = 4$. The detector measures rays from the source $S$ with an equal angular spacing. The FOV is a disk of radius $R_1 = 0.8$. The virtual source trajectory (in bold dashed line) is the arc at the FOV border outside the object.

## 3 Results

### 3.1 Simulations with noiseless projections

Figure 3 shows the reconstructed images and the corresponding profiles for the five formulae using the same noiseless sinogram. All reconstructions were satisfactory with minor differences. The reconstructed image obtained with formula (a) showed some artefacts (ripples close to the external white envelope) which were avoided with formula (b). The back-projection along the virtual source trajectory (formula (c)) produced more artefacts than the back-projection along the real source trajectory (formula (d)). The reason is probably due to the virtual trajectory being closer to the object than the real trajectory, as the lines contributing to the back-projection must all go through the vertices along the virtual trajectory, so these lines are irregularly sampled in case of a point close to the virtual trajectory (this caused the artefacts inside the object at the middle bottom and top left and right). Concerning the arc of circle of white artefacts at the vicinity of the virtual trajectory, it seems to be caused by the factor $1/||R\vec{\theta}_\lambda - \vec{x}||$ applied to this irregular sampling. We note that (b) and (d) seemed to have a similar accuracy. Finally the reconstruction obtained with formula (e) had a smaller exact reconstruction area because the filtering step in proposition 2 was only accurate for rays which cross the virtual source trajectory at a point where its tangent did not intersect the object. Moreover, the discretization of $h_H(\sin(\Delta_{R_1}^{R_2}(\gamma_2, \gamma_2')))$ required finer sampling so, for formula (e), we chose $\Delta\gamma_2' = \Delta\gamma_2/3$ (i.e. with 969 rays) instead of $\Delta\gamma_2' = \Delta\gamma_2$. The computation time was similar in all formulae except formula (e) for which it was much longer (about 10 times longer for the whole computation or 40 times longer for the part not involving the back-projection) due to the shift-variant "convolution" (which is not a true convolution so we cannot use the Fourier convolution theorem) and the finer sampling of $\gamma_2'$.

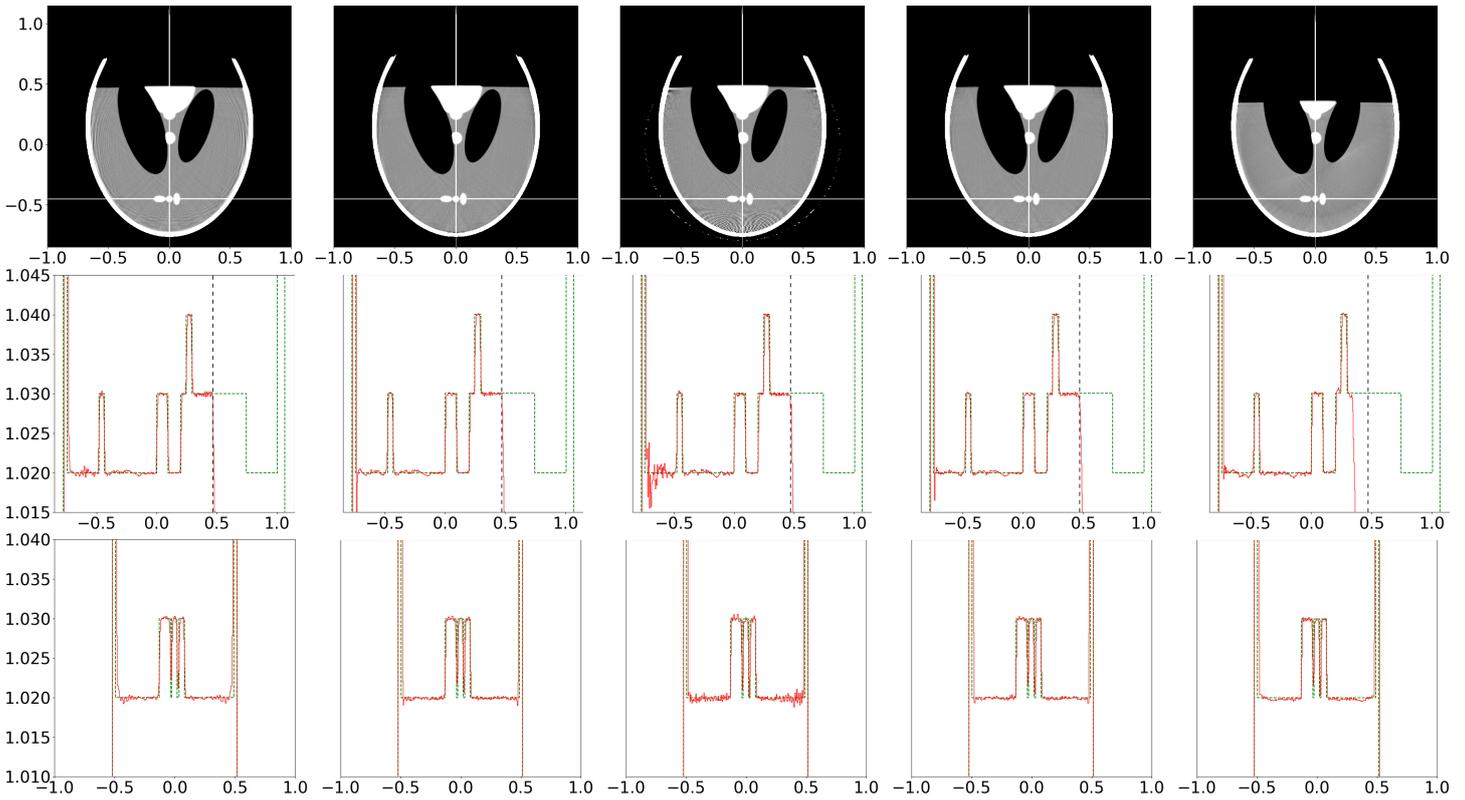

**Figure 3:** Top row: image reconstructed with, from left to right, formulae (a), (b) , (c), (d) and (e) when the data are measured with a source trajectory of radius $R = 4$. The plotting scale is [1.015 (black), 1.025 (white)]. The vertical and horizontal white lines correspond to the profiles plotted respectively in the middle row with scale [1.015, 1.045] and in the bottom row with scale [1.01, 1.04]. The ideal profiles are plotted in green dashed line and the real ones in red. The vertical black dashed line defines the boundary of the possible reconstruction area.

## 3.2 Simulations with noisy projections and variance study

Figure 4 shows the pixel-wise variance computed for $n = 100$ realizations in the case of Poisson noise simulated before taking the logarithm of the projections to obtain line integrals. Following [13], the Shepp-Logan densities were weighted by $1.879 \ a.u^{-1}$ , i.e., the linear attenuation coefficient of water at 75 keV with $1 \ a.u. = 100 \ mm$. The number of photons received per detector pixel without object in the beam was constant for all pixels and equal to $10^7$. Striking differences were observed in the spatial maps of the variance between the different formulae, with formula (c) the least homogeneous.

## 4 Conclusion

In this work, we compared five different implementations for ROI reconstruction from truncated fan-beam projections measured along a circular source trajectory. The first three formulae (a), (b) and (c) were already established in [8] and formulae (d) and (e) are, to our knowledge, new formulae. All reconstructions gave satisfactory results. Image quality was slightly better with formulae (b), (d) and (e) but the computation time was longer for method (e). Method (c) presented the worst satisfactory variance results, but we have not yet performed a control study of image resolution in the reconstructions.

## Acknowledgment

This work was supported by grant ANR-17-CE19-0006 (ROIdoré project) from the Agence Nationale de la Recherche (France).

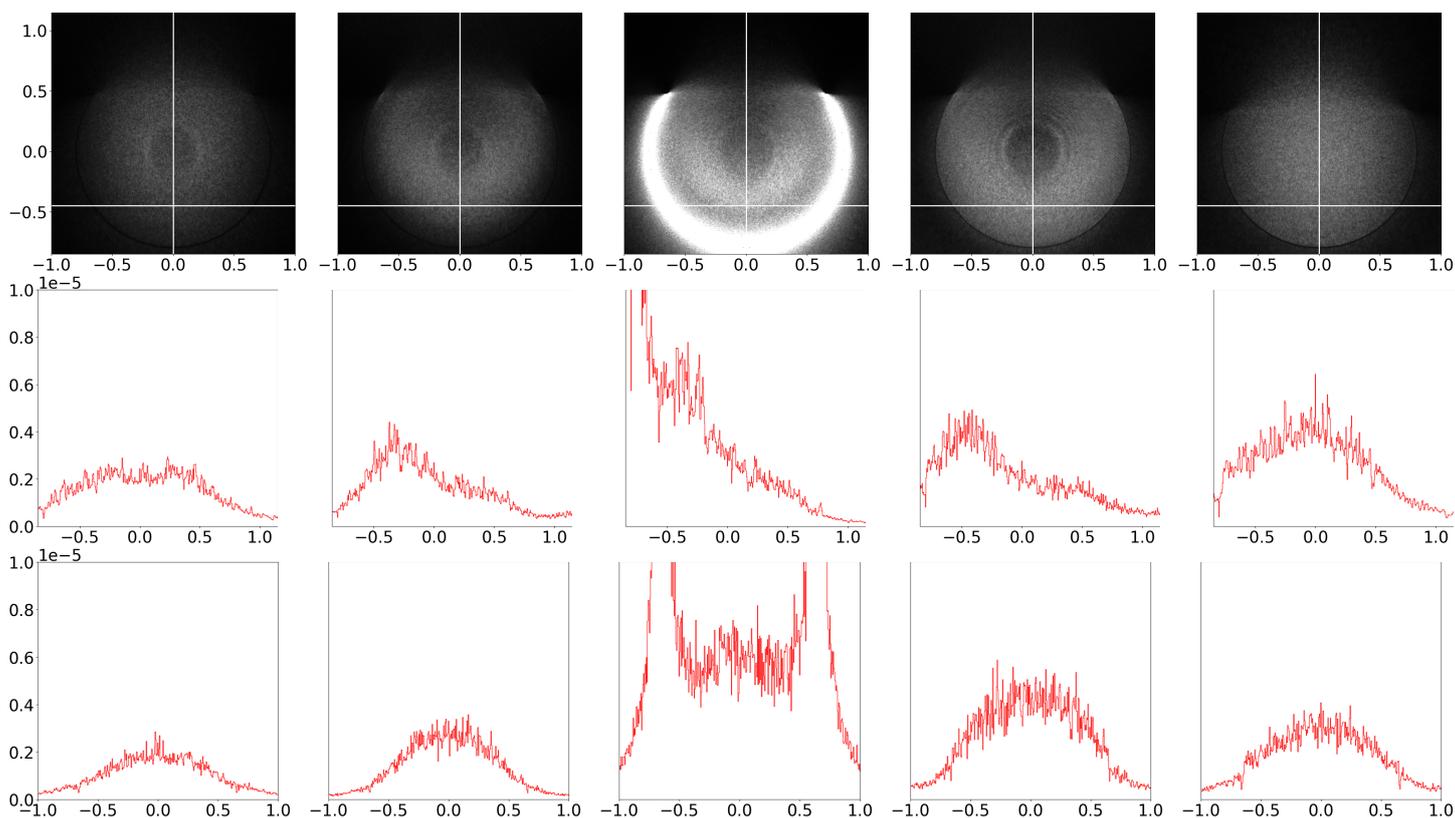

**Figure 4:** Top row: pixel-wise variance of the images reconstructed with, from left to right, formulae (a), (b) , (c), (d) and (e) when the data are measured with a source trajectory of radius $R = 4$. The plotting scale is [0 (black), $10^{-5}$ (white)]. The vertical and horizontal white lines correspond to the profiles plotted respectively in the middle row and in the bottom row with scale [0, $10^{-5}$].

# CB reconstruction for the 3-sin trajectory with transverse truncation

Nicolas Gindrier[1], Laurent Desbat[1], and Rolf Clackdoyle[1]

[1]TIMC-IMAG laboratory, CNRS UMR 5525 and Univ. Grenoble Alpes 38000 Grenoble, France

**Abstract** In cone-beam tomography Differentiated BackProjection method (DBP) is a suitable approach for image reconstruction from truncated projections. However, the reconstruction of a point with this method is possible only if the point lies on a chord connecting two source positions of the x-ray source trajectory. Using an approach initially proposed for the reverse helix with axial truncation, we present a configuration and its associated (theoretical) reconstruction method to deal with points which do not lie on any chord of the 3-sin trajectory (sine on a cylinder of period $2\pi/3$) and with transversely truncated projections.

## 1 Introduction

Cone beam (CB) geometry is an important part of the computed tomography. A main result of CB tomography comes from Tuy [1] and Finch [2]. They prove that for an X-ray source trajectory which is bounded and connected, an exact reconstruction is only possible within the convex hull of this trajectory. Moreover, in this case, the Tuy condition says exact reconstruction is possible if there is no data truncation. FOV is defined as follows in our article: the measured rays for each projection are exactly those that intersect the FOV. In this article, the FOV will be a $e_z$-axis cylinder and we deal with *transverse truncation*, appearing when the detector is not large enough (the FOV and the object intersect at their sides). To manage this kind of reconstruction, the Differentiated BackProjection method (DBP) [3] is suitable, for example in [4] for the helix trajectory. Yet this method requires that each point of the object $\Omega_O$ to be reconstructed is intersected by a chord (a line segment linking two source points of the X-ray source trajectory).

However, many trajectories have points within their convex hull which are not intersected by a chord. For example, this is the case for the reverse helix [5] and for the 3-sin trajectory, which is a sinusoid on a cylinder, defined by:

$$S \overset{\text{def}}{=} \{(R\cos\lambda, R\sin\lambda, H\cos(3\lambda)), \lambda \in [0, 2\pi)\} \quad (1)$$

with $R > 0$, $H > 0$, see Fig. 1, *left*. Nevertheless, [6] shows by numerical methods that exact reconstruction with transverse truncation appears to be possible even in some regions which are not intersected by chords. Moreover, $S$ has a convex hull bigger than that of the saddle trajectory (a 2-sin trajectory more extensively studied in the literature [7]), which is why we find it useful to study. We write $\Omega_S$ for the convex hull (Fig. 1, *right*) and $C_S (\subset \Omega_S)$ for the union of all chords $c$ for the 3-sin trajectory $S$, and $N_S \overset{\text{def}}{=} \Omega_S \setminus C_S$.

The article [5], treating the reverse helix case, explains how to perform reconstructions dealing with some points in the convex hull which are not lying on a chord and axial truncation

(the article [8], published at the same time, works on the same point and proposes a similar approach, except for the last step). Inspired by the method of [5], the goal of this article is to describe and to test one configuration for the trajectory $S$, where it is possible to reconstruct $\Omega_{\text{in}} \overset{\text{def}}{=} \text{FOV} \cap \Omega_O \cap N_S$ despite transverse truncation. To do this, the next section analyses and describes the regions $C_S$ and $N_S$. Section 3 describes the reconstruction principles and a computer simulation study is presented in section 4. We end with a short discussion and conclusion.

## 2 The 3-sin trajectory

### 2.1 Union of chords

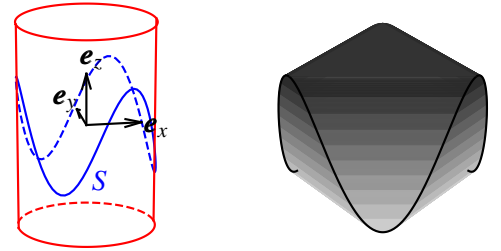

**Figure 1:** *Left:* the 3-sin trajectory $S$, which is a sinusoid on a cylinder. *Right:* the 3-sin trajectory with its convex hull $\Omega_S$. (The shades of grey vary according to the height.)

To build the union of chords $C_S$ of the 3-sin trajectory, it is useful to consider the intersection between the trajectory and a horizontal plane $\Pi_{\tilde{z}}$ with equation $z = \tilde{z}$, where $-H \le \tilde{z} \le H$, illustrated in Fig. 2. The angles in this figure are:

$$
\begin{aligned}
\lambda_A &= -\frac{1}{3}\arccos(\tilde{z}/H) & \lambda_B &= \frac{1}{3}\arccos(\tilde{z}/H) \\
\lambda_C &= \lambda_A + \frac{2\pi}{3} & \lambda_D &= \lambda_B + \frac{2\pi}{3} \\
\lambda_E &= \lambda_A + \frac{4\pi}{3} & \lambda_F &= \lambda_B + \frac{4\pi}{3}
\end{aligned}
\quad (2)
$$

We let $S_A$ denote $S_{\lambda_A}$. The aim is to show that each point of the hexagon of Fig. 2 (*right*) is intersected by a chord, except points in the central triangle (defined by the intersection of the line segments $[\boldsymbol{S}_C, \boldsymbol{S}_F]$, $[\boldsymbol{S}_D, \boldsymbol{S}_A]$ and $[\boldsymbol{S}_E, \boldsymbol{S}_B]$). Considering Fig. 3 and with equations (1) and (2), we see chords $c_1(\tilde{z})$ (linking $\boldsymbol{S}_D$ to $\boldsymbol{S}_E$ for $\tilde{z} \in [0, H]$) and $\tilde{c}_1(\tilde{z})$ (linking $\boldsymbol{S}_F$ to $\boldsymbol{S}_C$ for $\tilde{z} \in [0, H]$) move (and meet when $\tilde{z} = H$), continuously approaching with respect to increasing $\tilde{z}$, for $\tilde{z} \in [0, H]$. The chord $c_2(\tilde{z})$, where $\tilde{z} \in [-H, 0]$, moves continuously with respect to decreasing $\tilde{z}$ until $c_2(-H) = \boldsymbol{S}_\pi$. With the union of all chords $c_1(\tilde{z})$ and $\bar{c}_1(\tilde{z})$ for $\tilde{z} \in [0, H]$ and all chords $c_2(\tilde{z})$ for $\tilde{z} \in [-H, 0]$, a surface can be created (see Fig. 4, *right*). Let's note that the 3-sin trajectory is invariant through a rotation of $2\pi/3$ around the $\boldsymbol{e}_z$-axis and is invariant through a rotation of $\pi/3$ (around the same axis) then a symmetry with respect to the plane $(\boldsymbol{e}_x, \boldsymbol{e}_y)$. With these invariances it is possible to create six similar surfaces as described previously (see Fig. 4, *left*). By adapting the proof of [7] (appendix A.2), it is possible to prove that each point between these two surfaces lies on a chord, i.e. if it exists two points $\boldsymbol{x}' = (x, y, z')$ and $\boldsymbol{x}'' = (x, y, z'')$, $z' < z''$, each intersected by a chord, then each point $\boldsymbol{x} = (x, y, z)$, with $z' < z < z''$, is also intersected by a chord. To finish the construction of $C_S$ we must match $\boldsymbol{x}'$ and $\boldsymbol{x}''$ points, being at the union of the six surfaces, to ensure that each point within the volume of this union is intersected by a chord.

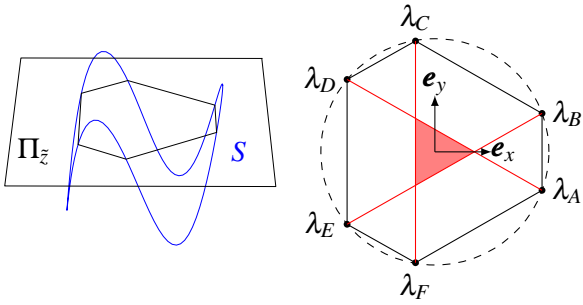

**Figure 2:** Intersection between a horizontal plane $\Pi_{\tilde{z}}$ and $S$ (with the intersection points linked). *Right:* The dashed circle is the projection of $S$ on $\Pi_{\tilde{z}}$. The red and black lines are chords of $S$ contained in this plane $\Pi_{\tilde{z}}$. This section shows that all points contained in the black polygon, but outside the red triangle, are intersected by a chord.

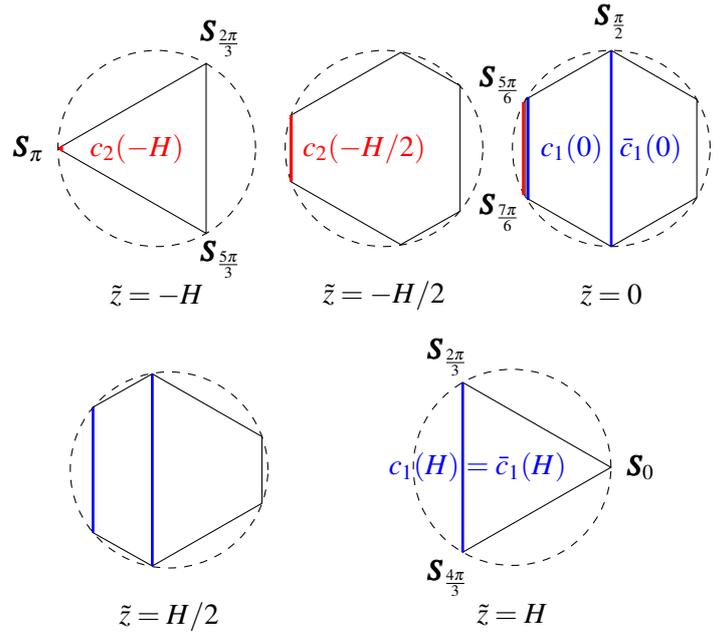

**Figure 3:** Different chords contained in some horizontal planes. The blue chords, defined for $\tilde{z} \geq 0$, are the chords $c_1(\tilde{z})$ (linking $\boldsymbol{S}_E$ to $\boldsymbol{S}_D$ for $\tilde{z} \geq 0$) and $\bar{c}_1(\tilde{z})$ (linking $\boldsymbol{S}_F$ to $\boldsymbol{S}_C$) and are parallel (and even merged for $\tilde{z} = H$.). The red chord is $c_2(\tilde{z})$ (linking $\boldsymbol{S}_E$ to $\boldsymbol{S}_D$ for $\tilde{z} \leq 0$). The union of these chords, for all $\tilde{z} \in [-H, H]$ is drawn Fig. 4.

We define the blue surfaces as the surfaces generated by the chords $c_1$ ($[\boldsymbol{S}_D\boldsymbol{S}_E]$, $[\boldsymbol{S}_C\boldsymbol{S}_B]$ and $[\boldsymbol{S}_F\boldsymbol{S}_A]$ for $\tilde{z} \in [0, H]$, $[\boldsymbol{S}_E\boldsymbol{S}_F]$, $[\boldsymbol{S}_B\boldsymbol{S}_A]$ and $[\boldsymbol{S}_C\boldsymbol{S}_D]$ for $\tilde{z} \in [-H, 0]$) and $\bar{c}_1$ ($[\boldsymbol{S}_C\boldsymbol{S}_F]$, $[\boldsymbol{S}_E\boldsymbol{S}_B]$ and $[\boldsymbol{S}_D\boldsymbol{S}_A]$). The red surfaces are defined by the surfaces generated by the chords $c_2$ ($[\boldsymbol{S}_D\boldsymbol{S}_E]$, $[\boldsymbol{S}_C\boldsymbol{S}_B]$ and $[\boldsymbol{S}_F\boldsymbol{S}_A]$ for $\tilde{z} \in [-H, 0]$, $[\boldsymbol{S}_E\boldsymbol{S}_F]$, $[\boldsymbol{S}_B\boldsymbol{S}_A]$ and $[\boldsymbol{S}_C\boldsymbol{S}_D]$ for $\tilde{z} \in [0, H]$)(see Fig. 4). For this section each projection will be an orthogonal projection onto the plane $(\boldsymbol{e}_x, \boldsymbol{e}_y)$ (which is the plane $\Pi_0$). The intersection of the projections of the blue surfaces covers the regular hexagon defined by the convex hull of the intersection between $\Pi_0$ and $S$ (see Fig. 3 and 5). Then a point "between" two blue surfaces (one above and one below $\Pi_0$) and whose projection is in the hexagon is intersected by a chord. A point whose projection is in the region between the hexagon defined above and the circle of radius $R$ is intersected by a chord if it is "between" a red surface and a blue surface on the same side of $\Pi_0$. The set of points intersected by a chord is $C_S$, illustrated in Fig. 7, *left*.

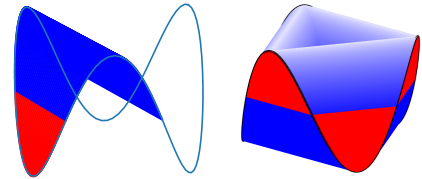

**Figure 4:** *Left:* One surface created for chords described Fig. 3. *Right:* The union of six surfaces from the left figure, using the invariances of the 3-sin trajectory.

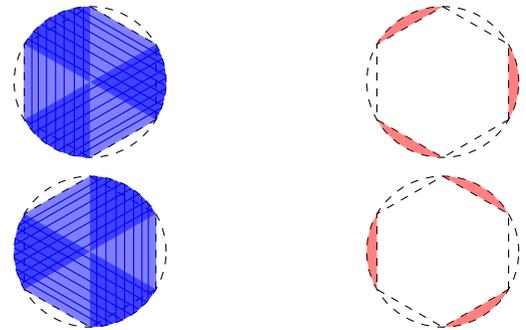

**Figure 5:** Orthogonal projections on the plane $(\boldsymbol{e}_x, \boldsymbol{e}_y)$ of the red and blue surfaces. *Top:* Surfaces defined for $\tilde{z} \geq 0$. *Bottom:* Surfaces defined for $\tilde{z} < 0$. The dashed hexagon links the points of the intersection between $S$ and $\Pi_0$.

## 2.2 Region without chords

We have defined $C_S$, but to be complete now we must be sure that no chords intersect a point in the central triangle or equivalently we construct region $N_S$. (see Fig. 2). Considering a horizontal plane $\Pi_{\bar{z}}$, $\bar{z} > 0$, we cut $S$ into several pieces $s$ (short pieces, above $\Pi_{\bar{z}}$) and $l$ (long pieces, under $\Pi_{\bar{z}}$), see Fig. 6 (for example $s_1$ is the piece of the trajectory linking $\boldsymbol{S}_A$ to $\boldsymbol{S}_B$). We study chords linking these pieces. There are four cases: chords linking $s_1$ to $l_1$ (directly opposite), $s_1$ to $l_2$ or $l_3$ ("$l$" to "$s$" but not directly opposite), $s_1$ to $s_2$ or $s_3$ ("$s$" to "$s$") and $l_1$ to $l_2$ or $l_3$ ("$l$" to "$l$"). It is clear that chords linking $s_1$ to $l_2$ do not intersect the central triangle. Chords linking $s_1$ to $s_2$ (resp. $l_1$ to $l_2$) are above (resp. below) $\Pi_{\bar{z}}$. The last case (chords linking $s_1$ to $l_1$), is more complicated, and an analytic approach would be tedious. We show some numerically calculated intersections between these chords and $\Pi_{\bar{z}}$ in Fig. 6 that suggest that all such intersections occur outside the triangle.

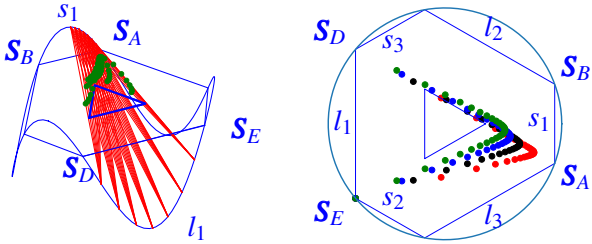

**Figure 6:** Chords for some values of $\lambda_1 \in [\lambda_A, \frac{\lambda_A + \lambda_B}{2}]$ (half of the $s_1$ piece) and $\lambda_2 \in [\lambda_D, \lambda_E]$ ($l_1$ piece) and intersections for the plane $\Pi_{\bar{z}}$, $\bar{z} = H/2$. *Right:* Intersections for four values of $\lambda_1$.

From equations (1) and (2) we are able to draw the central triangles for each $\bar{z} \in [-H, H]$ and build an illustration of $N_S$, as shown in Fig. 7, *right*.

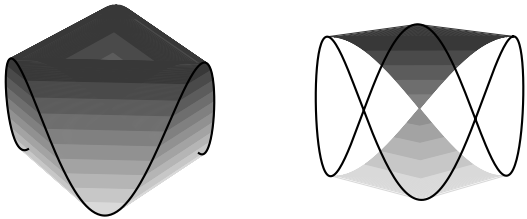

**Figure 7:** *Left:* The union of chords of $S$: $C_S$. *Right:* The set of points of $\Omega_S$ which are not intersected by a chord: $N_S$. (The shades of grey vary according to the height).

# 3 Reconstruction

## 3.1 General method

The regions $\Omega_O$ and FOV are assumed known, and $C_S$ and $N_S$ have been previously calculated. We can summarize the reconstruction approach in four steps:

1. Reconstruction of $\Omega_{DBP} \subseteq \text{FOV} \cap \Omega_O \cap C_S$ with the DBP

method, where $\Omega_{DBP}$ is the region where DBP is possible

2. Reprojection of reconstructed points (cone-beam projections of the new object reconstructed in the region $\Omega_{DBP}$)

3. Subtraction of reprojections from the original cone-beam data, to present a new reconstruction problem with a smaller object, defined on the region $\Omega_O \setminus \Omega_{DBP} = \Omega_{in} \cup \Omega_{out}$, with $\Omega_{out} \stackrel{\text{def}}{=} \Omega_O \setminus (\Omega_{DBP} \cup \Omega_{in})$ (the regions $\Omega_{DBP}$, $\Omega_{in}$ and $\Omega_{out}$ are mutually disjoint)

4. For reconstruction to be possible, the new configuration must be a problem without truncation satisfying Tuy condition: reconstruction of $\Omega_{in}$ by any of the various methods for cone-beam reconstruction from non-truncated projections (e.g. [1], [9], [10]...).

However, in order to apply this method, two points must be taken into account. Firstly, the DBP method does not generally allow reconstruction in the whole region FOV $\cap$ $\Omega_O \cap C_S$ because although the necessary Hilbert transforms can be formed along these chords, there are further geometric conditions required for Hilbert inversion (more precisely, we consider methods that guarantee the existence, stability and uniqueness of the inversion, so called 1-sided and 2-sided inverse Hilbert transforms [11]). Thus the $\Omega_{DBP}$ region must be carefully identified. Secondly, there must be no *contaminated lines*, which are defined as measured lines of $\Omega_{in}$ intersecting $\Omega_{out}$. The region $\Omega_{out}$ could then be removed from the reconstruction problem. Note that it would not be possible to reconstruct the part of the $\Omega_{out}$ region being outside the FOV. If there are contaminated lines this approach to reconstruction of $\Omega_{in}$ fails.

## 3.2 Configuration proposed

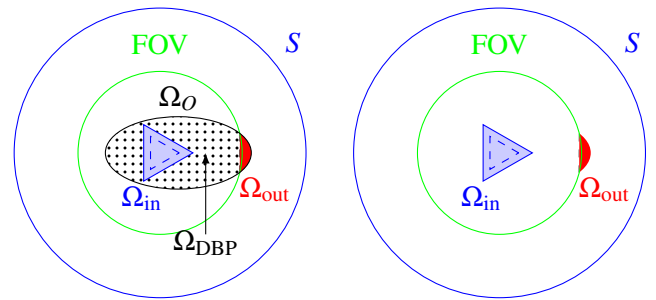

**Figure 8:** Top view of the considered configuration. The dashed blue triangle (resp. biggest blue triangle) delimits the intersection of $N_S$ with the horizontal plane $z = 12$mm (resp. $z = 20$mm). A zoom on $\Omega_{out}$ is done Fig. 9, *left*. *Left:* before the subtraction of the reprojection of $\Omega_{DBP}$ (dotted region) from the data. *Right:* after the subtraction.

We now propose an example configuration without contaminated lines. Other examples are also possible. The FOV

is a cylinder centered on the $\boldsymbol{e}_z$-axis of radius 90mm. The object support $\Omega_O$ is a cylinder of same direction with an elliptical defined by $\{(a\cos\lambda_e + c_o, b\sin\lambda_e), \lambda_e \in [0, 2\pi), a = 80\text{mm}, b = 40\text{mm}, c_o = 20\text{mm}\}$. Its axial extent (in the z-direction) is the interval $[12\text{mm}, 20\text{mm}]$ and the FOV is axially extended on a larger interval (no axial truncation). Finally, concerning $S$, we have $H = 60\text{mm}$ and $R = 160\text{mm}$. We present this configuration in top view, before and after subtraction of the reprojections from the data, see Fig. 8. It can readily be shown that, with the DBP method, we can reconstruct each point of $\Omega_{\text{DBP}} = \text{FOV} \cap \Omega_O \cap C_S \setminus A$ (the dotted region of Fig. 8, *left*), with $A \stackrel{\text{def}}{=} \text{conv}(\Omega_O \setminus \text{FOV}) \setminus (\Omega_O \setminus \text{FOV})$ (the orange region of Fig. 9, *left*). However it might be possible to reconstruct some points of the small region $A$ with the M-line methods [12], but this is not the central aim of this article, which is to prove that it is possible to reconstruct $\Omega_{\text{in}}$. A 3D illustration is given Fig. 9, *right*. We see from Fig. 8, *right*, that if there is no contaminated line, the configuration satisfies Tuy condition (there is "no longer any truncation"). Instead of drawing all measured lines (here they are the lines from a source point of $S$ and intersecting the FOV, especially $\Omega_{\text{in}}$), we choose to focus on lines intersecting $\Omega_{\text{in}}$ and $\Omega_{\text{out}}$ at the same time. These lines delimit two cones and a polyhedron (Fig. 10). We see from Fig. 10 that these cones (and the polyhedron) do not intersect $S$, so no contaminated lines exist, and thus reconstruction of $\Omega_{\text{in}}$ is possible.

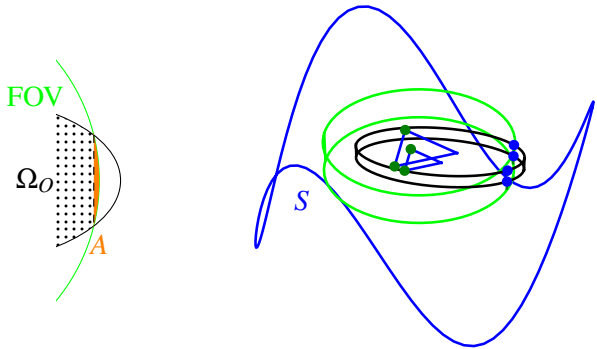

**Figure 9:** *Left:* Zoom on the right-side of $\Omega_O$: the dotted region is $\Omega_{\text{DBP}}$, the orange region is $A$ and the non-dotted (white and orange) region of $\Omega_O$ is $\Omega_{\text{out}}$. *Right:* The configuration proposed. The FOV is delimited by both green circles, $\Omega_O$ by both black ellipses and $N_S$ by the blue triangles (at $z = 12\text{mm}$ and $z = 20\text{mm}$). Dark green and blue dots are used to draw the limit lines of Fig. 10.

## 4   Simulation

We created a thin cylindrical phantom with an elliptical base as $\Omega_O$, and added some ellipsoid and balls, see Fig. 11. The configuration for the source trajectory and the FOV was the same as described in the previous section. The rectangular detector of $400 \times 430$ pixels is at a distance of 290mm from the source. A total of 360 cone-beam projections were simulated along the 3-sin source trajectory. The reconstruction

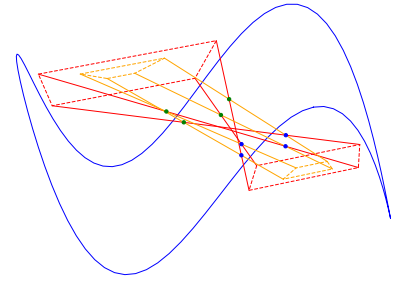

**Figure 10:** Cones (red) and the polyhedron (orange) demiliting the lines intersecting both $\Omega_{\text{in}}$ and $\Omega_{\text{out}}$. They do not intersect the trajectory $S$ so these lines are not contaminated lines.

volume consisted of $162 \times 82 \times 8$ voxels (pixels and voxels have a 1-mm side).

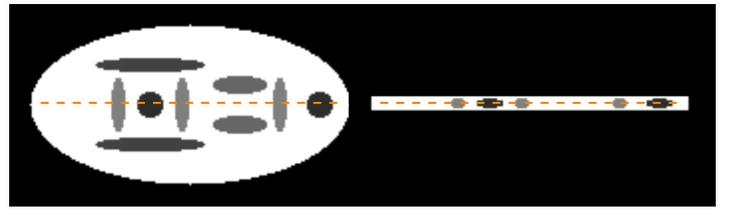

**Figure 11:** The phantom used for simulations. The orange lines indicate the location of the profile used in Fig. 13. *Left:* Top view. *Right:* Side view.

The objective was to verify the theory that the triangular region $N_S$ could be accurately reconstructed according to the theory established above. The goal was to investigate the results of [6], obtained by an iterative method, so we did not use the DBP method, with the 4 step approach outlined in section 3.1. We just used the conjugate gradient to minimize $\|Rf - p\|_2^2 + \gamma\|\nabla f\|_2^2$ with $R$ the forward projection operator and $p$ the measured projections. With $\gamma = 500$ we performed 60 iterations at which point we considered that convergence had been achieved. Some results are shown Figs. 12 and 13 (with another reconstruction performed without truncation, with a FOV of radius 102mm).

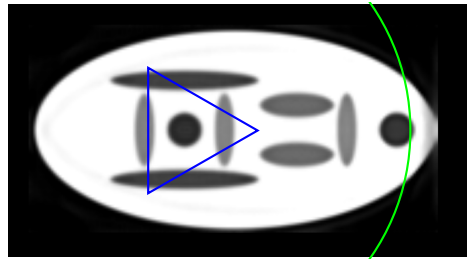

**Figure 12:** A cross-section at $z = 16\text{mm}$ of the reconstruction. The green circular arc delimits the FOV and the blue triangle is $N_S$ (for this cross-section).

## 5   Discussion and conclusion

We have adapted the scheme introduced in [5] for the reverse helix with axial truncation to the 3-sin trajectory with trans-verse truncation. To our knowledge, it is an original way to manage certain situations of transverse truncation for points lying in the Tuy-Finch region but not lying on a chord.

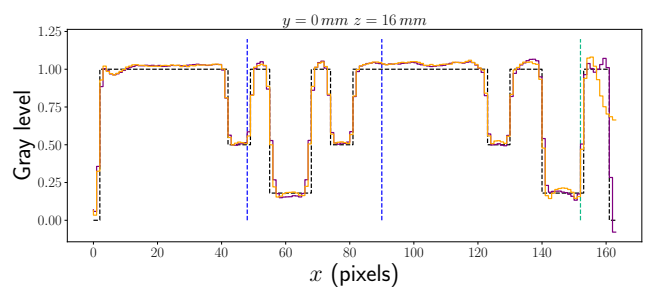

**Figure 13:** A profile of the reconstruction at $y = 0$mm and $z = 16$mm. The black line is the phantom, the purple line is the reconstruction without truncation and the orange line is the reconstruction with truncation. The dashed green line is the right limit of the FOV (for the reconstruction with truncation) and the dashed blue lines are the limits of $N_S$.

We performed a simulation (with an iterative method) which showed the same quality of reconstruction in the chord zone as well as the non-chord zone $N_S$. However, our example only involved very mild transverse truncation. On the other hand, in [6] we presented results showing good quality reconstruction for the same trajectory with much more transverse truncation but without theoretical results to justify it.

The configuration we have presented is rather limited in practice. For example, the object is quite flat. Nevertheless this represents a beginning of a lead, and other more general configurations could be found, for example by considering sub-trajectories of $S$ after subtraction of the reprojections, while guaranteeing Tuy's condition.

**Acknowlegdgements** The authors would like to thank F.Noo who pointed out the link between articles [5] and [6] and which led to this work. This work was supported by the "Fonds unique interministériel" (FUI) and the European Union FEDER in Auvergne Rhône Alpes (3D4Carm project) and by the ANR (ROIdoré project ANR-17-CE19-0006-01).

# Null space image estimator using dual-domain deep learning for Region-of-Interest CT reconstruction

Yoseob Han[1], Kyungsang Kim[1], and Quanzheng Li[1]

[1]Department of Radiology, Massachusetts General Hospital (MGH), Boston, MA, USA

**Abstract** Interior tomography for Region-of-Interst (ROI) imaging has various advantages in terms of reducing a number of detectors and decreasing X-ray radiation exposure dose. However, large patient or small field-of-view (FOV) detector can cause truncated sinograms, and then reconstructed images with truncated sinograms suffer from severe cupping artifacts. We proved that cupping artifacts are identified as null space images of truncated Radon transform, and the null space images are concentrated in low-frequency parts based on Bedrosian identity. To accurately reconstruct the ROI imaging, we propose a null space image estimator using deep learning with novel dual-domain framework. We demonstrate that the proposed method can outperform the conventional deep learning methods.

## 1 Introduction

X-ray computed tomography (CT) imaging provides high-quality and high-resolution images, but X-ray CT causes potential cancer risks due to radiation exposures. Thus, many researches have studied to reduce the radiation dose, where three approaches were widely used by reducing (1) photon counts of X-ray source (low-dose CT), (2) projection views (sparse-view CT), and (3) ROI (interior tomography). Unlike the low-dose CT reducing photon counts and the sparse-view CT undersampling projection views, the interior tomography retains these factors but uses small FOV detectors, which are useful for imaging of small target regions such as cardiac and dental imagings. In addition, portable C-arm CTs also use interior tomography imaging in order to miniaturize the hardware system. Therefore, the interior tomography not only reduce the radiation exposures, but also has a cost benefit due to a small size of detectors. While the interior tomography has advantages, truncated projection data has not been correctly reconstructed using analytic CT reconstruction algorithms such as filtered backprojection (FBP) and the reconstructed image suffers from severe cupping artifacts. Simple method to mitigate the cupping artifacts is projection extrapolation [1]. Even though the reconstructed image using extrapolated projection data shows moderated cupping artifacts, hounsfield units (HU) can be biased due to inaccurate extrapolation. Other researchers have developed model-based iterative reconstruction (MBIR) methods with several penalty teams such as total variation (TV) [2] and generalized L-spline [3, 4]. Specifically, Lee et al [4] showed that based on Bedrosian identity, signals of null space images remain at low frequency parts, while high frequency parts can be reconstructed using analytic functions such as Hilbert transform.

Recently, deep learning algorithms have been proposed as high-performance solutions for low-dose CT [5–7], sparse-view CT [8–10], and interior tomography [11, 12]. The solutions based on deep learning have surpassed the conventional MBIR methods in terms of image quality and reconstruction time. Deep learning methods demonstrated powerful performances for various applications, but limitations, such as blurring effect caused by $L2$ loss, still remain. To solve the blurring effects, many researchers use $L1$ loss and/or $GAN$ loss, but $L1$ loss increases mean square error (MSE) values and $GAN$ loss might generate realistic structures that do not exist.

In this paper, we propose a novel dual-domain deep learning method to solve interior tomography problem as shown in Fig. 1(b). Specifically, the proposed deep learning method estimates null space images rather than clean images. An advantage estimating the null space images is to avoid blurring effects because the null space images are composed of low frequency signals [4]. Even though previous works [11, 12] directly estimated clean images from truncated FBP images using image-domain CNN (See Fig 2(a)), but cupping artifacts are inherently caused by truncated projections, not image-domain. The novel dual-domain CNN (see Fig 2(c)), combined with projection-domain CNN (see Fig 2(b)) and image-domain CNN (See Fig 2(a)), is proposed to simultaneously handle the null space components in both domains. Numerical experiments show that the proposed method outperforms other CNNs and preserves high frequency parts in reconstructed images.

## 2 Theory

### 2.1 Problem Formulation

Here, we first describe Radon transform $\mathcal{R}$ and then extend to interior tomography problem using truncated Radon transform $\mathcal{T}_\mu \mathcal{R}$. Let $\theta$ denotes a vector on the unit sphere $\mathbb{S} \in \mathbb{R}^2$. The set of orthogonal vectors $\theta^\perp$ is described as

$$\theta^\perp = \{\mathbf{v} \in \mathbb{R}^2 : \mathbf{v} \cdot \theta = 0\}, \tag{1}$$

where $\cdot$ denotes an inner product. If an image is defined by $f(\mathbf{x})$ for $\mathbf{x} \in \mathbb{R}^2$, the Radon transform $\mathcal{R}$ of an image $f$ is formulated as

$$\mathcal{R}f(\theta, u) := \int_{\theta^\perp} f(u\theta + \mathbf{v})d\mathbf{v}, \tag{2}$$

where $u \in \mathbb{R}$ and $\theta \in \mathbb{S}$. Fig 1(a) shows a coordinate geometry for interior tomography and $\mu$ denotes a radius of a ROI (See green circle in Fig 1(a)). If the Radon transform $\mathcal{R}f$ is restricted by radius $\mu$ as $\{(\theta, u) : |u| < \mu\}$, then a truncated Radon transform is denoted as $\mathcal{T}_\mu \mathcal{R}f$, where $\mathcal{T}_\mu$ is truncation mask with radius $\mu$. Therefore, a interior tomography problem can be explained to find an unknown image $f(\mathbf{x})$ for $|\mathbf{x}| < \mu$ from the truncated Radon transform $\mathcal{T}_\mu \mathcal{R}f$.

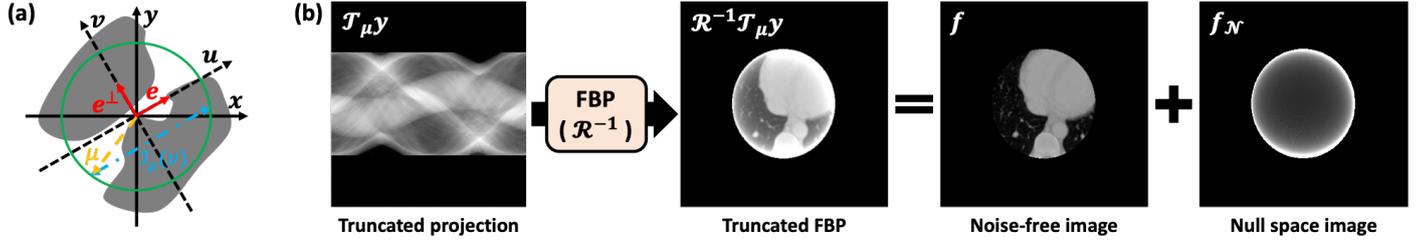

**Figure 1:** (a) A coordinate geometry for interior tomography, and (b) Truncated FBP consisting of noise-free image and null space image. $y$ (projection image) denotes $\mathcal{R}f$ (Radon transform $\mathcal{R}$ of unknown image $f(\mathbf{x})$).

## 2.2 Null Space Formulation

The truncated Radon transform $\mathcal{T}_\mu \mathcal{R}f$ described in Sec. 2.1 causes a null space $\mathcal{N}_\mu$, and the null space $\mathcal{N}_\mu$ makes the interior tomography problem a strong ill-posed problem. A mathematical analysis of the null space $\mathcal{N}_\mu$ follows Ward et al [3]. Specifically, an analytic inversion of the truncated Radon transform $\mathcal{T}_\mu \mathcal{R}f$ can be equivalently described a Hilbert transform $\mathcal{H}_\mu f(u) = \frac{1}{\pi} \int_{u \in \mathcal{I}_\mu} \frac{du'}{(u-u')} f(u')$ of a differentiated backprojection (DBP) along an restricted 1D chord lines $\mathcal{I}_\mu(u) := \{u' \in \mathbb{R} \mid \sqrt{(u')^2 + v^2} \le \mu\}$ (See blue line in Fig. 1(a)). Then, the null space $\mathcal{N}_\mu$ of the truncated Radon transform $\mathcal{T}_\mu \mathcal{R}f$ is represented by

$$\mathcal{N}_\mu := \left\{ f_{\mathcal{N}} \mid f_{\mathcal{N}}(u, v) = -\frac{1}{\pi} \int_{u' \notin \mathcal{I}_\mu(v)} \frac{du'}{(u-u')} \psi(u', v) \right\}, \quad (3)$$

for any functions $\psi(u, v)$ outside of a ROI. For example, a null space image $f_{\mathcal{N}}$ is illustrated in Fig. 1(b) and a cupping artifact is a common impact of the null space $\mathcal{N}_\mu$.

## 2.3 Spectral Decomposition with Bedrosian Identity

When high-passed components $h(x)$ restricted for $|\omega| > \omega_0$ and low-passed components $l(x)$ restricted for $|\omega| < \omega_0$ exist, the Bedrosian theorem [4] satisfies as:

$$\mathcal{H}\{l(x)h(x)\} = l(x)\mathcal{H}\{h(x)\}. \quad (4)$$

Here, the image $f(x) = f_H(x) + f_L(x)$ is restricted by the 1D chord lines $\mathcal{I}_\mu$ which are usually bounded at low frequency ranges. Then, according to the Bedrosian theorem:

$$\begin{aligned} \mathcal{I}_\mu(x)f(x) &= \mathcal{I}_\mu(x)(f_H(x) + f_L(x)) \\ &= \mathcal{I}_\mu(x)(\mathcal{H}\{g_H(x)\} + \mathcal{H}\{g_L(x)\}) \\ &= \mathcal{H}\{\mathcal{I}_\mu(x)g_H(x)\} + \mathcal{I}_\mu(x)\mathcal{H}\{g_L(x)\}. \end{aligned} \quad (5)$$

Therefore, only high frequency components $f_H(x)$ can be analytically reconstructed by

$$f_H(x) = \frac{\mathcal{H}\{\mathcal{I}_\mu(x)g_H(x)\}}{\mathcal{I}_\mu(x)}. \quad (6)$$

## 2.4 Null Space Estimator using Deep Learning

In previous works [11, 12], the researchers have solved the interior tomography problem by using image-CNNs (See Fig. 2(a)) to directly estimate noise-free images $f$ from truncated FBP images $\mathcal{R}^{-1}\mathcal{T}_\mu y$. However, Since we have verified that the high frequency components $f_H(x)$ can be accurately reconstructed by analytic methods, keeping the high frequency components $f_H(x)$ unprocessed can improve performance over dealing with the high frequency parts $f_H(x)$. In addition, there are blurring issues using $L2$ loss. To overcome the limitations, we proposed a null space estimator using deep learnings rather than clean image estimators [11, 12]. In addition, we developed a novel network architecture called as dual-domain CNN $\mathcal{Q}_{prj}$ (See Fig. 2(c)) combined with projection-domain CNN $\mathcal{Q}_{prj}$ (See Fig 2(b)) and image-domain CNN $\mathcal{Q}_{img}$ (See Fig 2(a)) to estimate null space images $f_{\mathcal{N}}$ in image-domain and projection-domain, simultaneously. Specifically, an optimization problem of the image-domain CNN $\mathcal{Q}_{img}$ is formulated as

$$\mathcal{L}_{img}(\mathcal{Q}_{img}) = \min_{\mathcal{Q}_{img}} \sum_{i=1}^{N} ||f_{\mathcal{N}}^{(i)} - \mathcal{Q}_{img}(FBP(\mathcal{T}_\mu \mathcal{R}f^{(i)}))||^2. \quad (7)$$

For the projection-domain CNN $\mathcal{Q}_{prj}$, an optimization problem is represented as

$$\mathcal{L}_{prj}(\mathcal{Q}_{prj}) = \min_{\mathcal{Q}_{prj}} \sum_{i=1}^{N} ||f_{\mathcal{N}}^{(i)} - BP(\mathcal{Q}_{prj}(\mathcal{T}_\mu \mathcal{R}f^{(i)}))||^2. \quad (8)$$

The proposed dual-domain CNN uses a loss $\mathcal{L}_{dual}$ such that

$$\mathcal{L}_{dual}(\mathcal{Q}_{img}, \mathcal{Q}_{prj}) = \mathcal{L}_{img}(\mathcal{Q}_{img}) + \mathcal{L}_{prj}(\mathcal{Q}_{prj}). \quad (9)$$

## 3 Method

### 3.1 Datasets

Ten subject datasets from the American Association of Physicists in Medicine (AAPM) Low-Dose CT Grand Challenge

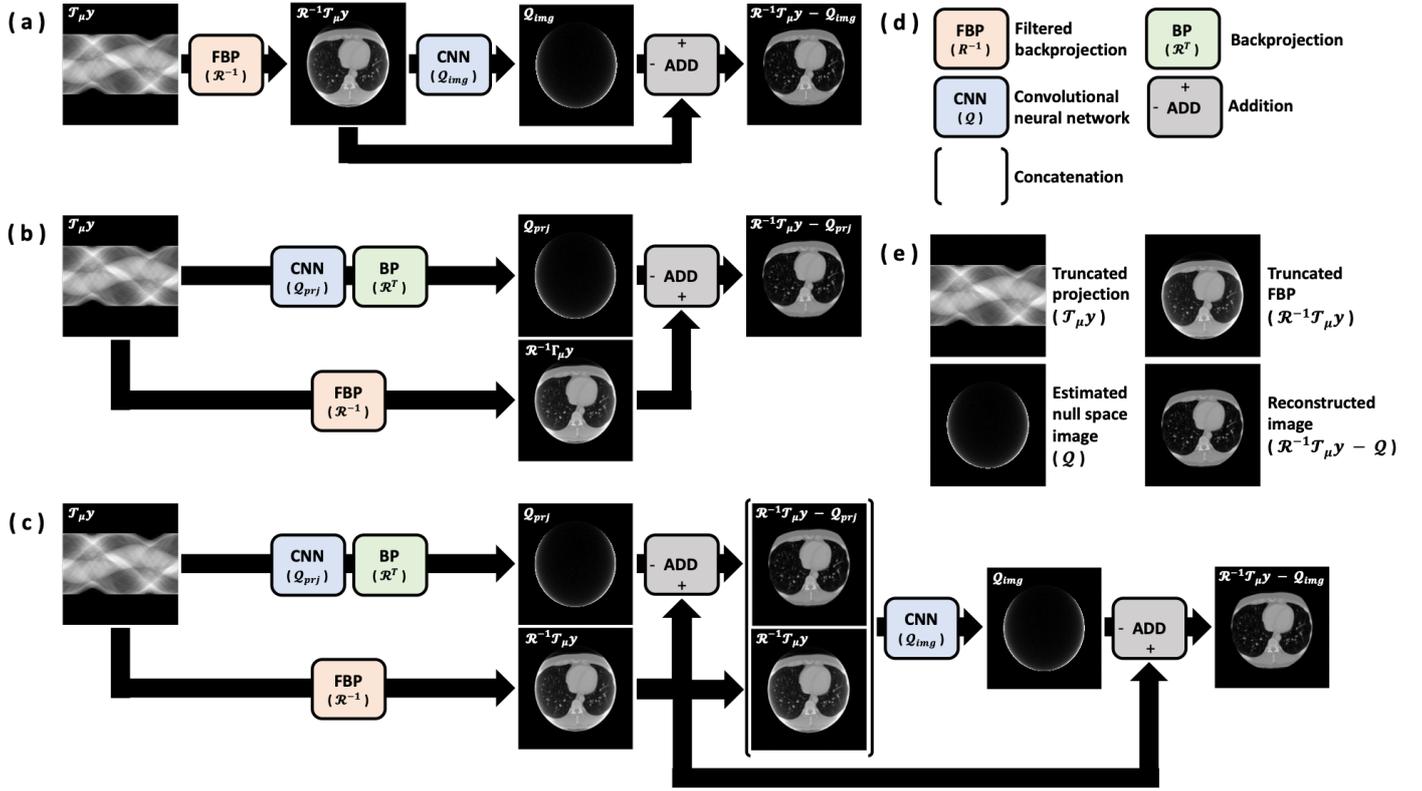

**Figure 2:** Several types of null space image estimators. (a) image-domain estimator $Q_{img}$, (b) projection-domain estimator $Q_{prj}$, and (c) dual-domains estimator $Q_{img}(Q_{prj})$ learnings. (d) describes modules shown in (a), (b), and (c), and (e) shows definitions of each image. $y$ (projection image) denotes $\mathcal{R}f$ (Radon transform $\mathcal{R}$ of unknown image $f(\mathbf{x})$).

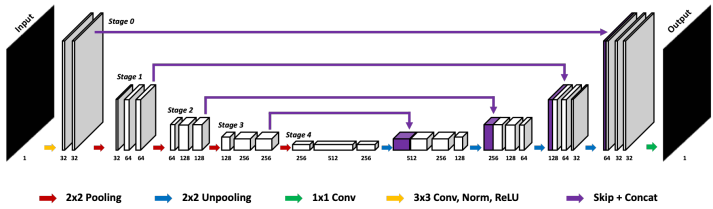

**Figure 3:** Backbone architecture for a neural network $Q$.

were used. Among ten subjects, nine subjects were used as train and validation datasets. Another subject was used to test dataset. From the datasets, projection data was numerically generated using a forward projection operator with parallel beam geometry. A size of images is $512 \times 512$ and its pixel resolution is $1mm^2$. The number of view is 720 views and a range of rotation for X-ray source is $[0°, 360°)$. The number of detectors is 720 and detector pitch is $1mm$. Truncation ratios were used as 0%, 50%, and 75%, so datasets were extended three times.

## 3.2 Architectures

Fig. 2(a-c) illustrate image-domain estimator, projection-domain estimator, and dual-domain estimator, respectively. Specifically, dual-domain estimator are consisted of a front part of projection-domain estimator and a back part of image-domain estimator. CNN module used in each estimator is U-Net as shown in Fig. 3.

## 3.3 Training

The proposed method was implemented using Pytorch. Also the proposed architecture was directly trained with end-to-end learning because FBP and BP modules were implemented in Pytorch's layer structures. Single graphic processing unit (GPU) as NVIDA Tesla V100 is used to train each estimator. CNN parameters are described as below. Adam optimizer was used and an initial learning rate was $10^{-4}$ and it was multiplied by 0.1 if validation loss did not decrease over 5 epochs. The number of batch size is 8. For data augmentation, vertical flipping was applied.

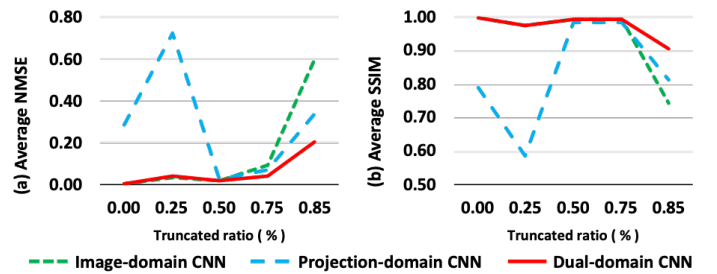

**Figure 4:** (a) Average NMSE and (b) Average SSIM with respect to various truncated ratios.

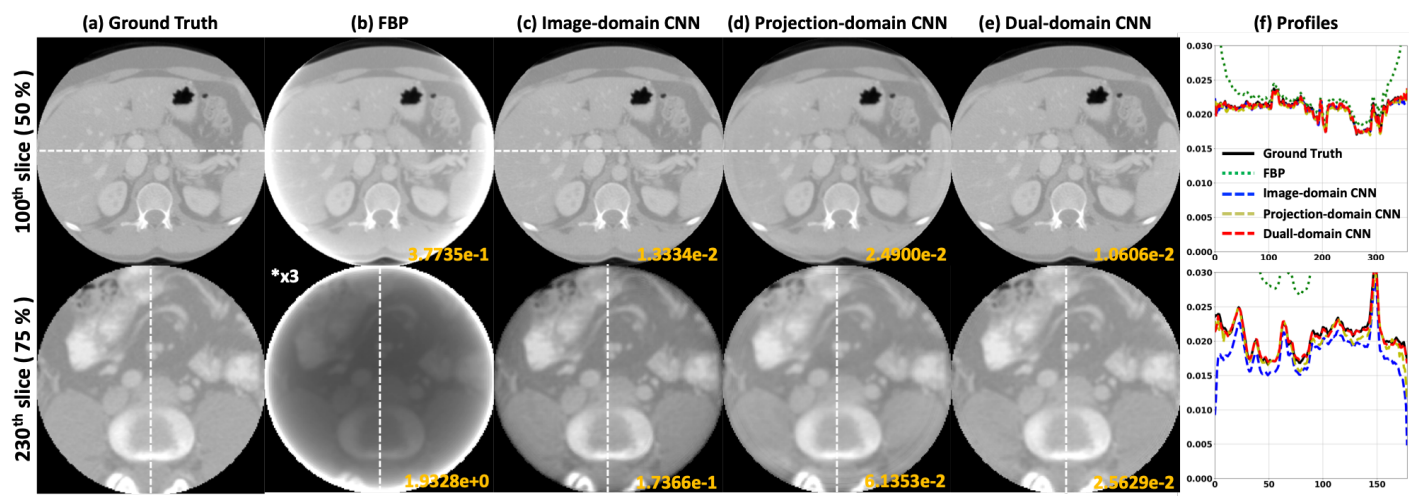

**Figure 5:** (a) and (b) are Ground truth and FBP image, respectively. (c-e) show reconstructed images by image-domain CNN in Fig. 2(a), projection-domain CNN in Fig. 2(b), and dual-domain CNN in Fig. 2(c). (f) illustrates profiles according to white line on the results. First and Second rows are 50% and 75% ROI situations, respectively. *x3 denotes that a window scale is magnified three times. NMSE values are written at the corner.

## 4    Results

Fig. 4 shows quantitative metrics such as normalized mean square error (NMSE) and structural similarity index measure (SSIM) according to various truncated ratios. Image-domain CNN in Fig. 2(a) and dual-domain CNN in Fig. 2(c) show similar quantitative values at small truncations range from 0% to 50%. However, as the truncation ratio increases, the dual-domain CNN outperforms the image-domain CNN. Interestingly, projection-domain CNN in Fig. 2(b) shows bad performance at the small truncation ranges, but outperforms than image-domain CNN at large truncation ratios. This is the reason that the projection-domain CNN is more efficient for large truncation ratios than the image-domain CNN, while the image-domain CNN is useful for small truncations. Fig. 5 compares the reconstructed results by image-, projection-, and dual-domain CNNs. The proposed dual-domain CNN shows lowest values in NMSE, and the smaller the truncated ratio, the larger the gap of NMSE between proposed method and other methods.

## 5    Conclusion

In this paper, we proposed the deep learning method to estimate the null space image, which is one of major problems in interior tomography. Because the null space image is caused by the truncated data in the projection domain, we developed a novel dual-domain network architecture. Numerical results showed that the proposed method is robust to the size of FOV, and outperforms single-domain CNNs.

# A multiple-energy-window projection-domain quantitative SPECT method for joint regional uptake quantification of $^{227}$Th and $^{223}$Ra

Zekun Li[1], Nadia Benabdallah[2], Daniel L. J. Thorek[1,2], and Abhinav K. Jha[1,2]

[1]Department of Biomedical Engineering, Washington University, St. Louis, MO, USA
[2]Mallinckrodt Institute of Radiology, Washington University, St. Louis, MO, USA

**Abstract** $\alpha$-particle radiopharmaceutical therapies ($\alpha$-RPTs) based on Thorium-227 ($^{227}$Th) isotope are currently in different stages of investigation. $^{227}$Th decays to $^{223}$Ra, another $\alpha$-particle-emitting isotope that redistributes inside the body. Reliable dose quantification with these therapies is clinically important. Since these isotopes also emit gamma-ray photons, SPECT provides a mechanism to perform this quantification. However, reliable quantification is challenging due to orders-of-magnitude lower administered activity and thus very low count levels compared to conventional SPECT, emission spectra of these isotopes containing multiple photopeaks, and significant overlap in the spectra of these isotopes. To address these issues, we propose a multiple-energy-window projection-domain quantification method that jointly estimates the regional activity uptake of both $^{227}$Th and $^{223}$Ra directly using the SPECT projection data from multiple energy windows. We evaluated the method with realistic simulation studies conducted with an anthropomorphic digital-phantom population. Our results demonstrated the convergence of the method, the ability of the method to yield reliable estimates of the regional uptake for both isotopes, and the reliability of the method for different lesion sizes. In conclusion, this study provides a method to perform reliable dose quantification for $^{227}$Th-based $\alpha$-RPTs.

## 1 Introduction

Thorium-227 ($^{227}$Th) base conjugates are an emerging $\alpha$-particle emitting isotope for $\alpha$-particle-based radiopharmaceutical therapies ($\alpha$-RPTs) that are being investigated in several clinical and preclinical studies [1]. Since these therapies distribute throughout the body, including radio-sensitive vital organs, there is an important need to quantify the dose distribution of these therapies. This could then help with post-therapy management, therapy outcomes prediction, and adverse events monitoring.

The $^{227}$Th isotope also emits gamma-ray photons that can be detected by a gamma camera. Thus, SPECT imaging may provide a mechanism to quantify the spatial distribution of absorbed dose in the patient. However, this quantification is challenging for several reasons. First, usually, the administrated activity in $\alpha$-RPTs is two-to-three orders lower compared with those in the conventional SPECT procedures. The emitted and thus the detected gamma-photon counts are many times lower than conventional SPECT studies. Further, the decay chain of $^{227}$Th is complicated. $^{227}$Th decays to Radium-223 ($^{223}$Ra), which can then disassociate with the antibody and form an independent biodistribution. Consequently, we have two independent isotope distributions, each of which has an overlapping spectrum and each of which is emitting a low number of photons. The inverse problem is to use these measurements to jointly quantify the regional uptake of both these isotopes in different organs and lesions.

One approach to solve this joint quantification problem is to jointly reconstruct the activity distribution with the two isotopes over a voxelized grid, and then quantify the activity within the VOIs [2, 3]. However, this reconstruction-based approach has been observed to yield limited accuracy in $\alpha$-RPTs, with bias values between 10-35% even with finely tuned reconstruction protocols [4–6]. We note here that reconstruction is only an intermediate step for quantification. Reconstructing the $^{227}$Th and $^{223}$Ra activity images require jointly estimating these activity distributions over a large number of voxels, a challenging task with a small number of detected counts. In contrast, our objective is only to jointly quantity the uptake in the lesion and few organs. A method that directly estimates the uptake from the projection data for only these small number of regions is a less ill-posed problem and may help achieve the goal of reliable quantification. Such a method may also help avoid reconstruction-related information loss.

Methods to directly estimate regional uptake from projection data have been proposed previously [7–10]. The maximum-likelihood region of interest (ML-ROI) approach [7] is particularly compelling due to the optimality of the ML estimator. We recently extended this method to quantify regional uptake in $\alpha$-RPT SPECT, addressing several unique challenges with imaging these isotopes [11, 12]. The resultant projection-domain quantification (PDQ) method was evaluated in the context of measuring regional uptake from SPECT for patients administered $^{223}$Ra-based therapy. We observed that the PDQ method yielded reliable regional uptake and consistently outperformed reconstruction-based quantification approaches. This provides the motivation to extend this method for the joint quantification task. The approach we take for this purpose is similar to a dual-isotope SPECT reconstruction approach [13], but we propose our method to directly perform quantification from the projection data, and skip the reconstruction step.

To further address the issue of low counts, we recognize that both $^{227}$Th and $^{223}$Ra emit gamma-ray photons over multiple photopeak energies. Thus, using photons from multiple energy windows corresponding to these different photopeaks provides a way to improve effective system sensitivity [4]. Recent studies have shown that using measurements from multiple energy windows can help improve quantification performance [14, 15]. We thus frame the inverse problem to use data from multiple energy windows.

To solve this inverse problem, we propose a multiple-energy-window projection-domain quantification (MEW-PDQ) method. The method jointly estimates the regional activity uptake of $^{227}$Th and $^{223}$Ra directly from the low-count SPECT projection data acquired over multiple energy windows. The proposed method is then evaluated with realistic simulation studies. We first describe the theory of the method.

## 2 Theory

Consider a SPECT system imaging an activity distribution that consists of both $^{227}$Th and $^{223}$Ra. The system acquires data over multiple energy windows and at multiple projection angles. Our goal is to estimate the regional uptake of both these isotopes within $K$ different volumes of interest (VOIs). Assume that the $^{227}$Th and $^{223}$Ra isotope distributions can be represented in terms of the VOI basis functions, with the underlying assumption that the activities within these VOIs are constant. Denote the $K$-dimensional vectors of regional activity uptake of $^{227}$Th and $^{223}$Ra by $\boldsymbol{\lambda}^{Th}$ and $\boldsymbol{\lambda}^{Ra}$, respectively. Denote the measured projection data by a $M$-dimensional vector $\boldsymbol{g}$. Each projection bin corresponds to data acquired within an energy window over a certain detector pixel and at a certain projection angle. Denote the $M \times K$ dimensional system matrices that describe the acquisition of the projection data from $^{227}$Th and $^{223}$Ra by $\boldsymbol{H}^{Th}$ and $\boldsymbol{H}^{Ra}$, respectively. The elements of these matrices, namely $H_{mk}^{Th}$ and $H_{mk}^{Ra}$, denote the response in the $m^{th}$ bin of the SPECT system to a decay event of $^{227}$Th and $^{223}$Ra from the $k^{th}$ VOI, respectively.

At the low count levels when imaging these isotope distributions, the stray-radiation-related noise is not negligible. We model this noise as Poisson distributed with the same mean for all projection bins, denoted by $\psi$. The imaging system equation can then be derived to be

$$\boldsymbol{g} = \boldsymbol{H}\boldsymbol{\lambda} + \psi + \boldsymbol{n}, \qquad (1)$$

where

$$\boldsymbol{H} = \begin{bmatrix} \boldsymbol{H}^{Th} & \boldsymbol{H}^{Ra} \end{bmatrix}, \ \boldsymbol{\lambda} = \begin{bmatrix} \boldsymbol{\lambda}^{Th} \\ \boldsymbol{\lambda}^{Ra} \end{bmatrix}, \qquad (2)$$

and $\boldsymbol{n}$ is a M-dimensional vector that denotes the Poisson distributed noise in the imaging system, with mean of $\boldsymbol{H}\boldsymbol{\lambda} + \psi$. Then, the probability of the measured projection data is given by

$$\Pr(\boldsymbol{g}|\boldsymbol{\lambda}) = \prod_{m=1}^{M} \exp[-(\boldsymbol{H}\boldsymbol{\lambda})_m - \psi] \frac{[(\boldsymbol{H}\boldsymbol{\lambda})_m + \psi]^{g_m}}{g_m!}. \qquad (3)$$

To estimate the regional activity uptake of each isotope, we take the partial derivative of the logarithm of Eq. 3 with respect to each element in $\boldsymbol{\lambda}$. This yields the following iterative estimates of activity uptake of $^{227}$Th and $^{223}$Ra in the $k^{th}$ VOI, denoted by $\lambda_k^{Th}$ and $\lambda_k^{Ra}$, respectively

$$\widehat{\lambda_k^{Th}}^{(t+1)} = \widehat{\lambda_k^{Th}}^{(t)} \frac{1}{\sum\limits_{m=1}^{M} H_{mk}^{Th}} \sum_{m=1}^{M} \frac{g_m}{[\boldsymbol{H}\hat{\boldsymbol{\lambda}}^{(t)}]_m + \psi} H_{mk}^{Th}, \qquad (4)$$

$$\widehat{\lambda_k^{Ra}}^{(t+1)} = \widehat{\lambda_k^{Ra}}^{(t)} \frac{1}{\sum\limits_{m=1}^{M} H_{mk}^{Ra}} \sum_{m=1}^{M} \frac{g_m}{[\boldsymbol{H}\hat{\boldsymbol{\lambda}}^{(t)}]_m + \psi} H_{mk}^{Ra}, \qquad (5)$$

where $\widehat{\lambda_k^{Th}}^{(t)}$ and $\widehat{\lambda_k^{Ra}}^{(t)}$ denote the estimates of $\lambda_k^{Th}$ and $\lambda_k^{Ra}$ at the $t^{th}$ iteration, respectively. We refer to this iterative approach as the MEW-PDQ method.

Implementation of the MEW-PDQ method required computing the elements of the two system matrices. These elements were computed using a Monte Carlo (MC)-based approach. More specifically, SIMIND, a well-validated MC-based simulation software was used to determine the response of the SPECT system to photon emission from each VOI. During this process, all relevant image-degrading processes in SPECT were modeled, including the effects of attenuation, scatter, collimator response, septal penetration and scatter, and finite energy and spatial resolution of the detector. Then, the mask of each VOI with unit activity uptake of each isotope, together with the attenuation map of the whole phantom, were input to SIMIND individually. Simulation of a large number of photons for each isotope yielded almost noiseless projection data from all relevant energy windows, which yielded columns of the system matrices for each isotope.

Next, the MEW-PDQ method required the mean stray-radiation-related noise in each energy window. This was estimated by acquiring a planar blank scan on the relevant SPECT system for a long duration. The mean stray-radiation-related counts in each energy window of each projection bin were calculated. These numbers were then normalized to generate the mean of this noise. The computed system matrices and mean stray-radiation-related noise were applied in Eq. 4 and 5 to estimate the regional uptake of each isotope simultaneously from the projection data acquired over different energy windows.

## 3 Evaluation of the proposed method

A realistic simulation study was conducted to evaluate the performance of the proposed MEW-PDQ method in clinically realistic scenarios that modeled patient-population variability. The pelvic region of 50 Digital 3D male patients with different anatomies was simulated using the Extended Cardiac-Torso(XCAT) [16]. The size of patients was sampled from a Gaussian distribution with a mean equal to the $50^{th}$ percentile male among US adults and a 10% standard deviation. The diameter of the lesions was sampled from a Gaussian distribution, with a mean of 33.75 mm and a standard deviation of 12.64 mm, as derived from clinical data. The phantom was divided into three primary VOIs, the lesion, bone, and gut. The rest of the isotope distribution within the patient was categorized as background, resulting in four VOIs in total. The mean activity uptake ratio of $^{227}$Th and $^{223}$Ra in each VOI is shown in Table 1. These ratios simulated the scenario where the patients were imaged after 120 h of administering $^{227}$Th. In this duration, approximately 17% of the $^{227}$Th had decayed to $^{223}$Ra. The bio-distribution of two isotopes was chosen to simulate a difference of uptake in different VOIs. The activity uptake ratios in each VOI were independently sampled from a Gaussian distribution with the mean as given in Table 1 and a 10% standard deviation.

**Table 1:** Activity uptake ratios of $^{227}$Th and $^{223}$Ra in VOIs.

| VOIs | Background | Bone | Gut | Lesion |
|------|-----------|------|-----|--------|
| $^{223}$Ra | 2 | 5 | 25 | 20 |
| $^{227}$Th | 12 | 30 | 100 | 300 |

Projection data corresponding to this patient population was generated using Monte Carlo modeling. A GE Optima 640 SPECT system with high energy general purpose (HEGP) collimator was simulated in SIMIND. As suggested in [17], the projection data was collected at four energy windows. The bounds of energy windows were 75 - 100 keV, 135 - 165 keV, 215 - 260 keV, and 260 - 285 keV. Projections were acquired from 60 angles spaced uniformly over 360°. The count level of these projections simulated the scenario where the pelvic region of a patient was administrated with 2 Mbq activity of $^{227}$Th and was imaged for 30 minutes with this dual-headed SPECT system. The MEW-PDQ method was applied to the acquired projection data. We quantified the accuracy and reliability of the MEW-PDQ method in estimating the regional uptake across all four VOIs using the absolute normalized bias and normalized root mean square error (RMSE).

To evaluate the convergence of the MEW-PDQ method, a patient with the average size of body and lesion was considered. One realization of the projection data was generated for that patient using the process described above. The proposed method was applied and the normalized error in the estimated uptake of each VOI after every 32 iterations was computed. 800 iterations were performed.

To evaluate the sensitivity of the MEW-PDQ method to lesion size, we simulated five patients with average body sizes. Each patient had a lesion of different diameter ranging from 15 mm to 35 mm. The activity uptake ratio in the different VOIs for both the isotopes was as shown in Table 1. Projection data for these patients was generated as described above. 50 noise realizations were generated for each patient to compute both the bias and the precision of the estimated activity uptake for different lesion sizes.

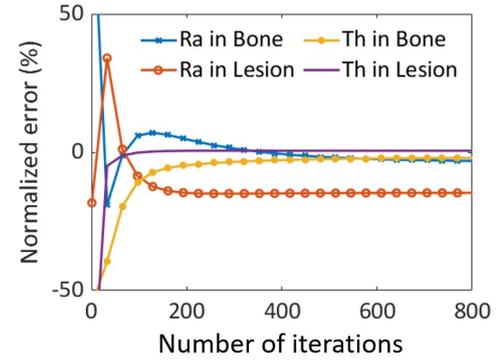

**Figure 1:** The normalized error of activity uptake estimates in bone and lesions regions of both isotopes as a function of iteration number using the MEW-PDQ method.

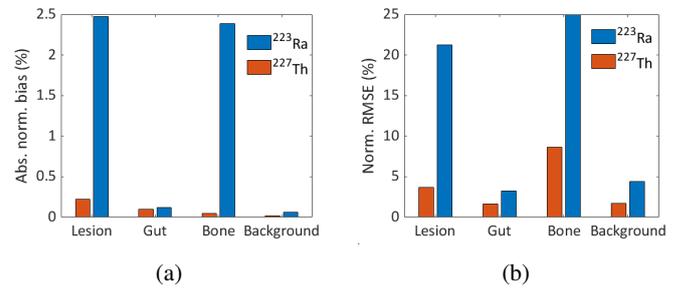

**Figure 2:** The (a) absolute normalized bias and (b) normalized RMSE of the estimated regional uptake for a realistic simulation study.

## 4 Results

The normalized error of the activity uptake estimates in the bone and lesion regions of both isotopes as a function of iteration number using the MEW-PDQ method is shown in Fig. 1. We observe that after 512 iterations, the change in error in all VOIs is less than 0.1%, showing the convergence of the method. Thus, we chose 512 as the number of iterations for the method for all experiments. Due to the low number of VOIs, 512 iterations took less than 5 minutes with an Intel(R) Core(TM) i7-6700 CPU with 8 cores and 16.0 GB RAM.

The absolute normalized bias and normalized RMSE of the activity uptake estimates of the two isotopes in the lesion, gut, bone, and background regions in the realistic simulation study are shown in Fig. 2. We observe that for all regions, the MEW-PDQ method yields a bias close to zero for both isotopes. Further, the normalized RMSE for both isotopes in all regions is low, with a mean of 8.7%.

The absolute normalized bias and normalized RMSE of the estimated activity uptake in the lesion as a function of the lesion diameter using the MEW-PDQ method are shown in Fig. 3. We observe that the estimates of the activity uptake of $^{227}$Th are almost unbiased for all the considered lesion sizes. For $^{223}$Ra, the bias and RMSE of estimates in the lesion region are higher for lesions with diameters below 25 mm. This is due to the very low number of detected counts emitted by $^{223}$Ra in that region considering the total count values as low as 10,000 counts per slice in the projections. However, beyond a lesion diameter of 25 mm, the bias from the $^{223}$Ra uptake is close to zero.

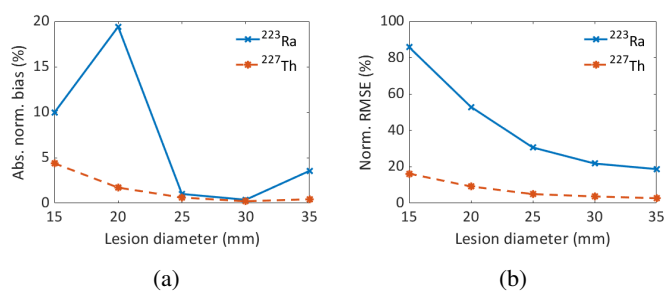

**Figure 3:** The (a) absolute normalized bias and (b) normalized RMSE of the estimated activity uptake in the lesion region as a function of the lesion diameter.

## 5 Conclusions

We proposed a multiple-energy-window projection-domain quantification method to jointly quantify the regional uptake of $^{227}$Th and $^{223}$Ra in different volumes of interest. Evaluation with realistic simulation studies provides evidence that the proposed method yields reliable absolute quantification of both $^{227}$Th and $^{223}$Ra regional activity uptake simultaneously at very low count levels, as is the case when imaging these isotopes in clinical scenarios. Further, the proposed method yields reliable absolute quantification of both isotopes across different lesion sizes. Our results suggest that the proposed method may provide a mechanism to perform reliable dose quantification with $^{227}$Th based $\alpha$-RPTs and motivate further evaluations.

### Acknowledgement

This work was supported in part by grants R21-EB024647 and R01-EB031051, awarded by National Institute of Biomedical Imaging and Bioengineering. We also thank the Washington University Center for High Performance Computing for providing computational resources for this project. The center is partially funded by National Institutes of Health (NIH) grants 1S10RR022984-01A1 and 1S10OD018091-01.

# Chapter 3

# Oral Session - Motion Management

**session chairs**
Roger Fulton, *University of Sydney (Australia)*
Simon Rit, *CREATIS (France)*

# Collision avoidance trajectories for on-line trajectory optimization in C-arm CBCT

Sepideh Hatamikia[1,2], Ander Biguri[3], Gernot Kronreif[1], Tom Russ[4], Joachim Kettenbach[5], Wolfgang Birkfellner[2]

[1]Austrian Center for Medical Innovation and Technology, Wiener Neustadt, Austria
[2] Center for Medical Physics and Biomedical Engineering, Medical University of Vienna, Austria
[3] Institute of Nuclear Medicine, University College London
[4]Computer Assisted Clinical Medicine, Medical Faculty Mannheim, Heidelberg University, Germany
[5]Institute of Diagnostic and Interventional Radiology and Nuclear Medicine, Landesklinikum Wiener Neustadt, Austria

**Abstract**
Kinematic constraints due to the additional medical equipment or patient size are common while acquiring C-arm cone beam computed tomography (CBCT). Such constraints cause collisions with the imager while performing a full circular rotation and therefore eliminate the chance for three dimensional (3D) imaging in CBCT-based interventions. In a previous paper, we proposed a framework to develop patient-specific collision-free trajectories for the scenarios where circular CBCT is not possible. However, the proposed method required kinematic constraints to be known beforehand. As collisions are mainly unpredictable in the operation theater, a framework which enables a real-time trajectory optimization is of great clinical importance. In this study, we introduce a new search strategy which has the potential to optimize trajectories on-the-fly. We propose an optimization procedure which identifies trajectories with the highest information to reconstruct a volume of interest (VOI) by means of maximizing an objective function; then a local search is performed around the best selected initial candidates and better trajectory solutions are investigated among newly created neighbors. The experimental results based on two imaging targets inside an Alderson Rando phantom showed that proposed trajectories achieve image quality comparable to that of the reference circular CBCT while simulating strong kinematic constraints. The overall time required for the whole optimization process was around three to four minutes using one GPU.

## 1 Introduction

Recently, cone beam computed tomography (CBCT) has become an important imaging modality in interventional radiology [1, 2]. One important feature of interventional radiology is that a prior knowledge of patient anatomy (e.g. high quality CT or pre-operative CBCT) is usually available. This gives the opportunity to incorporate such prior knowledge into image acquisition process by using a customized CBCT. Nowadays, robotic CBCT C-arms enable additional degrees of freedom and extend the scanning geometry possibilities beyond the standard circular source-detector trajectories. Several studies have demonstrated an improvement in image quality and/or reduction the radiation dose using noncircular trajectories. In these studies, trajectory parameters were computed in the way to maximize the imaging performance of particular imaging tasks [3-5]. Gang et al. [3] proposed a target-based imaging acquisition framework for robotic C-arm CBCT systems using a gradient-based optimization of the tube current, reconstruction kernel and orbital tilt. Noncircular source-detector trajectories have been introduced using periodic and B-spline-based functions for simulation studies, as well as in neuroradiology applications to increase the image quality in a volume of interest (VOI) [4, 5]. Recently, optimal sinusoidal trajectories were proposed in order to avoid the metal parts of the imaged object while still assuring a high coverage in Radon space and its vicinity

[6]. All the aforementioned researches [3-6] were effectively applied to C-arm CBCT trajectory optimization. However, in all these studies, hard constraints on the rotation angle were applied for the trajectory design; thus, the employed trajectories did not take patient-specific collisions into account. Furthermore, all these studies [3-6] calculated the optimal trajectory parameters in a (semi) offline manner.

In another study [7], patient-specific collision avoidance trajectories were proposed for linac-mounted CBCT devices using a virtual isocenter and variable magnification during data acquisition. Although their proposed trajectories could integrate case-specific collisions into the trajectory design, their method requires a high amount of computational time which hampers its usage for real-time trajectory optimization and therefore, it is not appropriate to react to unforeseen collisions which happen during interventions. To the best of our knowledge, the only study that introduced a real-time trajectory optimization was [8], in which the authors proposed optimizing the C-arm CBCT trajectories during the CBCT scan and performed the adjustments on-the-fly using a convolutional neural network and regressed an image quality measured over all possible next projections given the current X-ray image. However, the main focus of this research was metal artifact reduction and the trajectories introduced did not incorporate patient-specific collisions in their trajectory design. The research we present in this study is the first demonstration that proposes an on-the-fly trajectory optimization framework for customized CBCT acquisition that is able to react to scene-specific unforeseen collisions.

Our group has recently published a method to optimize imaging quality for CBCT using semi-circular scan trajectories which can also be arranged out-of-plane [9, 10]. A VOI is selected using a prior CT scan and a variety of possible trajectory combinations from short arcs is simulated while taking kinematic constraints in account. The optimal arc combination is designated based on the image quality within the VOI. The time needed for designing a patient specific trajectory was in the range of 80 minutes [9]. This required collisions and kinematic constraints to be known previously. As such constraints are mostly unpredictable in a clinical scenario, e.g. caused by additional medical devices or patient size (Fig.1), a real time trajectory optimization protocol is of great clinical importance even at the cost of losing a bit of image quality.

In the current study, we introduce a search strategy to overcome the aforementioned computational constraints.

Compared to our previous study [9], the major scientific novelty of this study lies on the introduction of the new search strategy that enables the on-the-fly feature for the trajectory optimization scheme; this finally brings a remarkably important clinical benefit for interventions where a 3D CBCT is otherwise not possible due to unforeseen collisions.

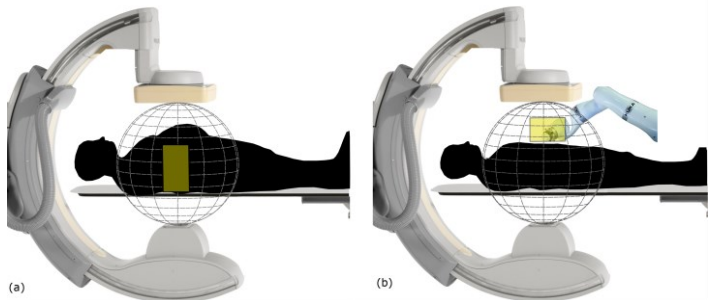

Figure 1. Two examples of common kinematic constraints during interventions. Collision due to the patient size (a), and due to other medical devices (b) [11].

## 2 Materials and Methods

### 2.A. Adaptation of workflow for the on-the-fly customized trajectory optimization

In this study, we modified our previous work to enable a dynamic optimization in the operation room which can integrate kinematic constraints emerging during the interventions into the trajectory optimization design.

We used the geometry of the Philips Allura FD20 Xper C-arm in order to define a set of possible arcs. The C-arm is able to perform two different types of rotations: 1) rotation by angle $\theta_1$ towards the Right Anterior Oblique (RAO)/Left Anterior Oblique (LAO) direction while having a tilt $\psi$ at various fixed Cranial (CRA)/Caudal (CAU) angles, 2) rotation by angle $\theta_2$ towards the CRA/CAU direction while having a tilt $\varphi$ at various fixed RAO/LAO angles. Different subset of arcs (each arc included around 80 projections) were defined similar to that in the previous work [9] (Fig 2. a, b). We simulated kinematic constraints as two forbidden areas on the geometry of the C-arm (represented as yellow rectangles in Fig. 2 c, d). The arcs which had more than 10% of their angular range in the two forbidden area were removed and those that had less than 10% in these areas were cropped (Fig. 2 e, f) [9]. In order to accelerate the optimization process in the current work, the previous approach was modified by sparsifying the initial subset of arcs (Fig. 2 e, f) to include just arcs for every six degrees (Fig. 2 g, h); this led to a significant reduction in the computation time. However, a reduction of the initial subset of arcs may introduce an unfavorable bias in the path selection process. To address this issue, we propose to perform a heuristic local search around the arcs with the largest amount of information. First, we selected the three arcs with the best objective function values as the arcs with highest amount of information. Then, we created new neighbor arcs for each of the three selected arcs and consequently, searched through such nearest neighbor arcs until an improvement in the objective function is observed.

Finally, we selected the arc with the highest objective function value (Fig. 3). We repeated this procedure for the arc subsets RAO/LAO and CRA/CAU one after the other, with the previous best arcs were still being used, until a predefined number of arcs was designated as the final trajectory. We used the value of Feature SIMilarity Index (FSIM) as the objective function, as in our previous study. The pseudocode for this procedure is presented in Algorithm 1.

### 2.B. Image reconstruction

A modified version of the Tomographic Iterative GPU-based Reconstruction (TIGRE) toolkit [12] for arbitrary trajectories was used [9], but the Adaptive Steepest Descent Projection Onto Convex Sets (ASD-POCS) reconstruction was limited to five iterations. For simulations, we sampled projections every four degrees, and therefore, 20, 40, and 60 projections were simulated for trajectories that included one, two, and three arcs, respectively. Projection number reduction was done only in simulations for a further acceleration of the process; however, for the real data, the full sampling projections were used for reconstruction. For projection simulations, we used a monoenergetic forward model with added Poisson noise. Bare-beam fluence was also modeled to approximate device exposure.

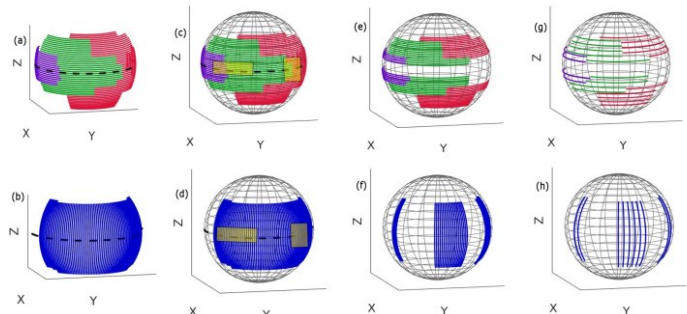

Figure 2. A) RAO/LAO arcs with CRA/CAU obliques shown in the purple, green, and red colors, (b) CRA/CAU arcs with RAO/LAO obliques shown in the blue color, (c) and (d) spherical plot of arcs with two forbidden areas, (e) and (f) spherical plot of the arcs after removing those that intersected the forbidden area, (g) and (h) spherical plot of these remaining arcs after sparsification. Only these arcs were in the search space for trajectory optimization. (Kinematic constraints are simulated as forbidden areas are shown as yellow rectangles) [11].

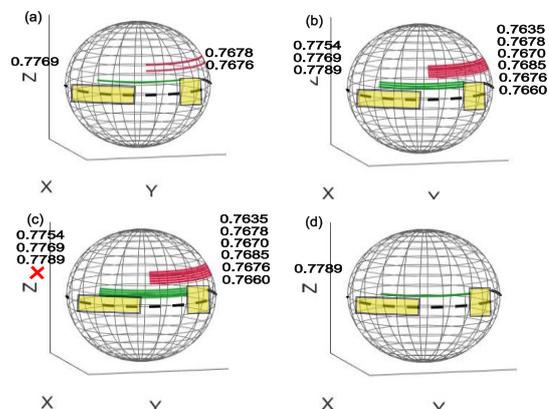

Figure 3. Illustration of the search strategy for optimizing the first best arc, (a) the three arcs with the highest objective function value are selected by searching through the RAO/LAO arcs sparsely sampled initial subset (Fig. 2 g), (b-d) the nearby arcs are searched until the objective function decreased. The sign (×) shows that the arc included more than 10% of its angular range in the forbidden area, and therefore, was rejected from the search space and FSIM was not calculated.

**Algorithm 1. Trajectory optimization**

**Input:** Search space, number of desired arcs
**Step 1:** Simulate projections for all defined arcs with the digital phantom
**Step 2: FOR** 1: number of subsets
**Step 3: FOR** 1: number of arcs in subset
- Reconstruct the image using the set of projections related to the corresponding arc
- Crop the reconstructed image at the VOI
- Calculate the objective function at the cropped area
   **END**
   **Step 4:** Select best three arcs from Step3
   **Step 5: WHILE** expanded arcs increase objective function
   **Step 6: FOR** 1:number of arcs to expand
- Create neighboring arcs at one degree each side
- Evaluate objective function in newly created neighbors
   **END**
   **END**
   **Step 7:** Select best arc and prepend to search space

## 2.B.1. Optimization of computational time

The implementation of ASD-POCS in the TIGRE toolbox was modified to remove CPU-GPU transfer functions and to run the reconstruction fully on the GPU. Our implementation takes approximately 1.4, 2.2, and 3.05 seconds for each ASD-POCS reconstruction (with five iterations), including 20, 40, and 60 projections using a computer with an NVIDIA GeForce RTX 2080 and a 32-core Advanced Micro Devices (AMD) processor. $256^3$ voxel volumes with $512^2$ projections were used for the reconstruction. The overall time required for the whole optimization process was around three to four minutes. The reported numbers in this study are using one GPU.

## 3 Results

In our experiments, two imaging targets in the thoracic spine (regions T3/T4 and T10/T11 for Target 1 and Target 2, respectively) of an Alderson-Rando phantom were evaluated. In the simulations, we optimized trajectories including three arcs for both imaging targets. 3D visualizations of the optimized trajectories compared to standard circular trajectory are shown in Fig. 4 a and Fig. 4 c for Target 1 and Target 2, respectively. The selected angular range and projection numbers related to optimized trajectories of both targets are shown in Table I. The (-) sign denotes rotation to the right/caudal directions and the (+) sign denotes rotation to the left/cranial directions. We implemented the optimal trajectories using a step-and-shoot protocol on C-arm to acquire real data. The reconstruction results were then compared to the C-arm circular trajectory (313 projections, 210° angular range). Furthermore, they were also compared with respect to a reconstruction from a partial circular trajectory with an angular range and projections equivalent to the optimized trajectory (Fig. 4 b, d). Reconstruction results using simulation data as well as real data for the optimized trajectories, standard C- arm circular, and partial circular trajectories for both targets are shown in Fig 5. The reconstruction results were evaluated by FSIM and Universal Quality Image (UQI). For both indexes, the image quality metric between the prior CT and

C-arm circular CBCT was considered the reference value. The quality index value between the prior CT and optimized/partial circular trajectory was also calculated as the measured value. The relative deviation between the reference and measured values was used for the image quality evaluation. According to the results of Table III, the optimized trajectories delivered relative deviations up to 9.47% and 4.06% in both image quality metrics for Target 1 and Target 2, respectively. A relative deviation up to 7.87% for Target 1 and 5.39% for Target 2, respectively, was also calculated for the reconstructed images related to partial circular trajectories. These results show a small decreased reconstruction performance (a slightly higher relative deviation) for Target 1, while an increased image quality (a slightly lower relative deviation) for Target 2 for both image quality metrics when using optimized trajectories compared to the partial circular trajectory. However, the differences observed are not significant and reconstructed images from optimized trajectories revealed a comparable image quality for both targets with regard to the partial circular trajectories.

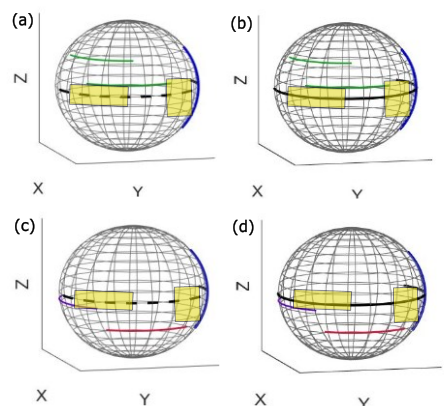

Figure 4. 3D visualization of the optimized trajectories (arcs shown in color) with respect to the C-arm circular trajectory (black dashed plot) and partial circular trajectory (black solid plot) for Target 1 (a, b) and Target 2 (c, d).

Table 1. The angular range and projection number of the three selected arcs for the optimized trajectories related to Target 1 and Target 2

| Trajectory | Arc | Angle | Projection number per arc | Total number of projections |
|---|---|---|---|---|
| **Target 1** | Arc 1 | $\theta_1$= -39:1:+39, $\psi$= -26 | 72 | 228 |
| | Arc 2 | $\theta_2$- -34:1:+40, $\varphi$= -60 | 75 | |
| | Arc 3 | $\theta_1$=+44:1:+124,$\psi$=-6 | 81 | |
| **Target 2** | Arc 1 | $\theta_1$= -22:1:+50, $\psi$= 10 | 73 | 227 |
| | Arc 2 | $\theta_2$- -40:1:+38, $\varphi$= -50 | 79 | |
| | Arc 3 | $\theta_1$= +9:1:+89, $\psi$=+32 | 75 | |

Table 1. Relative deviations (%) of image quality measures FSIM and UQI for Target 1 and Target 2 using both optimized and partial circular trajectories

| Target | Image quality metric | Trajectory | Relative deviation (%) |
|---|---|---|---|
| **Target 1** | FSIM | Opt. | 9.47 |
| | | Partial-circ. | 7.87 |
| | UQI | Opt. | 8.49 |
| | | Partial-circ. | 4.83 |
| **Target 2** | FSIM | Opt. | 3.90 |
| | | Partial-circ. | 5.39 |
| | UQI | Opt. | 4.06 |
| | | Partial-circ. | 5.38 |

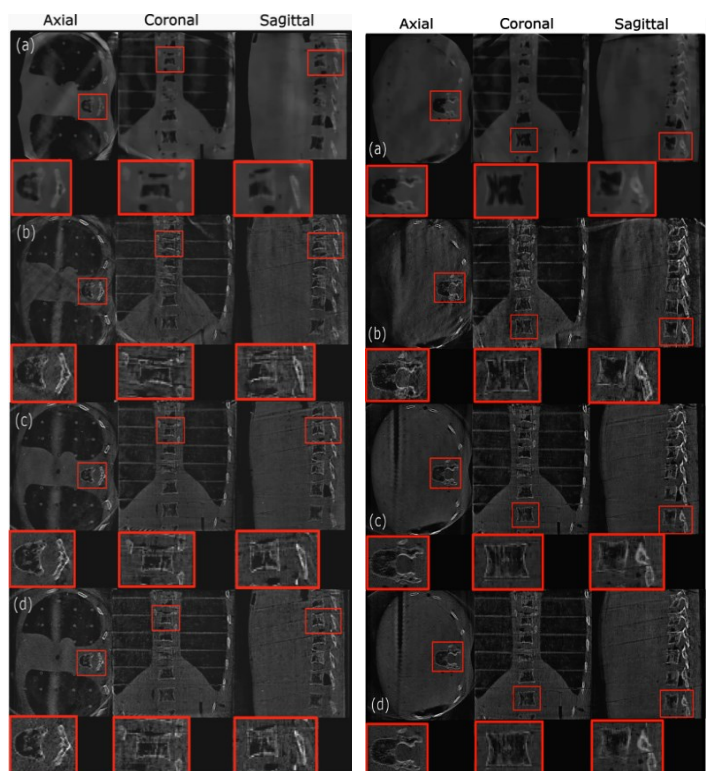

Figure 5. Reconstructions related to Target 1 (Left column) and Target 2 (Right column), (a) optimized trajectory based on simulation data, (b) optimized trajectory based on real data, (c) C-arm circular trajectory based on real data, and (d) partial circular trajectory based on real data. The display window has a range of 200-3000 HU for (a) and a range of 0-21 in gray values for (b-d), respectively.

## 4 Discussion and Conclusion

We proposed a framework for a patient-specific trajectory design for CBCT imaging which is suitable to react to unforeseen collisions. In fact, the major difference with the previous trajectory optimization approach [9] is that we now search for the optimal arcs within the most informative areas in 3D space to reconstruct the VOI (rather than searching among all plausible arcs as proposed in our previous study [9]), and consequently, we propose to perform a local search around the initially selected optimal arcs to find a better arc solution. Our results showed a slight decreased reconstruction performance for Target 1, while a small increase in image quality was seen for Target 2 using optimized trajectories compared to partial circular trajectories. Considering the fact that our approach is the first proposed protocol in literature that can facilitate CBCT for interventions in which a 3D circular CBCT would not be otherwise possible due to unpredictable collisions, our results show acceptable performance even if there is a slight reduction in the image quality for some targets compared to the partial circular trajectory. In this study, we achieved a considerably higher speed in comparison to our previous work [9], which required approximately 80 minutes for reconstruction. Our proposed trajectory optimization framework requires three to four minutes overall time on one GPU and a further reduction in time to about one minute is anticipated by using multiple GPUs. Our framework has the potential to be done on-the-fly; therefore, it can be considered suitable for interventions with unexpected and arbitrary collisions.

## ACKNOWLEDGEMNET

This work was supported by ACMIT - Austrian Center for Medical Innovation and Technology, which is funded within the scope of the COMET program and funded by Austrian BMVIT and BMWFW and the governments of Lower Austria and Tyrol. We also gratefully acknowledge the support of NVIDIA Corporation for the donation of the Titan Xp GPU used for this research. The support from the personnel of the Institute of Diagnostic and Interventional Radiology and Nuclear Medicine, Wiener Neustadt, Austria, for the performance of measurements is also gratefully appreciated.

# Reference-Free, Learning-Based Image Similarity: Application to Motion Compensation in Cone-Beam CT

Heyuan Huang[1], Jeffrey H. Siewerdsen[1,2], Wojciech Zbijewski[1], Clifford R. Weiss[2], Tina Ehtiati[3] and Alejandro Sisniega[1]

[1]Department of Biomedical Engineering, Johns Hopkins University, Baltimore, MD USA 21205

[2]Russell H. Morgan Department of Radiology, Johns Hopkins University, Baltimore, MD USA 21287

[3]Siemens Healthineers, Forchheim, Germany

**Abstract** Cone-beam CT (CBCT) is increasingly used in the interventional suite, providing three-dimensional imaging for intervention planning, guidance during the procedure, or post-procedure assessment. However, the moderately long acquisition time of CBCT (~4 to 20 s) makes it susceptible to artifacts from patient motion during image acquisition. Recent work has shown promising reduction of artifacts caused by patient motion in CBCT with purely image-based autofocus motion compensation approaches. However, conventional autofocus metrics (e.g., gradient entropy) are agnostic to the realistic presentation of anatomical structures in the image and can yield unrealistic solutions. Image structural similarity metrics (e.g., visual information fidelity, VIF) combine measures of image quality and structural similarity to a reference image, offering a potentially ideal basis for autofocus metrics. However, matched motion-free reference images are usually not available. In this work, we propose a reference-free, learning-based image similarity metric obtained using a deep neural network (denoted DL-VIF). A convolutional neural network was trained on simulated motion-corrupted CBCT data to extract features associated with the structural components contributing to VIF. The DL-VIF showed correlation with conventional VIF in motion corrupted abdominal CBCT for both rigid ($R^2 = 0.981$ and slope = 0.987) and deformable ($R^2 = 0.852$ and slope = 0.928) motion. The DL-VIF was incorporated in an autofocus motion compensation framework, and its performance was compared against a conventional metric (gradient entropy). Motion compensation with DL-VIF resulted in more robust motion compensation (pointing to a lower susceptibility to local minima), and in improved performance (SSIM = 0.943 compared to 0.900 for gradient entropy). The development of autofocus metrics that recognize the integrity of anatomical structures in the image is an important step toward reliable motion compensation in scenarios of complex soft-tissue deformable motion in CBCT.

## 1 Introduction

Involuntary patient motion remains one of the main challenges to image quality in cone-beam CT (CBCT). Motion artifacts arising from rigid motion in, e.g., head CBCT [1] or from deformable soft-tissue motion in, e.g., interventional CBCT [2], severely impact CBCT diagnosis and guidance capabilities.

Recent work has shown the feasibility of rigid [3] and deformable [4, 5] CBCT motion estimation with multi-region autofocus methods. Such approaches estimate the motion trajectory by optimizing an autofocus metric that emphasizes properties associated with motion-free images, such as image sharpness [3], piecewise constancy [6, 7], or sparsity of gradients [5]. However, such metrics do not guarantee the preservation of anatomical structures (e.g., shape and texture) in the image. For example, image entropy [6] can show similar values for in-focus images and unrealistic images formed by a nearly constant value [3]. Such degeneracies make the optimization susceptible to local minima within the non-convex space of the autofocus cost-function.

Contrary to conventional autofocus metrics, metrics including measures of structural similarity can capture the affinity between two input images beyond pixel intensity or gradient values, by estimating the correspondence between the (anatomical) structures present in the two images, comparable to recognition via the human visual system (HVS). For example, the structural similarity index (SSIM) [8] combines luminance and contrast with estimations of local spatial correlation between image patches. Similarly, the visual information fidelity (VIF) [9] estimates the loss of image information caused by a distortion process between the input image and its ideal counterpart. However, such similarity metrics require a reference image and are therefore not generally applicable to CBCT motion estimation where such a reference is not available.

In this work we hypothesize that the ability of deep convolutional neural networks (CNNs) to extract features representative of structural components of image data can be applied to the estimation of similarity metrics in images corrupted by patient motion *without a matching, motion-free reference*. This premise is supported by recent results that showed the capability of deep CNNs to estimate motion-induced image quality degradation in CBCT, and be used as basis for deep autofocus approaches. Recent developments in deep autofocus include deep CNNs for estimation of approximate motion amplitude in individual image patches [10, 11] for compensation of deformable motion in abdominal CBCT and approaches that built on multi-branch deep CNN architectures to automatically extract three-dimensional anatomical landmarks in motion corrupted brain/head CBCT images and to quantify motion via estimation of the reprojection error with respect to equivalent landmarks in the CBCT projection space [12].

In particular, we explored the potential of deep CNNs to reproduce VIF without a reference image in motion-corrupted CBCT data. A dedicated network architecture was designed and trained on pairs of simulated CBCT images with and without motion featuring soft-tissue and high-contrast structures in the abdomen. In the training stage, the VIF value was computed against the reference, motion-free, image and used to compute the CNN loss. In inference, only the motion-corrupted image was input to the network that outputs a learned VIF estimation (denoted DL-VIF). The validity of the DL-VIF as an autofocus metric was investigated within an autofocus framework for rigid motion estimation, with future work extending to deformable soft-tissue motion.

## 2 Materials and Methods

### 2.1 DL-VIF for Motion Quantification

VIF quantifies the similarity between a reference and a "distorted" image by estimating the loss of information attributable to the image degradation (distortion) process. The process to compute VIF is described in [9]. Briefly, the reference image is processed with a convolution channel that models the response of the HVS. The mutual information between the original and the HVS-processed reference quantifies the information preserved by the HVS. Application of an analogous processing channel to the test image yields the information preserved by the combination of both the perturbation (patient motion) and the HVS channels. The final VIF is computed as the ratio between the preserved information for the reference and distorted images. For motion-corrupted CBCT, VIF is computed as follows:

$$I_{MC} = \sum \log_{10}(1 + \frac{g^2 + \sigma_{MF}^2}{\sigma_v^2 + \sigma_n^2}) \qquad (1)$$

$$I_{MF} = \sum \log_{10}(1 + \frac{\sigma_{MF}^2}{\sigma_n^2}) \qquad (2)$$

$$VIF = \frac{I_{MC}}{I_{MF}} \qquad (3)$$

where $\sigma_{MF}^2 = H * (MF^2) - (H * MF)^2$ is given by the convolution of the reference, motion-free volume ($MF$) with an HVS channel ($H$) modelled as a set of four Gaussian kernels with size $N = 17, 9, 5,$ and 3 voxels (in the three volume dimensions of the volume) and standard deviation of $N/5$. The term $g$ estimates the loss of signal from patient motion and is computed as $g = \sigma_{MF-MC}/\sigma_{MF}^2 + 10^{-10}$, with $\sigma_{MF-MC} = H * (MF \cdot MC) - (H * MF) \cdot (H * MC)$, where $MC$ is the motion corrupted volume. The term $\sigma_v^2$ models the variance induced by patient motion, computed as $\sigma_v^2 = \sigma_{MC}^2 - g \cdot \sigma_{MF-MC}$, where $\sigma_{MC}^2$ is equivalent to $\sigma_{MF}^2$ but computed for the motion-corrupted volume. The term $\sigma_n^2$ is a scalar describing the noise level in the HVS and was set according to the method in [9].

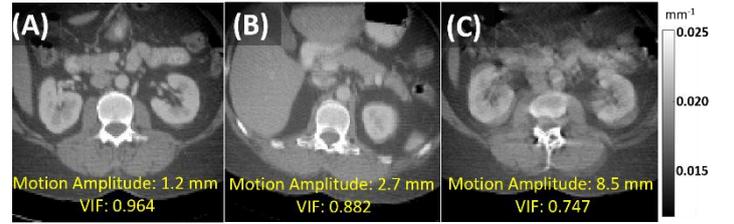

**Figure 1**: Images with various amount of motion artifacts and their corresponding motion amplitudes: 1.2 mm (A), 2.7 mm (B), and 8.5 mm (C). VIF values referenced to motion-free images show the correspondence between low VIF and severe motion artifacts, associated with large amplitude of motion.

Figure 1 illustrates the capability of VIF to capture image quality degradation caused by patient motion in CBCT images simulated as described below (Section 2.3). VIF values computed with the motion-free reference reduce in a manner that is strongly correlated with motion amplitude and qualitatively consistent with visual image appearance, making it a plausible autofocus metric if a reference-free estimation can be learned.

To extract the image features associated with VIF, a CNN was designed with topology illustrated in Fig. 2. The DL-VIF CNN is inspired by a previous CNN design for motion amplitude estimation [10] but uses 3D modules acting on 128×128×128 voxels volumes. An initial 3D convolution layer is followed by three residual blocks (ResBlock, similar to ResNet models [13]) that contain three convolution stages, each followed by an instance normalization layer and a rectified linear unit. Every ResBlock is followed by an average pooling layer. The final convolution layer is input to a dense linear layer that outputs the scalar DL-VIF.

### 2.2 Autofocus Motion Compensation with DL-VIF

The DL-VIF metric was integrated into a multi-region autofocus motion compensation framework able to compensate for rigid and deformable motion [3, 4, 5]. In its simplest form, the method considers a single region of interest subject to rigid motion. Motion estimation is then posed as an optimization problem that estimates a motion trajectory $T$ to minimize the following cost function:

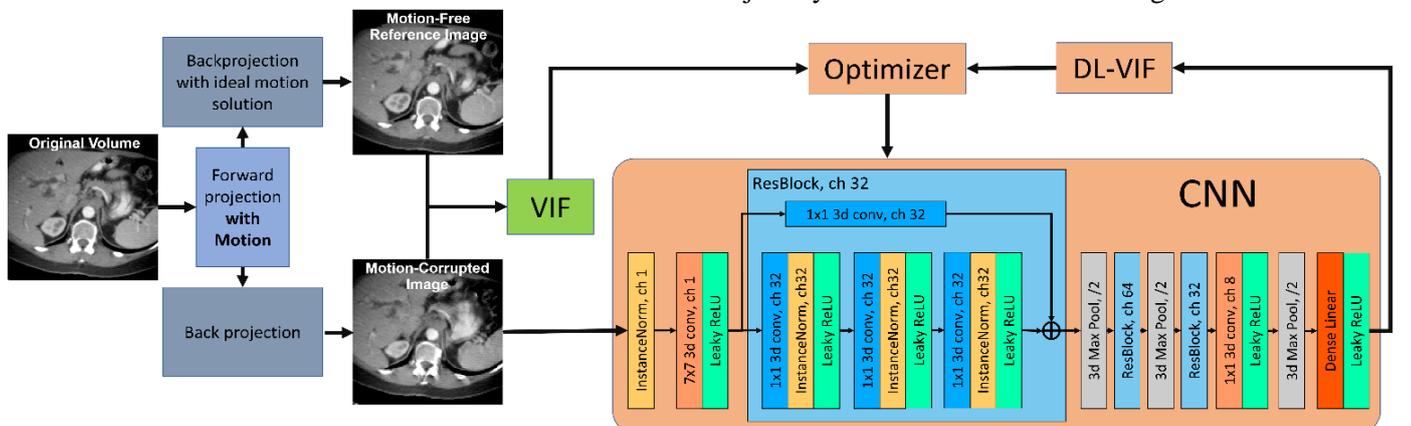

**Figure 2**: CNN training strategy and topology graph. Abdominal MDCT data were used as the basis for simulation of motion corrupted CBCT data with a high-fidelity forward projector. For each instance, a motion corrupted volume was subsequently obtained through backprojection, and a motion-free reference image was generated by applying the ideal solution during backprojection. VIF values were computed for each motion corrupted volume, using the corresponding motion-free reference. The CNN was trained to predict DL-VIF from the input motion-corrupted data with a loss function minimizing the difference between the predicted and reference VIF.

$$T = argmin_T[S(\mu(T)) + \beta_T \cdot R_T(T)] \quad (4)$$

where $\mu(T)$ is the reconstructed image for the candidate trajectory $T$, and $S(\mu(T))$ is an autofocus metric that (conventionally) maximizes image sharpness and penalizes residual motion artifacts while (ideally) preserving the fidelity of anatomical structures presented in the image. Abrupt temporal transitions in the motion trajectory are discouraged by the regularization term $R_T(T)$ that penalizes the norm-2 between the position difference for consecutive time points (viz. projections) in the acquisition. The contribution of the penalty and autofocus terms is balanced by the scalar hyperparameter $\beta_T$. To reduce the problem dimensionality, $T$ was modelled with $N_t$ cubic b-splines. In previous work[5], gradient entropy was used as the autofocus metric, denoted $S_{conventional}(\mu(T))$:

$$S_{conventional}(\mu)$$
$$= -\sum_{l=1}^{L} h_1\left(\sqrt{\nabla_x(\mu)^2 + \nabla_y(\mu)^2 + \nabla_z(\mu)^2}\right)$$
$$\cdot \log\left(h_1\left(\sqrt{\nabla_x(\mu)^2 + \nabla_y(\mu)^2 + \nabla_z(\mu)^2}\right)\right) \quad (5)$$

The proposed DL-VIF can be incorporated as the autofocus metric in Eq. (4), as $S_{DL-VIF}(\mu(T))$:

$$S_{DL-VIF}(\mu) = -\ln(DL\text{-}VIF(\mu)) \quad (6)$$

The negative logarithm in Eq. (6) provided basic conditioning to scale the DL-VIF value for optimization.

## 2.3 Data Generation and Training Process

Training and validation datasets were generated from 75 high quality abdominal MDCT volumes extracted from the cancer imaging archive (TCIA) "CT Lymph Nodes" collection. CBCT projection datasets were generated by forward projection of the MDCT volumes, incorporating both rigid and deformable patient motion in the forward projection process.

For each training instance, a volume was randomly selected from the dataset, and a 20 mm long sub-volume of interest was extracted at a random longitudinal position. CBCT projection data were generated by forward projection with a high-fidelity CBCT forward model and a scanner geometry pertinent to interventional C-arms, with source-to-detector distance (SDD) of 1200 mm and source-to-axis distance (SAD) of 785 mm. The detector was modelled as a flat-panel with 864×660 pixels and 0.64 mm isotropic pixel size. For rigid motion, translational motion was induced during the forward projection with amplitude varying randomly between 0 mm and 10 mm, and acting in a random direction. Deformable motion was induced with a motion field placed at a random location inside the abdomen. Motion amplitude of the ranged from 0 mm to 40 mm, set at the central point of the motion field, and faded from the center following an elliptical spatial pattern, with axes ranging from 61 mm to 182 mm. Both rigid and deformable motion followed a cosine temporal trajectory with random frequency set within a range covering 0.75 to 1.25 periods during the complete scan. For every dataset, a motion-free reference was generated analogously. Motion-corrupted and motion-free volumes were reconstructed on 128×128×128 voxels grids with isotropic 2 mm voxels. Training VIF values were calculated using the motion-free volume as a reference and were fed to the CNN along with the motion-corrupted volume, as illustrated in Fig. 2.

The network was trained with the Adam optimizer and a mean squared error (MSE) loss function between the DL-VIF and target VIF. Learning rate was $10^{-3}$ and training was performed with 10,000 rigid training instances and 1,000 deformable training instances. The network was trained on both dataset for 150 epochs. Training results were validated on 1,250 abdominal CBCT cases generated with the same method described above.

## 2.4 Validation of DL-VIF as an autofocus metric

The performance of DL-VIF for autofocus motion estimation was assessed with simulated abdominal CBCT data obtained similarly to the training and validation datasets. In this work, for simplicity, we experimented only on rigid motion patterns. To test the generalizability of the DL-VIF metric, the motion trajectory included random rotations around LAT, AP, and CC axes of ±5°.

Motion compensation was performed on 20 abdominal CBCT images with a conventional metric (gradient entropy) and with DL-VIF. Motion compensation results were evaluated in terms of SSIM, computed as in [8], using the motion-free volume as reference. A region of interest encompassing the entire volume was selected. A total of $N_t$ = 6 spline knots were used to model the motion trajectory, and $\beta_T$ was set to 50 for the conventional metric and 5 for DL-VIF to similarly scale the regularization term.

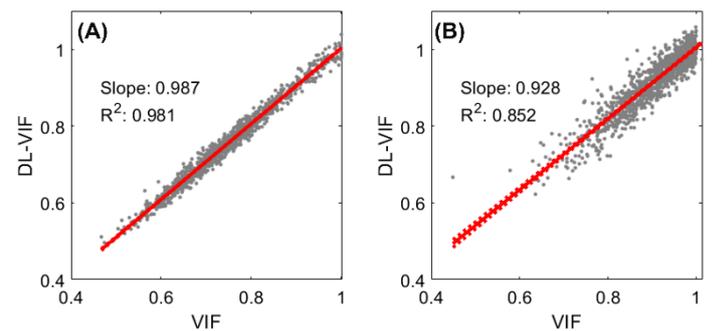

**Figure 3**: Agreement between DL-VIF and conventional VIF (computed with a reference motion-free image) in motion-corrupted CBCT volumes with rigid (A) and deformable (B) motion trajectories.

## 3 Results

### 3.1 Accuracy of DL-VIF

Figure 3 shows the agreement between the inferred DL-VIF and the reference VIF for the rigid and deformable motion abdominal test dataset after 150 epochs of training. Inferred DL-VIF showed good agreement with the reference, with a linear correlation ($R^2 = 0.981$ and slope = 0.987) and average MSE of $2.58 \times 10^{-4}$. Figure 4 shows the DL-VIF and corresponding VIF on deformable motion test data. Correspondence between DL-VIF and the reference VIF was similar as the one observed in rigid motion, with an approximated linear correlation ($R^2 = 0.928$ and slope = 0.852) and average MSE of $1.082 \times 10^{-3}$.

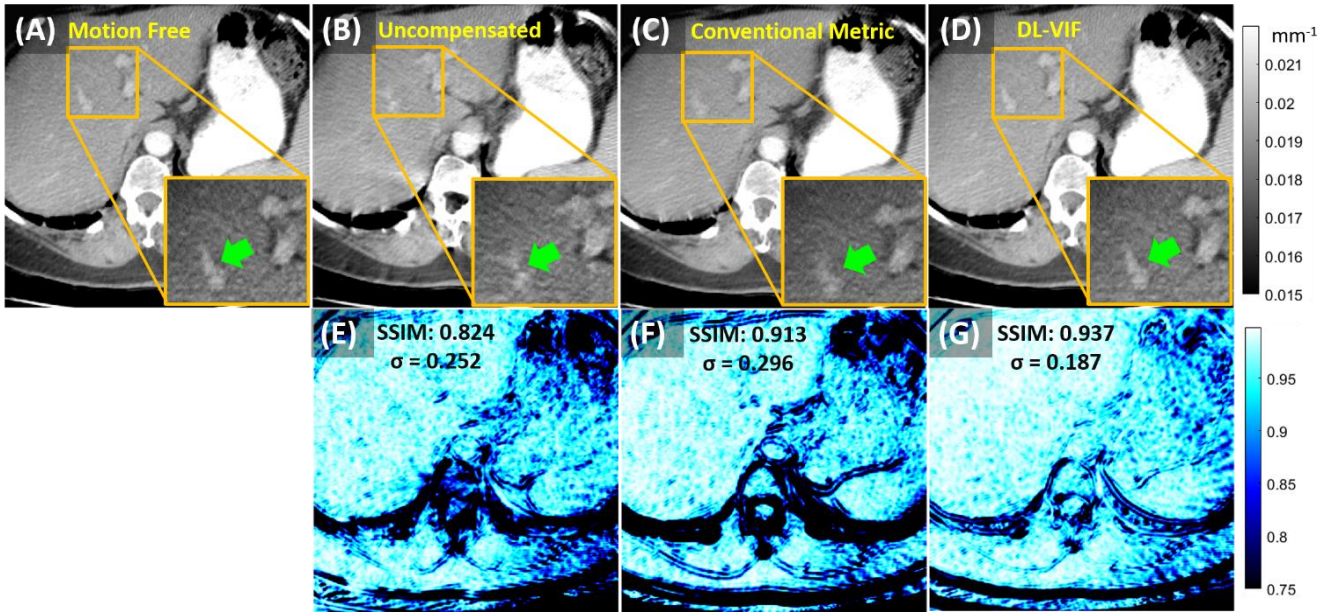

**Figure 4**: Motion compensation results in the abdomen (A-D) and their corresponding SSIM (E-G). (A) shows the reference motion-free reconstruction. (B and E) Motion artifacts are evident before motion compensation and are reflected in the reduced SSIM. (C and F) Motion compensation with a conventional autofocus metric resulted in reduced artifacts and increased SSIM. (D and H) Motion compensation with DL-VIF further improved results by preserving the structural content and penalizing residual artifacts in the liver.

## 3.2 Motion Compensation with DL-VIF

Figure 4 shows example results from motion compensation in the abdomen (Fig. 4A, 4B, 4C, and 4D) using the conventional metric and DL-VIF. While both methods achieved successful motion compensation, DL-VIF resulted in reduced residual artifacts and better delineation of anatomical structures (e.g., contrast-enhanced vessels in the liver, marked by green arrows). Motion compensation with DL-VIF yielded larger reduction of motion artifacts and better similarity with the motion-free reference image resulting in higher average SSIM (0.937 for DL-VIF vs 0.913 for conventional in the example in Fig. 4) and lower SSIM standard deviation ($\sigma_{SSIM} = 0.187$ vs $\sigma_{SSIM} = 0.296$ for conventional autofocus). The reduced standard deviation in SSIM suggests more consistent performance across different regions in the image when using DL-VIF. The spatial distribution of SSIM for the motion-corrupted image and compensated image with conventional and DL-VIF autofocus are shown in Figs. 4E-F-G. Consistent with the SSIM values, the motion-compensated images with DL-VIF exhibit more uniform (and overall higher) SSIM, with particular improvement in bone (e.g., spine) and soft-tissue features (e.g., liver vascularity).

Aggregated SSIM values for the 20 test cases are shown in Fig. 5. Both metrics yielded increased SSIM, but DL-VIF

resulted in larger SSIM improvement (average increase of 4.25%, compared to 0.56% for the conventional metric). The conventional metric allowed unrealistic images for certain motion trajectories and degenerate solutions evidenced by the red markers falling below the unity line.

The average SSIM across all cases was 0.943 for DL-VIF with standard deviation $\sigma_{SSIM} = 0.036$ compared to 0.900 ($\sigma_{SSIM} = 0.048$) for the conventional metric, and 0.895 ($\sigma_{SSIM} = 0.050$) for the uncompensated volume. The lower $\sigma_{SSIM}$ for DL-VIF suggests more consistent performance across cases. Fig. 5B illustrates the performance of conventional and DL-VIF autofocus as a function of motion amplitude. While the performance of both approaches slightly degrades with increasing amplitude, DL-VIF performed more consistently across the range investigated.

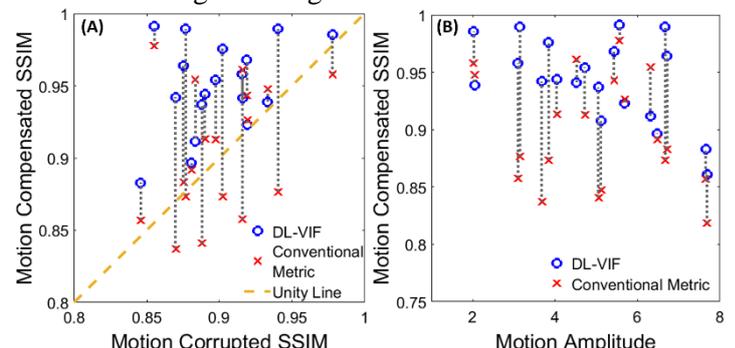

**Figure 5**: (A) SSIM for motion compensation with the conventional metric and with DL-VIF. SSIM values with DL-VIF are above the identity line for all cases, showing consistent SSIM improvement, while the conventional metric included degenerate solutions in many instances (a net decrease in SSIM). (B) SSIM as a function of motion amplitude illustrated stable compensation with DL-VIF over the full range of motion amplitude investigated.

## 4 Discussion and Conclusions

The feasibility of learning reference-free structural similarity metrics (VIF in this work) without a ground truth reference was demonstrated. The learned DL-VIF was generated by a combination of representative low and high-level features extracted with a deep CNN that contain information of both individual characteristics of the image (e.g., sharpness of edges) and structural content (the appearance of anatomical features) to match anatomical content present in the training data.

When integrated into an autofocus framework for CBCT motion compensation, DL-VIF was shown to be a more robust metric, penalizing implausible solutions that might otherwise be encouraged by conventional metrics.

Our preliminary results illustrated the capability of DL-VIF to provide better autofocus motion compensation for CBCT imaging of the abdomen, compared to a state-of-the-art autofocus metric. Ongoing work targets extension of the concept to rigid motion compensation in head CBCT and to deformable motion compensation in interventional CBCT of the abdomen.

Learning of reference-free structural similarity metrics poses a key step towards reliable autofocus motion compensation in challenging imaging scenarios, such as deformable motion compensation in soft-tissue CBCT interventional imaging.

## Acknowledgements

This research was funded by the National Institute of Health (NIH – R01EB030547) and by industry-academy research collaboration with Siemens Healthineers.

# DeepSLM: Image Registration Aware of Sliding Interfaces for Motion-Compensated Reconstruction

Markus Susenburger[1,2], Pascal Paysan[4], Ricky Savjani[5], Joscha Maier[1], Igor Peterlik[4], and Marc Kachelrieß[1,3]

[1]Division of X-Ray Imaging and Computed Tomography, German Cancer Research Center, Heidelberg, Germany
[2]Faculty of Physics, University of Heidelberg, Heidelberg, Germany
[3]Faculty of Medicine, University of Heidelberg, Heidelberg, Germany
[4]iLab, Varian Medical Systems, Baden-Dättwil, Switzerland
[5]iLab, Varian Medical Systems, Palo-Alto, CA

## Abstract

A common deep convolutional neural network architecture for deformable image registration is adapted to fit the needs of anatomical consistency for motion estimation in motion-compensated reconstruction. We introduce a sliding interface motion constraint to decompose the motion into perpendicular and tangential components in the vicinity of organ interfaces. Three separation schemes are evaluated. The results for the proposed approach, referred to as DeepSLM, show comprehensive motion adaption at the lung border for unseen test data. During inference, no additional input is needed to enable sliding lung motion registration. DeepSLM improves the registration quality and is able to learn anatomical features of the sliding lung border. The network is the basis for future investigations in motion-compensated reconstruction with deep learning techniques.

## 1 Introduction

Many motion-compensated (MoCo) methods in CT reconstruction rely on motion vector fields (MVF) estimated by deformable image registration [1]. Common deformable image registration algorithms utilize cost function optimization, featuring MVF regularization based on local smoothness [2, 3]. These common regularizers do not require any knowledge of anatomy, so the resulting fields lack physiological accuracy. Affected regions are sliding organ interfaces, where motion typically occurs in tangential direction to the organ boundary with different magnitude or even direction on each side of the border. An example for a sliding interface is the ventral body cavity. The sliding interface in the thoracic cavity — a part of the ventral body cavity — is denoted as sliding lung motion (SLM). Various methods have been published capable of dealing with SLM [4–10]. Our work is based on methods described in [4] and [9], however, instead of a conventional approach to deformable image registration, we employ convolutional neural networks (CNN) capable of dealing with SLM. The CNN in use is a modified version of VoxelMorph [11, 12], which has been successfully used in deformable image registration. For a given source and target image it predicts the MVF, which warps the source into the target. CNNs in deformable image registration has been recently explored as they speed up the registration time while achieving comparable registration accuracy [13]. VoxelMorph is built upon the U-Net architecture, which has shown to be capable of learning anatomical structures. We propose in this work a new method called DeepSLM, which modifies the VoxelMorph cost function so that it becomes capable of learning the correct estimate of SLM.

## 2 Materials and Methods

### 2.1 Deformable Image Registration

Consider a source image $f$ and a target image $g$. We search for a MVF $\boldsymbol{d}(\boldsymbol{r})$, which warps $f$ into $g$:

$$f(\boldsymbol{d}) = g$$

In conventional deformable image registration as well as in unsupervised deep learning-based methods, the MVF is found by optimizing a loss function $\mathscr{L}$ with respect to $\boldsymbol{d}$:

$$\underset{\boldsymbol{d}}{\operatorname{argmin}} \mathscr{L}(f, g, \boldsymbol{d}) = \mathscr{L}_{\mathrm{SIM}}(f, g, \boldsymbol{d}) + \lambda \mathscr{L}_{\mathrm{REG}}(\boldsymbol{d}) \quad (1)$$

$\mathscr{L}_{\mathrm{SIM}}$ measures the image similarity between $f(\boldsymbol{d})$ and $g$, while MVF regularization $\mathscr{L}_{\mathrm{REG}}$ typically enforces $\boldsymbol{d}$ to be smooth. $\lambda$ balances $\mathscr{L}_{\mathrm{SIM}}$ and $\mathscr{L}_{\mathrm{REG}}$. From a physiological point of view, smooth MVF are especially interesting in soft tissue, where the intensity differences are small. However, enforcing an isotropic smoothness is inadmissible in the case of sliding interfaces, as it generates an unphysical motion across the organ boundaries. In this case, $\boldsymbol{d}$ can be decomposed into components perpendicular and tangential with respect to the sliding organ interface:

$$\boldsymbol{d} = \boldsymbol{d}_\perp + \boldsymbol{d}_\parallel \quad (2)$$

The organ surface is conveniently described with a map of normal vectors $\boldsymbol{n}$ which point towards the organ boundary. Now we can explicitly decompose $\boldsymbol{d}$ into

$$\boldsymbol{d}_\perp = \frac{(\boldsymbol{n} \cdot \boldsymbol{d})\boldsymbol{n}}{\boldsymbol{n}^2}, \quad \boldsymbol{d}_\parallel = \boldsymbol{d} - \boldsymbol{d}_\perp. \quad (3)$$

Assuming vanishing gradient of the normal vector map $\nabla \boldsymbol{n} \approx 0$, it can be shown that

$$||\nabla \boldsymbol{d}||_2^2 \approx ||\nabla \boldsymbol{d}_\perp||_2^2 + \lambda_\parallel ||\nabla \boldsymbol{d}_\parallel||_2^2. \quad (4)$$

With a regularization constructed from this weighted decomposition, we can reward smooth MVF along the direction perpendicular to the organ boundaries while suppressing this reward along the direction tangential to the organ boundary. This allows sliding (e.g. to allow expansion of lungs during breathing) by the spatial variation of $\lambda_\parallel$, which is defined by:

$$\lambda_\parallel = \begin{cases} 1 & \text{if } |w| > 2\varepsilon \\ 0 & \text{if } |w| < \varepsilon \\ |w|/\varepsilon - 1 & \text{else.} \end{cases} \quad (5)$$

The map $w$ describes the distance of each point to its closest sliding organ boundary, it is derived from the signed Euclidean distance transform of a ventral cavity segmentation $m$ (Fig. 3). The parameter $\varepsilon$ defines the vicinity of the organ boundary; it is set to a value of 9 mm.

## 2.2  MVF Regularization Loss Functions

In this paper, we examine four networks based on the following MVF regularization loss functions:

$$\mathscr{L}_{\text{REG}} = \frac{1}{3N} \sum_r ||\nabla \boldsymbol{d}||_2^2 \quad (6a)$$

$$\mathscr{L}_{\text{REG}} = \frac{1}{3N} \sum_r (||\nabla \boldsymbol{d}_\perp||_2^2 + \lambda_\parallel ||\nabla \boldsymbol{d}_\parallel||_2^2) \quad (6b)$$

$$\mathscr{L}_{\text{REG}} = \frac{1}{3N} \sum_r \begin{cases} ||\nabla \boldsymbol{d}||_2^2 & \text{if } \lambda_\parallel = 1 \\ ||\nabla \boldsymbol{d}_\perp||_2^2 + \lambda_\parallel ||\nabla \boldsymbol{d}_\parallel||_2^2 & \text{else.} \end{cases} \quad (6c)$$

$$\mathscr{L}_{\text{REG}} = \frac{1}{3N} \sum_r \lambda_\parallel ||\nabla \boldsymbol{d}||_2^2 \quad (6d)$$

The weighting factor $3N$ refers to the number of entries in $\boldsymbol{d}$, connecting all $N$ voxels of $f$ to $g$. The first term is the conventional regularization to enforce smoothness of $\boldsymbol{d}$. The following three regularization loss functions are introduced in this study to regularize with respect to SLM. They implement the SLM constraint globally ($\boldsymbol{d}$ is decomposed for all spatial locations into $\boldsymbol{d}_\perp$ and $\boldsymbol{d}_\parallel$), locally (the decomposition occurs only close to the segmentation boundary) and strictly (the smoothness of $\boldsymbol{d}$ close to the border is not enforced).

## 2.3  Evaluation Metrics

For evaluation we used the Dice of the ventral cavity segmentation ($S$), the normalized cross-correlation coefficient (NCC) (applied window-wise) and the mean-square error (MSE) (also used during training):

$$\text{MSE} = \frac{1}{N} \sum_r ||f(\boldsymbol{d}) - g||_2^2 \quad (7)$$

$$\text{NCC} = \frac{1}{Q} \sum_q \frac{(\sigma_{f(\boldsymbol{d}),q} - \mu_{f(\boldsymbol{d}),q})(\sigma_{g,q} - \mu_{g,q})}{\sigma_{f(\boldsymbol{d}),q}\sigma_{g,q}} \quad (8)$$

$$\text{Dice} = \frac{2|S_{f(\boldsymbol{d})} \cap S_g|}{|S_{f(\boldsymbol{d})}| + |S_g|} \quad (9)$$

The NCC is calculated over small moving windows $q$ ($9 \times 9 \times 9$) and the values are averaged by the total number of windows Q, $\mu$ indicates the mean of a window and $\sigma$ its standard deviation. For each patient, each phase is registered towards the patient's first respiration phase and mean relative changes are reported for training and test data (Tab. 1).

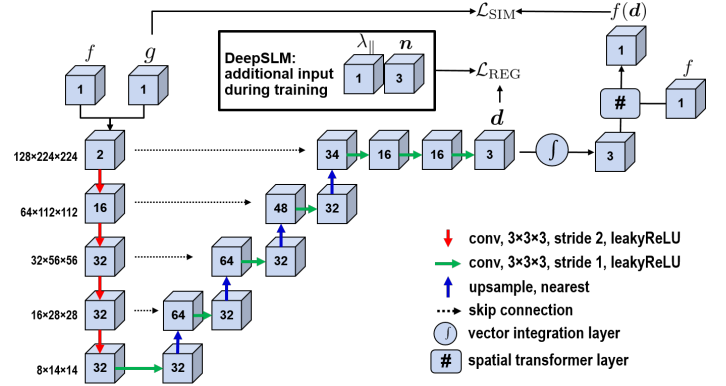

**Figure 1:** Scheme of the DeepSLM architecture. The cuboids indicate a volume with the number of feature channels on its outer face. The sizes of layers at each stage are shown on the left (z-y-x order). Source $f$ and target $g$ are concatenated and send in the U-Net encoder/decoder path. The output is reduced by three additional convolution layers to the MVF $\boldsymbol{d}$. The vector integration layer constraints $\boldsymbol{d}$ to diffeomorphic solutions [3] before it is used to warp $f$ towards $g$. The similarity loss is calculated between the warped source $f(\boldsymbol{d})$ and the target $g$. The gradient loss is calculated on the non-integrated $\boldsymbol{d}$. During inference, the additional DeepSLM input is not needed.

## 2.4  Data and Training

We trained the network on thoracic 4D CT data, consisting of 83 patients with 10 respiratory phases each. The data were acquired on a diagnostic CT system. Manual segmentations on the ventral body cavity have been performed by specialists and were then used to train a segmentation network for autosegmentation, providing autosegmentations for all respiratory phases. The patient data sets were split into 66 training and 17 test sets.

Each scan is transformed to a spatial grid of $3 \times 3 \times 3$ mm$^3$. Then the center of mass for the ventral cavity segmentation is calculated and each scan and segmentation is shifted such that the center of mass is in the middle of the final volume size of $224 \times 224 \times 128$ voxels. If the scan dimensions do not match the final size, the volume is either constantly padded or cropped. A signed Euclidean distance transform is calculated on each segmentation (see Fig. 2 for details), resulting in a map $w$ of positive distances inside the segmentation and negative distances outside. The map of normal vectors $\boldsymbol{n}$ is the gradient of $w$ and indicates for each voxel the direction towards the closest point on the segmentation border. The tangential weight map $\lambda_\parallel$ is calculated once on the distance map and forms together with normal vectors the additional input to the regularization loss of the DeepSLM network variants.

During training, a random patient is chosen at each iteration.

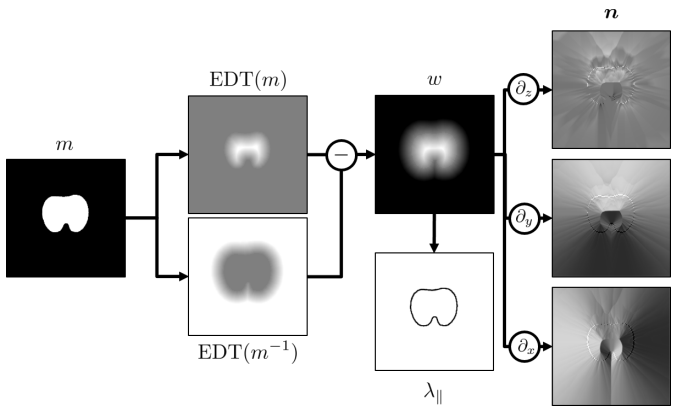

**Figure 2:** From an initial binary segmentation of the ventral body cavity $m(\boldsymbol{r})$, two Euclidean distance transforms (EDT) are performed to find the distance map on $m(\boldsymbol{r})$ and its inverse. The final distance map $w(\boldsymbol{r})$ is the difference of the outer and inner distance maps. The normal vector field $\boldsymbol{n}(\boldsymbol{r})$ is the gradient of $w(\boldsymbol{r})$. The tangential weight map $\lambda_{\parallel}(w(\boldsymbol{r}))$ is defined in Eq. 5.

From this patient two different randomly selected phases are used as input to the neural network. This results in $66 \times 90 = 5940$ combinations for training input. Each network is trained with stochastic gradient descent for 150000 iterations. Four networks are trained on an NVIDIA RTX 2080 GPU with MSE as the $\mathcal{L}_{\mathrm{SIM}}$ and the $\mathcal{L}_{\mathrm{REG}}$ defined in Eq. 6(a-d). The architecture (Fig. 1) is based on the VoxelMorph [12] PyTorch (v 1.7) implementation [14]. The MVF is finally applied within a spatial transformer layer on the source $f$. Our modifications to the default VoxelMorph configuration: The MVF are calculated at full resolution ($224 \times 224 \times 128$). The weight for the auxiliary term is set to $\lambda = 0.01$ (Eq. 1). The networks are trained with automatic mixed-precision, which results in a training time of 27 to 33 hours.

## 3  Results

All flavors of DeepSLM show better registration performance than the unmodified VoxelMorph in both training and testing. The mean registration results of one test patient are displayed in Fig. 3. The image similarity metrics are consistently increased by DeepSLM, the ranking of methods is from best to worse: local, strict, global DeepSLM, VoxelMorph (Tab. 1). The Dice for the ventral cavity segmentation before and after registration is decreased for all methods. The motion vectors have a lower magnitude outside the lung (Fig. 4) in DeepSLM. The strict DeepSLM shows less smooth MVF at the lung border.

## 4  Discussion

The increased registration performance of DeepSLM compared to VoxelMorph demonstrates the importance of proper handling of sliding interfaces in deformable image registration. From the different flavors of DeepSLM, the local method performs best in terms of similarity loss and in MVF appearance at the lung border on the test data. The strict variant of DeepSLM has its disadvantage of unregulated MVF at the segmentation border. The global variant decouples the normal and tangential motion components at all locations, i.e., also in the areas where the motion is coupled in all directions (e.g., the central area of ventral cavity). For this reason, the regularization is not sufficient and the training is dominated by the similarity term, which yields goods results in terms of MSE, but does not correspond to the physical motion of the tissue. The effect of the SLM loss is visible in the test data, which have not seen the segmentations during evaluation. This shows DeepSLM is capable of learning anatomical features such as the lung border. A downside of the deep learning approach is the coarse resolution of the registration volumes, which is limited by the memory of the used GPU. Unsupervised learning, employed in this paper, is necessary due to the lack of ground truth MVF. In this paper, we have shown that proper handling of sliding interfaces, necessary for modeling of organ motion, can be introduced into a deep learning-based deformable image registration suitable for MoCo. There are two main advantages of our approach: First, the registration time is significantly reduced by deploying the training of the network before the application of the network, and, secondly, SLM methods can work without additional segmentation in the application. Further research is needed in order to transfer the registration from the well-sampled 4D CT to the sparse 4D CBCT data used in 4D MoCo reconstruction.

| Training data | VM | global DeepSLM | local DeepSLM | strict DeepSLM |
|---|---|---|---|---|
| $\Delta$MSE | 81.57% | 82.10% | 82.86% | 82.53% |
| $\Delta$NCC | 5.09% | 5.53% | 5.40% | 5.24% |
| $\Delta$Dice | -0.28% | -0.20% | -0.27% | -0.33% |
| Test data | | | | |
| $\Delta$MSE | 86.66% | 87.29% | 88.35% | 88.05% |
| $\Delta$NCC | 7.91% | 8.45% | 8.59% | 8.40% |
| $\Delta$Dice | -0.19% | -0.12% | -0.20% | -0.23% |

**Table 1:** Mean relative change $\Delta$ for MSE, NCC and Dice evaluated on the training and test data for the networks: VoxelMorph (Eq. 6a), global DeepSLM (Eq. 6b), local DeepSLM (Eq. 6c) and strict DeepSLM (Eq. 6d). The relative change is evaluated by registering all respiratory phases of one patient to the first respiratory phase of that patient. Then, the values are averaged for all patients in the test and training data. The change is calculated as the relative difference to the preregistration measure. Positive/negative values indicate improvement/declination.

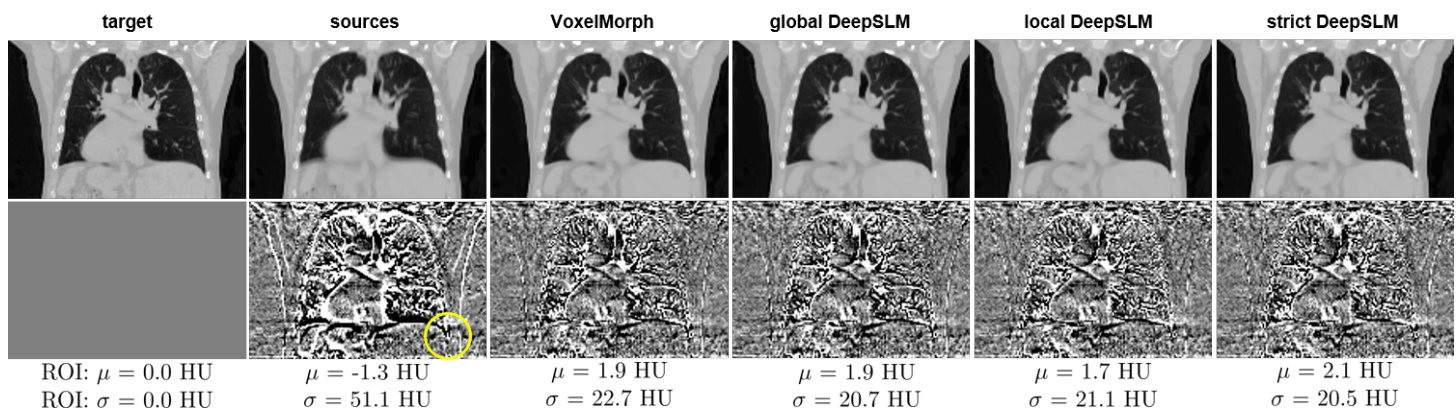

**Figure 3:** Ten respiration phases from one patient are registered using the models defined by Eq. 6–6d to the first phase of that patient data set. Upper row: The sharp target phase, the mean of the source images and the mean of the registration results for each network ($C = -250$ HU, $W = 1500$ HU). Lower row: The difference of target phase and registration mean ($C = 0$ HU, $W = 50$ HU) together with the mean and standard deviation of a SLM affected ROI.

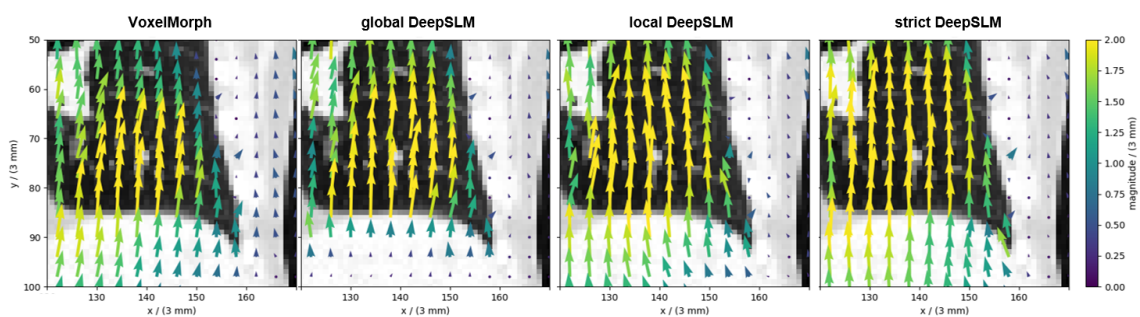

**Figure 4:** The resulting MVF for SLM show good motion decoupling of the lung interior and exterior. The DeepSLM methods decrease the motion transfer from the diaphragm to the lung exterior.

# Prior-aided Volume of Interest CBCT Image Reconstruction

Daniel Punzet[1,3], Robert Frysch[1,3], Oliver Speck[2,3], and Georg Rose[1,3]

[1]Institute for Medical Engineering, Otto-von-Guericke-University Magdeburg, Magdeburg, Germany
[2]Department of Biomedical Magnetic Resonance, Otto-von-Guericke-University Magdeburg, Magdeburg, Germany
[3]Research Campus *STIMULATE*, Magdeburg, Germany

**Abstract**  Typical CBCT acquisition protocols require the user to acquire two orthogonal fluoroscopic images for the purpose of positioning the patient ideally for the following acquisition. Since these fluoroscopic positioning images are fairly low-dose, they are typically not truncated even for volume-of-interest imaging acquisitions, except in the case of a too small detector. Yet, they share the positioning of the CBCT acquisition succeeding them and can therefore provide additional information to the reconstruction from truncated projections. We present a prior-aided reconstruction scheme which registers these fluoroscopic images to potentially available priors of the same patient in order to enhance CBCT volume-of-interest imaging acquisitions and ultimately reduce patient dose. We make use of a novel registration method which moves the computationally expensive steps of the registration to a timepoint prior to the volume-of-interest acquisition and therefore allows for fast registration of priors for usage in interventional volume-of-interest imaging settings.

## 1   Introduction

Imaging techniques like computed tomography (CT) enable physicians to look at slice images of a patient. These slices typically show the complete cross-section of a patient. In many clinical situations though, physicians are only interested in very limited regions of a patient. In this case volume-of-interest (VOI) imaging, which irradiates only small parts of a patient, can lead to considerable dose reduction. Inevitably though, the projection data will be incomplete since the projections are cut-off transversally. This problem is called truncation. The tomographic images reconstructed from truncated projections can be severely impaired by image artifacts depending on the degree of truncation. Therefore, the achievable dose reduction is limited by demands of a certain image quality. Typical strategies to counter the degradation in image quality stemming from truncation try to extrapolate the limited data heuristically [1] or try to minimize the impact of the filtering step [2] [3] in the reconstruction process which is one of the main contributing factors for truncation artifacts. Another strategy is to utilize available prior knowledge, e.g. in the form of prior images of the patient. Practical example scenarios where prior data is available are interventions such as the placement of a stent or coil. These are usually planned using an angiogram and performed under guidance of fluoroscopic images acquired with an angiographic C-arm system. Often prior CT or magnetic resonance images are available as well. Upon completion of the intervention, an additional cone-beam CT (CBCT) acquisition is performed in order to check for correct positioning and deployment of the stent structure. For this final acquisition however, generally no prior knowledge is used, even though the expected changes in the volume are minimal. We present a prior-aided reconstruction scheme which makes use of prior data to allow for strongly truncated acquisitions without suffering from the typical severe truncation artifacts. We make use of a novel registration method which moves the computationally expensive steps of the registration to a timepoint prior to the intervention and therefore allows for fast registration of priors for usage in the prior-aided image reconstruction method.

## 2   Method

All simulations have been performed with the Computed Tomography Library (CTL) tool [4] available at https://gitlab.com/tpfeiffe/ctl/.

### 2.1   Registration

A rigid registration of prior 3D volume data to 2D fluoroscopic images is performed using a recently published registration method [5] making use of Grangeat's relation. This method works by minimizing the registration error on Grangeat's intermediate function of both the 2D projection and the 3D volume. The advantage of this approach is that the computationally expensive steps, such as the computation of the 3D radon transform of the volume and the intermediate function of the projection data can be performed ahead of time and are then available in a pre-computed form during the actual registration process which then reduces to inexpensive re-sampling of the intermediate space. The computation speed, which is normally a limiting factor for the usage of registration methods, is therefore greatly increased compared to conventional projection-based registration methods. For further details on the Grangeat registration method we refer to the corresponding publication with the modification [eq. (11) [5]], in order to be more robust against the unavoidable axial truncation of the projection images.

### 2.2   Reconstruction

The prior-aided reconstruction is initialized with the prior volume $x^0$ registered to match the truncated projection data $g_{\text{trunc}}$. An OS-SART reconstruction algorithm then updates the initialized volume using the truncated projection images and corresponding projection matrices that have been corrected by the precedent registration. The update using the system matrix $\mathbf{A}_j$ of the $j$th subset reads

$$\boldsymbol{x}^{i+1} = \boldsymbol{x}^i - \omega\mathbf{A}_j^{\mathsf{T}}(\mathbf{A}_j\boldsymbol{x}^i - \boldsymbol{g}_{\text{trunc}}),$$

where $\omega \in (0, 1)$ denotes a relaxation parameter, which is determined using the Power method [6]. Note that no further regularization was used in this study.

## 2.3 Simulation & Evaluation

An interventional setting was simulated to evaluate the proposed method on (reprojected) clinical head and abdomen volumes from the low-dose CT grand challenge [7] [8] datasets. Figure 1 illustrates the simulation setup. First, the clinical volume data is used to create different volumes simulating the different states of the patient during the acquisitions of the prior, the interventional fluoroscopy and the final CBCT for outcome control upon completion of the intervention. Depending on the dataset, three volumes are created in a different fashion to simulate region-specific challenges for the registration procedure.

**Abdomen** The breathing deformation method published in [9] was applied to generate different breathing states for all three volumes. Furthermore, to account for additional changes in the VOI due to interventional instruments etc., the voxel volume of a stent is subsequently placed in volumes 2 and 3. This stent might not coincide with real anatomic structures and is only inserted to demonstrate the robustness of the proposed method to smaller changes within the VOI between the prior and interventional acquisitions.

**Head** The patient and patient table were manually segmented from the initial volume. A rigid displacement (translation and rotation) was then applied only to the patient while keeping the patient table stationary resulting in a combined dual-rigid i.e. non-rigid displacement of the total volume. Furthermore, the same stent voxel volume used already in the abdomen dataset was subsequently placed in volumes 2 and 3.

The resulting three different volumes are then forward-projected using different CBCT acquisition setups. A non-truncated shortscan acquisition (496 views) of volume 1 is performed and reconstructed (FDK or ART) to form the prior volume data. Of volume 2 fluoroscopic projections are acquired to simulate the fluoroscopic positioning images prior to a shortscan acquisition. The shortscan acquisition (496 views) is then performed on volume 3. The projection data of volume 3 is truncated by cropping the projection data in transversal and axial direction and modifying the projection matrices accordingly. Note that all scanned volumes differ either in their breathing states (abdomen volume) or some non-rigid displacement (head volume). Furthermore, there is a rigid displacement introduced between volume 1 and volumes 2 and 3 to account for the fact that the position of a prior is generally not available at the time of a later intervention. The displacement used in this study mimics the displacement derived from a real patient's interventional dataset. Volumes 2 and 3 share the same positioning since the fluoroscopic acquisitions (of volume 2) precede the following shortscan (of volume 3) in clinical routine with the same positioning.

The proposed reconstruction scheme now uses the non-truncated fluoroscopic images and registers them to the prior volume. As mentioned before, it can be assumed that the fluoroscopic images and the truncated shortscan acquisition share the same rigid translation parameters, i.e. positioning, and therefore the prior, when successfully registered to the fluoroscopic images, is also registered to the truncated acquisition. The reconstruction is then performed using a prior-initialized iterative reconstruction method, which reconstructs the volume of interest from the truncated projections. To estimate the performance of the prior-aided reconstruction, additionally, reconstructions from just the truncated projection data and from simple mirror-extrapolated projection data are performed and compared to the prior-aided reconstruction. A reconstruction from the untruncated projection data provides the groundtruth to compare to. Note that the initialization with the prior does not affect the reconstructed volume of interest in a negative way since the VOI is reconstructed from the truncated data. The initialization only aids the reconstruction with correctly distributing the excessive absorption values outside of the VOI correctly instead of forming the ring artifact and image offset typical for truncated reconstructions.

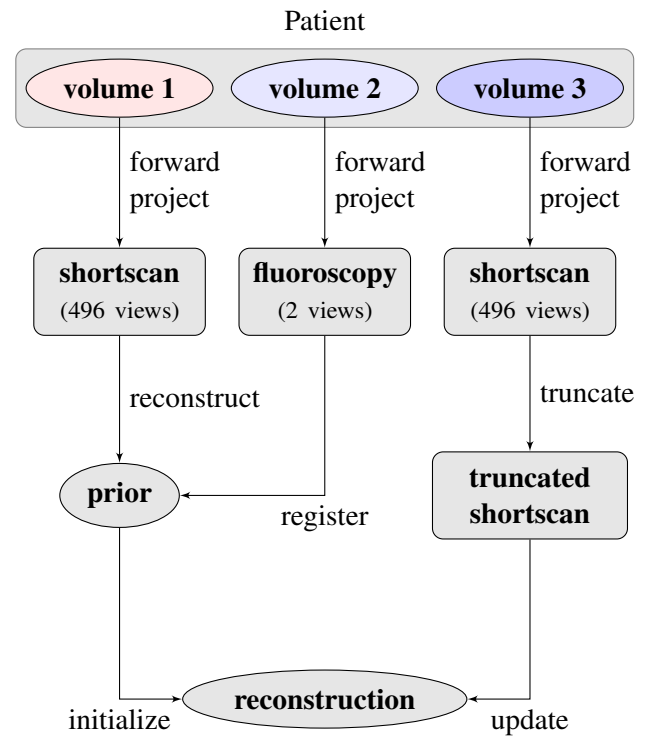

**Figure 1:** Reconstruction scheme evaluation setup.

## 3   Results

First, the accuracy of the Grangeat registration method compared to a conventional digitally reconstructed radiograph (DRR) method is demonstrated in Fig. 2. For this, instead of taking just two projections, a complete shortscan acquisition with 496 views was performed on volume 2 in order to have a larger sample base for assessing the registration quality. Shown are the registration parameters (translations $x$, $y$, $z$ and rotations $\alpha$, $\beta$, $\gamma$) obtained from each individual view for both methods applied on the head dataset. The corresponding dashed lines mark the true values for each parameter. The overall good fit of the Grangeat registration parameters in Fig. 2a reflects also in the calculated mean target registration errors (mTRE) shown in Table 1. This table shows the mTREs averaged over the whole 496 views.

**Table 1:** Mean Target Registration Errors

|  | Grangeat | DRR |
|---|---|---|
| average mTRE | 0.8662 mm | 2.6844 mm |
| min mTRE | 0.0470 mm | 0.1194 mm |
| max mTRE | 1.9659 mm | 5.6186 mm |

One thing to mention here is the susceptibility of the Grangeat registration method to truncation. Because the 3D Radon space is computed, consisting of plane integrals, strong truncation can pose problems to this registration method as the integrals become incomplete. In the case of the abdomen dataset this was more severe than for the head dataset as truncation in that case occurs in both axial directions of the patient instead of just one. Thus, the registration of the abdomen was found to be more demanding on the Grangeat registration than the registration of the head.

Figures 3 and 4 show the reconstruction results and corresponding profile plots as indicated in the reconstruction images. Shown is the central slice of the abdomen volume. The groundtruth reconstruction in Fig. 3a is the OS-SART reconstruction of volume 3 from untruncated projection data. The region of interest (ROI) in this slice is indicated by the green circle in Fig. 3a. The proposed prior-aided reconstruction method is shown in Fig. 3b. For comparison the OS-SART reconstruction from truncated projection data (Fig. 3c) and mirror-extrapolated FDK reconstruction (Fig. 3d) are shown. Figs. 3e to 3h show the magnified ROIs with the rest of the volumes masked out. From Fig.3 it is obvious that the proposed prior-aided reconstruction method achieves the best image quality. The anatomic details of the patient within the ROI are reconstructed correctly and the stent is clearly visible. Both the cupping artifact and the general offset of the reconstructions are reduced considerably. This shows also in the profile plots in Fig. 4 and the structural similarity index measures (SSIMs) indicated in Figs. 3f to 3h. Furthermore, the anatomic structures outside of the ROI visible in the prior-aided reconstruction in Fig. 3b might allow for better orientation for the operating surgeon. However, the validity of these structures cannot be guaranteed as they might originate from the prior volume and information from outside of the ROI should therefore be used with caution.

Figs. 5 and 6 show the reconstruction result for the head dataset. Shown are also the volumes of the prior before and after successful registration. The window $(C, W) = (-23, 38)$ HU in Fig. 5 was chosen to demonstrate that even some soft tissue regions can be discerned in the prior-aided reconstruction in Fig. 5b. Both the OS-SART reconstruction from truncated projection data and FDK reconstruction from mirror-extrapolated projection data fail to provide usable reconstructions at this windowing and are therefore not displayed as images. However, the profile is plotted in Fig. 6 in purple. Furthermore, from Figs. 5 and 6 it is apparent that the prior-aided reconstructions show a good agreement with the groundtruth within the ROI and slightly outside of it. The further away from the ROI, the more influences from the prior remain and do not agree with the groundtruth anymore. This is especially apparent looking at the alignment of the patient table.

## 4   Conclusion & Outlook

We have shown that by registering a prior to fluoroscopic images taken during an intervention, the reconstruction of the following CBCT acquisition can be enhanced by making use of the prior to initialize the iterative reconstruction algorithm and thereby suppressing the formation of the typical truncation ring artifact and erroneous offset. The registration performs well compared to established DRR-based registration methods while providing a clear advantage in computation speed. The resulting reconstructions show overall improved image quality. This also holds for far off-center slices and thereby would allow for further dose reductions by enabling stronger axial truncation while maintaining a comparable image quality. Future work will focus on applying the proposed method on real clinical interventional data and also for prior data from different modalities, e.g. conventional CT or even MRI. Furthermore, approaches to impose further constraints on the reconstruction method taking the prior data into account might prove beneficial and are worth investigating.

## Acknowledgment

This work was conducted within the context of the International Graduate School MEMoRIAL at Otto von Guericke University (OVGU) Magdeburg, Germany, kindly supported by the European Structural and Investment Funds (ESF) under the programme Sachsen-Anhalt WISSENSCHAFT Internationalisierung (project no. ZS/2016/08/80646). The work of this paper is also funded by the German Ministry of Education and Research (BMBF) within the Forschungscampus STIMULATE under grant number 13GW0473A.

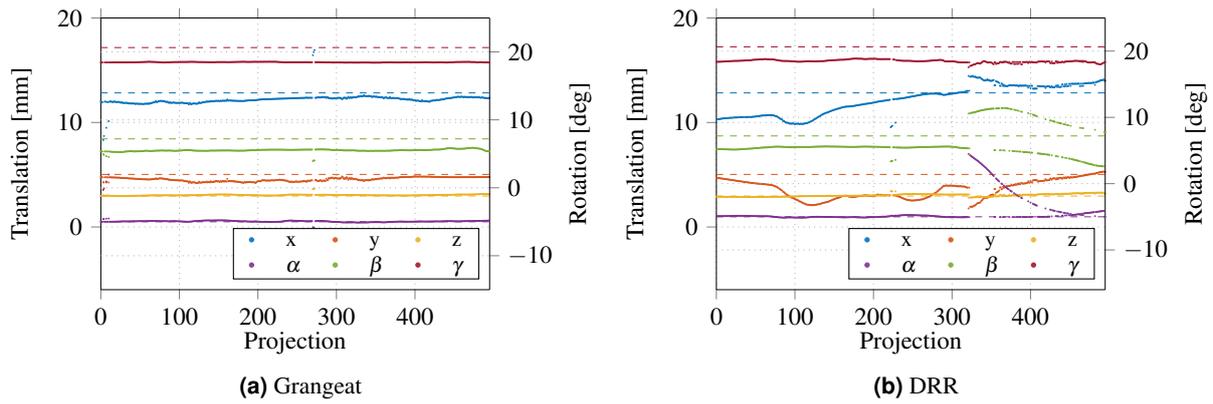

**(a)** Grangeat                          **(b)** DRR

**Figure 2:** Registration parameters as derived from Grangeat registration (Fig. 2a) and DRR registration (Fig. 2b).

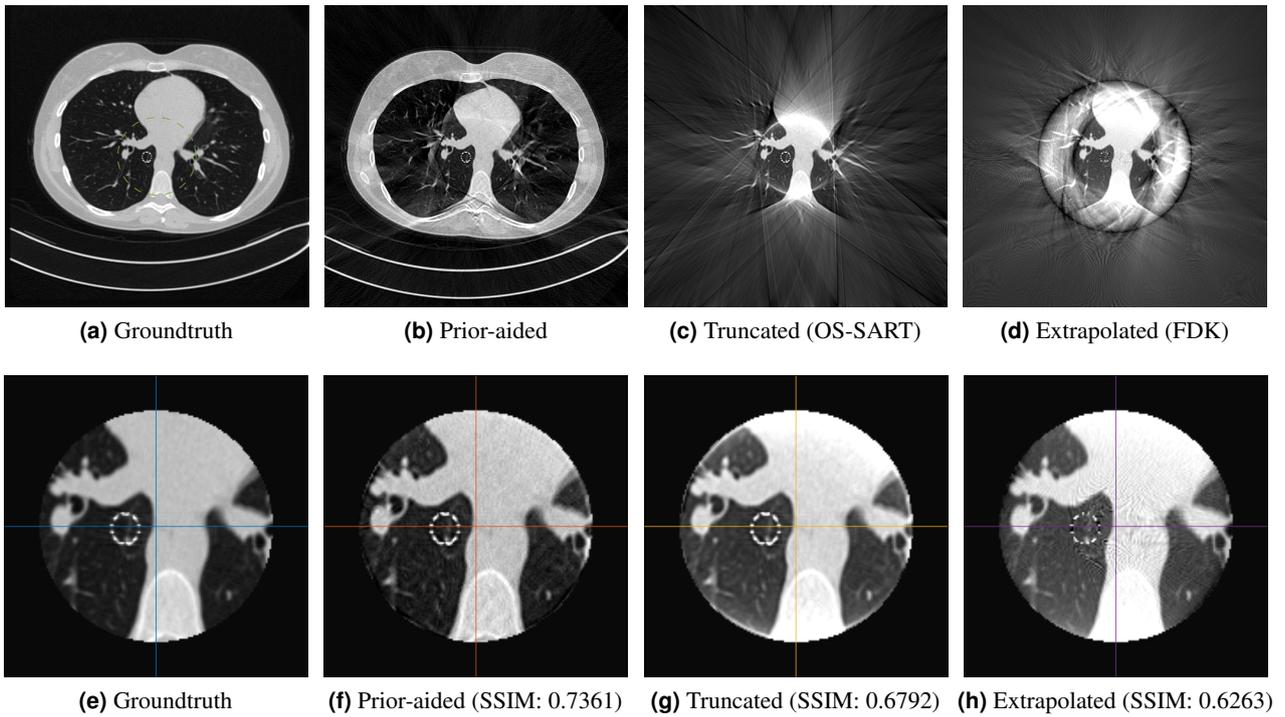

**(a)** Groundtruth          **(b)** Prior-aided          **(c)** Truncated (OS-SART)          **(d)** Extrapolated (FDK)

**(e)** Groundtruth     **(f)** Prior-aided (SSIM: 0.7361)     **(g)** Truncated (SSIM: 0.6792)     **(h)** Extrapolated (SSIM: 0.6263)

**Figure 3:** Reconstructions of the central slice of the abdomen dataset. (a): Groundtruth reconstruction from untruncated data with ROI marked in green, (b): the proposed prior-aided reconstruction, (c): OS-SART reconstruction from truncated data, (d): FDK reconstruction from mirror extrapolated data, (e)-(h): corresponding enlarged ROIs. Constant mutual window of (C, W) = (-330, 1445) HU.

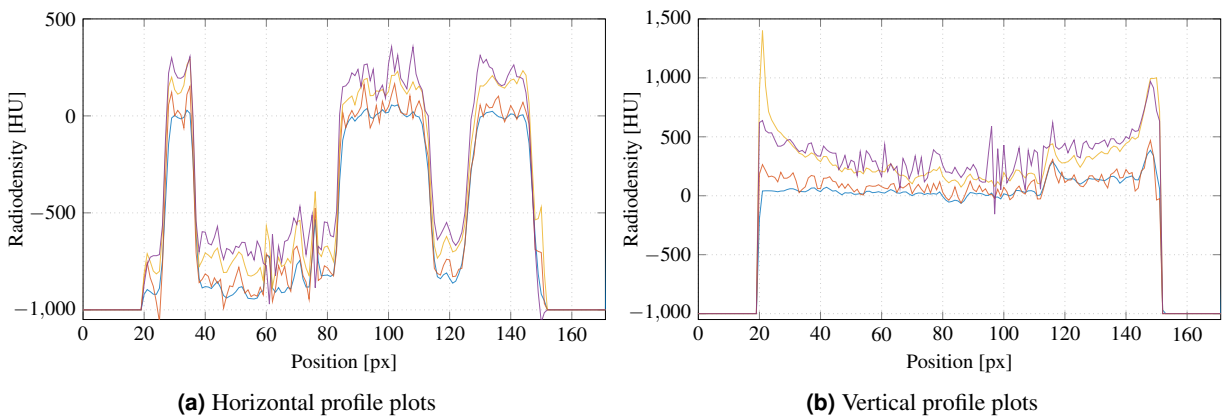

**(a)** Horizontal profile plots                    **(b)** Vertical profile plots

**Figure 4:** Profile plots of Figs. 3e (blue), 3f (red), 3g (yellow) and 3h (purple) as indicated.

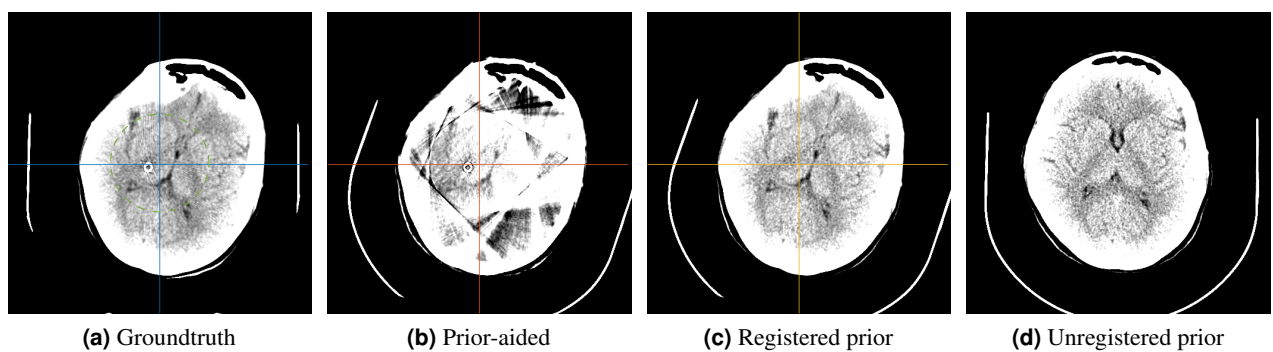

**(a)** Groundtruth    **(b)** Prior-aided    **(c)** Registered prior    **(d)** Unregistered prior

**Figure 5:** Reconstructions of the same slice of the head dataset. (a): Groundtruth reconstruction from untruncated data with ROI marked in green, (b): the proposed prior-aided reconstruction, (c): the prior volume after registration, (d): the prior volume before the registration. Constant mutual window of (C, W) = (23, 38) HU.

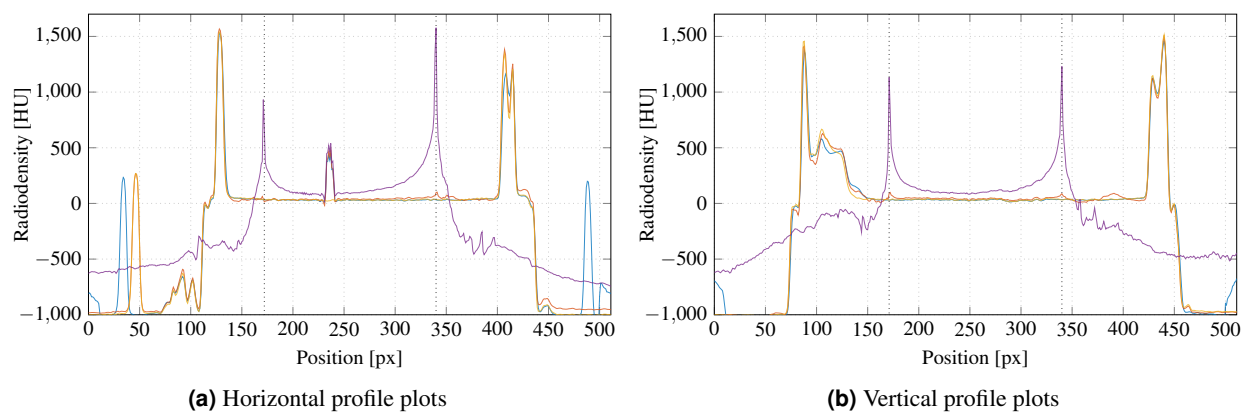

**(a)** Horizontal profile plots    **(b)** Vertical profile plots

**Figure 6:** Profile plots of Figs. 5a (blue), 5b (red) and 5c (yellow) as indicated. Additionally, the profile plot of an OS-SART reconstruction from truncated data (purple, no image shown) for comparison. The dotted lines mark the borders of the ROI.

# Data-driven motion compensated SPECT reconstruction for liver radioembolization

Antoine Robert[1,2], Simon Rit[1], Julien Jomier[2], and David Sarrut[1,3]

[1]Univ.Lyon, INSA-Lyon, Université Claude Bernard Lyon 1, UJM-Saint Etienne, CNRS, Inserm, CREATIS UMR 5220, U1206, F-69373, Lyon, France.
[2]Kitware SAS, 6 Cours André Philip, 69100 Villeurbanne
[3]Centre Léon Bérard, 28, rue Laennec, 69373 Lyon Cedex 08, France

**Abstract** The need for quantitative accuracy of single photon emission computed tomography (SPECT) image analysis is increasing with the emergence of targeted radionuclide therapies such as liver radioembolization. Breathing motion is a major issue for quantitation as it leads to misestimation of the tumor activity in the SPECT images. In this paper, we developed a data-driven motion compensated SPECT reconstruction algorithm to account for respiratory motion. A respiratory signal was retrospectively extracted from SPECT list-mode data with the Laplacian Eigenmaps algorithm and used to sort the projections into temporal bins of fixed phase width. A 2D affine motion was then estimated between projections at different phases. The transformation parameters were used to re-bin the list-mode data into one set of compensated projections that was then used to reconstruct a 3D motion-compensated SPECT image using all available events of the list-mode data. The method was evaluated on both simulated and real SPECT acquisitions of liver patients, and compared to respiratory-gated reconstruction. The motion-compensated reconstruction retrieved larger activity in the tumors compared to conventional 3D SPECT reconstruction with a better contrast-to-noise ratio than gated reconstruction.

## 1 Introduction

Single photon emission computed tomography (SPECT) is a key tool for imaging cancer, both for diagnosis and therapeutic purposes. The recent emergence of targeted radionuclide therapies, such as liver radioembolization or neuroendocrine tumors treated with $^{177}$Lu, increased the need for quantitative SPECT analysis, both for pre-treatment planning or per-treatment activity distribution monitoring. One key step in liver radioembolization is the pre-treatment $^{99m}$Tc SPECT/CT acquisition used to assess lung shunt and extrahepatic uptake. This acquisition can also be used to evaluate the planned dose delivered by therapeutic $^{90}$Y microspheres injected in the liver. The accuracy of this patient-specific treatment planning directly depends on the accuracy of the pre-treatment SPECT images.

Respiratory motion has a major impact on the image quality by blurring the SPECT image. For example, Bastiaannet *et al* [1] showed on simulated data that it may underestimate the SPECT activity in liver tumors. A widely used method to correct for breathing motion in tomography is respiration-gated reconstruction. It consists in using a respiratory signal to sort the measured projections in small temporal respiratory gated frames with minimal motion. Then, the sorted projections are reconstructed phase per phase yielding a series of motion-free volumetric images [2]. This method has proven effective to reduce the blur around moving tumours and improves the quantification of the tracer concentration. However, the lower photon count in the projection data of each individual time frame compared to conventional reconstruction leads to SPECT images with a poorer signal-to-noise ratio.

Other methods like motion-compensated reconstruction potentially allow to reconstruct images without motion artifacts while using all the available data. This type of method requires the knowledge of the motion of the patient during the whole acquisition. Then, the motion information can be used to combine all individual gated frames into a single motion corrected image either before [3], during [4] or after [5] the reconstruction.

A common method to retrieve the motion vector field is to use a previous 4D image of the same patient, e.g., a 4D CT [4]. However, a 4D image is rarely available in clinical practice and the respiratory motion changes from day to day. Another approach is to estimate the breathing motion field with image registration between the frames of the respiratory-gated PET or SPECT images [3, 6]. In that case, a prior gated reconstruction, which might be time consuming, is needed. Also, additional hardware is often used to get the respiratory signal needed for the gated reconstruction.

One way to avoid the previous reconstruction step is to estimate the movement in projection space. For example, Bruyant *et al* [7] tracked the center of mass of the gated projections to get the motion information. However, this method only assumes rigid motion and might not be suitable for complex tumor deformation.

In this paper, we correct for breathing motion in a conventional SPECT acquisition, without extra hardware or extra image. Affine deformation is estimated between gated projections and accounted for by re-binning list-mode data. The respiratory signal used for gating the projections is extracted from the list-mode data. The method was validated on simulated data and real patient acquisitions and compared to uncorrected and respiratory-gated reconstruction.

## 2 Materials and Methods

We assume a list-mode dataset of a conventional 3D SPECT acquisition, i.e., a list of *events* describing the spatial position, the energy and the time of each photon impinging on the detector for a set of detector positions around the patient.

## 2.1   Breathing signal extraction

A respiratory signal was first estimated using the algorithm described in [8]. The algorithm bins the list-mode data into low-count $256 \times 256$ pixel projections at a high framerate (200 ms) and uses Laplacian Eigenmaps to reduce the dimensionality to a 1D breathing signal sampled at 200 ms. The breathing phase was computed by assuming linearity from 0% to 100% between two consecutive end-inhale positions.

## 2.2   Motion estimation

The projections were then gated using the breathing phase. Projections were sorted into temporal bins sampling the breathing cycle regularly, eight for patient acquisitions and twenty for simulated data. The projections were acquired with a dual-head SPECT system (section 2.5), thus acquiring two opposite projections per gantry angle. For motion estimation, one of the two projections was laterally flipped and added to the other to improve the signal to noise ratio during the registration.

A reference bin was chosen and the other projection bins were registered to it using the Elastix image registration package [9]. Registration computed a 2D affine transformation (6 degrees of freedom) using the correlation coefficient as similarity measure and adaptive stochastic gradient descent for the optimisation. The resulting transformations were then applied to the list-mode data to bin a set of motion-compensated projections, keeping the two opposite projections separated. This step compensates the *apparent* 2D projective motion in the projections.

For each acquisition, we generated one set of compensated projections corresponding to the end-inhale position. This bin was chosen because it minimizes the attenuation correction error made by the use of a 3D attenuation map for liver tumors located close to the lungs.

## 2.3   SPECT reconstruction

All volumes were reconstructed using the Reconstruction Toolkit (RTK) [10]. The projections were reconstructed with 20 iterations and 4 subsets into a $128^3$ voxel matrix (voxel size 4.42 mm), using the OSEM algorithm with a quadratic penalization. All the reconstructions included scatter and attenuation correction. For the attenuation correction, the 3D CT image acquired by the SPECT/CT system (section 2.5) was used.

For each dataset, we reconstructed one conventional 3D image with no motion compensation, one respiratory-gated 4D image and one motion-compensated 3D image with the proposed method. In addition, for the simulation dataset, reference 3D images were reconstructed with the patient static at end-inhale.

## 2.4   SPECT simulation

The SPECT acquisition of a breathing patient was simulated with the Monte Carlo software Gate [11]. The simulated SPECT scanner was the same as the one of the patient acquisitions (section 2.5). Patient motion over a breathing cycle was estimated on the 4D CT image of a thoracic patient. The end-exhale phase was used as reference and registered to each of the nine other phases using 3D deformable image registration. The liver was delineated on the CT image of the reference phase and a spheroid activity source of 20 mm radius was positioned in the liver with of 1:9 ratio between the activity in the tumor and the liver activity in the background. Deformation vector fields were interpolated between consecutive frames to obtain 20 frames which were applied to the reference frame of the 4D CT and the activity sources (tumor and background). The resulting 20 positions in the respiratory motion were simulated individually. At the end of the simulation, the twenty list-mode files were gathered into a single list-mode file according to the input respiratory signal to mimic the acquisition of a breathing patient with a continuous rotation.

## 2.5   Patient acquisitions

The patient datasets were acquired with the dual-head General Electric Discovery NM/CT 670 of the Léon Bérard cancer center equipped with the Low Energy High Resolution (LEHR) collimator. Sixty projections were acquired over $360°$, each with $128 \times 128$ pixels, 4.42 mm isotropic spacing and an acquisition time of 25 s. Patient acquisitions used for this study came from the pre-treatment imaging of liver radioembolization procedure. This step consists in injecting around 350 MBq of $^{99m}$Tc macro aggregated albumin (MAA) inside the liver to assess lung shunt and extrahepatic uptake. Twelve distinct patients were included in this study. One patient was scanned three times but the results were analyzed separately, yielding fourteen patient samples.

## 2.6   Image analysis

On each reconstructed volume, a volume of interested (VOI) was delineated. For the simulations, the VOI of the reference 3D SPECT image (with no motion), of the motion-compensated reconstructions and of each frame of the reconstructed 4D image was the known tumor delineation at the respective position in the breathing cycle. The VOI of the blurred 3D image was the union of the 20 tumor positions. For the patient acquisitions, a fixed threshold at 42% of the maximum value was used to define the VOI [12].

The activity was evaluated by computing the mean activity A in the VOI. The activity recovery was defined as follows:

$$\Delta A = \frac{A_{evaluated} - A_{ref}}{A_{ref}} \times 100 \qquad (1)$$

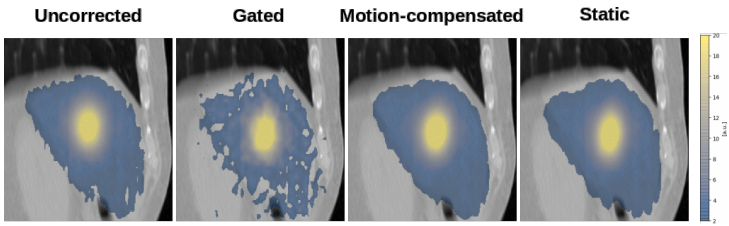

**Figure 1:** Sagittal slices of the images reconstructed from the SPECT simulation. The gated, motion-compensated and static images are images of the end-inhale position.

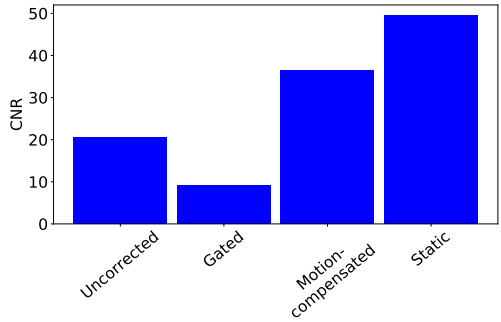

**Figure 2:** Contrast-to-noise ratio for the different reconstructions of the SPECT simulation.

where $A_{evaluated}$ was the mean activity in the VOI of either the gated, motion-compensated or blurred reconstruction and $A_{ref}$ the one in the reference image. The reference was the reconstruction with no motion for the simulation and the blurred reconstruction for the patient acquisition.

The contrast-to-noise ratio (CNR) was computed for each reconstruction as follows:

$$\text{CNR} = \frac{\mu_1 - \mu_2}{\sigma_2} \qquad (2)$$

where $\mu_1$ is the mean activity in the VOI, $\mu_2$ the one in the background and $\sigma_2$ the corresponding standard deviation.

The amplitude of the tumor motion was also measured in each acquisition by tracking the center of mass in the respiratory-gated reconstruction.

## 3 Results

### 3.1 Simulations

Sagittal slices of the images reconstructed from simulated data are shown in Figure 1. Motion-compensated reconstruction visually improved image quality compared to the gated or the blurred reconstructions.

This visual observation was quantitatively confirmed by the CNR value (Figure 2), which was higher for the motion-compensated reconstruction than the reconstruction without motion correction and the gated reconstruction and closer to the reference.

The mean activity in the tumor of the blurred 3D reconstruction was 24% lower than the reference. The gated and the motion-compensated reconstructions were closer to the reference: 1% below and 2% above, respectively.

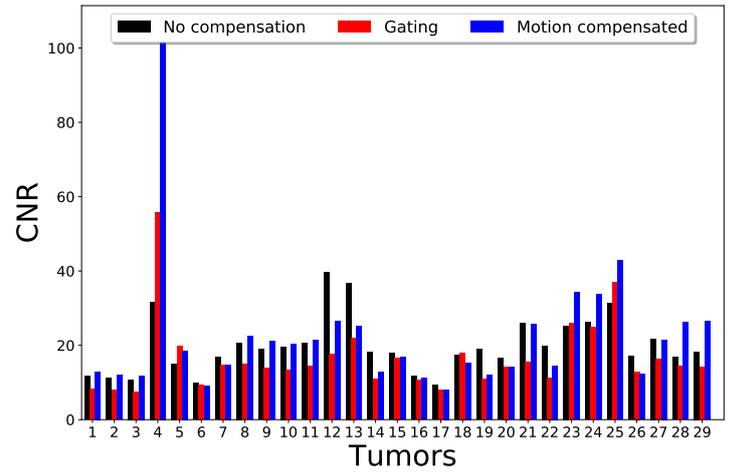

**Figure 3:** Contrast-to-noise ratio for the 29 tumors segmented on the patient reconstructions.

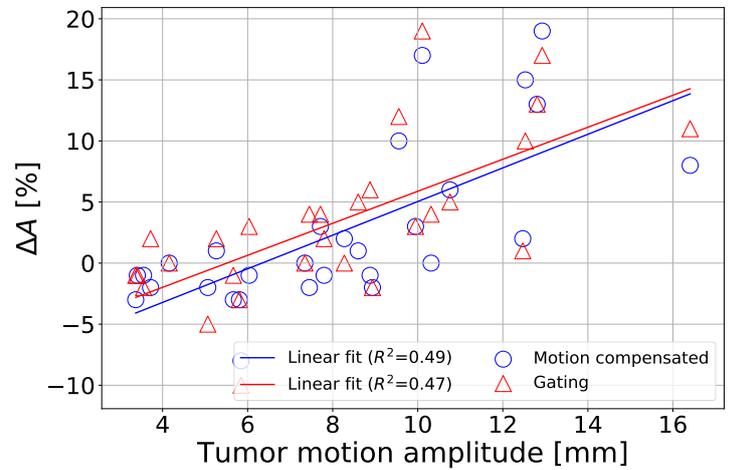

**Figure 4:** Activity recovery $\Delta A$ for the gated and motion-compensated images of the patient acquisitions at end-inhale. The blurred image was used as reference.

#### 3.1.1 Patients

Patient evaluation included 29 tumors in 20 SPECT acquisitions of different patients. The average motion amplitude was 8.1 mm with a 3.4–16.4 mm range. The CNR value is given in Figure 3. On average, the CNR in the motion-compensated SPECT image was 31.1% and 8.2% higher than the one obtained with the gated and the blurred SPECT images, respectively. The maximum gains were 90.1% and 234% compared to the uncorrected and the gated reconstructions.

The activity recovery $\Delta A$ was 3.4% on average with a -10%–19% range for the gated reconstruction. The average value was 2.3% for the motion-compensated reconstruction and the value range between -8% and 19% (Figure 4).

## 4    Discussion

The proposed motion-compensated reconstruction was applied to simulated and real liver radioembolization pretreatment $^{99m}$Tc SPECT acquisitions. The method is fully data-driven and does not require any extra-hardware or other image than the conventional 3D SPECT list-mode data.

The motion-compensated reconstruction recovered larger activity value in the tumor compared to the reconstruction without compensation. On the simulation, the activity recovery $\Delta A$ was closer to the reference 3D reconstruction obtained from a static simulation.

For the patient acquisition, the average increase was 2.3% compared to the uncorrected reconstruction. For five tumors, this value was above 10%. The activity recovery was slightly correlated to the tumor motion amplitude (fig. 4). Therefore, for tumors with substantial motion, the motion-compensated reconstruction should help to improve the predictive dosimetry for $^{90}$Y radioembolization, leading to more accurate patient-specific treatment planning. However, further studies are needed to assess the real impact on the predictive dosimetry, e.g. studying the impact of the method on the quantification of the tumor-to-normal liver ratio or the lung shunt fraction.

The use of a 3D attenuation map to perform the attenuation correction of the motion-compensated reconstruction might lead to mismatch between the attenuation map and the emission map. This can result in under- or over-estimation of the activity recovery depending on the tumor location. The most critical position is near the border between the liver and the lung, which have very different attenuation coefficients. In that case, part of the tumor can match the lung position in the attenuation map when reconstructed at end-exhale, whereas at end-inhale, most of the tumor is in the liver. In this study, we chose to reconstruct the motion-compensated images at end-inhale to mitigate this effect.

The activity recovery obtained with motion-compensated reconstruction was in the same range as the one obtained with respiratory-gated reconstruction. However, motion compensation improved the CNR with respect to gated reconstruction. Higher CNR could improve the detection of small tumors or regions with low uptake ratios, thus improving the diagnostic. In this study, we only considered tumors in the liver but the method can be easily adapted to other organs. However, it might be limited by the estimation of motion in the projections since two tumors might overlap but move differently in these images [3].

## 5    Conclusion

We have developed a fully data-driven motion-compensated SPECT reconstruction and evaluated it in the context of liver radioembolization. The method was compared to a respiratory-gated reconstruction in terms of activity recovery and CNR. The motion compensation recovered larger activity in the tumor compared to conventional 3D SPECT reconstruction with higher CNR than respiratory-gated reconstruction, which should eventually improve the treatment planning of liver radioembolization.

# Chapter 4

# Oral Session - Advanced CT

**session chairs**
Joseph Webster Stayman, *Johns Hopkins University (United States)*
Nicole Maass, *Siemens Healthcare GmbH (Germany)*

# Diffraction tomography inversion and the transverse ray transform

William R.B. Lionheart[1] and Alexander M. Korsunsky[2]

[1]Department of Mathematics, University of Manchester, UK
[2]Department of Engineering Science, University of Oxford, UK

**Abstract** We show that a reciprocal space squared intensity map of a material can be recovered, for each characteristic length scale, from diffraction tomography data by a simple slice-by-slice reconstruction method. Moreover if the reciprocal space map can be represented by a finite sum of spherical harmonic components for each length scale then the coefficients of that expansion can be recovered from inverting the transverse ray transform (TRT), where the data are polynomial coefficients of the azimuthal diffraction pattern for each length scale.

## 1 Introduction

X-ray diffraction experiments give information about the structure of a material on the length scale of the wavelength X-rays used. In X-ray crystallography a periodic crystal structure gives rise to a periodic diffraction pattern with distinct peaks. For less regular materials a less distinct diffraction pattern can never-the-less detect preferred orientations and nearly periodic structures.

If a narrow gauge volume is illuminated with a monochromatic X-ray beam the diffraction pattern is a sum of diffraction patterns in that volume [**korsunsky2011strain**]. Increasingly, not just for X-rays but also neutrons and electrons, we have the capability to raster scan a narrow beam measuring a diffraction pattern and perform a combination of tomography and diffraction, hoping to reconstruct a 3D diffraction pattern that that summarizes the properties of the material in each voxel. Small angle X-ray scattering (SAXS), see Figs 1 and 2, tomography is a particularly promising variant of this idea. However so far there is no theory for the reconstruction in this field, and although it has been assumed so, it is not yet proven that an isotropic average can be meaningful reconstructed from this data. In this paper we lay the theoretical foundation for diffraction tomography. We will demonstrate a theoretical reconstruction method using data from all directions. We will also show how a slice-by-slice approach can be used to reconstruct a diffraction pattern given by a finite sum of spherical harmonics. In this case the problem reduces to the transverse ray transform of symmetric tensor fields.

## 2 Physical model

We assume that at each point $\mathbf{x}$ in the object and for each three dimensional reciprocal space vector $\mathbf{q}$ there is a scattering intensity-squared map $f(\mathbf{x}, \mathbf{q})$. For a given ray direction $\xi \in S^2$ (the unit sphere) a diffraction pattern for the material near $\mathbf{x}$ would produce a 2D scattering pattern on a planar detector normal to $\xi$ with squared intensity $f(\mathbf{x}, \mathbf{q})$ for $\mathbf{q} \in \xi^{\perp}$

(the space of vectors perpendicular to $\xi$). Clearly there is an underlying assumption that the problem can be formulated on two length scales, and we will not make this explicit mathematically, but roughly that on the scale of the wave length of the X-rays (or particles) the 3D distribution of scatterers has $f(\mathbf{x}, \mathbf{q})$ as the square magnitude of its Fourier transform and this can be treated as a constant on a small length scale, but on a larger length scale, commensurate with the width of the beam and the spatial scanning increments, the Fourier transform is variable.

We treat intensity squared as the variable as we assume that the diffraction pattern observed from one ray is an incoherent average and so the result of the sum of squared intensities along the gauge volume. Note that as $f$ is magnitude squared Fourier transform of a real function it is even with respect to $\mathbf{q}$: $f(\mathbf{x}, \mathbf{q}) = f(\mathbf{x}, -\mathbf{q})$.

Our data then is the generalized transverse ray transform (GTRT)

$$g(\mathbf{x}, \xi, \mathbf{q}) = \int_{-\infty}^{\infty} f(\mathbf{x} + s\xi, \mathbf{q}) \, ds, \qquad (1)$$

for $\mathbf{x} \in \mathbb{R}^3, \xi \in S^2, \mathbf{q} \in \xi^{\perp}$. In the case where $f(\mathbf{x}, \mathbf{q}) = F \cdot \mathbf{q}^m$ where $F$ is a rank $m$ symmetric tensor field (the dot denotes contraction over $m$ indices) this coincides, after a small change in notation, with the transverse ray transform of symmetric tensor fields defined by Sharafutdinov [**sharafutdinov2012integral**].

## 3 Uniqueness and reconstruction for complete data

Suppose that we have diffraction data for all rays passing through the object (the support of $f$). For simplicity consider a single value of $|\mathbf{q}| = Q$, corresponding physically to one reciprocal length scale, and a circle on the detector plane of radius $Q$ centred on its intersection with the ray. We now follow the same argument used by [**sharafutdinov2012integral**] for the transverse ray transform of symmetric tensor fields. Choose a direction $\eta \in S^2$, which conceptually we think of as a rotation axis for the sample in an experiment. Now consider measurements of $g$ for all rays in directions $\xi \in \eta^{\perp}$.

For a given plane through $\mathbf{x}_0 + \eta^{\perp}$, through $\mathbf{x}_0$ normal to $\eta$, $g(\mathbf{x}, \xi, Q\eta)$ for $\mathbf{x} \in \mathbf{x}_0 + \eta^{\perp}, \xi \in \eta^{\perp} \cap S^2$ is the 2D X-ray transform of the scalar function $f(\mathbf{x}, Q\eta)$ on that plane. Hence it can be reconstructed using the inverse Radon transform. We see now that $f$ can be reconstructed from complete data $g$ for all rays. In practice only reciprocal length scales $0 < Q_0 < Q < Q_1$ in some fixed range would make physical sense. We note that this idea is already present in the SAXS tomography literature, for a single axis [**schroer2006mapping**], and for multiple axes [**schaff2015six**, **feldkamp2009recent**] without formally describing the generalized transverse ray transform. In the mathematical literature the extension of ray transforms to the *sphere bundle* (space with a sphere at each point) appear as the geodesic ray transform on a Riemannian manifold.

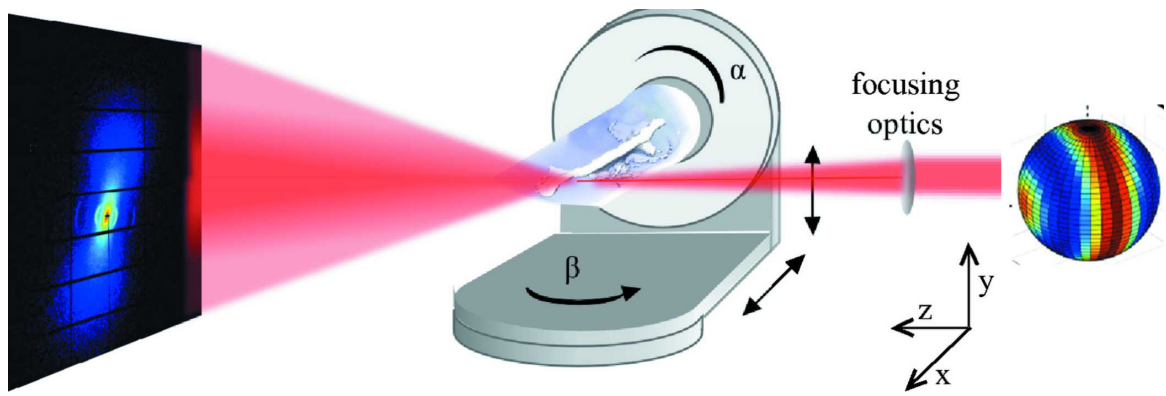

**Figure 1:** Small Angle X-ray Scattering tomography [**liebi2018small**](from which figure is taken) is a diffraction tomography method currently popular: its feasibility and application has been demonstrated and synchrotron radiation facilities are investing heavily in the necessary apparatus. The sample is placed on a tilt stage and for each direction $\xi$ on the sphere (relative to the sample) a narrow beam is raster scanned in a plane normal to $\xi$. For each such position a diffraction pattern is recorded on a planar detector to give the 6D data set. For each (range of) $|\mathbf{q}|$ and each voxel a reciprocal space map on a sphere is reconstructed. This gives information about the orientation of structures at that length scale.

## 4 Consistency conditions

In inverse problems in general and especially in tomography it is important to characterize data that is consistent with the assumed model: in mathematical terms, to describe the range of the operator. For some cases the singular value decomposition (SVD) gives an explicit orthogonal basis for the range and for its orthogonal complement. For the scalar x-ray transform in two dimensions see [**louis1984orthogonal**] and three dimensions see [**maass1987x**]. Consistency conditions are systems of equations that characterize the range. For the 2D Radon (X-ray) transform Helgason's range conditions characterize the range in terms of moments of the data [**natterer2001mathematics**]. For the 3D X-ray transform the range is characterized by satisfying John's ultrahyperbolic partial differential equation (PDE) [**john1938ultrahyperbolic**].

For the isotropic case $f$ independent of $\mathbf{q}$, the GTRT (1) reduces to the X-ray transform in 3D space. This is formally overdetermined as the space of lines in 3D space is four dimensional. The data is one function of four variables and we seek a function of three variables. So it is no surprise that the data satisfies one PDE. By contrast in the 2D isotropic problem (Radon transform) we seek one function of two variables and our data is one function of two variables —

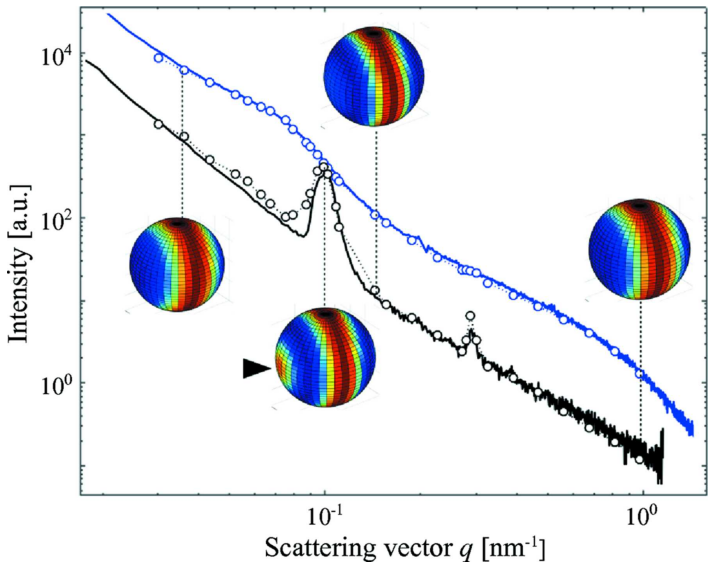

**Figure 2:** Figure from Liebi *et al* [**liebi2018small**] demonstrates how some reciprocal length scales have different orientation structure

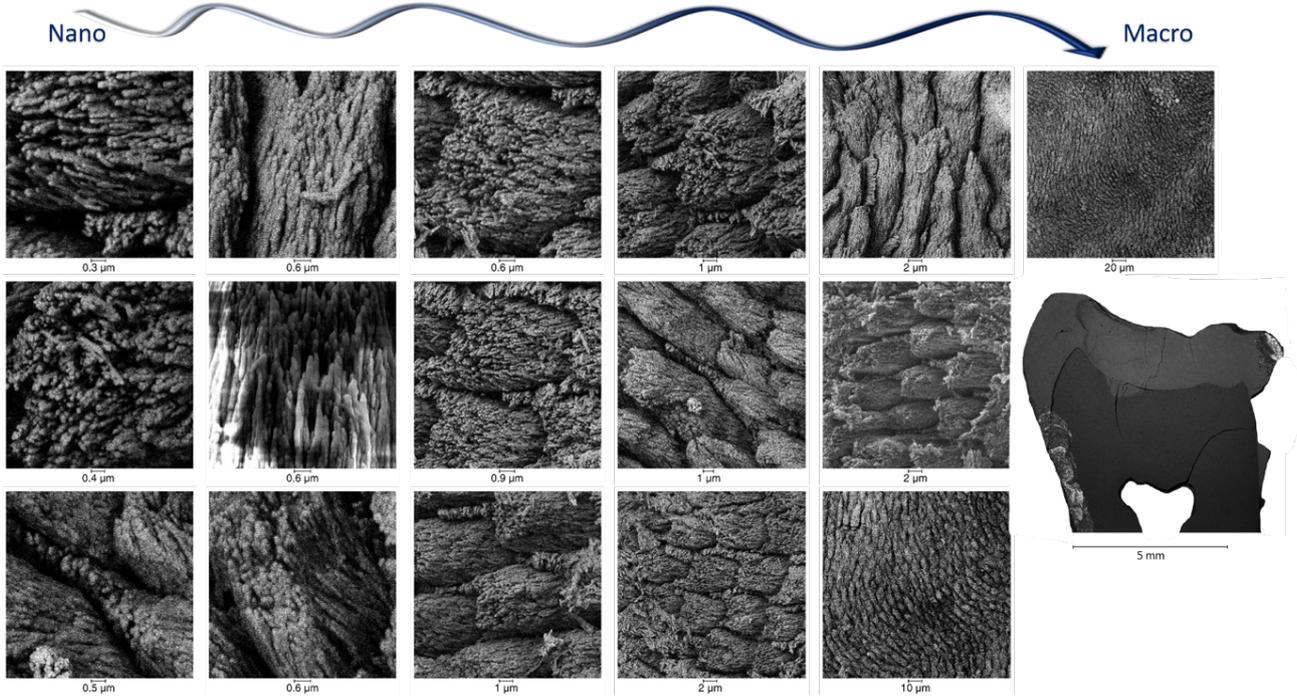

**Figure 3:** Scanning electron micrographs of human teeth from nano-to macro-scale showing of the hierarchy of the structure of artificially demineralised human enamel with secondary electron and backscattered electron images, and the last image on the right, the view of a slice of a human carious tooth showing enamel (light grey), dentine (dark grey), and the pulp chamber. These cross-sectional images do not provide direct evidence of the structural arrangement in 3D that can only be deduced for large volume, using high-resolution tomographic imaging. [**besnard20213d**]

formally correctly determined. In the case of the GTRT we seek one function of five variables (for fixed $Q$). Our data is a function on a circle for each line, and is also a function of five variables, so formally correctly determined.

In Sec 3 we saw how one can reconstruct $f(\mathbf{x}, \mathbf{q})$ on each plane normal to $\mathbf{q}$ as a scalar Radon transform on each plane. Notice for a fixed $\mathbf{q}$ the only data involving $f(\mathbf{x}, \mathbf{q})$ is exactly the lines in the plane through $\mathbf{x}$ normal to $\mathbf{q}$. So Helgason's range conditions are the only consistency conditions that apply, beyond that $g$ is even in $\mathbf{q}$. Data $g$ satisfying these consistency conditions is associated with a (unique) $f$.

## 5   The Transverse ray transform of tensor fields

The transverse ray transform of a symmetric tensor field is the integral along rays of the projection of the ray normal to that direction. Let $e_i$ be the unit Cartesian vectors in 3-space. We will denote the tensor product of tensors $a$ and $b$ by $a \otimes b$. A general rank two tensor has the form

$$a = \sum_{i,j=1,...,3} a_{ij} e_i \otimes e_j$$

We denote symmetric tensor product $a \odot b = (a \otimes b + b \otimes a)/2$ and the symmetric $k$-th tensor power by $a^k$. Let $\xi \in S^2$ be any unit vector, then the matrix

$$\Pi_\xi = I - \xi \xi^T$$

projects a vector on to the subspace $\xi^\perp$ or $(\Pi_\xi)_{ij} = \delta_{ij} - \xi_i \xi_j$ as a tensor. For a rank $k$ symmetric tensor $a$ the projection $P_\xi(a)$ is the $k$ fold contraction of $a$ with $\Pi_\xi$. In components

$$P_\xi(a)_{i_1 \cdots i_k} = \sum_{j_1 \cdots j_k} (\Pi_\xi)_{i_1 j_1} \cdots (\Pi_\xi)_{i_k j_k} a_{j_1 \cdots j_k}$$

for example as a matrix the components of the projection of a rank two symmetric tensor $P_{e_3} a$ are

$$\begin{bmatrix} a_{11} & a_{12} & 0 \\ a_{12} & a_{22} & 0 \\ 0 & 0 & 0 \end{bmatrix}$$

For a rank $k$ symmetric tensor field $a$ the Transverse ray Transform (TRT) is defined as

$$Ja(\mathbf{x}, \xi) = \int_{-\infty}^{\infty} P_\xi(a)(\mathbf{x} + s\xi) \, \mathrm{d}s,$$

note that the data for each ray defined by $\mathbf{x}, \xi$ is a symmetric rank $k$ tensor in three variables. However it is restricted to $\xi^\perp$ so it would be natural to express it in a two dimensional coordinate system for actual measurements (such as detector screen coordinates).

We can now review the known theory for sufficiency of data, characterization of consistent data and inversion for the TRT. Sharafutdinov [**sharafutdinov2012integral**] (and earlier Russian edition ) gives an inversion method for the TRT of a symmetric rank $k$ tensor field that is equivalent to the argument we gave in Sec 2 applied to the special case

$$f(\mathbf{x}, \mathbf{q}) = a(\mathbf{x}) \cdot \mathbf{q} \cdots \mathbf{q}$$

where the dots denote contraction and the result of the $k$ fold contraction is a scalar.

As before we consider a rotation axis $\eta$ and rays in directions $\xi \in \eta^\perp$, in each plane $\eta^\perp + z\eta$ the component $a(\mathbf{x}) \cdot \eta \cdots \eta$ ($k$-fold contraction) transforms as a scalar in the plane. We perform the reconstruction by application of the scalar inverse Radon transform to $Ja(x, \xi) \cdot \eta \cdots \eta$. One then has to repeat for at least $K = \binom{k+2}{2}$ (the dimension of the space of symmetric rank $k$ tensors, sometimes called the 'stars and bars' problem) choices $\eta^1, .., \eta^k$, such that the set of symmetric $k$ fold products $(\eta^i)^k$ is linearly independent. For example for $k = 2$ the six diagonals of the icosahedron is a suitable choice (in fact optimal as it maximizes the condition number of an associated linear system). See [**lionheart2015diffraction**] for a geometric criterion for $k = 2$.

For modest $k$ this still seems rather wasteful in that data is discarded and therefore more rotation axes are need than is strictly necessary. In [**lionheart2015diffraction**] we gave a filtered back projection formula for the reconstruction of a rank-2 tensor from *complete* TRT data, that is rotation about *every* axis. This is even more wasteful for small $k$ however it does use all the data that can be collected, averaging over the redundancy.

In an attempt to reduce the number of rotation axes needed, in [**desai2016explicit**] we showed that there is an explicit reconstruction algorithm for the TRT for a rank-2 tensor using only three rotation axes. However this has a certain type of instability compared to using six axes.

## 6 Spherical harmonic expansion

It has been suggested (see eg [**liebi2015nanostructure**],[**guizar2020validation**]) that the intensity squared reciprocal space map be expanded in spherical harmonics. Suppose we have

$$f(\mathbf{x}, \mathbf{q}) = \sum_{l \leq K, l \text{ even}, |m| \leq l} a(\mathbf{x}, |\mathbf{q}|)_{lm} Y_l^m(\mathbf{q}/|\mathbf{q}|). \quad (2)$$

Where we use the abbreviated notation for Laplace's spherical harmonics $Y_l^m(\hat{\mathbf{q}})$ for $Y_l^m(\theta, \phi)$ where $\hat{\mathbf{q}} = (\sin\theta\cos\phi, \sin\theta\sin\phi, \cos\theta)$ is a unit vector. In [**liebi2015nanostructure**] the complex reciprocal space map (not the square magnitude) is represented as a sum of spherical harmonics up to some order $K$. As the product of spherical harmonics can be expressed in spherical harmonics up to $K$ we lose no generality.

The question arises if we can deduce the coefficients $a(\mathbf{x}, Q)_{lm}$ from less than the full data $g(\cdot, \cdot, \mathbf{q})$ with $|\mathbf{q}| = Q$. In particular can the isotropic term $a_{00}$ be deduced from averages over the circles of radius $Q$ of the diffraction patterns? More generally can the components of each order be reconstructed separately by some form of preprocessing of diffraction pattern data? To answer these questions we need to consider the relationship between spherical harmonics and polynomials.

A homogeneous degree $k$ polynomial on $\mathbb{R}^3$ is a polynomial $p$ satisfying $p(c\mathbf{q}) = c^k p(\mathbf{q})$, In the discussion of polynomials we will use $\mathbf{q} = (q_1, q_2, q_3)$ as our general vector in reciprocal space. For example $p(\mathbf{q}) = q_1^3 - 2q_2^2 q_3$ is a 3-rd degree homogeneous polynomial. Homogeneous polynomials of degree $k$ are in one to one correspondence with symmetric $k$-th rank tensors, we just replace the $q_i$ by unit basis vectors $e_i$ and treat the product as the symmetric tensor product. For example $q_1^2 + q_2^2 + q_3^2$ corresponds to the Kroneker tensor with components $\delta_{ij}$.

A harmonic polynomial is $p(\mathbf{q})$ is simply a polynomial satisfying Laplace's equation

$$\Delta_{\mathbf{q}} p(\mathbf{q}) = 0, \quad \Delta_{\mathbf{q}} = \frac{\partial^2}{\partial q_1^2} + \frac{\partial^2}{\partial q_2^2} + \frac{\partial^2}{\partial q_3^2},$$

for example $q_1^2 - q_3^2$ is harmonic.

The dimension of the space of spherical harmonics of degree $l$ in 3 variables is $2l + 1$ [**axler2013harmonic**]. The Laplace spherical harmonics $Y_l^m(\mathbf{q})$ span the space of harmonics polynomials of degree $l$.

Our aim is to convert (2) to a tensor expression so that we can apply the known theory of tensor tomography. The problem is that while we can regard spherical harmonics as polynomials and polynomials as symmetric tensors we appear to have a sum of tensors of different ranks. To get around this first we impose the condition $|\mathbf{q}| = Q$, a reciprocal length scale.

In many applications the orientation structure depends very much on the reciprocal length scale $Q$ and it would be sensible to investigate this over specific ranges of interest, see Figs 2,3,4. The expression

$$f_Q(\mathbf{x}, \mathbf{q}) = \sum_{l \leq K, l \text{ even}, |m| \leq l} a(\mathbf{x}, Q)_{lm} |\mathbf{q}|^{K-l} Y_l^m(\mathbf{q}) \quad (3)$$

is the a homogeneous polynomial of degree $K$ in $\mathbf{q}$ at each $\mathbf{x}$. This polynomial has an associated rank $K$ symmetric tensor field we will call $F_Q(\mathbf{x})$, and $f_Q(\mathbf{x}, \mathbf{q}) = F_Q \cdot \mathbf{q}^K$.

Our task now is to show how the TRT data for $F_Q$ can be recovered from $g$ restricted to $|\mathbf{q}| = Q$. It is well known that a bilinear function $B(v, w)$ in two vector variables can be recovered from the quadratic form $P(v) = B(v, v)$ using the polarization identity

$$B(v, w) = \frac{1}{4} \left( P(v + w) - P(v - w) \right). \quad (4)$$

It is perhaps not surprising, but less well known, that a similar identity applies to symmetric multi-linear functions [**defant2017non**]. The relevance to us is that for each ray and a given $Q$ we know $g(\mathbf{x}, \xi, \mathbf{q})$ for $|\mathbf{q}| = Q, \mathbf{q} \in \xi^\perp$, the diffraction pattern around a circle of radius $Q$. This is the integral of $F_Q \cdot q^m$ along a ray and we can find the TRT $JF_Q(\mathbf{x}, \xi)$ by applying the multi-linear polarization identity to $g(\mathbf{x}, \xi, \mathbf{q})$.

As long as it is known (2) is valid for some $K$ one can attempt a reconstruction using the known reconstruction methods for the TRT detailed in Sec 5, or using regularized iterative methods widely used for large scale linear inverse problems: CGLS on an augmented matix for generalized Tikhonov and FISTA when a TV regularized term is included. These are implemented, for example, in our Core Imaging Library [**CIL1**] for scalar problems. If $K$ is not known *a priori* one can use a higher value than necessary, at the expense of higher computational cost, and then decide if the coefficients $a_{lm}$ are significant if including them results in a significantly better fit to the data. Of course for some values of $Q$ one may need a higher value of $K$ than others as is illustrated in the diffractions patterns in Fig 4, and is inherent in the micrographs in Fig 3.

One tempting approach that may well fail is to take an average in the detector plane over a circle of constant $Q$ and then attempt a slice by slice reconstruction assuming a scalar (that is isotropic) model with $K = 0$. The underlying problem is that the restriction of a harmonic polynomial in three variables to a plane is not necessarily harmonic. For example $q_1^2 - q^3$ is harmonic in three dimensional space but its restriction to $q_3 = 0$ is $q_1^1$ which is not harmonic. The projection on to spherical harmonic components of each order is not preserved by projection on to a plane.

In HAADF-STEM (High-Angle Annular Dark-Field Scanning Transmission Electron Microscope) tomography (see for example[**kubel2005recent**] [**leary2012quantitative**]) an azimuthal average of a diffraction pattern is used to reconstruct a scalar image from a single tilt-series. Our analysis suggests this is flawed where the electron diffraction pattern is anisotropic.

To some extent the danger of assuming isotropy, or indeed too small a $K$ in general, is reduced provided enough data is collected. For example if slice-by-slice data is collected for one rotation axis and a scalar reconstructed that best fits that data, one can then test if the same scalar reconstruction is consistent with reconstruction from rotation about a different axis.

On each plane normal to a vector $\eta$ the contractions of $Ja$ with $\eta$ appear as the TRT of lower rank tensor fields on $\eta^\perp$. The range of the 2D TRT is given completely by the SVD described by [**kazantsev2004singularA**].

## 7    Conclusions and further work

We have laid the theoretical framework for diffraction tomography including an explicit inversion procedure, for each reciprocal length scale, for *full data*, that is a sufficiently dense sampling of the four dimensional space of lines. We have also shown that assuming the reciprocal space map for each reciprocal length scale can be expanded in even spherical harmonics up to some fixed degree is equivalent to reconstructing a symmetric tensor field using the transverse ray transform data. The next steps practically are to do a full regularized algebraic reconstruction on real data, for both complete data, and with limited data assuming a finite spherical harmonic expansion. While an explicit reconstruction formula for limited TRT data is available for rank two tensors, none have been derived for higher rank tensors. Recent results on the TRT for higher rank tensors and limited data focus on the divergent beam case [**krishnan2020microlocal**] and this application may provide the impetus needed for further work on the parallel beam case relevant to synchrotron X-ray (SAXS and WAXS), electron (HAADF-STEM) and neutron (SANS) diffraction tomography[**treimer2008neutron**].

## Acknowledgements

WL would like to thank Marianne Liebi for helpful discussions, and to thank the Royal Society for a Wolfson Research Merit Award. Both authors gratefully acknowledge support from EPSRC grants EP/V007742/1 and EP/V007785/1 "Rich nonlinear tomography for advanced materials".

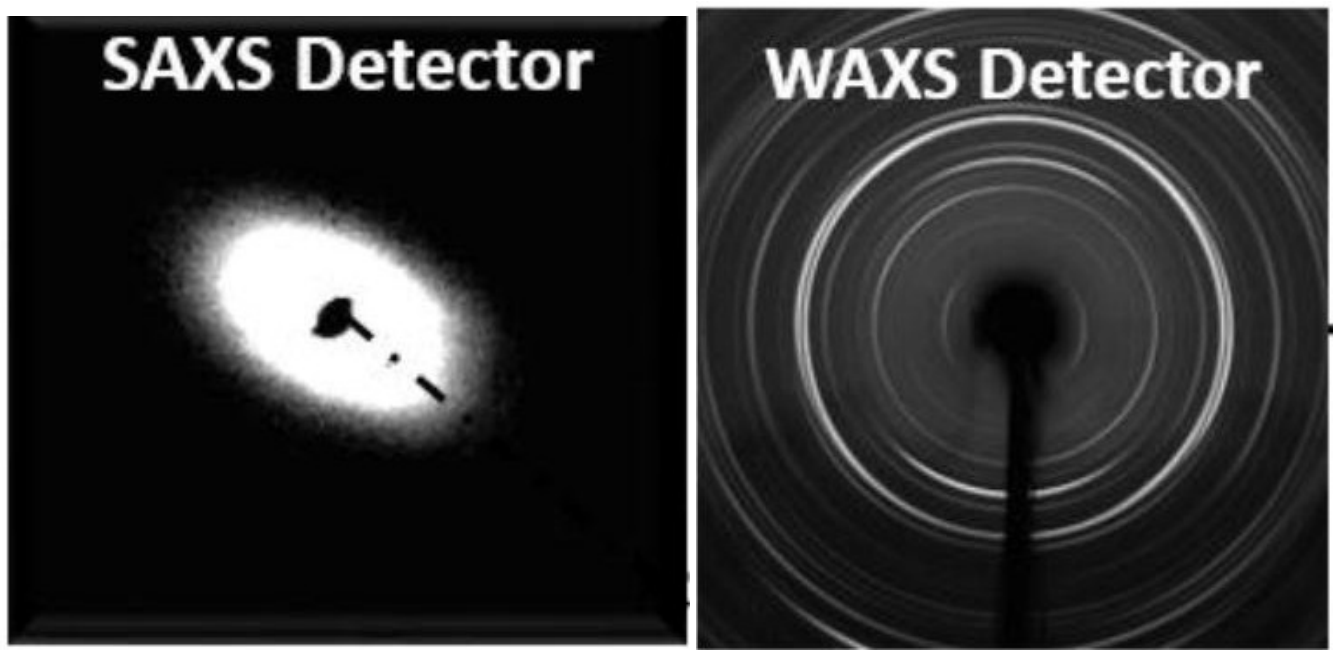

**Figure 4:** SAX, top, and WAX (wide angle X-ray scattering),bottom, diffraction patterns of dental enamel showing anisotropic structure on different length scales [**sui2018situ**]

# MeV Dual-energy CT Material Decomposition With Self-supervised Deep Learning Based Denoising: Simulation Results

Wei Fang, Liang Li and Zhiqiang Chen

Department of Engineering Physics, Tsinghua University, Beijing, China

**Abstract** MeV dual-energy CT can be used in customs as the next generation inspection tool to replace X-ray radiography for cargo/container imaging. Compared to X-ray radiography, MeV dual energy CT can give cross-section image, which is free of the overlapping problem. Besides, the recorded dual energy projection data can be used for material decomposition. Material decomposition can give more specific information on the scanned materials, such as the electron density and equivalent atomic number. Compared to keV dual-energy CT, MeV dual-energy CT material decomposition is much more difficult since the mass attenuation coefficients of different materials are very close in MeV energy range, making the decomposition problem more ill-conditioned. Besides, the heavy metals in the container will cause strong beam hardening artefacts, further degrading the image quality. In this paper, we used a projection domain-based method for material decomposition to reduce the effect of beam hardening. We also used a self-supervised Noise2Noise neural network for equivalent atomic number image denoising. The simulation results showed the effectiveness of our method for MeV dual-energy CT material decomposition.

## 1 Introduction

There are hundreds of millions of cargo transactions per year in customs, airports, stations and harbors all over the world. The giant size and complicated contents of cargo give the chance for concealing illicit materials, such as explosives, drugs, nuclear materials, smuggled goods and even stowaways. For homeland security, routine security screenings of these cargos are required. At present, high energy X-ray radiography with energy up to 9 MeV is the dominant technique for the inspection. However, the material discrimination capability of X-ray radiography technology is highly constrained by the overlapping problem. To overcome this problem, computed tomography technique can be introduced into cargo imaging as the next generation inspection tool. We have reported the design and the performance of the first commercial MeV dual energy CT system in our previous work [1-3].

For MeV dual-energy CT system targeted at cargo or container imaging, the material decomposition is a big challenge. On one hand, the spectrum generated by the X-ray tube or accelerator is polychromatic, making the dual energy attenuation equations nonlinear and much more complicated than the case of monochromatic X-ray imaging. On the other hand, the heavy metals often exist in the cargo or container, causing severe beam hardening artefacts. Besides, at MeV energy range, the mass attenuation coefficients of different materials are very close, making the decomposition problem more ill-conditioned and very sensitive to noise. The existed material decomposition methods can be generally classified into three types: the projection domain-based, image domain-based and one-step inversion methods. The projection domain-based decomposition methods can utilize the spectra information and maximumly reduce the beam hardening effect but with the cost of heavy computation on solving the nonlinear attenuation equations. The image-domain methods are easy to implement but with the beam hardening effect hard to eliminate. With the material decomposition finished, the electron density image and equivalent atomic number image can be calculated from the decomposed material coefficient images. Because of the ill-condition of the material decomposition process, the generated equivalent atomic number image can be very noisy. A specific denoising step can be taken to alleviate the noise in the equivalent atomic number image.

In this paper, we proposed a framework that includes a projection domain-based material decomposition process and a self-supervised deep learning-based method for equivalent atomic number image denoising. An optimization algorithm called Levenberg-Marquardt method [4] was used for solving the nonlinear attenuation equations. A self-supervised Noise2Noise network [5], which was trained by mapping one noise realization to another, was utilized for the equivalent atomic number image denoising. The simulation study showed the effectiveness of our proposed method.

## 2 Materials and Methods

### 2.1 The Data Simulation

The Monte Carlo simulated 6 and 9 MeV spectra we used for simulation are displayed in Figure 1. The spectra has been normalized, which means the photon counts have been divided by the total photon counts.

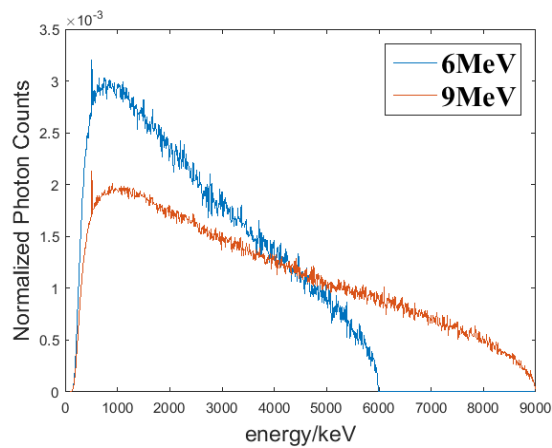

Figure 1. Monte Carlo simulated 6 and 9 MeV spectra used for simulation.

Figure 2 shows a water phantom with different pure material insertions. The atomic numbers of these materials are indicated in the small circles.

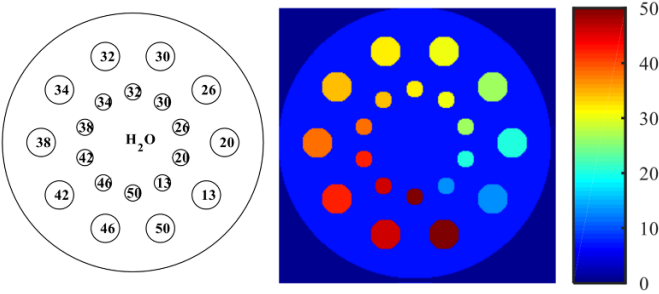

Figure 2. The water phantom with ten different material insertions used for simulation. The right part of the figure shows the water phantom with a colourful map that changing from blue to red along with increasing atomic numbers from 0 to 50.

A 2D fan-beam CT geometry was utilized in the simulation. The distance from the source to the center of rotation was 81.92cm and the distance from the detector to the center of rotation was 30.72cm. The number of views was 540, covering a full circle. The number of detector pixels was 736 and the fan angle was 42.17°. A total of $1 \times 10^6$ photons were used for simulation. The simulated low and high energy projection data and FBP reconstructed image are shown in Figure 3.

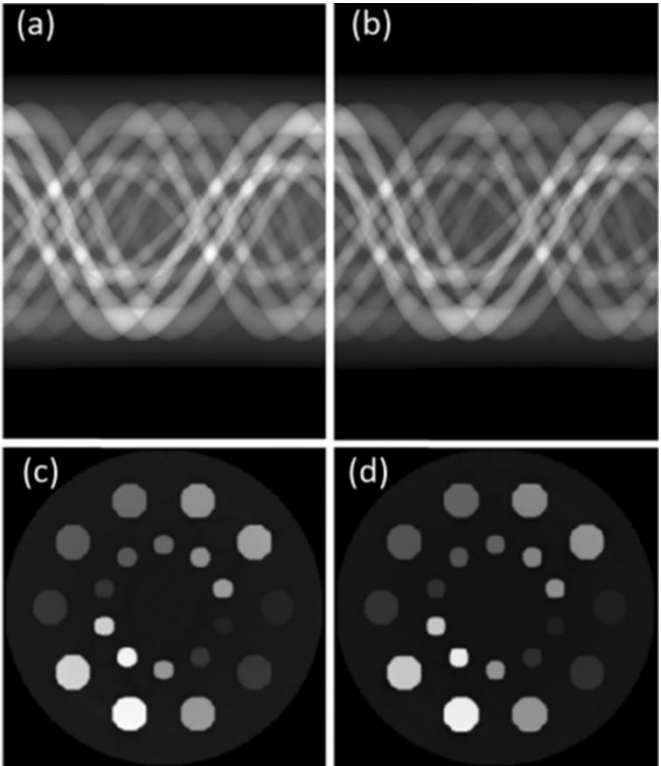

Figure 3. The left part is low energy sinogram and corresponding reconstruction. The right part is high energy sinogram and corresponding reconstruction. The display window for sinogram and reconstructed image are respectively [0, 6] and [0, 0.49].

## 2.2 The Projection Domain Decomposition

The projection domain-based decomposition tries to firstly estimate the line integral of material coefficients $B_1$ and $B_2$ from the nonlinear attenuation equations:

$$\begin{cases} p_L(ray_i) = -ln \int_0^{E_L} D_L(E) \exp[-B_1 \mu_1(E) - B_2 \mu_2(E)] \, dE \\ p_H(ray_i) = -ln \int_0^{E_H} D_H(E) \exp[-B_1 \mu_1(E) - B_2 \mu_2(E)] \, dE, \end{cases} \quad (1)$$

where $p_L$ and $p_H$ are respectively the low and high energy projection data. $D_L$ and $D_H$ are respectively the normalized low and high energy spectra. $\mu_1(E)$ and $\mu_2(E)$ are the attenuation coefficients of two basis materials at energy $E$. In this paper, We used Levenberg-Marquardt method for solving this two nonlinear equations. After the line integral of material coefficients $B_1$ and $B_2$ are estimated, the coefficient images can be reconstructed by using FBP method:

$$\begin{cases} b_1 = R^{-1}(B_1) \\ b_2 = R^{-1}(B_2) \end{cases} \quad (2)$$

where $R^{-1}$ represents inverse radon transform. After we get the decomposition coefficients, the electron density $\rho_e$ and equivalent atomic number $Z$ can be calculated by using the following formulas:

$$\rho_e = 2\left(b_1\rho_1\frac{Z_1}{A_1} + b_2\rho_2\frac{Z_2}{A_2}\right), \quad (3)$$

$$Z = \left[b_1\rho_1\frac{(Z_1)^2}{A_1} + b_2\rho_2\frac{(Z_2)^2}{A_2}\right], \quad (4)$$

where $\rho_i$ represents the mass density of the basis material $i$, $Z_i$ and $A_i$ respectively represent the atomic number and mass number of material $i$. More detailed derivation of formula (3) and (4) can be found in publication [6].

## 2.3 The Denoising of Atomic Number Image

The atomic number image can be very noisy due to the noise magnification characteristic of material decomposition process. It is necessary to conduct a further denoising step on the atomic number image. Deep learning has shown its great success in several image based tasks [7]. Traditional supervised deep learning method based on Noise2Clean mapping targets at minimizing such a loss:

$$\Theta^* = \underset{\Theta}{\mathrm{argmin}} \frac{1}{N} \sum_i \|f(\mathbf{z}_i + \mathbf{n}_i; \Theta) - \mathbf{z}_i\|_2^2 \quad (5)$$

where $\boldsymbol{z}_i$ is the ground truth of the $i_{th}$ image and $\boldsymbol{n}_i$ is the corresponding noise. $N$ represents the number of training samples. The function $f(\boldsymbol{z}; \Theta): \mathbb{R} \rightarrow \mathbb{R}$ represents the neural network mapping and $\Theta$ is its trainable parameters.

Different from Noise2Clean, Noise2Noise framework uses another independent noise realization as the training label, which can be illustrated as:

$$\Theta^* = \underset{\Theta}{\mathrm{argmin}} \frac{1}{N} \sum_i \|f(\boldsymbol{z}_i + \boldsymbol{n}_{i1}; \Theta) - (\boldsymbol{z}_i + \boldsymbol{n}_{i2})\|_2^2, \quad (6)$$

where $\boldsymbol{n}_{i1}$ and $\boldsymbol{n}_{i2}$ are two independent noise realizations. And the denoised result of the network can be given by:

$$y_i = \frac{f(\boldsymbol{z}_i + \boldsymbol{n}_{i1}; \Theta^*) + f(\boldsymbol{z}_i + \boldsymbol{n}_{i2}; \Theta^*)}{2}, \quad (7)$$

It has been demonstrated that under certain mild conditions that Noise2Noise training is equivalent to Noise2Clean training [8]. The process of obtaining two independent noisy equivalent atomic image implementations is like this. Firstly, we performed an angular separation to the projection data to get odd and even projections. We respectively performed material deconposition to these two separated datasets and calculated the equivalent atomic number images. The noise in the two calculated atomic number images can be approximately regarded as independent and zero-mean. In our case, the number of training samples N is 1. However, since the filters in convolutional neural network are shift-invariant, different parts of the training image actually serve as multiple training samples. Therefore the network can be well trained even with only one training image.

Figure 4 shows the architecture of the encoder-decoder network we used for Noise2Noise mapping. It had similar structures with U-net [9] but without resampling and doubled channels when moving to different stages.

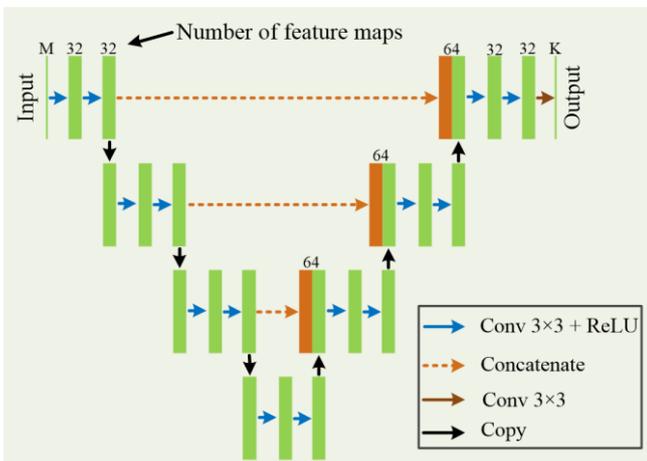

Figure 4. The architecture of the neural network for Noise2Noise mapping.

## 3 Results

The two basis materials we used for decomposition are carbon (Z=6) and tin (Z=50). Figure 5 shows the estimated line integral of basis material coefficients after solving the equations in (1). By performing FBP reconstruction to the decomposed projection, we obtained the material coefficient images, which are shown in Figure 6.

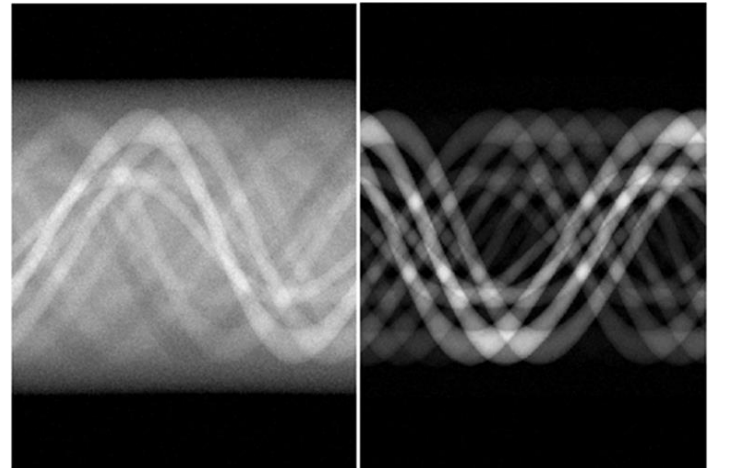

Figure 5. The decomposed projection. The left part is the decomposed carbon (C) projection and the right part is the deconposition tin (Sn) projection. The display window for carbon is [0, 4] and the display window for tin is [0, 0.15].

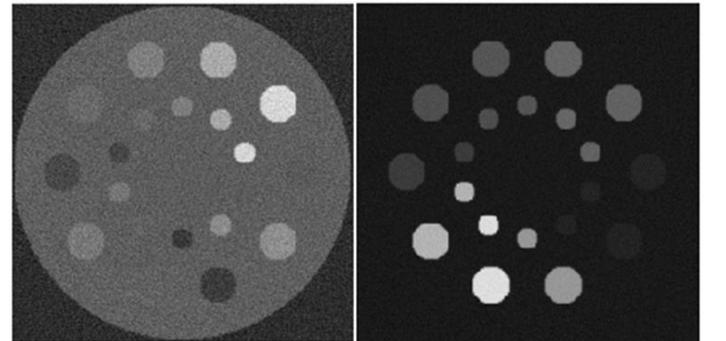

Figure 6. The reconstruction of material coefficients. The left part of the figure is the reconstructed carbon coefficient image with display window [-0.6, 2.8]. The right part of the figure is the reconstructed tin coefficient image with display window [-0.2, 1.8].

With the reconstructed material coefficients, we can calculate the electron density image and equivalent atomic number image according to formula (3) and (4). The two images are shown in Figure 7.

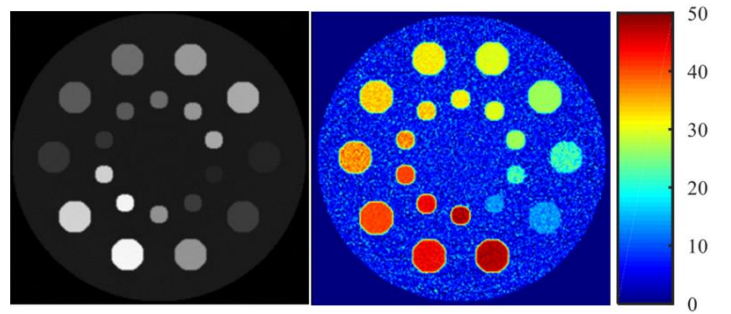

Figure 7. The electron density image (left) with display window [0, 11] and the equivalent atomic number image (right) with display window [0, 50].

To look at the performance of Noise2Noise network on atomic number image denoising. We put ground truth image, original calculated image and network denoised image together for comparison in Figure 8. A profile plot was shown in Figure 9.

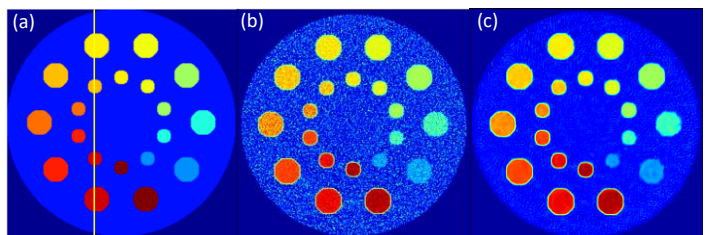

Figure 8. The equivalent atomic image. (a) the ground truth. (b) projection domain-based docomposition without denoising. (c) projection domain-based decomposition with Noise2Noise denoising. The display window is [0, 50].

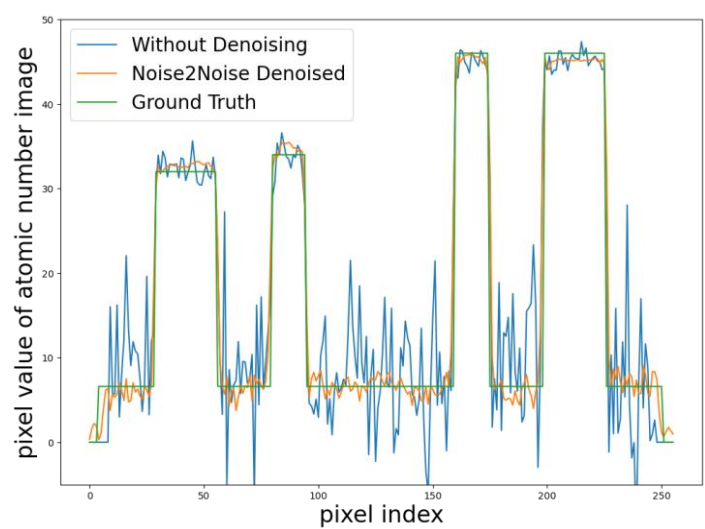

Figure 9. A profile plot (marked as yellow line in Figure 8) of equivalent atomic number images.

We can figure out that the noise is significantly suppressed after applying Noise2Noise network. Besides, the bias is also within a reasonable range.

## 4 Discussion and Future Work

The computation time of Levenberg-Marquardt algorithm for solving the two nonlinear attenuation equations in formula (1) is high. We plan to develop CUDA version of the algorithm running on GPU for solving this problem. Compared to image domain-based decomposition methods, the proposed projection domain-based method can reduce the effect of beam hardening for knowing the spectra information. Since now it is a simulation study, the access to the dual-energy spectra can be easily achieved. While in practical case, a spectrum estimation step is needed, which can be done by measuring the attenuation of incident spectra on some known calibration materials. The accuracy of spectra estimation is also very important to the final decomposition result. Regarding to the equivalent atomic number image denoising part, we also want to explore the architecture and property of the Noise2Noise network itself, especially on what benefits will be brought to denoising if we had a constrain on the sparsity or orthogonality of network parameters. Besides, this is an initial study currently. We will perform more quantitative evaluation on our method.

## 5 Conclusion

In this paper, we presented a simulation study of material decomposition of MeV dual-energy CT targeting for cargo or container imaging. A projection domain-based material decomposition method was applied for reducing the effect of beam hardening. A self-supervised neural network called Noise2Noise was used for equivalent atomic number image denoising. The simulation results showed the effectiveness of our method on beam hardening reduction and noise suppression of equivalent atomic number image.

## Acknowlegment

The corresponding author of this paper is Dr. Liang Li (email: lliang@tsinghua.edu.cn). This work was supported by NSFC 11775124.

# Chapter 5

# Oral Session - TOF-PET methods

**session chairs**
Michel Defrise, *Vrije Universiteit Brussel (Belgium)*
Yusheng Li, *University of Pennsylvania (United States)*

# Estimating the relative SNR of individual TOF-PET events for Gaussian and non-Gaussian TOF-kernels

Johan Nuyts[1], Michel Defrise[2], Emilie Roncali[3], Stefan Gundacker[4], Christian Morel[5], Dimitris Viskikis[6], and Paul Lecoq[7]

[1]Nuclear Medicine & Molecular Imaging, MIRC, KU Leuven, Belgium; [2] Laboratory for In vivo Cellular and Molecular Imaging, ICMI-BEFY, Vrije Universiteit Brussel, Brussels, Belgium, [3] Department of Biomedical Engineering and Department of Radiology, UC Davis, One Shields Avenue, Davis, CA 95616, United States of America; [4] Department of Physics of Molecular Imaging Systems, Institute for Experimental Molecular Imaging, RWTH Aachen University, Forckenbeckstrasse 55, 52074 Aachen, Germany; [5] Aix-Marseille Univ, CNRS/IN2P3, CPPM, Marseille, France; [6] LaTIM, INSERM UMR 1101, University of Brest, Brest, France; [7] Polytechnic University of Valencia, Spain

**Abstract** It is well known that measurement of the time-of-flight increases the information provided by coincident events in positron emission tomography. This information increase propagates through the reconstruction and decreases the variance in the reconstructed image for the same spatial resolution. T. Tomitani has analytically computed the gain in variance provided by a particular time-of-flight resolution, for the center of a uniform disk. This calculation is complicated, because it involves computing how the noise propagates through reconstruction with filtered backprojection. Here, we obtain the same result with a simpler image-based analysis of a single coincidence event. The proposed method assigns a relative signal-to-noise ratio to non-TOF and TOF events, as a function of the width and shape of the TOF kernel. This approach is also applicable to non-Gaussian TOF kernels. Such kernels are obtained for detectors in which different photon detection mechanisms are combined, such as detectors using both Cherenkov and scintillation photons and hybrid detectors. The proposed approach is verified mathematically by extending Tomitani's approach to the sum of shifted Gaussians, and it is verified with simple ray tracing simulation experiments.

## 1 Introduction

Tomitani provided a mathematical analysis of the variance in TOF and non-TOF images reconstructed with filtered backprojection (FBP) and post-filtered to the same image resolution [1]. This analysis only holds for the center of a uniform radioactive disk. In [2] we computed the TOF variance improvement for maximum-likelihood expectation maximization (MLEM) reconstruction numerically (using a Fisher information based analysis), and found excellent agreement with Tomitani's formula. Here we obtain that very same formula by directly analyzing the information in the data, thus avoiding the relatively complex mathematical treatment of the noise propagation through FBP or MLEM. This formula assigns a relative signal-to-noise (SNR) ratio to each event. It can account for some non-uniformity in the activity and for the contribution of randoms. More importantly, it can also deal with non-Gaussian TOF-kernels. This is of interest for detectors that make use of different photon detection mechanisms, e.g. detectors capturing both Cherenkov photons and scintillation photons [3–5], and detectors consisting of two (or more) scintillators with different characteristics [6, 7]. In some cases, each event can be assigned to a particular TOF-kernel, in other cases, this is not possible and the different TOF-kernels have to be combined into a single, averaged TOF-kernel. With the proposed approach, the value of a TOF event is quantified for both cases.

## 2 Methods

Following Tomitani, we consider the center of a disk shaped object with diameter $D$, filled with a uniform activity. We assume there is no attenuation. Due to symmetry, all the PET events along lines of response (LOR) intersecting the center have the same expectation and variance. The activity along the LOR equals $B$ per unit length.

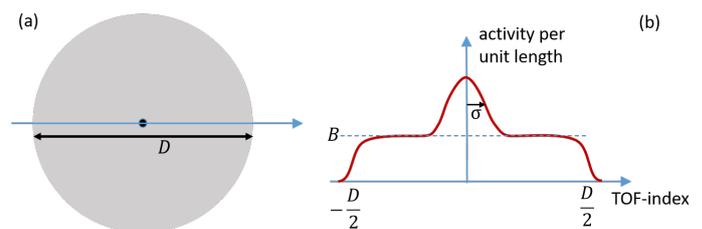

**Figure 1:** a) Disk phantom with diameter $D$ and a hot spot in the center. (b) The TOF profile as measured along a central LOR, which is a blurred version of the true profile. (Copied from [8]).

We propose to quantify the information provided by such an event as the SNR of an optimal (numerical) observer, who has to detect the presence or absence of a small spot of increased activity at the center, in a "signal known exactly, background known exactly" task [9]. The hypothesis is that this SNR is proportional to the information which the event contributes to the reconstruction of the central pixel value. If this is correct, then improvements to this SNR, e.g. due to changes to the TOF kernel, would imply identical improvements to the SNR in the reconstructed image.

Assuming that the Poisson noise on the data can be well approximated as Gaussian noise, the optimal observer uses the prewhitening matched filter [9] to compute a test statistic, which is compared to a threshold to decide if the hot spot is present or absent. In the cases considered here, the noise is already white and the observer reduces to a non-prewhitening matched filter. Further, we assume that the activity $S$ located at the "hot" spot is very small, such that it has a negligible effect on the variance of the measurements. The signal $t$ produced by the observer is the difference of the test statistic expectations when the spot is present and absent. The perfor-mance of the observer is degraded by the noise on the test statistic. As explained below, it is convenient to compute the squared SNR instead of the SNR.

### 2.1  SNR of a non-TOF event

For the non-TOF case, only a single value is measured for each LOR and the matched filter reduces to multiplication with an arbitrary constant, which can be set equal to 1. The observer signal $t$ is the difference between the expectations of the spot present and spot absent: $t_{\text{nonTOF}} = BD + S - BD = S$. The variance on the test statistic equals $BD$, because the data are subject to Poisson noise and $S << BD$. Consequently, the squared SNR equals

$$\text{SNR}^2_{\text{nonTOF}} = \frac{S^2}{BD} \qquad (1)$$

### 2.2  SNR of a Gaussian TOF event

Assume the same setup as above, except that now the acquisition is done with a TOF-PET system with spatial TOF uncertainty $\sigma$ (fig. 1). We assume that $\sigma << D$ such that we can ignore edge effects. Near the center, the expectation of the measurement with signal present $y(x)$ can be written as

$$\langle y(x) \rangle = B + \frac{S}{\sqrt{2\pi}\sigma} e^{-\frac{x^2}{2\sigma^2}}, \qquad F(x) = \sqrt{2}\, e^{-\frac{x^2}{2\sigma^2}}, \quad (2)$$

where $x$ is the 1D coordinate along the LOR (TOF-index converted to position). The corresponding matched filter $F(x)$ equals the difference between the expectations of the spot present and spot absent profiles, up to a constant. Using $F(x)$ as in (2), one obtains for the expectation of the difference of test statistics

$$t_{\text{TOF}} = \int_{-\infty}^{\infty} (\langle y(x) \rangle - B) F(x) dx = S, \qquad (3)$$

and for the variance

$$\text{var}_{\text{TOF}} = \int_{-\infty}^{\infty} B\, F^2(x) dx = 2\sqrt{\pi}\sigma B. \qquad (4)$$

Combining the above, the squared SNR associated with a TOF event equals

$$\text{SNR}^2_{\text{TOF}} = \frac{S^2}{2\sqrt{\pi}\sigma B}. \qquad (5)$$

Consequently, the variance gain (or SNR$^2$ ratio) obtained with TOF equals

$$\frac{\text{SNR}^2_{\text{TOF}}}{\text{SNR}^2_{\text{nonTOF}}} = \frac{D}{2\sqrt{\pi}\sigma} = \frac{\sqrt{2\ln 2}}{\sqrt{\pi}} \frac{D}{W} = 0.66 \frac{2D}{c\,\text{CTR}}, \quad (6)$$

where $c$ is the speed of light and CTR is the coincidence time resolution, i.e. the full width at half maximum of the Gaussian TOF kernel in units of time. $W = c\,\text{CTR}/2$ is the CTR converted to the corresponding distance. This result is identical to that obtained by Tomitani for FBP [1].

### 2.3  Accounting for non-uniform activity

Consider the same situation as above, except that now the activity of the disk is increased with $C$ per unit length over a distance $E$ near the edge of the disk (fig. 2). For a sufficiently narrow TOF kernel, nothing changes. In contrast, for the non-TOF measurement, the variance is increased due to the additional activity $2EC$. As a result, the TOF induced variance gain now becomes

$$\frac{\text{SNR}^2_{\text{TOF}}}{\text{SNR}^2_{\text{nonTOF}}} = 0.66 \cdot 2 \frac{D + 2EC/B}{c\,\text{CTR}}. \qquad (7)$$

This confirms that the gain due to TOF is higher for regions that are surrounded by more activity. For a decreased activity near the edge (negative $C$), the TOF gain would be reduced compared to the uniform activity case.

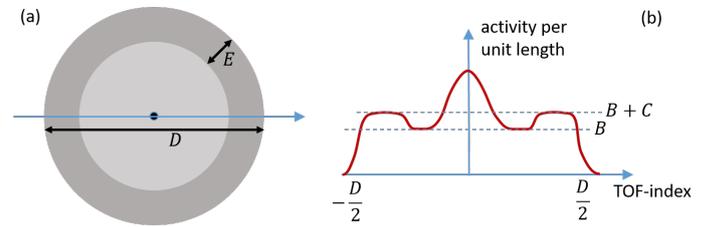

**Figure 2:** (a) Disk phantom with a hot spot in the center and an increased ring of activity (with width E) near the boundary. (b) The TOF profile as measured along an LOR through the hot spot. Near the center, the TOF profile is the same as in figure 1. (Figure copied from [8]).

### 2.4  Accounting for randoms

The randoms contribution is independent of the TOF index. Consequently, if the TOF acquisition would see $R$ randoms per unit length, the non-TOF acquisition would be contaminated by $RD_{\text{FOV}}$ randoms, where $D_{\text{FOV}}$ is the diameter of the field of view. Inserting this in eq. (6) results in

$$\frac{\text{SNR}^2_{\text{TOF}}}{\text{SNR}^2_{\text{nonTOF}}} = 0.66 \frac{2D}{c\,\text{CTR}} \frac{B + R/\beta}{B + R}, \qquad (8)$$

where $\beta = D/D_{\text{FOV}}$. Since $\beta < 1$, this result predicts that TOF reduces the variance more when the randoms fraction is higher, as has been observed e.g. in [10].

### 2.5  Non-Gaussian TOF kernels

Assume that the non-Gaussian TOF-kernel can be well approximated as a sum of (possibly shifted) Gaussians (see e.g. [7]):

$$k_{\text{TnG}}(x) = \frac{1}{\sqrt{2\pi}\sum_{i=1}^{n} A_i \sigma_i} \sum_{i=1}^{n} A_i e^{-\frac{(x-x_i)^2}{2\sigma_i^2}}, \qquad (9)$$

where the subscript TnG denotes a TOF kernel consisting of $n$ Gaussians, $\sigma_i$ is the standard deviation of Gaussian $i$ and

$x_i$ is its shift. Then, proceeding as above, we find for the variance gain over non-TOF PET:

$$0.66 \frac{2D}{c} \frac{1}{\left(\sum_{i=1}^{n} A_i \, \mathrm{CTR}_i\right)^2} \left( \sum_{i=1}^{n} A_i^2 \mathrm{CTR}_i \right.$$
$$\left. + 2\sqrt{2} \sum_{i=1}^{n-1} \sum_{j=i+1}^{n} A_i A_j \frac{\mathrm{CTR}_i \, \mathrm{CTR}_j}{\sqrt{\mathrm{CTR}_i^2 + \mathrm{CTR}_j^2}} e^{-\frac{(x_i - x_j)^2}{2(\sigma_i^2 + \sigma_j^2)}} \right). \quad (10)$$

where $\mathrm{CTR}_i$ represents the CTR of the Gaussian component $i$.

## 2.6   Combining events with different SNR$^2$

A non-Gaussian TOF kernel is observed when all events have a non-Gaussian probability distribution, or when mixed events are detected and each event is drawn from a Gaussian probability distribution. In the latter case, we have events with different SNR$^2$, and it would be convenient to estimate the expectation of the SNR$^2$ for such system. This raises the question how events with different SNR have to be combined. Assume that we have two measurements of the same quantity, producing the values $m_i, i = 1, 2$, corrupted by independent noise with standard deviations $s_i, i = 1, 2$. We can combine these measurements in a weighted sum, using weight $w$, with the following expectation and variance:

$$m = \frac{m_1 + w \, m_2}{1 + w} \qquad s^2 = \frac{s_1^2 + w^2 s_2^2}{(1 + w)^2}. \qquad (11)$$

The optimal weight $w = m_2 s_1^2 / (m_1 s_2^2)$ minimizes the variance, and the resulting SNR$^2$ of $m$ equals

$$\mathrm{SNR}^2 = \frac{m^2}{s^2} = \frac{m_1^2}{s_1^2} + \frac{m_2^2}{s_2^2} = \mathrm{SNR}_1^2 + \mathrm{SNR}_2^2. \qquad (12)$$

Thus, assuming that the reconstruction algorithm makes optimal use of the data, the SNR$^2$ of a set of independent events is the sum of the SNR$^2$ of each event. Accordingly, the mean SNR$^2$ of a single event is the sum of the SNR$^2$ of each type of event, weighted by their relative abundance. This shows also that if event type $a$ has an SNR$^2$ that is $N$ times higher than that of event type $b$, a single type $a$ event is equivalent to $N$ type $b$ events.

With the equations above, one can estimate the

- the variance improvement (or equivalently, the gain in SNR$^2$) obtained by going from non-TOF to TOF PET,

- the gain in SNR$^2$ obtained by improving the TOF resolution,

- the gain in SNR$^2$ obtained by associating each event with its proper TOF kernel, as compared to using the averaged TOF-kernel for all events,

- the equivalent Gaussian TOF resolution for a non-Gaussian TOF kernel.

Consequently, these equations should be useful for predicting and comparing the performance of detectors that make use of events with different characteristics.

Note that above we have considered the center of a radioactive disk in a 2D PET system, but the same symmetries and reasoning apply for a radioactive sphere positioned in a fully 3D PET system.

## 2.7   Comparison to Tomitani's approach

To verify if the agreement between Tomitani's result and (6) is fundamental, and not a particular feature of the Gaussian function, we apply Tomitani's approach to the sum of two shifted Gaussians. Eq (11) of [1] gives the variance $V$ in the center of the image, reconstructed with FBP from a noisy TOF-PET measurement of a uniform cylinder:

$$V = 2\pi a \int_0^\infty \frac{P^2(R)}{\mathscr{R}[K^2](R)} R \, dR \qquad (13)$$

where $a$ is the emission count in cm$^{-2}$, $P$ is the Fourier transform of the image PSF $p$ (imposed by post-smoothing with a Gaussian), $K$ is the Fourier transform of the TOF kernel $k$ and $\mathscr{R}$ is the rotational mean operator as defined in (1) of [1]:

$$\mathscr{R}[K^2](R) = \frac{1}{\pi} \int_0^\pi [K^2](R \cos \theta) d\theta. \qquad (14)$$

When calculating (14) for a TOF kernel consisting of Gaussians, one has to solve integrals of the form

$$Q(R) = \int_0^\pi e^{-c_1 R^2 \cos^2 \theta} f(R \cos \theta) d\theta. \qquad (15)$$

For a single Gaussian TOF kernel, $f(R \cos \theta) = 1$ and the integral produces a modified Bessel function of the first kind. To obtain a closed form solution, in [1], this Bessel function is replaced with its asymptotic form, which is a good approximation for large $R$. However, for a sum of shifted Gaussians, in some of the terms, the integral $f$ is more complicated, and that same approach cannot be applied. This problem can be avoided by using an alternative approximation. The integrand in (15) is the product of a Gaussian and the function $f$. For large $R$, the Gaussian is a narrow peak around the point where $\theta$ is close to $\pi/2$. Consequently, for large $R$, the integral can be approximated as

$$Q(R) \simeq \int_{-\infty}^{\infty} e^{-c_1 R^2 (\frac{\pi}{2} - \theta)^2} f\left(R\left(\frac{\pi}{2} - \theta\right)\right) d\theta. \qquad (16)$$

No further approximations are then needed to obtain a closed form solution. For a single Gaussian, the result is identical to that with Tomitani's derivation, i.e. (6); for a sum of shifted Gaussians, the result is identical to (10).

## 2.8 Equivalence to detection using the reconstructed image

As shown in section 2.7, by generalizing Tomitani's approach, the (relative) SNR$^2$ calculated from the variance of the reconstructed images coincides with the SNR$^2$ obtained directly from the data for a SKE/BKE task. It is known that data processing does not change the ideal observer's performance provided the operation is invertible (see [11], p. 855). Image reconstruction is not in general invertible, but the same result holds for the minimum norm weighted least-squares reconstruction of a discrete linear problem.

When using the data for detection, the task is to decide between $\langle y \rangle = Ab + As$ (signal present) and $\langle y \rangle = Ab$ (signal absent), from a noisy measurement $y$ with covariance $V$, where $A$ is the projection matrix and the covariance $V$ is assumed independent of the presence of the signal. The prewhitening matched filtering (PWMF) test statistic equals $t = (As)^T V^{-1} y$, where $s$ is the signal in the image and the superscript $^T$ denotes the transpose. For the signal and variance of the test statistic one finds:

$$\Delta \langle t \rangle = \langle t \rangle_{\text{spresent}} - \langle t \rangle_{\text{sabsent}} \tag{17}$$

$$= s^T A^T V^{-1} As = s^T F s \tag{18}$$

$$\text{var}(t) = (As)^T V^{-1} V V^{-1} As = s^T F s, \tag{19}$$

where $F = A^T V^{-1} A$ is the Fisher information matrix. Therefore, the SNR$^2$ for this task equals

$$\text{SNR}^2_{\text{data}} = \frac{(\Delta \langle t \rangle)^2}{\text{var}(t)} = s^T F s \tag{20}$$

The minimum norm least-squares reconstruction equals $x^\dagger = F^\dagger A^T V^{-1} y$, where $F^\dagger$ is the pseudoinverse of $F$. When detecting on the image, the task is to decide between $x = b + s$ (signal present) and $x = b$ (signal absent) from a noisy image $x$ with covariance

$$W = F^\dagger A^T V^{-1} V V^{-1} A F^\dagger = F^\dagger F F^\dagger = F^\dagger. \tag{21}$$

Denote with $\eta$ the filter used to compute the test statistic, i.e. $t = \eta^T x^\dagger$. The squared SNR equals

$$\text{SNR}^2_{\text{image}} = \frac{(\eta^T F^\dagger A^T V^{-1} As)^2}{\eta^T W \eta} = \frac{(\eta^T F^\dagger F s)^2}{\eta^T F^\dagger \eta}. \tag{22}$$

Maximizing the expression in the RHS yields the PWMF $\eta = Fs$. Consequently,

$$\text{SNR}^2_{\text{image}} = s^T F s, \tag{23}$$

where we used $F F^\dagger F = F$. This shows that the SNR for the SKE-BKE detection task is the same when the detection is done using the data or using the reconstructed image.

## 3 Experiments

To verify the results obtained above, five PET systems with the same geometry but different event types were simulated. The systems had either

1. non-TOF events

2. Gaussian TOF kernel events of 70 ps FWHM

3. Gaussian TOF kernel events of 400 ps FWHM

4. an equal probability for the two TOF events above, and for each event, the associated TOF kernel was known

5. the same two TOF events as above, but individual events could not be assigned to a particular TOF kernel.

Simple 2D simulations were performed for these systems. An image of $100 \times 100$ pixels, with pixel size of 2.5 mm $\times$ 2.5 mm, containing a uniform disk with diameter of 200 mm was generated. A finite system resolution was modeled with an image-based, shift invariant Gaussian point spread function of 3.75 mm FWHM. The field of view of the simulated PET system was circular with a diameter 450 mm.

Five noise-free measurements were created. Three were produced by forward projecting this image with a non-TOF projector, a projector with 70 ps and one with 400 ps TOF-resolution. Two additional measurements were produced, one by combining equal fractions of unlabeled events with 70 and 400 ps TOF resolution, and a second one where each event was labeled with its TOF uncertainty. All sinograms had $5 \times 10^6$ events. For each sinogram, 100 (Poisson) noise realizations were produced. The sinograms were reconstructed with MLEM in combination with a non-TOF projector, a Gaussian TOF projector, non-Gaussian TOF projector (eq. (9)) or a combination of two Gaussian TOF projectors, as required for optimal reconstruction of the respective sinograms.

From these simulations, SNR$^2$ gains were computed with (6) for the Gaussian kernels and with (10) for the unlabeled mixed events. For the labeled events, the two SNR$^2$ computed with (6) were averaged.

SNR$^2$ gains will translate into variance gains, if the reconstructed images have identical spatial resolution. To achieve that with good approximation, a high number of iterations was applied, and the resulting images were post-smoothed to suppress the effect of residual resolution differences caused by small differences in convergence. For the non-TOF reconstruction, 400 iterations were applied. Considering that the TOF-reconstructions converge much faster, they were done with 200 iterations. All reconstructions were post-smoothed with a Gaussian kernel of 3 pixels (7.5 mm) FWHM. An image of pixel variances was computed for each case from the 100 noise realizations. The mean variance value in a central region of interest with diameter of 60 mm was computed for each case (this region contains 462 pixels). The results of the 100 noise realizations were then used to estimate the mean variance value and the error on that estimate.

A similar simulation was done to evaluate the predicted TOF-gain in the presence of randoms. Non-TOF and 200 ps FWHM TOF events were simulated. The simulation parameters were as above, but the number of non-TOF iterations was increased to 800 because the presence of the randoms slows down convergence, in particular for non-TOF. In addition, the noise amplitude was decreased to reduce non-negativity effects, by increasing the total counts to $2.5 \times 10^8$.

## 4 Results

The results of the simulation experiment with different TOF kernels are shown in table 1. Table 2 lists the results from the randoms simulation experiment.

| SNR$^2$ ratio | gain | error | predicted |
|---|---|---|---|
| TOF1 / nonTOF | 12.5 | 0.27 | 12.5 |
| TOF2 / nonTOF | 2.18 | 0.049 | 2.21 |
| TOFmix / nonTOF | 5.24 | 0.12 | 5.21 |
| TOFlab / nonTOF | 7.39 | 0.15 | 7.34 |
| TOF1 / TOF2 | 5.75 | 0.13 | 5.63 |
| TOF1 / TOFmix | 2.40 | 0.054 | 2.39 |
| TOF1 / TOFlab | 1.70 | 0.035 | 1.70 |

**Table 1:** Results of the TOF kernel simulation experiment. TOF1: 70 ps, TOF2: 400 ps, TOFmix: unlabeled mixed events, TOFlab: mixed events, each labeled with its TOF resolution. The second column gives the inverse of the variance ratio (i.e. the SNR$^2$ ratio) of the events obtained from the simulation, the third column the error estimate, and the fourth column the corresponding value predicted by the proposed method.

| randoms/trues | gain | error | predicted |
|---|---|---|---|
| 0.0 | 4.35 | 0.085 | 4.37 |
| 1.0 | 5.16 | 0.11 | 5.12 |
| 2.5 | 5.47 | 0.11 | 5.69 |
| 5.0 | 6.06 | 0.14 | 6.14 |

**Table 2:** Results of the randoms simulation experiment. From left to right: ratio of randoms to trues, the gain due to 200 ps TOF seen in the simulations, the error on the gain and the predicted gain.

## 5 Discussion

The proposed method determines the "value" of a single PET event, assuming that this event contributes to the reconstruction of the center of a uniform disk. This value is computed as the SNR of the optimal linear observer for detecting a small signal in an SKE-BKE setting. As shown above, the SNR for detection using the data is the same as the SNR for detection using the image. Because the latter is strongly related to lesion detection tasks in clinical PET, we believe that the proposed method is clinically relevant. In addition, following Tomitani [1], we show that this SNR is a good approximation of the SNR of the central pixel in the post-smoothed reconstruction, which is relevant for PET image quantification.

Good agreement is obtained between the observed and predicted variance ratios (tables 1 and 2). The results show that for a 20 cm disk, a 70 ps FWHM TOF event is worth about 12.5 non-TOF events, and that the effective sensitivity is inversely proportional to the TOF-resolution. The results also confirm that when events with different timing resolution occur, the effective sensitivity is higher if each event can be labeled with its own TOF kernel. From these results, one can deduce that when the mixed 70 and 400 ps are not labeled, the system is equivalent to a system with Gaussian TOF kernel of 168 ps. If each event can be associated with its own TOF kernel, then the system is equivalent to a regular TOF system with 119 ps FWHM. The randoms simulation experiment confirms that the gain obtained with TOF increases with increasing randoms fractions, because the non-TOF reconstruction is corrupted by the randoms that end up outside the object, whereas the TOF reconstruction is not.

Tomitani's analysis was done for reconstruction with FBP, whereas in these experiments MLEM was used. FBP is approximately an unweighted least squares estimator. MLEM is well approximated with a weighted least squares estimator, and because of symmetry, all weights are identical here.

When comparing non-TOF to TOF events, the analysis holds for the center of a uniform cylinder. However, when comparing different TOF events, the analysis also holds for the center of a uniform region within an arbitrary object, provided that the diameter of that region is large compared to the widest TOF kernel involved.

The proposed method only considers individual events, assuming that the two systems are identical except for a change to the event type. In real situations, changes to the detectors would not only change the type of events, but probably also the system sensitivity, the energy resolution etc, which should also be considered when predicting the effective sensitivity.

## 6 Conclusion

The proposed method assigns a relative SNR$^2$ to TOF events of different types. Its predicted pixel variance ratios in the reconstructed images agreed well with those observed in the simulation.

# 2D study of a joint reconstruction algorithm for limited angle PET geometries

Marina Vergara[1,2], Ahmadreza Rezaei[1], Maria Jose Rodriguez-Alvarez[2], Jose Maria Benlloch Baviera[2], and Johan Nuyts[1]

[1]Department of Imaging and Pathology, Division of Nuclear Medicine, KU Leuven, Belgium
[2]Instituto de Instrumentacion para Imagen Molecular Centro Mixto CSIC—Universitat Politecnica de València, Valencia, Spain

**Abstract**  Recently, a wide interest on organ-dedicated PET systems has been shown. Some of those systems present geometries that produce an incomplete sampling of the tomographic data due to limited angular coverage and/or truncation, which lead to artifacts on the reconstructed image. Moreover, they are often designed as stand-alone systems, which implies the absence of anatomical information to estimate the attenuation factors. In this work, we propose a joint reconstruction algorithm for estimating the activity and the attenuation factors on a limited angle PET system with time-of-flight capabilities. This algorithm is based on MLACF and uses literature linear attenuation coefficients in a known tissue-class region to obtain an absolute quantification. We evaluate the algorithm through simple 2D simulations for different TOF resolutions and angular coverage. The results show that with good TOF resolution quantitative PET imaging can be achieved even with aggressive angular limitation.

## 1   Introduction

In recent years, organ-dedicated PET (positron emission tomography) systems have been proposed as an alternative to the whole-body scanner. These systems are focused on being less expensive, requiring less space and/or providing easier patient access, having higher resolution and/or better sensitivity [1]. They are typically stand-alone systems, which means the absence of supplementary CT or MR systems that could be used to provide the attenuation image. Many such systems use a geometry that can lead to limited angular coverage and, therefore, the acquisition of incomplete tomographic data. Because the reconstruction from these data does not have a unique solution, the reconstructed images usually suffer from artifacts. It has been shown that in PET, the availability of TOF (time-of-flight) information reduces these limited angle artefacts [2, 3].

When data are provided with TOF information, joint reconstruction algorithms can be used to estimate the attenuation sinogram from the emission data up to a global constant [4] for all the LORs where activity is present, so long as the spread of the tracer is wider than the TOF resolution. This constant can be determined if prior knowledge about attenuation (or activity) values is available [5–8].

In this work we consider the problem of jointly reconstructing the activity image and the attenuation sinogram from TOF-PET data suffering from limited angular coverage or truncation. We propose an approach based on the MLACF algorithm of [9]. For limited angle data, it is not guaranteed that all sinogram pixels are affected by the same constant, but below we show that in many cases this will be the case.

To determine the value of the constant, the attenuation image is reconstructed from the estimated attenuation coefficients. The problem is only studied in 2D. As shown below, the results indicate that for PET with high TOF resolution, quantitative image reconstruction can be achieved even for systems providing very limited angular coverage and severe truncation.

The work presented here has also been submitted to IEEE TRPMS [10]

## 2   Materials and Methods

### 2.1   System design

In order to examine the effect of limited angular coverage in TOF-PET image reconstruction, we use a 2D simulation of a partial arc of a circular PET. We consider limited angle effects similar to those seen by a pair of flat panels of size $W$, separated by distance $D$, by setting to zero the sensitivity value of all the LORs that are not seen by both flat panels (LOR 1 and 3 in figure 1).

### 2.2   Joint reconstruction

In order to exploit the TOF information in the joint estimation of activity and attenuation process, Maximum Likelihood estimation of the activity and the Attenuation Correction Factors (MLACF) [9] is applied.

In TOF PET, the expected count $\bar{y}_{it}$ for a certain line of response (LOR) $i$ and TOF-bin $t$ can be written as:

$$\bar{y}_{it} = a_i p_{it} + r_{it} \quad \text{with} \quad p_{it} = \sum_j c_{ijt} \lambda_j \qquad (1)$$

where $p_{it}$ is the unattenuated TOF projection of the activity image $\lambda_i$ for LOR $i$ in TOF-bin $t$, $a_i$ is the total linear attenuation coefficient along the LOR $i$, $c_{ijt}$ is the sensitivity of the measurement bin $(i,t)$ for activity in voxel $j$ in the absence of the attenuation and $r_{it}$ is an additive contribution made by randoms and/or scatter.

Then the MLACF algorithm [9] is given by:

$$\lambda_j^{(n+1)} = \frac{\lambda_j^{(n)}}{\sum_i c_{ij} a_i^{(m+1)}} \sum_{it} c_{ijt} a_i^{(m+1)} \frac{y_{it}}{\sum_k c_{ikt} a_i^{(m+1)} \lambda_k^{(n)} + r_{it}},$$

$$(2)$$

$$a_i^{(m+1)} = a_i^{(m)} \sum_t \frac{p_{it}^{(n)}}{p_i^{(n)}} \frac{y_{it}}{a_i^{(m)} p_{it}^{(n)} + r_{it}}. \qquad (3)$$

where $n$ and $m$ denote the iteration numbers for the activity and the attenuation correction factors (ACFs), respectively, and $p_i^{(n)} = \sum_t p_{it}^{(n)}$. Also for other TOF-dependent variables, we will drop the TOF-index to denote summation over the TOF-bins.

TOF data determine the attenuation sinogram up to a constant, which correspond to a multiplicative factor in the activity image. To obtain it, we propose to reconstruct an image from the attenuation sinogram and use prior information in image space (e.g. known tissue attenuation in a certain region) to obtain the constant. To reconstruct the attenuation sinogram we use the MLTR algorithm [11]:

$$\mu_j^{(n+1)} = \mu_j^{(n)} + \frac{\sum_i l_{ij} \frac{\bar{y}_i - r_i}{\bar{y}_i} (\bar{y}_i - y_i)}{\sum_i l_{ij} \sum_k l_{ik} (\bar{y}_i - r_i)(1 - \frac{y_i r_i}{\bar{y}_i^2})} \qquad (4)$$

where $\mu_j$ is the reconstructed attenuation value in voxel $j$, $\bar{y}_i = b_i \exp(-\sum_i l_{ij}\mu_j) + r_i$, $l_{ij}$ is the intersection length of LOR $i$ and pixel $j$. Equation 4 uses the measured activity $y_i = \sum_t y_{it}$ as transmission scan and the unattenuated forward projection of the reconstructed activity $b_i = \sum_{jt} c_{ijt} \lambda_j$ as the blank scan.

In order to determine the constant, agreement of some reconstructed attenuation values with the known attenuation coefficient of soft tissues at 511 keV was imposed [7, 12]. A region composed mostly of soft tissue was identified in the image and the mean attenuation coefficient in that region was computed. The region was determined by thresholding the central region of the image, keeping the values greater than the median over the region. Given the ratio of the correct tissue attenuation value to the extracted mean attenuation value ($\beta$), the following expression was used to estimate $\gamma$, the factor by which the activity image (and therefore the blank scan) had to be scaled:

$$\sum_i c_{ij} \gamma \lambda_j e^{-\mu\beta L} \simeq \sum_i c_{ij} \lambda_j e^{-\mu L} \rightarrow \gamma \simeq (e^{\mu L})^{\beta - 1} \qquad (5)$$

The blank scan was then rescaled with the factor $\gamma$, the attenuation image with $\beta$ and a new MLTR-iteration was computed. This sequence of MLTR reconstruction and rescaling was repeated until $0.99 < \gamma < 1.01$, which typically happens after a few iterations. Note that the relation between the attenuation image and the transmission sinogram is non-linear, which is why MLTR iterations are required to correctly propagate the effect of rescaling the activity with $\gamma$ into the attenuation image.

## 3  Simulation experiments

In this section, the performance of a circular PET system for scanning a heart phantom is studied for different system parameters, including: full or limited angular coverage, sinogram truncation to a width $W$ of 50 cm, a TOF resolution of 250 ps or 60 ps FWHM and presence or absence of a scatter contribution. The noise was modelled using 200 noise realizations with a moderate noise level of $8.5 \cdot 10^5$ counts in the activity sinogram. The 2D simulated randoms are scaled to obtain randoms to primary ratio of 50%.

The images were reconstructed with 20 iterations of the re-scaled MLACF algorithm outlined above, and post-smoothed with a 2D Gaussian with FWHM of 4 mm.

We analyze two types of systems. We will consider a limited angle system that we call "open configuration" (described in figure 1 left) and a "closed configuration" that is obtained by accepting all LORs seen by the panels if they were rotating continuously.

In order to challenge the reconstruction, four (three horizontal and one vertical) small Defrise phantoms were added to the heart phantom (see figure 1). These phantoms have been added with, from top to bottom, distances of 16, 10, 10 and 30 mm between the rods.

In the right part of figure 1 we compare the activity reconstructions for both re-scaled MLACF and MLEM with perfect attenuation map for the 250 ps in the case of limited angle for open and closed configuration

Figure 1 show that the performance of the re-scaled MLACF algorithm is close to the MLEM with perfect attenuation map as the difference between their values is low compared to the background activity.

For a scatter simulation study (figure ??), a scatter sinogram was generated. It was produced by convolving the trues sinogram with a 3D Gaussian with a FWHM of 120 mm in radial direction, 0.43 radians in angular direction and 94 mm in TOF direction. After smoothing, the scatter sinogram was multiplied with a scatter to trues ratio of 70%. In the final sinogram, the ratio of scatters to prompts was 0.38. No scatter correction was applied to the reconstructed images.

The two narrowest horizontal Defrise phantoms are poorly reconstructed in the open configuration, but this is solved if the TOF resolution is improved to 60 ps (9 mm). As expected, the vertical Defrise phantoms are in any case well reconstructed. The absolute differences between the reconstructions and ground truth images (also post-smoothed with a 4 mm gaussian) are small.

## 4  Discussion

In this work, joint reconstructions of the activity and attenuation were done using a re-scaled MLACF approach. The scale factor can be determined by imposing a-priori known attenuation values to regions in the attenuation image [12]. When the scale factor is obtained, the activity image can be re-scaled accordingly, no new activity reconstruction is required. For these 2D simulations, a good value of the scale factor could be obtained, despite the presence of limited angle

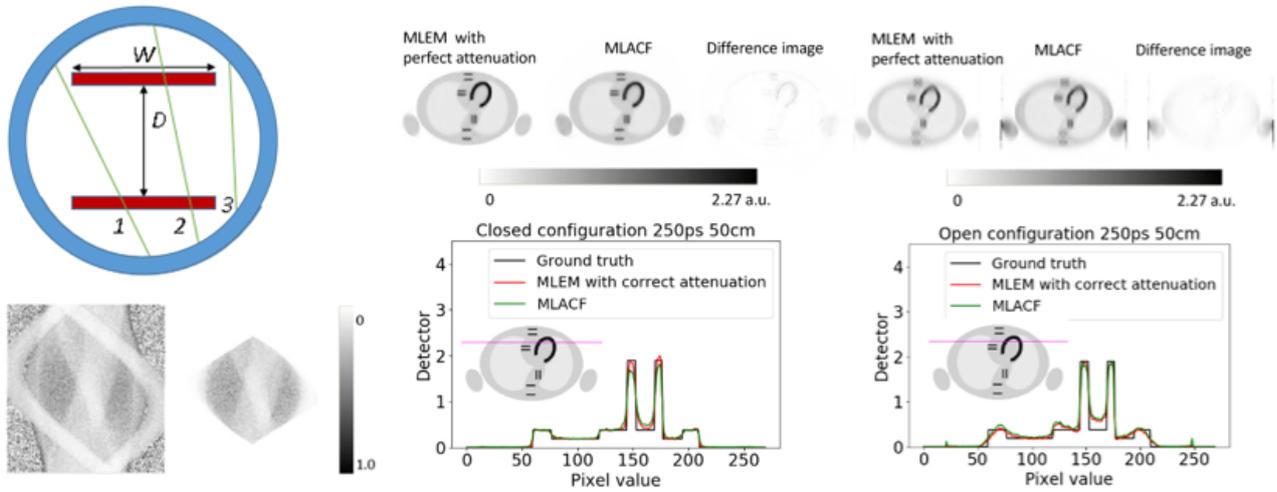

**Figure 1:** *Left up and down*: Description of the system and estimated attenuation sinograms for the closed and open configuration. *Up right*: Activity reconstructed images from perfect attenuation MLEM, re-scaled MLACF, and the difference image for closed and open configuration and 250 ps, respectively. *Bottom right*: Activity profiles along the line shown in the picture above. Even for the closed systems, the arms are outside the FOV and therefore not seen in all parallel projections.

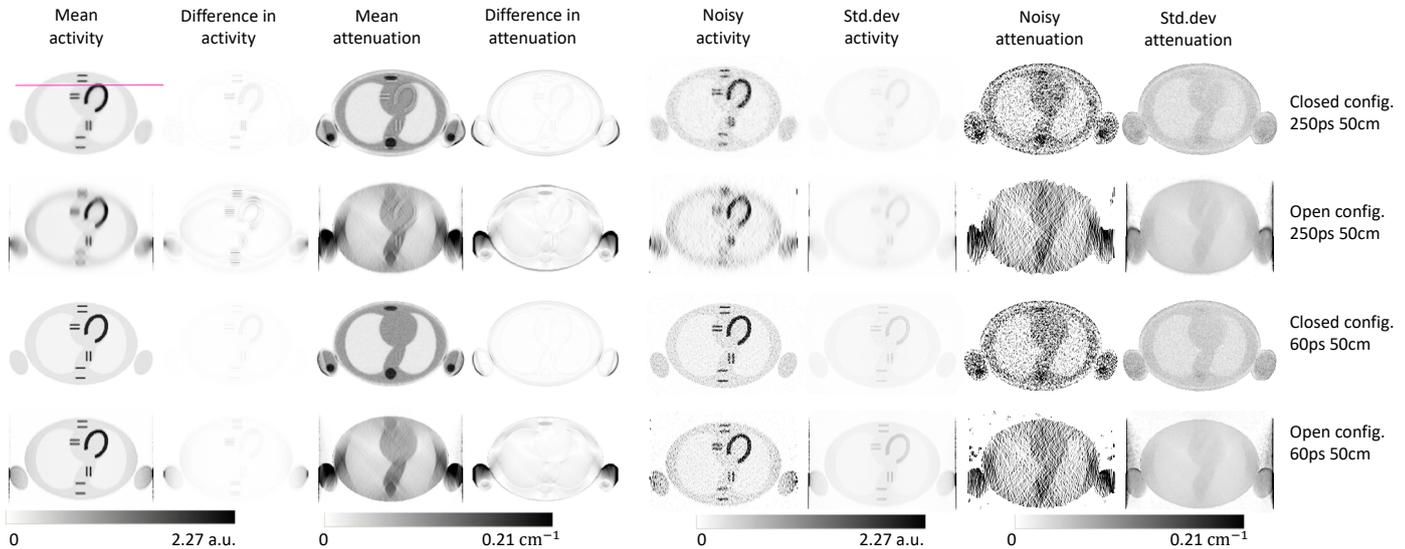

(a)

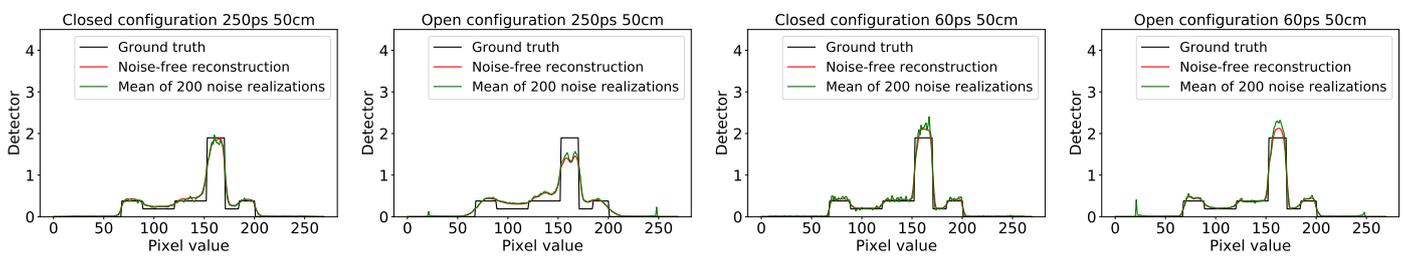

(b)

**Figure 2:** Results from 200 noise realisations for $W = 50$ cm without scatter contribution, (*from top to bottom*): closed configuration with 250 ps, open configuration with 250 ps, closed configuration with 60 ps and open configuration with 60 ps. In (a) and (*from left to right*): mean of the 200 noise realizations for the activity reconstruction, difference of this mean image with the ground truth, mean attenuation reconstruction, difference of this mean image with the ground truth, first noise realization of the activity, standard deviation for the 200 noise realizations of the activity, first noise realization of the attenuation and standard deviation of the attenuation. In (b) the line profiles of the activity image along a row through the apex of the heart (shown in (a) in purple) of the ground truth (black), the mean of 200 noise realizations (green) and the noiseless reconstructed image (red) for (*from left to right*) 250 ps closed configuration, 250 ps open configuration, 60 ps closed configuration and 60 ps open configuration.

artefacts.

In general, the images are worse for the open system, because of the increased amounts of missing data. For the closed system, the truncation is very limited, only part of the arms extend beyond the field of view and nearly artefact-free reconstructions are obtained. For the open system, the images are reconstructed from limited angle data. The horizontal Defrise phantoms, and in particular the two narrow ones, are poorly reconstructed when the TOF resolution is 250 ps, because the system has no projection data that "have seen" that there is a cold region between the hot rods. The vertical Defrise phantom is reconstructed accurately, because the two rods appear well separated in the acquired vertical projections. When the TOF resolution is improved to 60 ps, it provides a spatial resolution of 9 mm along the LORs. This enables the system to detect the cold region between the rods and to obtain accurate reconstruction of the Defrise phantoms.

Strong limited angle artifacts appeared in the attenuation and activity images near the edges of the phantom. MLACF is less performant if the activity is distributed along the LOR over a distance which is small compared to the TOF resolution. This is the case near the edges of the object. They produce an overestimation of the attenuation and activity estimates, which will adversely affect scatter estimates computed from the images, which, in turn, will propagate into the scatter corrected image. These effects will degrade image quality mostly near the edges of the object, but it can affect the central part to some extent. In addition, for the open system, the central part of the attenuation map is vertically blurred as a result of the missing data. Therefore, to ensure artefact-free and quantitative reconstruction of the activity near the center of the field of view, additional constraining of the attenuation image may be necessary, e.g. by imposing a maximum value to attenuation coefficients.

The analysis of the scatter estimation and correction problem is left for future research, but as a first exploration, the effect of a scatter estimation error on the performance of the re-scaled MLACF algorithm was investigated. For that purpose, re-scaled MLACF reconstructions without scatter correction were computed from scatter-contaminated sinograms. The reconstructed images are not exact, but they have a reasonably good visual quality. Based on these results, an iterative procedure alternating the estimation of the joint activity together with attenuation estimation and scatter estimation may work, even when initialized with a zero scatter estimate. We will evaluate this for fully 3D TOF-PET.

## 5 Conclusion

In this work, we investigated with simple 2D noise-free and noisy simulations the feasibility of obtaining attenuation corrected images from stand-alone limited angle TOF-PET systems. For the joint estimation of the activity and attenuation images, we proposed a re-scaled MLACF algorithm. We consider the results promising, warranting further investigation with more sophisticated fully 3D simulations and real PET measurements.

# First Results of HYPR4D Kernel Method with a Truly Four Dimensional Feature Vector on TOF PET Data

Ju-Chieh (Kevin) Cheng[1,2], Connor Bevington[2], and Vesna Sossi[2]

[1]Pacific Parkinson's Research Centre, University of British Columbia, Vancouver, BC, Canada
[2]Department of Physics and Astronomy, University of British Columbia, Vancouver, BC, Canada

**Abstract** We describe the initial results of our HYPR4D kernelized reconstruction method (i.e. PSF-HYPR4D-K-TOFOSEM) on dynamic 4D TOF PET data. The proposed HYPR4D kernel method was implemented for the GE SIGNA PET/MR system capable of high TOF resolution PET acquisition and compared to all the available TOF reconstructions with PSF resolution modeling on the system, namely PSF-TOFOSEM with and without standard post filter and PSF-TOFBSREM with various beta values. Results from experimental contrast phantom and human subjects injected with various PET tracers showed that more robust and accurate image features can be obtained from the proposed method compared to other regularization methods. For example, the preservation of contrast for the PSF-HYPR4D-K-TOFOSEM was observed to be better and less dependent on the contrast and size of the target structures as compared to PSF-TOFBSREM and PSF-TOFOSEM with filter. At the same contrast level, PSF-HYPR4D-K-TOFOSEM achieved better 4D noise suppression than other methods.

## 1 Introduction

Inspired by kernel methods in machine learning, kernelized reconstruction has shown promising results in PET in terms of noise suppression while preserving contrast [1][2]. In conventional kernel methods, high signal-to-noise ratio feature vectors which guide the de-noising process are typically pre-defined from either anatomical MRI or temporal sum of PET data. As a result, feature vectors contain very little or no temporal information of PET tracers and thus are not able to provide sufficient temporal noise reduction nor properly 'track' and preserve the temporal pattern of PET tracers. Moreover, the pre-defined feature vectors can introduce bias to PET images whenever there are mismatches in features between feature vectors and target PET images [2].

Recently, a spatiotemporal kernel method has been proposed to achieve high temporal resolution (HTR) by incorporating a single temporal kernel extracted from PET sinogram data [3]. More recently, we have proposed an intrinsic data-driven/prior-free 4D kernel method based on the 4D modified HighIY constrained backPRojection (HYPR4D), utilizing a truly 4D feature vector which applies voxel-specific temporal kernel generated directly within the reconstruction and demonstrated better preservation of spatiotemporal patterns while achieving 4D noise reduction as compared to the HTR kernel method as well as other standard noise reduction methods on conventional non-TOF PET data [4].

In this work, we implemented our HYPR4D kernel method (i.e. PSF-HYPR4D-K-TOFOSEM) using the PET toolbox for the GE SIGNA PET/MR system capable of high TOF resolution PET acquisition and compared its performance in terms of contrast recovery and noise suppression as well as time-activity curves in relatively small structures to that obtained from all the available TOF reconstructions with point-spread-function (PSF) resolution modeling on the system, namely PSF-TOFOSEM with and witout standard (3.5mm FWHM transaxial and 1-4-1 axial) post filter and PSF-TOFBSREM (Q.Clear) with various beta values using data acquired from contrast phantom and human subjects.

## 2 Materials and Methods

### PSF-HYPR4D-K-TOFOSEM

The HYPR4D kernel matrix consists of spatiotemporally variant convolutional basis functions, and the feature vector is computed as the sum of de-noised subset estimates which can be generated directly within the reconstruction at every time point. As a result, a truly 4D and purely data-driven feature vector can be obtained. The proposed PSF-HYPR4D-K-TOFOSEM is given by:

$$\alpha_{4D}^{m,s} = \frac{\alpha_{4D}^{m,s-1}}{(K_{H4D}^m)^T P_{S;t}^T 1} \cdot \left( (K_{H4D}^m)^T P_{S;t}^T \frac{y_{4D}^{s;t}}{P_{S;t} K_{H4D}^m \alpha_{4D}^{m,s-1} + b_{4D}^{s;t}} \right) \quad (1)$$

$$\lambda_{4D}^{m,s} = K_{H4D}^m \alpha_{4D}^{m,s} \quad (2)$$

$$K_{H4D}^m = diag[h^m] F_{4D}, where \ h^m = \frac{c_{4D}^m}{F_{4D} * c_{4D}^m} \quad (3)$$

$$C_{4D}^m = \sum_s K_{H4D}^{m-1} \alpha_{4D}^{m-1,s} \quad (4)$$

where $\alpha_{4D}^{m,s}$ is the 4D kernel coefficient at $s^{th}$ subset of $m^{th}$ iteration, $K_{H4D}^m$ is the HYPR4D kernel matrix which is decomposed into the self-normalized spatiotemporal weights extracted from the 4D feature vector ($C_{4D}^m$) for the preservation of 4D high frequency features ($h^m$) and the spatiotemporally invariant 4D Gaussian convolution ($F_{4D}$). The sparsity of the kernel matrix only depends on the width of the 4D Gaussian since the matrix which contains $h^m$ is diagonal. $P_{s;t}$ is the system matrix for the $t^{th}$ TOF bin within the $s^{th}$ subset; the projection based resolution modeling with spatially variant PSF and time spread function used for TOF reconstruction are embeded here along with normalization and attenuation corrections. $y_{4D}^{s;t}$ is the measured dynamic 4D TOF sinogram data, $b_{4D}^{s;t}$ is the estimate of background contamination such as randoms and scattered coincidences at $t^{th}$ TOF bin within the $s^{th}$ subset, and $\lambda_{4D}^{m,s}$ is the 4D (de-noised) PET image estimate at $s^{th}$ subset of $m^{th}$ iteration.

One iteration of PSF-TOFOSEM was used to initialize the 4D feature vector (i.e. sum of subset updates within the first iteration of PSF-TOFOSEM) in the kernel matrix. The one PSF-TOFOSEM iteration images are also used as the input initial 4D estimate for Eq. (1). After the 1st HYPR4D iteration, the feature vector is updated using the de-noised subset images from the previous iteration as shown in Eq. (4) and thus provides a highly constrained noise increment per update and allows the 4D high frequency features to be updated in a cleaner fashion as compared to conventional methods.

In short, the proposed method makes use of inconsistent noise patterns across subset data as well as the low noise property of the early updates of reconstruction to achieve noise constraint/reduction. The progressive update of the 4D feature vector ensures the extracted 4D high frequency features are adaptive to the measured PET data. As a result, better preservation of spatiotemporal patterns can be attained by the proposed method as compared to other methods while achieving 4D noise reduction. Additional benefits of the proposed method include reduction of zero trapping and limit cycle behaviour typically observed from OSEM reconstructions.

*Experimental Setups and Reconstructions*
A 16 cm diameter cylindrical contrast phantom with a 10 mm diameter sphere was filled with a 4:1 sphere-to-background ratio and injected with a total activity of 1.5 mCi of [18F]FDG. The phantom was scanned on the GE SIGNA PET/MR within the HNU coil for 15 minutes. The list-mode data were unlisted/framed according to the temporal count distribution (ranging from 5 million to 140 million counts) formed by our dynamic framing protocol for 11C human subject scans (i.e. 60s x 4, 120s x 3, 300s x 8, 600s x 1).

After unlisting, the dynamic 4D TOF sinogram data were reconstructed using PSF-TOFOSEM with and without the standard 3.5 mm FWHM transaxial and 1-4-1 axial filter, PSF-TOFBSREM with 8 different beta values ranging from 50 to 400 with an increment of 50, and PSF-HYPR4D-K-TOFOSEM with a 4D kernel size of 13 x 13 x 7 x 13 doxels which corresponds to 5.6 mm FWHM in the spatial domain and 4 frames FWHM in the temporal domain. The 4D kernel size used in PSF-HYPR4D-K-TOFOSEM was selected to achieve sufficient 4D noise reduction without making the kernel matrix too excessively non-sparse; i.e. the computation speed for the 4D kernel operations with this kernel size is 4 times faster than that of TOF projection operations and is thus making the HYPR4D kernel method practical in realistic scanning situations.

All reconstruction methods were run up to 10 iterations with 28 subsets except PSF-TOFBSREM. For each beta value used in PSF-TOFBSREM, 2 iterations of OSEM, 3 iterations of BSREM, and 8 iterations of PSF-TOFBSREM were used according to the GE protocol for PSF-TOFBSREM. All corrections such as normalization, scatter, randoms, and CT based attenuation correction of the phantom were applied for all reconstruction methods. The reconstructed image matrix size is 256 x 256 x 89 with voxel size of 1.39 x 1.39 x 2.78 mm³ for all methods. The average Contrast Recovery Coefficient (CRC) +/- STD across all dynamic frames for the 10 mm sphere was computed and plotted as a function of average voxel noise from the uniform background regions for each reconstruction method.

Human [11C]RAC and [11C]DTBZ scans with 10 mCi bolus injection were acquired for 60 minutes on GE SIGNA PET/MR. Lint-mode data were unlisted using the same dynamic framing protocol for the phantom mentioned above. The dynamic 4D TOF sinogram data were reconstructed using 4 iterations of PSF-TOFOSEM with and without the standard 3.5 mm FWHM transaxial and 1-4-1 axial post filter, PSF-TOFBSREM with various beta values, and 10 iterations of PSF-HYPR4D-K-TOFOSEM with the same 4D kernel size as mentioned above.

The number of iterations selected for each method was based on the CRC vs noise trade-off (see Figure 1) except for PSF-TOFBSREM which was run according to the GE protocol as mentioned above. All corrections were applied, and ZTE based MRAC was used for the attenuation correction of human subjects. In addition, a 5 mCi human [18F]FDG scan was also reconstructed using the same framing protocol for the evaluation of a more clinically relevant task. Time-Activity-Curve (TAC), image profile, and visual image quality comparisons were performed for the human scans.

## 3 Results and Discussion

The CRC vs voxel noise comparison for the 10 mm diameter sphere is shown in Figure 1. PSF-TOFOSEM had the highest noise increment per iteration compared to all other methods while the filtered PSF-TOFOSEM achieved noise reduction at the cost of lower CRC as expected. PSF-TOFBSREM achieved better CRC vs noise trajectory than PSF-TOFOSEM with and without filter. PSF-HYPR4D-K-TOFOSEM had the lowest noise increment per iteration and achieved even better CRC vs noise trade-off than PSF-TOFBSREM.

It can be observed that PSF-TOFBSREM with different beta values introduced different level of additional partial volume effect (PVE) and thus created different CRC values. PSF-TOFBSREM with high beta value (e.g. beta=400) was observed to have similar contrast vs noise trade-off but with higher variation in CRC (i.e. bigger error bar) as compared to the early iteration estimate of PSF-HYPR4D-K-TOFOSEM. Additionally, 10 iterations of PSF-HYPR4D-K-TOFOSEM had similar noise level as compared to 1 iteration of PSF-TOFOSEM (which was used as the input image estimate for PSF-HYPR4D-K-TOFOSEM) but with CRC similar to that of the later iterations of PSF-TOFOSEM.

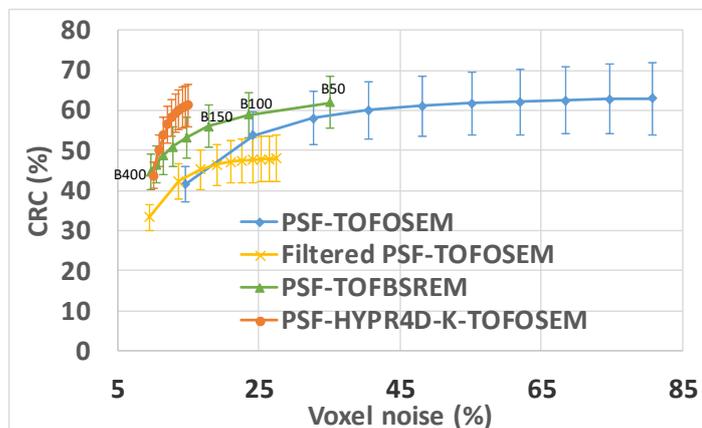

**Figure 1. Contrast recovery coefficient vs voxel noise for the 10 mm sphere obtained from various reconstruction methods. Each point represents an OSEM iteration except for PSF-TOFBSREM where each point represents a beta value ranging from 50 to 400. The beta value increases from right to left with an increment of 50 (see labels for guidance) while the number of OSEM iteration increases from left to right.**

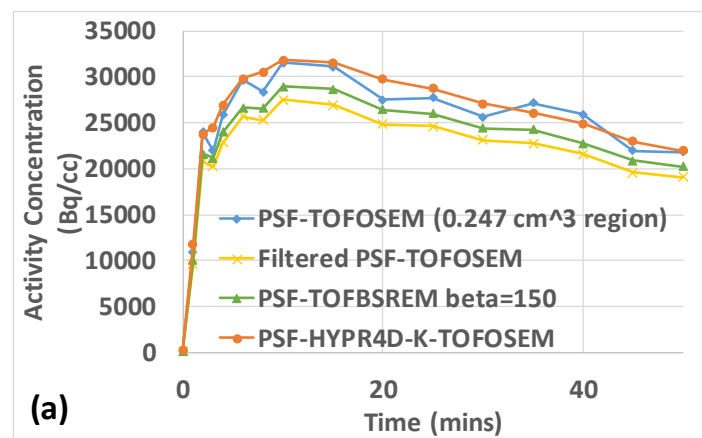

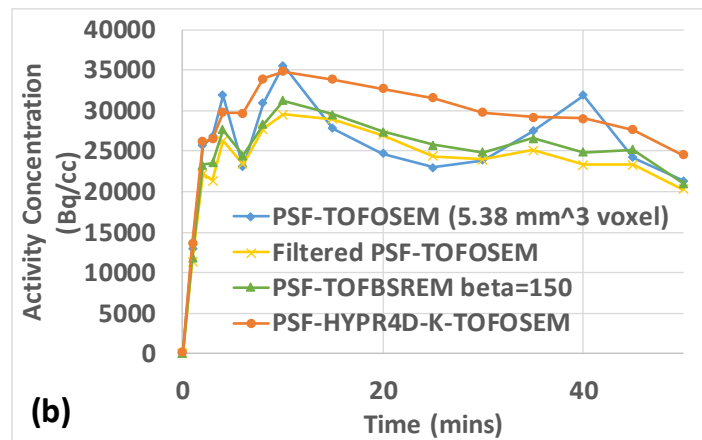

**Figure 2. (a) Regional-level TAC and (b) voxel-level TAC comparisons in the caudate of the human [11C]RAC scan reconstructed using various methods.**

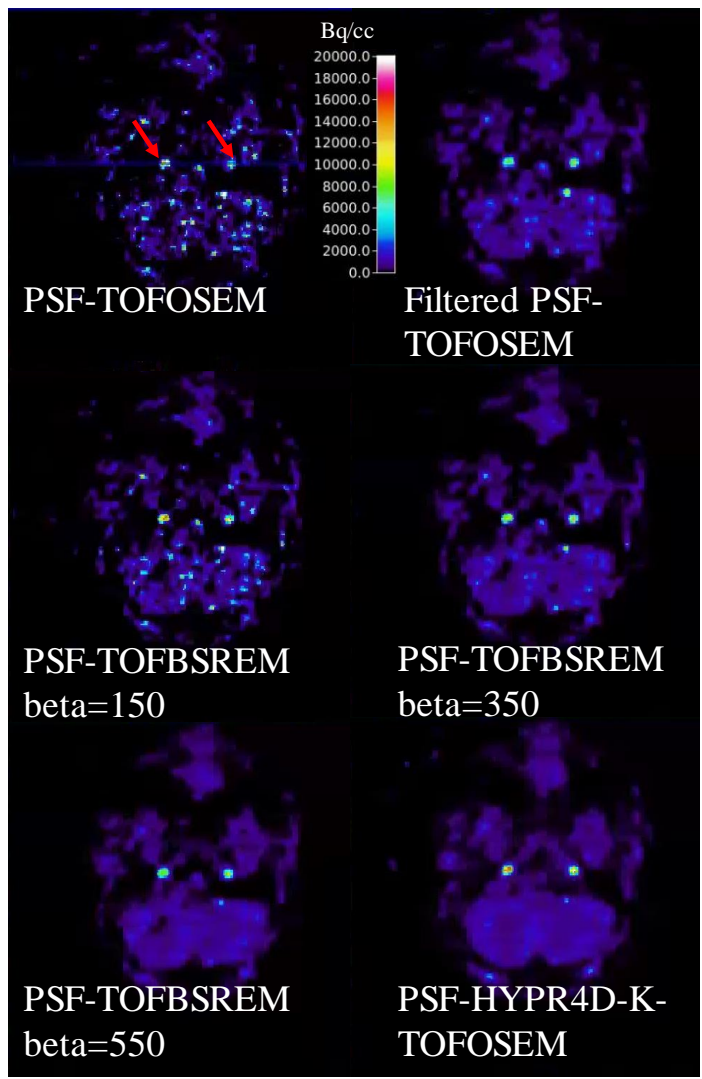

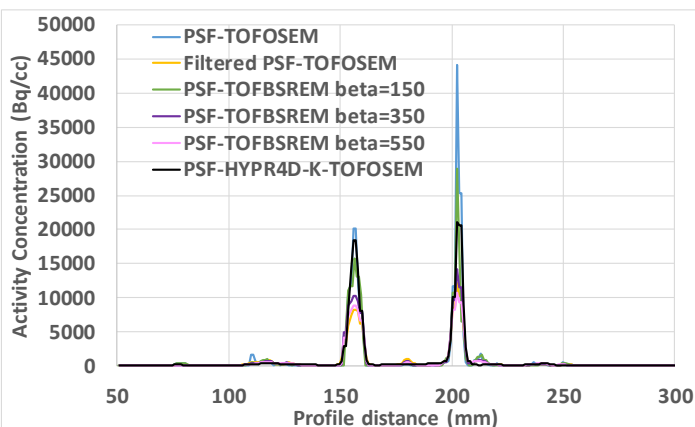

**Figure 3. (Top) A transaxial slice which contains very high contrast (much greater than 4:1 contrast ratio) in the carotid arteries with a size of 5-7 mm in diameter (see red arrows) and (Bottom) line profile across the carotid arteries in a low count frame of human [11C]DTBZ scan reconstructed using various methods. The location of the line profile can be seen from the PSF-TOFOSEM panel. Note that a substantially higher peak signal likely induced by noise can be observed from the right carotid artery than that from the left carotid in the PSF-TOFOSEM image though the ground truth is not known here.**

The TAC comparison for a relatively small region in the caudate and for a single voxel within the region from the human [11C]RAC scan is depicted in Figure 2. At both regional and voxel levels, PSF-HYPR4D-K-TOFOSEM showed the lowest temporal noise as compared to other methods. A fairly consistent difference in regional activity concentration or magnitude across time can be observed in Figure 2a between PSF-TOFOSEM and filtered PSF-TOFOSEM due to the consistent PVE introduced by the post filter given that the contrast in this region does not vanish over time for this tracer during the scan.

However, this difference was not observed at the voxel level as shown in Figure 2b due to the relatively high voxel noise in the PSF-TOFOSEM images. On the other hand, both regional and voxel TACs obtained from PSF-HYPR4D-K-TOFOSEM showed similar difference in magnitude as compared to filtered PSF-TOFOSEM. TACs obtained from PSF-TOFBSREM with beta=150 was observed to have a slightly higher magnitude than those from filtered PSF-TOFOSEM but with similar noise-induced temporal patterns. Higher beta values resulted in lower magnitudes in TACs as expected (not shown).

A low count frame which contains very high contrast signal in the carotid arteries (5-7 mm in diameter) from the human [$^{11}$C]DTBZ scan and a high count frame which contains moderate contrast in the colliculi (~4 mm in diameter) from the human [$^{18}$F]FDG scan are shown in Figures 3 and 4, respectively. As expected, post filter drastically reduced the contrast in small structures especially when structures contain very high contrast as shown in the line profile comparison in Figure 3 (Bottom); i.e. surrounding voxels contain much lower activity concentration values. The additional PVE introduced by the post filter became stronger with increasing contrast and/or decreasing size in the target structures and vice versa.

Interestingly, the reverse trend was observed for PSF-TOFBSREM. For a given beta value (see beta=150 for example) PSF-TOFBSREM was observed to preserve the contrast in small structure with very high contrast substantially better than PSF-TOFOSEM with post filter as shown in Figure 3 (Bottom) though the noise reduction was not sufficient for low count data with beta=150. However, for moderate contrast level as depicted in Figures 2 and 4, the preservation of contrast for PSF-TOFBSREM with beta=150 became similar to that of PSF-TOFOSEM with post filter. On the other hand, the preservation of contrast for PSF-HYPR4D-K-TOFOSEM showed less dependency on the contrast level and size of the target structures as compared to other regularization/noise reduction methods (i.e. more robust).

## 4 Conclusion

The preservation of contrast for the PSF-HYPR4D-K-TOFOSEM was observed to be better and less dependent on the contrast and size of the target structure as compared to other regularization methods such as PSF-TOFBSREM and

PSF-TOFOSEM with filter. At the same contrast level, PSF-HYPR4D-K-TOFOSEM also achieved better 4D noise suppression than other methods. These promising initial results on TOF PET data demonstrated that the proposed HYPR4D kernel method is likely suitable for all imaging tasks without requiring any prior information. Future work includes comparisons of other PET derived parameters as well as validations with more subjects.

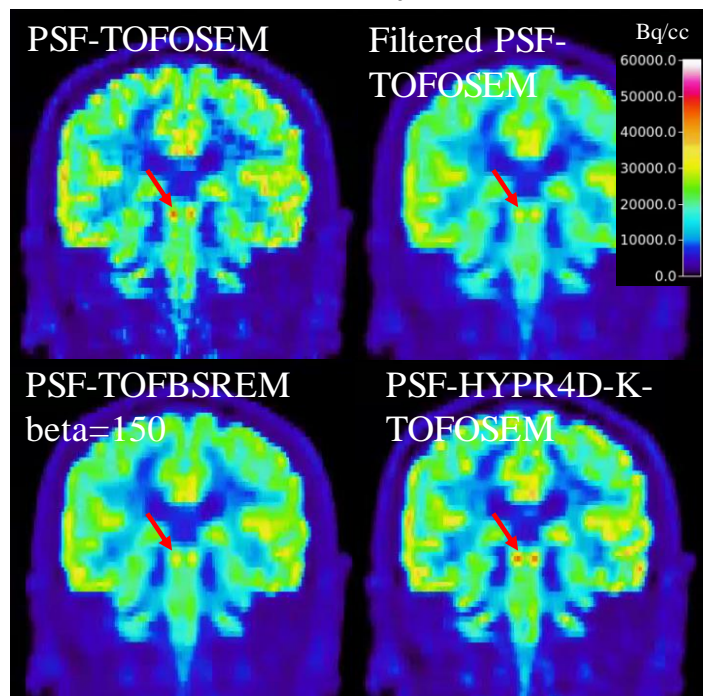

**Figure 4. A coronal slice which contains colliculi with a size of ~4 mm in diameter (see red arrows) in a high count frame of human [$^{18}$F]FDG (50 minutes post injection with 10 minutes frame duration) reconstructed using various methods. PSF-TOFBSREM with higher beta values were omitted since higher beta values only reduce contrast further without providing any benefit for high count data.**

## Acknowledgement

The authors would like to thank Kristen Wangerin and Timothy Deller from GE Healthcare for their helpful discussions about PET toolbox usage for the GE SIGNA PET/MR.

# DeepDIRECT: Deep Direct Image Reconstruction from Multi-View TOF PET Histoimages using Convolutional LSTM

Yusheng Li[1] and Samuel Matej[1]

[1]Department of Radiology, University of Pennsylvania, Philadelphia, United States

**Abstract**
Time-of-flight (TOF) PET scanners offer the potential of previously unachievable image quality in clinical PET imaging. The large sizes of TOF PET data are challenge for fully 3D image reconstructions. In our previous research, we proposed using multi-view TOF histoimages to efficiently store and process TOF PET data. In this work, we propose DeepDIRECT, a deep neural network (DNN) for TOF PET image reconstruction using convolutional long short-term memory (ConvLSTM). DeepDIRECT uniquely processes multi-view TOF histoimages in image space, which allows computationally efficient DNN reconstruction directly from 5D TOF histoimages to 3D reconstructed images. We train a ConvLSTM network using simulated datasets generated from a generic TOF PET with 250 ps timing resolution. We show improved image quality (in terms of PSNR and NRMSE) in DeepDIRECT reconstructed image with multi-view histoimage compared to single-view histoimage. We further show that DeepDIRECT reconstruction has substantially reduced noise performance at a similar contrast compared to TOF FBP and TOF OSEM reconstructions.

## 1 Introduction

Time-of-flight (TOF) PET scanners offer the potential of previously unachievable image quality in clinical PET imaging. TOF PET data are often stored in binned data known as TOF sinograms (with reduced number of TOF bins) or in list-mode data. There is no localized relation between TOF sinograms and images, let alone list-mode data. On the other hand, TOF histoimage proposed in our previous work exhibit localized properties leading to very efficient implementation of classical and deep learning reconstruction approaches involving convolutional operations.

Over the past few years, deep learning based image reconstruction using neural network has emerged as a fundamentally new approach with the advantages that it can dramatically reduce reconstruction time and incorporate more complicated priors given sufficient training datasets. Methods for deep image reconstruction generally fall into four categories: 1) image domain methods, e.g., post-processing reconstructions, using a convolutional neural network (CNN); 2) data-domain methods, e.g., pre-processing sinograms; 3) hybrid learning methods to incorporate priors; and 4) direct, end-to-end deep reconstruction methods, which produce reconstructed image directly from measured data. Among these methods, direct reconstruction methods, such as AUTOMAP [1] and deepPET [2], offer great potential for high-quality reconstructed images; however, they are very computationally expensive and difficult to train, and thus only been applied to small 2D slices. Deep learning image reconstruction using fully connected feed forward neural network can be infeasible or impractical for clinical PET data. CNN and RNN utilizing localized properties and relations between adjacent views of TOF histoimages are promising building blocks for the practically efficient deep image reconstruction for fully 3D PET data.

Thanks to our previously proposed TOF histoimage format, the TOF histoimage has the same coordinates at the reconstructed image, and the histoimage is essentially related with reconstructed image with a convolution operation with a localized TOF kernel with size depending on the timing resolution [3–5]. TOF histoimage can be deposited/partitioned from TOF sinograms or list-mode data with no information loss, or with little information loss using additional view grouping to significantly reduce data size. TOF histoimage is naturally suitable and ideal for deep image reconstruction for TOF PET. Recently, TOF backprojected images as the starting point for deep learning using a U-net demonstrated promising results [6], where the backprojected images is just a single-view histoimage. In this work, we propose the convolutional Long short-term memory (LSTM) based deep neural network (DNN) reconstruction using multi-view TOF histoimages, which allows for computationally efficient direct DNN reconstruction directly from TOF histoimages to reconstructed image.

## 2 Materials and Methods

### 2.1 TOF Histoimage Formation

For time-of-flight PET, each coincidence line-of-response (LOR) can be determined by two detectors located at $\vec{a}_1$ and $\vec{a}_2$. The TOF data generally can be formulated as

$$p(\vec{a}_1, \vec{a}_2, t) = \int_{-\infty}^{+\infty} dl \, h(t-l) f\left(\frac{\vec{a}_1 + \vec{a}_2}{2} + l\hat{n}\right), \quad (1)$$

where $f \in C_0(\mathbb{R}^3)$ is a 3-D tracer distribution, $h$ is a TOF profile, $\hat{n} = \frac{\vec{a}_2 - \vec{a}_1}{\|\vec{a}_2 - \vec{a}_1\|}$ is the direction, $t$ is the TOF parameter in a unit of length. The TOF profile is usually modeled as a Gaussian distribution with standard deviation $\sigma$,

$$h(t) = \frac{1}{\sqrt{2\pi}\sigma} \exp\left(-\frac{t^2}{2\sigma^2}\right). \quad (2)$$

TOF data are often stored in binned format known as TOF sinograms or in list-mode format. For TOF histoimages, we

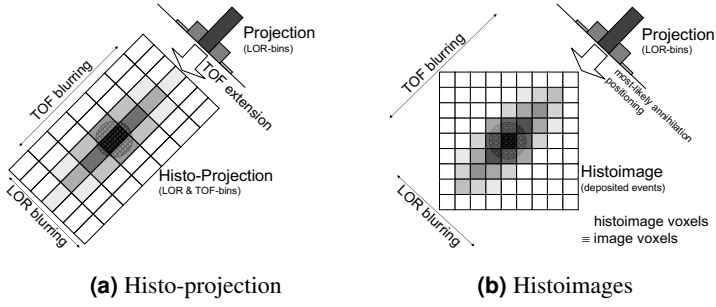

**(a)** Histo-projection      **(b)** Histoimages

**Figure 1:** Comparison of TOF data binned into histo-projections (a) and TOF histoimages (b). Histo-projections can be described as an extension of non-TOF projections (radial bins) along the TOF direction (time bins); the sampling intervals relate to the projection geometry and timing resolution, while histoimage sampling is given by the voxel size and geometry of the reconstructed image. (This figure is adapted from Fig. 1 in reference [4].)

first deposit the TOF list-mode data (or sinogram) into image space using the most likely annihilation (MLA) position

$$\vec{x}_{MLA} = \frac{\vec{a}_1 + \vec{a}_2}{2} + t\, \frac{\vec{a}_2 - \vec{a}_1}{\|\vec{a}_2 - \vec{a}_1\|}. \qquad (3)$$

The confidence weighted (CW) histogramming method can also be used with slightly increased computation. A TOF histo-image has the same coordinate as the reconstructed image with the geometry determined by the desired voxel size, as illustrated in figure 1 (right). The expectation of deposited TOF histo-image can be modeled as a convolution [5]

$$q(\vec{x}, \hat{n}) = \int_{-\infty}^{\infty} \mathrm{d}l\, h(l) f(\vec{x} - l\hat{n}). \qquad (4)$$

We can rewrite (4) as a 3D convolution in vector form

$$q(\vec{x}, \hat{n}) = f(\vec{x}) *** \kappa(\vec{x}, \hat{n}), \qquad (5)$$

where the 3D TOF kernel is given by

$$\kappa(\vec{x}, \hat{n}) = h(\vec{x} \cdot \hat{n})\, \delta(\vec{x} \cdot \hat{u})\, \delta(\vec{x} \cdot \hat{v}). \qquad (6)$$

The timing resolution is in the range of 200 ps to 400 ps for modern PET scanners using (digital) silicon photomultiplier. Thanks to the good timing resolution, the angular sampling requirements can be substantially reduced for TOF histoimages [4, 7]. We can dramatically reduce TOF histoimage storage using view grouping by grouping events into a set of transverse views and axial tilts, satisfying the TOF angular sampling requirements.

## 2.2 Neural Network Architecture

We propose the convolutional long short-term memory (Conv-LSTM) for computationally efficient direct deep leaning image reconstruction, as conceptualy illustrated in figure 2. TOF multi-view histoimage has 5 dimensions, 3 spatial dimension $(n_x, n_y, n_z)$ and 2 angular dimensions $(n_\phi, n_\theta)$ [5].

Rather than just using convolutional neural network (CNN), we propose to use CNN in spatial (image) domain, and recurrent neural network (RNN) to process the sequence along the angular dimensions.

A recurrent neural network (RNN) has a memory/state that stores the information pertaining to what it has observed/processed, and it processes sequential data through a number of iterations. RNN is a generalization of Markov chain and has much stronger processing capacities. The RNN, however, suffers from the problem of vanishing gradients. The long short-term memory (LSTM) is one of the most popular RNNs developed by Hochreiter and Schmidhuber [8] that adds a way to carry information across sequences, which prevents older signals from vanishing gradually. Figure 3 shows detailed implementation using basic LSTM cells, where $t$ denotes the view index (with a slight abuse of notation). In the figure, $i_t$ is the input gate, $f_t$ is the forget gate, $c_{t-1}$ is the previous cell output, $o_t$ is the output gate, and $h_t$ is the final state. LSTM processing updates for view $t$ given input $x_t$, and the previous state $h_{t-1}$, and previous cell output $c_{t-1}$.

Our approach of combining LSTM and CNN is based on the idea of ConvLSTM proposed by Shi et al. [9]. Since then, ConvLSTM has been applied to other different applications, e.g., processing stacked back projections for sparse-view CT reconstruction [10]. Our ConvLSTM structure differs from a simple CNN plus LSTM structures, where the convolution structure (CNN) is applied as the first few layers and sequentially LSTM layer is applied in the later layers. In the proposed ConvLSTM, the convolution structures are applied to the spatial domain, and the convolutional operations are nested into an LSTM structure to learn across multi-view hitoimages along the angular domain, as shown in figure 4. We also stack a few ConvLSTM blocks/layers to form a deep structure across spatial domain to fully learn the information. Currently, we used 4 ConvLSTM layers, and each layer has 4 filters ($n_f = 4$) with kernel size of $3 \times 3$. The 'same' padding option is used in convolution since the array size in spatial domain of histoimage is kept the same during learning. Batch normalization layers were included between ConvLSTM layers. The 4 channels of output from the last ConvLSTM can be further processed using a simple $1 \times 1$ convolutional layer or a simple U-net for refined reconstruction.

## 2.3 Simulated Datasets and Network Training

As preliminary work, we perform numerical simulations with a generic 2D TOF PET system to generate training datasets. Based on the positive results, our studies will be expanded to involve full 3D TOF dataset from clinically PET scanners. The generic system has timing resolution 250 ps and crystal size of 4 mm. We used random phantoms with warm background of 35 cm disk discretized in $144 \times 144$ with 4 mm pixel size. Each image contains 24 mixed ellipses (2/3 in probability) and rectangles (1/3 in probability) with random locations and orientations, with semi-axes 10, 13, 17, 22, 28, 37 and 50 mm with both hot and cold concentrations. As shown in figure 5, we generated simulated histo-images from the random phantoms in an array size of $144 \times 144 \times 48$ with 4 mm pixel and 48 views, which was group from TOF data with 288 azimuthal angles uniformly spaced over 180°. There were 384 datasets: 320 for training, 32 for validation, 32 for testing. Both noise-free and noisy TOF histoimages were used for training.

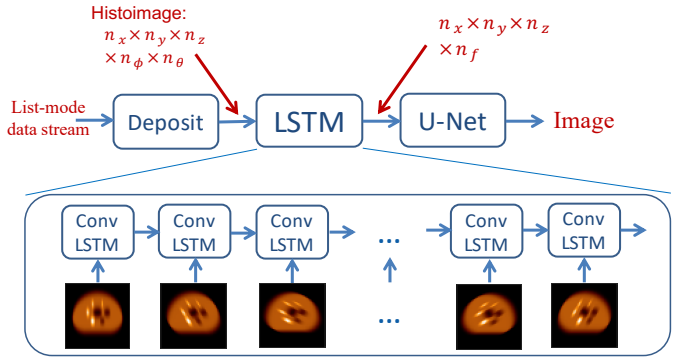

**Figure 2:** Diagram for deep image reconstruction from multi-view TOF histoimages using ConvLSTM. The PET list-mode data stream is deposited into multi-view TOF histoimages as input to the ConvLSTM followed by a U-net for refined processing.

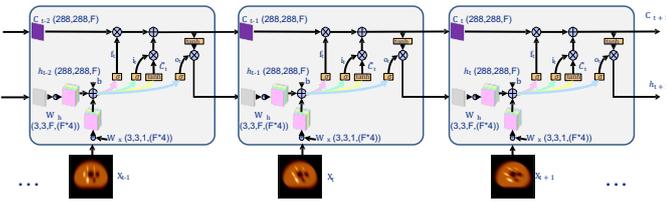

**Figure 3:** Detailed implementation of the ConvLSTM using basic LSTM cells with convolutional operations. $\sigma$ is sigmoid function, and tanh is hyperbolic tangent.

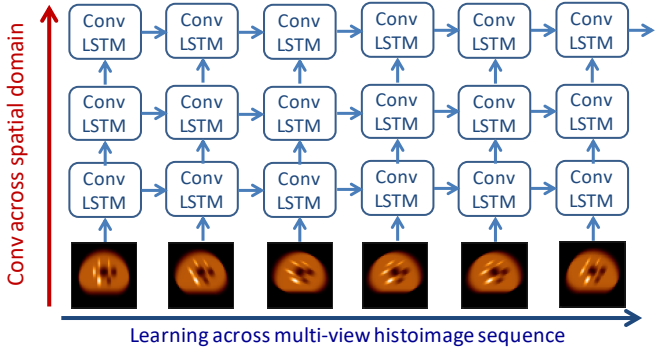

**Figure 4:** Deep ConvLSTM architecture.

The ConvLSTM neural network was implemented using `TensorFlow` and `Keras`, and was trained using a Nvidia Titan RTX GPU. The mean squared error (MSE) loss function was used.[1] The Adam optimizer, a gradient-based optimization based on adaptive estimates of lower-order moments,

---

[1] We plan to investigate Kullback–Leibler divergence loss function for reconstruction with nonnegative values to incorporate Poisson statistics.

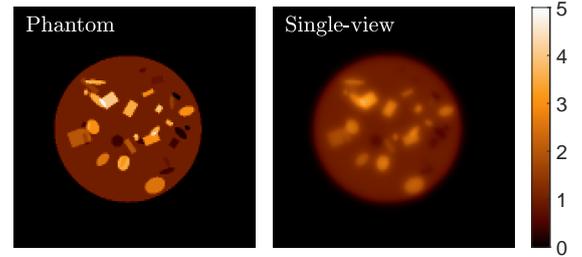

**(a)** Phantoms and single-view histoimage

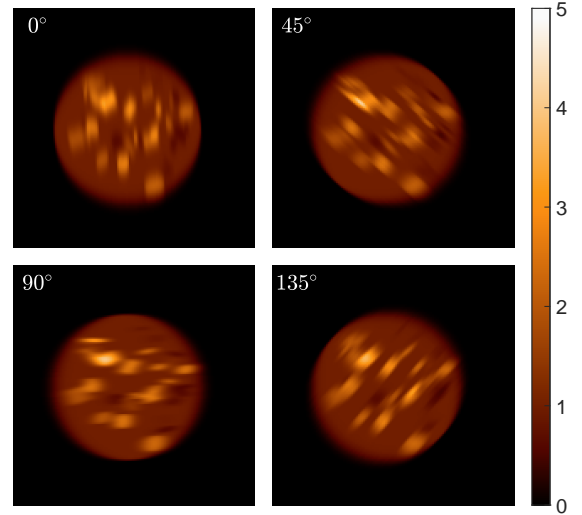

**(b)** Multi-view hisotimage

**Figure 5:** A sample training dataset contains a random phantom and the corresponding histoimages. A random phantom generated from ellipses and rectangles and the corresponding single-view histoimage (summing of all views) are shown in (a). Four views from a multi-view TOF histoimage along 4 directions along angles at 0°, 45°, 90° and 135° are shown in (b).

was used for training. We ran up to 200 epochs with mini batch size of 8 to ensure the convergence of training with very small training and validation error (below $1 \times 10^{-3}$).

## 3 Results

We separately trained the ConvLSTM network with TOF histoimages with different number of views. The 48-view TOF histoimages in the training dataset were rebinned into TOF histoimages with 1, 6, 12 and 24 views by averaging adjacent views. Figure 6 shows sample difference images between ConvLSTM reconstructed images and true images. We also computed peak signal-to-noise ratio (PSNR) and root-mean-square error (RMSE) from the 32 reserved test datasets, and show the results in table 1.

We also trained the ConvLSTM using 384 noisy datasets. A TOF histoimage with 12 views with 1M counts was used for each dataset. We then tested the trained ConvLSTM network using a NEMA phantom, never seen by the network. We generated 60 noise realizations of TOF histoimages from the NEMA phantom with the same 1M counts in each realization. The DeepDIRECT reconstructed images were com-pared with TOF FBP and TOF OSEM reconstructed images. There were 10 iterations and 12 subsets in TOF OSEM reconstructions. Figure 7 shows a comparison of the sample, mean and variance of TOF FBP, TOF OSEM and DeepDIRECT reconstructions. The horizontal profiles of the sample and variance images through the centers of the 22 mm hot rod and the 37 mm cold rod are shown in figure 8 for a quantitative comparison.

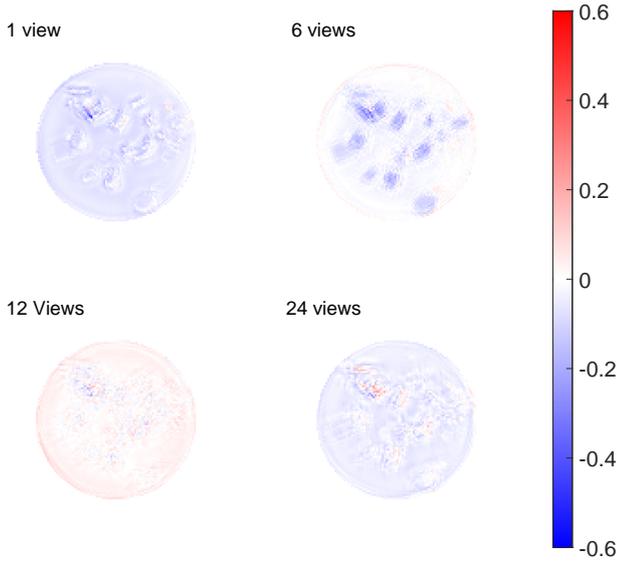

**Figure 6:** Difference images between DeepDIRECT reconstructed images and true images for different number of views.

| View | PSNR | RMSE |
|------|------|------|
| 1 | 29.77 | 0.0326 |
| 6 | 32.20 | 0.0250 |
| 12 | 35.31 | 0.0172 |
| 24 | 32.25 | 0.0245 |

**Table 1:** The evaluation of PSNR and RMSE of ConvLSTM reconstructed images with different number of views of histoimages.

## 4 Discussion and Conclusion

Generally, a TOF histoimage with more views has more information than one with fewer views. The angular information can become plateaued, or the angular information gain becomes diminished as the number of views increases. The view sampling requirements were investigated elsewhere for DIRECT reconstruction approaches [4]. The view smapling requirements can be different for different reconstruction methods and can be much less stringent for TOF data with improved timing resolution. We showed worse performance for histoimages with 24 views compared to that with 12 views in table 1, which may due to the network capacity for histoimages with a large number of views. And we expect to improve the performance for histoimages with 24 views by tuning hyperparameters and training with more datasets.

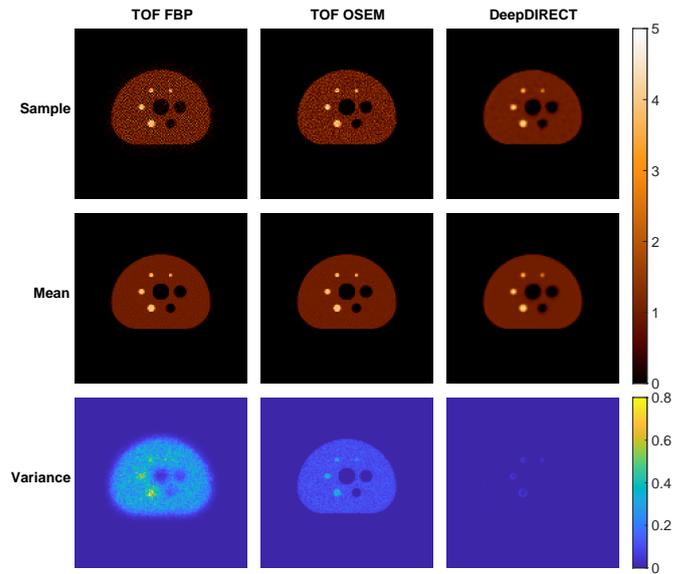

**Figure 7:** Comparison of TOF FBP, OSEM and DeepDIRECT reconstructed images of the NEMA phantom. The mean and variance reconstructions were calculated from 60 noise realizations.

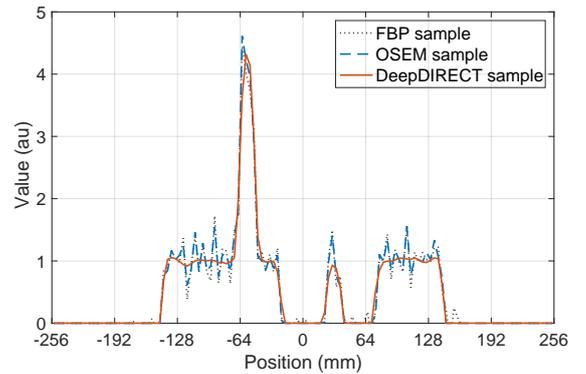

**(a)** Sample profiles

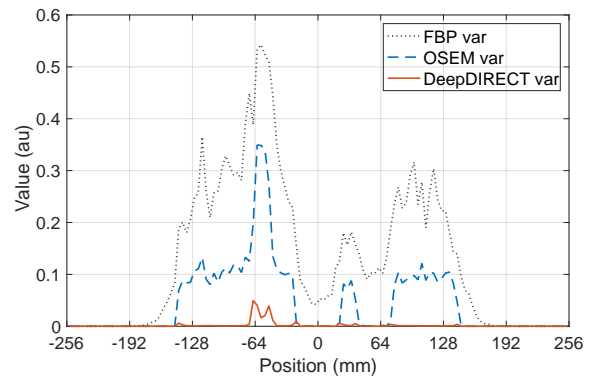

**(b)** Variance profiles

**Figure 8:** Comparison of the horizontal sample and variance profiles for the DeepDIRECT and OSEM reconstructed images of the NEMA phantom.

In summary, we developed DeepDIRECT deep image reconstruction using multi-view TOF hisotimages, which allows for computationally efficient direct DNN reconstruction directly from TOF histoimages to reconstructed images. The ConvLSTM network will be expanded with 3D convolutional operations to directly reconstruct 3D images from fully 3D multi-view TOF histoimages for clinical PET studies.

## Acknowledgment

Research reported in this work was supported by the National Institute of Biomedical Imaging and Bioengineering of the National Institutes of Health under award numbers R01EB031806 and R01EB023274. The content is solely the responsibility of the authors and does not necessarily represent the official views of the National Institutes of Health.

# Kernel MLAA Using Autoencoder for PET-enabled Dual-Energy CT

Siqi Li[1] and Guobao Wang[1]

[1]Department of Radiology, University of California Davis Medical Center, Sacramento, CA 95817, United States.

**Abstract** PET-enabled dual-energy CT combines a low-energy x-ray CT image with a high-energy γ-ray CT (GCT) image reconstructed from time-of-flight PET emission data to enable dual-energy CT multi-material decomposition on a time-of-flight PET/CT scanner. The maximum-likelihood attenuation and activity (MLAA) algorithm has been used for GCT reconstruction but suffers from noise. Kernel MLAA exploits x-ray CT image prior and the kernel framework to guide GCT reconstruction and has demonstrated substantial improvements on noise suppression. However, similar to other kernel methods for image reconstruction, the existing kernel MLAA uses image intensity-based feature vector for constructing the kernel representation, which is not always robust and may lead to suboptimal reconstruction with artifacts. In this paper, we propose a modified kernel method by using autoencoder convolutional neural network (CNN) to extract intrinsic feature set from the x-ray CT image prior for kernel construction. Computer simulation results show that the autoencoder kernel MLAA method can achieve a significant image quality improvement for GCT and multi-material decomposition as compared to the existing algorithms.

## 1 Introduction

Combined use of PET and dual-energy (DE) CT provides a multi-parametric characterization of disease states in cancer and other diseases [1]. Nonetheless, the integration of DECT with existing PET/CT would not be trivial, either requiring costly CT hardware upgrade or significantly increasing CT radiation dose. We have proposed a new dual-energy CT imaging method that is enabled using a standard time-of-flight (TOF) PET/CT scan without change of scanner hardware or adding additional radiation dose or scan time [2]. Instead of using two different x-ray energies as commonly used by conventional DECT, the PET-enabled dual-energy CT method combines a radiotracer annihilation-generated high-energy "γ-ray CT (GCT)" at 511 keV with the already-available low-energy x-ray CT (usually ≤ 140 keV) to produce a pair of dual-energy CT images on PET/CT for multi-material decomposition.

The reconstruction of GCT image from the PET emission scan can be achieved using the maximum likelihood attenuation and activity (MLAA) method [3]. However, standard MLAA reconstruction is commonly noisy because the counting statistics of PET emission data is limited. To suppress noise, a kernel MLAA approach [2] has been developed by use of x-ray CT as image prior and has demonstrated substantial improvements over standard MLAA.

In the kernel methods for image reconstruction (e.g. [4, 5, 6]), a set of features need to be defined for constructing the kernel representation of the image to be estimated. Existing kernel methods have mainly used image pixel intensities of a small patch (e.g., for MR-guided PET reconstruction [5, 6]) or temporal sequence (e.g., for dynamic PET reconstruction [4]). However, the intensity-based features do not always provide satisfactory results. As shown later in this paper, the reconstructed GCT image by such a method suffers from artifacts.

In this paper, we propose to use a convolutional neural network (CNN) feature set that is adaptively learned on the prior image to build the kernel representation for MLAA reconstruction. Deep learning with CNN has a strong ability to derive a latent feature representation in different tasks [7]. While it is often impractical to collect a large amount of training data for supervised deep learning, here we utilize the concept of autoencoder [8], an unsupervised representation learning technique, for intrinsic feature extraction from the x-ray CT image prior for the kernel construction. We expect the autoencoder-derived feature set to provide a more robust kernel representation for the GCT image reconstruction than the conventional intensity-based features. [1]

## 2 PET-enabled Dual-Energy CT

### 2.1 PET-enabled GCT by MLAA

The measurement $y$ in TOF PET can be well modeled as independent Poisson random variables using the log-likelihood function,

$$L(\boldsymbol{y}|\boldsymbol{\lambda},\boldsymbol{\mu}) = \sum_{i=1}^{N_d}\sum_{m=1}^{N_t} y_{i,m}\log\bar{y}_{i,m}(\boldsymbol{\lambda},\boldsymbol{\mu}) - \bar{y}_{i,m}(\boldsymbol{\lambda},\boldsymbol{\mu}), \quad (1)$$

where $i$ denotes the index of PET detector pair and $m$ denotes the $m$th TOF bin. The expectation of the PET projection data, $\bar{y}_m$, is related to the radiotracer activity image $\boldsymbol{\lambda}$ and GCT attenuation image $\boldsymbol{\mu}$ at 511 keV via

$$\bar{\boldsymbol{y}}_m(\boldsymbol{\lambda},\boldsymbol{\mu}) = \text{diag}\{\boldsymbol{n}_m(\boldsymbol{\mu})\}\boldsymbol{G}_m\boldsymbol{\lambda} + \boldsymbol{r}_m, \quad (2)$$

where $\boldsymbol{G}_m$ is the PET detection probability matrix and $\boldsymbol{r}_m$ accounts for the expectation of random and scattered events. $\boldsymbol{n}_m(\boldsymbol{\mu})$ is the normalization factor with the $i$th element being $n_{i,m}(\boldsymbol{\mu}) = c_{i,m} \cdot \exp(-[\boldsymbol{A}\boldsymbol{\mu}]_i)$, where $c_{i,m}$ represents the multiplicative factor excluding the attenuation correction factor and $\boldsymbol{A}$ is the system matrix for transmission imaging.

The MLAA method [3] jointly estimates the attenuation image $\boldsymbol{\mu}$ and the activity image $\boldsymbol{\lambda}$ from the projection data $\boldsymbol{y}$ by maximizing the Poisson log-likelihood,

$$\hat{\boldsymbol{\lambda}}, \hat{\boldsymbol{\mu}} = \arg\max_{\boldsymbol{\lambda} \geq 0, \boldsymbol{\mu} \geq 0} L(\boldsymbol{y} | \boldsymbol{\lambda}, \boldsymbol{\mu}). \tag{3}$$

---

[1]An extended version of this work is available on arXiv (2010.07484) and has been submitted to the Philosophical Transactions of the Royal Society A for a theme issue on synergistic reconstruction.

Previous use of MLAA was mainly for PET attenuation correction [3, 9]. In our PET-enabled dual-energy CT method [2], the estimated high-energy GCT image $\boldsymbol{\mu}$ is combined with the low-energy x-ray CT image $\boldsymbol{x}$ to form dual-energy imaging for multi-material decomposition.

## 2.2  Kernel MLAA

The GCT image estimate by standard MLAA is commonly noisy due to the limited counting statistics of PET emission data. To suppress noise, the kernel MLAA approach [2] incorporates the x-ray CT image as *a priori* information to guide the GCT reconstruction in the MLAA. It describes the intensity of the GCT $\mu_j$ in pixel $j$ as a linear representation in a transformed feature space [2]. Thus, we can obtain he equivalent matrix-vector form for the GCT image $\boldsymbol{\mu} = \boldsymbol{K}\boldsymbol{\alpha}$, where $\boldsymbol{K}$ is the kernel matrix built upon the x-ray CT image prior and $\boldsymbol{\alpha}$ denotes the corresponding kernel coefficient image. Substituting $\boldsymbol{\mu} = \boldsymbol{K}\boldsymbol{\alpha}$ into the MLAA formulation in Eq. (3) gives the following kernel MLAA optimization formulation,

$$\hat{\boldsymbol{\lambda}}, \hat{\boldsymbol{\alpha}} = \arg\max_{\boldsymbol{\lambda} \geq 0, \boldsymbol{\alpha} \geq 0} L(\boldsymbol{y} | \boldsymbol{\lambda}, \boldsymbol{K}\boldsymbol{\alpha}). \tag{4}$$

Once $\hat{\boldsymbol{\alpha}}$ is obtained, the final estimate of the GCT image is obtained by $\hat{\boldsymbol{\mu}} = \boldsymbol{K}\hat{\boldsymbol{\alpha}}$.

## 2.3  Material Decomposition for PET-enabled DECT

For each image pixel $j$, the GCT attenuation value $\mu_j$ and x-ray CT attenuation value $x_j$ jointly form a pair of dual-energy measurements $\boldsymbol{u}_j \triangleq [x_j, \mu_j]^T$, which can be modeled by a set of material bases, such as air (A), soft tissue (S) or equivalently water, and bone (B):

$$\boldsymbol{u}_j = \boldsymbol{U}\boldsymbol{\rho}_j, \quad \boldsymbol{U} \triangleq \begin{pmatrix} x_A & x_S & x_B \\ \mu_A & \mu_S & \mu_B \end{pmatrix}, \boldsymbol{\rho}_j \triangleq \begin{pmatrix} \rho_{j,A} \\ \rho_{j,S} \\ \rho_{j,B} \end{pmatrix}, \tag{5}$$

subject to $\sum_k \rho_{j,k} = 1$. The coefficients $\rho_{j,k}$ with $k = A, S, B$ are the fraction of each basis material in pixel $j$. The material basis matrix $\boldsymbol{U}$ consists of the linear attenuation coefficients of each basis material measured at the low and high energies. Finally, $\boldsymbol{\rho}_j$ is estimated using the least-square optimization.

## 3  Modified Kernel Method Using Autoencoder

### 3.1  Building Kernels Using CNN Features

In the kernel MLAA and other kernel methods for image reconstruction (e.g., [4, 5, 6]), the formation of the pixel-wise feature vector is a key factor to build the kernel matrix $\boldsymbol{K}$ and directly impacts the reconstruction result. Conventionally, the feature vector $\boldsymbol{f}_j$ at pixel $j$ is commonly defined by the intensity values of a pixel or its surrounding small patch in the prior image. However, such an approach may lead to suboptimal feature representation because of the simplification of feature attributes and the small size of receptive field. Artifacts were observed in the reconstructed GCT images of the kernel MLAA [2].

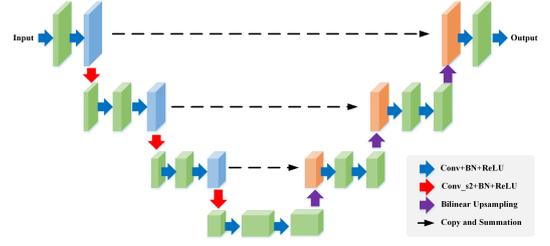

**Figure 1:** U-Net model used for CNN feature extraction.

To alleviate the issues, we propose to exploit deep learning with CNN [7] to extract intrinsic feature set. A radial Gaussian kernel between pixel $j$ and $l$, i.e. the $(j, l)$th element of $\boldsymbol{K}$, will then have the form,

$$\kappa(\boldsymbol{f}_j^{\text{CNN}}, \boldsymbol{f}_l^{\text{CNN}}) = \exp\left(-\left\|\boldsymbol{f}_j^{\text{CNN}} - \boldsymbol{f}_l^{\text{CNN}}\right\|^2 / 2\sigma^2\right), \tag{6}$$

where $\sigma$ is a hyper-parameter which can be set to 1 if the feature set is normalized [4]. $\boldsymbol{f}_j^{\text{CNN}}$ denotes the the feature set of pixel $j$ obtained by

$$\boldsymbol{f}_j^{\text{CNN}} = [\mathscr{F}_\ell(\boldsymbol{\theta}; \boldsymbol{I})]_j, \tag{7}$$

where $\mathscr{F}_\ell$ denotes the output of the $\ell$th layer of a CNN model $\phi(\boldsymbol{\theta}; \boldsymbol{I})$ with $\boldsymbol{\theta}$ the model parameters and $\boldsymbol{I}$ the input. We specifically consider a Unet model shown in Fig. 1, which is widely used for image segmentation and reconstruction (e.g. [10]). For general consideration, $\ell$ can be set to the penultimate layer which generally provides pixel-wise multi-channel feature sets. Similar to the previous kernel method [4], $\boldsymbol{K}$ is built to be sparse using k-nearest neighbors.

### 3.2  Unsupervised Learning Using Autoencoder

A natural choice for CNN feature extraction is by use of supervised deep learning which has shown a strong potential for feature extraction in image recognition tasks. One major challenge with supervised deep learning is it commonly requires a large number of training data sets, which are not always available or the data acquisition is costly.

An autoencoder is an unsupervised technique for deep representation learning using neural networks [8]. The corresponding optimization problem for applying autoencoder for our kernel method is defined by,

$$\hat{\boldsymbol{\theta}} = \arg\min_{\boldsymbol{\theta}} \|\boldsymbol{x} - \phi(\boldsymbol{\theta}; \boldsymbol{x})\|^2, \tag{8}$$

where both the input and output are set to the x-ray CT image $\boldsymbol{x}$. The optimization essentially seeks an adaptive

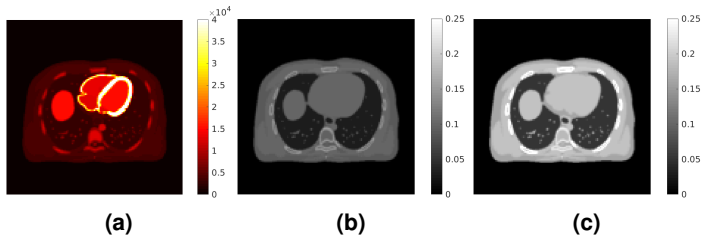

**Figure 2:** The digital phantom used in the PET/CT computer simulation. (a) PET activity image in Bq/cc; (b) PET attenuation image at 511 keV in cm$^{-1}$; (c) x-ray CT image at 80 keV.

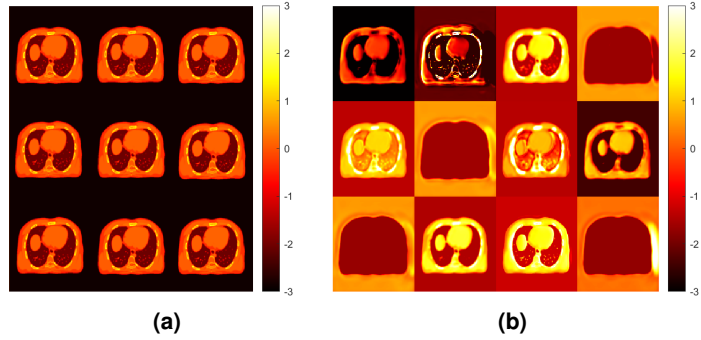

**Figure 3:** Maps of the feature set used by (a) standard kernel and (b) proposed Unet kernel.

CNN representation of the image $\boldsymbol{x}$, without requiring a large training database. Once the model is trained, the penultimate layer is used to extract $\{\boldsymbol{f}_j^{\mathrm{CNN}}\}$.

## 4 Simulation Results

We simulated a GE Discovery 690 PET/CT scanner in 2D with a TOF resolution of 550 ps. The true PET activity image and 511 keV attenuation image are shown in Fig. 2(a) and (b), respectively. The images were first forward projected to generate noise-free sinogram of 11 TOF bins. A 40% uniform background was included to simulate random and scattered events. Poisson noise was generated using 5 million expected events. The x-ray CT image at a low-energy 80 keV was also simulated and is shown in Fig. 2(c).

Three types of reconstruction were compared, including (1) standard MLAA [3], (2) existing kernel MLAA [2] with $\boldsymbol{f}_j$ being the pixel intensities of x-ray CT image $\boldsymbol{x}$ in a 3×3 image patch centered at pixel $j$, and (3) Unet kernel MLAA: proposed autoencoder kernel method with $\boldsymbol{f}_j$ extracted using the Unet. The 511 keV attenuation map converted from the x-ray CT image was used as the initial estimate of $\boldsymbol{\mu}$. All different kernel matrices were built using 50 nearest neighbors based on the distance of feature vectors in a way similar to [4]. We used the Adam optimization algorithm to train the Unet of the x-ray CT image with 300 epochs. The learning rate of Unet was chosen to be $10^{-2}$ for approximately optimal performance. All MLAA reconstructions were run for 3000 iterations for the purpose of studying convergence, with one inner iteration for the $\boldsymbol{\lambda}$-estimation step and five inner iterations for the $\boldsymbol{\mu}$-estimation step. Different MLAA

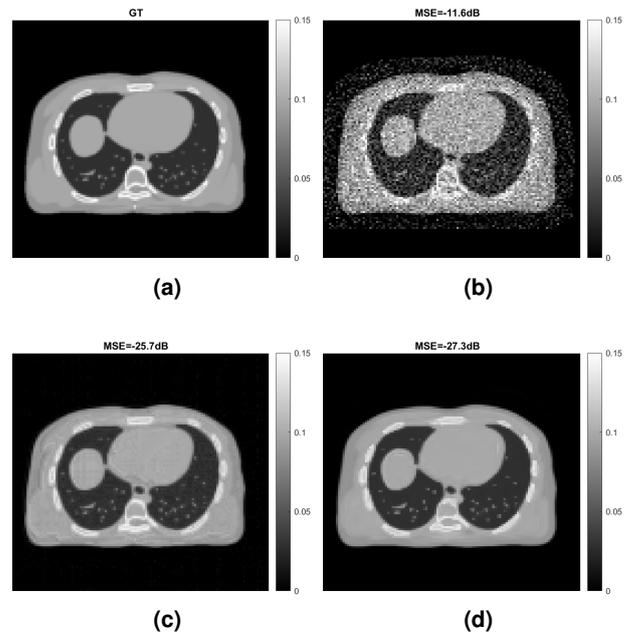

**Figure 4:** GCT images by different reconstruction algorithms. (a) Ground truth, (b) standard MLAA, (c) standard kernel MLAA, and (d) Unet kernel MLAA.

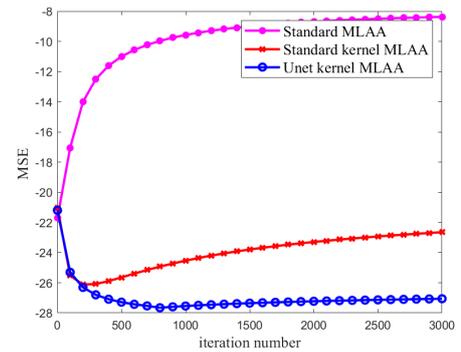

**Figure 5:** Plot of image MSE as a function of iteration number for different MLAA reconstruction algorithms.

methods were first compared for the image quality of GCT using the mean squared error (MSE).

To demonstrate the differences between the Unet kernel and standard kernel, Fig. 3(a) and (b) visualize the maps of the feature set used by the two types of kernels extracted from the x-ray CT image. Each subimage corresponds to one element of the feature vector $\boldsymbol{f}_j$ or $\boldsymbol{f}_j^{\mathrm{CNN}}$ at all different pixels. The standard intensity-based kernel was formed from $3 \times 3$ neighboring patches, which explains why the feature maps look similar. The Unet-derived CNN features was learned using an autoencoder as the output of the penultimate layer (12 channels).

Figure. 4 shows examples of the reconstructed GCT image $\boldsymbol{\mu}$ by different MLAA algorithms with a specific iteration number 400. While the standard MLAA reconstruction was noisy, the two kernel MLAA reconstructions significantly improved the result according to both visual quality and image MSE. The Unet kernel MLAA demonstrated least artifacts with good visual quality and achieved the lowest

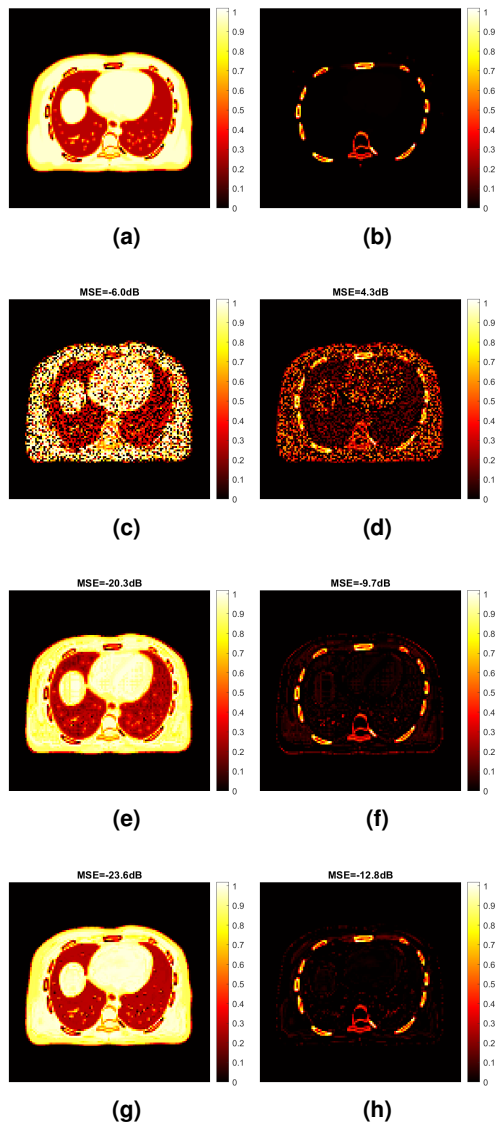

**Figure 6:** True and estimated fractional images of two basis materials using different reconstruction algorithms: soft tissue (left column) and bone (right column). (a-b) Ground truth,(c-d) standard MLAA, (e-f) standard kernel MLAA, (g-h) proposed Unet kernel MLAA.

MSE among different algorithms. Figure. 5 further shows image MSE as a function of iteration number for different algorithms. The Unet kernel MLAA outperformed the other two algorithms, though its convergence rate was slightly slower.

Figure. 6 shows the fractional basis images of soft tissue and bone from multi-material decomposition of the PET-enabled dual-energy CT images. The results were obtained from the MLAA reconstructions with the best GCT image MSE for each method. Compared to the standard MLAA reconstruction, the standard kernel MLAA significantly suppressed the noise but still with artifacts. The Unet kernel MLAA reconstruction led to better visual quality and decreased image MSE. Fig. 7 further shows image MSE as a function of iteration number for each basis fractional image. The two kernel MLAAs were significantly superior to the standard MLAA, with the best MSE performance from the Unet kernel MLAA.

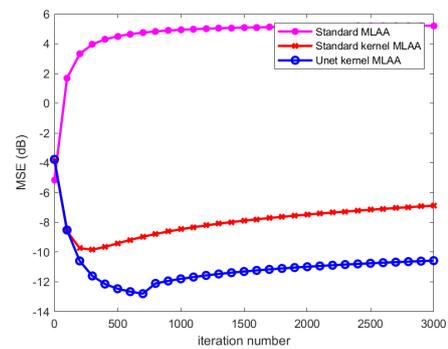

**Figure 7:** Plot of image MSE as a function of iteration number for bone fractional image.

## 5 Conclusion

We have developed an autoencoder kernel MLAA reconstruction method for PET-enabled dual-energy CT. The autoencoder-derived feature set can provide an improved kernel representation to incorporate x-ray CT as the image prior for GCT image reconstruction from PET emission data. Computer simulation results have demonstrated the improvement of the autoencoder kernel MLAA over existing MLAA algorithms for GCT image quality and dual-energy CT material decomposition. The proposed method can suppress noise efficiently and reduce image artifacts.

# Chapter 6

# Poster Session 1

**session chairs**
Guobao Wang, *University of California Davis Medical Center (United States)*
Kjell Erlandsson, *University College London (United Kingdom)*

# Posterior image distribution from misspecified PET image reconstruction estimators

Marina Filipović [1], Thomas Dautremer[2], and Éric Barat[2]

[1]Université Paris-Saclay, CEA, CNRS, Inserm, BioMaps, Service Hospitalier Frédéric Joliot, Orsay, France
[2]CEA, LIST, Laboratory of Systems Modelling and Simulation, Gif-sur-Yvette, France

**Abstract** PET image reconstruction methods are based on some possibly misspecified modelling assumptions: the model relating the unknown image to the acquired data and the regularization assumptions may not fully match the reality. Here we provide a new point of view on the uncertainty in PET image reconstruction. We use a statistical framework that allows to: 1) state explicitely that the modelling assumptions are misspecified, 2) define explicitely how much we trust our prior beliefs, 3) given any image reconstruction estimator (e.g. penalized ML), update our prior beliefs upon acquiring some data into a posterior image distribution which remains true in the case of estimator misspecification, and 4) provide uncertainty information about some modelling assumptions in addition to the uncertainty due to the data noise. Initial results are shown and explained on simulated and real data for some common PET image reconstruction methods. Future work will focus on including the uncertainty of more diverse modelling assumptions.

## 1   Introduction

Image estimators in optimization-based (iterative) PET image reconstruction are based on some modelling assumptions (e.g. Poisson data noise, system matrix, image smoothness/roughness) and on some empirical adjustments (e.g. parameter values for spatial/temporal regularization). The loss (objective) function to be minimized (maximized) may have a statistical interpretation (e.g. data likelihood distribution, image posterior distribution) and it may contain some empirical customizable components. The assumptions implied by this loss function approximate reality only to some extent and so the entire model that generates the data is necessarily somewhat misspecified. Usually, efforts are put into showing that the approximation is good enough. An other approach is possible: we can use statistical tools, [1], [2], to assert openly that we are not entirely sure about the postulated model (e.g. assumption that smooth areas in a PET image and in an associated MRI image match). The aim is to provide a more honest statistical interpretation of the reconstruction process and to take into account the uncertainty of some modelling assumptions in addition to the uncertainty due to the noise in the acquired data.

## 2   Theory

First, a vocabulary reminder: a "parametric" statistical model assumes that the noisy measured data (e.g. PET sinogram) follow a specific probability distribution type (e.g. Poisson, Gaussian), whose parameters are unknown (e.g. PET image of radiotracer emission concentration); a "non parametric" statistical model assumes that the measured data follow a probability distribution which is itself a realization drawn from some chosen probability distributions capable of generating probability distributions (e.g. a Dirichlet process), so the unknown of interest (PET image) is no longer an explicit parameter.

Optimization-based (iterative) PET image reconstruction methods produce a single image estimate by minimizing a loss function based on a model relating the unknown image to the acquired data. This model often has a parametric statistical interpretation: the data are assumed to follow a Poisson distribution (likelihood) whose unknown parameter is the PET image. If the model is parametric Bayesian, an image prior distribution can be specified, including some assumptions about image smoothness/roughness, similarities with an associated MRI image, etc, to produce a posterior maximum image estimate. Any such image estimator (e.g. regularized/penalized ML, MAP) hence relies on the assumption that the underlying model is true.

From now on, we are no longer concerned with the specific contents of the loss function nor with the truthfulness of the assumptions it is based on: it suffices that minimizing the loss function produces practically useful image estimates from an acquired PET dataset. Let's consider a list-mode dataset composed of $K$ counts, where each detected count $r_k$ and its attributes (e.g. LOR, TOF) are a realization from an independent and identical (unknown) data distribution $F$. Let $\lambda$ be the (unknown) PET image, and $\hat{\lambda}$ an image estimate obtained using any chosen loss function defined independently of any considerations about $F$. Then, the image estimate can be defined as [1], [2]:

$$\hat{\lambda} = argmin_{\lambda} \sum_k \text{loss}(\lambda, r_k) F(r_k). \qquad (1)$$

We then build a nonparametric Bayesian model focused on the data distribution $F$, independently of the loss function. We express our prior assumptions about the data distribution and how much we believe in them using a generic customizable data prior, a Dirichlet process (DP), whose realizations represent probability distributions. Then, upon acquiring (observing) some data, we update these prior beliefs by producing a posterior probability distribution of the data $F_{post}$, which turns out to be an other Dirichlet process, see [1] for more details.

We use this nonparametric posterior data distribution $F_{post}$

to produce a nonparametric posterior probability distribution of the image $\lambda$, which is meaningful with respect to the nonparametric Bayesian model for the data distribution and with respect to the chosen image estimator. It has a different meaning from posterior image distributions obtained using parametric Bayesian models (e.g. [3], [4]). If $f_{post}$ is a realization drawn from $F_{post}$, a realization $\lambda^\star$ from the nonparametric posterior image distribution is obtained as follows:

$$\lambda^* = argmin_\lambda \sum_k \text{loss}(\lambda, r_k) f_{post}(r_k). \tag{2}$$

Hence, a sample of B realizations from the nonparametric image posterior is obtained using the algorithm 1: we repeatedly produce posterior data distribution realizations $f_{post}$ and include them into image estimator runs.

---

**Algorithm 1** General algorithm for list-mode datasets

---

1: **for** $b = 1$ to $B$ **do**
2:     draw $f_{post_b}$
3:     $\lambda_b^* = \arg\min_\lambda \sum_k \text{loss}(\lambda, r_k) f_{post_b}(r_k)$
4: **end for**

---

To actually apply this method, we have to choose a prior on the data distribution $F$. For these preliminary tests, we chose the simplest prior which does not carry any assumptions/information (except the iid assumption about list-mode counts), a Dirichlet process with a null parameter, DP($\alpha = 0$). Hence, the posterior data distribution $F_{post}$ becomes a uniform Dirichlet distribution of K unitary parameters, Dir($1, 1, \ldots, 1$), [1]. Drawing a realization $f_{post}$ from this posterior data distribution may be interpreted as assigning a random probability to each detected count. The algorithm for drawing a sample of $B$ realizations from the nonparametric image posterior from a list-mode dataset is given in Algorithm 2.

Under some assumptions about the loss function (see the Appendix), that are satisfied in most existing PET image reconstruction estimators, an equivalent algorithm exists for histogram datasets. If the list-mode data are histogrammed into some kind of bins (sinogram, time-of-flight), resulting in a histogram dataset $y$, it can be shown (see the Appendix) that drawing realizations from the posterior data distribution is equivalent to drawing randomized histograms, where the number of counts for each bin $i$ is drawn from $Gamma(y_i, 1)$. The algorithm for drawing a sample of $B$ realizations from the nonparametric image posterior from a histogram dataset is given in Algorithm 3.

---

**Algorithm 2** List-mode dataset

---

1: **for** $b = 1$ to $B$ **do**
2:     draw $(w_1, w_2, \ldots, w_K)$ from Dir($1, 1, \ldots, 1$)
3:     $\lambda_b^* = \arg\min_\lambda \sum_k \text{loss}(\lambda, r_k) w_k$
4: **end for**

---

**Algorithm 3** Histogram dataset

---

1: **for** $b = 1$ to $B$ **do**
2:     draw posterior histogram realization $y^*$, where $y_{i_b}^* \sim$ Gamma($y_i, 1$)
3:     $\lambda_b* = \arg\min_\lambda \sum_i \text{loss}(\lambda, y_{i_b}^*)$
4: **end for**

---

A sample from the nonparametric posterior image distribution remains true even if the image estimator contains assumptions that do not match reality perfectly. In this initial setting, the nonparametric data prior does not include any modelling assumptions, so the nonparametric image posterior contains mostly the uncertainty due to the data noise propagation through the image estimator. However, the presented algorithms are directly expandable to include data priors that contain some of our beliefs about the data distribution, see [1], and so include the uncertainty of some modelling assumptions. There are many modelling assumptions that could be taken into account, e.g. Poisson expectation parameterization, system matrix, relevance of an MRI image for the spatial regularization of the PET image, and many ways to define them, so it is material for further work.

There is a direct link between the described nonparametric image posterior and the more classical parametric image posterior. As shown in [1], if the loss function is based on a parametric Bayesian model, as in MAP PET image reconstruction methods, the nonparametric posterior (using the noninformative Dirichlet data prior chosen here) becomes approximately equivalent to the parametric posterior. The algorithms shown here for nonparametric image posteriors are identical to the algorithms shown in [3] for parametric image posteriors. In other words, the same posterior sample can have two different interpretations, according to the context (well-specified vs misspecified model in the loss function), and according to which Bayesian model the word "posterior" refers to. If some prior assumptions are included into the nonparametric data prior, the nonparametric image posterior will deviate from the parametric image posterior and express richer uncertainty information.

## 3 Methods and Results

All the image estimators (iterative image reconstruction methods) were implemented using the CASToR (Customizable and Advanced Software for Tomographic Reconstruction) platform in C++, [5], [6]. The methods were fully quantitative, contained image PSF resolution modelling and corrections (random/scattered coincidences, attenuation, normalization).

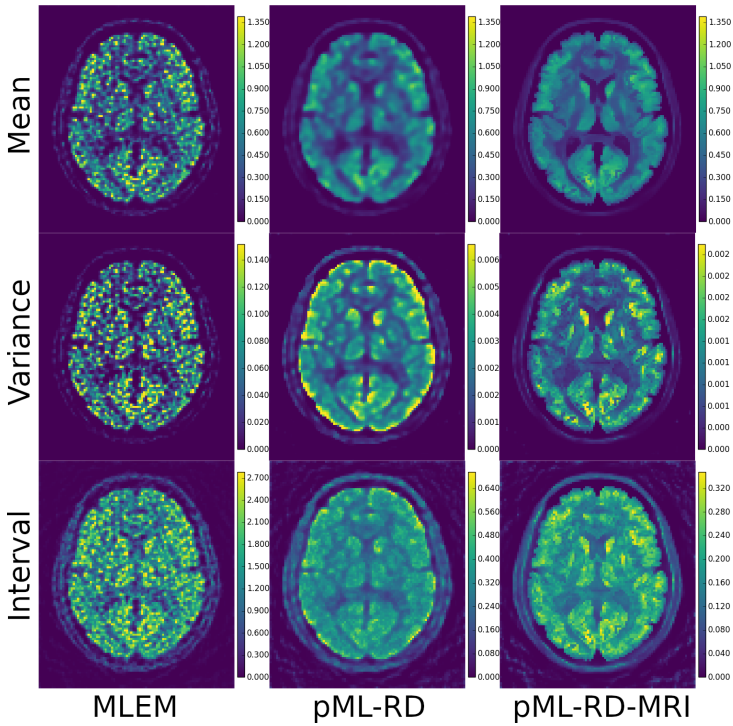

**Figure 1:** Nonparametric posterior mean, variance and interval size for several image estimators. Colorscale maximum differs across estimators.

### 3.1 Simulated data

Realistic simulations were performed using a 2D slice of the high-resolution heterogeneous $^{18}$F-FDG PET/MRI brain phantom [7]. The nonparametric posterior image distribution was obtained for several common image estimators: MLEM, pML-RD (ML penalized/regularized with a Markov Random Field with relative differences potential function, [8]), pML-RD-MRI (same including an additional MRI image using the asymmetric Bowsher method [9]). Here we refer to these image estimators as regularized ML instead of MAP, to stress the fact that the loss function can include any (possibly empirical) components and does not have to comply with a parametric Bayesian model. Regularization parameters were chosen to minimize the RMSE compared to the true phantom and each image reconstruction estimator method was run for 1000 iterations, which resulted in relatively early stopping for MLEM and a reasonable convergence for pML. The number of nonparametric posterior image realizations was 1000.

Figure 1 shows some distribution characteristics (mean, variance, interval size). The mean of the nonparametric posterior was visually indistinguishable from and quantitatively close to the corresponding standard pML image estimate (not shown). Hence, the nonparametric posterior mean and the pML estimate had similar properties.

The voxel-wise nonparametric posterior variance has to be interpreted with respect to the posterior mean. This variance is higher for MLEM and decreases for estimators with stronger spatial regularization. For pML-RD, it presents some local smoothness and is relatively higher near strong edges, e.g.

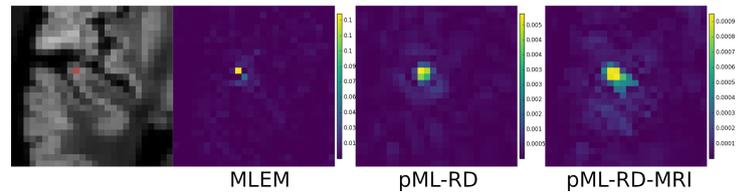

**Figure 2:** Nonparametric posterior covariance images for a gray matter voxel (left) for several image estimators.

the edge between the gray matter and the skull. For pML-RD-MRI, it presents clearer edges, because of the spatially regularizing influence of the MRI image. This nonparametric posterior uncertainty conveys mostly the propagation of data noise through the image estimator for the given dataset.

Each voxel has a nonparametric posterior covariance image, as illustrated in Figure 2. The absolute covariance values decrease with stronger regularization. For pML-RD, the covariance is lower in edges and higher in smoother areas in the voxel neighbourhood, while for pML-RD-MRI it is high in a larger neighbourhood that appears smooth in the MRI image.

TODO intervals assessment

### 3.2 Real data

Data were obtained from a GE Signa PET/MR scanner for the brain bed step of a whole body $^{18}$F-FDG oncological exam, with an associated 3D T1 weighted fast spin echo MRI acquisition after Gadolinium injection. A nonparametric posterior image distribution was obtained for pML-RD-MRI, using 28 subsets and 16 iterations, with empirically adjusted regularization parameters (neighbourhood sphere radius = $6mm$, RD$\gamma = 3$, $\beta = 0.005$, Bowsher percentage=30%).

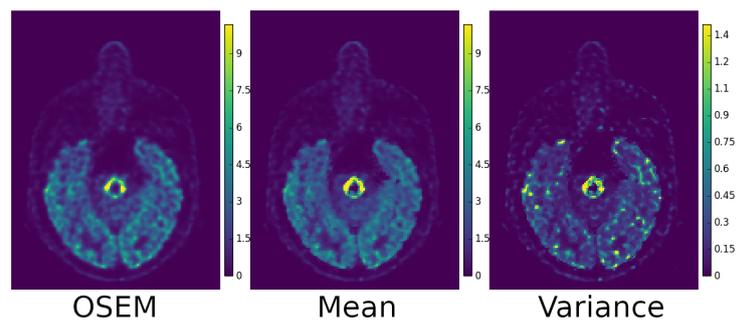

**Figure 3:** Clinical reconstruction (OSEM) and the nonparametric posterior mean and variance for pML-RD-MRI

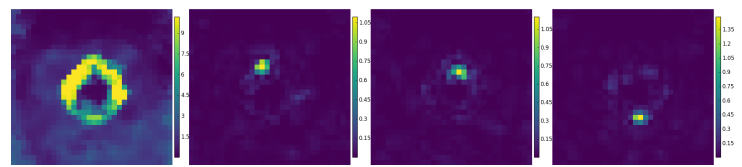

**Figure 4:** Zoomed lesion area (left) and nonparametric posterior covariance for several voxels in the lesion for pML-RD-MRI

Figure 3 shows a clinical reconstruction (OSEM with 28

subsets and 8 iterations) and the nonparametric posterior voxel-wise mean and variance. The posterior mean is visually indistinguishable from the corresponding pML image estimate (not shown). The posterior variance is highest in the lesion area and in some high-contrast areas in the gray matter. The covariance of lesion voxels is higher in the nearest voxel neighbourhood and shows also in some non adjacent lesion voxels, see Figure 4.

There are no well-established standard validation methods for these posterior distributions, but some characteristics can be explored and validated using for instance notions that link frequentist and Bayesian points of view. In a nutshell, if the PET examination is repeated on the same patient in the exact same conditions several times, producing several datasets corresponding to the same PET image, we can check whether the intervals and percentiles on the posterior distribution obtained from a single dataset match the intervals and percentiles on the frequentist (estimator) distribution over multiple datasets. As the applied algorithm presented here are identical to the ones in [3], but with a different interpretation, the detailed description and satisfactory results regarding intervals validation provided in [3] apply also to the nonparametric posterior distributions presented here.

## 4    Discussion

The proposed point of view on uncertainty in PET image reconstruction is new and there are no available methods for direct comparison to our knowledge. There are no standard validation methods for this kind of nonparametric posterior distributions, but some characteristics can be explored and validated as exposed in the results. The proposed algorithms build on the already available iterative PET image reconstruction methods and do not impose any requirements on their performance, e.g. regarding tuning of parameters, achieving convergence, empirical modifications. The computation time depends on the number of required posterior realizations but the multiple runs of image estimators can be entirely parallelized. The interplay between parametric and nonparametric Bayesian models and between estimation/optimization and full probability distributions may appear confusing and redundant, but this statistical framework may be viewed as a versatile generalization of all these points of view. There are many possibilities for the choice of the loss function and of the nonparametric prior data distribution. It should be noted that most usual assumptions in PET image reconstruction can be included either in the loss function or in the data prior, though the interpretation is different: if an assumption is included in the loss function, it is considered as a more or less misspecified though empirically useful approximation of reality, and if a assumption is included in the nonparametric data prior, we can explicitely state how much we believe in it. The resulting nonparametric posterior distribution must be interpreted accordingly. The data noise and its propagation always show in the posterior uncertainty but depend on the properties of the image estimator.

## 5    Conclusion

A different and generalized probabilistic point of view on the uncertainty in PET image reconstruction is provided. Modelling assumptions implied in iterative PET image reconstruction methods are viewed as an imperfect approximation of reality. Easy-to-implement algorithms are provided for producing realizations from a type of posterior image distributions that remains true if the image estimators are misspecified. Such posterior distributions were produced and explained for several usual images estimators on simulated and on real data. The algorithms can be readily extended to honestly state our prior beliefs about some modelling assumptions (e.g. match of smooth areas between PET and MRI) and thus include some model uncertainty into the posterior image distribution. This is material for further work.

## 6 Appendix

Here, we show how the nonparametric posterior bootstrap with the chosen data distribution prior can be applied on a histogram dataset (Algorithm 3) starting from the application on a list-mode dataset (Algorithm 2). Drawing a realization from the K-dimensional Dirichlet distribution $(w_1, w_2, \ldots, w_K)$ in Algorithm 2 can be implemented using $K$ Gamma distributions, with their shape parameters equal to the parameters of the Dirichlet distribution, as:

$$p_k \sim \text{Gamma}(1, 1) \tag{3}$$

$$w_k = p_k / \sum_m p_m \tag{4}$$

$$\tag{5}$$

To apply the proposed method to histogram PET datasets, let's replace $w_k$ with $p_k$ in Algorithm 2, as this does not change the produced realizations $\lambda^*$:

$$\lambda^* = argmin_\lambda \sum_k \text{loss}(\lambda, r_k) p_k. \tag{6}$$

Let's histogram the counts into some sort of bins (detection LOR, TOF) $i$, where the number of counts $y_i = \sum_{k \in S_i} 1_k$, where $S_i$ is the set of counts that belong to the bin $i$. Let's assume that the loss function is linear with respect to the detected counts, i.e. $\sum_{k \in S_i} \text{loss}(\lambda, r_k) = \text{loss}(\lambda, y_i)$, and with respect to the multiplicative weigths, i.e. $\text{loss}(\lambda, r_k) p_k = \text{loss}(\lambda, r_k p_k)$, which is the case for all the loss functions that rely on the Poisson assumptions for PET data [10]. Then, the Equation 6 becomes:

$$\lambda^* = argmin_\lambda \sum_i \text{loss}(\lambda, \sum_{k \in S_i} r_k p_k) \tag{7}$$

$$\lambda^* = argmin_\lambda \sum_i \text{loss}(\lambda, y_i^*) \tag{8}$$

Hence, this is equivalent to applying an optimization-based image estimator to a new histogram dataset $y^*$. Given the choice for $p_k \sim \text{Gamma}(1, 1)$, the number of counts in each histogram bin $y_i^*$ can be drawn from $\sum_{k \in S_i} \text{Gamma}(1, 1) = \text{Gamma}(y_i, 1)$, as in Algorithm 3.

# Image reconstruction from tissue scattered events for $\beta^+\gamma$ coincidences in Compton-PET

Satyajit Ghosh[1] and Pragya Das[1]

[1]Department of Physics, Indian Institute of Technology Bombay, Mumbai, INDIA

**Abstract** For long time non-pure beta emitters are avoided from PET imaging due to extra dose and increase in background from Compton scattering. But advent of high-resolution Compton camera system opens up new domain of imaging. Various non-pure beta emitters are formed as beam irradiation byproduct in therapy which can be used in online beam range verification. In this case, though, the number of usable counts for imaging is generally 1-3 order lesser than normal PET scan. On the other hand, we know that in human PET scanner, 30-60% can be tissue scattered coincidences in 3D case containing 80% single scattered events. In this work, we have investigated feasibility of imaging using only single scattered coincidences for non-pure beta emitters in a Compton-PET system. The locus of tissue scatter point can be reduced to in generally two points after using Compton cone from both ends of 511 keV detections. Finally, annihilation point is estimated using Compton cone of 1157 keV gamma and time-of-flight information for the 511 keV. We believe independent assessment of underlying activity from single scattered data sets will increase confidence in image interpretation.

## 1 Introduction

For long time non-pure beta emitter radioisotopes (e.g., $^{44m}$Sc,$^{94}$Tc,$^{14}$O,$^{68}$Ga,$^{124}$I,$^{10}$C) are not used in PET imaging. This is because of extra dose and Compton scattering background that the quasi-simultaneously emitted extra gamma ray produces. But with the development of excellent resolution Compton camera systems this situation had changed. New concept of imaging using triple coincidence data was proposed [1]. In this new imaging, Compton cone drawn using extra gamma interaction points was used to estimate original annihilation point on the LOR, similar to TOF-PET imaging [2]. Application of these type of radioisotopes is in generally of two types. It is used as conventional radiopharmaceutical, e.g., [44Sc]Sc-PSMA-617 [3] in prostate cancer imaging. Besides, various non-pure beta emitters are formed as beam irradiation byproduct in ion therapy [4]. Hence, it is online or offline beam range monitoring agent. But in this case, generally the emitted count is 1-3 orders magnitude smaller than conventional PET scan [5]. On the other hand, it is known that tissue scattering can contribute to 30-60% of coincidences in human 3D-PET [6] in which 80% are single scattered [7]. So, in this work, we have investigated the feasibility of image reconstruction from those tissue scattered events in Compton-PET system. Our aim is to produce a physically meaningful image from the single scattered data which is independent from unscattered data. We believe having two independent image of same underlying activity distribution will assists us in better diagnosis. In this context, it is worth to mention that the motivation of WGI imaging concept is indeed to use all types data independently [8]. We performed GATE [9] simulation with finite resolution parameters for a Compton-PET system with silicon as scatterer ring and LaBr$_3$:Ce as absorber ring. Geometrical arrangement and parameters were chosen keeping in mind the sensitivity and resolution. Line sources of $^{44}$Sc was used. And a cylindrical water phantom of diameter 10 cm was placed axially. At first, we had shown that the locus of tissue scattering point of a single scattered coincident (single scatter surface) is prolate spheroid (for scattering angle, $\theta_s < 90^0$) and spindle toroid (for $\theta_s > 90^0$) where to acquire these single scattered coincidences photo-peak and off-peak energy windows positioned in accordance with detector resolution were used. Data acquisition was performed using an appropriately defined trigger logic. Compton cones from both end of 511 keV detection were projected on the single scatter surface to obtain two 3D curves which cut each other in generally at two points forming two possible broken LORs. Finally, annihilation point was estimated by projecting Compton cone of 1157 keV and TOF information was used to choose between the two. The image we obtained is physically meaningful and proves the feasibility of single scattered imaging in non-pure beta emitter cases.

## 2 Materials and Methods

At first, we have discussed the locus of single scattering point in case of Compton-PET system. For proving the feasibility a GATE simulation was performed. Trigger logic was developed for data extraction. Finally the image reconstruction algorithm was proposed for single scatter imaging.

### 2.1 Locus of scattering point

We have drawn a typical Compton-PET set in figure (1). From here on, we named the locus as single scatter surface to avoid any confusion. Now, we assume a single scattered coincident event where annihilation happened at point O and tissue scattering at C. To find the locus of the scattering point C, at first, we write down the equation of locus depending only on scattering angle in tissue ($\theta_s$),

$$\vec{AC}.\vec{CB} = \left|\vec{AC}\right|\left|\vec{CB}\right| \cos\theta_s$$

$$\Rightarrow (\vec{r} - \vec{r_A}).(\vec{r_B} - \vec{r}) = |\vec{r} - \vec{r_A}||\vec{r_B} - \vec{r}| \cos\theta_s \quad (1)$$

where tissue scattering angle is calculated using this equation

$$\theta_s = \arccos\left(2 - \frac{511}{E_1 + E_2}\right)$$

where $E_1$ and $E_2$ are energies deposited in scatterer and absorber by the tissue scattered photon and here we have assumed a full energy deposition.

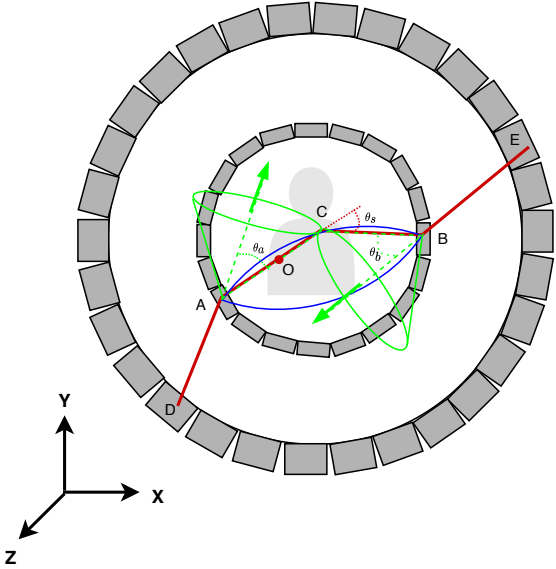

**Figure 1:** Compton-PET set up with a single scatter event get detected at points A, B (scatterer ring detection points) and points D, E (absorber ring detection points) where annihilation happended at O and tissue scattering happended at point C; tissue scattering angle is $\theta_s$ and scattering angles in scatterer ring are $\theta_a$, $\theta_b$; locus of scattering point shown in blue curve and Compton cones from both ends of 511 keV detection are shown in green colored cones.

Now applying Compton cone constraint from both side of 511 keV detection the tissue scattering point can be further localised. The equations of the Compton cones are

$$(\overrightarrow{r} - \overrightarrow{r_A}) \cdot \widehat{n_A} = |\overrightarrow{r} - \overrightarrow{r_A}| \; cos \; \theta_a \qquad (2)$$

and,

$$(\overrightarrow{r} - \overrightarrow{r_B}) \cdot \widehat{n_B} = |\overrightarrow{r} - \overrightarrow{r_B}| \; cos \; \theta_b \qquad (3)$$

where $\theta_a$ and $\theta_b$ are scattering angles from scatterer ring and $\widehat{n_a}$ and $\widehat{n_b}$ are unit vectors along line joining from absorption to scattering point respectively.

It is known that eq. (1), which represents single scatter surface (blue curve in figure 1), is a surface equation of a prolate spheroid for $\theta_s < 90^0$ and of spindle toroid for $\theta_s > 90^0$. For further constraining the locus of scattering point, we have solved eq. (1) with eq. (2) and (3) which means solution between single scatter surface and cones. We found that the solution to be closed contour 3D curves on the single scatter surface. And two curves from both end of 511 keV detection cut each other in generally at two points. Further discussion about this is given in section 3.

## 2.2 GATE simulation

We performed a GATE [9] simulation of a Compton-PET system. Silicon scatterer ring of thickness 2.5 cm and radius 20 cm was chosen. Radius was chosen larger since we are working with human scanner. And LaBr$_3$:Ce absorber of ring radius 28 cm and thickness 3 cm was used. Axial width of each ring was 28 cm. Energy resolutions of scatterer and absorber were 2.5% and 5% @511 keV and time resolutions were 1 ns and 200 ps respectively. Finally the spatical resolution was chosen to be 2 mm and 5 mm respectively. For image resolution study a [44]Sc line source, situated at the centre of the scanner, of activity 1 MBq was used. Activity was chosen low to have a smaller number of random events. A cylindrical water phantom of diameter 10 cm and height 28 cm was defined axially. The decision of working with a human scale Compton-PET set up was due to the fact that the scatter fraction in human PET scan is significant enough to interest us in the proposed idea whereas in small animal imaging scatter fraction is not so high.

## 2.3 Trigger logic

After generating the data from GATE simulation, we had defined a trigger logic to select out usable valid triple gamma single scattered events. A coincidence time window of 10 ns was used for data selection. At first, two different energy windows, for 511 keV, the energy window was from 10-255 keV and for 1157 keV gamma it is 255-818 keV were used to select out scatterer detector interactions. If three hits in the above specified energy windows (two for 511 keV and one for 1157 keV) for the scatterer were obtained then we collected all the events in absorber ring falling in that coincidence time window and sorted out events with only three hits in absorber ring. In next stage, correspondence between individual scatter hit and absorber interactions were made. At first, the absorber hit corresponding to 1157 keV is identified depending on closeness of summed energy to 1157 keV. Then remaining two absorber hits were allocated depending on closeness from scatter hits. Finally, single scattered coincidences were acquired using photo-peak and off-peak energy windows of 495-525 keV and 250-495 keV respectively.

## 2.4 Image reconstruction

Image reconstruction was performed without applying any typical algorithm (e.g., MLEM, OSEM [7]). Rather the annihilation points were estimated independently for each event. At first, we had calculated two possible scattering points on the single scatter surface as explained in section 2.1. Then Compton cone of 1157 keV was projected on the two separate broken LORs to obtain at most four possible annihilation points (figure 2). One point among those was selected depending on FOV constraint and TOF information.

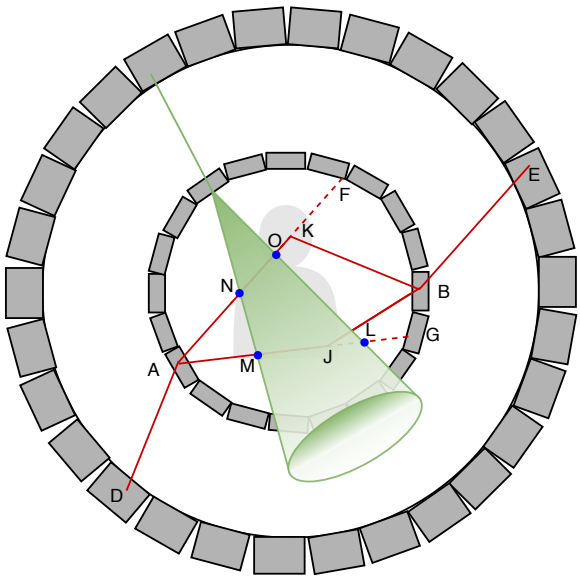

**Figure 2:** A single scattered event is detected at points A, B in scatterer ring and at point D, E in absorber ring; the intersection between single scatter surface, two Compton cones for 511 keV lefts us with two possible scattering points K, J; the intersection between Compton cone of 1157 keV and broken LORs gives us four possible annihilation points; one among those are chosen depending on FOV constraint and TOF information.

## 3　Results

As described in the section 2.1, using Compton cones from both sides of 511 keV detection, the locus can be further constrained. The cross section between single scatter surface and the cone in these cases is a type of 3D curves (red and green curve in figure 3) such that two such 3D curves cut each other in generally at two points. It is worth to mention here that the generation of two cross points is not due to finite resolution of detectors and hence those can be quite a distant apart (figure 3).

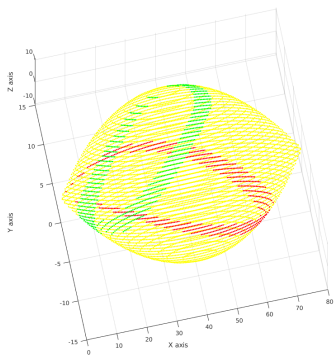

**Figure 3:** The solution between single scatter surface (yellow envelope) and Compton cones from each end of 511 keV detection is shown as green and red curve, in generally these two curves cut each other at two points, it is to be noted that two point is not due to finite resolution of detector.

We had performed the GATE simulation of Compton-PET

system with parameters described in section 2.2. Then root output data was processed using the trigger logic described in section 2.3. The trigger logic was implemented through MATLAB scripts. Figure 4-6 shows the 2D energy histogram plot between scatter energy deposition vs. absorber energy deposition for 511 and 1157 keVs. For unscattered photons we can find x+y=511 keV line where x and y are absorber and scatter deposition respectively which shows that the proposed trigger logic is able to collect 511 keV data (figure 4). The line is discontinued at scatter energy 10 keV and 255 keV, because of energy window on scatter deposition (see section 2.3). Besides the width of the x+y=511 line is decided by photo-peak width chosen in trigger logic. On the other hand, for 1157 keV detection similar x+y=1157 keV line can be seen (figure 5). Here we have discontinuity on scatterer energy at 255 keV and 818 keV due to energy window applied in trigger logic. Point to be noted, here unlike 511 keV we have count below the x+y=1157 line. This is because there is no window applied on total energy like photo-peak energy window for 511 keV. We have assumed a full deposition of energy of 1157 keV gamma in scatterer and absorber. Finally, for single scattered events, rather than having a line we have area bounded by x+y=250 keV, x+y=495 keV, x=10 keV, and x=255 keV (figure 6). First two bounds are due to off-peak window and last two are applied in initial stage of trigger logic.

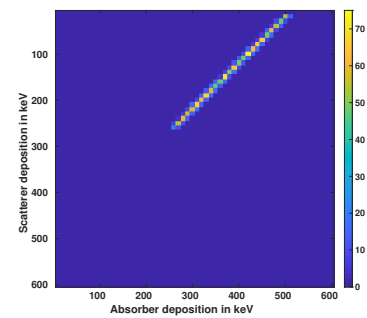

**Figure 4:** The count histogram color plot between scatterer and absorber energy deposition for 511 keV unscattered photon detection.

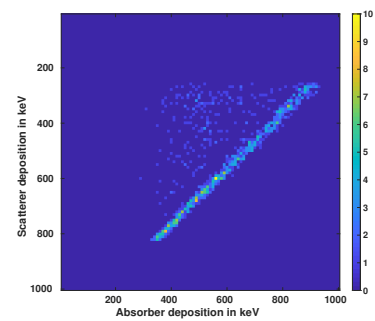

**Figure 5:** The count histogram color plot between scatterer and absorber energy deposition for 1157 keV photon detection.

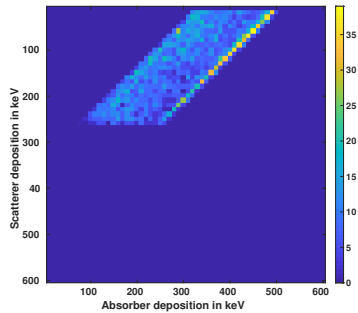

**Figure 6:** The count histogram color plot between scatterer and absorber energy deposition for 511 keV single scattered photon detection.

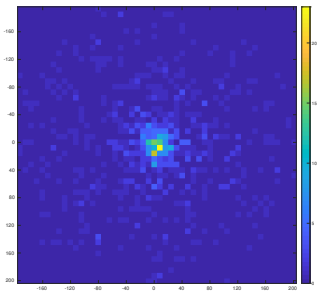

**Figure 7:** Single scattered image (cross sectional) of line source, pixel size was chosen to be $8 \times 8$ $mm^2$.

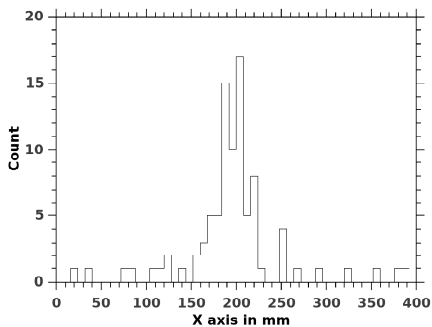

**Figure 8:** Intensity line profile (horizontal) of the single scatter image through (0,0) point, pixel size was chosen to be $8 \times 8$ $mm^2$.

Finally, we had produced image from single scattered data set (figure 7). For image resolution study, we had calculated FWHM of intensity line profile (horizontal) of the image. The histograms were shown in figure (8) with FWHM calculated to be 35.864 mm. To sum up, we were able to produce physically meaningful single scatter images. This proves the feasibility of single scatter imaging for Compton-PET system with triple gamma source.

## 4  Conclusion

We have proposed the idea of feasibility of imaging from single scattered (inside tissue) data in triple gamma imaging.

Tissue scattered data in human PET scan can go up to 40-60%. On the other hand, triple gamma imaging suffers from low count specially in online ion range verification in ion therapy and hence in that context the idea of imaging from scattered data is relevant. Although a better resolution image than unscattered image can't be expected from scattered data due to inherent resolution effects, we believe that producing image from two independent data sets – unscattered and single scattered – will improve our diagnosis ability. We have shown the feasibility of the proposed concept. Analysing GATE simulation data, we are able to produce physically meaningful images. The trigger logic used here is not claimed to be perfect. Rather simplicity is invoked to make the task computationally simple as this work is related to only feasibility. We believe that the idea proposed can be beneficial in triple gamma imaging based beam range monitoring and late point imaging in case of Scandium DOTA-TOC imaging.

## 5  Acknowledgement

Authors wish to gratefully acknowledge the Center for Development of Advanced Computing (C-DAC), Pune, India, for providing the supercomputing facilities [10] for data analysis.

# Analytic Continuation and Incomplete Data Tomography

Gengsheng L. Zeng[1,2] and Ya Li[3]

[1]Department of Computer Science, Utah Valley University, Orem, USA
[2]Department of Radiology and Imaging Sciences, University of Utah, Salt Lake City, USA
[3]Department of Mathematics, Utah Valley University, Orem, USA

**Abstract** A unique feature of medical imaging is that the object to be imaged has a compact support. In mathematics, the Fourier transform of a function that has a compact support is an entire function. In theory, an entire function can be uniquely determined by its values in a small region, using, for example, power series expansions. Power series expansions require evaluation of all orders of derivatives of a function, which is an impossible task if the function is discretely sampled. In this paper, we propose an alternative method to perform analytic continuation of an entire function, by using the Nyquist–Shannon sampling theorem. The proposed method involves solving a system of linear equations and does not require evaluation of derivatives of the function. Noiseless data computer simulations are presented. Analytic continuation turns out to be extremely ill-conditioned.

## 1 Introduction

It is known that for stable image reconstruciton using noisy data, measurements must be sufficiently acquired. There are many data sufficiency conditions that are proposed. For example, in cone-beam imaging, Tuy's condition must be satisfied [1]. In PET (positron emission tomography), Orlov's condition must be satisfied [2]. In MRI (magnetic resonance imaging), the k-space (i.e., the Fourier space) preferably should be fully sampled.

As theoretical curiosity, one would wonder whether it is possible to reconstruct the image using incomplete data. This subjuct has been systematically discussed in Natterer's book [3], where the incomplete data situations are classified in to 3 categories: limited angle problems, exterior problems, and truncated problems. For the limited angle problems and exterior problems, the inversion is so seriously ill-posed that it is hopeless to have any practical value.

The goal of this paper is *not* to develop an algorithm that can be used immediately in practice. The motivation of this paper is purely theoretical. Assuming that we live in an ideal world without any noise around, we investigate whether it is possible to reconstruct an image using incomplete measurements.

## 2 Methods

### 2.1 Mathematical foundation

A function has compact support if it is zero outside of a compact set that is closed and bounded. The Fourier transform of a compactly supported function is an entire function. An entire function is a complex-valued function that is holomorphic at all finite points over the whole complex plane. A holomorphic function is complex differentiable at every point of its domain. Any holomorphic function is infinitely differentiable and equal, locally, to its own Taylor series. A holomorphic function whose domain is the whole complex plane is called an entire function [4].

The possibility of imaging with incomplete data is established as follows, using a one-dimensional (1D) example. The object $f(x)$ is compactly supported, for example, defined on [-½, ½] and $f(x) = 0$ elsewhere. Let $F(\omega)$ be the Fourier transform of $f(x)$. Assume that $f(x)$ is unknown, $F(\omega)$ is partially known. Without loss of generality, $F(\omega)$ is assumed to be known in a small region around the point $\omega = 0$. One can evaluate derivatives of $F(\omega)$ at all orders, and thus construct the Taylor series of $F(\omega)$ at $\omega = 0$. Since $F(\omega)$ is an entire function, this Taylor series converges to $F(\omega)$ in the entire complex plane. In other words, $F(\omega)$ becomes known in the entire complex plane through analytic continuation.

This analytic continuation of $F(\omega)$ is actually of no use in practice, because most real-world measurements are discrete. This fact inhibits the evaluation of derivatives of $F(\omega)$, and thus the Taylor series of $F(\omega)$ cannot be obtained.

### 2.2 Lagrange interpolation method

Other than the Taylor series expansion, the Lagrange interpolation formula can be an alternative method to perform analytic continuation [5]. The essential idea of the Lagrange interpolation formula is to find the lowest order polynomial that passes through given points.

We do not believe that the Fourier transform of a compactly supported function in medical imaging behaves like a polynomial. The energy of a compactly supported function $f(x)$ is finite. Parseval's Theorem tells us that the energy of its Fourier transform $F(\omega)$ is the same and finite. As $|\omega| \to \infty$, we must have $F(\omega) \to 0$. Hence, $F(\omega)$ cannot behave as a polynomial, because the magnitude of a polynomial tends to infinity as the magnitude of the variable tends to infinity.

### 2.3 Nyquist-Shannon method

If a spatial-domain function is band-limited, then this signal can be represented by its discrete samples, and the sampling interval is inversely proportional to the bandwidth. If we switch the roles of these two domains, the spatial-domain function $f(x)$ is spatially bounded and the corresponding Fourier-domain function $F(\omega)$ can be represented by its discrete samples. The Fourier-domain sampling interval $\Delta\omega$ is inversely proportional to the spatial-domain object size. According to the Nyquist-Shannon Theorem, we can express the complex Fourier-domain function $F(\omega)$ by its own samples $F(n\Delta\omega)$ as

$$F(\omega) = \sum_{n=-\infty}^{\infty} F(n\Delta\omega) \cdot sinc\left(\frac{\omega - n\Delta\omega}{\Delta\omega}\right), \qquad (1)$$

where the sinc function is defined as

$$sinc(x) = \begin{cases} \dfrac{sin(\pi x)}{\pi x} & if \quad x \neq 0 \\ 1 & if \quad x = 0. \end{cases} \qquad (2)$$

Formula (1) is referred to as the Whittaker-Shannon interpolation formula. Formula (1) implies that the function $F(\omega)$ is sufficently determined by its discresete values $F(n\Delta\omega)$, where $n \in \mathbb{Z}$ (integers). Because $f(x)$ is the inverse Fourier transform of $F(\omega)$, the spatial-daumin compactly supported function $f(x)$, in turn, is determined by the samples $F(n\Delta\omega)$.

According to the fact that $F(\omega) \to 0$ as $|\omega| \to \infty$, we can obtain an approximate expression of (1) by using only a finite number of terms in the summation:

$$F(\omega) \approx \sum_{n=-N}^{N} F(n\Delta\omega) \cdot sinc\left(\frac{\omega - n\Delta\omega}{\Delta\omega}\right). \qquad (3)$$

It is reasonable to further assume that the function $f(x)$ is real and $F = F_r + iF_i$, and then the real part $F_r(\omega)$ is even and the imaginary part $F_i(\omega)$ is odd. Therefore, (3) can be written as

$$F_r(\omega) \approx F_r(0) + 2\sum_{n=1}^{N} F_r(n\Delta\omega) \cdot [sinc\left(\frac{\omega - n\Delta\omega}{\Delta\omega}\right) + sinc\left(\frac{\omega + n\Delta\omega}{\Delta\omega}\right)] \qquad (4a)$$

$$F_i(\omega) \approx 2\sum_{n=1}^{N} F_i(n\Delta\omega)[sinc\left(\frac{\omega - n\Delta\omega}{\Delta\omega}\right) - sinc\left(\frac{\omega + n\Delta\omega}{\Delta\omega}\right)]. \qquad (4b)$$

## 2.4 Proposed method

This part presents the main result of current paper. We consider a Fourier transform pair: $f(x)$ and $F(\omega)$, where $f(x)$ is real and has a compact support, and $F(\omega)$ is complex and entire. The spatial-domain function $f(x)$ is unknown. The Fourier-domain function $F(\omega)$ is measured at discrete points, $\omega_k$, $k = 1, 2, \ldots, M$, which are in a smaller interval than $[-N, N]$. Let us form 2 real column vectors:

$$p = [F_r(\omega_1), F_r(\omega_2), \ldots, F_r(\omega_M), F_i(\omega_1), F_i(\omega_2), \ldots, F_i(\omega_M)]^T \qquad (5)$$

and

$$u = [F_r(0), F_r(\Delta\omega), \ldots, F_r(N\Delta\omega), F_i(\Delta\omega), \ldots, F_i(N\Delta\omega)]^T. \qquad (6)$$

The vector $p$ contains the measurements, and the vector $u$ contains the unknowns. It is allowed that some of the unknowns are the measurements. The approximations (4a) and (4b) can be written in the matrix form as

$$Au \approx p, \qquad (7)$$

where the real $(2M) \times (2N + 1)$ matrix $A$ is determined according to (4a) and (4b). A numeric algorithm is required to solve the unknown vector $u$ from (7). Once the vector $u$ is obtained, the spatial-domain function $f(x)$ is constructed by $u$ as follows.

If we consider the compactly supported function $f(x)$ as one period of a periodic function, then $f(x)$ has a Fourier series expansion

$$f(x) = \sum_{n=-\infty}^{\infty} c_n e^{\frac{i2\pi nx}{T}} \approx \sum_{n=-N}^{N} c_n e^{\frac{i2\pi nx}{T}} \qquad (8)$$

with Fourier coefficients

$$c_n = \frac{1}{T} \int_{-T/2}^{T/2} f(x) e^{-\frac{i2\pi nx}{T}} dx, \qquad (9)$$

where $T$ is the period and can be same as (or larger than) the span of $f(x)$. Since the Fourier transform of $f(x)$ is defined as

$$F(\omega) = \int_{-T/2}^{T/2} f(x) e^{-i2\pi\omega x} dx, \qquad (10)$$

we have

$$c_n = \frac{1}{T} F\left(\frac{n}{T}\right). \qquad (11)$$

If we choose $\Delta\omega = 1/T$, the Fourier series coefficients can be obtained by solving (7). When the Fourier series (8) is truncated, the summation can be implemented by the $(2N+2)$-point inverse discrete Fourier Transform (IDFT) or the inverse fast Fourier transform (IFFT).

$$\sum_{n=0}^{2N+1} c_n e^{\frac{i2\pi nx}{2N+2}} \qquad (12)$$

where $c_{N+1} = 0$ and $c_{N+1+k} = c_{N+1-k}^*$, $k = 1, \ldots, N$.

Solving for $u$ from (7) is challenging, because the system is seriously ill-posed. In the computer simulations in this paper, no noise is added to the measurements $p$. The computer roundoff errors are already serious enough to make the solution deviate from the true solution. The following approaches can be used to find the vector $u$:

Approach 1:

The Moore-Penrose pseudoinverse with a tolerance $tol$. This approach finds the singular value decomposition (SVD) of the matrix $A$ and replaces the singular values that are smaller than $tol$ by zeros before calculating the generalized inverse of $A$.

$$u = A^{\dagger} p. \qquad (13)$$

Approach 2:

$$\min_{u}(\|Au - p\|_2 + \alpha\|u\|_2). \qquad (14)$$

Approach 3:

$$\min_u(\|Au - p\|_1 + \alpha\|u\|_2). \tag{15}$$

Approach 4:

$$\min_u(\|Au - p\|_\infty + \alpha\|u\|_2). \tag{16}$$

Approach 5:

$$\min_u(\|u\|_2) \tag{17}$$

Subject to

$$Au = p. \tag{18}$$

Approach 6:

$$\min_u(\|u\|_1) \tag{19}$$

Subject to (18).

Approach 7:

$$\min_u(\|u\|_\infty) \tag{20}$$

Subject to (18).

### 2.5 Applications to medical imaging

One application of the proposed method is in limited angle tomography, where the Radon transform is only available in an angular range smaller than 180º. According to the Central Slice Theorem, in the two-dimensional (2D) Fourier domain, two angular sections are measured, and two remaining angular sections are not, as illustrated in Fig.1.

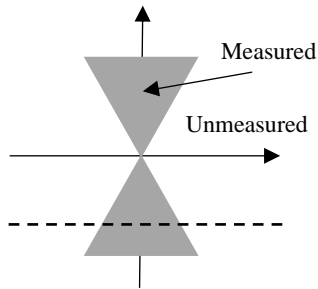

Fig. 1. Illustration of a 2D Fourier space when the angular sampling is less than 180º. The shaded regions are measured, while the unshaded regions are not measured. In theory, one can use analytic continuation to estimate the data in the unmeasured regions.

The measured Fourier components are in the shaded regions. In theory, it is possible to complete the unmeasured Fourier components by line-by-line (which can be row-by-row, or column-by-column) analytic continuation. One analytic continuation method is suggested in Section 2.4.

Another application of the proposed method is in fast MRI (magnetic resonance imaging), where the k-space is not completely measured. The unmeasured k-space data can be estimated from measured data. If this analytic continuation technology works, MRI procedures can be sped up significantly.

## 3 Results

The first computer simulation considered a 1D function $f(x)$ that was composed of two boxcars. The Fourier transform of a boxcar function is a sinc function. Therefore, the closed-form of $F(\omega)$ in this case was known. It was assumed that 64 uniform discrete samples of $F(\omega)$ were sufficient to represent the function.

We measured the first 16 frequency components, and measured additional 135 components within the measured range. The condition number of $A^T A$ was $6.0325\times10^{17}$. The strategy of the proposed method is to over-sample the region where data is available. However, when we used additional 1350 components (instead of 135 components) within the measured range, the condition number of $A^T A$ worsened to $3.4168\times10^{18}$.

The following parameters were used for this simulation: Approach 1: $tol = 10^{-14}$. Approach 2: $\alpha = 10^{-9}$. Approach 3: $\alpha = 10^{-8}$. Approach 4: $\alpha = 0$.

The second simulation was with a Shepp-Logan phantom, for which we did not have a closed-form expression for its Fourier transform. To work around this problem, we first used a computer simulated digitized Shepp-Logan phantom in a 256×256 array that was column-by-column zero-padded so that each column had 2560 pixels. After taking 2560-point 1D DFT, we obtained an over-sampled Fourier spectrum. Among these 2560 samples, we chose the first 100 samples as our measurements. These low-frequency 100 components were used to estimate the unmeasured frequency components using the method proposed in Section 2.4.

The DFT assumes discrete and periodic $f(x)$, as well as discrete and periodic $F(\omega)$. The actual $F(\omega)$ is aperiodic, because the actual $f(x)$ is continuous. The errors introduced by discretization of $f(x)$ can be reduced by using smaller sampling intervals. For example, using an array size of 1024×1024 or 2048×2048 to represent the Shepp-Logan phantom.

Fig. 2 shows the Fourier-domain signals and their associated inverse DFT reconstructions. All computer simulation results for the simulations are shown in Figs. 3 and 4. In the first simulation, it was assumed that 64 samples were good enough to represent the original signal. Frequency components were available only lower than sample #15. In the second simulation, frequency components lower than #10 were available.

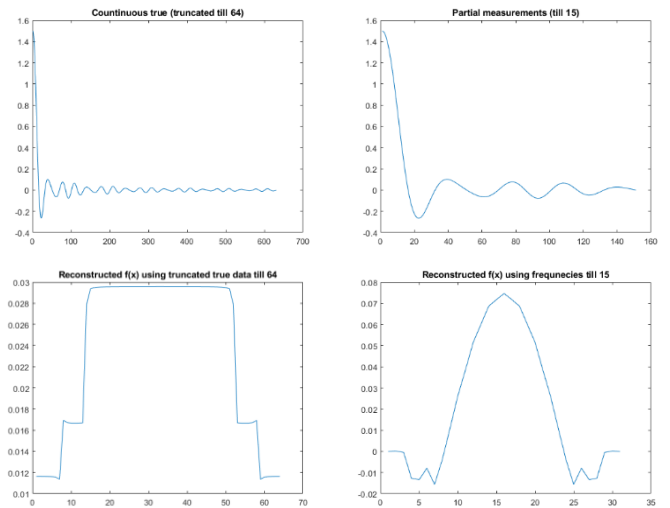

Fig. 2. Fourier-domain signal $F(\omega)$ and its reconstruction $f(x)$, up to "64" and up to "15", respectively.

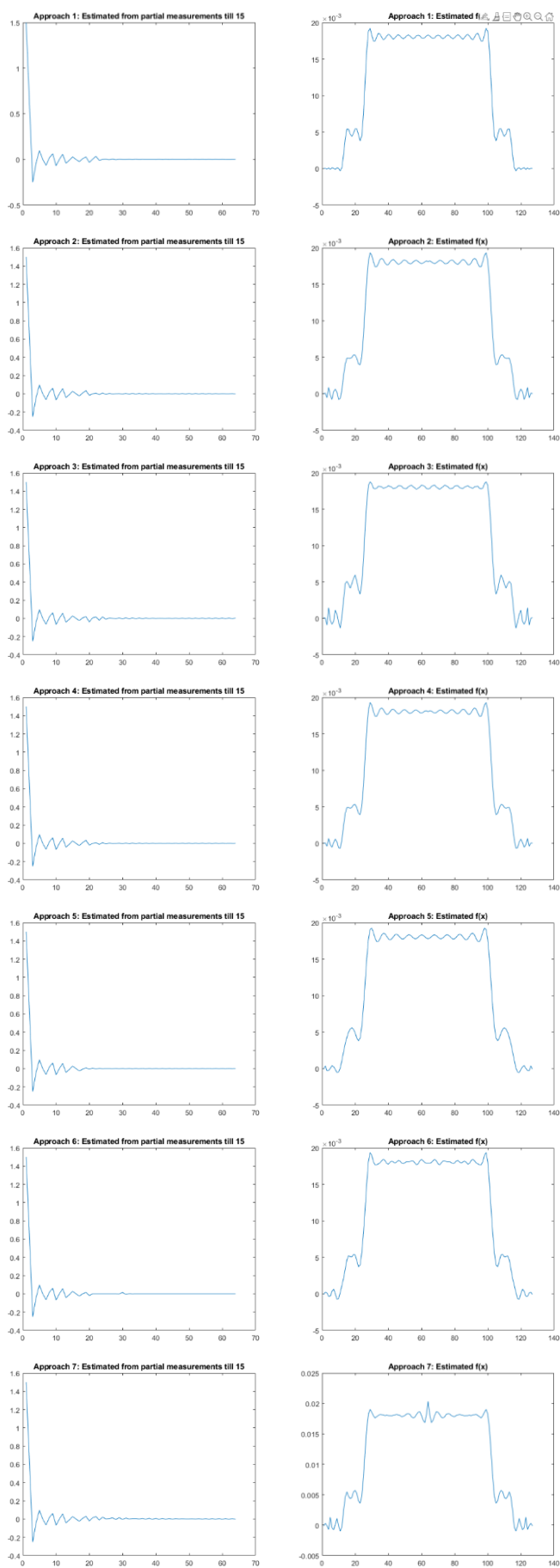

Fig. 3. Results for the first simulation. Estimated Fourier componets $F(\omega)$ and their reconstructions $f(x)$ using proposed 7 approaches.

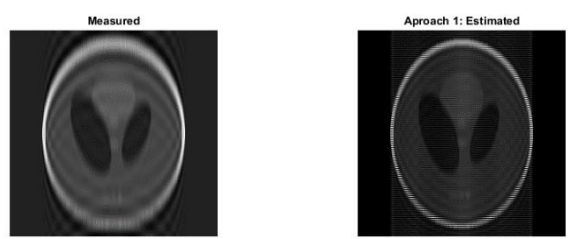

Fig. 4. Left: Reconstructed result from measured data. Right: Reconstructed result for the second simulation, using approach #1.

## 4 Discussion

Analytic continuation is a powerful tool in mathematics to determine the values of an entire fucntion in a wider region. This paper has developed an analytic continuation method by over-sampling the 'known' region and solving a system of linear equations. The system turns out to be seriously ill-posed. Our computer simulations cannot obtain exact estimation even though no noise is added to the measurements. The computer rounding errors are already too large to handle. Seven approaches have been tested. It is interesting to notice that Approach #4 with the infinity norm allows $\alpha = 0$, while Approaches #1-#3 with $L_1$ or $L_2$ norms require some regularization. The $L_1$ norm forgives outliers, the $L_2$ norm manages the error energy, and the $L_\infty$ norm controls the maximum error.

Our results do not imply that the analytic continuation is useless in the real world. Our simulations used sampled data. The analytic continuation requires 'continuous' data, which is difficult to implement with today's computers. It is still an open problem whether analytic continuation is helpful if 'continuous' and 'rounding error free' computers are available. We believe that denoising must be performed prior to analytic continuation.

# Fast and memory-efficient reconstruction of sparse TOF PET data with non-smooth priors

Georg Schramm[1] and Martin Holler[2]

[1]Department of Imaging and Pathology, Division of Nuclear Medicine, KU Leuven, Belgium
[2]Institute of Mathematics and Scientific Computing, University of Graz, Austria

**Abstract** In this work, we propose and analyze a modification of the stochastic primal-dual hybrid gradient (SPDHG) algorithm which substantially reduces its memory requirements for reconstruction of sparse time-of-flight (TOF) PET data with non-smooth priors. Moreover, we study the influence of the ratio of the primal and dual step sizes on the convergence of SPDHG. The performance of the optimization algorithm is investigated based on simulated 2D TOF data using a brain-like software phantom. We find that the memory requirement of SPDHG for sparse TOF PET data can be substantially reduced by a better initialization without noticeable losses in the convergence speed. Moreover, a careful choice of the ratio of the primal and dual step sizes, depending on the magnitude of the image to be reconstructed, is crucial to obtain fast convergence.

## 1 Introduction

Due to limitations in acquisition time, injectable dose and scanner sensitivity, acquired data in positron emission tomography (PET) suffer from high levels of Poisson noise that is transferred into the reconstructed image, necessitating noise suppression during or post reconstruction. One possible way of noise suppression is the maximum a posteriori approach where a smoothing prior is added next to the data fidelity term (the negative Poisson loglikelihood) in the cost function optimized in iterative image reconstruction. Unfortunately, many advanced smoothing priors such as e.g. Total Variation (TV) [1], Total Generalized Variation (TGV) [2], Joint T(G)V [3, 4] or Parallel Level Sets [5, 6] are non-smooth functions which permits the use of simple and efficient purely gradient-based optimization algorithms. Moreover, due to the large number of data bins in a (time-of-flight) sinogram of modern PET scanners, the computation time for a single evaluation of the complete forward (and adjoint) model is usually slow, favoring optimization algorithms that use only a subset of the data in every update step like maximum expectation maximization with ordered subsets (OSEM).

Recently, Chambolle et al. [7] and Ehrhardt et al. [8] introduced the stochastic primal-dual hybrid gradient (SPDHG) algorithm which is a provably convergent algorithm that allows to solve the PET reconstruction problem including many non-smooth priors with only a few iterations. Algorithm 1 shows SPDHG using a typical PET forward model. As seen in line 6 and 7, in every update step only a forward and backprojection of a subset of the data is required. Using two clinical FDG and Fluorbetapir data sets from the Siemens mMR, it was shown in [8], that approximately 10 iterations, meaning 10 complete forward and back projections of the data are

---

**Algorithm 1** SPDHG for PET reconstruction [8]

1: **Initialize** $x(=0), y(=0), (S_i)_i, T, (p_i)_i,$
2:    $\bar{z} = z = P^T y$
3: **repeat**
4:      $x = \text{proj}_{\geq 0}(x - T\bar{z})$
5:      Select $i \in \{1, \ldots, n+1\}$ randomly according to $(p_i)_i$
6:      **if** $i \leq n$ **then**
7:        $y_i^+ \leftarrow \text{prox}_{D_i^*}^{S_i}(y_i + S_i(P_i x + s_i))$
8:      **else**
9:        $y_i^+ \leftarrow \text{prox}_{D_i^*}^{S_i}(y_i + S_i P_i x)$
10:      **end if**
11:      $\delta z \leftarrow P_i^T (y_i^+ - y_i)$
12:      $y_i \leftarrow y_i^+$
13:      $z \leftarrow z + \delta z$
14:      $\bar{z} \leftarrow z + (\delta z / p_i)$
15: **until** stopping criterion fulfilled
16: **return** $x$

---

sufficient to reach reasonable convergence for clinical purposes when using preconditioning and proper sampling of the subsets.

In this work, we focus on time-of-flight (TOF) PET reconstruction using TV regularization, noting, however, that generalizations to other non-smooth priors as mentioned above are possible within the same framework. The TV regularized TOF PET reconstruction method requires to solve the optimization problem

$$\arg\min_{x \geq 0} \sum_j (Px)_j - d_j \log((Px)_j + s_j) + \beta \|\nabla x\|_1, \quad (1)$$

where $x$ is the PET image to be reconstructed, $P$ is the TOF forward projector including the effects of attenuation and normalization, $d$ are the acquired prompt TOF coincidences (the emission sinogram), and $s$ are additive contaminations including random and scattered coincidences. The operator $\nabla$ is the gradient operator, $\|\nabla u\|_1$ is sum over all entries of the pointwise Euclidean norm of $\nabla u$, and $\beta$ is a scalar controlling the level of regularization.

In the application of Algorithm 1 we follow the approach of [8] by splitting the data into n non-overlapping subsets with the corresponding sequence of partial PET forward operators denoted as $(P_i)_{i=1}^n$. To simplify notation, we set $P_{n+1} = \nabla$ and choose the probabilities $p_1 = \ldots = p_n = 1/(2n)$ and $p_{n+1} = 1/2$. For $\rho < 1$ and $\gamma > 0$, we define preconditioned step sizes for the partial PET operators, for $i = 1, \ldots, n$

$$S_i = \gamma \, \mathrm{diag}\left(\frac{\rho}{P_i 1}\right) \qquad T_i = \gamma^{-1} \mathrm{diag}\left(\frac{\rho \, p_i}{P_i^T 1}\right)$$

and for the gradient operator

$$S_{n+1} = \gamma \frac{\rho}{\|\nabla\|} \qquad T_{n+1} = \gamma^{-1} \frac{p_i \rho}{\|\nabla\|} \ .$$

As mentioned in [8], if we set $T = \min_{i=1,\ldots,n+1} T_i$ pointwise, SPDHG converges.

The proximal operator for the convex dual of $D_j(y) := y_j - d_j \log(y_j)$ is given by

$$(\mathrm{prox}_{D_j^*}^{S_i}(y))_j = \frac{1}{2}\left(y_j + 1 - \sqrt{(y_j - 1)^2 + 4(S_i)_j d_j}\right) \quad (2)$$

and the proximal operator for the convex dual of the TV term is given by

$$(\mathrm{prox}_{D_{n+1}^*}(y))_j = \beta \, y_j / \max(\beta, |y_j|) \ . \quad (3)$$

As discussed in Remark 2 of [8], a potential drawback of SPDHG is that it requires to keep at least one more complete (TOF) sinogram ($y$) in memory. Moreover, if the proposed preconditioning is used, a second complete (TOF) sinogram (the sequence of step sizes $(S_i)_{i=1}^n$) needs to be stored. In general, this is less of a problem for static single-bed non-TOF PET data, where sinogram sizes are relatively small. However, for simultaneous multi-bed, dynamic or TOF PET data, the size of complete sinograms can be become problematic, especially when using GPUs. E.g., for modern PET TOF scanners with 25 cm axial FOV and a TOF resolution of ca. 400 ps, a complete unmashed TOF sinogram in single precision for one bed position has approximately $4.4 \cdot 10^9$ data bins, requiring ca. 17 GB of memory. Note that the memory required to store a complete TOF sinogram will further increase with better TOF resolution. Due to the large number of data bins and the limitations in injected dose and acquisition time, modern TOF sinograms are usually very sparse, meaning that in most data bins no data is acquired. E.g., for a typical 3 min-per-bed-position whole-body FDG scan with an injected dose of around 200 MBq acquired 60 min p.i. on a state-of-the-art TOF PET/MR scanner, more than 95% of the data (TOF sinogram) bins are empty. For short early frames in dynamic brain scans, this fraction is even higher. And even for "high count" late static 20 min FDG brain scans with an injected dose of 150 MBq acquired 60 min p.i., still around 70% of the data bins are empty.

Considering the very sparse nature of TOF emission sinograms, in this work, we propose and analyze a modification of SPDHG for sparse PET data which substantially reduces its memory requirements. Moreover, we also analyze the influence of the scalar hyperparameter $\gamma$, that determines the ratio between the primal and dual step sizes, on the convergence of SPDHG.

## 2 Materials and Methods

### 2.1 Memory efficient TOF PET SPDHG through better initialization

In [8], the authors propose to initialize $x$ and $y$ with zeros everywhere. However, we can observe from Eq. (2) that for data bins $j$ where $d_j = 0$ (empty TOF sinogram bins), $(\mathrm{prox}_{D_j^*}(a))_j = 1$ for $a_j \geq 1$ and $(\mathrm{prox}_{D_j^*}(a))_j = a_j$ otherwise. Moreover, we see that $a_j = (y_i + S_i(P_i x + s_i))_j \geq 1$ provided that $(y_i)_j \geq 1$ since all other quantities are positive. Hence, if we initialize all bins of $y$ where the data $d$ equals zero with 1, these bins remain equal to 1 during all iterations. This in turn means that these bins do not contribute to the solution, since only the change in $y$ is backprojected in line 7 of algorithm 1. Consequently, this implies that these bins do not need to be kept in memory after the initialization of the $z$ and $\bar{z}$ with the backprojection of $y$, which dramatically reduces the memory requirement and also the number of projections that need to be calculated, keeping in mind that for most acquisitions with modern TOF PET scanners, most of the data bins are 0 as discussed before. To differentiate between SPDHG with the originally proposed initialization and the initialization proposed above, we call the latter SPDHG-S.

### 2.2 Numerical experiments

To compare the convergence of SPDHG and SPDHG-S, we performed reconstructions of simulated TOF PET data from a virtual 2D scanner mimicking the TOF resolution (ca. 400 ps FWHM) and geometry of one ring (direct plane) of the GE SIGNA PET/MR (sinogram dimension: 357 radial bins, 224 projection angles, 27 TOF bins). A software brain phantom with a typical gray to white matter contrast of 4:1 was created based on the brainweb phantom and used to generate simulated data including the effects of attenuation and flat contamination (scattered) coincidences with a simulated scatter fraction of 16%. Noisy simulated prompt emission TOF sinograms were generated for $10^5$, $10^6$, and $10^7$ counts. In the case of $10^5$, 80% of the bins in the 2D TOF emission sinogram are 0. The simulated data were reconstructed with SPDHG and SPDHG-S using 50 iterations, 112 subsets, a fixed level regularization ($\beta = 0.6$ for $10^5$ counts, and $\beta = 0.2$ for $10^6$ and $10^7$ counts), and different values for $\gamma$. As in [8], convergence was monitored by tracking the relative cost function

$$c_{\mathrm{rel}}(x) = (c(x) - c(x^*))/(c(x^0) - c(x^*)).$$

and the peak signal to noise ratio

$$\mathrm{PSNR}(x) = 20 \log_{10}\left(\|x^*\|_\infty / \sqrt{\mathrm{MSE}(x, x^*)}\right)$$

compared to an approximate minimizer $x^*$. which was calculated using the deterministic PDHG with 5000 iterations without subsets. In all reconstructions, the TOF PET operator $P$ was renormalized such that the norm of $P$ equaled the number of projection angles and the subsets were defined via equidistant projection angles. The gradient operator was implemented as the finite forward difference.

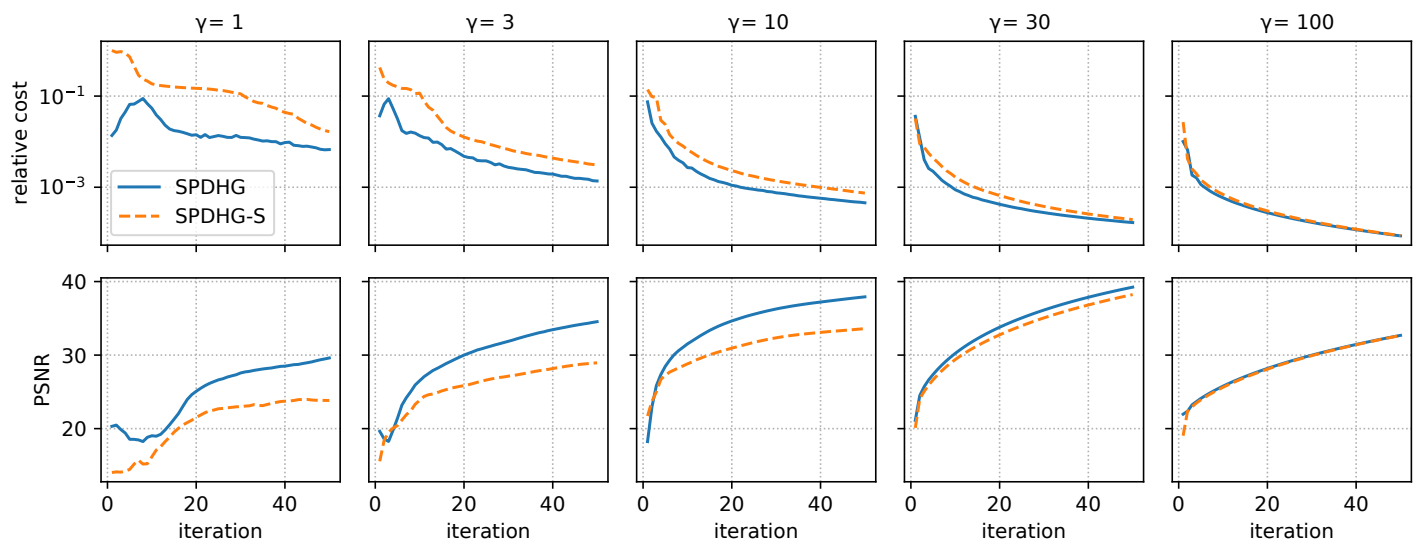

**Figure 1:** Cost and PSNR relative to the approximate minimizer of SPDHG and SPDHG-S for different $\gamma$ values, using $10^5$ counts and 112 subsets.

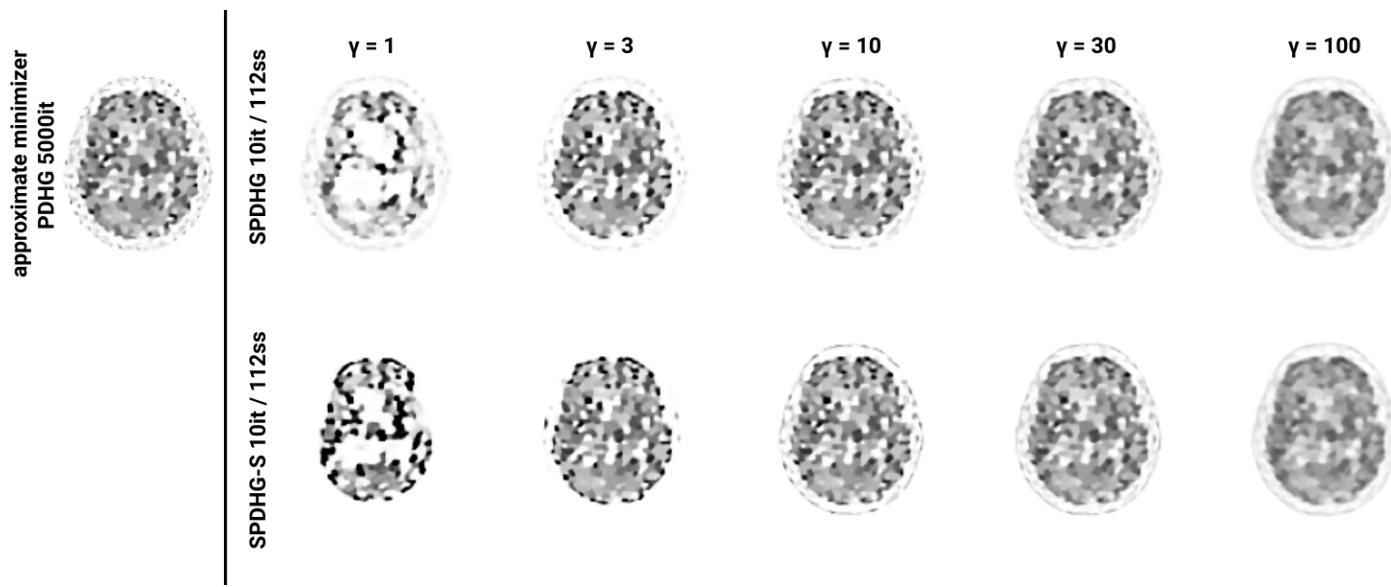

**Figure 2:** Reconstruction results of SPDHG (top) and SPDHG-S (bottom) after 10 iterations with 112 subsets for different $\gamma$ values indicated above the reconstructions using $10^5$ counts.

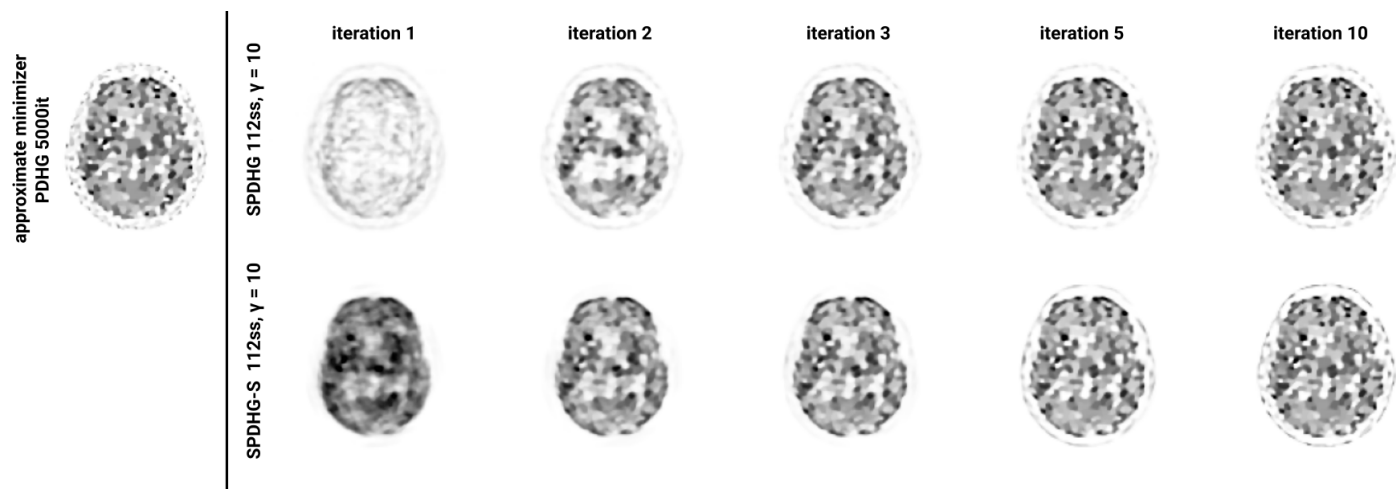

**Figure 3:** Comparison of convergence of SPDHG (top) and SPDHG-S (bottom) in early iterations using $10^5$ counts, $\gamma = 10$ and 112 subsets.

## 3   Results

Figure 1 shows the relative cost and PSNR to the approximate minimizer of SPDHG and SPDHG-S for different $\gamma$ values, using $10^5$ counts and 112 subsets. First of all, we see that the choice of $\gamma$ has a strong influence on the convergence for SPDHG and SPDHG-S which can be also seen in Fig. 2 where the reconstructions after 10 iterations are shown. Moreover, it can be observed in both figures that, with increasing $\gamma$, SPDHG-S performs more and more similar to SPDHG in terms of PSNR and the relative cost. Figure 3 shows a comparison of SPDHG and SPDHG-S in the very early iterations for $\gamma = 10$.

## 4   Discussion

All three figures in this work demonstrate that the difference in the convergence between SPDHG and SPDHG-S for appropriate $\gamma$ values (e.g. 10) after ca. 3 iterations is very minor. As reported in [8], we can confirm that after ca. 10 iterations the visual difference in the image quality compared to the approximate minimizer is very small. The remaining slight differences are mainly in low uptake regions in the skin around the brain. Major differences in the convergence between SPDHG and SPDHG-S are only seen after the 1st iterations which is explained by the impact of the initialization of $y$ on the initialization of $z$ and $\bar{z}$ as shown in line 2 in algorithm 1.

The finding reported for $10^5$ counts were also confirmed for the higher count levels of $10^6$ and $10^7$ counts not shown here. However, note that at $10^7$ the emission sinogram is of less sparse which naturally reduces the influence of our proposed initialization. In contrast to [8], we find that in general $\gamma = 1$ is not optimal in terms of convergence speed. Comparing the results with different simulated count levels, we find that the optimal value for $\gamma$ is inversely proportional to the magnitude of the image to be reconstructed, or, in other words, inversely proportional to the number of acquired counts.

So far, the proposed SPDHG algorithm uses binned data (sinograms). However, the promising results shown in this work might help to develop a convergent algorithm that allows to directly reconstruct list-mode TOF PET data with non-smooth priors.

## 5   Conclusion

The memory requirement for the SPDHG algorithm for sparse TOF PET data, can be substantially reduced by a better initialization of the dual variable $y$ without noticeable losses in the convergence speed. Careful choice of the step size ratio parameter $\gamma$ depending on the magnitude of the image to be reconstructed is crucial to obtain fast convergence.

# Ratio of Multi-Channel Representation for Spectral CT Material Decomposition

Yongyi Shi[1], Qiong Xu[2], Yanbo Zhang[3], Zhengrong Liang[4] and Xuanqin Mou[1]

[1] Institute of Image Processing and Pattern Recognition, Xi'an Jiaotong University, Xi'an, Shaanxi 710049, China.

[2] Beijing Engineering Research Center of Radiographic Techniques and Equipment, Institute of High Energy Physics, Chinese Academy of Sciences, Beijing 100049, China.

[3] PingAn Technology, US Research Lab, Palo Alto, CA 94306, USA.

[4] Departments of Radiology, Electrical and Computer Engineering, Computer Science and Biomedical Engineering, State University of New York at Stony Brook, Stony Brook, NY 11794, USA

**Abstract:** Photon counting spectral computed tomography (PCCT) produces attenuation maps at different narrow energy windows simultaneously, being a promising technique due to its ability of material decomposition. However, because of its low signal-to-noise (SNR) and nonideal detector response in each individual energy channel, PCCT encounters the challenge of reconstructing distorted data for material decomposition. Generally, material decomposition methods aim to obtain the product of two sets of unknown variables (i.e., material densities and material composition maps). In this paper, we proposed a PCCT material decomposition method, which uses the ratio of multi-channel representation (RMCR) to achieve material decomposition in a one-step procedure. With the help of RMCR, the distortion caused by nonideal detector response was corrected. Hence, we estimated the material composition maps from the post-log data directly, avoiding knowing spectrum information which is necessary for existing one-step decomposition methods. Given the calculated material composition maps, the material densities could be accurately recovered. In addition, the noise could be reduced by strengthening prior information into the iterative framework. Experimental results show higher accuracy of material decomposition of our proposed method compare with traditional methods.

## 1 Introduction

Photon counting spectral computed tomography (PCCT) can separately obtain the incident photons at multiple energy bins via pulse-height analysis, enabling more than three materials decomposition and K-edge imaging. PCCT also provides a relatively high signal-to-noise ratio (SNR) compared to conventional energy-integration detection, because the electronic noise is eliminated. However, in each individual channel, the SNR is relatively low due to the limited counting rate. On the other hand, the PCCT measurements are also corrupted by complicated noises and artifacts caused by nonideal detector response, such as detector elements response variation, fluorescence x-ray effects, charge sharing, K-escape, and pulse pileups, in each individual energy channel [1].

Aimed to improve the accuracy of material decomposition, numerous methods were proposed to reconstruct basis material map images which are usually employed to quantify material composition and density. These methods can be divided into two categories: indirect (two-step procedure) (e.g., [2-4]) and direct (one-step procedure) (e.g., [5-7]) methods. Indirect methods estimate the material composition maps by two steps. In the two-step procedure, image reconstruction and materials decomposition are independent. The information lost in the first step cannot be compensated in the second step, which may result in aberrant decomposition. Direct methods estimate basis material maps from the energy-windowed measurements in a one-step procedure. The direct methods combine the statistical properties of measured projections, prior information in the basis material maps, and even parameters of the imaging system into one unified objective function, which can greatly improve decomposition accuracy.

However, Mory et al. compared the convergence speeds of five direct methods and mentioned that these methods are relatively time consuming [6]. In addition, these one-step methods are typically relied on detailed knowledge about the spectral response of the detectors. However, it's difficult to obtain the spectral information in real clinical practice. To solve the spectrum dependency of these one-step methods, Chang et al. proposed a spectrum estimation-guided iterative reconstruction algorithm for dual-energy CT [7]. Nevertheless, this method is hard to be introduced into PCCT because the decomposition results are sensitive to the spectrum.

In this study, we proposed a PCCT material decomposition method that is based on the previously proposed RMCR [8]. The improvement lies in that we proposed a material decomposition method with one-step procedure, named as decomposition based on RMCR (DRMCR), while the previous work adopted a two-step procedure with RMCR only as a constraint. Specifically, RMCR is defined as the ratio of each individual channel image with the reconstructed broad-spectrum image using all available photons. With the help of the RMCR operation, beam-hardening artifacts and ring artifacts are suppressed. And also, the density variation of same materials is eliminated because it's energy independent. As a result, the RMCR image is sparser than linear attenuation coefficient image, with which the proposed DRMCR may lead to better noise reduction ability by strengthening prior information into iterative framework. On the other hand, since the decomposition is not affected by density, we estimated the material composition maps from the post-log data directly, rather than calculate the product of material densities and material composition maps. After that, given the material composition maps, the density of materials could be accurately recovered.

## 2    Materials and Methods

### A.   Ratio of Multi-Channel Representation

In PCCT imaging, the linear attenuation coefficient of the reconstructed image can be denoted as $\boldsymbol{\mu} \in \mathbb{R}^{S \times J}$, where $S$ is the number of energy channels, $J$ is the number of image pixels, and its element $\mu_{sj}$, $s = 1, 2, \cdots S$, $j = 1, 2, \cdots J$, can be expressed as the product of density and mass attenuation coefficient:

$$\mu_{sj} = \rho_j \sigma_{sj} \tag{1}$$

where $\rho_j$ is the density at the $j^{th}$ pixel, $\sigma_{sj}$ denotes the mass attenuation coefficient of the $j^{th}$ pixel at energy channel $s$. However, same materials may have different densities, such as the iodine solution with different concentrations, muscle and fat. To eliminate the effect of density, which is energy independent, on materials. We define the RMCR as:

$$r_{sj} = \frac{\mu_{sj}}{\bar{\mu}_j} = \frac{\rho_j \sigma_{sj}}{\rho_j \bar{\sigma}_j} = \frac{\sigma_{sj}}{\bar{\sigma}_j} \tag{2}$$

where $r_{sj}$ denotes the RMCR of the $j^{th}$ pixel at energy window $s$, $\bar{\mu}_j$ is the linear attenuation coefficient reconstructed by the broad-spectrum projection using all available photons.

To illustrate the benefits of RMCR, a physical phantom is scanned in a PCCT system. One representative reconstructed linear attenuation coefficient image and its corresponding RMCR image are shown in Fig. 1. Both the linear attenuation coefficient and RMCR images are energy dependent which used to describe the same object. The mean values in the red box region are marked in the images. In linear attenuation coefficient image, it can be observed that iodine solution with different concentrations are also different linear attenuation coefficient. In RMCR images, it does not include density information, resulting in the same RMCR value of Iodine solution with different concentrations. In linear attenuation coefficient images, the streak artifacts caused by beam-hardening are indicated by arrow "1", which are not observed in the RMCR images. Similarly, the ring artifacts caused by inconsistent detector response are indicated by arrow "2" in the linear attenuation coefficient images, which are eliminated or suppressed in RMCR images.

Given M basic materials for decomposition, the RMCR can be decomposed into two parts:

$$r_{sj} = \sum_{m=1}^{M} r_{sm} f_{mj} \tag{3}$$

where $r_{sm}$ is the RMCR of material $m$ at energy channel $s$, $f_{mj}$ denotes the material composition value at $j^{th}$ pixel of basis material $m$, $M$ is the total number of material types. Obviously, density is not embedded in material composition value. In this study, we first estimate the material composition maps. After that, given the material composition maps, the densities of materials can be recovered by the following equation:

$$\rho_{mj} = \frac{f_{mj} \times \bar{\mu}_j}{\bar{\sigma}_j} \tag{4}$$

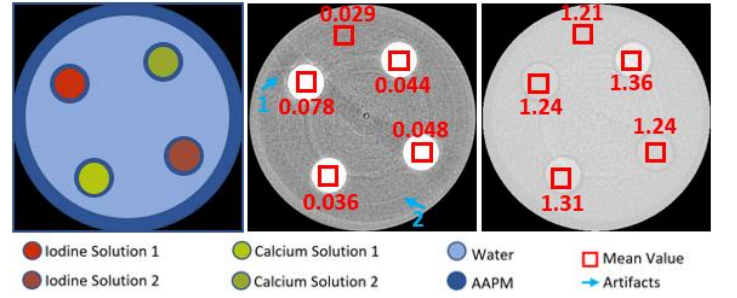

Fig. 1: The reconstructed images of a physical phantom. From left to right are abridged general view, a reconstructed linear attenuation coefficient image (energy bin information), and one RMCR image (energy bin information), respectively. The corresponding markers are list at the bottom of the figure. The display window for linear attenuation coefficient images is [0.02 0.035] $mm^{-1}$. The display window for RMCR image is [0 1.5].

### B.   PCCT Image Reconstruction Framework

Considering the basis material decomposition, the measured photons of PCCT at photon counting detector (PCD) can be estimated as follows:

$$I_{si} = b_{si} exp \left( -\sum_{m=1}^{M} \sum_{j=1}^{J} a_{ij} r_{sm} \bar{\mu}_j f_{mj} \right) \tag{5}$$

where $b_{si}$ and $I_{si}$ are incident photons and the expected value of measured photons in energy channel $s$ along the $i^{th}$ x-ray path, respectively. $a_{ij}$ denotes the length of the intersection between ray $i$ and pixel $j$.

After negative logarithmic operation, we obtain the post-log data:

$$\bar{y}_{si} = -\log \left( \frac{I_{si}}{b_{si}} \right) = \sum_{m=1}^{M} \sum_{j=1}^{J} a_{ij} r_{sm} \bar{\mu}_j f_{mj} \tag{6}$$

With the post-log data, the solution for spectral CT material decomposition can be solved by minimizing following objective function:

$$\underset{f}{argmin} \sum_{s=1}^{S} \sum_{i=1}^{I} \frac{I_{si}}{2} (\bar{y}_{si} - y_{si})^2 + \beta R(\boldsymbol{f}) \tag{7}$$

where $y_{si}$ is the measurement along the $i^{th}$ x-ray path at energy channel $s$. $R(\boldsymbol{f})$ is the regularization term on the material composition maps. $\beta$ represents a parameter to balance the strength between fidelity term and the regularization term.

In this paper, the regularization term in Eq. (7) is described by normally used TV, without loss of generality. Using the TV regularization, the material composition value $f_{mj}$ can be denoted in triple subscripts as

$$f_{mj} = f_{mpq}, \tag{8}$$

where $W$ and $H$ are respectively the width and height of the each two-dimensions material composition maps, and $J = W \times H$, where $j = (p-1) \times W + q, p = 1, 2, \cdots H, q = 1, 2, \cdots, W$. Then, the TV of material composition maps can be expressed as:

$$\text{TV}(\boldsymbol{f}) = \sum_{m=1}^{M} \|\nabla \boldsymbol{f_m}\|_1 \qquad (9)$$

where $\nabla \boldsymbol{f_m} = \left(\nabla f_{m1}, \nabla f_{m2}, \cdots, \nabla f_{mJ}\right)^T$ and

$$\nabla f_{mj} = \nabla f_{mpq}$$
$$= \sqrt{\left(f_{mpq} - f_{m(p-1)q}\right)^2 + \left(f_{mpq} - f_{mp(q-1)}\right)^2} \quad (10)$$

When the TV regularization is introduced to Eq. (7), the problem can be solved by minimizing the following objective function

$$\underset{\boldsymbol{f}}{\arg\min} \sum_{s=1}^{S} \sum_{i=1}^{I} \frac{I_{si}}{2} \|\bar{y}_{si} - y_{si}\|_2^2 + \beta \, \text{TV}(\boldsymbol{f}) \qquad (11)$$

### C. Optimization Via Alternating Minimization

We present an alternating minimization strategy to minimize the objective function Eq. (11). Introducing an auxiliary variable $\boldsymbol{v_m} = \nabla \boldsymbol{f_m}$, we can obtain the following unconstraint objective function:

$$\underset{\boldsymbol{f}}{\arg\min} \sum_{s=1}^{S} \sum_{i=1}^{I} \frac{I_{si}}{2} (\bar{y}_{si} - y_{si})^2 + \beta \sum_{m=1}^{M} \|\boldsymbol{v_m}\|_1$$
$$+ \alpha \sum_{m=1}^{M} \|\boldsymbol{v_m} - \nabla \boldsymbol{f_m}\|_2^2 \qquad (12)$$

Eq. (12) can be solved by three steps in an alternating manner until the stopping criterion is met.

**Step 1)** *Minimize the fidelity term.*

In this paper, we utilize the separable paraboloid surrogate method to update $f_{mj}$:

$$f_{mj}^{t+1} = f_{mj}^{t} - \frac{\left.\frac{\partial H}{\partial f_{mj}}\right|_{f=f^t}}{\left.\frac{\partial^2 H}{\partial f_{mj}^2}\right|_{f=f^t}} \qquad (13)$$

where

$$\frac{\partial H}{\partial f_{mj}} = \sum_{s=1}^{S} \sum_{i=1}^{I} r_{sm} a_{ij} \left( \sum_{m=1}^{M} \sum_{j=1}^{J} a_{ij} r_{sm} \bar{\mu}_j f_{mj} - y_{si} \right) \quad (14)$$

$$\frac{\partial^2 H}{\partial f_{mj}^2} = \sum_{s=1}^{S} \sum_{i=1}^{I} r_{sm} a_{ij} \left( \sum_{m=1}^{M} \sum_{j=1}^{J} a_{ij} r_{sm} \bar{\mu}_j \right) \qquad (15)$$

**Step 2)** *Update $v_{mj}$:*

$$v_{mj}^{t+1} = max\{\nabla f_{mj}^{t+1} - \beta/2\alpha, 0\} \qquad (16)$$

This process is a soft-threshold filtration of the TV with a threshold $\beta/2\alpha$.

**Step 3)** *Minimize $\alpha \sum_{m=1}^{M} \|\boldsymbol{v_m} - \nabla \boldsymbol{f_m}\|_2^2$.*

This problem can be solved by inverted $\nabla \boldsymbol{f_m}$. Because $\nabla \boldsymbol{f_m}$ is not uniquely invertible, we use a pseudo-inverse of $\nabla \boldsymbol{f_m}$ to solve this problem.

In summary, the workflow of the proposed method can be described as following.

## 3 Results

### A. Numerical Simulation Study

In the numerical simulation study, a two-dimensions (2D) numerical circle phantom was employed. Fig. 2 shown the circle phantom consists of water background with diameter of 200 mm, four circular inserts with a diameter of 35 mm. The materials were indexed, and the corresponding densities are listed in Table I. The bases materials are selected as water, calcium and iodine. Gaussian blurs are applied to iodine to simulate the permeation of iodine, where the Gaussian window is $20 \times 20$ and the standard deviation of Gaussian filter is 5. It worth noting that iodine solution consists of iodine and water and the density of the water remains constant.

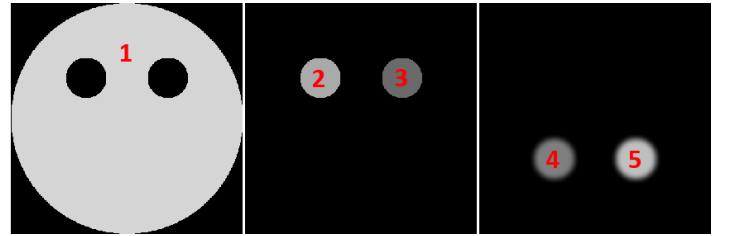

Fig. 2 Numerical circle phantom. From left to right are water, calcium and iodine, respectively. The display window for water and calcium is [0 1.2] g/$cm^3$. The display window for iodine is [0 0.02] g/$cm^3$. The materials were indexed, and the corresponding densities are listed in Table I.

TABLE I: LIST OF THE MATERIALS AND DENSITIES.

| Index | Material | Density (g/$cm^3$) |
|---|---|---|
| 1 | Water | 1 |
| 2 | Calcium | 0.8 |
| 3 | Calcium | 0.5 |
| 4 | Iodine | 0.01 |
| 5 | Iodine | 0.015 |

Four monochromatic images (30 keV, 40 keV, 50 keV, and 60 keV) are used to simulate the multi-energy projection data. A 90 kVp x-ray spectrum was assumed, which was generated from the SpectrumGUI software. Poisson noise is superimposed onto the measurement by assuming that there are 5000 photons emitted from each x-ray path. The emitted photons were distributed to each energy bin with the weights calculated from the x-ray spectrum. An equidistant fan-beam geometry is assumed for a PCCT scanner. 640 post-log projections are collected over a full scan range, the detector is composed of 512 detector elements, and each element size is 0.1 *mm*. The reconstructed images are $512 \times 512$ with in-plane resolution of $0.075\ mm \times 0.075\ m$. The linear attenuation coefficient images for each window reconstructed by simultaneous algebraic reconstruction technique (SART) method are shown in Fig. 3.

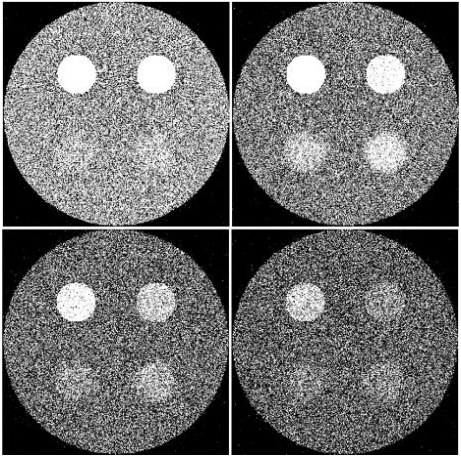

Fig. 3. Numerical circle phantom reconstructed linear attenuation coefficient images at noise scenarios by SART. The display window is [0 0.05] $mm^{-1}$.

For comparison study, one is the post-reconstruction method which first reconstruct the linear attenuation coefficient images in each individual energy channel by SART. Then decompose the reconstructed images into material fraction maps [9] (DSART). Another two-step method is also employed. This method first reconstructs the linear attenuation coefficient images by tensor dictionary learning. Then decompose the reconstructed images into material fraction maps [3] (DTDL). The other one is direct method which estimate material fraction maps from the energy-windowed measurements in a one-step procedure (DOS). This method is similar with our proposed method but without using RMCR.

Fig. 4 show the decomposition results. We can observe that the results of DSART method contains severe noise, especially the iodine is hard to identify. DTDL method can suppress noise tremendously. However, in region "3", part of calcium is decomposed to water using DSART and DTDL methods. In addition, a circle edge, which indicate by the arrow, introduces to iodine fraction maps, this is because the error in the reconstruction step cannot be compensated by decomposition step. DOS and DRMCR methods can avoid these errors because the material fraction maps are estimated from measurements directly. The noise also well suppressed thanks to TV constraints. The material fraction values in the indexed regions are listed in TABLE II. The mean values of DSART are significantly different than that in the noise free scenario because of the severe noise. It worth noting that selecting different density regions to calculate the decomposition matrix will lead different results in DSAR, DTDL and DOS methods. Our proposed method excludes the effect of density. Hence, the result is not affected by density. We will further discuss that in the discussion section.

The recovered density maps are shown in Fig. 5. The density can be recovered well with a good performance in noise reduction. The quantitative evaluation of the ROIs in Fig. 5 is listed in Table. III. The mean values are close to the reference.

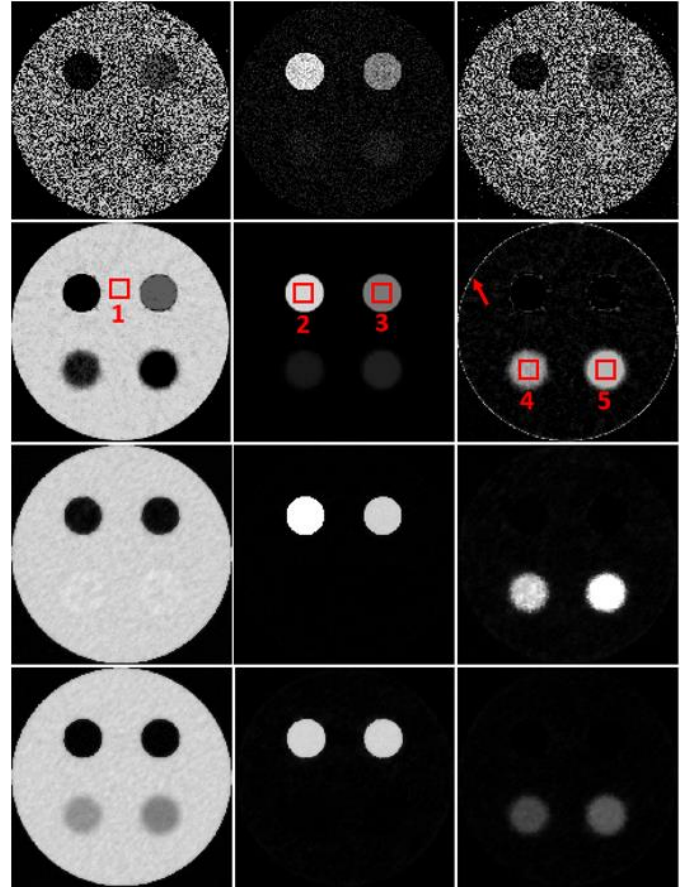

Fig. 4. From top to bottom are the material composition maps estimated by DSART, DTDL, DOS and DRMCR methods. From left to right are water, calcium, iodine and the corresponding color images. The display window is [0 1.2].

TABLE II: The Mean and STD Values of the ROIs in Figure 4.

|  | Region | 1 | 2 | 3 | 4 | 5 |
|---|---|---|---|---|---|---|
| Mean | DSART | 0.4522 | 0.9838 | 0.5749 | 0.5521 | 0.6361 |
|  | DTDL | 0.9656 | 0.9997 | 0.5681 | 0.7729 | 0.8530 |
|  | DOS | 0.9624 | 0.9868 | 0.6174 | 0.6553 | 0.9622 |
|  | DRMCR | 0.9710 | 0.9911 | 0.9912 | 0.3153 | 0.4157 |
| STD (e-4) | DSART | 1660.5 | 291.06 | 196.94 | 1856.7 | 1421.9 |
|  | DTDL | 16.648 | 0.2373 | 0.1853 | 23.469 | 0.1018 |
|  | DOS | 3.7734 | 1.4544 | 1.0466 | 19.287 | 18.297 |
|  | DRMCR | 11.973 | 0.7579 | 1.3520 | 3.6589 | 2.3214 |

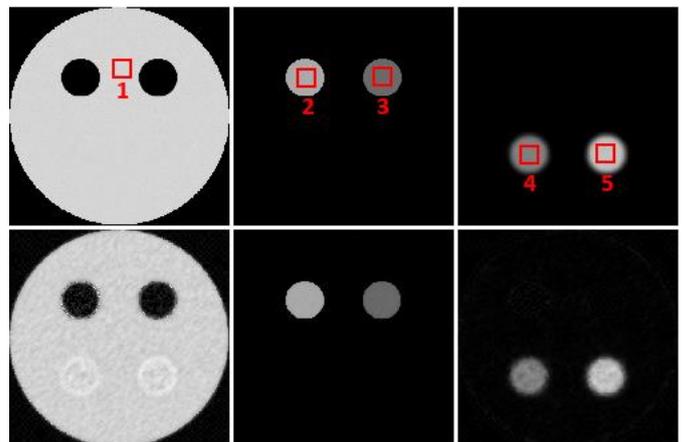

Fig. 5. Recovered density maps for numerical study. The display window for (a) and (b) is [0 1.2] $g/cm^3$. The display window for (c) is [0 0.02] $g/cm^3$

TABLE III: THE MEAN AND STD VALUES OF THE ROIs IN FIGURE 5.

|  | Region | 1 | 2 | 3 | 4 | 5 |
|---|---|---|---|---|---|---|
| MEAN | Reference | 1.0000 | 0.8000 | 0.5000 | 0.0100 | 0.0150 |
|  | DRMCR | 0.9710 | 0.7929 | 0.4956 | 0.0099 | 0.0151 |
| STD(e-5) | DRMCR | 120.00 | 4.8508 | 3.3799 | 0.0358 | 0.0306 |

## B. Physic Phantom Study

Fig. 6 shows a photography of a cylindrical physical phantom. The cylindrical phantom is a polymethylmethacrylate (PMMA) container filled with water, which contains two calcium solution rods with concentrations of 100 mg/ml and 50 mg/ml, and iodine solution rods with concentrations of 5 mg/ml and 15 mg/ml. The diameters of PMMA background and four circular inserts are 35 mm and 10 mm, respectively. The physical phantom was scanned in a PCD based micro-CT system in Institute of High Energy Physics (IHEP) of Chinese Academy of Sciences (CAS). For the micro-CT system, an X-ray source with 4 mm aluminum filter is operated at 90 kVp with a PCD detector. The dimension of the linear detector is 1027 with size of 0.2 mm. 2160 projections are collected per rotation. Six energy windows are used to collect projection with the set energy thresholds at 28keV, 34 keV, 40 keV, 47 keV, 56 keV, and 67 keV, respectively. The CT images of experimental materials for each window reconstructed by SART method are shown in Fig. 7. The reconstructed images are $512 \times 512$ with in-plane resolution of $0.25\ mm \times 0.25\ mm$.

The decomposition results for cylindrical physical phantom study are shown in Fig. 8. Since the basis material is selected as water with density of 1 g/$cm^3$, calcium with density of 0.1 g/$cm^3$ and iodine with density of 0.015 g/$cm^3$. The regions including basis materials, e. g. water, can be accuracy decomposed using all the four methods. In water fraction maps, the results of DSART method contain noise. DTDL method can suppress noise in some extent. DOS and DRMCR methods suppress noise significantly. However, DOS method includes beam-hardening artifacts and ring artifacts, which are indicated by arrow "1" and "2", respectively. And also, the PMMA and water have different fraction values because of the density variation. RMCR method could eliminate both the artifacts and density variation. In calcium fraction maps, because we select the calcium with density of 0.1 g/$cm^3$ instead of calcium solution as the basis material, both DSART and DTDL methods failed at calcium decomposition. And a ring artifact in the edge of the phantom are introduced. DOS and DRMCR method can avoid this error decomposition. In iodine fraction maps, while DSART and DTDL methods decompose calcium into iodine, DOS and DRMCR obtain accuracy decomposition. The mean value and STD are listed in Table. IV. The quantitative evaluation is consistent with the visual inspection.

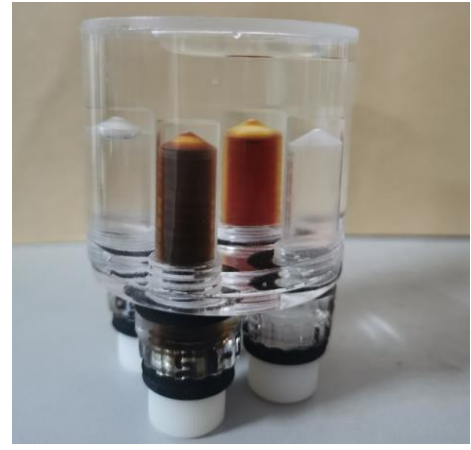

Fig. 6. Photography of physical phantom.

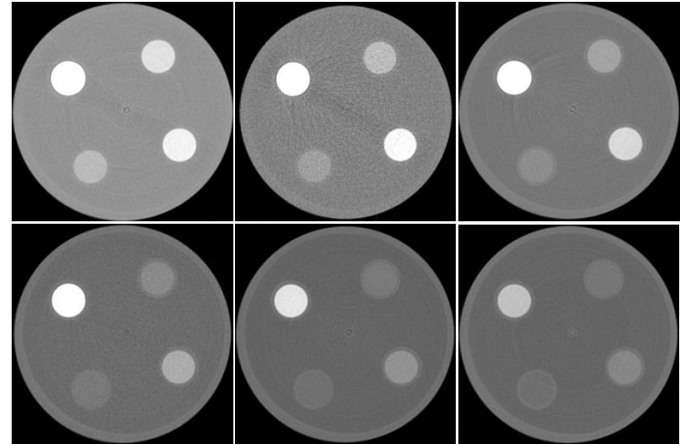

Fig. 7. Physical phantom reconstructed LAC images by SART. The display window is [0 0.035] $mm^{-1}$.

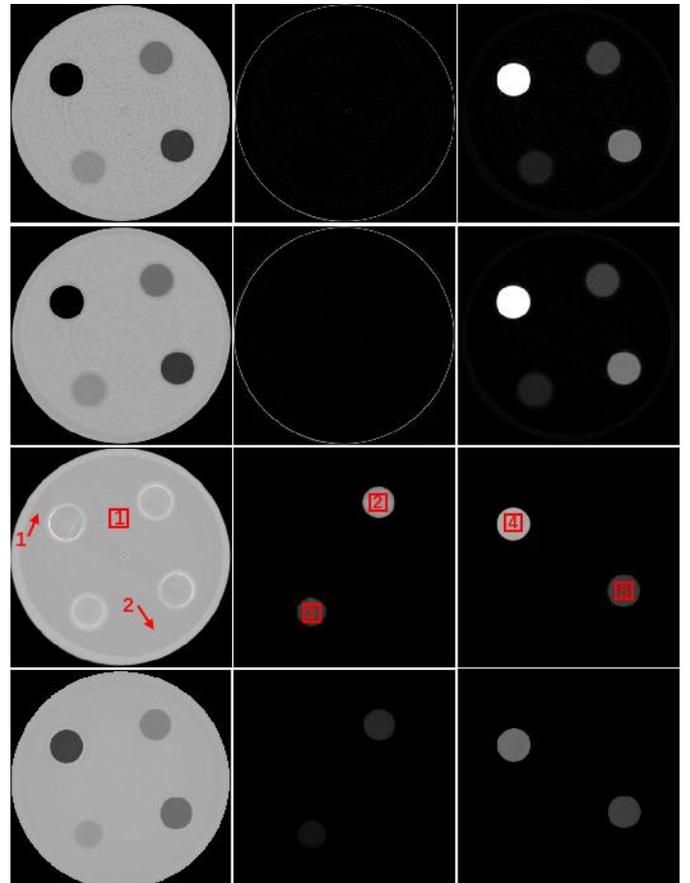

Fig. 8. From top to bottom are the images estimated by DSART, DTDL, DOS and DRMCR, respectively. From left to right are the fraction maps of water, calcium, and iodine, respectively. The display window is [0 1.5].

TABLE IV: THE MEAN AND STD VALUES OF THE ROIs IN FIGURE 8.

|  | Region | 1 | 2 | 3 | 4 | 5 |
|---|---|---|---|---|---|---|
| Mean | DSART | 0.9767 | 0.0000 | 0.0000 | 1.5049 | 0.6792 |
|  | DTDL | 0.9930 | 0.0000 | 0.0000 | 1.5047 | 0.6798 |
|  | DOS | 1.0039 | 0.9817 | 0.3755 | 0.9817 | 0.3517 |
|  | DRMCR | 1.0039 | 0.6170 | 0.1120 | 0.6170 | 0.3648 |
| STD | DSART | 0.0231 | 0.0000 | 0.0000 | 0.0106 | 0.0167 |
|  | DTDL | 0.0130 | 0.0000 | 0.0000 | 0.0086 | 0.0077 |
|  | DOS | 0.0072 | 0.0786 | 0.0355 | 0.0228 | 0.0130 |
|  | DRMCR | 0.0037 | 0.0193 | 0.0084 | 0.0088 | 0.0084 |

TABLE V: THE MEAN VALUES OF ROIs IN FIGURE 9.

| Region | 1 | 2 | 3 | 4 | 5 |
|---|---|---|---|---|---|
| Reference | 1.0000 | 1.6000 | 1.0000 | 0.0000 | 0.0000 |
| DOS | 1.0000 | 1.6000 | 1.0000 | 1.0004 | 1.5006 |
| DRMCR | 1.0000 | 1.0000 | 1.0000 | 0.3192 | 0.4141 |

## 4. Discussions

The decomposition results are usually affected by the selection of basis materials. In the simulation study, the densities of calcium in region "2" and "3" are 0.8 g/$cm^3$ and 0.5 g/$cm^3$, respectively. The concentration of iodine solution in region "4" and "5" are 0.01 g/$cm^3$ and 0.015 g/$cm^3$, respectively. In section 3.A, we select calcium with density of 0.8 g/$cm^3$, water, and iodine with density of 0.015 g/$cm^3$ as basis materials. Here, we select calcium with density of 0.5 g/$cm^3$, water, and iodine with density of 0.01 g/$cm^3$ as basis materials. The decomposition results are shown in Fig. 9, both of DSART and DOS methods produce different material composition maps because of the difference of density of basis materials. Our proposed method excludes the effect of density. Hence, the results are same as section 3.A. TABLE V shows the mean values in the ROIs. In the selected basis material regions "1" and "3", all the three methods could obtain same decomposition value. In region "2", the decomposition values of DSART and DOS methods become 1.6, because the density in this region is 1.6 times than region "3". In region "4" and "5", the composition values of DSART are close to zeros. For DOS method, the composition values in region "4" is close to 1. In region "5", the composition values are close to 1.5, because the concentration in this region is 1.5 times than region "4". Our proposed DRMCR method obtain an exactly similar value compare with section 3.A, thanks to the density excluded from the reconstruction procedure. Hence, our proposed method was more robustness about the selection of basic materials.

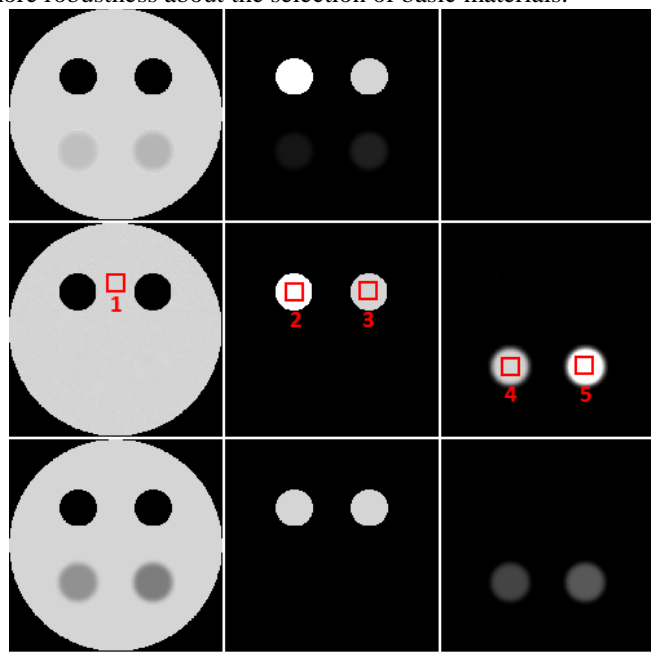

Fig. 9. From top to bottom are the material composition maps estimated by DSART, DOS and DRMCR method. From left to right are water, calcium, iodine and the corresponding color images. The display window is [0 1.2].

## 5. Conclusions

In this paper, we present a RMCR based spectral CT material decomposition method with one-step procedure. The presented method estimates the material composition maps from post-log data directly, followed by densities recovery. With the help of RMCR operation, the beam-hardening artifacts and ring artifacts are suppressed. In addition, the noise could be reduced by strengthening prior information in iterative framework. Both numerical and physic phantom results demonstrate that the presented method can achieve attractive decomposition accuracy with high computational efficiency.

## Acknowledgments

This work was partially supported by National Nature Science Foundation of China (NSFC) (No. 62071375 and No.11975250). Dr. Liang was partially supported by the NIH/NCI grant #CA206171.

# Joint activity and attenuation reconstruction for the alignment of hardware attenuation

Ahmadreza Rezaei, Donatienne Van Weehaeghe, Georg Schramm, Johan Nuyts, and Koen Van Laere

KU Leuven - University of Leuven, Department of Imaging and Pathology, Nuclear Medicine & Molecular imaging; Medical Imaging Research Center (MIRC), B-3000, Leuven, Belgium.

**Abstract**

A method was implemented to increase the comfort during PET/MR imaging for patients diagnosed with amyotrophic lateral sclerosis (ALS). The system head holder and the system software require appropriate positioning of the head, and for many of these patients, this required head pose is not possible. Some hardware was developed to enable a more flexible positioning, and new software was developed to ensure that accurate quantitative images are obtained independent of the head position.

## 1 Introduction

Amyotrophic Lateral Sclerosis (ALS) is a rare neuro-degenerative disease in which patients experience among other symptoms, increasing weakness and difficulty in breathing. In a recent study conducted at UZ Leuven [1] patients underwent dynamic Positron Emission Tomography / Magnetic Resonance (PET/MR) imaging for which the typical patient positioning in the scanner was either not possible or very uncomfortable. To increase patient comfort, additional hardware (a wedge shaped apparatus) was designed which the MR coils were placed on. This setup allowed patient positioning and data acquisition in an elevated head position. However, since the PET/MR system assumes that the MR coils are anchored to a fixed position on the bed, additional processing was required for accurate quantitative tracer reconstruction.

For quantitative reconstructions of tracer distributions in PET imaging, an accurate correction for photon attenuation of the emission data is critical. In hybrid PET/MR scanners the problem of attenuation correction can at times still be challenging. It has been shown that joint activity and attenuation reconstructions from TOF-PET can provide accurate tracer distribution reconstructions, comparable to the state of the art [2]. The study, demonstrates that accurate attenuation estimation (and hence correction) can be obtained for regions with in the tracer activity support. Joint reconstruction of activity and attenuation has been used previously for the completion of truncated attenuation images [3] or for the reconstruction of flexible attenuating hardware [4] (e.g. MR headphones for ear protection).

In this work, we describe our pipeline for processing of patient scans with the MR head coils being (un)intentionally improperly positioned according to the scanner. Clearly as there are no ground truth reconstructions to compare the results against, examples of the alignment will be shown for brain and neck/thorax patient datasets.

## 2 Materials and Methods

### 2.1 Data Acquisition and Processing

Brain and neck/thorax $^{18}$F-FDG patient scans were acquired on the GE SIGNA TOF-PET/MR scanner [5] at UZ Leuven [1]. The local institutional review board approved this study and informed consent was obtained from all subjects. The emission data were acquired in 1 and 2 bed positions for the brain and the neck/thorax datasets, respectively. The data were collected in 5D sinograms consisting of 357 radial bins of 2.016 mm width, 224 azimuth angles over 180 degrees, a combined 1981 planes of 2.658 mm width for sinogram planes and 27 TOF-bins of 169 ps width. Duetto v02.03 provided by GE Healthcare was used to process the raw data and to generate the expected scatter and randoms contribution of the emission measurements independently for each bed position.

### 2.2 Reconstruction

Using in-house reconstruction software data collected from multi-bed scans were reconstructed simultaneously in a single volume. Our in-house projector works on a bed-by-bed basis, however the (back/) projections are computed (onto) from a single whole-body volume. Activity and attenuation images were reconstructed in a 210 × 210 pixel grid of 3.125 mm width transaxially and up to 154 planes of 2.78 mm width axially. The TOF resolution of the scanner was modelled as a Gaussian having a 400 ps full width at half maximum (accounting for the TOF-binning of the data).

The activity reconstruction was initialized by applying 1 iteration and 32 subsets of the OSEM algorithm, taking into account all attenuating components except the wedge and coil attenuations. Using the patient attenuation in combination with the OSEM activity reconstruction a sinogram mask of the activity support was generated which was used to encourage zero attenuation outside the activity support [6], [7].

The attenuation reconstruction was initialized with an image of all zeros. Since all but the wedge and coil attenuation components have been accounted for during OSEM (and will be accounted for during OSAA) updates, the attenuation reconstruction will contain these hardware attenuating volumes. To further control any attenuation buildup (and to minimize changes to the scale problem of joint estimation) an intensity prior favoring zero attenuation was active on the entire attenuation volume as well as a quadratic smoothing prior. Activity and attenuation reconstructions (with a 1:2 update activity to attenuation ratio) were generated using 1 iteration of 32 subsets of OSAA.

## 2.3 Registration

A template of the head/body coil was subsequently rigidly aligned to the reconstructed attenuation image. The registration was done using in-house developed tools, initially using normalized mutual information as the registration cost function, and then using normalized cross-correlation to refine the 6 estimated degrees of freedom of the rigid transform.

## 3 Results

Figure 1 and Figure 2 show the results of our processing pipeline for a brain and neck/thorax dataset. Since the patient and bed attenuation were accounted for during reconstruction, the OSAA attenuation reconstructions only contain the missing coil attenuation structures. We see in the reconstruction that although parts of the coil attenuation have been "smeared", the high attenuating components and high spatial frequency regions are well resolved. These are the "control-points" which the subsequent template registration operate on.

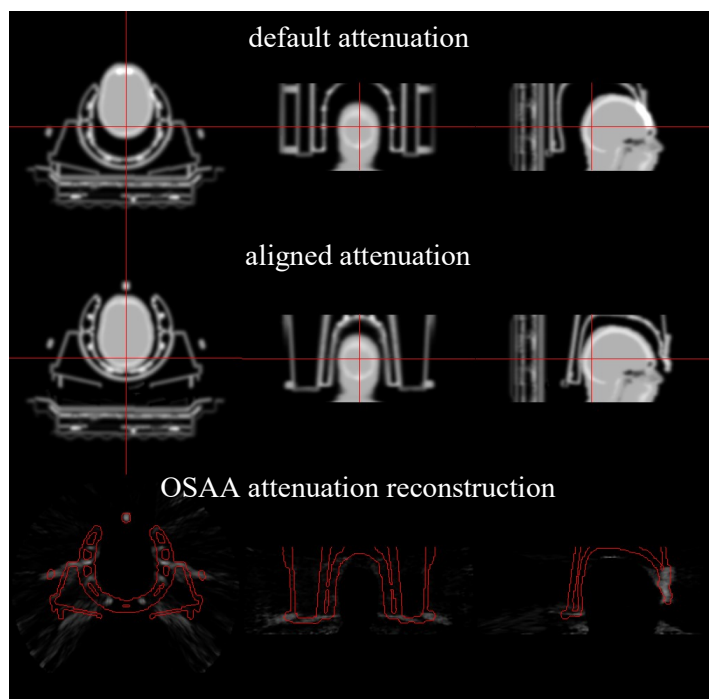

*Figure 1 Results of a brain scan dataset; top: original attenuation image, center: aligned attenuation image, and bottom: OSAA attenuation reconstruction. The red contour is showing the position of the aligned coil template.*

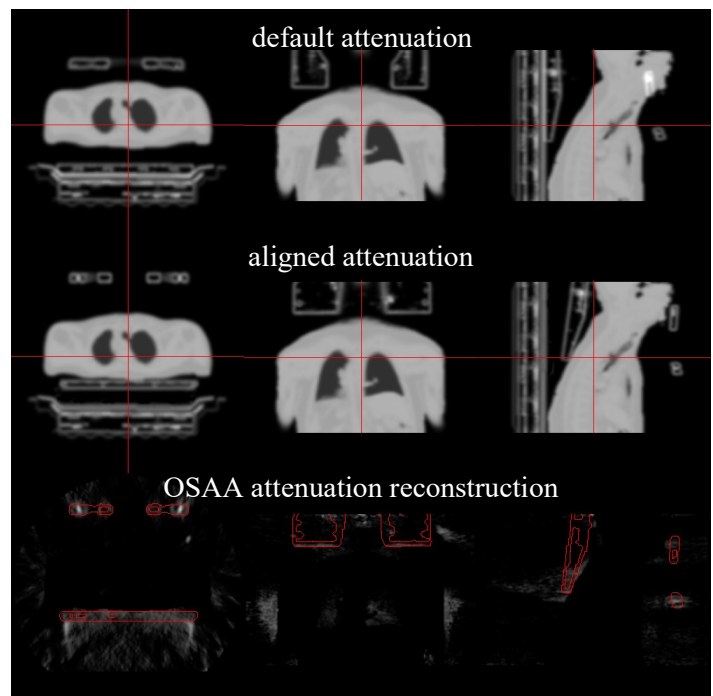

*Figure 2 Same results as in Figure 1 for a two bedposition neck and chest scan dataset.*

## 4 Discussion

PET/MR scans were collected for an ALS research study at our institution, where depending on the stage of the disease the typical patient positioning was not possible for all patients. These patients were scanned with MR coils placed on an in-house developed (low attenuating) wedge sitting on the scanner bed, which made the scan possible for some and slightly more convenient for others. Since the head and neck MR coils are assumed to be anchored and fixed to the bed by the scanner, additional alignment of the coils was required for this study.

We had access to the PET/MR coil templates used in this study, and hence we opted for aligning the coils using the TOF-PET data as opposed to reconstructing the additional hardware attenuation as proposed in [4]. As a consequence no additional scale correction strategies were required during OSAA reconstructions, since quantitatively accurate attenuation images are not necessary for the alignment. Furthermore, we believe that this approach would allow for a slightly more accurate scatter simulation, which requires further quantification.

## 5 Conclusion

In this work we demonstrate how we have utilized the maximum likelihood activity and attenuation reconstruction algorithm for the alignment of harware atterution. The processing pipeline was created for data collected from patients diagnosed with ALS for which normal patient positioning was not possible. We find that although exact reconstruction of external attenuating hardware outside the activity support is challenging, in cases where a template of the hardware attenuation is available, reliable alignment of the attenuating medium is possible using the TOF-PET emission data.

# Solid angle non-uniformity correction in frequency domain for X-ray diffraction computed tomography

Kaichao Liang[1,2], Li Zhang[1,2], and Yuxiang Xing* [1,2]

[1]Department of Engineering Physics, Tsinghua University, Beijing, China.
[2]Key Laboratory of Particle & Radiation Imaging (Tsinghua University), Ministry of Education, China.
*Corresponding author: Yuxiang Xing, E-mail: xingyx@mail.tsinghua.edu.cn

**Abstract** X-ray diffraction (XRD) has shown its great performance in distinguishing molecular level structures of different materials which is hopefully to be applied in clinical diagnostics. In the literatures, pencil-beam X-ray diffraction computed tomography (XRDCT) has been studied to improve the spatial-resolution of point-wise XRD scan. Line-integral XRD projections are reconstructed analytically similar to transmission CT. However, in XRDCT imaging, the XRD projection is weighted by the scattering solid angle. For a fixed size detector pixel, the scattering solid angle is non-uniform along X-ray transmission paths which causes artifacts after direct FDK reconstruction. The non-uniform solid angle can be modelled by squared-distance weights mathematically. In this work, we proposed a correction method for the squared-distance weights in the frequency domain of XRD projection data. For validation, we simulated the XRDCT scan weighted by non-uniform solid angle analytically based on a manually designed digital phantom, the proposed method corrected the artifacts caused by solid angle non-uniformity well.

## 1 Introduction

XRD measurs the coherent scattering signal of materials which is sensitive to molecular level structures. XRD has been a powful component analysing tool in crystallography, material science and security screen. Many studies have also shown that XRD has the ability distinguishing different types of biological tissues. For instance, there are significant differencies in the XRD patterns of fat and glandular tisssue, which indicates that the XRD inspection can be a useful method in breast cancer diagnosis [1-3]. Oliveira et. al. used a commercial powder diffractor for nueplasias classification [3]. Moss et. al. adopted a pixelated energe-dispersive dectecor for breast sample XRD inspection, the XRD results were in correspondance with histopalogical classification [2]. However, point-wise XRD inspection has shown a poor spatial resolution along the X-ray transmission direction. With the hole size of collimator on the detector side being 0.5 mm, the spatial resolution is worse than10 mm caused by the small diffraction angle [4], which leads to significant partial volume effects for biological tissues. Thus point-wise XRD are mainly suitable for thin sample inspection.

XRDCT have been proposed over the decades for crystallography as well as diagnostics [5-7]. Generally, an XRDCT scan follows the first generation CT scan mode. For each measurement in the Radon space, XRDCT collects photons scattered along the X-ray transmission path with collimator on the detector side removed. With analytical CT reconstruction, XRDCT provided XRD pattern information accompanied by structural information of improved spatial resolution, which is very valuable for non-invasive diagnosis. However, one of the differences between transmission CT and XRDCT is that the XRDCT projection is weighted by the scattering solid angle [7]. For a fixed size detector pixel, the solid angle is non-uniform along X-ray transmission paths. In most studies, the sizes of inspected samples were assumed far smaller compared to the center-to-detector distance, thus the solid angle can be approximated to a constant. However, larger center-to-detector distance weakens the coherent scattering signal greatly. In applications for luggage check or cancer ROI imaging in vivo, the center-to-detector distance may be reduced to about only twice the ROI radius to increase the intensity of scattering signals. When the center-to-detector distance is reduced to the magnitude of object size, the influence of non-uniform solid angle is not ignorable. The non-uniformity of solid angle can be expressed by a squared-distance weighting function mathematically. In this work, we address the influence of non-uniform solid angle in the frequency domain of the XRD projection data, and further proposed a mathematical correction method in the frequency domain, the artifacts caused by solid angle non-uniformity were removed in our results of simulation experiments.

## 2 Methods

### 2.1 XRDCT physical model

The linear differential coherent scattering coefficient $\dfrac{\partial \mu_{\mathrm{coh}}(E,\theta_{\mathrm{S}})}{\partial \theta_{\mathrm{S}}}$ of an amorphous material is given by:

$$\frac{\partial \mu_{\mathrm{coh}}(E,\theta_{\mathrm{S}})}{\partial \theta_{\mathrm{S}}} = \frac{r_{\mathrm{e}}^{2}\left(1+\cos^{2}\theta_{\mathrm{S}}\right)}{2}\frac{N_{\mathrm{A}}\rho}{M}F^{2}_{\mathrm{IAM}}(q)m(q) \quad (1)$$

Here, $E$ denotes the incidence X-ray photon energy, $\theta_{\mathrm{S}}$ the scattering angle which is generally in the range from three to eight degree, $r_{\mathrm{e}}$ the classical radius of electron, $N_{\mathrm{A}}$ the Avogadro constant, $\rho$ the material density, $M$ the mean relative molecular mass of material, $F^{2}_{\mathrm{IAM}}$ the molecular form factor function determined by the atomic composition with independent atomic model (IAM), $m$ the molecular interference function, $q$ the momentum transfer which is a function of both $E$ and $\theta_{\mathrm{s}}$ given by:

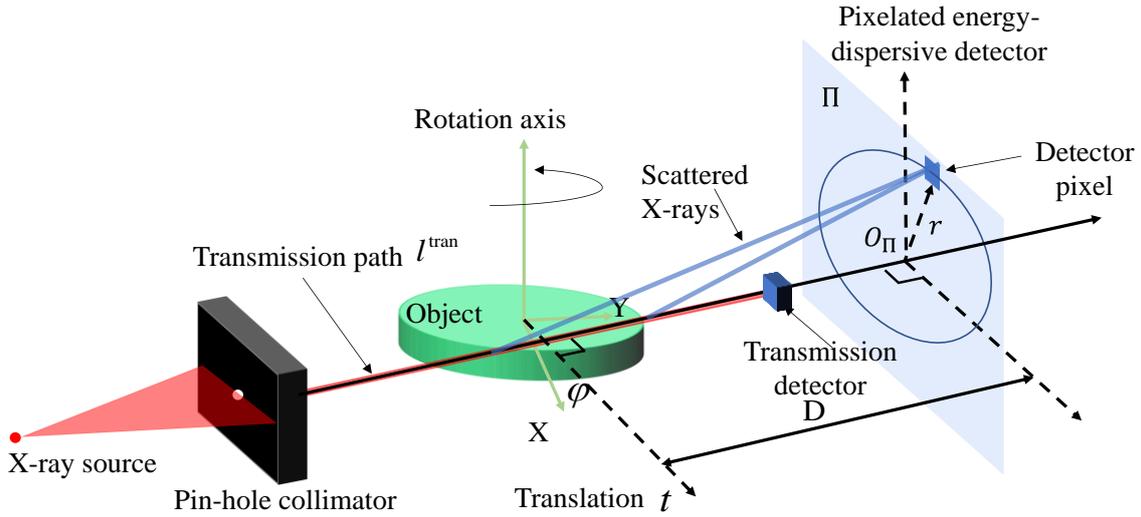

Figure 1: Perspective view of a pencil-beam XRDCT system with pixelated energy-dispersive detector.

$$q = \frac{E \sin (\theta_s / 2)}{hc} \qquad (2)$$

where $h$ is the Planck constant and $c$ the speed of light. Here, we define the material specific factor in $\frac{\partial \mu_{\text{coh}} (E, \theta_s)}{\partial \theta_s}$ as relative scattering intensity: $f(q) \triangleq \frac{\rho}{M} F_{\text{IAM}}^2 (q) m(q)$. The curve $f(q)$ forms the specific XRD pattern of a material. For a two-dimensional slice of a non-uniform object, the $f(q)$ is also a function of spatial location $(x, y)$. For convenience, we denote $f(x, y, q) \triangleq \frac{\rho}{M} F_{\text{IAM}}^2 m(x, y, q)$ to address its spatial distribution.

For a pencil-beam energy-dispersive XRDCT system, a pin-hole collimator installed in front of an X-ray source. The incident pencil-beam X-ray photons are attenuated and scattered along transmission paths. Photons scattered by materials along X-ray transmission paths are collected by a pixelated energy-dispersive photon-counting detector. The transmission signals are detected by an additional transmission detector placed before the energy-dispersive pixelated detector for attenuation correction. The XRDCT scanning process is the same as the first-generation parallel-beam CT scan. The X-ray source, collimator and detector of XRDCT system rotate around the iso-center together. At each view, the X-ray source, collimator and detector translate perpendicular to the X-ray transmission path. Thus a full set of XRDCT data is acquired. Each XRD transmission path is determined uniquely by the rotation angle $\varphi$ and translation position $t$. We denote the transmission path as $l^{\text{tran}}$, the plane that pixelated detector for scattered photons as $\Pi$ that is perpendicular to $l^{\text{tran}}$, their intersection point as $O_\Pi$, the distance from iso-center to $\Pi$ as D. An illustration of a pencil-beam XRDCT system is in Fig. 1.

The diffraction signal reaches a unit area in $\Pi$ at a distance $r$ from $O_\Pi$ can be modelled as:

$$I^{\text{XRD}} (t, \varphi, E, r) = \iint I^{\text{T}} (t, \varphi, E) \frac{r_e^2 \left(1 + \cos^2 \theta_s\right) N_A}{2}$$
$$\bullet \Omega_r (x, y, \varphi) f (x, y, q) \delta (x \cos \varphi + y \sin \varphi - t) \mathrm{d}x \mathrm{d}y \qquad (3)$$

Here, $I^{\text{T}} (t, \varphi, E) = I^0 (E) T (t, \varphi, E)$ measures the influence of incident spectrum $I^0 (E)$ and diffraction signal attenuation ratio $T (t, \varphi, E)$. $I^{\text{T}} (t, \varphi, E)$ is recorded by the transmission detector. The scattering angle $\theta_s = \arctan(\frac{r}{D - y \cos \varphi + x \sin \varphi})$ is at position $x, y$ is computed from system geometry. $\Omega_r (x, y, \varphi) = \frac{\cos \theta_s}{r^2 + (D - y \cos \varphi + x \sin \varphi)^2}$ accounts for the non-uniform scattering solid angle at position $x, y$ toward a unit area in $\Pi$. In small diffraction angle situations, $\Omega_r (x, y, \varphi) \approx \frac{1}{(D - y \cos \varphi + x \sin \varphi)^2}$.

We define the diffraction signal without incident spectrum weighting and attenuation as intrinsic XRD projection denoted by $s^{\text{XRD}} (t, \varphi, E, r)$. With the small angle approximation, the momentum transfer $q = \frac{E \sin (\theta_s / 2)}{hc} \approx \frac{Er}{2hc (D - y \cos \varphi + x \sin \varphi)}$, $s^{\text{XRD}} (t, \varphi, E, r)$ can be written as:

$$s^{\text{XRD}} (t, \varphi, E, r) \triangleq \frac{I^{\text{XRD}}}{AI^{\text{T}} (t, \varphi, E)} = \iint \frac{D^2}{(D - y \cos \varphi + x \sin \varphi)^2}$$
$$\bullet f \left(x, y, \frac{Er}{2hc (D - y \cos \varphi + x \sin \varphi)}\right) \delta (x \cos \varphi + y \sin \varphi - t) \mathrm{d}x \mathrm{d}y$$

$$(4)$$

Here $A \equiv r_e^2 N_A D^2$ is a constant. From Eq. 4, the intrinsic XRD projection $s^{\mathrm{XRD}}(t,\varphi,E,r)$ is the function of $Er$ rather than the function of independent variables $E$ and $r$. Thus, we introduce a new variable $k \triangleq Er$, the $s^{\mathrm{XRD}}(t,\varphi,E,r)$ can be averaged according to $k$ weighted by transmission signal intensity for data dimension reduction and noise reduction:

$$g(t,\varphi,k) \triangleq \bar{s}^{\mathrm{XRD}}(t,\varphi,E,r)\big|_{k=Er,w=I^{\mathrm{T}}(t,\varphi,E)}$$

$$= \frac{\int_{E_{\min}}^{E_{\max}} \int_{R_{\min}}^{R_{\max}} s^{\mathrm{XRD}}(t,\varphi,E,r) I^{\mathrm{T}}(t,\varphi,E)\delta(k-Er)r\,\mathrm{d}r\mathrm{d}E}{\int_{E_{\min}}^{E_{\max}} \int_{R_{\min}}^{R_{\max}} I^{\mathrm{T}}(t,\varphi,E)\delta(k-Er)r\,\mathrm{d}r\mathrm{d}E} \quad (5)$$

Here, $g(t,\varphi,k)$ denotes the mean XRD projection from $s^{\mathrm{XRD}}(t,\varphi,E,r)$. Combining Eq. 4 and 5, $g(t,\varphi,k)$ can be acquired from raw XRD data $I^{\mathrm{XRD}}(t,\varphi,E,r)$ directly by:

$$g(t,\varphi,k) = \frac{\int_{E_{\min}}^{E_{\max}} \int_{R_{\min}}^{R_{\max}} I^{\mathrm{XRD}}(t,\varphi,E,r)\delta(k-Er)r\,\mathrm{d}r\mathrm{d}E}{A\int_{E_{\min}}^{E_{\max}} \int_{R_{\min}}^{R_{\max}} I^{\mathrm{T}}(t,\varphi,E)\delta(k-Er)r\,\mathrm{d}r\mathrm{d}E} \quad (6)$$

Eq. 6 forms the pre-processing method of raw XRD data with attenuation correction and signal average. And according to Eq. 4, the final XRDCT model is:

$$g(t,\varphi,k) = \iint \frac{D^2 f(x,y,q)}{(D-y\cos\varphi+x\sin\varphi)^2}\delta(x\cos\varphi+y\sin\varphi-t)\mathrm{d}x\mathrm{d}y$$

$$(7)$$

Here, $q = \dfrac{k}{2hc(D-y\cos\varphi+x\sin\varphi)}$. The XRDCT reconstruction is to reconstruct $f(x,y,q)$ from XRD projection $g(t,\varphi,k)$. As displayed in Eq. 7, the XRDCT reconstruction is similar to cone-beam transmission CT. Previous works adopt either FBP reconstruction assuming $q \approx \dfrac{k}{2hcD}$ [5, 6] or FDK reconstruction with Taylor approximation [7]. The reconstruction method is not discussed in this work. Here we only focus on the non-uniform solid angle weights $\dfrac{D^2}{(D-y\cos\varphi+x\sin\varphi)^2}$ which is different from transmission CT reconstruction.

### 2.2 Solid angle non-uniformity correction in frequency domain

To solve the solid angle non-uniformity weighing problem, we consider the 1D Fourier transform of $g(t,\varphi,k)$ at dimension $t$:

$$G(\omega,\varphi,k) \triangleq \mathcal{F}\big(g(t,\varphi,k)\big)$$

$$= \iint \frac{D^2 f(x,y,q)}{(D-y\cos\varphi+x\sin\varphi)^2} \exp^{-j\omega(x\cos\varphi+y\sin\varphi)} \mathrm{d}x\mathrm{d}y \quad (8)$$

Taking partial derivative of $G(\omega,\varphi,k)$ with respect to $\varphi$ gives:

$$\frac{\partial G(\omega,\varphi,k)}{\partial \varphi} = j\omega(x\sin\varphi-y\cos\varphi)G-$$

$$2\iint f(x,y,q)\frac{D^2(x\cos\varphi+y\sin\varphi)}{(D-y\cos\varphi+x\sin\varphi)^3}\mathrm{e}^{-j\omega(x\cos\varphi+y\sin\varphi)}\mathrm{d}x\mathrm{d}y \quad (9)$$

$$-\iint \frac{k\partial f(x,y,q)}{\partial k}\frac{D^2(x\cos\varphi+y\sin\varphi)}{(D-y\cos\varphi+x\sin\varphi)^3}\mathrm{e}^{-j\omega(x\cos\varphi+y\sin\varphi)}\mathrm{d}x\mathrm{d}y$$

We further define the Fourier transform of XRD projection with first-order distance weights in denominator as $G^1(\omega,\varphi,k)$:

$$G^1(\omega,\varphi,k) \triangleq \iint \frac{Df(x,y,q)}{(D-y\cos\varphi+x\sin\varphi)} \exp^{-j\omega(x\cos\varphi+y\sin\varphi)} \mathrm{d}x\mathrm{d}y$$

$$(10)$$

The first term in Eq. 9 becomes $j\omega D(G^1 \cdot G)$. The second and third terms are actually the residual terms between $G^1$ and $G$. We approximate the denominators of the second and third term in Eq.9 as:

$$\frac{1}{(D-y\cos\varphi+x\sin\varphi)^3} \approx \frac{1}{D(D-y\cos\varphi+x\sin\varphi)^2} \quad (11)$$

With the approximation, Eq. 9 can be simplified as:

$$G^1 = G + \frac{\partial G}{j\omega D\partial\varphi} + 2\frac{\partial G}{\omega D^2 \partial\omega} + \frac{k\partial^2 G}{\omega D^2 \partial\omega\partial k} \quad (12)$$

With Eq. 12, the second order distance weights in denominator is reduced to first order.

Further define the Fourier transform of XRD projection without non-uniform solid angle as $P(\omega,\varphi,k)$:

$$P(\omega,\varphi,k) \triangleq \iint f(x,y,q)\exp^{-j\omega(x\cos\varphi+y\sin\varphi)} \mathrm{d}x\mathrm{d}y \quad (13)$$

Similar to Eq. 12, $P(\omega,\varphi,k)$ can be acquired with:

$$P = G^1 + \frac{\partial G^1}{j\omega D\partial\varphi} + \frac{\partial G^1}{\omega D^2 \partial\omega} + \frac{k\partial^2 G^1}{\omega D^2 \partial\omega\partial k} \quad (14)$$

The ideal XRD projection without non-uniform solid angle weights can be acquired through I-Fourier transform on $P(\omega,\varphi,k)$.

The non-uniform solid angle correction process can be concluded as:

(1) Calculate the Fourier transform of $g(t,\varphi,k)$ to acquire $G(\omega,\varphi,k)$.

(2) Estimate the $G^1(\omega,\varphi,k)$ with Eq. 12.

(3) Estimate the $P(\omega,\varphi,k)$ with Eq. 14.

### 3 Experiments and results

In our simulation experiments, we adopted a 90 mm diameter manually designed phantom containing water, glandular, fat and bone as shown in Fig. 2 (a). The energy-dispersive detector was set of 100 mm×100 mm with detector bin size 1 mm and the energy bin was set 1keV. The detector was placed at 150 mm away from the iso-center. We compared direct FDK reconstruction of the original XRD projection (referred as FDK), FDK reconstruction of the XRD projection corrected with the proposed method (referred as Corrected-FDK). We also simulated FDK reconstruction of XRD projection without solid angle weights for reference (referred as Reference-FDK. As the XRDCT physical model does not satisfy data completeness condition, cone-beam artifacts are not avoidable at XRD pattern peak positions. In this work, we do not discuss the cone-beam artifacts, and Reference-FDK is treaded as the up-limit performance for the solid angle correction step. The cone-beam artifacts and discrete artifacts in Corrected-FDK are similar to those in Reference-FDK.

### A.  Simulation study of ideal configuration

To evaluate the accuracy of the proposed correction method with the influence of discrete error and noise avoided, we set simulation sampling rate during scan high enough and the projection noise free. In this ideal configuration study, there were 320 translation steps under each view with translation step interval 0.3375 mm, and there were 360 views uniformly distributed over $2\pi$. The reconstruction image was of $256\times256$ grids with pixel size 0.36 mm. The reconstruction results at fat XRD pattern peak $q=1.12$ nm$^{-1}$ were displayed in Fig. 2.

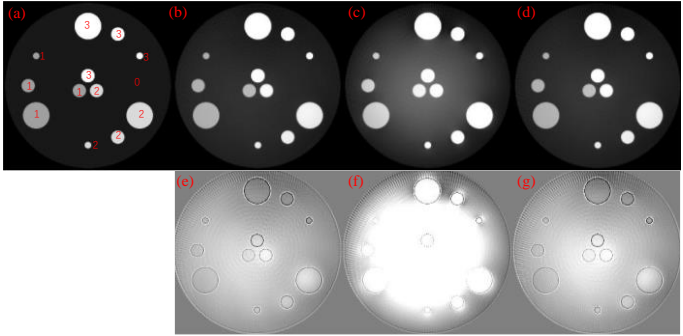

Figure 2: FDK reconstruction results of different XRD projections under ideal configuration. Martial 0: water, material 1: glandular, material 2: bone. (a) Ground-truth, (b) Reference-FDK, (c) FDK, (d) Corrected-FDK, display window: [0.5,4.5] mm$^{-1}$; (e) is the difference between (a) and (b), (f) is the difference between (a) and (c), (g) is the difference between (a) and (d), display window: [-0.5, 0.5] mm$^{-1}$.

In Fig. 2, the artifacts caused by non-uniform solid angle are majorly low frequency bright artifacts which is structure related, while there are also weak dark artifacts near fat. The proposed method corrected the non-uniform solid angle well, and the Corrected-FDK reconstruction was similar to Reference-FDK. The relative root mean square error (RRMSE) between Reference-FDK and Corrected-FDK was 0.0117, while the RRMSE between Reference-FDK and FDK was 0.1313. Quantitative results also confirmed the accuracy of the proposed correction method.

### B.  Simulation study of realistic configuration

In practical XRDCT scans, the sampling rate is lower limited by the pin-hole size on the collimator and scan time. For the realistic configuration, we set 80 translation steps under each view with step interval 1.35 mm, and 90 views uniformly distributed over $2\pi$. The pin-hole was a square hole with height and width 0.5 mm. The X-ray source was placed 150 mm from the collimator, the source voltage was 80kVp and incident tube current time was 24 mAs for each beam. Poisson noise was added to $I^{\mathrm{XRD}}(t,\varphi,E,r)$ with mean photon counts estimated analytically. The reconstruction image was of $64\times64$ grids with pixel size 1.44 mm. When calculating projection differential operation in Eq. 12 and Eq. 14, we smoothed the noisy XRD projections at the other two dimensions except the differential dimension for noise reduction. Ground-truth and the results of Reference-FDK, FDK and Corrected-FDK were displayed in Fig. 3. And the profiles of different methods at the red line in Fig. 3 were compared in Fig. 4.

In this realistic configuration study, although the sampling rate was low, and the noise was not-ignorable, the results were still in consistent with ideal simulation results suggesting the robustness of the proposed method.

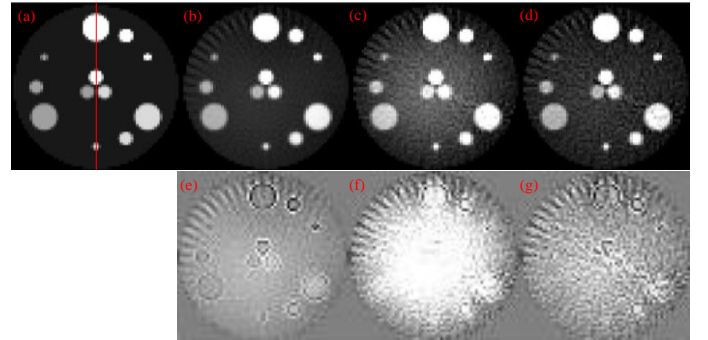

Figure 3: FDK reconstruction results of different XRD projections under realistic simulation condition. (a) Ground-truth, (b) Reference-FDK, (c) FDK, (d) Corrected-FDK, display window: [0.5,4.5] mm$^{-1}$; (e) is the difference between (a) and (b), (f) is the difference between (a) and (c), (g) is the difference between (a) and (d), display window: [-0.8, 0.8].

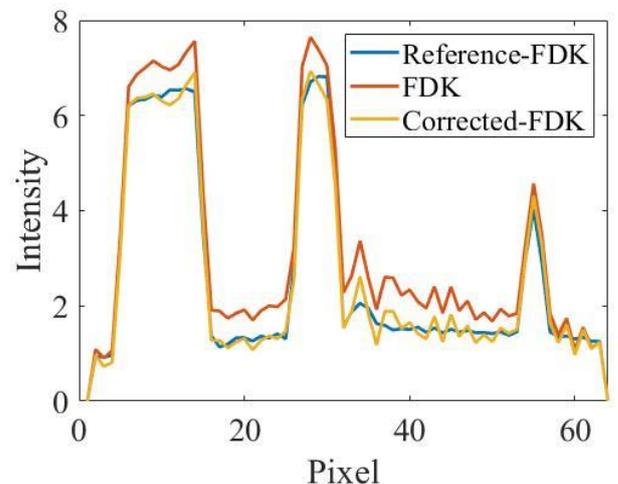

Figure 4: Comparison of profiles of different reconstruction results.

## 4 Conclusion

In XRDCT, the XRD projection differs from a general cone-beam transmission CT projection by a weighting term. The weighting term is due to the scattering solid angle according to its physical model. The scattering angle is non-uniform along the X-ray transmission path and it introduces aditional artifacts to analytical reconstruction results if treated as a general cone-beam projection reconstruction problem. In mathematics, the solid angle weights are in a squared-distance form. In this work, we addresed the influence of the weights in the frequecy domain of the projection data. The residual terms were well corrected in frequency domain, and the artifacts caused by the solid angle non-unifomity were restored after reconstruction in simulation studies in both ideal and realistic situations. The solid angle non-uniformity problem is one problem in XRDCT reconstruction. The scattering angle also varies at different locations, it transforms the XRDCT reocnstruction to a more complicated 3D reconstruction problem. Reconstructed with traditional FDK methods, cone-beam artifacts may further degrades the results significantly. We will further work on the XRDCT analytical reconstruction methods to improve the reconstruction accuracy.

## Acknowledgements

This work is partially supported by National Natural Science Foundation of China (No. 61771279)..

# Ultra-fast Monte Carlo PET Reconstructor

P. Galve[1], F. Arias-Valcayo[1], A.L. Montes[1], A. Villa-Abaunza[1], P. Ibanez[1], J.L. Herraiz[1,2], J.M. Udías[1,2]

[1]Grupo de Física Nuclear, EMFTEL & IPARCOS Universidad Complutense de Madrid, CEI Moncloa, Madrid, Spain
[2] Instituto de Investigación Sanitaria del Hospital Clínico San Carlos (IdISSC)

**Abstract** The Ultra-fast Monte Carlo PET simulator (UMC-PET) is an accurate, fast and flexible PET simulator which has been developed for multiple purposes. The UMC-PET was tested against other MC PET simulators such as PeneloPET, obtaining similar results while being more than 3000 times faster. These features allow applying the UMC-PET for a 3D iterative reconstruction, with a projection step based on, on-the-fly, raw, MC calculations and thus avoiding physics simplifications in the system response matrix. We compared this novel reconstruction scheme with traditional projection techniques combined with Monte Carlo based scatter correction. On a single common GPU (500 USD) these fully MC reconstructions require a few hours for a scanner with > 1 billion lines of response. This provides not only a useful and flexible gold standard method, but may become a practical reconstruction approach if it is combined with variance reduction methods and/or high performance multi-GPU systems.

## 1 Introduction

Positron Emission Tomography (PET) is a complex medical imaging technique that involves many processes related to nuclear and particle physics, optics inside the detectors, and biological processes [1]. In order to predict the scanner performance before manufacture, it is important to have fast and accurate models capable of reproduce all the nature of the technique. Monte Carlo (MC) methods allow us to precisely model all the physics involved, such as positron range, scatter and attenuation inside the patient, photon interaction with the scanner, and detector response [2]. Furthermore, the current development of parallel computing with GPU affords to speed up the simulation codes several orders of magnitude [3]. Simulators must also be prepared to be flexible since state-of-the-art scanners are being focused on dedicated geometries that require arbitrary morphologies [4]-[5].

MC simulators may also be used to improve image quality via corrections implemented in the reconstruction software. The ordered subsets expectation maximization (OSEM) [6] is the most commonly used algorithm for PET reconstruction, and its success depends on the accuracy of the System Response Matrix (SRM) in the projection kernels. Furthermore, scatter and attenuation corrections are usually approximated instead of using a realistic MC approach.

We present a fast, accurate, and flexible Ultra-fast Monte Carlo PET simulator (UMC-PET), a GPU-based code that implements all the aforementioned physics in a flexible framework that allows to simulate any kind of scanner. We also propose to use the UMC-PET directly in the reconstruction, for scatter estimation, and implement it as a realistic projector in the OSEM algorithm.

## 2 An Ultra-Fast Monte Carlo PET simulator

The UMC-PET simulator has been developed focusing in three aspects: accuracy, flexibility and speed.

### 2.1 Accuracy

The UMC-PET simulator included the most relevant physics related to the emission, transport and detection of photons inside a PET acquisition. The positron range was modeled using the convolution kernels developed by J. Cal et al. [12]. Non-collinearity is modeled with a gaussian distribution for the deviation angle 0.5 degrees FWHM. The photon interaction cross sections of photoelectric effect, Compton scatter, and Rayleigh scatter were taken from PENELOPE [13]. Only annihilation photons with 511 keV energy were considered. The energy resolution and time resolution are implemented using a gaussian distribution. For multiple crystal events, the Anger logic has been implemented in the crystal pixel identification.

The time evolution was not implemented, and therefore the exponential decay rule for activity and random or multiple events are not included.

### 2.2 Speed

The code was implemented using PGI CUDA Fortran Compiler, which gives the possibility to define CUDA kernels to be run in NVIDIA GPUs.

### 2.3 Flexibility

The UMC-PET easily adapts to any kind of scanner geometry. The scanner geometry, patient composition, and source distribution are described in voxellized images that are loaded in the GPU global memory, thus allowing to simulate any kind of predefined geometry. With this framework, several scanners have already been successfully simulated, like the 6R-SuperArgus (see section 3), and other spherical morphologies [7].

## 3 Materials

Two preclinical scanners based on the GE healthcare eXplore VISTA scanner [8] were implemented in this work, the 2R-Argus PET/CT, and the 6R-SuperArgus PET/CT (Sedecal Medical Imaging). Both were based on 13×13 double layer modules of LYSO-GSO pixellized scintillator phoswich of 1.55 mm crystal pitch. The 2R-Argus scanner consisted of two rings of 18 modules with 118 mm diameter, a transversal field of view (FOV) of 70 mm and axial FOV of 50 mm. The 6R-SuperArgus is made of six rings of 24 modules with 178 mm diameter, 130 mm transversal FOV and 150 mm axial FOV.

An acquisition following the NEMA NU4-2008 [9] protocol was used to measure the percent standard deviation of the uniform region against the recovery coefficient (RC) in an image quality (IQ) phantom for the preclinical 6R-SuperArgus PET/CT scanner.

## 4 Validation

We have tested the UMC-PET accuarcy against the latest version of PeneloPET [2], a validated PET simulator. In this section we have simulated the NEMA NU 4-2008 [9] rat phantom for the scatter fraction (SF) study inside the 6R-SuperArgus scanner. Since the UMC-PET does not simulates the time evolution, the scanner parameters for the PeneloPET simulation were set to avoid random coincidences (ideal time resolution).

As stated in the NEMA protocol, the sinograms for each simulation were centered to the maximum of each angular bin, and angularly and axially collapsed. In fig. 3, the resultant radial profiles for two energy windows are shown. The SF calculated is shown in table I.

|  | Rat SF (%) | |
|---|---|---|
| (keV) | 425-600 | 100-700 |
| UCM-PET | 11.2 | 24.7 |
| PeneloPET | 11.5 | 24.9 |

Table I. Scatter Fraction (SF) for the rat-like phantoms with UMC-PET and PeneloPET for two different energy windows.

The code was executed in an Intel(R) Xeon(R) W-2155 CPU @ 3.30GHz with a single 11 Gb GeForce RTX 2080 Ti GPU, with 4352 cores. We compared the timing performance against PeneloPET [2] running in a single Intel(R) Xeon(R) CPU E5-2650 0 @ 2.00GHz, obtaining a ratio of $3.25 \cdot 10^7$ decays per second for the UMC-PET, whereas PeneloPET achieved $1.7 \cdot 10^4$ decays per second.

## 5 Image reconstruction implementations

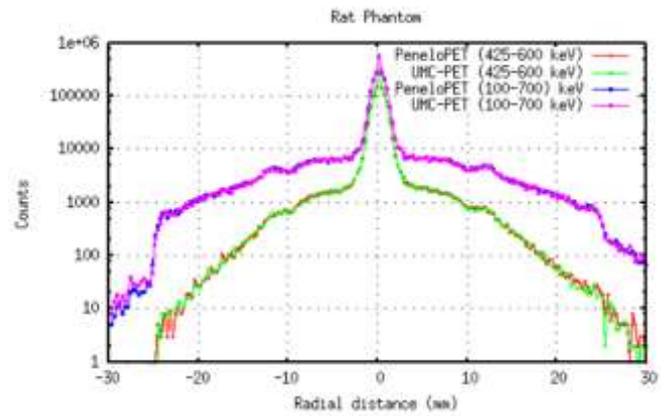

Fig. 3. Comparison of the axially and angularly collapsed profiles for the NEMA NU 4-2008 rat-like phantom for two different energy windows (425-600 keV and 100-700 keV).

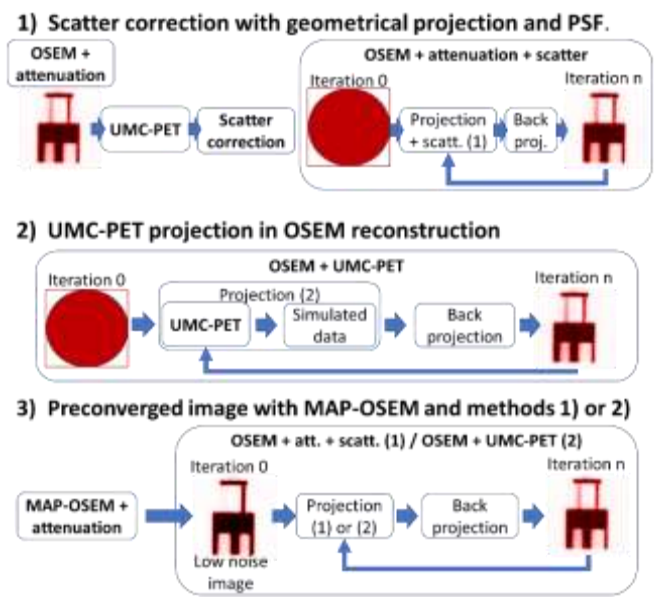

Fig. 4. Schematic representation of the suggested reconstruction methods.

### 5.1 Geometrical projection scheme with Spatially Variant PSF (SV-PSF)

The SRM is separated in geometrical projections combined with a gaussian Point Spread Function (PSF) [10]. We will compare the use of a homogeneous PSF and a variable PSF based in point source simulations to correct DOI effect. The attenuation correction is estimated based on the CT image, and the scatter correction is estimated using the UMC-PET simulator (fig. 4.1). Both factors are included in the projection scheme.

### 5.2 UMC-PET projection

A homogeneous object with the correspondent CT was simulated to obtain a list of events emission voxel-coincidence LOR. To estimate the projections, in each OSEM iteration we run over the list and accumulate the image values of the emission event for each coincidence LOR (fig. 4.2). Since the projection statistics are limited, this methodology was combined with a MAP-OSEM [11] preconverged image with very low noise and resolution (fig. 4.3).

### 5.3 symUMC-PET projection

We use the projection scheme in subsection 5.2 without the CT to explode the scanner symmetries, and the attenuation and UMC-PET scatter corrections in subsection 5.1. The corrections are incorporated to the symUMC-PET projection in the OSEM algorithm.

## 6 Results

### 6.1 Simulation of a micro Derenzo phantom in the 2R-Argus scanner.

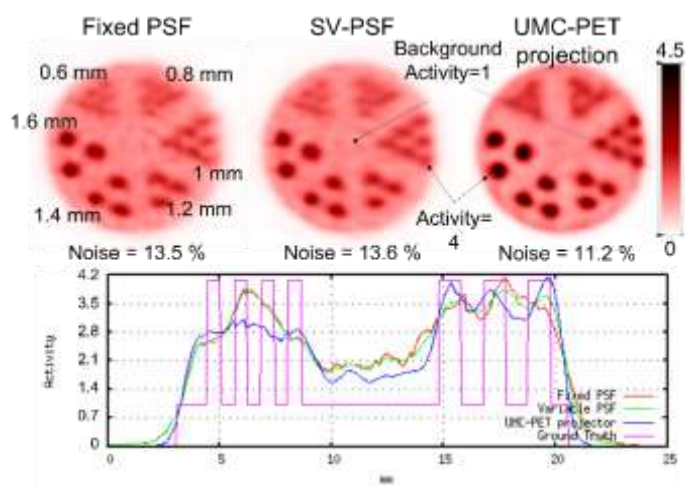

Fig. 5. Reconstruction of the simulated micro Derenzo phantom with fixed PSF, variable PSF, and the UMC-PET projection. The line profile along the 0.6 mm rods and 1.0 mm rods is shown.

The simulated micro Derenzo phantom has been located at the edge of the FOV (25 mm off-center) to increase the DOI effects. The rods diameter were 0.6 mm, 0.8 mm, 1.0 mm, 1.2 mm, 1.4 mm and 1.6 mm, and the rods to background activity ratio was 4:1. In this case, no object was inserted in order to test the image quality separately to the scatter and attenuation corrections. The UMC-PET simulation with ~$1.78 \cdot 10^9$ coincidences took less than 20 minutes, and each iteration took 24 seconds, resulting in an image reconstruction in less than 2 hours.

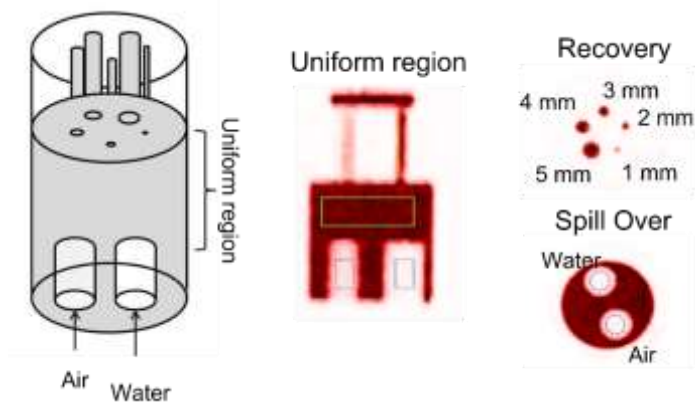

Fig. 6. Scheme of the image quality phantom of the NEMA NU 4-2008 [9] (left) and regions for quantification of noise (center, green region in the uniform region of the phantom), recovery coefficients (right top, yellow regions in the rods; transversal top view), and spill over (right bottom, blue regions in the non-active inserts filled with water and air; transversal bottom view).

### 6.2 Image Quality phantom in the 6R-SuperArgus scanner

The IQ phantom described in fig. 6 ($1.08 \cdot 10^8$ coincidences) was reconstructed with all the methods in section 5. The noise, recovery coefficients and spill over were measured as stated in the NEMA NU 4-2008 protocol (the regions under study are schematized in fig. 6), and the results are summarized in fig. 7 and table II. Figure 7 shows the evolution of the recovery and spill over against noise when we increase the number of iterations.

The simulation of the events list used for the UMC-PET projector took 12 hours to generate ~$6.4 \cdot 10^{10}$ events. The reconstruction took 4 minutes and 37 seconds per iteration, resulting in a total reconstruction of 17 hours (60 iterations).

| | Noise (%) | Recovery coefficient (%) | | | | | Spill over (%) | |
|---|---|---|---|---|---|---|---|---|
| | | 1 mm | 2 mm | 3 mm | 4 mm | 5 mm | Water | Air |
| SV-PSF (wo inp.) | 5.23 | 33.5 | 85.5 | 82.2 | 87.3 | 92.6 | 5.86 | 6.18 |
| SV-PSF | 5.08 | 31.4 | 68.0 | 76.0 | 84.5 | 92.0 | 3.49 | 3.45 |
| UMC-PET | 5.01 | 38.1 | 104.6 | 103.0 | 105.5 | 105.0 | 2.85 | 3.11 |
| UMC-PET (filt.) | 5.08 | 35.2 | 93.2 | 99.5 | 101.3 | 101.7 | 2.53 | 2.82 |
| symUMC-PET | 5.08 | 40.2 | 97.3 | 98.0 | 101.9 | 101.7 | 3.00 | 3.04 |

Table II. Noise, spill over, and recovery coefficients for the IQ phantom (fig. 6) in the 6R-Super Argus scanner ($1.08 \cdot 10^8$ coincidences) for all the methods in section 5. The SV-PSF is tested with and without MAP-OSEM preconverged input image (all the UMC-PET based methods include the input image). The UMC-PET (filt) was post-filtered with a gaussian FWHM of 0.75 mm.

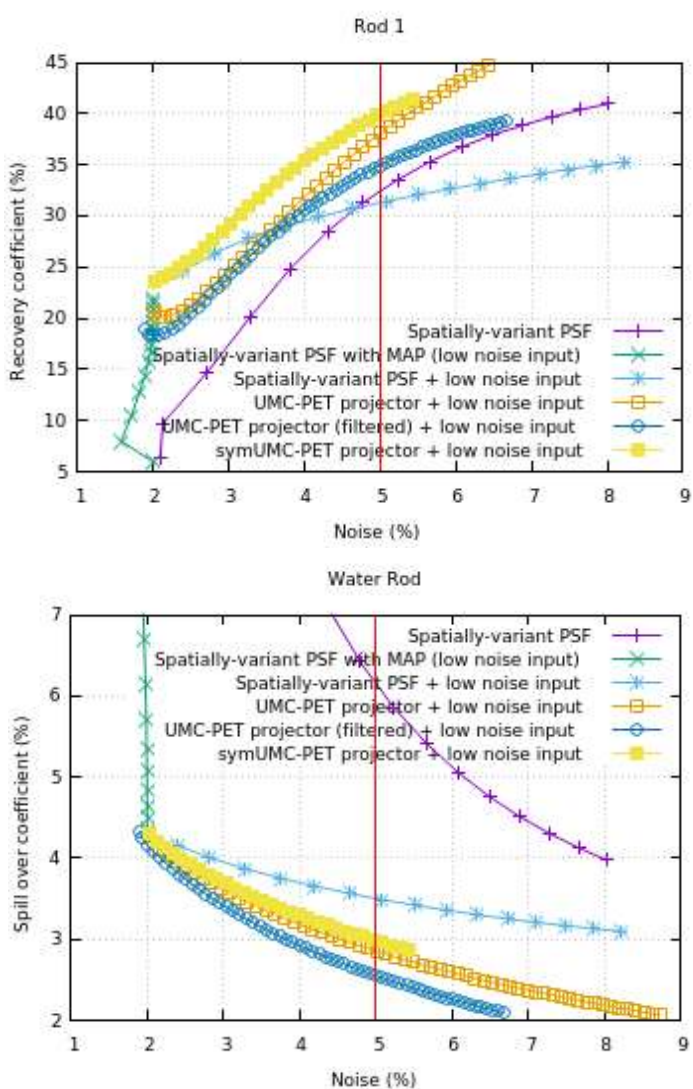

Fig. 7. (Up) Comparison of the recovery-noise curve in the 1 mm rod and (down) the spill over-noise curve in the water insert. The red line shows the 5 % noise level.

In the symUMC-PET projection, the events list took 4 hours to generate ~$5.1 \cdot 10^{10}$ events, and each iteration took 7 minutes. Since the events list can be generated in advance, the image reconstruction with 60 iterations took than 7 hours.

## 7 Discussion

The UMC-PET has been tested against Penelopet, a realistic and validated PET simulator, showing great agreement in the sinogram distribution of detected activity. Further validation would require testing it against a real scanner acquisition.

One of the main goals of this simulator was to accurately predict the SRM. For this reason, the physics related to time evolution were not implemented. Furthermore, the implementation of a global timing in the simulation may require processing the single events sequentially, complicating the computational parallelization of the code in the GPU. However, if necessary, a list of single events generated with the UMC-PET simulator might be post processed to include the global time stamps.

The line profile shown in fig. 5 proves the image quality gain when the SRM is optimized. In table II, the results for the recovery and spill-over for the UMC-PET projector outperform the traditional projection scheme. The recovery coefficients show promising improvements against the standard methods, although the overshoot should be reduced. In the other hand, the noise in the projections must be reduced as well since the UMC-PET projector itself raised the noise without increasing image quality and regularization methods were necessary to achieve practical results (see fig. 7 and table II). The symmetries helped to increase the statistics in the projection and reduce the overshoot. The spill over obtained the best results with the UMC-PET projector without symmetries. Since the scatter is obtained in both cases using the UMC-PET simulator, we assume that the differences may relay in the evolution of the scatter estimation within each image update.

The reconstruction times of the UMC-PET projector were very high for practical use (17 hours and 7 hours for each of the presented examples) and will vary depending on the scanner and image size. For this reason, the proposed method is presented as a gold standard that might not be available for daily use. The use of multiple GPUs will help to reduce the computation time, and further strategies to optimize the PSF implementation will be studied.

## 8 Conclusion

The UMC-PET simulator is a flexible, fast and accurate PET simulator based on Monte Carlo methods and GPU acceleration. The simulator might be used to simulate any kind of scanner geometry. Its high-performance allows to use it for real time scatter prediction and gold standard reference for reconstruction. However, more optimizations in the code and variance reduction methods might be necessary to reduce the computation time to practical applications.

# CBCT Estimation from Limited-Angle Projections by an End-to-End Unsupervised Deformable Registration Network (2D3D-RegNet)

You Zhang[1]

[1]Department of Radiation Oncology, University of Texas Southwestern Medical Center, Dallas, TX, 75235, USA

**Abstract** Limited-angle based cone-beam computed tomography (CBCT) acquisition can reduce the imaging dose, shorten the scan time, and allow fast and continuous tumor/target localizations throughout arc-based radiotherapy treatments. The lack of sufficient scan angle spans, however, leads to severe distortions/artifacts in the reconstructed CBCT images by traditional reconstruction algorithms. In comparison, 2D-3D deformable registration can deform a prior fully-sampled CT/CBCT volume to estimate a high-quality CBCT, by a deformation vector field (DVF) solved using only limited-angle projections. The CBCT images estimated by 2D-3D deformable registration can successfully suppress the distortions and artifacts by incorporating prior information, and reflect up-to-date patient anatomy through deformation. However, currently the 2D-3D deformable registration algorithm is limited by its computational speed, which can take up to several hours to converge to an accurate DVF. In this study, an end-to-end, unsupervised 2D-3D deformable registration framework (2D3D-RegNet) was developed using convolutional neural networks to address the speed bottleneck of the conventional iterative 2D-3D deformable registration algorithm. 2D3D-RegNet can solve the DVFs in < 5 seconds for 90 orthogonally-arranged projections covering a combined 90° scan angle, with DVF accuracy superior to 3D-3D deformable registration, and on par with the conventional 2D-3D deformable registration algorithm.

## I Introduction

For radiotherapy treatments, fast and accurate imaging-guided tumor localization is often needed to account for anatomical motion/deformation, and to pinpoint the radiation to the tumors and avoid surrounding healthy organs. Limited-angle CBCT, which is acquired through a partial arc rotation, can allow 3D imaging at a high temporal resolution [1]. A limited-angle acquisition also naturally reduces the overall imaging dose to patients, and may allow continuous tumor localizations through arc-based radiotherapy deliveries. However, the image quality of limited-angle CBCT is severely affected by the poor resolution along the direction perpendicular to the scan angle due to partial Fourier domain sampling [2].

2D-3D deformable registration, which is a technique that solves a DVF to map a previously acquired fully-sampled CT/CBCT (source) to a new on-board CBCT (target) via 2D projection matching, can be particularly effective under the limited-angle sampling scenario [1]. The combination of *a priori* high-quality information from the source image, and on-board information from limited-angle projections, can effectively mitigate the under-sampling issue to render high-quality on-board CBCT images. Instead of measuring the similarity directly between a deformed source image and the artifacts-ridden limited-angle CBCT image, 2D-3D deformable registration calculates the similarity between the projected 2D digitally reconstructed radiographs (DRRs) of the deformed 3D source image and the 2D artifacts-free on-board projections. 2D-3D deformable registration has been investigated in recent years for its potential in sparse-view and limited-angle projection based CBCT estimation with very promising results [1, 3]. However, the current 2D-3D deformable registration algorithms involve a very computationally-expensive optimization scheme, with considerable runtimes up to hours needed to derive a high-accuracy DVF even with GPU acceleration.

To address this issue, we developed an unsupervised, end-to-end, 2D-3D deformable registration network (2D3D-RegNet) on the basis of a core U-net structure, which proved effective in handling various image domain tasks [4]. A simple Feldkamp-Davis-Kress (FDK) reconstruction module was included into the 2D3D-RegNet to align the 2D projections with the source image to feed into the U-net as parallel channels [5]. Forward projection module was also included into the network to generate 2D DRRs from the deformed 3D source images to assess their match to 2D target cone-beam projections. A DVF inversion module was included in the 2D3D-RegNet to invert the forward DVF to promote inverse deformation consistency, which also adds additional constraints and regularizations for the ill-conditioned 2D-3D deformable registration problem. Different limited-angle sampling scenarios were simulated to evaluate the accuracy of 2D3D-RegNet against the traditional 2D-3D deformable registration algorithm, and against a mainstream, open-source 3D-3D deformable registration package (Elastix) [6].

## II Materials and Methods

### II.A. Network structures

Fig. 1 shows the overall workflow of 2D3D-RegNet. The function and design of each of the modules were introduced as following:

### II.A.1. Reconstruction layer:

The input source 3D images and the cone-beam projections are of different physical properties, on different image reference frames, and also of different dimensionality and resolution. To align the two inputs as parallel channels, we added a GPU-enabled, non-trainable reconstruction layer on the basis of the 3D FDK algorithm to convert the 2D projections into the 3D image domain [7]. The FDK-reconstructed target image and the source image were concatenated as two channels to input into a following U-net core structure to solve the DVFs.

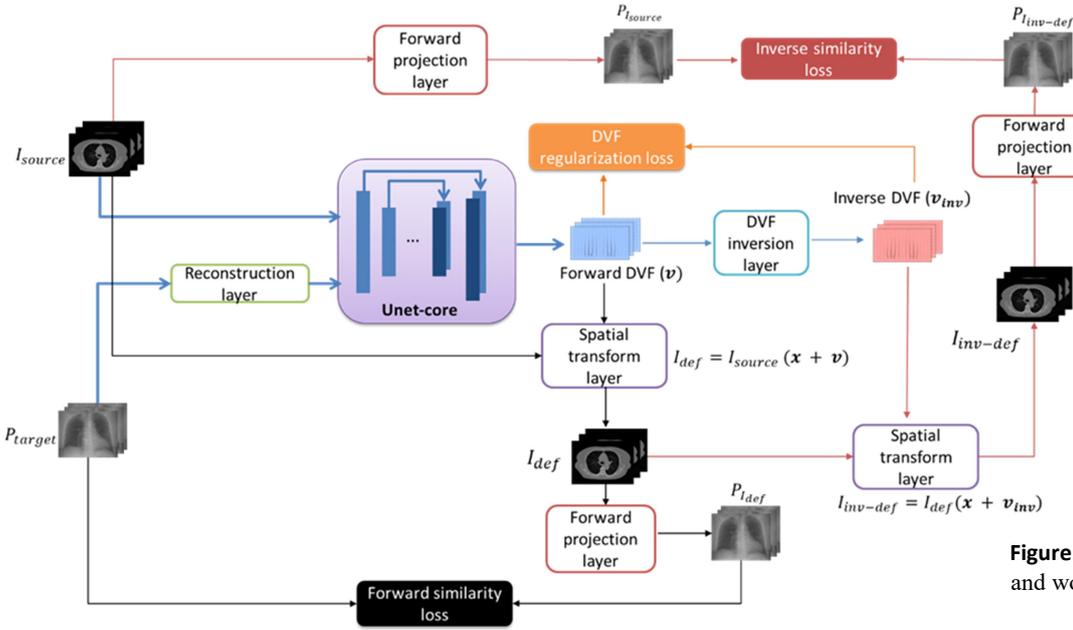

**Figure 1.** The overall 2D3D-RegNet structure and workflow.

### II.A.2. U-Net core:

In this study, we used U-net as the core structure to generate the DVFs [4]. The U-net structure starts with a 16-filter convolutional layer, which was followed by four down-sampling convolutional layers (stride of two) of 16, 32, 32, and 32 filters, respectively. The expansive path features four layers symmetric to the contractive path, with each layer composed of up-sampling, concatenating skip connection, and convolution (32 filters) operations. The output of the expansive path was further convolved by three additional layers of 16, 16 and 3 filters, respectively. The final 3-filter convolutional layer yields a DVF output the same size as the original image, with three channels each representing the DVF along one Cartesian direction (x, y and z). All convolutional filters are of size 3 x 3 x 3, and each convolution operation was followed by a LeakyRelu activation with parameter 0.2.

### II.A.3. DVF inversion layer:

The U-net core structure outputs the forward DVF that maps the source image to the target image. Contrary to the forward DVF, inverse DVF maps the target image back to the source image. The simultaneous solution of the inverse DVF along with the forward DVF is desired for many applications [8]. The solution of inverse-consistent DVF pairs also further regularizes and improves the DVF accuracy. In 2D3D-RegNet, we incorporated a DVF inversion layer based on the iterative fixed-point method to generate the inverse DVFs [9].

### II.A.4. Forward projection layer:

To compute the similarity metric in the 2D projection domain, a GPU-enabled, non-trainable forward projection layer was incorporated as a network layer, which computes 2D cone-beam projections from the deformed images using the Siddon's ray-tracing algorithm [7]. The conjugate filtered backprojection operation (FDK) was registered as the gradient of the network layer when back-propagating the gradient of the designed network.

### II.B. Loss function design

In 2D3D-RegNet, we defined three different loss functions to train the network. The 1st loss term ($Loss_1$) measures the similarity between the input 2D cone-beam projections and the DRRs projected from the deformed source image (Equation 1, Fig. 1). It serves as the main data fidelity term to drive the optimization of the DVF ($v$).

$$I_{def} = I_{source}(x + v),$$
$$Loss_1 = D\,(A * I_{def}, P_{target}) \qquad (1)$$

$I_{source}$ indicates the source image to be deformed. $x$ indicates the Cartesian coordinates of $I_{source}$ voxels, and $v$ indicates the corresponding DVF at each voxel coordinate. $I_{def}$ represents the deformed source image. $P_{target}$ indicates the on-board acquired target 2D projections. $A$ is the system matrix that maps $I_{def}$ onto the $P_{target}$ reference frame. $D$ indicates the image similarity metric (mean squared error for this study).

The 2nd loss term ($Loss_2$) is the inverse similarity loss, based on the inverse DVF generated from the DVF inversion layer (II.A.3). In the DVF inversion layer, the inverse DVF ($v_{inv}$) was calculated via an iterative fixed-point conversion scheme as [9]:

$$v_{inv}^0 = 0,$$
$$v_{inv}^n = -v\,(x + v_{inv}^{n-1}), \quad n = 1, \dots, N \qquad (2)$$

$n$ indicates the iteration number, with its maximum $N$. To strike a balance between the computational/memory load and accuracy, we set $N = 7$. $v_{inv}$ yields the corresponding $Loss_2$ as shown in Equation set 3:

$$I_{inv-def} = I_{def}(x + v_{inv}),$$

$$Loss_2 = D\left(A * I_{inv-def}, A * I_{source}\right) \quad (3)$$

This loss term serves to measure the similarity between the inversely-deformed $I_{inv-def}$ and the original $I_{source}$, enforcing the inverse-consistency of the DVF.

The 3rd loss term ($Loss_3$) for the network training calculates the DVF energy and enforces the DVF smoothness:

$$E(v) = \sum_{x=1}^{n_i}\sum_{y=1}^{n_j}\sum_{z=1}^{n_k}\sum_{m=x,y,z}\left(\left(\frac{\partial v_m}{\partial x}\right)^2 + \left(\frac{\partial v_m}{\partial y}\right)^2 + \left(\frac{\partial v_m}{\partial z}\right)^2\right) \quad (4)$$

In Equation 4, $m$ indicates one of the three Cartesian directions $x$, $y$ and $z$. $v_m$ indicates the DVF along the corresponding $m$ direction. $n_i$, $n_j$ and $n_k$ indicate the image dimension along the three Cartesian directions.

**II.C. Training and testing scheme**

To train the 2D3D-RegNet, we used 4D-CT lung datasets from two public libraries: the cancer imaging archive (TCIA) [10] and the CREATIS laboratory [11]. A total of 26 4D-CT sets were used as our training dataset. Each 4D-CT set has 10 respiratory phase volumes. For each 4D-CT, we extracted the end-expiration phase as the 3D source image, and simulated 2D cone-beam projections from all phases (including the end-expiration phase) for 2D-3D registration. The projection matrix was simulated in full-fan mode with 512 x 512 pixels, with each pixel measuring 0.8 mm x 0.8 mm in dimension. The projections were simulated under two limited-angle acquisition scenarios: (1). Single-view: projections distributed over a single-angle spanning around the anterior-posterior (AP) direction of the patient; and (2). Ortho-view: projections distributed over two orthogonally-arranged angles, one along AP, and the other along the left lateral direction. A total of 50000 iterations were used to train 2D3D-RegNet, which took ~72 hours on a NVIDIA V100 GPU card (NVIDIA Corporation, Santa Clara, CA). Independent models were developed and trained for each of the angular acquisition scenarios.

We used an independent in-house 4D-CT lung dataset to test the 2D3D-RegNet. The corresponding dataset has 12 lung patient cases, and each has 10-14 respiratory phase volumes. The end-expiration phase of each case was selected as the source image, and cone-beam projections were simulated from the end-inspiration phase volume. The end-inspiration phase was used for testing as it has the largest extent of deformation from the source image, and could better assess the accuracy of 2D3D-RegNet. Both the relative error (RE) of the deformed images and the target registration error (TRE) of the solved DVFs were assessed to compare different algorithms [3].

**III Results**

As shown in Fig. 2, the limited-angle projections from a single-view yielded significant artifacts and structure distortions in the reconstructed FDK images (Target-FDK). Correspondingly, direct 3D-3D deformable registration by Elastix is error-prone, which generated severe distortions in the deformed images (Elastix). For the Elastix registration, we used a region-of-interest mask to exclude regions outside the imaging field-of-view to account for the limited projection size, which however is unable to suppress the strong distortion artifacts from limited-angle sampling. In comparison, 2D3D-RegNet has preserved the image integrity and did not introduce the distortions into the deformed image (2D3D-RegNet). It also deformed the lungs to well match with the ground-truth target images (Target-GT). Quantitative results shown in Table 1 echoed the images presented in Fig. 2. The 2D3D-RegNet substantially improved the accuracy of the estimated CBCT images through the deformation-driven approach, as compared to that of the CBCT images directly reconstructed by the FDK algorithm. The 2D3D-RegNet also outperformed the Elastix in terms of the accuracy of the deformed images and the accuracy of the DVFs. Increasing the scan angle expectedly improves the accuracy of 2D3D-RegNet. With the same total scan angles, projections acquired from an orthogonal-view setting yielded better results than those acquired from a single direction, due to the complimentary information offered from the orthogonal directions. Comparing the traditional 2D-3D deformable registration with 2D3D-RegNet, under the single-view scan angles their results are similar (2D3D-RegNet performed better on the TRE metric), while under the orthogonal-view scan angles the traditional 2D-3D deformable registration algorithm performed better in the RE metric. Speed-wise, the 2D3D-RegNet solved DVFs for image volumes of size 256 x 256 x 256 in < 5 seconds at test time. In comparison, the iterative 2D-3D registration technique takes ~1.5 hours for 30 projections, and 4-5 hours for 90 projections.

| Metrics | | Target Registration Error (TRE, **mm**) | | | | Relative Error (RE, **%**) | | | |
|---|---|---|---|---|---|---|---|---|---|
| Scan Angle Scenarios | | Elastix | 2D-3D Def | 2D3D-RegNet | | FDK | Elastix | 2D-3D Def | 2D3D-RegNet |
| single -view | 0º | 28.4 ± 17.7 | 6.4 ± 4.7 | 5.4 ± 4.1 | | 155.3 ± 20.0 | 53.7 ± 8.9 | 21.3 ± 3.4 | 20.4 ± 3.0 |
| | 15º | 29.3 ± 18.2 | 5.5 ± 4.1 | 4.8 ± 3.5 | | 113.0 ± 15.2 | 53.4 ± 8.7 | 19.6 ± 3.1 | 18.2 ± 2.8 |
| | 30º | 25.0 ± 16.8 | 5.2 ± 4.0 | 4.6 ± 3.5 | | 95.3 ± 12.9 | 53.5 ± 9.9 | 18.1 ± 2.9 | 17.7 ± 2.8 |
| | 60º | 14.0 ± 9.3 | 4.6 ± 3.7 | 4.3 ± 3.4 | | 72.5 ± 8.5 | 44.6 ± 8.5 | 15.5 ± 2.5 | 16.1 ± 2.5 |
| | 90º | 7.0 ± 5.3 | 5.0 ± 3.9 | 4.3 ± 3.3 | | 57.1 ± 5.7 | 41.0 ± 7.9 | 15.0 ± 2.8 | 15.3 ± 2.4 |
| ortho-view | 0º (0º each) | 9.4 ± 6.0 | 6.0 ± 4.4 | 5.4 ± 3.7 | | 133.2 ± 19.0 | 36.8 ± 5.1 | 17.8 ± 2.7 | 19.2 ± 2.4 |
| | 15º (7.5º each) | 9.5 ± 6.2 | 4.7 ± 3.5 | 4.8 ± 3.4 | | 105.3 ± 15.6 | 36.3 ± 5.4 | 13.8 ± 2.5 | 16.3 ± 2.8 |

| 30º (15º each) | 8.8 ± 6.2 | 4.3 ± 3.4 | 4.5 ± 3.4 | 90.3 ± 13.1 | 34.1 ± 5.4 | 12.4 ± 2.1 | 15.4 ± 2.5 |
| 60º (30º each) | 7.3 ± 5.8 | 4.0 ± 3.3 | 4.1 ± 3.1 | 67.8 ± 10.6 | 29.3 ± 4.9 | 11.2 ± 1.6 | 14.8 ± 2.4 |
| 90º (45º each) | 5.4 ± 4.6 | 3.8 ± 3.1 | 3.9 ± 3.0 | 52.0 ± 9.9 | 24.3 ± 4.6 | 11.1 ± 1.5 | 14.4 ± 2.1 |

**Table 1.** Comparison between the target registration errors (TREs) of DVFs, and the relative errors (REs) of images solved by different methods, using different scan angle schemes. FDK: Feldkamp-Davis-Kress (reconstruction algorithm). Def: deformation.

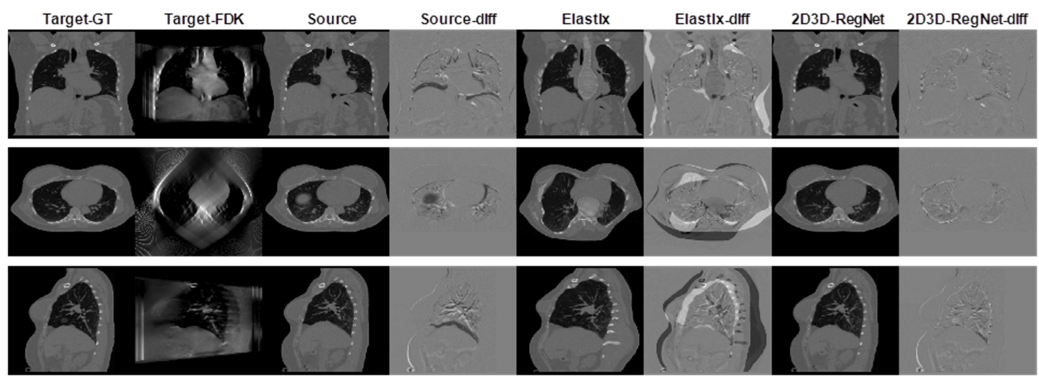

**Figure 2.** Three-view image comparison between 'ground-truth' target images (**Target-GT**), FDK-reconstructed target images (**Target-FDK**) by limited-angle projections, source images (**Source**) before deformable registration, difference images (**Source-diff**) between **Source** and **Target-GT**, Elastix-deformed images (**Elastix**) with deformation between **Source** and **Target-FDK**, difference images (**Elastix-diff**) between **Elastix** and **Target-GT**, and 2D3D-RegNet-deformed images (**2D3D-RegNet**), and difference images (**2D3D-RegNet-diff**) between **2D3D-RegNet** and **Target-GT**. The results are based on projections from a 90º scan angle (single-view).

## IV Discussion

To address the speed bottleneck of current iterative 2D-3D deformable registration techniques, we developed an end-to-end, unsupervised 2D-3D deformable registration network to allow near real-time DVF solution. Visual comparisons of the deformed images (Fig. 2), and corresponding quantitative evaluations of these images and the solved DVFs (Table 1), demonstrated the superiority of 2D3D-RegNet over the conventional FDK reconstruction algorithm, and a 3D-3D registration method (Elastix). In the current 2D3D-RegNet structure, we reconstructed the 2D projections at the beginning to a 3D volume via FDK to align with the source 3D image, such that both can be conveniently fed into a subsequent U-net core as parallel channels. The FDK reconstruction is a degenerative process, especially considering the 2D projections are limited-angle, which leads to severe under-sampling artifacts in the reconstructed images (Fig. 2). However, since the image similarity metric is measured by 2D3D-RegNet on re-projected 2D DRRs of the deformed image, instead of directly on artifact-ridden 3D images, the final deformation results are superior to direct 3D image-domain registration, with artifacts and distortions successfully suppressed.

The results also showed that 2D3D-RegNet performed similarly to the traditional iterative 2D-3D registration algorithm (Table 1), while with substantially improved efficiency (< 5 seconds as compared to hours). However, the iterative 2D-3D registration algorithm generated comparatively smaller REs than 2D3D-RegNet (Table 1) under ortho-view scan angles. The discrepancy is potentially due to the artifacts and the distortions presented in the FDK input channel when the projections from two distinct, orthogonally-arranged scan angles were mixed together in reconstruction, as well as the loss of information from the degenerative FDK reconstruction process (Fig. 2). Potential solutions to further improve the accuracy of 2D3D-RegNet include adding additional image filtration or enhancement layers after the reconstruction module, or directly inputting the 2D projections into the network without explicit reconstruction, or feeding the projections into the network multiple times through a recurrent or cascaded pattern.

## V Conclusion

We developed an unsupervised 2D-3D deformable registration network, 2D3D-RegNet, for efficient and accurate CBCT estimation from limited-angle projections. 2D3D-RegNet solves DVFs with similar accuracy as the traditional iterative 2D-3D deformable registration algorithm, while only takes a few seconds as compared to several hours of the latter method.

# Imaging Resolution Analysis of Compton camera for 20-80 keV X-ray Fluorescence Photons

Chuanpeng Wu, Liang Li

Department of Engineering Physics, Tsinghua University, Beijing, 100084, China

**Abstract**

X-ray fluorescence CT(XFCT) is a novel medical imaging modality for molecular imaging. The structure of XFCT can be pinhole collimation structure or Compton camera(CC) structure. In this work, we study the imaging resolution upper-limit of Compton camera imaging with 20-80keV X-ray fluorescence(XF) photons through theoretical analysis. We choose two-layer detectors structure CC and four elements (I, Ba, Gd, and Au) which are most commonly used in XF contrast agents. The imaging spatial resolution of CC depends strongly on detection distance. So angular resolution measurement(ARM) is commonly used to evaluate the performance of CC. Three mainly affecting factors of ARM are taken into consideration: energy resolution, spatial resolution and Doppler broadening. From the results, when incident photons are low-energy, spatial resolution of detectors has minimal impact, while Doppler broadening is the most significant influencing factor. The ARM upper-limit of the four elements is 12.1971°, 10.7426°, 7.9424° and 4.9747° respectively. As a conclusion, through theoretical analysis, we believed that the feasibility of X-ray fluorescence Compton camera is not high because of the significant negative impact of Doppler broadening effect.

## 1 Introduction

XFCT is a novel medical imaging(MI) modality that can present molecular and functional information in organisms [1]. Compared with some traditional MI modalities such as CT and MRI, XFCT has higher molecular sensitivity due to the specific characteristic energy of XF photons. Compared with nuclear MI modalities, such as SPECT and PET, the tracers used in XFCT are not spontaneously decayed. So the synthesis, storage of contrast agents is more convenient and the administration time of patients is more unlimited. The facility cost of XFCT is lower as well[2].

Among MI modalities mentioned above, PET can only image with 511keV photons. While SPECT and XFCT, which can use a variety of traces, usually use mechanical collimation methods to obtain the direction of incident X-rays. Collimation results in the reduction of collected photons and the decline of detection efficiency.

Compton camera (CC) is a double-layer detector structure imaging system without mechanical collimation[3]. Compton imaging has been used for photon detection in many application fields[4, 5]. However previous researches were all focused on radioisotopes with high photon energy. For example, ZhongHe used 3D position-sensitive large-volume CZT detectors for CC imaging[6], which has excellent performance in high-energy and wide-energy-spectrum imaging, but was not optimal under tens keV.

Therefore, combining the advantages of X-ray fluorescence and CC imaging is a brand new idea. In world, the Monte Carlo simulation work carried out by Vernekohl in 2016 is the only exploration to X-ray fluorescence Compton camera (XFCC) MI modality[2]. 82keV monochromatic X-rays is incident into gold nanoparticle (GNPs) solution, and

Si/CdTe double-layer detector CC is used to collect photons. This work verified the clinical feasibility of XFCC.

The purpose of this work is to theoretically analyze the imaging resolution upper-limit of CC imaging under 20-80keV low-energy XF photons. The imaging SR of CC depends strongly on reconstruction method and detection distance. Thus, ARM is commonly used to evaluate the performance of CC. In order to examine the ARM upper-limit, we choose two-layer detectors structure CC which has two state-of-the-art detectors with high ER and SR. We selected four elements (I, Ba, Gd, and Au, whose typical fluorescence peak energy is 28.610keV, 32.191keV, 42.983keV, 68.794keV respectively.) which are most commonly used in XF contrast agents.

## 2 Methods

Although the 3D position-sensitive CZT detector CC system mentioned above has good performance, we hope to use low-Z material suchas Si detectors due to imaging with 20-80keV low energy XF photons. The main reason is that low-Z detectors will lead to lower Doppler broadening and better energy resolution for low-energy photons. Therefore, Therefore, we abandon the single large-volume CZT crystal structure, and choose the traditional double-layer detector CC structure to carry out the research.

In CC imaging , as shown in Figure1, photons are incident into the first layer of detector and has Compton scattering interaction with detector atoms. The scattering decector will record the interaction position $r_1$ and the deposited energy of the recoil electron $E_1$. Then the scattered photons are emitted out of the first detector, and absorbed by the second layer of detector. The absorption position $r_2$ and the deposited energy $E_2$ are recorded by the absorption detector. The sum of the two deposited energy is the energy of the incident photon, that is, $E_0 = E_1 + E_2$. According to Compton kinematics, with $E_0$ and $E_2$ substituted into Eq. (1), the scattering angle θ of Compton scattering can be calculated:

$$cos\theta = 1 - m_e c^2 \left( \frac{1}{E_2} - \frac{1}{E_0} \right) \qquad (1)$$

Where $E_0$ is the energy of incident photon. $E_2$ is the energy of scatted photon. θ is the scattering angle of Compton scattering. $m_e$ is the rest mass of the electron and $m_e c^2 = 511 keV$.

After calculating the scattering angle $\theta$, we are still not sure where the specific incident direction of the incident photon is. But we can build a cone surface with $r_1 r_2$ as the axis and $\theta$ as the cone angle, on which we can find the incident direction. When enough Compton scattering events are detected, each event can be inversely calculated to a cone surface. The intersection of these cone surfaces is theoretically the spatial position of the radioactive source.

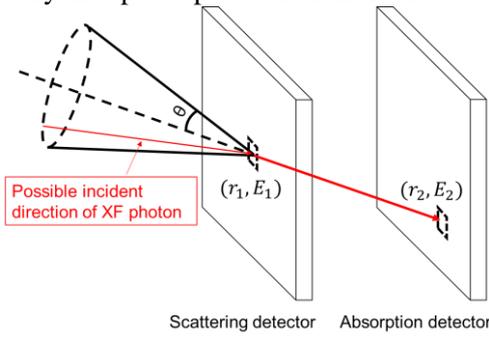

Possible incident direction of XF photon

Scattering detector          Absorption detector

Figure1 diagram of Compton camera principle

ARM is most important when evaluating the performance of a CC system, which is defined as the deviation between the calculated scatter angle θ (cone angle) and the real scatter angle. There are three main factors affecting ARM: ER of detector, SR of detector, and Doppler broadening effect[4]. ARM = $\theta_{ALL}$, and $\theta_{ALL}$ is shown as Eq. (2):

$$\tan(\theta_{ALL})^2 = \tan(\Delta\theta_E)^2 + \tan(\Delta\theta_r)^2 + \tan(\Delta\theta_D)^2 \quad (2)$$

Where $\Delta\theta_E$ is the angle uncertainty caused by the ER of detector. $\Delta\theta_r$ is the angle uncertainty caused by erros of detecting interaction position, which is related to SR of detector. $\Delta\theta_D$ is the angle uncertainty caused by Doppler broadening. Next, we will analyze the angle uncertainty caused by each factor separately.

**2.1 Angle uncertainty due to energy resolution: $\Delta\theta_E$**
From Eq. (1), the calculated scattering angle θ is related to the detected energy. Bad energy resolution of the detector will lead to a larger angle uncertainty. In real detection process, not all scattered photons will completely deposit all of its energy in the detector. Therefore, electronic coincidence is used to filter out Compton scattering events that have deposited all energy in the two-layer detector system to reconstruct the image. This will reduce the background noise caused by incomplete-deposition events. Then Eq. (1) can be transformed into the form of Eq. (3), which brings another advantage that we can calibrate the specific XF peak energy $E_0$, which can be obtained by summing th detected energy $E_1$ and $E_2$. After energy calibration, $E_0$ is unbiased and the angle uncertainty is mainly determined by the energy resolution of the first layer detector.

$$\cos\theta = 1 - m_e c^2 \left( \frac{1}{E_0 - E_1} - \frac{1}{E_0} \right) \quad (3)$$

According to the law of error propagation, the uncertainty of scattering angle caused by ER $\Delta\theta_E$ is shown as Eq. (4):

$$(\Delta\theta_E)^2 = \left(\frac{d\theta}{dE_0}\right)^2 (\Delta E_0)^2 + \left(\frac{d\theta}{dE_1}\right)^2 (\Delta E_1)^2 \quad (4)$$

Where the derivative is calculated as eqution (5) and (6):

$$\frac{d\theta}{dE_0} = \frac{m_e c^2}{\sin\theta} \left[ \frac{1}{E_0^2} - \frac{1}{(E_0 - E_1)^2} \right] \quad (5)$$

$$\frac{d\theta}{dE_1} = \frac{m_e c^2}{\sin\theta} \frac{1}{(E_0 - E_1)^2} \quad (6)$$

The uncertainty of $E_0$ after energy calibration is extremely small, $\Delta E_0 \ll \Delta E_1$ Finally $\Delta\theta_E$ can be expressed as Eq.(7):

$$\Delta\theta_E \approx \frac{m_e c^2}{\sin\theta} \frac{1}{(E_0 - E_1)^2} \Delta E_1 \quad (7)$$

**2.2 Angle uncertainty due to spatial resolution: $\Delta\theta_r$**
For double-layer detector CC imaging problem, Ordonerz et al. gave the analytical form of the angular uncertainty caused by the spatial resolution of the detector $\Delta\theta_r$ through theoretical calculations in 1999[7]. There are three parts of angle uncertainty contributing to $\Delta\theta_r$ :(a) $\Delta\theta_1$ caused by horizontal-direction SR of the first layer detector $\Delta r_1$ ; (b) $\Delta\theta_2$ caused by depth-direction SR of the first layer detector $\Delta z_1$; (c)$\Delta\theta_3$ caused by horizontal-direction SR of the second layer detector $\Delta r_2$. $\Delta\theta_r$ is expressed as Eq. (8):

$$\Delta\theta_r = \sqrt{\Delta\theta_1^2 + \Delta\theta_2^2 + \Delta\theta_3^2}$$
$$= \frac{1}{d}\sqrt{2cos^2\theta(\Delta r_1)^2 + sin^2\theta(\Delta z_1)^2 + 2(\Delta r_2)^2} \quad (8)$$

Where $\theta$ is scattering angle. $d$ is the distance between two layers of detectors.

**2.3 Angle uncertainty due to Doppler broadening: $\Delta\theta_D$**
Eq. (1) describes the classic Compton effect, in which the scattering angle formula is under the assumption that the inner electron interacting with the incident photon is at rest. However in reality, the electrons constantly move around the nucleus and have momentum. As a result, for specific scattering angles and incident photons with specific energy, the energy of recoil electrons and scattered photons is diffused and no longer a specific value. This phenomenon is called Doppler broadening effect[8]. According to Matscheko's research in 1989[9], The broadening of the recoil electron energy $E_1$ caused by Doppler broadening follows Eq.(9):

$$\Delta E_{1D} = \frac{E_0 - E_1}{E_0}\sqrt{E_0^2 + (E_0 - E_1)^2 - 2E_0(E_0 - E_1)cos\theta} \frac{\Delta p_z}{m_e c} \quad (9)$$

Where $\Delta p_z$ is the FWHM of the Compton profile of detector material, which reflects the distribution of the initial momentum $p_z$ of the electrons around the nucleus. The relationship between $\Delta\theta_D$ and $\Delta E_{1D}$ can be derived form the law of error propagationas as Eq.(7).

**3 Calculation and Results**
We select state-of-the-art detectors with high ER and SR in our theoretically analysis. About the material of detectors, we choose low-Z Silicon material for the first layer detector. Despite of lower detection efficiency, low-Z detectors have lower Doppler broadening and better energy resolution. Besides, for low-Z elements Compton scattering cross-section accounts for larger proportion of the total cross-section. Besides, the energy resolution of Si detectors is better when detecting low-energy X-rays. As for the second layer detector, we chose CdTe detector in order to absorb as many scattered photons as possible.

**3.1 Calculate $\Delta\theta_E$**
The novel array silicon drift detector (SDD) is able to detect low-energy X-rays with high energy resolution, such as 127eV (@5.9keV) [10]. Its electronic noise can reach about $(FWHM)_{noise} = 0.09 keV$. For Silicon, best Fano-factor is $F = 0.084$ and mean ionization energy is $W = 3.8eV$.

With these parameters we calculate the FWHM of statistical fluctuations as Eq. (10) and energy resolution as Eq. (11):

$$(FWHM)_{statistic} = 2.35\sqrt{FWE}$$
$$= 2.35\sqrt{0.132 \times 0.0038E} \ keV \quad (10)$$

$$\Delta E = (FWHM)_{total}$$
$$= \sqrt{(FWHM)_{statistic}^2 + (FWHM)_{noise}^2} \quad (11)$$
$$= \sqrt{0.002770E + 0.0081} keV$$

According to Eq. (11), Figure2 shows the energy resolution of silicon detector. Combining Eq.(11) and Eq.(7), we can calculate $\Delta\theta_E$ when the XF rays of 4 elements are the incident rays. The results are shown in Figure3.

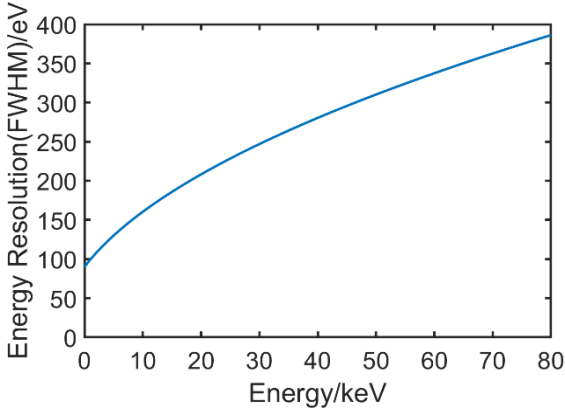

Figure2 ER as a function of Energy for Silicon drift detector

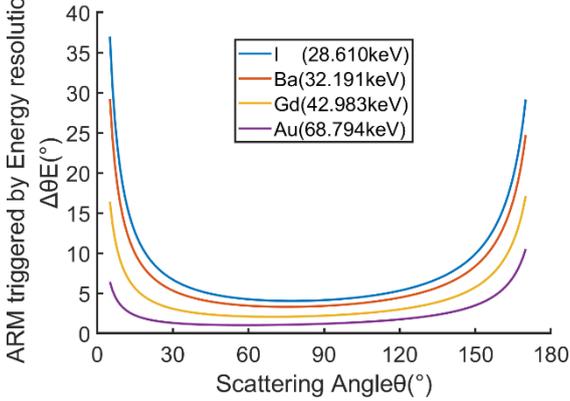

Figure3 $\triangle\theta_E$ as a function of Scattering Angle $\theta$ for four XF elements

From Figure3, when the scattering angle is close to 0° or 180°, $\Delta\theta_E$ is too large because $sin\theta$ is close to 0. Therefore, large-angle scattering and small-angle scattering events should be discarded in actual measurements. Considering backscattering events are discarded as well, we select moderate angles in the range of 20°~80° and calculate average value of $\Delta\theta_E$. The number of scattered photons in each scattering angle $\theta$ direction is not isotropic, but conforms to the description of the Klein-Nishina formula as Eq. (12). So we take this factor into consideration and calculate the weighted average value of $\Delta\theta_E$.

$$f_{KN}(E_0, \theta) = \frac{d\sigma}{d\theta} = 2\pi sin\theta \frac{d\sigma}{d\Omega}$$
$$= \frac{2\pi sin\theta r_e^2}{(1+\alpha(1-cos\theta))^2}\left(\frac{1+cos^2\theta}{2}\right)\Bigg\{1 \quad (12)$$
$$+ \frac{\alpha^2(1-cos\theta)^2}{(1+cos^2\theta)(1+\alpha(1-cos\theta))}\Bigg\}$$

Where $f_{KN}(E_0, \theta)$ is Compton scattering differential cross-section. $\alpha = E_0/m_e c^2$ . $r_e$ is the electronic classical radius, $r_e = e^2/m_e c^2 = 2.818 \times 10^{-13} cm$.

With Eq. (12), we can plot the curve of Compton scattering cross-section as a function of scattering angle $\theta$, as shown in Figure4. Use $f_{KN}$ under different $\theta$ as the weight and calculate the weighted average of $\Delta\theta_E$ in the range of 20°~80°. We finally get the results as shown in Table1.

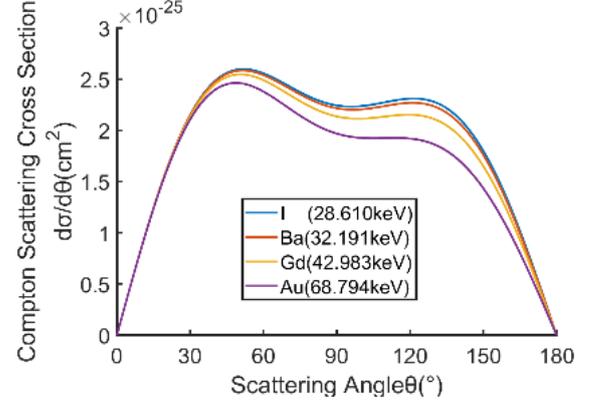

Figure4 KN Cross-Section as a function of Scattering angle $\theta$

Table1 The results of the average $\Delta\theta_E$ for 4 different elements

| Incident XF energy/keV | I (28.620) | Ba (32.191) | Gd (42.983) | Au (68.794) |
|---|---|---|---|---|
| $\Delta\theta_E/°$ | 4.5835 | 3.7102 | 2.2554 | 1.0892 |

### 3.2 Calculate $\Delta\theta_r$

According to Eq. (8), there're three factors affecting $\Delta\theta_r$. For horizontal SR $\Delta r_1$ and $\Delta r_2$, we select the performance index of advanced array SDD in world, which can reach 0.5mm[10].About $\Delta z_1$, with the decrease of detector thickness $\Delta z_1$ will be better, but the detection efficiency will be worse. Considering these two factors, the thickness is set to 0.5mm. As for the distance between detectors $d$, too small $d$ makes $\Delta\theta_r$ terrible, while too large $d$ makes the number of scatted photon detected by the second detector decrease. Referring to previous works we set $d = 60mm$,. Substituting these performance parameters into Eq. (8), the curve of $\Delta\theta_r$ can be plotted, as shown in Figure5.

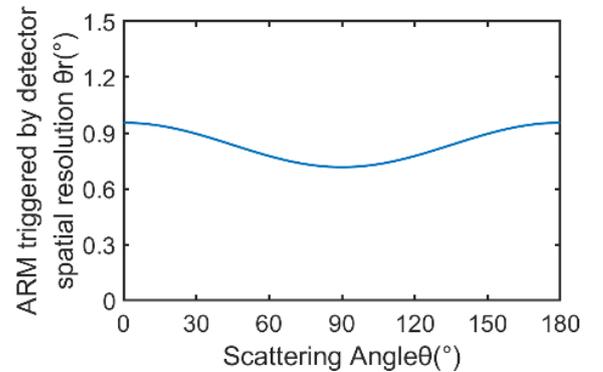

Figure5 $\triangle\theta_r$ as a function of Scattering Angle $\theta$

Similarly, calculate the weighted average of $\Delta\theta_r$ in the range of 20°~80° according to Eq. (12), and the calculation results are shown in Table2.

Table2 The results of the average $\Delta\theta_r$ for 4 different elements

| Incident XF energy/keV | I (28.620) | Ba (32.191) | Gd (42.983) | Au (68.794) |
|---|---|---|---|---|

| $\Delta\theta_r/°$ | 0.8142 | 0.8143 | 0.8148 | 0.8160 |
|---|---|---|---|---|

### 3.3 Calculate $\Delta\theta_D$

In this section we calculate the scattering angle uncertainty caused by Doppler broadening when XF photons of four elements are incident on Si detector. According to Eq. (9) and (7), the curve how $\Delta\theta_D$ varies with the scattering angle $\theta$ is shown in Figure6. It is worth noting that in Eq.(9) $\Delta p_z$ is constant for specific material. For silicon, $\Delta p_z/m_e c = 9.50 \times 10^{-3}$. Result of the average value of $\Delta\theta_D$ according to Eq. (12) is shown in Table3.

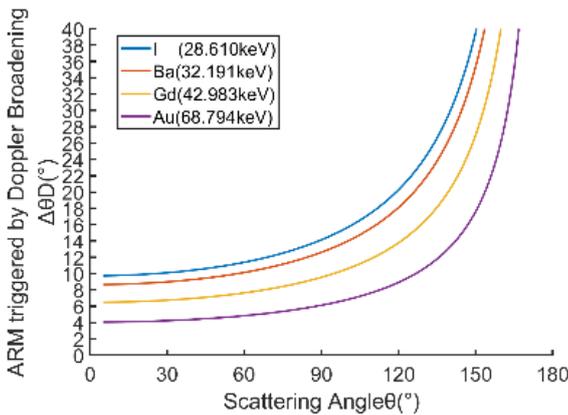

Figure6 $\Delta\theta_D$ as a function of Scattering Angle $\theta$ for four XF elements

Table3 The results of the average $\Delta\theta_D$ for 4 different elements

| Incident XF energy/keV | I (28.620) | Ba (32.191) | Gd (42.983) | Au (68.794) |
|---|---|---|---|---|
| $\Delta\theta_D/°$ | 11.3229 | 10.0779 | 7.5806 | 4.7867 |

### 3.4 Calculate $ARM$

Combining all the calculation results and Eq. (2), we can calculate ARM upper-limit of CC imaging under 4 types of 20-80keV low-energy XF photon. The results are shown in Table4. Table4 also shows SR value corresponding to ARM at a typical imaging detection distance of 10 cm.

Table4 ARM upper-limit of CC imaging for 4 elements

| Incident XF energy/keV | I (28.620) | Ba (32.191) | Gd (42.983) | Au (68.794) |
|---|---|---|---|---|
| $ARM/°$ | 12.197 | 10.743 | 7.9424 | 4.9747 |
| SR/mm | 21.62 | 18.97 | 13.95 | 8.70 |

## 4 Discussion and Conclusion

In this paper, four most commonly used XF elements, I, Ba, Gd and Au, are selected to explore the upper-limit of the ARM of CC for 20-80keV XF photon imaging through theoretical analysis. From the results, with the increase of incident photon energy, the angular error caused by energy resolution $\Delta\theta_E$ and Doppler broadening $\Delta\theta_D$ will decrease, while the angular error caused by spatial resolution $\Delta\theta_r$ has little relationship with incident photon energy. As for the results of ARM as shown in Table4, with the incident light energy increasing, ARM will gradually decrease. Among the four elements, Au has the best performance 4.97°, which corresponding to SR 8.70mm at the detection distance of 10cm. So Au has a certain imaging potential.

As a conclusion, Compton imaging has very limited feasibility under the low-energy incidence of 20-80keV fluorescent photons. Because the negative impact of the Doppler broadening effect is too significant, which makes the angular and spatial resolution unsatisfactory. Through analysis, we believed that the feasibility of XFCC imaging modality is not very well.

**Acknowledgements:**
This work was partially funded by grants from the National Key Research and Development Program of China 2018YFC0115502, and NNSFC 11775124.
Corresponding Author: lliang@tsinghua.edu.cn

# 90°-plus-90° DECT imaging

Buxin Chen[1], Zheng Zhang[1], Dan Xia[1], Emil Sidky[1], and Xiaochuan Pan[1,2]

[1]Department of Radiology, The University of Chicago, Chicago, USA
[2]Department of Radiation and Cellular Oncology, The University of Chicago, Chicago, USA

**Abstract** In this work, we investigate the dual-energy CT imaging with limited-angular-range data. A scan configuration is set up with two $\sim 90°$ arcs of low- and high-kVp spectra that are next to each other. Low- and high-kVp images of a digital phantom containing different materials are reconstructed separately from their respective $\sim 90°$ data of limited-angular range by use of the directional total-variation (DTV) algorithm previously developed. The images are assessed visually for artifacts reduction, and decomposed into interaction-type bases for estimating the effective atomic number. The results suggest that the DTV algorithm can effectively reduce the artifacts in reconstructed images from the limited-angular-range data. The estimated values of the effective atomic number from the decomposed dual-energy images reconstructed from the limited-angular-range data by use of the DTV algorithm are comparable to those from the full-scan data by use of the standard FBP algorithm.

## 1 Introduction

Dual-energy CT (DECT) imaging is able to correct for beam-hardening artifacts, improve image contrast, and differentiate different materials by collecting data from two spectrally distinct scans [1]. A standard method to acquire and process the dual-energy data, commonly referred to as the image-domain decomposition method, takes two full-rotation data sets with low- and high-kVp X-ray spectra for each set, reconstructs them separately, and accesses the dual-energy information with image post-processing steps [2]. As a result, the reconstruction step for either the low- or high-kVp scan data is no different than that for the conventional CT scan data.

Meanwhile, CT imaging with limited-angular-range data is of high level of interest in clinical and industrial applications, for its ability to reduce radiation dose, increase scanning throughput, and avoid collision. Considering the severe ill-conditionedness of the reconstruction problem from the limited-angular-rage data, recent works have used optimization-based reconstructions with different designs of objective functions and image constraints [3–6]. We have previously developed a directional total-variation (DTV) algorithm for image reconstruction from limited-angular-range data, based on an optimization problem with separate constraints along two orthogonal axes in the image array [7, 8]. The DTV algorithm has been demonstrated to reduce artifacts in images, especially for very small angular ranges, and also yield smaller minimum angular range than the isotropic TV method in terms of accurately inverting the discrete X-ray transform (DXT) data model [3]. In this work, we aim to bring the potential of reducing the angular range to DECT imaging by applying the DTV algorithm.

In this work, the low- and high-kVp images are reconstructed separately from their corresponding limited-angular-range data of low- and high-kVp scans. As a result, the limited-angular ranges can, but do not necessarily need to, overlap, and can also cover an arbitrary range of degrees. In this work, we present results from one particular type of scan configurations of interest, as they have been previously investigated and reported [9, 10]. It consists of two arcs of $\sim 90°$, that are next to each other either with or without a gap in between, for low- and high-kVp scans. In our approach, no effective extrapolation from limited-angular range into full- or short-scan range is needed for either kVp data set. The image reconstruction from each kVp data set, despite being of limited-angular range, is carried out individually and separately by use of the DTV algorithm.

Numerical studies with a digital suitcase phantom are carried out, focusing on industrial applications such as luggage screening. Dual-energy data with limited-angular ranges of $\sim 90°$ for each kVp scan, as well as full-scan data with two full rotations, are generated using a non-linear data model considering polychromatic X-ray spectra. Reconstructed images from limited-angular-range data are assessed with visual inspection and quantitative metrics, and compared against the reference images reconstructed from the full-scan data.

## 2 Materials and Methods

### 2.1 Data generation

The digital phantom, as shown in Fig. 1a, is designed as a suitcase containing four elliptical features, representing water, ANFO (a type of explosives), PVC, and Teflon, as well as three calibration bar structures of elements C, Al, and Ca. It is discretized on an image array of size $175 \times 256$, where each pixel is labeled with a specific material type associated with a spectral response, i.e., the linear attenuation coefficient as a function of energy. The image shown in Fig. 1a, for example, represents the linear attenuation coefficients for each material at 40 keV, i.e., the monochromatic image of the suitcase phantom at 40 keV.

Schematic drawings of the scanning configurations studied in this work is also shown in Fig. 1. Data are collected over two $\sim 90°$ circular arcs that are next to each other. For the case without a gap in between two arcs, as shown in Fig. 1b, the limited-angular ranges covering the low- and high-kVp scans are 90° each, i.e., $\alpha_1 = \alpha_2 = 90°$. For the case with a gap of 5°, as shown in Fig. 1c, the limited-angular ranges are

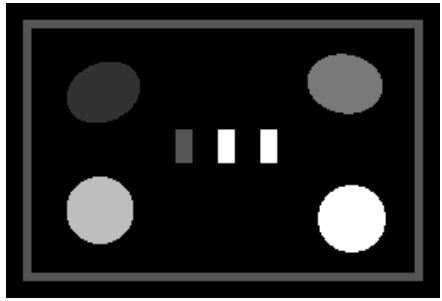

**(a)** suitcase phantom

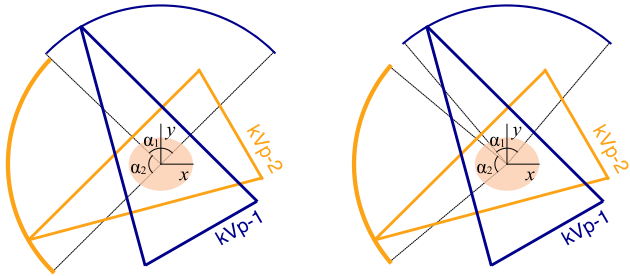

**(b)** 90-plus-90 scan w/o a gap    **(c)** 90-plus-90 scan with a 5° gap then 85° each, i.e., $\alpha_1 = \alpha_2 = 85°$. The two configurations simulate an instantaneous or 5°-delayed operation of switching the kVp [11], and are thus referred to as the *90-plus-90 scans* without and with a gap, respectively. As a reference, a scanning configuration with two full rotations, i.e., 360° angular ranges for both kVp scans, are also generated and referred to as the *full scan*. With an angular interval of 1° per view, 360° or 90° scans collect 360 or 90 views in the data, respectively. Other scanning geometry parameters include source-to-rotation and source-to-detector distances of 100 and 150 cm, respectively, and a linear detector of 32 cm consisting of 512 bins. The image array is set up such that the $x$- and $y$-axes divides evenly the high- and low-kVp scan angular ranges, respectively.

**Figure 1:** Top (a): the monochromatic image of the suitcase phantom at 40 keV, with a displaying window of [0.15, 0.75] $cm^{-1}$; Bottom: schematic drawings of the 90-plus-90 scanning configurations without a gap (b) and with a 5° gap (c) between the two arcs of low- and high-kVp scans.

The dual-energy data are generated using the non-linear, energy-integrated data model [12], with the 80 and 140 kVp spectra generated using the TASMIC model with an additional 5-mm Al filter [13]. A series of monochromatic images of the phantom at different energy levels are generated from the material label map of the phantom and used for forward projection with the DXT, followed by negative exponential transform, energy integration with the normalized X-ray spectrum, and negative logarithmic transform. As a linear data model is considered in the reconstruction with either the FBP or the DTV algorithm, the generated data in this work necessarily contain inconsistencies including the beam-hardening effect and possibly decomposition errors with a limited number, i.e., 2 in this work, of basis functions used in the image post-processing.

## 2.2 Image reconstruction

Images are first reconstructed by use of the standard FBP algorithm. The FBP algorithm is applied with a Hanning kernel and a cutoff frequency of 0.5. It is used in this work for providing a benchmarking reference with the full-scan dual-energy data and for demonstrating typical limited-angular-range artifacts in the image, if not accounted for.

Image are then reconstructed by use of the DTV algorithm, which has been previously proposed and developed for conventional CT reconstruction with limited-angular-range data [7, 8]. The DTV algorithm is developed based on the first-order primal-dual (PD) algorithm [14, 15]. The detailed derivation and its pseudo-codes of the algorithm can be found in Ref. [8], while a brief summary is provided as below. The DTV algorithm is based on a linear data model, $\mathbf{g} = \mathcal{H}\mathbf{f}$, and a constrained optimization problem with a data-$\ell_2$-distance as its objective function to minimize and two directional TV constraints, together with an image non-negativity constraint. The directional TVs are $\ell_1$ norms of image's partial derivatives along the $x$ and $y$ directions, and they are upper-bounded by two constraint parameters, referred to as $t_x$ and $t_y$, forming two inequality constraints. The constrained optimization problem is adapted into the framework of the general PD algorithm and then solved by deriving an instance, particularly by solving analytically the proximal-mapping problems, of the PD algorithm. Just like any other algorithms, the DTV algorithm involves parameters, most importantly, the DTV constraints parameters $t_x$ and $t_y$, which have impact on the reconstruction results. We will report at in the conference how these parameters are selected.

In this work, for either the FBP or DTV algorithm, it is applied separately and directly to each of the low- and high-kVp data with limited-angular-ranges, thus reconstructing two individual images of low- and high-kVp scan, referred to as low- and high-kVp images, or $\mathbf{f}^L$ and $\mathbf{f}^H$, respectively.

## 2.3 Image post-processing

### 2.3.1 Basis decomposition

For the two images, $\mathbf{f}^L$ and $\mathbf{f}^H$, an interaction-based decomposition model is used to decompose the images into basis images of Photoelectric effect (PE) and Compton scattering (KN), as below

$$\begin{pmatrix} \mathbf{f}^L \\ \mathbf{f}^H \end{pmatrix} = \begin{pmatrix} \mu^L_{PE}, \mu^L_{KN} \\ \mu^H_{PE}, \mu^H_{KN} \end{pmatrix} \begin{pmatrix} \mathbf{b}_{PE} \\ \mathbf{b}_{KN} \end{pmatrix}, \qquad (1)$$

where $\mathbf{b}_{PE}$ and $\mathbf{b}_{KN}$ are the basis images of Photoelectric effect and Compton scattering, and $\mu^s_k$ ($s = L$ or $H$, $k = $ PE or KN) are effective linear attenuation coefficients of basis PE or KN for the low-(L) and high-kVp (H) spectra. While the spectral responses of the two basis interaction types, PE and KN, are known to be $1/E^3$ and the Klein-Nishina formula [16, 17], respectively, the effective linear attenuation attenuation coefficients used in this work are the spectrum-weighted average of the spectral responses. We can then obtain the basis images from the low- and high-kVp images by inverting the $2 \times 2$ matrix in Eq. (1).

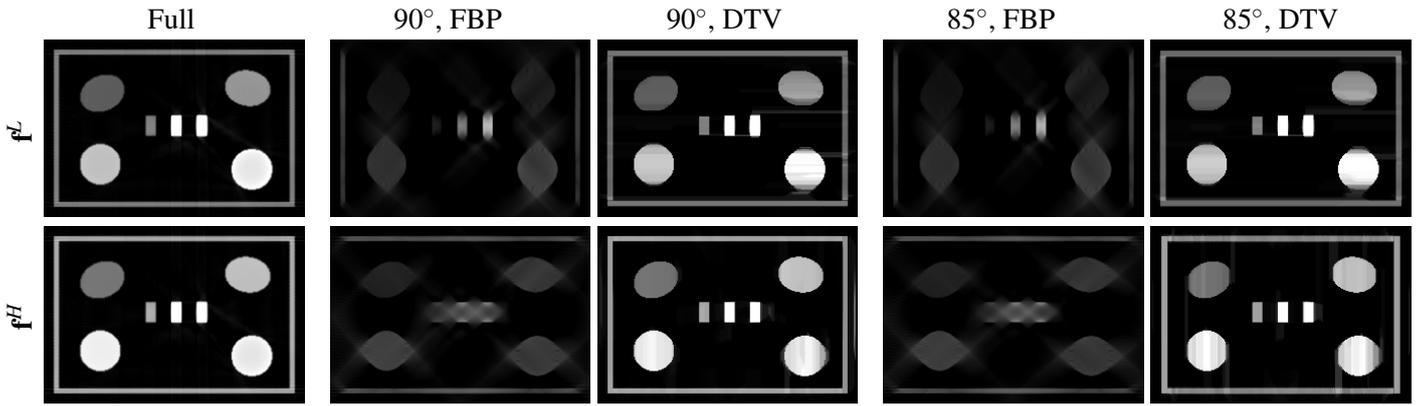

**Figure 2:** Reconstructed low- (row 1) and high-kVp (row 2) images from full-scan data (column 1), 90-plus-90-scan data without a gap and with a 5° gap by use of the FBP algorithm (columns 2 & 4) and by use of the DTV algorithm (columns 3 & 5). The displaying windows are [0, 0.65] $cm^{-1}$ and [0,0.43] $cm^{-1}$ for rows 1 and 2, respectively.

**Table 1:** Estimated effective atomic numbers for the materials in the suitcase phantom. The standard values are either the true atomic number for the elements or calculated using Murty's formula for the compounds.

|                | Carbon | Aluminum | Calcium | Water | ANFO | Teflon | PVC   |
| -------------- | ------ | -------- | ------- | ----- | ---- | ------ | ----- |
| Standard value | 6.00   | 13.00    | 20.00   | 7.42  | 7.40 | 8.43   | 13.86 |
| Full, FBP      | 6.04   | 12.79    | 20.20   | 7.28  | 7.22 | 8.19   | 13.72 |
| 90°, DTV       | 5.89   | 13.91    | 19.03   | 7.39  | 7.02 | 8.01   | 13.24 |
| 85°, DTV       | 5.88   | 14.14    | 18.77   | 6.71  | 6.32 | 7.47   | 12.80 |

### 2.3.2 Effective atomic number estimation

An important application for dual-energy CT imaging in luggage screening is to identify different materials, especially with explosive detection, by estimating the effective atomic number of materials. It can be estimated by exploring the dependence on the atomic number, $Z$, of the basis components in the interaction-based decomposition model [16, 18], as

$$Z = K(\frac{b_{PE}}{b_{KN}})^{1/n},  \qquad (2)$$

where $K$ and $n$ are two coefficients that can be theoretically defined and approximated. Alternatively, in this work we choose to calibrate $K$ and $n$ with a linear fitting in the log-log domain using the three calibration slabs of C, Al, and Ca ($Z = 6, 13,$ and $20$, respectively) in the suitcase phantom. The estimated $K$ and $n$ are then used to estimate the effective atomic numbers of the other compounds, namely, water, ANFO, teflon, and PVC, in the phantom.

## 3 Results

### 3.1 Image results

We show in Fig. 2 reconstructed low- and high-kVp images from the 90-plus-90-scan data without and with a 5° gap by use of the FBP and DTV algorithms, as well as the reference images from the full-scan data by use of the FBP algorithm. It can be observed that the limited-angular-range artifacts in the FBP images, mostly notably the "invisible boundaries" [19] that are parallel to the source's scanning directions, are effectively compensated and corrected for in the DTV images, which are visually similar to the reference images from the full-scan data, except for some minor remaining artifacts. Moreover, the displayed gray-scale values in the DTV images are closer to those in reference images, indicating comparable quantitative accuracy. The reconstructed images from the full-scan data by use of the DTV algorithm are not shown here, as they are visually similar to the reference FBP images.

### 3.2 Estimation of effective atomic numbers

We show in Table 1 estimated values of the effective atomic numbers for the 7 different materials in the suitcase phantom. The standard values are either the true atomic numbers for the elements or calculated using Mutry's formula [20] for the compounds. It can be observed that the estimated effective atomic numbers from the 90-plus-90-scan data by use of the DTV algorithm are comparable to those from the reference images. Admittedly, biases exist, as compared to the standard values, in the estimated values from the 90-plus-90-scan images, as well as the reference images. They are likely due to the beam-hardening effect in the data. Further, for the task of explosive detection, while the estimated effective atomic numbers of ANFO from the full- or 90-plus-90-scan data are not the same as the standard value, they are sufficiently different from those of, and thus can be differentiated from, water. On the other hand, the FBP images of 90-plus-90 scans are filled with negative values, which led to numerical errors in the estimation, and are thus not shown in the table.

## 4  Discussion and conclusion

In this work, we have investigated dual-energy CT imaging with limited-angular-range data by applying the DTV algorithm previously developed. Data were collected from two $\sim 90°$ arcs that are next to each other either with or without a gap, and reconstructions were performed separately and directly on each kVp data set of limited-angular range. Numerical studies with a suitcase phantom were carried out, and the evaluation is based on image visualization and the estimation of effective atomic numbers. Results have suggested that the DTV algorithm can reconstruct dual-energy images that are visually similar to the reference images from full-scan data and also obtain estimated values of the effective atomic numbers that are comparable to those from the reference images.

Some beam-hardening artifacts can be observed in the DTV images from the 90-plus-90-scan data, as well as the reference images, since they are not accounted for in the data model used in the reconstruction in this work. Such inconsistency is likely the source of discrepancies between the standard and estimated values of the effective atomic number. Further investigation will be focused on correcting for the beam-hardening effect in the image reconstruction, e.g., using the data-domain decomposition or the one-step reconstruction method.

In this work, as the reconstructions are performed separately and directly from the limited-angular-range data from either low- or high-kVp scan, the two scanning arcs for the low- and high-kVp spectra do not need to be next to each other or within a gap, as demonstrated in the 90-plus-90 scanning configurations. We have carried out additional studies with other scanning configurations for DECT with limited-angular-range data, e.g., two overlapping arcs, and also studies with more angular ranges, especially those considerably smaller than $90°$, and will report these results in the conference.

## 5  Acknowledgments

This work is supported in part by NIH R01 Grant Nos. EB026282 and EB023968, and the Grayson-Jockey Club Research.

# Symmetric-Geometry CT with Linearly Distributed Sources and Detectors in a Stationary Configuration: Projection Completion and Reconstruction ROI Expansion

Tao Zhang[1,2], Zhiqiang Chen[1,2], Li Zhang[1,2], Yuxiang Xing[1,2], and Hewei Gao *[1,2]

[1]Department of Engineering Physics, Tsinghua University, Beijing 100084, China

[2]Key Laboratory of Particle & Radiation Imaging (Tsinghua University), Ministry of Education, Beijing 100084, China

**Abstract** With linearly distributed sources and detectors in a stationary configuration, the symmetric-geometry computed tomography (SGCT) has great potential for fast CT imaging. In this work, we propose a projection completion approach to suppress the truncation artifacts and expand the reconstruction region of interest (ROI) for both filtered backprojection (FBP) and Linogram reconstruction methods in a dual-SGCT scan mode. The preliminary results of numerical simulations and physical experiments validated the effectiveness and feasibility of our proposed methods. For the Catphan phantom reconstruction with a truncation problem, our proposed completion method can decrease the relative standard deviation of uniform regions by 40.4% for the FBP algorithm, and by 38.3% for the Linogram algorithm.

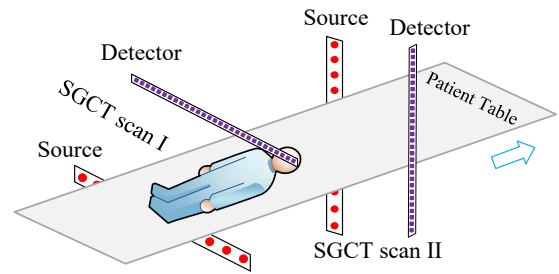

**Figure 1:** The diagram of the daul-SGCT scan mode. The arrow indicates the direction of the patient table movement.

## 1 Introduction

The concept of multi X-ray sources is being investigated in computed tomography (CT) system to accelerate data acquisition [1]. Recently, a symmetric-geometry computed tomography (SGCT) has been presented, where the X-ray sources and detectors are both linearly distributed [2, 3]. In the object scanning of SGCT, the X-ray sources distributed equally along a straight-line will sequentially fire in one side, and the detectors continuously collect data in the opposite side. As a result, no spinning of X-ray source or detector is involved in the data acquisition of SGCT, which simplifies the system design and equipment manufacturing. In addition, with such a stationary configuration, SGCT has great potential for fast CT imaging.

In practical applications, with finite length of source and detector arrays, only one SGCT scan segment is not enough. Therefore, dual-SGCT scan mode was introduced to supply sufficient projection data. As illustrated in Fig. 1, there are two SGCT scan segments (i.e., SGCT scan I and SGCT scan II) in the dual-SGCT scan mode. Each SGCT scan segment contains a linearly distributed source array and a linear detector array parallel to it. And the source array of the SGCT scan I is perpendicular to that of the SGCT scan II. A filtered backprojection (FBP) reconstruction method [2] and a Linogram reconstruction method [3] were both derived to achieve image reconstruction from the scanning data of SGCT. The fourier transform of the projection along the detector direction is needed for both FBP and Linogram methods. Thus, the constrain that the projection data is not truncated along the detector direction should be satisfied to obtain accurate reconstructed images, which limits the size and location of the reconstruction region of interest (ROI) of dual-SGCT scan mode.

In this work, we propose a projection completion approach to suppress the truncation artifacts and expand reconstruction ROI for both FBP and Linogram methods in dual-SGCT scan mode. The preliminary results of numerical simulations and physical experiments validate the effectiveness and feasibility of our proposed methods.

## 2 Methods

When using FBP or Linogram methods for reconstruction from CT projection data, to obtain accurate reconstructed image, the size and location of reconstruction ROI is usually limited by the following two constrains:

- CONS-I: *The passing rays cover at least 180 degrees for each and every points of the scanned object;*

- CONS-II: *No truncation occurs along the filtering direction of the projection data.*

The CONS-I is the data sufficiency constrain. With the lengths of the source and detector array being finite in real applications, only one SGCT scan segment cannot satisfy the CONS-I as expected. Fortunately, the CONS-I can be met in the introduced dual-SGCT scan mode if carefully designed. Specifically, the SGCT scan I is responsible for supplying $-45^0 \sim 45^0$ projection data, and the SGCT scan II is for $-135^0 \sim -45^0$. The CONS-II is to avoid truncation artifacts,

---

which will further reduce the size of the reconstruction ROI after meeting the CONS-I.

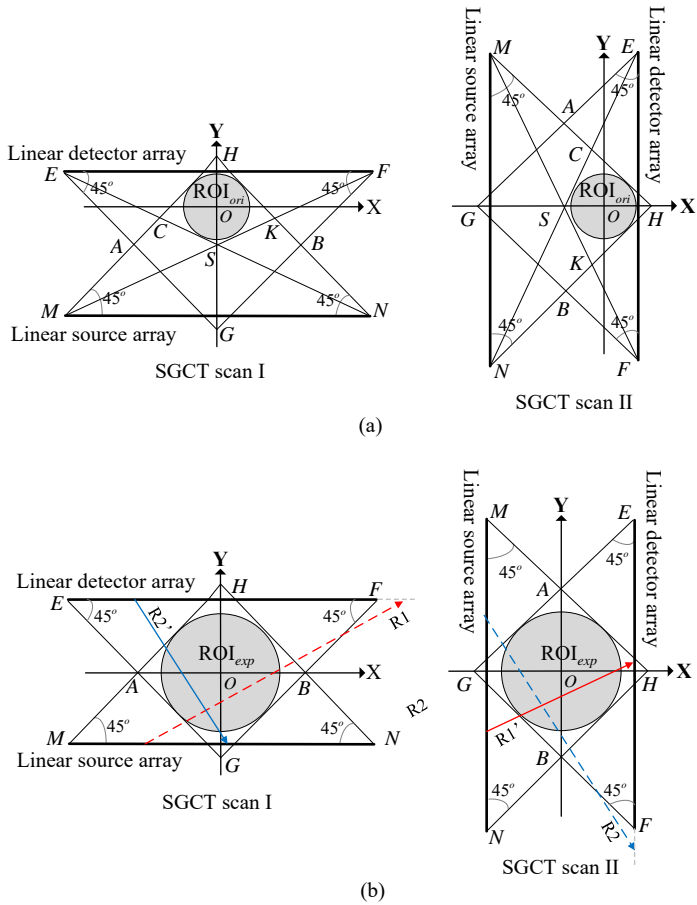

**(a)**

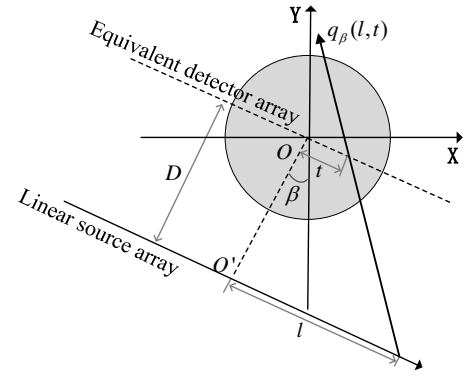

**(b)**

**Figure 2:** Reconstruction ROI analysis for the dual-SGCT scan mode. (a) The maximal circular reconstruction ROI meeting both the CONS-I and CONS-II, denoted as $\mathrm{ROI}_{ori}$; (b) the maximal circular reconstruction ROI only meeting the CONS-I, denoted as $\mathrm{ROI}_{exp}$.

In Fig. 2 (a), we demonstrate the maximal circular reconstruction ROI for dual-SGCT scan mode when using FBP or Linogram algorithms. The reconstruction ROI should locate in the quadrilateral region $AHBG$ for meeting the CONS-I and in the triangle region $ESF$ for the CONS-II. Thus the maximal circular reconstruction ROI is identically the largest inscribed circular of the intersection between the quadrilateral region $AHBG$ and the triangle region $ESF$, which is denoted as $\mathrm{ROI}_{ori}$.

However, when only taking the CONS-I into consideration but ignoring the CONS-II, the maximal circular reconstruction ROI can be expanded to $\mathrm{ROI}_{exp}$ as shown in Fig. 2(b). The $\mathrm{ROI}_{exp}$ is the largest inscribed circular of the quadrilateral region $AHBG$, of which the size greatly exceeds that of $\mathrm{ROI}_{ori}$. It is worth noting that the centers of $\mathrm{ROI}_{ori}$ and $\mathrm{ROI}_{exp}$ are different. Thus the relative spatial position of SGCT scan I and SGCT scan II should be adjusted according to the reconstruction ROI center in real implementation of the dual-SGCT scan. A projection completion method was proposed in this work to obtain accurate reconstructed image for $\mathrm{ROI}_{exp}$ through FBP or Linogram methods while avoiding truncation artifacts.

The imaging geometry of a SGCT scan is defined in Fig. 3. The center of the object coordinates (X,Y) is denoted as $O$. An equivalent detector array parallel to the real one is introduced, which passes through the object center $O$. The element on the equivalent detector is indexed by $t$, its offset from $O$. The projection of the object center $O$ on the source array is $O'$, and we use the offset from $O'$ (i.e., denoted as $l$) to represent the element of the source array. And $D$ is the distance from the source array to the equivalent detector array. $\beta$ is the overall scanning angle of a SGCT scan, which is the angle between the line $OO'$ and the axis Y. Thus, the projection from a SGCT scan of the overall scanning angle being $\beta$ is defined as $q_\beta(l,t)$, with the corresponding elements on the source array and equivalent detector array being $l$ and $t$, respectively. According to the geometry relation, projection $q_\beta(l,t)$ can be written from the object function $f(x,y)$ [3],

$$q_\beta(l,t) = \int_{-\infty}^{+\infty}\int_{-\infty}^{+\infty} \mathrm{d}x\mathrm{d}y \, \frac{\sqrt{(l-t)^2+D^2}}{x\sin\beta+y\cos\beta+D} f(x,y)$$
$$\times \delta\left(\frac{(x\cos\beta-y\sin\beta)D+(x\sin\beta+y\cos\beta)l}{x\sin\beta+y\cos\beta+D}-t\right). \quad (1)$$

In the dual-SGCT scan mode, the overall scanning angle $\beta$ is set to 0 and $\frac{\pi}{2}$ for the SGCT scan I and II, respectively. Therefore, according to Eq. (1), the two projection segments in dual-SGCT scan mode be written as, respectively,

$$q_\mathrm{I}(l_1,t_1) = \int_{-\infty}^{+\infty}\int_{-\infty}^{+\infty} \mathrm{d}x\mathrm{d}y \, \frac{\sqrt{(l_1-t_1)^2+D^2}}{y+D} f(x,y)$$
$$\times \delta\left(\frac{xD+yl_1}{y+D}-t_1\right). \quad (2)$$

$$q_\mathrm{II}(l_2,t_2) = \int_{-\infty}^{+\infty}\int_{-\infty}^{+\infty} \mathrm{d}x\mathrm{d}y \, \frac{\sqrt{(l_2-t_2)^2+D^2}}{x+D} f(x,y)$$
$$\times \delta\left(\frac{xl_2-yD}{x+D}-t_2\right). \quad (3)$$

Here, $q_\mathrm{I}(l_1,t_1)$ and $q_\mathrm{II}(l_2,t_2)$ are the projection from SGCT scan I and II. The combined projection of SGCT scan I and

**Figure 3:** Imaging geometry a SGCT scan segment. The $\beta$ is the overall scanning angle, which is different for different SGCT scan segments.

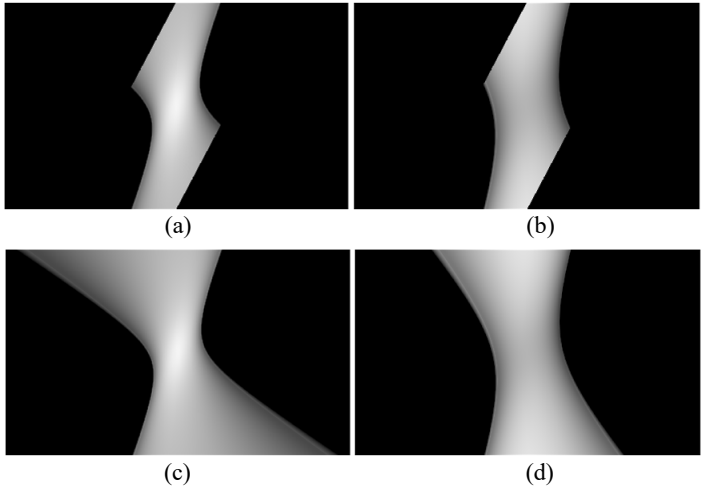

**Figure 4:** Projection of Shepp-Logan head phantom from dual-SGCT scan mode. (a) and (b) are the projection data obtained from SGCT scan I and II respectively, where truncation occurs along the detector direction; (c) and (d) are the completed projection for SGCT scan I and II using our proposed method.

II is sufficient for every points in ROI$_{exp}$ due to the CONS-I. Thus, for the ROI$_{exp}$ in the dual-SGCT scan mode, the projection ray truncated in a SGCT scan can be completed using the complementary one in another SGCT scan. According to the Eqs. (2) and (3), the projection completion method for ROI$_{exp}$ in dual-SGCT scan mode can be described as follows,

1) When the projection is truncated in SGCT scan I, completion is done using the complementary one in SGCT II through Eq. (4);

$$q_{\mathrm{I}}(l_1, t_1) = q_{\mathrm{II}}(l_2, t_2) \begin{cases} l_2 = \frac{-(D+t_1)D}{l_1 - t_1} \\ t_2 = \frac{-t_1 D}{l_1 - t_1} \end{cases} \quad (4)$$

2) When the projection is truncated in SGCT scan II, completion is done using the complementary one in SGCT I through Eq. (5);

$$q_{\mathrm{II}}(l_2, t_2) = q_{\mathrm{I}}(l_1, t_1) \begin{cases} l_1 = \frac{(t_2 - D)D}{l_2 - t_2} \\ t_1 = \frac{t_2 D}{l_2 - t_2} \end{cases} \quad (5)$$

For example, as shown in Fig. 2(b), one can utilize the existing projection ray R1' in SGCT scan II to complete the truncated projection ray R1 in SGCT scan I, and use the existing projection ray R2' in SGCT scan I to complete the truncated projection ray R2 in SGCT scan II.

## 3 Results

### 3.1 Simulation study

The dual-SGCT scan mode was simulated with the parameters listed in Table 1. With the same length of source array and detector array, and the same distance from source array to detector array, the ROI$_{ori}$ has a radius of 98.73 mm which satisfies the CONS-I and CONS-II as analyzed in Ref. [2]. However, when only satisfying the CONS-I, the reconstruction ROI can be expanded to ROI$_{exp}$ with a radius of 176.78 mm. To simulate such expanded ROI, the distance from source to isocenter need to be 250 mm. The 2D Shepp-Logan head phantom was used, of which projections from simulated SGCT scan I and II was demonstrated in Fig. 4 (a) and (b), respectively. The truncation occurs along the detector direction of the projection, which can be completed using our proposed method. And the projection after completion are shown in Fig. 4 (c) and (d). The original FBP method in Ref. [2] and Linogram method in Ref. [3] were both implemented to reconstruct images from dual-SGCT scan. Also the above two methods were combined with our proposed projection completion method, denoted as FBP* and Linogram*, respectively. It is seen form the results in Fig. 5 and profiles in Fig. 6 that, there are severe truncation artifacts in the reconstructed images without projection completion techniques, while our proposed projection completion method can greatly improve the image quality for both FBP and Linogram reconstructions.

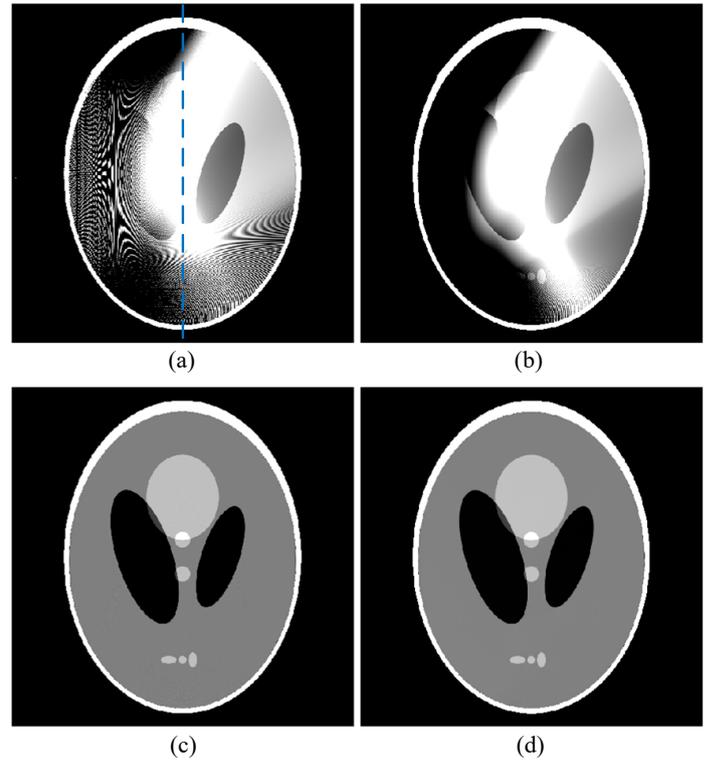

**Figure 5:** Reconstructed images of the Shepp-Logan head phantom. Reconstruction size: 512 x 512 pixels with 0.625 x 0.625 mm$^2$ for a pixel. (a) and (b) are the FBP and Linorgam reconstruction from dual-SGCT scan, respectively, with no projection completion techniques applied; (c) and (d) are the FBP* and Linorgam* reconstruction from dual-SGCT scan, respectively, where our proposed projection completion method were used.

**Table 1:** Simulation Experiments parameters for the dual-SGCT scan mode

|  | Parameters | Values |
|---|---|---|
|  | Distance from source to detector | 500mm |
|  | Distance from source to isocenter | 250mm |
| Geometry | Interval of source array | 0.5mm |
|  | Length of source array | 1000mm |
|  | Interval of detector array | 0.5mm |
|  | Length of detector array | 1000mm |
| Oveall scan angle | SGCT scan I | 0 |
|  | SGCT scan II | $\frac{\pi}{2}$ |

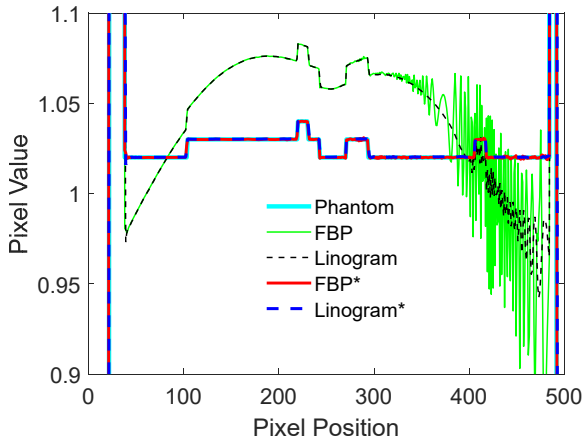

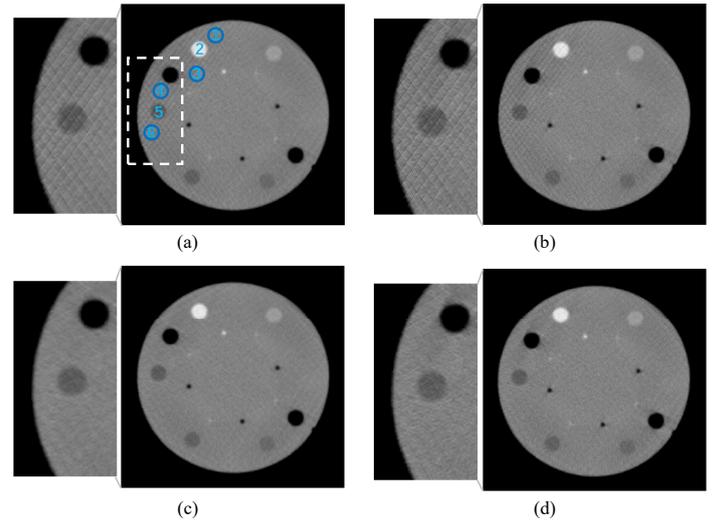

**Figure 6:** 1D profiles of of Fig. 5(a), (b), (c), (d) and the digitalized Shepp-Logan head phantom along the vertical line indicated in Fig. 5(a).

## 3.2  Real experiments

To further demonstrate the effectiveness of the projection completion method, a dual-SGCT scan was performed for the Catphan phantom on a SGCT prototype, where the truncation occurs in the reconstruction area where the phantom located. Fig. 7 are the reconstruction results of FBP and Linogram methods without and with the proposed projection completion. The FBP* and Linogram* methods can suppress truncation artifacts, surpassing the original FBP and Linogram methods. For a quantitative comparison, we selected six uniform regions indicated in Fig. 7(a), and calculated the relative standard deviation (*RSD*) of every regions for the reconstructed images in Table 2. In addition, a factor was used to quantitatively measure the improvement by our proposed projection completion approach, which is defined as follows,

$$f_{imp} = \frac{1}{N} \sum_{i=0}^{N} \frac{RSD_i - RSD_i^*}{RSD_i} \times 100\% \qquad (6)$$

here, $i$ is the index of the selected regions. $RSD_i$ represents the relative standard deviation of the region $i$ in the reconstructed images by the analytic method (i.e., FBP or Linogram) without the projection completion, while $RSD_i^*$ is that with the projection completion. The results in Table 2 indicate that our proposed completion method can decrease the relative standard deviation of uniform regions by 40.4% for the FBP algorithm, and by 38.3% for the Linogram algorithm.

**Figure 7:** Reconstructed images of the Catphan phantom scanned on an SGCT prototype. Reconstruction size: 350 x 350 pixels with 0.5 x 0.5 mm² for a pixel. Display window: [0.05, 0.3]/cm. (a) and (b) are the FBP and Linorgam reconstruction from dual-SGCT scan, respectively, with no projection completion techniques applied; (c) and (d) are the FBP* and Linorgam* reconstruction from dual-SGCT scan, respectively, where our proposed projection completion method were used. Dotted rectangle in (a) indicates the region of zoomed-in displays.

**Table 2:** Relative standard deviation of selected uniform regions of the Catphan phantom in Fig. 7.

| Regions | 1 | 2 | 3 | 4 | 5 | 6 | $f_{imp}$ |
|---------|-----|-----|-----|-----|-----|-----|------|
| FBP | 0.030 | 0.023 | 0.039 | 0.044 | 0.059 | 0.051 | \ |
| FBP* | 0.021 | 0.015 | 0.023 | 0.023 | 0.032 | 0.029 | 40.4% |
| Linogram | 0.037 | 0.030 | 0.048 | 0.054 | 0.070 | 0.064 | \ |
| Linogram* | 0.025 | 0.022 | 0.028 | 0.031 | 0.039 | 0.037 | 38.3% |

## 4  Discussion and Conclusion

In this work, a projection completion method is proposed for both FBP and Linogram methods in the dual-SGCT scan to suppress the truncation artifacts and expand the reconstruction ROI. The preliminary results of numerical simulations and physical experiments validate the effectiveness and feasibility of our proposed methods.

# Tabu-DART: a dynamic update strategy for the Discrete Algebraic Reconstruction Technique based on tabu-search

Daniel Frenkel, Jan De Beenhouwer, and Jan Sijbers

imec-Vision Lab, Universiteit Antwerpen, Antwerp, Belgium

**Abstract** In X-ray Computed Tomography (XCT), the Discrete Algebraic Reconstruction Technique (DART) has been proposed as a practical method for reconstructing images measured of an object that is composed of only a small number of different materials. For such objects, DART has shown the potential to reconstruct high quality images even in the case of a low number of radiographs or a limited angular range. To this end, DART follows a set of rules to enforce the material discreteness prior knowledge. However, these rules are static in that they remain unchanged throughout the entire reconstruction process, which limits the full potential of the DART concept. To increase flexibility during the reconstruction process, we introduce an update framework that dynamically adjusts update rules throughout the iterations. Our experiments show that such dynamic update strategy leads to increased reconstruction quality and lower computational burden.

## 1 Introduction

In X-ray Computed Tomography (XCT), prior knowledge about the object to be reconstructed is often exploited to improve the quality of images reconstructed from limited data. A specific class of prior knowledge is the assumption that the object consists of only a small number of different materials. The domain of Discrete Tomography (DT) studies algorithms that reconstruct objects adhering to this assumption. In 2011, the Discrete Algebraic Reconstruction Technique (DART) was proposed as a practical algorithm that provides high reconstruction quality in tomographic reconstruction problems with limited X-ray projection data [1]. Since then, many variations of the DART algorithm have been reported [2–7].

The DART algorithm iteratively interchanges a reconstruction step, where the image is updated by minimizing the projection distance, and a segmentation step, where the image pixels are classified into the few different material classes. However, the rules used by DART to attribute labels to the pixels to be updated, are rigid in the sense that they do not exploit knowledge gained about the intermediate reconstructed images throughout the iterations. This slows down the algorithm or causes it to converge to a local minimum [8].

To improve upon the rigid DART update rules, we propose a generalization of the DART update strategy by introducing a dynamic update probability map of the image throughout the reconstruction. We express update strategies as changes to the update probability map and we exploit the probability map sequence by using a tabu-search framework. We show that this approach improves both convergence speed and reconstruction quality.

## 2 Materials and Methods

### 2.1 The DART algorithm

DART assumes that the object to be scanned consists of a small number (typically $k < 5$) of different materials. Let $\{\rho_1 < ... < \rho_k\}$ be the gray values representing the different materials present in the object and $\mathbf{x} \in \mathbb{R}^n$ the representation of the pixel grid of attenuation values of the object. Given the measured projection data $\mathbf{p} \in \mathbb{R}^m$ and the system matrix $\mathbf{W} \in \mathbb{R}^{m \times n}$, the reconstruction problem comes down to solving the linear system

$$\mathbf{W}\mathbf{x} = \mathbf{p}, \quad \text{such that } \mathbf{x} \in \{\rho_1, .., \rho_k\}^n. \quad (1)$$

To this end, the following steps are performed in the DART algorithm: First, an initial reconstruction is calculated with the use of an Algebraic Reconstruction Method (ARM), such as ART, SART or SIRT [9]. Without loss of generality, we will use the SIRT algorithm as the ARM. The output vector is denoted as $\mathbf{x}^{(0)}$. Since the output of an ARM has continuous gray values, which violates the discreteness assumption, a segmentation step is performed to enforce discreteness. Similar to [1], we use a global thresholding step with the following mapping function:

$$S(\mathbf{x}, \rho) : \mathbb{R}^n \longrightarrow \{\rho_1, \rho_2, ..., \rho_k\}^n \ \mathbf{x} \to \mathbf{s},$$

$$s_i = \begin{cases} \rho_1, & x_i < \tau_1 \\ \rho_2, & \tau_1 \le x_i < \tau_2 \\ \vdots & \\ \rho_k, & \tau_{k-1} \le x_i, \end{cases} \quad i = 1, ..., n,$$

where the thresholds $\tau_j$ are calculated as

$$\tau_j = \frac{\rho_j + \rho_{j+1}}{2}, \ j = 1, ..., k-1. \quad (2)$$

The resulting discrete image is denoted as $\mathbf{s}^{(0)} = S(\mathbf{x}^{(0)})$.
Let $\mathbf{s}^{(\ell)}$ be the segmentation from the $\ell$-th iteration of DART. First, all pixels in $\mathbf{s}^{(\ell)}$ classified either as boundary or interior pixels. A pixel is considered interior when it belongs to the same material class as its neighbours. All other pixels are considered boundary pixels. Only the boundary pixels are updated in the next ARM iteration while interior regions are kept fixed. Let $(\mathbf{w}_1, ..., \mathbf{w}_n)$ be the columns of the system matrix $\mathbf{W}$. The boundary pixels are reconstructed on the

residual data:

$$\mathbf{W}_{(\ell)}\mathbf{x}^{(\ell)} = \mathbf{p} - \mathbf{w_i}s_i^{(\ell)}, \tag{3}$$

$$\mathbf{W}_{(\ell)} = (\mathbf{w_1}, ..., \mathbf{w_{i-1}}, \mathbf{w_{i+1}}, ..., \mathbf{w_n}), \tag{4}$$

$$\mathbf{x}^{(\ell)} = (x_1, ..x_{i-1}, x_{i+1}, ..., x_n)^T \tag{5}$$

To minimize the risk of local minima, each interior pixel also has a probability $p$ of being included in the next reconstruction step. Solving the reduced system (3) yields a new vector containing updated values for the boundary pixels and the fixed pixels that were randomly selected. Along with the fixed pixels, a new image $\mathbf{x}^{(\ell+1)}$ is computed. Finally, a smoothing operation is performed by convolving the image with a $3 \times 3$ mean kernel.

The above process is repeated until a convergence criterion is met or a predetermined maximum number of DART iterations is reached.

## 2.2 The Tabu-search concept

Tabu-search is a mathematical optimization method that employs a memory structure to improve local search methods. By manipulating adaptive memory structures, Tabu-search methods can reach parts of the solution space that would otherwise have been left unexplored by more traditional methods. There exist many variations that characterize the memory structure in Tabu-search [10]. However, one of specific interest for the DART update rules is frequency based memory. This variant contains and uses information on the amount of times a certain attribute has appeared in recent solutions. If the presence of a property is correlated to good solutions, then remembering search directions where many solutions with this property exist increases the probability of finding an optimal solution. There are various metrics that we can track about a reconstruction that change once the DART algorithm nears convergence. An example of that would be how many pixels still change their material class. By measuring the class change for each pixel individually, we essentially create a frequency based memory structure related to the material labeling of the image. We can exploit changes in this structure to adapt the DART algorithm update step. In this way, the solution guiding process becomes more refined over time.

In the next section, we will generalize the DART update step as a framework which uses a probability map to function as a frequency based memory structure for the update step inside the algorithm as shown in Figure 1. We will also describe an algorithm called Tabu-DART, which uses a dynamic set of rules to update the probability map. By changing the values of the probability map, we directly influence the frequency with which individual pixels are updated in the following iterations.

## 2.3 Tabu-DART: using a probability map to function as memory for DART

Tabu-search is a heuristic technique which uses the concept of memory to increase control of the solution space. We implemented this concept in the DART update step because it directly relates to both convergence speed and reconstruction quality of the DART algorithm. In [1], this step is based on a boundary criterion and a probability parameter $p$ for each pixel. Instead of one parameter $p$ describing the probability that an interior pixel is updated in the next iteration, Tabu-DART uses a map:

$$P : \mathbb{R}^n \longrightarrow [0,1]^n, \mathbf{x} \longrightarrow \mathbf{p_x} \tag{6}$$

As such, each pixel in the image has its own unique probability and for each pixel it is individually decided whether or not it is updated in the next iteration. The Tabu-DART can be summarized as follows:

1. After an initial segmentation, the probability map $\mathbf{p_x}^{(\ell)}$ is initialized.

2. During the partitioning step, a random value $r_i^{(\ell)}$ between 0 and 1 is generated for each pixel $x_i^{(\ell)}$. If $r_i^{(\ell)} \le p_{x_i}^{(\ell)}$, then the pixel is selected for update.

3. At the end of every DART iteration, a feedback step is added that updates the probability map based on changes between the new segmented image and the one found in the previous DART iteration. In this way, the probability map adapts quickly to changes in the reconstructed image.

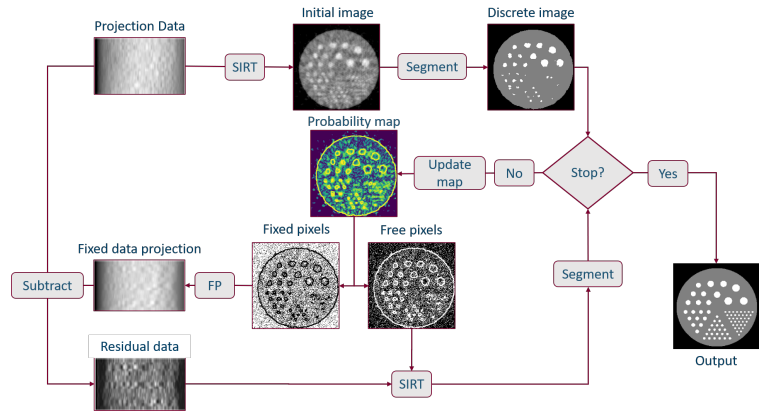

**Figure 1:** A flowchart of the Tabu-DART algorithm.

Note that this framework encloses the original DART algorithm [1] with random parameter $p$ and segmentation $s^{(\ell)}$ as follows:

$$p_{x_i}^{(\ell)} = \begin{cases} 1, & \text{if } s_i^{(\ell)} \text{ is boundary} \\ p, & \text{if } s_i^{(\ell)} \text{ otherwise} \end{cases} \quad i = 1, ..., n. \tag{7}$$

We decided on a different approach in Tabu-DART. Each pixel $x_j$ is linked to a probability vector of length $k$ representing the probabilities that $x_i$ belongs to each material class.

We denote this vector by $\mathbf{v}_{x_j}$. By using the entropy

$$\mathscr{H}(x_j) = -\mathbf{v}_{x_j}^T \log_2(\mathbf{v}_{x_j}), \qquad (8)$$

a single value representing uncertainty of the pixel $x_i$ can be calculated. Varga et al. [11] described a method to calculate the entropy for binary images. Let $y_j$ be the value of pixel $x_j$ as a result of the initial ARM iterations. Then, the probability vector $\mathbf{v}_{x_j}$ for a pixel $x_j$ is defined

$$\mathbf{v}_{x_j} = [y_j, 1 - y_j],$$

and this becomes the input into (8). However, this approach is applicable only to binary images. As to generalize it to more than two classes, we suggest the following for the output $\mathbf{y}$ from the ARM. Let

$$\mathbf{d}_{x_i} = \left[ \frac{1}{|y_i - \rho_0|}, \dots, \frac{1}{|y_i - \rho_k|} \right]$$

$$\mathbf{v}_{x_i} = \frac{\mathbf{d}_{x_i}}{\|\mathbf{d}_{x_i}\|_1}$$

The resulting vector $\mathbf{v}_{x_i}$ is input into (8) to yield a single value $\mathscr{H}(x_i)$ measuring uncertainty for the pixel $x_i$. These uncertainty values are used to initialize the probability map. The probability map update step is also different and based on three rules:

1. A pixel changing class during the last DART iteration, indicates that the uncertainty of which class it belongs to is still high. To ensure that the pixel will be updated again in the next iteration, its update probability is set to 1.

2. When a pixel did not change material classes compared to the last DART iteration, its corresponding update probability is halved instead. In this way, stable regions are iteratively removed from the reconstruction problem.

3. As was pointed out in [1], the boundary plays a key role as it holds the most uncertainty in the image. the update probability of each boundary pixel is set to 1 as in [1].

## 3   Experiments and Results

### 3.1   Simulation experiments

To evaluate the effect of the proposed dynamic update strategy on the reconstruction quality, we simulated projection datasets of a laminate profile phantom (Figure 2) with decreasing angular range with a geometry that represents the one used when scanning objects in the UAntwerp FlexCT scanner [12]. We assumed a monochromatic beam with fan-beam geometry with a phantom size of $200 \times 400$, a Source-Object-Distance (SOD) of 360 mm and a Source-Detector-Distance (SDD) of 90 mm. The voxel size was set to 0.120 mm. We varied the angular range from 40 to 140 degrees, with the number of projections taken varying from 20 up to 70. The simulation was performed with the ASTRA toolbox [13]. Simulated Poisson noise with an average photon count of 25000 was added to the projection data. The reconstruction was performed using both the DART and Tabu-DART algorithms described in Section 2. In addition, we implemented the ADART algorithm [2] and a variant of it employing the Tabu-DART based map update. The update step for ADART is given by:

$$P^k(x_i, s) = \begin{cases} 1 & \text{if } i \in B_s^k \\ p & \text{if } i \notin B_s^k, \end{cases}$$

where the boundary set $B_s^{(k)}$ changes over time. A total of 50 initial SIRT iterations were run, followed by 95 DART iterations. Following the original paper [1], each DART iteration contained a subroutine of 10 masked SIRT iterations, the value for $p$ for DART and ADART was set to 0.15, and the smoothing factor was set to 0.1. This amounts to 1000 SIRT iterations for each method. To counteract the effects of noisy data, a relaxation factor $\lambda$ was introduced to the SIRT algorithm in the following way:

$$\mathbf{x}^{(k+1)} = \mathbf{x}^{(k)} + \lambda \mathbf{CW}^\top \mathbf{R}(\mathbf{p} - \mathbf{Wx}^{(k)}), \qquad (9)$$

where each DART iteration $\lambda$ was set to the number of free pixels divided by the total number of pixels. To measure the performance of the methods, we calculate the number of misclassified pixels, denoted as the *pixel error*.

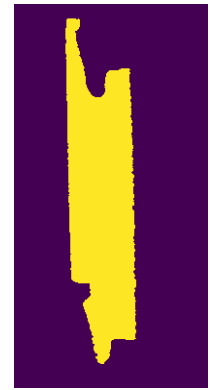

**Figure 2:** The Laminate phantom used in the experiment

### 3.2   Simulation results

Figure 3 shows the pixel error as a function of the angular range for the four DART methods described in the previous section and the average relaxation factor $\lambda$. These errors have been averaged over 50 repetitions with different seeds for the generation of the Poisson noise. We observe a lower pixel error for the algorithms based on Tabu-search compared to the original methods (DART and ADART) for each choice of angular range. The reconstructions are shown in Figure 4. The visual difference, however, is negligible. The relaxation factor in this experiment was set to reflect the number of freed pixels. These directly influence the computational cost

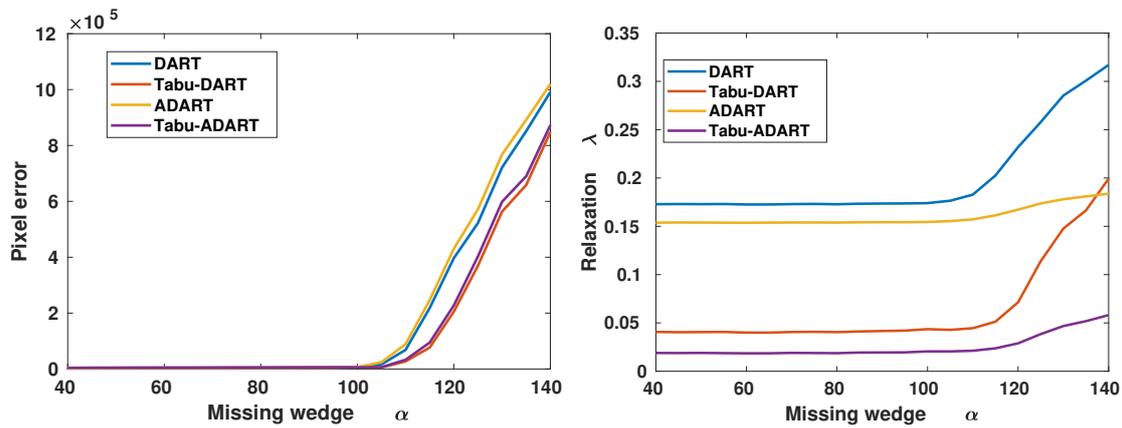

**Figure 3:** The pixel error as a function of the angular range for the laminate phantom (left) and the average relaxation factor (right). Tabu-DART and Tabu-DART start outperforming DART and ADART once the missing wedge becomes large. The relaxation factor represents the system size and indicates where the algorithm performance starts deteriorating.

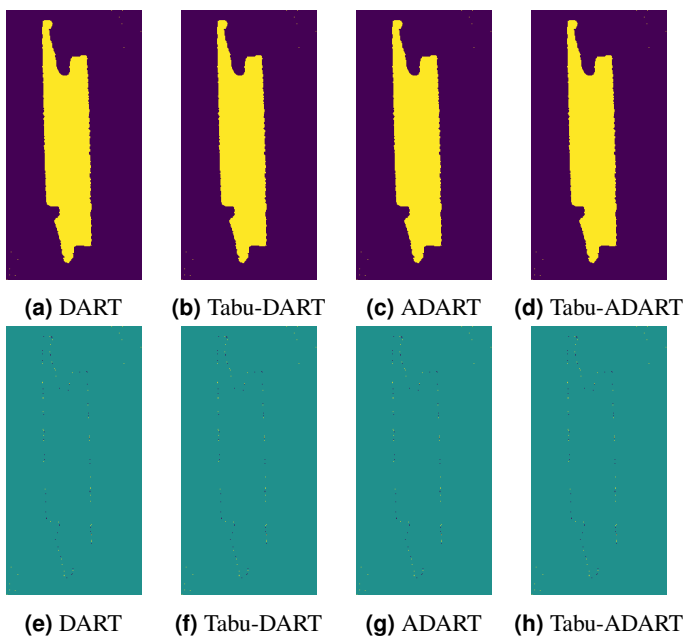

**(a)** DART  **(b)** Tabu-DART  **(c)** ADART  **(d)** Tabu-ADART

**(e)** DART  **(f)** Tabu-DART  **(g)** ADART  **(h)** Tabu-ADART

**Figure 4:** Reconstruction for a 100 ° angular range. The first row (a-d) shows the laminate image resulting from the methods. The second (e-h) row shows difference images with the phantom. While there is a difference in pixel error, the visual difference is negligible.

of performing the DART iterations, and are lowest for Tabu-DART and Tabu-ADART. Our approach allows for a lower pixel error and similar visual quality at a lower computational cost. A sudden increase of $\lambda$ can be observed for all methods once the removed wedge increases past 110°. This indicates the breaking point of the DART algorithm, reconstruction becoming more and more unreliable past this point.

## 4   Conclusion

We have introduced a new update strategy which generalizes the rigid update rules that DART and some of its variants use in subsequent iterations. By representing the update strategy with a probability map we yield more dynamic control of the reconstruction regions and even singular pixels. The specific example of the framework that we presented is however far from optimal. Because of the flexibility of the framework it is possible to introduce complex selection methods that are based on priors already used in other methods such as Total Variation minimization (TV) algorithms, statistical reconstruction methods, or even learned priors. This is a subject of our further work.

## 5   Acknowledgements

This research is funded by the FWO SBO project MetroFlex (S004217N).

# Correction of low-frequency artifacts in CBCT images

Sébastien Brousmiche[1] and Guillaume Janssens[1]

[1]Ion Beam Applications SA, Louvain-la-Neuve, Belgium

**Abstract** Cone-beam CT images are mostly used today in proton therapy to improve patient positioning, which only requires sufficient high-contrast resolution between bones and soft-tissues. However, adaptive therapy requires much higher image quality, with reduced noise and exceptional soft-tissue contrast.

Random noise, residual scatter, or beam hardening artifacts deservedly receives particular attention for improving image quality. An additional challenge in proton therapy systems is the couch induced artifacts, the most notable of which being truncation. All these errors manifest to some extend as low-frequency variations impacting the contrast. This study aims at addressing it.

We first introduce a new model for generating many contrast-impaired CBCT scans from a single CT and demonstrate how a deep-learning network can efficiently be trained based on it. A 5-layers UNet has then been used following this method to correct clinical pelvis and head-and-neck CBCT images. The results show substantial contrast improvements.

## 1 Introduction

Today in proton therapy, cone-beam CT (CBCT) images are mostly used for accurate patient positioning through rigid registration with the planning CT. As it primarily relies on high contrast regions such as soft-tissues-bones interfaces, this processing neither requires accurate CT numbers nor a good soft-tissue contrast. However, the lack of soft-tissue contrast is a significant impediment to detecting anatomical changes between the planning CT and the CBCT of the day. It also limits its use for monitoring target motion, delineating anatomical structures, and computing dose, the latter two being prerequisites for online adaptive proton therapy [1–4]. Several synthetic CT methods have been proposed to enhance CBCT quality and enable daily adaptive proton therapy [5]. The most common method to generate a synthetic CT from a CBCT is through deformable image registration [6–9]. Although the obtained result is less noisy and with much better contrast, there is little guarantee that the resulting soft-tissue structures and contours are accurately matching those of the CBCT, which are usually drowned in noise.

In CBCT reconstruction, the most common sources of low-spatial-frequency contrast loss are residual patient scatter, inaccurate beam hardening correction (water equivalence assumption), data incompleteness, or patient truncation. Moreover, proton therapy systems include couches, which are typically more rigid and more attenuating than those used in CT scanners to limit the impact of couch deflection on the treatment delivery accuracy. Couch truncation in addition to couch-induced scatter and beam hardening have shown to be pretty difficult artifacts to correct with enough accuracy. Fully mitigating some of these sources has shown to be very expensive in modeling, system characterization, and/or algorithm development. Without removing the need for a complete system understanding, deep-learning strategies allow for more straightforward image quality improvements [10–12].

We propose a deep convolution neural network based on the UNet architecture [13] for contrast improvement and CT number offsets correction of a CBCT image. We also developed a noise model to generate a large number CBCT-like images from planning CT, used for the network training and validation. The network has been successfully tested on real CBCT from pelvis and head-and-neck.

Section 2 describes the network architecture and the CBCT image generator used for its training. The results obtained on clinical CBCT images are presented in Section 3. The paper ends with a discussion on the potential improvements in Section 4.

## 2 Materials and Methods

The soft-tissue contrast of a typical medium-dose CBCT is mostly impacted by smooth spatial variations. In this work, we designed and trained a network to accurately estimate them without the need for CBCT and synthetic CT pairs to be used respectively as noisy inputs and ground truth outputs. One common alternative to these pairs is to generate realistic CBCTs from a number of planning CT by simulating the various physical sources of contrast-loss involved. However, to reduce the simulation cost and drastically speed up the training time, we decided to develop a generator that applies random low-spatial frequency perturbations to planning CTs. The necessary assumption for this is that such a random distribution can draw realistic CBCT variations.

### 2.1 Contrast-loss and HU deviation model

A CBCT slice is modelled by a pixel-wise linear transformation of a CT slice $f(\mathbf{x})$ by a scalar low-spatial frequency map $M(\mathbf{x})$, such that

$$g(\mathbf{x}) = f(\mathbf{x})\,M(\mathbf{x}) + n(\mathbf{x}), \qquad (1)$$

where $\mathbf{x}$ is the pixel position and $n(\mathbf{x})$ is a stationary zero-mean random Gaussian noise with variance $\sigma_n^2$. The map $M(\mathbf{x})$ is generated randomly so to create an almost infinite number of CBCTs from a single CT. This can efficiently be done through spectral factorization in which a particular power spectral density is defined. We naturally chose a 2D

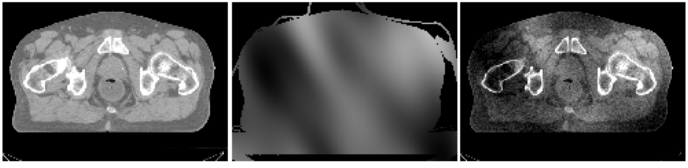

**Figure 1:** Generation of a CBCT slice $g(\mathbf{x})$ (right) from a CT image $f(\mathbf{x})$ (left) with map $M(\mathbf{x})$ (center), using $\sigma_x$=22mm, $\sigma_y$=58mm, $\theta$=0.65$\pi$, $a$=0.3, $b$=0.95, and $\sigma_n$=30 HU.

Gaussian density $G(\mathbf{k})$ defined in the frequency domain $\mathbf{k} = \{k_x, k_y\}$ by its standard deviations, $\sigma_{kx}$ and $\sigma_{ky}$, and a rotation angle $\theta$. The spatial correlation lengths of $M(\mathbf{x})$ are therefore given by $\sigma_x = 1/\sigma_{kx}$ and $\sigma_y = 1/\sigma_{ky}$.

The factorization of a complex spatial map $S(\mathbf{x}|\sigma_x, \sigma_y, \theta)$ is obtained by

$$S(\mathbf{x}|\sigma_x, \sigma_y, \theta) = F^{-1}\{r(\mathbf{k})\, G(\mathbf{k}|\sigma_{kx}, \sigma_{ky}, \theta)\}, \quad (2)$$

where $r(\mathbf{k})$ is a sequence of random complex numbers drawn from a Gaussian distribution $N(0, 1)$ and $F^{-1}$ is the inverse discrete Fourier transform. This process allows to create simultaneously two orthogonal maps in the real and imaginary parts of $S(\mathbf{x})$. These maps are further normalized in the $[-1, 1]$ range. We then compute the scaled map $M$ from, e.g., the real part, with

$$M(\mathbf{x}|\sigma_x, \sigma_y, \theta, a, b) = a\; \text{Re}(S(\mathbf{x}|\sigma_x, \sigma_y, \theta)) + b, \quad (3)$$

where the parameters $a$ and $b$ are respectively the map scaling factor, representing the dynamic of the map, and the CT numbers constant deviation factor.

All the six model parameters ($\sigma_x$, $\sigma_y$, $\theta$, $a$, $b$, $\sigma_n$) are sampled randomly in their typical range, i.e. $\sigma_{x,y}$ in [10, 500] mm, $\theta$ in [0, 2$\pi$], $a$ in [0, 0.4], $b$ in [0.7,1.1] and $\sigma_n$ in [0, 20] HU. These values have been determined by analyzing the available CBCT data set. An illustration of the CBCT generation process is given in Fig. 1.

## 2.2 Noise estimation and network architecture

The network architecture is the commonly used UNet [13]. This is a convolution-based multiresolution network composed of an encoding and a decoding branch, as illustrated in Fig. 2. It takes the noisy image $g(\mathbf{x})$ as input and estimates the noise map $M(\mathbf{x})$. The corrected image $\hat{f}(\mathbf{x})$ can be obtained from the estimated noise map $\hat{M}(\mathbf{x})$ with

$$\hat{f}(\mathbf{x}) = g(\mathbf{x})/\hat{M}(\mathbf{x}). \quad (4)$$

The number of kernels $N_l$ at a given layer $l$ is $N_l = 2^l N_0$ with $l = 0 \ldots L-1$, $L$ is the number of layers and $N_0$ the number of kernels at the first and last layers (excluding the 1x1 convolution layer). Zero padding is done before each convolution step, and batch normalization is applied before changing the layer on both the encoding and decoding branch. The activation function is a rectified linear one known as ReLU.

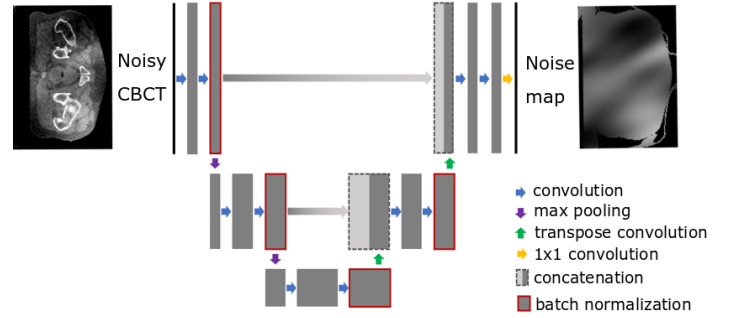

**Figure 2:** Example of a UNet with 3 layers processing the noisy input $g(\mathbf{x})$ to estimate the noise map $M(\mathbf{x})$. The gray box widths are proportional to $N_l$.

- convolution
- max pooling
- transpose convolution
- 1x1 convolution
- concatenation
- batch normalization

The network exhibiting the best performance is a 6-layers UNet, with $N_0$=32 and a 5x5 kernel size. The mini-batches contains 16 slices downsampled to $256^2$ pixels. The maps are upsampled to the CBCT original size before correction.

## 2.3 Data set and network training

As much as 71 planning CT scans of all available anatomical regions are used as training and development data sets. About two-thirds of the data were used for training, the other third for testing. The network architecture and parameters tuning were done on CT images exclusively and the validation on clinical CBCT images. As the CT scans already contain random noise, the maximum variance $\sigma_n^2$ of the additive noise has been computed downwards accordingly. Moreover, we checked that these scans do not contain any strong streaking artifacts that could impair the network training.

The loss function is the addition of a $L_2$ norm between $\hat{M}(\mathbf{x})$ and $M(\mathbf{x})$, and a regularization term, proposed in [14], which measures the edge coherence between them, i.e.

$$L_s = \|SF(\hat{M}(\mathbf{x})) - SF(M(\mathbf{x}))\|^2, \quad (5)$$

where $SF$ is an operator summing the absolute values of the Sobel filtering about the two directions. The loss function to be minimized is therefore $L = L_2 + \lambda_s L_s$, where $\lambda_s$ was tuned to 0.5. The network was optimized with the Adam optimization method with 32 mini-batches of 16 slices per epoch and over 20k epochs. The learning rate was constant and set to 0.001. The network was implemented with the TensorFlow Library.

## 3 Results on real CBCT data

The network performance was mainly evaluated on real CBCT images of pelvis and head-and-neck scans obtained from the IBA compact gantry system and exhibiting severe contrast reduction. The corrected images are compared with synthetic CT images obtained after deformable registration of the planning CT onto the acquired CBCT before contrast correction. The entire 3D image is processed so to target potential inter-slices inconsistencies in the correction and to assess the performance for slices at high cone angles.

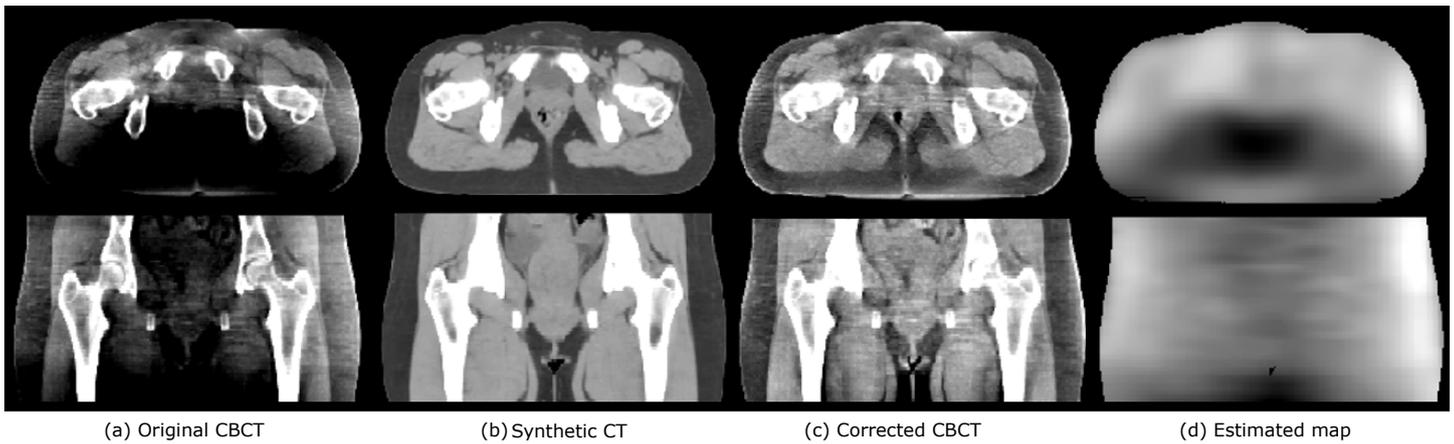

(a) Original CBCT        (b) Synthetic CT        (c) Corrected CBCT        (d) Estimated map

**Figure 3:** Results for a real pelvis CBCT (a) and comparison with the synthetic CT (b). The corrected image (c) is obtained by dividing the original image (a) by the estimated map (d) using Eq. 4. The W/L=400/0 HU for the CTs and 0.5/0.85 for the map.

The results for a pelvis scan are plotted in Fig. 3. The original CBCT shows substantial contrast reduction in the prostate as well as in posterior regions. The impact of the couch has been demonstrated to be dominant there and can hardly be corrected upstream in the processing pipeline. Horizontal streaking artifact through the hip bones can also be observed, mainly due to photon starvation. The corrected image shows an evident improvement over the original one, both in terms of contrast and CT numbers offset correction, with minimal impact on resolution. The coronal views also demonstrate image restoration at higher cone angles, especially at the tights level. This has for advantage to increase the effective scan length of the system. The SSIM metric has been computed relatively to the synthetic CT and improved from 0.4 to 0.75 for the central slice. These numerical results must be weighted by the fact that the CBCT image has a much higher noise and that, due to the model used, the network is not meant to accurately correct bony regions. As required by the model, the estimated correction map is globally at low-spatial frequencies in the axial plane. However, it shows inter-slice variations in the coronal view, which has to be expected due to the slice-by-slice processing. Nevertheless, the impact of those variations on the corrected images looks rather limited compared to the noise. The correction of the skin-line is globally insufficient, probably because of the high gradients to be corrected. Despite the global contrast improvement between, e.g., muscles and fat, their relative CT number distance varies locally. This issue is caused by the linearity assumption used in our model, which does not allow for non-linear artifacts, such as those linked to the poly-energetic nature of the beam, to be corrected.

Fig. 4 depicts the results for a head-and-neck scan. The main difference with the pelvis one is the predominance of bones and, therefore, bone-soft-tissues interfaces, insufficiently corrected bone beam hardening, and metal artifact. The contrast is again well improved, especially inside the skull, with no impact on resolution. The streaks caused by beam hardening at the base of the skull are removed as can be seen in the coronal and sagittal views. The jaw region is also given a low-frequency correction despite the severe streaks due to metal artifacts. This indicates a relatively good tolerance to strong inconsistencies. The last observation is that a partial or inaccurate correction of some streaking artifacts may still produce small structures in the corrected images as we can see for instance in the axial plane. The computed SSIM metric improved from 0.35 to 0.65 at the base of the skull and from 0.5 to 0.8 for a slice at the middle of the skull.

## 4   Discussion and conclusions

The good results obtained on real CBCT images demonstrate the overall adequacy of the CBCT generator for the task as no CBCT image has been used during the training. The risk of overfitting is therefore almost non-existent. However, its main limitation is to only partially correct for non-linear effects. This fact was actually predictable as our main requirement was to be able to provide a scalar low-spatial frequency correction map for inspection. We therefore naturally chose a linear model. This point will need to be reevaluated depending on the new developments made to improve the network. A question that arises is whether separate training should be done for each anatomical region. Even if the correction accuracy would most probably be improved, it would make the correction practically more complex as usual scan lengths are such that multiple regions are covered.

In addition to further improving the results, our future work will focus on evaluating how much this CBCT image quality enhancement makes any subsequent processing, such as contouring or deformable registration, more accurate.

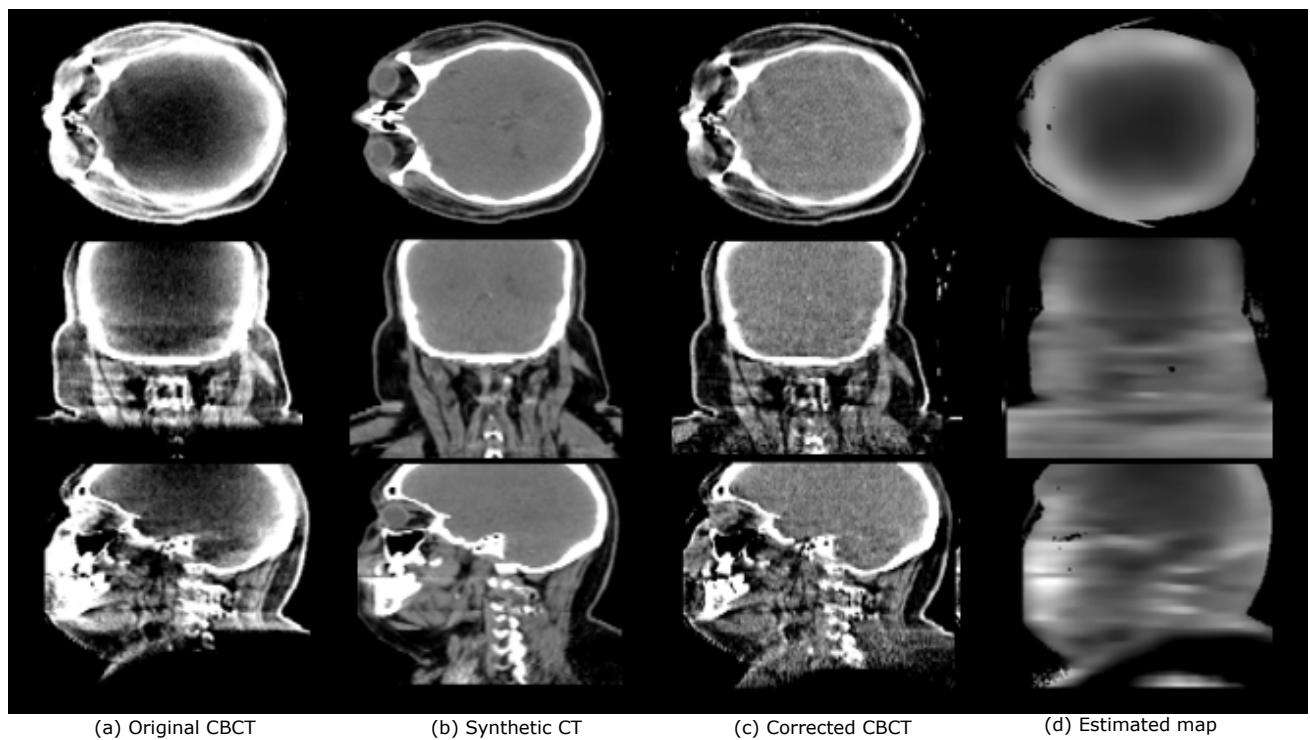

    (a) Original CBCT          (b) Synthetic CT          (c) Corrected CBCT         (d) Estimated map

**Figure 4:** Results for a real head-and-neck CBCT (a) and comparison with the synthetic CT (b). The corrected image (c) is obtained by dividing the original image (a) by the estimated map (d) using Eq. 4. The W/L=400/0 HU for the CTs and 0.5/0.85 for the map.

# Image reconstruction for ion imaging with the TIGRE software framework

S. Kaser[1], T. Bergauer[1], W. Birkfellner[2], D. Georg[3,4], S. Hatamikia[2,5], A. Hirtl[6], C. Irmler[1], F. Pitters[1], and F. Ulrich-Pur[1]

[1]Institute of High Energy Physics, Austrian Academy of Sciences, Vienna, Austria
[2]Center for Medical Physics and Biomedical Engineering, Medical University of Vienna, Vienna, Austria
[3]Department of Radiation Oncology, Div. Medical Radiation Physics, Medical University of Vienna, Vienna, Austria
[4]MedAustron Ion Therapy Center, Wiener Neustadt, Austria
[5]Austrian Center for Medical Innovation and Technology, Wiener Neustadt, Austria
[6]Atominstitut, TU Wien, Vienna, Austria

**Abstract** For ion therapy, an accurate estimate of the ion energy deposition per path length (stopping power) in the patient is essential. Ion computed tomography (iCT) allows to directly measure this quantity. However, as a result of multiple Coulomb scattering, ions pass through the patient on a curved trajectory. Considering each ion path separately in the reconstruction process adds complexity to the problem and often results in long reconstruction times. In this work, a simple and fast approach for iCT reconstruction with the GPU-based open-source software toolkit TIGRE is presented. Since the framework was initially intended for x-ray CT, a straight line approach is used to approximate ion paths to use the framework without modification.

With this simplified approach, imaging data obtained from Monte Carlo simulations and measurement data from an ion CT demonstrator are reconstructed in TIGRE. The accuracy of the demonstrated reconstruction approach is limited by the straight line approximation of the ion path. However, reconstruction results could be improved with additional data cuts. The structure of TIGRE and possibilities for its improvement for iCT reconstruction are discussed.

## 1 Introduction

Proton computed tomography (pCT) was already discussed by Cormack [1] in the 1960ies. Since protons and other ions are affected by multiple Coulomb scattering, they do not pass through a material on a straight line which leads to reduced image quality. An accurate path estimate is therefore necessary in the reconstruction process. Here, most likely path [2, 3] and cubic spline [4] have been shown to achieve better results than a straight line approach [5]. Although the reconstruction problem is more complex than conventional CT, ion CT (iCT) gained interest in the context of ion therapy for cancer treatment [6, 7] where treatment planning is based on the relative stopping power (RSP) values within a patient. So far, a treatment plan is based on conventional CT, where Hounsfield units (HU) have to be converted to RSP via a calibration curve [8] which is the main source of range uncertainties [9]. In iCT, the RSP can be directly obtained from the measurement, thus offering the potential for improved treatment planning in ion therapy.

A typical iCT setup, as it is displayed in Figure 1, was introduced by Schulte, Bashkirov, Li, et al. [10] and consists of a particle tracker and a device to measure the residual energy of each particle (calorimeter). The information from the tracking system is used to reconstruct the ion's path through the medium, while the residual energy is used to calculate the projection value, which is then back projected along the path estimate in the reconstruction.

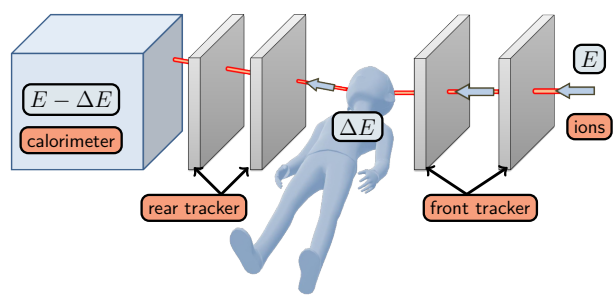

**Figure 1:** iCT setup consisting of two trackers, to measure the ion's position and direction and a calorimeter to measure the ion's residual energy.

Although there have already been promising ion CT reconstruction approaches [11, 12] and frameworks [13, 14], iCT reconstruction often faces problems such as long reconstruction times. The aim of the present study is to demonstrate iCT image reconstruction with the open-source framework TIGRE [15] (Tomographic Iterative GPU-based REconstruction toolbox). The layered structure of the framework allows for an iCT reconstruction approach which is fast (GPU-based reconstruction) and easy to apply for the user (Matlab user layer). Although the framework was originally intended for X-ray CT reconstruction, iCT reconstruction along straight ion paths was demonstrated to be possible.

## 2 Materials and Methods

### 2.1 TIGRE reconstruction toolkit

TIGRE is an open source cone beam CT reconstruction framework. It introduces a large set of iterative reconstruction methods, mainly algorithms using total variation (TV) regularization, which allow efficient reconstruction from sparse view and limited angle projection data. While the forward and backward projectors are fully implemented in CUDA (hence run on one or multiple GPUs), the reconstruction algorithms and user layer are written in Matlab. These layers are communicating via C++ scripts.

## 2.2 Ion CT demonstrator and Monte Carlo simulations

An ion CT demonstrator, consisting of four 300 μm silicon strip tracking detectors and a range telescope, has been tested at MedAustron and yielded first RSP images of a phantom [16, 17]. Due to the limited size of the tracking detectors, only small phantoms can be imaged. 80 non-equidistant projections of an aluminum cube with a side length of 1 cm and a stair profile were measured over a range of 360 degrees using protons with an initial energy of 100.4 MeV.

For larger phantoms, imaging data were generated from Monte Carlo simulations with Geant4 [18] and GATE [19] using the physics list *QGSP_BIC*. To analyze line pair resolution and RSP accuracy of the investigated reconstruction method, two Catphan modules (CTP528 and CTP404 [20]) were used as phantoms. They are made of a cylindrical acrylic body with a diameter of 15 cm and specific inserts. While the CTP528 module contains aluminum strip inserts to determine the spatial resolution of a reconstruction, the CTP404 module contains different cylindrical inserts that can be used to determine the RSP accuracy of the reconstruction (the central slice was analyzed for 200 MeV protons in [21]). In this work, reconstruction results with helium ions (200 MeV/u) are presented. Furthermore, to investigate biological materials, a CT image of the CIRS head phantom [22] was imported to a GATE simulation and a reconstruction was performed using 200 MeV protons.

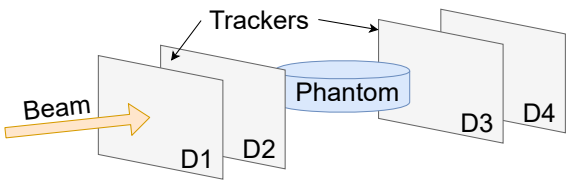

**Figure 2:** Monte Carlo simulation setup.

The Monte Carlo simulation setup for all simulations performed in the scope of this work (see Figure 2) contained two trackers, each consisting of two 300 μm silicon detectors. One tracker was located upstream (detector D1 and detector D2) and one downstream (detector D3 and detector D4) the phantom to measure entry and exit position and direction of ions to the phantom. Between the two detectors of a tracker, a 10 cm distance was set while the distance between tracker and phantom was always kept greater than 10 cm. The residual energy of the ions was determined at detector D4 in this idealized setup (no additional calorimeter). The setup was located within an air volume and a fluence of 800 ions/mm² (Catphan modules) and 200 ions/mm² (CIRS head) was used in the simulations. In each simulation, 90 projections were generated over an angular span of 178°.

## 2.3 iCT projection definition and reconstruction

To use the framework TIGRE without modification, a straight line approach was used for the ion path. In order to remove ion paths with a strong curvature, position cuts were introduced in addition to the standard 3σ cuts [3]: The difference of proton hit positions in *x*- and *y*-direction between detector D2 and detector D3 was calculated and, if it exceeded a certain threshold, the track was rejected (this method was adapted from Cirrone, Bucciolini, Bruzzi, et al. [23]). The optimal position cut threshold hereby depended on the phantom thickness and material. It was set to 0.5 mm for the measurement data, 2 mm for the simulation of Catphan modules and 4 mm for the simulation of the CIRS head since this was found to be the ideal compromise between optimized spacial resolution, RSP accuracy and amount of rejected ion paths. If a track passed the cut condition, the ion was assigned to the pixel corresponding to the average of the hit positions on detectors D2 and D3.

To obtain the RSP in the reconstructed image, the water equivalent path length (WEPL) had to be calculated for each ion and further used as projection value. While for the measurements from the iCT demonstrator, the range telescope measurement was directly converted to the WEPL, the Donahue [24] definition of the ion range *R*,

$$R = \frac{1}{\kappa}[\beta E_{\text{in}}^q + \alpha E_{\text{in}}^p + \frac{h}{g}(\exp(-gE_{\text{in}}) + gE_{\text{in}} - 1)]u, \quad (1)$$

was used for simulated data. Here, the stopping power *S* depends on the material's ionization potential $I_{\text{material}}$ and the ion energy *E*. $E_{\text{in}}$ is the initial ion energy and *u* is the atomic mass number of the ion. *α*, *β*, *g*, *h*, *p*, and *q* are material-dependent parameters, which were already defined for protons in water in Donahue, Newhauser, and Ziegler [24]. For helium ions, the material parameters for the Donahue model were calibrated with NIST data [25] between 5 MeV and 250 MeV using a least squares algorithm provided by Python's scipy module [26]. To obtain the WEPL for each ion, the range at the initial and residual ion energy $E_{\text{in}}$ and $E_{\text{out}}$ are subtracted [21]

$$\text{WEPL} = R_{\text{water}}(E_{\text{in}}) - R_{\text{water}}(E_{\text{out}}). \quad (2)$$

In the projection, the average WEPL per pixel was calculated for all ions assigned to the pixel.

Adaptive-Steepest-Descent Projection Onto Convex Sets (ASD-POCS) [27] of the Total Variation (TV) algorithm family was selected as the main reconstruction method in this study due to its highly demonstrated performance under limited angle scanning trajectories. Algorithms of this family have also shown promising results for iCT reconstruction problems [28, 29]. In [21] it was shown that especially for limited data, ASD-POCS outperforms other algorithms implemented in TIGRE, such as Ordered-Subset Simultaneous

Algebraic Reconstruction Technique (OS-SART) [30]. However, OS-SART was used to reconstruct the measurement data from the iCT demonstrator at MedAustron. Due to the phantom size, the data cuts did not have such a strong influence and the better statistics per pixel allowed for the faster reconstruction with OS-SART.

## 3  Results

### 3.1  Measurement data from the iCT demonstrator

In the reconstructed 3D view of the phantom (Figure 3, right), the stair profile is clearly visible. Furthermore, the reconstructed RSP was analyzed within each stair (see Figure 4). Edge voxels have been excluded from this analysis. The relative error of the median RSP within each step was ranging from 1.4% to 11% (thinnest stair), while the relative error of the average values was ranging from 2.7% to 11.6%.

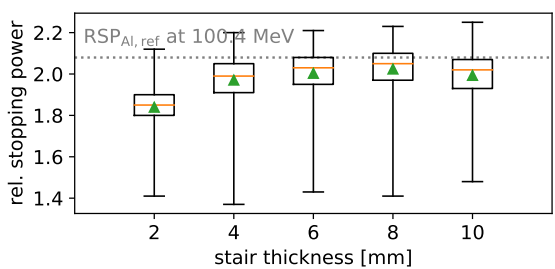

**Figure 3:** Photo of aluminum phantom (left) and 3D view of its reconstruction (right). 3D view was created with Slicer [31].

In Ulrich-Pur, Bergauer, Burker, et al. [17], the RSP values within the phantom stairs (again, after using position cuts of 0.5 mm) were analyzed while including edge voxels. It could be seen that the relative error of the most probable value (MPV) within a stair could be lowered to 0.28 - 1.56% with these position cuts. The reason for the smaller errors compared to the values stated before lies in the shape of the RSP distribution observed within a stair: Rather than being Gaussian-shaped, the distribution showed a significant tail towards lower values. While this influences the average and median value within a stair and shifts it to a lower value, the MPV could still be found closer to the expected reference value.

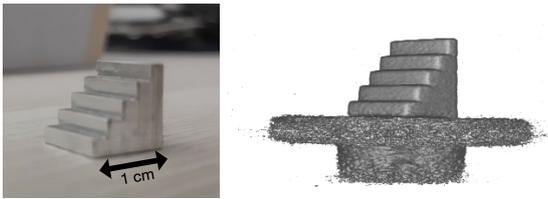

**Figure 4:** Reconstructed RSP values within the stairs.

### 3.2  Monte Carlo simulations – Catphan modules

Based on the work in Kaser, Bergauer, Birkfellner, et al. [21] reconstruction results using helium ions are summarized in Figure 5 for the central slices of the CTP528 and CTP404 modules.

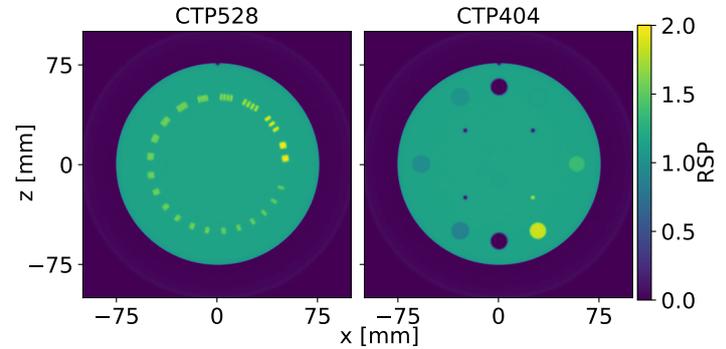

**Figure 5:** Reconstructed central slices of CTP528 (left) and CTP404 (right) using 200 MeV/u helium ions.

The contrast – corresponding to the value of the modulation transfer function as defined as in Volz, Piersimoni, Bashkirov, et al. [32] – of the first three line pair inserts within the CTP528 from a reconstruction using protons or helium ions and a 2 mm position cut is shown in Figure 6. While the 3 lp/cm insert could be distinguished with a contrast of 26% for protons, the contrast of higher line pair inserts was below 10%. Using helium ions, the contrast for the first three line pair inserts was higher than for protons, for example, 40% for the 3 lp/cm insert.

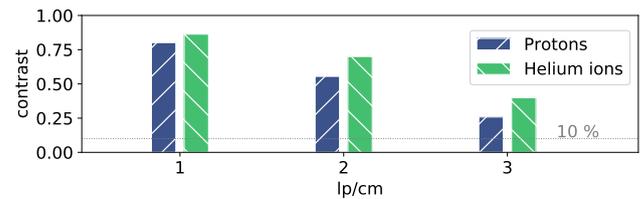

**Figure 6:** Contrast values in the line pair inserts of the CTP528.

For the CTP404, RSP values of three from the outer inserts, namely LDPE (Low Density Polyethylen), PMP (Polymethylpentene) and Delrin, can be found in Figure 7 for protons and helium ions. Edge pixels have been excluded from the analysis.

For helium ions, the spread of values was smaller than for protons within a region of interest. Furthermore, the average RSP did correspond very well to the the reference RSP, which was defined using an R80 calibration as described in Kaser, Bergauer, Birkfellner, et al. [21]. For example, the reconstructed average RSP for the Delrin insert yielded 1.369 which is 0.2% above the reference value of 1.366. For protons, the average RSP value of 1.357 was approximately 1% below the reference value (1.371).

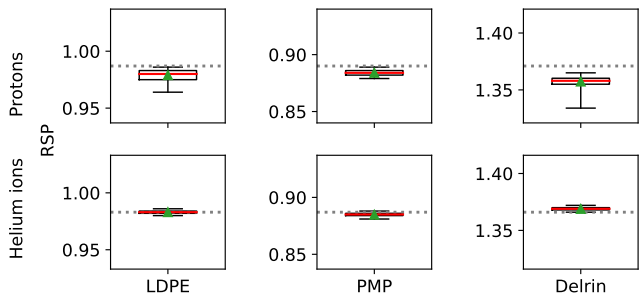

**Figure 7:** Reconstructed RSP values in the LDPE, PMP and Delrin insert.

### 3.3 Monte Carlo simulations – CIRS head phantom

To visually demonstrate the effect of position cuts on the reconstruction result, a CT image of the CIRS head phantom was inserted as phantom to a GATE simulation. Figure 8 shows the reconstruction result from 90 projections using 200 MeV protons.

The effect of the position cuts can be visually observed regarding the transition between bone and tissue within the phantom.

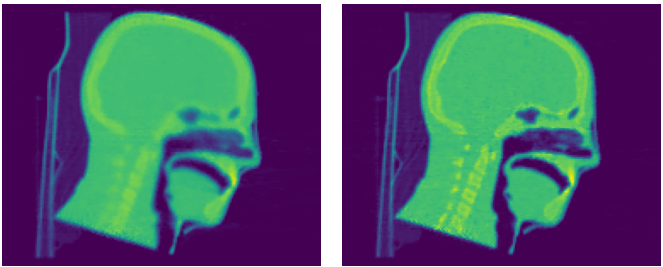

**Figure 8:** Reconstructed slices of CIRS head phantom without (left) and with (right) position cuts.

### 4 Discussion

To apply the OS-SART and the ASD-POCS algorithm (based on total variation) to the iCT reconstruction problem, no modification of the TIGRE software framework was necessary (only a redefinition of projection values).

Measurement data from an ion CT demonstrator could be successfully reconstructed and position cuts could be used to increase RSP accuracy. The reconstruction time for a volume of $256 \times 256 \times 128$ voxels was smaller than 10 s on a standard GPU using OS-SART.

For larger phantoms, Monte Carlo simulations were used to generate projection data. The reconstruction of the CTP528 using protons showed the expected limitations in spatial resolution due to the straight line approximation of the proton path. Nevertheless, inserts with 1 to 3 lp/cm could be distinguished using the additional position cut. Comparing reconstruction results to other straight line reconstructions using protons, a similar spatial resolution can be found (for example approx. 2 lp/cm for 180 projections and using ART with 120 iterations and 200 protons $mm^{-2}$ in Li, Liang, Singanallur, et al. [5]). Using helium ions instead of protons did increase the contrast of the reconstruction result.

For the CTP404, RSP values within inserts were analyzed (transitions between materials were neglected in the analysis for this phantom). Regarding the average RSP values obtained in the inserts, the Donahue approximation seems to be an adequate option for the projection definition since only minor deviations from the literature values [33] were found. For helium ions, reconstructed RSP values were closer to the reference value than for proton ions.

For the CIRS head phantom, the effect of position cuts could be well observed regarding the transitions between phantom/air and bone/tissue in Figure 8.

To further test the applicability of TIGRE for limited projection data, the number of projections or the particle fluence used in this study have to be further reduced. However, the 800 protons/$mm^2$ that were used in the present work lie in the typical range for iCT: For example, Rit, Dedes, Freud, et al. [11] used 900 protons per $mm^2$ to investigate a 3 lp/cm insert (720 projections) while Giacometti, Bashkirov, Piersimoni, et al. [34] reported a contrast above 10 % for 3 lp/cm for 100 protons per $mm^2$ and 90 projections.

The main limitations of the presented method arise from the effect of multiple Coulomb scattering, which leads to decreased spatial resolution and RSP inaccuracies if a straight ion path is assumed. The position cut used in the present study allowed to compensate for this effect to some part, however, a large number of primary particles were filtered. For the Catphan modules, 70-90% of primary protons were rejected by the 2 mm position cut, depending on the phantom thickness (lower data rejection at the edges than in the central region of the cylindrical phantom). Using helium ions, 34-75% of primary ions were rejected by the same position cut. To optimize TIGRE for iCT, two main requirements have to be addressed: The already binned projection data have to be replaced by list-mode data. This step is crucial to treat each ion path separately in the reconstruction process. Here, the straight line approximation has to be replaced by cubic spline or most likely path estimation.

The structure of TIGRE allows to keep a Matlab header for the user while changes in the projection and back projection operators have to be done in CUDA. Such changes in the CUDA implementations have already been proposed in Hatamikia, Biguri, Kronreif, et al. [35], where reconstruction from arbitrary rotation scan trajectories were added to the framework. In addition, the CUDA layer was modified to speed up the implementation of the total-variation based algorithms. The TIGRE toolbox offers multiple incentives to perform the proposed adaptions for iCT: It offers a wide range of algorithms which have already been shown to generate promising results with low input data [15, 35], the use of multi-GPUs is possible and the layered structure makes the framework a promising candidate for a user-friendly iCT reconstruction framework. This layer structure already allows the use of multi-GPUs in the reconstruction, further speeding up the reconstruction time.

## 5 Conclusion

The applicability of the TIGRE reconstruction framework to the ion CT reconstruction problem was shown using simulated and measured projection data. Further improvements needed to optimize ion CT reconstruction were discussed. Most importantly, a sophisticated path estimate has to be implemented to the framework.

## 6 Acknowledgements

The authors would like to thank Barbara Knäusl (MedUni Wien) and Markus Stock (MedAustron) for providing the CT image of the CIRS head phantom. This project received funding from the Austrian Research Promotion Agency (FFG), grant number 875854.

# New method for correcting beam-hardening artifacts in CT images via deep learning

C. Martínez[1,2], C. F. Del Cerro[1,2], M. Desco[1,2,3,4], M. Abella[1,2,3]

[1] Depto. Bioingeniería e Ingeniería Aeroespacial, Universidad Carlos III de Madrid, Spain (crismart, carlosfe)@pa.uc3m.es, (mdesco, mabella)@ing.uc3m.es

[2] Instituto de Investigación Sanitaria Gregorio Marañón, Madrid, Spain,

[3] Centro Nacional de Investigaciones Cardiovasculares Carlos III (CNIC), Madrid, Spain

[4] Centro de investigación en red en salud mental (CIBERSAM), Madrid, Spain

**Abstract** Beam-hardening is the increase of the mean energy of an X-ray beam as it traverses a material. This effect produces two artifacts in the reconstructed image: cupping in homogeneous regions and dark bands among dense areas in heterogeneous regions. The correction methods proposed in the literature can be divided into post-processing and iterative methods. The former methods usually need a bone segmentation, which can fail in low-dose acquisitions, while the latter methods need several projections and reconstructions, increasing the computation time.

In this work, we propose a new method for correcting the beam-hardening artifacts in CT based on deep learning. A U-Net network was trained with rodent data for two scenarios: standard and low-dose. Results in an independent rodent study showed an optimum correction for both scenarios, similar to that of iterative approaches, but with a reduction of computational time of two orders of magnitude.

## 1 Introduction

The origin of the beam-hardening effect lies in the polychromatic nature of the X-ray sources. It is defined as the process whereby the mean energy increases its value when traversing a material. This energy shift is due to the fact that low-energy photons are more easily absorbed than high-energy photons. The beam-hardening effect produces two artifacts on the reconstructed image: cupping in homogeneous regions and dark bands among dense areas in heterogeneous regions [1].

We can find multiple correction schemes in the literature. It is common to pre-harden the beam by using a physical filter that eliminates most of the low-energy photons [1]. However, this is not enough to completely eliminate the artifacts, making it necessary to use image processing methods. The method implemented in most of the scanners is the water linearization. It assumes that the sample is homogeneous, correcting only the cupping artifacts [2]. To correct both cupping and dark bands, the beam-hardening effect can be modeled using the spectra knowledge and an estimation of the tissue thicknesses [3, 4]. The spectra knowledge was substituted with a beam-hardening model using information either from a calibration phantom [5] or the sample itself [6]. Other works avoid the characterization of the beam-hardening model by maximizing the flatness [7] or the entropy [8] of the reconstructed image. However, all the previous methods need a segmentation that can fail in low-dose acquisitions. In these scenarios, the use of iterative algorithms allows for the improvement of the segmented masks with successive iterations. The work proposed by Elbakri et al. [9] included a polychromatic model of the source, but required the spectra knowledge to incorporate the energy effect into the projection matrix. This requirement was eliminated in the method proposed by Abella et al. [10], called bhSIR, with a simplification of the polychromatic model based on two parameters and the same calibration step of the water-linearization method. However, the use of iterative methods leads to an increase in the execution time.

Over recent years, deep learning has been widely used in CT images for segmentation and classification [11, 12] or to improve the quality of low-dose acquisitions [13, 14]. U-net [15], originally used for image segmentation and one of the most known architectures, has already been used to reduce the sparse-view artifacts in CT images [16], metal artifacts [17] or ring artifacts [18]. To the best of our knowledge, there are no deep learning approaches to reduce the beam-hardening artifacts on CT images.

In this work, we proposed a new method to obtain images free of beam-hardening artifacts in CT. We compensate the artifacts by using deep-learning techniques based on a U-net architecture in low and standard-dose scenarios.

## 2 Materials and Methods

The proposed method uses a modification of the original U-net architecture [15], eliminating the sigmoid layer that normalizes the resulting image to allow the restoration of the monochromatic values. We use the mean squared error (MSE) as the cost function. Figure 1 shows the network architecture.

The training was performed during 100 epochs using the Adam optimizer [19] with axial slices of four rodent studies acquired with the micro-CT scanner ARGUS/CT (SEDECAL) [20]. Two scenarios, standard dose (360 projections covering 360 degrees) and low dose (180 projections covering 360 degrees), were acquired and reconstructed with the software FUX-SIM [21], obtaining projections of 512×375 pixels and 0.2×0.2 mm of pixel size. Reconstruction was performed with the FDK algorithm [22], resulting in volumes of 512×512×375 voxels and 0.121×0.121×0.121 mm of voxel size. In both scenarios, images obtained with bhSIR [10] from standard dose data were used as reference (Figure 2).

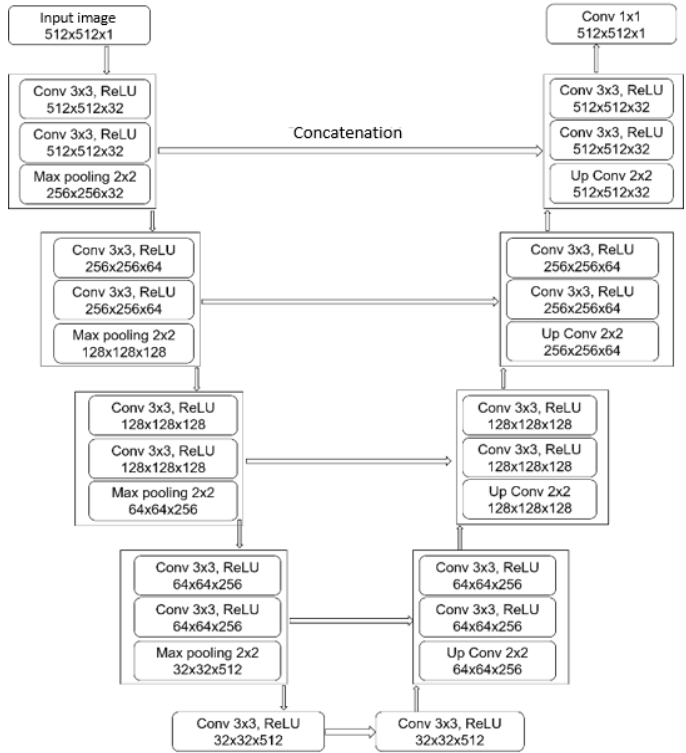

*Figure 1: Modified Architecture of the U-net*

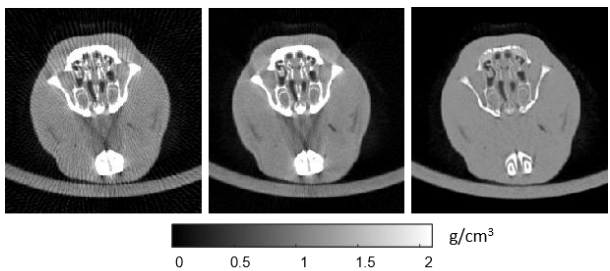

*Figure 2: Axial slice of the rodent study for the low (left) and standard-dose (right) scenario and the reference obtained with the iterative method (right)*

Images obtained with bhSIR [10] were used as reference. To select the appropriate learning rate, we used the Leslie N. Smith test [23], resulting in $10^{-5}$ (Figure 3).

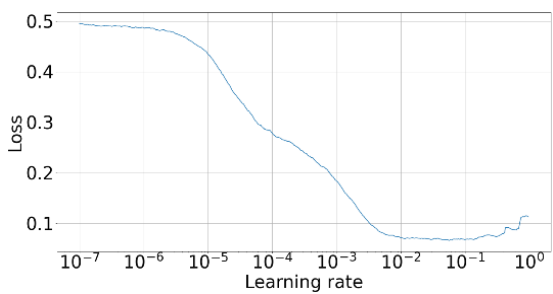

*Figure 3: Results of the Leslie N. Smith test to determine the optimum learning rate.*

## 3 Evaluation and results

The network was applied to a fifth rodent study, also acquired in standard- and low-dose scenarios. We compared the proposed method with the FDK, FDK+2DLinBH [5] and bhSIR [10] visually and in terms of execution time.

Figure 4 shows the two axial slices of the standard dose scenario obtained with the different methods. We can observe a reduction of the dark bands with all the methods but with a slight noise increase with the analytical approach FDK+2DLinBH. The image corrected with the proposed method is very similar to the one obtained with the iterative algorithm bhSIR, with higher SNR and a complete reduction of the beam-hardening artifacts.

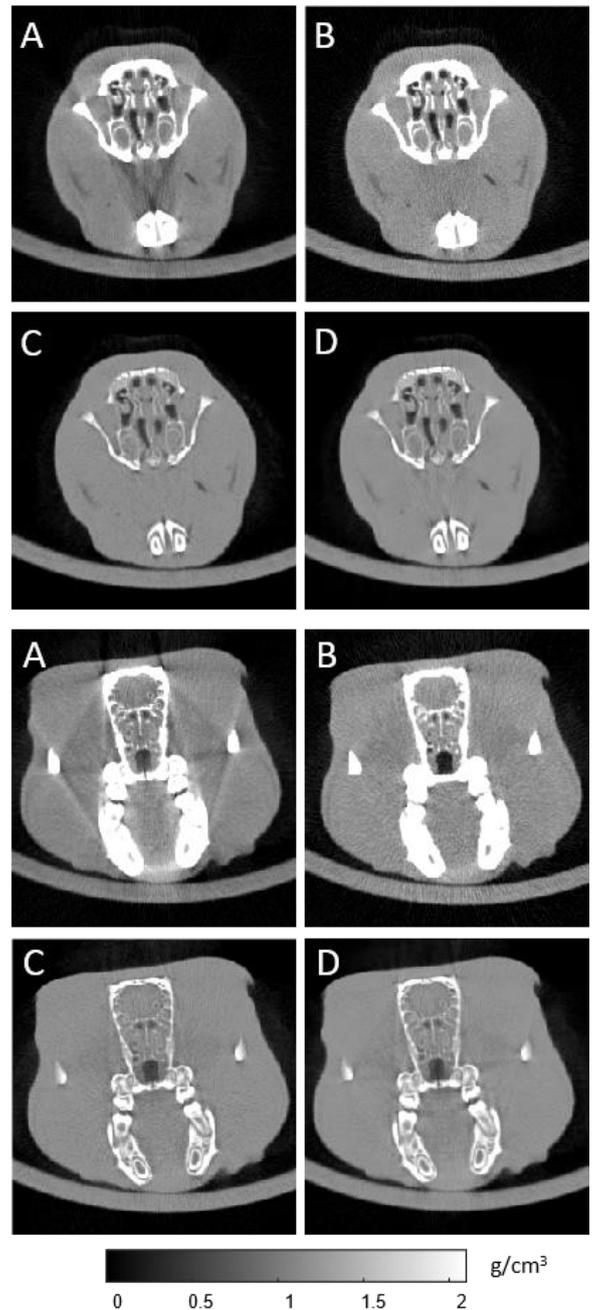

*Figure 4: Standard-dose scenario for two different axial slices 1 (top) and 2 (bottom) obtained with the FDK (A), FDK+2DLinBH (B), bhSIR (C) and the proposed method (D)*

Figure 5 shows the results for the low-dose scenario. FDK+2DLinBH shows streak artifacts because of the low angular sampling. The proposed method reduces these low-sampling artifacts and compensates the beam-hardening artifacts similar to that in the reference.

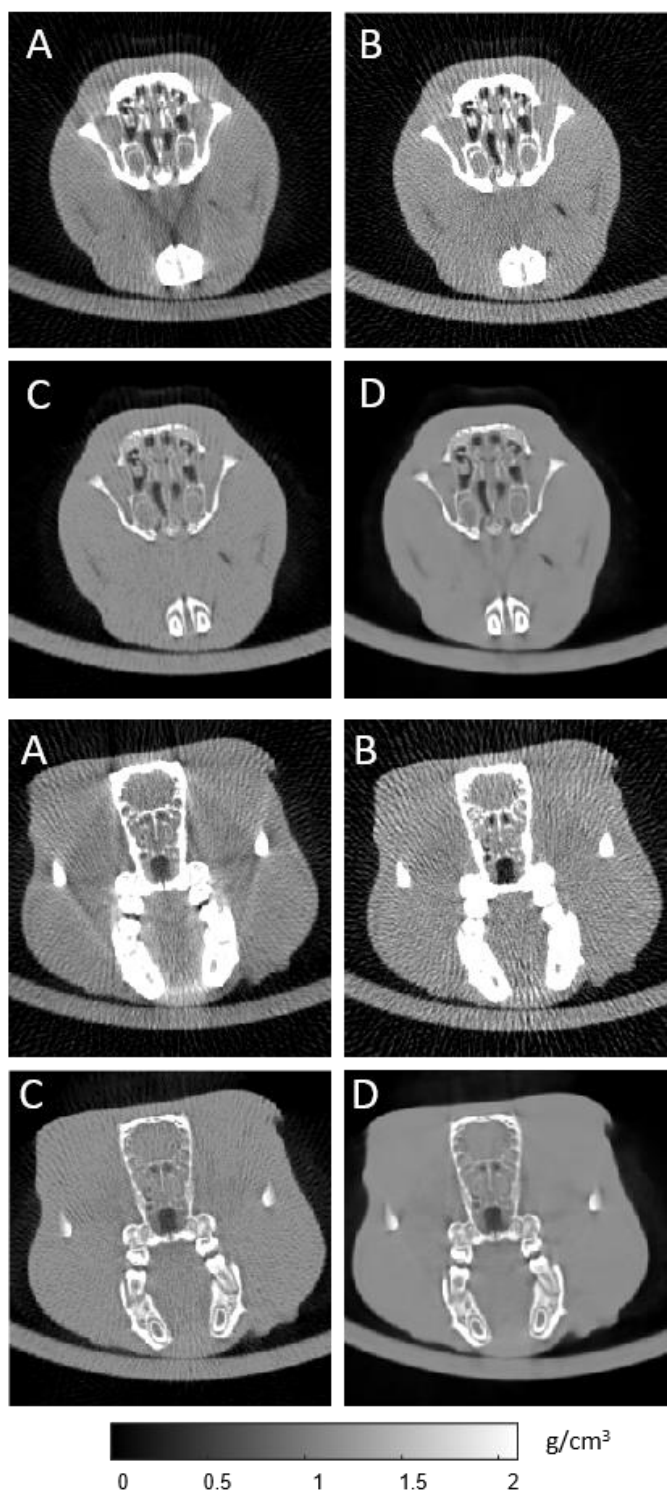

g/cm³
0    0.5    1    1.5    2

*Figure 5: Low-dose scenario for the slices 1 (top) and 2 (bottom) obtained with FDK (A), FDK+2DLinBH (B), bhSIR (C) and the proposed method (D)*

Table I shows the computational time of the complete volume for the different methods. We can observe that the lowest time corresponds to the proposed method.

TABLE I
EXECUTION TIME OF EACH METHOD (SECONDS)

|  | FDK | FDK+2DLinBH | bhSIR | FDK+DL |
|---|---|---|---|---|
| Standard dose | 10.4 | 10.4+59.9 | 28800 | 10.4+28.7 |
| Low dose | 6.3 | 6.3+42.7 | 28080 | 6.3+28.7 |

## 4 Discussion

We have proposed a new method to compensate the beam-hardening artifacts on CT images based on the combination of conventional reconstruction and deep learning. Our method outperforms classical post-processing methods in low-dose data, showing a similar performance to a polychromatic iterative method (bhSIR) but with a considerable reduction of computational time.

Evaluation performed on real data showed a good correction of the beam-hardening artifact but a slight loss of spatial resolution. The selection of the simple cost function MSE for these preliminary results may be responsible for this loss of spatial resolution. Future work will evaluate the use of more sophisticated cost functions, such as SSIM or perceptual loss, or architectures like GAN (Generative Adversarial Networks).

Due to the impossibility of acquiring the rodent studies with a monochromatic source, an iterative method was used as the gold standard.

We focused on head studies, creating a different model depending on the number of projections. Further work will evaluate the performance of the method when other anatomical parts, such as the abdomen or thorax, are included in the dataset. We also expect that this increase in the amount of training data would enable a single model to work independently of the number of acquired projections.

## 5 Conclusion

The proposed method based on deep learning corrects the beam-hardening artifacts in CT images with a reduction of noise and low-sampling streaks similar to iterative methods but with a significant reduction of computational time. This reduction allows the method to be used in real-time applications like intraoperative imaging. The method can be easily implemented in real systems, since it involves only an extra processing step right after a conventional reconstruction.

## Acknowledgment

This work has been supported by project "DEEPCT-CM-UC3M", funded by the call "Programa de apoyo a la realización de proyectos interdisciplinares de I+D para jóvenes investigadores de la UC3M 2019-2020, Convenio Plurianual CAM – UC3M" and project "RADCOV19", funded by CRUE Universidades, CSIC and Banco Santander (Fondo Supera). The CNIC is supported by the

# Feasibility study of residual U-Net based low-dose dual-energy imaging for a gantry-type CT scanner

Sanghoon Cho[1], Taejin Kwon[1], Jongha Lee[1], and Seungryong Cho[1,2]

[1]Department of Nuclear and Quantum Engineering, KAIST, Daejeon, South Korea
[2]KAIST Institutes for ICT, HST, and AI, KAIST, Daejeon, South Korea

**Abstract** We have earlier demonstrated the feasibility of realizing low-dose dual-energy imaging on top of an existing single-energy fast-rotating gantry-type CT scanner by use of a multi-slit beam-filter placed between the x-ray source and the patient. By sliding a beam filter with its slits being at a slanted angle and exploiting associated image processing and reconstruction algorithm, we have overcome hardware- and software- challenges. However, our previous method has limitations of not zero-loss of structural information due to notch-filtering and image domain-denoising, and not short-computation time in iterative image reconstruction framework. In this work, we propose a CNN-based dual-energy imaging method to overcome these challenges. We firstly deployed a residual U-Net neural network to make fully restored low- and high-energy sinogram out of original streaky sinogram, followed by filtered-backprojection (FBP) of two output sinograms. Using these two images as a prior and initial guess, we have processed 5 steps of the compressed-sensing-inspired iterative reconstruction (CS-inspired IR) algorithm to improve the fidelity of the reconstructed image value. We conducted a simulation study using the anthropomorphic digital phantom, and showed its successful results in image reconstruction and material decomposition.

## 1 Introduction

Spectral CT imaging techniques can provide material decomposed images or variable contrast images for better diagnosis or distinction of lesions, reducing the burden of uptake of contrast agents of patients. High-end gantry-type CT systems are equipped with these dual-energy techniques such as dual-source technology [1,2], fast kV-switching [3], and detector-based spectral CT [4,5]. Very high-end CT systems are equipped with deep-learning-based dual-energy imaging technology enabling better material decomposition performance in a short computation time [6].

We have earlier shown a many-view under-sampling (MVUS) technique that uses a multi-slit beam filter [7-10]. It enables dual-energy imaging in a low-dose manner [9-10]. Recently, we proposed to use a linear motion of a beam-filter with its slits being at a slanted angle with the rotation axis to implement the MVUS technique on top of existing single energy fast-rotating gantry-type CT scanner [10]. Out of streaky sinogram, a notch-filter was applied to come up with restored sinogram. By filtered-backprojection (FBP) and image-based smoothing, an initial image without a streaky pattern was made. In turn, compressed-sensing inspired image reconstruction (CS-inspired IR) was processed by selectively using specific energy data out of the original streaky sinogram and by exploiting the structural information of the initial image as a prior [10,11]. Despite its successful demonstration, it takes not short-computation time having a difficulty in clinical use. In addition, the conventional notch-filtering process and image-based denoising can lead to not zero loss of structural information.

In this work, we propose a convolutional neural network (CNN) based low- and high-energy sinogram restoration method out of the original streaky sinogram to overcome these challenges. After FBP of two output sinograms, 5 steps of CS-inspired IR are processed to improve the fidelity of the reconstructed image. To test the feasibility, a simulation study has been conducted using anthropomorphic digital phantom.

## 2 Materials and Methods

### A. MVUS scanning and CNN-based dual-energy imaging

As shown in Fig. 1, a multi-slit beam filter slides along the rotation axis while CT scanning so that one can acquire sparsely sampled data. Photons through the filter attenuate leading to x-ray mean-energy shift to a higher level, enabling dual-energy imaging, and it leads to dose reduction to the patient as well. As shown in Fig.2, in step 1, sparsely sampled sinogram data go through the neural network, resulting in fully recovered low- and high-energy sinograms, and both sinograms are fed into the FBP framework. In step 2, both images go through a CS-inspired IR process by selectively using specific energy data regions in the original streaky sinogram to improve the fidelity of the reconstructed image value, followed by image-based material decomposition.

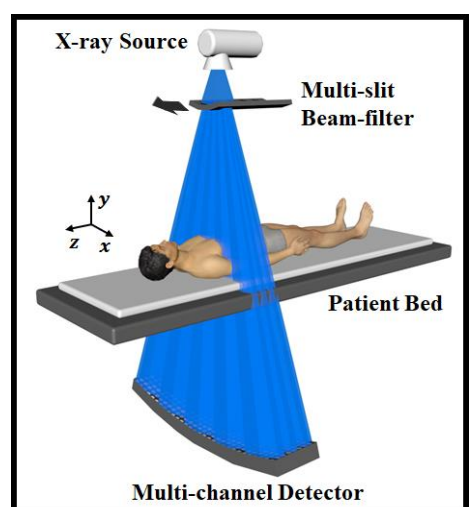

**Figure 1. Schematic of MVUS scanning**

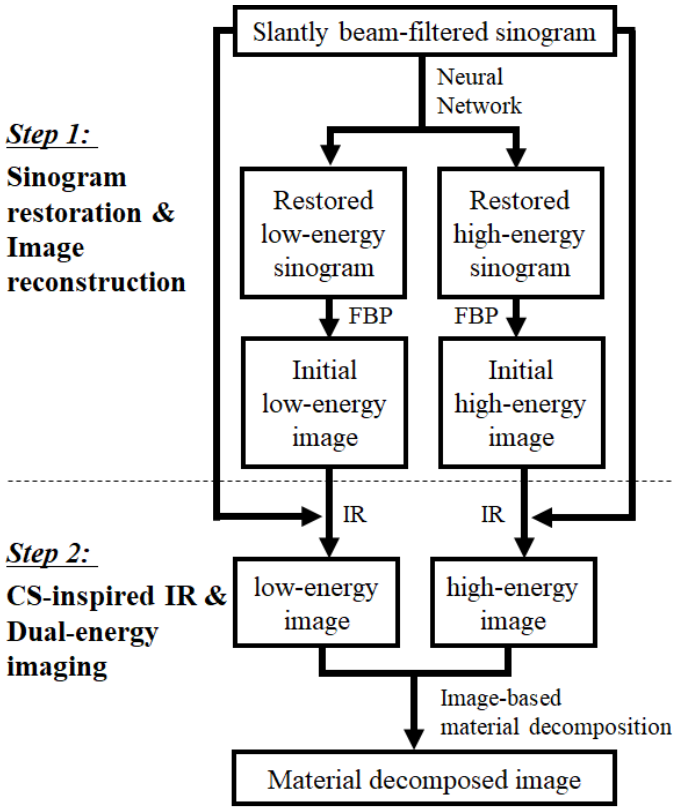

**Figure 2. Proposed imaging algorithm flowchart.**

## A1. Step 1: CNN based sinogram restoration and image reconstruction

We propose a sinogram domain-based deep learning framework to generate fully restored low- and high-energy sinograms. Figure 3 shows the input and output of the proposed deep learning network. Input data is composed of 3 channels sinogram patches which are the original sinogram patch in channel 1, the binary mask patch of the low-energy sinogram in channel 2, and that of the high-energy sinogram in channel 3. Please note that penumbra was considered in sinogram and masks. Output data is composed of 2 channles which are fully restored low-energy sinogram and high-energy sinogram. Residual U-Net network was exploited as shown in Fig. 4, where the residual of the network is difference of label and input streaky sinogram [12]. In turn, output sinograms are fed into FBP framework. In this study, all the source codes were programmed in Python and the Pytorch library on an RTX 2080-ti GPU. The Adam method was employed to optimize the network. The learning rate starts from 1e-4 and is multiplied by 0.99 times every epoch. For the loss, MSELoss was used and a total of 15 epochs were trained.

## A2. Step 2: CS-inspired IR & Dual-energy imaging

5 steps of CS-inspired IR are processed for each energy image by selectively using low- and high-energy data regions in the original sinogram. Each initial guess of IR is FBP of low- and high-energy output sinogram in Step 1. A constrained total-variation minimization algorithm was exploited where it minimizes the l2 norm of the difference between gradient magnitude image (GMI) of the reconstructed image in Step 1 and the image to be reconstructed as following equations:

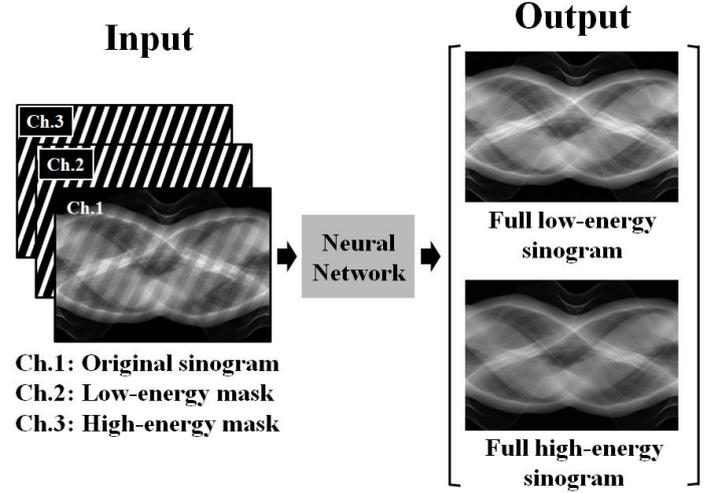

**Ch.1: Original sinogram**
**Ch.2: Low-energy mask**
**Ch.3: High-energy mask**

**Figure 3. Input and output of our proposed deep-learning framework.**

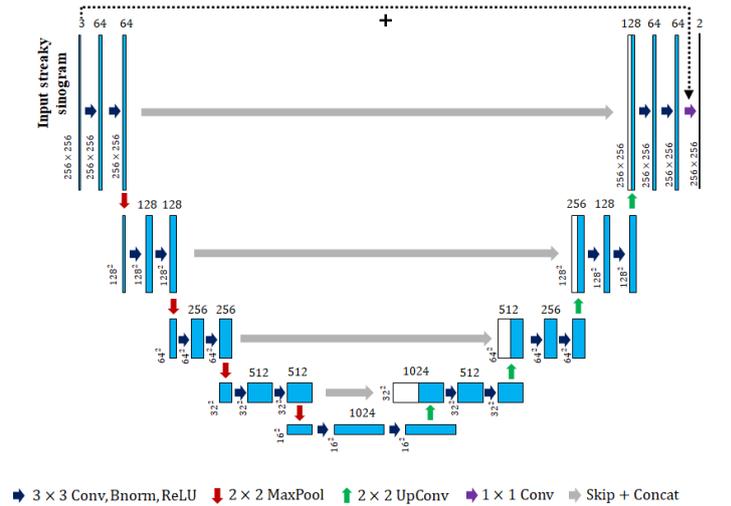

**Figure 4. Proposed residual U-Net backbone.**

$$\vec{f}^{low} = \underset{\vec{f}^*}{\text{argmin}} \left( \left\| \vec{f}^{low} \right\|_{TV} + \beta \left\| GMI^{low\ prior} - w_1 GMI^{low} \right\|_2 \right)$$
$$such\ that\ \left\| A\vec{f}^{low} - \vec{g}^{low} \right\| < \varepsilon_1\ and\ \vec{f}^{low} \geq 0\ \cdots (1)$$

$$\vec{f}^{high} = \underset{\vec{f}^*}{\text{argmin}} \left( \left\| \vec{f}^{high} \right\|_{TV} + \beta \left\| GMI^{high\ prior} - w_1 GMI^{high} \right\|_2 \right)$$
$$such\ that\ \left\| A\vec{f}^{high} - \vec{g}^{high} \right\| < \varepsilon_1\ and\ \vec{f}^{high} \geq 0\ \cdots (2)$$

, where $\vec{g}^{low}$ and $\vec{g}^{high}$ correspond to the low- and high-energy projection data. $GMI^{low\ prior}$ and $GMI^{high\ prior}$ are GMI of FBP images of low- and high-energy output sinograms in Step 1, respectively. $w_1$ and $w_2$ correspond to the ratio of $l_1$ norm of $GMI^{low\ prior}$ to that of $GMI^{low}$ and the ratio of $l_1$ norm of $GMI^{high\ prior}$ to that of $GMI^{high}$, respectively, to normalize GMI scales between them [13]. In last, image-based material decomposition is then followed [14].

## B. Beam-filter parameters optimization

Optimization of beam-filter parameters such as the number of slits and the cycle number of the streaks in the sinogram domain is important in image quality and also in the accuracy of material decomposition. We have calculated sampling density (SD) of reconstruction image domain for the optimization considering penumbra region [15], and 4 strips and 4 cycles were found to be optimum as shown in Fig 5, which were used in the simulation.

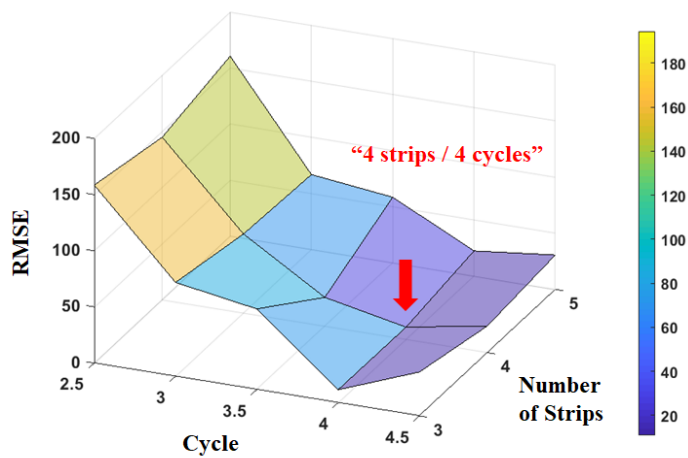

**Figure 5. RMSE of SD from the ideal uniform case for each streak pattern.**

## C. Simulation study

A simulation study was conducted using the 3D digital anthropomorphic phantom. Acquired projection data were composed of 80 kVp and 140 kVp energy spectrum information, for the unfiltered and filtered regions, respectively. 67,500 data set was used for the training, and fan-beam circular scanning geometry was used for the feasibility study. Detailed simulation parameters are summarized in Table 1.

**Table 1. Simulation & Deep learning parameters.**

| Parameters | Values |
| --- | --- |
| Views per rotation | 720 |
| Detector pixel number | 512 × 1 |
| Detector pixel pitch | 0.8 mm |
| Distance of source to detector | 1300 mm |
| Distance of source to object | 1100 mm |
| X-ray source | 140 kVp / 80 kVp |
| Beam-filtered region ratio | 50 % |
| Beam-filter shape | 4 strips / 4cycles |
| Patch size | 256 × 256 |
| Patch stride | 58(view), 64(detector) |
| Training data number | 1,500 case |
| (Patch number) | (67,500 patches) |
| Learning rate | 1e-04 |
| Optimizer | ADAM |
| Loss function | MSELoss |
| Library | Pytorch |

## 3 Results

Figure 6 shows (a) original, (b) low-energy label, (c) high-energy label, (d) low-energy output, and (e) high-energy output sinogram. In turn, the figure shows (f) difference of (b) and (d), and (g) difference of (c) and (e). Overall structural and energy information is well recovered as seen in Fig. (d) and (e), however, the error remains in counterpart energy region in output sinograms as shown in Fig. (f) and (g). To handle this issue, after FBP of (d) and (e), 5 steps of CS-inspired IR using the original sinogram are processed to improve the fidelity of the reconstructed image value.

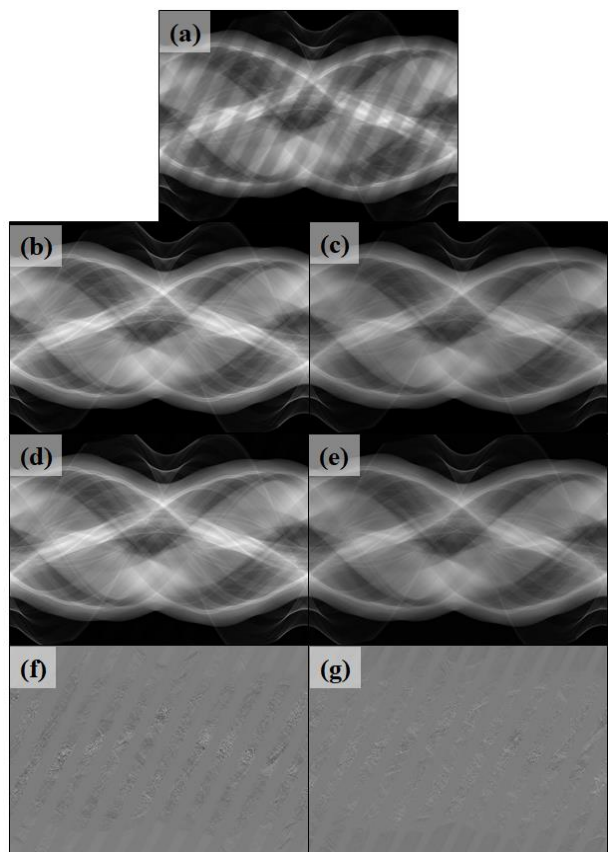

**Figure 6. (a) original, (b) low-energy label, (c) high-energy label, (d) low-energy output, and (e) high-energy output sinogram. (f) difference of (b) and (d), and (g) difference of (c) and (e).**

Figure 7 shows FBP of the label (a) low- and (b) high-energy data, FBP of output (c) low- and (d) high-energy data, and FBP + 5 steps of CS-inspired IR of output (e) low- and (f) high-energy data. Structural similarity (SSIM) was evaluated for the cases of Fig. 7 (c) and (e) from (a) for low-energy cases, and (d) and (f) from (b) for the high-energy cases. As result, SSIM value has been improved in (e) and (f) as summarized in Table 2. Figure 8 shows corresponding material decomposed images of (a) soft tissue, (b) bone, and (c) air. The results show that the proposed method can successfully provide the reconstruction image for both energies and decompose an image into the specific-material map in a shorter computation time than our previous method.

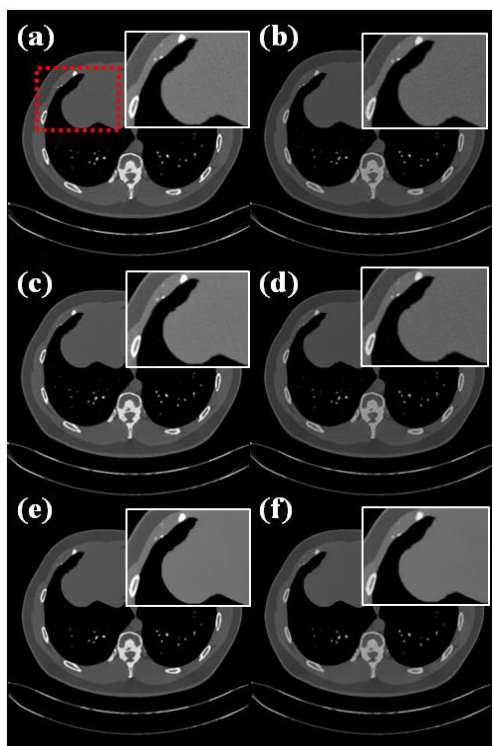

**Figure 7. FBP of label (a) low- and (b) high-energy data. FBP of output (c) low- and (d) high-energy data. FBP and 5 CS-inspired IR of output (a) low- and (b) high-energy data.**

**Table 2. Calculated SSIM values for the cases of (c) and (e) for (a) in low-energy case, and SSIM values for the cases of (d) and (f) for (b) in high-energy case.**

| SSIM | FBP | FBP + 5 CS-inspired IR |
|---|---|---|
| **Low-energy** | 0.9918 | 0.9970 |
| **High-energy** | 0.9949 | 0.9986 |

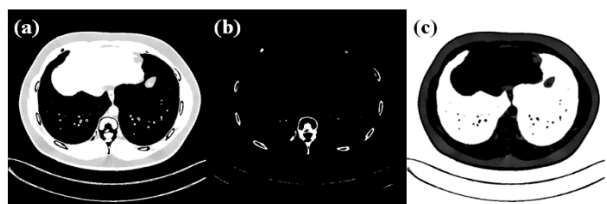

**Figure 8. Material decomposed images of (a) soft tissue, (b) bone, and (c) air.**

## 4 Conclusion

In this preliminary simulation study, we extended our previous MVUS method of using sliding multi-slit beam-filter for low-dose dual-energy imaging to a deep learning-based imaging method. We exploited the residual U-Net network to make fully restored low- and high-energy sinograms out of the original streaky sinogram. 5 steps of CS-inspired IR were then followed to improve the fidelity of the reconstructed image value. Through a simulation study, we have successfully demonstrated the feasibility of our proposed method.

**Chapter 7**

# Oral Session - Denoising methods for low dose imaging

**session chairs**

Johan Nuyts, *KU Leuven (Belgium)*
Liang Li, *Tsinghua University (China)*

# Low-Dose CT Denoising with Multi-scale Residual Attention Convolutional Neural Network

Zhan Wu[1,2], Yikun Zhang[3], Yongshun Xu[2], Yang Chen[1,3,4,5,*], Limin Luo[1,3,4,5], and Hengyong Yu[2,*]

[1]School of Cyberspace Security, Southeast University, Nanjing, Jiangsu, China
[2]Department of Electrical and Computer Engineering, University of Massachusetts Lowell, Lowell, MA, USA
[3]School of Computer Science and Engineering, Nanjing, Jiangsu, China
[4]Centre de Recherche en Information Biomedicale Sino-Francais (LIA CRIBs) Rennes, France
[5]Key Laboratory of Computer Network and Information Integration, Southeast University, Nanjing, Jiangsu, China
*email: chenyang.list@seu.edu.cn, Hengyong-yu@ieee.org

**Abstract: Low-dose computed tomography (LDCT) has emerged as a powerful tool for clinical diagnosis. However, the reduction of radiation dose severely degrades the reconstructed CT image quality with noise and artifacts. In the past decade, deep learning (DL) based methods have made rapid progress for CT image denoising driven by convolution neural networks (CNNs). Nevertheless, the contextual feature representation for LDCT denoising has not been fully investigated. In this paper, we proposed a novel multi-scale residual attention network (MSRANet) for LDCT denoising. To enrich the contextual information, a dilated convolution is introduced. To incorporate different scales of spatial features and enhance spatial feature dependencies, a multi-scale attention mechanism is designed. The proposed MSRANet is evaluated on the AAPM-Mayo Clinic Low Dose CT Grand Challenge dataset, and it outperforms the state-of-art competing methods.**

## 1 Introduction

Low-dose X-ray Computed tomography (CT) is of great significance for high resolution imaging that is increasingly applied in clinical diagnosis such as lung cancer screening. Many algorithms for low-dose CT (LDCT) denoising are proposed to improve the image quality and avoid the abuse of radiation dose. Considering the fact that the utilization of inaccurate prior knowledge and low-level handcrafted features may limit the robustness, supervised deep learning based methods were developed to learn hierarchical and complicated representation of data features for LDCT denoising [1, 2]. Existing deep learning-based methods contain two strategies for CT denoising: **encoder-decoder** and **high-resolution feature processing**.

**Encoder-decoder:** The encoder-decoder based models extract the hierarchical LDCT features to low-dimension representation and back to the original dimension using gradual reverse mapping. Hu *et al.* proposed to combine the auto-encoder, deconvolution network, and shortcut connections into a residual encoder-decoder CNN for LDCT denoising [3]. Mao *et al.* combined symmetrically convolutional and deconvolutional layers with skip-layer connections for much faster training convergence and attained a higher-quality local optimum [4]. Even so, this strategy cannot focus on the detailed artifacts and noise when complicated and hierarchical features are learned via dimension downsampling. This tends to produce false positive samples and cause clinical misdiagnosis [5].

**High-resolution feature processing:** The high-resolution feature processing based strategy keeps pixel-to-pixel correspondence with the input LDCT images and do not employ any downsampling units [6-8]. Yi *et al.* combined an adversarially trained network and a sharpness detection network for LDCT denoising. They obtained small resolution loss and excellent denoising performance [6]. Yang *et al.* introduced a generative adversarial network (GAN) with Wasserstein distance and perceptual similarity for LDCT denoising. They effectively reduced the image noise level and kept the critical information.

All the aforementioned methods show great potentials for LDCT image denoising. However, these methods do not sufficiently capture comprehensive and enriched information and make full use of contextual features including low- and high-dimension features. It causes ineffective feature representation for LDCT image denoising. In this paper, we propose a multi-scale residual attention network (MSRANet) based on the deep convolution neural networks (DCNNs) for LDCT image denoising. A dilated convolution is introduced in the proposed MSRANet to enlarge the receptive fields. By explicitly using multi-layer dilated convolution operations, wider contextual information are captured for feature representation with no extra computation cost. Besides, multi-scale attention mechanism is designed to generate low- and high- dimension feature representations for multi-scale spatial information integration and suppress irrelevant information. It effectively increases model representation power and improves the performance of LDCT image denoising.

## 2 Materials and Methods

The LDCT denoising task can be described as a noise reduction model in the image domain. Let $x \in \mathbb{R}^{N \times N}$ denotes an input LDCT image and $y^* \in \mathbb{R}^{N \times N}$ denotes the corresponding NDCT image. The target of this model is to find a function $G(\cdot)$ that can synthesize a new output image $y$ close to the NDCT image $y^*$ from the LDCT image $x$:

This work was supported in part by the State's Key Project of Research and Development Plan under Grant 2017YFA0104302, Grant 2017YFC0109202 and 2017YFC0107900, in part by the National Natural Science Foundation under Grant 61801003, 61871117and 81471752, in part by the China Scholarship Council under NO. 201906090145.

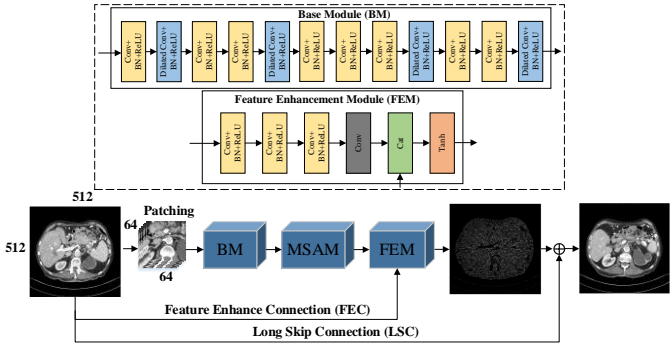

**Fig. 1.** Overview of the proposed MSRANet for LDCT image denoising. BM alternately utilizes the dilated and standard convolution layers for feature representation. MSAM employs the attention mechanism and multi-scale feature fusion for comprehensive contextual aggregation. FEM effectively combines original input LDCT image and feature maps in deeper layer for feature enhancement. The final denoised CT images are obtained by combining the original noisy LDCT images and predicted residual images.

$$y = \mathrm{G}(x) \rightarrow y^*. \qquad (1)$$

The LDCT denoising task can be transformed to build a function $\mathrm{G}(\cdot)$ for optimal approximation:

$$\underset{G}{argmin} \left\| G(x) - y^* \right\|_2^2 . \qquad (2)$$

In this paper, we propose a multi-scale residual attention network (MSRANet) for LDCT denoising. It is composed of three interactional functional components: **Base module (BM)**, **Multi-scale attention module (MSAM)**, and **Feature enhancement module (FEM)**. The overall framework of the proposed network architecture is illustrated in Fig. 1.

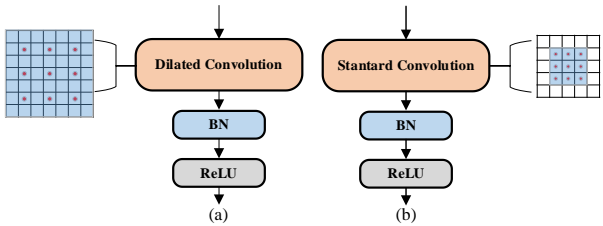

**Fig. 2.** The representative types of (a) dilated convolution layer with kernel size of $3 \times 3$ and dilated factor of 2, and (b) standard convolution layer with kernel size of $3 \times 3$.

## 2.1 Base module (BM)

Serving as the base of the proposed MSRANet, the base module (BM) consists of 12 convolution layers and performs contextual feature representation. Four dilated convolution layers and eight standard convolution layers are alternately utilized in the BM.

Inspired by the work [9], the dilated convolution operation is utilized. Different from standard convolution operation, it is a convolution applied to input with defined gaps in the kernels, which can effectively increases receptive fields and linear parameter accretion of the convolution layer to integrate wider context information with less cost for LDCT image denoising. Compared with standard convolution operation, the size of receptive field F can be expressed as:

$$\mathrm{F}_{i+1} = (2^{i+r} - 1) \times (2^{i+r} - 1) , \qquad (3)$$

where $i$ is the kernel size of the dilated convolution and $r$ is the dilated convolution rate. The dilated convolution operation can be transformed into standard convolution operation when $r$ is 1. Fig. 2 demonstrates the difference between the dilated and corresponding standard convolution layers. In this paper, the dilated convolution layer denotes that a dilated convolution with dilated factor of 2 and kernel size of $3 \times 3$, batch normalization (BN), and Rectified Linear Unit (ReLU) activation function are connected. The standard convolution layer denotes that a standard convolution with kernel size of $3 \times 3$, BN and ReLU are connected.

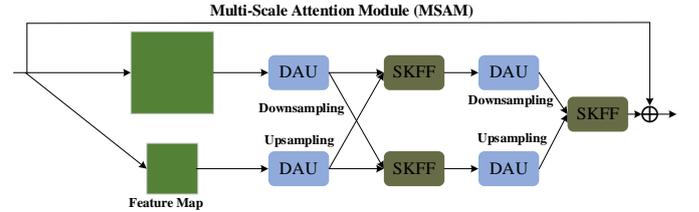

**Fig. 3.** The architecture of multi-scale attention module (MSAM)

## 2.2 Multi-scale attention module (MSAM)

The multi-scale attention module (MSAM) generates spatially-precise multi-scale feature representation, which effectively maintains high-resolution representations and receives rich contextual information from low-resolution noisy LDCT feature maps. This module consists of selective kernel feature fusion unit (SKFFU), dual attention unit (DAU), which is illuminated in Fig. 3.

Dual attention unit (DAU) extracts the attention-guided feature representation to refine the denosed LDCT image, which includes the channel and spatial attention operations. The channel attention operation focuses on identifying channel information to represent meaningful channel features. The spatial attention operation figures out where should be paid attention to and provides inter-spatial information for a given feature map. The outputs of two attention operations are combined via concatenation layer and recovered via the convolution layer into the feature map same with original channel to extract enriched features.

Selective kernel feature fusion unit (SKFFU) nonlinearly fuses the multiple scale features to capture more robust features. First, two scales of feature maps ($L_1, L_2$) from DAUs are aggregated via element-wise summation to form one feature map $L \in \mathbb{R}^{C \times H \times W}$ :

$$L = \sum_{i=1}^{2} L_i . \qquad (4)$$

The combined feature map $L$ is fed into global average pooling (GAP) to compute channel-wise features $L \in \mathbb{R}^{C \times 1 \times 1}$ for dimension reduction. One convolution layer is utilized for channel-wise feature extraction. The extracted channel-wise feature vector $\hat{L}$ is employed to form the feature map $\hat{L}_n \in \mathbb{R}^{C \times 1 \times 1}$, $n \in \{1, 2\}$ with channel number of C=2, and these two scales of feature maps are normalized using Softmax function into $y_n \in \mathbb{R}^{C \times l \times l}$ with the range of $(0, 1)$.

Finally, these two scales of normalized features are combined with the original three feature maps by element-wise product:

$$y = \sum_{n=1}^{3} y_n \cdot L_n , \qquad (5)$$

where '·' stands for element-wise product and y is the combined feature maps with multiply scales of information.

For these two different scales of feature maps, the dimension unification of feature maps between layers is performed by down-sampling and up-sampling operations. When the size of upper feature map is greater than that of latter layer, the upper feature map is down-sampled. When the size of upper feature map is less than that of latter layer, the upper feature map is up-sampled.

### 2.3 Feature enhancement module (FEM)
Feature enhancement module (FEM) highlights the features in the deep layers. It is known that the deep networks are more vulnerable to weaken influences in feature representation compared with shallow networks for LDCT image denoising [10]. The FEM effectively extracts the global and local features through combination between original images and deep feature maps for more robust residual features.

In this paper, we choose $L_1$ loss as objective function for our network to accelerate training [11]. The utilized $L_1$ loss function is given by:

$$loss = |\hat{y} - y| , \qquad (6)$$

where y is the labeled vector and $\hat{y}$ is the predicted probability vector.

## 3 Experiment and Results
In this section, the public dataset is employed to train and evaluate the proposed method. The related data preprocessing, implementation details, and evaluation criteria are discussed. The ablation experiments are carried out, and quantitative and qualitative results of the proposed network for LDCT image denoising are presented. Furthermore, the proposed method is compared with the state-of-the-art methods. Our experiments show that the proposed MSRANet network outperforms other state-of-the-art methods, which indicates its effectiveness for LDCT image denoising.

### 3.1 Data Sources
The evaluation is performed on a publicly available dataset that was authorized by Mayo Clinics for "the 2016 NIH-AAPM-Mayo Clinic Low Dose CT Grand Challenge". This dataset contains 2378 pairs of CT image slices from 10 anonymous patients, each pair includes 3mm thickness normal dose CT (NDCT) and quarter dose CT (LDCT) $512 \times 512$ images [12]. For fairness, nine patients including

LDCT scans and the corresponding NDCT scans are employed for training, and the rest of patient L506 is used for testing the proposed MSRANet network. It is noted that there is no data overlapping between the training images and testing images.

### 3.2 Data preprocessing
At the training stage, to effectively increase the sample number and reduce the computation complexity, the small patches are extracted to serve as the training dataset. In this paper, we randomly extracted eight patches with size of $64 \times 64$ pixels in each pair of LDCT and NDCT images, and they are fed into the proposed MSRANet. In the testing stage, the whole CT images are used. Two metrics, including the root mean square error (RMSE) and structural similarity index measure (SSIM), were chosen for quantitative assessment of image quality.

### 3.3 Ablation Study of Proposed Methods

**Table 1.** List of all trained networks for ablation study.

| Experiments | Descriptions |
|---|---|
| LDCT | The images reconstructed by FBP from quarter-dose CT projection data. |
| CNN-Normal | No Dilated Convolution and no Multi-scale Attention Module |
| CNN-Dilated | With Dilated Convolution, no Multi-scale Attention Module |
| MSRANet | With Dilated Convolution and Multi-scale Attention Module |

In this section, we perform model ablation studies on different neural networks with different settings (see Table 1) to validate the effectiveness of the employed dilated convolution and multi-scale attention module (MSAM), respectively. The evaluation results with the metrics of SSIM and RMSE values are summarized in Tables 2 for all the slices in case L506. In our experiments, the normal dose CT images are used as the reference.

As shown in Table 2, the performances of different networks are computed on the testing case L506, including the CNN-Normal, CNN-Dilated, and MSRANet networks. It can be found that CNN-Dilated network outperforms the CNN-Normal network in SSIM of 0.0268 and RMSE of 2.3385. MSRANet can obtain the best denoising performance in these two metrics.

### 3.4 Comparison with the state-of-the-art methods
To visualize the denoising performance, we carried out experiments to test the case L506. We depict the results of CNN-Normal, the proposed CNN-Dilated and MSRANet networks. For comparison, we also show the results of several state-of-the-art methods, including RED-CNN [3], WGAN-VGG [7], and MAP-NN [13]. Fig. 4 shows the visualization results on the representative slice for case L506. Our proposed MASNet is effective in removing noise and produces perceptually-pleasing and sharp images. Furthermore, it is capable of maintaining the spatial smoothness without introducing artifacts.

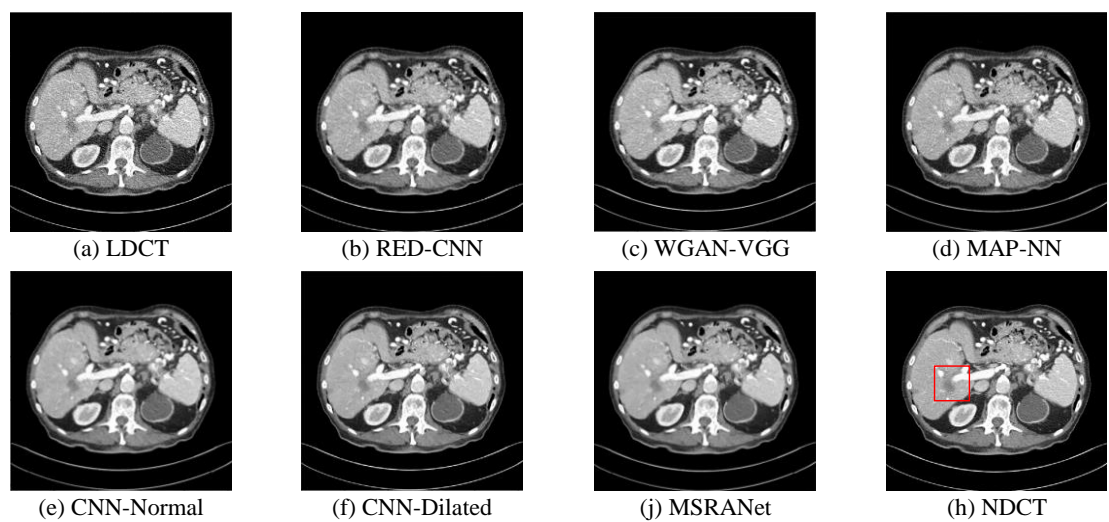

**Fig. 4.** Comparison study for case L506 with lesion 575. (a) and (h) are the LDCT and corresponding ground-truth NDCT image; (b) - (j) are the denoised images generated by different networks. The display window is [-160, 240] HU.

For the quantitative comparison with respect to NDCT, we used all slices of case L506 as testing dataset to compare the denoising performance in the terms of SSIM and RMSE. As shown in Table 2, our proposed MSRANet achieved the most excellent performance, which indicates the advantage of the proposed network architecture.

**Table 2.** Quantitative comparison results with the state-of-the-art methods for case L506.

| Method | SSIM | RMSE |
|---|---|---|
| LDCT | 0.8759 | 14.2416 |
| RED-CNN | 0.8952 | 11.5926 |
| WGAN-VGG | 0.9008 | 11.6370 |
| MAP-NN | 0.8941 | 11.5848 |
| CNN-Normal | 0.8908 | 10.9982 |
| CNN-Dilated | 0.9176 | 8.6597 |
| MSRANet | 0.9187 | 8.6722 |

## 4 Discussion

In this paper, for the first time, we proposed a novel multi-scale residual attention network (MSRANet) for LDCT denoising. The motivation for this paper includes two parts: 1) the traditional convolution operations do not sufficiently capture the contextual information across the whole CT images; 2) most of methods do not make full use of different scales of spatial features, which can neglect some key features hidden in the complex background. The contributions of this work are as follows: 1) multilayer dilated convolution operation in the proposed MSRANet to enlarge the receptive fields for capturing wider contextual information; 2) multi-scale attention mechanism is designed to generate different dimensions of features for multi-scale spatial information integration. The proposed MSRANet is trained and evaluated on a publicly available dataset released as the 2016 NIH-AAPM-Mayo Clinic Low Dose CT Grand Challenge and achieves an excellent CT denoising performance.

# MAGIC: Manifold and Graph Integrative Convolutional Network for Low-Dose CT Reconstruction

Wenjun Xia[1], Zexin Lu[1], Yongqiang Huang[1], Zuoqiang Shi[2], Yan Liu[3], Hu Chen[1], Yang Chen[4], Jiliu Zhou[1], and Yi Zhang[1,*]

[1]College of Computer Science, Sichuan University, Chengdu 610065, China
[2]Department of Mathematical Sciences, Tsinghua University, Beijing 100084, China
[3]School of Electrical Engineering Information, Sichuan University, Chengdu 610065, China
[4]Laboratory of Image Science and Technology, Southeast University, Nanjing 210096, China

**Abstract** Low-dose computed tomography (LDCT) scans, which can effectively alleviate the radiation problem, will degrade the imaging quality. In this paper, we propose a novel LDCT reconstruction network that unrolls the iterative scheme and performs in both image and manifold spaces. Because patch manifolds of medical images have low-dimensional structures, we can build graphs from the manifolds. Then, we simultaneously leverage the spatial convolution to extract the local pixel-level features from the images and incorporate the graph convolution to analyze the nonlocal topological features in manifold space. The experiments show that our proposed method outperforms both the quantitative and qualitative aspects of state-of-the-art methods. In addition, aided by a projection loss component, our proposed method also demonstrates superior performance for semi-supervised learning. The network can remove most noise while maintaining the details of only 10% (40 slices) of the training data labeled.

## 1 Introduction

Low-dose X-ray computed tomography (LDCT) can effectively reduce the risk of radiation exposure and thus plays an important role in radiology. However, a lower-dose scan will degrade the signal-to-noise ratio (SNR) of the reconstructed images and compromise the diagnosis accuracy. It is very difficult to meet the diagnostic demands with LDCT images reconstructed via the classic analytical method, i.e., filtered back-projection (FBP). To balance the radiation dose and imaging quality, a number of algorithms have been developed for LDCT reconstruction. With the very recent technological innovations, these algorithms can generally be divided into two categories: 1) regularization-based methods and 2) learning-based methods.

The regularization-based methods formulate the prior knowledge into a reconstruction model. Appropriate prior information, such as total generalized variation (TGV) [1], which efficiently characterizes the target image, can maintain the critical details of the reconstructed result while eliminating unexpected noise and artifacts. However, the regularization-based methods are difficult to be applied to clinic because of the expensive time consumption.

Inspired by the success of deep learning in many related fields, learning-based methods have become the mainstream of medical imaging. By using skip connections, Chen et al. developed a residual encoder-decoder convolutional neural network (RED-CNN) for LDCT denoising [2]. Chen et al. unrolled the steepest gradient descent algorithm and proposed the learned experts' assessment-based reconstruction network (LEARN) for sparse-view CT [3]. Adler and Öktem generalized the primal-dual hybrid gradient (PDHG) algorithm by replacing both the primal and dual proximal operators with learned operators, which were implemented by a trained CNN [4]. However, spatial convolution is a local operator only focused on adjacent pixels, ignoring the fact that CT image data are located on a low-dimensional manifold, which accommodates rich topological structure information.

In this paper, to simultaneously extract the pixel-level and topological features of LDCT data, we propose a manifold and graph integrative convolutional (MAGIC) network that performs in both image and manifold spaces for LDCT reconstruction. First, we unroll the gradient descent algorithm into a neural network and use a CNN module to replace the handcrafted regularization terms. Then, to introduce the low-dimensional manifold features, overlapped patches with a small size are extracted from the image to form a patch set. This operation is based on a well-accepted assumption that the patch set is located on a low-dimensional smooth manifold referred to as a patch manifold [5]. Since spatial convolution cannot process such data, inspired by the success of a graph convolution [6], we construct a graph using the points sampled from the patch manifold, and a graph convolution is applied to extract the topological features from the graph. In addition, since it is difficult to obtain a large amount of paired low-dose and normal-dose data in clinical practice, our proposed method alleviates this drawback by introducing a projection loss, which enables our semi-supervised learning model.

## 2 Methods

A general model for regularized reconstruction is as follows:

$$\min_{x} \frac{1}{2}\|Ax - y\|_2^2 + \lambda R(x), \qquad (1)$$

where $x \in \mathbf{R}^{M_2}$ denotes the vectorization of image $f \in \mathbf{R}^{m \times n}$ ($M_2 = m \times n$), $y \in \mathbf{R}^{M_1}$ represents the measured projection data, and $A \in \mathbf{R}^{M_1 \times M_2}$ is the system matrix. $R(x)$ denotes the regularization term reflecting the prior knowledge of the image to reconstruct, and $\lambda$ is a weight to balance the measurement and regularization term.

A simple gradient descent algorithm can be used to solve the model, and a classic method to unroll the iterative algorithm into CNN model can be formulated as:

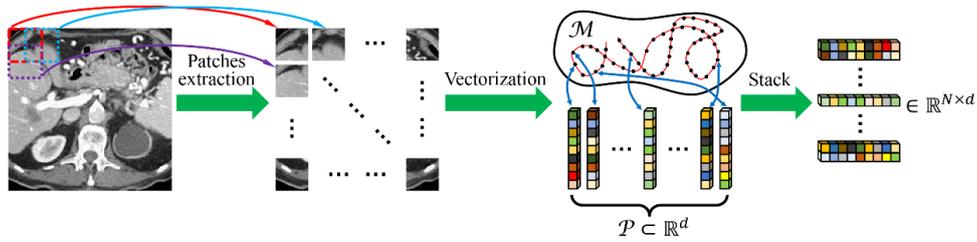

Fig. 1. Diagram of the linear transform to obtain $X$ from $f$. Each vectorized patch corresponds to (blue arrows) a black point on the patch manifold. The patch set $P$ (black points) has a trivial 2D parameterization (red curve) on the patch manifold $M$.

$$x^{t+1} = x^t - \alpha^t A^T (Ax^t - y) + \Phi(x^t), \qquad (2)$$

in which

$$\Phi(x^t) = w_3^t * \sigma\big(w_2^t * \sigma(w_1^t * x^t)\big), \qquad (3)$$

where $w$ is the trained kernel, $*$ denotes the convolution operator and $\sigma(\cdot)$ is the activation function.

In our method, we attempt to simultaneously extract the pixel-level and topological features by incorporating both spatial and graph convolutions. In Eq. (2), a three-layer CNN module is used to extract the local pixel-level features of $x^t$. To impose the nonlocal topological features from the low-dimensional manifold space, we modified Eq. (2) and added a graph convolutional network (GCN) term. First, a patch set $P(f^t)$ is built, we extract a small rectangular patch $p_{ij}(f^t)$, which has pixel $f^t(i,j)$ as the top-left corner and a size of $s_1 \times s_2$. $P(f^t)$ can be seen as a point cloud sampled from a low-dimensional manifold $M(f^t)$ embedded in $\mathbf{R}^d$, referred to as the patch manifold associated with $f^t$. Then we construct a graph $G^t(V, E)$ with $N$ nodes, each of which corresponds to a certain element of $P(f^t)$. The adjacency matrix $W \in \mathbf{R}^{N \times N}$ of the graph can be calculated with a Gaussian function [6]:

$$W_{ij} = \exp\left(-\frac{\|v_i - v_j\|_2^2}{\sigma(V)^2}\right), \qquad (4)$$

where $v_i, v_j \in V$ are the two nodes in the graph and $\sigma(V)$ is the standard deviation of the nodes. The diagonal degree matrix $D$ is defined as $D_{ii} = \sum_j W_{ij}$. Considering the autocorrelation, we can obtain $\widetilde{W} = I + W$ and $\widetilde{D}_{ii} = \sum_j \widetilde{W}_{ij}$. Then, the nodes are stacked to obtain the matrix signal $X^t \in \mathbf{R}^{N \times d}$, and two successive graph convolutions are applied on it. Fig. 1 illustrates the main steps to obtain $X$ from $f$. Our model is modified from Eq. (2) to:

$$x^{t+1} = x^t - \alpha^t A^T (Ax^t - y) + \Phi(x^t) + \Psi(X^t) \quad (5)$$

where $\Psi(X^t) = \widetilde{D}^{-\frac{1}{2}} \widetilde{W} \widetilde{D}^{-\frac{1}{2}} \sigma\left(\widetilde{D}^{-\frac{1}{2}} \widetilde{W} \widetilde{D}^{-\frac{1}{2}} X^t \Theta_1^t\right) \Theta_2^t$, $\Theta_1 \in \mathbf{R}^{d \times F}$ and $\Theta_2 \in \mathbf{R}^{F \times d}$ are the graph convolutional kernels. Notably, the computation of the adjacency matrix is time-consuming if we update it in each iteration. Based on this consideration, we divide the whole iteration procedure into two stages: coarse and fine stages. Fig. 2 shows the flowchart of our proposed unrolled iteration network MAGIC. In the coarse stage, $x^0$ (initial reconstruction with FBP) and projection data $y$ are fed into the network. Compared with LEARN, one parallel path, which performs graph convolution, is added into each iteration block. The adjacency matrix of coarse stage $W_C$ is calculated based on $x^0$ and kept fixed in each iteration block during the entire coarse stage. The graph transform in Fig. 2 is the linear transform to obtain $X$ from $f$, and the inverse graph transform denotes the inverse operator. In the coarse stage, the result of FBP usually suffers from heavy noise, which makes $W_C$ inaccurate. After the $t+1$ iteration, once the noise of $x^{t+1}$ has been basically removed, the network enters the fine stage. We recalculate the adjacency matrix $W_F$ based on $x^{t+1}$ and leave it unchanged during the entire fine stage.

Mean square error (MSE) is adopted as the loss function:

$$L_{MSE} = \frac{1}{N_s} \sum_{i=1}^{N_s} \|x_i - \hat{x}_i\|_2^2, \qquad (6)$$

where $x_i$ is the predicted reconstruction result and $\hat{x}_i$ is the corresponding label. $N_s$ is the total number of samples. In addition, we apply our proposed MAGIC to only part of the labeled samples. The projection loss is proposed as:

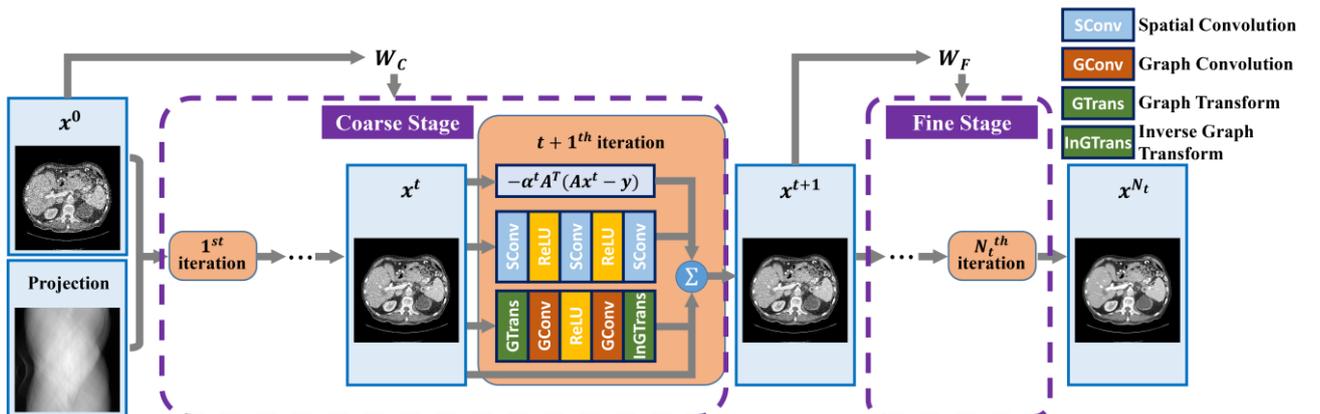

Fig. 2. Illustration of our proposed MAGIC.

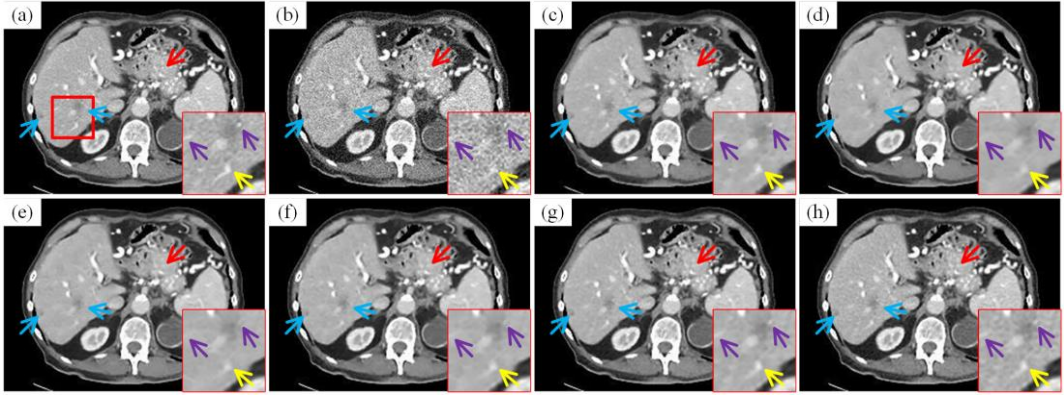

Fig. 3. Abdominal reconstruction with 10% dose data by different methods. (a) Ground truth, (b) FBP (25.02/0.7084), (c) TGV (30.84/0.8925), (d) RED-CNN (31.20/0.8955), (e) LPD (31.38/0.9040), (f) LEARN (31.80/0.9078), (g) MAGIC (**34.00**/0.9356) and (h) MAGIC-Semi (33.55/**0.9360**). The display window is [-160, 240] HU.

$$L_{Pro} = \frac{1}{N_S} \sum_{i=1}^{N_S} \|Ax_i - y_i\|_2^2, \qquad (7)$$

where $y_i$ is the corresponding measured projection data. In the case of semi-supervised learning, the loss function can be formulated as:

$$L = \frac{1}{|S_1|} \sum_{i \in S_1} \|x_i - \hat{x}_i\|_2^2 + \frac{1}{|S_2|} \sum_{i \in S_2} \|Ax_i - y_i\|_2^2 \quad (8)$$

where $S_1$ and $S_2$ are the sets of labeled and unlabeled samples, respectively. $|S_1|$ and $|S_2|$ denote the numbers of elements in $S_1$ and $S_2$, respectively, and $|S_1| + |S_2| = N_s$. While dealing with the unlabeled data in the training set, the projection loss can be leveraged to avoid overfitting.

## 3 Results

To evaluate the performance of our MAGIC, the dataset "*the 2016 NIH-AAPM-Mayo Clinic Low-Dose CT Grand Challenge*" was used in our experiments. In our experiments, 400 images were randomly selected from 8 patients as the training set, and 100 images were chosen from the remaining 2 patients as the test set. The size of the image was $256 \times 256$. The distances of the X-ray source and detector to the rotation center were both 25 cm. The physical height and width of a pixel were both 0.6641 mm. The detector had 512 elements, each of which had a length of 0.72 mm. On average, 1024 projection views were sampled in the 360 degree range. To simulate a realistic clinical environment, Poisson noise and electronic noise were added into the measured projection data as:

$$y = \ln \frac{I_0}{\text{Poisson}(I_0 \exp(-\hat{y})) + \text{Normal}(0, \sigma_e^2)}, \qquad (9)$$

where $I_0$ is the number of photons before the X-rays penetrate the object, $\sigma_e^2$ is the variance of electronic noise, and $\hat{y}$ represents the noise-free projection. In our experiments, the X-ray intensity of a normal dose was set to $I_0 = 10^6$. Three different dose levels were simulated as low-dose cases, including 10%, 5% and 2.5%, i.e., $I_0 = 10^5$, $5 \times 10^4$, and $2.5 \times 10^4$, respectively. We fixed the electronic noise variance at $\sigma_e^2 = 10$.

The size of the spatial convolution kernels was set to $3 \times 3$. When building the graph, the extracted patch size was set to $6 \times 6$. While calculating the adjacency matrix, 8 nearest neighbors of each node were included to make the adjacency matrix sparse and reduce the computational complexity. The sizes of the graph convolution parameters $\Theta_1$ and $\Theta_2$ were $36 \times 64$ and $64 \times 36$, respectively. The number of iterative blocks was fixed to 50. The coarse and find stages had 25 blocks. In the semi-supervised learning experiments, only 10% of the training data, which means only 40 images have labels. Four state-of-the-art methods were involved for comparison, including TGV [1], RED-CNN [2], learned primal-dual (LPD) [4] and LEARN [3]. The semi-supervised learning version of MAGIC is referred to as MAGIC-Semi.

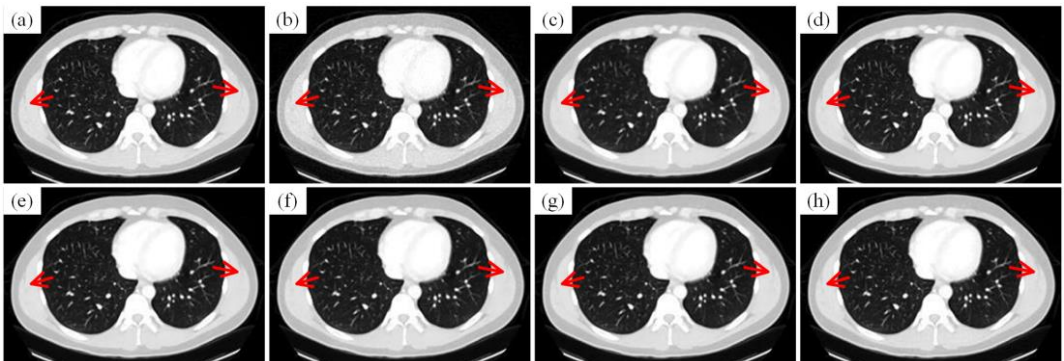

Fig. 4. Thoracic reconstruction with 10% dose data by various methods. (a) Ground truth, (b) FBP (27.93/0.7762), (c) TGV (31.49/0.9299), (d) RED-CNN (32.98/0.9423), (e) LPD (33.24/0.9521), (f) LEARN (33.74/0.9517), (g) MAGIC (**36.26/0.9696**) and (h) MAGIC-Semi (35.58/0.9692). The display window is [-1000, 200] HU.

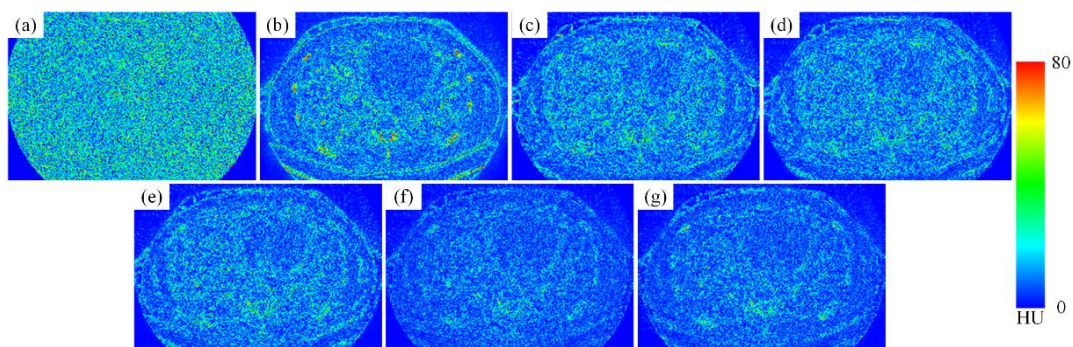

Fig. 5. Absolute difference images associated to the ground truth (a) FBP, (b) TGV, (c) RED-CNN, (d) LPD, (e) LEARN, (f) MAGIC and (g) MAGIC-Semi.

Table I Quantitative results (Mean±SD) using different methods. The best scores are marked in red, and the second best scores are marked in blue.

| dose | 10% | | 5% | | 2.5% | |
|---|---|---|---|---|---|---|
| | PSNR | SSIM | PSNR | SSIM | PSNR | SSIM |
| FBP | 26.35 ±0.68 | 0.6969 ±0.0321 | 23.56 ±0.74 | 0.6160 ±0.0363 | 20.67 ±0.77 | 0.5381 ±0.0374 |
| TGV | 31.91 ±0.50 | 0.9210 ±0.0097 | 31.12 ±0.52 | 0.9103 ±0.0119 | 29.41 ±0.62 | 0.8721 ±0.0207 |
| RED-CNN | 32.89 ±0.58 | 0.9251 ±0.0124 | 31.57 ±0.60 | 0.9123 ±0.0144 | 29.93 ±0.63 | 0.8911 ±0.0169 |
| LPD | 33.12 ±0.59 | 0.9356 ±0.0117 | 31.59 ±0.61 | 0.9194 ±0.0140 | 30.05 ±0.62 | 0.8981 ±0.0165 |
| LEARN | 33.51 ±0.60 | 0.9363 ±0.0112 | 32.18 ±0.61 | 0.9299 ±0.0116 | 30.38 ±0.61 | 0.9090 ±0.0142 |
| MAGIC | **35.89 ±0.66** | **0.9587 ±0.0092** | **34.18 ±0.64** | **0.9460 ±0.0107** | **32.72 ±0.64** | **0.9335 ±0.0120** |
| MAGIC-Semi | 35.18 ±0.59 | 0.9548 ±0.0092 | 33.70 ±0.59 | 0.9425 ±0.0111 | 32.16 ±0.57 | 0.9275 ±0.0133 |

Fig. 3 shows the results of an abdominal image reconstructed by different methods with a 10% dose. The TGV method removes most of the noise while preserving the details. Two possible metastases, which are indicated by the blue arrows, are apparent in all the results of Fig. 3. Without the help of the measured data, the detailed distortion in the result of the RED-CNN is obvious. LEARN and LPD achieved similar performance. The proposed MAGIC and MAGIC-Semi obtained the best visual result and preserved most details. In the region indicated by the red arrow, the vascular structures in our results are more complete and have a higher contrast than the other methods. To better visualize the performance of different methods, we magnify the region indicated by the red rectangle in Fig. 3 (a). Two purple arrows indicate two minute vessels, and only TGV and our methods recovered them well. All the other methods blurred these details to varying degrees. In the area indicated by the yellow arrow, TGV result produced piecewise smooth result. Although the result of MAGIC-Semi has more mottle-like noise than that of MAGIC, the visual effect is more similar to the ground truth. Fig. 4 demonstrates the reconstructions of a thoracic slice using different methods. Two red arrows indicate that two edges can visually differentiate the performance of different methods. Only MAGIC preserved these structures well, and other methods smoothed them to varying degrees. To better visualize the denoising performance of different methods, we show the absolute difference images associated with the ground truth in Fig. 5. It is clear that our proposed methods yielded the smallest difference from the ground truth, eliminating most noise and maintaining more details.

The statistical quantitative results of the whole testing set using different learning-based methods are shown in Table I, which gives the means and standard deviations (SDs) of PSNR and SSIM. It is clear that our methods obtained higher scores than all the other methods.

## 4 Conclusion

In this paper, we propose a novel manifold and graph integrative convolutional network for LDCT reconstruction. This method not only uses spatial convolution to extract local pixel-level features in image space but also utilizes graph convolution to analyze the nonlocal topological features in manifold space. Compared with other methods, our method can capture the self-similarity during local pixels and nonlocal patches simultaneously. The experimental results prove that our method outperforms other methods in both visual and quantitative aspects.

# Two-layer Clustering-based Sparsifying Transform Learning for Low-dose CT Reconstruction

Xikai Yang[1], Yong Long[1], and Saiprasad Ravishankar[2]

[1]University of Michigan - Shanghai Jiao Tong University Joint Institute, Shanghai Jiao Tong University, Shanghai 200240, China

[2]Department of Computational Mathematics, Science and Engineering and Department of Biomedical Engineering, Michigan State University, East Lansing, MI 48824, USA

**Abstract** Achieving high-quality reconstructions from low-dose computed tomography (LDCT) measurements is of much importance in clinical settings. Model-based image reconstruction methods have been proven to be effective in removing artifacts in LDCT. In this work, we propose an approach to learn a rich two-layer clustering-based sparsifying transform model (MCST2), where image patches and their subsequent feature maps (filter residuals) are clustered into groups with different learned sparsifying filters per group. We investigate a penalized weighted least squares (PWLS) approach for LDCT reconstruction incorporating learned MCST2 priors. Experimental results show the superior performance of the proposed PWLS-MCST2 approach compared to other related recent schemes.

## 1 Introduction

Low-dose computed tomography (LDCT) has received much interest in clinical and other settings. A predominant challenge in LDCT is to obtain high-quality reconstructions despite the reduced intensity of radiation. Traditional methods such as analytical filtered back-projection (FBP) perform poorly for LDCT reconstruction, and produce substantial streak artifacts. Model-based image reconstruction [1] methods have been especially popular for LDCT. In particular, penalized weighted least squares (PWLS) approaches incorporating the edge-preserving (EP) regularizer [2] significantly reduce the artifacts present in FBP images.

Other works have proposed an adaptive regularization term for statistical iterative reconstruction. In particular, there has been growing interest in designing data-driven regularizers that capture complex sparse representations of signals from training datasets [3]. Sparsifying transform learning [4] is a generalization of analysis dictionary learning and is an approach for learning models that when applied to images approximately sparsify them. Compared to conventional synthesis dictionary learning methods [3] that are often NP-Hard and involve expensive algorithms, sparsifying transform learning methods are computationally very efficient due to closed-form sparse code and transform updates. In particular, the optimal sparse coefficients in the transform model are typically found by thresholding operations.

Due to their low computational cost, several transform learning-based methods have been studied for image reconstruction in recent years including the union of transforms approach based on data clustering (ULTRA) [5] and multi-layer sparsifying transform (MRST) models [6], where the transform domain feature maps (filter sparsification residuals) are sequentially sparsified over layers. Although, both ULTRA and learned MRST models offer benefits for LDCT reconstruction, the MRST model tends to oversmooth image details [6]. On the other hand, the union of transforms (ULTRA) model can flexibly capture a diversity of image edges and subtle details and contrast by learning a transform for each class of features, which motivates combining its benefits with the richness of deep transform models.

In this paper, we propose *unsupervised* learning of a two-layer clustering-based sparsifying transform model (referred to as MCST2) for images, where image patches and their feature maps (filtering residuals) in the transform domain are clustered into different groups, with a different learned transform per group. The image patches or features in each group are assumed sparse under a common transform. We derive an exact block coordinate descent algorithm for both transform learning and for image reconstruction with a learned MCST2 regularizer. We investigate the performance of PWLS with MCST2 regularization for LDCT reconstruction. Our experimental results show that MCST2 achieves improved image reconstruction quality compared to several recent learned sparsity-based approaches. PWLS-MCST2 also significantly outperforms conventional methods such as FBP and PWLS-EP.

## 2 Algorithm for Model Training

### 2.1 Problem formulation

Our proposed method is patch-based. The underlying cluster optimization over two layers groups the training patches and their corresponding filter residuals (feature maps) into different classes. The transform domain residuals in the second layer contain finer structures that are sparsified with a union (collection) of transforms. Our formulation for training the MCST2 model is as follows, with $\mathscr{H}_0 = \{\mathbf{\Omega}_{1,k}, \mathbf{\Omega}_{2,l}, \mathbf{Z}_{1,i}, \mathbf{Z}_{2,j}, C_{1,k}, C_{2,l}\}$ denoting the set of *all* optimized variables:

$$\min_{\mathscr{H}_0} \sum_{k=1}^{K} \sum_{i \in C_{1,k}} \|\mathbf{\Omega}_{1,k}\mathbf{R}_{1,i} - \mathbf{Z}_{1,i}\|_2^2 + \eta_1^2 \|\mathbf{Z}_{1,i}\|_0$$
$$+ \sum_{l=1}^{L} \sum_{j \in C_{2,l}} \|\mathbf{\Omega}_{2,l}\mathbf{R}_{2,j} - \mathbf{Z}_{2,j}\|_2^2 + \eta_2^2 \|\mathbf{Z}_{2,j}\|_0, \quad \text{(P0)}$$

s.t. $\mathbf{R}_{2,i} = \mathbf{\Omega}_{1,k}\mathbf{R}_{1,i} - \mathbf{Z}_{1,i}, \forall i \in C_{1,k}, \forall k, \quad \{C_{1,k}\} \in \mathscr{G}_1,$
$\{C_{2,l}\} \in \mathscr{G}_2, \quad \mathbf{\Omega}_{1,k}\mathbf{\Omega}_{1,k}^T = \mathbf{\Omega}_{2,l}\mathbf{\Omega}_{2,l}^T = \mathbf{I}, \forall k, l,$

where $\mathbf{R}_1 \in \mathbb{R}^{p \times N'}$ denotes the training matrix, whose columns $\mathbf{R}_{1,i}$ represent vectorized patches extracted from images, and $\mathbf{R}_2$ denotes the residual map obtained by subtracting the transformed patches and their sparse approximations. In particular, $\mathbf{Z}_1$ and $\mathbf{Z}_2$ denote the sparse coefficient maps for the two layers, with $\eta_1$ and $\eta_2$ denoting sparsity controlling parameters. We assume that the image patches and residual matrix columns can be grouped into $K$ and $L$ disjoint classes, respectively, with $\{\mathbf{\Omega}_{1,k}\}$ and $\{\mathbf{\Omega}_{2,l}\}$ denoting the learned sparsifying transforms in the first and the second layer, respectively. We let $C_{1,k}$ and $C_{2,l}$ denote the sets containing the indices of the columns of $\mathbf{R}_1$ and $\mathbf{R}_2$ that belong to the $k$th class in the first layer and the $l$th class in the second layer, respectively. The sets $\mathscr{G}_1$ and $\mathscr{G}_2$ include all the possible disjoint partitions of $[1:N']$ into $K$ and $L$ sets, respectively.

## 2.2 Transform Learning Algorithm

We propose an exact block coordinate descent (BCD) algorithm to optimize (P0) by alternatively updating $\{\mathbf{Z}_{1,i}, C_{1,k}\}$, $\{\mathbf{Z}_{2,j}, C_{2,l}\}$, and the transforms $\mathbf{\Omega}_{1,k}$ and $\mathbf{\Omega}_{2,l}$. When updating each set of variables, the other variables are kept fixed. Since the solutions to the subproblems are computed exactly, the objective function in (P0) converges over the BCD iterations.

### 2.2.1 Update Coefficients and Clusters in the First Layer

Here, we solve (P0) with respect to the coefficients and cluster memberships in the first layer ($\{\mathbf{Z}_{1,i}, C_{1,k}\}$), with the other variables fixed. This leads to the following subproblem (1), whose exact solution is shown in (2) and (3) (can be derived similar to [5]), with $H_\eta(\cdot)$ denoting the hard-thresholding operator that sets vector elements with magnitude less than $\eta$ to zero.

$$\min_{\substack{\mathbf{Z}_{1,i}, \\ C_{1,k}}} \sum_{k=1}^{K} \sum_{i \in C_{1,k}} \|\mathbf{\Omega}_{1,k}\mathbf{R}_{1,i} - \mathbf{Z}_{1,i}\|_2^2 + \eta_1^2 \|\mathbf{Z}_{1,i}\|_0$$
$$+ \sum_{l=1}^{L} \sum_{j \in C_{2,l}} \|\mathbf{\Omega}_{2,l}\mathbf{R}_{2,j} - \mathbf{Z}_{2,j}\|_2^2. \quad (1)$$

$$\hat{k}_i = \arg\min_{1 \le k \le K} \|\mathbf{\Omega}_{1,k}\mathbf{R}_{1,i} - \tilde{\mathbf{Z}}_{1,i}\|_2^2 + \eta_1^2 \|\tilde{\mathbf{Z}}_{1,i}\|_0$$
$$+ \|\mathbf{\Omega}_{2,\hat{l}_i}(\mathbf{\Omega}_{1,k}\mathbf{R}_{1,i} - \tilde{\mathbf{Z}}_{1,i}) - \mathbf{Z}_{2,i}\|_2^2 \quad (2)$$

where $\tilde{\mathbf{Z}}_{1,i} = H_{\eta_1/\sqrt{2}}(\mathbf{\Omega}_{1,k}\mathbf{R}_{1,i} - 0.5\mathbf{\Omega}_{2,l}^T\mathbf{Z}_{2,i})$ and $\hat{l}_i$ denotes the *fixed* cluster membership in the second layer. The optimal $\mathbf{Z}_{1,i}$ is then given as follows:

$$\hat{\mathbf{Z}}_{1,i} = H_{\eta_1/\sqrt{2}}(\mathbf{\Omega}_{1,\hat{k}_i}\mathbf{R}_{1,i} - 0.5\mathbf{\Omega}_{2,\hat{l}_i}^T\mathbf{Z}_{2,i}), \quad \forall i. \quad (3)$$

### 2.2.2 Update of Transforms in the First Layer

In this step, we solve subproblem (4) for the transforms $\{\mathbf{\Omega}_{1,k}\}$ with the other variables fixed.

$$\min_{\{\mathbf{\Omega}_{1,k}\}} \sum_{k=1}^{K} \sum_{i \in C_{1,k}} \|\mathbf{\Omega}_{1,k}\mathbf{R}_{1,i} - \mathbf{Z}_{1,i}\|_2^2$$
$$+ \sum_{k=1}^{K}\sum_{l=1}^{L} \sum_{j \in C_{2,l} \cap C_{1,k}} \|\mathbf{\Omega}_{2,l}(\mathbf{\Omega}_{1,k}\mathbf{R}_{1,j} - \mathbf{Z}_{1,j}) - \mathbf{Z}_{2,j}\|_2^2. \quad (4)$$

The problem decouples into $K$ parallel updates, one for each $\mathbf{\Omega}_{1,k}$. Let $\mathbf{R}_{1,C_{1,k}}$ and $\mathbf{Z}_{1,C_{1,k}}$ be matrices with columns $\mathbf{R}_{1,i}$, $\mathbf{Z}_{1,i}$, $i \in C_{1,k}$, respectively, and let $\mathbf{Z}_{2,C_{2,l} \cap C_{1,k}}$ be matrix defined similarly. Then, denoting the full singular value decomposition (SVD) of $\mathbf{R}_{1,C_{1,k}}(\mathbf{Z}_{1,C_{1,k}}^T + 0.5\mathbf{Q}_k^T)$ as $\mathbf{U}_{1,k}\mathbf{\Sigma}_{1,k}\mathbf{V}_{1,k}^T$, where $\mathbf{Q}_k$ is defined in (5), the optimal solution is $\hat{\mathbf{\Omega}}_{1,k} = \mathbf{V}_{1,k}\mathbf{U}_{1,k}^T$ [4].

$$\mathbf{Q}_k = [\mathbf{\Omega}_{2,1}^T\mathbf{Z}_{2,C_{2,1} \cap C_{1,k}}, \mathbf{\Omega}_{2,2}^T\mathbf{Z}_{2,C_{2,2} \cap C_{1,k}}, \cdots, \mathbf{\Omega}_{2,L}^T\mathbf{Z}_{2,C_{2,L} \cap C_{1,k}}] \quad (5)$$

### 2.2.3 Update Coefficients and Clusters in the Second Layer

Next, we solve subproblem (6) with $\{\mathbf{Z}_{1,i}, C_{1,k}, \mathbf{\Omega}_{1,k}, \mathbf{\Omega}_{2,l}\}$ fixed. The (joint) optimal solution for $\{\mathbf{Z}_{2,j}, C_{2,l}\}$ can be exactly computed as shown in (7) and (8).

$$\min_{\substack{\mathbf{Z}_{2,j}, \\ C_{2,l}}} \sum_{l=1}^{L} \sum_{j \in C_{2,l}} \|\mathbf{\Omega}_{2,l}\mathbf{R}_{2,j} - \mathbf{Z}_{2,j}\|_2^2 + \eta_2^2 \|\mathbf{Z}_{2,j}\|_0. \quad (6)$$

$$\hat{l}_j = \arg\min_{1 \le l \le L} \|\mathbf{\Omega}_{2,l}\mathbf{R}_{2,j} - \tilde{\mathbf{Z}}_{2,j}\|_2^2 + \eta_2^2 \|\tilde{\mathbf{Z}}_{2,j}\|_0, \; \forall j \in C_{2,l}, \quad (7)$$

where $\tilde{\mathbf{Z}}_{2,j} = H_{\eta_2}(\mathbf{\Omega}_{2,l}\mathbf{R}_{2,j})$. Then the optimal solution to $\mathbf{Z}_{2,j}$ is as follows:

$$\hat{\mathbf{Z}}_{2,j} = H_{\eta_2}(\mathbf{\Omega}_{2,\hat{l}_j}\mathbf{R}_{2,j}), \quad \forall j. \quad (8)$$

### 2.2.4 Update of Transforms in the Second Layer

In this step, we solve the following subproblem for $\mathbf{\Omega}_{2,l}$, with the other variables kept fixed:

$$\min_{\{\mathbf{\Omega}_{2,l}\}} \sum_{l=1}^{L} \sum_{j \in C_{2,l}} \|\mathbf{\Omega}_{2,l}\mathbf{R}_{2,j} - \mathbf{Z}_{2,j}\|_2^2 \quad (9)$$

Problem (9) decouples into $L$ different updates, one for each transform. Let $\mathbf{R}_{2,C_{2,l}}$ and $\mathbf{Z}_{2,C_{2,l}}$ be matrices with columns $\mathbf{R}_{2,j}$ and $\mathbf{Z}_{2,j}$, $j \in C_{2,l}$, respectively. Then, denoting the full SVD of $\mathbf{R}_{2,C_{2,l}}\mathbf{Z}_{2,C_{2,l}}^T$ as $\mathbf{U}_{2,l}\mathbf{\Sigma}_{2,l}\mathbf{V}_{2,l}^T$, the optimal $\hat{\mathbf{\Omega}}_{2,l} = \mathbf{V}_{2,l}\mathbf{U}_{2,l}^T$.

## 3 Approach for Image Reconstruction

### 3.1 CT Reconstruction Formulation

After learning the collections of transforms $\{\mathbf{\Omega}_{1,k}\}$ and $\{\mathbf{\Omega}_{2,l}\}$, the learned MCST2 model is incorporated into the reconstruction problem via a data-driven regularizer. We then reconstruct the (vectorized) image $\mathbf{x} \in \mathbb{R}^{N_p}$ from noisy sinogram measurements $\mathbf{y} \in \mathbb{R}^{N_d}$ by solving the following problem:

$$\min_{\mathbf{x} \succeq \mathbf{0}} \frac{1}{2}\|\mathbf{y} - \mathbf{A}\mathbf{x}\|_{\mathbf{W}}^2 + \beta S(\mathbf{x}), \quad (P1)$$

where the regularizer $S(\mathbf{x})$ is defined as follows, with $\mathscr{H}_1 = \{\mathbf{Z}_{1,i}, \mathbf{Z}_{2,j}, C_{1,k}, C_{2,l}\}$ again denoting the set of *all* optimized variables:

$$\min_{\mathscr{H}_1} \sum_{k=1}^{K} \sum_{i \in C_{1,k}} \|\mathbf{\Omega}_{1,k}\mathbf{R}_{1,i} - \mathbf{Z}_{1,i}\|_2^2 + \gamma_1^2 \|\mathbf{Z}_{1,i}\|_0$$
$$+ \sum_{l=1}^{L} \sum_{j \in C_{2,l}} \|\mathbf{\Omega}_{2,l}\mathbf{R}_{2,j} - \mathbf{Z}_{2,j}\|_2^2 + \gamma_2^2 \|\mathbf{Z}_{2,j}\|_0, \quad (10)$$

s.t. $\mathbf{R}_{1,i} = \mathbf{P}_i\mathbf{x}$, $\mathbf{R}_{2,i} = \mathbf{\Omega}_{1,k}\mathbf{R}_{1,i} - \mathbf{Z}_{1,i}$, $\forall i \in C_{1,k}, \forall k$.

In problem (P1), $\mathbf{A} \in \mathbb{R}^{N_d \times N_p}$ denotes the CT measurement matrix, and $\mathbf{W} \in \mathbb{R}^{N_d \times N_d}$ is a diagonal weighting matrix, whose diagonal elements are the estimated inverse variances of elements of $\mathbf{y}$. The operator $\mathbf{P}_i$ extracts the $i$th vectorized patch of $\mathbf{x}$ as $\mathbf{P}_i \mathbf{x}$. The parameter $\beta$ denotes the regularizer weighting, and $\gamma_1$ and $\gamma_2$ are non-negative parameters that control the sparsity levels of the sparse coefficients.

### 3.2 Image Reconstruction Algorithm

Similar to the learning algorithm, we use an exact block coordinate descent (BCD) algorithm to optimize (P1). The algorithm cycles over updates of the image $\mathbf{x}$, sparse coefficients $\mathbf{Z}_1$ and $\mathbf{Z}_2$, and cluster memberships $\{C_{1,k}\}$, $\{C_{2,l}\}$. The algorithm enforces monotone decrease of the underlying objective.

#### 3.2.1 Image Update Step

In this step, we update $\mathbf{x}$ in (P1) with the other variables fixed, which leads to the subproblem (11). We use the efficient relaxed LALM (rLALM) algorithm [7] to solve (11). The detailed description of this algorithm can be found in [6].

$$\min_{\mathbf{x} \succeq \mathbf{0}} \frac{1}{2} \|\mathbf{y} - \mathbf{A}\mathbf{x}\|_{\mathbf{W}}^2 + \beta S_1(\mathbf{x}), \quad (11)$$

where $S_1(\mathbf{x}) \triangleq \sum_{k=1}^K \sum_{i \in C_{1,k}} \|\mathbf{\Omega}_{1,k} \mathbf{P}_i \mathbf{x} - \mathbf{Z}_{1,i}\|_2^2 + \sum_{l=1}^L \sum_{j \in C_{2,l}} \|\mathbf{\Omega}_{2,l} (\mathbf{\Omega}_{1,k} \mathbf{P}_j \mathbf{x} - \mathbf{Z}_{1,j}) - \mathbf{Z}_{2,l,j}\|_2^2$.

#### 3.2.2 Sparse Coding and Clustering Step

With $\mathbf{x}$ fixed, (P1) is reduced to the same subproblems as (1) and (6). Then $\{\mathbf{Z}_{1,i}, C_{1,k}\}$ and $\{\mathbf{Z}_{2,j}, C_{2,l}\}$ are updated in the same manner as in (2), (3), (7), and (8).

## 4 Experiments

### 4.1 Experiment Setup

We study the performance of MCST2 for the XCAT phantom and Mayo Clinic data. For XCAT phantom case, the low-dose measurements are simulated from the groundtruth image with GE 2D LightSpeed fan-beam geometry corresponding to a monoenergetic source. For Mayo Clinic data case, we simulated the low-dose measurements from the regular-dose images with a fan-beam CT geometry corresponding to a monoenergetic source. The width of each detector column is 1.2858 mm, the distances from source to detector, source to rotation center are 1085.6 mm and 595 mm, respectively. We set the incident photon intensity $I_0 = 1 \times 10^4$ per ray and with no scatter. The "Possion + Gaussian" noisy model is used to generate synthesized low-dose measurements of size $888 \times 984$ for the XCAT phantom and $736 \times 1152$ for Mayo Clinic data, respectively. Two types of metrics (RMSE and SSIM) are applied for evaluating image reconstruction quality. We compute the root mean square error (RMSE) and the structural similarity index measure (SSIM) in a circular central region of the images, which includes all the tissues.

### 4.2 Transform Learning

For the learning stage, we used five $420 \times 420$ XCAT phantom slices to train the MCST2 model. We also used seven slices of size $512 \times 512$ from the Mayo Clinic data set to learn transforms. We ran 1000 iterations of the BCD algorithm to ensure convergence. The number of clusters in the two layers were 5 and 2, respectively. We set $(\eta_1, \eta_2) = (125, 70)$ and $(60, 10)$ for the XCAT phantom and Mayo Clinic data, respectively. Fig. 1 shows the transforms in the MCST2 model that were learned from the XCAT phantom data. Each row of the transform matrices is displayed as an $8 \times 8$ square patch. The pre-learned transforms in the first layer (blue box) show oriented and gradient-like features that sparsify the training image patches. For the second layer, the pre-learned transforms (red box) capture finer level features that further sparsify the filtering residuals.

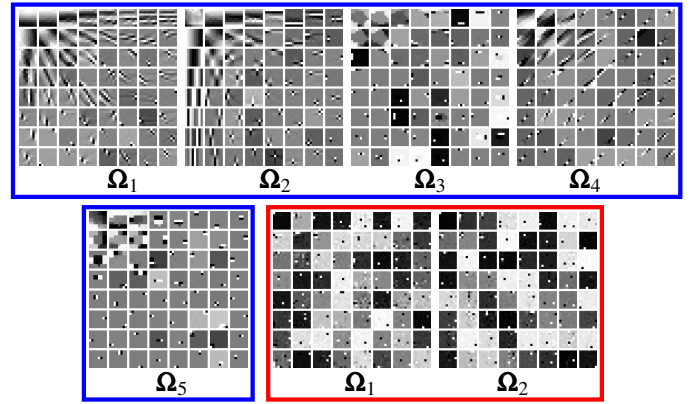

**Figure 1:** Pre-learned transforms from XCAT phantom for the MCST2 model with **5** clusters in the first layer (shown in the blue box) and **2** clusters in the second layer (shown in the red box). Each row of the transform matrices is displayed as a square $8 \times 8$ patch.

### 4.3 Reconstruction Results

We compare our method with the conventional FBP and PWLS-EP algorithms. PWLS-MRST2 [6] and PWLS-ULTRA [5] are also included to verify the usefulness of the proposed MCST2 model. We fine-tuned the hyperparameter $\beta = 2^{15.5}$ for PWLS-EP and ran 1000 iterations of the algorithm to ensure convergence. We ran 1500 iterations for the other iterative algorithms. The parameters for the three transform learning-based were tuned to achieve the best reconstruction quality (i.e., RMSE and SSIM) in the XCAT phantom and Mayo Clinic data experiments, and are as $(\beta, \gamma_1, \gamma_2) = (7 \times 10^4, 30, 10)$ and $(2 \times 10^4, 30, 12)$ for PWLS-MRST2; $(\beta, \gamma) = (2 \times 10^5, 20)$ and $(5 \times 10^4, 20)$ for PWLS-ULTRA; $(\beta, \gamma_1, \gamma_2) = (1.5 \times 10^5, 20, 5)$ and $(4.5 \times 10^4, 25, 5)$ for PWLS-MCST2. Fig. 2 and 3 show the reconstructions of slices of the XCAT phantom and Mayo Clinic data, respectively. We use modified Hounsfield units, where air is 0 HU and water is 1000 HU. Apart from significantly outperforming the traditional FBP and PWLS-EP methods, the proposed PWLS-MCST2 method performs the best in terms of both RMSE and SSIM compared to the recent MRST2 and ULTRA schemes. Furthermore, PWLS-MCST2 improves the image reconstruction quality by removing more notorious artifacts in the margin regions and preserving critical details in the central region.

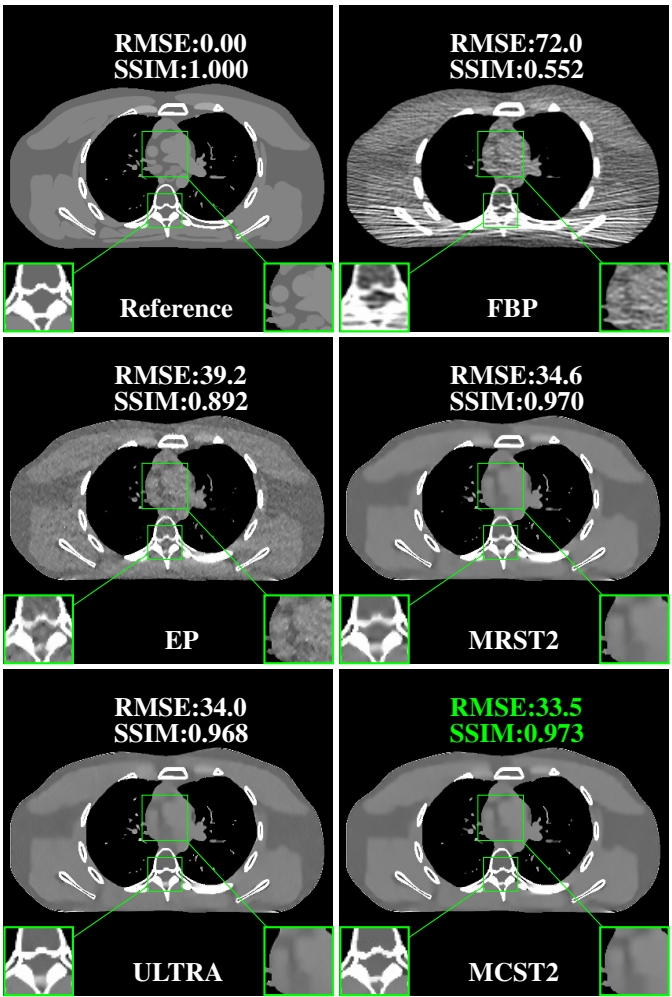

**Figure 2:** Comparison of reconstructions of one slice of the XCAT phantom with the FBP, PWLS-EP, PWLS-MRST2, PWLS-ULTRA, and PWLS-MCST2 methods, respectively, at incident photon intensity $I_0 = 1 \times 10^4$. The display window is [800, 1200] HU.

## 5   Conclusion

This paper proposes learning a two-layer clustering-based sparsifying transform model (MCST2) for CT images, wherein both the image data and feature (transform residual) maps are divided into multiple classes, with sparsifying filters learned for each class. We present an exact block coordinate descent algorithm to train the MCST2 model from limited unpaired (clean) training data. Our experimental results with simulated XCAT phantom and Mayo Clinic data illustrate that the PWLS approach incorporating the learned MCST2 regularizer outperforms recently proposed MRST2 and ULTRA models. It also provides a significant improvement compared to conventional FBP and PWLS-EP methods. Future work will incorporate and explore deeper MCST models as well as other imaging applications.

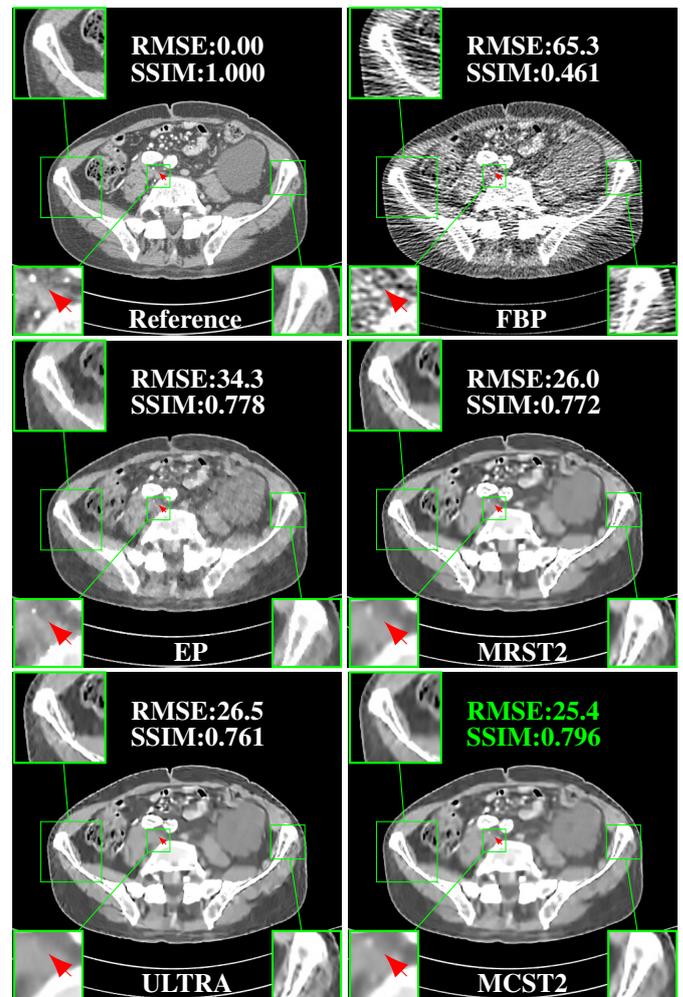

**Figure 3:** Comparison of reconstructions of one slice from the Mayo Clinic data with the FBP, PWLS-EP, PWLS-MRST2, PWLS-ULTRA, and PWLS-MCST2 schemes, respectively, at incident photon intensity $I_0 = 1 \times 10^4$. The display window is [800, 1200] HU.

# Adaptive Ensemble of Deep Neural Networks for Robust Denoising in Low-dose CT Imaging

Dufan Wu[1,2], Kyungsang Kim[1,2], Ramandeep Singh[3], Mannudeep K. Kalra[3], and Quanzheng Li[1,2]

[1]Center for Advanced Medical Computing and Analysis, Massachusetts General Hospital and Harvard Medical School, Boston, MA 02114, USA
[2]Gordon Center for Medical Imaging, Massachusetts General Hospital and Harvard Medical School, Boston, MA 02114, USA
[3]Department of Radiology, Massachusetts General Hospital and Harvard Medical School, Boston, MA 02114, USA

**Abstract** Deep learning-based denoising/reconstruction algorithms have been widely studied for low-dose CT imaging. However, the trained neural networks tend to have deteriorated performance on noise levels that are different from the training noise level. In this work, we propose to tackle this problem by ensembling different pretrained denoisers at testing time. The ensemble is done by linear combination of denoisers trained for different doses. The combining coefficients are calculated to reduce the noise to the level of a normal-dose scan. The proposed method was validated by training on the Mayo Clinic Low-dose CT Challenge dataset, and testing on the Mayo dataset as well as low-dose dual-energy CTs from our institution. It demonstrated improved robustness to noise levels compared to single networks trained using multiple dose levels. Better texture preservation was also achieved than the comparing deep-learning methods.

## 1 Introduction

Low-dose CT (LDCT) imaging is often accompanied with increased noise in the images, which needs to be compensated by denoising and reconstruction algorithms. In recent years, deep learning-based denoising and reconstruction algorithms have attracted a lot of attention because of their simplicity and good performance. However, most deep neural networks are designed to work under one noise level, and their performance will be deteriorated when applied under a different noise level than the training noise level. Although training data can include multiple noise levels, it cannot eliminate the noise-dependency performance of the trained network. In practice, the noise in the image is associated with various factors such as kVp, tube current, patient size, etc. For example, in dual energy CT, the noise in the low energy images changes drastically with the patient size [1]. Hence, it is desirable to have a denoiser that could robustly work for a wide range of noise levels. Most existing solutions are based on combining deep learning with iterative reconstruction, where the data fidelity could compensate for under/over-smoothing of the network [2]. A recurrent network, MAP-NN, was proposed where each module progressively denoise the image, so the user could manually choose from outputs with different denoising strength [3].

In this work, we propose an automatic and non-iterative methods to adapt deep learning-based denoisers to different noise levels. The proposed approach is based on ensemble of multiple denoisers which are pretrained under different noise levels. We used a simple linear combination of the denoisers for each slice, where the combining coefficients are calculated to reduce the noise to a given level, such as normal-dose level. To validate the proposed method, we trained 5 denoisers at 5 different simulated dose levels on the Mayo Clinic Low-dose CT Challenge dataset and tested on various dose levels without changing any hyperparameters. The method was further tested on another dataset which consists of low-dose dual-energy CT (DECT) scans at Massachusetts General Hospital (MGH). The performance was compared to denoising networks trained on each specific dose level as well as on all the dose levels.

## 2 Materials and Methods

An overview of the proposed testing-time algorithm is given in figure 1. It consists of two major parts: noise estimation (blue), which estimates noise in the LDCT and the corresponding normal-dose CT (NDCT) noise level; network ensemble (orange), which calculate the linear combination coefficients based on an LDCT-NDCT pair generated by the noise estimation algorithm.

### 2.1 Noise Estimation

The projection data $\mathbf{p}$ is separated to odd and even projections $\mathbf{p}_{odd}$ and $\mathbf{p}_{even}$, which leads to two filtered backprojection (FBP) reconstructions $\mathbf{x}_{odd}$ and $\mathbf{x}_{even}$ [4]. $\mathbf{x}_{odd}$ and $\mathbf{x}_{even}$ have almost identical underlying structures but independent noises. Subtracting them will eliminate the common structures and leave the noises only. Since both $\mathbf{x}_{odd}$ and $\mathbf{x}_{even}$ take half the projections, their noise level is $\sqrt{2}$ of the original noise. Subtracting will further increase the noise by a factor of $\sqrt{2}$. Hence, the final noise estimation of the FBP image $\mathbf{x}$ is:

$$\mathbf{n}_s = (\mathbf{x}_{odd} - \mathbf{x}_{even})/2. \qquad (1)$$

Note that $\mathbf{n}_s$ shares the same distribution with the original noise, but they are not identical.

To estimate the noise level of an NDCT, we first denoisd $\mathbf{x}$ with a relatively strong denoiser to acquire a smooth image $\mathbf{x}_{est}$. In this work we used the network trained under 16x dose reduction as the smoothing denoiser.

Then Poisson noise was added in the forward projection of $\mathbf{x}_{est}$ to simulate a normal-dose scan. To make it adaptive across patients, the initial photon number per ray, $N_0$, was determined by fixing the average number of photons received by the detectors, $N_{\text{ref}}$. The added noise is then reconstructed via FBP to get the noise estimation of NDCT $\mathbf{n}_r$.

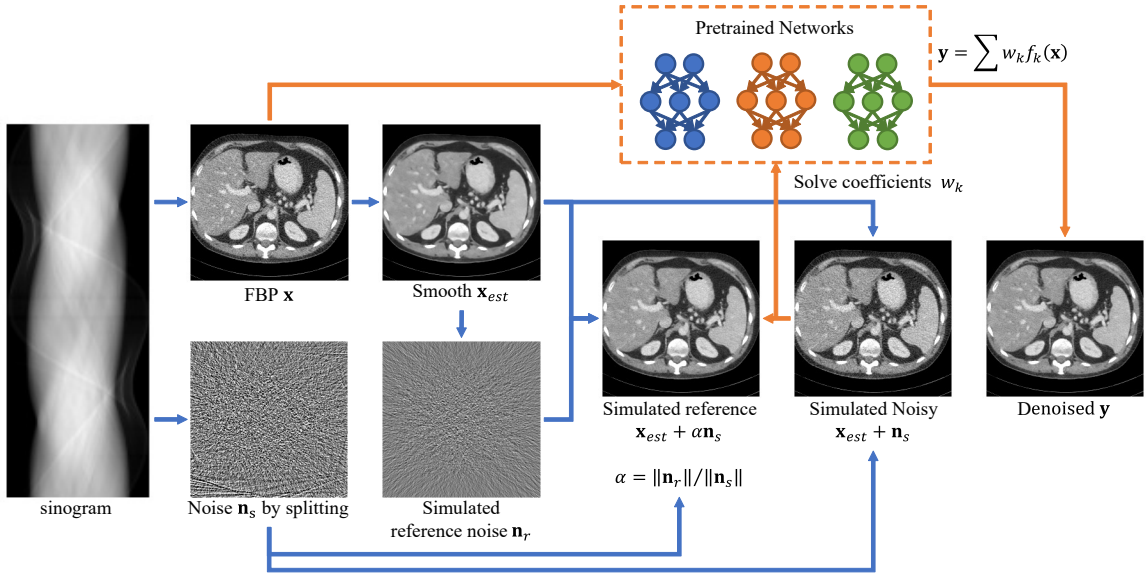

**Figure 1:** Testing-time workflow of the proposed method. Blue lines are noise estimation and orange lines are network ensemble.

With the help of $\mathbf{n}_r$, we could shrink $\mathbf{n}_s$ to the level of NDCT with a scalar:

$$\alpha = \|\mathbf{n}_r\| / \|\mathbf{n}_s\|. \tag{2}$$

If we "train" a model to denoise $\mathbf{x}_{\text{est}} + \mathbf{n}_s$:

$$f^* = \underset{f}{\operatorname{argmin}} \|f(\mathbf{x}_{\text{est}} + \mathbf{n}_s) - (\mathbf{x}_{\text{est}} + \alpha \mathbf{n}_s)\|, \tag{3}$$

because $\mathbf{x}_{\text{est}} + \mathbf{n}_s$ shares similar structure and noise pattern with the original LDCT $\mathbf{x}$, $f^*(\mathbf{x})$ is expected to bring $\mathbf{x}$ to the NDCT level.

Note that we cannot use $\mathbf{n}_r$ instead of $\alpha \mathbf{n}_s$ in the target image, otherwise the independent noises will lead to smoothing images due to the "Noise2Noise" effect [5].

### 2.2 Network Ensemble

We use simple linear combination of pretrained denoisers. Denote pretrained denoisers as $f_1(\mathbf{x}), f_2(\mathbf{x}), \ldots, f_K(\mathbf{x})$, the linear combination coefficients are optimized via (3) as:

$$\mathbf{w} = \underset{w_1, w_2, \ldots, w_K}{\operatorname{argmin}} \|\sum_k w_k f_k(\mathbf{x}_{\text{est}} + \mathbf{n}_s) - (\mathbf{x}_{\text{est}} + \alpha \mathbf{n}_s)\|_2^2, \tag{4}$$

which can be solved by taking derivative with respect to $\mathbf{w}$ and set it to zero:

$$\mathbf{A}\mathbf{w} = \mathbf{b}, \mathbf{A} \in R^{K \times K}, \mathbf{b} \in R^K \tag{5}$$

where

$$A_{ij} = f_i(\mathbf{x}_{\text{est}} + \mathbf{n}_s)^T f_j(\mathbf{x}_{\text{est}} + \mathbf{n}_s), \tag{6}$$

and

$$b_i = f_i(\mathbf{x}_{\text{est}} + \mathbf{n}_s)^T (\mathbf{x}_{\text{est}} + \alpha \mathbf{n}_s). \tag{7}$$

After $\mathbf{w}$ is solved, the final denoised image can be obtained by combining denoising results on $\mathbf{x}$:

$$\mathbf{y} = \sum_k w_k f_k(\mathbf{x}) \tag{8}$$

### 3 Experimental Setups

The denoisers $f_k$ were trained on the Mayo Clinic Low-dose CT Challenge dataset [6]. We rebinned the projection data to multi-slice fanbeam with 3mm slice thickness and used FBP with Hann window to reconstruct images to 0.75 mm pixel size. 7 patients were used as the training dataset, where 100 slices randomly extracted from each patients for the training. We injected noises equivalent to 2, 4, 8, 12, 16 times dose reduction to construct 5 levels of noises. A denoiser $f_k$ was trained for each dose level using L2 loss.

During testing, the rest 3 patients in the Mayo dataset were used. We estimated the performance on dose reduction rates of 2, 4, 6, 8. SSIMs inside liver window [-160, 240] HU were computed against the original NDCT. To estimate the noise level difference between the denoised images and NDCT, we further extracted 100 flat patches of $32 \times 32$ pixels inside the liver for each patient. For each patch the standard deviation (std) was calculated, and the absolute difference between the stds of corresponding denoised and NDCT patches were further calculated to estimate the noise level difference.

The trained networks were further tested on 6 low-dose DECT images acquired at MGH. The patients had a normal DECT scan followed by a low-dose scan with at least 50% dose reduction. Multiple $32 \times 32$ patches were selected inside flat liver regions for each patient, in order to calculate and compare the noise power spectrum (NPS) between the LDCT and NDCT. The NPS was normalized by the mean intensity within the patches to compensate for the change of iodine concentration between the two scans. We only calculated the NPS for the 140kVp images due to the excessive inference from vessels and contrast changes in the 80kVp images.

A target noise level of $N_{\text{ref}} = 7.5 \times 10^4$ was used for the Mayo dataset and $N_{\text{ref}} = 3.75 \times 10^4$ was used for the DECT dataset. For comparing methods, we trained denoising networks using

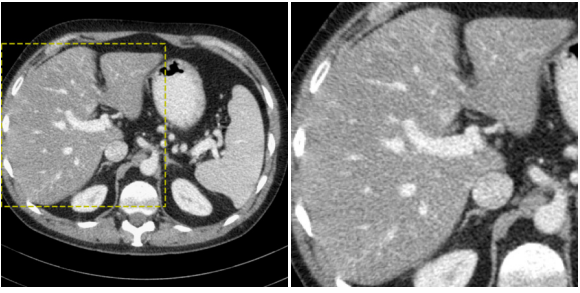

**Figure 2:** The reference image and the zoom-in region corresponding to figure 3. The display window is 40±350 HU.

L2-loss and WGAN [3] using data from all the 5 dose levels (L2 (all) and WGAN (all)). We also trained dose-specific networks using L2-losses (L2 (matched)). All the networks have the same structure of REDCNN [7], with 3 encoders and 3 decoders, and 2 convolution layers within each encoder/decoder. All the convolutional layers has $3 \times 3$ kernels, 64 featuremaps and leaky ReLU activation ($\alpha = 0.2$) except for the last one. Adam algorithm with step size of $10^{-4}$ was used for all the training. We followed [3] for the WGAN training setup, with an additional 0.5-dropout layer in the discriminator to stabilize the training. The single-dose networks were trained for 25 epochs and the multi-dose networks were trained for 5 epochs.

## 4 Results

### 4.1 Mayo dataset results

Figure 2 shows an NDCT slice where the denoising results under different dose levels are given in figure 3. It can be observed that as the dose reduction rate increases, the noise level also increases for the denoising results of L2(all) and WGAN(all), which indicates that the they cannot well adapt to different noise levels despite being trained using multiple dose levels. Furthermore, instead of increasing uniformly, the noise appears as structure-mimicking spikes. On the other hand, the denoising images from the proposed Ensemble shared similar textures and noise levels at different dose levels. There is also little spikes in the Ensemble results. The L2(matched) results also have similar noise levels because dose-specific networks were used. But it has more spikes and less textures compared to the ensemble results.

Figure 4 shows quantitative analysis of the denoising results compared to the original NDCT. L2(all) and WGAN(all) have faster drops on SSIM as the dose reduction rate increases, whereas Ensemble had similar trend with L2(matched), which requires different configurations under different dose rates. The mean patch std distance plot demonstrated that Ensemble had consistent and closest noise levels to the NDCT across all the dose reduction rates.

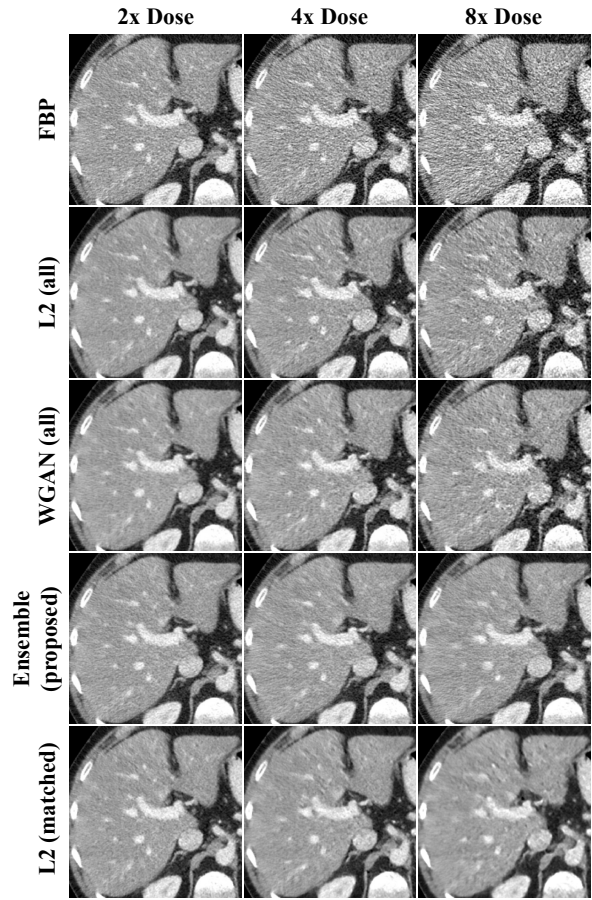

**Figure 3:** The denoising results of a testing ROI under different dose levels. The display window is 40±350 HU.

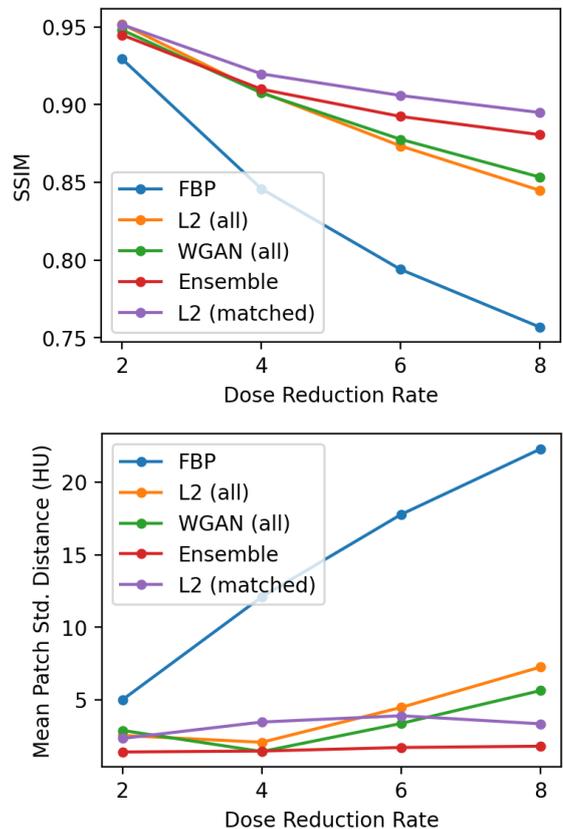

**Figure 4:** SSIMs and mean patch std distances against the NDCT.

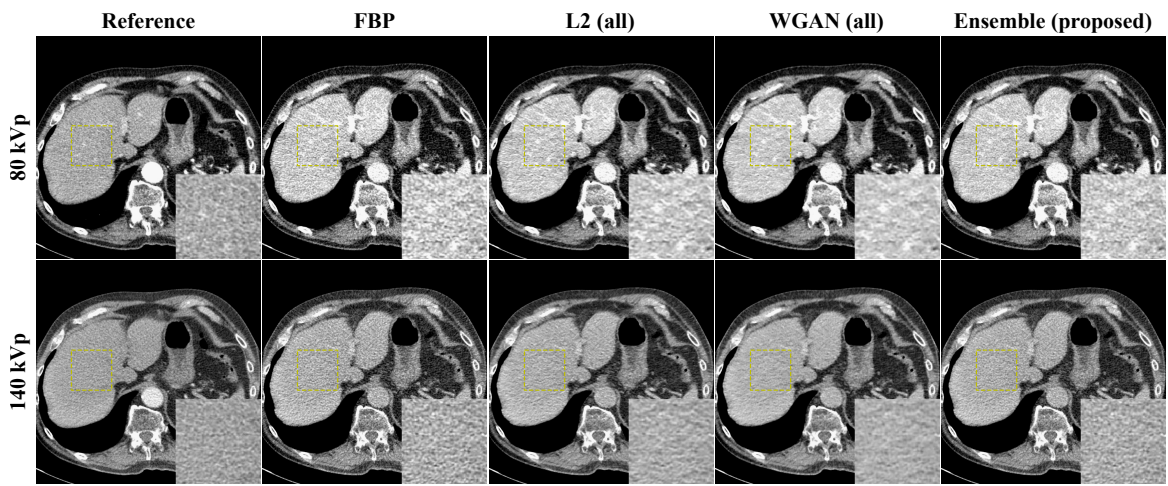

**Figure 5:** The denoising results of a DECT slice. Reference is FBP from NDCT. The display window is 40±350 HU.

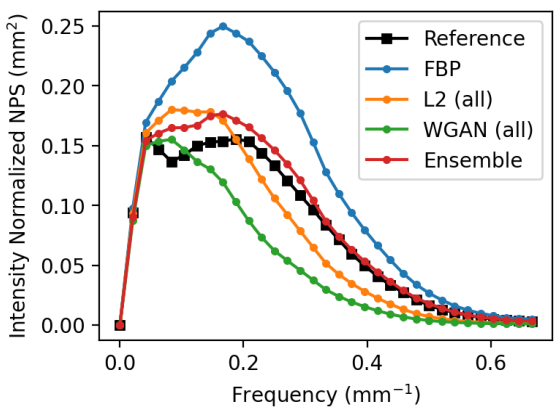

**Figure 6:** The mean intensity-normalized NPS of 140kVp images. Reference stands for FBP from NDCT.

## 4.2 DECT dataset results

Figure 5 shows the denoising results on a DECT slice for both 80kVp and 140kVp images. The reference images are FBP results from the normal-dose scan, which happened before the low-dose scan so there is less iodine take-up in the liver. It can be observed from the zoom-in that L2(all) and WGAN(all) have less high-frequency textures compared to the reference images and Ensemble.

Further NPS analysis verified the observation as shown in figure 6. The intensity-normalized NPS of Reference and Ensemble are very close, whereas the NPS of L2(all) and WGAN(all) shows less high-frequency components.

## 5 Discussion and Conclusion

We proposed a testing-time ensemble method which adapts pretrained denoisers to different dose levels. The method showed improved robustness and texture preservation compared to networks trained on multiple dose levels.

The proposed method has one important hyperparameter $N_{ref}$ which controls the target noise level. $N_{ref}$ has clear physical meaning and can be determined empirically and should be stable across patients. Larger $N_{ref}$ would give less noisy results at the risk of oversmoothing. More sophisticated models can be used to more accurately estimate the noise level of NDCT.

A drawback of the current method is the L2-loss used for the ensemble (4) may encourage smoothing. A potential solution is to select different $f_k$ based on the noise level indicator $\alpha$, so that strong denoisers will only be used when the noise level is really high.

# Fine-tunable Supervised PET Denoising

Jianan Cui*[1], Kuang Gong*[1], Ning Guo[1], Huafeng Liu[2], Scott Wollenweber[3], Floris Jansen[3], and Quanzheng Li[1]

[1]Center for Advanced Medical Computing and Analysis, Massachusetts General Hospital and Harvard Medical School, Boston, USA
[2]State Key Laboratory of Modern Optical Instrumentation, College of Optical Science and Engineering, Zhejiang University, Hangzhou, China
[3]GE Healthcare, Waukesha, USA

**Abstract** Denoising is of vital importance for Positron emission tomography (PET) imaging to improve its diagnostic merits. Recently deep neural networks (DNNs) have been successfully applied to PET image denoising through supervised or unsupervised learning methods. In this work, considering the advantages of supervised and unsupervised learning methods, we propose a deep learning framework jointly employing them together for PET image denoising. Firstly, a supervised neural network is trained by a group of low-quality and high-quality training pairs. During the testing phase, the pre-trained network is finetuned by using the test noisy image itself as the training label. Though finetuning, the supervised result can be further optimized according to the inherent information in the testing data, thus avoiding the pitfall of mismatches between training and testing. Quantification results based on a clinical PET/CT dataset containing 47 $^{18}$F-FDG, 4 $^{18}$F-Fluciclovine, and 1 $^{68}$Ga-DOTATATE scans demonstrated that the proposed framework had better performance than using the supervised method or single-image-based unsupervised method alone.

## 1 Introduction

Positron emission tomography (PET) is a powerful molecular imaging technique that can reveal physiological and pharmacological processes *in vivo*. Currently, PET has been widely applied in oncology due to its sensitivity for early detection of tumor and occult recurrences [1, 2]. However, due to the dose-safety concerns and the acquisition-time limitation, PET images still suffer from low signal-to-noise ratios (SNR). Noise in the PET image can seriously compromise its lesion detectability, especially for small tumors. Developing effective and efficient denoising methods for PET imaging is of vital importance.

Recently, with the development of deep neural networks, deep learning-based methods have made outstanding achievements in natural image denoising and are gradually being applied to medical imaging. For PET denoising, there are two categories of deep learning-based methods. One is the supervised method [3–5]. Through pre-training deep neural networks by low-quality and high-quality image pairs, the network can learn latent features and mapping relations between low-quality and high-quality image pairs. However, in real clinical practice, it is hard to collect a large enough number of high-quality images. Once the test image has different characteristics from the training dataset, the network performance would drop dramatically. Another category is unsupervised learning method[6, 7], particularly the single-image-based unsupervised learning method [6] which learns features from the noisy image itself. However, for the single-image-based unsupervised method, the network is fed with the random initialization, easier to be trapped into the local optimum. In addition, a new network needs to be trained for each image to learn its intrinsic features during denoising, which is time-consuming.

In this paper, we combined the supervised and unsupervised methods together to leverage their advantages and achieve optimum performance. The proposed method consists of two steps: supervised training and unsupervised finetuning. In the supervised training step, a neural network was trained by a group of low-quality and high-quality training pairs. In the unsupervised finetuning step, the pre-trained network was finetuned by using the specific low-quality test image as the training label. Through finetuning, the neural network can optimize the supervised result and output a denoised image with higher quality. The co-registered CT images were supplied as input to the second channel to provide additional anatomical information. Contributions of this work include:

- Compared to the supervised method, the proposed method is more robust for data with different characteristics. The fine-tuning step can recover denoised image quality even if the pre-trained supervised network gets an outlier input.

- Compared to the single-image based unsupervised method, the proposed method needs less training time. When the single-image-based unsupervised method[6] is implemented, for each image, the network usually needs several hundred epochs' training. In the proposed method, based on the pre-trained network, fine-tuning step converges much fast and only needs training for several additional epochs.

## 2 Methods and Materials

### 2.1 Methods

The proposed method includes two steps: supervised training and unsupervised finetuning. During the supervised training step, training was performed by a widely used supervised denoising framework. In this framework, an Unet was trained from $N$ low-quality and high-quality training pairs as

$$\hat{\theta} = arg\min_{\theta} \frac{1}{N} \sum_{i=1}^{N} \left\| X_i^{high} - f\left(\theta | X_i^{low}, \alpha_{X_i}\right) \right\|^2, \quad (1)$$

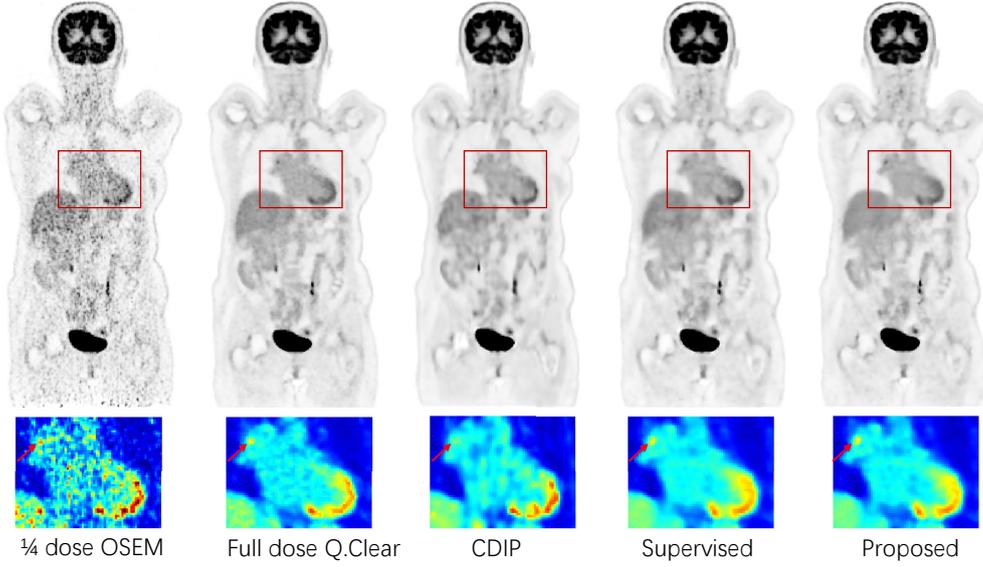

**Figure 1:** [18]F-FDG results of the quarter dose OSEM PET image, the full dose Q.Clear image, the post-filtered image using CDIP method, the supervised method, and the proposed method. Lesions are pointed out using arrows. The subfigures show the view in the red box using the Jet colormap.

where $f$ stands for a neural network, $\theta$ are the parameters of the neural network, $X_i^{low}$ and $X_i^{high}$ are the $i$th low-quality and high-quality training pair. The network inputs are low-quality images $X^{low}$ and the training labels are the corresponding high-quality image $X^{high}$. $\alpha_{X_i}$ is the co-registered anatomical image of the $i$th patient and serves as the second channel input to provide additional anatomical information. Through training, the network can learn the mapping relationship from low-quality images to high-quality images. Once given a test image $X_{test}$, the well-trained neural network can directly output a corresponding high-quality image $f\left(\hat{\theta}|X_{test},\alpha_{X_{test}}\right)$. However, in clinical practice, it is hard to acquire a perfect training dataset that is large enough to include data with different characteristics. When the characteristics of the test image do not match the training data set, the network would fail to output a good result.

In this method, we added a unsupervised finetuning step after the supervised training. The finetuning step can be generalized as

$$\hat{\phi} = arg\min_{\phi}\left\| Y^{low} - f\left(\phi|Y^{low},\alpha_Y,\hat{\theta}\right)\right\|^2, \quad (2)$$

where $f$ and $\hat{\theta}$ are the pretrained network and its parameters in the first step, respectively. $Y^{low}$ is the low-quality image in test data set $Y$. In this step, the two-channel network inputs are the low-quality image $Y^{low}$ and its co-registered anatomical image $\alpha_Y$, respectively. Here, the low-quality image itself was served as the network training label $Y^{low}$ to optimize the results of directly supervised output. The denoising effect of using noisy image itself as the training label has already been verified in our previous work [6]. After finetuning for several epochs, the network can output a denoised image

$$Y^{high} = f\left(\hat{\phi}|Y^{low},\alpha_Y,\hat{\theta}\right). \quad (3)$$

A modified 3D Unet[8] was utilized in the proposed method. The detailed structure of the modified 3D Unet is shown in the reference[6]. The optimization algorithm is ADAM[9] for both supervised training and unsupervised finetuning.

## 2.2 Experiment Set Up

The experiment was performed on a PET/CT dataset. The patient scans were acquired with a single-site Discovery MI (DMI) PET/CT system (GE Healthcare). Low-dose CT scans (120 kV; 59 mA; pitch 0.984:1; matrix, $512 \times 512 \times 345$; voxel size, $1.3672mm \times 1.3672mm \times 2.8mm$; FOV, 70 cm) were obtained for PET attenuation correction. PET images were acquired at 60 min post-injection with a matrix size of $256 \times 256 \times 345$, and a voxel size of $2.7344mm \times 2.7344mm \times 2.8mm$. Full dose Q.Clear PET images reconstructed using time-of-flight contrast recovery with Bayesian penalized likelihood method (TOF-Q.Clear, beta=350) [10] were used as the high-quality image to supervise the training. Low-quality images are quarter-dose OSEM PET images reconstructed using the time-of-flight ordered subset expectation maximization method(TOF-OSEM, 2 iterations, 34 subsets, 25% duration).

There is a dataset from 47 patients injected with [18]F-FDG for evaluation. We used 35 datasets for training, 6 datasets for validation, and 6 datasets for testing. To further validate the performance of the proposed method, 4 [18]F-Fluciclovine and 1 [68]Ga-DOTATATE datasets were additionally tested by the proposed method to test its robustness regarding different radiotracers. The results of the supervised training in step one and a single-image-based unsupervised method (conditional deep image prior, CDIP)[6] were presented for reference. Epochs of supervised training were 500, cho-

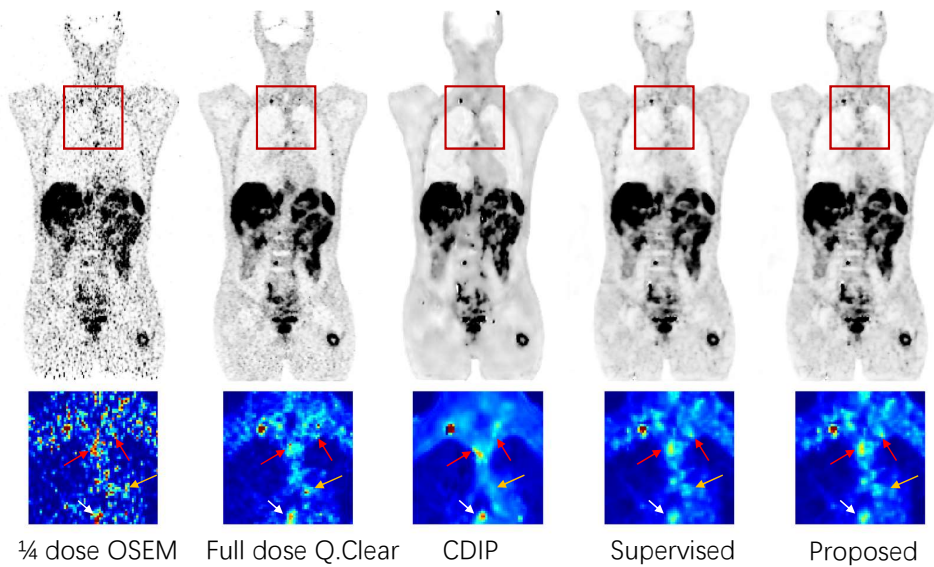

¼ dose OSEM     Full dose Q.Clear     CDIP     Supervised     Proposed

**Figure 2:** $^{68}$Ga-DOTATE results of the quarter dose OSEM PET image, the full dose Q.Clear image, the post-filtered image using CDIP method, the supervised method, and the proposed method. The subfigures show the zoomed view from the red box using the Jet colormap.

sen based on the validation data. Unsupervised finetuning epochs were around 5 to 10 epochs. Training epochs of CDIP were 1400. Quantitative analysis was based on the peak signal-to-noise ratio (PSNR) and structural similarity index measure (SSIM), with full dose Q.Clear PET images as the ground truth.

## 3  Results

### 3.1  $^{18}$F-FDG Results

Figure 1 shows the coronal view of the quarter-dose OSEM $^{18}$F-FDG PET images processed using different methods. Full-dose Q.Clear image is shown as the ground truth. Both the supervised method and the proposed method preserved the tumor uptake (pointed out using arrows) in the lung, but CDIP failed. This is because both the supervised and the proposed methods had prior information learned from the full-dose Q.Clear images. CDIP only learned information from the quarter-dose OSEM image itself. Compared to the supervised method, the proposed method further improved the uptake of the tumor and the structure of different organs. The PSNRs and SSIMs calculated from the results of the different methods shown in Figure 3 (a,b) also verify that the proposed method has the best performance. One interesting thing is that for dataset 3, the supervised method's SSIM is even worse than the original noisy OSEM. With the help of unsupervised finetuning, the proposed method got a better SSIM.

### 3.2  $^{18}$F-Fluciclovine and $^{68}$Ga-DOTATATE Results

Figure 2 shows the coronal view of the quarter-dose OSEM $^{68}$Ga-DOTATATE PET images processed using different

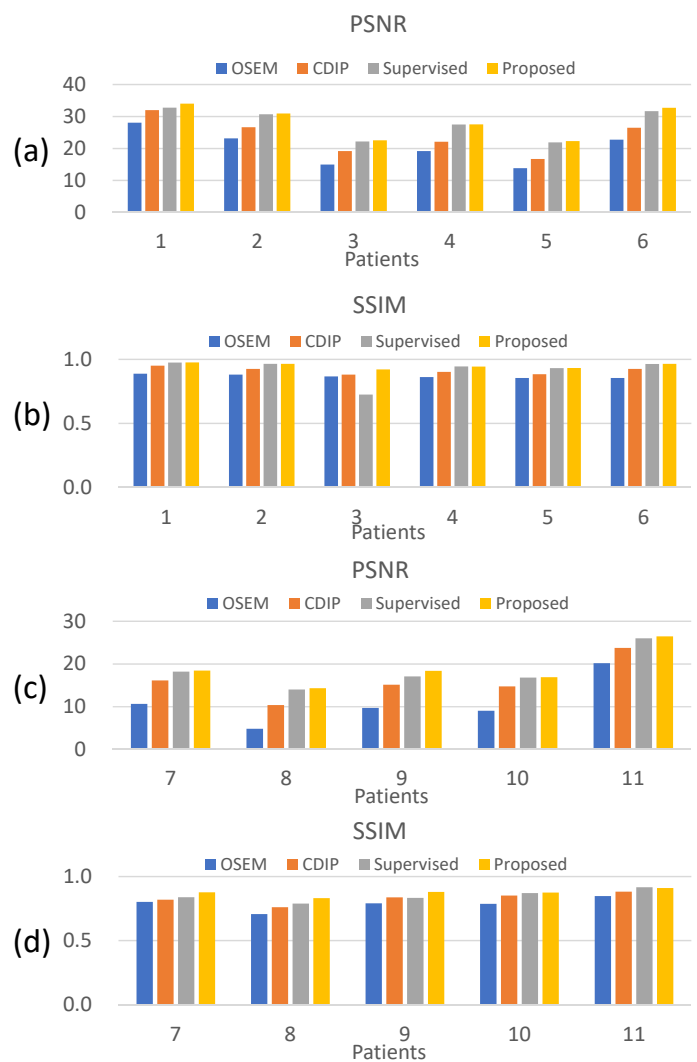

**Figure 3:** PSNRs and SSIMs of the test datasets with different tracers using different methods. Dataset 1-6: $^{18}$F-FDG; Dataset 7:10: $^{18}$F-Fluciclovine; Dataset 11: $^{68}$Ga-DOTATATE.

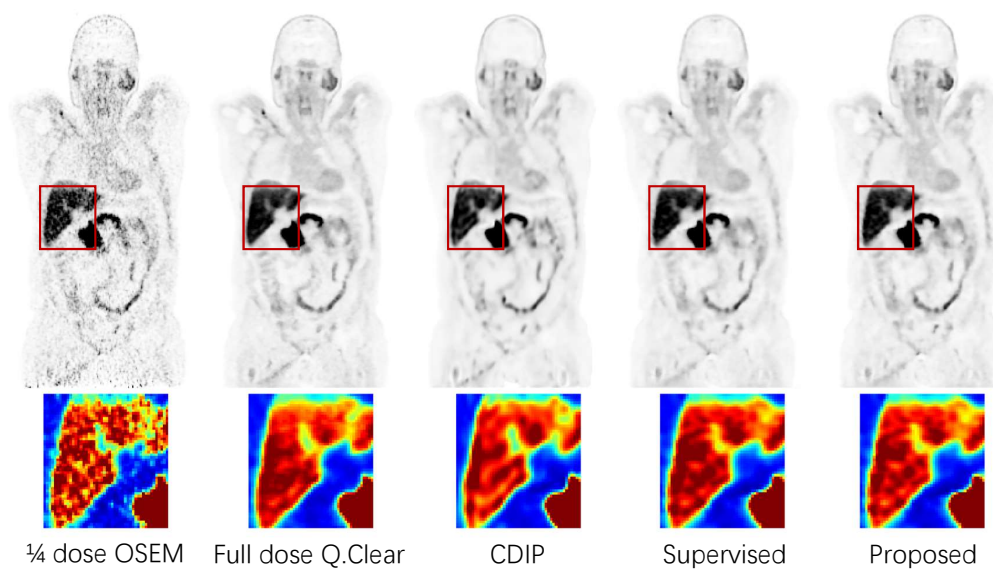

¼ dose OSEM    Full dose Q.Clear    CDIP    Supervised    Proposed

**Figure 4:** $^{18}$F-Fluciclovine results of the quarter-dose OSEM PET image and the full-dose Q.Clear image, as well as the post-filtered image using the CDIP method, the supervised method, and the proposed method. Lesions are pointed out using arrows. The subfigures show the zoomed view from the red box using the Jet colormap.

methods. Full-dose Q.Clear image is shown as the ground truth. From the subfigures, we can see that the proposed method can improve the tumor contrast comparing to supervised results (pointed out using red and white arrows). CDIP is over-smoothing in some regions (pointed out using orange arrows). However, in the white arrow position, it has superior performance than the supervised method and the proposed method. Figure 4 displays the coronal view of the quarter-dose OSEM $^{18}$F-Fluciclovine PET images processed using different methods. All the denoising methods work well for this tracer. In the subfigure shown using the Jet colormap, the recovered structures of the proposed method are the closest to the full-dose Q.Clear image. The PSNRs and SSIMs are shown in Figure 3(c,d). The proposed method achieves the highest PSNRs and SSIMs for all the $^{18}$F-Fluciclovine datasets. As for the $^{68}$Ga-DOTATATE (Patient 11), the SSIM value of the proposed method is a lit bit lower than the supervised method but it has a higher PSNR value.

## 4 Conclusion

In this work, we proposed a deep learning-based denoising method, combining both supervised and unsupervised learning for PET image denoising. A PET/CT $^{18}$F-FDG dataset containing 47 patients were utilized for evaluation. The proposed method was further validated using $^{68}$Ga-DOTATATE and $^{18}$F-Fluciclovine datasets. Quantitative results show that the proposed method outperforms the supervised method and the single-image-based unsupervised method. Our future work will focus on more clinical evaluations for the proposed method.

# Chapter 8

# Oral Session - PET acquisition modelling

**session chairs**
Claude Comtat, *CEA (France)*
Georgios Angelis, *University of Sydney (Australia)*

# Singles-Based Random-Coincidence Estimation for Noise Reduction on Long Axial Field-of-View PET Scanners

Stephen C. Moore, Margaret E. Daube-Witherspoon, Varsha Viswanath, Matthew E. Werner, Joel S. Karp

Department of Radiology, University of Pennsylvania, Philadelphia, USA

**Abstract** While long axial field-of-view (FOV) PET scanners provide significantly increased true-coincidence sensitivity compared with most commercial systems, their greater axial detector coverage can also lead to higher random-coincidence rates and randoms fractions. Because these can adversely affect the precision, as well as the accuracy of reconstructed PET images, it is important to consider noise-reduction techniques for the randoms estimate, e.g., by using detector singles count rates to estimate the randoms. We implemented a fast Monte Carlo (**FMC**) program to evaluate expected true (**T**), scatter (**S**), and random (**R**) count rates for a scanner similar in design to the PennPET Explorer and compared its predictions to those from GATE simulations. Randoms computed from delays (**RD**) and from singles rates (**RS**) were compared over a wide range of singles rates with values simulated for a NEMA 70-cm line source; accuracy and precision of randoms estimates from low-count scans were then assessed using a simulated 70-cm IEC phantom containing $^{18}$F and, separately, $^{89}$Zr. T, S, and R count rates from FMC simulations of the NEMA 70-cm line source all agreed to <5% with those from GATE over the range of activities; and RS and RD from FMC both agreed to <0.6% with RD from GATE. FMC-simulated late acquisitions of the IEC phantom showed for both $^{18}$F and $^{89}$Zr that images of the RS estimates using most likely positioning from time-of-flight information agreed closely with those from the simulated randoms, while reducing image noise by a factor > 50. NEMA count-rate physical phantom data acquired on the PennPET Explorer system (112-cm axial FOV) also confirmed that RS yields accurate and precise estimates of randoms for this system; additional validation studies using phantoms and human subjects are in progress.

## 1 Introduction

Long axial field-of-view (FOV) PET scanners [1,2] have demonstrated their potential in clinical and research applications due to their high sensitivity and long axial coverage. Their excellent performance has enabled studies over a wide range of count rates and count densities (e.g., very fast dynamic studies, delayed imaging of slow biology, and imaging with reduced doses to limit radiation burden to the subject), allowing studies that would not be possible on a standard axial FOV system. However, because of their larger acceptance angle, these systems utilize a wider coincidence window, $\tau$, which leads to more random coincidences compared to systems with smaller axial FOV. This can be especially challenging under high count-rate conditions, such as when imaging the blood input function; however, image noise arising from randoms is also exacerbated at low count rates that may only be slightly above the $^{176}$Lu background rate from LYSO detectors, e.g., for very delayed imaging of $^{18}$F-fluorodeoxyglucose (**FDG**). Radiopharmaceuticals with long half-lives and/or low positron yield (e.g., $^{64}$Cu-DOTATATE) and/or with accompanying gamma photons (e.g., $^{89}$Zr-labeled tracers) are also often scanned under low count-rate conditions to keep the radiation dose to the subject acceptably low. For this reason, conventional randoms-from-delays (**RD**) estimates can be quite noisy, often requiring smoothing or other noise-reduction techniques. Also important is that the total number of lines-of-response (**LOR**) in, for example, the PennPET Explorer systems, increases from 140-million with one detector ring up to 3.5-billion for the 5 detector-ring system. This implies that the RD sinogram becomes even more sparse as the axial field increases, so most LORs have just 0 or 1 count; this can introduce bias and noise into RD estimates. Fig. 1 shows images from a research study on the prototype PennPET Explorer (64-cm axial FOV) following injection of 550 MBq FDG; the total system randoms fraction (**RF**) increased to 75% when imaging ~19 hours post injection.

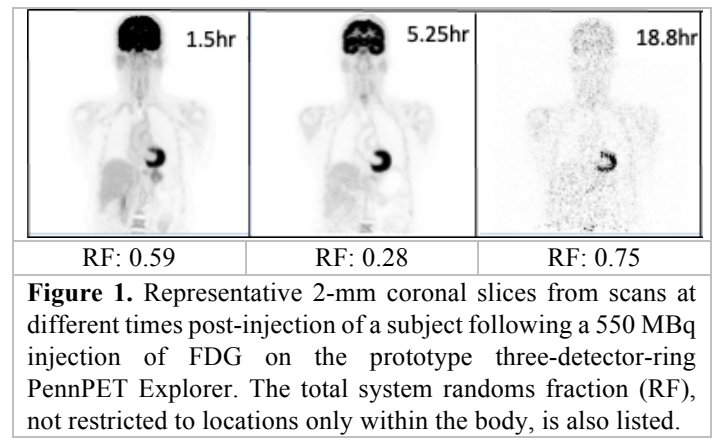

**Figure 1.** Representative 2-mm coronal slices from scans at different times post-injection of a subject following a 550 MBq injection of FDG on the prototype three-detector-ring PennPET Explorer. The total system randoms fraction (RF), not restricted to locations only within the body, is also listed.

While the RD approach can provide accurate estimates of random-event count rates, there are practical challenges dealing with the very large singles list-mode files obtained from a long axial FOV system, in addition to the noise-related issues described above. Alternatively, the randoms-from-singles (**RS**) algorithm used in some early PET scanners (e.g., [3]) has been well validated for volume imaging on standard axial FOV systems [4,5]. However, the wide range of count rates and count densities seen on long axial FOV systems makes it important to revisit this approach for such scanners.

To test the performance of a RS approach, we have recently implemented a fast Monte Carlo (**FMC**) simulation program to evaluate the expected true (**T**), scatter (**S**), and random (**R**) coincidence count rates for a geometry similar in design to the prototype three-detector-ring (~70-cm-long) configuration of the PennPET Explorer [1]. To check the accuracy of the FMC program over a range of singles rates, we compared FMC count rates to those obtained from the detailed and well-known (but significantly slower) GATE Monte Carlo software [6,7] when simulating a NEMA 70-cm line source [8] (count-rate phantom). After validating accuracy with GATE [9], FMC was then used for the rest of our simulations because of its faster run-time.

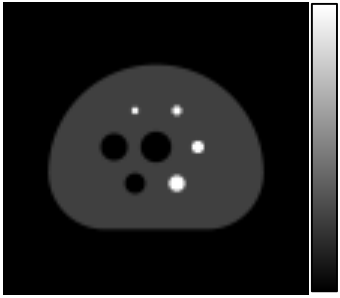

**Figure 2.** Central transverse slice through the NEMA image quality (IQ) phantom [8], extended axially for this simulation to be 70 cm long. The 'hot' spheres (10, 13, 17, 22-mm in diameter) contained activity concentration values 4-times higher than that of the uniform background; the 28- and 37-mm-diameter 'cold' spheres and the central cylindrical 'lung' contained only water with no activity. Several different total-activity values, yielding detector singles rates up to 90 Mcps, were simulated.

## 2 Materials and Methods

The FMC program was developed to simulate a continuous cylindrical PET scanner to approximate the geometry of the PennPET Explorer with 3 rings (~70-cm), containing the same number of LYSO crystal elements: 576 (azimuthal) x 168 (axial). The crystal pitch of the simulated system was almost the same as the scanner's (4.1 mm x 4.1 mm); its energy resolution was 11% FWHM at 511 keV, and changed with incident photon energy, E, according to:

$$FWHM_E = FWHM_{511} \bullet (511 \text{ keV/E})^{1/2}. \quad (1)$$

When processing singles events to form coincidences, the "keep all goods" coincidence policy was used for both the FMC and GATE simulations. For the NEMA phantom simulations with [18]F, the lower-level discriminator (LLD) was set to 444 keV; for the simulations of the IEC image-quality phantom, the energy window used to qualify all single events was 451 to 649 keV. The simulated time-of-flight (TOF) resolution was 250 ps; the detector pulse width, $\tau$, was 2.01 ns for the [18]F NEMA line-source simulations and 2.5 ns for most NEMA IQ phantom simulations. Because each of the PennPET Explorer's small LYSO crystals is coupled 1:1 to its own silicon photomultiplier (SiPM) element, we simplified the FMC simulation by assuming zero deadtime and no pulse pileup in the detectors. The intrinsic [176]Lu background count-rate was simulated -- based on a measured 'blank' scan -- to be 1.613 Mcps for the three-detector-ring system.

Compton and coherent photon scatter were simulated within the phantom, but not within the detectors, for which the probability of photon absorption depended on each photon's energy and its angle-of-incidence on the detector. Each single qualified detector "hit" was recorded in a singles list-mode file, which also contained a unique nuclear decay number, a detection-time 'stamp', and a code identifying the history of the detected photon. The singles list-mode file was processed to compute total singles rates in each detector element, as well as the numbers of T, S, and R coincidence events simulated. The RS event rate for a line-of-response (LOR) connecting detectors i and j was computed from the following well-known formula,

$$R_{i,j} = 2 \tau S_i S_j, \quad (2)$$

where $S_i$ and $S_j$ are the singles rates of detectors i and j, and $2\tau$ is the full width of the coincidence time window.

RD events were estimated by creating a duplicate copy of a singles list-mode file and adding a constant time offset (greater than the coincidence time window) to the time stamp of every event in the duplicate file, before searching for coincidence events between the original and time-shifted files.

To study the impact of singles rate on the accuracy and precision of the RS and RD methods, the NEMA count rate performance measurement was simulated using both FMC and GATE. The 70-cm-long NEMA line-source was positioned 4.5 cm below the center of the scanner within a 20.3-cm-diameter polyethylene attenuator. Consistent with the usual NEMA count-rate data analysis procedure, all random coincidences intersecting a centered 24-cm-diameter cylindrical volume-of-interest were included. We then used FMC to compare the performance of RD and RS estimates of randoms from a simulated 70-cm-long NEMA image quality phantom (Fig. 2) filled with [18]F, and then (separately) with [89]Zr -- both under delayed imaging (low activity) conditions. For [89]Zr, we also optimized the upper-level discriminator (ULD) setting by computing noise-equivalent count-rates (NECR) over a range of ULD values in order to minimize the impact of random coincidences from scattered high-energy gammas on NECR.

Finally, to assess RS performance under real imaging conditions, we compared the RS and RD count rates obtained experimentally from the extended 5-detector-ring PennPET Explorer system (112-cm axial FOV, [10]) when performing the NEMA count rate study using a 70-cm-long line source of [18]F, located 4.5 cm below the center of the scanner within 20- and 35-cm-diameter polyethylene attenuators, as well as a single measurement with a 140-cm-long line source in a 20-cm diameter attenuator. Last, total system RD and RS rates were compared for a research study of a human subject who had been injected 24 h earlier with a new [89]Zr-labeled radiopharmaceutical.

## 3 Results

For the simulated NEMA count-rate line-source phantom, the T, S, and R count rates from the FMC program agreed to better than 5% with those from the more detailed GATE simulation software over the entire range of activity concentration values simulated (Fig. 3).

Comparison of the FMC RS and RD rates and GATE RD rates (Table 1) showed that all three agreed within <0.6% for all activity conditions.

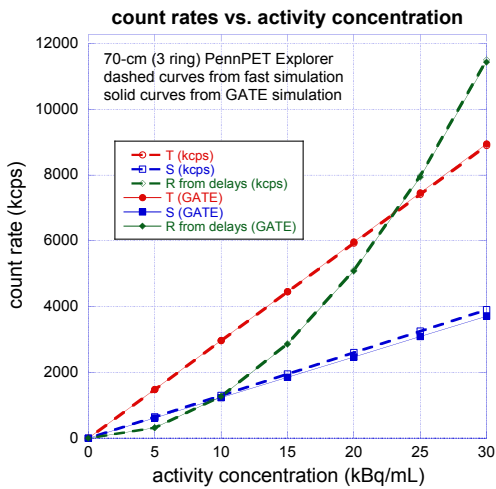

Figure 3. True, scatter, and random count rates obtained from the FMC simulation program (dashed curves), compared to GATE simulations (solid curves) for the same off-center NEMA count-rate line-source phantom and geometry of the PennPET Explorer.

**Table 1.** Comparison of randoms rates (kcps) obtained from the two programs used to simulate the NEMA line source.

| kBq/mL | GATE RD(delayed) | Fast Sim. RD(delayed) | Fast Sim. RS(from singles) |
|---|---|---|---|
| 5 | 318 | 320 | 319 |
| 10 | 1,271 | 1,276 | 1,278 |
| 15 | 2,857 | 2,878 | 2,876 |
| 20 | 5,083 | 5,113 | 5,111 |
| 25 | 7,937 | 7,981 | 7,988 |
| 30 | 11,436 | 11,506 | 11,503 |

For evaluating RS estimation under very low [18]F-activity conditions (corresponding to a ~18.5 hour delay after injecting a patient with 550 MBq of FDG), we simulated a one-hour acquisition of a NEMA image quality phantom with the background extended to cover 70 cm axially, containing just 165 kBq (4.5 μCi) of [18]F. For this case, many of the randoms arose from coincidences involving one or two [176]Lu photons detected in the LYSO detector array. The total singles rate for this activity distribution was 1.65 Mcps, just above the intrinsic [176]Lu background rate for the three-ring configuration of the PennPET Explorer. 29.2-million prompt coincidences were recorded during the one-hour simulated scan; of these, 24.3 million were randoms, i.e., 83% randoms fraction (RF). When including only LORs traversing an elliptical cylinder larger than the phantom (Fig. 4), RF was 0.48, 0.53, and 0.60, respectively, for simulated 1-, 2,- and 3-detector-ring systems utilizing τ values of 1.93, 2.16, and 2.5 ns, chosen to allow a fixed larger (56-cm-diam.) circular reconstruction FOV for all three axial detector coverages.

The singles distribution from the 3-ring simulation was used with equation (2) to compute the RS MLP image shown in Fig. 4b, and delayed events were used to compute the noisier RD MLP image in Fig. 4c. Events were added into these images at their most likely positions (**MLP**) within the elliptical cylinder, based on the TOF difference between the two detectors defining each coincidence LOR.

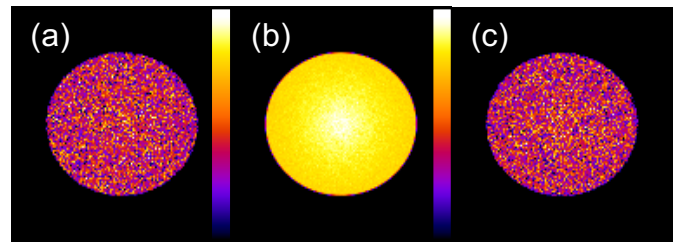

**Figure 4.** (a) MLP image of simulated random-coincidence events from very late [18]F acquisition of the image-quality phantom, (b) distribution calculated using the RS method, and (c) noisier MLP image of randoms from delayed events (RD).

Horizontal profiles (Fig. 5) across Figs. 4a and b show that the RS approach provided very accurate and precise random estimates under these challenging imaging conditions characterized by low counts but a very high randoms fraction.

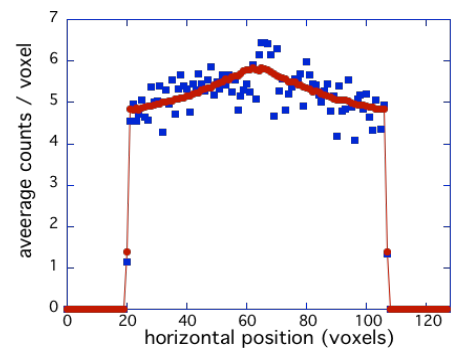

**Figure 5.** Profiles across the randoms images shown in Figs. 4a and 4b. Blue squares: simulated randoms; red circles: randoms-from-singles estimates.

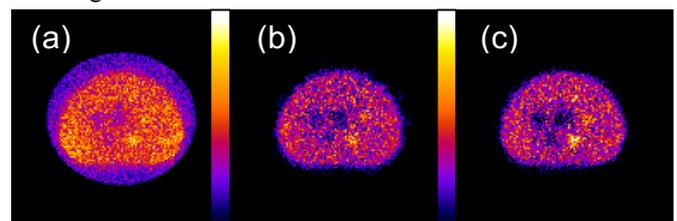

**Figure 6.** (a) MLP images from all (T, S, and R) coincidence events for the simulated late FDG acquisition; (b) after subtracting S events and random events from delays (RD); (c) same, but using randoms-from-singles (RS) estimate of randoms. Attenuation and sensitivity corrections were used for b and c.

The cold and hot spheres were visualized better with RS subtraction (Fig. 6c) than with RD subtraction (Fig. 6b). The relative noise in a large volume-of-interest (VOI) drawn on a uniform region within 5 coronal slices, away from all spheres and phantom boundaries, was 58.3% ± 1.2% for RD, and less (51.6% ± 1.3%) with RS subtraction.

The results of simulated 1-min acquisitions of the same phantom, filled with 750 µCi of $^{89}$Zr, showed that despite the large abundance of high-energy (>900 keV) decay gammas, their overall contribution to the random-coincidence rate was relatively small. The mean ± standard deviation of the counts/voxel in a large volume of interest on the randoms MLP image (from the simulation) was just 0.313 ± 0.541, as compared to 0.320 ± 0.010 for the same region in the corresponding RS image; the noise of the RS-estimated randoms was thus reduced by a factor of ~50.

For late acquisitions of $^{89}$Zr -- as mentioned earlier for late $^{18}$F studies -- a major source of random events arises from intrinsic $^{176}$Lu decay photons in the LYSO detectors. Using a measured $^{176}$Lu energy spectrum from a "blank scan" on the PennPET Explorer, we modeled the singles detector rates in the FMC program to simulate the extended IQ phantom containing low levels of $^{89}$Zr activity ranging from 100 to 1600 µCi. With a fixed lower level discriminator (LLD) setting of 451 keV, we varied the upper-level discriminator (ULD) and computed the noise-equivalent count-rate (NECR) values from the simulated T, S, and R rates. The optimal ULD setting for MLP events located inside the elliptical cylinder was ~585 keV over most of the activity range considered (e.g., Fig. 7).

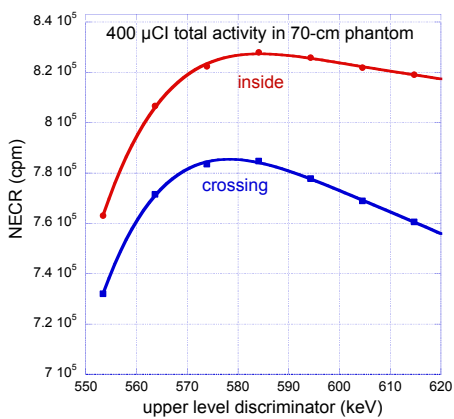

**Figure 7.** NECR vs. ULD setting for LORs crossing the elliptical cylinder volume containing the IQ phantom (**blue curve**), and for MLP events located inside the same volume (**red curve**).

Using real data acquired on the 5-ring PennPET Explorer from the NEMA count-rate phantom, Fig. 8 shows that the total randoms rates estimated by the RS method were within 10% of the corresponding RD rates at all count rates, and for all three activity/attenuator distributions.

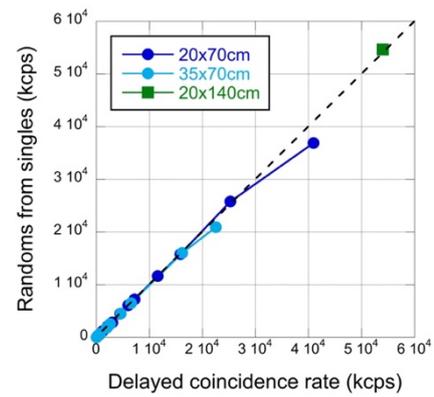

**Figure 8.** Comparison of randoms estimated for the NEMA count-rate phantoms measured on the 5-ring PennPET Explorer for the RS and RD methods. The dashed line is the line of identity.

These results are for the total system randoms rate, and the RS estimate was based on the total singles rate, although the singles rate was not constant across the axial FOV for the 70-cm-long line source; this may account for the deviation of the 70-cm RS estimates from the line of identity at the highest activity values.

Finally, the total RS rate was also estimated for a one-hour scan of a human subject acquired on the PennPET Explorer ~24 hours following the injection of 22.2 MBq of a new $^{89}$Zr-labeled anti-CD8 minibody and compared with the RD approach (Fig. 9). The total RS rate was within 5% of the RD rate, even for this low count-rate study (5 Mcps singles rate, which was just above the 3.2 Mcps $^{176}$Lu singles rate).

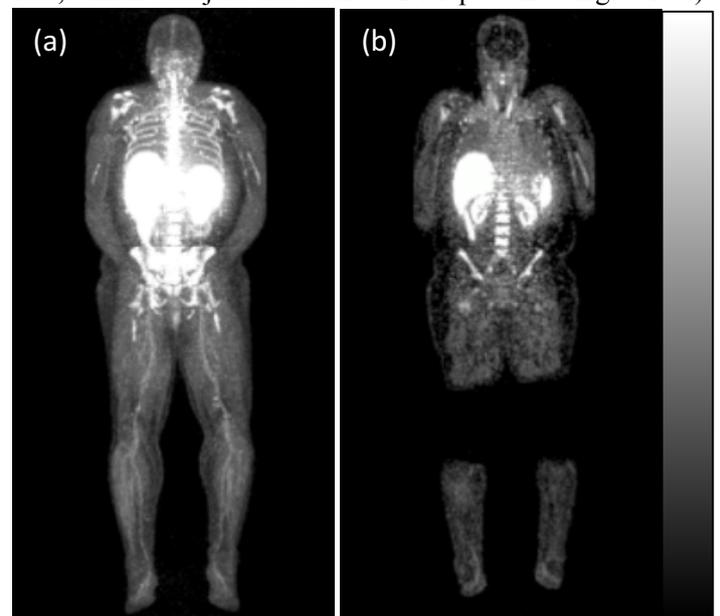

**Figure 9.** **(a)** Maximum-intensity projection (MIP) image at 0 degrees acquired for one hour on the 5-detector-ring PennPET Explorer (112-cm AFOV) from a human subject, 24 h after administration of 22.2 MBq of a new $^{89}$Zr-labeled anti-CD8 minibody. **(b)** 4-mm-thick TOF list-mode OSEM-reconstructed coronal slice through the same subject, whose knees were elevated during the acquisition.

Even after full reconstruction using TOF list-mode OSEM, the residual image noise seen in, e.g., the thigh and calf regions of Fig. 9b suggest that the RD-corrected reconstruction could benefit from use of the RS approach for randoms correction.

## 4 Discussion

The FMC simulation program demonstrated comparable delayed randoms rates to the slower GATE simulation for both count-rate and image-quality phantom studies. This has allowed us to investigate the RS and RD approaches for a variety of imaging conditions efficiently. While the FMC program does not model the exact geometry or all details affecting the spatial-resolution of the PennPET Explorer system, nevertheless, the program models the T, S, and R count rates quite accurately and is thus useful for understanding the count rate behavior of the system.

In this study, we have chosen to use most-likely-position (MLP) images, obtained using TOF-weighted backprojection of each coincidence event along its LOR. MLP images are linear 'reconstructions', i.e., they do not depend on characteristics such as the number of iterations or subsets or the degree of possibly nonstationary smoothing used in a full reconstruction; nevertheless, it was still straightforward in this study to compare imaging performance under different conditions using MLP images.

Oliver and Rafecas have advocated use of the uncorrelated singles rates for estimating the randoms [11]; however we found that using the uncorrelated singles rates significantly underestimated the correct random rates, whereas the total (correlated + uncorrelated) detector singles rates provided accurate randoms estimates over a wide range of total activity (Fig. 10).

The PennPET Explorer acquires data as single events and stores them in a list for off-line coincidence processing. Up to now, a delayed coincidence window has been used to estimate the random coincidences. However, because the data are acquired as singles, the RS method is a natural choice for the PennPET Explorer, and it is currently being incorporated into the data processing stream for evaluation in future studies. Initial phantom tests (with [18]F) have shown comparable levels and spatial distributions of randoms estimated using the RS method to the measured delayed coincidence rate (RD). Planned studies include application to dynamic scans with both [18]F and [11]C-labeled radiotracers, and studies with non-standard radionuclides in addition to [89]Zr, including those with prompt gammas.

## 5 Conclusion

The randoms-from-singles approach is a practical method to provide low-noise, almost unbiased estimates of randoms from long axial field-of-view scanners such as the PennPET Explorer. It is shown to be capable of accurately and precisely measuring randoms arising from a variety of event types, including [176]Lu background radiation, high-energy single photons from radionuclides like [89]Zr, and more conventional randoms from positron-annihilation events within the subject. The RS method works accurately and reliably at both high and low count rates.

## Acknowledgments

We acknowledge support from NIH R01-CA225874, NIH R01-CA113941 and a Siemens Research Agreement.

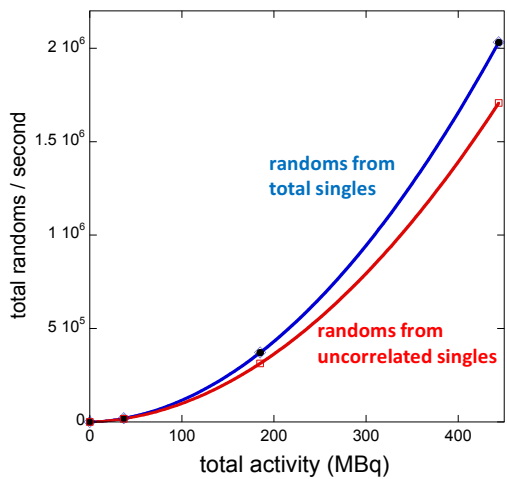

**Figure 10.** Total randoms rates simulated for [18]F in the 70-cm IQ phantom (**black data points**); randoms estimated using all singles (RS) in each detector (**blue curve**), and estimated using only uncorrelated singles (**red curve**) obtained after subtracting singles comprising true- and scattered-coincidence events from the total singles in each detector.

# Practical Energy-based method for correction of Scattered events in Positron Emission Tomography

Nikos Efthimiou[1], Joel S. Karp[1], and Suleman Surti[1]

[1]Department of Radiology, Perelman School of Medicine, University of Pennsylvania, USA

**Abstract** Scattered events lead to bias in PET images that requires an accurate estimation of their distribution in order to produce quantitative images. This paper presents a practical energy-based (EB) scatter estimation method that uses the marked difference between the energy distributions of the unscattered and scattered events in PET data in the presence of random events. The method is fast, efficient, robust, and the scatter estimate is generated in scale with the emission data; therefore, tail-fitting that is common in single scatter simulation (SSS) is unnecessary. We evaluated the method using Monte Carlo (MC) simulations and measured data from a single ring of the PennPET Explorer (PPEx) scanner, a long axial field-of-view (AFOV) scanner currently configured with 5 rings with 112 cm AFOV. We show that EB estimate of the scatter sinogram is in good agreement with the MC ground truth. Also, compared to correction with Single Scatter Simulation (SSS), there was a marked improvement in the reconstructed images, both in the background and cold regions. An ideal reconstructed image is given for reference. The method is efficient and fast, making it attractive for use in standard as well as long axial field-of-view (FOV) scanners.

## 1 Introduction

Scattered and Random events (background) lead to bias in the PET images. The contribution of randoms is commonly estimated using the delayed window method, while scatter estimation is commonly performed with the Single Scatter Simulation (SSS) [1, 2]. Extended versions of SSS including time-of-flight (TOF) [3, 4] information and modeling double scattering [5, 6] have also been presented. However, while SSS is generally robust for routine clinical imaging, it underperforms in certain imaging scenarios. Therefore alternative methods such as Monte Carlo simulations [7, 8] and, more recently, deep learning based scatter estimation [9–11] have been explored.

In this paper, we present an extension to a method initially proposed by Popescu et al. [12, 13] for estimating the scatter distribution based on photons' energy properties (energy-based, or EB, scatter estimation). The proposed method utilizes difference between the unscattered and scattered events energy distributions, to obtain a total scatter estimate for each line-of-response (LOR) [14]. To simplify the extraction of the energy probability density function (PDF), we consider the energies of the two coincident photons to be uncorrelated, which allows us to avoid fitting a 2D energy plane [15]. In addition to accuracy, our method focuses on algorithm efficiency, robustness and practicality for routine use. We evaluated its performance using Monte Carlo simulations of the PennPET Explorer (PPEx) scanner, comparing the results to ground truth available in simulations as well SSS estimates.

We also tested our method on NEMA phantom data acquired on a the PPEx scanner where we compare the estimated EB scatter sinograms with those generated by the 3D-TOF-SSS that is routinely used with PPEx.

## 2 Materials and Methods

Each PET event includes the measurement of two coincident $\gamma$-photons with energies $E_1$, $E_2$ at a line-of-response (LOR, $l$) between two detectors. Let $T(E)$ be the distribution of $\gamma$-photons that have not undergone Compton scattering and $S(E)$ that of scattered photons, then the total energy histogram is the sum of three types of events; i) with both photons unscattered, ii&iii) with one of the two photon scattered, and iv) with both photons scattered. The above can be expressed as:

$$C(E_1, E_2) = \sigma_0 T(E_1)T(E_2) + \sigma_1 T(E_1)S(E_2) + \quad (1)$$
$$\sigma_2 S(E_1)T(E_2) + \sigma_3 S(E_1)S(E_2)$$

with $\sigma_0 \ldots \sigma_3$ their respective weights. Assuming that the energies of the two coincident photons per event are uncorrelated, in Figure 1 we show the total 1D energy histogram for all events, and categorized as the prompt (or total), unscattered, and scattered events (simulation data). The sum of the unscattered events curve is the $\sigma_0$ from 1. Likewise, the scattered events curve sum equals to $\sigma_T = \sigma_1 + \sigma_2 + \sigma_3$ and can be decomposed as:

$$\sigma_0 = kT(E) \quad (2)$$
$$\sigma_T = p_0 T(E) + p_1 S(E) \quad (3)$$

where $\sigma_T$ is illustrated in the bottom row of Fig. 1.
Our hypothesis is that if we are able to extract the energy distribution of scattered photons (red curve in Fig. 1(bottom)) then we will be able to solve for $\sigma_1 \ldots \sigma_3$ by appropriately scaling a Gaussian function by $\sigma_0$. Since the prompt data also include random events, we also estimate and subtract the contribution of random events in the energy spectrum by using the data collected in the delayed coincidence window.

### 2.1 Estimation of models for scattered and unscattered photons energy distributions

The unscattered photons global energy spectrum, T(E), has a peak at 511 keV, broadened by the detector energy resolu-tion, while the scattered photons have a continuous energy spectrum with energies lower than 511 keV. The detector's has an energy response, in principle, can be modeled as a Gaussian distribution scaled to the total integrated counts collected within a high energy window (HEW). In the HEW, the number of scattered photons would be negligible to the total number of unscattered events. However, due to inter-crystal scattering, partial energy deposition, $^{176}$Lu background activity, coarse bin size of energy in the collected data, and statistical noise etc., the shape of T(E) is not a perfect Gaussian distribution. This discrepancy of the unscattered photon's peak from a Gaussian can lead to an underestimation of the model's scaling factor, resulting in an influx of unscattered events in the scatter estimated sinogram (overestimation of scatter). Or in overestimation of the scaling factor, resulting in a reduction in the number of high energy scattered photons.

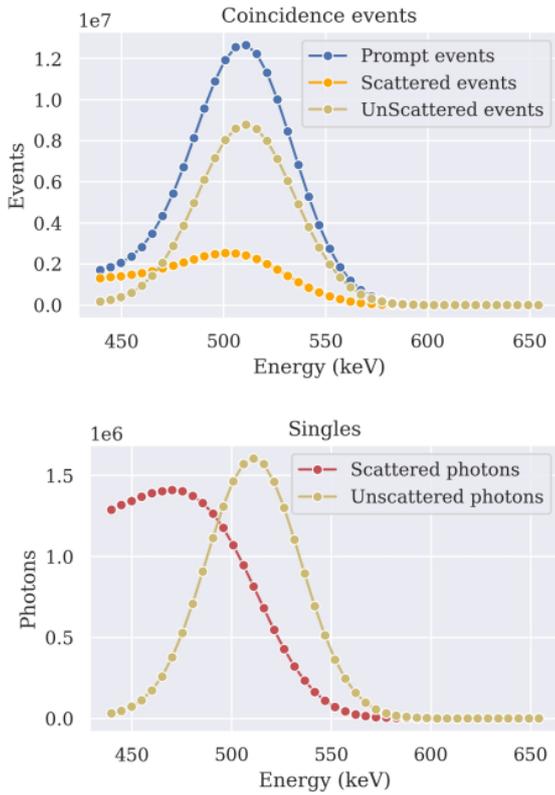

**Figure 1:** (top) Prompt minus delays energy histogram (C), and (bottom) decomposed energy distributions of scattered and unscattered photons in the scattered events.

In order to work around this limitation we designed an algorithm that scans several lower bounds for the HEW and checks which of the estimated scatter photon energy distribution shapes meet the following three conditions: i) it does not have negative values, ii) it should not present a photopeak at 511 keV, and iii) above, 570 keV, for the current detector energy resolution, they do not assign higher probabilities for delayed events than prompts. This creates a small *pool* of scattered photon energy distribution shapes and applies them on a row-by-row basis in the sinogram, choosing the one that minimizes the root square mean error (RMSE) between the estimated scatter profile and the tails of the emission data.

## 2.2 Fitting of the energy distribution models to the data to estimate scatter

The energy distribution models for scattered and unscattered events as obtained from the global energy spectrum are fitted to the measured data in an LOR or a group of LORs, using the method of moments. The moments of order $(m, n)$ of the energy distribution can be calculated from the listmode events in each LOR as $\hat{q}_{mn} = \sum_i^M E_{1,i}^m E_{2,i}^n$ with $M$ being the number of events in the LOR and should match the theoretical value:

$$q_{mn} = \int_\varepsilon E_1^m E_2^n C(E_1, E_2) dE_1 dE_2 \qquad (4)$$

$$= \sigma_0 \alpha_m \alpha_n + \sigma_1 \alpha_m \beta_n + \sigma_2 \beta_m \alpha_n + \sigma_3 \beta_m \beta_n \qquad (5)$$

where $a_m = \int_\varepsilon E^m T(E) dE$, $\beta_m = \int_\varepsilon E^m S(E) dE$ and likewise for the $n$ moment. Solving for $\sigma_0 \ldots \sigma_3$ was done with the Cramer method. In order to remove the random events we subtract from $\hat{q}_{mn}$ a term $\hat{w}_{mn}$ calculated using the same formula, but with the delayed events.

## 2.3 Data generation and image reconstruction

### 2.3.1 Simulations

For the generation of the simulated data we used the GATE Monte Carlo simulation toolbox(v.8.1.p01) [16]. We simulated the geometry of a single ring of the PPEx scanner [17]. In brief, the geometry consists of $3.86 \times 19$mm$^3$ LYSO crystals arranged in an $8 \times 8$ to form a detector tile. Each scanner module is composed of a $4$(radial) $\times 7$(axial) array of these detector tiles with 18 such modules arranged to form a single scanner ring. In total, the scanner ring has 56 crystal rings and 576 crystal per crystal ring. We set the energy resolution to 11% and the timing resolution to 250 ps, binning in 161 TOF bins of 25 ps.

We simulated the NEMA phantom (Fig. 2) (hot sphere activity ratio of 4:1 with respect to the background) for a 150 s acquisition with total activity 43.1 MBq in the phantom. We collected $79.3 \times 10^6$ prompt events with a randoms ratio (delayed / prompts events) of 13%. We also simulated a cold slab phantom 35 cm diameter x 30 cm long cylinder with 1 inch thick polyethylene slab placed centrally in the phantom (Fig. 2), for a 500 s acquisition with 60 MBq, and we recorded $7.6 \times 10^7$ prompts with a randoms ratio of 46%. For scatter estimation, the aforementioned scanner geometry was downsampled to 36 (**.d1**) and 72 (**.d2**) detectors per ring, over 7 rings and 17 TOF bins (272 ps). For the simulated data, all image reconstruction and various data space operations were performed using the STIR image reconstruction toolkit (v.4) [18–21]. Image reconstruction was performed using ordered subsets, expectation maximization (OSEM) with 11 subsets and 20 full iterations in the projection space. The voxel size of the reconstructed images was $1.93 \times 1.93 \times 1.98 mm^3$. In order to reduce the noise of the NEMA phantom, Gaussian post-filtering ($1 \times 1 \times 1$ mm$^3$), was applied.

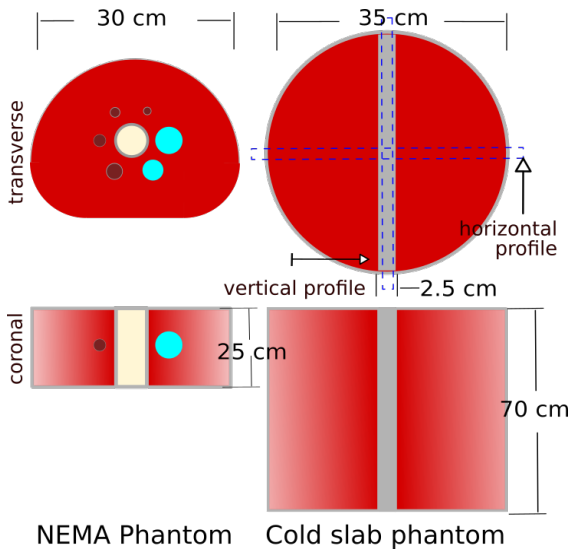

**Figure 2:** Illustrations of the NEMA and cold slab phantoms used in this study.

### 2.3.2 Measurements

We collected data on a single ring of the PPEx scanner using the NEMA phantom with hot sphere activity uptake ratio of 9.4:1. The total activity in the phantom was 59.2 MBq and data were collected for 1800 s. Scatter was estimated using our EB scatter estimation method as well as the standard 3D-TOF-SSS algorithm as implemented on this scanner.

## 3   Results

### 3.1   Data-space

#### 3.1.1   Simulated data

In Figure 3.A we show sinograms followed by relevant profiles from the MC simulated data (Figures 3.B and 3.C). In Figures 3.D and 3.E we compare sinogram profiles from 2D-SSS and EB scatter estimates to the MC ground truth for the NEMA and cold slab phantoms. As we can see in Figures 3.B and 3.C, in the downsampled sinogram space the EB estimated scatter profiles are in very good agreement with the ground truth profiles, which are created by directly selecting the appropriate events in the simulated datasets. Especially in the tails of the emission data, the agreement is excellent. The good performance also extends to the estimated unscattered events. When we use the less coarse down-sampling (d2), the estimation follows more closely the phantom's higher spatial frequency structures. However we also see increased noise, possibly due to the smaller number of events per bin.

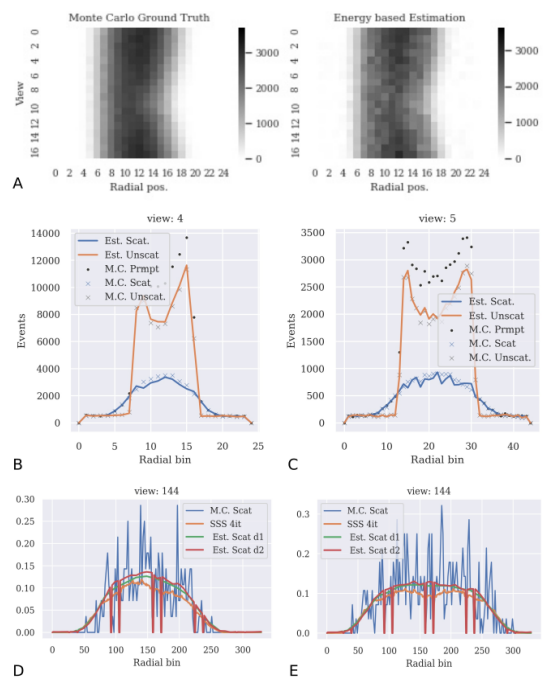

**Figure 3: A.** (left) Ground truth scatter sinogram from the MC simulated NEMA phantom and the sinogram from EB scatter estimate using d1 down-sampling (right). **B.** Profiles drawn across sinograms shown in (A) and (B) above for the fourth view. The MC values are marked with "x" and the EB estimates are shown with solid lines. **C.** Same profiles as in (B) but with less coarse down-sampling (d2). **D.** Profiles across the prompts, 2D_SSS estimated scatter sinogram and the EB scatter estimate sinograms for the simulated NEMA phantom, all after up-sampling to the original scanner geometry. **E.** Same as in (D) but for the simulated cold slab phantom.

In Figures 3.D and 3.E we demonstrate how the EB method compares with 2D-SSS and the ground truth for the NEMA and the cold slab phantoms, respectively. From the NEMA MC simulations we know that the scatter fraction (SF) should be 28.2%. The 2D-SSS gave 25.3% and the EB 26.5% - 27.8% depending on the down-sampling and usage of TOF. These differences are reflected in Figures 3.D and 3.E where we see an underestimation of scatter with SSS near the center of the phantoms. However, it should be noted that both scatter estimation methods are in good agreement with tails of the phantom emission data.

#### 3.1.2   Measured data

In Fig. 4 we show sinograms from the measured NEMA IQ phantom and scatter estimations using 3D-TOF-SSS algorithm as implemented on PPEx and the proposed EB method. In addition, we also show two radial profiles. In general, we can see that for most part the SSS and EB methods are in good agreement. However, as the measured sinogram shows there was a deactivated detector tile (empty diagonal) and the EB scatter sinogram reflects that. In addition, in the EB sinogram we can see a second diagonal with higher values. The acquisition of this dataset was performed while the ring was being calibrated prior to integration with the PPEx scanner. In this situation one detector tile presented increased operating temperature at the time of data acquisition, leading to a change in local energy resolution that affects the EB scatter estimation. Under normal operating conditions this calibration issue will not be present.

## 3.2 Image-space

Quantitative analysis of the reconstructed images for the simulated NEMA phantom showed that the ideal (no scattered events) $\Delta_{lung}$ residual was 6.2%, while those corrected with 2D-SSS had 10.0% and those with EB 6.6% and 6.2%, for down-samplings d1 and d2, respectively. We can see that the results from EB scatter estimation are much closer to the ideal results. In addition, using a less coarse down-sampling (d2) we show a $2 - 2.5\%$ improvement in the cold sphere contrast recovery coefficient (CRC), possibly a result of the reduced interpolation between the cold and background regions. However, we also saw an equivalent reduction in the CRC of the smallest (10 mm) hot sphere, possibly due to noise introduced by the additive correction. Optimization of the level of downsampling should account for statistics and will depend on the type of imaging protocol (dynamic or static).

NonTOF reconstructed images and profiles of the challenging cylindrical phantom with the cold slab are illustrated in Fig. 5. One can see that the scatter correction in the cold slab region is excellent with the EB method. It is also clear that SSS under corrects for scatter in the cold region that is also confirmed by the sinogram profiles as shown in Figure 3. In addition, we cannot observe any effect from the out-of-FOV activity.

## 4 Discussion

This paper describes the implementation, validation, and initial evaluation of an energy-based scatter estimation method. The method was evaluated both in data-space and image-space, using Monte Carlo generated data of a standard NEMA phantom and a challenging cylinder with a cold slab. The evaluation was against the MC ground truth and the 2D-SSS, as implemented in STIR. In addition, we compared the energy-based estimated sinogram generated from measured data acquired on a single ring of the PPEx scanner.

This method's accuracy depends on the accuracy in estimating the energy distribution functions of the scattered and unscattered events for the scanned object and the effectiveness in rejecting the randoms from the estimation of the scattered. As the detector is downsampled, the energy response of the individual crystals and other effects from the obliqueness of the LORs are not significant. However, it is critical for our method to have the correct energy resolution for each downsampled group of detectors. As Fig 4 demonstrates, in the case that this does not hold, then the estimated scatter will not be accurate. Ideally, all individually read-out crystals should also have approximately similar energy resolution. This suggests that in older PET scanners built with light-sharing detectors that lead to some spatial variability in energy resolution as well as pulse pile-up effects at higher count-rates, the proposed EB method may be limited in its accuracy. However, this method should be more effective in the latest generation of scanners using SiPMs that have minimal to no light sharing.

In addition, although the method of moments is a swift and efficient way to fit data when the moments of the two models offer good separation, an inappropriate selection of the used moments can affect the results. However, as we have seen with the first two moments, we get good discrimination between the scattered and unscattered distributions. The excellent performance of all the above points has been successfully demonstrated as shown by our results in this paper (Figure 5).

We have shown that the EB estimation is very close to the MC ground truth. In nonTOF reconstructed images, the $\Delta_{lung}$ residual values are closer to the ideal (without scattered events) than those corrected with 2D-SSS. In addition, there might be some benefits in using a less coarse scanner downsampling for the scatter estimation as it can improve the convergence of cold regions. However, this will reduce the number of events per bin, and the effect of noise in the additive reconstruction term has to be considered.

The cold slab phantom presented several challenges as higher attenuation, out-of-FOV activity, and high SF, especially due to multiple scatter. The EB method performed well in this situation, and the reconstructed images closely matched the ideal image (with a rejection of scattered events from the data). The excellent performance of the EB-based correction was demonstrated by the horizontal and vertical profiles selected through the center of the cylinder's image.

An additional advantage of the EB method is the total execution time of the algorithm. On an i7 6700HQ processor, the EB-based method finished in 8 s the nonTOF, fully 3D, scatter estimation, and 35 s the TOF case. In addition, we did not see any significant impact (change) in performance with the number of events present in the data set all the way up to $500 \times 10^6$ prompts. For comparison, the 2D-SSS, as implemented in STIR, needs approximately 12 min for four scatter estimation iterations.

### 4.1 Future work

In the future, we plan to use this energy-based scatter estimation method with additional measured data, including phantom and human studies from the PennPET Explorer scanner.

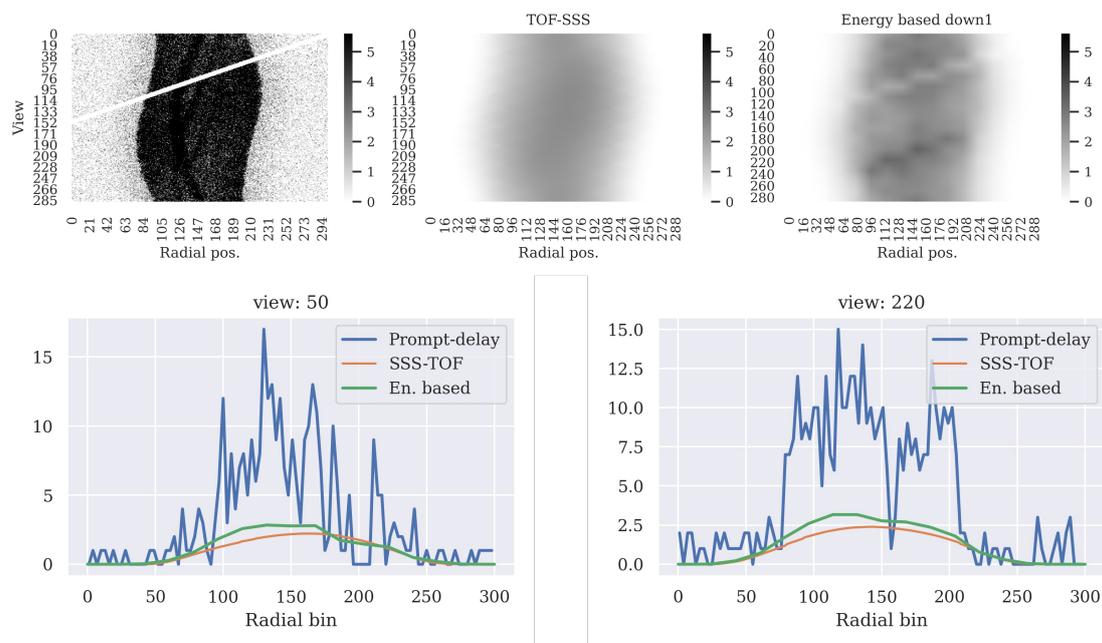

**Figure 4:** (top) NEMA-IQ sinograms from (left to right): i) Prompts-delays ii) 3D-TOF-SSS scatter estimation iii) EB scatter estimation. (bottom) Profiles across the sinograms of the top row.

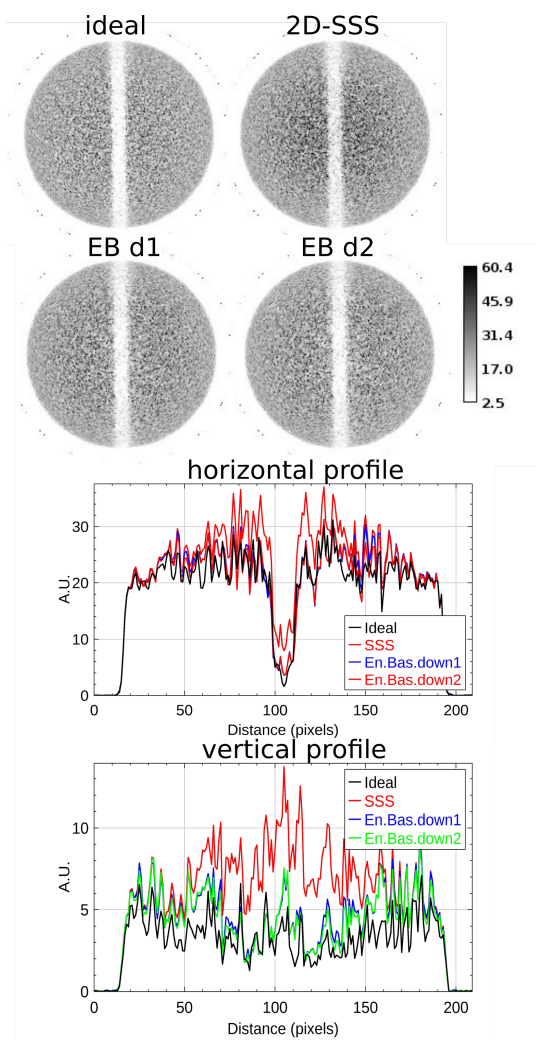

**Figure 5:** NonTOF reconstructed images of the cylindrical phantom with a cold slab. The ideal dataset does not include scattered events. In all cases the randoms correction is the same. In addition, horizontal profiles and vertical, are illustrated.

## 5 Conclusion

We demonstrated through realistic studies of a typical NEMA phantom and a challenging cylindrical phantom with a cold slab that our energy-based scatter estimation method provides accurate scatter estimation, leading to quantitatively accurate images. Additionally, using measured data for a NEMA phantom we showed good agreement with 3D-TOF-SSS in most regions of the scatter sinogram. Overall, the method is much faster than SSS and MC based methods, does not need prior training data sets, and does not rely on a transmission image.

## 6 Acknowledgments

This work was supported in part by NIH grants R21-CA239177, R01-EB028764, R01-CA196528, R01-CA113941 and the Siemens Research agreement.

**Chapter 9**

# Oral Session - Advanced image reconstruction methods 2

**session chairs**
Laurent Desbat, *University Grenoble Alps (France)*
Abhinav Jha, *Washington University in St. Louis (United States)*

# Novel approaches to reconstruction of highly multiplexed data for use in stationary low-dose molecular breast tomosynthesis

Kjell Erlandsson[1], Andrew Wirth[2], Ian Baistow[2], Kris Thielemans[1], Alexander Cherlin[2] and Brian F Hutton[1]

[1]Institute of Nuclear Medicine, University College London, London, UK
[2]Kromek Ltd, County Durham, UK

**Abstract** Molecular breast imaging (MBI) can be a useful complement to conventional X-ray mammography. Currently, planar imaging is normally used for this purpose. We are developing a stationary tomosynthesis system for MBI, based on CZT detectors with depth-of-interaction (DOI) capability, and multi-pinhole collimation, which could offer significantly improved contrast compared to planar imaging. A large number of pinholes are used in order to obtain high sensitivity and also improved sampling to compensate for the lack of detector motion. This results in multiplexing (MX), which leads to ambiguity regarding the direction of incidence of the detected γ-photons. We have developed various novel approaches to address this problem by performing de-MX either before or during the image reconstruction, aided by the DOI information. We have shown that, by optimising the system geometry, it is possible to gain a factor of 2 in effective sensitivity as compared to a system without MX.

## 1 Introduction

Conventional X-ray mammography has limited sensitivity in patients with dense breasts. Molecular imaging is advantageous in these situations, the main drawbacks being the relatively high radiation dose and long imaging times [1]. Dedicated cameras for molecular breast imaging (MBI), operating in planar mode, have been in use for years [2]. Improved image quality in planar MBI can be obtained using CZT detectors with depth-of-interaction (DOI) capability [3]. A tomosynthesis (limited angle tomography) system for MBI, based on multi-pinhole (MPH) collimators, has been proposed [4]. This system was designed to avoid multiplexing (MX), i.e. overlap of the individual pinhole projections on the detectors, and requires scanning detector motion to cover the whole field-of-view (FoV).

We are developing a stationary tomosynthesis system for MBI, based on CZT detectors with DOI and MPH collimation. Our basic idea is to use a large number of pinholes, allowing for MX, resulting in higher sensitivity and improved sampling. With MX, there is some degree of ambiguity regarding the direction of incidence of the detected γ-photons, which can lead to artefacts in the reconstructed images. However, it has been shown in the past that artifact-free images can be obtained by combining multiplexed and non-multiplexed data [5-9]. DOI information has the potential to provide data with variable amounts of MX, which could therefore aid in de-multiplexing.

We have investigated various design configurations in a multi-parameter space in order to optimize the system performance. We have also developed a novel de-MX approach that can be applied to the projection data before reconstruction. Here we compare this approach with direct reconstruction that incorporates MX in the system matrix as well as a hybrid approach.

## 2 Materials and Methods

### Data generation

The system consists of two planar CZT detector arrays placed opposite each other (Fig. 1). We assume the use of mild breast compression for a mean thickness of 6 cm [2]. We performed simulations for a 16x16 cm detector size with a pixel-size of 1x1 mm and DOI estimation in 1-mm layers. For the system optimisation, we investigated the following parameters: Number of pinholes, pinhole aperture size, pinhole opening angle and collimator-to-detector distance.

We first used analytical calculations of contrast-to-noise ratio (CNR) to narrow down the parameter space. Next we performed analytical simulations generating projection data corresponding to a phantom containing one layer of spherical lesions in four quadrants of 36 spheres each. The sphere diameter was 6 mm and the sphere-to-background ratios were 5, 10, 15 and 20 in the four quadrants, respectively. Simulations were also performed with four layers of spheres separated by 15 mm. The simulations, which included attenuation but not scatter, represented 10-min patient scans after injection of 150 MBq of $^{99m}$Tc-MIBI. We estimated that this would result in a background activity concentration of 760 Bq/mL, based on information in [10].

### De-multiplexing and reconstruction

As MX is a process occurring by definition in the detector domain, it depends only on the line-integrals and de-MX can in principle be applied to the projection data without knowledge of the 3D activity distribution. We developed a new approach for de-multiplexing the MPH projection data, utilising the DOI information provided by the CZT detectors. The algorithm consists of an iterative procedure where data are forward and back-projected between virtual 2D planes, representing each pinhole, and 3D detector blocks (Fig. 2). The following steps are repeated for a number of iterations:

$$\boldsymbol{V}_{i,j}^k = \boldsymbol{A} \cdot \boldsymbol{P}_{i,j}^k \ ; \ i = 1..N_d \ ; \ j = 1..N_p$$

$$\boldsymbol{V}_i^k = \left[ \boldsymbol{V}_{i,1}^k, \boldsymbol{V}_{i,2}^k, ... \boldsymbol{V}_{i,N_p}^k \right]^T \ ; \ i = 1..N_d$$

$$\boldsymbol{C}_i^k = \left[ \boldsymbol{C}_{i,1}^k, \boldsymbol{C}_{i,2}^k, ... \boldsymbol{C}_{i,N_p}^k \right]^T = \boldsymbol{B}^T \frac{\boldsymbol{Q}_i}{\boldsymbol{B} \cdot \boldsymbol{V}_i^k} \ ; \ i = 1..N_d$$

$$\boldsymbol{P}_{i,j}^{k+1} = \frac{\boldsymbol{P}_{i,j}^k}{\boldsymbol{A}^T \cdot \boldsymbol{1}} \boldsymbol{A}^T \cdot \boldsymbol{C}_i^k \ ; \ i = 1..N_d \ ; \ j = 1..N_p$$

where $\boldsymbol{P}_{i,j}^k$ and $\boldsymbol{V}_{i,j}^k$ are the 2D virtual data plane and 3D detector data, respectively, for detector $i$ and pinhole $j$ after $k$ iterations, $N_d$ and $N_p$ are the number of detectors and the number of pinholes per detector, respectively, $\boldsymbol{A}$ is a matrix for transformation from the 2D to the 3D data representation, $\boldsymbol{B}$ is a matrix representing the multiplexing operator, and $\boldsymbol{Q}_i$ is the measured data for detector $i$. The MX process was implemented by combining data from different pinholes in detector pixels with overlap.

This de-MX method differs from the one presented in [11] as it is entirely independent of the tomographic reconstruction process.

For the tomographic reconstruction, we have implemented three different approaches: 1) 1-step: direct image reconstruction, incorporating MX in the system matrix; 2) 2-step: de-MX is applied to the projection data before tomographic reconstruction; and 3) a combination of the two methods, in which, at each iteration, the image is updated using the average of the correction factors obtained from the MX data and the de-MX data (Fig. 3). For the reconstruction we used a MAP algorithmm [12] with a prior obtained by distance dependent smoothing for resolution equalisation.

Here we compare the three approaches in terms of contrast and noise. We also compare the results with images reconstructed from ideal projection data for the same geometry but without MX (not possible in practice).

## 3 Results

Target-to-background ratios (TBR) were calculated for the spheres in the single layer phantom and the coefficient-of-variation was calculated in the uniform region away from the sphere-plane. Figure 4 shows TBR vs. CoV curves with different MPH configurations from 8x8 to 16x16 pinholes per head with separations in the range 10-20 mm. The 8x8 configuration with 20 mm separation corresponds to the actual MX-free case. The graph resembles a "bow and arrow", with the "arrow" corresponding to the ideal no-MX situation, and the other three curves corresponding to the different reconstruction approaches. Starting from the MX-free case on the right side of the graph, all three curves initially move more or less in the same direction as the ideal curve. They then seem to hit an invisible barrier and bounce off in different directions, due to unresolved MX or noise-amplification. Along the "invisible barrier" there are multiple solutions, which are essentially equivalent, but have different bias vs.

noise trade-offs. The "bow" crosses the "arrow" at a point corresponding to ~14.3 mm pinhole separation. Compared to the MX-free case, this corresponds to a pinhole density increase by a factor of $(20/14.3)^2 \approx 2$, which represents the effective increase in sensitivity.

The MPH configuration with 14x14 pinholes with 12 mm separation was chosen for further evaluation. Reconstructed images are shown in Fig. 5 for the different reconstruction approaches. Fig. 6 shows TBR vs. CoV curves for the different reconstruction approaches in the single layer phantom and in the multi-layer phantom. It can be seen that the best approach (apart from the ideal no-MX case) is different for the two phantoms; the 1-step for the single layer phantom, and the 2-step for the multi-layer phantom. In both cases, the results of the hybrid method are between the other two, representing a good compromise.

## 4 Discussion

In terms of the different reconstruction approaches, we found that the method giving the best results depended on the activity distribution. The method of choice could therefore be the hybrid approach, but more work is required in order to evaluate this option. Although this work was performed for MBI, the same ideas could be applied in other imaging situations.

These results have important implications for clinical MBI. Tomosynthesis provides significantly improved contrast compared to planar imaging but to achieve sufficient angular sampling without MX normally requires detector movement. The approach we have developed not only solves this sampling limitation but also provides potential for significant gain in sensitivity compared to non-MX systems. The possible options are to improve performance, decrease imaging time or reduce dose (or to choose some combination depending on goals of the clinical study). For example, this may stimulate more widespread use of MBI in screening applications.

## 5 Conclusion

We have designed a novel stationary MBI tomosynthesis system, incorporating multiplexing for improved image quality. We have developed various de-multiplexing and reconstruction approaches, specifically for this system. Using simulations we found that it is possible to obtain an effective increase in sensitivity by a factor of 2 by utilising multiplexing.

## Acknowledgements

Staff at the Institute of Nuclear Medicine at UCLH also receive support from the NIHR University College Hospitals Biomedical Research Centre.

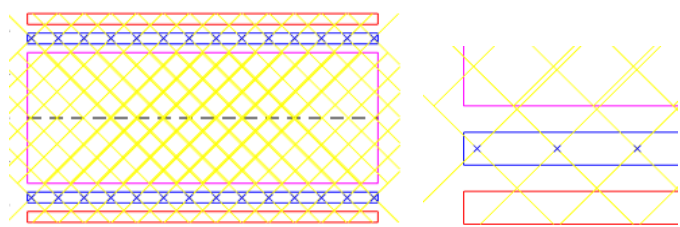

**Figure 1:** Full system geometry (left), blow-up of corner, showing multiplexing (right).

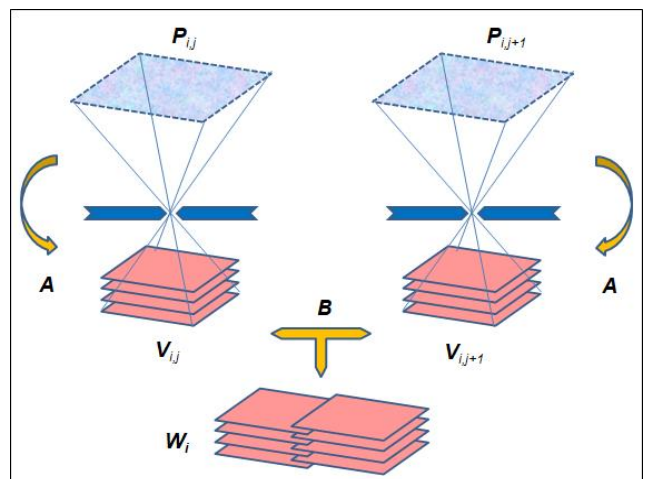

**Figure 2:** Illustration of the forward transformation process in the de-multiplexing algorithm. Virtual planes are projected (FP) onto multi-layered detector blocks, which are then merged with multiplexing (MX).

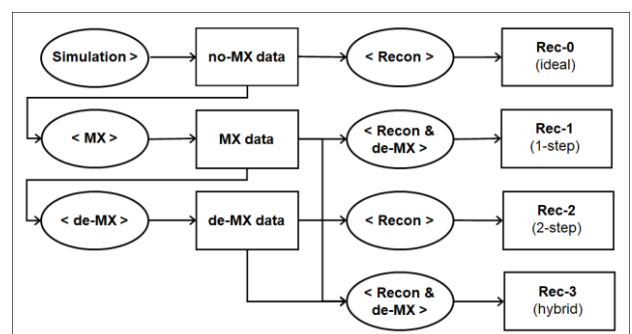

**Figure 3:** Schematic description of hybrid reconstruction algorithm. The correction factors are calculated based on both MX and de-MX data.

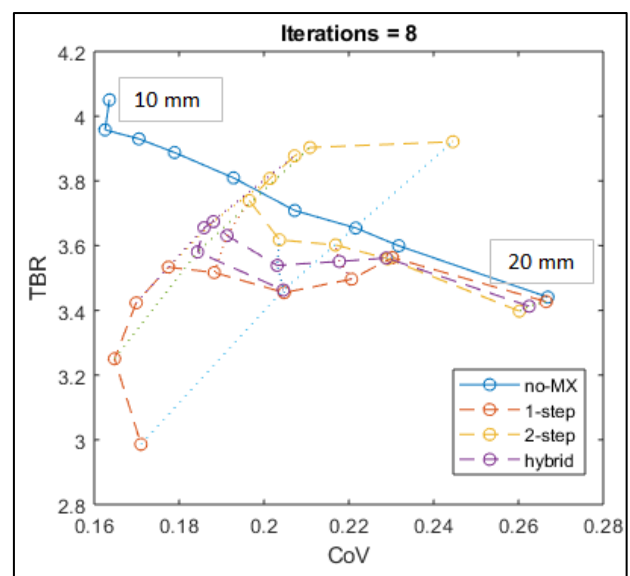

**Figure 4:** TBR vs. CoV curves for different pinhole configurations, corresponding to pinhole separations of 10-16, 18 and 20 mm. The solid line with circles represents the ideal (but unachievable) no-MX case, while the dashed lines with circles represents different reconstruction approaches for the MX data. Dotted lines join points with the same pinhole separation.

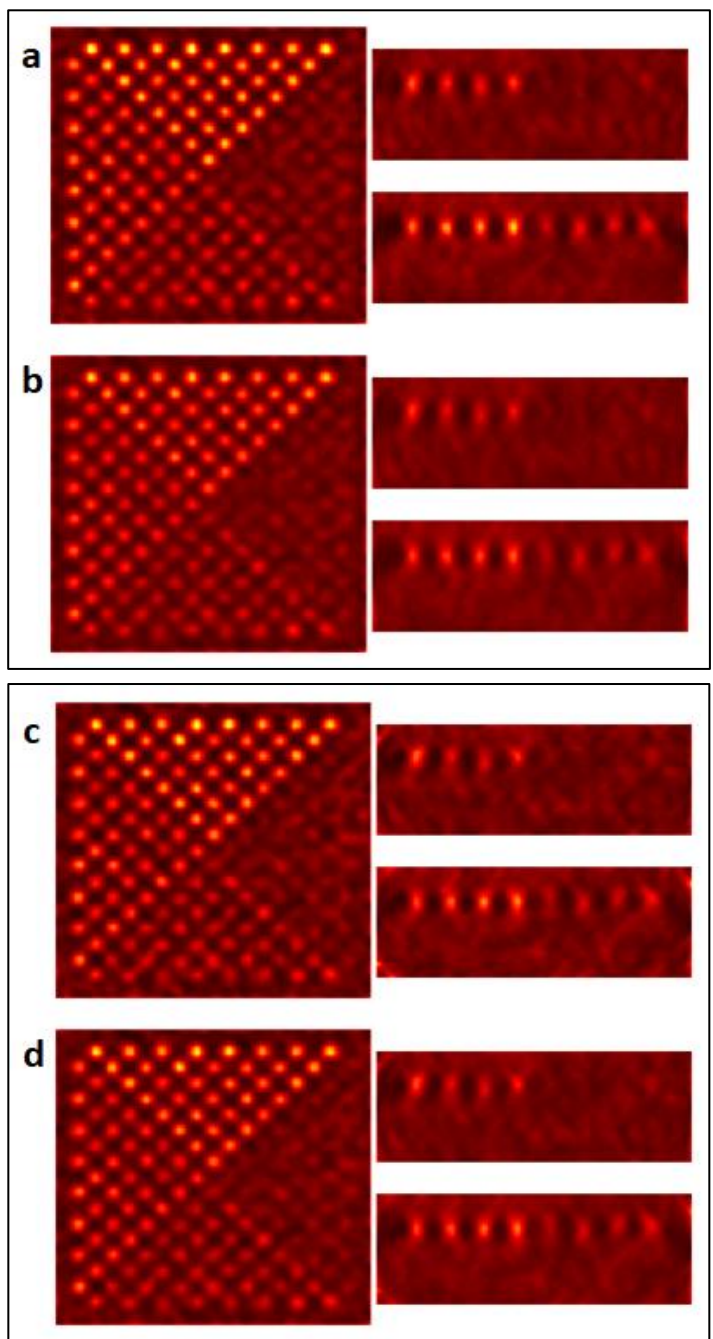

**Figure 5:** Reconstructed images (trans-axial, coronal and sagittal) for a 14x14 pinhole configuration (12 mm separation) with different reconstruction approaches: a) ideal case without MX, b) 1-step, c) 2-step, and d) hybrid reconstruction.

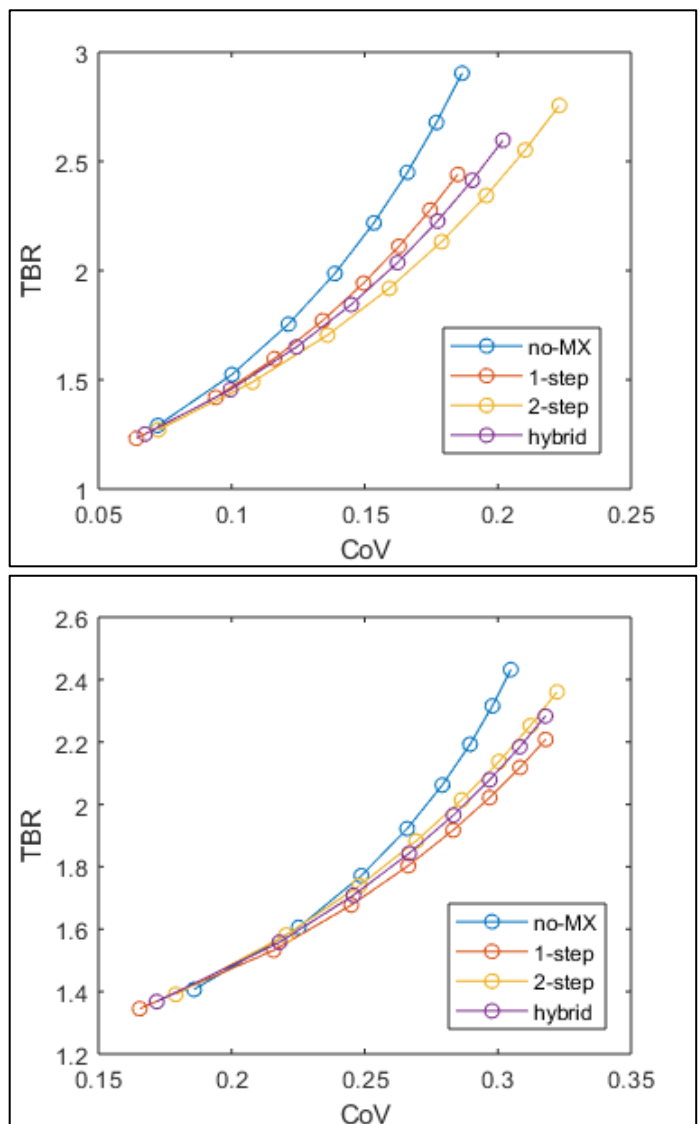

**Figure 6:** TBR vs. CoV curves with different number of iterations (1-8) for different reconstruction approaches in a single layer (top) and a multi-layer phantom (bottom). The no-MX curve is always best, as it represents an ideal case without MX which is not possible in practice.

# Analytical Covariance Estimation for Iterative CT Reconstruction Methods

Xiaoyue Guo[1,2], Yuxiang Xing[1,2], and Li Zhang[1,2]

[1]Department of Engineering Physics, Tsinghua University, Beijing, China
[2] Key Laboratory of Particle & Radiation Imaging, Tsinghua University, Beijing, China

**Abstract** Model observers are often used to simulate the way of doctors' diagnosis in medical imaging for task-based performance evaluation. However, a big number of image samples are needed to estimate statistical mean and covariance of reconstructions necessary in the construction of traditional model observer templates, which is hard to realize in practical medical applications. For CT imaging, since the statistical distribution of the sinogram is known, we propose a new method to propagate the covariance from projection domain to image domain. For iterative reconstruction methods, the gradient of objective function disappears at the point of convergent solution. Making use of this condition, one can deduce an analytical covariance estimator accordingly. We considered in this paper three typical cases of penalty in iterative reconstruction, 1) no penalty, 2) quadratic penalty, and 3) non-quadratic penalty. We mainly focus on case 1 and 3 since case 2 is exactly same as the published work. For case 1, we mainly deal with the ill-condition and solve it by widely used Tikhonov-Phillips method in regularization area. For case 3, we mainly consider a linear approximation for the gradient of non-quadratic penalty, and analyze popular WLS-TV and WLS-qGGMRF reconstruction. Analytical estimation of the covariance matrices of a simulated phantom are compared with results estimated from the early published Taylor expansion based method. MAPE is calculated for quantitative analysis. Results indicate that our proposed method outperforms the published work for TV penalty and is comparable for qGGMRF penalty. However, MAPEs of analytical covariance for TV are still different from statistical covariance by more than 20%, which needs further improvement.

## 1 Introduction

Traditional model observer methods require the mean vector and covariance matrix of images to be observed to construct observer templates [1]. The mean vector and covariance matrix can be calculated statistically by repeatedly obtaining numerous images under the same condition. However, it is impossible to acquire such quantities of images in real situation, especially in medical imaging since patients' radiation dose should be minimized. In CT imaging, it is widely accepted that a sinogram can be approximated as Gaussian with known mean and covariance [2]. Image reconstruction is a transformation function of projections so that the mean and covariance of the sinogram can be transmitted from projection domain to image domain. Thus, analytical methods for image mean and covariance calculation are proposed that merely require a few images. For analytical reconstruction, Noo et al [3] get the covariance matrix of image from that of the corresponding sinogram based on the linear property of filter back projection (FBP) method. For iterative reconstructions, Fessler [4] studies mean and variance estimation based on the Taylor expansion of the converged reconstructed image. Considering the diversity of iterative methods, we focus on covariance matrix estimation of the iterative reconstruction images in this work.

## 2 Materials and Methods

Iterative CT reconstruction methods with Gaussian noise in projection data normally can be expressed as following:

$$\hat{\boldsymbol{\mu}} = \underset{\boldsymbol{\mu}}{\arg\min}\, \Phi(\mathbf{g}, \boldsymbol{\mu}) = \underset{\boldsymbol{\mu}}{\arg\min}\, \frac{1}{2}\|\mathbf{H}\boldsymbol{\mu} - \mathbf{g}\|_{\mathbf{W}}^2 + \alpha R(\boldsymbol{\mu}) \quad (1)$$

where $\hat{\boldsymbol{\mu}}$ is the reconstruction, $\mathbf{g}$ the projection, $\mathbf{H}$ the system matrix, $R$ the penalty function, and $\alpha$ the hyper-parameter. Theoretically, $\hat{\boldsymbol{\mu}}$ is the point where the gradient of the cost function in Eq. (1) equals to 0:

$$\frac{\partial \Phi(\mathbf{g}, \boldsymbol{\mu})}{\partial \boldsymbol{\mu}}\bigg|_{\boldsymbol{\mu}=\hat{\boldsymbol{\mu}}} = \mathbf{H}^T \mathbf{W}(\mathbf{H}\boldsymbol{\mu} - \mathbf{g}) + \alpha \nabla R(\boldsymbol{\mu})\big|_{\boldsymbol{\mu}=\hat{\boldsymbol{\mu}}} = \mathbf{0} \quad (2)$$

By rearranging Eq. (2), we have:

$$\mathbf{H}^T \mathbf{W} \mathbf{H}\hat{\boldsymbol{\mu}} + \alpha \nabla R(\hat{\boldsymbol{\mu}}) = \mathbf{H}^T \mathbf{W} \mathbf{g} \quad (3)$$

Hence, the covariances for both sides of Eq. (3) is related by:

$$\mathbf{H}^T \mathbf{W} \mathbf{H}\,\mathrm{Cov}(\hat{\boldsymbol{\mu}})\mathbf{H}^T \mathbf{W} \mathbf{H} + \alpha^2 \mathrm{Cov}\big[\nabla R(\hat{\boldsymbol{\mu}})\big]$$
$$+ 2\alpha \mathbf{H}^T \mathbf{W} \mathbf{H}\,\mathrm{Cov}\big[\hat{\boldsymbol{\mu}}, \nabla R(\hat{\boldsymbol{\mu}})\big] = \mathbf{H}^T \mathbf{W}\mathrm{Cov}(\mathbf{g})\mathbf{W} \mathbf{H} \quad (4)$$

To calculate $\mathrm{Cov}\big[\hat{\boldsymbol{\mu}}, \nabla R(\hat{\boldsymbol{\mu}})\big]$ and $\mathrm{Cov}\big[\nabla R(\hat{\boldsymbol{\mu}})\big]$, we express $\nabla R(\hat{\boldsymbol{\mu}})$ in a linear form:

$$\nabla R(\hat{\boldsymbol{\mu}}) = \mathbf{L}(\bar{\boldsymbol{\mu}})\hat{\boldsymbol{\mu}} + \mathbf{c}(\hat{\boldsymbol{\mu}}) \quad (5)$$

where $\mathbf{L}(\bar{\boldsymbol{\mu}})$ is a coefficient matrix and $\mathbf{c}(\hat{\boldsymbol{\mu}})$ a constant matrix. Therefore, Eq. (4) is simplified as:

$$\big[\mathbf{H}^T \mathbf{W} \mathbf{H} + \alpha \mathbf{L}(\bar{\boldsymbol{\mu}})\big]\mathrm{Cov}(\hat{\boldsymbol{\mu}})\big[\mathbf{H}^T \mathbf{W} \mathbf{H} + \alpha \mathbf{L}(\bar{\boldsymbol{\mu}})\big]^T$$
$$\cong \mathbf{H}^T \mathbf{W}\mathrm{Cov}(\mathbf{g})\mathbf{W} \mathbf{H} \quad (6)$$

Then the covariance matrix of the reconstruction image $\hat{\boldsymbol{\mu}}$ can be estimated from the covariance of sinogram $\mathbf{g}$ by:

$$\mathrm{Cov}(\hat{\boldsymbol{\mu}}) \cong \mathbf{A}^{-1}\mathbf{H}^T \mathbf{W}\mathrm{Cov}(\mathbf{g})\mathbf{W} \mathbf{H}\big(\mathbf{A}^T\big)^{-1} \quad (7)$$

with $\mathbf{A} = \mathbf{H}^T \mathbf{W} \mathbf{H} + \alpha \mathbf{L}(\bar{\boldsymbol{\mu}})$.

Many penalty functions $R(\boldsymbol{\mu})$ has been studied for iterative CT reconstructions. In this work, we study three cases of iterative CT reconstruction for their covariance estimation with this method by Eq. (7).

### 2.1 No penalty term

One typical iterative reconstruction method is weighted least square (WLS) method with $\alpha = 0$ in Eq. (1). Correspondingly:

$$\text{Cov}(\hat{\mathbf{\mu}}) = (\mathbf{H}^T \mathbf{W} \mathbf{H})^{-1} \mathbf{H}^T \mathbf{W} \text{Cov}(\mathbf{g}) \mathbf{W} \mathbf{H} (\mathbf{H}^T \mathbf{W} \mathbf{H})^{-1} \quad (8)$$

Because the condition number of the muplication term $\mathbf{H}^T \mathbf{W} \mathbf{H}$ is very high so that it leads to unstable and inaccurate estimation of $\text{Cov}(\hat{\mathbf{\mu}})$.

To solve the ill-posed Eq. (8), we adopt the Tikhonov-Phillips method commonly used for solving ill-posed equations [5]. A weighted identity matrix $\mathbf{I}$ is added to reduce the condition number of $\mathbf{H}^T \mathbf{W} \mathbf{H}$:

$$\text{Cov}(\hat{\mathbf{\mu}}) =$$
$$(\mathbf{H}^T \mathbf{W} \mathbf{H} + \beta \mathbf{I})^{-1} \mathbf{H}^T \mathbf{W} \text{Cov}(\mathbf{g}) \mathbf{W} \mathbf{H} (\mathbf{H}^T \mathbf{W} \mathbf{H} + \beta \mathbf{I})^{-1} \quad (9)$$

which results in a more stable and accurate covariance estimation. In fact, the term $\beta \mathbf{I}$ in Eq. (9) is equivalent to adding a L2-norm penalty in WLS:

$$\hat{\mathbf{\mu}} = \underset{\mathbf{\mu}}{\text{argmin}} \, \Phi(\mathbf{g}; \mathbf{\mu}) = \underset{\mathbf{\mu}}{\text{argmin}} \, \frac{1}{2} \|\mathbf{H}\mathbf{\mu} - \mathbf{g}\|_{\mathbf{W}}^2 + \beta \|\mathbf{\mu}\|_2^2 \quad (10)$$

which indicates $\hat{\mathbf{\mu}}$ of minimum potential is preferred.

## 2.2 Quadratic penalty

The commonly used quadratic penalty prior can be represented as:

$$R(\hat{\mathbf{\mu}}) = \frac{1}{2} \sum_i \sum_{j \in \mathcal{N}_i} w_{ij} (\hat{\mu}_i - \hat{\mu}_j)^2 \quad (11)$$

where $w_{ij}$ is the inverse of the spatial distance, and $\mathcal{N}_i$ the neighborhood of pixel $i$. In addition, $\nabla^2 R(\hat{\mathbf{\mu}})$ is the second derivative of $R(\hat{\mathbf{\mu}})$:

$$\nabla^2 R(\hat{\mathbf{\mu}}) = \begin{cases} \sum_j w_{ij}, & i = j \\ -w_{ij}, & i \neq j \end{cases}$$

Since $\nabla^2 R(\hat{\mathbf{\mu}})$ is a constant indepent of $\hat{\mathbf{\mu}}$, $R(\hat{\mathbf{\mu}})$ can be expressed as $\frac{1}{2} \hat{\mathbf{\mu}}^T \nabla^2 R(\hat{\mathbf{\mu}}) \hat{\mathbf{\mu}}$. Obviously, the first derivative of $R(\hat{\mathbf{\mu}})$ is linear:

$$\nabla R(\hat{\mathbf{\mu}}) = \nabla^2 R(\hat{\mathbf{\mu}}) \hat{\mathbf{\mu}} \quad (12)$$

Hence, the covariance matrix estimator becomes:

$$\text{Cov}(\hat{\mathbf{\mu}}) = \mathbf{A}^{-1} \mathbf{H}^T \mathbf{W} \text{Cov}(\mathbf{g}) \mathbf{W} \mathbf{H} (\mathbf{A}^T)^{-1} \quad (13)$$

with $\mathbf{A} = \mathbf{H}^T \mathbf{W} \mathbf{H} + \alpha \nabla^2 R(\hat{\mathbf{\mu}})$.

Eq. (13) is in accordance with covariance estimator proposed by [6] in case of quadratic penalties.

## 2.3 Non-quadratic penalty

For non-quadratic penalties, covariance extimation is more difficult. Eq. (12) dose not hold. Different non-quadratic penalty functions have different expressions so that have their unique linear approximation of $\nabla R(\hat{\mathbf{\mu}})$. One natural way is to perform 1$^{st}$ order Taylor expansion on $\nabla R(\hat{\mathbf{\mu}})$, which means the coefficient matrix $\mathbf{L}(\bar{\bar{\mathbf{\mu}}}) = \nabla^2 R(\bar{\bar{\mathbf{\mu}}})$, and it is exactly the Taylor approximation based method (TAM) published in [4].

In this work, we propose a linearity approximation based covariance estimation method (LAM). Rewrite $\nabla R(\hat{\mathbf{\mu}})$ as a linear function and the linear coefficients are extracted as coefficient matrix $\mathbf{L}(\bar{\bar{\mathbf{\mu}}})$, i.e. $\mathbf{c}(\hat{\mathbf{\mu}}) = \mathbf{0}$ in Eq. (5), and get:

$$\nabla R(\hat{\mathbf{\mu}}) = \mathbf{L}(\bar{\bar{\mathbf{\mu}}}) \hat{\mathbf{\mu}} \quad (14)$$

Apply Eq. (14) to two penalty terms, Total Variance (TV) and q-Generalized Gaussian Markov Random Field (qGGMRF). These two penalties are representative because TV is an approximate function of linear function, and qGGMRF is a function between the linear and quadratic functions.

### 2.3.1 TV penalty

For reconstruction image $\hat{\mathbf{\mu}} \in \mathbb{R}^{N^2 \times 1}$, TV function is defined as:

$$R(\hat{\mathbf{\mu}}) = \sum_i \varphi(\hat{\mu}_i)$$
$$\varphi(\hat{\mu}_i) = \sqrt{(\hat{\mu}_i - \hat{\mu}_{i-N})^2 + (\hat{\mu}_i - \hat{\mu}_{i-1})^2 + \varepsilon} \quad (15)$$

where small quantity $\varepsilon$ is added to avoid zero value of the gradient of $\varphi(\mu_i)$. The gradient of Eq. (15) is:

$$\frac{\partial R(\hat{\mathbf{\mu}})}{\partial \hat{\mu}_i} = \left[ \frac{2}{\varphi(\hat{\mu}_i)} + \frac{1}{\varphi(\hat{\mu}_{i+N})} + \frac{1}{\varphi(\hat{\mu}_{i+1})} \right] \hat{\mu}_i$$
$$- \frac{\hat{\mu}_{i-N} + \hat{\mu}_{i-1}}{\varphi(\hat{\mu}_i)} - \frac{\hat{\mu}_{i+N}}{\varphi(\hat{\mu}_{i+N})} - \frac{\hat{\mu}_{i+1}}{\varphi(\hat{\mu}_{i+1})} \quad (15)$$

Thus, the $i^{th}$ row of $\mathbf{L}_{\text{TV}}(\bar{\bar{\mathbf{\mu}}})$ is:

$$\mathbf{L}_{ii}^{\text{TV}} = \frac{2}{\varphi(\bar{\bar{\mu}}_i)} + \frac{1}{\varphi(\bar{\bar{\mu}}_{i+N})} + \frac{1}{\varphi(\bar{\bar{\mu}}_{i+1})}$$
$$\mathbf{L}_{i,i-N}^{\text{TV}} = -\frac{1}{\varphi(\bar{\bar{\mu}}_i)}, \, \mathbf{L}_{i,i-1}^{\text{TV}} = -\frac{1}{\varphi(\bar{\bar{\mu}}_i)} \quad (16)$$
$$\mathbf{L}_{i,i+N}^{\text{TV}} = -\frac{1}{\varphi(\bar{\bar{\mu}}_{i+N})}, \, \mathbf{L}_{i,i+1}^{\text{TV}} = -\frac{1}{\varphi(\bar{\bar{\mu}}_{i+1})}$$

Consequently, the covariance matrix for TV constrained WLS (WLS-TV) method can be computed as in Eq. (7).

### 2.3.2 qGGMRF penalty

For reconstruction image $\hat{\mathbf{\mu}} \in \mathbb{R}^{N^2 \times 1}$, qGGMRF function is defined as:

$$R_q(\hat{\mathbf{\mu}}) = \sum_i \sum_{j \in N_i} \omega_{ij} \varphi(\hat{\mu}_i - \hat{\mu}_j)$$
$$\varphi(\Delta) = \frac{|\Delta|^p}{1 + \left| \frac{\Delta}{\delta} \right|^{p-q}}, \, (1 \leq q \leq p \leq 2) \quad (17)$$

where $\omega$ is a distance dependent weight factor, and $\delta$ is a threshold to adjust the degree of denoising. The $1^{st}$ derivative of Eq. (17) is:

$$\frac{\partial R_q(\hat{\boldsymbol{\mu}})}{\partial \hat{\mu}_i} = \sum_{j \in N_i} \omega_{ij} \frac{\partial}{\partial \hat{\mu}_i} \varphi(\hat{\mu}_i - \hat{\mu}_j)$$

$$\nabla \varphi(\Delta) = l(\Delta) \Delta \tag{18}$$

$$= \frac{|\Delta|^{p-2}}{1 + \left|\frac{\Delta}{\delta}\right|^{p-q}} \left( p - \frac{(p-q)}{\delta^{p-q}} \frac{|\Delta|^{p-q}}{1 + \left|\frac{\Delta}{\delta}\right|^{p-q}} \right) \Delta$$

Therefore, the $i^{th}$ row of $\mathbf{L}_q(\overline{\boldsymbol{\mu}})$ is:

$$\mathbf{L}_{ii}^q = \sum_{j \in N_i} \omega_{ij} l(\overline{\mu}_i - \overline{\mu}_j)$$

$$\mathbf{L}_{ij}^q = -\omega_{ij} l(\overline{\mu}_i - \overline{\mu}_j), \ j \in N_i \tag{19}$$

Smilarly, the covariance matrix for qGGMRF constrained WLS (WLS-qGGMRF) method can be calculated according to Eq. (7).

## 3 Experimental Results

A $64 \times 64$ phantom is simulated to validate the predicted covariance of WLS method, and another larger phantom of size $128 \times 128$ is simulated to validate the covariance estimations of TV- and qGGMRF-WLS methods. Incident photons $I_0$ are set to be $10^5$, $10^4$, and $10^3$ to model different noise level.

Since the case of quadratic penalty term has already been studied by [7], only the results of WLS and TV/qGGMRF methods, are presented and discussed in this section. Only one noisy sinogram $\mathbf{g}$ is used to estimate the covariance of the sinogram $\text{Cov}(\mathbf{g}) = \exp(-\mathbf{g})/I_0$. In addition, Eq. (8) and (9) are calculated to get ill-conditioned and regularized covariance results of WLS. Meanwhile, for TV/qGGMRF reconstruction, covariance matrices are predicted by TAM as well as LAM covariance estimation method. Besides, statistical estimation of covariance matrix $\text{Cov}_{\text{sta}}(\hat{\boldsymbol{\mu}})$ is calculated as reference for validation:

$$\text{Cov}_{\text{sta}}(\hat{\boldsymbol{\mu}}) = \frac{1}{M-1} \sum_{m,n=1}^{M} (\hat{\boldsymbol{\mu}}_m - \overline{\overline{\boldsymbol{\mu}}})(\hat{\boldsymbol{\mu}}_n - \overline{\overline{\boldsymbol{\mu}}}) \tag{20}$$

where $M$ is the total number of noise realizations. We use mean absolute percent error (MAPE) as the metric for comparison.

### 3.1 Covariance estimation for WLS reconstruction

In this experiment, the condition number of the matrix $\mathbf{H}^T \mathbf{W} \mathbf{H}$ is $10^6$. Regularized by the identity matrix, the condition number of the matrix $\mathbf{H}^T \mathbf{W} \mathbf{H} + \beta \mathbf{I}$ reduced by two order of magnitude to $10^4$ so that the stability of the image covariance is significantly improved.

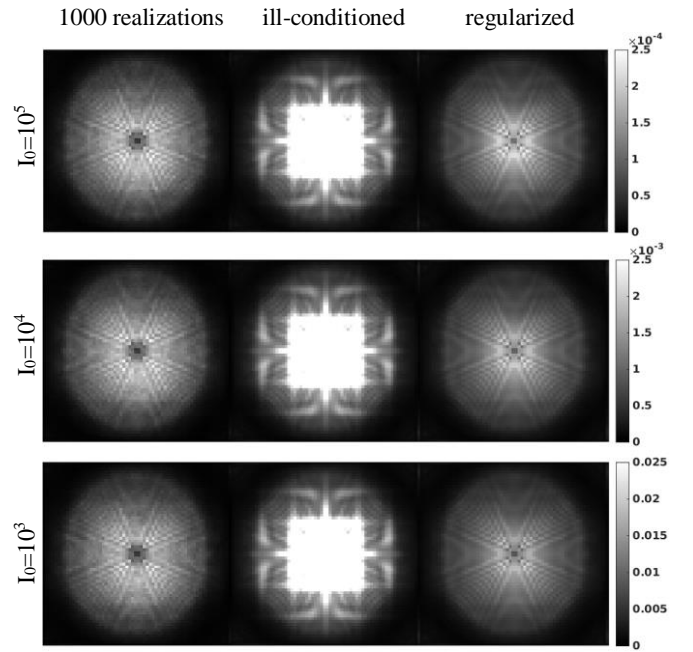

Fig.1 Variance images of WLS.

Table 1 MAPEs of variance estimation for WLS.

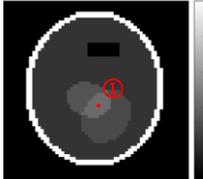

| MAPE | ill-conditioned | regularized | Phantom |
|---|---|---|---|
| $I_0=10^5$ | 137.94% | 25.34% | |
| $I_0=10^4$ | 138.08% | 26.07% | |
| $I_0=10^3$ | 149.25% | 26.80% | |

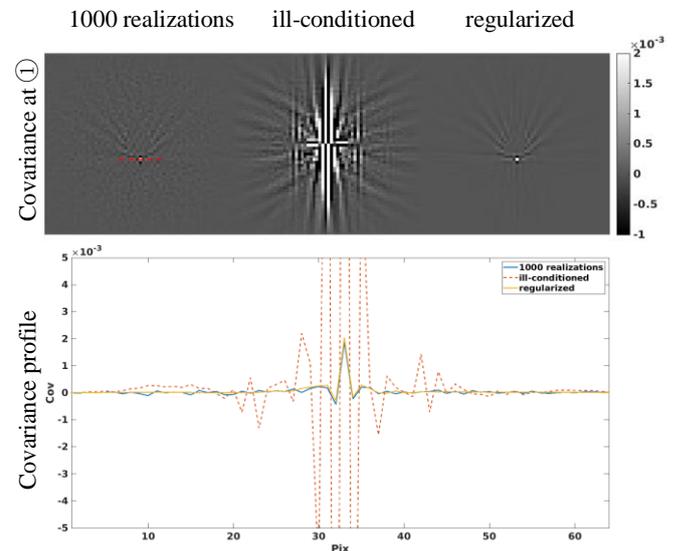

Fig.2 WLS covariance images at point ① when $I_0=10^4$. The profile of the red dashed line in covariance image is plotted.

Corresponding variance image estimations are displayed in Fig.1. Compared with variance images statistically computed from 1000 noise realizations, the effect caused by the ill-conditioned matrix is clearly shown. Once the ill-conditioned matrix regularized with identity matrix, the effect is significantly suppressed. As photons decrease, the conclusion still holds. Identity matrix Regularization helps improve the accuracy and stableness. The MAPE of image variance decreases about 80% from ill-conditioned one to regularized one, and results close to reference is obtained.

### 3.2 Covariance estimation for WLS-TV reconstruction

We set $\alpha$ of TV to be 0.008 when $I_0 = 10^5$, 0.02 when $I_0 = 10^4$ and 0.12 when $I_0 = 10^3$ respectively. 200 noisy image samples are used to calculate the reference covariance.

Table 2 MAPEs of variance estimation for WLS-TV.

| MAPE | TAM | LAM | Phantom |
|---|---|---|---|
| $I_0=10^5$ | 278.87% | 24.14% | |
| $I_0=10^4$ | 279.45% | 33.97% | |
| $I_0=10^3$ | 1610.45% | 29.24% | |

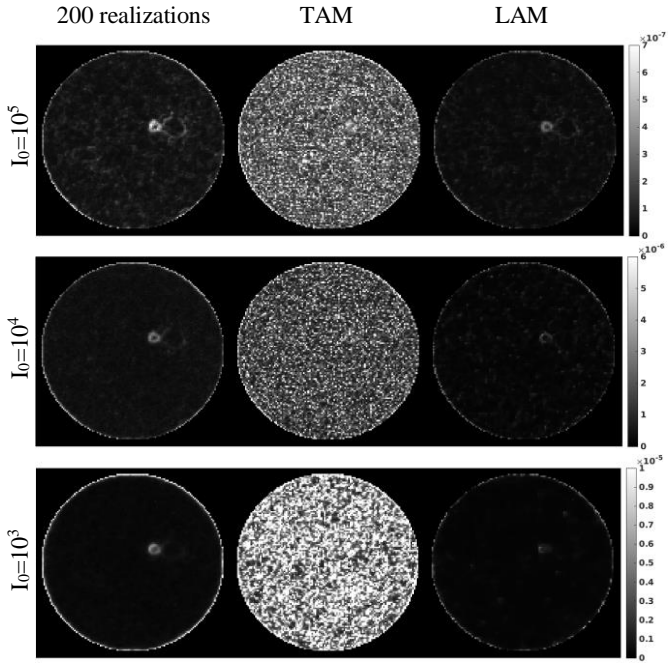

Fig.3 Variance images of WLS-TV.

The variance image results are displayed in Fig.3. Obviously, the LAM outperforms the TAM. Especially in smooth regions, the TAM results in much higher variance compared with the reference. And the LAM keeps a similar structure as the reference but with lower values.

The quantitative analysis results are listed in Table. 2. In case of the $10^5$ incident photons, the MAPE of Taylor is about 280%, which is almost 10 times that of linearity. When incident photons is $10^4$, the MAPE drops about 88% from Taylor to linearity. With even lower incident photons, the MAPE of TAM gets even higher to around 1000%, while the MAPE of linearity still around 30%. At all noise levels, the linearity based method has better performance than the TAM.

Covariance images of two points (shown in the phantom image in Table. 2) are displayed in Fig. 4, one with relatively high attenuation (marked by red ①) while the other with relatively low attenuation (marked by red ②). For the high attenuating point, although the central value (variance) of the LAM covariance map is of larger error than TAM compared with the reference, the total MAPE of LAM is about 40% lower than that of TAM. The absolute covariance values of the low attenuating point obtained by TAM are extremely large, which is around 10 times the reference values on average. The covariance map of Taylor

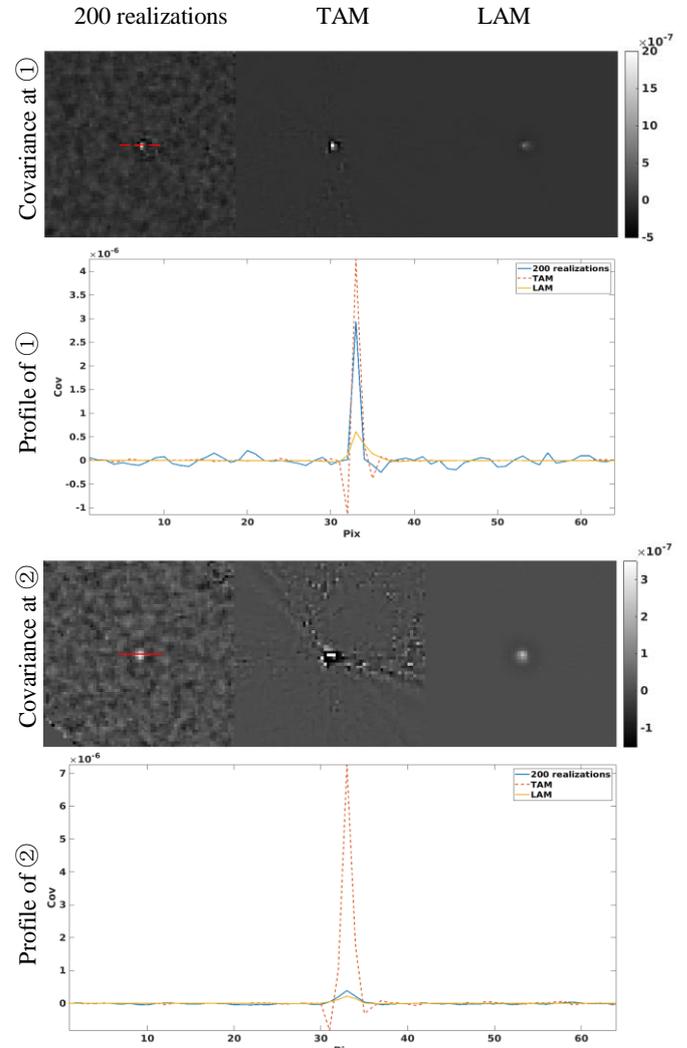

Fig. 4 covariance images when $I_0=10^4$ for TV penalty case. Profiles of the red dashed line in covariance image ① and the red solid line in covariance image ② are plotted.

is distorted, and the covariance map of LAM is more accurate.

### 3.3 Covariance estimation for WLS-qGGMRF reconstruction

A common parameter setting of qGGMRF with p=2, q=1.2 and δ=0.002 is studied. The penalty parameter $\alpha$ is set to be 3 when $I_0 = 10^5$, 9 when $I_0 = 10^4$ and 35 when $I_0 = 10^3$ respectively. Again, we use the statistical covariance of 200 noisy reconstructed images as the reference.

The corresponding variance images are displayed in Fig.5. For the situation of $10^5$ incident photons, both TAM and LAM covariance estimation methods show relatively high accuracy of covariance. Variance maps of both methods have similar structure as the variance map of reference. However, TAM gives more accurate variance map. As the number of incident photons reduces to $10^4$, TAM as well as LAM has similar performance, the MAPEs of both methods are about 10%. The difference is that the variance of TAM is slightly higher than the variance of reference, while the variance of LAM is slightly lower. When incident photons set to $10^3$, the LAM performs better, since the variance of Taylor based method is slightly higher than the variance of reference. Linearity gets more accurate variance with about 8% MAPE.

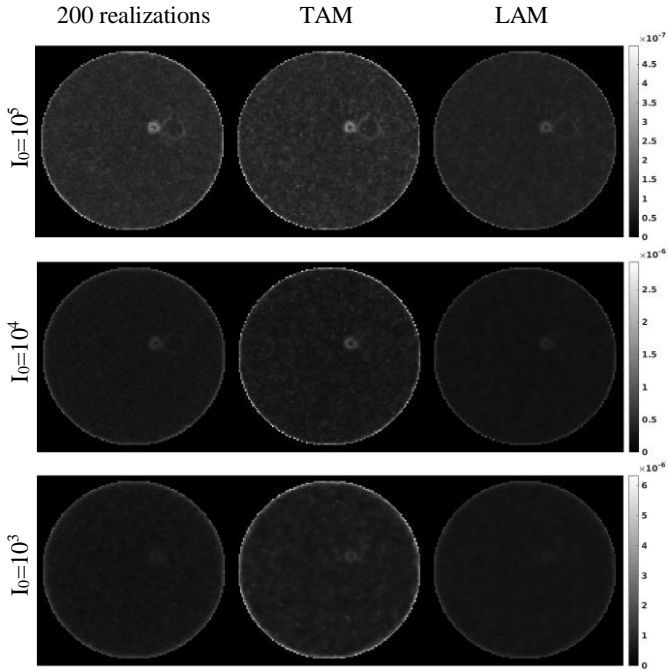

Fig.5 Variance images of WLS-qGGMRF reconstruction.

As shown in Table.3, covariance maps of two points ① and ② are presented. For both point ① and ②, TAM results in slightly higher covariance values compared with reference, while LAM results in slightly lower values, which indicates that both methods perfoms similarily when $I_0 = 10^4$.

## 4 Conclusion

We propose a method to estimate the covariance of reconstructions analytically from sinogram covariacne, which avoid using a large amount of noisy image samples for covariance estimation. Three cases of penalized iterative methods are studied. For the case without penalty term, the ill-codition problem is solved by Tikhonov-Phillips method to give a stable estimation of covariance. For the case with non-quadratic penalty terms, the typical case of a TV and qGGMRF constraints are studied. The gradient of TV and qGGMRF can be rewritten as a linear function despite that the linear coefficient matrices $\mathbf{L}_{TV}(\bar{\hat{\boldsymbol{\mu}}})$ and $\mathbf{L}_q(\bar{\hat{\boldsymbol{\mu}}})$ are $\bar{\hat{\boldsymbol{\mu}}}$ dependent. The linear coefficient matrix is applied to the covariance estimation. For WLS-TV reconstruction, the proposed LAM method is more accurate compared with

TAM. For WLS-qGGMRF reconstruction, it is of higher accuracy at high noise level compared with TAM, and of relatively low accuracy at low noise level.

However, the mean absolute percent error of image variance of WLS-TV calculated by the proposed method is still over 20% compared with the statistical estimation. Besides, the covariance is hard to estimate when image size is large because of the computational complexity. Our future work is to further improve covariance accuracy and efficiency for large dimention images.

Table 3 MAPEs of variance estimation for WLS-qGGMRF.

| MAPE | TAM | LAM | Phantom |
|---|---|---|---|
| $I_0=10^5$ | 9.32% | 14.74% | |
| $I_0=10^4$ | 13.67% | 12.10% | |
| $I_0=10^3$ | 23.63% | 7.85% | |

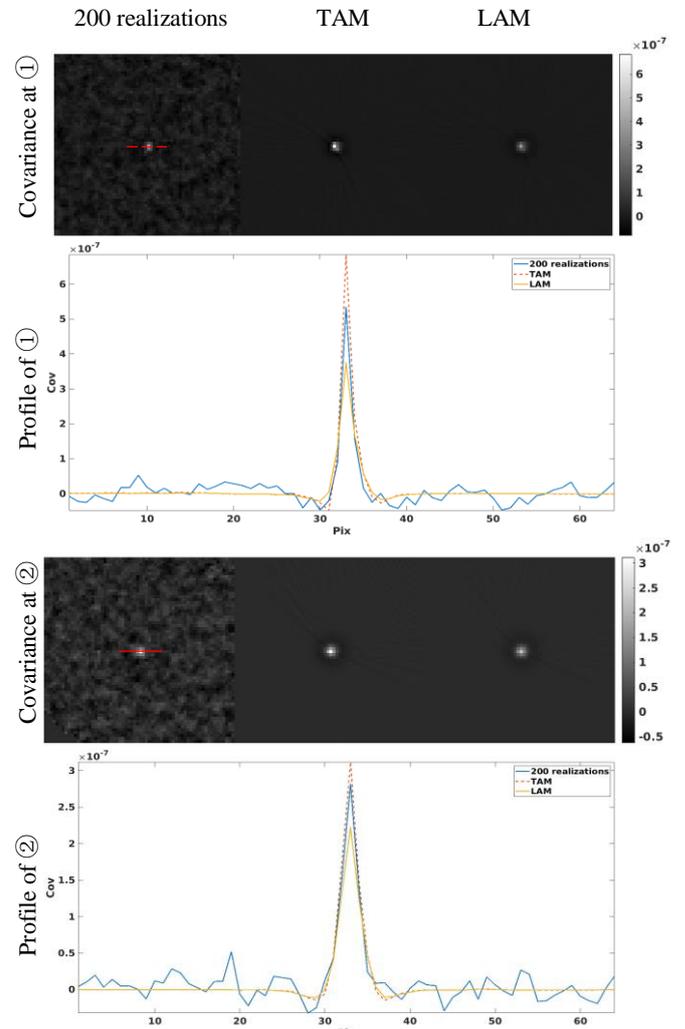

Fig.6 covariance images when $I_0=10^4$ for qGGMRF penalty case. Profiles of the red dashed line in covariance image ① and the red solid line in covariance image ② are plotted.

## Acknowledgement

This work is supportted by National Natural Science Foundation of China (Grant No. 62031020 and 61771279).

# Investigation of Subset Methodologies Applied to Penalised Iterative PET Reconstruction

Robert Twyman[1], Simon Arridge[2], and Kris Thielemans[1]

[1]Institute of Nuclear Medicine, University College London, London, UK
[2]Department of Computer Science, University College London, London, UK

**Abstract** Subset PET image reconstruction algorithms accelerate reconstruction during early iterations. However, at later iterations, many subset algorithms exhibit limit cycle behaviour leading to undesirable variations between subsequent images, resulting in non-convergence in the absence of step size relaxation. A class of variance reduction algorithms address this issue by incorporating previous subset gradients into the update direction computation. This generates an update direction that is a better approximation of the full objective function direction than standard subset algorithms while maintaining similar low computational cost. In this work, the impact on reconstruction performance when using a deterministic ordered subset method and two stochastic subset methods is investigated. These subset selection methods are applied to a preconditioned gradient ascent algorithm and three variance reduction algorithms. The ordered subset methodology resulted in superior performance for both subset gradient ascent and two of the variance reduction algorithms during early passes through the data. Yet, at later iterations, the stochastic subset variance reduction algorithm reconstructions converged closer to the solution.

## 1 Introduction

Iterative algorithms are commonly used in Positron Emission Tomography (PET) image reconstruction. A selection of these algorithms iteratively improve a discretised estimated PET tracer distribution by adding a preconditioned gradient of the objective function to the current estimate. These are known as gradient-based optimisation algorithms. The computation of the gradient requires forward and backward projections between measured data and image spaces. For large scale PET reconstruction problems, these projection operations are computationally demanding [1].

Projecting the estimated distribution into only a subset of the measured data significantly reduces the computational demand. A sub-class of gradient-based algorithms utilise these subsets during optimisation of the objective function by cycling through each of the unique and distinct subsets to compute an approximation of the full gradient to be applied at each update. These ordered subset (or 'block sequential') algorithms in PET, e.g. Ordered Subset Expectation Maximisation (OSEM) and Block Sequential Regularised Expectation Maximisation (BSREM), utilise this methodology with $M$ subsets to realise accelerated convergence rates during early updates by applying a single subset at each iteration [2, 3]. However, at later iterations, as the estimate approaches the (unique) solution, discrepancies between subset gradients due to variations in noise and geometric projection sensitivities may be observed [3]. This results in behaviour commonly known as the limit cycle and it prevents algorithm convergence. While linear acceleration may be realised initially with respect to the number of subsets, image quantification and lesion detectability may be inhibited when larger numbers of subsets are used [4]. Authors have attempted to address the non-convergence of subset algorithms by utilising step size relaxation or a Complete-data OSEM (COSEM) algorithm [3, 5]. An issue with these methods is the impact of their hyper-parameters on convergence rates.

Stochastic optimisation methods are commonly used in a number of fields, such as the training of deep learning models, but it remains largely unexplored in tomography. The simplest algorithm is Subset Gradient Ascent (SGA), which uses random sample selection together with step-size relaxation to accelerate convergence [6]. A contemporary class of stochastic first order optimisation algorithms, known as stochastic Variance Reduction Methods (VRM), aim to reduce the impact of variations between subsets by applying a step that incorporates the current subset gradient as well as previously computed subset gradients [7–9]. In our previous work, we applied three stochastic variance reduction algorithms, Stochastic Average Gradient (SAG), SAGA and Stochastic Variance Reduced Gradient (SVRG), to the PET reconstruction problem in a preconditioned form [10]. This preliminary investigation constructed subsets using the same methodology as the BSREM algorithm but selected a random subset index at each algorithm iteration. The stochastic algorithms allowed for the use of a larger number of subsets than the comparison BSREM algorithm. They performed similarly to BSREM during early passes through the data and converged to the Maximum A Posteriori (MAP) solution within numerical tolerance.

In this work, we expand on this previous study by investigating a set of alternative subset sampling methodologies for a Preconditioned SGA (PSGA) algorithm and for the three previously investigated preconditioned VRMs [10]. The aim of this study is to accelerate image reconstruction, reduce the variations between sequential image estimates, and allow the optimisation algorithms to converge closer to the solution.

## 2 Methodology

In this study, we investigated three subset construction and sequence methodologies for subset optimisation. A standard subset construction method in tomographic image reconstruction involves the construction of $M$ subsets that are unique and complete [2]. Equally spaced rows of a sinogram are binned into a subset and subsequent subsets are constructed similarly but are offset from one another by their respective subset index $m$. The aim of this construction method is to balance the subsets, i.e. to minimise the variance in the detection probability of each image voxel between different subsets [1, 3]. The first two investigated subset sequence methodologies in this work use this subset construction method.

The first methodology, known as Ordered Subset (OS), is intuitively described by Herman et al. [11]. This methodology attempts to apply a subset with a direction that is as orthogonal as possible to the space generated by recently used subsets. A deterministic and cyclical sequence of subset indices is generated and it applies each subset to the algorithm before the sequence returns to the beginning. This sequence methodology is commonly used in the BSREM algorithm [3].

The second subset sequence methodology, denoted as stochastic subsets, uses the same construction method as described above. However, at each iteration of the reconstruction algorithm, a subset index is selected at random from a uniform probability distribution [10].

The final subset methodology, designated as randomised batches, does not utilise the aforementioned subset construction method. Instead, a number of projection angles are selected at random (without replacement) and applied to the reconstruction at each iteration of the algorithm [10].

To evaluate the subset methodologies, an XCAT torso phantom (Figure 1a and 1b) scan was simulated in a scanner with 280 projection angles and two rings using STIR [12, 13]. Poisson noise was added and scattered events were simulated in the measured data. The respective data corrections were included within the forward model and, to further suppress noise in the images while encouraging edges, the Relative Difference Prior (RDP) was used to penalise the objective function [14]. Image reconstructions were performed using STIR-python and the penalty weighting factor was tuned visually for this data set.

To assess the performance of the subset methodologies, global image reconstruction performance was measured as a distance from convergence using the following metric:

$$\Delta_k(x_k, \hat{x}) = \frac{\|x_k - \hat{x}\|_2}{\|\hat{x}\|_2} \cdot 100\%, \qquad (1)$$

where $x_k$ is the image at the $k$th iteration of the algorithm, $\|\cdot\|_2$ is the $\ell_2$-norm, and $\hat{x}$ is the unique solution, which is computed using the Limited memory Broyden Fletcher Goldfarb Shanno Bounded Pre-Conditioned (L-BFGS-B-PC) reconstruction algorithm [15] and shown in Figure 1c.

The previously mentioned subset methods were applied to both a PSGA algorithm and three VRMs. The preconditioned update equation is given by:

$$x_{k+1} = P_+ \left[ x_k + \alpha_k D \tilde{\nabla}_{k,m} \right], \qquad (2)$$

where $P_+[\cdot]$ is a non-negativity projection operation, $\alpha_k = 1$ is a scalar step size that is fixed for this work, $D$ is a diagonal preconditioner, and $\tilde{\nabla}_{k,m}$ may be interpreted as an objective function gradient approximation. The preconditioner used in this work is given by

$$D = \frac{x_{\text{init}} + \delta}{A^T 1}, \qquad (3)$$

where $\delta$ is a small positive constant that allows voxels with zero value to be updated, and $A^T 1$ is the backprojection of a uniform sinogram. This preconditioner is inspired by the Expectation Maximisation (EM) preconditioner used in BSREM. We used $x_{\text{init}} = x_{\text{OSEM}}$, which is the resulting image after 1 epoch of OSEM with 20 subsets.

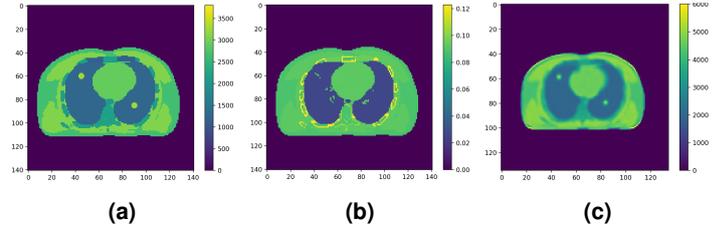

**Figure 1:** Transaxial slices of the: (a) simulated source distribution of the XCAT volume with two inserted lung lesions, (b) simulated attenuation map of the XCAT volume, (c) converged estimate that is computed using L-BFGS-B-PC [15].

For PSGA, the approximate objective function gradient is given as the $m$th subset objective function gradient, i.e. $\tilde{\nabla}_{k,m} := \nabla\Phi_m(x_k)$. The VRM gradient approximations are given by:

$$\tilde{\nabla}_k^{SAG} := \frac{\nabla\Phi_m(x_k) - g_m}{M} + \frac{1}{M}\sum_{n=1}^{M} g_n, \qquad (4a)$$

$$\tilde{\nabla}_k^{SAGA} := \nabla\Phi_m(x_k) - g_m + \frac{1}{M}\sum_{n=1}^{M} g_n, \qquad (4b)$$

$$\tilde{\nabla}_k^{SVRG} := \nabla\Phi_m(x_k) - g_m + \mu, \qquad (4c)$$

where $g_m$ are the previously computed subset gradients (stored in memory) [7–9]. Both SAG and SAGA update these variables after each iteration with the latest computed $m$th subset gradient. However, SVRG recomputes each $g_m$ from the current estimate every three epochs and resets $\mu = \frac{1}{M}\sum_{n=1}^{M} g_n$. It should be noted that SAG is a stochastic modification of the deterministic cyclical Incremental Aggregated Gradient (IAG) algorithm [7, 16].

Due to the stochastic nature of two of the subset selection methods investigated, we define an *epoch* as equivalent to the computational cost of computing a full objective function gradient in terms of projection operations. Furthermore, as the VRMs may be sensitive to initial conditions, they are initialised from $x_{\text{OSEM}}$ [6, 10].

## 3 Results and Discussion

In Figure 2a, the OS method exhibited slightly improved performance compared to the stochastic subsets method. However, this was only realised after approximately 25 epochs

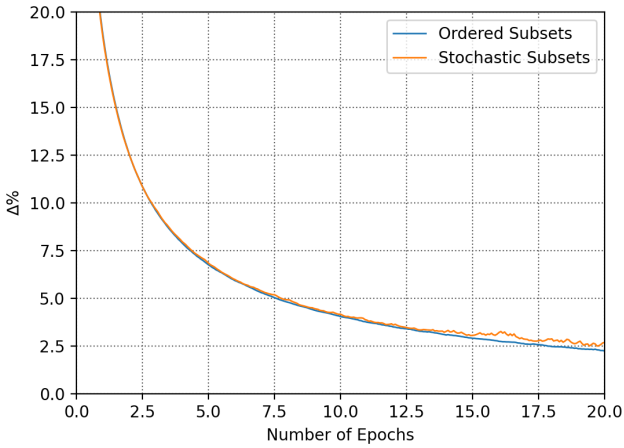

**(a)** 14 Subsets

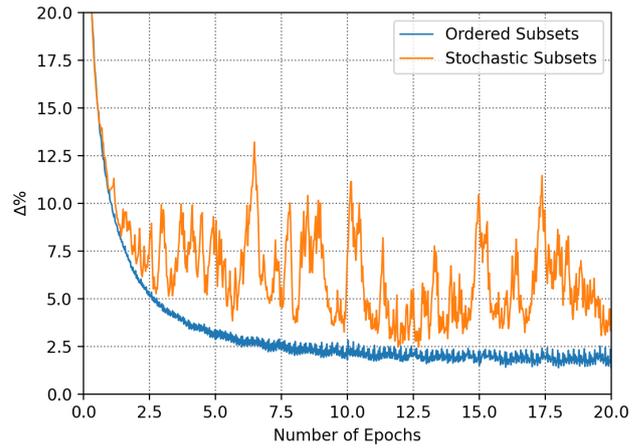

**(b)** 40 Subsets

**Figure 2:** Global convergence performance plotted using two subset methodologies. The Stochastic Batches plots are not included.

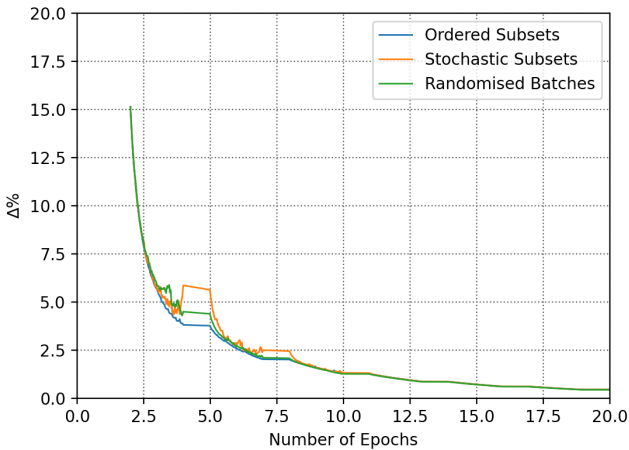

**Figure 3:** Global convergence performance plotted throughout three 70 subset SVRG reconstructions using different subset selection methodologies.

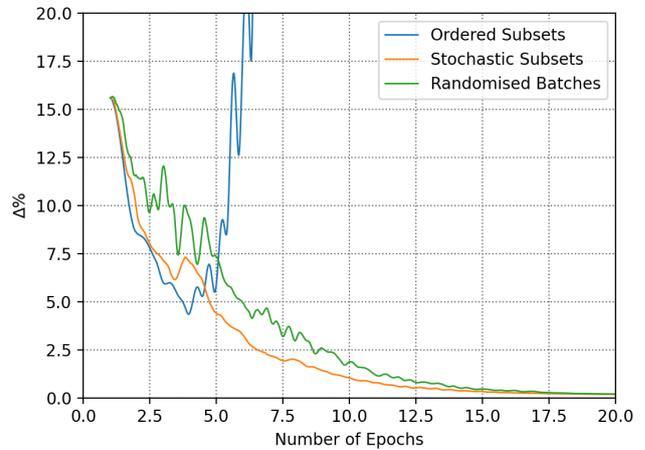

**Figure 4:** Global convergence performance plotted throughout three 70 subset SAG reconstructions using different subset selection methodologies.

where the impact of the randomly selected subsets began to affect the reconstruction and the Δ% values began to significantly vary.

Increasing the number of subsets, Figure 2b, improved the convergence rate during the first few epochs for both methods. Yet, the stochastic subsets method demonstrated large fluctuations in Δ% values and performed poorer on average than the OS method. The randomised batches reconstructions are not included in these figures as the Δ% metric evaluations resulted in significantly poorer performance that is not competitive with either of the presented subset methods.

In Figure 3, the variations of SVRG are shown and all three methods appear to perform well. However, the OS subset sequence method outperformed both of the stochastic methods during early iterations. At later iterations, all three methods performed equally as the reconstructions approached convergence. This indicates that the SVRG algorithm reconstructions may be improved with the OS sampling methodology. The Δ% metric performance of SAG, Figure 4, indicates that the stochastic subsets method optimised the reconstruc-tion problem at a faster rate than random batches and with with fewer fluctuations. During the first four epochs the OS method outperformed both of the other methods. However, the algorithm appears to demonstrate divergent behaviour as the Δ% values increased rapidly. Voxel values in the image oscillated from zero to significantly larger than those of the converged image and similar behaviour was observed when fewer subsets were used (results not shown).

For the SAGA variant reconstructions, Figure 5, the OS method outperformed the stochastic methods, which exhibited large fluctuations during the first 6-7 epochs. The stochastic subsets method's reconstruction performance became comparable at this point and then outperformed the OS method at later iterations. The randomised batches method demonstrated a similar trend in behaviour but with larger variations during early iterations and reduced fluctuations at later iterations.

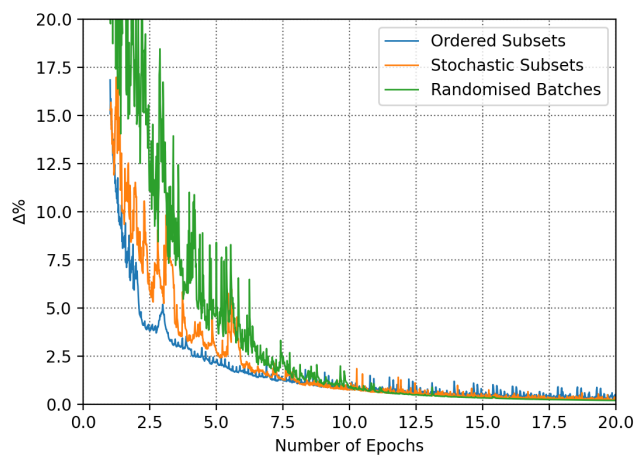

**Figure 5:** Global convergence performance plotted throughout three 70 subset SAGA reconstructions using different subset selection methodologies.

## 4    Discussion

The OS method outperformed the stochastic methods when applied to the PSGA algorithm, particularly when a greater number of subsets was utilised, but did not converge to the MAP solution. However, all VRM reconstructions were able to converge to the MAP solution, except for the OS SAG reconstruction, an algorithm comparable to IAG. Additionally, the VRMs allowed for the use of a greater number of subsets, resulting in fewer epochs required to reach a certain value of $\Delta\%$, an observation consistent with our earlier work [10].

The OS method applied to the SVRG and SAGA algorithms lead to faster convergence than the stochastic subsets during early iterations. As the algorithms approached convergence, the stochastic methods exhibited either comparable or superior performance. Therefore, the OS method may be utilised during the first few epochs before a heuristic switch to a stochastic subset selection methodology is made. However, for the SAG algorithm, the OS method is not recommended, based on our preliminary results. Furthermore, the randomised batches method was consistently outperformed by the stochastic subsets method and therefore, deliberate construction of subset structure is important for improved algorithm performance.

This preliminary study has many limitations, including the use of a single object and noise level. It should be noted that repeated stochastic reconstructions may lead to different results. Yet, we found that the stochastic results and conclusions presented represent the general behaviour observed over other reconstructions of this data set. Another limitation is the use of a fixed step-size. PSGA has been shown to converge when using a suitable relaxation scheme. Relaxation will decrease the fluctuations, although it may decrease the overall convergence rate. Use of too large a step-size might also explain the divergent behaviour observed for OS SAG. Despite this limitation, it is encouraging that convergence was observed for the two other OS VRMs.

In future work, we aim to address these limitations and in-vestigate the relationships between the presented variance reduction, IAG and COSEM algorithms [5, 16].

## 5    Conclusion

While the variance reduction methods were developed for stochastic optimisation, our preliminary results indicate their promise for adapting to ordered subset methods. A careful selection of subset structure and/or sequence is advantageous at initial iterations, while stochastic selection helps convergence at later stages.

## Acknowledgements

This research is supported by: GE Healthcare, the NIHR UCLH Biomedical Research Centre, the EPSRC Centre for Doctoral Training i4Health (EP/S021930/1) and the EPSRC-funded UCL CDT in Medical Imaging (EP/L016478/1).

# On Stochastic Expectation Maximisation for PET

Zeljko Kereta[1], Robert Twyman[2], Simon Arridge[1], Kris Thielemans[2], and Bangti Jin[1]

[1]Computer Science Department, University College London, London, UK
[2]Institute of Nuclear Medicine, University College London, London, UK

**Abstract** Ordered subset variants of statistical iterative reconstruction algorithms for PET can improve the performance in early iterations and thus are popular. However, they suffer from convergence issues, e.g., entering limit cycles. This work considers a stochastic variant of the maximum likelihood expectation maximisation. We adapt the algorithm to PET MAP reconstruction, and for a non spatially separable prior, we combine it with the separable surrogate approach to facilitate the computation of the M-step. Preliminary numerical results indicate that the method is competitive with traditional approaches and enjoys excellent convergence behaviour.

## 1 Introduction

Iterative reconstruction methods have been used to solve the positron emission tomography (PET) problem since the 70s. A statistical reformulation of the problem by Shepp and Vardi allows to compute the tracer distribution through the maximum likelihood (ML) estimate [1]. The resulting PET expectation maximisation (EM) algorithm has been widely used due to its simple form and desirable properties, e.g., nonnegativity preservation, and consists of two steps: the E-step and M-step. The former computes the complete data sufficient statistic, given the current distribution estimate, and the latter updates the estimate by maximising the complete-data log-likelihood. Due to the large data sizes associated with modern scanners, the full batch computation of the algorithm, i.e., using all of the measurement data to compute the sufficient statistic, is often infeasible or inefficient. To improve convergence rates, one established procedure is the ordered subset EM method (OSEM), whereby, at each iteration, only a subset (mini-batch) of the given data is used [2]. The ordered subsets strategy greatly reduces the per-update computational cost, and it has been observed to provide significant acceleration in initial iterations.

However, the acceleration comes at a cost since most ordered subset algorithms do not converge to a maximising solution but rather enter a limit cycle [3]. This has led to the development of several variants of OSEM that maintain the speedup in early iterations but also guarantee convergence to the ML solution by suitably adjusting the step-size schedule [3–6]. Another difficulty for EM methods is the incorporation of a prior distribution, often employed for combating the ill-posedness of the reconstruction problem, which can be cumbersome for the M-step, making it not analytically solvable for most spatially non-separable priors [7, p. R561]. Additional approximations have to be employed, which tend to further slow down the convergence.

In this paper we address these issues by adapting the Online-EM algorithm [8] to maximum *a posteriori* (MAP) PET reconstruction. Online-EM was developed for conducting EM on exponential latent models in the setting of streaming data. Since not all the data is available at the start, Online-EM instead aims to approximate the conditional statistic by exponentially moving averages as the data streams in. This reduces the computational complexity of the E-step for exponential latent models (compared to standard MLEM) since the full conditional expectation is never computed.

In this work we develop a stochastic EM algorithm by adapting the online EM algorithm to PET image reconstruction. The proposed algorithm utilizes ordered subsets of the measurements as in OSEM, and an exponentially moving average as in Online-EM. To handle priors, we employ separable parabolic surrogates [9] to facilitate an explicit solution of the M-step for a range of popular priors. Thus, the resulting algorithm is mathematically principled and easy to implement.

We study the performance of our method, and compare it to BSREM, using a relaxed step-size regime that ensures it converges [3], on reconstruction of a simulated PET scan of a torso, using the STIR library [10]. The numerical results indicate that the proposed algorithm enjoys steady convergence and is competitive with existing approaches.

## 2 Stochastic Expectation Maximisation

EM algorithms are designed for probabilistic models with observed quantities, but which depend on latent, unobserved quantities. Let $\mathbf{f} = (f_1, \ldots, f_N)^\top$ be the discretisation of the tracer distribution organised into $N$ discretisation boxes, and $\mathbf{g} = (g_1, \ldots, g_M)^\top$ be the measured data collected at $M$ detector bins. Statistical formulations of PET model emission measurements in the $i^{\text{th}}$ detector bin using the Poisson model, $g_i \sim \text{Poisson}(\mathbb{E}[g_i])$. In this paper we use the complete data framework for PET proposed by Shepp and Vardi [1]. We denote the complete data by $\mathbf{G}$, with entries $g_{ij}$ denoting the number of emissions detected in bin $j$ that originated from voxel site $i$, which satisfy $\mathbb{E}[g_{ij}|\mathbf{f}] = a_{ij} f_j$ and $\mathbb{E}[g_i] = \sum_j \mathbb{E}[g_{ij}]$. The complete data likelihood can then be written as

$$p(\mathbf{G}|\mathbf{f}) = C_{AG} \exp(\boldsymbol{\eta}(\mathbf{f})^\top \mathbf{T}(\mathbf{G}) - \mathcal{U}(\mathbf{f})), \qquad (1)$$

where $C_{AG}$ is a constant independent of $\mathbf{f}$, and

$$\mathcal{U}(\mathbf{f}) = \sum_{i=1}^{M} \mathbf{a}_i^\top \mathbf{f}, \qquad \boldsymbol{\eta}(\mathbf{f}) = \left(\log(f_j)\right)_{j=1}^{N},$$

$$\mathbf{T}(\mathbf{G}) = \left(\sum_{i=1}^{M} g_{ij}\right)_{j=1}^{N}. \qquad (2)$$

## 2.1 Maximum Likelihood

Maximising the likelihood leads to MLEM iterates

$$\mathbf{f}^{(k+1)} = \underset{\mathbf{f} \geq 0}{\operatorname{argmax}} \, \boldsymbol{\eta}(\mathbf{f})^\top \mathcal{T}(\mathbf{f}^{(k)}) - \mathcal{U}(\mathbf{f}),$$

where $\mathcal{T}(\mathbf{f}^{(k)}) = \mathbb{E}_{\mathbf{G}|\mathbf{g},\mathbf{f}^{(k)}}[\mathbf{T}(\mathbf{G})]$. Instead of all the measured data, OSEM uses only subsets of the measurements for computing conditional expectations.

Online-EM [8] is an EM type algorithm, developed for computing on-line estimates of the (full) conditional expectation $\mathcal{T}(\mathbf{f}^{(k)})$ from data coming in a stream, that continuously updates approximations $\hat{\mathbf{s}}^{(k)}$ of the (full) conditional expectation $\mathcal{T}(\mathbf{f}^{(k)})$ using an exponentially moving average. Consider an OS decomposition with $N_s$ subsets of the form $\mathcal{T}(\mathbf{x}) = \frac{1}{N_s}\sum_{r=1}^{N_s} \tau_r(\mathbf{x})$. In PET reconstruction, for a subset $S_r$, we have

$$\tau_r(\mathbf{x}) = N_s \mathbf{x} \odot \mathbf{A}_r^\top (\mathbf{g}_r \oslash (\mathbf{A}_r \mathbf{x})), \qquad (3)$$

where $\odot$ and $\oslash$ denote entry-wise multiplication and division, respectively. Then we compute

$$\hat{\mathbf{s}}^{(k+1)} = (1 - \alpha_k)\hat{\mathbf{s}}^k + \alpha_k \tau_{t_k}(\hat{\mathbf{f}}^{(k)}), \qquad (4)$$

where $\alpha_k \geq 0$ is a sequence of decaying step-sizes[1], and $t_k \in \{1, \dots, N_s\}$. The $\mathbf{s}^{(k)}$ are sufficient statistic estimates of the conditional expectation $\mathcal{T}(\mathbf{f}^{(k)})$. Note that the original Online-EM considers streaming statistic updates $\tau_{t_k}$, which is different from PET construction. Here $t_k$ denotes the subset index used at the $k^{\text{th}}$ iteration, which can be either cycled through deterministically (i.e. shuffling) or chosen at random. This algorithm is called Stochastic Expectation Maximisation (SEM) below. The M-step is then given by

$$\hat{\mathbf{f}}^{(k)} = \underset{\mathbf{f} \geq 0}{\operatorname{argmax}} \, \boldsymbol{\eta}(\mathbf{f})^\top \hat{\mathbf{s}}^{(k)} - \mathcal{U}(\mathbf{f}). \qquad (5)$$

For the Poisson likelihood in PET, solving (5) admits an explicit solution. Next we discuss the extension of SEM to MAP estimation, where the use of the prior is of fundamental importance.

---

[1] For Online-EM choosing $\alpha_k = \mathcal{O}(1/k)$ ensures the convergence [8]

## 2.2 Regularisation

The goal is to develop a method within a (generalised) EM framework, in the sense that each iterate is a maximiser of the log posterior density, and it admits explicit solutions at the M-step. MAP maximises $p(\mathbf{G}|\mathbf{f})q(\mathbf{f})$, where $q(\mathbf{f}) = \exp(-\beta \mathcal{R}(\mathbf{f}))$ is the pre-defined prior. The resulting M-step then follows

$$\hat{\mathbf{f}}^{(k)} = \underset{\mathbf{f} \geq 0}{\operatorname{argmax}} \, \boldsymbol{\eta}(\mathbf{f})^\top \hat{\mathbf{s}}^{(k)} - \mathcal{U}(\mathbf{f}) - \beta \mathcal{R}(\mathbf{f}). \qquad (6)$$

Maximising (6) through iterative schemes (which is typically needed if $\nabla \mathcal{R}(\mathbf{f})$ is not separable) incurs computational costs and numerical inaccuracies [7]. To mitigate these issues, and explicitly solve the M-step, we shall instead use a separable surrogate of the prior. Surrogates for the prior have been used in many works [3, 11–13]. We consider priors of the form

$$\mathcal{R}(\mathbf{f}) = \frac{1}{2} \sum_{i=1}^{N} \sum_{j \in \mathcal{N}_i} w_{ij} \rho(f_i - f_j),$$

where $\rho$ is a potential function, $\mathcal{N}_i$ is a neighbourhood of the voxel $f_i$, and $w_{ij}$ are the weights. We employ a surrogate defined as (up to an additive constant) [9]

$$\hat{\rho}^{(k)}(f_i; f_j) = \gamma_\rho(f_i^{(k)} - f_j^{(k)}) \left( \left(f_i - \frac{f_i^{(k)} + f_j^{(k)}}{2}\right)^2 \right. \qquad (7)$$
$$\left. + \left(f_j - \frac{f_i^{(k)} + f_j^{(k)}}{2}\right)^2 \right),$$

with $\gamma_\rho(f) = \frac{\rho'(f)}{f}$. Accordingly, the surrogated prior is given by

$$\hat{\mathcal{R}}(\mathbf{f}; \mathbf{f}^{(k)}) = \frac{1}{2} \sum_{i=1}^{N} \sum_{j \in \mathcal{N}_i} w_{ij} \hat{\rho}(f_i; \mathbf{f}^{(k)}).$$

Some admissible potential functions are in Table 1.

**Table 1:** Admissible potential functions

| | $\rho(t)$ | $\rho'(x)$ | $\gamma_\rho(x)$ |
|---|---|---|---|
| quadratic | $\frac{x^2}{2}$ | $x$ | $1$ |
| log cosh | $\delta^2 \log \cosh(x/\delta)$ | $\delta \tanh(x/\delta)$ | $\delta \frac{\tanh(x/\delta)}{x}$ |
| hyperbola | $\delta(\sqrt{1 + (x/\delta)^2} - 1)$ | $\frac{x}{\sqrt{1+(x/\delta)^2}}$ | $\frac{1}{\sqrt{1+(x/\delta)^2}}$ |

The surrogate M-step for MAP-SEM is thus given by

$$\hat{\mathbf{f}}^{(k)} = \underset{\mathbf{f} \geq 0}{\operatorname{argmax}} \, \boldsymbol{\eta}(\mathbf{f})^\top \hat{\mathbf{s}}^{(k)} - \mathcal{U}(\mathbf{f}) - \beta \hat{\mathcal{R}}(\mathbf{f}; \mathbf{f}^{(k)}). \qquad (8)$$

Equating the gradient of the objective with zero we arrive at a quadratic equation, with a single (and explicit) non-negative solution.

## 3   Experimental Results

Now we illustrate the performance of the algorithm on simulated data. We use PET scan of a torso, obtained as an XCAT simulated phantom using 2 rings and 280 projection angles. Image reconstruction was performed using the Software for Tomographic Image Reconstruction STIR [10], via a python environment [10], available at https://github.com/UCL/STIR.

The log cosh prior is used to penalise the reconstruction (cf. Table 1), with the regularisation parameter $\beta = 0.001$ and $\delta = 1.0$. The methods examined in this section are initialised with 5 epochs of standard OSEM using 35 subsets.

As a performance metric we use the value of the objective function

$$\Phi(\mathbf{f}) = \mathcal{L}(\mathbf{f}) - \beta \mathcal{R}(\mathbf{f}),$$

where $\mathcal{L}(\mathbf{f})$ is the log-likelihood and $\mathcal{R}(\mathbf{f})$ is the prior. We compare the performance of our method, which uses (4) and solves (8) in each iteration, with step-size relaxed BSREM [3].

The sinogram data was binned into 10 and 20 subsets, respectively, using geometric projections of the scanner. To facilitate a fair comparison, MAP-SEM and BSREM use the same step-size relaxation regime $\alpha_k = \frac{1}{\gamma k+1}$, where the step-size decay factor $\gamma$ depends on the number of subsets. For this stepsize schedule both BSREM and the original Online-EM converge to the maximising solution. In the case of 10 subsets, we set $\gamma = 0.002$, and for 20 subset we use $\gamma = 0.008$. For a given number of subsets, the subsets are created by partitioning the data into equally spaced rows of the sinogram. Then, for each iteration $k$ the subset index $\tau_k$ is selected uniformly at random.

## 4   Discussion and Conclusion

The proposed algorithm provides performance competitive with BSREM (given the same step-size regiment), ensuring convergence. Like OSEM-based approaches we maintain the speed-up in early iterations. MAP-SEM shows a better performance than BSREM, particularly in early stages. Regarding the later epochs, the performance of the two methods is largely indistinguishable, although, MAP-SEM can provide more stable improvements, see Figure 2.

In the first panel of Figure 1, we show the reconstructed solution computed using 1500 epochs of BSREM with line-search for the step-sizes (which converges to the maximising solution within numerical accuracy). The other two panels show the corresponding reconstruction errors of iterates computed using 100 epochs of MAP-SEM and BSREM. The figures show that MAP-SEM and BSREM are, at this point, visually indistinguishable, and the largest errors with both methods are observed at boundaries.

Comparing the behaviour with a different number of subsets in Figure 2, two observations can be made. First, using a lower number of subsets provides a more steady behaviour. This is expected since a smaller subset number gives a sufficient statistics estimate of lower variance. Meanwhile, increasing the number of subsets increases both the acceleration in early iterations but also the variability in the objective function value. Second, asymptotically we observe stabilisation as the step-size decreases in later iterations and the two subset regimes show comparable behaviour.

As observed in [9], in case of online-EM, the step-size schedule used for MAP-SEM requires some tuning in order to ensure convergence and optimise the acceleration in early iterations. Moreover, empirical investigations suggest that the optimal step-size decay factor depends on the number of subsets, but the precise dependence remains unknown.

Furthermore, as observed in Figure 2, increasing the number of subsets increases the variability between the updates which requires an adjustment in the step size decay constant to stabilise the iterations. This behaviour is due to several factors, e.g., imbalances between subsets and larger variances of the sufficient statistic estimates.

Future work will aim to address both of these issues by investigating a MAP-SEM variance-reduction technique, e.g., [14, 15].

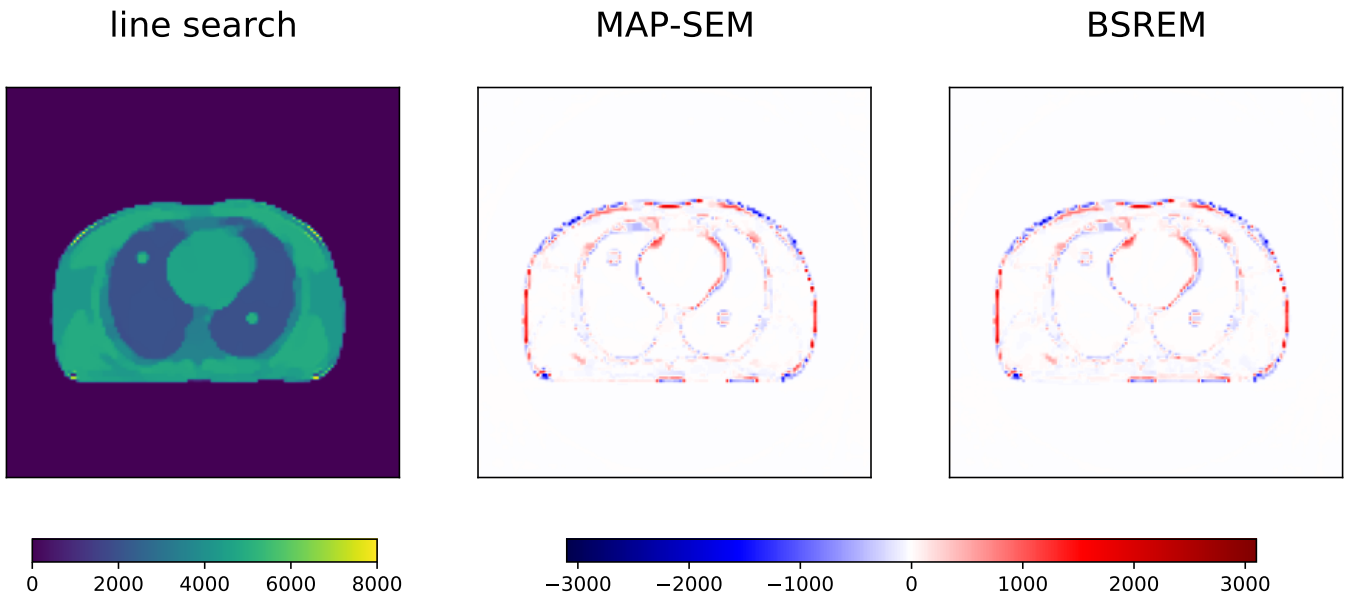

**Figure 1:** The left-most panel shows the (up to numerical precision) converged maximising solution computed with 1500 epochs of line search BSREM. The panels to its right show respective errors of $100^{th}$-epoch updates of MAP-SEM and relaxd BSREM, computed using 20 subsets

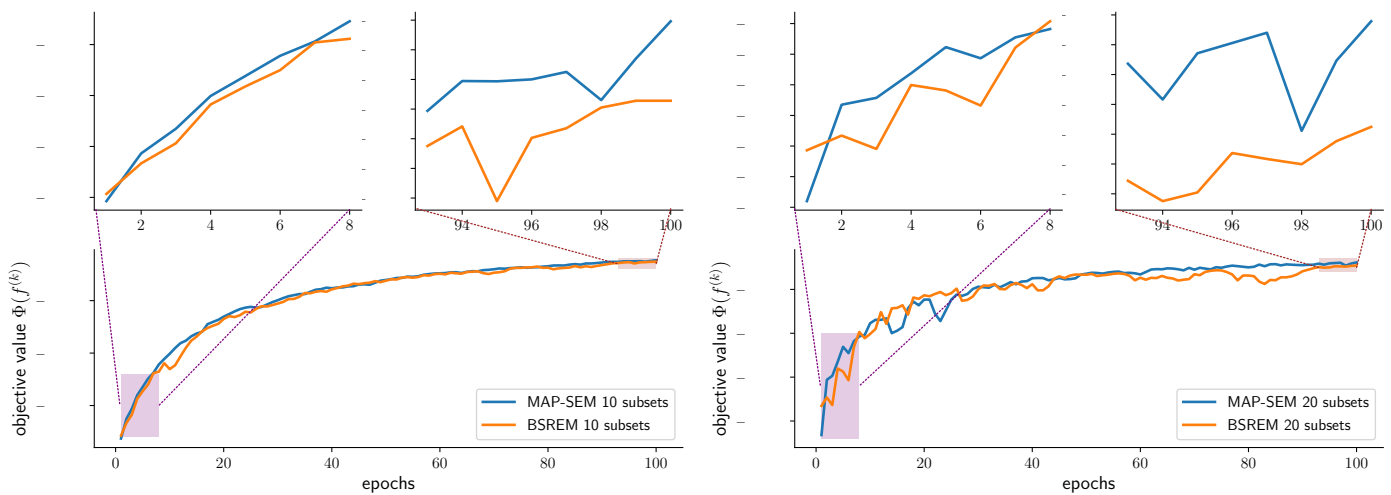

**Figure 2:** Comparison of the values of the objective function. The figure on the left corresponds to the results for the 10 ordered subsets of the data, and the figure on the right corresponds to 20 subsets of the data. The highlighted sections on the top show the behaviour in the first, respectively last, 7 epochs

# Image Properties Prediction in Nonlinear Model-based Reconstruction using a Perceptron Network

Wenying Wang, J. Webster Stayman, and Grace J. Gang

Department of Biomedical Engineering, Johns Hopkins University, Baltimore, MD, 21205

**Abstract** Nonlinear reconstruction algorithms have demonstrated superior resolution to noise tradeoffs compared to traditional linear reconstruction methods. However, their nonlinear, shift variant, and data-dependent nature complicates performance analysis. Furthermore, there usually lacks a predictive framework for image properties that allows efficient control and optimization of imaging performance. In this work, we quantify the system response of general nonlinear reconstructions using a quantitative perturbation response metric and develop a data-driven approach for prospective prediction of such properties as a function of varying perturbations (size, shape, contrast, and contrast profile), patient anatomy, and algorithmic parameter. The feasibility of prediction framework is demonstrated for a penalized-likelihood reconstruction algorithm with a Huber penalty (PLH). We incorporated a compact representation of the imaging system and the perturbation as the input to the network and used a three-layer perceptron network for image property prediction. The predicted perturbation response shows good agreement with those obtained from empirical measurements. The prediction accuracy is generalizable to all perturbations, anatomical locations, and regularization parameters investigated. Results in this work suggest that the data-driven method and training strategies developed herein is a promising approach for prospective image property prediction and control in nonlinear reconstruction algorithms.

## 1 Introduction

The recent proliferation of nonlinear reconstruction algorithms have presented tremendous opportunities for image quality improvement and dose reduction. However, despite promising results in the research setting, clinical translation of these algorithms have met a number of challenges. Due to their nonlinear, shift-variant, and data-dependent nature, traditional image quality assessment metrics rooted in linear system analysis (e.g., impulse response, noise power spectrum) may no longer apply. For example, the appearance of a lesion of interest in an MBIR reconstructed image can be highly dependent on its location in the anatomy. Lesions of different contrast may also result in different edge profiles [1]. Furthermore, the performance of nonlinear algorithms often rely on careful tuning of algorithmic parameters (e.g., regularization strength). The relationship between these parameters and image properties, however, is often opaque. As a result, image properties are often analyzed in a *retrospective* fashion via empirical measurements. Optimization of nonlinear algorithms therefore frequently relies on exhaustive evaluations overs the parameters of interest, which is time consuming due to the large number of dependencies mentioned previously.

In previous work [2, 3], we proposed a novel image quality analysis framework capable of *prospective* predictions of image properties in general nonlinear reconstruction algorithms. Leveraging the universal approximation theorem, we trained an artificial neural networks model to map the nonlinear transfer functions of an example model-based reconstruction algorithm. In this work, we present further development of the framework focusing on efficient training strategies that allows the predictive capability of the model to be generalized to arbitrary stimuli, anatomy, and imaging conditions.

## 2 Materials and Methods

### 2.1 Generalized system response

In linear shift-invariant imaging systems, the system response can simply be characterized by the impulse response function which is dependent on the system parameters, $\mathbf{S}$. For general nonlinear algorithms, the system response carries additional dependencies on the measurement data, $y$, and the stimulus/perturbation, $\mu_s$. Following Ahn and Leahy[4], we define the generalized system response of a reconstruction algorithm, $\mathcal{H}(\mu_s; \mu, \mathbf{S})$, as the difference between the mean reconstructions ($\hat{\mu}$) with and without the perturbation:

$$\mathcal{H}(\mu_s; \mu) = \overline{\hat{\mu}(y(\mu + \mu_s; \mathbf{S}))} - \overline{\hat{\mu}(y(\mu; \mathbf{S}))}. \quad (1)$$

The generalized system response extends the characterization of the dependencies on the background anatomy $\mu$ and the stimuli $\mu_s$ introduced in nonlinear algorithms in addition to the dependencies on imaging system characterizations and reconstruction approach including regularization designs demonstrated in locally linearizable algorithms.

### 2.2 Penalized-likelihood reconstruction with a Huber penalty

In this work, we demonstrate methods for developing the predictive analysis framework on an example MBIR algorithm based on a penalized-likelihood objective with a Huber penalty. The objective function is given by:

$$\Phi(\mu, y) = L(\mu; y) - R(\mu, \beta, \delta) \quad (2)$$

where $L(\mu; y)$ is the log-likelihood term that presumes the measurements follow an independent Poisson distribution, and $R(\mu, \beta, \delta)$ is the Huber penalty active in the 4-nearest neighborhood:

$$R(\mu, \beta, \delta) = \beta \sum_j \sum_{k \in \mathcal{N}_j} \phi_H(\mu_j - \mu_k; \delta) \quad (3)$$

$$\phi_H(x; \delta) = \begin{cases} \frac{x^2}{2\delta}, & |x| \leq \delta \\ |x| - \frac{\delta}{2}, & |x| > \delta \end{cases} \quad (4)$$

The term contains two regularization parameters, where $\beta$ controls the overall regularization strength and $\delta$ controls the threshold in voxel differences where the potential function transitions from quadratic to linear. The interaction between $\beta$ and $\delta$ results in a complex tradeoff between overall smoothness and edge preservation. Examples of such dependencies

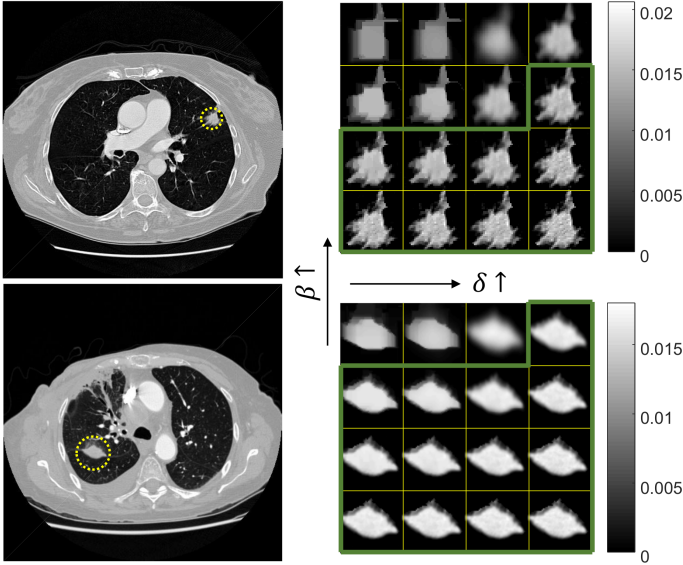

**Figure 1:** Generalized system response of two lung nodules with different regularization values. The green line circumscribes the "good" perturbation responses. (Unit: mm$^{-1}$)

are illustrated in Figure 1 with two lung nodules in the Lung Image Database Consortium (LIDC) [5]. Each subplot on the right shows the generalized system response (Eq.1) corresponding to PLH reconstructions at different combinations of $\beta$ and $\delta$. From visual inspection, the perturbation response can be distorted with poor regularization parameters combinations. Moreover, because the system is nonlinear, a set of regularization parameters that can achieve good response with one perturbation may not work for other perturbations. The spiculated lung nodule (top) shows less tolerance of high regularization strength compared to the smooth lung nodule (bottom). As a quantitative measure of how faithfully nodules are represented in the reconstructions, we circumscribed the region of "good" ($\beta$, $\delta$) where the perturbation response has less than 30% relative root mean square error (rRMSE) compared to the ground truth (rRMSE is defined as the normalized RMSE over the root mean square of the ground truth ). While the regions in Fig.1 were identified through empirical measurements, the following sections aim to establish a model to predict perturbation responses without the need for reconstructions.

## 2.3 Prediction Framework Implementation

The prediction framework leverages the universal approximation theorem which states that a fully-connected neural network with a single hidden layer that is of arbitrary number of nodes or an arbitrarily deep fully-connected network with a finite number of nodes in each hidden layer can approximate any well-behaved continuous function $f : \mathbb{R}^d \to \mathbb{R}^D$ with a arbitrarily small residual distance.[6, 7] In this section, we discuss the efficient information for perturbation response prediction, space sampling strategy, and prediction neural network architecture setup.

### 2.3.1 Efficient information for prediction

We leveraged prior knowledge of the image properties of PLH to devise efficient network inputs. Ahn and Leahy [4]

derived an explicit closed-form expression of the perturbation response in locally linearizable algorithms:

$$\mathcal{H}(\mu_s) = [\mathbf{A}^T\mathbf{W}\mathbf{A} + \mathbf{R}]^{-1}\mathbf{A}^T\mathbf{W}\mathbf{A}\mu_s \qquad (5)$$

where $\mathbf{W}$ is the covariance matrix of the measurements and $\mathbf{R}$ denotes the Hessian of the regularizer. For Huber penalty, the Hessian term is image dependent and therefore difficult to evaluate. However, the Fisher information term $\mathbf{A}^T\mathbf{W}\mathbf{A}\mu_s$ efficiently characterizes the dependencies on system geometry (through $\mathbf{A}$), data statistics (through $\mathbf{W}$), and perturbation (through $\mu_s$). Therefore, as a compact representation of the imaging system, the anatomical background, and the perturbation, the Fischer information term $\mathbf{A}^T\mathbf{W}\mathbf{A}\mu_s$ along with the regularization parameters are used as inputs to the network to provide sufficient information to determine the generalized system response of the PLH. This expression also informs the range of training data required to achieve generalizable predictive capability.

### 2.3.2 Parameter space sampling in training data

In a data-driven method, the range of training data is directly related to network performance. In this work, we seek to build a predictive model for the generalized system response as a function of (A) perturbation, (B) background anatomy and locations, and (C) regularization. We propose the following sampling strategies for each parameter in the context of lung imaging:

**(A) Sampling the perturbation ($\mu_s$):** The perturbations, or lung lesions, have large variabilities in terms of their size, contrast, shape, and contrast profiles. To efficiently sample the perturbations, we adopted a parametric model for realistic lesion simulation developed by Solomon and Samei [8]:

$$c(\theta, r) = C(1 - (r/R_\theta)^2)^n \qquad (6)$$

where $C$ is the peak contrast value, $n$ describes the steepness of the profile, and $R_\theta$ is the distance from the center to the edge along the radial direction $\theta$. This type of model allows us to systematically represent perturbations by sampling combinations of ($C, n, R_\theta$). We sampled scalars $C$ and $n$ with the range described in [8]. For the vector $R_\theta$, we sampled $R_\theta$ along 8 radial directions according to a normal distribution of mean $\bar{R}$ and variance $\sigma_R$), and used interpolation to achieve smoothly varying $R_\theta$ for arbitrary $\theta$.

**(B) Sampling the anatomy and locations (W):** We used the XCAT chest digital phantom [9] as the background anatomy in the simulation study and manually selected locations to insert lesions. These locations represent various profiles of statistical weights $\mathbf{W}$ pertaining to lung imaging.

**(C) Sampling the regularization ($\beta, \delta$):** We sampled the different combinations of regularization parameters ($\beta$, $\delta$) using a 2D sweep. The range of the regularization parameters are selected to sufficiently include a variety of reconstruction outcomes illustrated in Fig.1.

### 2.3.3 Efficient network architecture

With the efficient input proposed in Sec. 2.3.1, we seek to approximate the following function with a neural network:

$$\mathcal{H}(\mu_s, \mu) = f(\text{ROI}[\mathbf{A}^T\mathbf{W}\mathbf{A}\mu_s], \beta, \delta). \qquad (7)$$

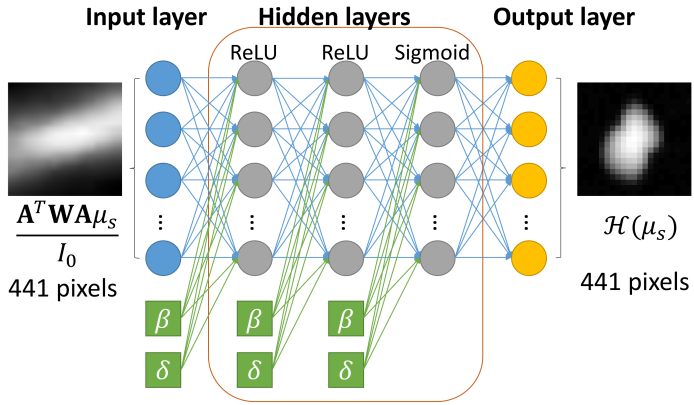

**Figure 2:** Structure of the multi-layer perceptron neural network.

| Dataset | $C$ (HU) | $n$ | $\bar{R}$(mm) | $\sigma_R / \bar{R}$ |
|---|---|---|---|---|
| Training | 600:200:1400 | 0.5,0.75,1.0 | 3,4,5,6 | 0.2,0.5 |
| Testing | 700,1100 | 0.6,0.9 | 3.5,4,5,5.5 | 0.35 |

| Dataset | $\log(\beta)$ | $\log(\delta)$ |
|---|---|---|
| Training | 2.0:0.8:6.0 | 2.4:0.8:5.6 |
| Testing | -5.0:0.8:-1.0 | -4.6:0.8:-1.4 |

**Table 1:** Lesion synthesis and regularization parameters in data generation.

According to the universal approximation theorem, a perceptron network with a single hidden layer can approximate arbitrary functions give enough nodes in the hidden layer. When using more hidden layers, each layer requires a smaller number of nodes. Through empirical experimentation, we adopt a perceptron network that has three hidden layers as shown in Figure 2. The input stacks the Fisher information term the regularization parameters. The corresponding output is the measured system response as the difference between PLH reconstructions with the perturbation and without the perturbation. Furthermore, for this investigation, we assumed a well-sampled imaging geometry where the perturbation response is contained in a local region. This assumption allows us to truncate $\mathbf{A}^T\mathbf{W}\mathbf{A}\mu_s$ and $\mu_s$ to reduce the dimensionality of the network. All Fisher information term ($\mathbf{A}^T\mathbf{W}\mathbf{A}\mu_s$) and outputs ($\mathcal{H}(\mu_s,\mu)$) of the network is contained in a 21x21 grid. Each hidden layer has 441 nodes that is of the same size as the output layer. To feed the regularization parameters to the network, we concatenate $(\beta,\delta)$ to the first two hidden layer. All nodes are fully connected. The first two layer is activated with a rectified linear unit (ReLU), and the last layer with a sigmoid function.

### 2.4 Experiment setup

Following the sampling strategies proposed in Sec. 2.3.2, we generated training and testing datasets according to the various dependent parameters. The lung nodule parameters and $(\beta,\delta)$ are shown in Table 1. With each set of parameters, we generated 50 nodules for training and 10 nodules for testing, resulting in 6000 nodules in the training dataset and 120 nodules in testing dataset. Figure 3 illustrates example nodules corresponding to nodule parameters in both training and testing. The simulated lung lesions are inserted in 19 locations on a 2D slice in the chest phantom as shown in

Fig.4. For initial investigation, we performed training and testing on the same anatomical background. Generalizing the prediction framework to varying anatomical background is the subject of ongoing work.

The perceptron neural network was trained by minimizing the mean square error between the predicted perturbation responses and the measurements using the ADAM optimizer. Prediction feasibility was validated through qualitative comparison and quantitative evaluation using structural similarity index measure (SSIM). We raised one example application of the proposed prediction framework in efficient prospective regularization selection, where the boundaries of proper regularizations that can produce "good response" with small rRMSE (akin to Fig. 1) were determined using measured response through a retrospective exhaustive parameters sweep or prospective evaluation using the prediction model. The comparison between the measured and predicted boundaries demonstrated the efficacy of prospective approach.

### 3 Results

Prediction accuracy of the proposed framework was validated through comparisons between predicted and measured perturbation responses. Pairs of measurement and prediction are shown in Figure 5 with varying regularization parameters ($\beta$, $\delta$), perturbations, and locations, respectively. The predictions show good agreement with the measurements and are capable of characterizing all dependencies investigated. The agreement between prediction and measurement is further quantified in terms of the SSIM metric. The mean SSIM among all testing cases is 0.9991. Over 99% of the predictions achieves 0.995 of SSIM when compared to the measured ground truth.

An example application of the proposed predictor is demonstrated in efficient selection of "good" regions of regularization parameters, i.e., a quantitative alternative to Fig.1 without the need for additional reconstructions. Figure 6 shows two maps of rRMSE with varying regularization combinations. The ground truth plot (top left) were computed from measured perturbation response through an exhaustive sweep. The bottom left plot shows the predicted rRMSE map computed with finer sampling of regularization parameters. The green lines circumscribed the "good response" areas where the rRMSE is smaller than 30%. The plots on the right show two examples of perturbation responses from the "good response" region circumscribed in green and the "bad response" region circumscribed in red. We notice that despite the predicted rRMSE values deviate from the measured values in the highly-regularized region, the predicted "good response" region shows great agreement with the outcome of retrospective evaluation, demonstrating the capability of using this predictor for regularization parameters selection.

### 4 Discussion and Conclusion

In this work, we propose a prediction framework that quantifies the perturbation response of a nonlinear reconstruction algorithm, where a multi-layer perceptron network is used to approximate the perturbation response in a data-driven fashion. We establish a sampling strategy to guarantee good characterization of the perturbation dependencies on regularization parameters, perturbation features, and locations. We demonstrate the feasibility of the prediction framework in simulation study, and show the potential application of the tool towards regularization tuning for reliable system response. Ongoing work includes incorporating variability in background anatomy and refine the perturbation model to achieve good agreement for more realistic perturbations in clinical dataset.

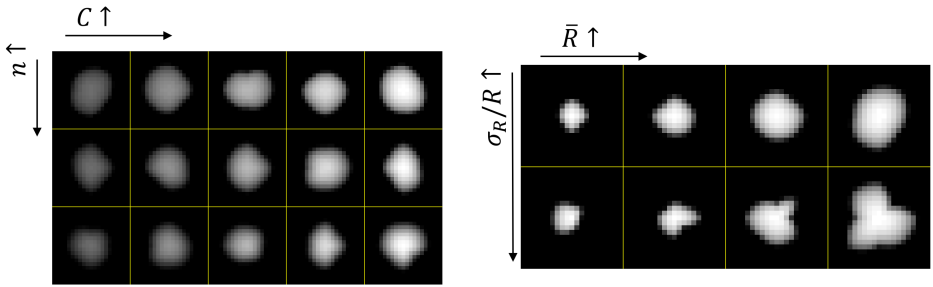

**Figure 3:** Examples of synthesized lung nodules for training.

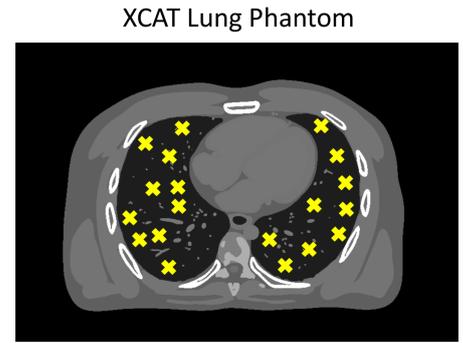

**Figure 4:** XCAT simulated anatomical background. The yellow crosses indicate the locations to insert perturbations.

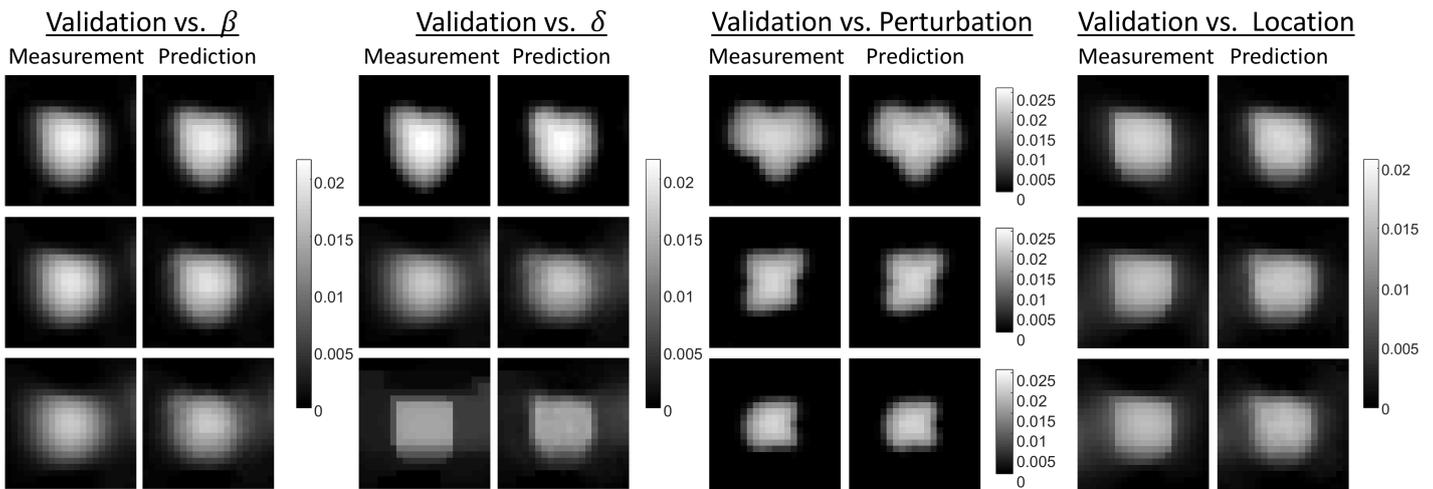

**Figure 5:** Comparisons between measurements and predictions with varying regularizations, perturbations, and locations. Unit: mm$^{-1}$.

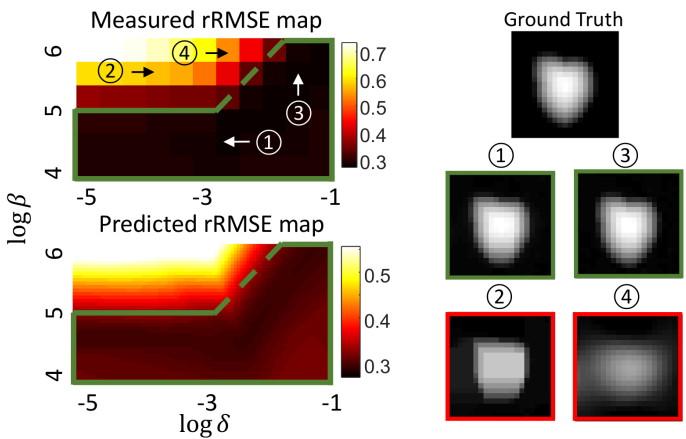

**Figure 6:** Measured and predicted rRMSE map with varying $\beta$ and $\delta$. The regularization parameter combinations are circumscribed with green lines. Examples of perturbation responses are shown on the right.

### Acknowledgement

This work is supported, in part, by NIH grants R01CA249538 and R01EB027127.

# Chapter 10

# Poster Session 2

**session chairs**
Samuel Matej, *University of Pennsylvania (United States)*
Vesna Sossi, *University of British Columbia (Canada)*

# One-step spectral computed tomography image reconstruction for subjects containing metal

Emil Y. Sidky[1], Rina Foygel Barber[2], Taly Gilat-Schmidt[3], and Xiaochuan Pan[1]

[1]Department of Radiology, The University of Chicago, Chicago, IL, USA
[2]Department of Statistics, The University of Chicago, Chicago, IL, USA
[3]Department of Biomedical Engineering, Marquette University, Milwaukee, WI, USA

**Abstract** A one-step spectral computed tomography (CT) image reconstruction algorithm is developed that is tailored to scanning subjects that contain metal. The algorithm is based on a large-scale nonconvex and non-smooth optimization problem that includes a transmission Poisson likelihood (TPL) term and a regularizing total-variation constraint. From an algorithmic standpoint, metal presents a challenge due to low X-ray transmission through such objects, and the combination of photon starvation with the inherent nonconvexity of the optimization problem can make it difficult to formulate a convergent iterative algorithm. In this work, this problem is addressed by considering a new convexification of the TPL term. In addition, one of our previous preconditioning strategies is seen to be effective in improving algorithm efficiency. The algorithm is demonstrated on inversion of simulated noiseless spectral CT data of a test phantom of a human pelvis with metal prosthesis.

## 1 Introduction

Photon-counting detectors (PCD) enable the possibility of acquiring X-ray transmission data in a number of energy windows simultaneously. When PCDs are incorporated in a CT system, the resulting spectral CT device provides sufficient information to determine the energy dependent linear X-ray attenuation map, assuming that the attenuation energy dependence can be represented by few basis functions. This capability has use for quantitative CT and particularly quantitative imaging with K-edge contrast agents [1]. Spectral CT can also be useful to overcome beam-hardening artifacts, which can be particularly severe for highly-attenuating materials. An extreme example of such materials are metal objects. Metal not only poses a challenge due to beam-hardening, it can also completely block X-ray transmission causing photon starvation.

One-step image reconstruction for spectral CT [2, 3], where the linear attenuation map is estimated directly from the transmission data, allows for a unique opportunity to address objects with metal. One-step algorithms do not need all energy windows to be collected for every transmission ray. This fact is useful in particular for the situation where a given transmission ray has useful information for higher energy windows while it is photon-starved for lower energy windows. When a given transmission ray is completely blocked for all energy windows, it can also be discarded, and this strategy leads to various "missing data" or in-painting problems. There is, however, a degree of arbitrariness in this approach because a cut-off strategy needs to be specified that determines which data to throw out and whether to have a sharp transition or a smooth data-weighting scheme. This involves specifying addition parameters, increasing complexity of the image reconstruction algorithm.

In this work, we present an algorithm that accepts spectral CT data in the form of photon counts along each transmission ray resolved in multiple energy-windows. In particular, it can be applied to data from a subject that contains metal where the corresponding transmission data is photon-starved, enabling new ways to deal with CT scanning of patients with metal. In Sec. 2, we briefly describe the image reconstruction algorithm; in Sec. 3 the algorithm is demonstrated on a challenging model inversion problem for a pelvis phantom with metal hip implants; and the conclusion and outlook is provided in Sec. 4.

## 2 Methods

We write the continuous spectral CT data model as

$$I_{w,\ell} = \int S_{w,\ell}(E) \exp\left[-\int_\ell \mu(E, \vec{r}(t))dt\right] dE, \qquad (1)$$

where $I_{w,\ell}$ is the transmitted X-ray photon fluence along ray $\ell$ in energy window $w$; $t$ is a parameter indicating location along $\ell$; $S_{w,\ell}(E)$ is the spectral response; and $\mu(E, \vec{r}(t))$ is the energy and spatially dependent linear X-ray attenuation map. This unknown function is four dimensional, and the dimensionality can be reduced by employing a standard material-expansion decomposition

$$\mu(E, \vec{r}(t)) = \sum_m \left(\frac{\mu_m(E)}{\rho_m}\right) \rho_m f_m(\vec{r}(t)), \qquad (2)$$

where $\rho_m$ is the density of material $m$; $\mu_m(E)/\rho_m$ is the mass attenuation coefficient of material $m$; and $f_m(\vec{r})$ is the spatial map for material $m$.

Combining Eq. (1) with Eq. (2), normalizing the spectral response, and discretizing the integrations leads to the data model used for image reconstruction

$$\hat{c}_{w,\ell}(f) = N_{w,\ell} \sum_i s_{w,\ell,i} \operatorname{qexp}\left(-\sum_{m,k} \mu_{m,i} X_{\ell,k} f_{k,m}\right), \qquad (3)$$

where $\hat{c}w, \ell(f)$ and $N_{w,\ell}$ are respectively the mean transmitted photon count and the total number of incident photons along ray $\ell$ in energy window $w$; $s_{w,\ell,i}$ is the normalized spectral response, i.e. $\sum_i s_{w,\ell,i} = 1$; $i$ indexes the sum over energy, which replaced the energy integration; $X_{\ell,k}$ represents X-ray projection along the ray $\ell$; and $f_{k,m}$ is the pixelized material map with $k$ and $m$ indexing pixel and expansion-material, respectively. The function qexp is a modification of the exponential

$$\text{qexp}(x) = \begin{cases} \exp(x) & x \leq 0 \\ \frac{1}{2}x^2 + x + 1 & x > 0 \end{cases},$$

that is identical to $\exp(\cdot)$ for non-positive physical argument values. The softer quadratic dependence for positive arguments is helpful for iterative image reconstruction where unphysical, negative attenuation values may occur at intermediate iterations. The goal of spectral CT image reconstruction is to invert Eq. (3), obtaining the material maps $f$ from measured counts data $c$.

Maximizing the transmission Poisson likelihood (TPL) is equivalent to minimizing the following data divergence between the photon count data, $c$, and mean photon count model, $\hat{c}(f)$,

$$D_{\text{TPL}}(c, \hat{c}(f)) = \\ \sum_{w,\ell} \left[ \hat{c}_{w,\ell}(f) - c_{w,\ell} - c_{w,\ell} \log \frac{\hat{c}_{w,\ell}(f)}{c_{w,\ell}} \right], \quad (4)$$

where $c_{w,\ell}$ are the measured counts in energy window $w$ along ray $\ell$. We consider both minimization of $D_{\text{TPL}}$ alone and constrained minimization

$$f^\star = \arg\min_f D_{\text{TPL}}(c, \hat{c}(f)) \text{ such that } \text{GTV}(f) \leq \gamma, \quad (5)$$

where

$$\text{GTV}(f) = \sum_{\text{pixels}} \sqrt{\sum_m |Df_m|_{\text{mag}}^2};$$

$D$ is a numerical gradient operator; $|\cdot|_{\text{mag}}$ computes the spatial vector magnitude and accordingly $|Df_m|_{\text{mag}}$ is the gradient magnitude image of the material map $f_m$. The parameter $\gamma$ is the constraint value for the image GTV.

The challenge for minimizing $D_{\text{TPL}}$ is that it is a large-scale nonconvex optimization problem, and introducing the constraint in Eq. (5) adds non-smoothness to the optimization problem. We discuss optimization of the smooth nonconvex objective function $D_{\text{TPL}}$. Most strategies for solving nonconvex problems involve "convexification", where the complete nonconvex problem is approximated by a convex problem that depends on the current value of the image estimate. The convex approximation is used to generate a descent step, and if the convexification is well-designed the descent step will be an efficient descent step for the original non-convex problem. In our previous work [2], convexification is achieved by forming a convex quadratic expansion that is guaranteed to be an upper bound to the smooth part of the nonconvex problem locally. While this approach to convexification has been effective for numerous spectral CT studies, it encounters difficulty in dealing with data sets containing photon starvation.

We have recently been pursuing a different form of convexification for spectral CT image reconstruction. First, we note that

$$D_{\text{TPL}}(c, \hat{c}(f)) = \sum_{w,\ell} \left[ \hat{c}_{w,\ell}(f) - c_{w,\ell} \log \hat{c}_{w,\ell}(f) \right] + C,$$

where $C$ is independent of $f$. Accordingly, we consider minimization of

$$L(f) = L_1(f) + L_2(f) \quad (6)$$

$$L_1(f) = \sum_{w,\ell} \hat{c}_{w,\ell}(f) \quad (7)$$

$$L_2(f) = -\sum_{w,\ell} c_{w,\ell} \log \hat{c}_{w,\ell}(f). \quad (8)$$

The first term $L_1(f)$ is convex, and it is the second term $L_2(f)$ that is nonconvex. The form of convexification employed for this work involves linearization of the second term

$$L_c(f, f_0) = L_1(f) + f^\top \nabla L_2(f_0), \quad (9)$$

where $f_0$ is the expansion point for the linearization of the second term, and $L_c$ is a convexification of $L$. In the proposed iterative algorithm, the value of $f$ from the previous iteration is used as the expansion point, $f_0$. This form of convexification is analyzed for spectral CT in [4], where it is shown that convergence is proved under an assumption of restricted strong convexity. The details of the algorithm are presented in [4].

For this work, we consider $D_{\text{TPL}}$ minimization with and without introducing a constraint on the material map GTV. We also investigate the impact of $\mu$-preconditioning ($\mu$-PC), described in [2], where a transformation of the material linear attenuation functions is performed that orthonormalizes these functions. The same transformation is applied to the material maps so that we have

$$\sum_m \mu_{m,i} f_m = \sum_m \mu'_{m,i} f'_m \text{ and } \sum_i \mu'_{m,i} \mu'_{m,j} = \delta_{i,j},$$

where $\delta_{i,j}$ is the Kronecker delta function, and the primed quantities are transformed.

## 3 Results

The presented reconstruction problem is an idealized spectral CT set-up where we investigate the solution of the inverse problem corresponding to noiseless 4-window spectral CT data for a simulated pelvis phantom with metal hip implants. The spectral sensitivities used in the study are shown in Fig. 1. The object is exactly represented in a two-material basis expansion using bone and water for expansion materials. The test phantom is shown in Fig. 2 in two grayscale windows; the high contrast window is displayed so that the cross section of the Co-Cr-Mo metal rods is visible, and the low contrast window is shown so that soft-tissue can be resolved. The CT setup involves 512 X-ray projections over a 360 degree scanning angular range.

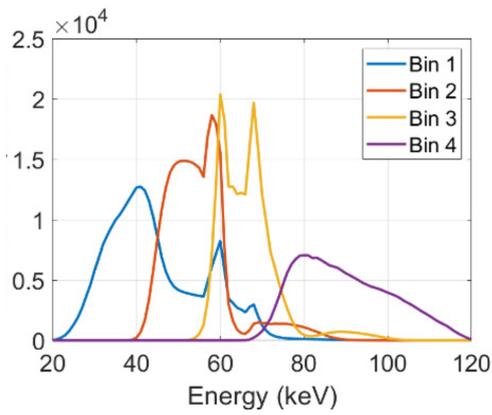

**Figure 1:** Photon energy distributions used for the simulated 4-energy bin spectral CT system. While the transmission counts data is simulated, these distributions result from calibration measurements of an actual DxRay PCD.

Due to the metal rods, the mean transmitted photon counts has a high dynamic range. The modeled X-ray fluence is $10^6$ photons incident on each detector pixel, and the maximum photon count on a single energy window for the unattenuated beam is $2.85 \times 10^5$ photons. The mean photon count for a ray crossing both metal rods and detected in the lowest energy window is as low as $6.71 \times 10^{-5}$ – ten orders of magnitude lower than the maximum count value. For the study presented here, we consider ideal noiseless data and investigate the ability to recover the pelvis phantom. Accordingly, we use the mean photon counts as the input data for image reconstruction. The large dynamic range of the transmitted photon count data makes this inverse problem challenging, and our previous algorithm [2] could only be applied to this system by masking out the measurements with a mean photon count of 10 or less.

The approach specified by the convexification described in Eq. (9) and explained in [4] can yield convergent iteration without masking the photon count data. This is demonstrated by the $D_{TPL}$ plots in Fig. 3, where progress over 1000 iterations is shown for minimization based on $D_{TPL}$ is shown. Also shown are the use of the GTV constraint and $\mu$-preconditioning. Because this study uses noiseless consistent data it should be possible to drive the $D_{TPL}$ to zero, and we note that all curves show a downward trend over the 1000 iterations. The results also show that use of the GTV constraint and $\mu$-preconditioning can substantially improve convergence.

To examine convergence of the algorithm more closely, reconstructed images by use of GTV-constrained $D_{TPL}$-minimization with $\mu$-preconditioning are shown in Fig. 4. Corresponding line profiles are shown in Fig. 5 for a line in the image that crosses both metal implants. The images show that the soft tissue regions are recovered to a high degree of accuracy. There is noticeable discrepancy in the reconstructed water map when viewed in a narrow gray scale window; small bright patches appear on either side of the hip bone regions. Interestingly, the recovery of the metal objects themselves appears to be quite challenging in a water-bone basis. There is significant discrepancy between reconstructed profiles and the truth for the water and bone profiles in the metal regions. When the profile comparison is shown for the monochromatic image, however, the agreement appears to be quite good even for the metal rods. This may be an indication that a different material basis would allow better phantom recovery performance.

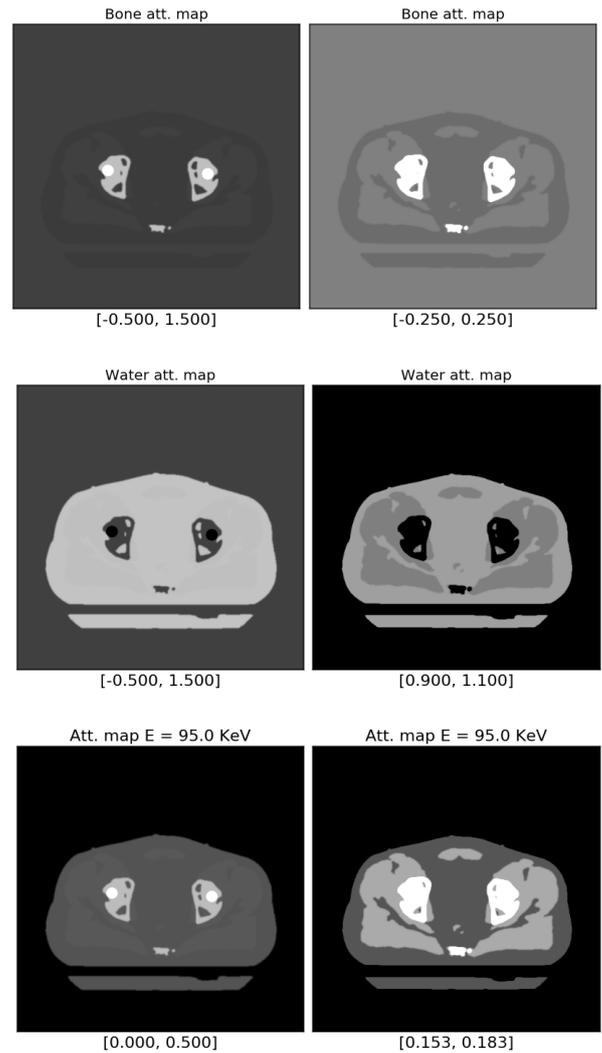

**Figure 2:** Pelvis phantom shown in high contrast and low contrast gray scales on the left and right columns, repsectively. The gray scale window values appear below each panel. The bone and water material maps are shown along with a monochromatic image corresponding to linear attenuation at 95 keV. The shown image array is 512x512, and the pixel size is 1 mm².

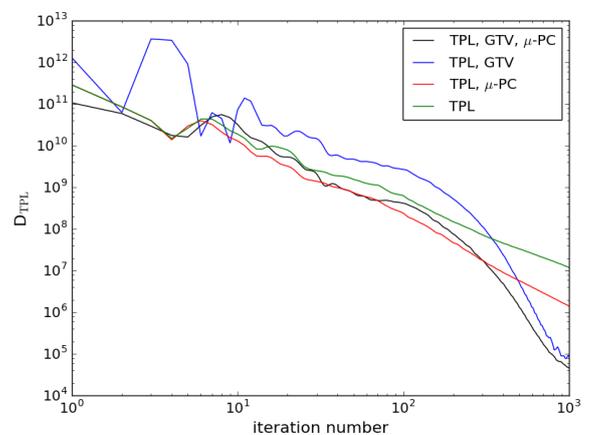

**Figure 3:** Convergence of $D_{TPL}$ as a function of iteration number. The green curve shows iteration based solely on minimization of $D_{TPL}$. The label "GTV" indicates use of the GTV constraint with the constraint parameter $\gamma$ set to the true phantom value. The label $\mu$-PC indicates use of $\mu$-preconditioning.

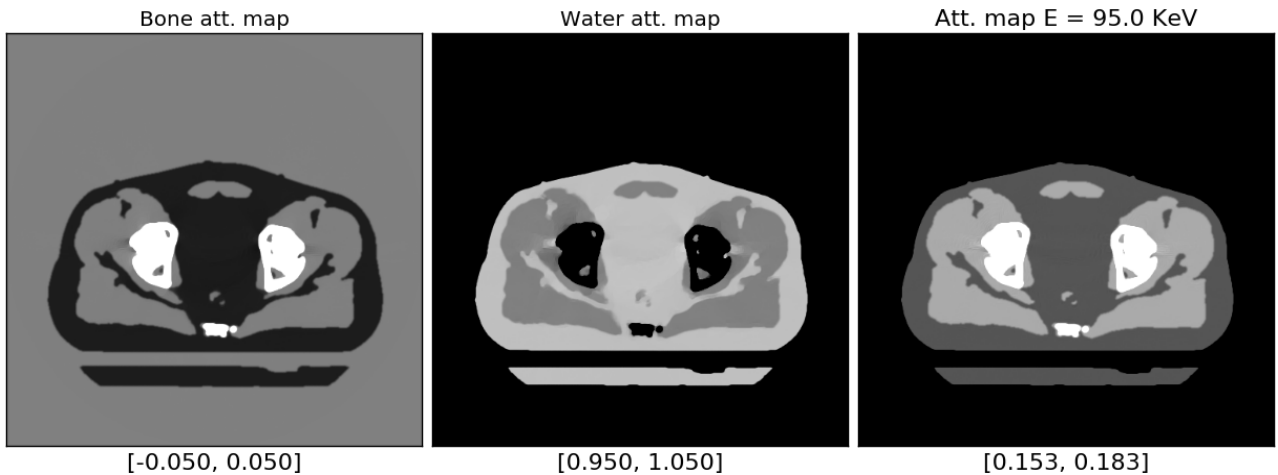

**Figure 4:** Results from reconstruction using GTV-constrained, $D_{\text{TPL}}$-minimization with $\mu$-preconditioning. The gray scale windows for the material maps is set to the 10% level so that subtle discrepancy from the ground truth can be appreciated.

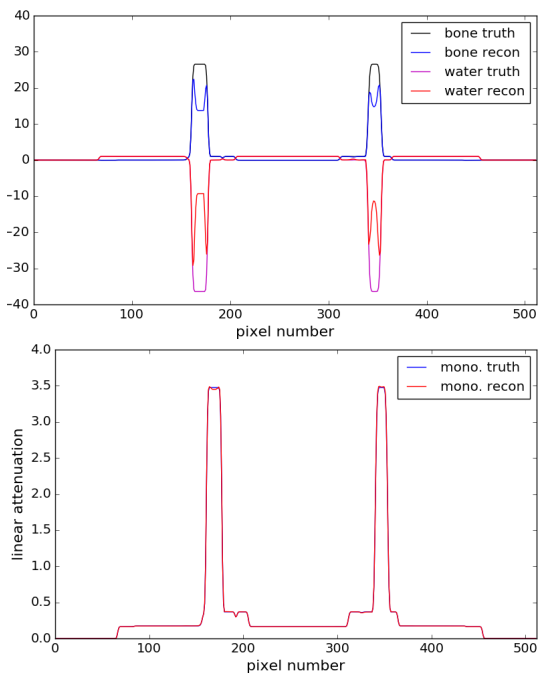

**Figure 5:** Profile comparisons for a horizontal line going through both metal implants for the images shown in Fig. 4. The top graph shows the comparison for the material maps and the bottom graph shows the same for the monochratic image at 95 keV.

## 4 Conclusion

We have presented preliminary results on one-step image reconstruction for spectral CT for the situation where the subject contains metal and the transmitted photon count data is photon starved. The corresponding nonconvex optimization problem is approached by a new convexification strategy, which enables image reconstruction from spectral CT data that has a high dynamic range. The results show promise in solving the corresponding inverse problem. Further results at the meeting will focus on exploring the boundaries of accurate image recovery. Also, we will present results focussing on inconsistent data where noise is present.

### Acknowledgements

Work supported by the National Institutes of Health via grants R01-023968 and R01-026282. R.F.B. was also supported by the National Science Foundation via grant DMS–1654076 and by the Office of Naval Research via grant N00014-20-1-2337. The contents of this article are solely the responsibility of the authors and do not necessarily represent the official views of the National Institutes of Health.

# Few-shot learning with Light-weight Neural Network for limited-angle computed tomography reconstruction

Ping Yang[1, 2], Yunsong Zhao[1, 2], and Xing Zhao[1,2, *]

[1] School of Mathematical Sciences, Capital Normal University, Beijing, China
[2] Beijing Advanced Innovation Center for Imaging Technology, Capital Normal University, Beijing, China.

**Abstract** Traditional reconstruction algorithms for limited-angle computed tomography (CT) have shortcomings including excessive iterations and prolonged reconstruction time. With the rapid development of deep learning, several deep learning-based limited-angle CT reconstruction algorithms emerged to improve image quality and to overcome the disadvantages of traditional algorithms. However, due to the small sample size and complex learning models, many algorithms can't be widely applied. Against such a backdrop, in this work, a new network structure based on traditional optimization imaging model is built. By fully utilizing the characteristics of limited-angle CT, the new network has low complexity, thus the number of samples needed for training is small. In addition, an attention mechanism is introduced to learn the hyperparameters in the network. The results show that the proposed method has good performance in improving image quality and the limited-angle artifacts in the images are alleviated effectively.

## 1 Introduction

Computed tomography (CT) technology has been widely used in medical diagnosis, industrial nondestructive testing (NDT), and other fields [1]. In its practical applications, limited by geometric shapes of scanned objects and the scanning system, problems of limited-angle CT imaging due to missing projection data have emerged. Fields like circuit board tomography and breast imaging have been research focus and difficult issues in CT reconstruction algorithms. In recent years, artificial intelligence (AI) and deep learning have achieved rapid development, and it is widely recognized that deep learning may provid new ideas and means to tackle CT imaging problems [2-5]. In terms of limited-angle CT imaging, multiple deep learning-based limited-angle CT reconstruction algorithms have also emerged with good performance, such as AirNet proposed by Gaoyu Chen et al. [6].

Although many deep learning-based CT reconstruction algorithms have been built, their effective use is still hindered by a lack of training samples [7-8]. According to the latest research, to improve the generalization performance of few-shot learning, one should focus on two aspects: more effective feature decoupling and integration of physical information with network design [9]. Images reconstructed with limited-angle CT reconstruction algorithms have such distinct features as image blurring and serious streak artifacts along the directions perpendicular to the missing scanning angles, and images of acceptable quality in directions are not perpendicular to the scanning angles. Some researchers have conducted research based on the above characteristics, and improved image quality by introducing them to the reconstruction model [10].

In this paper, a light-weight deep learning model-based method for limited-angle CT reconstruction is proposed. The model is based on traditional methods of optimization which integrates features of images reconstructed with limited-angle CT into network design. The proposed algorithm combines a neural network with the existing limited-angle CT optimization model, it can obtain effective image features and reduce the number of required samples. In addition, it uses the attention mechanism to learn hyperparameters in the optimization model, we can avoid manual adjustment. It should be noted that, ray angle information is integrated into the model to design a specific neural network structure, thus enhancing the generalization ability of few-shot learning.

## 2 Methods

### A. Image Reconstruction Models

The CT reconstruction model is shown in (1):

$$A\vec{x} = p \qquad (1)$$

where $A$ is a large sparse matrix, $\vec{x}$ is the reconstructed image, $p$ is the projection image. The limited-angle CT reconstruction is a typical ill-posed problem, and the regularization is an effective solution[11-14]. It is usually described as:

$$\min_{\vec{x}} \frac{1}{2} \|A\vec{x} - p\|_w^c + \lambda \sum_{k=1}^{K} \lambda_k \phi_k (G_k \vec{x}) \qquad (2)$$

where $\|A\vec{x} - p\|_w^c$ is a measure of data fidelity, and $\lambda$ is used to balance the weight of the fidelity term and the regularization term. The second item is expressed as the sum of different regularization terms, in which K is the number of different regularisation items, $\lambda_k$ is the weight of each regular item, and $\phi(\cdot)$ is a measure, like $l_0$, $l_1$ and $l_2$, $G_k$ is a transformation type like sparse transform and wavelet transform. When solving (2), both the Alternating Direction Method of Multipliers and Split Bregman will inevitably increase the number of hyperparameters, and the determination of hyperparameters is very difficult.

In view of the difficulty in choosing regularization terms and its hyperparameters, many researchers have combined traditional methods with deep learning to learn the unknowns in (3). It can be expressed as:

$$x = \arg\min_{\vec{x}} \frac{1}{2} \|A\vec{x} - p\|_w^c + \sum_{k=1}^{K} \lambda_k NN_k (\vec{x}) \qquad (3)$$

where $c$ and $w$ are forms of measurement of the fidelity terms of network learning; $\lambda_k$ is the weight of each regular item, $NN_k(x)$ is a regularization term using neural network.

## B. Prior Regularized Neural for Limited-angle CT

In the field of industrial CT scanning, it cannot be ignored that the number of training sets for deep learning-based reconstruction or denoising is insufficient. To tackle the problem, methods such as training set expansion and introduction of prior knowledge are often adopted. By contrast, the second method is a more feasible approach [15]. In addition to the piecewise constant, the limited-angle CT reconstructed image also feature blurring and streak artifacts along the direction perpendicular to the missing rays. In other words, image blurring and streak artifacts are correlated to the direction of ray. Combining the characteristics of the limited-angle CT reconstruction image and the deep learning method, (3) is further expressed as:

$$\min_{\vec{x}} \frac{1}{2}\|A\vec{x}-p\|_2^2 + \lambda_h HNN(\vec{x}) + \lambda_v VNN(\vec{x}) + \lambda_r CNN(\vec{x}) \quad (4)$$

where $HNN(\vec{x})$ and $VNN(\vec{x})$ are used for edge-preserving extension and edge-preserving smoothing; $CNN(\vec{x})$ is the regularization term remainder; $h$ and $v$ stand for the horizontal and vertical directions; $\lambda_h$, $\lambda_v$ and $\lambda_r$ are to adjust the role of $HNN(\vec{x})$, $VNN(\vec{x})$, and $CNN(\vec{x})$ in the optimization function. (4) is the solution to a typical coupling problem, which can be broken down to the following steps:

$$\vec{x}^{k+1/4} = \arg\min_{\vec{x}}\left\{\|\vec{x}^k - \vec{x}\|_2^2 + \|A\vec{x}-p\|_2^2\right\} \quad (5)$$

$$\vec{x}^{k+2/4} = \arg\min_{\vec{x}}\left\{\|\vec{x}^{k+1/4} - \vec{x}\|_2^2 + \lambda_h HNN(\vec{x}^{k+1/4})\right\} \quad (6)$$

$$\vec{x}^{k+3/4} = \arg\min_{\vec{x}}\left\{\|\vec{x}^{k+2/4} - \vec{x}\|_2^2 + \lambda_v VNN(\vec{x}^{k+2/4})\right\} \quad (7)$$

$$\vec{x}^{k+1} = \arg\min_{\vec{x}}\left\{\|\vec{x}^{k+3/4} - \vec{x}\|_2^2 + \lambda_r CNN(\vec{x}^{k+3/4})\right\} \quad (8)$$

In this paper, (5) is solved approximately using Simultaneous Algebraic Reconstruction Technique(SART). (6-8) are approximated with Prior Regularized Neural Network (PRNN). The algorithm is presented in Table 1.

### Table 1. PRNN Algorithm steps

| Algorithm： |
| --- |
| 1.    **Inputs**： projection data $p$ |
| 2.    **Output**： $\vec{x}^N$ |
| 3.    initial image $\vec{x}^0 = 0$, iteration $N$ |
| 4.    **for** $k = 1: N$ **do** |
| 5.    $\vec{x}^{k+1/4} = SART(\vec{x}^k, p)$  # step 1. solving the fidelity term |
| 6.    $H^k = HNN(\vec{x}^{k+1/4})$  # step 2. Horizontal regularization |
| 7.    $\lambda_h^k = AttentionHNN(H^k, x^{k+1/4})$ |
| 8.    $\vec{x}^{k+2/4} = \lambda_h^k * H^k$ |
| 9.    $V^k = VNN(\vec{x}^{k+2/4})$  # step 3. Vertical regularization |
| 10.   $\lambda_v^k = AttentionVNN(V^k, x^{k+2/4})$ |
| 11.   $\vec{x}^{k+3/4} = \lambda_v^k * V^k$ |
| 12.   $R^k = CNN(\vec{x}^{k+3/4})$  # step 4. Other regularization |
| 13    $\lambda_r^k = AttentionCNN(R^k, x^{k+3/4})$ |
| 14.   $\vec{x}^{k+1} = \lambda_r^k * R^k$ |
| 15.   **end for** |

The overall network architecture of PRNN is shown in Fig. 1. It mainly includes 5 modules: traditional reconstruction (REC) module, HNN module, VNN module, CNN module, and Attention Mechanism module. HNN and VNN are shown in Fig. 2. The VNN module regularizes results of the previous module along the direction parallel to the missing rays, and it is used for image smoothing. The design is roughly the same with that of the HNN module, but different in taking each column of data as one set to input. The number of iterations $N$ is set to 10.

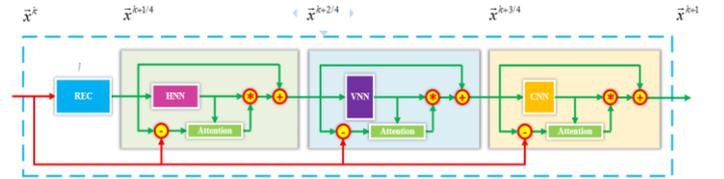

Fig 1. Integral network framework.

With the REC module reconstructed images, the HNN module uses a fully connected neural network to regularize along the direction perpendicular to the missing rays. The neural network is applicable for the actual research problem. For one thing, according to features of limited-angle CT images, we can reduce the nodes and network parameters of the input layer by learning line by line; for another, the traditional reconstruction algorithm has been obtained abundant information, so we can enhance network expression without increasing the number of layers.

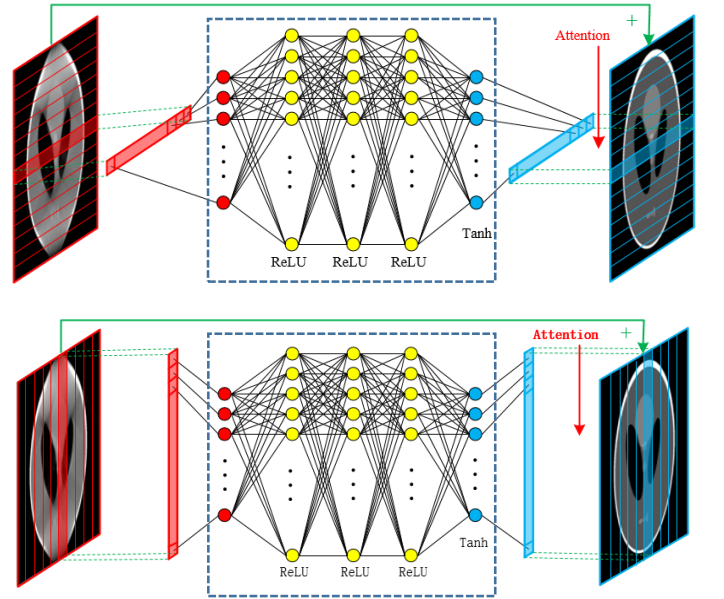

Fig 2. HNN-Net and VNN-Net

The module contains a 4-layers neural network. Each row of data in the horizontal direction (direction perpendicular to the ray) of the reconstructed images is input as one set of the network. When the image width is W, the number of neurons and image width in the input layer and the output layer will also be W, and the number of neurons for the hidden layers is 2W+1 [16]. Residuals between input images and tag images are used as the network label for training. The sum of output and input of the network is taken as corrected images.

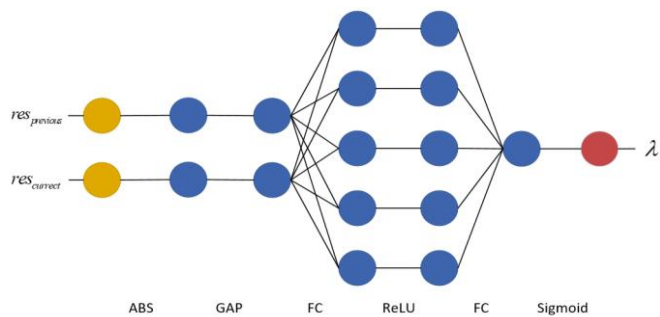

Fig 3. Attention-Net

The CNN module is a 7-layers residual network using 64*3*3 convolution filters, where residuals between input images and label images are used for training. The module trains weighted parameters in (4). According to actual demand, we modified the basic attention network [17]. As shown in Fig. 3, there are two inputs: I. the difference between preposition module output and SART module input; II. The residual of the previous module output. First, we perform a non-negative operation on the input, and then enter the results to the Squeeze-and-Excitation module, eventually obtain the residual coefficient λ of the current term.

## 3 Experiment

In this section, a simulation experiment is used to verify the feasibility and validity of the proposed algorithm, and to evaluate its performance through comparison with the following algorithms: SART, SART-TV, and Air-Net.

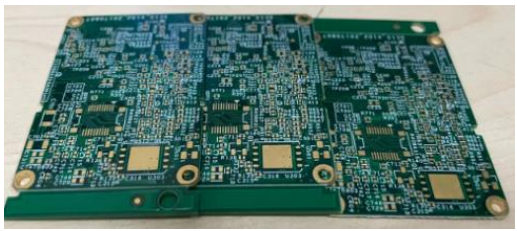

Fig 4. PCB

The experiment adopts two datasets, TCIA Collections and real PCB. The TCIA Collections dataset includes data of 20 patients, from which 10 patients are sampled at the same interval to form the training set, and the remaining patients sampled in the same way to form the test set, image size being 256*256. The PCB dataset includes 7 real PCB omnidirectional data, in which 5 data sampled at the same interval to form the training set, and the remaining sampled in the same way to form the test sets. The experimental environment is specified as: 1) Hardware: CPU: Intel Core Processor 2.59GHz (16 cores); memory: 64GB; GPU: Tesla P100 16GB; 2) Software: MatConvNet deep learning framework and Matlab R2019b.

To better evaluate the proposed algorithm, peak signal-to-noise ratio (PSNR) and structural similarity index measure (SSIM) are adopted for a quantitative evaluation [18].

In this study, we select two representative images from two test sets for display. Fig. 5 shows results from different methods at the sample size of 100 in the projection range of [-45°,45°]. The first row represents reconstruction results of the LUNG using different methods. SART shows obvious artifacts; SART-TV does reduce some artifacts but fails to restore the image structure accurately; Air-Net restores the structure of the image of the lung, the border of the result is blurred, but for PCB, SART-TV is better than Air-Net. Air-Net is not suitable for reconstructing this type of data. While the proposed algorithm better restores structural information, the smoothness of the reconstructed images still needs to be improved compared with the reference. The second row shows reconstruction results of the PCB using different methods. Compared the above methods, the algorithm herein better reconstructs edge information, reduces artifacts, and improves image quality. Besides, we calculate the SSIM and PSNR of the reconstructed images, and extract a row of pixel values for comparison. The SSIM and PSNR of PRNN are higher than other methods. The number of parameters of PRNN is reduced. They are shown in Table 3. PRNN's pixel curve is closer to the reference curve, as in Fig. 6.

In the PCB dataset, different methods are adopted and experiments are conducted for comparison at different scanning ranges. It is known to all that as an important test method of PCB, X-ray can detect through-hole copper fractures, soldering joint quality, and cracks, among others. It can be found from Fig. 7 that the images of PRNN show PCB through-hole copper more clearly within [-75°,75°], and the image quality is better.

In addition to the above experiments, this paper also examines the effect of the sample size on the algorithm. The roles of the HNN, VNN, CNN module in the algorithm are also analyzed. In the learning process, the sample size is closely related to an abundance of image feature information. The effect of sample sizes of 100 and 200 on PCB data in the paper is tested within the scanning range of [-45°,45°], it is shown in Fig. 8. In Fig. 9, two curves of pixel values obtained from training are not completely consistent with the reference; the curve of the sample size 100 and that of the sample size 200 are different, but the difference is acceptable because they have similar waveforms. $\lambda_h$, $\lambda_v$ and $\lambda_r$ balance the fidelity term and the regular term. In other words, they are also used to control the introduction of errors. In this article, the error value is set not higher than sum of all pixels $*10^{-2}$.

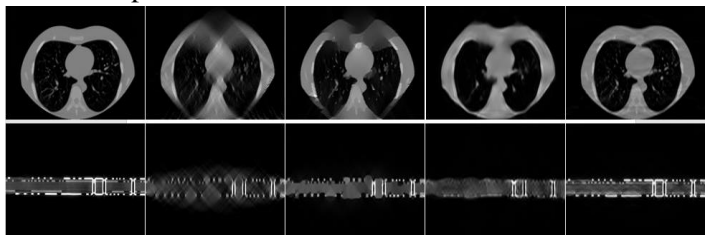

Fig 5. 1st column is the reference image, and 2nd -5th column are the images reconstructed by SART, SART-TV, Air-Net and PRNN.

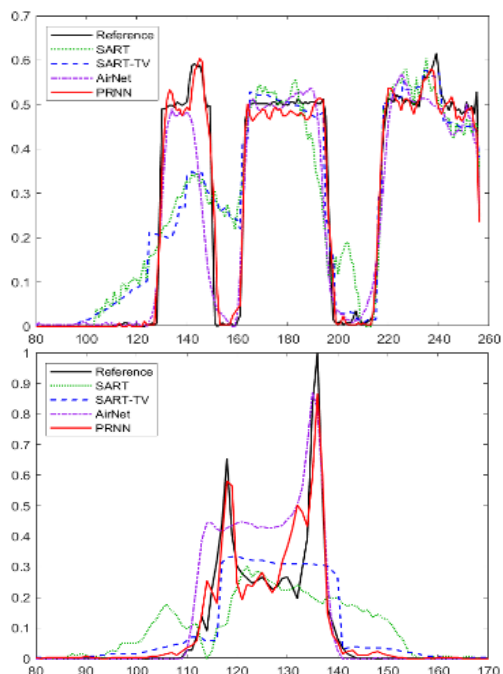

Fig 6. The two pictures are the line graphs of the pixel values of a certain row in the extraction result: LUNG(left) and PCB(right).

**Table 3. PSNR and SSIM :when the scanning range is [-45°,45°]**

| Category | Algorithm | Number of parameters | PSNR | SSIM |
|---|---|---|---|---|
| LUNG | SART | | 21.35330 | 0.91925 |
| | SART-TV | | 21.43434 | 0.92127 |
| | Air-Net | 3,000,000 | 25.26701 | 0.97020 |
| | **PRNN** | **1,854,720** | **31.8804** | **0.9933** |
| PCB | SART | | 23.18693 | 0.81345 |
| | SART-TV | | 24.98306 | 0.88787 |
| | Air-Net | 3,000,000 | 24.23246 | 0.89024 |
| | **PRNN** | **1,854,720** | **30.58110** | **0.97000** |

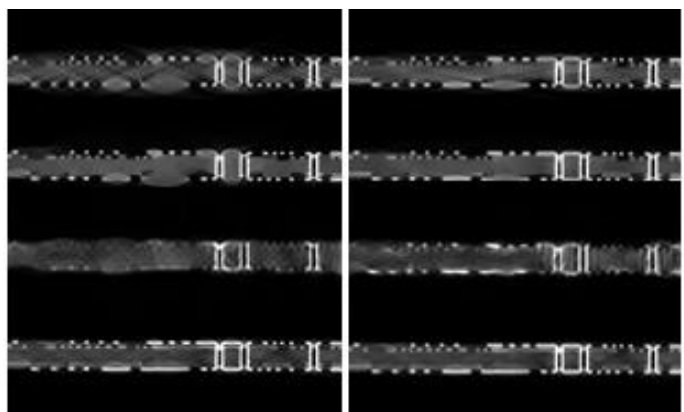

Fig 7. When the scanning range are [-60°,60°] and [-75°,75°], the PCB uses SART, SART-TV,Air-Net and PRNN to obtain the results. 1st column is the result of [-60°,60°], and 2nd column is the result from [-75°,75°].1st~4th rows are the results of the above four methods.

The REC module plays an important role in providing rich information to facilitate other modules' input and reducing the number of required samples. In the PRNN, we adopt a fully connected network alone to implement HNN and VNN modules, reduce parameters of each operation, and simplify the network structure. In the meantime, the introduction of Attention and CNN modules improves network generalization to a certain extent. The experimental results show that the CNN module not only learns regularization term remainders, but also overcomes the drawback of a fully connected network that ignores the spatial structure of images. The Attention module can reasonably control hyperparameters values.

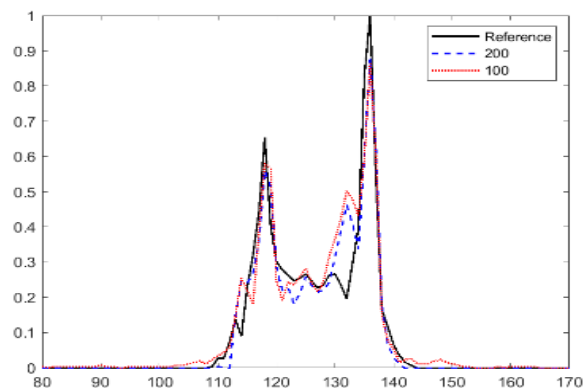

Fig 8. This is the curve when we use different capacity.

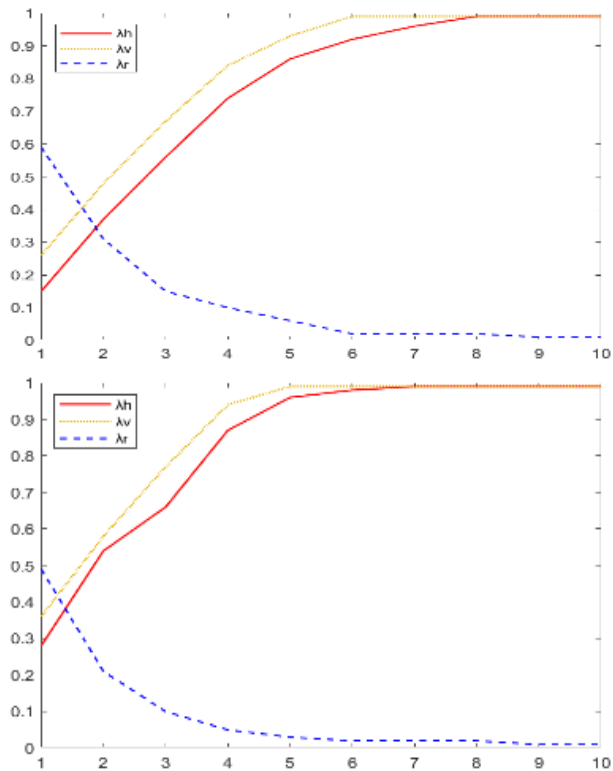

Fig 9. The curve of the weight parameters: LUNG(left) and PCB(right).

## 4 Conclusion

In this paper, a limited-angle CT reconstruction method is proposed based on a few-shot lightweight learning model, whose effectiveness is verified with both simulated and real data. We can conclude that PRNN reduces the number of samples required, simplifies the network structure, and enhances model generalization and improves image quality. Combined with traditional methods of optimization and reconstruction and characteristics of limited-angle CT reconstructed images, the modularized network structure improves network adaptability while making full use of features of limited-angle CT reconstructed images. PRNN, whose experiment is conducted on two datasets in the paper, reduces the number of required samples, simplifies the network structure, and enhances model generalization and image quality. However, due to limited types of training samples, more data are needed to test the performance of the algorithm.

# Application of Time Separation Technique to Enhance C-arm CT Dynamic Liver Perfusion Imaging

Hana Haseljić[1], Vojtěch Kulvait[1], Robert Frysch[1], Bennet Hensen[2], Frank Wacker[2], Georg Rose[1], and Thomas Werncke[2]

[1]Institute for Medical Engineering and Research Campus STIMULATE, University of Magdeburg, Magdeburg, Germany
[2]Institute of Diagnostic and Interventional Radiology, Hannover Medical School, Hannover, Germany

**Abstract** Perfusion imaging is an interesting new modality for evaluation and assessment of the liver cancer treatment. C-Arm CT provides a possibility to perform perfusion imaging scans intra-operatively for even faster evaluation. The slow speed of the C-Arm CT rotation and the presence of the noise, however, have an impact on the reconstruction and therefore model based approaches have to be applied. In this work we apply the Time separation technique (TST), to denoise data, speed up reconstruction and improve resulting perfusion images. We show on animal experiment data that Dynamic C-Arm CT Liver Perfusion Imaging together with the processing of the data based on the TST provides comparable results to standard CT liver perfusion imaging.

## 1 Introduction

Perfusion CT imaging is an important modality for the treatment of the liver cancer, see [1]. It would be the additional benefit to have C-arm CT available as part of the interventional suite, see [2, 3]. The ability of C-arm systems to measure parenchymal blood volume (PBV) has been the subject of research during the last few years, see [2, 4–7]. The study in [8] has used animal model to evaluate the dynamic reconstruction algorithm [9] and measure Arterial Liver Perfusion using C-arm CT data.

Using C-Arm CT perfusion imaging in the liver cancer management could provide an option to evaluate the performed embolization intraoperatively. The data moreover could be used to plan ablation. The slow rotation time, limited number of projections and time gap between rotations causes undersampling and artifacts in the contrast agent dynamics reconstruction. Neglecting these limitations by static reconstruction i.e. reconstruction of each rotation individually as if it was a native CT scan, will cause loss of accuracy and could result in incorrect perfusion measurements, see [10]. To overcome these problems we use the model-based approach to describe time attenuation curves (TAC) as a weighted sum of temporal basis functions, so called Time separation technique (TST), see [11, 12].

In this paper we use the data from the C-Arm CT perfusion reconstruction of the swine liver. We are solving the aforementioned problems by using an analytical basis derived from Fourier analysis and applying TST. Then we use the deconvolution based algorithms together with Tikhonov stabilization to compute perfusion parameters, see [13]. Using this approach we reduce noise in the data and show that produced perfusion maps are comparable to the CT perfusion data.

## 2 Materials and Methods

### Procedure

We used 2 anaesthetized domestic pigs to perform C-Arm CT perfusion scans of the liver using iodinated contrast agent after the embolization that induced the area of decreased perfusion. To induce areas of hypoperfusion in the swine liver model we embolized branches of the right hepatic artery with tantalum-based embolization material (Onyx) and coils. Two matching C-arm and CT perfusion scans were acquired using Siemens ARTIS pheno C-arm and SOMATOM Force CT. A 15ml of contrast material Imeron 300 was injected with the duration of 5s. The tube voltage was set to 90 kVp. Each C-arm scan consisted of five forward-backward sweep pairs. Each sweep covered rotation of 200° and with angular step of 0.8° acquired 248 projections. The scans with the two scanners were performed ten minutes apart to insure the contrast material has washed out.

### Time separation technique

Let the interval $\mathscr{I} = [0, T]$ represent the duration of the scan. The TACs are modeled as a linear combination of a defined set of $N$ orthogonal functions

$$\mathscr{B} = \{\Psi_1, \ldots, \Psi_N\}, \tag{1}$$

where for each $i \in \{1, \ldots, N\}$ $\Psi_i = \Psi_i(t), t \in I$ are scalar functions of the time. We call the set $\mathscr{B}$ basis and refer to functions $\Psi_i$ as to the basis functions. In practice these functions might be analytical functions and therefore $I = \mathscr{I}$ or $\Psi_i$ can be represented as a vector of its values in the $M$ time points $I = \{0, T/(M-1), \ldots, T\}$. According the TST, the time attenuation curve in a particular volume point $x_v$ is given by the linear combination of the basis functions

$$x_v(t) = \sum_{i=1}^{N} w_{v,i} \psi_i(t). \tag{2}$$

Under the assumption of the orthogonality of the basis functions, we can transform the contrast agent dynamic reconstruction problem to the $N$ standard CT reconstruction problems to reconstruct weight coefficients $w_{v,i}$, see [12] for the details.

To do so, we assume that the projection data for any C-Arm spatial configuration, namely angle and pixel position, encoded by index $k$ satisfies

$$p_k(t) = \sum_{i=1}^{N} \omega_{k,i} \psi_i(t). \qquad (3)$$

Note that by means of equations (2) and (3) we separate time development, encoded by basis functions $\Psi_i(t)$, from the spatial configuration encoded by weighting coefficients $w$ and $\omega$ respectively, thus the method is called time separation technique.

In order to reduce noise in the data and extract important information about the contrast agent dynamics, we use trigonometric functions as a basis. Based on the numerical experiments we figured out that $N = 5$ provides a good tradeof between the number of reconstructions and the image quality. Therefore we have chosen $N = 5$ and $\Psi_i$ to be

$$\Psi_0 = 1, \quad \Psi_1 = \sin(\frac{2\pi t}{T}), \quad \Psi_2 = \cos(\frac{2\pi t}{T}),$$
$$\Psi_3 = \sin(\frac{4\pi t}{T}), \quad \Psi_4 = \cos(\frac{4\pi t}{T}). \qquad (4)$$

These functions are orthogonal with respect to the scalar product $< \Psi_i, \Psi_j > = \int_0^T \Psi_i(t)\Psi_j(t)\,\mathrm{d}t$. To find the weighting coefficients of the projection data $w^p$, we performed a least squares fitting. This was followed by the 40 iterations of the algebraic reconstruction, see [14] [1] and the TAC data was obtained using (2).

In case of CT perfusion data, the acquisition on SIEMENS SOMATOM Force was followed by the analytical reconstruction using Br36 kernel in syngo CT VA50A software. The reconstructed volume data was interpolated by means of cubic splines to obtain TAC data. The estimation of the perfusion maps by our software is identical for both modalities by the method that follows.

**Perfusion Parameters Estimation**

The artery input function (AIF) is needed to compute the blood flow through the organ. It describes the contrast agent flow over time. To generate comparable perfusion maps artery was detected as suggested in [13].

We estimate TAC in every voxel as a convolution of AIF with residual function

$$\mathrm{tac}(t) = \mathrm{aif}(t) * f_r(t). \qquad (5)$$

We discretize the function $aif(t)$ by its values in $k = 100$ time points. Then we apply the pseudoinverse with Tikhonov regularization to (5) in order to recover function $f_r(t)$.

---

[1] Source code of the reconstruction technique, namely CGLS, which has been used can be found at **https://bitbucket.org/kulvait/kct_cbct**.

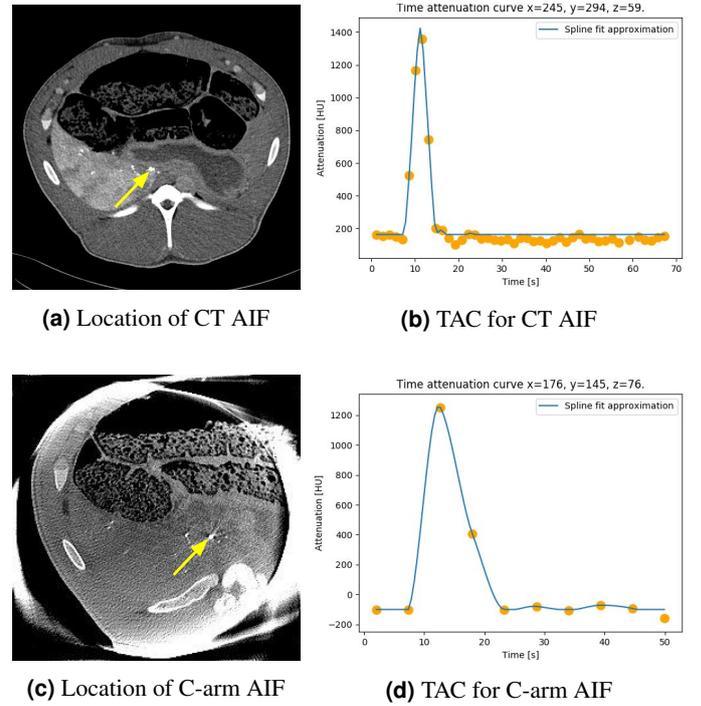

**(a)** Location of CT AIF          **(b)** TAC for CT AIF

**(c)** Location of C-arm AIF          **(d)** TAC for C-arm AIF

**Figure 1:** C-arm and CT artery input function

We compute four perfusion parameters, blood flow (BF), blood volume (BV), mean transit time (MTT) and time to peak (TTP) using the following formulas from [13]:

$$BF = \max f_r(t), \quad BV = \sum_{i=1}^{n} f_r(\mathrm{i}),$$
$$MTT = \frac{BV}{BF}, \quad TTP = \arg\max_t f_r(t). \qquad (6)$$

## 3  Results

We processed the data from the swine liver perfusion using methods described in M&M section. To compute perfusion parameters, first we have to locate the arterial inlet in order to derive AIF function. In Figure 1 the detected location of AIF is shown together with the time attenuation profile of this voxel from both modalities. Due to the injection duration, an undersampling of the contrast material peak in C-arm CT compared to CT can occur.

Using the model (5) and (6) we have computed perfusion maps. The results for a selected C-arm slice and corresponding CT slice are shown in Figure 2. The perfusion maps generated by the means of described TST technique are given in the first row, in the second row are the results from the TAC obtained by spline interpolation of the static reconstructions and the third row contains the CT perfusion maps.

To compare the results we mainly look for the hypoperfused areas. They are easily distinguishable in the reconstruction image, see Figure 3. We can clearly observe these areas in the perfusion maps as they have different color, see Figure 2.

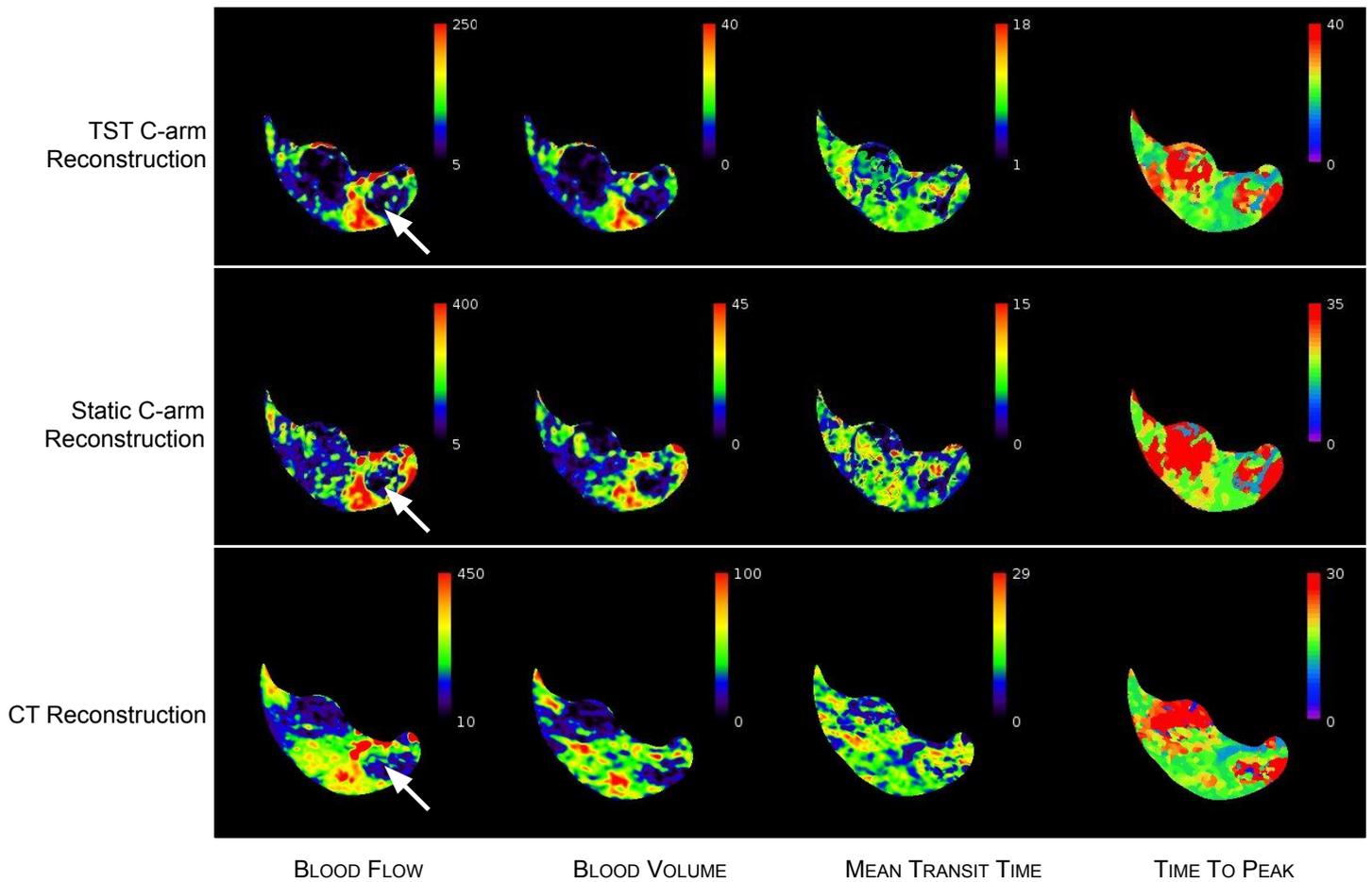

BLOOD FLOW    BLOOD VOLUME    MEAN TRANSIT TIME    TIME TO PEAK

**Figure 2:** Perfusion maps, BF in mL/100g/min, BV in mL/100g, MTT in s and TTP in s, white arrow is pointing to the area of reduced perfusion induced by embolization.

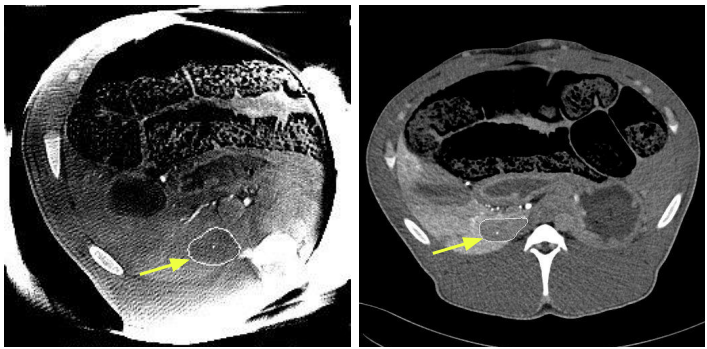

**Figure 3:** Hypoperfused area in C-arm (left) and CT (right).

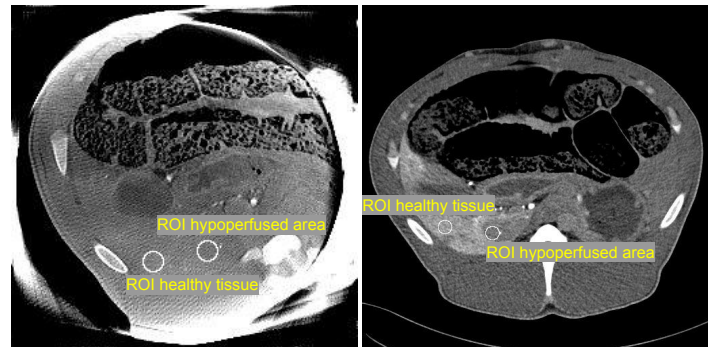

**Figure 4:** C-arm (left) and CT (right) regions of interest.

The unhealthy tissue induced by embolization is very well distinguishable namely in the BF and BV maps of the TST reconstruction, where they have dark blue color code.

We observe that in the TST C-arm reconstruction perfusion maps the hypoperfusion area is much more pronounced than in the static C-arm reconstructions.

To better evaluate the differences between hypoperfusion area and healthy tissue, we selected regions of interest in both CT and C-Arm CT images, see Figure 4. We then computed mean value and standard deviation of the respective perfusion coefficient.

We have found that not only the means are different, but when performed T-test to find out the significance of that difference, we found a p-value of $< 10^{-10}$. The mean value and standard deviation of BF and BV for selected regions are given in Tables 1 and 2.

## 4 Discussion

In this paper we have shown that C-Arm CT liver perfusion imaging can provide similar results as CT perfusion imaging when we use an adequate model based approach.

| Reconstruction | Hypoperfused area | Healthy tissue |
| --- | --- | --- |
| TST | $20.7 \pm 20.4$ | $95.6 \pm 24.3$ |
| Static | $38.8 \pm 39.6$ | $133.4 \pm 41.4$ |
| CT | $91.0 \pm 32.9$ | $1162 \pm 122$ |

**Table 1:** BF measurements in units of mL/100g/min. Mean and the standard deviation taken over the selected regions of interest.

| Reconstruction | Hypoperfused area | Healthy tissue |
| --- | --- | --- |
| TST | $2.3 \pm 3.0$ | $13.4 \pm 3.1$ |
| Static | $1.9 \pm 2.4$ | $19.1 \pm 5.0$ |
| CT | $18.8 \pm 8.6$ | $58.9 \pm 14.1$ |

**Table 2:** BV measurements in units of mL/100g. Mean and the standard deviation taken over the selected regions of interest.

It can be seen that proposed basis set of the TST provides a perfusion maps that reduce the noise and clearly separates the hypoperfusion areas from healthy tissues, see Figure 2 and Tables 1 and 2. Thus, this method enables detection of hypoperfusion regions and has the potential to be introduced into clinical practice.

C-arm perfusion maps show also some differences in well perfused areas when compared to CT perfusion maps. The difference in values are mostly noticeable on BF and BV. This can be mainly attributed to the fact that the positioning of the animal in two different modalities is different but the different setup of the reconstruction for the two modalities can also play its role, see [6]. C-Arm and CT devices also were not calibrated to provide equal attenuation values.

## 5    Conclusion

From the results it can be seen that model based reconstruction of the C-Arm CT perfusion scans outperforms methods based on individual static reconstructions. The data are less noisy and the area with reduced perfusion is more visible on perfusion maps. Therefore it provides the results comparable to the CT perfusion imaging. We plan to perform clinical evaluation of these data to assess whether this approach should be part of the clinical setup.

Additional improvement of the TST results is expected by including dedicated perfusion basis functions based on CT data as prior knowledge, as studied for brain data in [12, 15].

## Acknowledgments

*This work was partly funded by the European Structural and Investment Funds (International Graduate School MEMo-RIAL, project no. ZS/2016/08/80646) and the German Ministry of Education and Research (Research Campus STIMULATE, grant no. 13GW0473A and 13GW0473B).*

# Laplace-Beltrami Regularization for Anisotropic X-Ray Dark-field Tomography

Ana Radutoiu[1], Theodor Cheslerean-Boghiu[1], Franz Pfeiffer[2], and Tobias Lasser[1]

[1]Computational Imaging and Inverse Problems, Department of Informatics and Munich School of BioEngineering, Technical University of Munich, Germany
[2]Chair of Biomedical Physics, Department of Physics and Munich School of BioEngineering, Technical University of Munich, Germany

**Abstract** Anisotropic X-ray Dark-field Tomography (AXDT) enables visualization of microstructure orientations via Talbot-Lau grating interferometers. The measured dark-field signal produced by the small-angle scattering of X-rays passing through the sample can be used to reconstruct spherical scattering functions inside the sample, which in turn allows the extraction of the orientation of microstructures that are otherwise invisible. Dark-field measurements are prone to high levels of noise, indicating the use of regularization in the AXDT reconstruction process. In this work, we introduce Laplace-Beltrami operator for spherical harmonics as a regularization method for the AXDT inverse problem. To validate our model, we present two different experiments, one performed on synthetic data and one executed on experimental data acquired by measuring a wooden sample.

## 1 Introduction

Anisotropic X-ray Dark-field Tomography (AXDT) enables the reconstruction of spherical scattering functions from the dark-field measurements [1]. The dark-field signal is produced by the ultra-small angle scattering effects of X-rays when they pass through structures smaller than the voxel size and can be measured using a Talbot-Lau grating interferometer [2]. In order to perform tomographic reconstruction of the spherical scattering functions, AXDT discretizes them using spherical harmonics (SH) and seeks to compute their respective spherical harmonics coefficients.

In previous works, the AXDT inverse problem has been solved using conventional non-regularized least squares-based algorithms [1, 3]. As the measured dark-field contrast obtained with a grating interferometer is very noisy, adjustments to stop the iterative non-regularized reconstruction process before the solution overfits to noise are advantageous. A regularization strategy to minimize the impact of measurement noise would, therefore, be highly desirable, but so far there is a lack of successful developments in this direction.

For this work, we will look to regularization methods for Q-Ball imaging (QBI) in the context of Diffusion MRI for inspiration. In QBI, spherical functions are used to represent orientation distribution functions (ODF) at each voxel location. One such approach was introduced by Descoteaux et al. [4] who introduced the Laplace-Beltrami operator, an extension of the Laplacian operator to functions defined on the unit sphere to penalize the high order spherical harmonics terms which are usually only modelling noise in the system and to leave those that are necessary to describe the underlying ODF.

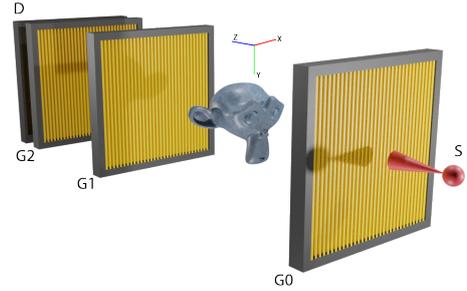

**Figure 1:** Schematic of an X-ray grating interferometer. The stream of X-ray photons generated by source S illuminates the sample through the source grating G0. The phase grating G1 creates an interference pattern that is sampled by the analyzer grating G2 in front of the X-ray detector D. For AXDT, either the sample or the grating interferometer have to be rotated around all three axes in order to fully sample the scattering functions.

As the tomographic reconstruction of the ODFs in QBI is equivalent to the one of the spherical scattering functions in AXDT, in this work we build upon the works of Wieczorek et al. [3] and Descoteaux et al. [4], introducing a new regularized tomographic reconstruction approach in the context of AXDT based on the Laplace-Beltrami operator. Additionally, we perform a suite of tests with different regularization parameters in order to find the optimal one for the reconstructions.

## 2 Laplace-Beltrami Regularization for AXDT

### 2.1 AXDT Forward Model Discretization

In order to allow the extraction of multiple fiber directions inside each voxel, the forward model assumes the integration over all scattering directions on the unit sphere. The continuous model can be expressed as follows:

$$-\ln d_j = \frac{1}{4\pi} \int_{\mathbb{S}^2} h(\widehat{\varepsilon}, \widehat{\mathbf{t}}_j, \widehat{\mathbf{s}}_j) \left[ \int_{L_j} \eta(\widehat{\varepsilon}, \mathbf{r}) \, ds \right] \mathrm{d}^2 \widehat{\varepsilon} \quad (1)$$

for $j = 1, \ldots, J$ where we aim to reconstruct the scattering amplitude

$$\eta_k(\mathbf{r}_i) = |\varepsilon_k(\mathbf{r}_i)|^2 : \mathbb{R}^3 \to \mathbb{R}$$

for each direction $\widehat{\varepsilon}_k$ and at every voxel $\mathbf{r}_i$, $i = 1, \ldots I$, where $I$ is the number of voxels. Here, $d_j$ denotes the $j$th scalar dark-field measurement, $L_j$ the corresponding X-ray with the normalized incident direction $\widehat{s}_j \in \mathbb{S}^2$, and $\widehat{t}_j \in \mathbb{S}^2$ the sensitivity direction of the grating interferometer, and $h \colon \mathbb{S}^2 \times \mathbb{S}^2 \times \mathbb{S}^2 \to \mathbb{R}$ is a weighting function representing the physical interaction process of the X-ray with the sample, as defined in [3]. In order to discretize our continuous problem, let us consider the spherical harmonics expansion of the scattering amplitudes:

$$\eta(\widehat{\varepsilon}, \mathbf{r}) = \sum_{n=0}^{N} \sum_{m=-n}^{n} \eta_n^m(\mathbf{r}) Y_n^m(\widehat{\varepsilon}). \qquad (2)$$

Next, we discretize the continuous model from eq. (1) and we obtain the following forward model:

$$\mathbf{A}\eta = \mathbf{b}, \qquad (3)$$

where $\eta$ represents the scattering functions vector $\eta = \begin{bmatrix} \eta_0^0 & \dots & \eta_N^{-N} & \dots & \eta_N^N \end{bmatrix}$ and $\eta_n^m$ is a vector that stacks the evaluation of the spherical harmonic coefficient of order $n$ and phase $m$ at each voxel, $\mathbf{A}$ is the forward operator encoding weighting functions and projection matrix, and $\mathbf{b}$ is the vector of dark-field measurements. For more details about the forward model we refer to ref. [3].

## 2.2 The Regularization Term

If we want to perform reconstruction on the AXDT inverse problem, we will be facing a discrete ill-posed problem. To mitigate the accuracy issues caused by this, we employ regularization with the aim of damping the noise. For this, we start by considering the well-know method of Tikhonov regularization. This approach consist in finding a minimum for the objective function

$$F_\lambda(\eta) = \frac{1}{2}||\mathbf{A}\eta - \mathbf{b}||_2^2 + \lambda R(\eta), \qquad (4)$$

where

$$||\mathbf{x}||_2^2 = [\mathbf{x}]_1^2 + \dots + [\mathbf{x}]_N^2 \qquad (5)$$

is the $L_2$-norm of the vector $\mathbf{x} \in \mathbb{R}^N$, $\lambda$ the regularization parameter, and $R(\eta)$ the regularization term.

## 2.3 Laplace-Beltrami-Based Regularization

Because the reconstruction is performed in the spherical harmonics coefficient space, we have to choose an appropriate regularization term that controls the magnitude and/or the smoothness of the obtained solution. Consequently, we enforce smoothness of $\eta(\widehat{\varepsilon}, r)$ with respect to the angular variable $\widehat{\varepsilon}$ using the Laplace-Beltrami operator, which is equivalent to the Laplace operator in spherical coordinates. Another important aspect is the simplicity of evaluating the Laplace-Beltrami operator, denoted by $\triangle_b$, on spherical harmonics, which is simply

$$\triangle_b Y_n^m = -n(n+1) Y_n^m, \qquad (6)$$

where $Y_n^m$ represents the spherical harmonic of order $n$ and phase $m$. Using this information, we can now define the regularization term in a continuous setting as

$$R(\eta) := \frac{1}{2} \int_{\mathbb{S}^2} \int_D |\triangle_b \eta(\widehat{\varepsilon}, \mathbf{r})|^2 \, \mathrm{d}V \mathrm{d}^2 \widehat{\varepsilon}, \qquad (7)$$

where $\triangle_b$ acts on $\widehat{\varepsilon} = (\theta, \varphi)$, while the corresponding discrete version is

$$R(\eta) = \frac{1}{2} \sum_{i=1}^{I} \int_{\mathbb{S}^2} (\triangle_b \eta(\widehat{\varepsilon}, \mathbf{r}_i))^2 \, \mathrm{d}^2 \widehat{\varepsilon}. \qquad (8)$$

We apply the Laplace-Beltrami operator on Eq. 2:

$$\triangle_b \eta(\widehat{\varepsilon}, \mathbf{r}_i) = \sum_{n=0}^{N} \sum_{m=-n}^{n} \eta_n^m(\mathbf{r}_i)(\triangle_b Y_n^m(\widehat{\varepsilon}))$$
$$= -\sum_{n=0}^{N} \sum_{m=-n}^{n} n(n+1) \eta_n^m(\mathbf{r}_i) Y_n^m(\widehat{\varepsilon}).$$

We can now define the regularization term, making use of the orthogonality property of spherical harmonics:

$$R(\eta) = \frac{1}{2} \sum_{i=1}^{I} \int_{\mathbb{S}^2} (\triangle_b \eta)^2 \, \mathrm{d}^2 \widehat{\varepsilon}$$
$$= \frac{1}{2} \sum_{i=1}^{I} \sum_{mn} \sum_{m'n'} nn'(n+1)(n'+1) \eta_n^m(\mathbf{r}_i) \eta_{n'}^{m'}(\mathbf{r}_i)$$
$$\times \left[ \int_{\mathbb{S}^2} Y_n^m(\widehat{\varepsilon}) Y_{n'}^{m'}(\widehat{\varepsilon}) \, \mathrm{d}^2 \widehat{\varepsilon} \right]$$
$$= \frac{1}{2} \sum_{n=0}^{N} \sum_{m=-n}^{n} n^2(n+1)^2 \left[ \sum_{i=1}^{I} \eta_n^m(\mathbf{r}_i) \eta_n^m(\mathbf{r}_i) \right]$$
$$= \frac{1}{2} \sum_{n=0}^{N} \sum_{m=-n}^{n} n^2(n+1)^2 (\eta_n^m)^T \eta_n^m$$
$$= \frac{1}{2} \eta^T D^T D \eta$$
$$= \frac{1}{2} ||D\eta||_2^2,$$

where $D$ represents a diagonal matrix with entries $n(n+1)$, where $n$ is the order of the corresponding spherical harmonics coefficient, and $\eta$ is the quantity that we aim to reconstruct, the spherical harmonics coefficients [5].

Finally, we rewrite the objective function from eq. (4):

$$\mathscr{F}_\lambda(\eta) = \frac{1}{2}(||\mathbf{A}\eta - \mathbf{b}||_2^2 + \lambda ||D\eta||_2^2), \qquad (9)$$

and we define its corresponding normal equation

$$(\mathbf{A}^T \mathbf{A} + \lambda D^T D)\eta = \mathbf{A}^T \mathbf{b}. \qquad (10)$$

As a linear problem, the solution can be computed in a straightforward manner, for example via the conjugate gradient method.

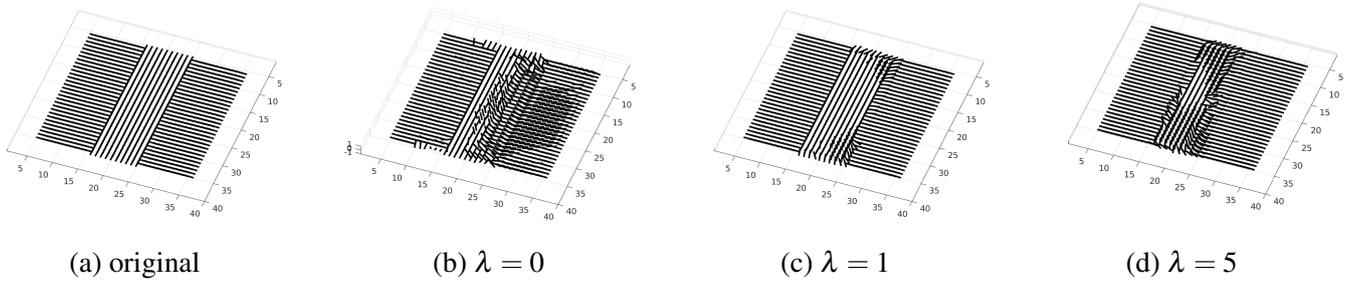

(a) original          (b) $\lambda = 0$          (c) $\lambda = 1$          (d) $\lambda = 5$

**Figure 2:** A slice of the synthetically generated volume for the first experiment. (a) is extracted from the original volume, while (b), (c) and (d) are slices of different reconstructed volumes using the proposed regularization method: (b) with $\lambda = 0$ (i.e. no regularization), (c) with $\lambda = 1$ and (d) with $\lambda = 5$. In (c) the fibers are much closer to the original (a) compared to (b) where noise is still present, while for $\lambda = 5$ the solution already appears to over-regularize.

## 3 Experiments and Results

### 3.1 Reconstructing the data

To test the effects of the regularization term on the reconstructed solution, the functionality of the tomographic reconstruction framework *elsa* [6] has been extended accordingly to include the Laplace-Beltrami-based regularization matrix. After the reconstruction, fiber extraction was performed using the Funk-Radon transform [1]. We ran our experiments on a machine with two Intel Xeon E5-2687W v2, 128GB of memory and two RTX 2080Ti.

We performed tests on two different data sets: one simulated measurements set from synthetically generated volume, and one experimental measurements set of two crossed wooden sticks obtained using the hardware setup from Fig. 1. The reconstructions were all obtained running 25 iterations of the conjugate gradient method, and for different values of the regularization parameter.

### 3.2 Assessing the quality of the synthetic data reconstruction

In order to evaluate the effects of the regularization parameter $\lambda$ on the reconstruction, we generated a $40^3$ voxel sample. The volume has been created starting from two different scattering profiles and populating three distinct regions with them, as presented in Fig. 2-a. We simulated dark-field measurements using *elsa* [6] and added normally distributed noise with mean 0 and variance $10^{-2}$. After that, we performed the reconstructions using 15 different regularization parameters and eventually extracted the fiber directions from the reconstructed data. To assess the quality of the data, we computed two different metrics: the mean absolute difference between the spherical harmonics coefficients vectors of the original and of the reconstructed data, which is a measure that shows the similarity between two spherical harmonics, and the mean dot product between the fibers of the original volume and the fibers of the reconstructed volume, which indicates the angle between the fibers and ranges between 1 (fibers are parallel) and 0 (fibers are perpendicular). Considering this, a good reconstruction then corresponds to a mean absolute difference score close to 0 and to a mean dot product score close to 1.

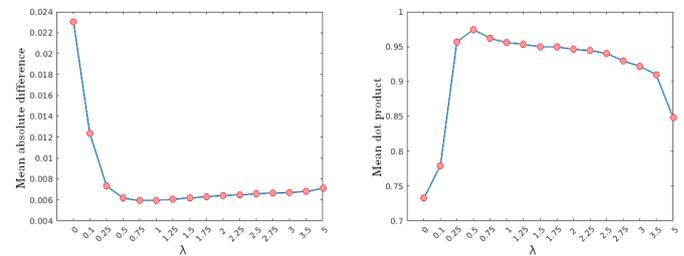

**Figure 3:** Fifteen different regularization parameters from 0 to 5 plotted against the mean absolute difference of the spherical harmonics coefficients *(left)* and the mean dot product scores of the extracted fibers *(right)* over the slice shown in Fig. 2. The best scores were obtained for $\lambda$ values close to 1 and represent a considerable improvement compared to the reconstructions performed without regularization.

Plotting the mean absolute difference and the mean dot product against the regularization parameters $\lambda$ for each of the 15 reconstructions, we observed optimal results for the reconstructions corresponding to values for $\lambda$ close to 1 (see Fig. 3).

### 3.3 Assessing the Quality of a Real Data Reconstruction

In the same manner as presented above, we performed reconstructions for the same $\lambda$ values of a $320^3$ volume from experimental data acquired using the setup showed in Fig. 1, containing two crossing wooden sticks (more detail on the experimental setup can be found in [3]).

In this case, the comparison between the original and the reconstructed data was not possible, as we do not have ground truth data of the crossed sticks volume. Instead, we assess the quality of the reconstruction through visual inspection of the results obtained with different values for $\lambda$ based on a priori assumptions about the sample: we can make the supposition that the wooden fibers from each stick are parallel along the length of the stick, thus we want to also observe this

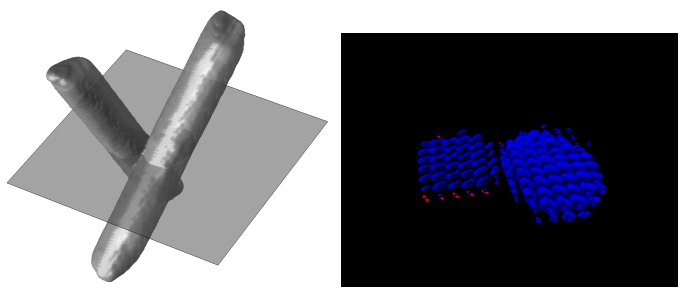

**Figure 4:** *Left:* 3D rendering of the two wooden sticks used for acquiring the data and the slice selected for the visual assessment of the result. *Right:* Respective rendering of the spherical scattering functions.

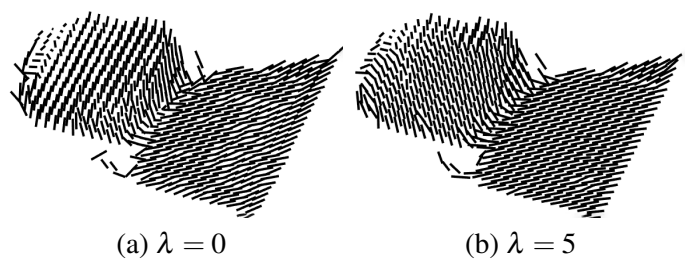

(a) $\lambda = 0$ · · · · · · · · · · · · · · (b) $\lambda = 5$

**Figure 5:** Two representative slices from the reconstructed wooden sticks volume. We extracted the dominant fiber orientations in each voxel by applying the Funk-Radon transform on the reconstructed scattering function. The area between the two sticks contains fibers facing in approximately opposite directions while fibers inside a particular stick are mostly facing the same direction. (a) reconstruction with $\lambda = 0$ (i.e. no regularization), (b) reconstruction with $\lambda = 5$. As the regularization term tries to reduce the impact of high order spherical harmonics in the reconstruction, the direction of the fibers can be imaged in a more accurate fashion, as the noise that correlates to high-order coefficients is reduced.

phenomenon in the visualized slice [7]. In other words, we visualize the reconstructed data and decide whether the observation meets our expectations. For this, we selected a slice around the region where the two sticks cross, as presented in Fig. 4. Two representative reconstructions are shown in Fig. 5 for $\lambda = 0$ (no regularization) and $\lambda = 5$. We remark that in the reconstruction slice corresponding to $\lambda = 5$, the fibers are arranged in a parallel manner, while the reconstruction with $\lambda = 0$ has fibers oriented in uneven directions. We can thus conclude that the regularized result closer resembles our expectations compared to the non-regularized one.

## 4 Conclusion

In this work we introduced a regularized approach to solving the AXDT inverse problem based on the Laplace-Beltrami operator for spherical harmonics. We showed that given an well-chosen regularization parameter we can obtain higher quality results compared to a conventional non-regularized method. We can conclude that both in the simulation and experimental study the addition of the regularization term improved the results. In the first case, good scores for the mean absolute difference and mean dot products metrics can be observed, while in the second case by employing the results we get a reconstruction that is closer to our expectations regarding the experimental sample. In future work, we will seek to implement a more robust algorithm of finding an optimal regularization parameter while also using a quantitative method to assess results when ground truth data is not present.

## 5 Acknowledgments

The authors would like to thank C. Jud and S. Seyyedi for recording the wooden sticks data set. We acknowledge financial support through the Munich Centre for Advanced Photonics (MAP), the DFG (Gottfried Wilhelm Leibniz program) and the European Research Council (AdG 695045). This work was carried out with the support of the Karlsruhe Nano Micro Facility (KNMF, www.kit.edu/knmf), a Helmholtz Research Infrastructure at Karlsruhe Institute of Technology (KIT).

# Learned energy-flexible algorithm using joint sparsity for dual-energy CT: a preliminary study

Donghyeon Lee[1] and Seungryong Cho[1]

[1]Department of Nuclear and Quantum Engineering, Korea Advanced Institute of Science and Technology, Daejeon, Republic of Korea

**Abstract** In dual-energy computed tomography (DECT), it is often the case that an x-ray energy pair differ from site to site and from even scan to scan. In this study, in an attempt to come up with a flexible material decomposition engine, we propose a deep learning (DL)-based algorithm that can be easily applicable for various scanning conditions of DECT. The proposed network is trained to find the optimal energy-flexible sparsifying transform based on the training data, and enhance image quality by using joint sparsity between dual-energy images. We demonstrate the performance of the proposed method not only for the pairs of energies (80kVp and 140/80kVp) used for training, but also for the other unused pair (100/140kVp). We expect that the proposed algorithm can greatly increase the material decomposition accuracy in DECT and improve the overall diagnostic performance of DECT.

## 1 Introduction

Dual-energy computed tomography (DECT) is a widely used imaging modality that can increase the diagnostic performance through providing material-specific information or monoenergetic image in various clinical practices. Applications of DECT are diverse depending on the purposes including: virtual non-contrast imaging [1], bone removal in vascular imaging [2], characterization of renal stones [3], differentiation of hemorrhage in neuroimaging [4], etc. Material decomposition process is based on the nonlinear attenuation characteristics of the objects for two different x-ray energy spectra.

Recently, deep learning (DL) approaches have received much attention in many fields including medical imaging due to their successful performance and potential, and several DL-based methods for dual-energy or spectral CT have also been proposed [5]–[7]. However, the existing DL-based methods that are aimed at achieving both reconstruction and material decomposition have several difficulties in applying to real systems. The performance of the DL-based methods largely depend on the amount and the quality of training data, but it is usually challenging to obtain enough data for training in DECT. This is because not only does the combination of materials vary greatly according to the purposes (i.e., iodine/tissue [1], bone/tissue [2], liver tissue/fat [8], xenon or krypton/lung tissue [9], etc.), but also the pair of x-ray tube energies can differ depending on the scanning anatomical target organ or on the patient obesity. The two standard energies that have been typically employed in DECT are 80 kVp and 140 kVp, but 80/120 kVp and 100/140 kVp have also been used [10]. It should be noted that, even with the same system setting, there may be drift in measured energy spectra as the equipment ages.

The purpose of this research is to reduce the aforementioned difficulties by proposing an energy-flexible algorithm that uses joint sparsity between spectral images to improve image quality. The proposed DL network is given one of the sinograms as an input and enhances image quality by incorporating another sinogram acquired at different energies. We named it the energy-flexible algorithm since other sinograms measured at "different" energy settings can go through the same network that has been trained using sinogram data acquired at a preset energy pairs and the algorithm can still enhance the image quality. The energy-flexible algorithm was inspired by the previous studies in the iterative image reconstruction framework that use several sparsifying transforms to find joint sparsity of the images obtained at different energy settings and improve image quality [11]–[13]. In these studies, minimizing the differences in sparsifying transform of different energy images was adopted as a regularizer in the iterative reconstruction algorithms. On the other hand, in this study, the proposed energy-flexible algorithm is learned not only to find the optimal sparsifying transform based on training data but also to minimize the differences in the extracted joint sparsity between the images.

We demonstrate the feasibility of the proposed method by a numerical simulation study that includes the polychromatic nature of x-rays, and conduct quantitative evaluation between the ground truth and our results.

## 2 Methods

### A. Learned Energy-Flexible Algorithm

A schematic workflow of the proposed energy-flexible algorithm is summarized in Fig. 1. The framework of this network was inspired by Adler's learned primal-dual algorithm [14]. One of the sinograms in the given pair is fed into the layers 'in the projection domain', and the other sinogram is first fed into image reconstruction by the filtered-backprojection (FBP) and then is sent to the layers 'in the image domain'. Both sinograms are repeatedly used in each subnet as shown in Fig. 1.

In all the previous studies [11]–[13], minimizing the differences of a sparsifying transform has been performed in the image domain. Likewise in our method, the convolutional layers in the image-domain play a role of calculating the difference in sparsifying transform between different energy images. The batch normalizations support learning sparsifying transform by normalizing the scale of different energy images. The output of the layers in the image domain at each subnet goes through forward-projection and enters the projection domain as an input in the following subnet.

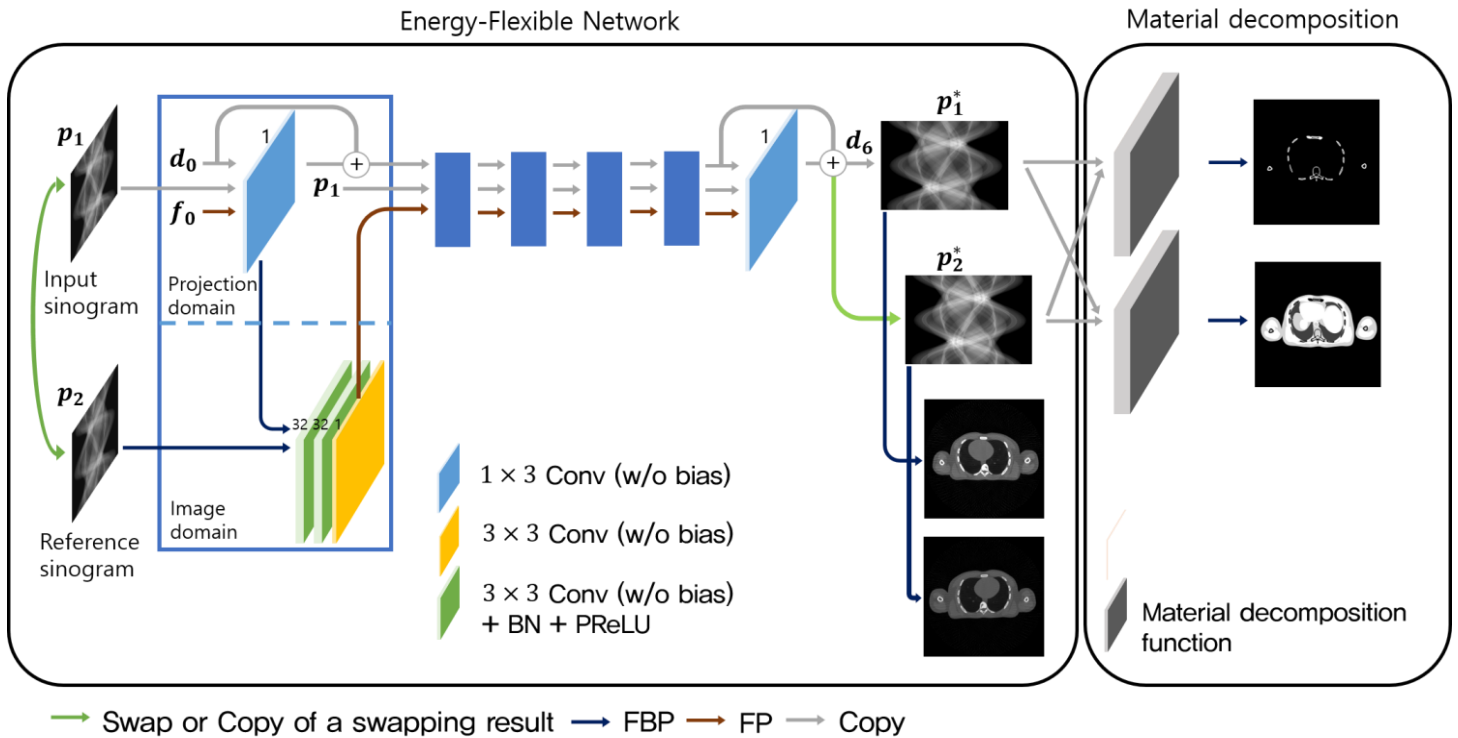

Fig. 1. Architecture of the proposed algorithm. The sinogram $p_1$ is given to the layer in the projection domain, and the sinogram $p_2$ is given to the layer in the image domain after reconstructed by FBP. The small blue boxes have the same architecture as the layers in the area bordered by the blue lines. The arrows and layers have different meanings for each color and the meanings are indicated in the figure. The green arrow stands for sinogram swap. $p_1^*$ and $p_2^*$ are the results of the network before and after the swap, respectively.

A reconstructed image of the ground truth corresponding to the input sinogram is given as a label of the energy-flexible algorithm. A data array, $d_0$, having the same size with the sinograms is initialized with zero and is updated at each subnet in the projection domain. After a total of five subnets and the last convolutional layer, the final $d_6$, which is nothing but $p_1^*$ (a desirable output after all), is reconstructed by FBP to calculate the error with respect to the label during the training. Additionally, after the training is done, we utilize the final $d_6$ to perform material decomposition in the projection domain.

For each training pair, we swap the input sinogram and the reference sinogram so as to train the algorithm to be energy-flexible. In case the sinograms are swapped, the final $d_6$ is $p_2^*$.

### B. Data description

To examine the performance of the proposed method, we conducted a numerical simulation with an XCAT phantom. We generated a total of 600 material maps from the XCAT phantom, which have four materials (air, adipose tissue, soft tissue, and bone) indices. Using a lab-made numerical simulation tool, sinograms measured at 80kVp, 100kVp, and 140kVp were obtained. We utilized an open X-ray spectra simulator and the database of X-ray mass attenuation coefficients from the National Institute of Standard Technology (NIST) [15], [16]. A fan-beam CT system was simulated, and 180 views for each spectrum were acquired. The number of detector pixels in a row is 512. The size of the reconstructed image is $256 \times 256$. For all data, Poisson noise corresponding to the background intensity $5 \times 10^3$ is added. The ground truth reconstructed images for the energy-flexible algorithm were made from the corresponding noise-free input sinogram.

### C. Implementation details

Among the simulated sinograms, we used randomly selected 580 sinograms for training, 10 for validation, and the other 10 for testing. Both pairs of 80/140kVp and 140/80kVp were used for training. Each energy of the pair means the input sinogram and the reference sinogram in order. For testing, the data corresponding to 80/140kVp and 140/80kVp were given in the algorithm, and the 100/140kVp pair were additionally tested.

### 3 Results

We visualize the reconstructed images of the proposed method and the other methods (FBP and SART-TV) from the sinograms taken at 80kVp and 140kVp in Fig. 2. The FBP were performed with a ramp filter, and a TV weight and a total number of iterations for the SART-TV were empirically chosen. The reconstructed images of the FBP severely suffer from noise. In case of the SART-TV, the noise was considerably reduced, but the reconstructed images were excessively smoothed. The images of the proposed method show the best performance among them in terms of both noise and resolution. The region pointed by the red arrow is only clearly visible in the proposed method.

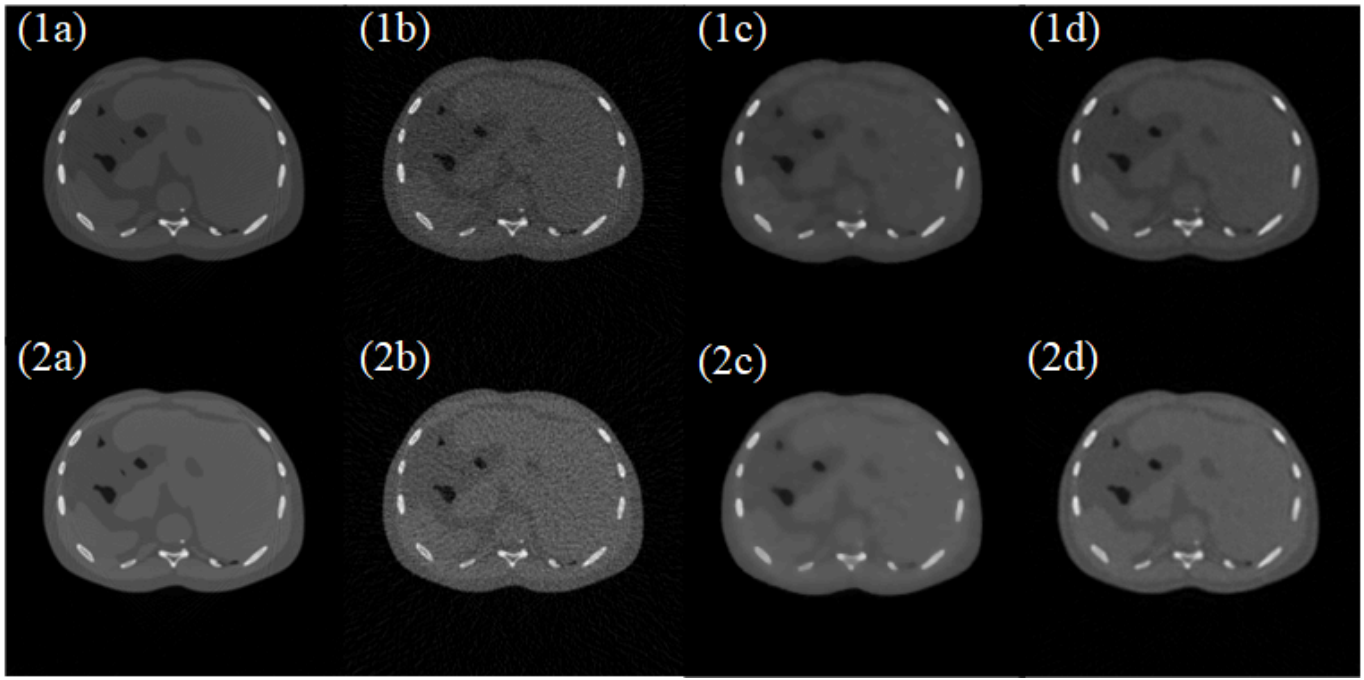

Fig. 2. Reconstructed images by (a) FBP with the ground truth, (b) FBP with the noisy sinogram, (c) SART-TV with the noisy sinogram, and (d) the proposed algorithm with the noisy sinogram with. (1) and (2) are 80kVp and 140kVp, respectively. The display windows for 80kVp and 140kVp are set to [0.0, 0.09] and [0.0, 0.06].

Quantitative results of the reconstructed images in Fig. 2 are shown in Table 1. We adopted the root-mean-square error (RMSE) and the structural similarity index measure (SSMI). The proposed method outperfo rms the other reconstruction methods w.r.t both RMSE and SSIM.

Additionally, we tested our algorithm with the 100/140kVp pair and show the results in Fig. 3. The noise level of the reconstructed image of FBP in Fig. 3(a) is substantially reduced in Fig. 3(b). We also evaluated the RMSE and the SSIM between the reconstructed results and the ground truth in Table. 2. The result also convinces an outperformance of the proposed method over the conventional methods in the 100kVp data. It verifies that our algorithm is applicable to data taken at another level of energy that has not been used for training.

Results of the material decomposition in the 80kVp/140kVp setting reconstructed by FBP from the original noisy sinograms and those reconstructed by the proposed method are shown in Fig. 4. The material decomposition was conducted in the projection domain, and the same decomposition function and parameters were used for both cases. The noise level in the original sinograms was largely amplified in the results of material decomposition. On the other hand, most of the noise are removed in the material-specific images in the proposed method.

Table 1. RMSE (in units of $1\,cm^{-2}$) and SSIM of reconstructed images by the proposed method and other methods at low (80kVp) and high (140kVp).

|          | Low (80kVp) | | High (140kVp) | |
|----------|-------------|------|---------------|------|
|          | RMSE | SSIM | RMSE | SSIM |
| FBP      | 0.0284 | 0.966 | 0.0209 | 0.980 |
| TV       | 0.0212 | 0.986 | 0.0122 | 0.992 |
| **Proposed** | **0.0133** | **0.994** | **0.0087** | **0.997** |

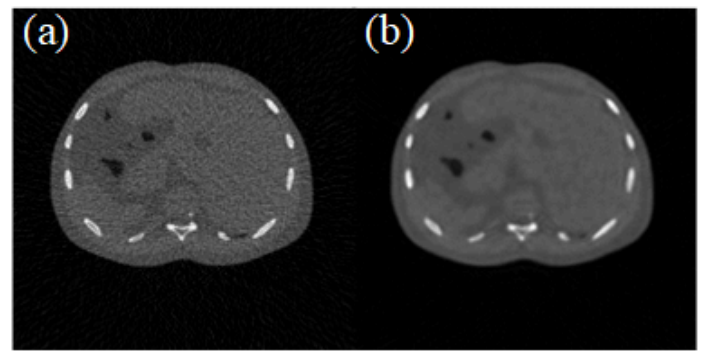

Fig. 3. Reconstructed images by (a) FBP and (b) the proposed method with the noisy 100kVp sinogram. The display is set to [0.0, 0.07].

Table 2. RMSE (in units of $1\,cm^{-2}$) and SSIM of reconstructed images at 100kVp.

|          | 100kVp | |
|----------|--------|------|
|          | RMSE | SSIM |
| FBP      | 0.0244 | 0.974 |
| TV       | 0.0137 | 0.992 |
| **Proposed** | **0.0120** | **0.995** |

## 4 Discussion

We have developed a data-driven energy-flexible algorithm using joint sparsity. The algorithm shows the quantitative improvement compared to the conventional methods in the numerical experiments. Additionally, we confirmed that the algorithm is applicable not only to the data measured at the energies used for training, but also data measured at different energies. We conjecture this is because the dependence of the network parameters on the energy dramatically decreases by training both 80/140kVp and 140/80kVp pairs alternately or in a swapping manner.

In the furue work, we will perform additional simulation studies and real experiments for further verification. Also, we will test wider energy range to check the flexbility on energy of the proposed method.

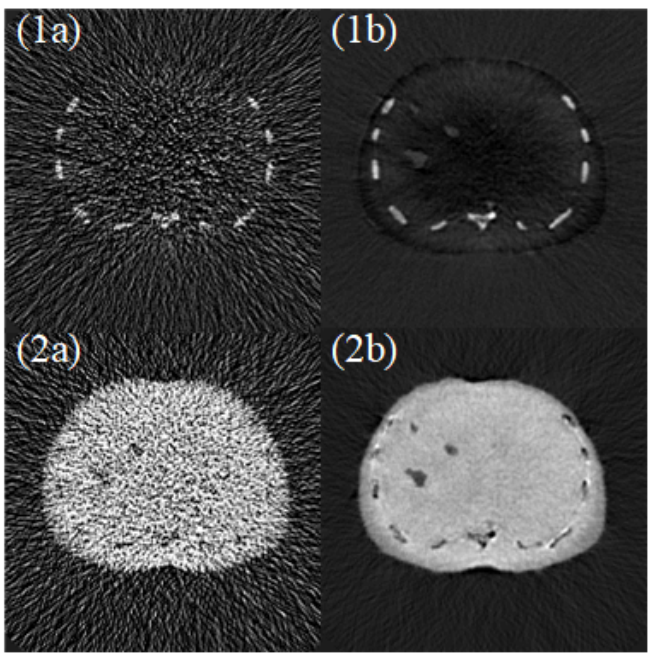

Fig. 4. Results of material decomposition from (a) the original sinograms and (b) the quality improved sinograms by the proposed method. (1) and (2) represent bone-only images and tissue-only images. The display for the bone-only images and tissue-only images are set to [-0.2, 1.2] and [-0.2, 1.5].

## 5 Conclusion

In this study, we proposed a learned energy-flexible algorithm using joint sparsity of dual-energy images. The proposed algorithm greatly improves the reconstruction performance in terms of noise and resolution. We have successfully obtained the improved material-specific results by the algorithm. Particularly, the flexibility of this algorithm against energy is verified by applying data measured at different energy that is unused for training.

# 4D iterative HYPR-based denoised reconstruction and post-processing improves PET image quality

Connor W. J. Bevington[1], Ju-Chieh (Kevin) Cheng[1,2], and Vesna Sossi[1]

[1]Department of Physics and Astronomy, University of British Columbia, Vancouver, Canada
[2]Pacific Parkinson's Research Centre, University of British Columbia, Vancouver, Canada

**Abstract** High-resolution dynamic PET images are useful for detecting small neurophysiological changes in the brain, but often suffer from low acquired counts per voxel, leading to high spatiotemporal voxel noise. Denoising—within reconstruction and/or as post-processing—is thus required to bring such images to acceptable levels of accuracy and precision for these applications. HYPR is a denoising operator that reduces noise using a high signal-to-noise composite image without compromising resolution. Traditionally a 3D post-processing operator, more recently 4D iterative formulations of HYPR have been proposed within reconstruction (HYPR4D-K-OSEM) and for post-processing (IHYPR4D). Both have shown increased precision without reduced accuracy, yet the information used to denoise in both algorithms is distinct; the combination HYPR4D-K-OSEM + IHYPR4D is thus proposed to maximize denoising while preserving high accuracy. This is demonstrated through phantom, simulation, and human studies, with HYPR4D-K-OSEM + IHYPR4D reducing noise up to 50% more than HYPR4D-K-OSEM alone without affecting accuracy.

## 1 Introduction

High resolution dynamic PET images are becoming increasingly sought-after to identify small, often rapid neurophysiological changes in response to intervention or disease. This includes studying metabolism changes from to a variety of tasks (task-based [$^{18}$F]FDG [1-3]) and inferring neurotransmitter release from changes in radiotracer binding [4-6]. However, high spatiotemporal resolution images contain a low number of acquired counts per voxel, leading to extremely high noise in the spatial and temporal dimensions when using standard image reconstruction methods. This in turn decreases image precision and thus limits the sensitivity of these applications. This motivates a 4D denoised reconstruction and post-processing combination that increases image precision with little to no reduction in image accuracy.

HighlY constrained backPRojection (HYPR) is a denoising operator that filters a target image, $I$, (i.e., the dynamic frames) to reduce noise, then uses a higher signal-to-noise (SNR) composite image, $C$, to return high frequency features to the filtered target. Its first application in PET was as a post-processing method where the composite image is a weighted sum of all temporal frames and thus the operator is 3D, applied separately to each frames (HYPR3D) [7]

$$I_{H3D}(x, t_j) = C(x) \frac{I(x, t_j) * F(x)}{C(x) * F(x)} \quad (1)$$

$$C(x) = \sum_j w_j I(x, t_j) \quad (2)$$

where $j$ indexes the dynamic frames, $F$ is the filtering kernel, $w_j$ is the weight for the $j^{th}$ frame, and "*" denotes convolution. This formulation implies all temporal information is lost and thus only spatial high frequency features can be extracted from the composite. As a result, if there is any contrast mismatch between the composite and the target bias will be introduced. This is especially relevant for PET tracers manifesting different contrasts during tracer uptake, progression to equilibrium, and washout periods, such as [$^{11}$C]raclopride (RAC).

Recently, a 4D formulation of HYPR has been incorporated into a kernelized Ordinary Poisson OSEM reconstruction (HYPR4D-K-OSEM [8]) as well as an iterative post-processing method (IHYPR4D [9]). By 4D we mean the composite and filtering kernels are 4D; 4D filtering allows for greater noise reduction while a 4D composite allows for preservation of high frequency spatiotemporal image features. In both HYPR4D-K-OSEM and IHYPR4D this 4D composite is iteratively updated to better match the contrast between the composite and target at each frame.

In HYPR4D-K-OSEM, the composite is constructed by summing over all subset images separately for each frame; initially one iteration of OSEM is required to initialize the algorithm. The subset images have unique noise properties, thus allowing noise to add decoherently during summation. The first OSEM iteration and thus the initial composite has relatively low noise but underestimates contrast; subsequent iterations improve the composite contrast by updating the composite to be the sum of the denoised subset images from the previous iteration. Consequently, HYPR4D-K-OSEM increases contrast recovery with a significantly smaller noise increment per update than OSEM. Spatiotemporal features from the 4D composite for the $m^{th}$ iteration ($C_{4D}^m$) are extracted using the 4D HYPR operator into a kernel matrix, $K$

$$K^m = diag[h^m]F_{4D} \quad (3)$$

$$h^m = \frac{C_{4D}^m}{F_{4D} * C_{4D}^m} \quad (4)$$

where $F_{4D}$ is the 4D filtering kernel. The OSEM algorithm is applied to the kernel coefficients, $\alpha$

$$\alpha^{m,s} = \frac{\alpha^{m,s-1}}{(K^m)^T (P^s)^T 1} \cdot \left( (K^M)^T (P^s)^T \frac{y^s}{P^s K^m \alpha^{m,s-1} + b^s} \right) \quad (5)$$

where $s$ indexes the subsets, $P$ is the system matrix, $y$ is the measured projection data, $b$ includes scatter and randoms contributions, and "T" denotes matrix transpose. The subset image estimates are given as

$$\lambda^{m,s} = K^m \alpha^{m,s} \tag{6}$$

allowing updating of the composite for the next iteration

$$C_{4D}^{m+1} = \sum_s \lambda^{m,s} \tag{7}$$

While HYPR4D-K-OSEM produces images substantially less noisy than OSEM images—especially in high resolution PET—the spatiotemporal noise at the voxel level may be further decreased by post-processing denoising. The post-processing method IHYPR4D is an iterative 4D analog of (1) that uses a set of spatial regions of interests (ROI) to define the initial 4D composite: for a given ROI $\mathcal{R}$, each voxel time activity curve (TAC) within the ROI is set to be the regional TAC. The composite is then filtered with the system/scanner point spread function to match the composite and target resolution

$$C'(x_i, t_j) = \frac{1}{|\mathcal{R}|} \sum_{r \in \mathcal{R}} I(x_r, t_j), \tag{8}$$

$$C(x, t_j) = C'(x_i, t_j) * PSF(x) \tag{9}$$

After running one iteration of HYPR, in subsequent iterations the composite is replaced by the denoised image of the previous iteration

$$I_{H4D}^{n=1}(x, t) = C(x, t) \frac{I(x,t)*F(x,t)}{C(x,t)*F(x,t)} \tag{10}$$

$$I_{H4D}^{n+1} = I_{H4D}^n(x, t) \frac{I(x,t)*F(x,t)}{I_{H4D}^n(x,t)*F(x,t)} \tag{11}$$

By averaging over the many voxels of each ROI, the initial composite has extremely low noise. However, any nonuniform image features within an ROI are evidently not present within the initial composite. Like HYPR4D-K-OSEM, the contrast of these features will be initially reduced, but subsequent iterations improve contrast with a low noise increment per iteration.

HYPR4D-K-OSEM has already been shown to offer substantial improvements in image precision over OSEM as well as some improvements in accuracy through a reduction of zero-trapping and being less biased to the last subset of the data [8]. Initial work on OSEM + IHYPR4D has shown improvements in image precision over OSEM + HYPR3D at the same the same accuracy level of OSEM [9] (unlike HYPR3D, which introduces bias through composite-target contrast mismatch). Thus, we hypothesize that HYPR4D-K-OSEM + IHYPR4D will further improve precision over HYPR4D-K-OSEM while maintaining high accuracy in a high resolution PET setting. This will likely increase the detection sensitivity of task-based neurophysiological changes such as metabolism and neurotransmitter release.

## 2 Materials and Methods

To test HYPR4D-K-OSEM + IHYPR4D in a high resolution PET setting, we perform phantom, simulation, and human studies on the Siemens High Resolution Research Tomograph (HRRT), currently the highest resolution brain-dedicated PET scanner. For HYPR4D-K-OSEM, we use 16 subsets, set $F_{4D}$ to be a Gaussian kernel with a $(5 \text{ mm})^3$, four frame FWHM, and perform 10 iterations. This combination was found to balance contrast recovery and noise suppression for data acquired on the HRRT. For IHYPR4D, we use a segmented anatomical MRI image to derive the ROIs. Gaussian spatial kernel sizes of 1x, 1.5x, and 2x the PSF of the HRRT (i.e. $(2.5 \text{ mm})^3$ FWHM) were tested with a Gaussian temporal kernel size of four frames FWHM. For comparison, we also test HYPR4D-K-OSEM + HYPR3D with the HYPR3D spatial kernel sizes set to match those of IHYPR4D, and a simple 4D Gaussian filtering with a 1x PSF, two frame kernel size to keep the bias introduced by filtering relatively low.

*Phantom study:* A modified Esser phantom with 4, 6, 8, 12, and 16 mm hot inserts filled with $^{18}$F in a 4:1 hot-to-background ratio was scanned. The data were binned to match the temporal count distribution of a RAC scan with a 4 x 1, 3 x 2, 8 x 5, 1 x 10 min framing protocol; the short half-life of $^{11}$C in general produces lower SNR dynamic images and thus presents a greater challenge for a given denoising operator. The IHYPR4D "segmentation" was a single hand-drawn ROI encompassing the entire phantom, such that the hot inserts serve as nonuniform image features within an ROI. Percentage contrast recovery (%CRC) in each hot insert versus voxel noise (standard deviation divided by mean) in a background ROI was evaluated.

$$\%CRC = 100\% \times \frac{C_H/C_B - 1}{A_H/A_B - 1} \tag{12}$$

where $C_H$ and $C_B$ are the estimated concentrations in the hot and background regions and $A_H$ and $A_B$ are the corresponding ground truth concentrations.

*Simulation study:* 20 noisy realizations of a realistic HRRT RAC scan, based on real human data, with the same framing protocol as above were simulated as described in [8]. The simulation includes uniform RAC binding in the caudate as well as a small (~30 voxel) and large (~100 voxel) hot spot in the putamen—corresponding to elevated binding—to simulate nonuniform image features, not separately segmented, within an anatomical ROI. To evaluate accuracy, percentage bias was calculated across all noisy realizations, then averaged separately across all voxels of the caudate, small hot spot, and large hot spot.

$$\%bias = 100\% \times \frac{1}{N} \sum_{n=1}^{N} \frac{C_n - A}{A} \tag{13}$$

where $C_n$ is the estimated concentration for the $n$th noisy realization and $A$ is the corresponding ground truth concentration. To evaluate precision, percentage standard deviation was calculated across all noisy realizations then averaged across all voxels of the cerebellum, a uniform low binding structure in RAC

$$\%std = 100\% \times \frac{1}{\bar{C}} \sqrt{\frac{1}{N-1} \sum_{n=1}^{N} (C_n - \bar{C})^2} \tag{14}$$

To jointly analyze accuracy and precision, for each region we take the mean of the absolute bias over all frames and plot it against the mean %std across frames to yield bias-noise trajectories.

*Human study:* we visually compare the images and TACs from the simulation study and a real RAC scan of a healthy volunteer on the HRRT using the same framing protocol.

## 3 Results

*Phantom study:* %CRC vs. voxel noise curves in Fig. 1 demonstrate that adding IHYPR4D to HYPR4D-K-OSEM causes a loss of %CRC initially since the hot inserts are not included in the IHYPR4D segmentation. With further iteration, the %CRC returns to that of HYPR4D-K-OSEM at a significantly decreased noise level, i.e. provided enough iterations are run, IHYPR4D reduces noise by ~25% without changing the %CRC of the input data. More iterations are required for a larger IHYPR4D kernel size as well as for smaller structures. The trajectories for different kernel sizes are nearly identical for all inserts except the smallest (4 mm) where now the 2x PSF kernel size lies systematically below the others. Since 2x PSF is ~(5 mm)$^3$, this suggests unsegmented features smaller than the kernel size cannot be fully recovered. Thus, we use the 1x PSF kernel size for IHYPR4D going forward.

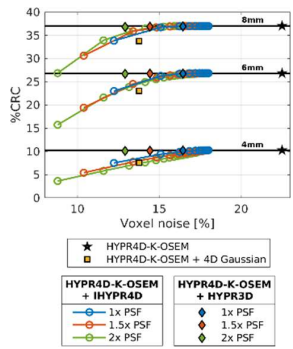

Fig. 1: %CRC versus voxel noise for the three smallest phantom inserts. Each marker represents an IHYPR4D iteration, increasing left-to-right. The non-iterative post-processing methods are shown as single markers.

The 4D Gaussian filter induces a small loss in %CRC. At equivalent voxel noise, (1x PSF) IHYPR4D has higher %CRC. Additionally, at equivalent %CRC IHYPR4D has lower voxel noise. We thus drop the 4D Gaussian from future analyses. Like IHYPR4D, applying HYPR3D after HYPR4D-K-OSEM significantly decreases voxel noise without a decrease in %CRC—even with increasing kernel size. However, in this experiment the contrast in the phantom is not time-dependent, so the composite and target contrast are perfectly matched at each frame. In a more realistic scenario with time-dependent contrast, any contrast mismatch between the composite and a particular dynamic frame will be amplified by a larger kernel (see simulation study results below).

*Simulation study:* the %bias vs. %std plots of Fig. 2a are similar to the %CRC vs. voxel noise plots; in the unsegmented hot spots, adding IHYPR4D after HYPR4D-K-OSEM increases bias, but after an appropriate number of iterations the bias reduces to that of HYPR4D-K-OSEM at a ~50% reduced noise level. In the segmented caudate,

IHYPR4D introduces little bias initially since the concentration is uniform in this ROI. HYPR3D also reduces noise, but now—unlike the phantom study—it also introduces bias through composite-target contrast mismatch; examining the %bias at each frame in Fig. 2b, the composite overestimates contrast during early (uptake) frames—resulting in positive bias—and underestimates contrast during later (washout) frames—resulting in negative bias. These effects are exacerbated by a larger HYPR3D kernel size. A sample uptake frame in Fig. 3a provides visual evidence of the HYPR3D overestimation. IHYPR4D, with its 4D composite, can better match the composite and target contrast in each frame to yield reduced bias at equivalent noise as HYPR3D, or reduced noise at equivalent bias.

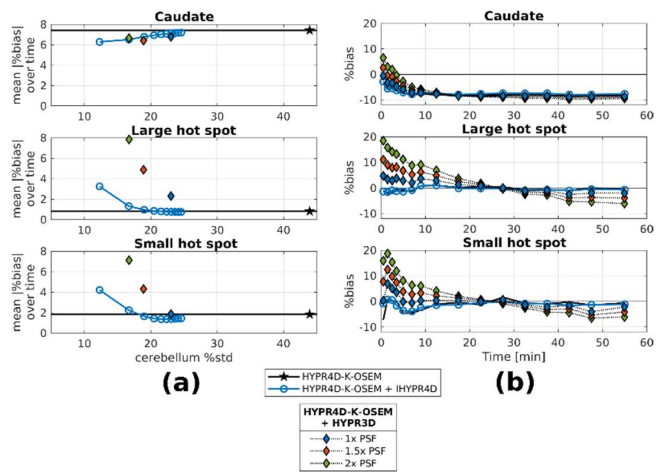

Fig. 2: (a) bias-noise trajectories for three regions of interest in the RAC simulation; (b) %bias at each frame for the same three regions. In this case, IHYPR4D is shown after three iterations.

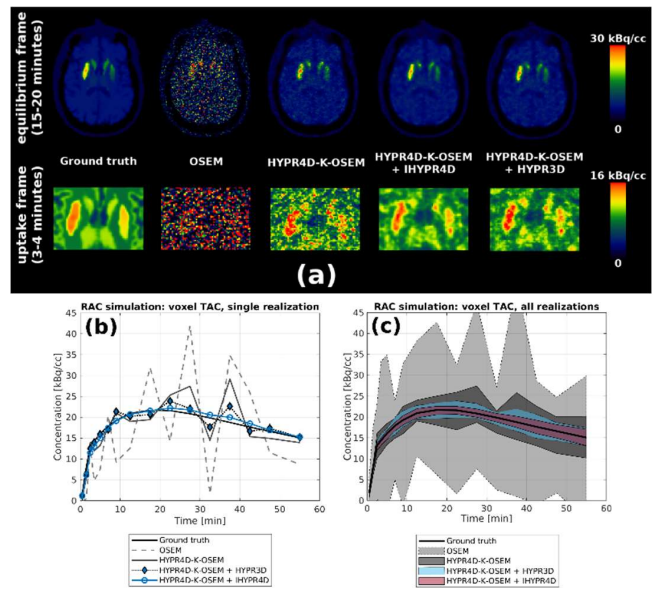

Fig. 3: (a) An axial slice of a representative realization of the RAC simulation for an equilibrium frame and an uptake frame (focused on the striatum); (b) a representative putamen voxel TAC from a sample noisy realization; (c) the mean ± standard deviation over all realizations of the same voxel TAC. Both IHYPR4D (three iterations) and HYPR3D use a 1x PSF spatial kernel size.

From the images of Fig. 3a and the sample voxel TACs of Fig. 3b, a clear reduction in spatiotemporal noise is seen going from the traditional reconstruction OSEM to HYPR4D-K-OSEM. Adding IHYPR4D further decreases spatial and especially temporal noise—more so than HYPR3D, whose noise reductions primarily manifest in the spatial domain. This gives IHYPR4D higher voxel-level precision, as seen by the smaller standard deviation in the TAC values over all noisy realizations (Fig. 3c).

*Human study:* the human study RAC images and TACs appear very similar to the simulated images (Fig. 4); the spatiotemporal noise is noticeably less in HYPR4D-K-OSEM versus OSEM, and the highest degree of denoising is achieved when IHYPR4D is added in post-processing. Though the ground truth is unknown, the HYPR4D-K-OSEM + HYPR3D uptake frames appear to have higher contrast than HYPR4D-K-OSEM alone, which would agree with the simulation study finding that HYPR3D overestimates contrast during RAC uptake.

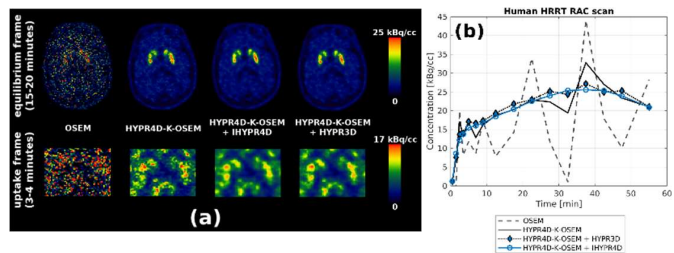

Fig. 4: (a) An axial slice of the human RAC scan for an equilibrium frame and an uptake frame (zoomed in on the striatum); (b) a representative putamen voxel TAC. Both IHYPR4D (three iterations) and HYPR3D use a 1x PSF spatial kernel size.

## 4 Discussion

While HYPR4D-K-OSEM offers substantial reductions in spatiotemporal noise compared to traditional PET image reconstruction, in high resolution PET the acquired counts per voxel are very low and thus the summation over subset images during reconstruction produces a composite image that may not sufficiently increase voxel-level precision for applications requiring sensitivity to small spatiotemporal signal changes. Ideally, a bias-free post-processing denoising algorithm should be added to further increase precision without affecting the high accuracy of HYPR4D-K-OSEM. We have shown through phantom, simulation, and human studies that a 4D iterative HYPR-based post-processing algorithm, IHYPR4D, approaches this ideal by adding little to no additional bias to the reconstructed images while reducing voxel noise by up to 50%.

IHYPR4D uses regional averaging over a set of ROIs to generate an extremely low noise 4D composite image for a 4D implementation of the HYPR operator. Initially these HYPR-denoised images are biased, but by iterating the HYPR operator one can essentially remove this bias—provided the image feature of interest is not smaller than the kernel sized used with IHYPR4D. By comparison, the previously proposed HYPR3D uses a 3D composite which is a weighted sum of all temporal frames, thus limiting temporal denoising and introducing bias in frames where the contrast differs from that of the composite.

## 5 Conclusion

HYPR4D-K-OSEM produces images of high voxel-level accuracy and precision compared to traditional reconstruction methods. Adding IHYPR4D as a post-processing step further improves precision without sacrificing accuracy. Future research will use these denoised images in applications where high sensitivity to small voxel-level signal changes is required, such as detecting neurotransmitter release and task-based FDG.

# Combining conditional GAN with VGG perceptual loss for bones CT image reconstruction

Théo Leuliet, Voichiţa Maxim, Françoise Peyrin, Bruno Sixou

Univ Lyon, INSA-Lyon, Université Claude Bernard Lyon 1, UJM-Saint Etienne, CNRS, Inserm, CREATIS UMR 5220, U1206, F-69621, LYON, France
*{theo.leuliet,voichita.maxim,francoise.peyrin,bruno.sixou}@creatis.insa-lyon.fr*

**Abstract** Reducing both the radiation dose to patients and the reconstruction time is key for X-ray computed tomography. The imaging of bone microarchitecture at high spatial resolution is all the more challenging as noisy data can severely deteriorate structural details. Deep Learning based algorithms are efficient for post-processing poor-quality reconstructions obtained with Filtered BackProjection, though MSE-trained networks hardly capture the structural information relevant for bones. Instead, conditional GANs allow to generate very realistic volumes that correspond to their corrupted FBP. Moreover, perceptual losses are efficient to capture key features for the human eye. In this work we combine both concepts within a new framework called CWGAN-VGG that is designed for the reconstruction of bones at high spatial resolution, with an emphasis put on the preservation of their structural information. We show on simulated low-dose CT bones data that our CWGAN-VGG outperforms state-of-the-art methods that involve GANs and/or perceptual losses in terms of PSNR and other metrics.

## 1 Introduction

Bone microstructure study with X-ray computed tomography (CT) is a challenging task due to the complexity of the underlying structures [1], [2]. Physical limitations of scanners and the need for reducing the patient's radiation dose may lead to noisy data, which need to be corrected to help practitioners get relevant parameters. When the number of projections is sufficiently large with a reasonable amount of noise, analytical algorithms like the Filtered Back-Projection (FBP) can offer satisfying results. When this is no longer the case, iterative methods [3] [4] can be considered. A major drawback of such algorithms is the reconstruction time and the need for tuning parameters for every reconstruction.

Deep Learning based algorithms have the potential to enhance the quality of images by learning patterns from ground-truth data, while significantly reducing the reconstruction time compared to iterative algorithms. A solution is to use neural networks to improve a poor-quality analytically obtained reconstruction [5] [6], e.g the FBP obtained from low-dose projections.

A critical point to address is the way the network should be trained. A Mean Squared Error (MSE) loss between predicted images and the corresponding ground-truths as in [7] might lead to slight oversmoothing that deteriorates some important structural details and thus affects the study of bone microarchitecture.

Instead, the use of a generative adversarial network (GAN) [8] allows to capture the probability distribution of the ground-truth images. In [5], such a network is trained with the Wasserstein distance along with a perceptual loss that compares the network output against the ground truth in a feature space designed to match the human eye perception, thus preserving key structural information. The resulting WGAN-VGG achieves impressive results on noise removal and artifacts correction. A similar architecture was proposed in [9] for underwater image restoration.

Nevertheless in both cases, the Wasserstein distance might be low even if the output does not correspond to the FBP it has been generated from. This is not the case when considering a conditional GAN [10] where the discriminator also takes the conditional information as an input. Such a framework was proposed in [11] and for medical image reconstruction in [6]. In [12], authors propose a conditional Wasserstein GAN (CWGAN) to capture the probability distribution of some volume conditionally to the FBP obtained from low-dose projections.

Combining such a conditional GAN with a perceptual loss has never been performed, though it seems to be perfectly adapted to bone microarchitecture imaging in order to capture their structural information. We then propose the CWGAN-VGG network that learns a probability distribution conditionally to the FBP obtained from low-dose projections, with a perceptual loss that is added to the generator loss function in order to preserve bone microstructure information.

In section 2, we present our CWGAN-VGG algorithm. In section 3 we detail the numerical experiments that we performed on simulated low-dose projections of $\mu$CT bone data. In section 4 we discuss the impact of both the conditioning and the perceptual loss on the quality of the reconstructions.

## 2 CWGAN-VGG framework

### 2.1 Conditional GAN and perceptual loss

Let $y$ be the FBP reconstructed volume from low-dose projections and $x$ the reference volume. We recall the CWGAN introduced in [12], where the aim is to approximate the posterior distribution $\pi(x|y)$ with a parametrized generator $G_\theta(y)$. Knowing this distribution allows to generate a number of

The authors acknowledge financial support of the French National Research Agency through the ANR project LABEX PRIMES (ANR-11-IDEX-0007) of Université de Lyon. The authors thank Andrew Burghard from University of California, San Francisco, USA, for providing the experimental $\mu$CT data.

volumes that can be responsible for data $y$. To approximate such a posterior distribution, the objective is to find $\theta^*$ that minimizes $d(G_\theta(y), \pi(x|y))$, with $d$ some distance between probability distributions. A now commonly used method to improve neural networks convergence is to consider the Wasserstein distance [13]. Denoting the probability distribution associated to the generator $G_\theta(y)$ by $P_\theta(y)$ - remember that $y$ is the condition here -, the dual characterization of this distance writes

$$W(\pi(x|y), P_\theta(y)) = \sup_{||f||_L \le 1} \mathop{\mathrm{E}}_{x \sim \pi(x|y)}[f(x)] - \mathop{\mathrm{E}}_{v \sim P_\theta(y)}[f(v)] \quad (1)$$

where the supremum is taken over all the 1-Lipschitz functions. Since it is not feasible to cover the entire space of these functions, they are parametrized with a neural network $D_w$ called a discriminator, with parameters $w$. Also, the generator $G_\theta(y)$ takes as input realizations $z$ drawn from a simple probability distribution $\eta$. We ensure the Lipschitz condition by adding a gradient penalty term to the distance function as in [14]. Minimizing the Wasserstein distance approximated by neural networks finally gives the optimization problem

$$
\begin{aligned}
\theta^* \in \mathop{\mathrm{argmin}}_\theta \sup_w L_{\mathrm{CWGAN}}(D, G) = & \mathop{\mathrm{E}}_{(x,y) \sim \mu}[D_w(x, y)] \\
& - \mathop{\mathrm{E}}_{\substack{z \sim \eta \\ y \sim P_y}}[D_w(G_\theta(z, y), y)] \\
& - \lambda \mathop{\mathrm{E}}_{\hat{x} \sim P_{\hat{x}}}[(||\nabla_{\hat{x}} D_w(\hat{x}, \hat{y})||_2 - 1)^2]
\end{aligned}
\quad (2)
$$

where $\mu$ is the joint distribution of $(x, y)$ corresponding to paired low-dose FBPs and high-dose ground truths, $P_y$ the unknown distribution of low-dose FBP data, $\hat{x} \sim P_{\hat{x}}$ are sampled along straight lines between samples from both $\pi(x|y)$ and the generated distribution $P_\theta(y)$, $\hat{y}$ are sampled along straight lines between the corresponding FBPs, and $\lambda$ is the weighting term for the gradient penalty. During training, expectations are replaced by their empirical counterpart obtained with paired data.

The resulting network generates stochastic samples conditionnally to the FBP of low-dose projections, according to a probability distribution that approximates the true distribution $\pi(x|y)$. Also, one can take $\eta$ as a Dirac distribution. In that case, the network is deterministic and generates a single output from the low-dose FBP. We call this network Det-CWGAN.

In [5], authors propose the WGAN-VGG framework which consists in training a generator by adding a perceptual loss to an unconditioned WGAN objective function, in order to better fit human perception of images, given as

$$L_{\mathrm{VGG}}(G_\theta) = \frac{1}{n} \mathop{\mathrm{E}}_{(x,y) \sim \mu}[||VGG(G_\theta(y)) - VGG(x)||_F^2] \quad (3)$$

where $n$ is the total number of voxels and $VGG$ is the $16^{th}$ output of the pre-trained VGG-19 model [15], $||.||_F$ is the Frobenius norm, and in their case $G_\theta$ only takes $y$ as input.

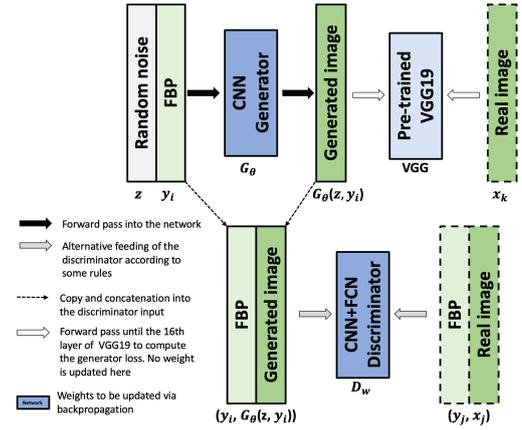

**Figure 1:** Scheme of the proposed CWGAN-VGG model. The FBP data is taken as input of the generator $G_\theta$. Both the FBP and the generated image are concatenated to produce the input of the discriminator. The network is trained according to (4).

It is shown in [16] that such a loss better suits human perception compared to pixel-wise based losses. In this framework, the output is deterministic and the discriminator is not conditioned on the FBP input, which amounts to taking $\eta$ as a Dirac distribution and $x \sim \pi(x)$ instead of $x \sim \pi(x|y)$ in (1).

## 2.2 Proposed architecture

In this work, we make use of the FBP computed from the acquired low-dose projections, to learn a conditional probability, by adding this FBP as an input to the discriminator. The benefits of conditioning the discriminator were already shown in [11] for natural images. Though authors used pixel-wise based additional losses, we propose to use the VGG perceptual loss since retrieving structural information on data is of major importance in bone microarchitecture imaging. Thus we propose the CWGAN-VGG framework that is trained as

$$\min_\theta \max_w L_{\mathrm{CWGAN}}(D_w, G_\theta) + \lambda_1 L_{\mathrm{VGG}}(G_\theta) \quad (4)$$

with $\lambda_1$ a weighting parameter. The scheme of the resulting network is presented in Fig. 1. In WGAN-VGG, the discriminator is not fed with the low-dose FBP, which results in a different paradigm compared to conditional GANs; the distribution that is learned is $\pi(x)$, and the generator is a mapping between the space of low-dose FBPs and the space of high-dose images. In conditional GANs, the low-dose FBP $y$ is the conditional data and the generator is a mapping between the latent space $Z$, where samples $z$ are drawn from $\eta$, and the space of high-dose images. To our knowledge this is the first time that the CWGAN-VGG architecture is proposed.

Moreover, both [11] and [12] pointed out the difficulties of CWGAN to generate stochasticity, as the network tends to ignore the input noise. Thus in our tests, we also implemented a deterministic CWGAN-VGG (Det-CWGAN-VGG) that only learns a Dirac distribution, for comparison.

## 3 Numerical Experiments

### 3.1 Materials and methods

The ground-truth data consist of human bone volumes reconstructed from acquisitions of radius and shin structures obtained on a SCANCO $\mu$-CT 100 with a 24-$\mu$m resolution. We create 180 2D projections of these volumes - corresponding to a low-dose acquisition - with ASTRA Toolbox [17] in Python. To simulate counting noise, random values are drawn from a Poisson law with mean the projections pixels. To simulate detectors noise, we then add a zero-mean Gaussian noise with standard deviation $\sigma = 0.8\%$ of the mean value in the projections. We then take the FBP - with Hann filter - as the input of the neural networks.

The dataset is composed of 13 volumes from different patients, 3 of which are only taken for evaluation. These 3 volumes have respectively a number of slices, height and width of $164 \times 882 \times 752$, $194 \times 466 \times 372$ and $180 \times 824 \times 702$. The trained networks are first evaluated with the Peak Signal to Noise Ratio (PSNR) and the Structural SIMilarity index (SSIM). Then, we post-process the reconstructed volumes with Otsu segmentation [18], and we compute the DICE index between the segmented reconstructed volumes and the segmented ground-truth data. Also we compute the ratio between the segmented bone volume and the total volume (BV/TV) that we compare with the one of ground-truth data. These metrics help better reflect the capability of the networks to preserve bone microstructure information.

Since CWGAN and CWGAN-VGG produce stochastic outputs, we average each voxel of 10 generated outputs to produce the volume for evaluation. Note that in our tests, increasing this number does not improve the performance.

Training is performed on 64x64 patches from 1,992 different 2D slices for a total of 297,976 patches, 20% of which are used for validation. The evaluation is performed by averaging metrics on the 3 test volumes.

The generator is a 16-layer Convolutional Neural Network (CNN) with 128 3x3 filters in each layer, except for the last layer which has only one since the output is the generated image. We used the same discriminator structure as in [5]. For both the discriminator and the generator, Leaky ReLU activations are used with parameter 0.3 and He initialization [19], except for the output of the discriminator that has no activation function. Optimization is performed with Adam algorithm [20] with $\beta_1 = 0.9$, $\beta_2 = 0.999$. The learning rate is $10^{-6}$ - except for WGAN-VGG where it is $10^{-5}$ -, with a batch size of 128 and 7,000 epochs. We took $\lambda_1 = 10$ for all the algorithms that include a VGG loss. For one update of the discriminator, we update the generator 4 times.

For a fair comparison, the kernel size, batch size, learning rate, number of generator updates and $\lambda_1$ have all been optimized for every single network, on the validation set. Computations are performed on a NVIDIA Tesla V100 GPU, and training of one network takes approximately 30 hours.

| | PSNR | SSIM | DICE | BV/TV |
|---|---|---|---|---|
| FBP | 15.96 | 0.491 | 0.880 | 0.2317 |
| Det-CWGAN | 23.05 | 0.634 | 0.928 | 0.1951 |
| CWGAN | 25.41 | 0.739 | 0.939 | 0.2043 |
| WGAN-VGG | 25.10 | 0.739 | **0.952** | 0.2154 |
| Det-CWGAN-VGG | 25.20 | 0.696 | 0.948 | 0.2096 |
| CWGAN-VGG | **26.00** | **0.753** | 0.951 | **0.2091** |

**Table 1:** Metrics computed on the 3 test volumes. PSNR, DICE and BV/TV were computed by stacking the 3 - potentially segmented - volumes, SSIM is the average value of the metric computed on each of them. The ground-truth BV/TV is 0.2077.

### 3.2 Results

Reconstructions of one of the three testing volumes are shown in Fig 2, along with a region of interest. Note that the 2 other testing volumes as well as the 10 training volumes all have a significantly different shape, which attests for the ability of the networks to generalize. In the second row of Fig 2, we notice that Det-CWGAN is the only one that fails to recover some continuous structure of the bone, which is a key feature for bones imaging. However, for the others there is no clear indication that one reconstruction outperforms the others.

Table 1 allows to better distinguish between the obtained reconstructions. Results show that CWGAN-VGG performs the best in terms of PSNR, SSIM and BV/TV ratio, which is an important metric for bone microarchitecture. The algorithm presents a DICE index that is only slightly inferior to the one of WGAN-VGG, which outperforms the other algorithms for this metric. We also note that both CWGAN and CWGAN-VGG perform better than their deterministic version for the tested metrics.

## 4 Discussions

Our results suggest that using a perceptual loss for training our generator as in [5] allows the network to produce volumes that are closer to the real ones, in terms of pixel-wise metrics and structural-specific evaluation methods. Indeed, CWGAN-VGG outperforms CWGAN, whether it is on the deterministic or stochastic version.

We also argued that conditioning the discriminator would produce outputs that better match the FBP they are conditioned on. This is the case in our tests, where CWGAN-VGG produced better results than WGAN-VGG for 3 out of the 4 tested metrics, and the DICE index for both methods is very close.

We also find that it is less optimal for the network to learn a Dirac conditional distribution. Indeed, the strategy of averaging several stochastic outputs gave a significant improvement compared to using a deterministic network for both the CWGAN and CWGAN-VGG networks. Along with improvements on those metrics, the non-deterministic outputs that CWGAN and CWGAN-VGG produce might be very

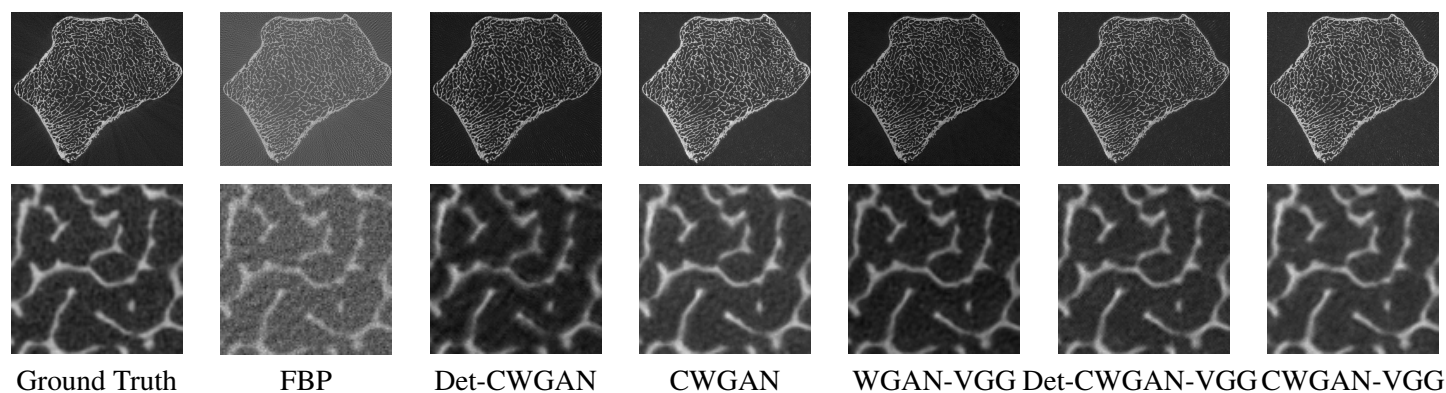

Ground Truth    FBP    Det-CWGAN    CWGAN    WGAN-VGG   Det-CWGAN-VGG   CWGAN-VGG

**Figure 2:** Entire slice (first row) and zoom on this slice (second row) of the bone volume reconstructed with different architectures, with pixel intensities between 0 and 1

useful for practitioners in order to get a level of confidence for specific regions of interest in the reconstruction.

In order to fully show the potential of CWGAN-VGG, work is under progress to train and test it on different noise configurations to get a more robust model and evaluate it on more realistic data for even more metrics.

## 5 Conclusion

We proposed a new framework called CWGAN-VGG for the task of enhancing the quality of a FBP acquired from low-dose projections. It combines both the ability of GANs to learn conditional probabilities and the preservation of key structural information provided by perceptual losses. We showed the benefits provided by both conditioning the discriminator with the low-dose FBP and adding a perceptual loss to train the generator. We also showed the improvement on the evaluated metrics when using a non-deterministic network. Our resulting architecture thus outperformed state-of-the-art ones that rely on similar methods, for PSNR and other metrics, on CT bones data.

# Trajectory Upsampling for Sparse Conebeam Projections using Convolutional Neural Networks

Philipp Ernst[1,2], Marko Rak[1,2], Christian Hansen[1,2], Georg Rose[2,3], and Andreas Nürnberger[1]

[1]Faculty of Computer Science, University of Magdeburg, Germany
[2]Research Campus *STIMULATE*, University of Magdeburg, Germany
[3]Institute for Medical Engineering, University of Magdeburg, Germany

**Abstract** In this paper, we present an approach based on a combination of convolutional neural networks and analytical algorithms to interpolate between neighboring conebeam projections for upsampling along circular trajectories. More precisely, networks are trained to interpolate the angularly centered projection between the input projections of different angular distances. Experiments show that an analytical interpolation as additional input is more beneficial than adding more neighboring projections. Using our best model, we achieve an x8 upsampling by repeating the interpolation three times. Though not depending on a specific reconstruction algorithm, we show that FDK reconstructions substantially benefit from this upsampling for removing streak artifacts. Using this FDK reconstruction as initialization for ART is also superior to other initializations but comes with a higher computation time and therefore cannot be considered as an option in an interventional setting.

## 1 Introduction

Conebeam X-ray CT (CBCT) is a helpful tool for surgeons to guide them during interventions. The downside is that the patients as well as the surgeons are exposed to harmful X-radiation. Reducing it is possible by acquiring fewer projections or applying less radiation while keeping the number of projections high. In both cases however, the image quality of the reconstructed volumes using the commonly used FDK algorithm is severely impaired by streak artifacts or noise. Many algorithms have been proposed to overcome these artifacts [1] but are too costly in terms of computation time to be applicable in an interventional setting, especially iterative methods that need to compute both forward and backprojections in each iteration for the entire 3-D volume.
Convolutional neural networks (CNNs) and deep learning have found their way into medical imaging [2] and CT image reconstruction [3]. Owing to their trainable nature, they directly incorporate domain knowledge to approximate the reconstruction more closely. Despite the high computational power of modern PC systems, training CNNs on whole 3-D data sets is usually not feasible due to the large memory requirements and is often stripped down to 2-D problems or patch-based 3-D approaches. The inherent two-dimensionality of conebeam projections suggests using them in combination with CNNs. The method proposed here will be used to interpolate between these projections which enables an upsampling along a circular trajectory around the scanned subject. Since the interpolation is carried out in projection space, the method does not rely on any reconstruction algorithm and preserves data consistency.

## 2 Method

### 2.1 Analytical Projection Interpolation

As described in [4], conebeam projections can be approximately interpolated by using (Eq. 24 in [4])

$$g(\lambda + \varepsilon\Delta\lambda, \underline{\alpha}) \simeq (1 - \varepsilon)g(\lambda, \underline{b}(\lambda + \varepsilon\Delta\lambda, \underline{\alpha}) - \underline{a}(\lambda)) + \varepsilon g(\lambda + \varepsilon\Delta\lambda, \underline{b}(\lambda + \varepsilon\Delta\lambda, \underline{\alpha}) - \underline{a}(\lambda + \Delta\lambda)) \quad (1)$$

for projections $g(\lambda, \underline{\alpha})$ from source positions $\underline{a}(\lambda)$ in directions $\underline{\alpha}$ and points of interest $\underline{b}(\lambda, \underline{\alpha})$ that are closest to the rotation axis on the line through $\underline{a}(\lambda)$ with direction $\underline{\alpha}$. Unlike [4], the directions $\underline{\alpha}$ here are chosen to coincide with the projection lines of the projection to be interpolated. This only requires interpolating on the given projections.

### 2.2 CNN Approach

Assuming an equiangular sampling of conebeam projections along a circular trajectory, the presented approach upsamples along the trajectory by subsequently interpolating projections angularly centered between neighboring projections. Simple algorithms like linear interpolation are not applicable because of the sinusoidal structure and perspective distortions caused by the conebeam. A U-Net [5] is used to approximate this highly complex interpolation because of its large receptive field that is able to capture and trace larger translations in the projections compared to flat CNN architectures. (1) Networks are trained to predict the projection angularly centered between two projections from only its direct neighbors for 2°, 4° and 8° of angular distance (referred to as nn2). (2) The number of neighboring input projections is increased from 2 to 4 and 8 neighbors to provide more angular information (referred to as nn4, nn8). (3) Instead of increasing the number of neighboring projections, the analytical interpolation described in Sec. 2.1 with $\varepsilon = 0.5$ is used as an additional input which is supposed to guide the network closer to the true interpolation (referred to as nn2+ana).

### 2.3 Datasets and Training

The data of 22 subjects from the CT Lymph Nodes collection [6] of The Cancer Imaging Archive [7] is used, consisting of reconstructed volumes of the abdomen with different in-plane spacings that serve as ground truth. Conebeam projec-tions were generated using the CTL toolkit [8] equiangularly along a circular trajectory with a source to detector distance (SDD) of 1000 mm and a source to isocenter distance (SID) of 750 mm. The flat panel detector consists of $256 \times 256$ elements with a pixel size of $4 \text{ mm}^2$ (cone angle of $54.2°$). The values were chosen such that most projections were not truncated and to enable a faster training.

| Up | Method | NMSE ($\times 10^{-5}$) | PSNR [dB] | SSIM [%] |
|----|--------|------|------|------|
| x2 | ana | 10.88 | 50.28 | **99.26** |
| x2 | nn2 | **8.63** | **50.74** | 99.01 |
| x2 | nn4 | 11.49 | 49.68 | 98.80 |
| x2 | nn8 | 17.32 | 47.57 | 97.87 |
| x2 | nn2+ana | 10.91 | 49.80 | 98.74 |
| x4 | ana | 32.60 | 45.48 | 97.84 |
| x4 | nn2 | **17.15** | **47.56** | **98.24** |
| x4 | nn4 | 27.63 | 45.52 | 97.53 |
| x4 | nn8 | 40.20 | 44.10 | 96.34 |
| x4 | nn2+ana | 18.24 | 47.44 | 98.01 |
| x8 | ana | 93.52 | 41.09 | 94.44 |
| x8 | nn2 | 58.64 | 42.70 | 96.03 |
| x8 | nn4 | 79.94 | 41.32 | 94.39 |
| x8 | nn8 | 114.83 | 39.76 | 92.61 |
| x8 | nn2+ana | **32.45** | **45.17** | **96.98** |

**Table 1:** Projection errors for different upsampling methods.

| Method | NMSE [%] | PSNR [dB] | SSIM [%] |
|--------|------|------|------|
| full | 4.95 | 28.76 | 99.14 |
| sparse | 16.09 | 20.86 | 97.72 |
| ana | 7.43 | 25.03 | 98.81 |
| nn2 | 6.51 | 26.07 | 98.97 |
| nn4 | 6.74 | 25.77 | 98.92 |
| nn8 | 7.28 | 25.21 | 98.83 |
| nn2+ana | **6.00** | **26.81** | **99.03** |

**Table 2:** Reconstruction errors of FDK reconstructions for different upsampling methods from 45 available projections.

The U-Net [5] has a depth of 5 and is slightly modified. The encoder doubles the number of layers after each average pooling, whereas the decoder halves the number of layers after each nearest neighbor upsampling. The optimizer is SGD with a weight decay of $1 \times 10^{-4}$ and a learning rate of $6 \times 10^{-3}$ that gradually drops to $1 \times 10^{-6}$ by a factor of 0.8 after every 10 epochs of no improvement in validation loss. Every network was trained for 300 epochs using mean squared error (MSE) and another 300 epochs using equally weighted $l_1$ and MS-SSIM loss similar to [9] to focus more on general structures and edges. 16, 4 and 2 datasets were used for training, validation and testing, respectively. For faster convergence, the projections were normalized between 0 and approximately 1 by dividing by the 99th percentile of all projections of all datasets.

## 3 Results

### 3.1 Projections

The different interpolation methods are evaluated on the projections first. Except for the analytical upsampling described in Sec. 2.1, all methods interpolate the projection angularly centered between the input projections, which is repeated for x4 and x8 upsampling using the corresponding trained networks. For the analytical upsampling, the parameter $\varepsilon$ is chosen to directly resemble the positions of the projections to be interpolated. Tab. 1 shows the results for the error metrics normalized mean squared error (NMSE), peak signal-to-noise ratio (PSNR) and structural similarity index measure (SSIM) averaged over all projections. The calculation of the metrics obviously excludes the non-interpolated projections. Interestingly, the results are quite different for the different upsampling stages.

For the single interpolation (x2, angular difference of $2°$), nn2 gives the best results for NMSE and PSNR. The analytical interpolation however results in the highest SSIM.

Interpolating twice (x4, angular difference of $4°$) is done best by nn2, this time for all metrics.

Finally, the optimal method for carrying out the interpolation three times (x8, angular difference of $8°$) is using nn2+ana. A patch of an exemplary x8 interpolation created with the different methods is shown in Fig. 1. Compared to the ground truth patch, the other patches are more blurry. The patch created with the analytical interpolation looks like the superimposition of two projections. The nn4 and nn8 patches seem to have more high frequencies than nn2 and consequently look less blurry. nn2+ana is visually closest to the ground truth and the least blurred.

### 3.2 Reconstructions

Evaluating in projection space only does not fully show the benefits of the proposed method. It is also necessary to compare the reconstructions. We decided for the commonly used FDK [10] algorithm as well as ART [11] without interpolated projections initialized with the FDK reconstruction using all interpolated projections.

All reconstructions are created with the CTL toolkit [8]. The ART reconstructions run for 5 iterations with enabled positivity constraint.

Since the number of projections is still relatively small and the resolution of the detector is quite low, the reconstructions will also be compared to the FDK reconstruction using all 360 projections to find lower bounds for the error metrics. As described previously, though not depending on any recon-struction algorithm, the interpolated projections are supposed to increase the quality of the reconstructions by providing a more appropriate sampling of projections.

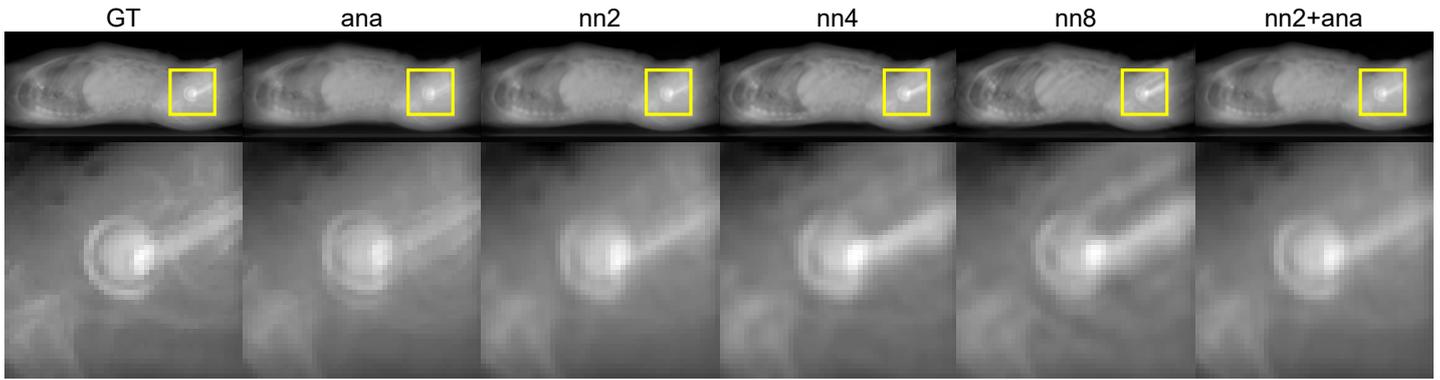

**Figure 1:** Top: Interpolated projections (central part) of x8 upsampling of different interpolation methods compared to ground truth projection (GT). Bottom: Zoomed patches around a hip implant.

This hypothesis is evaluated using the FDK reconstruction algorithm, first. For brevity, only the reconstructions of the highest upsampling (x8, 45 available projections) are investigated. Tab. 2 shows the error metrics for the different methods averaged over all axial slices. For reference, the first two rows serve as lower/upper bounds: values for the full FDK describe the errors between the ground truth volume and the volume reconstructed from 360 projections, whereas values for the sparse FDK describe the errors between ground truth and reconstruction from 45 projections. All interpolation methods optimize the sparse FDK reconstruction and are quantitatively closer to the full FDK. nn2+ana works best, followed by nn2, nn4, nn8 and using only the analytical interpolation. This closely resembles the errors on the projections described in the previous section.

The left column of Fig. 2 shows exemplary FDK reconstructions using the different methods. Compared to the direct FDK reconstruction from 45 projections (sparse), every method reduces the streak artifacts. The analytical upsampling (ana), however, basically results in a radially blurred reconstruction. None of the CNN-based reconstructions suffers from streak artifacts or radial blur, but they appear slightly more blurred than the sparse FDK reconstruction. As expected from the quantitative analysis, nn2+ana also creates the best visual result.

ART provides another simple reconstruction algorithm. Due to its iterative nature, it is inherently slower than FDK but enables simply adding additional constraints resulting in reconstructions of higher quality. For a better convergence, ART is initialized with another reconstruction. In our experiments, we use the FDK reconstructions of the different interpolation methods and run ART with only the 45 available projections, which results in the best compromise between reconstruction time and quality. Tab. 3 shows the error metrics. Zero-initialized ART and sparse-FDK-initialized ART are shown for reference. In all cases, ART outperforms FDK. Again, nn2+ana works best, followed by the other methods

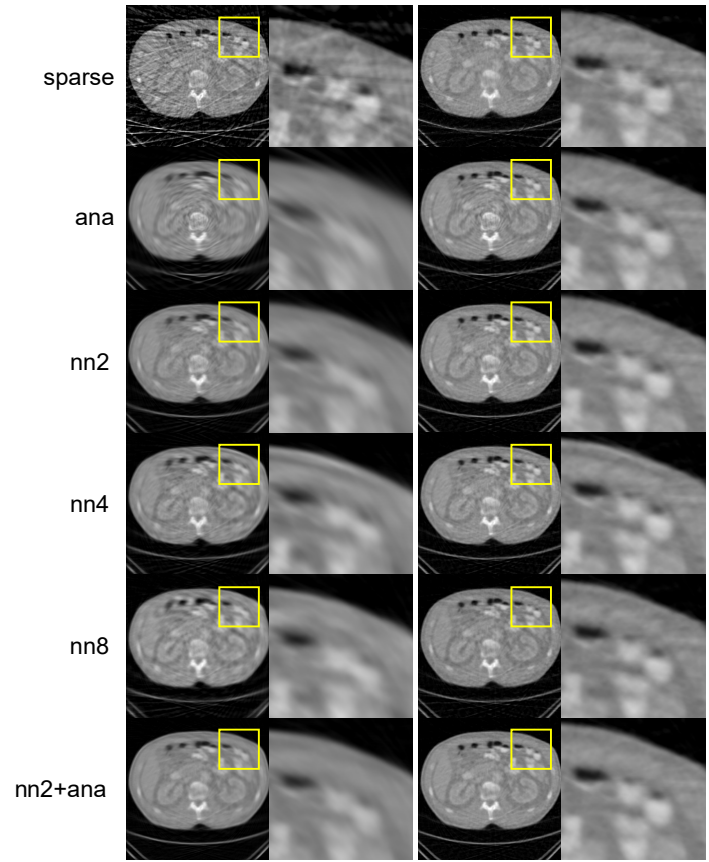

**Figure 2:** Reconstructions for different upsampling methods. Left column: FDK. Right column: ART initialized with FDK.

| Init. | NMSE [%] | PSNR [dB] | SSIM [%] |
|---|---|---|---|
| zero | 2.72 | 28.16 | 99.73 |
| sparse | 2.28 | 28.91 | 99.78 |
| ana | 2.19 | 29.06 | 99.80 |
| nn2 | 1.88 | 29.79 | 99.84 |
| nn4 | 1.96 | 29.57 | 99.83 |
| nn8 | 2.12 | 29.23 | 99.81 |
| nn2+ana | **1.65** | **30.40** | **99.86** |

**Table 3:** Reconstruction errors of ART reconstructions for different upsampling methods from 45 available projections.

in the same order as in the FDK reconstructions.

The right column of Fig. 2 shows exemplary ART reconstructions using the different methods. They are not only quantitatively closer to the ground truth but also qualitatively outperform their FDK counterparts. There are only slight visual differences of the ART initialized with the different FDK reconstructions. For the `sparse` case, edges are preserved well but tissues of the same absorption coefficient appear noisy. `nn2+ana` has the best visual quality with the least noise and the best edge preservation compared to the other methods.

## 4 Discussion

Increasing the number of neighboring projections does not increase the quality of the interpolated projections. Since the additional projections are only provided to the CNN as input channels and the convolutions are carried out per channel, it is possible that (without any special weight initialization) the information from more distant neighbors is not local enough to be considered as helpful knowledge during backpropagation. Moreover, increasing the number of input projections even impairs the prediction quality. Further tests need to investigate why different interpolation methods work best for certain upsampling stages.

The simulated projections do not contain noise, are almost not truncated, have a low resolution and a rather large pixel spacing. Further experiments need to focus on more realistic detector and gantry parameters and the method needs to be tested on real data, especially including interventional instruments and other artifact creating influences.

The used error metrics only give a rough impression of the quality. Due to the blurring of edges caused by the interpolation, future work needs to focus on how exactly mappings of edges are changed as well as how the reconstructions compare to other state-of-the-art methods.

Using the neighboring projections as input channels of the U-Net is a rather straightforward way. As with other deep learning methods, it is conceivable that another network architecture can extract more information from the input data and thus improve the quality even further, which will be part of future experiments. The code is available on Github[1].

## 5 Conclusion

It was shown that conebeam projection interpolation using CNNs applied to trajectory upsampling significantly reduces streak artifacts from FDK reconstructions and provides a strong prior for iterative reconstruction algorithms when used for the initialization in ART. Providing further knowledge about the interpolation to the network in terms of the analytical interpolation approach similar to [4], the quality can be improved even further. This allows for a dose reduction by a factor of at least eight while still providing a good quality of the reconstructions. Compared to an FDK reconstruction from 45 projections, our best interpolation method increases the PSNR by almost 6 dB. Though not applicable in interventions due to time requirements, initializing an ART with the FDK reconstruction further increases the PSNR to 30.40 dB.

## Acknowledgements

This work was conducted within the International Graduate School MEMoRIAL at OVGU Magdeburg, supported by the ESF (project no. ZS/2016/08/80646).

---

[1] https://github.com/phernst/conebeam_interpolation

# Efficient prior image generation for normalized metal artifact reduction (NMAR) using normalized sinogram surgery with the beam-hardening correction

Sungho Yun[1], Donghyeon Lee[1], Hyeongseok Kim[2], Rizza pua[1], Sanghoon Cho[1], and Seungryong Cho[1]

[1]Department of Nuclear and Quantum Engineering, KAIST, Daejeon, Republic of Korea
[2]KAIST Institute for Artificial Intelligence, KAIST, Daejeon, Republic of Korea

**Abstract** A clean and anatomically informative prior image is important in normalized metal artifact reduction (NMAR) method. To fully utilize the advantages of NMAR in terms of speed and computational cost, the prior image generation should desirably be fast and least case-dependent. In this study, we propose an efficient sinogram surgery method for computing prior image that does not require iteration or optimization. The beam-hardening corrector was employed to remove the artifact-corrupted region in the sinogram domain. Also, the normalization was used to minimize the inconsistencies of the corrected-sinogram instead of using iterative fidelity check. The results show well-preserved anatomical structures while substantially reducing the metal artifacts. With the use of the image prior, as proposed in this work, for NMAR, it is expected to preserve anatomical structures and increase the soft-tissue contrast without compromising the benefits of NMAR.

## 1 Introduction

Metal artifacts in CT imaging constitute one of the major causes that hinder accurate diagnosis of and effective treatment planning of various diseases in the patients who have metallic inserts. Continuing research efforts have been exerted on reducing the metal artifacts for decades. Normalized metal artifact reduction (NMAR) [2] is one of the popularly used techniques and is in focus of this work.

NMAR has desirable features for practical applications in that it is fast and effective with a relatively low computational cost. The key idea exists in the usage of a prior image for sinogram normalization. The normalization using this prior sinogram before interpolation can minimize the inconsistencies of the sinogram and prevent introducing additional artifacts. Also, the anatomical structures of the prior image such as bone can be preserved after the de-normalization step. However, the results heavily depend on the quality of the prior image, which means additional secondary artifacts may be introduced by an inaccurate or artifact-corrupted prior image.

Thus, acquiring a clean and anatomically informative prior image is critical in NMAR. Furthermore, the prior image generation should be fast and least case-dependent in order not to undermine the advantages of NMAR in terms of speed and computational cost. In the literature, the image fusion based approach [3, 4] and iterative algorithms [5, 6] were proposed for prior image generation. Recently, neural network based approach [7] was also investigated. However, they are still computationally rather expensive and time-consuming in a relative sense to the one-shot analytic approaches due to the optimization or training nature.

In this study, we propose an efficient sinogram surgery-based method[8, 9] for computing prior image that does not require iterative processes or optimization. The sinogram surgery method enables to fully utilize the anatomical information of the metal-trace compared to a simple interpolation. A mathematically derived beam-hardening corrector [10] was utilized in the sinogram domain for the removal of the artifacts. The normalization was also used to minimize the inconsistencies of the corrected-sinogram after the surgery.

## 2 Materials and Methods

### A. Reprojection of metal-replaced FBP image

Initially reconstructed FBP image from the original sinogram $S^{orig}$ has severe streaks and noises due to the metallic objects. To create a metal-replaced FBP image, a smoothing filter (Gaussian used in this study) was applied to mitigate the noise, and the values of metal regions were replaced by a constant value $P^{soft}$ that has a similar scale with surrounded soft tissues. The metal region was segmented using threshold and $P^{soft}$ was chosen from averaged values of soft tissues. By forward projecting of metal-replaced FBP image, the reprojection sinogram $S^{reproj}$ was obtained.

The metal trace of $S^{reproj}$ shows important differences when compared with the original sinogram $S^{orig}$. In $S^{orig}$, the metal-overlapped region $M^o$ is the main cause of the beamhardening artifacts, because, the bright streaks and dark shade artifacts correlate with the high-frequency and low-frequency change of sinogram consistency in $M^o$ [1]. (In sigle metal case, only cupping artifacts occur inside of the metal [1]. In this study, only cases that considered which have at least two metal objects.) Also, the anatomical information tend to be buried due to the high attenuation value of the metals. However, in $S^{reproj}$, the reprojected bright streaks relocated in the extended boundary region of $M^o$, and the reprojected dark shade artifacts are centered in $M^o$ (in Fig. 1). Because, the forward projection of the metal-replaced FBP image just relocate the artifacts into more extensive position in the $S^{reproj}$. More, due to the replacement of metal objects, the anatomical information could be excavated from the large metal value in $S^{reproj}$.

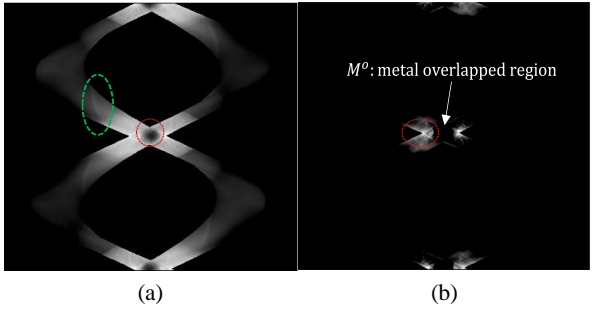

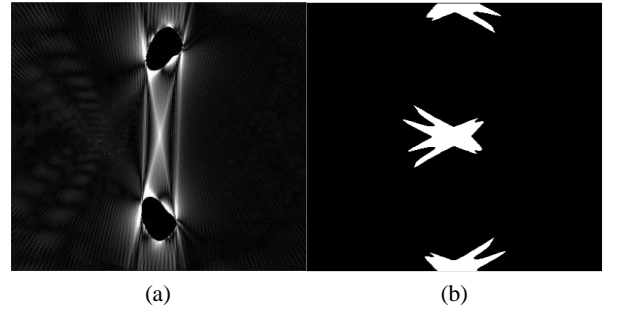

(a)                              (b)

Figure 1. (a) is the only metal trace image of $S^{reproj}$. (b) is highlighted $M^o$ of (a). In (a), the green circle represents excavated anatomical data, and red circle represents the relocated dark shade artifacts. The red circle of (b) represents relocated bright streaks artifacts.

## B. Specifying artifact-corrupted region $M^{corrupt}$ using the beam-hardening corrector $\Phi_{D,\lambda}$

In section $A$, the relocation of the beamhardeing artifacts in $S^{reproj}$ was explained. To compensate them, *K.Y Jeong et al* was proposed directional interpolation of $M^o$, after the total-variation smoothing operation [8]. However, only interpolating $M^o$ is not sufficient to cover the invading artifacts without pre-iterative treatment. To cover the artifacts more effectively, the beam-hardening corrector $\Phi_{D,\lambda}$ in [10] was used. The mathematically derived $\Phi_{D,\lambda}$ can generate the modeled beamhardening artifacts for given metal image. It was supposed to use in the image domain through optimization of energy-related parameter λ.

In this study, the metal trace of $R\left\|\Phi_{D,\lambda}^{nometal}\right\|$ was used to generate a mask for sinogram correction ($R$ indicates forward projection operator, and the superscript indicates metal removed image). $\Phi_{D,\lambda}$ guarantees the streaks generation for any λ value, so it does not require optimization for generating a mask in the sinogram domain. The mask for beamhardening correction $M^{corrupt}$ was created following the equation (1). The purpose of the mask $M^{corrupt}$ is to properly cover only the neighboring regions of $M^o$ for selective interpolation. $M^{corrupt}$ was calculated by following equation (1).

$$M^{corrupt} = \begin{cases} NaN & \text{if metal trace of } R\left\|\Phi_{D,\lambda}^{nometal}\right\| > T* \\ 1 & \text{else} \end{cases} \quad (1)$$

$T*$ is a threshold value. The *NaN* indicates regions that will be interpolated. $M^{corrupt}$ was possible to cover more extensive areas of artifacts region more than $M^o$ (in Fig. 2(b)).

Figure 2. (a) is the image of $\left\|\Phi_{D,\lambda}^{nometal}\right\|$, and (b) is the image of $M^{corrupt}$ which indicates the artifact-corrupted region that will be interpolated (white).

## C. Normalization of $S^{reproj}$

In the sinogram surgery approaches, directly replacing the $S_{metal}^{orig}$ to $S_{metal}^{reproj}$ (subscript indicates the only metal trace part) will be introduced secondary artifacts, due to the mismatch of two different sinogram consistencies (in Fig. 3 (a)). In [8], they constructed a new sinogram to be transplanted using the high-frequency of denoised $S_{metal}^{reproj}$ (iteration used), and low-frequency of linear-interpolation (LI) data of $S_{metal}^{orig}$. This method may work well with iterative smoothing algorithms, however, noise amplification emerges as a problem in iteration-free cases.

As an alternative, low-frequency normalization was used. By the normalization step (in eq. 2, 3), the low frequency of the $S_{metal}^{reproj}$ was converged into the LI-data of $S_{metal}^{orig}$ (in Fig. 3 (b) and (d)).

$$N = G*\left(\frac{LI\,data\,of\,S_{metal}^{orig}}{S_{metal}^{reproj}}\right) \quad (2)$$

$$\mathbb{N}(S_{metal}^{reproj}) = N \cdot S_{metal}^{reproj} \quad (3)$$

$N$ is a normalization factor, $G$ is a smoothing filter (Gaussian was used in this study), and $\mathbb{N}(S_{metal}^{reproj})$ indicates normalized $S_{metal}^{reproj}$.

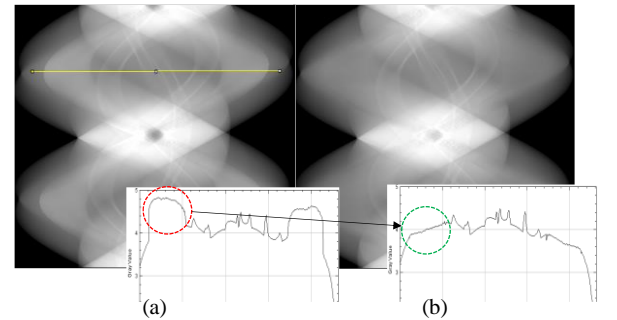

(a)                              (b)

Figure 3. (a) is a synthesized sinogram from the sinogram surgery, (b) is a synthesized sinogram after normalization. The line-profile of (a), (b) was presented following the yellow line of (a) in the bottom-left of images, respectively.

*D. Sinogram surgery and prior image generation*

The sinogram surgery was simply done by replacing $S_{metal}^{orig}$ to $\mathbb{N}(S_{metal}^{reproj})$ with the interpolation using $M^{corrupt}$ (in Fig. 4(b)). For the smooth connection in the boundary, the marginal gap of the metal trace was also interpolated. The corrected-sinogram was reconstructed using FBP, and the simple flat prior image was generated by bone segmentation using a threshold. The over all process is shown at Fig. 5.

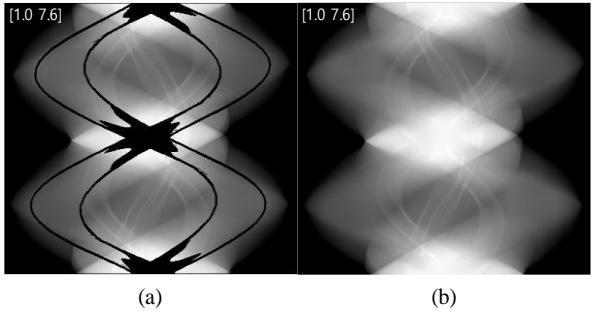

(a)           (b)

Figure 4. (a) represents the masked sinogram after surgery by $M^{corrupt}$, and the boundary of the metal trace. Then, the masked region was interpolated for artifacts removal.

*E. Materials*

Polychromatic (80kVp) x-ray measurement data of XCAT phantom were acquired using fan-beam CT simulation. The poisson random noises were added to create beam-hardening and photon starvation artifacts. Phantom 1 has two insertions of titanium with 360 views of projection data (SDD = 1300mm, SOD = 900mm). Phantom 2 has two insertions of titanium with 720 views of projection data (SDD = 2750mm, SOD = 2500mm). Also, a real sinogram of 120 kVp of helical CT projection data for monkeys with multiple metalic objects was rebinned and used to show the robustness of the proposed algorithm [11].

## 3 Results

In Fig. 6 the reconstructed results of phantom 1 and 2 are shown in the first and second row, respectively. The FBP results in (a) and (e) are severely corrupted by beam-hardening artifacts. The LI-MAR results in (b) and (f) show reduced artifacts, whereas they lost the structure of bones and introduced secondary artifacts. The proposed method

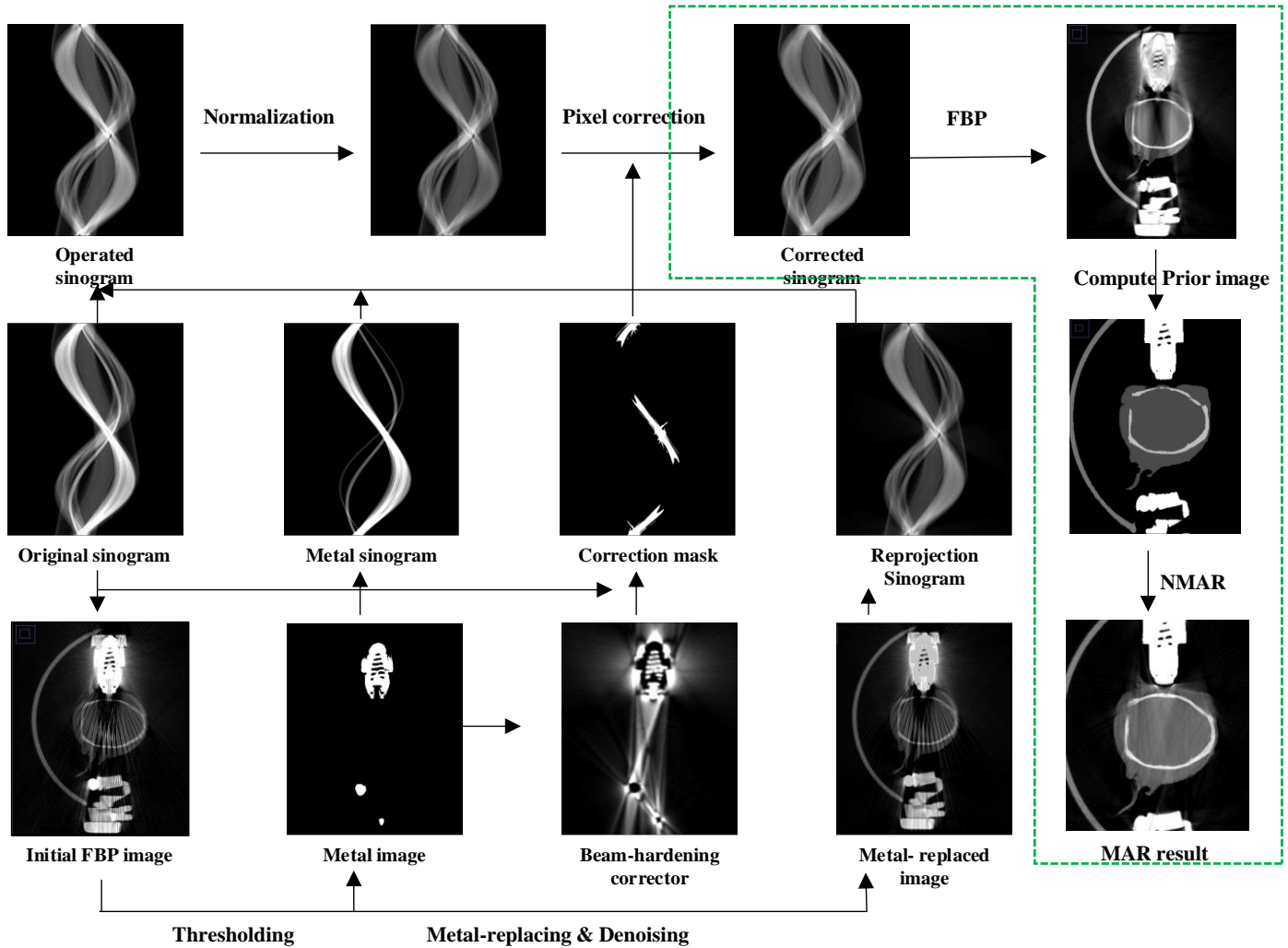

Figure 5. The overall procedure of proposed method. The green dashed lined procedure indicates the steps required for NMAR method.

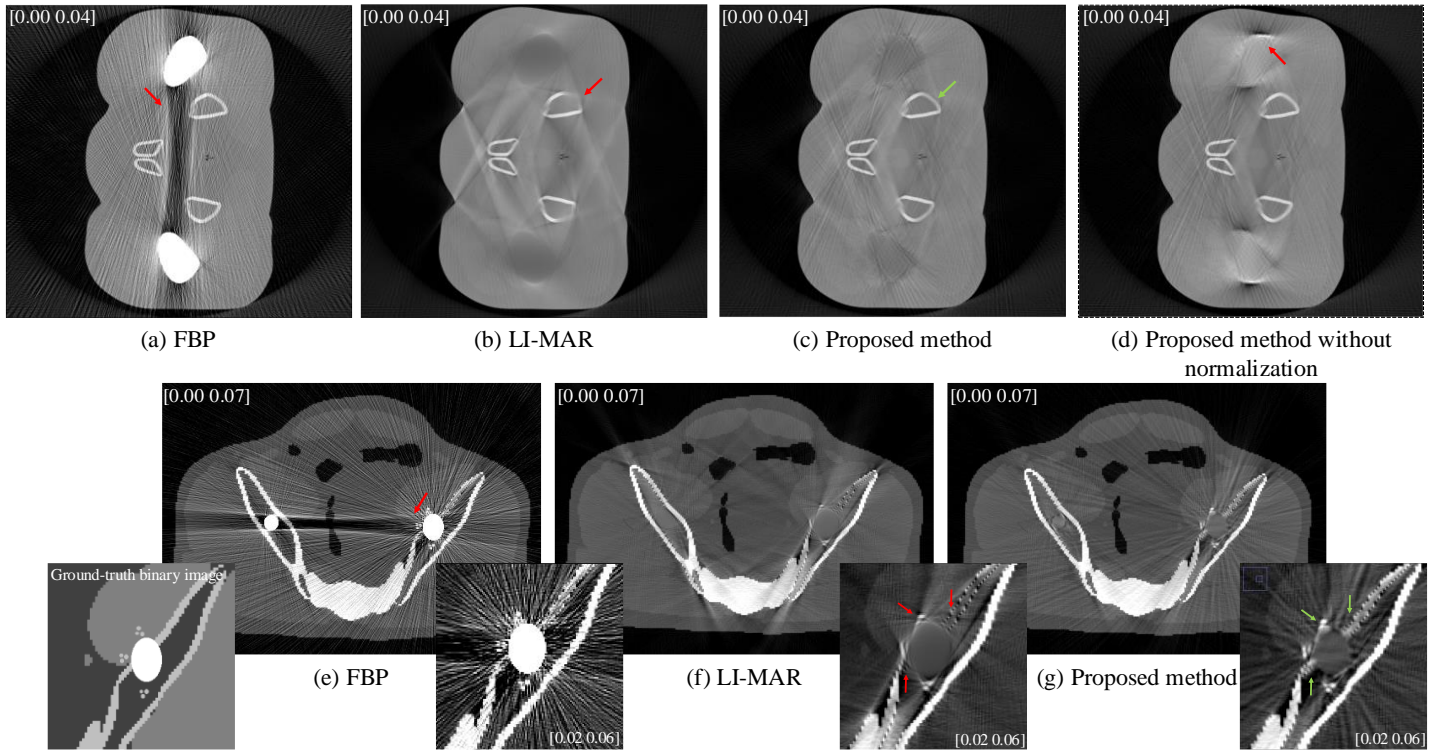

Figure 6. Reconstruction results of FBP, LI-MAR, and proposed method. (a), (b), and (c) indicates the results of phantom 1. (d) is the reconstrucition result of proposed method without normalization. (e), (f), and (g) indicates the results of phantom 2.

results in (c) and (g) show reduced artifacts with well-preserved bone.

In Fig. 7, the generated prior images are shown. It was possible to successfully describe the complete shape of bones without artifacts using the proposed method results.

In Fig. 8, the NMAR results of phantom 1 and 2 are shown in the first and second row, respectively. The results using FBP prior image in (a) and (d) introduced the streak artifacts. The results using LI-MAR prior image in (b) and (e) lost the structure of bones and introduced secondary artifacts. The results using the proposed prior image show well-preserved bone with the least secondary artifacts.

In Fig. 9, the NMAR results of real monkey head data are shown in last row, respectively. The results using FBP prior image introduced the streak artifacts and highlighted by red-dashed box. The results using LI-MAR prior lost the structure of bones and highlighted by red-dashed box. The results using the proposed prior image show well-preserved bone structure.

## 4 Discussion

As shown in Fig. 6 (c) and (g), the results of the proposed method contained more anatomical information than LI-MAR. This is because the proposed surgery method enables to fully utilize the anatomical data of $\mathbb{N}(S_{metal}^{reproj})$ than simply to use linearly interpolated data of the metal trace. Also, the proposed method greatly reduced the beam-hardening artifacts. It indicates $M^{corrupt}$ was effective to remove the artifact-corrupted region of the sinogram, even there are no additional iterative procedures. Meanwhile, the mitigated noises were presented in the proposed method due to the presence outside of $M^{corrupt}$. Yet, it was easily removed using the threshold in the prior image generation.

Furthermore, the strong secondary artifacts of Fig. 6 (d) were eliminated by the proposed normalization while maintaining the edges of bones, because it minimized the mismatch between $S^{orig}$ and $S_{metal}^{reproj}$ by treating only the low-frequency. Hence, these features serve as an advantage for computing a clean and anatomically informative prior image in Fig 7.

Figure 8 shows that the artifact-corrupted FBP prior image and incomplete LI-MAR prior image affect NMAR results to have streaks or lose the anatomical information. In contrast, the proposed prior image affects NMAR results to have well-preserved bones with the least secondary artifacts. For that reason, the accurately estimated prior image contributes to a proper normalization, and it minimizes the inconsistency in the interpolation. As a result, they also have the best soft-tissue contrast by avoiding artifacts corruption.

Also in Figure 9, they show similar aspects with above simulation studies. The NMAR result with FBP prior mage introduced streak artifacts and the case with LI-MAR prior image lose the bone structure information. However, the proposed method was successfully conserve the bone while reducing the artifacts. These results supports that the beamhardening corrector also works well in multiple metal cases. Thus, the robustness of algorithm was well described in this real data caes.

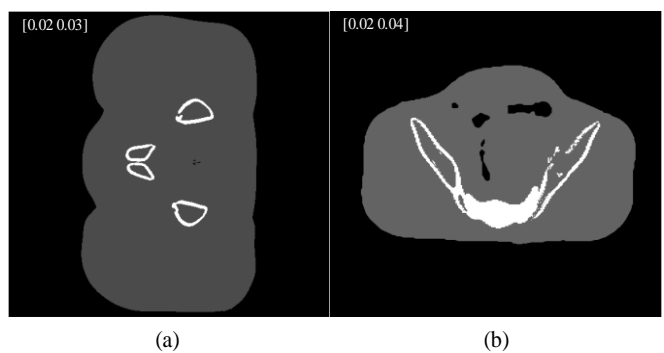

Figure 7. (a) is a prior image of phantom 1. (b) is a prior image of phantom 2. They are created from the image of Fig. 4 (c) and (g)

## 5 Conclusion

In our study, an efficient way for computing the high-quality prior image that does not require any iteration or optimization was proposed. Therefore, the complementary use of this technique for NMAR is possible to preserve anatomical structures, and increase the soft-tissue contrast without compromising their benefits in terms of speed, and computational cost.

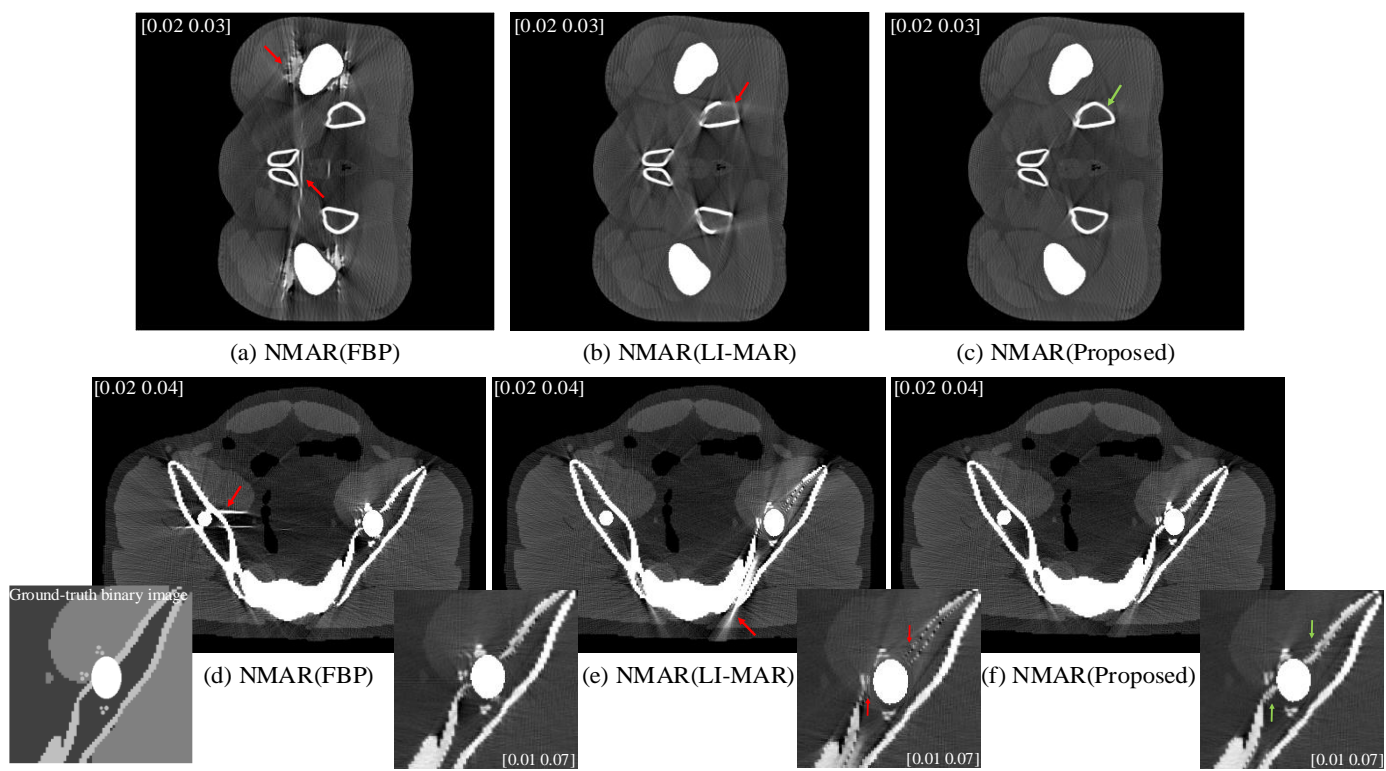

Figure 8. NMAR results of FBP, LI-MAR, and proposed method. (a), (b), and (c) indicates the results of phantom 1. (d), (e), and (f) indicates the results of phantom 2.

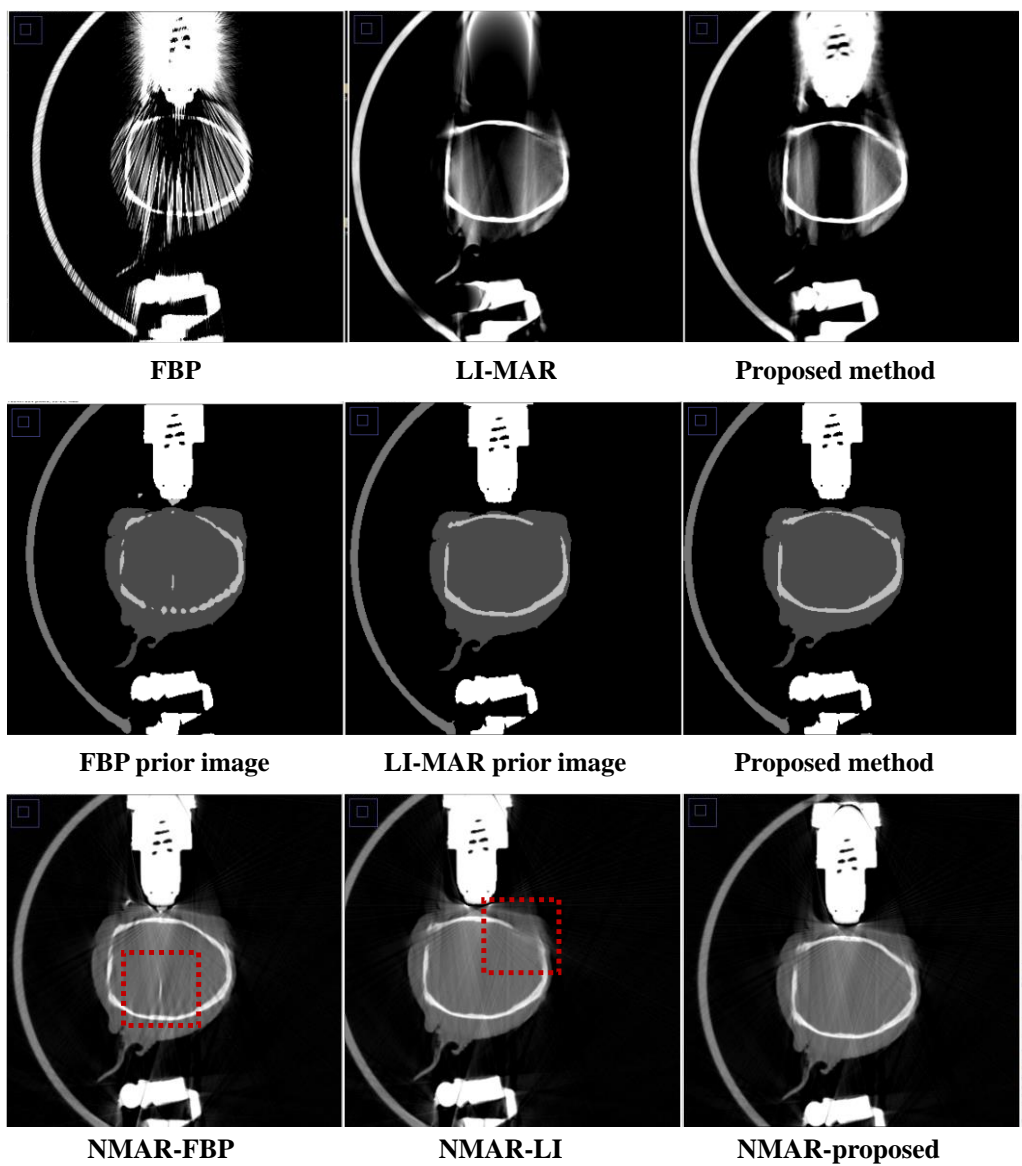

Figure 9. First row indicates the reconstructed monkey projection data by FBP, LI-MAR and  proposed method. Second row indicates corresponding prior image with different reconstruction methods. Last row indicates NMAR results of each methods/

# Feasibility of CBCT Scatter Estimation Using Trivariate Deep Splines

Philipp Roser[1,2], Annette Birkhold[2], Norbert Strobel[3], Markus Kowarschik[2], Andreas Maier[1], and Rebecca Fahrig[2]

[1]Pattern Recognition Lab, Department of Computer Science, Friedrich-Alexander Universität Erlangen-Nürnberg, Erlangen, Germany
[2]Advanced Therapies, Innovation, Siemens Healthcare GmbH, Forchheim, Germany
[3]Institute of Medical Engineering Schweinfurt, University of Applied Sciences Würzburg-Schweinfurt, Schweinfurt, Germany

**Abstract** Learning-based projection-wise scatter correction for cone-beam computed tomography can replace previous hardware- or software-based approaches. Recently, a learning-based approach has been proposed that constrains the estimated scatter to bivariate B-splines. While being on par with purely data-driven methods performance-wise, this method has introduced compelling benefits such as decreased parameter and runtime complexity, increased interpretability and improved data integrity. However, issues such as inconsistent scatter estimates among successive X-ray projections taken along a 3D scan remained. To obtain more consistent scatter estimates over 2D projections, this work seeks to extend the spline constraint to trivariate ones. In a simulation study using nested cross-validation, we show the overall feasibility. Nonetheless, quantitatively and qualitatively, the trivariate approach falls short of the bivariate approach, mainly due to limited training samples.

## 1 Introduction

X-ray cone-beam computed tomography (CBCT) is an important imaging technique in the interventional suite. It provides three-dimensional (3-D) image information, e.g., to assess outcomes during otherwise two-dimensional (2-D) fluoroscopically-guided procedures using flat-panel detector angiography systems. One major drawback of such systems is a large amount of scattered radiation due to the relatively large field of view and consequently irradiated volume. Besides the contrast deterioration in 2-D projection imaging, 3-D reconstructions may suffer from streaking or smearing artifacts due to scatter-induced inconsistencies. To reduce scatter-related artifacts, today's C-arm systems are typically equipped with a detector-side anti-scatter grid, which physically blocks large parts of the scatter. Unfortunately, such grids inevitably attenuate the primary radiation and lead to increased patient and, as a consequence, occupational dose. Furthermore, anti-scatter grids need to be precisely manufactured and they represent a considerable cost factor, especially for low-budget systems.

Consequently, much research has been devoted to investigating alternative, in particular, software-based approaches to scatter compensation [1]. Over the last decade, deep learning and, most notably, convolutional neural networks (CNN) found their way to modeling physical effects [2, 3]. As in many other fields, CNNs surpassed the state-of-the-art in various CBCT artifact correction tasks from a quantitative perspective, such as metal artifact reduction, denoising, as well as scatter compensation [4]. However, at the same time, questions arose how much such approaches might violate data integrity [5, 6]. For projection-wise deep-learning-based scatter compensation, we found in previous studies that constraining the co-domain of a CNN to bivariate B-splines yields considerable benefits over purely data-driven approaches [7, 8]. Besides reducing parameter and runtime complexity, the estimated scatter signal is ensured to be low-frequency, making it unlikely that the CNN introduces spurious artifacts such as edges or new, hallucinated details. First, by estimating the scatter for each projection independently, inconsistencies between successive projections can occur resulting in streaking artifacts. Second, CNNs tend to overestimate the scatter in low-intensity regions leading to highly inaccurate attenuation coefficients after applying the log-transform.

To increase consistency, we extend our previous bivariate B-spline based approach to trivariate B-splines in this work. This allows us to directly estimate the scatter for each projection image in a one-shot procedure. In this simulation study, we investigate the overall feasibility of constraining the projection stack scatter to a trivariate spline and evaluate potential pitfalls for future developments. As potential advantages, we consider two aspects. First, by constraining each projection's scatter to lie in the convex hull of its supporting spline coefficients, the overall consistency of the scatter-corrected projections is inherently increased. Second, estimating the scatter for each projection at once lays the foundation to tailor potentially better-suited loss functions to train the CNN, e.g., calculating the loss in different domains. In Sec. 2, we give a short primer on B-splines and introduce our notation for multivariate splines before discussing the learning-based scatter estimation and the experimental setup. Sec. 3 summarizes the results and compares them to our previous approach [8]. The report concludes with a discussion of the main findings in Sec. 4 before pointing out future research plans in Sec. 5.

## 2 Material and Methods

### 2.1 Image Formation Model

Neglecting typical photon shot noise, the X-ray projection scatter signal shows low-frequency characteristics within the 2-D projections, but also along the stack assuming a small enough angular increment. The flat-field normalized X-ray projection stack $\mathbf{I} \in \mathbb{R}^{p \times h \times w}$ with $p$ projections of width $w$ and height $h$ in pixels can be expressed by $\mathbf{I} = \mathbf{I}_p + \mathbf{I}_s$ with its primary component $\mathbf{I}_p$ and scatter component $\mathbf{I}_s$. While the primary component can be described quite accurately by the Beer–Lambert law, we introduce a trivariate spline as surrogate for the scatter component.

## 2.2 Trivariate B-splines

A non-uniform univariate B-spline series $s_{n,t}(u)$ of degree $d \in \mathbb{N}$ and order $n = d + 1$ in the parameter $u \in \mathbb{R}$ is defined by

$$s_{n,t}(u) = \sum_{i=1}^{N} c_i B_{i,n,t}(u) \quad , \quad (1)$$

with $N$ spline coefficients, comprising the coefficient vector $\boldsymbol{c} \in \mathbb{R}^N$. The knot vector $\boldsymbol{t} \in \mathbb{R}^{N+n}$ with $t_{i-1} \leq t_i \leq t_{i+1} \forall i$ recursively defines the B-spline in $u$ via convex combinations

$$B_{i,n,t}(u) = \frac{u - t_i}{t_{i+n} - t_i} B_{i,n,t}(u) + \frac{t_{i+n+1} - u}{t_{i+n+1} - t_{i+1}} B_{i+1,n-1,t}(u) \quad , \quad (2)$$

with

$$B_{i,1,t}(u) = \begin{cases} 1, & t_i \leq u < t_{i+1} \\ 0, & \text{else} \end{cases} \quad . \quad (3)$$

In the following, when there is no chance for ambiguity, we reduce the number of indices to improve readability, i.e., $s(u) = s_{n,t}(u)$ and $B_i(u) = B_{i,n,t}(u)$. Introducing vector notation for B-splines $\boldsymbol{B}_i(u) = (B_{i-d}(u), B_{i-d+1}, \ldots, B_i)^T$ and the supporting coefficients $\boldsymbol{c}_i = (c_{i-d}, c_{i+1-d}, \ldots, c_i)^T$, a non-uniform univariate B-spline series can also be represented as

$$s(u) = [\boldsymbol{B}_i(u)]^T \boldsymbol{c}_i \quad . \quad (4)$$

Multivariate B-splines series are constructed from univariate ones via the tensor product, e.g., a trivariate B-spline of degree $\boldsymbol{d} \in \mathbb{N}^3$ and order $\boldsymbol{n} = \boldsymbol{d} + 1$ in $\boldsymbol{u} \in \mathbb{R}^3$ series is defined by

$$s(\boldsymbol{u}) = \sum_{i_1=1}^{N_1} \sum_{i_2=1}^{N_2} \sum_{i_3=1}^{N_3} c_i B_{i_1}(u_1) B_{i_2}(u_2) B_{i_3}(u_3) \quad , \quad (5)$$

with the coefficient tensor $\mathbf{C} \in \mathbb{R}^{N_1 \times N_2 \times N_3}$ and $c_i = c_{i_1, i_2, i_3}$. Similar to the univariate case, the multivariate B-spline can also be defined in tensor notation $\mathbf{B}_i(\boldsymbol{u})$ via the outer product $\otimes$ of univariate B-spline vectors $\boldsymbol{B}_i(u)$

$$\mathbf{B}_i(\boldsymbol{u}) = \boldsymbol{B}_{i_1}(u_1) \otimes \boldsymbol{B}_{i_2}(u_2) \otimes \boldsymbol{B}_{i_3}(u_3) \quad . \quad (6)$$

Consequently, $s(\boldsymbol{u})$ is evaluated via tensor contractions $\langle \cdot, \cdot \rangle$

$$s(\boldsymbol{u}) = \langle \mathbf{B}_i(\boldsymbol{u}), \mathbf{C}_i \rangle \quad , \quad (7)$$

where $\mathbf{C}_i$ is constructed similarly to the univariate case.

## 2.3 Convolutional Spline Encoders

The tensor-based representation of multivariate B-splines can be seamlessly integrated into deep learning frameworks. To this end, we employ a lean convolutional encoder $f_\theta^{(k)} : \mathbb{R}^{w_1} \otimes \cdots \otimes \mathbb{R}^{w_k} \mapsto \mathbb{R}^{N_1} \otimes \cdots \otimes \mathbb{R}^{N_k}$, where $\theta$ comprises the parameters to train, $k$ is the dimensionality of the image data, $w_i$ are the dimensions of the image data, and $N_i$ are the dimensions of the spline coefficients. The encoder consists of four $k$-D convolutional blocks of two $3^k$ convolutional layers with 16 feature channels followed by a rectified linear unit activation [9]. Each convolutional block is preceded by a $2^k$ average pooling layer. The encoder is completed with a $1^k$ convolution to sum the feature channels followed by a global weighting layer, which establishes a global context [7, 8]. The architecture is depicted in Fig. 1. The network expects either a single 2-D X-ray projection or a 3-D projection stack $\mathbf{l}^{(k)}$ as input and outputs the corresponding coefficient tensor $\mathbf{C}^{(k)}$. This coefficient tensor is then evaluated to a discretely sampled spline $\tilde{\mathbf{l}}_s^{(k)}$ approximating the true scatter signal $\mathbf{l}_s^{(k)}$:

$$\tilde{\mathbf{l}}_s^{(k)} = \langle \mathbf{B}^{(k)}, f_\theta^{(k)}(\mathbf{l}^{(k)}) \rangle. \quad (8)$$

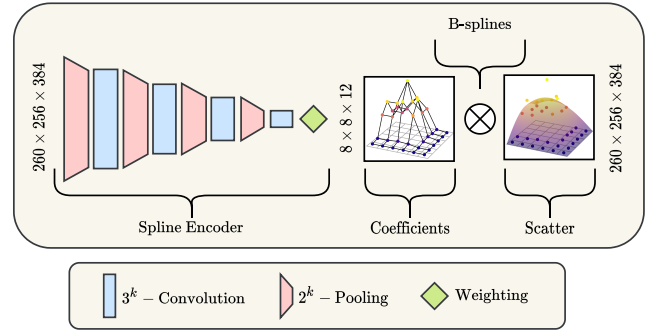

**Figure 1:** Architecture of the employed convolutional spline encoder with corresponding projection and spline dimensions for our proposed approach with $k = 3$. As baseline, we consider the exact same configuration with $k = 2$ [8]. As a consequence, for the 2-D case, the associated spline dimensions are $8 \times 12$, but one spline is estimated and evaluated for each projection separately.

The primary component is then calculated by subtracting the scatter estimate.

## 2.4 Simulation Study

For training, validation, and testing, we used the same synthetic data as in a previous study [8] comprising 20 head and 15 thorax scans, respectively. The data was simulated using MC-GPU [10] and openly available computed tomography scans from The Cancer Imaging Archive [11]. Each scan comprises $p = 260$ X-ray projections ($h = 768$, $w = 1152$) over an angular range of $200°$ with 785 mm source-to-isocenter distance and 1300 mm source-to-detector distance. Each projection image was simulated using $10 \times 10^9$ photons sampled from a 85 kV tungsten spectrum. To accelerate the training procedure, and reduce simulation noise, we applied Gaussian filtering and down-sampled the projections to $256 \times 384$ pixels. CBCT volumes were reconstructed on a 1 mm$^3$ spaced, isotropic $256^3$ voxel grid. We separately trained the networks in a nested 4*3 or 5*4 cross-validation for the head and thorax data, respectively. The network parameters are optimized using adaptive moments [12] with an initial $10^{-5}$ learning rate with respect to minimizing the mean absolute percentage error (MAPE) $\varepsilon_{\text{MAPE}}$

$$\varepsilon_{\text{MAPE}} = \frac{|\mathbf{B}^{(k)} f_{\theta}^{(k)}(\mathbf{I}^{(k)}) - \mathbf{I}_{\text{s}}^{(k)}|}{\mathbf{I}_{\text{s}}^{(k)}} \cdot 100\% \quad . \qquad (9)$$

The performance of the networks is assessed based on the structural similarity index (SSIM) with respect to the scatter-free ground truth simulations.

## 3 Results

Figure 2 shows boxplots for each test fold for both datasets. For all folds and datasets, the 3-D approach is considerably outperformed by the 2-D one. This is confirmed by the quantitative comparison displayed in Fig. 2, which shows exemplary error maps for scatter estimates and reconstructed central slices in Hounsfield units (HU). While, for the second thorax case, the 3-D approach yields more consistent results than the 2-D one, this is merely an exception. Even for the overall less challenging head dataset, the 3-D method's inferior scatter compensation leads to a few streaking artifacts and causes more severe HU inaccuracies.

## 4 Discussion

We presented an extension from bivariate B-splines to trivariate B-splines to constrain learning-based scatter estimation. This way, X-ray scatter can be inferred from the whole projection stack in a one-shot procedure. We investigated the performance for synthetic head and thorax data in a nested cross-validation procedure and compared the results to a bivariate baseline, shown to be on par with a purely data-driven method previously [8]. While we could demonstrate the overall feasibility of including trivariate splines, the results fell short of expectations. Throughout all experiments, the 3-D approach was outperformed by the 2-D baseline. However, we identify a likely reason responsible for these somewhat subpar results. Since the whole projection stack is processed in its entirety when applying the 3-D approach, it merely represents a single data point. Therefore, the total amount of data points is decreased 260-fold (i.e., by the number of projections). This implies that slightly misaligned anatomies in the test folds are perceived to lie out-of-distribution for the 3-D network. We believe that by increasing the training data tremendously, this issue can be resolved. However, we hasten to add that it is difficult to generate such an amount of training data. This is why we prefer, at least initially, the use of more sophisticated loss or regularizing functions, such as the absolute error in the reconstruction domain or epipolar consistency conditions. Future studies will show if the shortcoming of limited training data can be circumvented. As a compromise, merging the 2-D and 3-D approach could combine the advantages of both worlds. This could be realized by including a forced consistency check after the 2-D scatter estimation or using a stack of 2-D scatter estimates as an additional input to the 3-D network.

## 5 Conclusion

Trivariate splines can be used to constrain learning-based scatter estimation. In a first proof-of-concept study, projection-wise bivariate splines, however, outperformed the projection stack-wise trivariate splines. A combination of both approaches appears promising.

**Disclaimer:** The concepts and information presented are based on research and are not commercially available.

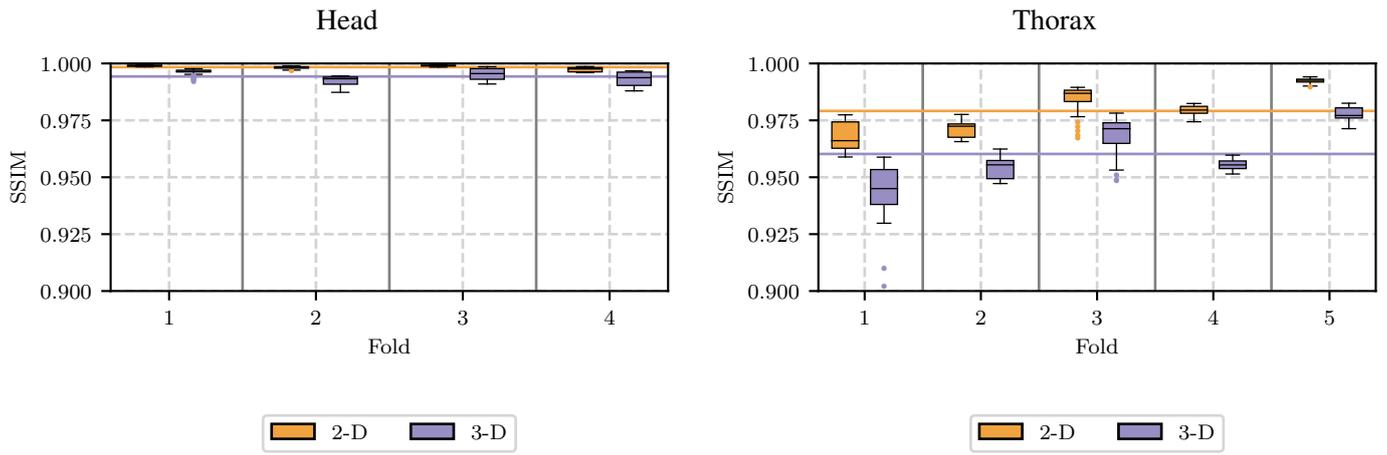

**Figure 2:** Quantitative results of the cross-validation for head (left) and thorax (right) reconstructions after scatter correction. The structural similarity index (SSIM) is given with respect to the simulated scatter-free ground truth data. The vertical lines encode the respective average values.

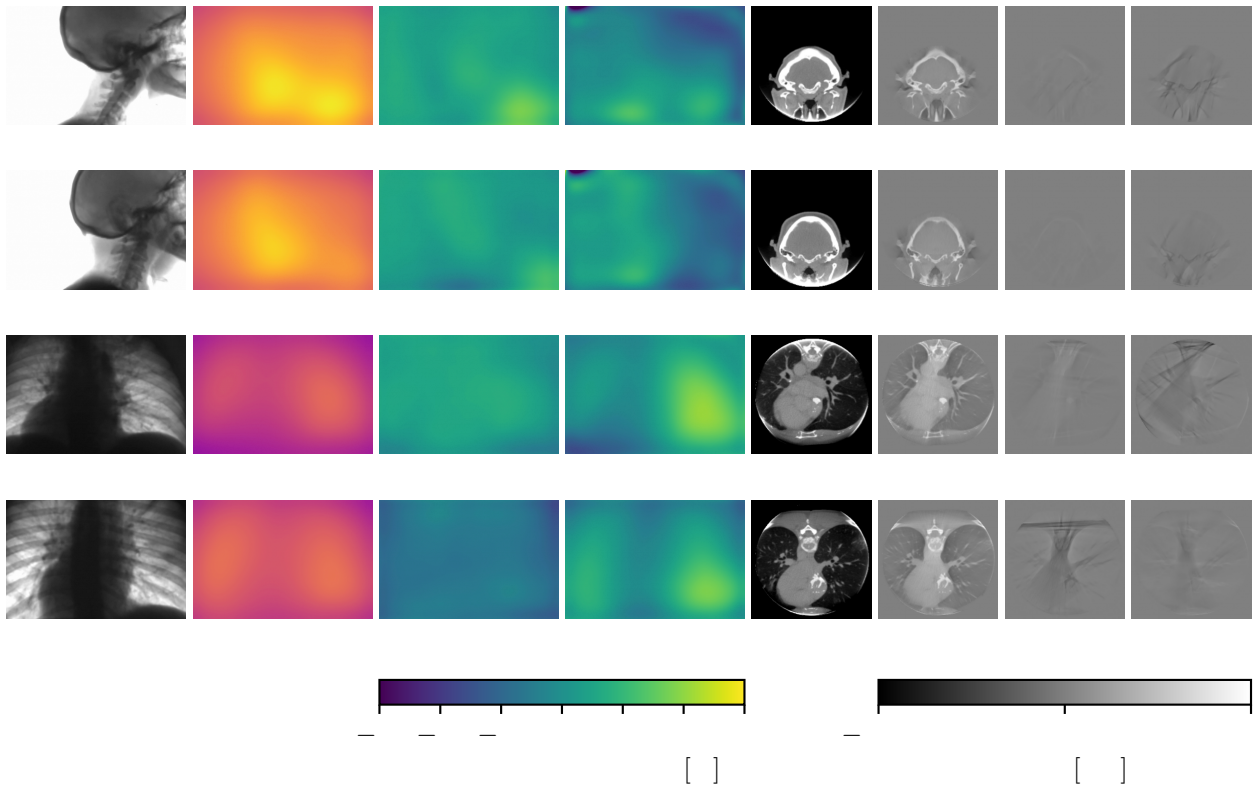

**Figure 3:** Qualitative error map examples calculated from predictions of the 2-D and 3-D networks for scatter estimates and respective reconstructed central slices. Below the error maps, the mean absolute percentage error (MAPE) and the absolute error in Hounsfield units (HU) is given for the scatter and the reconstructed central slices, respectively.

# Non-uniqueness in C-arm calibration with bundle adjustment

Anastasia Konik[1], Laurent Desbat[1], and Yannick Grondin[2]

[1]TIMC-GMCAO, Univ. Grenoble Alpes, Grenoble, France
[2]SurgiQual Institute, Meylan, France

**Abstract** In this work, we consider the problem of C-arm geometric calibration with the bundle adjustment (BA) method. This method was initially developed and used in computer vision. It's based on markers, but with unknown geometry. We don't know the 3D positions of the markers from which we have to estimate the calibration parameters. Thus, the geometric calibration is only based on the positions of the projected markers on 2D projection images. Pollefeys et al. [1] have shown that such calibration from BA can be performed up to a similarity transformation. In our work, we present a numerical solution to the C-arm geometric calibration with the BA method. We show the non-uniqueness of the geometric calibration with the BA method for cone-beam tomography. Just like in computer vision, we can't find the solution better than up to a similarity transformation.

## 1 Introduction

A C-arm X-ray imaging system is designed as a C-shaped arm which connects a X-ray source and a X-ray detector. We consider isocentric C-arms rotating around their isocenter. Usually projection images of a patient placed at the isocenter are collected. In the figure 1, we present some geometric parameters of a C-arm in a schematic view. This geometric model is classical in the cone-beam (CB) geometry.

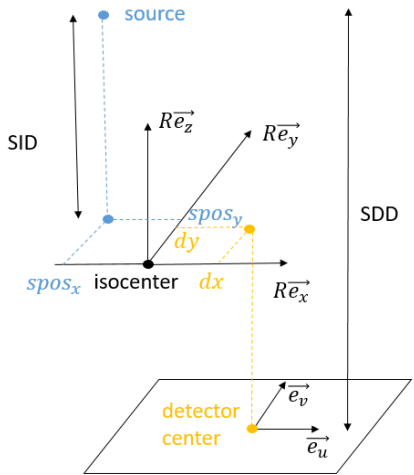

**Figure 1:** Some geometric parameters of a C-arm in a schematic view.

In the section 2 we recall the geometric model of a C-arm system. This projection geometric model maps 3D patient points to 2D detector points. The model contains geometric parameters. In general, these parameters need to be calibrated for each projection. The identification of these parameters is necessary for an accurate 3D reconstruction [2]. For most C-arms, in order to take into account mechanical vibrations over time, it is necessary to periodically perform the C-arm geometric calibration. In this work, we want to discuss bundle adjustment (BA) geometric calibration.

This C-arm geometric calibration process is similar to the camera geometric calibration in computer vision [3]. By analogy with computer vision, we can divide all image-based calibration methods for C-arms into two groups: calibration with markers and without markers as in [4]. For the first group, the calibration problem is solved with specific scans of a calibration object, usually based on few opaque markers. Either the 3D coordinates of marker centers are known in the world coordinate system or not. In the second case, both these 3D coordinates of marker centers and the geometric calibration parameters need to be identified. In this work, we consider this bundle adjustment problem and the method to solve this problem described in [5]. Just as for the calibration without markers, BA only uses the projection data and thus belongs to self-calibration methods.

## 2 Geometric calibration of a C-arm

As we know from computer vision [3], a camera can be modelled by a projection matrix $P$ mapping $Q$, a 3D point in the world coordinate system, to $q$, its corresponding projection onto the image plane. We usually have the decomposition of the projection matrix $P$:

$$P \sim \begin{pmatrix} K & 0 \end{pmatrix} \begin{pmatrix} R & -Rt \\ 0^T & 1 \end{pmatrix} = KR \begin{pmatrix} I & -t \end{pmatrix}. \qquad (1)$$

To be more precise, we can connect homogeneous coordinates of $q = (u, v, 1)^T$ in the pixel coordinate system and homogeneous coordinates of $Q = (X^w, Y^w, Z^w, 1)^T$ in the world coordinate system with special matrices $K$, $R$ and $t$:

$$\begin{pmatrix} u \\ v \\ 1 \end{pmatrix} \sim KR \begin{pmatrix} I & -t \end{pmatrix} \begin{pmatrix} X^w \\ Y^w \\ Z^w \\ 1 \end{pmatrix}, K = \begin{pmatrix} f_x & s & u_0 \\ 0 & f_y & v_0 \\ 0 & 0 & 1 \end{pmatrix}. \qquad (2)$$

The matrix $K$ consists of intrinsic calibration parameters. For the classical pinhole camera model the skew $s = 0$, $f_x = f_y = f$ is the focal length, $u_0$ and $v_0$ are the coordinates in the camera image of the orthogonal projection of the optical center. The rotation matrix $R$ and the translation vector $t$ are extrinsic calibration parameters; they describe the orientation and the position of the camera in the world coordinate system. Let us use here a IEC 61217 standard in order to describe the C-arm used in our simulation in the same way as cameras. In order to do this, 9 parameters are used, see the table 1.

Usually, we don't know exactly these parameters, so we need to calibrate them. We present some of these parameters in the figure 1.

| Parameter | Value | Noise bounds |
|---|---|---|
| $SDD$ (mm): source-detector distance | 1000 | $\pm 3.5$ |
| $SID$ (mm): source-isocenter distance | 700 | $\pm 6.9$ |
| $spos_x$ (mm): $x$-coordinate of the position of the source in the rotated frame | 0 | $\pm 6.9$ |
| $spos_y$ (mm): $y$-coordinate of the position of the source in the rotated frame | 0 | $\pm 6.9$ |
| $dx$ (mm): $x$-coordinate of the center of the detector in the rotated frame | 0 | $\pm 13.9$ |
| $dy$ (mm): $y$-coordinate of the center of the detector in the rotated frame | 0 | $\pm 13.9$ |
| $\theta_x$ (degrees): orientation of the rotated frame relative to the world frame along the $x$ axis | 0 | $\pm 1.4$ |
| $\theta_z$ (degrees): orientation of the rotated frame relative to the world frame along the $z$ axis | 0 | $\pm 1.4$ |
| $\theta_y$ (degrees): angle of scan | $i\delta$ | $\pm 0.7$ |

**Table 1:** C-arm calibration parameters: initial values and noise bounds used to simulate mechanical vibrations (we show in the third column bounds for the uniform distribution, we added a small uniform noise to initial values except the case of the SDD parameter for which we have completely different initial value, its realistic values for all projections are around 1300 mm). Here $i \in \mathbb{N}$ is the projection index, $\delta$ is the angular step.

We can describe the C-arm with approximately the same projection matrix as the basic pinhole camera. As in computer vision, we can build the intrinsic geometric calibration matrix $K$ with zero skew ($s = K_{12} = 0$). The rotation matrix $R$ is here the rotation around the isocenter defined by the position of the rotated frame. The translation $-t$ is the source position in the rotated frame:

$$P = K \begin{pmatrix} R & t \end{pmatrix}, t = \begin{pmatrix} -spos_x \\ -spos_y \\ -SID \end{pmatrix}, K = \begin{pmatrix} -f & 0 & u_0 \\ 0 & -f & v_0 \\ 0 & 0 & 1 \end{pmatrix}, \quad (3)$$

$$f = \frac{SDD}{dim_{\text{pixel}}}, u_0 = \frac{spos_x - dx}{dim_{\text{pixel}}}, v_0 = \frac{spos_y - dy}{dim_{\text{pixel}}}, \quad (4)$$

$$R = \begin{pmatrix} c_z & -s_z & 0 \\ s_z & c_z & 0 \\ 0 & 0 & 1 \end{pmatrix} \begin{pmatrix} 1 & 0 & 0 \\ 0 & c_x & -s_x \\ 0 & s_x & c_x \end{pmatrix} \begin{pmatrix} c_y & 0 & s_y \\ 0 & 1 & 0 \\ -s_y & 0 & c_y \end{pmatrix}, \quad (5)$$

$$c_\alpha = \cos(-\theta_\alpha), s_\alpha = \sin(-\theta_\alpha), \alpha \in \{x, y, z\}. \quad (6)$$

So, during the calibration we want to identify the elements of the projection matrix $P$ or the calibration parameters which define the projection matrix.

## 3 Bundle adjustment

In tomographic situations, we assume that we have collected many X-ray projections. For each X-ray projection $i$, we want to estimate both projection matrices $\hat{P}^i, i = 1, \dots, N_{\text{projections}}$ and unknown 3D marker points $\hat{Q}_j, j = 1, \dots, N_{\text{markers}}$ from known image points $q_j^i$. We minimize the mean of Euclidean distances between the projected points and the measured image points for all X-ray images, i.e.

$$\min_x D(x) \overset{\text{def}}{=} \min_{P^i, Q_j} \frac{1}{N_{\text{markers}} N_{\text{projections}}} \sum_{i,j} d(P^i Q_j, q_j^i)^2, \quad (7)$$

where $d(q_1, q_2)$ is a geometric image distance between homogeneous points $q_1$ and $q_2$, $D$ is the cost function, $x$ is a vector containing the parameters of $P^i, i = 1, \dots, N_{\text{projections}}$, $Q_j, j = 1, \dots, N_{\text{markers}}$, so in our case $x$ contains $9N_{\text{projections}} + 3N_{\text{markers}}$ parameters to be identified. This is the general formulation of the BA problem.

In [5] authors described basic local optimization methods for differentiable functions to solve the BA problem. Let us try to minimize the cost function $D(x)$ over $x$ with the initial estimate $x_0$. We want to find a displacement $\delta x$ which locally minimizes $D(x)$. This cost function can be replaced by an approximate local model. The quadratic local model is based on the Taylor expansion:

$$D(x + \delta x) \approx D(x) + g^T \delta x + \frac{1}{2} \delta x^T H \delta x, \quad (8)$$

where $g$ is a gradient vector of $D$ at $x$ and $H$ is a Hessian matrix at $x$. In [5] authors proposed methods to optimize such as the damped Newton methods which solve the following regularized system:

$$(H + \lambda W)\delta x = -g, \quad (9)$$

where $\lambda$ is a weighting factor and $W$ is a positive definite weight matrix. This is the basis for trust region methods, for example, the popular Levenberg-Marquardt method. We use this method in our numerical experiments.

## 4 Numerical experiments: calibration with BA

In order to solve numerically the optimization problem (7) we used the C++ package Ceres [6]. We simulated data for our numerical experiments. Firstly, we started with 20 markers, see the 3D plot in the figure 2. We call these points true values $Q_{j,\text{true}}$. Then we fixed $dim_{\text{pixel}} = 0.5$ mm, $N_{\text{projections}} = 181$ with the angular step as 2 degrees. We computed from the table 1 the initial values for the 9 calibration parameters $f^i$, $u_0^i$, $v_0^i$, $\theta_x^i$, $\theta_y^i$, $\theta_z^i$, $t_x^i$, $t_y^i$, $t_z^i$ for each projection $i$. These values were used as initial estimations for our optimization algorithm. In order to simulate realistic values for calibration parameters and $P_{\text{real}}^i$ which correspond to clinical situations with mechanical vibrations of the C-arm, we added a uniform noise to each calibration parameter (see the table 1 for details).

We simulated $q_j^i$ by the multiplication of $P_{\text{real}}^i$ by $Q_{j,\text{true}}$. We usually don't know exactly $q_j^i$, because we detect key points on X-ray images using specific algorithms. In order to simulate this process, we also added a uniform noise with bounds $\pm 0.3$ pix to the image points $P_{\text{real}}^i Q_{j,\text{true}}$. Thus, as inputs for the optimization algorithm we had noisy image points $q_j^i$.

In order to start the minimization algorithm, we also need initial estimations for 3D points $Q_j$ (we described before just the initial calibration parameters). We did our first guess with the basic triangulation algorithm of Python's OpenCV from two known initial projection matrices and known projections for 0 and 90 degrees. A full description of the basic triangulation algorithm could be found in [3]. After this simulation and initialisation, the optimization algorithm was launched. We started from the initial cost 1337. With the Levenberg-Marquardt method we achieved the final cost 0.003.

We show in the figure 2 estimated 3D points obtained by this algorithm. We calculated the reprojection error as $\frac{1}{N_{\text{markers}} N_{\text{projections}}} \sum_{i,j} ||\hat{P}^i Q_{j,\text{true}} - q_j^i||_2$. It is equal to 34.41 pix. The maximum errors for the estimated calibration parameters through all projections are: 9.78 pix for $f$, 10.28 pix for $u_0$, 9.65 pix for $v_0$ (1 pix is 0.5 mm), 102.09 mm for $t$. In order to compare rotation matrices, we calculated the error rotation matrix for each projection $i$ as $R_{\text{err}}^i = (R_{\text{real}}^i)^{-1} \hat{R}^i$. According to the Euler rotation theorem, each rotation $R_{\text{err}}^i$ in three dimensions is defined by its axis and its angle $\psi_i$. We found the absolute value of $\psi_i$ from the error rotation matrix $R_{\text{err}}^i$ for each projection $i$ with $|\psi_i| = \arccos\frac{\text{tr}(R_{\text{err}}^i) - 1}{2}$. The maximum error for angles $|\psi_i|$ through all projections was 0.79 degrees. We observed the high reprojection error and high errors in calibration parameters. The same for 3D points: true and estimated 3D points are far to be exactly the same. Thus, we found the solution of the calibration problem and it differs from the true solution. But what is the reason?

## 5 Theoretical explanation of non-uniqueness

### 5.1 Computer vision BA limits

Let us start with different classes of transformations of 3D space. Let us remind that in computer vision we usually use homogeneous coordinates of the point, so for the point in 3D we have four coordinates. We identify each transformation by its matrix form. Moreover, these transformations form a hierarchy. So, we start with the general one.

**Definition 5.1.** A projective transformation is a transforma-

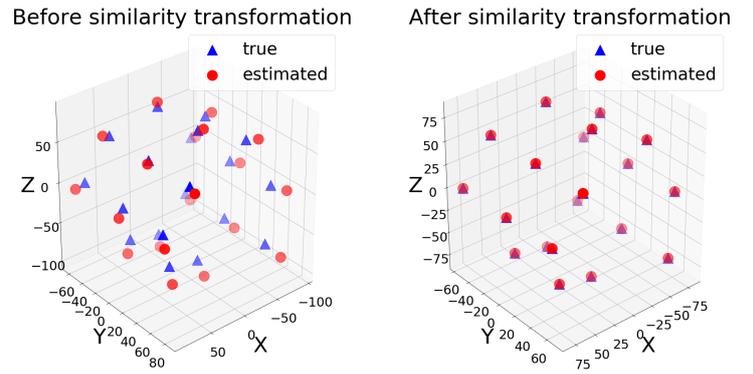

Before similarity transformation    After similarity transformation tion of the form $\begin{pmatrix} A & t \\ h^T & v \end{pmatrix}$, where $A$ is an invertible $3 \times 3$ matrix, $h$ is a general 3-vector.

**Figure 2:** True and estimated 3D points for Levenberg-Marquardt method.

**Definition 5.2.** An affine transformation is a transformation of the form $\begin{pmatrix} A & t \\ 0^T & 1 \end{pmatrix}$, where $A$ is an invertible $3 \times 3$ matrix.

**Definition 5.3.** A similarity transformation is a transformation of the form $\begin{pmatrix} \sigma R & t \\ 0^T & 1 \end{pmatrix}$, where $R$ is a $3 \times 3$ rotation matrix and $\sigma \neq 0$.

**Definition 5.4.** An Euclidean transformation is a transformation of the form $\begin{pmatrix} R & t \\ 0^T & 1 \end{pmatrix}$, where $R$ is a $3 \times 3$ rotation matrix.

From the literature [3] we know that we have a solution of the BA problem up to a projective transformation. We can take an invertible matrix $H$ and have as a solution also $\hat{P}^i H^{-1}$, $H\hat{Q}_j$. Moreover, we found in [1] that for special calibration matrices with zero skews ($s = 0$) the solution could be found up to a similarity transformation. The following theorem from [1] is true:

**Theorem 5.1.** The class of transformations which preserves the absence of skew is the group of similarity transformations.

If we have projection matrices as solutions of our calibration problem, they differ by some projective transformation. With this theorem, if the sequence of views is general enough and if in decompositions of the projection matrices we have zero skews, this projective transformation should be a similarity transformation. We build projection matrices for the C-arm BA problem exactly such that in decompositions they have zero skews. In essentially all digital X-ray CB systems the skew is zero because the lines and columns of digital X-ray detectors are perpendicular. Unfortunately, the sequence of C-arm positions often couldn't be general enough, which complicates the application of the theorem.

## 5.2 Cone-beam geometric self-calibration limits

Any CB system can be also described by an integral model. We consider cone-beam data in the form

$$d(\vec{s}_\lambda, \vec{\zeta}) = Df(\vec{s}_\lambda, \vec{\zeta}) = \int_0^{+\infty} f(\vec{s}_\lambda + l\vec{\zeta})dl, \qquad (10)$$

where $\lambda \in \Lambda \subset \mathbb{R}$ is a trajectory parameter of the source, $\vec{s}_\lambda$ is the 3D position of the source at $\lambda$, $\vec{\zeta}$ is a unit vector in $\mathbb{R}^3$, the direction of the integration line.

**Theorem 5.2.** Let $f_{R,\vec{t}}(\vec{x}) \overset{\text{def}}{=} f(R\vec{x} + \vec{t})$ for any $\vec{x} \in \mathbb{R}^3$, for any rotation $R$ and any translation vector $\vec{t} \in \mathbb{R}^3$, then

$$Df_{R,\vec{t}}(\vec{s}_\lambda, \vec{\zeta}) = Df(R\vec{s}_\lambda + \vec{t}, R\vec{\zeta}). \qquad (11)$$

*Proof.* We have $Df_{R,\vec{t}}(\vec{s}_\lambda, \vec{\zeta}) = \int_0^{+\infty} f_{R,\vec{t}}(\vec{s}_\lambda + l\vec{\zeta})dl = \int_0^{+\infty} f(R\vec{s}_\lambda + lR\vec{\zeta} + \vec{t})dl = Df(R\vec{s}_\lambda + \vec{t}, R\vec{\zeta})$. □

Thus, the cone-beam data $Df_{R,\vec{t}}$ of $f_{R,\vec{t}}$ from the source $\vec{s}_\lambda$ in the direction $\vec{\zeta}$ is nothing but the cone-beam data $Df$ of $f$ from the source position $R\vec{s}_\lambda + \vec{t}$ toward the direction $R\vec{\zeta}$. Conversely, let $\vec{v}_\lambda = R\vec{s}_\lambda + \vec{t}$ or $\vec{s}_\lambda = R^T(\vec{v}_\lambda - \vec{t})$, let $\vec{\eta} = R\vec{\zeta}$, thus $\vec{\zeta} = R^T\vec{\eta}$. Then for all $\vec{v}_\lambda$ and all unit $\vec{\eta}$ Eq. (11) is equivalent to

$$Df_{R,\vec{t}}\left(R^T(\vec{v}_\lambda - \vec{t}), R^T\vec{\eta}\right) = Df(\vec{v}_\lambda, \vec{\eta}). \qquad (12)$$

Thus, the projection data $Df$ of $f$ acquired from the source position $\vec{v}_\lambda$ toward the direction $\vec{\eta}$ is equal to the projection $Df_{R,\vec{t}}$ acquired from the source position $R^T(\vec{v}_\lambda - \vec{t})$ toward the direction $R^T\vec{\eta}$ for any rotation $R$ and translation vector $\vec{t}$. This shows that for the cone-beam geometry the geometric self-calibration problem can not be solved better than up to an Euclidean transformation. Moreover, we have

**Theorem 5.3.** Let $f_{\sigma R,\vec{t}}(\vec{x}) \overset{\text{def}}{=} f(\sigma R\vec{x} + \vec{t})$ for any $\vec{x} \in \mathbb{R}^3$, rotation $R$, translation $\vec{t} \in \mathbb{R}^3$ and scaling $\sigma > 0$, then

$$D\left(\sigma f_{\sigma R,\vec{t}}\right)\left(\vec{s}_\lambda, \vec{\zeta}\right) = Df\left(\sigma R\vec{s}_\lambda + \vec{t}, R\vec{\zeta}\right). \qquad (13)$$

*Proof.* We have $Df_{\sigma R,\vec{t}}(\vec{s}_\lambda, \vec{\zeta}) = \int_0^{+\infty} f_{\sigma R,\vec{t}}(\vec{s}_\lambda + l\vec{\zeta})dl = \int_0^{+\infty} f(\sigma R\vec{s}_\lambda + \sigma lR\vec{\zeta} + \vec{t})dl = \int_0^{+\infty} f(\sigma R\vec{s}_\lambda + nR\vec{\zeta} + \vec{t})d\frac{n}{\sigma} = \frac{1}{\sigma}Df(\sigma R\vec{s}_\lambda + \vec{t}, R\vec{\zeta})$. □

Thus, the cone-beam data $D(\sigma f_{\sigma R,\vec{t}})$ of $\sigma f_{\sigma R,\vec{t}}$ from the source $\vec{s}_\lambda$ in the direction $\vec{\zeta}$ is nothing but the cone-beam data $Df$ of $f$ from the source position $\sigma R\vec{s}_\lambda + \vec{t}$ toward the direction $R\vec{\zeta}$. Conversely, let $\vec{v}_\lambda = \sigma R\vec{s}_\lambda + \vec{t}$ or $\vec{s}_\lambda = \frac{1}{\sigma}R^T(\vec{v}_\lambda - \vec{t})$ and let $\vec{\eta} = R\vec{\zeta}$, thus $\vec{\zeta} = R^T\vec{\eta}$, then for all $\vec{v}_\lambda$ and all unit vector $\vec{\eta}$ Eq. (13) is equivalent to

$$D\left(\sigma f_{\sigma R,\vec{t}}\right)\left(\frac{1}{\sigma}R^T(\vec{v}_\lambda - \vec{t}), R^T\vec{\eta}\right) = Df(\vec{v}_\lambda, \vec{\eta}). \qquad (14)$$

Thus, for the cone-beam geometry the geometric self-calibration problem can not be solved better than up to a similarity transformation.

## 6 Numerical experiments: similarity error identification

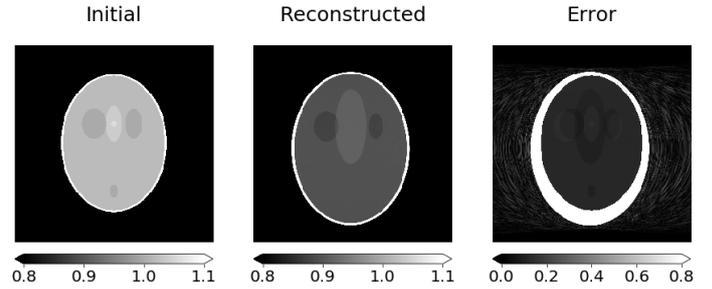

**Figure 3:** Slices $z = 6.5$ mm of the initial 3D Shepp–Logan phantom $f(\vec{x})$ (left), the reconstruction $g(\vec{x})$ from the estimated acquisition geometry (center) and $|f(\vec{x}) - g(\vec{x})|$ (right).

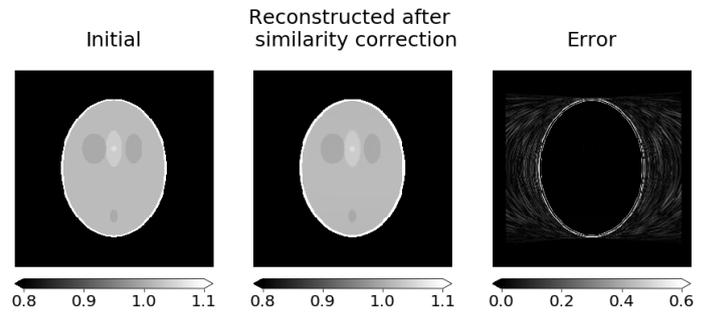

**Figure 4:** Slices $z = 6.5$ mm of the initial 3D Shepp–Logan phantom $f(\vec{x})$ (left), the reconstruction from the estimated acquisition geometry after the similarity correction $\frac{1}{\sigma}g\left(\frac{1}{\sigma}R^T(\vec{x} - \vec{t})\right)$ (center) and $\left|f(\vec{x}) - \frac{1}{\sigma}g\left(\frac{1}{\sigma}R^T(\vec{x} - \vec{t})\right)\right|$ (right).

The numerical experiments from the section 4 provided a scaling, a rotation and a translation, i.e. a similarity transformation, for computing the set of the true 3D marker coordinates from the set of the estimated 3D marker coordinates. Firstly, we computed the scaling factor. We computed the barycenters $b_{\text{true}}$ and $b_{\text{est}}$ of the true and the estimated 3D points. The mean of $\frac{||Q_{j,\text{true}} - b_{\text{true}}||_2}{||Q_j - b_{\text{est}}||_2}$ is a simple (and sufficient) estimation of the scaling. It is equal to 0.88. Then, after the scaling correction, we numerically found the rotation and the translation with the algorithm described in [7]. We can then apply to the set of scaled estimated points the rotation approximately equal to the identity matrix and the translation $(0.96, -3.14, 5.18)^T$ in mm. We show the result of such transformation of the estimated 3D points to the true 3D points in the figure 2.

The estimated calibration parameters from the section 4 can be used to perform a reconstruction. We started with $f(\vec{x})$ being the 3D Shepp–Logan phantom. We computed projections with the true acquisition geometry. From these data we performed a FDK reconstruction denoted $g(\vec{x})$ with the estimated acquisition geometry using the Python package RTK [8]. According to Eq. (13), the reconstructed image corresponds to the function $\sigma f_{\sigma R,\vec{t}}(\vec{x})$. In the figure 3, we show the same slice of both $f$ and $g$, and of $|f - g|$.

We then computed the similarity correction applied to $g$. The reconstruction $g(\vec{x})$ obtained from the estimated geometric parameters should be equal to $\sigma f_{\sigma R,\vec{t}}(\vec{x}) = \sigma f(\sigma R\vec{x} + \vec{t})$, thus $f(\vec{x})$ should be equal to $\frac{1}{\sigma} g\left(\frac{1}{\sigma} R^T\left(\vec{x} - \vec{t}\right)\right)$. We used $\sigma, R, \vec{t}$ estimated at the beginning of this section from the BA results obtained in the section 4 in order to compute the similarity correction, thus an estimation of the original image $f$. Note that an interpolation is needed for the image grid computation: we used the linear interpolation method `interpolate.RegularGridInterpolator()` from SciPy. After such similarity correction applied to the image $g$ we obtained an estimation $\frac{1}{\sigma} g\left(\frac{1}{\sigma} R^T\left(\vec{x} - \vec{t}\right)\right)$ of the initial image $f(\vec{x})$ (see the figure 4).

The widely used root-mean-square error (RMSE) between the initial 3D image and the reconstructed 3D image after the similarity correction was 0.08. For example, RMSE between the initial 3D image and the reconstructed with the true acquisition geometry 3D image was 0.09, which is normal for the numerical reconstruction implemented in RTK. Thus, we showed that the reconstruction with the estimated acquisition geometry can be performed and we verified that then the reconstructed image has the form $\sigma f_{\sigma R,\vec{t}}(\vec{x})$.

# 7 Conclusion

We have presented a numerical solution to the C-arm geometric calibration problem with the BA method. In simulations, we have observed the following phenomena: there are high errors in the estimated 3D marker positions and calibration parameters, but true and estimated 3D marker points differ almost by a similarity transformation. We analyzed the existing computer vision theory and translated it to the X-ray cone-beam geometry. Cone-beam geometric self-calibration problems can not be solved better than up to a similarity transformation. In the previous section we also presented the numerical verification of our theoretical result.

# Acknowledgement

This work is supported by the Univ. Grenoble Alpes IDEX grant CQFD, the "Fonds unique interministériel", the European Union FEDER in Auvergne-Rhône-Alpes (3D4Carm), the ANR ROIdoré (ANR-17-CE19-0006-01) and the ANR CAMI LABEX (ANR-11-LABX-0004-01).

# L1/L2 Minimization Based Method for Sparse-view CT Reconstruction

Xiaohuan Yu[1], Ailong Cai[1], Zhizhong Zheng[1], Lei Li[1], Fagui Zhang[2] and Bin Yan[1]

[1]Information Engineering University, Zhengzhou, China
[2]No. 988 Hospital of Joint Logistic Support Force, Zhengzhou, China
(Corresponding author: Bin Yan. Information Engineering University, Zhengzhou, China)
(Email: yu_xiao_huan@163.com; cai.ailong@163.com; zhengzz81@163.com; leehotline@163.com; 303864034@qq.com; ybspace@hotmail.com)

**Abstract** In this paper, we propose a nonconvex L1/L2 ratio model on the image gradient for sparse-view CT reconstruction. Based on this ratio model, we design an iterative algorithm under the forward-backward splitting framework and optimize the corresponding augmented Lagrangian function by minimizing each sub-problem alternatively. In this paper, efficient solvers are developed for each sub-problem. Ideal numerical verification indicates that the proposed algorithm behaves convergent and stable for under-sampled views. The experimental results verify that the proposed method is promising compared with classical methods for sparse-view CT reconstruction.

## 1 Introduction

Computed tomography (CT) is an advanced technique in many fields, such as medicine, industry, materials science, and so on. In practical application, the projection data is often incomplete because of various reasons. Sparse-view is one of the incomplete projection problems and it has great significance in the reduction of radiation dose. However, sparse-view CT reconstruction always has high level of noise and artifacts and is difficult to obtain high quality images because of incomplete projections.

The emergence of compressed sensing (CS) theory [1-2] brings a new idea to solve the incomplete projection data reconstruction problems and can be applied to ill-posed problem of sparse-view CT reconstruction. Total variation (TV) has become a powerful method in exploring the sparsity of gradient image and preserving edges [3]. Sidky and Pan [4] designed a TV and projection onto convex set based method for incomplete projection data reconstruction and achieved a huge success. Since then, many other variants of TV-based method have been proposed, such as the fractional-order TV [5], total generalized variation [6] and anisotropic TV by introducing direction information [7], et al. Although these TV based method have many successful applications, they still suffer from lack of contrast and staircase artifacts while prefers piece-wise constant images.

The ratio model L1/L2 is a sparse measure to approximate the desired L0 norm first appeared in [8] and was further proved to be a scale invariant model in [9]. Motivated by recent L1/L2 model based works [10, 11] for sparse signal recovery, we utilize the ratio model of L1/L2 on the gradient as the regularization term and design an iterative algorithm for sparse-view CT reconstruction. The experiments illustrate the promising effectiveness of our proposed method when the projection views are insufficient.

The rest of the paper is organized as follows. In section 2, we present the ratio model and explain the proposed algorithm in detail. In section 3, a simulation experiment is tested to evaluate the performance of the proposed method. Discussions and conclusions are given in section 4 and 5.

## 2 Materials and Methods

### A. Preliminaries

In this study, we consider the discreted system for CT reconstruction:

$$Au = f \tag{1}$$

where $u \in R^{N^2}$ denotes the unkonwn image to be reconstructed. $A$ is the system matrix. $f$ represents the measured projection data. However, problem (1) is usually ill-posed because the obtained projection data are incomplete in sparse-view CT reconstruction. So to solve this problem, we utilize the total variation (TV) as the regularization term that contains the prior information of the reconstructed image, and the model is as follows

$$\min_u \|\nabla u\|_1 \quad s.t. \ Au = f, \tag{2}$$

where $\nabla u := (\nabla_x u, \nabla_y u)$, and the TV term, i.e., $\|\nabla u\|_1$ is defined by Eq. (3),

$$\|\nabla u\|_1 = \|\nabla_x u\|_1 + \|\nabla_y u\|_1. \tag{3}$$

### B. L1/L2-TV model and algorithm

Now, we consider a constrained model of L1/L2 on the gradient image (L1/L2-TV) for sparse-view CT reconstruction as

$$\min_u \frac{\|\nabla u\|_1}{\|\nabla u\|_2} \quad s.t. \ Au = f. \tag{4}$$

The optimal conditions of (4) are

$$\begin{cases} 0 \in \frac{\partial \|\nabla u\|_1}{\|\nabla u\|_2} - \frac{\|\nabla u\|_1}{\|\nabla u\|_2^2} \partial \|\nabla u\|_2 + A^T \lambda, \\ 0 = Au - f, \end{cases} \tag{5}$$

where $\lambda$ is some vector. By introducing $\tilde{\lambda} = \|\nabla u\|_1 \lambda$, we further get

$$\begin{cases} 0 \in \partial \|\nabla u\|_1 - \frac{\|\nabla u\|_1}{\|\nabla u\|_2} \partial \|\nabla u\|_1 + A^T \tilde{\lambda}, \\ 0 = Au - f. \end{cases} \tag{6}$$

The conditions (6) are also the optimal conditions of another optimization problem:

$$\min_u \; g(u) - h(u), \tag{7}$$

where

$$g(u) = \|\nabla u\|_1 + \delta_{Au=f}(u), \tag{8}$$

and the function $h(u)$ satisfies

$$\partial h(u) := \frac{\|\nabla u\|_1}{\|\nabla u\|_2} \partial \|\nabla u\|_2. \tag{9}$$

Note that $h(\cdot)$ does not have an explicit expression from (9) and $\partial h(\cdot)$ represents the sub-differential of $h(\cdot)$.

And the indicator function $\delta_{Au=f}(u)$ is defined as

$$\delta_{Au=f}(u) = \begin{cases} 0, & \text{if } Au = f, \\ +\infty, & \text{otherwise.} \end{cases} \tag{10}$$

From the problem (7) and its optimal conditions (6), we utilize the forward-backward splitting method to get the iterative scheme

$$(\tau u^k + v^k) \in \tau I + \partial g \; (\bar{u}^k), \tag{11}$$

where $g$ is defined in (8) and $v^k \in \partial h(u^k)$. And then we design two descent steps according to the obtained point $\bar{u}^k$. The details of the proposed algorithm are summarized in Algorithm1.

$\bar{u}$ -sub-problem of (11) can be written as

$$\bar{u}^k = \arg\min_u \left\{ g(u) + \frac{\tau}{2} \left\| u - \frac{\tau u^k + v^k}{\tau} \right\|_2^2 \right\}$$

$$= \arg\min_u \left\{ \|\nabla u\|_1 + \frac{\tau}{2} \left\| u - \frac{\tau u^k + v^k}{\tau} \right\|_2^2 \; s.t. \; Au = f \right\}.$$

By introducing an auxiliary variable $z$, the augmented Lagrangian function is

$$L_A(u, z; \tilde{\lambda}, w) = \|z\|_1 + \frac{\tau}{2} \left\| u - \frac{\tau u^k + v^k}{\tau} \right\|_2^2 - \tilde{\lambda}^T (Au - f)$$

$$+ \frac{\rho_1}{2} \|Au - f\|_2^2 - w^T (z - \nabla u) + \frac{\rho_2}{2} \|z - \nabla u\|_2^2.$$

where $\tilde{\lambda}, w$ represent the Lagrangian multipliers, and $\rho_1, \rho_2 > 0$ are the penalty parameters.

Then the ADMM framework is applied to solve this sub-problem, which goes as follows

$$\begin{cases} z^{j+1} = \arg\min_z L_A(\bar{u}^j, z; \tilde{\lambda}^j, w^j), \\ u^{j+1} = \arg\min_u L_A(u, z^{j+1}; \tilde{\lambda}^j, w^j), \\ \tilde{\lambda}^{j+1} = \tilde{\lambda}^j - \rho_1 \xi(Au^{j+1} - f), \\ w^{j+1} = w^j - \rho_2 \xi(z^{j+1} - \nabla u^{j+1}), \end{cases} \tag{12}$$

where superscript $j$ is the inner loop index, which is the opposite of the outer loop $k$.

For the $z$-sub-problem in the inner loop has a closed-form solution via the *soft shrinkage*, i.e.,

$$z^{j+1} = shrink(\nabla u^j + \frac{w^j}{\rho_2}, \frac{1}{\rho_2}), \tag{13}$$

where $shrink(v, r) = sign(v) \max\{|v| - \frac{1}{r}, 0\}$.

---

**Algorithm 1: The L1/L2 Minimization on Image Gradient Via Forward-Backward Splitting (L1/L2-TV)**

Input: $A, f, \tau > 0$ and $\varepsilon > 0$ .

Initialize: $u^0, v^0 \in \partial h(u^0)$ .Set $k = 0, k \le \max outer$ .

1.  While $k \le \max outer$ or $\|u^k - u^{k-1}\| / \|u^k\| > \varepsilon$

2.  $\bar{u}^k = \tau I + \partial g^{-1} (\tau u^k + v^k)$ ,

3.  $u^{k+1} = u^k - \frac{\tau^2}{\tau^2 + 1}(u^k - \bar{u}^k)$ ,

4.  $v^{k+1} = v^k - \frac{\tau}{\tau^2 + 1}(\bar{u}^k - u^k)$ ,

5.  End while

Output: $u^{k+1}$ .

---

We first linearize $\frac{1}{2}\|Au - f\|_2^2$ at the current point $u^j$ and add a proximal term, i.e.,

$$\frac{1}{2}\|Au - f\|_2^2 \approx \frac{1}{2}\|Au^j - f\|_2^2 + \langle p^j, u - u^j \rangle + \frac{1}{2\beta}\|u - u^j\|_2^2,$$

where $p^j = A^T(Au^j - f)$ denotes the gradient of $\frac{1}{2}\|Au - f\|_2^2$ at $u^j$. Then the $u$-sub-problem in the inner loop can be further reformulated as

$$\min_u \; \frac{\tau}{2} \left\| u - \frac{\tau u^k + v^k}{\tau} \right\|_2^2 + \frac{\rho_2}{2} \|z - \nabla u\|_2^2$$

$$- w^T(z - \nabla u) + \frac{\rho_1}{2\beta} \left\| u - \left[u^j - \beta(p^j - \frac{A^T \tilde{\lambda}^j}{\rho_1})\right] \right\|_2^2. \tag{14}$$

Taking derivation on (14) with respect to $u$ and forcing the result to zero, we get

$$\left( \frac{\tau + \rho_1/\beta}{\rho_2} + \nabla^T \nabla \right) u = c^j, \tag{15}$$

where

$$c^j = \frac{\tau u^k + v^k}{\rho_2} + \frac{\rho_1}{\rho_2 \beta}[u^j - \beta(p^j - \frac{A^T \tilde{\lambda}^j}{\rho_1})] + \nabla^T(z^{j+1} - \frac{w^j}{\rho_2}).$$

Under the periodic boundary conditions for $u$, we utilize fast Fourier transforms (FFT) to solve (15). The algorithm for solving $\bar{u}$ -sub-problem is shown in Algorithm 2.

---

**Algorithm 2: The ADMM for Solving $\bar{u}$ -Sub-problem**

Input: $A, f, \tau, \beta, \rho_1, \rho_2 > 0, \varepsilon > 0$ and $\xi \in (0,1)$ .

Initialize: $u^0, z^0$ .Set $j = 0, j \le \max inner$ .

1.  While $j \le \max inner$ or $\|u^j - u^{j-1}\| / \|u^j\| > \varepsilon$

2.  update $z^{j+1}$ via Eq.(13),

3.  update $u^{j+1}$ via Eq.(15),

4.  update $\lambda^{j+1}, w^{j+1}$ via Eq.(10),

5.  End while

Output: $\overline{u}^k = u^{j+1}$.

## 3 Results

In this section, the performance of the proposed L1/L2-TV method is tested on the two-dimensional Shepp-Logan phantom with size 512 by 512. And the compared algorithms are classical FBP and TVAL3 [12] methods. The peak signal to noise ratio (PSNR), the structural similarity index (SSIM) and the root of mean square error (RMSE) are choosen as the quality metrics to measure the reconstructed resluts among different methods. In the experiments, all the parameters in the three methods are determined empirically. We set $\tau = 5$, $\beta = 1$, $\rho_1 = 100$, $\rho_2 = 10$, $\xi = 0.55$ and $\varepsilon = 10^{-5}$. The maximum outer and inner iterations are 100 and 200, respectively. What's more, we further verify the performance of the algorithm on the three-dimensional moby data with size $256 \times 256 \times 256$, compared with FDK and TVAL3-3D. The parameters are same as Shepp-Logan reconstruction except for $\xi = 0.95$ and the maximum outer and inner iterations are 100 and 2, respectively.

### A. Numerical Verification

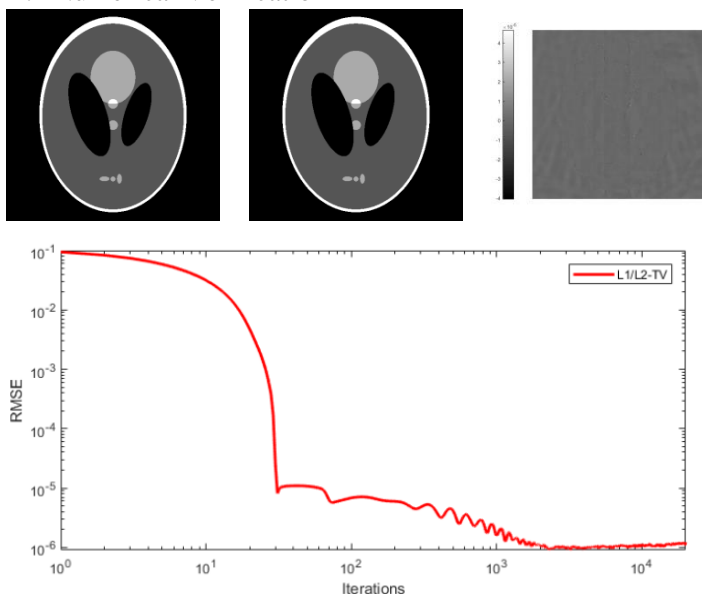

Fig.1 Convergence behavior of the proposed method with 36 views under 20000 iterations. The figures at the top row from left to right represent the ground truth, the reconstructed image of the proposed method and the difference of the former two images with 36 views, respectively.

We first give the numerical verification of our L1/L2-TV method as shown in Fig.1, whose top row consists of three images: the ground-truth image of Shepp-Logan phantom, the image of L1/L2-TV method reconstructed from 36 projection views and their difference image, respectively. To visibly illustrate the convergence behavior of the proposed method, we draw the logarithmic curve from 36 views with 20000 iterations which shown in the second row of Fig.1. It is clear that the proposed method generates a decreasing sequence, which is stable at the order of magnitude of $10^{-6}$.

### B. Comparison with FBP and TVAL3

The projection views in this experiment is selected as 19, 21, 23 uniformly distributed in $360^0$. The distance of source to object and detector are $164.86 \, mm$ and $1052 \, mm$, respectively. The number of detector bins is 512. The size of each pixel of reconstruction image is $0.0116 \, mm \times 0.0116 \, mm$.

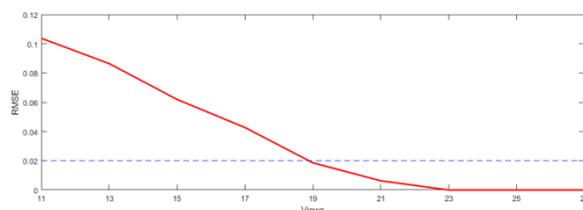

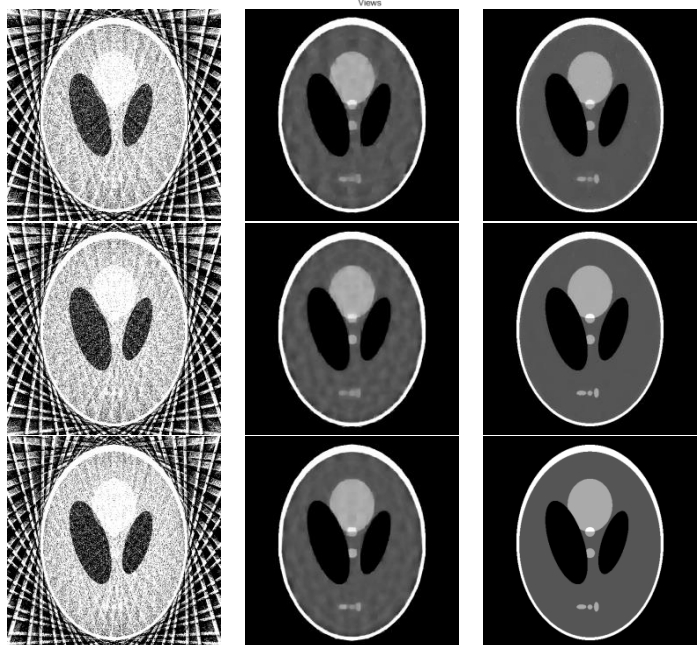

Fig. 2 Comparison of Shepp-Logan phantom reconstructed results between FBP, TVAL3 and the proposed method. The columns from left to right represent the results of FBP, TVAL3 and the proposed method. The rows from 1 to 3 represent the views of 19, 21, 23.The display windows are all [0.1, 0.4]. The top figure represents RMSE values under 9 number of views totally from 11 to 27.

We first choose the views from 11 to 27 to obtain corresponding reconstruction results and RMSE values, and set 0.02 as the maximum threshold according to the reconstructed image quality, whose curve is shown in the first row of Fig.2. When the number of views is 19, the reconstruction quality is relatively better, and when the number of views is greater than or equal to 23, the reconstruction accuracy has reached $10^{-6}$. As shown in the rest of Fig.2, whose row and column represent the projection views and comparison method, respectively, the results of FBP suffer from serious artifacts and have the worst quality because it is an analytic algorithm and can't be accurately reconstructed with a large lack of projection views. The reconstructions of TVAL3 method has a better image quality compared with FBP in preserving shapes, but has a poor ability to maintain details. The proposed method is superior to the other methods not only in preserving edges but also in suppressing artifacts.

To further evaluate the performance of the proposed method, we list the quantitative results in Table 1. Clearly, our algorithm has the lowest RMSEs, and the highest PSNRs under different projection views. In particular, when views increase from 21 to 23, the RMSE dropped sharply to the order of magnitude of $10^{-6}$. Although the SSIM of our method is lower than that of TVAL3 at 19 views, but it is close to 1 at 23 views.

Table 1 Quantitative results among different methods in different sparse views.

| Views | Methods | RMSE | PSNR | SSIM |
|---|---|---|---|---|
| 19 | FBP | 5.1758 | -14.280 | 0.0151 |
| | TVAL3 | 3.886e-2 | 28.210 | 0.952 |
| | L1/L2-TV | **1.861e-2** | **34.604** | 0.838 |
| 21 | FBP | 4.457 | -12.982 | 0.0126 |
| | TVAL3 | 4.138e-2 | 27.767 | 0.948 |
| | L1/L2-TV | **6.336e-3** | **43.964** | **0.973** |
| 23 | FBP | 3.949 | -11.930 | 0.0118 |
| | TVAL3 | 3.546e-2 | 29.004 | 0.969 |
| | L1/L2-TV | **6.813e-6** | **103.334** | **1** |

Additionally, Fig.3 visibly illustrate the convergence rate of the TVAL3 and our L1/L2-TV method under 2500 iterations. It is not difficult to find that the latter has a faster descent rate.

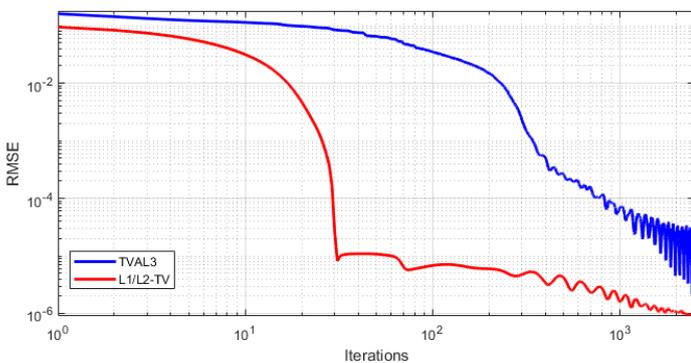

Fig.3 Convergence behaviors of TVAL3 and L1/L2-TV methods under 2500 iterations.

### C. 3D Reconstruction

In 3D reconstruction, the distance of source to object and detector are 1000mm and 1536mm, respectively. The detector consist of 512×512 bins at size of each pixel is 0.8mm×0.8mm. The size of each pixel of reconstruction image is 1mm×1mm×1mm.

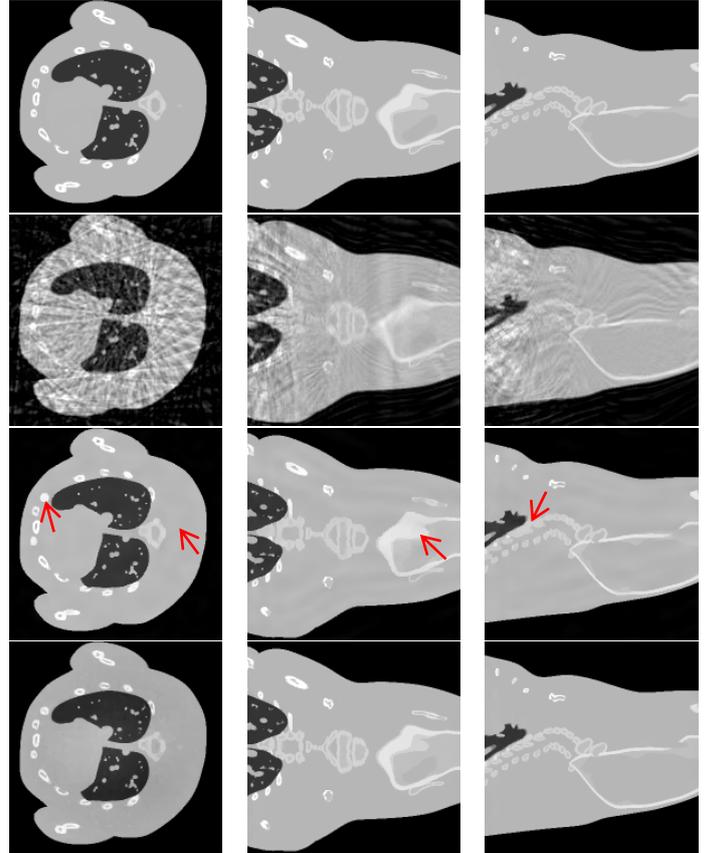

Fig.4 Comparison of 3D moby phantom reconstructed results, where rows represent the ground truth, FDK, TVAL3-3D and the proposed method, respectively. The columns from left to right represent x-y sections, x-z sections, and y-z sections, respectively and the display window is [0, 1].

We choose the minimal projections uniformly distributed in $360^{0}$ are 30 like the 2D reconstruction above to validate the performance of the proposed algorithm on 3D reconstruction. To clearly evaluate the moby image, the reconstruction images of all methods are orientated horizontally and present as 31×220 pixels. As shown in Fig.4, the rows and columns represent different methods (ground truth、FDK、TVAL3-3D and the proposed method) and three dimensional sections (x-y section、x-z section and y-z section), respectively. The results of FDK obviously have serious artifacts and lost many details in the image. The reconstructions of TVAL3-3D preserve a lot of details in three-dimensional sections, but some areas are over-smoothed out as shown by the red arrows. Compared with other methods, the reconstruction results of the proposed method are of the best quality and close to the ground truth.

## 4 Discussions

There are several parameters involved in the execution of the proposed algorithm. The penalty parameters $\rho_1, \rho_2$ are manually selected form a candidate set $10^{-3}, 10^{-2}, 10^{-1}, 1, 10, 10^2, 10^3$ , and finally fixed as $\rho_1 = 100$, $\rho_2 = 10$. And the parameter $\tau$ is choosen from the interval $0.1, 10$ empirically and fix as $5$. Determining the optimal adaptive strategy of the parameters mentioned above might be an interesting topic for future research. As for the parameter $\beta$ introduced in the proximal term of linearized process, its choice is theoretically assured under the condition of $0 < \beta < \frac{1}{A^T A}$, which has deduced in our subsequent work on the convergent proof of the proposed algorithm.

The proposed algorithm consists of one forward-backward step and two descent steps, computational cost will increase rapidly as the scale of the data larger. We utilize graphics processing units (GPU) acceleration technology to speed up our algorithm and the cost time of 20000 iterations is approximately 750 seconds. And when the reconstruction extended to three-dimensional, the cost time increases to 820 seconds approximately for 100 iterations. These results also encourage us to make further improvement.

## 5 Conclusion

Sparse-view CT reconstruction has practical significance in the medical and industrial fields. However, reducing the projection views is often accompanied by poor image quality. To make fully use of the edge preserving ability of TV regularization and the characterization of image sparsity by L1/L2, we propose the ratio model of L1/L2 on the image gradient under forward-backward splitting framework for sparse-view CT reconstruction. Then we design an iterative algorithm and an efficient solvers when solving the subproblem alternatively. The simulation Shepp-Logan phantom is used to validate the effectiveness of the proposed method. The reconstructed results show that the proposed method is superior to FBP and TVAL3 methods not only in the image quality but also in the descent rate when the projection views less than 30. Three-dimensional moby reconstructions also show that the proposed method is superior to other comparison methods at 30 projection views.

## Acknowledgments

This work was supported by the National Key Research and Development Project of China (Grant No. 2020YFC1522002) and the China Postdoctoral Science Foundation (Grant No. 2019M663996). This work was also supported by the National Natural Science Foundation of China (Grant No. 62101596).

# Helical Dark-field Fiber Reconstruction

Lina Felsner[1,2], Christopher Syben[1], Andreas Maier[1,2], and Christian Riess[1]

[1]Pattern Recognition Lab, Department of Computer Science, Friedrich-Alexander-University Erlangen-Nürnberg, Germany
[2]International Max Planck Research School - Physics of Light, Erlangen, Germany

**Abstract** X-ray dark-field imaging provides information about small angle scattering from objects that consist of micrometer-sized structures, such as fibers. The measured dark-field signal strength dependents on the orientation of the fiber relative to the setup, which prohibits the use of standard filter-based reconstruction algorithms. Hence, existing reconstruction algorithms use complex acquisition protocols to sample the object with multiple trajectories.
In this work, we propose a direct 3-D dark-field fiber reconstruction algorithm for data from just one single continuous trajectory. We describe a generic 3-D iterative reconstruction algorithm for the dark-field fiber directions and show experimentally that the reconstruction with a helical trajectory is possible. The results of our simulations show an excellent agreement with the ground truth, both quantitatively and qualitatively.

## 1 Introduction

The X-ray dark-field signal can provide complementary information to conventional X-ray attenuation imaging, as it relates to small angle scattering and unresolved edges. Examples for improved diagnostic information in medicine include structural information and fracture detection of bones [1] and improved lung diagnosis [2]. In a tomographic setup, it can enable the reconstruction of nerve fibers [3]. In this work, we consider dark-field small angle scattering, which is produced by objects that consist of micrometer-sized elongated structures such as fibers. The scattering distribution of a fiber can be modeled with a 3-D Gaussian [4]. The eigenvalues of the scatter distribution define the amount of scatter in the respective direction. A fiber **f** surrounded by its associated scatter spheroid is shown in the center of Fig. 1.

One prominent setup to measure the X-ray dark-field signal is the Talbot-Lau Interferometer (TLI) [5], which is sketched around the fiber in Fig. 1. The TLI consists of three gratings that are placed between the source and detector. The TLI measures changes in the X-ray wavefront. Variations in the wavefront below pixel resolution contribute to the dark-field signal. The dark-field signal samples the scatter spheroid along the sensitivity direction **s**, i.e., perpendicularly to the gratings (see Fig. 1). This reduces the measured signal to one dimension.

Due to the anisotropic 3-D scattering function and the one dimensional sampling, the strength of the measured dark-field signal depends on the orientation of the fiber relative to the setup. In particular, this dependency leads to the effect that the dark-field signal of a fiber varies under rotation. These properties make the reconstruction of the object challenging, such that standard tomographic reconstruction algorithms can not be used. Moreover, the one dimensional sampling imposes an additional constraint to the tomographic reconstruction, since it requires that the object is sampled in all three dimensions to obtain a fully three-dimensional reconstruction.

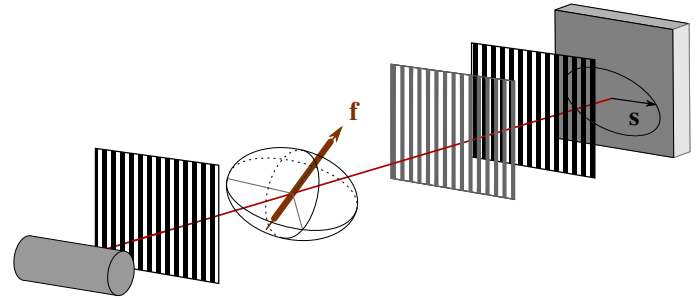

**Figure 1:** Sketch of a Talbot-Lau Interferometer and dark-field measurement for one fiber. The TLI consists of three gratings between source and detector. The 3-D fiber **f** has a 3-D Gaussian scatter function shown as a spheroid. The dark-field signal is measured in the sensitivity direction **s**.

Current dark-field reconstruction methods sample the 3-D distribution by combining scans from several different trajectories. Unfortunately, this leads to complex acquisition protocols and specially designed measurement setups. Malecki *et al.* used a three-circle Eulerian cradle to measure scatter directions along seven circular, planar trajectories [6]. Hu *et al.* showed that under ideal conditions, the estimation of the 3-D fiber direction from only two 2-D reconstructions is in principle possible [7]. However, the method requires perpendicular trajectories and an additional object registration. Schaff *et al.* used the reconstructions of seven circular 2-D trajectories to represent the scatter distribution in a 3-D ellipse [8].

In this paper, we present a proof of concept for a direct 3-D fiber reconstruction with a helical trajectory. To our knowledge, this is the first direct 3-D dark-field reconstruction algorithm for data from one continuous trajectory. We propose and describe a generic 3-D iterative reconstruction algorithm for the dark-field fiber directions. The high quality of the reconstruction is shown in qualitative and quantitative evaluations. The superiority of a helical trajectory over a circular trajectory is shown in an experiment that compares both setups. Furthermore, the feasibility of a helical 3-D fiber reconstruction is shown for a more complex phantom that consists of two rods. Finally, we discuss the results and conclude the paper.

---

**Algorithm 1:** Iterative reconstruction algorithm

---

**1**    initialization
**2**    project volume
**3**    compute difference sinogram
**4**    normalize with mask sinogram
**5**    compute initial error
**6**    **while** *iterations < nr of iterations* **do**
**7**      iterations ++
**8**      backproject the differenceSinogram
**9**      vectorfield($t-1$) + stepsize · vectorfield($t$)
**10**     normalize vectors
**11**     projection of new estimate
**12**     compute and nomalize difference sinogram
**13**     compute new error
**14**     **while** *new error > old error && step > 0.1* **do**
**15**       decrease stepsize
**16**       repeat lines **9** to **13**
**17**     **end**
**18**     **if** *stepsize < 0.1* **then**
**19**       iterations = nr of iterations
**20**       break
**21**     **end**
**22**    **end**

---

## 2   Materials and Methods

We use the 3-D dark-field projection model from [4]. The fiber is represented as a 3-D vector $\mathbf{f} \in \mathbb{R}^3$ and the amount of scattering consists of an isotropic and anisotropic term. These values are related to the eigenvalues of the 3-D Gaussian scatter distribution. The dark-field signal for a single fiber $\mathbf{f}$ measured along the sensitivity direction $\mathbf{s} \in \mathbb{R}^3$ is

$$D = d_{\text{iso}} + d_{\text{aniso}} \left( \mathbf{s}^\top \left( \mathbf{Rf} \right) \right)^2 \quad , \tag{1}$$

where $d_{\text{iso}}$ and $d_{\text{aniso}}$ are the isotropic and anisotropic coefficients, the rotation matrix $\mathbf{R} \in \mathbb{R}^{3 \times 3}$ encodes the rotation angle and the fan- and cone-beam angle of a X-ray in the given geometry. The quadratic part in the model implements the symmetry in the scatter distributions and ensures the compatibility to previous 2-D scatter models. The derivation and complete projection model can be found in [4]. In this work, we consider the measured dark-field signal as the integral over the dark-field signal from Eq. 1 of all fibers along a ray. The projection model is truly three-dimensional and allows to image arbitrary 3-D trajectories.

The general reconstruction problem is to estimate the quantities $d_{\text{iso}}$ and $d_{\text{aniso}}$ and the fiber direction. However, the reduction of a five-dimensional vector of unknowns ($d_{\text{iso}}$, $d_{\text{aniso}}$ and three dimensions for the fiber vector) to a one dimensional signal is a very difficult reconstruction problem. Hence, we slightly relax the problem statement in this work: we assume fixed and homogeneous isotropic and anisotropic values and only aim to reconstruct the fiber direction.

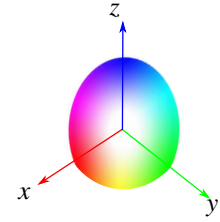

**Figure 2:** The vector visualization is using a RGB color-coded coordinate system in three dimensions.

The solution to this task can be formulated as a least-squares solution, analogously to state-of-the-art iterative computed tomography reconstruction algorithms. Hence, the objective function is

$$\hat{\mathbf{f}} = \arg\min_{\mathbf{f}} \left\| \left| D - d_{\text{iso}} + d_{\text{aniso}} \left( \mathbf{s}^\top \left( \mathbf{Rf} \right) \right)^2 \right| \right\|^2 \quad . \tag{2}$$

The derivative of the objective function with respect to $\mathbf{f}$ is

$$\nabla_{\mathbf{f}} \left( D - d_{\text{iso}} - d_{\text{aniso}} \left( \mathbf{s}^\top \left( \mathbf{Rf} \right) \right)^2 \right)^2 \tag{3}$$

$$= \mathbf{R}^\top \mathbf{s} \, 2 d_{\text{aniso}} \left( \mathbf{s}^\top \left( \mathbf{Rf} \right) \right)^2 2 \left( d - d_{\text{iso}} - d_{\text{aniso}} \left( \mathbf{s}^\top \left( \mathbf{Rf} \right) \right)^2 \right) \tag{4}$$

$$= \mathbf{R}^\top \mathbf{s} \, m \quad , \tag{5}$$

where we combined between Eqn. 4 to Eqn. 5 all scalar values into the value $m$ for improved readability. As a consequence, Eqn. 5 shows that the gradient vector-direction is encoded by the sensitivity direction and the transpose of the rotation matrix that encodes the projection ray. Equation 5 together with symmetry considerations of the scatter distribution enables the reconstruction of the true 3-D fiber direction.

The reconstruction itself follows a state-of-the-art iterative computed tomography algorithm, an algorithm listing as pseudo code is shown in Alg. 1.

We use the helical trajectory, since it is a well-defined, true 3-D trajectory and commonly used in medical applications. One general benefit of a 3-D trajectory over a circular trajectory is that its central slice is not limited to in-plane fiber rotations.

## 3   Experiments and Results

We perform two experiments to show the feasibility of the direct 3-D reconstruction of fiber vectors with a helical trajectory. All experiments are implemented in the CONRAD framework [9]. The helical trajectory contains seven rotations with 270 projections along each rotation, with a source-isocenter distance of 20 mm and a source-detector distance of 160 mm. The rotation axis is set to the $z$-axis, i.e. $(0, 0, 1)$. The grating bars are aligned with the rotation axis, which results in a sensitivity direction within the $xy$-plane. More specifically, the sensitivity direction is set to $\mathbf{s} = (1, 0, 0)$. The isotropic and anisotropic parameters are fixed and set to 2 and 1, respectively. The initial step size for the iterative reconstruction is 0.25 with a maximum number of 100 iterations. The vector directions in the experiments are color-coded for an easier visual verification of the directions in 3-D. We use for the color coding the RGB values representing the three coordinate axes. The color scheme is shown in Fig. 2.

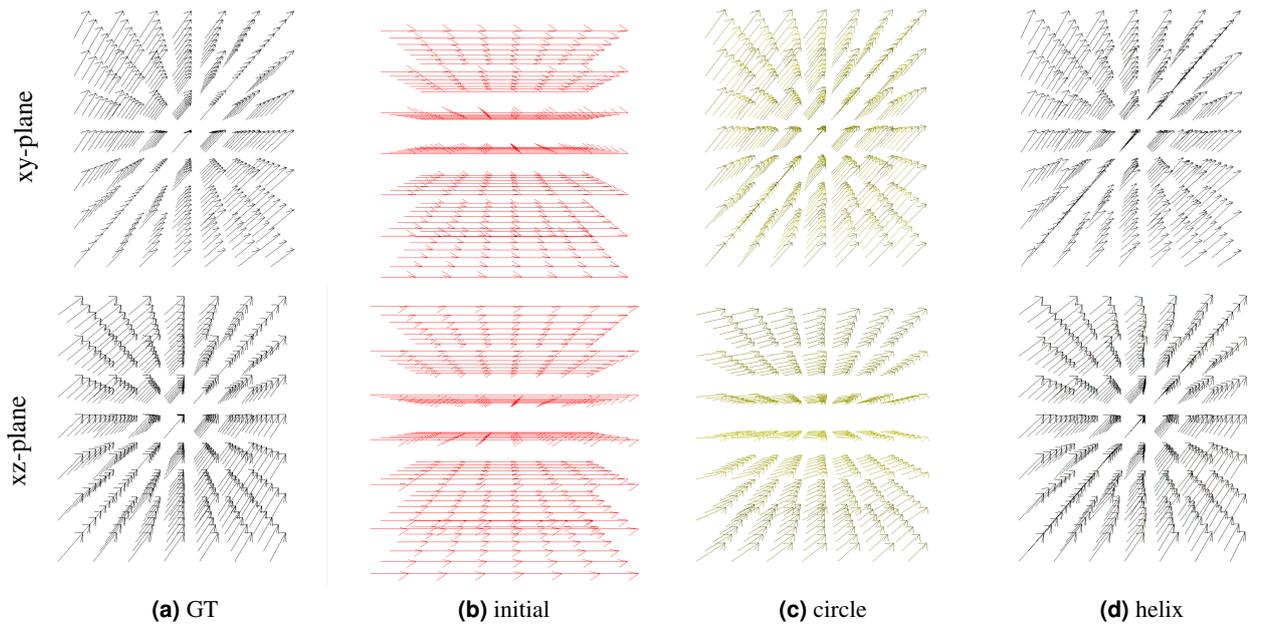

**Figure 3:** 3-D visualization of the volume and fiber directions for Experiment 1, using the color-coding of Fig. 2.

**Experiment 1:**

This experiment shows the feasibility of a direct 3-D reconstruction with a helical trajectory and compares the reconstructions visually and quantatively with a circular trajectory using the same reconstruction algorithm. The circular trajectory consists of 360 projections over $2\pi$ with the same magnification as the helical trajectory. The object consists of $7 \times 7 \times 7$ voxels (voxel spacing of 1 mm), and the helical pitch is chosen as 2 voxels per rotation.

Figure 3 shows the vector directions in the *xy*- and *xz*-plane for the ground truth phantom, the initialization of the reconstruction volume, and reconstructions from the circular and helical trajectory. The homogeneous phantom has a vector direction of $\mathbf{f} = (1, 1, 1)$ (Fig. 3a). The dimensions of the reconstruction volume are set identically to the ground truth. Voxels are initialized with a constant fiber direction of $\mathbf{f}^0 = (1, 0, 0)$ (Fig. 3b). The reconstructed fiber directions with the helical trajectory are very close to ground truth. The discrepancy of the circular reconstruction can be best recognized from the color-coding of the fiber direction, which is olive to yellow instead of gray. The divergence is particularly large in the central plane (bottom of Fig 3c).

A quantitative evaluation confirms the high-quality reconstruction of the helical trajectory. We compute the angle in 3-D between the ground truth and the reconstructed fiber direction as a performance metric. The mean angular error between the fiber direction of the GT and helical reconstruction is about $2°$, whereas the circular trajectory has a mean angular error of $25°$.

Figure 4 shows the mean angular error and standard derivation separately for all *xy*-slices. The error for the circular trajectory strongly depends on the vertical slice, i.e., the cone angle. The largest deviation from the ground truth is in the central plane. The fiber directions from the helical reconstruction are consistently accurate across all slices.

**Experiment 2:**

This experiment shows the feasibility to reconstruct a non-homogeneous object with a helical trajectory. The object consists of $15 \times 15 \times 15$ voxels (voxel spacing of 1 mm), which results in a helical pitch of 4 voxels. The object itself consists of two rods with different vector directions defined as $(0.5, 1, 0.5)$ and $(1, -0.5, 0)$ (see Fig. 5). This object geometry is similar to the object used in [10]. The iterative

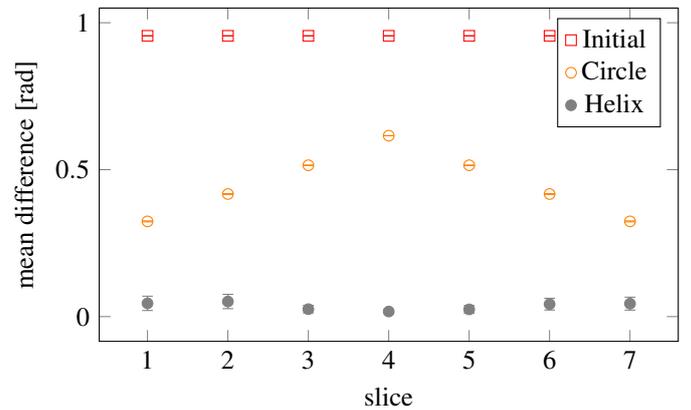

**Figure 4:** Mean and standard deviation of the angular error in 3-D per slice.

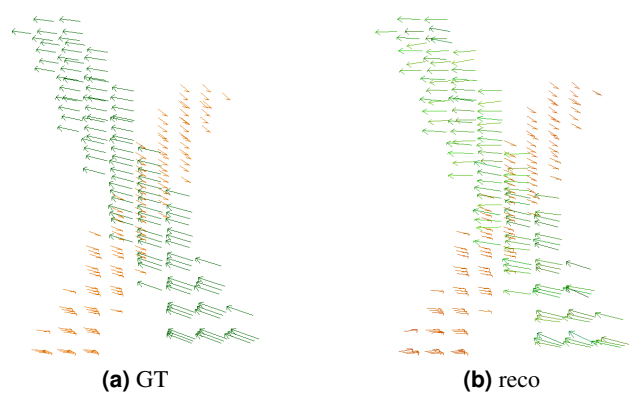

**(a)** GT          **(b)** reco

**Figure 5:** 3-D visualization of the volume and fiber directions for the two rods in experiment 2, using the color-coding of Fig. 2.

reconstruction is initialized as $\mathbf{f}^0 = (1, 0, 0)$.

The ground truth phantom and resulting reconstruction are shown in Fig. 5. The reconstruction agrees very well with the ground truth, which shows that a reconstruction of more than one fiber direction is possible. Quantitatively, the mean angular error of the non-zero voxels is $7.57°$ with a standard deviation of $4.58°$.

## 4    Discussion

The experiments show that the proposed reconstruction algorithm is able to estimate 3-D fiber directions from a single, continuous helix trajectory. The reconstruction of the homogeneous square (Exp. 1) shows the superiority of a helical trajectory over a circular trajectory. The reconstruction with the circular trajectory shows that diverging cone rays can lift the fiber out of the rotation plane. However, in the central plane, only a rotation within the plane is possible. The fiber directions that are reconstructed with a helical trajectory are overall very accurate, and the result is very homogeneous, with only minor deviations to the ground truth. Interestingly, the reconstruction error has a minimum in the central slice. The second experiment shows that the reconstruction of two rods with different fiber directions is possible. While the fiber of one rod is restricted to the *xy*-plane, the second rod has fibers that point in an out-of-plane direction. The reconstruction shows the ability to locally differentiate between fiber directions within the same object volume. We consider the results as a promising proof of concept and believe that further investigations with the proposed model allow for improved fine-tuning of the reconstruction parameters or update steps to further improve the results. Furthermore, for more complex objects, it may be beneficial to include an additional regularization term as shown in other settings [10].

Both experiments use a large magnification to exploit large cone angles and therefore a rotation matrix that allows to easily rotate the fiber direction out-of-plane. One major point that is not discussed so far is the feasibility of large cone angles in a real TLI setup. In current real setups, the fan angles are relatively small due to self-shadowing of rays at the grating bars. This may be overcome in future designs, e.g., with curved gratings. However, the cone angle is not substantially limited along the direction of the rotation axis, which is more important to the successful application of a helix trajectory. Overall, this simulation study demonstrates the feasibility of helical dark-field fiber reconstruction. Furthermore, it might enable the imaging of larger objects with a simpler setup, such that a full 3-D dark-field reconstruction becomes feasible in medical applications.

## 5    Conclusion and Outlook

This paper presents a reconstruction algorithm for a direct 3-D fiber reconstruction from a X-ray dark-field signal. It operates on a single, true 3-D trajectory, and is demonstrated on a helical trajectory. To our knowledge, this is the first work that enables a direct reconstruction of directional X-ray dark-field scatter.

In future work, we will investigate the stability of the algorithm itself, and its stability with respect to the trajectory parameters. We will also investigate a combined reconstruction of fiber direction, isotropic, and anisotropic coefficients.

# Software Implementation of the Krylov Methods Based Reconstruction for the 3D Cone Beam CT Operator

Vojtěch Kulvait[1] and Georg Rose[1]

[1]Institute for Medical Engineering and Research Campus STIMULATE, University of Magdeburg, Magdeburg, Germany

**Abstract** Krylov subspace methods are considered a standard tool to solve large systems of linear algebraic equations in many scientific disciplines such as image restoration or solving partial differential equations in mechanics of continuum. In the context of computer tomography however, the mostly used algebraic reconstruction techniques are based on classical iterative schemes. In this work we present software package that implements fully 3D cone beam projection operator and uses Krylov subspace methods, namely CGLS and LSQR to solve related tomographic reconstruction problems. It also implements basic preconditioning strategies. On the example of the cone beam CT reconstruction of 3D Shepp-Logan phantom we show that the speed of convergence of the CGLS clearly outperforms PSIRT algorithm. Therefore Krylov subspace methods present an interesting option for the reconstruction of large 3D cone beam CT problems.

## 1   Introduction

Krylov subspace methods, see [1], are attractive in the context of the solving tomographic problems as they do not require direct storage of the system matrix of the corresponding problem. This is especially true for 3D tomographic problems, where the size of the system matrix prohibits its efficient storage and the computation of the projection and back-projection operator on GPUs or dedicated hardware might be orders of magnitude faster than using stored precomputed matrix.

While most of the literature regarding algebraic reconstruction is based on Kaczmarz algorithm and related classical iterative schemes, see [2], there is also a growing body of the works on application of various Krylov subspace methods. Important topics are optimal preconditioning strategies, see [3] and the enforcing of properties such as non-negativity of the solution [4].

Until circa 2010 the direct application of the Krylov subspace methods to the 3D cone-beam CT operator (CBCT), was very rare and the works regarding Krylov methods were solving just smaller 2D problems. The block-wise algorithm to divide the tomographic reconstruction into the smaller sub-problems to apply CGLS and LSQR was proposed in [5]. Currently the implementation of CGLS for CBCT operator can be found in the two MATLAB based frameworks, see [6, 7]. The Split Bregman algorithm for CBCT TV norm minimization using Krylov BiCGStab was published in [8, 9]. There is however still lack of works to systematically study Krylov subspace methods or to compare their performance with the classical ART based approaches.

In this work we present the software package, which contains an open-source C++ and OpenCL implementation of the Krylov subspace methods for the CBCT reconstruction. We show that these methods poses a viable option for a fast and accurate reconstruction on recent GPU hardware.

Moreover on simple test problem using CBCT we compare performance of PSIRT, an advanced technique based on ART like algorithms, to the CGLS, implementation of conjugate gradients on the normal equations. We show that convergence speed of the Krylov method is much faster and to achieve the same accuracy, we need a lot less iterations.

## 2   Materials and Methods

Cone beam CT operator can be understood as a sparse matrix $\mathbf{A} \in \mathbb{R}^{m \times n}$ acting on discretized attenuation data $\mathbf{x} \in \mathbb{R}^n$ in the volume of interest to produce projection data $\mathbf{b} \in \mathbb{R}^m$. In the matrix form

$$\mathbf{b} = \mathbf{A}\mathbf{x}. \qquad (1)$$

As the matrix $\mathbf{A}$ is non square and often over-determined with $m > n$, we typically solve the least-squares problem by means of normal equations to find attenuation $\mathbf{x}$ such that

$$\mathbf{A}^\top \mathbf{A} \mathbf{x} = \mathbf{A}^\top \mathbf{b}. \qquad (2)$$

The matrix $\mathbf{A}^\top \mathbf{A}$ is symmetric, positive definite and it is possible to apply directly method of conjugate gradients on such system. Direct method to do so is referred as CGLS. There is also LSQR, mathematically equivalent method with improved numerical stability, see [10, 11]. For the sake of completeness, we include here the iterative scheme of CGLS implemented in our software as Algorithm 1. We restructured the algorithm in a way that at the end of each iteration we compute the update of $\mathbf{x}$ and postpone the update of the residuals to the beginning of the next iteration. By doing so, we save one projection and backprojection at the end of the algorithm.

## 3   Software

The software package was developped in C++ and OpenCL. The project implements various CBCT projection and backprojection operators, namely Siddon ray-caster [12], footprint methods [13] and also so called Cutting voxel projector. Cutting voxel projector uses the volume of the cuts of the voxels by the rays to particular pixel for the computation of projections, details are yet to be published. From the reconstruction **input :** Projection data $\mathbf{b}$, initial vector $\mathbf{x}_0$, relative discrepancy tolerance ERR, maximum number of iterations $K$.

**begin**

    allocate $\mathbf{x}$, $\mathbf{d}_x$ and $\mathbf{r}_x$;

    allocate $\mathbf{e}_b$ and $\mathbf{p}_b$;

    $\text{NB}_0 = \|\mathbf{b}\|_2$;

    $\mathbf{x} = \mathbf{x}_0$;

    $\mathbf{p}_b = \mathbf{A}\mathbf{x}$;

    $\mathbf{e}_b = \mathbf{b} - \mathbf{p}_b$;

    $\mathbf{r}_x = \mathbf{A}^\top \mathbf{e}_b$;

    $\mathbf{d}_x = \mathbf{r}_x$;

    $\text{NR2}_{\text{old}} = \|\mathbf{r}_x\|_2^2$;

    $\mathbf{p}_b = \mathbf{A}\mathbf{d}_x$;

    $\text{NP2} = \|\mathbf{p}_b\|_2^2$;

    $\alpha = \text{NR2}_{\text{old}}/\text{NP2}$ ;

    $\mathbf{x} = \mathbf{x} + \alpha \mathbf{d}_x$;

    $\mathbf{e}_b = \mathbf{e}_b - \alpha \mathbf{p}_b$;

    $\text{NB} = \|\mathbf{e}_b\|_2$;

    $i = 1$;

    **while** $\text{NB}/\text{NB}_0 > \text{ERR}$ & $i < \text{K}$ **do**

        $\mathbf{r}_x = \mathbf{A}^\top \mathbf{e}_b$;

        $\text{NR2}_{\text{now}} = \|\mathbf{r}_x\|_2^2$;

        $\beta = \text{NR2}_{\text{now}}/\text{NR2}_{\text{old}}$;

        $\mathbf{d}_x = \mathbf{d}_x + \beta \mathbf{r}_x$;

        $\text{NR2}_{\text{old}} = \text{NR2}_{\text{now}}$;

        $\mathbf{p}_b = \mathbf{A}\mathbf{d}_x$;

        $\text{NP2} = \|\mathbf{p}_b\|_2^2$;

        $\alpha = \text{NR2}_{\text{old}}/\text{NP2}$ ;

        $\mathbf{x} = \mathbf{x} + \alpha \mathbf{d}_x$;

        $\mathbf{e}_b = \mathbf{e}_b - \alpha \mathbf{p}_b$;

        $\text{NB} = \|\mathbf{e}_b\|_2$;

        $i = i + 1$;

    **end**

**end**

**Result:** Vector $\mathbf{x}$, number of iterations $i$, final norm of discrepancy NB.

**Algorithm 1:** CGLS with delayed residual computation.

techniques, the software contains CGLS and LSQR implementation with the option of basic Jacobi preconditioning and Tikhonov regularization. It is possible to select initial vector or guess of the solution, e.g. the result of analytical reconstruction or apriori knowledge. It is also possible to do a off-center reconstruction, where the volume to reconstruct can be positioned outside the center of rotation. For the purpose of the comparison of the different CBCT reconstruction methods, two classical ART like methods, SIRT and PSIRT [2] are also implemented.

The package also contains methods to project volumes or backproject projections without reconstruction. This could be useful when e.g. simulating acquisition of the given volume with particular geometry setting of a concrete CT device.

The program is distributed under the terms of GNU GPL3 license and its Git repository is available at https://bitbucket.org/kulvait/kct_cbct. The results presented were obtained using git commit f2bf01a.

## 4  Results

Here we present a test to compare convergence of the Krylov method (CGLS) with the classical scheme (PSIRT) when applied on CBCT operator. We have chosen 3D Shepp-Logan phantom with the 256x256x52 voxels of the dimensions $0.86\,\text{mm} \times 0.86\,\text{mm} \times 3.44\,\text{mm}$.

The geometry configuration is similar to the clinical C-Arm CT systems for the brain tomographic scanning, where the distance from the source to the isocenter is 749 mm and the distance from source to the detector is 1198 mm. Detector matrix have pixel size of $0.616\,\text{mm} \times 0.616\,\text{mm}$. The trajectory consists of 496 scanning angles.

We have first projected the 3D Shepp-Logan phantom using this geometry to obtain projection data. We have used for the projections and the reconstructions the implementation of TT projector and backprojector, see [13]. We run the tests to compare classical method PSIRT to Krylov subspace method CGLS, both implemented in our software. The tests were performed on computer with the AMD Ryzen 7 1800X and GPU Vega 20 Radeon VII with 16GB HBM2 Memory and 1TB/s bandwidth. Projectors and backprojectors are implemented in OpenCL and the computations were run on the GPU. Speed of the both methods in terms of average time per iteration was comparable, circa 12.9 s. To compare the speed of convergence, we have measured relative norm of discrepancy of the solution

$$e = \frac{\|\mathbf{A}\mathbf{x} - \mathbf{b}\|_2}{\|\mathbf{b}\|_2}. \tag{3}$$

during the iterative process.

Initially we run fixed number of 40 iterations of the both methods, the relative norm of discrepancy (3) after 40 iterations was then $e_{\text{CGLS}} = 0.18\%$ versus $e_{\text{PSIRT}} = 3.64\%$. The visualization in Figure 1 shows that PSIRT reconstruction is still blurry while CGLS has converged without visible problems.

Second, we test how many iterations every method needs to achieve norm of discrepancy under $e < 1\%$. For PSIRT it is $N = 112$ iterations compared to CGLS with $N = 20$. This means that the CGLS is circa five times faster than PSIRT when we would like to achieve the same accuracy. In figs. 2 and 3 can be found the graphs comparing the speed of convergence for both methods that illustrates clear advantage of the CGLS for the test problem.

## 5  Discussion

The library we present in this article was developed to test different approaches to implement cone beam CT operator and to test various reconstruction techniques. Focus was on solving problems that arise from the model based perfusion reconstructions, where we can have negative values in reconstructed volumes and classical solvers might often diverge. The accuracy is often prioritized over speed and many double precision computations could be substituted by float precision counterparts to increase speed. From the nature of the open-source solution and the C++ modular implementation it is straightforward to extend the program for the particular reconstruction algorithm or preconditioning strategy.

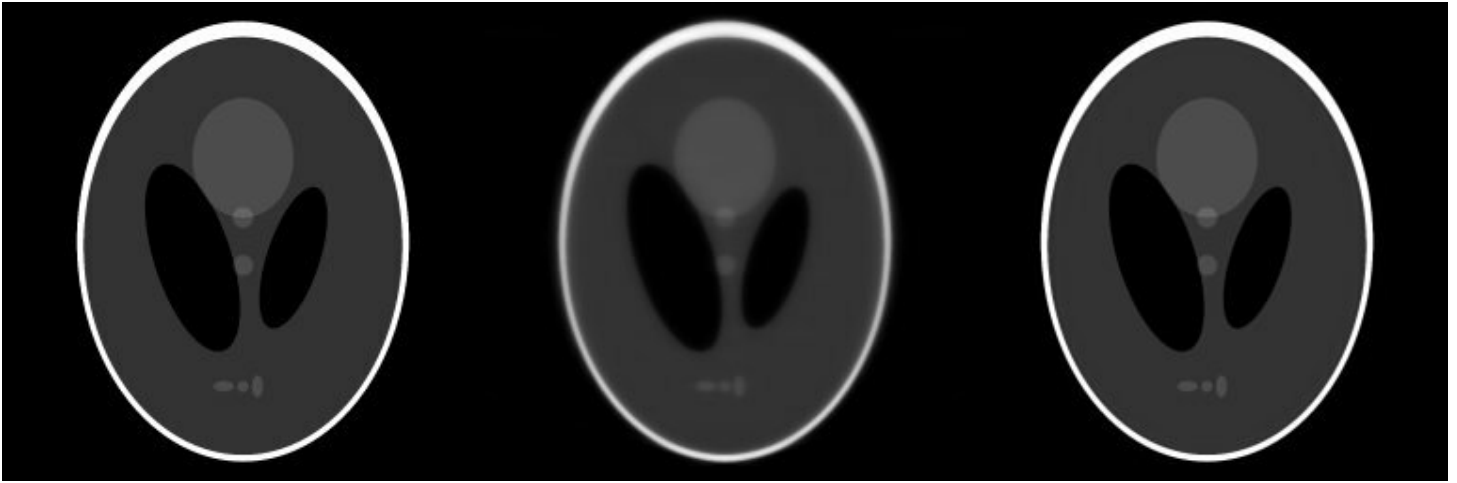

**Figure 1:** Center slice of the test volume of 256x256x52 voxels, all images have the same window [0,1], ground truth data on the left. PSIRT reconstruction after 40 iterations in the middle. CGLS reconstruction after 40 iterations on the right.

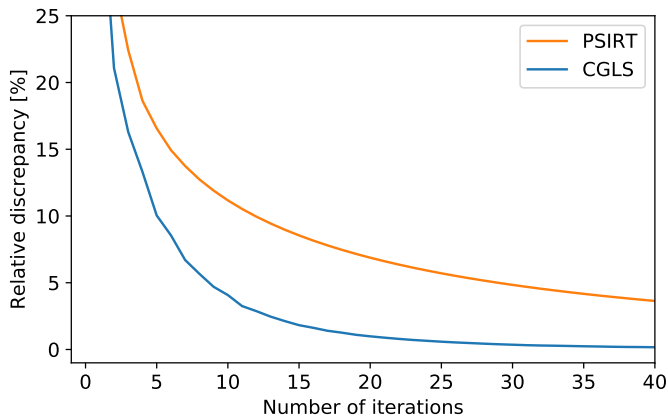

**Figure 2:** Comparison of the speed of convergence in terms of the relative norm of discrepancy (3) for CGLS and PSIRT for the test CBCT problem, 40 iterations.

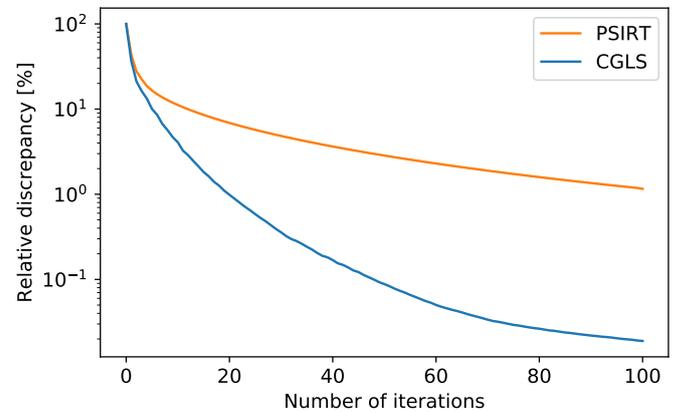

**Figure 3:** Comparison of the speed of convergence in terms of the relative norm of discrepancy (3) for CGLS and PSIRT on the test CBCT problem. Y axis in logarithmic scale, 100 iterations.

On a test problem we have demonstrated that the CGLS has much faster convergence characteristic than PSIRT. It can be seen from figs. 1 to 3. We have tested also SIRT as alternative classical reconstruction method, but its convergence was slower when compared to PSIRT.

When using Krylov methods, it is not well addressed how to enforce conditions such as non-negativity of the solution. While in the context of classical ART, this is an easy task, for Krylov methods we can not simply apply the box conditions as the regularized solution would not fit to the underlying

Krylov subspace and we would lose the convergence properties of the Krylov methods. To address this multiple schemes for restarted methods within this framework were introduced, see [4]. Implementing these techniques is also one of further goals for the development of presented software package. Main obstacle to do so is a larger memory footprint of such methods as compared to CGLS as we usually need to store more vectors of underlying Krylov space.

The method PSIRT could be used with the box conditions as for the test problem values lies within a range [0, 1]. The relative norm of discrepancy after 40 iterations was $e_{\text{PSIRT}} = 3.64\%$ without box conditions and $e_{\text{PSIRT}} = 3.62\%$ with them. Therefore for simplicity, we report only the results of PSIRT without box conditions as they did not have any meaningful effect on the convergence. In real world applications the upper bound can be hard to estimate.

The memory footprint of SIRT and CGLS is the same. Our implementation of CGLS, see Algorithm 1, needs to store three times the volume data and three times the right hand side data. It is necessary for storing the residual and Krylov subspace vectors on which we project the error. Our implementation of SIRT has the same memory footprint as we need to store the vectors of the row and column sums of the system matrix and update vectors. In PSIRT we do not store column sums vector so we need to store right hand side data only twice. However, current GPU hardware provides enough memory such that the memory footprint of these methods is not an issue in practical applications even for much bigger problems than the one presented.

There is also implemented Jacobi preconditioning strategy in the software. It is most simple preconditioning, where we approximate the matrix of the normal equations by its diagonal. It seems however that this strategy alone does not work very well especially due to the presence of very small diagonal values in system matrix on the cone boundary. Therefore better preconditioning strategies have to be found in order to speed the convergence.

Although LSQR should in theory be numerically more stable than CGLS, from our experiments for the tomographic reconstruction the two methods are producing practically identical results. So the type of instability that makes LSQR numerically more stable method is probably not present in a typical CT data. Due to smaller memory footprint, using CGLS might therefore be preferred.

In the CGLS algorithm the discrepancy and residual vectors are not computed directly but they are iteratively updated. Therefore it is proposed to reorthogonalize this vector once in $k$ iterations to avoid accumulation of errors. When we applied such scheme, we figured out that the difference between iteratively updated discrepancy and discrepancy computed from solution vector is less than 0.0001% for 10 iterations. Therefore we omit this reorthogonalization step in a default configuration.

# 6  Conclusion

Analytical reconstruction methods are still a gold standard in a CT reconstruction. Main argument for using them is their speed. As Krylov subspace methods provide much faster convergence when compared to the ART like methods, together with the hardware speed improvements, their wider application could change the speed narrative to widely adopt algebraic reconstruction techniques in practice.

The results show on a phantom problem very high advantage of the CGLS over PSIRT in terms of convergence and they shall be validated for practical problems and other setups. We believe that there will be still very strong advantage of Krylov solvers in practice. Further development should focus on adopting a good preconditioning strategies for Krylov solvers as it has potential to further reduce the run time.

# 7  Remarks

Just before the conference, we managed to achieve a significant speedup of some projectors and backprojectors implemented in the **https://bitbucket.org/kulvait/kct_cbct**, especially the so-called Cutting voxel projector, which will be introduced in a separate article. Because of this speedup, significantly better projection, backprojection and reconstruction times were presented on the poster. The main subject of this contribution, the convergence speed of Krylov methods as a function of the number of projections and backprojections, is not affected. However, it may significantly improve the usability of our software for fast algebraic reconstruction of moderate sized problems in minutes. Last stable commit as of writing this sentence is `0d7d6`.

After presenting a poster at the Fully3D 2021 conference, we received very important feedback. Namely, Simon Rit, the lead developer of the Reconstruction Toolkit, see [14], mentioned that methods derived from the Kaczmarz algorithm using the entire projection and backprojection operator at once are slow compared to ordered subsets methods, see [15]. This also applies to the PSIRT algorithm, which does not use ordered subsets and which we compare Krylov methods with. We take this criticism very seriously and have started working on implementing OS algorithms in our package to be able to reliably compare the methods. Unfortunately, this comparison is not yet complete and cannot be presented here. It is also worth mentioning that, compared to ordered subset schemes, it is easy to add L2 regularization directly in the problem formulation for Krylov methods. As recently shown, L2 regularization can be a faster alternative to L1 regularization, such as TV norm minimization, with similar results, see [16].

The previous question however, leads to the following consideration: if ordered subsets methods have yielded significant speedups of classical CT reconstruction schemes by using only a subset of the rows of the CT operator matrix in each step, can a similar approach be used for Krylov methods? A naive implementation of methods like CGLS with this strategy would very likely suffer from a rapid loss of orthogonality and convergence speed. On the other hand, it is possible that orthogonalization with respect to a larger number of vectors proportional to the number of subsets could stabilize the method. Although this would lead to longer recurrences and a larger memory footprint of the algorithm, it seems promising to investigate such methods in the future.

# Acknowledgments

*The work of this paper is partly funded by the Federal Ministry of Education and Research within the Research Campus STIMULATE under grant number 13GW0473A.*

# A Deep Residual Dual Domain Learning Network for Sparse X-ray Computed Tomography

Theodor Cheslerean-Boghiu[1], Daniela Pfeiffer[2], and Tobias Lasser[1]

[1]Computational Imaging and Inverse Problems, Department of Informatics and Munich School of BioEngineering, Technical University of Munich, Germany

[2]Department of Radiology, Klinikum Rechts der Isar, Munich, Germany

**Abstract** Sparse sampling in X-ray Computed Tomography (CT) is an interesting option towards reducing dose. Recently, the success of deep learning techniques in a lot of medical imaging applications has prompted an increased interest in data-driven tomographic reconstruction algorithms. Nevertheless, most of the proposed solutions boil down to denoising steps performed either pre-reconstruction on the sinogram or post-reconstruction on the reconstructed image, with no reconstruction being performed as part of the network model. In this work, we introduce an improved deep neural network architecture (DNN) which combines a fine-tuned Filtered Backprojection (FBP) operation with a dual domain filtering step applied on both the sinogram and reconstructed image to produce artifact-free reconstructions from sparse data and we present its performance potential in experiments with clinical data.

## 1 Introduction

In many works on data-driven tomographic reconstruction for sparse X-ray Computed Tomography (CT), the *reconstruction* part is done outside the network. Examples are using a UNet in the context of sinogram completion from sparse data and then applying a FBP operation to deliver an artifact-free reconstruction [1], or first reconstructing the sparse-view CT data and then using a UNet trained in the framelet domain (to enforce sparsity in said domain) to perform what boils down to a denoising step in the image domain to remove sparse data artifacts from the reconstruction [2]. However, pre-reconstruction sinogram completion or post-reconstruction denoising approaches all have the disadvantage of performing on either the sinogram or the image domain (or an equivalent wavelet representation of said domain) such that a neural model is unable to learn to distinguish features from both the measurements and the reconstruction. Additionally, the actual reconstruction step plays no role in the training or inference process.

More recently, new methods of dealing with the inverse problem in a more controlled data-driven fashion have been introduced by Würfl et al., who combined a back-projection operation with convolutional and fully connected layers that mimic the filtering operation of the FDK algorithm [3], or Jin et al., who developed an unrolled loop algorithm of an iterative reconstruction approach using a DNN [4]. The advantage of these methods is that the reconstruction step can now be used during training, which removes the single domain training limitation of the methods mentioned earlier.

In this work, we are building on the works of Würfl et al. [3] and Lin et al. [5], developing an improved architecture for a deep neural network that is able to learn both from the raw data (sinogram-domain learning) and from the reconstructed data (image-domain learning). In order to enable end-to-end to training, we introduce a fixed differentiable backprojection layer based on a fixed operator with an additional filtering step as a convolutional layer, mimicking an FBP operation in between the two denoising modules. We call the overall architecture *WNet*.

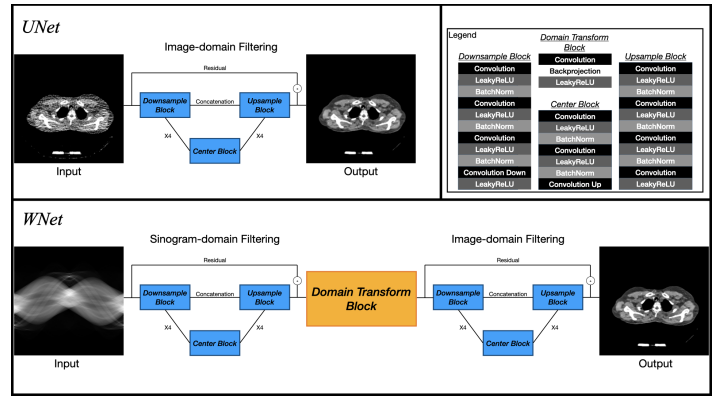

**Figure 1:** (*UNet*) In this work we use a slightly modified version of the UNet network. (*WNet*) Proposed network architecture for concurrent dual domain filtering and filtered backprojection as the domain transform. The domain transform block could in theory be replaced with any inversion operation implemented as a differentiable layer.

## 2 Materials and Methods

### 2.1 Mathematical Background

Given a set of sparse measurements $y_{\text{sparse}} \in \mathbb{R}^{M_1}$ and a forward operator $A_{\text{sparse}} \in \mathbb{R}^{M_1 \times N}$ describing a sparse CT configuration, we define the following forward model:

$$A_{\text{sparse}}\mu = y_{\text{sparse}} \tag{1}$$

where $\mu \in \mathbb{R}^N$ is the voxelized volume we seek to reconstruct. As in every tomographic reconstruction problem, starting from a set of measurements $y_{\text{sparse}}$ and a known system matrix $A_{\text{sparse}}$, we seek to invert the forward model in eq. (1). Therefore, we compute the solution $\mu_{\text{sparse}}$ as the result of an

FBP operation with a ramp filter:

$$\mu_{\text{sparse}} = P_W^{\text{sparse}} y_{\text{sparse}} \tag{2}$$

where $P_W^{\text{sparse}} = A_{\text{sparse}}^t W$ with $A_{\text{sparse}}^t$ being the equivalent back-projection operation to the forward one ($A_{\text{sparse}}$), and $W$ a filtering operator applying a convolution with the ramp kernel. Nonetheless, since the FBP result is only a solution to the inverse X-ray transform iff $N \to \infty$ and $M_1 \to \infty$, the solution $\mu_{\text{sparse}}$ will not only contain sampling artifacts but also sparse-data artifacts. Therefore, we are interested in eliminating such artifacts by finding a denoising operator which satisfies:

$$\hat{D} = \arg\min_D d(D\mu_{\text{sparse}} - \mu) \tag{3}$$

given a well defined distance metric $d(\cdot)$. To do this, we introduce a second operator $A \in \mathbb{R}^{M_2 \times N}$ with $M_2 > M_1$, which describes a fully sampled CT configuration to generate a set of measurements $y \in \mathbb{R}^{M_2}$:

$$A\mu_{\text{sparse}} = y + \varepsilon \tag{4}$$

The sampling and sparse-data artifacts, denoted as a combined $\varepsilon$, that are present in $\mu_{\text{sparse}}$ are also contained in the generated measurements. Thus, we need to find a pair of operators $(D_1, D_2)$ for which:

$$\hat{D}\mu_{\text{sparse}} = D_1 P_W D_2(y + \varepsilon) = \hat{\mu} \approx \mu \tag{5}$$

where $P_W = A^t W$ is the FBP operation working with fully-sampled measurements. In other words, we seek to eliminate the artifacts from the final reconstruction $\hat{\mu}$ in the measurement domain ($D_2$) as well as in the image domain ($D_1$).

## 2.2 Model Architecture

In this work we investigate two different approaches to convert a sparse reconstruction into an artifact-free one. Both image-domain and dual-domain filtering DNN models take advantage of a slightly modified, smaller version of the standard UNet architecture. A graphical representation of both architectures can be seen in Fig. 1. Both networks were implemented in Pytorch, while the domain transform layer is a differentiable Pytorch module using as a backend for the forward and back-projection operations of the C++ framework *elsa* [6, 7].

### 2.2.1 Image-domain Filtering

The first approach consists of a post-processing step performed after the FBP reconstruction from sparse-data sinograms, in which a network containing a single UNet module (taking the role of $\hat{D}$ from eq. (3)), called *UNet*, is trained towards eliminating both sampling and sparse-angle artifacts from the reconstructed image. Our version of the encoder-decoder network only has four levels instead of the five layers present in the original UNet architecture [8]. Instead of performing the downsampling operations using max pooling operations we used 2D convolution operations with stride 2. We also added an additional 2D convolution operation at each downsampling and upsampling level.

In denoising operations, the output of the network has to retain similarities to its input. Therefore, we appended a residual branch to our UNet module implementation. Lastly, all activation operations are leaky ReLUs.

### 2.2.2 Dual-domain Filtering

The second approach upgrades the FBP reconstruction into a module containing a convolution with a fine-tuned filter followed by a fixed back-projection operation, and combines it with a pair of UNets for the goal of removing artifacts pre- and post-reconstruction both in the sinogram and the reconstructed image domain. This is equivalent to finding the pair of operators $(D_1, D_2)$ which satisfy eq. (5), for which $D_2 = U_s$ where $U_s$ is a UNet module dealing with sinogram data, and $D_1 = U_r$ where $U_r$ is a UNet module dealing with the reconstruction data. Both of them have the same specifications as the one introduced in the previous section.

We propose an improved neural network architecture, *WNET*, as shown in Fig. 1, which consists of the FBP operation as a domain transform layer combined with a filtering operation sandwiched between the two UNet modules ($U_s$ and $U_r$) mentioned before. The first UNet module ($U_s$) takes as an input a set of noisy fully-sampled measurements and produces a denoised version of them. Next a convolution with a learnable filter combined with the back-projection operation in a domain transform layer ($T$) model the FBP operation and yield a reconstruction which is then fed to a second UNet module ($U_r$). The result of $U_r$ is the output of the network and consequently the sparse-angle artifact-free reconstruction.

## 3 Results and Evaluation

Our goal is to confirm that the proposed cascaded network approach *WNet* can produce superior results compared to the *UNet* used as a post-reconstruction denoising step.

### 3.1 Dataset Preparation

For the experiments presented in this paper we use five clinical investigative high-dose full-view scans obtained ex-vivo from two patients. Four of the five datasets where obtained of one patient and used for training and validation purposes while the fifth dataset was of the second patient and kept for testing. Each dataset consists of the 150 slices containing chest scans extracted from full-body CT reconstructions which were obtained at a resolution of 512-by-512 pixels. Each slice was downsampled to a resolution of 256-by-256 pixels and normalized between 0 and 1. Then, to avoid overfitting data augmentation was carried out on the 750 images by flipping each image both horizontally and vertically and rotating them by 90 degrees clockwise and counter-clockwise.

We used the tomographic reconstruction framework *elsa* [6, 7] to simulate two different sinograms in a parallel-beam configuration for each of the 3750 images obtained after the data augmentation step: one fully-sampled sinogram consisting of 512 pixels by 512 views over 359 degrees and one sparse sinogram with 512 pixels over 64 and 128 angles over 359 degrees. No additional detector noise was added to the sinograms.

A set of reconstructions were then obtained from these sinograms applying the FBP method with a ramp filter: the fully-sampled sinogram was used to obtain the "*ground-truth*" reconstruction required for training and evaluation, while with the sparse sinogram we produced sparse reconstructions of the original CT scan (called "*64/128-view*"). Furthermore, we simulated another sinogram, of the same size and properties as the fully-sampled one, but of the sparse reconstruction. We call this sinogram "*64/128-view full-sparse*".

### 3.2 Training Procedure

For training and validation we had a set of 3000 data points coming from the first patient. We use a 80%-20% split for training and validation purposes. For both architectures we use the same loss, the Mean Squared Error.

We train the *WNet* for 100 iterations on pairs of ("*64/128-view full-sparse*" sinogram, "*ground-truth*" reconstruction) using individual Adam optimizers ($\beta_1 = 0.9, \beta_2 = 0.999$) for each of the three parts: $U_s$, $T$, $U_r$. We initialize the weights of $U_s$ using Normal Kaiming distribution, while the weights of $U_r$ were initialized with the normal distribution. Besides, the convolution filter from $T$ is initialized with the values of the ramp filter. Both $U_s$ and $U_r$ were trained starting from a learning rate of $10^{-3}$ while the learning rate for the $T$ was initialized with $10^{-10}$, which can be viewed as a fine-tuning step of the ramp filter. Moreover, we employed a scheduler to exponentially decrease the learning rate of all three optimizer by a factor of 0.95 every 5th iteration. The batch size was set to 4 for all training stages and all models.

We use the same number of iterations to train the *UNet* on pairs of ("*64/128-view*" reconstruction, "*ground-truth*" reconstruction). As an optimizer we also use the Adam optimizer ($\beta_1 = 0.9, \beta_2 = 0.999$), and we initialize the learning rate with $10^{-4}$. The initialization of the weights is left the default one, and we use a batch size of 4.

Both *WNet* and *UNet* were showing learning saturation in the validation loss after 100 iterations. The need for a scheduler in the case of *WNet* was highlighted in an earlier ablation study we performed, and we found out that without a scheduler, the $U_s$ and $U_r$ modules were learning against the $T$ module and convergence would not be achieved.

### 3.3 Evaluation of the Networks

We compare the results obtained using the two networks, *WNet* and *UNet*, to reconstructions performed using conventional FBP with a ramp filter by computing PSNR and SSIM metrics over the test set (see Table 1) and we show example reconstructions in Fig. 2 and 3 .

In the red boxes from Fig. 3 there is a small thin feature located on the left side of the vertebra which in the case of *UNet* for both "*64-*" and "*128-view*" reconstructions has almost vanished. On the other hand, the *WNet* result matches more accurately the ground truth label and also manages in both "*64-*" and "*128-view*" cases to keep the blurring amount to a minimum compared to *UNet*.

In the blue box, a soft tissue feature with clear delimitations can be seen in the ground truth image. While for "*64-view*" data both methods lack some structural similarity to the ground truth, *WNet* manages to keep the blurring to a minimum and to not introduce "fake" soft-tissue. On the other hand, on the "*128-views*" data, both methods perform at closer level, yet the *WNet* result tends to have more contrast and definition compared to the *UNet* result.

| Metrics<br>Methods | PSNR (dB) | | SSIM | |
|---|---|---|---|---|
| | avg. | std. | avg. | std. |
| FBP (64-view) | 34.2034 | 1.8802 | 0.9419 | 0.0190 |
| *WNet* (64-view) | **40.9175** | **1.8999** | **0.9873** | **0.0048** |
| *UNet* (64-view) | 38.7319 | 1.7915 | 0.9793 | 0.0072 |
| FBP (128-view) | 39.2255 | 1.8400 | 0.9781 | 0.0078 |
| *WNet* (128-view) | **43.9907** | **1.7610** | **0.9937** | **0.0023** |
| *UNet* (128-view) | 42.5084 | 1.8967 | 0.9898 | 0.0039 |

**Table 1:** Mean and Standard deviation values of PNSR and SSIM for the area inside the volume. (bold) best PSNR and SSIM value pairs which correspond to the *WNet* results. Interestingly, the results obtained with the *WNet* trained on "*64-view*" data are superior to the "*128-view*" reconstructions produced with standard FBP.

### 4 Conclusion

We propose an improved cascaded deep neural network composed of a module containing a learnable filter combined with a differentiable fixed back-projection operation layer sandwiched between two UNet modules performing denoising in both the sinogram and the image domain. Our evaluation results show that the proposed *WNet* performs better than a conventional *UNet* applied as a post-reconstruction denoising step and is more robust to artifacts in the case of low-number-view sparse reconstruction. Moreover, we have seen an increase of around $1.5 - 2$dB in the PSNR value inside the volume over the *UNet* of the same size.

We seek to expand this architecture in the future with a compound loss module based on a combination of pixel-wise and perceptual loss functions. Part of a future study is also the development of variational modules based on more accurate iterative reconstruction algorithms as the domain transfer.

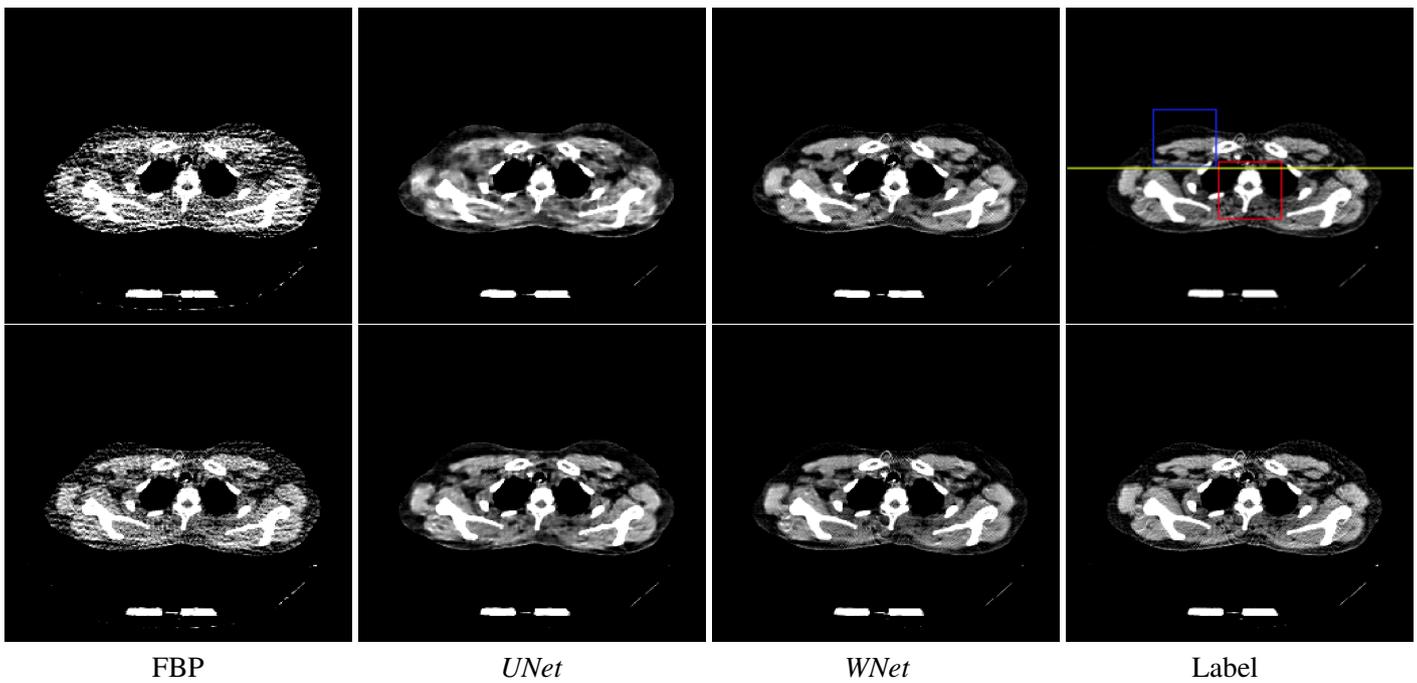

FBP                    UNet                    WNet                    Label

**Figure 2:** Reconstructions obtained using the different methods applied on "*64-view*" sinograms (first row) and on "*128-view*" sinograms (second row). (top-right) two squares (red and blue) highlighting two regions of interest which are visualized in Fig. 3. All images are visualized with soft tissue windowing.

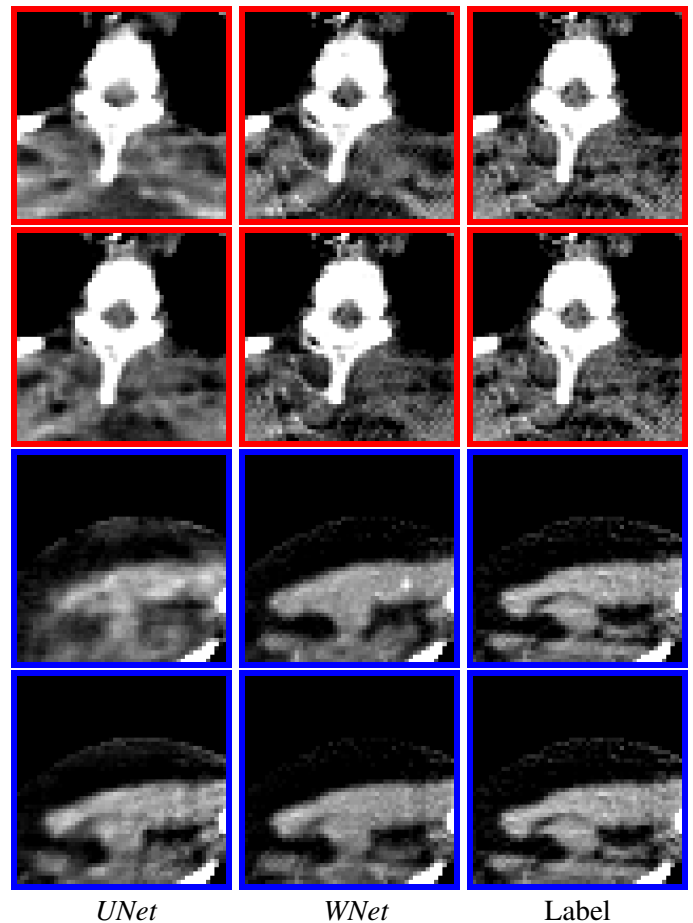

UNet                    WNet                    Label

**Figure 3:** Magnified regions highlighted in the top-right image from Fig. 2. The first and the second row of each color correspond to the reconstruction obtained from the "*64-view*" sinogram (top row) and the one from the "*128-view*" sinogram (bottom row).

# Tomographic diffusion filter

Cyril Riddell

GE Healthcare, Buc, France

**Abstract:** We introduce a diffusion filtering scheme called tomographic because it is integrated into the backprojection operation. Backprojection is decomposed into the sum of the images of the backprojection of each single projection (BSP). The BSP image is filtered in the directions parallel to the detector axes. The diffusion strength is simultaneously modulated by a priori ray-ray noise information and a priori edge information. The sum of the filtered BSP images gives the final FBP-like reconstruction image. The properties of diffusion make that the filtering is applied separately along all the directions of the tomographic acquisition. Simulations in cone-beam geometry illustrate how ray-wise statistics and an edge map can simultaneously reshape the signal-to-noise ratio of the output of an analytical reconstruction, providing a non-iterative alternative to diffusion regularized weighted least-squares optimization.

## Introduction

Analytical reconstruction is the most convenient way of reconstructing tomographic data. The processing pipeline is pre-filtering of the data, backprojection and post-filtering of the reconstructed image. There exists an infinite variety of ways to combine these filtering operations, based on a priori information about the noise of the data and of features to preserve in the image. Backprojection however, is an obstacle in-between. Backprojection at the voxel level is a sum of measurements (or rays) of different statistical quality; if the expectation of each measurement were the same, a weighted sum according to the noise statistics would be optimum. Obviously, the point of acquiring tomographic data is that the measurements do have a different expectation at each angle, and the correct summation is the non-weighed one; statistically weighted backprojection does not fit into analytical reconstruction. What is allowed is the translation of the noise statistics into a modulation of the pre-filtering of the data. The issue is that it cannot account for the edge information. Edge information is available in the image domain only, where, however, the filtered backprojection has created complex statistical correlations. To overcome these issues, one either combine multiple analytical reconstructions [1] or formulate the problem as a regularized weighted least-squares optimization problem [2]. The first approach, while quite practical, does not handle the noise simultaneously with the edges. On the contrary, model-based iterative reconstruction (MBIR) brings together the statistical description, assuming uncorrelated noise, which is a good-enough approximation at the detector level, with an edge-preserving regularization term. However, handling the interplay between the noise model and the image regularization requires a complex parameterization to deliver the right image for a given clinical task [3].

We propose instead to consider the case of the backprojection of a single projection. It is a very special image because it only replicates the projection over the image space. The ray-by-ray noise model can thus be propagated into the image space where the edge information already belongs. It becomes possible to perform a filtering that is guided by both noise and edge information. The filtering having no impact in the direction where the signal remains constant, it is implemented in the remaining orthogonal directions only. The final summation yields an image that is filtered over all sampled directions. Not all filters are separable into a sum of 1D filtering steps, but diffusion is, and will be considered exclusively in the following. Diffusion filtering is very flexible and benefits from a solid theoretical background regarding its discretization and fast computation, such as additive splitting schemes [4].

## Materials and Methods

We first expose the mathematics of tomographic diffusion in the 2D case of reconstructing a single slice, in parallel- or fan-beam geometry. We use the variational form of diffusion filtering.

### Noise-driven 1D-diffusion:

Let us use coordinates $(u, \theta)$ for the detector and angular axes of the projection domain respectively. We denote $\Theta$ the set of the $M$ sampled directions and $p_\theta(u) = p(u, \theta)$ the projection having direction $\theta$, i.e. such that the direction of the central measurement ray is $\theta^\perp = \theta - \frac{\pi}{2}$. The sinogram of all projections is denoted $p_\Theta$. Uniform diffusion of $p_\Theta$ is obtained by 1D-diffusion along $u$ of each sinogram row $p_\theta(u)$. We denote $\nabla_\theta$ the gradient along $u$ for direction $\theta$ (note that this is also the direction of the ramp filtering of the FBP process). We denote $G_\Theta(p)$ such filtering:

$$G_\Theta(p_0) = \min_p \left\{ \frac{1}{2}\|p - p_0\|^2 + T \sum_\theta p^t (\nabla_\theta^t \nabla_\theta) p \right\}$$

The penalty weight is called the diffusion time $T$. The reason for this is that, for $T$ not too large, $G_\Theta(p_0)$ is also the solution of the problem of uniform heat diffusion at time $t = T$ if the $p_0$ values are taken as the temperature values at time $t = 0$ [4]. For the same reason, diffusion filtering will be equivalent to a Gaussian filtering of full width at half maximum (FWHM) close to $3.333\sqrt{T}$. Note that time $T$ is thus homogeneous to the variance of the Gaussian while the FWHM is homogeneous to its standard deviation.

The interest of diffusion schemes with respect to Gaussian filtering in Fourier space is that a non-uniform filtering is easily obtained by inserting a diagonal matrix of diffusion weights $W_\theta$ called diffusion tensor. It modulates the diffusion at each sample $u$ for filtering direction $\theta$ according to:

$$G_{W_\Theta}(p_0) = \min_p \left\{ \tfrac{1}{2}\|p - p_0\|^2 + T \sum_{\theta \in \Theta} p^t \left( \nabla_\theta^t W_\theta \nabla_\theta \right) p \right\}$$

with $W_\Theta$ gathering the modulation $W_\theta$ for all directions in $\Theta$. A straightforward statistical interpretation comes by rewriting the above equation as:

$$G_{W_\Theta}(p_0) = \min_p \left\{ \tfrac{1}{2}(p - p_0)^t W_\Theta^{-1}(p - p_0) \right.$$
$$\left. + T \sum_{\theta \in \Theta} p^t (\nabla_\theta^t \nabla_\theta) p \right\}$$

A classical choice for weights $W_\Theta$ is to set them equal to the a priori known variances of the measurements. This fits well with time $T$ that is homogeneous to a variance, while the modulation of the diffusion strength in terms of FWHM will therefore follow the local variations of the standard deviation. Note that this is but one choice, made a priori. The output filtered image is optimal with respect to that choice. The optimality of the choice itself is not discussed in this work.

While $W_\Theta$ provides a ray-by-ray control of noise propagation, it is blind to the image features.

### Edge-preserving 2D-diffusion

In the image space, we denote a voxel location $\bar{x} = (x_1, x_2)$. The post-filter applies diffusion along each axis $x_i$ of the Cartesian grid where image $f$ is sampled. The variational definition using gradient $\nabla_{\bar{x}} = (\nabla_{x_1}, \nabla_{x_2})$ is:

$$G(f) = \min_g \left\{ \tfrac{1}{2}\|g - f\|^2 + T \sum_i g^t \left( \nabla_{x_i}^t \nabla_{x_i} \right) g \right\}$$

Uniform diffusion over the data is equivalent to uniform diffusion over the reconstructed image. But image diffusion can now be modulated by a diagonal diffusion tensor $E_{\bar{x}}$ that modulates scalar $T$ according to:

$$G_{E_{\bar{x}}}(f) = \min_g \left\{ \tfrac{1}{2}\|g - f\|^2 + T \sum_i g^t \left( \nabla_{x_i}^t E_{\bar{x}} \nabla_{x_i} \right) g \right\}$$

In this context, matrix $E_{\bar{x}}$ reflects the edge information, locally switching in a spatially continuous manner from 1 for full diffusion to 0 when an edge must be preserved. The modulation brought by $E$ is edge-preserving but blind to the noise propagation from the data. We here assume that the edge map is known a priori. The edge-based modulation is the same for all directions so that the diffusion is isotropic but non-stationary.

### Filtered backprojection of a single projection

Noise in analytical reconstruction is amplified by the ramp filtering and propagated to the reconstructed image by backprojection. Let us decompose the backprojection of acquired data $p_0$ through:

$$R^t p_0 = \sum_{\theta \in \Theta} R_\theta^t p_0$$

We denote $f_\theta(\bar{x}) = R_\theta^t p_\theta(u)$ the backprojection of $p_\theta(u)$. Image $f_\theta$ is a "backprojection-of-a-single-projection" (BSP) image. Since the BSP image is constant along $\theta^\perp$, 2D filtering is equivalent to 1D filtering along direction $\theta$. It can be written in variational form as:

$$G_\theta(f_\theta) = \min_g \left\{ \tfrac{1}{2}\|g - f_\theta\|^2 + T \sum_i g^t \left( \nabla_\theta^t \nabla_\theta \right) g \right\}$$

Because, in parallel geometry, gradient $\nabla_\theta f_\theta$ is the same as $\nabla_\theta p_\theta$, the filtering commutes with backprojection: post-filtering $G_\theta(f_\theta)$ of the BSP image is equal to the backprojection of the corresponding projection of the pre-filtered sinogram $G_\Theta(p_0)$. Therefore, it can be modulated by the same noise model through:

$$G_{W_\theta}(f_\theta) = \min_g \left\{ \tfrac{1}{2}\|g - f_\theta\|^2 + T \sum_i g^t \left( \nabla_\theta^t W_\theta \nabla_\theta \right) g \right\}$$

Because $f_\theta$ is in the same coordinate framework as the reconstructed image, it can alternatively be modulated by the edge information:

$$G_{\theta, E_{\bar{x}}}(f_\theta) = \min_g \left\{ \tfrac{1}{2}\|g - f_\theta\|^2 + T \sum_i g^t \left( \nabla_\theta^t E_{\bar{x}} \nabla_\theta \right) g \right\}$$

Most importantly, as illustrated on Fig. 1, it can be modulated by both:

$$G_{W_\theta E_{\bar{x}}}(f_\theta) = \min_g \left\{ \tfrac{1}{2}\|g - f_\theta\|^2 \right.$$
$$\left. + T \sum_i g^t \left( \nabla_\theta^t W_\theta E_{\bar{x}} \nabla_\theta \right) g \right\}$$

Product $W_\theta E_{\bar{x}}$ brings together the angle-independent edge map and the angle-dependent ray-by-ray noise model. The resulting filter is thus a linear non-uniform 1D diffusion

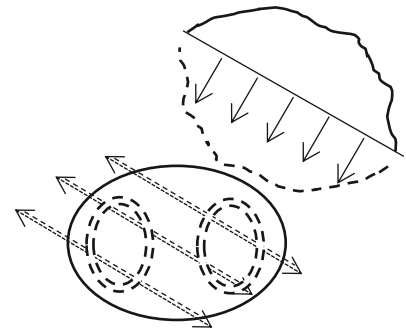

Tomographic diffusion

Figure 1: BSP image brings noise and edge information in the same space

filter along direction $\theta$, driven in intensity by the noise model $W_\theta^{-1}$ while the edges are preserved according to $E_{\bar{x}}$.

## Analytical reconstruction with tomographic diffusion

We denote $\acute{f}_\theta = R_\theta^t \acute{p}_\theta$ the BSP image equal to the backprojection of $\acute{p}_\theta(u)$, the ramp filtered version of $p_\theta(u)$. The filtered backprojection reconstruction is the sum of all BSP images:

$$f_{FBP} = \frac{\pi}{M} \sum_{\theta \in \Theta} \acute{f}_\theta$$

Tomographic diffusion is obtained by substituting $\acute{f}_\theta$ by its filtered version:

$$f_{W_\theta E_{\bar{x}}} = \frac{\pi}{M} \sum_{\theta \in \Theta} G_{W_\theta E_{\bar{x}}}(\acute{f}_\theta)$$

The filter can be seen as an apodization of the ramp filter in the image space that controls noise propagation while preserving edges. It must be noted that there is no assumption regarding the angular sampling.

## Implementation

By moving the apodization from the projection space to that of BSP images, the computation becomes more complex, especially as direction $\theta$ almost never corresponds to the axes of the cartesian grid. In [5-6], it was shown that the rows and columns of BSP image $f_\theta(\bar{x})$ are resampled versions of projection $p_\theta(u)$, the resampling being a 1d magnification in parallel geometry or a 1d homography in fan-beam geometry. Therefore, any convolution of $p_\theta(u)$ can be connected to the convolution of a row or column $f_\theta(x_i)$. Within diffusion, it is straightforward to replace $\nabla_\theta f_\theta$ by $\nabla_{x_i} f_\theta$, making its implementation aligned with axes $x_i$ along which the edges are defined. One must not forget the ratio between the sampling step $d\theta$ along direction $\theta$ and the sampling step $dx_i$ along axis $x_i$ to properly scale the gradients and get the same filtering output.

The computation may be further simplified by the linearity of the filtering: if $W_\theta$ is kept constant or averaged over a range of angles, one can limit the computation from $M$ to $M/K$ images made of the backprojection of $K$ projections with $K \ll M$. This latter simplification was not used in the following experiments.

## Experiments

A simulation was conducted for circular CBCT over 360° with 360 projections. The distance from the focal point to the center of rotation was set to 1500 voxels. The ground-truth image was made of a CT slice of an abdomen supplemented with a needle-like shape set at 1000 Hounsfield units (HU) above the background. For simplicity, the same slice was duplicated in the z direction so that FDK reconstruction remains exact along z. We used the same noise model and edge map for all slices. Cone-beam forward projection was performed, and pure Poisson noise was generated with value $10^6$ counts in air. After log-transform, the variance of the data was computed per column as the estimate of the input noise. All reconstructions were performed with the same implementation of the tomographic diffusion filter within the cone-beam backprojector. The projections were first Fourier-filtered with a non-apodized ramp filter; then all cone-beam BSP volumes were computed and filtered with diffusion. The diffusion was thus 2D: along z and along x or y, depending on which one was closer to angle $\theta$. Gradient scaling was applied as in parallel geometry: according to angle $\theta$ and neglecting the small extra angular variation within the cone. We used the FWHM to parameterize the diffusion time through $T = FWHM^2 / 11.09$. The filtered BSP volumes were then summed into the final reconstruction.

Four diagonal tensors were compared: 1) identity, 2) noise-weighted, 3) edge-weighted and 4) simultaneously weighted by noise and edges. Only the last case cannot be obtained by the association of FDK with a pre- or post-diffusion filter.

For diffusion with the identity tensor and diffusion with the edge-preserving tensors, the FWHM was set to 3 voxels. The edge weights were computed from the gradient of the ground-truth image. Denoting $n_g(\bar{x})$ the norm of the gradient at voxel $\bar{x}$, the weight at voxel $\bar{x}$ is equal to $1 - \exp\left(\frac{-3.315}{(n_g(\bar{x})/\tau)^8}\right)$ with a threshold $\tau = 50$ HU. Uniform and edge-preserving tensors are independent from angle $\theta$. For the noise-weighted tensors, the ray-by-ray noise variances were used for $W_\Theta$ and value $T$ was set so that the bulk of the ray-by-ray diffusion FWHM values were between 1.5 and 4 voxels. This range of FWHM was chosen empirically as the one matching the output SNR of uniform filtering on average. The values of $TW_\Theta$ were then clipped to yield a maximum FWHM of 8 voxels for the noisiest measurement locations.

For the last tensor choice, the previously described noise and edge tensors were multiplied by one another.

The ground-truth image, one non-apodized FDK noisy reconstruction, the edge map and the noise weights are shown on Fig. 2 in clockwise order.

The mean and standard deviation of the reconstructed volumes were computed along z, excluding the top and bottom slices affected by "long-object" truncation.

## Results

Fig. 3 shows one reconstructed slice for each of the four different tensor choices. The images are displayed with a common windowing of width 100. Uniform diffusion (top left image) reduces noise overall but blur the needle-like structure. The diffusion is not strong enough to remove the horizontal noise streaks. With the edge information only (bottom left image), edges are preserved, but there is no improvement regarding the noise streaks. Noise-weighted diffusion (top right image) allows a stronger filtering where the projection of the needle-like structure yields the highest noise, which results in the nearly complete removal of the noise streaks, but also into a strong blurring of the needle-like structure. Tomographic diffusion with a tensor capturing both noise and edges (bottom right image) provides the best of both worlds. Regarding the needle-like structure, the noise streak artifacts are removed but the needle is not blurred.

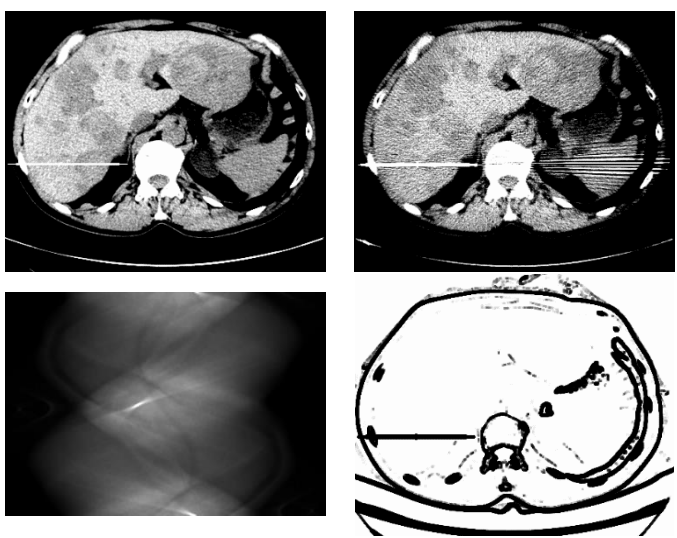

Figure 2: Top: Ground truth image (left) and non-apodized ramp filtered FDK reconstruction (right). Bottom: noise variance (left) and edge map (right)

Fig. 4 shows the corresponding SNR images displayed with a common windowing (150-400). The SNR is greater if the noise in the data is comparatively lower or if it is filtered out. Uniform filtering of projections with non-uniform noise yields a radially increasing SNR from the most attenuated center towards the outer parts of the body (top left image). It also increases the SNR for thin high-intensity structures, at the cost of a strong bias since they are blurred. With the edge information (bottom left image), the noise properties are unchanged with respect to uniform filtering except where the edge map is null. There the SNR is now much lower: no bias is introduced but noise is high. Use of the noise model (top right image) decreases the filtering on each side of the projections which yields SNR images with opposite properties than uniform diffusion: the SNR now (slightly) decreases when moving away from the center. The diffusion is now stronger where noise propagation would be stronger. Overall, the SNR is rather uniform except for even stronger biases at the needle-like structure and at the edges of the vertebra. Tomographic diffusion merging noise and edge information (bottom right image) is also a merge of the SNR image obtained with the ray-by-ray noise model with the edge map that avoids biases at the edges.

## Discussion

Tomographic diffusion filtering is a linear multi-directional anisotropic non-uniform diffusion filter that accounts for a priori known noise and edge models that can handle ray-by-ray noise variations. This provides an analytical alternative to diffusion regularized least-squares optimization that requires a forward projection model to achieve the same objective.

The parameterization requires an explicit translation of the noise information into a diffusion blur. Our setting directly translated the variation of the noise standard deviations into the modulation of the diffusion FWHM. This allowed us to illustrate how the SNR of the reconstructed image could be modulated and made more uniform. It is known that FBP reconstruction propagates the noise uniformly yielding a non-uniform SNR, while Poisson statistical models can be a better alternative because they allocate the reconstructed noise so that the output has a more uniform SNR [7]. Tomographic diffusion, which is linear, unfortunately remains radically different from Poisson models, as the latter, which are non-linear, achieve a uniform SNR by modeling mean and variance as a single parameter, not through filtering. In our experiments, tomographic diffusion modulated by the standard deviations of the noise yielded a rather uniform SNR, away from the edges. At edge locations, diffusion was simply stopped and did not act on the SNR anymore.

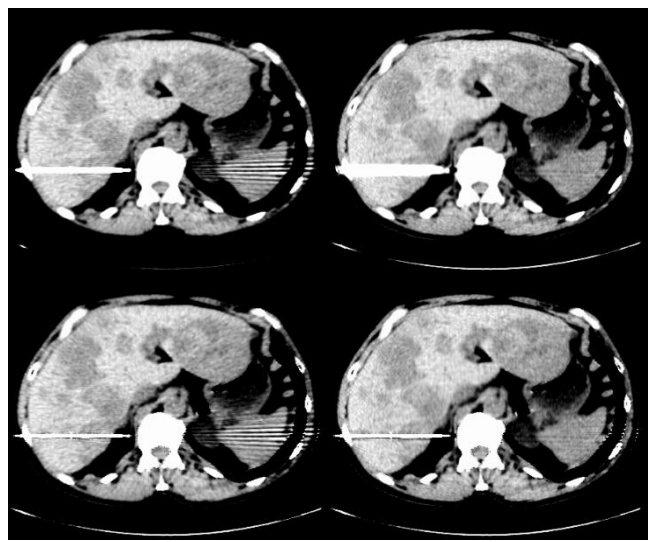

Figure 3: Tomographic diffusion reconstruction with uniform (top left), noise-modulated (top right) edge preserving (bottom left) and noise and edge preserving diffusion tensors (bottom right).

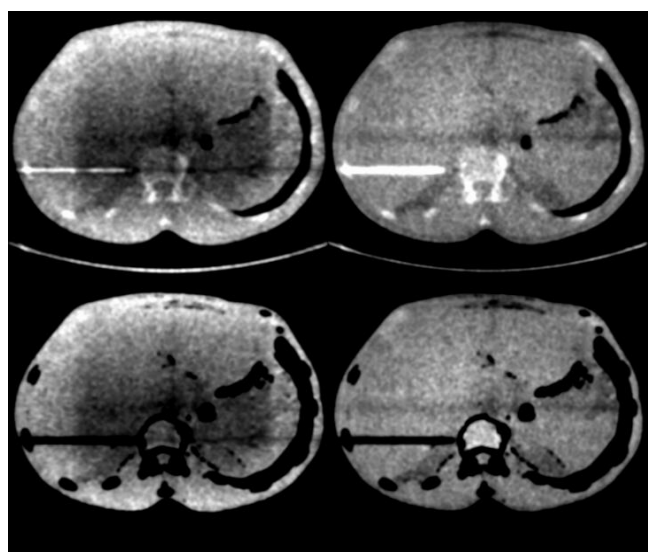

Figure 4: SNR images of tomographic diffusion reconstruction with uniform (top left), noise-modulated (top right) edge preserving (bottom left) and noise and edge preserving diffusion tensors (bottom right).

As we already mentioned, the interplay between a noise model and regularization is not an easy task. We believe that tomographic diffusion is a simple tool to make the transition between the input a priori knowledge and the desired properties of the output reconstruction. Our goal here was to demonstrate the flexibility and clarity of the diffusion parameterization, not to produce the best image. Note that diffusion filtering efficiency is high when the data is not correlated. Even if tomographic data is usually uncorrelated, as the diffusion goes, the signal becomes more and more correlated and diffusion less and less efficient.

We therefore see tomographic diffusion as a real improvement over pre-filtering, to accommodate for strong variations, especially when concentrated on a narrow angular range, as illustrated by the needle-like example, without altering the resolution of dense but fine structures, whose edges would be easily detected on a prior reconstruction. An image with more uniform SNR is expected to be a better starting point for a more sophisticated post-filtering. In particular, in conjunction with deep-learning based denoising, tomographic diffusion may alleviate the need to retrain a network for the very peculiar noise patterns induced by standard filtered backprojection; second, tomographic diffusion happens over intermediate images that are not accessible to a post-filter, it thus exploits better information than what can be given to a post-reconstruction network.

Finally, tomographic diffusion filtering within analytical reconstruction is an alternative to iterative reconstruction with least-square noise models only. Forward projectors capturing additional physical effects [8] or subsampling [9] still lead to least-square formulations that have no analytical counterpart. The tomographic diffusion filter being a variational filter, a possible next step is to use it as a regularization. In that case, the filter modulation replaces the noise model, which simplifies the data fidelity term into a non-weighted least-square term. This simplification further allows for integrating ramp filtering in the data fidelity term to both accelerate the convergence and ensure that constraining the square norm of the gradient does behave as a diffusion filter. Note that diffusion regularization can also be turned into a sparse regularization [9].

## Conclusion

A diffusion filtering scheme, called tomographic because it is built within the backprojection process, has been shown to control noise propagation and preserve edges within an FBP-like reconstruction by taking simultaneously into account an a priori ray-by-ray noise model and an a priori edge map.

# Learning projection matrices for marker free motion compensation in weight-bearing CT scans

Valentin Bacher[1], Christopher Syben[1], Andreas Maier[1], and Adam Wang[2]

[1]Pattern Recognition Lab, Friedrich-Alexander-University Erlangen-Nuremberg, Germany
[2]Departments of Radiology and Electrical Engineering, Stanford University, Stanford, California

**Abstract** Weight-Bearing Computed Tomography is gaining popularity due to its ability to generate a three-dimensional reconstruction of joints under weight-bearing condition. The major problem regarding scanning are motion artifacts which are inevitable, because of the standing position of the patient. State of the art methods overcome these problems using aids, such as external tracking devices, prior knowledge, or fiducial markers. Those methods require knowledge which might not be exist or are tedious or computationally expensive. Therefore, we investigate the possibility of using trainable CT operators to compensate rigid motion without any kind of aids or prior knowledge.

The motion is estimated and corrected using a TensorFlow based API (PYRO-NN) to incorporate the learning of projection matrices into an iterative comparison of the original sinogram with digitally reconstructed radiographs (DRRs) obtained from the motion-corrupted backprojected volume. The loss function used is a squared loss. This error is then used to calculate a gradient using a finite difference, which is used in an Adam optimizer to iteratively reduce the loss. This approach is able to estimate and correct the shifts, although it is unable to correct small rotations. The results of our simulation study show that this is still sufficient to obtain a very good reconstruction with a low level of motion artifacts.

## 1 Introduction

Weight-Bearing Computed Tomography (WBCT), based on Cone-Beam CT (CBCT) geometries, has established itself as a core modality in orthopedic imaging of the lower extremities [1]. It allows the imaging of the joints under weight-bearing conditions and thus valuable insights in the pathology of the bones, cartilage, and other soft tissue in proximity of the joints [1]. The greatest challenge associated with this imaging setup is motion artifacts. These are inevitable due to the upright standing position of the patient during scan time. Combined with relatively long scan times associated with cone beam scanners this causes severe motion artifacts.

Most state of the art methods use either physical aids or prior knowledge to reduce motion artifacts. Several approaches use fiducial markers placed on the skin as orientation [2–4]. However the placement of these markers is tedious and can cause errors due to movement of the skin relative to the bone. Thus Berger et al. proposed to use prior scans of the respective bone structure to register the projections [5]. While this is promising, it requires prior scans which are not always available. Algorithms based on sharpness metrics or image consistencies have been developed for other use cases with different approaches. One such, which originated in emission tomography, uses subdivision of the sinogram into almost motion-free subsets [6–8]. Other approaches use sharpness metrics on the motion corrupted reconstruction as a loss function [9]. Bodensteiner et al. proposed to evaluate a least squares difference between digitally reconstructed radiographs (DRRs) and the original projections to iteratively reduce motion artifacts [10].

Integrating trainable trainable layers into a fixed weight pipelines has proven it self as very beneficial [11]. We propose to use the differential CT operators introduced by PYRO-NN [12]. PYRO-NN implements forward and backward projectors as differential layers in deep learning frameworks. However, in the current version the weights of the CT operator layers are not trainable. In this proof-of-concept work we extend PYRO-NN to trainable CT operators and therefore allow to optimize for geometrical misalignment, e.g. occurring motion or miscalibration by learning compensated projection matrices specific to each scan.

## 2 Definition of the Model

For the proof-of-concept study we propose a model very similar to an iterative reconstruction pipeline. For this we make use of the differentiable CT layers of PYRO-NN [12]. Their theoretical description can be introduced by the system matrix $\mathbf{A}$ for the forward projection operator and by $\mathbf{A}^\top$ for the back-projection operand. As shown in the paper, the filtered back-projection (FBP) algorithm to reconstruct tomographic images $\mathbf{v}$ from sinogram data $\mathbf{p}$ can be fully described within the context of neural networks. The fundamental reconstruction problem can be described by performing a pseudo inverse of the forward model $\mathbf{p} = \mathbf{Av}$, resulting in $\mathbf{v} = \mathbf{A}^\top \mathbf{F}^\mathbf{H} \mathbf{CF}\mathbf{p}$, where $\mathbf{C}$ denotes the reconstruction filter in Fourier domain and $\mathbf{F}, \mathbf{F}^\mathbf{H}$ are the Fourier and inverse Fourier transform, respectively. As for realistic settings the system matrix $\mathbf{A}$ is infeasible to store in memory, the layers make use of the concept of projection matrices $\mathbf{R}^{3 \times 4}$, where one projection matrix describes a certain projection and a set of projection matrices forms the scanning trajectory acquiring sinogram data for tomographic reconstruction. However, up to now these CT operator layers are non-trainable and differentiable with respect to their input. In this proof-of-concept work we investigate the possibility to extend the CT operator layers to trainable layers, where the projection matrices are the weights of the layers. This can be described by $\mathbf{A}(w_i)$ and $\mathbf{A}^\top(w_i)$, where $w_i$ are the newly introduced trainable weights.

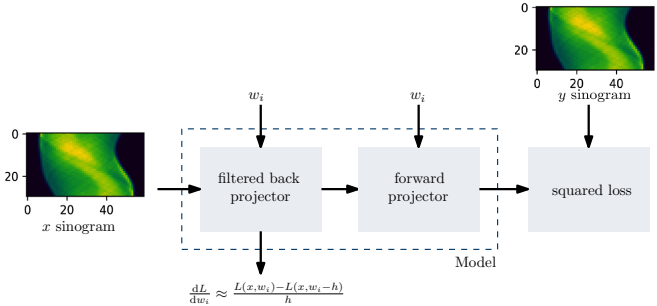

**Figure 1:** This model using a complete pipeline is used to train the projection matrices.

Trainable layers allow setups to update the geometric situation of the projection process and therefore, allow to setup pipelines for motion compensation. A naive implementation where we learn the projections matrices directly would lead to $N \times 3 \times 4$ trainable weights for $N$ projections in the trajectory. To reduce the amount of trainable parameters we used a static definition of the intrinsic camera parameters and a non static description of the extrinsic parameters consisting of three rotations and three translations. With this, we define the $x$-axis parallel to the detector plane horizontal, the scanner rotation is about the $y$-axis, and the $z$-axis points towards the detector. Linear motion parallel to the central ray is neglected based on some initial experiments and the findings of Gullberg et al. [13]. Scaling effects caused by this kind of motion are very small due to small cone angles. Consequently our model for motion compensation takes a motion-corrupted sinogram, which is reconstructed by the first layer. The output of the first layer is directly forward projected using the same geometric parameters as the first layer. The resulting re-projected sinogram is evaluated by a loss against the original sinogram. The complete model with the loss is shown in Fig. 1. The mathematical description of the layers results in

$$\hat{\mathbf{y}} = \mathbf{A}(w_i)\mathbf{A}^{\top}(w_i)\mathbf{F}^{\mathbf{H}}\mathbf{C}\mathbf{F}\mathbf{x} \ . \tag{1}$$

The loss is determined using a simple squared difference $L(\mathbf{x}, w_i) = (\hat{\mathbf{y}}(\mathbf{x}, w_i) - \mathbf{y})^2$. For the proof-of-concept study, we compute the gradient with respect to the weights of the layers using a finite backward difference resulting in

$$\frac{\partial L(\mathbf{x}, w_i)}{\partial w_i} = \frac{L(\mathbf{x}, w_i) - L(\mathbf{x}, w_i - h)}{h} \ . \tag{2}$$

The training is conducted using the Adam optimizer and the stepsize for the backward difference is set to $h = 0.002$.

## 3 Experiments Description

The experiments aim to evaluate the possibility of iteratively reducing motion artifacts in cone-beam CT data using a PYRO-NN based trainable reconstruction pipeline. The first experiment investigates the possibility to achieve meaningful gradients for the layer weights. For this, the model shown in Fig. 1 is further simplified to just do a reconstruction and compute a loss with a motion-free label reconstruction. The second experiment is conducted with the presented model to be free of supervision and is inspired by iterative reconstruction methods.

The simulated scanner setup consists of a $120 \times 120$ pixel flat-panel detector with a pixel size of $1 \times 1$ mm$^2$. The source is located at a distance of 1200 mm, while the source-to-isocenter distance measures 750 mm. The scanner simulated 30 projections obtained in a scan range of $180° + 2 \cdot 2.84°$. The data used in the experiments are simulated 3D Shepp-Logan phantoms. The phantom measures $60 \times 60 \times 60$ mm$^3$ with a pixel size of $0.5 \times 0.5 \times 0.5$ mm$^3$. Rigid motion is modeled by uniformly random shifts and rotations in the range of $-2$ to 2 [mm] or [°], respectively, for each projection with which the Shepp-Logan phantom is forward projected to obtain the motion-corrupted sinogram to reconstruct. The range of motion is oriented at the range of motion of a standing person [14]. To gain independence of the initial position of the phantom, the phantom is uniformly randomly shifted and rotated ensuring that the phantom stays within the field of view. The learning rate for the first experiment is chosen to be $\eta = 9 \cdot 10^{-2}$. The second experiment is conducted with a learning rate of $\eta = 1 \cdot 10^{-1}$. The trainable variables consisted of the five motion parameters $w_i = \{\phi_x, \phi_y, \phi_z, t_x, t_y\}$ for each of the 30 projections. Note that in the proof-of-concept study we seek to find the true weights to a specific corrupted scan and therefore only one sinogram is used for training.

## 4 Results

The results of the first experiment are shown in Figure 2 and show the training results for a single scan of a randomly initialized phantom, where these results are representative of other randomly initialized phantoms and motion. In Fig. 2a the variance of the difference over all 30 projections of the layer weights with respect to the ground truth parameter over the training procedure are plotted. For all parameters, except $\phi_y$, which describes the axis of the scanner's rotation, the ground truth parameter can be learned. Fig. 2c indicates that the high variance is due to the axis symmetrical error of the rotation angle. The loss (cf. Fig. 2b) correlates well with the variance of the shifts, while the rotations do not have a large effect on the loss. Fig. 3 shows the results of the second experiment. While for the shifts the true parameter could be found, this is not true for the rotations. Fig. 3c plots the difference between the learned parameters and the ideal ones. The shifts approach nearly zero, with a small negative constant offset for $t_y$. The rotations show random behaviour. Fig. 4 shows multi-planar views of the reconstruction of the phantom before and after the motion correction. While the transformation of the phantom stays the same, the sinograms used for the reconstruction differs in their type of motion corruption. A distortion just caused by rotations does not lead

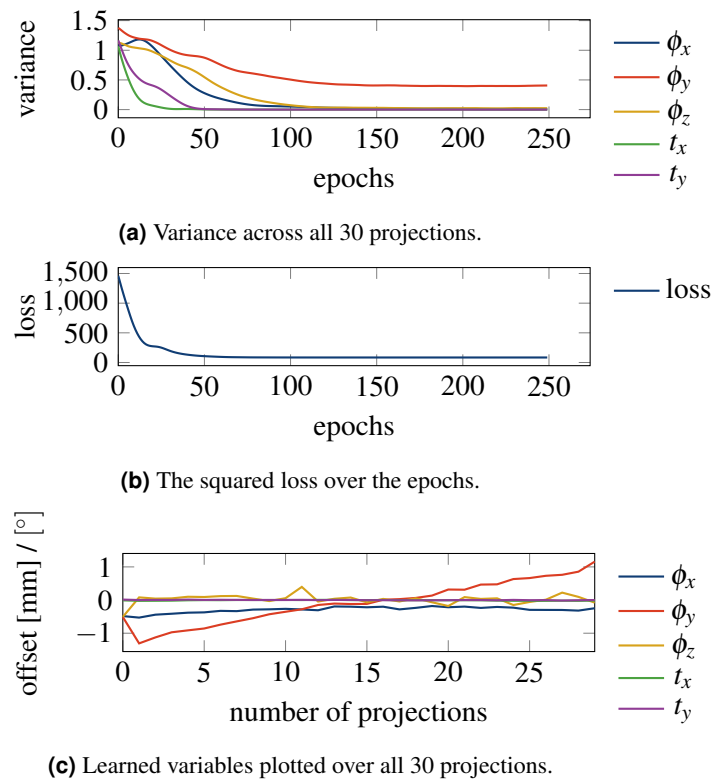

**(a)** Variance across all 30 projections.

**(b)** The squared loss over the epochs.

**(c)** Learned variables plotted over all 30 projections.

**Figure 2:** All results above are obtained from training on a randomly chosen single phantom, assuming a known reconstruction.

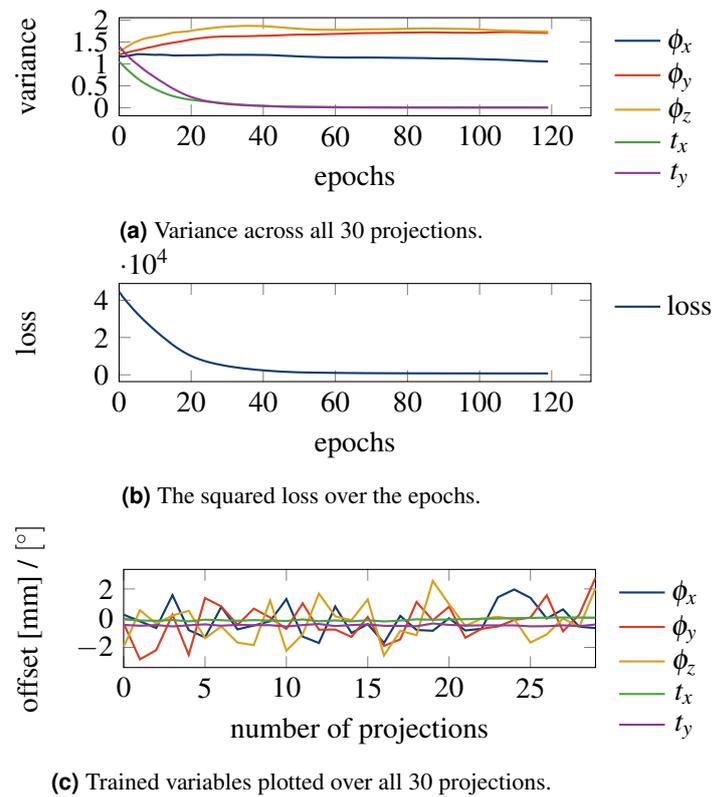

**(a)** Variance across all 30 projections.

**(b)** The squared loss over the epochs.

**(c)** Trained variables plotted over all 30 projections.

**Figure 3:** Results from training on a randomly chosen single phantom, without prior knowledge. For better visualization, the expected value are subtracted in sub figure (c).

to a substantial reduction in image quality. The shifts however cause severe motion artifacts. In general, all reconstructions using the learned parameters are very well defined, and the main artifacts are caused by the relatively low number of projections used.

## 5   Discussion

With both experiments, we were able to show that it is possible to interpret the geometric parameters of the CT operators as trainable layers and successfully update these layer weights in a training setting. With the learned parameters the image quality of the tomographic can be improved, even though the true value of all parameters could not be learned. While we were able to learn the shifts, this is not true for the rotations. In the first experiment, two out of three rotation parameter could be learned, while in the second experiment no stable solution for all three could be learned. The final learned rotation parameters do not have a substantial effect on the image quality of the reconstruction nor the loss. This indicates that the gradient response of artifacts imposed by rotation are neglected in the training process compared to the artifacts introduced by the translation parameters. This shortcoming could be overcome by introducing a weighted loss, focusing more on the rotation parameters or by choosing a composite loss function including metrics more sensitive to the artifacts introduced by the rotation parameters, e.g., incorporating sharpness metrics of the reconstructed image such as total variation. In 2017, Sisniega et al. published promising results using such a metric. They also added a term penalizing unrealistic jumps in motion parameters. Also, the object to reconstruct has an impact on the sensitivity of the loss function w.r.t changes in rotations, with the Shepp-Logan phantom particularly insensitive. Another observation to discuss is that for both experiments the reconstruction with only 30 projections leads to strong sparse sampling artifacts in the tomographic domain and therefore a substantial portion of the loss reflects artifacts which cannot be compensated by the proposed setup. This can negatively effects the training process, e.g., situations where motion artifacts can cancel out or reduce sparse sampling artifacts would lead to minima in the cost function which do not correspond with ideal layer weights. We are aware that this limitation is introduced by the limited amount of data, which is due to memory limitations in the context of 3D tomography and the use of finite difference for the gradient computation. However, the general results are promising and this does not diminish their insights as even with the limited data, the algorithm estimated the trajectory well enough to obtain good results. For this proof-of-concept study we use a finite difference, which is a computationally expensive estimation of the gradient. In the future, approximate or analytical partial gradients, which would utilize the back propagation algorithm, are desirable. The robustness with respect to a low amount of data may allow use of a Gaussian pyramid, which not only reduces spatial resolution but also the number of projections to estimate motion.

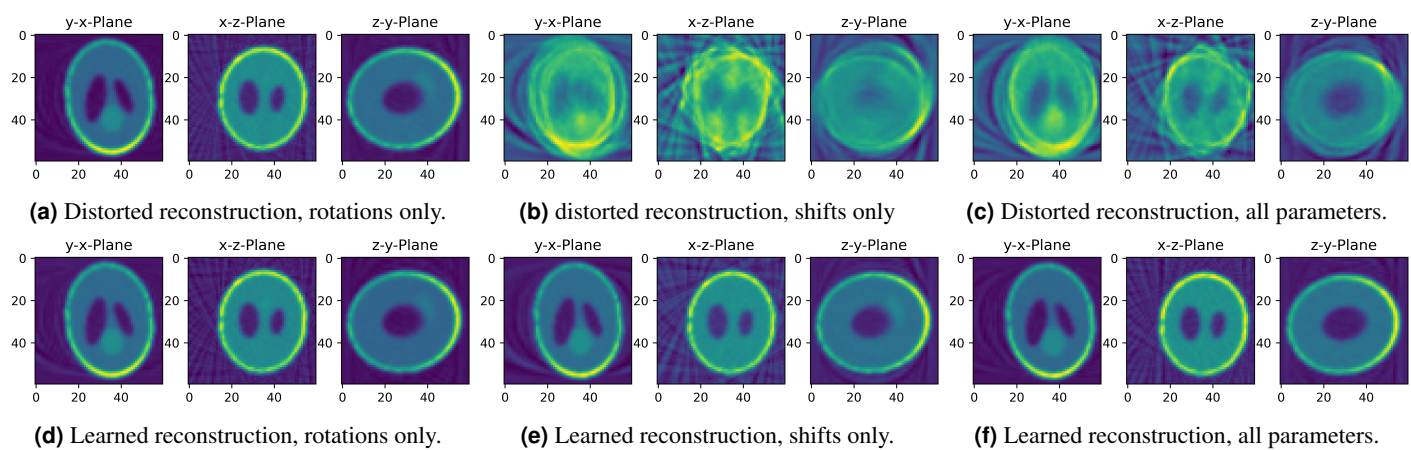

**(a)** Distorted reconstruction, rotations only.    **(b)** distorted reconstruction, shifts only    **(c)** Distorted reconstruction, all parameters.

**(d)** Learned reconstruction, rotations only.    **(e)** Learned reconstruction, shifts only.    **(f)** Learned reconstruction, all parameters.

**Figure 4:** Visualized are the reconstructions before (a-c) and after (d-f) the learning of the projection matrices. All graphs are the result of a training without any prior knowledge.

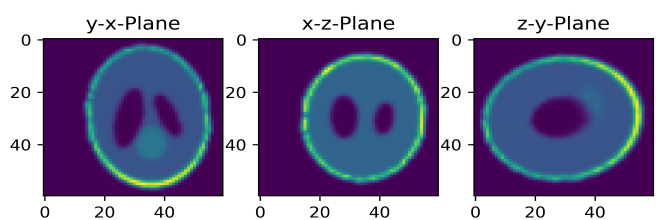

**Figure 5:** The randomly transformed phantom used for the results shown in Figure 4.

## 6 Conclusion

We were able to show that the general concept of trainable CT operators in neural networks are feasible. Further, an iterative reduction of motion artifacts by defining the trajectory parameters as trainable weights is possible. To make our approach applicable to clinical use, gradient computation needs to be improved and the loss function needs to be further investigated. The inability of the rotations to converge did not have a significant effect on the image quality. Still this approach is promising since it has proven itself as robust with respect to a small amount of data. Thus it could be combined with multi-scale algorithms increasing the performance.

## 7 Acknowledgements

This work was partially supported by NIH S10RR026714, NIH R01AR065248, and Siemens Healthineers.

# Chapter 11

# Oral Session - CT imaging 1

**session chairs**
Karl Stierstorfer, *Siemens Healthcare GmbH (Germany)*
Harald Schoendube, *Siemens Healthineers (Germany)*

# Region-specific Texture Prior-based Low-dose Cone-beam CT Bayesian Reconstruction for Image-guided Radiation Therapy

Shaojie Chang[1], Siming Lu[2], Yongfeng Gao[1], Hao Zhang[3], Hao Yan[4], and Zhengrong Liang[2*]

[1] Department of Radiology, Stony Brook University, Stony Brook, NY 11794, USA.
[2] Departments of Radiology and Biomedical Engineering, Stony Brook University, Stony Brook, NY 11794, USA
[3] Department of Medical Physics, Memorial Sloan Kettering Cancer Center, New York, NY 10065, USA
[4] OUR UNITED CORPORATION, Xi'an, Shaanxi, 710018, China

**Abstract:** In image-guided radiation therapy (IGRT), on-board cone-beam computed tomography (CBCT) provides volumetric information of a patient at treatment position and localizes the target of treatment. However, the repeated CBCT scanning during a treatment course delivers an excessive dose to the patient. Meanwhile, a planning CT is always available for treatment planning purposes, which has superior image quality. To reduce dose, we propose a region-specific texture prior-based low-dose CBCT reconstruction algorithm, which explores a prior strategy to connect the planning CT information with the low-dose CBCT reconstruction. The proposed method extracts the regional tissue-specific textures from the planning CT images to determine the Markov random field (MRF) weights on the neighborhood and uses this constructed tissue-specific MRF prior model as a priori knowledge to perform Bayesian reconstruction of the low-dose CBCT images with enhanced tissue-specific textures for improved IGRT. It is shown that the proposed method can better preserve structural details while effectively suppressing noise. Quantitatively, our proposed method shows the best performance and achieves 0.0149 and 0.5822 in terms of root mean square error (RMSE) and structure similarity index (SSIM) metrics.

## 1. Introduction

Image-guided radiation therapy (IGRT) has been widely used in radiotherapy clinics. An on-board cone-beam computed tomography (CBCT) in the IGRT system provides volumetric information of a patient at treatment position and localizes the target of treatment, which has been considered as a gold standard for IGRT. However, during the treatment, over 25 times CBCT scanning are applied to a patient [1]. It will deliver too much dose to the patient. In addition, Kan *et al.* reported that the repeated CBCT scanning for IGRT could increase the secondary cancer risk by 2% up to 4% [2]. Hence, the development of low-dose CBCT has raised a great concern in the field.

A simple way to achieve low-dose CBCT imaging is lowering the x-ray exposure level in a scan. However, it would inevitably lead to increased noise in projection data, which degrades the reconstructed image quality. Tremendous efforts have been devoted to developing effective low-dose CBCT reconstruction methods to reduce the radiation dose while maintaining the clinical image quality. Among them, the Bayesian theorem-based iterative reconstruction algorithms have been shown success to improve image quality for low-dose CT imaging, which consider the statistical properties in the projection domain and *priori* information in the image domain [3-6]. In particular, many types of *priori* model have been extensively studied, such as Markov random field (MRF), total variation (TV), dictionary learning (DL), and so on. The MRF plays an important role in preserving edge sharpness while suppressing noise, whose weights on the neighborhood are always fixed in traditional CT. However, for different tissues of CT images, the weights are variant to preserve the tissue-specific textures. As tissue textures have been realized as important imaging biomarkers for various clinical tasks, Zhang *et al.* proposed an MRF based texture preserving prior to improve the low-dose CT (LdCT) image via learning region-specific textures, such as muscle, fat, bone and lung, from the same patient's previous full-dose CT (FdCT) scans [3,4]. After that, Gao et al. [7] further reported the feasibility of learning region-specific textures from an FdCT database, which did not require previous FdCT scans for the current LdCT imaging.

Inspired by these works, to reduce dose, we proposed a region-specific texture prior-based (RSTP) low-dose CBCT reconstruction method for IGRT. In IGRT, a planning CT is always available for treatment planning purposes, which has superior image quality. In this work, we would like to propose a prior strategy to connect the planning CT information with the low-dose CBCT reconstruction in IGRT. Hence, we first extract the regional tissue-specific textures from the planning CT images to determine the Markov random field (MRF) weights on the neighborhood. Specifically, it captures the image textures of muscle, fat, bone and lung. Second, we incorporate this constructed tissue-specific MRF prior model as a *priori* knowledge for Bayesian reconstruction of the corresponding regions in the low-dose CBCT slice image. Finally, the enhanced low-dose CBCT images are iteratively reconstructed.

The remainder of this paper is organized as follows. Section 2 will describe the formulations of Bayesian reconstruction for CBCT, followed by an introduction to our proposed RSTP method and the overall workflow. Section 3 presents the experiment design and results. Discussion and conclusions are drawn in Sections 4 and 5.

## 2. Materials and Methods

### 2.1. Bayesian Formulation for Low-dose CBCT Reconstruction

Given a set of acquired line integral data, denoted by a vector $y \in \Re^{I \times 1}$, where I is the number of data elements. The solution $\mu \in \Re^{J \times 1}$, where J is the number of image voxels, is desired to maximizes the posterior probability $p(\mu|y)$. Based on the Bayes' theorem:

$$p(\mu|y) = \frac{p(y|\mu)p(\mu)}{p(y)} \approx p(y|\mu)p(\mu) \qquad (1)$$

where $p(y)$ becomes a constant when maximizing the posterior probability and is ignored.

The log data fidelity term $\log p(\mu|y)$ can be described as a re-weighted least squares (RWLS). And the log prior term, $\log p(\mu)$ is described by a Markov random field (MRF) model. The desired solution can be expressed as:

$$\mu^* = arg \min_{\mu}\{(y - A\mu)^T D(y - A\mu) + \beta U(\mu)\} \qquad (2)$$

where $A$ is the projection matrix with the size $I \times J$ and its element $A_{ij}$ is calculated as the intersection length of projection ray i with voxel j. $A\mu$ denotes the mean value vector $\bar{y}$ of the acquired $y$. $D$ is a diagonal matrix, where each diagonal element is called a weight for its corresponding datum and also called the variance of the detecting datum in statistics. $\beta$ is a parameter that controls the smoothing strength or the balance between the data fidelity term and the prior term.

## 2.2. Data Fidelity Term

Without considering the electronic noise, the transmission data $N_i$ along $i^{th}$ ray is assumed to follow the Poisson noise as follows:

$$N_i \sim \text{Poisson}(\overline{N}_i) \qquad (3)$$

where $\overline{N}_i$ denotes the mean of transmission data. Here, we introduce $y_i$ as a random variable describing the line integral $\sum_j A_{ij}\mu_j$, which is the mean of $y_i$, $\bar{y}_i = \sum_j A_{ij}\mu_j$. $i = 1,\dots,I$, $j = 1,\dots,J$, where $I$ is the total number of measurements in the scan and $J$ is the total number of pixels. And $\mu_j$ denotes the linear attenuation coefficient at $j^{th}$ pixel.

By Beer's law, we have

$$\overline{N}_i = N_0^i \exp\left(-\sum_j A_{ij}\mu_j\right) = N_0^i \exp\left(-\bar{y}_i\right) \qquad (4)$$

where $N_0^i$ represents the incident radiation intensity. With the analysis in [9], the variance of line integral is expressed by,

$$\sigma_{y_i}^2 = \frac{1}{\overline{N}_i}. \qquad (5)$$

Hence, the data fidelity term becomes,

$$L(y | \mu) = (y - A\mu)^T D(y - A\mu), \qquad (6)$$

where $D = diag\{\frac{1}{\sigma_{y_i}^2}\} = diag\{\overline{N}_i\}$. And it can be obtained by $\overline{N}_i = N_0^i exp(-\bar{y}_i) = N_0^i \exp(-[A\mu^{(n)}]_i)$ with one-step late approximation in iterations.

## 2.3. Region-specific Texture Prior

To extract the regional tissue-specific textures from the planning CT images to determine the Markov random field (MRF) weights on the neighborhood from different types of tissues, a vector quantization (VQ) automatic segmentation algorithm [7] is applied to this study. For this study, the chest CT image is segmented into four tissue types including lung, fat, bone and muscle. Morphological operations are adopted to enlarge the segmented lung parenchyma and bone region boundaries slightly so that the final lung region for MRF coefficients prediction would include both the blood vessels inside the lung and the juxta-pleural nodules attached to the pleural wall. The bone marrow with relatively lower intensities was also included in the refined bone region for MRF coefficients prediction of the bone tissue. An example for region-specific texture prior extraction is shown in Fig.1. Fig.1(a) shows the segmented regions by VQ and Fig.1(b) presents the MRF coefficients for each region. In this method, we extract the region-specific texture prior information from the planning CT image slice-by-slice in a 2D way.

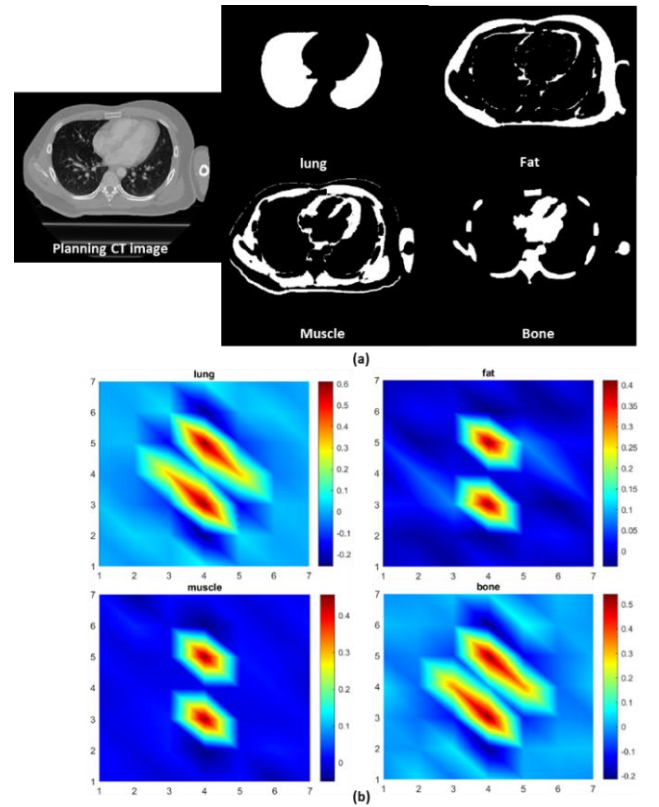

Fig.1 Region-specific texture prior extraction from Planning CT image. (a) Segmented regions. (b) MRF coefficients for each region.

As stated in Hao's paper [3], It is interesting to see that the MRF model coefficients of the lung and bone regions have some similarity while the coefficients of the fat and muscle regions also have some similarity, but the coefficients of the group of lung/bone are different from the coefficients of the group of fat/muscle. The former group has a large intensity variation while the latter group has a small intensity variation. All four tissue regions exhibit different spectral patterns corresponding to different image textures.

To preserve the tissue texture features, we proposed a tissue region-based texture-preserving regularization which can be given as:

$$U(\boldsymbol{\mu}) = \sum_{r=1}^{R} \sum_{j \in Region(r)} \sum_{m \in \Omega_j} b_{jm}^{pCT\_predict} (\mu_j - \mu_m)^2 \quad (7)$$

where R represents the different tissue regions and the index r will run through all types of regions. Index j goes over every voxel in the specific tissue. $\Omega_j$ denotes the MRF window (typically 48 neighbors in a 2D case) around voxel j. Index m runs over every voxel in the MRF window. $\left\{ b_{jm}^{pCT\_predict} \right\} = \boldsymbol{b}_r^{pCT\_predict}$ means the MRF coefficients of the specific tissue region r predicted from the planning CT scan.

With a planning CT image and an MRF window size, a linear regression strategy is applied to determine the set of MRF model coefficients corresponding to a tissue region. Among all the linear regression estimation algorithms, the least-squares algorithm is adapted because of its computational efficiency. With this method, every image voxel inside the MRF window can be predicted from a linear combination of its clique-mates (the pixels that position around the to-be predicted pixel and are bounded by the MRF window). The least-squares predicted MRF coefficients can be formulated as:

$$\boldsymbol{b}_r^{pCT\_predict} = arg \min_{\boldsymbol{b}_r} \sum_{k \in Region(r)} (\mu_k^{pCT} - \boldsymbol{b}_r^T \boldsymbol{\mu}_{\Omega k}^{pCT})^2 \quad (8)$$

where vector $\mu_k^{pCT}$ represents the planning CT image. It is expected that the sum of the predicted MRF coefficients for each region shall be close to one.

By substituting Eqs. (4), (6) into (2), the final objective function becomes,

$$\boldsymbol{\mu}^* = arg \min_{\boldsymbol{\mu}} \Big\{ (\boldsymbol{y} - \boldsymbol{A}\boldsymbol{\mu})^T \boldsymbol{D} (\boldsymbol{y} - \boldsymbol{A}\boldsymbol{\mu}) + \\ \beta \sum_{r=1}^{R} \sum_{j \in Region(r)} \sum_{m \in \Omega_j} b_{jm}^{pCT\_predict} (\mu_j - \mu_m)^2 \Big\}. \quad (9)$$

The denoised image can be obtained by optimizing the Eq. (9) in iterations.

## 2.4. Overall Workflow

As shown in Fig.2, we present a workflow for implementation of the RTSP-based low-dose CBCT Bayesian reconstruction for IGRT. First, a series of planning CT images are available for treatment planning, which are introduced for RSTP extraction. In this step, we segment the planning CT images into four regions (lung, fat, muscle, bone) and the regional MRF coefficients can be estimated. After that, the low-dose CBCT projection and RSTP are incorporated into the Bayesian reconstruction framework with Eq. (9). Finally, the enhanced CBCT images are iteratively reconstructed.

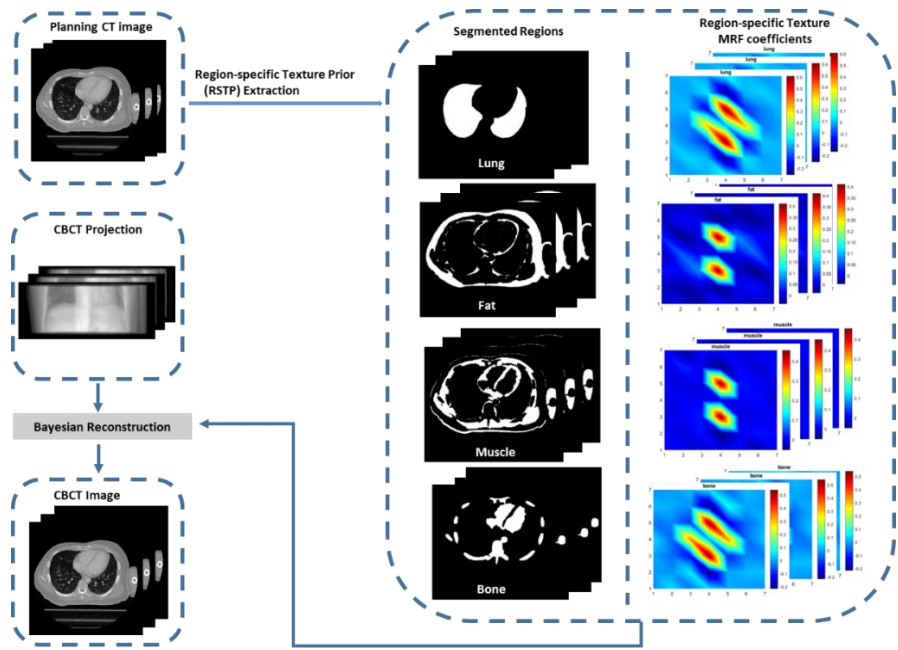

Fig.2 Workflow of the proposed region-specific texture prior-based low-dose CBCT image reconstruction for IGRT.

## 3. Experiments and Results

A full dose patient volume data under planning CT scan was used for numerical simulation. The volume size is $512\times512\times101$ with the voxel size of $0.5\times0.5\times1$ mm$^3$. We first performed a simulated CBCT scan with CT patient data volume by adding Poisson noise with levels of $10^4$ photon incidents per ray. The geometry was set to the same as the TrueBeam medical accelerator (Varian Medical System), where the source-to-axis and source-to-detector distances were 1000mm and 1500mm, respectively. The flat detector had $512\times384$ pixels with $0.776\times0.776$ mm$^2$ resolution. Projections were simulated uniformly in $360°$ by a full-fan scan.

For the image reconstruction, we use order subsets and GPU to accelerate the computation. In this work, we set 6 subsets and 10 iterations for the Bayesian reconstruction.

To show the performance of the proposed method, we compared the image reconstruction quality obtained with the proposed RSTP-based Bayesian reconstruction method with those of the conventional Feldkamp-Davis-Kress (FDK) method and total variation (TV) based reconstruction method. For quantitative comparison, the root means squared errors (RMSE) and the structure similarity (SSIM) with the planning CT image were used. The $\beta$ used in the reconstruction with each prior term is empirically selected to achieve the best denoising performance.

Fig.3 presents a single transversal slice of the reconstructed images of different methods. It is found that both TV regularization and RSTP method can efficiently remove the noise in comparison with the conventional FDK result. To further evaluate the texture preserving performance of the proposed method, the zoomed-in region of interests (ROIs) selected by the red rectangle are shown in Fig.4. The fine structures of the TV method are blurred to some extent, and the proposed method can enhance the textures and preserve the details. Table I shows the quantitative results of the selected ROI. It is indicated that our proposed method achieves the best performance in all metrics.

Table I:  RMSE and SSIM of reconstructed images of each method compared to the Planning CT image in Fig.4.

| Method | RMSE | SSIM |
|---|---|---|
| FDK | 0.0202 | 0.4770 |
| TV | 0.0150 | 0.5778 |
| **The proposed method** | **0.0149** | **0.5822** |

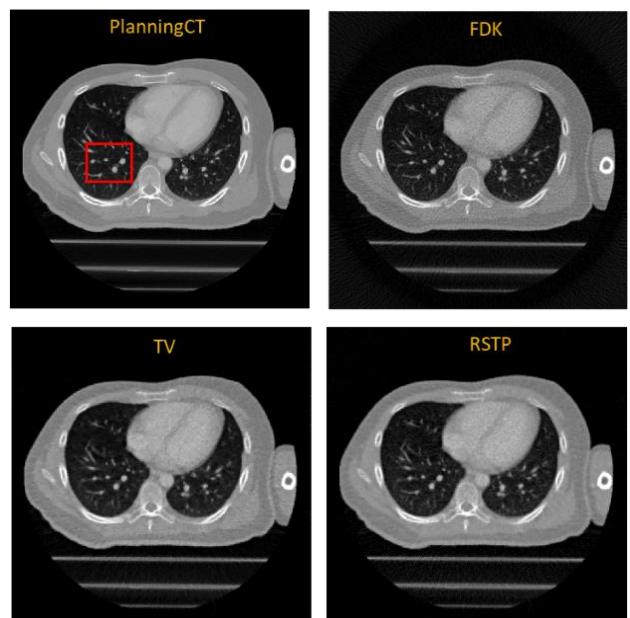

Fig.3 A single transversal slice of the reconstructed images of different methods. The display window is [0,0.35] cm$^{-1}$.

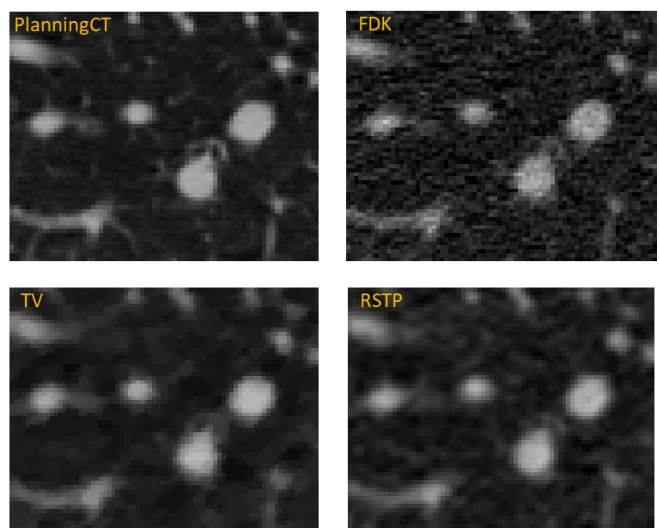

Fig.4 The zoomed-in ROIs selected by the red rectangle in Fig.3 of different methods. The display window is [0,0.35] cm$^{-1}$.

## 4. Discussion

In this study, we use a 2D slice-by-slice texture to present the feasibility of RSTP-based CBCT reconstruction in IGRT. The results indicate that our proposed method can reduce the noise while maintaining the clinical image quality, especially preserving the structural details. For the Bayesian reconstruction, there are different L_2 norm image penalty terms, such as Huber and Gaussian MRF. In the future, we will compare the results of the above penalty terms to verify the performance of our proposed method. Meanwhile, we will develop a 3D texture model for low-dose CBCT reconstruction in IGRT, which brings more prior information and is believed more efficient to represent the corresponding tissue.

In addition, the ordered subset strategy and parallel computing are applied to speed up the whole program in both segmentation and image reconstruction processing. It has a great potential to make our proposed method clinically practical.

Moreover, utilizing the planning CT images' region-specific texture prior is an effective way to improve the current low-dose CBCT reconstructed images in IGRT. The region-specific textures are modeled by the MRF theorem as a *priori* knowledge and applied adaptively as a regularizer to the corresponding regions in the CBCT reconstruction, instead of matching the patches between the planning CT image and the corresponding CBCT slice image. In the future, more clinical experiments at different dose levels will be implemented to verify the performance of the proposed method.

## 5. Conclusion

In summary, we proposed a region-specific texture prior-based low-dose CBCT reconstruction algorithm, which explored a prior strategy to connect the planning CT information with the low-dose CBCT reconstruction in IGRT. The proposed method extracted the regional tissue textures from the planning CT images before treatment as *priori* knowledge to enhance the reconstructed CBCT images for IGRT. It is shown that the proposed method can better preserve structural details while effectively suppressing noise as compared to the conventional FDK reconstruction and TV-based method.

### Acknowledgement

This work was partially supported by the NIH/NCI grant #CA206171.

# Sam's Net: A Self-Augmented Multi-Stage Neural Network for Limited Angle CT Reconstruction

Changyu Chen[1,2], Yuxiang Xing[1,2*], Hewei Gao[1,2], Li Zhang[1,2], and Zhiqiang Chen[1,2*]

[1]Department of Engineering Physics, Tsinghua University, Beijing, China

[2]Key Laboratory of Particle & Radiation Imaging (Tsinghua University), Ministry of Education, Beijing, China

**Abstract** In this paper, we presented a self-augmented multi-stage neural network (Sam's Net) for limited angle CT reconstruction. The network includes forward projection (FP) and filtered back-projection (FBP) layers as domain transformers and utilizes a feedback mechanism to formulate a self-augmented learning procedure. We employ two networks for sinogram completion and artifact reduction. Besides, a weighting layer is introduced to adjust the redundant weight in FBP reconstruction through learnable pixel-wise weights. In the training phase, the feedback mechanism serves as online augmentation which ensures information and errors can propagate between the sinogram domain and image domain conveniently and enhances data consistency. The online augmentation is a general framework which can be extended to other networks. In the inference phase, sinograms may run through the network several times to achieve better results. Numerical experiments under 90-degree fan-beam configuration are executed to evaluate the proposed method. The results indicate that Sam's Net can significantly improve the image quality compared with a simple dual-domain learning and it is stable and robust for limited angle tomography.

## 1 Introduction

As a non-invasive imaging method, X-ray computed tomography (CT) has been widely used in many fields, such as clinical applications and security inspections. Given projections under complete angular coverage (180 degrees for parallel-beam scanning and 180 degrees plus fan-angle for fan-beam scanning), analytical and iterative algorithms can achieve high-quality reconstruction. However, in some practical applications, data acquisition angle is limited and reconstructions with conventional algorithms may suffer from severe artifacts and structural distortions[1].

In recent decades, a lot of work has gone into limited angle tomography. There are mainly two strategies: **1) Improving reconstruction algorithms.** Such methods include iterative reconstruction-reprojection[2], wavelet decomposition[3], and projections onto convex sets (POCS) [4]. **2) Employing additional prior knowledge.** To enhance sparsity in the gradient domain, total variation (TV) regularized iterative reconstruction algorithm was proposed[5] and achieved great success. In limited angle tomography, the shape and orientation of streak artifacts are closely related to the missing angular range. Employing such prior information, anisotropic TV (ATV)[6] methods were proposed, which has been proved to be more efficient in reducing artifacts and recovering structures.

The resurgence of deep neural networks has yielded many new approaches. Some focus on sinogram completion[7], while others work as image post-processing[8]. But the capability of such single-domain methods is limited.

This work was supported in part by the National Natural Science Foundation of China (Grant No. 61771279 and 62031020).

*Corresponding author: Zhiqiang Chen (e-mail: czq@tsinghua.edu.cn); Yuxiang Xing (e-mail: xingyx@mail.tsinghua.edu.cn).

To integrate optimization in dual domains, end-to-end networks were proposed. Some researchers proposed analytical networks[9][10], which map analytical algorithms (such as FBP and linogram) to neural networks and learn the redundant weight of projections. Others presented more comprehensive cross-domain optimization[11][12] methods. However, due to the poor ability of sinogram completion, cross-domain methods still highly rely on the strength of image-domain optimization, which may be harmful to robustness and data consistency.

In this paper, we proposed a self-augmented multi-stage neural network (Sam's Net) for limited angle CT reconstruction. The network incorporates a self-augmented feedback procedure for consensus optimization and has the capability of improving robustness and data consistency for reliable prediction outputs.

## 2 Materials and Methods

Fig. 1 gives an overview of Sam's Net. The input of Sam' Net is pre-estimated full-size sinograms $\hat{\mathbf{p}} = [\mathbf{p}_{LA}, \hat{\mathbf{p}}^{\varnothing}]$ with $\mathbf{p}_{LA}$ being the acquired limited angle projections and $\hat{\mathbf{p}}^{\varnothing}$ an estimate of missing projections. A sinogram completion network (SCNet) learns to map $\hat{\mathbf{p}}$ to sinogram labels in a supervised residual learning manner. Then, a pixel-wise weighting layer and a filtered back-projection (FBP) layer transform data to the image domain. An artifact suppression network (ASNet) further recovers context information. Besides this main branch, the recovered image $\hat{\boldsymbol{\mu}}_{ASN}$ returns to the sinogram domain by forward projection (FP), which forms a feedback mechanism by feeding a new input to SCNet in the next epoch of training. More details of our method are explained in following sections.

### 2.1 Domain Transformer

The FBP layer and FP layer work as domain transformers in our net, which enables interaction between dual domains. They are implemented in a matrix format based on analytical operations for fan-beam CT:

$$\text{FBP Layer:} \quad \boldsymbol{\mu} = \mathbf{H}_W^T \mathbf{F} \mathbf{W}_{cos} \mathbf{W}_{red} \odot \mathbf{p}$$
$$\text{FP Layer:} \quad \mathbf{p} = \mathbf{H}\boldsymbol{\mu} \tag{1}$$

Here, $\boldsymbol{\mu}$ refers to images and $\mathbf{p}$ sinograms, $\mathbf{W}_{red}$ stands for redundant Parker weights, $\mathbf{W}_{cos}$ for cosine weighting, $\mathbf{F}$ for filtration, and $\mathbf{H}_W^T$ for weighted back-projection. FP layer is simply a multiplication with the system matrix $\mathbf{H}$. Inspired by [9][10], we formulated a pixel-wise learnable weighting layer $\tilde{\mathbf{W}}_{\text{red}}$ to adjust the inconsistency enforced by SCNet. It is initialized to conventional Parker weights $\mathbf{W}_{\text{red}}$ and the weight for the data missing region will be updated in the learning procedure. To apply a data consistency constraint to the whole network, especially ASNet, relative root mean square error (RRMSE) between the forward projection $\hat{\mathbf{p}}_{\text{FP}}$ and sinogram labels $\mathbf{p}_{\text{GT}}$ within the data acquired region is calculated:

$$L_{\text{FP}} = \frac{\left\| \hat{\mathbf{p}}_{\text{FP}} \odot (\mathbf{1} - \mathbf{M}^{\varnothing}) - \mathbf{p}_{\text{GT}} \odot (\mathbf{1} - \mathbf{M}^{\varnothing}) \right\|_2}{\left\| \mathbf{p}_{\text{GT}} \odot (\mathbf{1} - \mathbf{M}^{\varnothing}) \right\|_2} \quad (2)$$

where $\mathbf{M}^{\varnothing}$ denotes the mask of the data missing region.

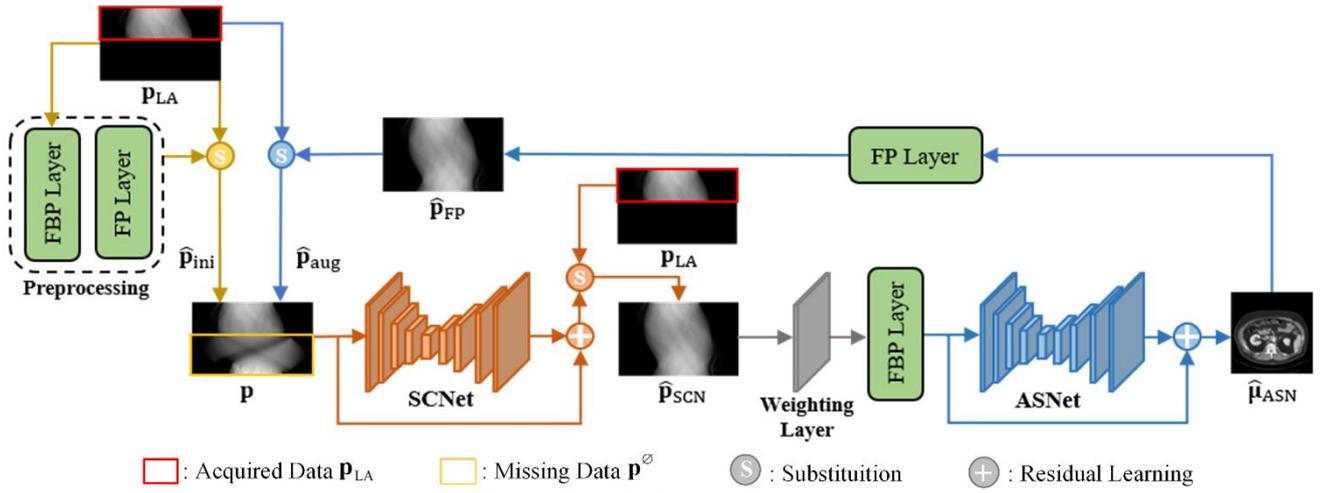

**Fig. 1** Schematic diagram of Sam's Net

## 2.2 Sinogram Completion Network (SCNet)

A five-stage U-Net with residual learning is used for SCNet where the number of each channel is reduced by half compared to the original U-Net. To reflect the full geometry of CT scan, SCNet takes complete size sinograms as input and outputs a more accurate version guided by labels. Initially, we set:

$$\hat{\mathbf{p}}_{\text{ini}} = [\mathbf{p}_{\text{LA}}, (\mathbf{H}\mathbf{H}_{\mathbf{w}}^{\text{T}} \mathbf{F} \mathbf{W}_{\cos} \mathbf{W}_{\text{red}} \odot \mathbf{p}_{\text{LA}}) \odot \mathbf{M}^{\varnothing}] \quad (3)$$

Denoting the output of SCNet as $\hat{\mathbf{p}}_{\text{SCN}}$. We employ RRMSE to minimize the difference between $\hat{\mathbf{p}}_{\text{SCN}}$ and $\mathbf{p}_{\text{GT}}$ in the data missing region:

$$L_{\text{SCNet}} = \frac{\left\| \hat{\mathbf{p}}_{\text{SCN}} \odot \mathbf{M}^{\varnothing} - \mathbf{p}_{\text{GT}} \odot \mathbf{M}^{\varnothing} \right\|_2}{\left\| \mathbf{p}_{\text{GT}} \odot \mathbf{M}^{\varnothing} \right\|_2} \quad (4)$$

To fully use the given knowledge, we formulated a substitution operator which sends the output sinogram $\hat{\mathbf{p}}_{\text{SCN}}$ with the data acquired region substituted by $\mathbf{p}_{\text{LA}}$ as $[\mathbf{p}_{\text{LA}}, \hat{\mathbf{p}}_{\text{SCN}} \odot \mathbf{M}^{\varnothing}]$ to the next layer.

## 2.3 Artifact Suppression Network (ASNet)

To further suppress the artifacts, we incorporate ASNet to work in the image domain. Similar to SCNet, ASNet also utilizes U-Net with the number of channels reduced by quarter of the original U-Net. Besides, it has been proved that perceptual loss helps learn deep features. Our group proposed a CT image feature space (CTIS) loss[13] which is defined by an autoencoder trained on many normal dose CT images. The latent space in the autoencoder is used to represent the deep features of CT images. To integrate pixel-wise precision and domain property supervision, we chose the bottom dense layer of the autoencoder to measure the feature space loss and combined it with RRMSE as the loss function:

$$L_{\text{ASNet}} = \frac{\left\| \hat{\boldsymbol{\mu}}_{\text{ASN}} - \boldsymbol{\mu}_{\text{GT}} \right\|_2}{\left\| \boldsymbol{\mu}_{\text{GT}} \right\|_2} + \lambda \frac{\left\| \psi(\hat{\boldsymbol{\mu}}_{\text{ASN}}) - \psi(\boldsymbol{\mu}_{\text{GT}}) \right\|_2}{\left\| \psi(\boldsymbol{\mu}_{\text{GT}}) \right\|_2} \quad (5)$$

where $\psi$ denotes the pre-trained CTIS loss model and $\lambda$ is set to 0.1.

## 2.4 Self-augmented Feedback Mechanism and Loss Function

Theoretically, sinogram completion networks are designed to learn a mapping from the data acquired region to the data missing region directly. However, convolutional networks tend to learn local mapping and object features rather than global sinogram properties, which may lead to the weak capability of sinogram completion and generalization. Thus, artifact suppression in the image domain is strongly relied on. But without enough guidance of physical imaging processes, some context information and data consistency condition will be abandoned by image-domain networks. To further enhance the consensus in sinogram and image domains, we proposed the self-augmented feedback mechanism.

Specifically, at the $t^{\text{th}}$ epoch in the training phase, the initial sinogram $\hat{\mathbf{p}}_{\text{ini}}$ (Eq. 3) and the augmented sinogram $\hat{\mathbf{p}}_{\text{aug}}^{(t)}$ (Eq. 6) are fed into SCNet as independent samples. As $\hat{\mathbf{p}}_{\text{aug}}^{(t)}$ originates from the output of the network at the previous epoch $\hat{\mathbf{p}}_{\text{FP}}^{(t-1)}$, it is a self-augmented sinogram.

$$\hat{\mathbf{p}}_{\text{aug}}^{(t)} = \mathbf{p}_{\text{LA}} \odot (\mathbf{1} - \mathbf{M}^{\varnothing}) + \hat{\mathbf{p}}_{\text{FP}}^{(t-1)} \odot \mathbf{M}^{\varnothing} \quad (6)$$

We can find that both sinograms are the same in the data acquired region but $\hat{\mathbf{p}}_{\text{aug}}^{(t)}$ provides a quite different pixel-to-pixel interpretation in the data missing region. SCNet needs to map these two different samples to the same sinogram label, which guides it to be more attentive to the information from data acquired region and learn sinogram properties to predict the missing data. And this mechanism provides a more efficient channel for spreading data consistency information and errors through the two domains.

The overall loss function can be interpreted as:

$$L_{\text{Total}} = \frac{1}{2N} \sum_i (0.5 \times L_{\text{SCNet}}^i + L_{\text{ASNet}}^i + 0.1 \times L_{\text{FP}}^i) \quad (7)$$

$$(\forall i \in [\hat{\mathbf{p}}_{\text{ini}}; \boldsymbol{\mu}_{\text{GT}}] \cup [\hat{\mathbf{p}}_{\text{aug}}; \boldsymbol{\mu}_{\text{GT}}])$$

where $[\hat{\mathbf{p}}_{\text{ini}}; \boldsymbol{\mu}_{\text{GT}}]$ refers to the training set and $[\hat{\mathbf{p}}_{\text{aug}}; \boldsymbol{\mu}_{\text{GT}}]$ the augmented. $N$ is the number of training samples.

## 3 Results

### 3.1 Experimental Set-up

We evaluated our method based on datasets from the American Association of Physicists in Medicine (AAPM) "Low Dose Grand Challenge" and The Cancer Imaging Archive (TCIA) "Low Dose CT Image and Projection Data". In our experiments, we used CT images from 28 patients for training, 5 patients for validation, and another 6 patients for inference. The limited angle CT scan is simulated under a 90-degree equidistant fan-beam configuration and the detailed geometry parameters are shown in Table 1. Under this setting, the angular coverage of a complete short scan is 208 degrees with [90, 180] degrees used for limited angle projections.

**Table 1** Parameters of the fan-beam geometry

| Parameter | Value |
| --- | --- |
| Distance between the source and isocenter (cm) | 100 |
| Distance between the detector center and isocenter (cm) | 80 |
| Number of detector elements | 640 |
| Size of each element (cm) | 0.14375 |
| Dimension of reconstruction grids (pixels) | 512 |
| Voxel size (cm) | 0.0625 |
| Sampling interval of projection (deg) | 0.5 |

For quantitative evaluation, we employed structural similarity index (SSIM), and peak signal-to-noise ratio (PSNR) between the prediction $\hat{\mathbf{x}}$ and ground truth $\mathbf{x}^*$ as evaluation metrics. The SSIM is defined as

$$\text{SSIM}(\mathbf{x}^*, \hat{\mathbf{x}}) = \frac{(2\mu_{\hat{\mathbf{x}}}\mu_{\mathbf{x}^*} + C_1)(2\sigma_{\hat{\mathbf{x}}, \mathbf{x}^*} + C_2)}{(\mu_{\hat{\mathbf{x}}}^2 + \mu_{\mathbf{x}^*}^2 + C_1)(\sigma_{\hat{\mathbf{x}}}^2 + \sigma_{\mathbf{x}^*}^2 + C_2)} \quad (8)$$

where $\mu$ and $\sigma$ denote the mean and standard deviation of the vector in subscripts correspondingly, and $\sigma_{\hat{\mathbf{x}}, \mathbf{x}^*}$ is the covariance between $\mathbf{x}^*$ and $\hat{\mathbf{x}}$. $C_1 = (0.01 \times L)^2$ and $C_2 = (0.03 \times L)^2$ with $L$ being the dynamic range of the pixel intensity that was set to be $\|\mathbf{x}^*\|_\infty$. The PSNR is defined as

$$\text{PSNR}(\mathbf{x}^*, \hat{\mathbf{x}}) = 10 \log_{10} \frac{\|\mathbf{x}^*\|_\infty^2}{\frac{1}{\dim(\mathbf{x}^*)} \|\mathbf{x}^* - \hat{\mathbf{x}}\|_2^2} \quad (9)$$

## 3.2 Experimental Results

In the training phase, the feedback mechanism enables Sam's Net to be trained on both original input sinograms and self-augmented sinograms. To evaluate the effect of the feedback mechanism, we adopted the dual-domain network architecture from [13] plus the proposed weighting layer as baseline, denoted as "DD_LW". Fig. 2 displays the result of sinogram completion from Sam's Net and DD_LW. We can find that Sam's Net is more capable of sinogram completion. It is worth noting that both Sam's Net and DD_LW have the same architecture and trainable parameters. Theoretically, the performance of both networks should have the same upper limit. The difference indicates that the feedback mechanism can significantly help networks to converge to a better stage and achieve better performance.

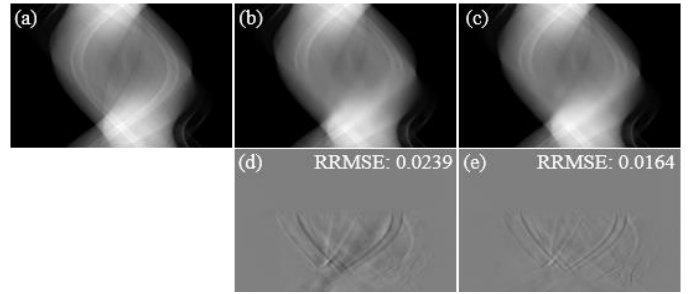

**Fig. 2** Comparison of the performance on sinogram completion between DD_LW and Sam's Net: (a) Ground Truth, (b)(d) DD_LW (residual), and (c)(e) Sam's Net (residual). The display window is [0,7] for sinograms and [-1.5,1.5] for residuals.

In the inference phase, Sam's Net operates in a multi-stage way, where input sinograms go through the whole network for several iterations and achieve different outputs. Fig. 3 shows the result of one slice from the test set. We can find that predictions get refined and more context information is recovered as the number of iterations increases. Obvious progress is achieved between the first two iterations, suggesting the gap between the training and inference data after the first iteration is smaller than that of initial inputs. After that, data consistency enhancement helps for detail promotion, and no significant improvement after 5 times. Thus, we chose Iter 4 for subsequent experiments.

To provide more comprehensive evaluations, we compared Sam's Net (Iteration=4) with ART regularized by ATV[6] (ART-ATV), FBPConvNet[15], DD_LW and dual-domain method[13]. The same loss function and parameter setting is utilized. Two test slices are displayed in Fig. 4 and the corresponding evaluation metrics are shown in Table 2. We may find that ART-ATV failed to handle such a severe ill-posed problem, while FBPConvNet can suppress artifacts to some extent. However, images generated by FBPConvNet are still with obvious artifacts and structural distortions. Incorporating optimization in the sinogram and image domains, the dual-domain method can further reduce artifacts. But detailed information is not well recovered. With enhanced data consistency and multi-stage interaction, more context information and tiny structures are disclosed by Sam's Net. From quantitative evaluations, it is also confirmed that Sam's Net outperforms the other methods in all metrics.

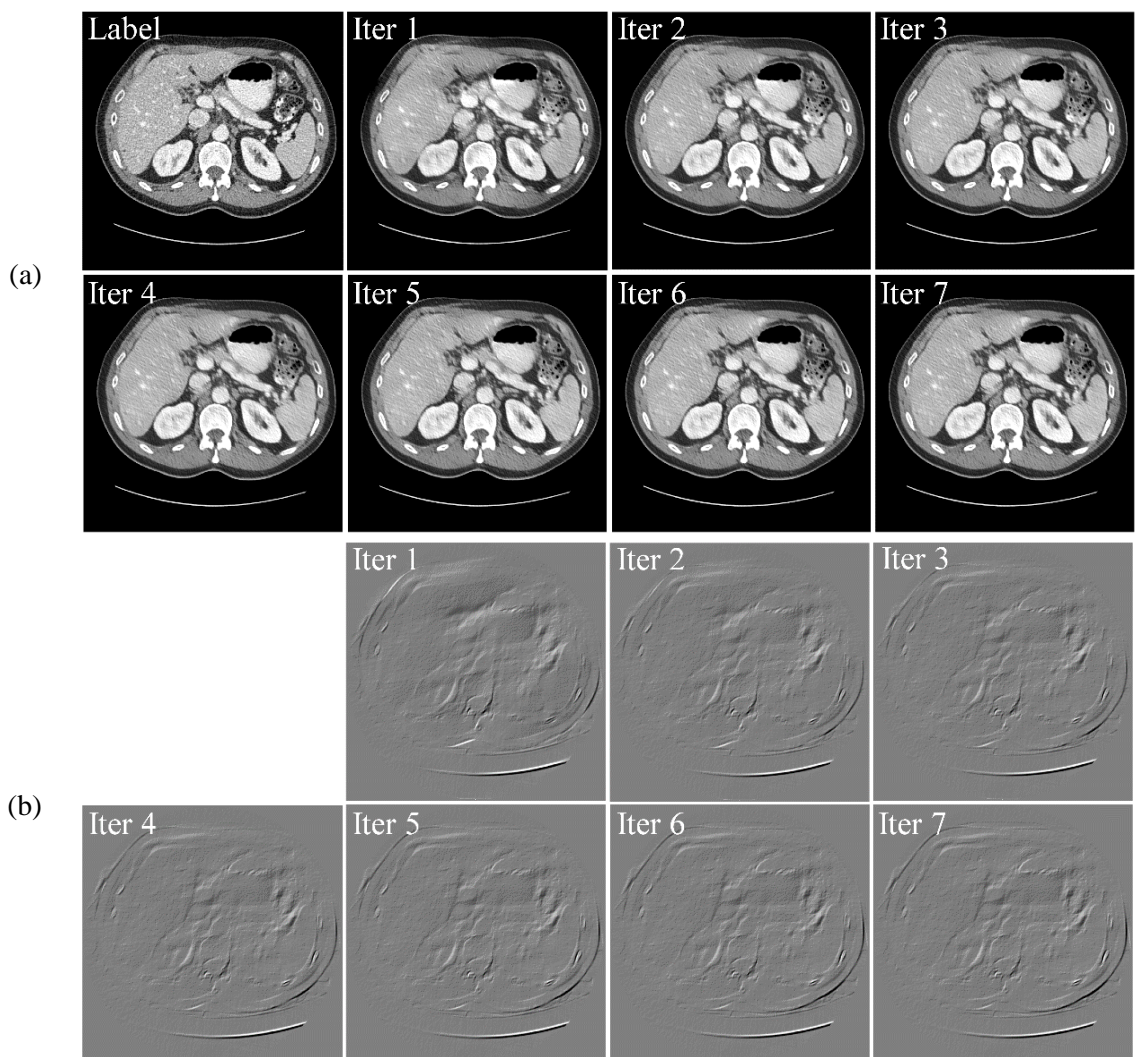

**Fig. 3** Images (a) and residual images (b) after different iterations from Sam's Net. The display window is [0.016, 0.024] for images and [-0.01,0.01] for residuals.

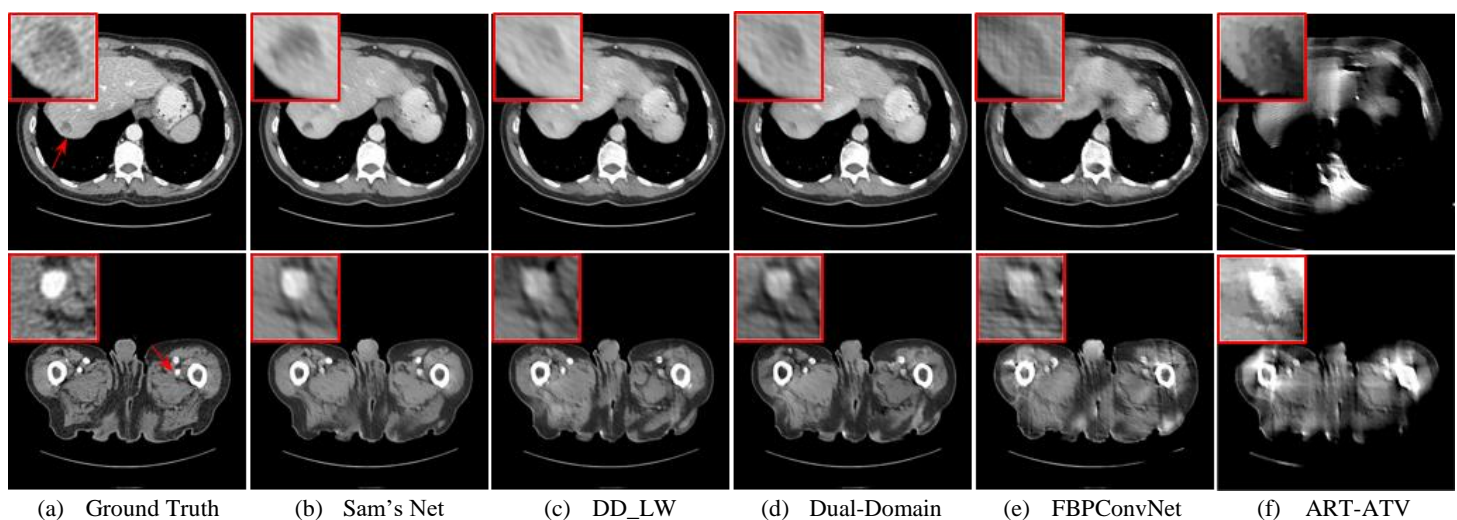

(a)    Ground Truth    (b)    Sam's Net    (c)    DD_LW    (d)    Dual-Domain    (e)    FBPConvNet    (f)    ART-ATV

**Fig. 4** Comparison among the five different methods. The display window is [0.016,0.024].

**Table 2**    Quantitative Evaluation

| Method | Slice 1 | | | Slice 2 | | |
|---|---|---|---|---|---|---|
| | SSIM | PSNR | RRMSE | SSIM | PSNR | RRMSE |
| ART-ATV | 0.7701 | 22.03 | 0.2631 | 0.7812 | 25.00 | 0.2680 |
| FBPConvNet | 0.8728 | 27.38 | 0.1422 | 0.8978 | 30.36 | 0.1468 |
| Dual-Domain | 0.9071 | 28.18 | 0.1297 | 0.9301 | 33.35 | 0.1041 |
| DD_LW | 0.9037 | 28.08 | 0.1311 | 0.9406 | 34.85 | 0.0876 |
| Sam's Net | **0.9122** | **29.73** | **0.1085** | **0.9496** | **36.42** | **0.0731** |

## 4 Discussion

Limited angle tomography is a challenging ill-posed problem. By bringing in data-driven prior knowledge, neural networks may improve reconstructions significantly. However, generalization caused by the bias between training data and test data is a big problem. Sam's Net presents two enhancements in this matter: 1) In the training phase, the self-augmented learning procedure feeds both original sinograms and self-augmented sinograms to SCNet. The two types of sinograms are with same acquired data in the data acquired region. In the data missing region, they provide different interpretations of the pixel-to-pixel relationship with the ground truth. SCNet learns to map these two versions of sinograms to the same sinogram label, which promotes it to utilize the data acquired region more and learn more about the true sinogram properties (see Fig. 2) instead of local mapping. 2) In the inference phase, multi-stage processing narrows the gap between the training data and test data, especially after the first iteration. It is because the whole network may encode original inputs from both datasets to more similar distributions and achieve continuous improvement after each iteration (see Fig. 3). Besides, the specially designed self-augmentation generates augmented sinograms from the model updated in the previous epoch. This is beneficial to promoting interaction and information propagation between the two domains. Additionally, the weighting layer adjusts the bias enforced by SCNet to improve the data consistency. Moreover, we introduced supervision on forward projection and CTIS loss, which give ASNet further guidance in data consistency and imaging processes. From the preliminary results from 90-degree fan-beam scanning, Sam's Net has shown great potential of artifact suppression and structure recovery with enhanced data consistency and robustness. Sam's Net presents a general framework for limited angle tomography. Yet more carefully designed network architectures may further improve the performance. Theoretical framework needs to be established to enable Sam's Net to work in a more general manner. And more experiments under different configurations and datasets are also needed.

## 5 Conclusion

We presented a general framework (Sam's Net) for limited angle tomography. Sam's Net adopts a self-augmented learning mechanism to improve generalization and robustness. Unlike single-domain or dual-domain methods, it works in a multi-stage manner with convenient propagation of information and errors between the sinogram domain and image domain. Preliminary evaluation has shown that Sam's Net outperforms other state-of-the-art methods without increasing the network scale. In the future, the theoretical framework will be established and more experiments will be conducted.

# Beam-hardening correction through a polychromatic projection model incorporated into an iterative reconstruction algorithm

Leonardo Di Schiavi Trotta[1], Dmitri Matenine[4], Philippe Letellier[2], Mathieu des Roches[2], Margherita Martini[2], Pascal Bourgault[2], Karl Stierstorfer[3], Pierre Francus[2], and Philippe Després[1]

[1]Department of Physics, Physical Engineering, and Optics, Université Laval, Québec, Canada
[2]Centre Eau Terre Environnement, Institut National de la Recherche Scientifique, Québec, Canada
[3]Siemens Healthcare, Computed Tomography, Forchheim, Germany
[4]Département de génie des systèmes, École de technologie supérieure, Montréal, Canada

**Abstract**

Typical reconstruction algorithms assume monochromatic attenuation, while the X-ray beam used in CT scanners are polychromatic, giving rise to beam hardening (BH) artifacts in the reconstructed image (e.g. dark streaks and cupping artifacts). In this work a novel, physics-rich beam-hardening (BH) correction algorithm was developed for X-ray computed tomography. This method uses the spectrum information, the detector response, the filter geometry and a calibration curve. The correction, which does not require prior material knowledge, is applied in an iterative reconstruction algorithm, and simulates the beam-hardening by estimating the X-ray spectrum at each voxel in the forward projection step. As a result, BH artifacts are inherently reduced in reconstructed images. Processing times of roughly 10 minutes per volume (or less depending on the number of projections), are achieved by using multiple GPUs. This method was also compared to the dual-energy beam-hardening correction method proposed by Alvarez and Macovski, which it outperforms when high-Z elements are involved.

## 1 Introduction

X-ray Computed Tomography (CT) is now ubiquitous in medicine for diagnosis purposes. This technology is also increasingly used for non-medical purposes in several fields, notably for high-resolution, non-destructive analysis [1]. The presence of high-density materials in scanned objects causes deterioration of CT image quality, where the polychromatic nature of the X-ray beam used in CT scanners is at the origin of image artifacts (e.g. streaks and cupping artifacts) [2]. Physical and non-physical models for beam hardening correction (BHC) were proposed to tackle this problem. This includes: the use of physical filters to pre-harden the beam, X-ray absorption considerations in the iterative reconstruction (IR) algorithm [3], effective energy shift of the X-ray spectrum in each voxel in the forward projector step of the IR algorithm [4] and dual-energy (DE) methods which inherently corrects for such artifacts [5, 6]. This latter method is known for producing images with amplified noise [5] in the CT energy range due to the nature of the photoelectric effect. Most methods require the knowledge of the material composition, which is not ideal for non-medical applications, since characterization is often the main objective. The heterogeneity of most samples requires physics-rich algorithms, capable of modelling the X-ray attenuation in the image formation process, without having to rely on prior material information.

In a clinical environment, filtered back-projection algorithms such as the one described by Feldkamp, Davis, and Kress [7] are often used. Advances in computing power have driven the development of iterative reconstruction algorithms (IR), which allow acquisitions with reduced dose, noise and number of projections [8]. This class of reconstruction algorithms are numerically intensive, typically requiring GPU computing to get results in reasonable time [9]. In principle, the inclusion of physics phenomena into the reconstruction algorithm would allow incorporating beam hardening correction in the reconstruction process by simulating the polychromatic behavior of the X-ray beam in the forward projector step [4]. A polychromatic reconstruction model that uses spectrum information, detector response, filter geometry of the CT scanner and a calibration curve to properly model the physics in the IR algorithm was developed, requiring no prior knowledge of the material composition. The numerical burden associated with such advanced modeling is offloaded by the use of multiple GPUs. With this approach, we aim at inherently reduce beam-hardening artifacts in the reconstruction process through a polychromatic forward projection model. We compared this approach with the dual-energy beam-hardening correction method of Alvarez and Macovski (DE-AM) [10].

## 2 Materials and Methods

### 2.1 Polychromatic forward projection

For a polychromatic beam traversing a heterogeneous material, the projection value $P_i$ in the sinogram is given by the following expression [11]:

$$P_i = -\ln\left\{\frac{\int s(E)d(E)\exp\{-\int_L \mu(E,r)dr\}\,dE}{\int s(E)d(E)dE}\right\}, \quad (1)$$

where $s(E)$ is the X-ray spectrum of the source, $d(E)$ is the energy-dependent detector response, $\mu(E,r)$ the linear attenuation coefficient at position $r$ along the path $L$ evaluated at the energy $E$ of the X-ray spectrum. In the forward projector step of most IR algorithms, a much simpler model is generally used, where the total attenuation is calculated as follows:

$$P_i = \sum_{j \in i} l_{ij} \mu_j, \qquad (2)$$

where $\mu_j$ is the linear attenuation coefficient in voxel $j$, traversed by the ray $i$, and $l_{ij}$ is the intersection length.

We posit that the total attenuation can be transformed from monochromatic to polychromatic by introducing the attenuation coefficient averaged over the local spectral response [12]:

$$\mu_{s_j} = \frac{\sum_{k=0}^{K} s'_{jk} \mu_{jk}}{\sum_{k=0}^{K} s'_{jk}} \qquad (3)$$

where $k$ is the energy index, $K$ the total number of energies, $\mu_{jk}$ the linear attenuation coefficient in pixel $j$ at energy $k$, and $s'_{jk}$ is the spectral response ($s'_{jk} = d_k \cdot s_{jk}$) in pixel $j$ at energy $k$. The local spectral response in $j$ is attenuated by $j-1$ voxels, and is calculated by the following expression:

$$s'_{jk} = s'_{0k} \exp\left\{ -\sum_{j' \in i}^{j-1} l_{ij'} \mu_{j'k} \right\}, \qquad (4)$$

where $s'_{0k}$ is the unattenuated spectral response at energy $k$. Thus, a polychromatic projection have the following form:

$$P_i = \sum_{j \in i} l_{ij} \mu_{s_j}. \qquad (5)$$

For simplification purposes, the linear attenuation coefficient at energy $E_k$ can be decomposed into the photoelectric effect and Compton scattering [10, 13]:

$$\mu_k = a_p E_k^{-3} + a_c f_{KN}(E_k), \qquad (6)$$

where $a_p$ and $a_c$ are constants related to each attenuation effect, $f_{KN}$ is the Klein-Nishina function and $E_k$ corresponds to the energies of the discretized spectrum. If we suppose that the uncorrected linear attenuation coefficient $\mu_j$ is evaluated at the effective energy of the X-ray spectral response

$$E_0 = \frac{\sum_{k=0}^{K} s'_{k0} E_k}{\sum_{k=0}^{K} s'_{k0}}, \qquad (7)$$

we can estimate the attenuation in voxel $j$ at any energy $k$ using the following relation:

$$\mu_{jk} \approx \frac{\left(\frac{a_p}{a_c}\right)_j E_k^{-3} + f_{KN}(E_k)}{\left(\frac{a_p}{a_c}\right)_j E_0^{-3} + f_{KN}(E_0)} \mu_j = f_{jk} \cdot \mu_j, \qquad (8)$$

where $f_{jk}$ is the conversion factor. If the spectral response is known, the only quantity yet to be determined is the ratio $(a_p/a_c)_j$, which gives the contribution of each physical effect in each voxel. This quantity can be estimated through a calibration curve of the form:

$$\left(\frac{a_p}{a_c}\right)_j = \sum_m b_m \mu_{E_0}^m \approx \sum_m b_m \mu_j^m \qquad (9)$$

where $\mu_{E_0}$ is the linear attenuation coefficient evaluated at $E_0$, which is roughly equal to the uncorrected $\mu_j$ in our approximation. The curve is calibrated against $\mu_{E_0}$, however, during the reconstruction, $(a_p/a_c)_j$ is determined by applying $\mu_j$ in Equation 9.

## 2.2 Implementation strategies

The strategy used to implement our polychromatic projection model is summarized in Figure 1. Within the forward projection step of an IR algorithm, and for each energy $E_k$ of the X-ray spectrum, given an arbitrary voxel $j$, where $j \in i$, the estimated and uncorrected attenuation $\mu_j$ is used first to calculate the Compton and photoelectric coefficients by using the calibration curve given by Equation 9. Once these terms are defined, one can obtain the conversion factor $f_{jk}$, which can be used to estimate $\mu_{jk}$ (see Equation 8). The total attenuation for the energy $E_k$, henceforth defined as $T$, is then accumulated over $j-1$ voxels and used to calculate the spectrum response for each energy bin at the voxel $j$. Once all energy bins are processed, one can calculate the local attenuation coefficient by applying Equation 3 and so the total attenuation, given by $t_i^{(n)} = \sum_{j \in i} l_{ij} \mu_{s_j}$.

## 2.3 Dual-energy decomposition

Following the DE-AM method [10], with two scans acquired with different tube voltages, one obtains two sets of logarithmic projections, $P_L$ and $P_H$. In order to calculate $A_p = \sum a_p l_{ij}$ and $A_c = \sum a_c l_{ij}$, the photoelectric absorption and Compton scattering line integrals, the Multivariate Newton-Raphson method was used to minimize the following system of Equations [5, 6]:

$$f_L = -\ln \sum_k s_{Lk} d_k \exp\left[ -A_p E_k^{-3} - A_c f_{KN}(E_k) \right] + \\ + \ln \sum_k s_{Lk} d_k - P_L, \quad (10)$$

$$f_H = -\ln \sum_k s_{Hk} d_k \exp\left[ -A_p E_k^{-3} + A_c f_{KN}(E_k) \right] + \\ + \ln \sum_k s_{Hk} d_k - P_H, \quad (11)$$

where $s_{Lk}$ and $s_{Hk}$ are the low- and high-energy X-ray spectrum. With the line integrals, $A_p$ and $A_c$, calculated from $P_L$ and $P_H$, the iterative algorithm OSC-TV, a combination of Ordered Subsets Convex (OSC) algorithm and the Total Variation minimization (TV) regularization technique [8], is used to reconstruct the photoelectric and Compton images, $a_p$ and $a_c$, respectively. This algorithm runs on multiple GPUs, which delivers a computation time of a few minutes (or seconds).

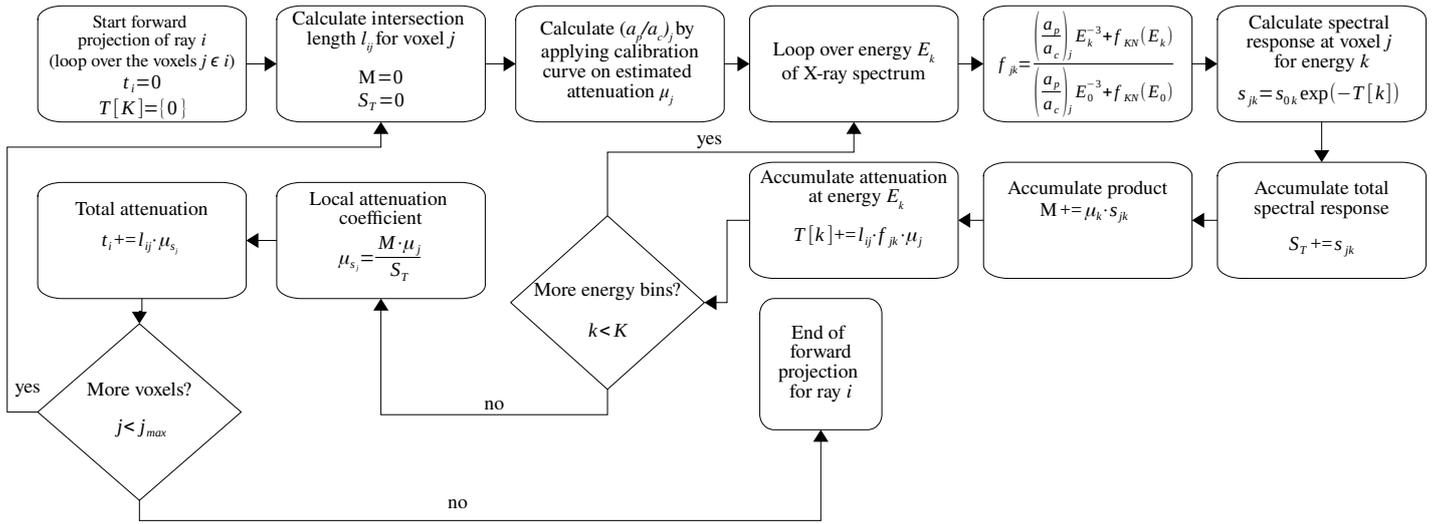

**Figure 1:** Flowchart depicting the strategy used to implement the polychromatic forward projection in the IR algorithm.

### 2.4   Detector dependent spectrum

The methods detailed in Sections 2.1 and 2.3 require the prior knowledge of the X-ray spectrum. Given that the spectrum is typically "pre-hardened" by the bowtie filter in medical CT scanners, one must take into account such effect. By modelling the bowtie filter, one can calculate the spectrum as a function of the CT fan angle $\theta$, and incorporate the detector dependent spectrum into the previous models through the following equation:

$$s_k(\theta) = s_k \exp\left[-x(\theta)\mu_{Al}(E_k)\right], \qquad (12)$$

where $x$ is the thickness of aluminum traversed by the beam and $\mu_{Al}$ is the linear attenuation coefficient for aluminum. The spectra and the detector response used in this work, both necessary for applying our BHC method and the DE-AM method, are the ones provided by the manufacturer.

### 2.5   Scanning process

The scans were performed using a Siemens SOMATOM Definition AS+ 128 CT scanner, installed at INRS Eau Terre Environnement in Quebec City [14]. In this work, we used vendor-provided binaries to remove the proprietary beam-hardening correction (BHC) preprocessing and to convert raw data into convenient image file format. All samples were scanned in sequential mode.

#### 2.5.1   Numerical simulations

In order to validate the proposed method, simulations of sequential acquisitions were conducted for 100 and 140 kVp with a virtual phantom, taking into account the geometry of the Siemens SOMATOM Definition AS+ 128 CT scanner as well as the same X-ray spectra and detector response provided by the manufacturer. The bowtie filter was not considered for this case and neither any noise model. The virtual phantom, illustrated in Figure 2, is composed of a water cylinder of 200 mm of diameter, with 4 cylinder rods of 30 mm of diameter. The material properties of the virtual phantom are reported in Table 1.

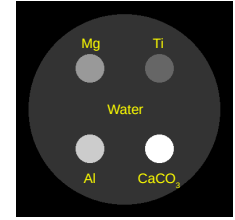

**Figure 2:** Virtual phantom.

| material | $Z_{eff}$ | $\rho_e$ ($\frac{electrons \cdot mol}{cm^3}$) | $\rho$ ($\frac{g}{cm^3}$) |
|---|---|---|---|
| Water ($H_2O$) | 7.42 | 0.555 | 1.000 |
| Titanium (Ti) | 22 | 2.071 | 4.506 |
| Magnesium (Mg) | 12 | 0.858 | 1.738 |
| Aluminum (Al) | 13 | 1.301 | 2.700 |
| Marble ($CaCO_3$) | 15.08 | 1.354 | 2.711 |

**Table 1:** Composition of the virtual phantom.

#### 2.5.2   Real-samples application

Three samples were used to illustrate the performance of both algorithms in a real scenario (see Figure 3): (i) a water phantom with a diameter of 200 mm; (ii) an aluminum cylinder with a diameter of 75 mm; (iii) a sodium iodide solution (NaI) with at 50 % concentration in a 13.5 mm diameter recipient. For the first sample, we show how the medical CT scanner can produce images presenting 'capping' artifacts [2] when beam hardening preprocessing is removed and how both BHC methods with detector dependent spectra can correct such effects. This artifact is due to the bowtie filter. At last, we show the cupping artifact generated in the aluminum sample and in the 50 % NaI solution, as well as their respective corrections using both algorithms.

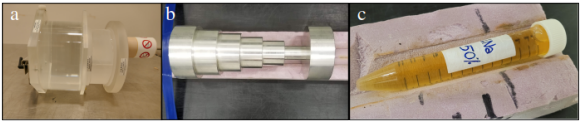

**Figure 3:** Samples used to test the proposed beam-hardening correction algorithm: (a) Water phantom for Definition AS, (b) aluminum sample and (c) 50 % NaI solution.

Each sample was scanned with a tube voltage of 100 and 140 kVp, so that the dual-energy method (DE-AM) can be applied.

We used the OSC-TV both with and without our polychromatic projection model, henceforth called OSC-TV-poly (Section 2.1), to reconstruct the images of the virtual phantom and the real samples acquired at an X-ray tube voltage of 140 kVp. For the dual-energy method, the 100 and 140 kVp projections were transformed into photoelectric and Compton components (see Equations 10 and 11), which are later reconstructed with the OSC-TV algorithm. Finally, these dual-energy images were combined (see Equation 6) to produce a virtual monoenergetic image (VMI) at the effective energy $E_0$ of the 140 kVp beam.

The OSC-TV-poly is computationally intensive. Hence, in order to evaluate the applicability of our method, reconstructions with both OSC-TV and OSC-TV-poly were performed on a computing node with four nVidia V100 Volta GPUs.

In order to evaluate potentially cupping and capping artifacts generated in the reconstructed images, a beam-hardening ratio is defined as [5]:

$$BHR = \frac{|P_{edge} - P_{center}|}{P_{edge}} \times 100\%, \quad (13)$$

in which $P_{edge}$ and $P_{center}$ are the pixel values at the edge and at the center of the reconstructed image, obtained from the line profiles illustrated in Figs. 6, 7 and 8.

## 3  Results and discussion

### 3.1  Calibration curve

Equation 6 was used to fit, with the least square method and data from the NIST XCOM database [15], the linear attenuation coefficient of a large list of materials (e.g. 3,000 materials and constrained by $Z_{eff} \leq 27$ and $\rho \leq 5.2 \ g/cm^3$), considering the energy range from 20 to 140 keV, thus obtaining each pair $(a_p, a_c)$. Only those fits where the coefficient of determination $R^2$ was higher than 0.999 were used. This constraint removes elements that present K-edges within the energy range of the CT scanner (outliers). Finally, the ratio $a_p/a_c$ is fitted against $\mu(E_0 = 83 \ keV)$, where $E_0$ is the effective energy of the 140 kVp beam, using a polynomial of order 6, with $b_0 = 0$, so $a_p/a_c \geq 0$ for low values of $\mu(E_0)$, obtaining $R^2 = 0.893$ (Figure 4).

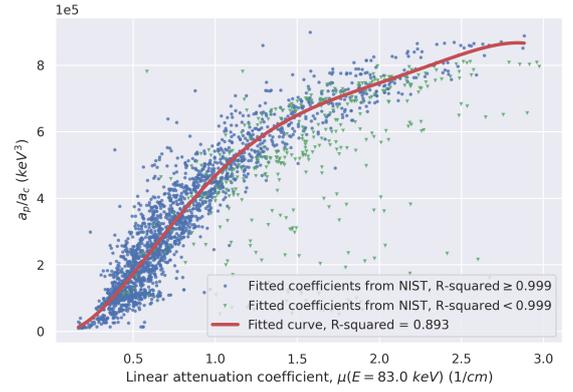

**Figure 4:** Calibration curve used to estimate $a_p/a_c$ from the uncorrected linear attenuation coefficient.

### 3.2  Simulation results

Figure 5 shows the comparison between the reconstruction of the virtual phantom with no BHC (a), the DE-AM method (b) and the proposed BHC method (c). As the X-ray spectra and the detector response are well known in the simulation framework, the dual-energy method efficiently removes beam-hardening artifacts, such as dark streaks and cupping. The OSC-TV-poly method also produces images with reduced BH artifacts, for both water and other materials. These results suggest that the OSC-TV-poly algorithm can handle a wide range of $Z_{eff}$ and $\rho_e$.

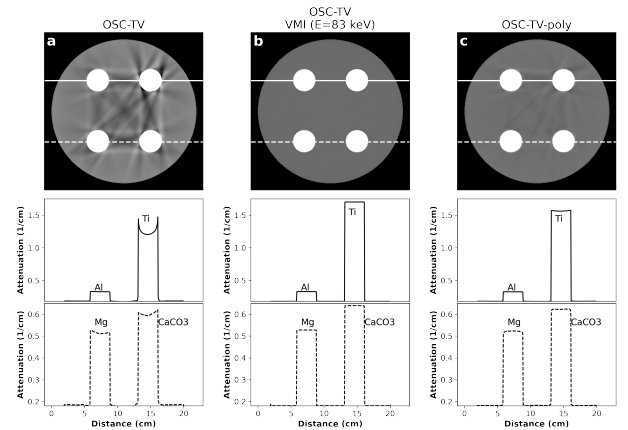

**Figure 5:** Virtual phantom: (a) OSC-TV (140 kVp), (b) virtual monoenergetic image (100/140 kVp) for $E = 83 \ keV$, (c) OSC-TV-poly (140 kVp). Window: [0.160, 0.200] $cm^{-1}$.

### 3.3  Real-samples application

Figures 6, 7 and 8 show line profiles of the water phantom, the aluminum sample and the 50 % NaI solution, without any BHC (a), the VMI for $E = 83 \ keV$, which is a linear combination of the photoelectric and the Compton images, both reconstructed with OSC-TV (b), and finally, the reconstruction with the BHC from the proposed model (c). Due to the presence of the bowtie filter, an important 'capping' artifact is produced in the water phantom. The modelling of

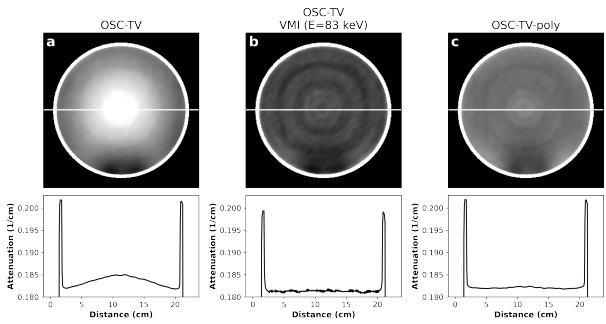

**Figure 6:** Water phantom: (a) OSC-TV (140 kVp), (b) virtual monoenergetic image (100/140 kVp) for $E = 83\ keV$, (c) OSC-TV-poly (140 kVp). Window: $[0.180, 0.185]\ cm^{-1}$.

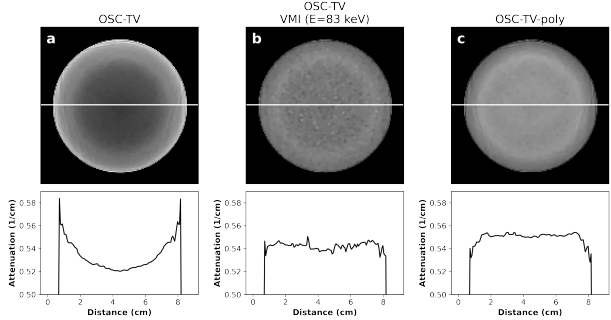

**Figure 7:** Aluminum sample: (a) OSC-TV (140 kVp), (b) virtual monoenergetic image (100/140 kVp) for $E = 83\ keV$, (c) OSC-TV-poly (140 kVp). Window: $[0.50, 0.59]\ cm^{-1}$.

this filter implemented in the DE-AM and the in the polychromatic algorithms were able to significantly reduce this effect. For higher density, higher Z samples, an important cupping artifact is observed in the uncorrected image of the aluminum sample and the 50 % NaI solution. Both BHC methods were able to reduce it.

Along the solid lines shown in Figure 6, one can see that the maximum and minimum value for the water phantom, when no BHC is applied (a), is 0.185 $cm^{-1}$ and 0.182 $cm^{-1}$, respectively. The correspondent *BHR* is only 1.6 %. Both BHC methods reduced the *BHR* to only 0.3 %, meaning the capping artifact was greatly reduced.

Concentric ring artifacts appears in the water phantom in Fig-ure 6 for all the cases studied, but mainly on the VMI image. These artifacts are the result of TV regularization, and so they can be reduced or removed by decreasing the regularization constant, at the expense of images reconstructed with more noise. The magnitude of these artifacts is increased by the level/window width used, which is the same for all figures. For the aluminum sample, the variations are more important, going from 0.52 $cm^{-1}$ to 0.58 $cm^{-1}$ in the image with no BHC, which represents a *BHR* of 10.3 %. The dual-energy and the OSC-TV-poly methods led to variations ranging from 0.53 $cm^{-1}$ to 0.55 $cm^{-1}$, equivalent to a *BHR* of 3.6 %.

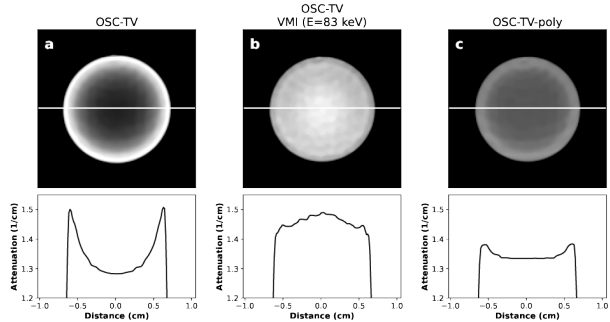

**Figure 8:** 50 % NaI solution: (a) OSC-TV (140 kVp), (b) virtual monoenergetic image (100/140 kVp) for $E = 83\ keV$, (c) OSC-TV-poly (140 kVp). Window: $[1.25, 1.50]\ cm^{-1}$.

The OSC-TV-poly approach was able to reduce the cupping artifact importantly for the NaI solution (*BHR* of 3.6 %), even if the sample has an effective atomic number higher than the ones used to define the calibration curve, and a K-edge at 33.2 keV leading to a poor fit through Equation 6 ($R^2 = 0.45$). This result is much better than the $BHR = 14.9$ % obtained without correction, and also lower than the dual-energy approach (*BHR* = 5.9 %). This last method led to an over-correction, as it can be seen in Figure 8 (b), where a capping artifact is produced. The presence of a K-edge is poorly compatible with the two basis functions of the Alvarez and Macovski attenuation model (see Equation 6), leading to this type of behavior [6, 16].

### 3.4    Pre-processing time

For the Siemens detector grid of 736 x 64, and a total of 2,304 projections, the developed Python script, allied with the Numba compiler [17], is able to perform the dual-energy decomposition (Section 2.3) of such configuration in half a minute. That is, once the low- and high-voltage sinograms are acquired, it takes roughly 30 seconds to generate the Compton and photoelectric pair. For the 50 % NaI solution protocol, where twice the number of projections are used, the decomposition takes around 80 seconds.

### 3.5    Processing time

The OSC-TV-poly is computationally intensive. Hence, in order to evaluate the applicability of our method, we calculated the reconstruction time (min) of the ordinary OSC-TV against the OSC-TV-poly, using the 140 kVp images for benchmarking. As reported in Table 2, OSC-TV-poly is able to reconstruct the same images in a few minutes. For the water phantom, the reconstruction takes roughly 20 times longer than the OSC-TV, where it takes 3.3 and 8.3 times longer for the aluminum sample and the virtual phantom, respectively. This difference is more important for the water phantom due to its larger size (200 mm), where the X-ray spectrum needs to be estimated for more voxels compared to the aluminum sample. The reconstruction time of the virtual phantom takes longer compared to the water phantom and the aluminum cylinder due to the high number of subsets used in the reconstruction, which increases the number of iterations within a full OSC step.

| Sample | Reconstruction time (min) | | Time increase factor |
|---|---|---|---|
| | OSC-TV | OSC-TV-poly | |
| Virtual phantom | 1.86 | 6.13 | 3.30 |
| Water phantom | 0.21 | 4.36 | 20.76 |
| Aluminum | 0.47 | 3.87 | 8.23 |
| 50 % NaI | 6.97 | 113.99 | 16.35 |

**Table 2:** Comparison of reconstruction time using 4x nVidia Tesla V100 16 GB.

For the 50 % NaI solution, processing time is longer due to its reconstruction matrix size (four times larger): almost two hours for the OSC-TV-poly and roughly 7 minutes for the OSC-TV. Such issue with the reconstruction time could be easily solved by making use of strategies to reconstruct regions-of-interests (ROI) through iterative reconstruction algorithms [18], hence, a smaller grid size could be used to avoid including the whole FOV into the image matrix and at the same time set smaller voxels sizes.

## 4  Conclusion

In this paper, we presented a novel, physics-rich algorithm to reduce beam hardening artifacts. The modelling of the physics allows the correction of beam hardening artifacts (capping, cupping and streaks) with no prior knowledge of the material information. The code can be implemented in the forward projection of an IR algorithm. Reasonable processing times were achieved by using multiple GPUs (less than 10 minutes). Both the dual-energy method and the IR algorithm with a polychromatic acquisition model are able to deliver satisfactory results, where our approach was able to outperform when it comes to high-Z elements.

## 5  Acknowledgements

This research was supported by the Sentinel North program of Université Laval, made possible by the Canada First Research Excellence Fund. This work was also supported by a grant from *Fonds de recherche du Québec - Nature et technologies*. We would like to thank Jérôme Landry and prof. Dr. Damien Pham Van Bang for all the support provided during the conception of this work. This research was made possible in part by support provided by WestGrid, Compute Canada and Calcul Québec.

# Image Synthesis for Data Augmentation in Medical CT using Deep Reinforcement Learning

Arjun Krishna[1], Kedar Bartake[1], Chuang Niu[2], Ge Wang[2], Youfang Lai[3], Xun Jia[3], and Klaus Mueller[1]

[1]Computer Science Department, Stony Brook University, Stony Brook, NY USA
[2]Department of Biomedical Engineering, Center for Biotechnology & Interdisciplinary Studies, Rensselaer Polytechnic Institute, Troy, NY USA
[3]Department of Radiology Oncology, UT Southwestern Medical Center, Dallas, TX USA

**Abstract** Deep learning has shown great promise for CT image reconstruction, in particular to enable low dose imaging and integrated diagnostics. These merits, however, stand at great odds with the low availability of diverse image data which are needed to train these neural networks. We propose to overcome this bottleneck via a deep reinforcement learning (DRL) approach that is integrated with a style-transfer (ST) methodology, where the DRL generates the anatomical shapes and the ST synthesizes the texture detail. We show that our method bears high promise for generating novel and anatomically accurate high resolution CT images at large and diverse quantities. Our approach is specifically designed to work with even small image datasets which is desirable given the often low amount of image data many researchers have available to them.

## 1 Introduction

One of the key challenges in unlocking the full potential of machine and deep learning in radiology is the low availability of training datasets with high resolution images. This scarcity in image data persists predominantly because of privacy and ownership concerns. Likewise, publicly available annotated high resolution image datasets are also often extremely small due to the high cost and small number of human experts who have the required amount of medical knowledge to undertake the labeling task. With insufficient data available for model training comes the inability of these networks to learn the fine nuances of the space of possible CT images, leading to the possible suppression of important diagnostic features and in the worst case making these deep learning systems vulnerable to adversarial attacks. We present an approach that can fill this void; it can synthesize a large number of novel and diverse images using training samples collected from only a small number of patients.

Our method is inspired by the recent successes of Deep Reinforcement Learning (DRL) [1, 2] in the game environments of Atari [3], Go and Chess [4] which all require the exploration of high-dimensional configuration spaces to form a competitive strategy from a given move. It turns out that this is not too different from generating plausible anatomical shapes in medical CT images. Our methodology combines the exploratory power of Deep Q Networks [5] to optimize the parameter search of geometrically defined anatomical organ shapes, guided by medical experts via quick accept and reject gestures. This need for feedback eventually vanishes, as the network learns to distinguish valid from invalid CT images.

During the generation, once the anatomical shapes for a novel CT image have been obtained from the DRL module, we use a style transfer module, designed for the texture learning of component organs and tissues [6], to generate the corresponding high resolution full-sized CT image. To the best of our knowledge, our proposed approach is the first attempt to incorporate DRL networks for the synthesis of new diverse full-sized CT images.

## 2 Methods

We adopt a two-step approach for synthesizing the full-resolution CT images. The first step consists of creating an anatomically accurate semantic mask (SM) for the image; this is the focus of this paper's discussion. The second step uses our existing style transfer network [6] to render anatomically accurate texture into the different portions of the generated SM.

As shown in Figure 1 (next page), step 1 consists of two phases. The first phase includes data pre-processing and training of a classifier following a traditional Convolutional Neural Network architecture [7] for classifying images. The data pre-processing stage produces the SMs of the high-resolution CT training images; it represents the annotated segmentations of the various anatomical features, such as organs and skeletal structures, as a set of 2D curves which are then geometrically parameterized as B-splines of order n for n+1 control points $\{(x_i, y_i)\}_{i=1}^n$. The control points of the anatomical features are stored as sequences of coordinates into vectors and then embedded into a lower dimensional space obtained via PCA. PCA is attractive since it preserves the spatial relationships of the SMs, has a linear inverse transform, and identifies a reduced orthogonal basis that approximates the shape of the SM statistical distribution well. Next, to train the classifier sufficiently, we generate a large number (on the order of 10,000) new semantic masks by interpolating in this PCA space and group these images into clusters via k-means. The clusters are then manually labeled by experts as good and bad image sets and the classifier is then trained on these clusters. The classifier thus represents an approximation of control points that could serve as valid semantic masks.

Phase 2 uses this trained classifier as the reward predictor in our Reinforcement Learning Environment (RLE). DRL networks learn by optimizing on results via a reward mech-anism that derives from the rules of the environment. This environment serves to stimulate the learning of an effective strategy for exploring the anatomical shape space to facilitate a diversified yet accurate image generation. Our specific environment for DRL involves a user-feedback interface that consists of a front-end where linear interpolations between the semantic masks of two distinct valid SMs are corrected by the agents of the RLE followed by the expert user marking them as good or not. This feedback is then used to further train the classifier/reward predictor such that it can give better predictions of the actual rewards to the agents as they try to correct future interpolations. Hence the agents in RLE and the reward predictor are trained asynchronously. As the reward predictor gets better, so do the actions of the agents and consequently we gain more semantic masks representing valid plausible anatomy.

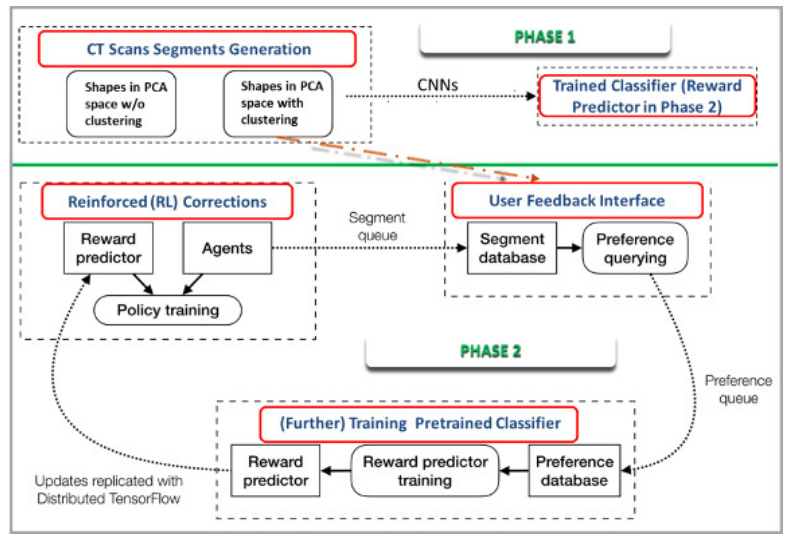

**Figure 1:** Two-Phase box diagram for training RL agents. The pre-trained classifier in Phase 1 is used as reward predictor in Phase 2. Segment refers to the resulting SM from agents' actions. Preference refers to the user preference of one segment (SM) over other.

Our contributions are as follows:

- We discuss a robust way of learning anatomical shapes via their geometrical representations of B-splines and their interpolations / samplings in PCA space.

- We define an environment where the true image space of the anatomical shapes could be discovered without the supporting dataset via Reinforcement Learning.

- We build a visual user-interface where users can control and guide the generation process. Once sufficiently trained, users have the option to add the generated images to the training dataset.

## 2.1 General Interpolation Framework: B-Splines and PCA Interpolation

Curvature is a central morphological feature of organs, tissues, cells, and sub-cellular structures [8]. Hence we represent the curve shapes by the set of control points with strongest curvatures between some predefined distances across the whole curves depicting organs, skeletal structures, etc., we shall refer to it as *anatomical shapes*. These control points also integrate easily with B-spline curves to decode them back into full curves. B-spline curves provide flexibility to represent these anatomical curves [9] since the degree of a B-spline curve is separated from the number of control points. Hence lower degree B-spline curves can still maintain a large number of control points and the position of a control point would not change the shape of the whole curve (local modification property). Since B-splines are locally adjustable and can model complex shapes with a small number of defined points, they are an excellent choice to model anatomical shapes with control points selected based on strong curvatures.

Since each semantic mask (SM) is expressed as a set of control points, we embed the training data SMs in a lower dimensional space via Principal Component Analysis (PCA). The PCA model is used to reconstruct the anatomical shapes of the training dataset giving us a repository of coefficients for eigen-vectors that make plausible anatomy for lung CT SMs. We can then reconstruct new anatomy curves by sampling these coefficients. Each type of anatomical shape, such as left lung, right lung, torso, spinal cord, esophagus, and heart, forms a dedicated subspace of SM vectors and is represented as a multivariate Gaussian with mean (for each coefficient of the corresponding eigen-vector) and co-variance matrix. The set of anatomical shapes for a specific SM are interlinked so they can be jointly used in the interpolation procedure. In our initial implementation we represented all anatomical shapes of the training SMs as a single vector to form a single multivariate Gaussian. In practice, however. this approach does not work well and fails to generate SMs with correlated anatomical shapes.

One way to generate a novel SM is to take any two available SMs and linearly interpolate between the two. One problem

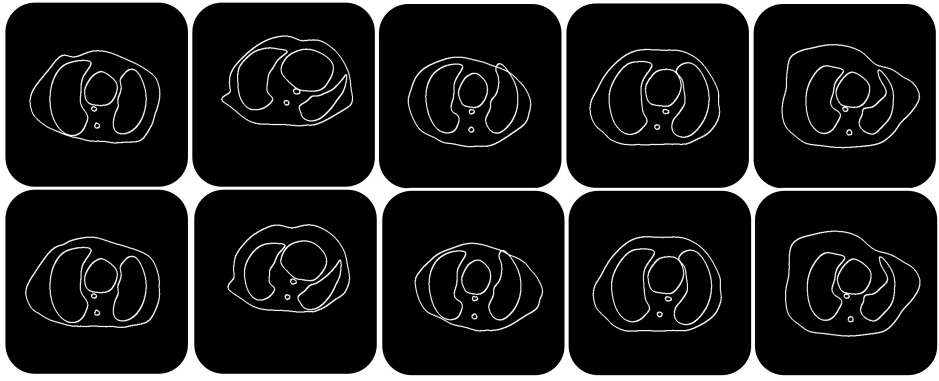

**Figure 2:** The first row shows linearly interpolated SMs for a lung CT image. The second row shows their improved counterparts from RL agents. In the first three columns, the agents tries to make them more symmetric and remove intersections. For anatomically accurate interpolated SMs, agents don't make much change as seen in the fourth column. The fifth column represents the anatomical space in our PCA for which agents have not yet been trained on and would improve with incoming user feedback

with this approach is that with small training datasets there is not enough variety to construct an accurate PCA decomposition. leading to noise and subsequently to erroneous features in the generated SM. Also, accurate anatomical shapes do not occupy a perfectly linear space even in heavily reduced dimensions and the interpolation on the eigen-vectors still limits the number of novel anatomical shapes that can be generated since the set of images between which the interpolation is being done is small. To overcome these limitations, we introduce the powerful mechanism of DRLs within our environment which we describe in the next section.

## 2.2 User Assisted Deep Reinforcement Learning

We propose to solve the aforementioned problem with PCA space exploration using Deep Reinforcement Learning, obtaining user feedback via a dedicated user interface. We ask a user to interpolate between two generated anatomies by moving a slider. We then present small perturbations made by the agents in the Deep Q Learning environment to the linear interpolation and present these to the user as alternative results. The user picks which ones are better and which ones are worse and submits his or her feedback via the interface. The submitted preferences train a CNN (Convolutional Neural Network) based image classifier that is simultaneously used as a reward predictor for training the agents in the Deep-Q Learning algorithm. Our approach of using a reward predictor to predict rewards based on user feedback mainly borrows from the work of Christiano et al.[10] who utilize user feedback on video clips of game play to train a reward predictor.

As shown in Figure 1, we pre-train the reward predictor during the data processing stage. By modifying the parameters in the clustering (via k-means), we can visibly alter the quality and anatomical accuracy of the generated SMs when interpolating in PCA space. These groups of SMs can be used to pre-train the reward predictor that is used in our DRL environment where it is further fine-tuned with the help of user feedback. The trained reward predictor on submitted user preferences then help the agents in learning the perturbations that need to be applied to the coefficients of eigen-vectors representing a SM while interpolating in between any two random SMs. Note that because of this setup once agents are trained, they can also be used to "fix" any generated SM interpolated on the PCA space. With the help of user verification, we add perfectly generated SMs in the training dataset that are then used to interpolate more novel SMs hence expanding the known PCA space representing valid anatomy. This helps our SM generating interface get better with the usage by the users.

## 2.3 Loss Function, Input/Output and Network Architecture of Deep-Q Agents

We follow the Deep-Q DRL algorithm used by the authors of Atari [3]. We maintain a policy $\pi$ that takes the observation state O as input and gives an action A to be performed; $\pi : O \rightarrow A$. The reward predictor takes the resulting image as input and gives a reward estimate R; $\hat{r} : O \times A \rightarrow R$. For training our policy $\pi$ we use the traditional Deep-Q loss:

$$y_i = E_{s' \sim \varepsilon}[\hat{r} + \gamma max_{a'} Q(s', a'; \theta_{i-1}))^2] \quad (1)$$

$$L_i(\theta_i) = E_{s,a \sim \rho(\cdot)}[(y_i - Q(s, a; \theta_i))^2] \quad (2)$$

where $y_i$ represents the discounted reward estimate from iteration i and $\rho(s,a)$ represents the distribution of all states and actions applicable on those states. Since our states are sequences of coefficients for representing the control points of every organ (thereby representing the set of anatomical shapes constituting SMs), we use a neural network using six fully connected layers to estimate the second term; $Q(s, a; \theta_i)$ in equation (2). The parameters from the previous iteration $\theta_{i-1}$ are held fixed when optimising the loss function $L_i(\theta_i)$ and are estimated via stochastic gradient descent.

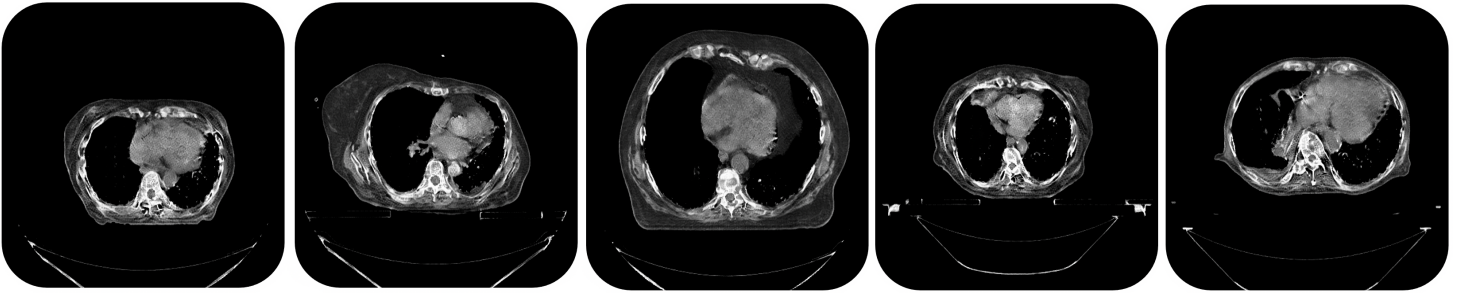

**Figure 3:** Some stylized CT images, generated by linear SM pair interpolation, and corrected with the RL framework.

## 2.4 Loss Function, Input/Output and Network Architecture of Reward Predictor

Once the agents modify the contributions of the eigen-components, the resulting anatomical shapes are assembled into a SM and sent to a six layer CNN with batch normalization layers and relu activations [7]. The CNN classifies the SM image in one of five or six categories indicative of their anatomical accuracy according to which a reward is assigned to the action of agent. The policy $\pi$ interacts with the environment to produce a set of trajectories $\{\tau^1...\tau^i\}$. A pair of such trajectory results (SMs) are selected and are sent to our front-end for user feedback. To fine-tune the reward predictor further we use the cross entropy loss between the predictions of the reward predictor and user feedback $v$ [10].

$$loss(\hat{r}) = \sum_{\tau^1,\tau^2,v} v(1)log\hat{P}[\tau^1 \succ \tau^2] + v(2)log\hat{P}[\tau^2 \succ \tau^1]$$

(3)

where under the assumption that user's probability of preferring a SM over other should depend exponentially on the true total reward over the SM's trajectory; $\hat{P}[\tau^1 \succ \tau^2]$ could be expressed as:

$$\hat{P}[\tau^1 \succ \tau^2] = \frac{exp\sum\hat{r}(s_t^1,a_t^1)}{exp\sum\hat{r}(s_t^1,a_t^1) + exp\sum\hat{r}(s_t^2,a_t^2)}$$

(4)

As evident from figure 1, the above two networks are trained asynchronously. With increasing data from the user's feedback, the reward predictor gets better which helps better train the RL agents.

## 3 Results, Future Work and Conclusion

Figure 2 shows corrected SMs via RL agents from badly formed counterparts which were interpolated linearly between two generated SM images. In most cases, our RL agents are able to correct the obvious errors like the intersections between the organ curves or the sharp unnatural bends in the boundaries of torsos, but as evident from the example in the last column of the figure, for some badly formed SMs the agents are unable to make better SMs. That's because we need more user feedback for training the reward predictor enough to make agents respond to a wide range of generated SMs. With more feedback that the reward predictor would receive, the agents could be trained better for responding to the generated SMs. Figure 3 shows stylized CT images on corrected SMs.

For future work, we intend to modify the user-interface to enable faster user interaction hence enabling larger feedback collection quickly for more efficient training of the reward predictor and the RL agents. We also plan to make the texture learning more robust on varied SMs and not just lung CT SMs. We also intend to extend our framework for learning and generating pathology which should integrate well with our two step approach. At the current time, we generate volumes slice by slice. For better continuity across slices, we plan to learn anatomical curves directly in 3D volumes, using B-spline patches.

# Spherical Ellipse Scan Trajectory for Tomosynthesis-Assisted Interventional Bronchoscopy

Fatima Saad[1,2], Robert Frysch[1,2], Tim Pfeiffer[1,2], Jens-Christoph Georgi[3], Torsten Knetsch[3], Roberto F. Casal[4], Andreas Nürnberger[5], Guenter Lauritsch[3], and Georg Rose[1,2]

[1]Institute for Medical Engineering, Otto-von-Guericke University, Magdeburg, Germany
[2]Research Campus STIMULATE, Otto-von-Guericke University, Magdeburg, Germany
[3]Siemens Healthcare GmbH, Forchheim, Germany
[4]Department of Pulmonary Medicine, The University of Texas MD Anderson Cancer Center, Houston, TX, USA
[5]Data and Knowledge Engineering Group, Faculty of Computer Science, Otto-von-Guericke University, Magdeburg, Germany

**Abstract** Accurate nodule location identification is a cornerstone in the diagnostic yield of transbronchial needle biopsy procedures. Due to the overlapping structures, depiction of lung nodules is challenging with chest radiography (CR). While cone-beam computed tomography (CBCT) might provide exact 3D information, its use in real practice is limited by the radiation dose involved and by the space restrictions in the operating room. Since digital chest tomosynthesis (DCT) provides a compromise between image quality and space requirements, it appears as a potential candidate for guiding bronchoscopy procedures. To mitigate the limited depth resolution in conventional linear DCT, the use of a spherical ellipse scan trajectory for interventional DCT is proposed in this work. The proposed trajectory is evaluated using numerical simulation of real chest CBCT volumes and is compared to CR, linear DCT and circular DCT. Analytic (FBP) and algebraic (ART) reconstructions are evaluated qualitatively and quantitatively. Compared to CR and linear DCT, the proposed trajectory demonstrates an improved visualization of the tool-in-lesion and the different chest structures. With respect to circular DCT, the proposed orbit provides a good compromise between image quality and workspace requirements.

## 1 Introduction

Lung cancer is one of the most common and fatal cancers worldwide [1]. Since a higher likelihood of successful treatment is linked to an early-stage diagnosis [2], improved screening methods and subsequently an increased rate of minimally invasive nodule biopsies are expected. One of the common methods for lung biopsies is transbronchial needle biopsy (TBNbx) [3]. Using this method, the bronchoscopist relies on pre-procedural CT images and on chest radiography (CR) to navigate the needle to the target nodule. Several limitations of this method have been reported. First, the location of the nodule in the prior CT may be different than its actual location, known as CT-to-body divergence [4]. Second, nodules are obscured in the radiographic image by the overlapping anatomical structures, only the needle and the ribs can be resolved. Consequently, poor positioning and inaccurate diagnosis are highly probable. While intra-operative cone-beam CT (CBCT) may seem as an ideal candidate to overcome the aforementioned issues, its use as a real-time image guidance tool is limited due to many barriers. Many hundreds of projections are required to reconstruct a non-ambiguous image with the standard reconstruction algorithms. Therefore, the patient as well as the bronchoscopist will be exposed to a high dose of excessive ionizing radiation. In addition, to perform a scan during the intervention, the C-arm has to be fully rotated around the patient (~200°), thus, some considerable logistic efforts are required. In the operating room, the available space is limited: a robotic arm holding the bronchoscope and many entangled cables block the C-arm trajectory and need to be rearranged before performing the scan. Moreover, these operations are time-consuming and time is a critical factor during interventions.

Recently, Aboudara et al. [5] proposed digital chest tomosynthesis (DCT) technology as a potential alternative for navigational bronchoscopy guidance. During this procedure, the C-arm performs a fluoroscopic sweep over a limited angular range and a limited set of projection images are acquired. The reconstruction of these images provides quasi 3D information and captures the location of both the needle tip and the nodule. However, these boundaries cannot be resolved in 3D using the conventional linear tomosynthesis due to the limited depth resolution. Multi-directional scan orbits might improve the depth resolution, nevertheless, some devices, in particular robotic bronchoscope holders, are placed in a way that the angulation in the cranial/caudal direction are limited. To avoid collision, either a small circular trajectory or a spherical ellipse trajectory obtained by stretching the small circle into a certain direction where space is not limited, could be applied. Therefore, this paper investigates which of the two options is more appropriate.

## 2 Materials and Methods

### 2.1 Proposed source-detector motions

In this study, three classes of source-detector trajectories are investigated using numerical simulations. A flat-panel detector composed of $616 \times 480$ pixels with a 0.616 mm pixel pitch is used. An X-ray source and the flat-panel detector are mounted on a robotic C-arm. For all the tested trajectories, the source to iso-center distance $d$ is constant by design and is set to 785 mm. The source to detector distance is fixed to 1200 mm. The detector mounted in opposite to the X-ray source performs an in-plane rotation in a way its rows are kept tangential to the scan trajectory. $N$ projection views were acquired on each of the proposed trajectories.
The three candidate scan orbits are shown in Figure 1:

- a linear scan trajectory "⌢" where the source and the detector rotate around the patient in one axial plane with a limited angular range of $\pm\alpha$. The source positions are evenly spaced on the linear trajectory. The source sampling locations are defined by:

$$\vec{T}_i^l = (d\sin\theta_i^l, d\cos\theta_i^l, 0) \qquad (1)$$

where

$$\theta_i^l = -\alpha + i\Delta\theta^l \qquad i = 0, ..., N-1 \qquad (2)$$

and $\Delta\theta^l = \frac{2\alpha}{N-1}$,

- a circular scan trajectory where the source and the detector rotate each in one coronal plane along a circular path. For this orbit, two cases with different limited angular ranges are considered: a small circle "○" with an angular range of $\pm\beta$ and a large circle "◯" with an angular range of $\pm\alpha$ ($\alpha > \beta$). The source sampling locations are defined by:

$$\vec{T}_i^c = (d\sin\gamma\cos\theta_i^c, d\cos\gamma, d\sin\gamma\sin\theta_i^c) \qquad (3)$$

where $\gamma = \alpha$ for the "◯" and $\gamma = \beta$ for the "○", $\theta_i^c = i\Delta\theta^c$, $i = 0, ... N-1$ and $\Delta\theta^c = \frac{360°}{N}$,

- a spherical ellipse scan trajectory "⬯" where the source and the detector rotate each along a spherical ellipse path defined by two angles $\pm\alpha$ and $\pm\beta$. To construct this orbit, the source positions are evenly spaced per arc length on a 2D ellipse located in a coronal plane. The ellipse has $a = d\tan\alpha$ as a large radius and $b = d\tan\beta$ as a small radius. Since a fixed iso-center is preferred, the equally spaced source points are then projected on the surface of a sphere with radius $d$. The source sampling locations on the 2D ellipse are given by:

$$\vec{T}_i^s = (r(\theta_i^s)\cos\theta_i^s, d, r(\theta_i^s)\sin\theta_i^s) \qquad (4)$$

where

$$r(\theta_i^s) = \frac{b}{\sqrt{1 - (e\cos\theta_i^s)^2}} \qquad (5)$$

and $e = \sqrt{1 - \frac{b^2}{a^2}}$ is the ellipse eccentricity. To find $\theta_i^s$ at position $i$, the circumference $C$ of the ellipse is computed by:

$$C = 4aE(e) \qquad (6)$$

where $E$ is the complete elliptic integral of the second kind. The arc length between each two consecutive source locations on the plane ellipse is given by:

$$\Delta l = \frac{C}{N} \qquad (7)$$

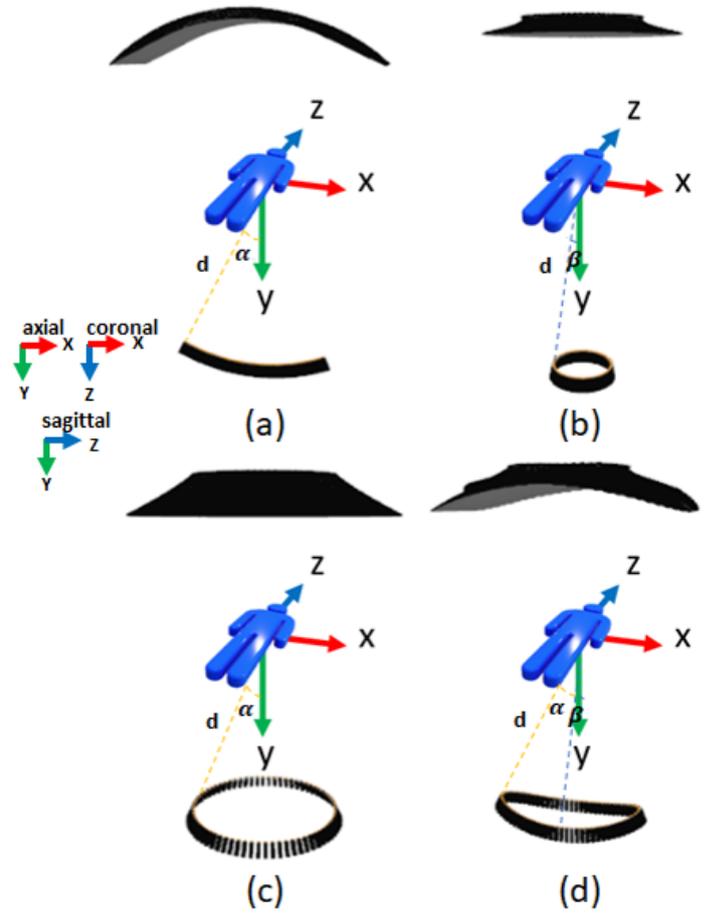

**Figure 1:** (a) Linear "⌢", (b) small circular "○", (c) large circular "◯" and (d) spherical ellipse "⬯" DCT scan orbits. The source trajectory is shown below the patient and the detector trajectory is shown above the patient.

The arc length from the source at position 0 (situated on the large radius) to the source at position $i$ is given by:

$$L(\theta_i^s) = i\Delta l = aE(e) - \varepsilon\left(\frac{r(\theta_i^s)\cos\theta_i^s}{a}, e\right) \qquad (8)$$

where $\varepsilon$ is the incomplete elliptic integral of the second kind. Using (8), one can write:

$$\varepsilon\left(\frac{r(\theta_i^s)\cos\theta_i^s}{a}, e\right) = aE(e) - i\Delta l \qquad (9)$$

$\theta_i^s$ can be found by computing the inverse of $\varepsilon$. The iterative procedure proposed by Boyd in [6] was used to compute this inverse.

Different number of projection views are investigated in this work, ranging from 36 to 180 views. The small tomographic angle $\beta$ was fixed to 15° and the large tomographic angle $\alpha$ was fixed to 23°.

## 2.2 Numerical simulation studies

All the tomosynthesis scan trajectories discussed above have been implemented in the Computed Tomography Library

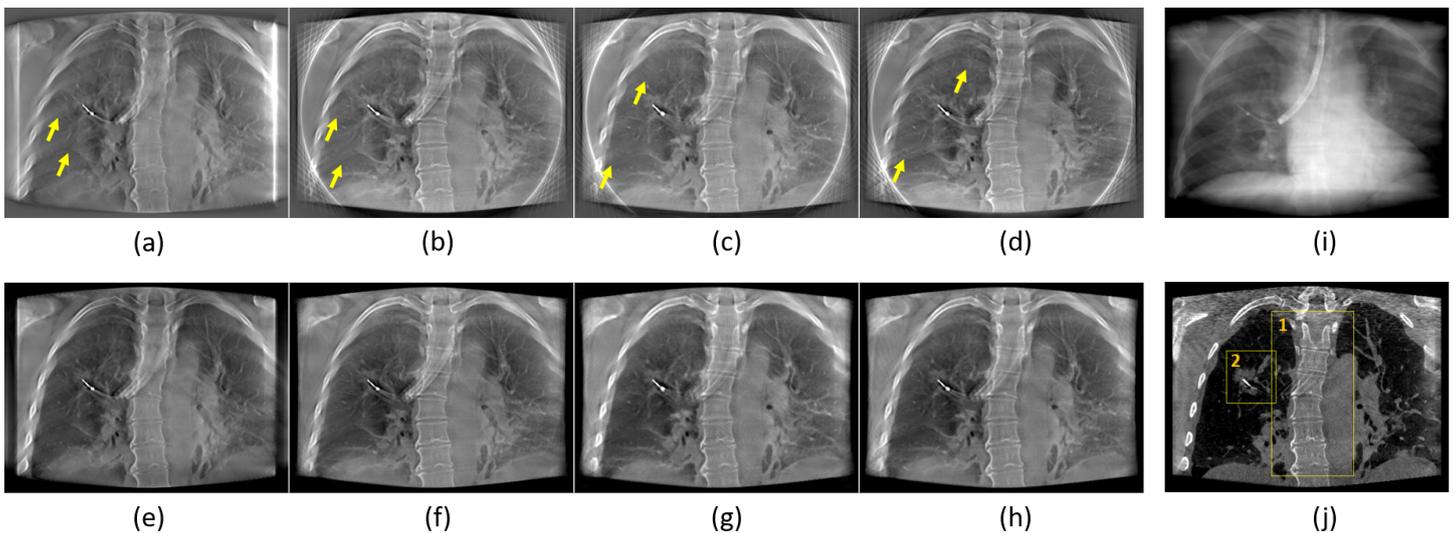

**Figure 2:** DCT coronal slice reconstructions with the different scan trajectories. (a), (b), (c) and (d) show the FBP reconstructions respectively with "⌒", "○", "◯" and "⬭" (displayed gray-scale window of [0.03,0.53] after a min-max normalization). (e), (f), (g) and (h) show the ART reconstructions respectively with "⌒", "○", "◯" and "⬭" (displayed gray-scale window of [0.00,0.05]). For comparison, (i) and (j) show respectively the radiographic image and the corresponding reference CBCT coronal slice.

CTL[1] [7]. In order to take the complex chest anatomy into account, real chest CBCT images acquired using a C-arm (Axiom Artis dTA, Siemens Healthcare GmbH, Erlangen, Germany) during an interventional bronchoscopy procedure (held at MD Anderson Cancer Center, Houston) were used in this study. The images show a bronchoscope and a transbronchial needle inserted within a target lesion into a patient lung. The DCT projection images were simulated by forward projecting the CBCT volumes. Poisson noise was added to the projection data. The photon flux was set to $4.75 \times 10^8$ photons per $cm^2$. Analytic and algebraic reconstructions were adopted in this study. A filtered backprojection (FBP) reconstruction using a Hann apodized ramp filter (cutoff at the Nyquist frequency) was used. The ramp filter was applied in the direction tangential to the detector motion trajectory. Moreover, an iterative algebraic reconstruction (ART) with an ordered-subsets scheme [8] was tested as well. Only a positivity constraint was incorporated into the ART reconstructions with no further regularization. ART iterations were stopped once the normal equation was numerically satisfied [7]. The size of the reconstructed images was set to $512 \times 512 \times 382$ voxels with a voxel size of $0.5003^3 mm^3$. To evaluate the different orbits, visual inspection as well as quantitative assessment have been conducted. Pearson correlation was computed in different regions of interest (ROI), focusing on the ability of the different orbits to detect the biopsy needle and the target lung nodule (ROI#2 in Figure 2(j) composed of $90 \times 9 \times 96$ voxels) and to resolve the different spine details (ROI#1 in Figure 2(j) composed of $149 \times 31 \times 300$ voxels).

## 3 Results

Figure 2 shows one coronal slice of the reconstructed DCT images using the "⌒" (a,e), the "○" (b,f), the "◯" (c,g), and the proposed "⬭" (d,h) orbits. FBP (top row) and ART (bottom row) reconstructions are shown for each case. 72 projection views were used in these reconstructions. For comparison, the corresponding CBCT coronal slice (j) is shown as a reference image and the conventional radiograph (i) is shown as well. Overall, DCT shows an improved visibility of the chest structures (normal pulmonary vasculature, spine) compared to plain radiography. In the radiographic image (i), the target nodule, the intervertebral disks and the pulmonary vasculature are completely obscured by the overlapping ribs and the bronchoscope. Compared to FBP, ART shows a better removal of out-of-plane artifacts (yellow arrows). Compared to the multi-directional orbits, the out-of-plane ribs are more visible with the uni-directional orbit ("⌒") due to the poor spatial resolution. Figure 3 shows an enlarged region of the spine (ROI#1 in Figure 2(j)) for the various scan orbits. Horizontal structures are better recovered with the multi-directional orbits. The horizontal edges of the intervertebral discs appear sharper in "○", "◯" and "⬭", but are hidden by some shading artifacts in "⌒" (black arrows). Despite the larger angular range with the "◯", strong ripple artifacts appear along the spine (red arrows). This is mainly due to a lower projection density on the "◯" compared to the other trajectories. Strong out-of-focus artifacts (white arrows) produced by the high-attenuation object (bronchoscope) appear in the "⌒" and the "○" as multiple ghosting copies of the bronchoscope (especially at the edge).

Figure 4 illustrates the enlarged regions around the needle and the target nodule (ROI#2 in Figure 2(j)). The nodule is better distinguished from the lung background with "○", "◯" and "◯" orbits compared to the "⌣" orbit. However, out-of-focus artifacts coming from the bronchoscope are slightly stronger with the "○" trajectory. For a quantitative assessment of the results, Figure 5 and Figure 6 show the plots of the Pearson correlation (PC) between the reconstructed DCT images and the reference CBCT image as a function of the number of projection views in ROI#1 and ROI#2 respectively. ART reconstructions show higher PC than the FBP ones for all the four trajectories. Just for clarity of illustration and due to space limitations, only the ART reconstruction results are shown in the plots. In accordance with the visual inspection and for the different number of views tested, the "◯" reconstruction has the highest PC in both the regions of interest. PC is higher in the "◯" than in the "⌣" and the "○". Interestingly, PC in the "◯" is just slightly smaller than the PC in the "◯" for ROI#1. In ROI#1, PC is higher in the "○" than in the "⌣" but the opposite occurs in ROI#2. PC could be object-dependent. There are more horizontal edges in ROI#1 which cannot be resolved with the "⌣".

---

[1]code available at: **https://gitlab.com/tpfeiffe/ctl**

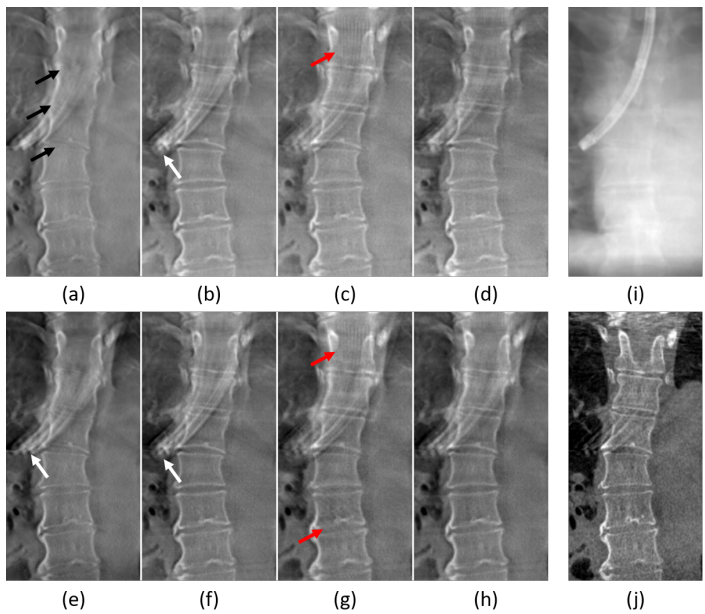

**Figure 3:** Enlarged views around the spine (ROI#1) corresponding to the images shown in Figure 2.

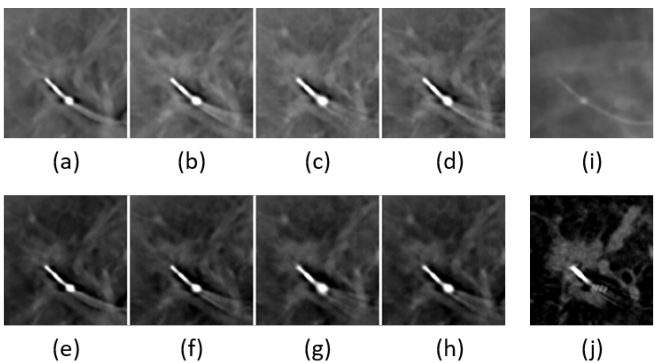

**Figure 4:** Enlarged views around the needle tip and the target nodule (ROI#2) corresponding to the images shown in Figure 2.

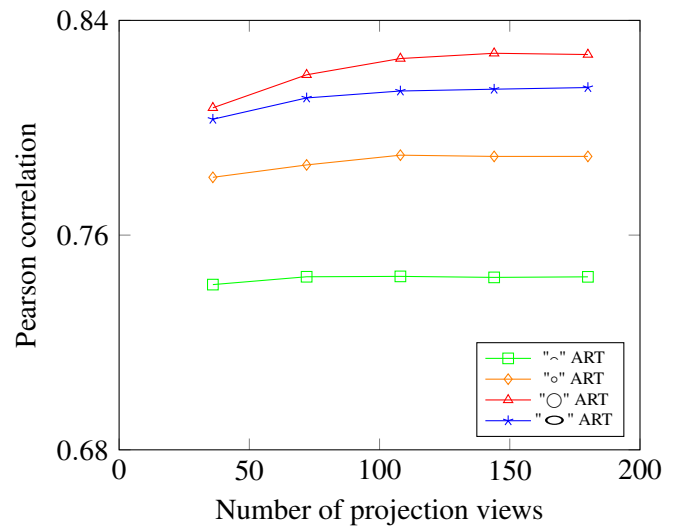

**Figure 5:** Pearson correlation in ROI#1 as a function of the number of projection views computed for the different scan orbits.

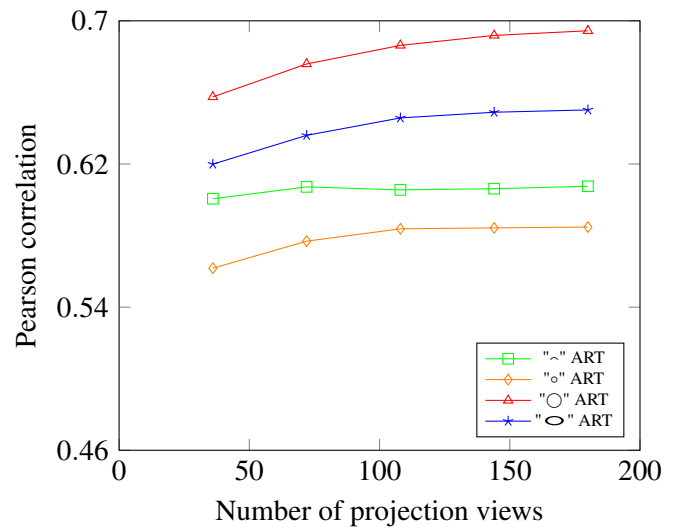

**Figure 6:** Pearson correlation in ROI#2 as a function of the number of projection views computed for the different scan orbits.

## 4    Discussion and Conclusion

In this paper, the benefits of a spherical ellipse scan trajectory "◯" for tomosynthesis-guided navigational bronchoscopy are investigated. It is compared to linear "⌣", small circular "○", and large circular "◯" scan orbits. Moreover, its benefits compared to conventional chest radiography (CR) are shown. Digital chest tomosynthesis (DCT) yields improved bony and soft structures visibility compared to CR. This is highly crucial in navigational bronchoscopy in order to confirm the tool-in-lesion and to collect the biopsy samples from the accurate positions. Compared to the "⌣" and the "○", the "◯" shows an improvement in the mitigation of out-of-focus artifacts. The "◯" demonstrates a very slight improvement in terms of image quality, however this tiny benefit is at the cost of a much more larger space required to perform the scan. In the operating room, the available space is highly limited, performing a "◯" is impractical and requires many logistic efforts. If one is restricted to circular trajectories, just the "∘" can be used, however having the "⌵" at hand, it can be used for an improved image quality. In addition, it provides flexibility: the larger aperture of the "⌵" can be chosen in the direction where more space is available. While intra-operative CBCT might provide better image quality, it is at the cost of a much larger radiation dose and space requirements. Therefore, the "⌵" appears to be potentially suitable and optimized for interventional DCT. Further investigations will focus on the impact of the angular range and the projection density on the reconstruction. Moreover, reconstructions in more challenging conditions (e.g. nodules in the periphery of the lungs) will be addressed.

## Acknowledgments and Disclaimer

This work was in part conducted within the context of the International Graduate School MEMoRIAL at the Otto von Guericke University (OVGU) Magdeburg, Germany, kindly supported by the ESF (project no. ZS/2016/08/80646) and was partly funded by the Federal Ministry of Education and Research within the Forschungscampus STIMULATE under grant number 13GW0473A. The authors would like to thank Gouthami Chintalapani, Siemens Medical Solutions USA, Inc., for data selection and organizing data transfer. The concepts and information presented in this paper are based on research and are not commercially available.

# Chapter 12

# Oral Session - Deep Learning Methods

**session chairs**
Jerome Liang, *Stony Brook University (United States)*
Kuang Gong, *Massachusetts General Hospital (United States)*

# Fast Reconstruction of non-circular CBCT orbits using CNNs

Tom Russ[1], Wenying Wang[2], Alena-Kathrin Golla[1], Dominik F. Bauer[1], Matthew Tivnan[2], Christian Tönnes[1], Yiqun Q. Ma[2], Tess Reynolds[3], Sepideh Hatamikia[4,5], Lothar R. Schad[1], Frank G. Zöllner[1], Grace J. Gang[2], and J. Webster Stayman[2]

[1]Computer Assisted Clinical Medicine, Heidelberg University, Heidelberg, Germany
[2]Department of Biomedical Engineering, Johns-Hopkins University, Baltimore, USA
[3]ACRF Image X Institute, University of Sydney, Australia
[4]Austrian Center for Medical Innovation and Technology, Wiener Neustadt, Austria
[5]Center for Medical Physics and Biomedical Engineering, Medical University of Vienna, Austria

**Abstract** Non-circular acquisition orbits for cone-beam CT (CBCT) have been investigated for a number of reasons including increased field-of-view, minimal interference within an intraoperative environment, and improved CBCT image quality. Fast reconstruction of the projection data is essential in an interventional imaging setting. While model-based iterative reconstruction can reconstruct data from arbitrary geometries and provide superior noise suppression for a wide variety of non-circular acquisitions, such processing is particularly computationally intensive. In this work, we present a scheme for fast reconstruction of arbitrary non-circular orbits based on Convolutional Neural Networks (CNNs). Specifically, we propose a processing chain that includes a shift-invariant deconvolution of backprojected measurements, followed by CNN processing in a U-Net architecture to address artifacts and deficiencies in the deconvolution process. Synthetic training data is produced using orbital specifications and projections of a large number of procedurally generated objects. Specifically, attenuation volumes are created via randomly placed Delaunay tetrahedrons. We investigated the reconstruction performance for different sets of acquisition orbits including: circular, sinusoidal and randomized parametric trajectories. Our reconstruction scheme yields similar image quality when compared to simultaneous algebraic reconstruction technique (SART) reconstructions, at a small fraction of the computation time. Thus, the proposed work offers a potential way to utilize sophisticated non-circular orbits while maintaining the strict time requirements found in interventional imaging.

## 1 Introduction

The advent of robotic interventional x-ray systems has opened the door to dramatically increased flexibility in the design of CBCT acquisition trajectories. Such orbits have been used to increase the imaging field-of-view and to minimally interfere with the other equipment in the interventional suite; but also to improve image quality. For example, a large variety of non-circular orbits has been investigated to improve data completeness, metal artifacts, and task-based detectability [1, 2, 3, 4]. Typically, reconstruction algorithms for non-circular data have relied on both analytical and model-based methods. Analytical solutions exist for specific classes of non-circular orbits such as saddle trajectories [5]. Model-based iterative reconstruction (MBIR) implicitly handles arbitrary geometries (providing a "best" estimate based on the available data). These algorithms, however, are computationally expensive, which poses a major limitation particularly for interventional applications. The recent proliferation of data-driven and machine-learning-based reconstruction methods provides opportunities for superior reconstruction speed *and* image quality comparable to MBIR.

In this work, we propose a reconstruction scheme that leverages Convolutional Neural Networks (CNNs). In particular, we develop a processing chain where data backprojection is followed by a shift-invariant deconvolution step followed by CNN processing. The deconvolution is based on the or-bital trajectory and the intrinsic system response but is only approximate. The CNN step is trained to mitigate deficiencies in this approximate deconvolution. Each of these steps is computationally efficient and non-iterative leading to a fast processing chain. The following sections detail this processing chain and its application to five different sets of orbit geometries. For comparison, an iterative reconstruction scheme, the simultaneous algebraic reconstruction technique (SART), is also applied and quantitative performance measures (relative to truth) are computed.

## 2 Materials and Methods

### 2.1 The Tomographic Reconstruction Problem

Presuming log-transformed projection data, tomographic reconstruction seeks to solve the following inverse problem:

$$y = \mathbf{A}(\Omega)\mu, \qquad (1)$$

where $y$ denote the measured line integrals of attenuation (e.g., projections) and $\mu$ is the distribution of attenuation values in the object. Here, we identify the dependence of the projection matrix, $\mathbf{A}$, on some parameterization of the acquisition orbit $\Omega$. Classic inversion approaches often seek to find the pseudo-inverse:

$$\mu = (\mathbf{A}^T \mathbf{A})^{-1} \mathbf{A}^T \mathbf{y}. \qquad (2)$$

The pseudo-inverse has the advantage that solutions can be found for non-square and rank-deficient $\mathbf{A}$ that are possible for arbitrary trajectories.

We note that $\mathbf{A}^T$ represents a backprojection operation. Thus, the operator $(\mathbf{A}^T \mathbf{A})^{-1}$ represents a kind of generalized filtering operation. In fact, under idealized imaging conditions (parallel beam, sufficient sampling, etc.) and a circular acquisition geometry, $(\mathbf{A}^T \mathbf{A})$ represents the operator that applies the well-known intrinsic response of tomography - a $1/r$ blur function. Thus, in the ideal case, $(\mathbf{A}^T \mathbf{A})^{-1}$ is the inverse filter that removes $1/r$ blur. For non-circular orbits, divergent beams, etc., the blur induced by $(\mathbf{A}^T \mathbf{A})$ is not generally shift-invariant nor of the form $1/r$. However, these observations suggest a potential scheme for fast reconstruction using similar processing stages.

### 2.2 Proposed Reconstruction Pipeline

Motivated by the above observations, we propose a new reconstruction pipeline using neural networks but leveraging what we already know about the required reconstruction process. Specifically, we maintain the backprojection step and address the operator $(\mathbf{A}^T \mathbf{A})^{-1}$.

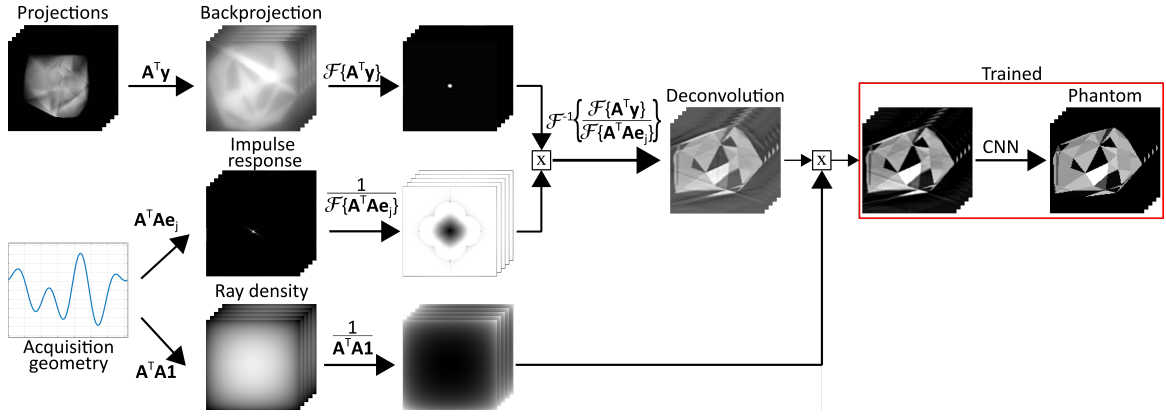

**Figure 1:** Flow chart illustrating the proposed reconstruction pipeline. We first deconvolve an approximation of the system response then deploy a CNN to remove residual artifacts.

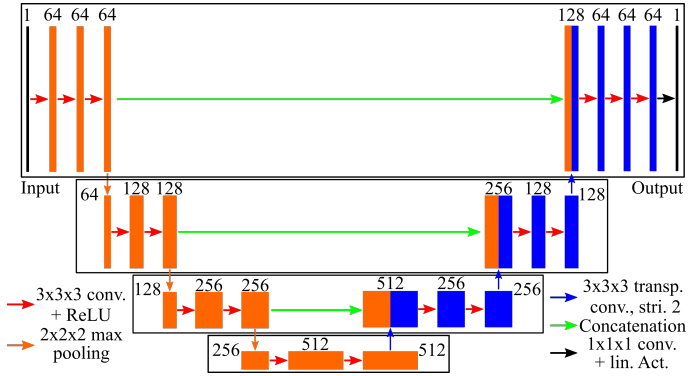

**Figure 2:** U-Net in the last step of the propose pipeline. Numbers over feature channel blocks indicate the number of channels. Max pooling halves the size of each dimension, whereas transposed convolution with stride two doubles the size of each dimension.

While one could develop a neural network to learn this inverse transform, there is an opportunity to provide a better network input. We will presume that the geometry, and hence $\mathbf{A}$ and $\mathbf{A}^T$ of non-circular acquisitions are known *a priori*. We can therefore devise network inputs to leverage such prior information. Specifically, if we can first deconvolve (in the case of a shift-invariant system) the system response, $\mathbf{A}^T\mathbf{A}$, from the backprojection, $\mathbf{A}^Ty$, we can effectively remove the dependency on geometry in the reconstruction process. Of course, such deconvolution procedure is noise amplifying and prone to artifacts. We therefore deploy a post-deconvolution CNN to remove residual artifacts.

Towards this end, we developed a processing chain illustrated in Fig.1. For initial investigation in this work, we assumed the system response to be approximately shift-invariant (true for small objects and/or long geometries) and approximated the system response as $\mathbf{A}^T\mathbf{A}e_j$ where $e_j$ denotes an impulse at the center of the image. We first deconvolved $\mathbf{A}^T\mathbf{A}e_j$ from $\mathbf{A}^Ty$ via direct Fourier inversion, i.e.:

$$\mathscr{F}^{-1}\left\{ \frac{\mathscr{F}\left\{\mathbf{A}^Ty\right\}}{\lfloor\mathscr{F}\left\{\mathbf{A}^T\mathbf{A}e_j\right\}\rfloor} \right\}. \qquad (3)$$

We adopted several techniques to mitigate artifacts associated with the deconvolution process. First, a threshold operator was used in the denominator to avoid division by zero. (Specifically, a value of 0.0025 was applied.) Second, the backprojection volume was expanded to approximately four times the reconstruction volume to mitigate spurious frequencies as a result of the fast Fourier transform of signals with discontinuities at the boundaries. Third, to mitigate artifacts in $\mathscr{F}\left\{\mathbf{A}^T\mathbf{A}e_j\right\}$ due to the combined effect of voxel sampling and ray-based projector, we computed $\mathbf{A}^TAe_j$ at eight voxel locations around the central voxel of the image and averaged the responses.

After the deconvolution, we additionally corrected for sampling density by performing an element-wise division of the volume by $\mathbf{A}^T\mathbf{A}1$, where $\mathbf{1}$ denotes a volume of 1s. We truncated the image to the same size as the reconstruction image volume to save memory. The resulting image volume was used as input to the CNN. In summary, the input to the CNN is represented mathematically as:

$$\mathbf{x} = \mathscr{F}^{-1}\left\{ \frac{\mathscr{F}\left\{\mathbf{A}^Ty\right\}}{\lfloor\mathscr{F}\left\{\mathbf{A}^T\mathbf{A}e_j\right\}\rfloor} \right\} \frac{1}{\mathbf{A}^T\mathbf{A}1}. \qquad (4)$$

For the CNN processing step, we chose a U-Net architecture consisting of seven convolutional blocks illustrated in Figure 2. The U-net architecture was chosen due to its successful application in image deconvolution and CT reconstruction.

The input of the network detailed above consists of 128x128x128 voxel volumes. The network is trained to predict the ground truth phantom images of the same size. For training, the root-mean-square error (RMSE) between the prediction and the ground truth image was chosen as the loss function. Optimization was performed using an Adam optimizer with a learning rate of 0.001 and terminated after 100 training epochs. Among the 1000 phantom images, 800 images were used for training, 100 for validation, and 100 for testing. Details of the training and evaluation data follow.

### 2.3 Phantom and Data Generation

For imaging phantoms used in training and evaluation, we procedurally generated 1000 random realizations of Delaunay tetrahedrons. We randomly sampled 40 vertex locations in 3D, then created a tetrahedron mesh by connecting these vertices using the 3D Delaunay triangulation algorithm in MATLAB. Within each tetrahedron, a uniform attenuation coefficient was randomly assigned based on the distribution of voxel values in an abdomen CT scan (*sans* background). The phantoms were then formed by voxelizing the meshes on a 128x128x128 grid with 0.5x0.5x0.5 mm$^3$ voxel spacing. Data were simulated using the ASTRA toolbox [6, 7]. The imaging geometry used a source-axis distance and source-detector distance of 1 m and 0.5 m, respectively. The reconstruction volume matched the voxel size and spacing of the ground truth. The projection data were simulated on a 256x256 detector with pixel size 0.75 mm x 0.75 mm. The detector size was large enough to avoid data truncation. Noiseless projection data were simulated.

## 2.4 Experimental Design

We exercised the proposed reconstruction pipeline on five different sets of acquisition geometries. For each geometry, 512 rotation angles, $\theta$, are evenly distributed between 0° and 360°. The elevation angles, $\phi$, are parameterized as sinusoidal functions of $\theta$ at varying frequencies. The amplitude has been set to 25° for all cases (except the circular geometry). Four networks were trained on data of only one orbit, while one network was trained on data with two different geometries in a common pool. This was done to investigate if our proposed approach is able to reconstruct data of more than one geometry.

The five acquisition geometries are as follows:

- circular, $\phi = 0$ for all $\theta$
- $\phi = \sin(2\theta)$,
- $\phi = \sin(3\theta)$,
- $\phi = \sin(2\theta)$ and $\phi = \sin(3\theta)$,
- one linear combination of sinusoidal basis functions with randomly generated coefficients.

## 2.5 Evaluation Metrics

Reconstruction performance was evaluated in terms of the the normalized RMSE (nRMSE), the feature similarity index (FSIM) and the structural similarity index (SSIM) between the network output and the ground truth phantom images. To compare the proposed reconstruction pipeline with state-of-the-art algorithms, we additionally performed reconstructions using 50 iterations of the SART algorithm (using the GPU-based TIGRE toolkit for arbitrary trajectories [8, 9]).

## 3 Results

Intermediate images and final reconstruction outputs from the proposed reconstruction pipeline are illustrated in Figure 3. Note the residual artifacts in the deconvolved volumes. The calculated evaluation metrics are compared in Table 1. The CNN-based approach consistently outperforms the SART reconstructions in terms of nRMSE and FSIM. This is also the case for SSIM except for the network trained on two sinusoidal geometries. While SART performs comparably for all geometries with only slight deviations, the performance of the CNNs show noticeable differences for the different geometries. Specifically, reconstruction performance decreases with increasing orbital complexity (possibly due to increased shift-variance). This is apparent in the slightly decreasing evaluation metrics as well as in the magnified areas and the difference images in Figure 3. The magnified region in particular contains fine-grain details, which every CNN struggles to reconstruct accurately.

The majority of the computation time for the proposed method is spent on the calculation of the impulse response (5 minutes), the ray density (30 seconds), and ultimately for the deconvolution operation (20 seconds). The CNN prediction takes around 1 second. In comparison, SART reconstructions take approximately 50 minutes for 50 iterations on a workstation with comparable specifications. Aside from the CNN, the mentioned implementations have not been optimized for runtime.

## 4 Discussion and Conclusion

In this work, we proposed a novel pipeline for fast reconstruction of non-circular geometries. The pipeline consists of an initial deconvolution step to remove an approximation of the system response followed by an artifact removal step using a CNN. We tested the pipeline in five sets of imaging geometries of single and mixed sinusoidal orbits. Our proposed method offers $\sim 90\%$ reduction in computation time and is comparable or superior to SART in terms of the nRMSE, FSIM, and SSIM. These results suggest that the pipeline offers a promising approach to reconstruct data acquired with non-circular orbits when time is of the essence.

This work has several limitations that are being addressed in ongoing work. First, the pipeline was only trained and assessed on piecewise-constant phantoms. Extending the reservoir of phantoms to include non-piecewise-constant phantoms will help to improve the generalizability of the proposed approach. Second, the case where two sinusoidal orbits were trained simultaneously illustrates some capacity of the method to accommodate multiple geometries within the same class. We plan to extend the reconstruction capability of the method to arbitrary orbits within classes (e.g., sinusoids).

## Acknowledgements

The authors acknowledge support, in part, by the US National Institutes of Health through grant R01EB027127; and by the state of Baden-Wuerttemberg through bwHPC and the German Research Foundation (DFG) through grant INST 35/1134-1 FUGG. This research project is part of the Research Campus M$^2$OLIE and funded by the German Federal Ministry of Education and Research (BMBF) within the Framework 'Forschungscampus - Public-Private Partnership for Innovation' under the funding code 13GW0388A.

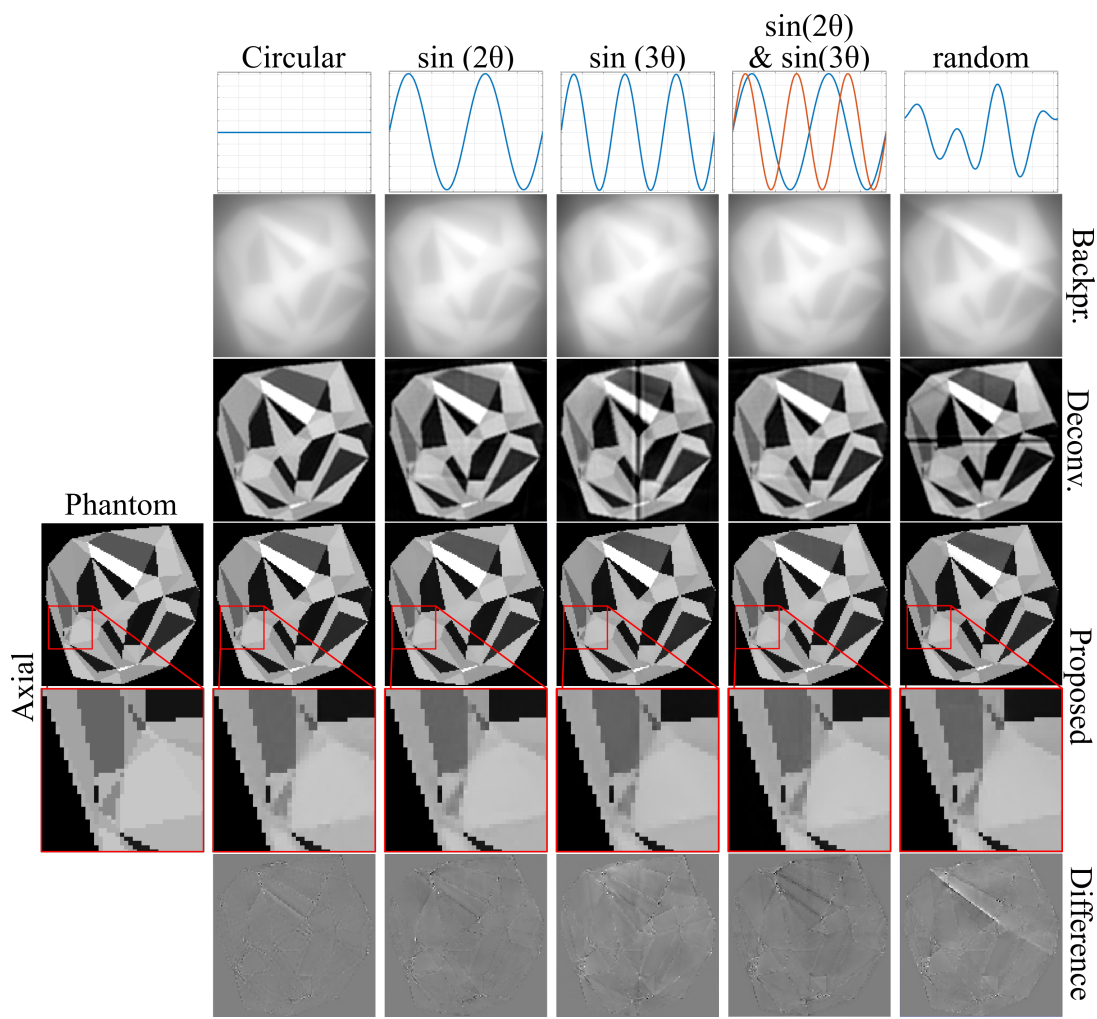

**Figure 3:** Intermediate image volumes and final reconstruction outputs from the reconstruction pipeline. Each column corresponds to a set of imaging geometries. Rows from top to bottom: elevation angle $\phi$ as a function of rotation angles $\theta$; backprojected volume; volume after deconvolution; CNN reconstruction (axial slice); zoomed in ROI within the slice; difference image between the reconstructions and ground truth phantom images.

| | | circular | sin(2θ) | sin(3θ) | sin(2θ)&sin(3θ) | random |
|---|---|---|---|---|---|---|
| **Proposed** | nRMSE ↓ | **0.033 ± 0.005** | **0.048 ± 0.007** | **0.060 ± 0.008** | **0.062 ± 0.007** | **0.061 ± 0.009** |
| | FSIM ↑ | **0.991 ± 0.005** | **0.983 ± 0.010** | **0.979 ± 0.010** | **0.977 ± 0.013** | **0.979 ± 0.013** |
| | SSIM ↑ | **0.994 ± 0.002** | **0.987 ± 0.004** | **0.984 ± 0.003** | 0.944 ± 0.013 | **0.985 ± 0.007** |
| **SART** | nRMSE ↓ | 0.116 ± 0.013 | 0.105 ± 0.016 | 0.109 ± 0.015 | 0.107 ± 0.016 | 0.108 ± 0.015 |
| | FSIM ↑ | 0.937 ± 0.019 | 0.943 ± 0.015 | 0.940 ± 0.015 | 0.942 ± 0.015 | 0.941 ± 0.015 |
| | SSIM ↑ | 0.941 ± 0.011 | 0.963 ± 0.011 | 0.956 ± 0.011 | **0.960 ± 0.011** | 0.958 ± 0.010 |

**Table 1:** Evaluation metrics for the propose pipeline compared with SART. All metrics were evaluated between the reconstructions and ground truth phantom images. The better of two values is marked **bold**.

# Noise Entangled GAN For Low-Dose CT Simulation

Chuang Niu[1], Ge Wang[1], Pingkun Yan[1], Juergen Hahn[1], Youfang Lai[2], Xun Jia[2], Arjun Krishna[3], Klaus Mueller[3], Andreu Badal[4], Kyle J. Myers[4], and Rongping Zeng[4]

[1]Department of Biomedical Engineering, Center for Biotechnology & Interdisciplinary Studies, Rensselaer Polytechnic Institute, Troy, NY USA
[2]Department of Radiation Oncology, UT Southwestern Medical Center, Dallas, TX USA
[3]Computer Science Department, Stony Brook University, Stony Brook, NY USA
[4]Division of Imaging, Diagnostics and Software Reliability, OSEL, CDRH, U.S. Food and Drug Administration, Silver Spring, MD USA

**Abstract** We propose a Noise Entangled GAN (NE-GAN) for simulating low-dose computed tomography (CT) images from a higher dose CT image. First, we present two schemes to generate a clean CT image and a noise image from the high-dose CT image. Then, given these generated images, an NE-GAN is proposed to simulate different levels of low-dose CT images, where the level of generated noise can be continuously controlled by a noise factor. NE-GAN consists of a generator and a set of discriminators, and the number of discriminators is determined by the number of noise levels during training. Compared with the traditional methods based on the projection data that are usually unavailable in real applications, NE-GAN can directly learn from the real and/or simulated CT images and may create low-dose CT images quickly without the need of raw data or other proprietary CT scanner information. The experimental results show that the proposed method has the potential to simulate the realistic low-dose CT images.

## 1 Introduction

An excess of x-ray exposure from computed tomography (CT) examinations could lead to the development of cancer, and thus optimizing CT protocols according to the as low as reasonably achievable (ALARA) principle has become important. Low-dose CT (LDCT) simulation techniques have developed as an effective tool to help determine the lowest dose in accordance with the ALARA principle, thereby circumventing the repetition of CT examinations with different exposure conditions for the same patients. However, reducing the radiation dose will inevitably increase the noise level in the reconstructed CT images and may compromise the accuracy of a radiologist's diagnostic decision. To this end, a lot of LDCT denoising methods have been proposed to improve the image quality. Recently, deep-learning-based denoising methods have been shown a potential to achieve the superior denoising performance, if properly trained with a large number of CT images. In this context, the results with LDCT simulation methods can help train and test the robustness of denoising methods or other image analysis models applied to the LDCT images.

Traditionally, LDCT simulation tools insert random noise to the raw sinogram data and reconstruct the noisy data to simulate LDCT images [1]. However, neither raw data nor the precise parameters of a CT imaging system are generally accessible without an established collaboration with the CT vendor. To circumvent the use of raw data, projection data can be approximated by forward projecting from the CT image, which are then added with noise and reconstructed using CT simulation software [2]. However, these sinogram-based methods are usually time-consuming and the simulated projection data may not truly reflect the real conditions so the simulated LDCT noise is likely still not perfect. Recently, Shan et al. designed a specific GAN with a conditional batch normalization layer to simulate LDCT noise from a random 2-dimensional Gaussinan noise vector in the latent space [3]. However, it is difficult for this method to generate realistic LDCT images from the Gaussian noise without explicit prior information of the LDCT noise.

In this work, we treat the LDCT simulation as a transformation from a higher dose CT (HDCT) image to the LDCT images. Specifically, we first generate a clean CT image and a high-dose noise image from the HDCT image, and then train a noise entangled GAN (NE-GAN) to generate different levels of LDCT images via entangling the high-dose noise image scaled by different noise factors into the clean CT image. The advantages of the proposed framework for LDCT simulation are: 1) The generated high-dose noise image explicitly contains the prior of noise and imaging system. 2) The NE-GAN can learn from both the simulated and real CT images, so that it has the potential to generate realistic LDCT images. 3) Once the model has been trained, the simulation speed for LDCT images is very fast.

## 2 Methods

Ideally, before simulating the LDCT image from the HDCT image, the noise component should be removed from the HDCT image and then low-dose noises are simulated and added to the denoised image. In practice, the HDCT images are usually regarded as the clean image and the high-dose noises are ignored. Although the magnitudes of high-dose noises are low, they do contain the prior information of CT noises and the imaging system to some extent. Based on above observations, we propose to simulate an LDCT image through two steps: the fist step is to generate a clean image and a high-dose noise image from the HDCT image, and the second step is to generate different levels of LDCT images by entangling the high-dose noise component scaled with a specific noise factor into the clean CT image.

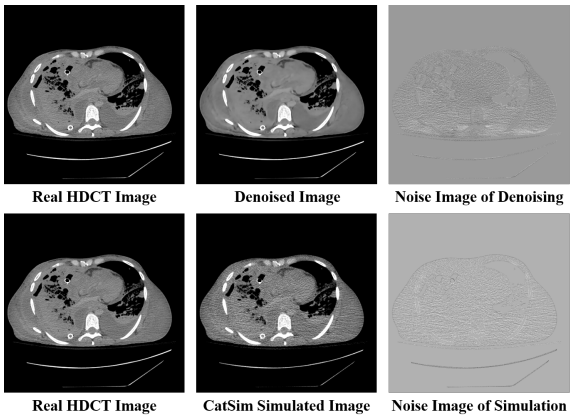

**Figure 1:** Generation of noise image.

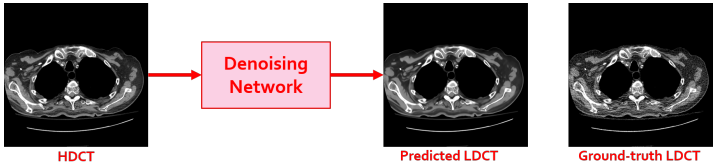

**Figure 2:** A denoising network trained using HDCT as input and LDCT as target. The predicted "LDCT" is a cleaner CT image rather than a noisier.

## 2.1 Generation of high-dose noise image

For generating the high-dose noise image that preserves the prior information of the CT noise and imaging system, we present two schemes, i.e., through a denoising network or through CT simulation, as illustrated in Fig. 1. For the denoising scheme, the HDCT image is first forwarded into a denoising model to obtain a clean CT image, which is then subtracted from the input HDCT image to generate the high-dose noise image. The denoising model is trained by directly mapping high-dose CT image to the low-dose CT image, as shown in Fig. 2. By doing this, the trained model can generate the denoised image instead of the images with more noises, which is consistent to the findings of Noise2Noise [4]. The denoising scheme can extract the real prior information from the real HDCT images, which are then transformed to LDCT images with specific noise level by NE-GAN. The CT simulation scheme is to use traditional sinogram-based methods to simulate a set of higher-dose noise images by virtually scanning the real HDCT image, and the real HDCT image is regarded as the clean CT image, as shown in Fig. 1. Then, NE-GAN takes the simulated noise image and the HDCT image as inputs to generate a set of LDCT images with different levels of noise. In this scheme, the sinogram-based method is only used to simulate a single dose of images, and other lower dose of images can be generated by NE-GAN to save computation time.

## 2.2 Noise Entangled GAN

In this Subsection, we describe the details of the proposed noise entangled GAN (NE-GAN). As shown in Fig. 3, NE-

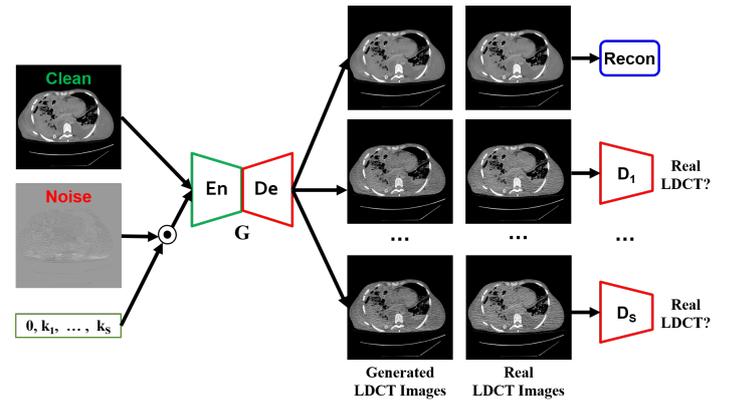

**Figure 3:** Framework of NE-GAN.

GAN consists of a generator $G$ and a set of discriminators $\mathbf{D} = \{D_j\}$, $j = 1, \cdots, S$, where the number of discriminators $S$ is equivalent to the number of lower-dose levels in the training set. Specifically, the generator $G$ is a encoder-decoder network that takes a clean CT image and a noise image scaled with a noise factor as inputs, and outputs a LDCT image corresponding to the input noise factor. All discriminators share the same network architecture. Each discriminator is to determine whether the generated image of a predefined level is real.

To train NE-GAN, we need a set of training samples $\{x_i^0, n_i^0, k^j, x_i^j\}$, $i = 1, \cdots, N$, $j = 1, \cdots, S$, where $x_i^0$ and $n_i^0$ denote the clean CT image and high-dose image respectively, $N$ is the total number of HDCT images, $x_i^j$ denotes the corresponding LDCT image, and $S$ is number of noise levels or discriminators, $k^j$ denotes the noise factor that is a positive real number and a larger value corresponds to a higher noise level or lower image quality. The loss function is:

$$L = \sum_{j=1}^{S} E_{X^j}[\log D_j(x^j)] + E_{X^0}\left[log(1 - D_j(G(x^0, n^0 \cdot k_j)))\right]$$
$$+ |x^0 - G(x^0, n^0 \cdot k_j)| + |x^0 - G(x^0, 0)|\Big].$$

The first two items in the loss function are the adversarial losses that train $\mathbf{D}$ to maximize the probability of assigning the correct label to both real LDCT images and the generated ones from $G$ and train G to minimize probability of assigning the correct label for $D$, the third item is a data fidelity loss to constrain the generated LDCT images to keep the same contents as those in the input images, and the fourth item is a reconstruction loss to ensure that the generated CT images are exactly the clean images when the noise factor is zero. After training, only the generator $G$ is retained to simulate different levels of LDCT images given the clean CT image, the high-dose noise image, and the specific noise factor, i.e., $\hat{x}^j = G(x^0, n^0 \cdot k_j)$. It is noted that although the noise factor in the training stage is predefined as a limited number of fixed values according to the training dataset, it could be any value in the testing stage beyond the predefined values in the training stage. With increasing the value of noise factor, the noise level of the simulated LDCT image will increase.

## 2.3 Implementation details

We adopted the same generator and discriminator networks as those in CycleGAN [5]. The architecture of the denoising network was the same as the generator network. During training, we used the Adam method to optimize the NE-GAN model with a batch of 8 $128 \times 128$ randomly cropped image patches. The initial learning rate was set to 0.0002 during the first 200 epochs and the learning rate was linearly decay to zero in the following 200 epochs. The momentum terms of Adam were set to 0.5 and 0.999. The noise factor $k_j$ is set to the ratio of the input dose level to the target dose level, see Subsections 3.2 and 3.3 for details.

## 3 Experiments and results

### 3.1 Dataset

In this study, we used a multi-dose of real CT image dataset from [6], in which the CT images were collected from anonymous cadavers and each of them was repeatedly scanned four times using four different radiation doses. In our experiments, we selected a sub-dataset that contains 261 groups of CT images for training and 251 groups of CT images for testing, each group includes four $512 \times 512$ FBP reconstructed images that have the same contents but different noise indices of 10, 20, 30, and 40. Here the noise index is approximately equal to standard deviation of CT number in the central region of the image of a uniform phantom, and used to define the image quality.

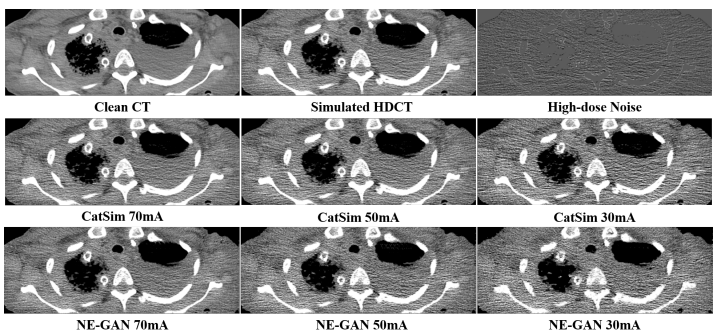

**Figure 4:** Results of NE-GAN on simulated dataset.

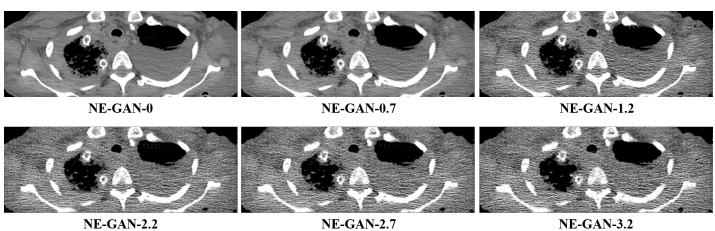

**Figure 5:** Results of NE-GAN on simulated dataset with additional noise factors beyond training.

## 3.2 Results on simulated dataset

In this subsection, we used the simulation scheme as described in Subsection 2.1 to generate high-dose noise images. Specifically, we used the CatSim [7] simulator to simulate LDCT images of four different dose levels corresponding to X-ray tube currents of 90 mA, 70 mA, 50 mA, and 30 mA, as shown in Fig. 4. The simulated CT image of 90 mA is used as the HDCT image and the real CT image with noise index of 10 is regarded as the clean CT image, thus the high-dose noise image is the difference between them. With these images, NE-GAN was trained and noise factors corresponding to 70 mA, 50 mA, and 30 mA were set to 1.3, 1.8, and 3.0 respectively. The results in this setting are shown in Fig. 4, we can see that the proposed method can simulate the different levels of LDCT images and the learned noise levels are similar to those simulated with CatSim. The NE-GAN simulated results with different noise factors that were not used in the training stage are shown in Fig. 5, where the number indicates the noise factor. Particularly, NE-GAN-0 means that the scale factor is zero and in this case no noises are added, consistent with the constraint in the loss function as described in Subsection 2.2. Also, when increasing the noise factor, the noise magnitude of simulated LDCT image increases and the image looks more noisier.

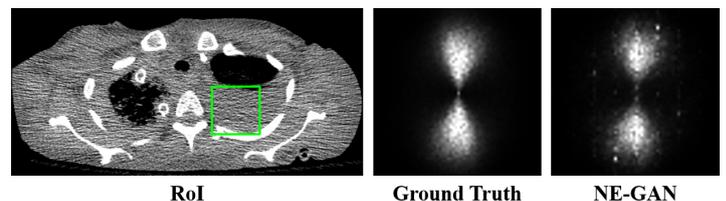

**Figure 6:** Results of noise power spectrum.

In addition, we evaluated the statistical property of noise power spectra (NPS) of the NE-GAN generated LDCT images [8]. Specifically, we repeatly generated HDCT noise image with CatSim and simulate the LDCT images with NE-GAN by 50 times. Then, the $64 \times 64$ image patches (green box) were cropped to calcuate the NPS, as shown in Fig. 6. The NPS of the NE-GAN generated LDCT images is similar to that of the targets, which indicates the proposed deep-learning-based method has the ability to preserve the statistical properties of noise.

### 3.3 Results on real dataset

In this subsection, the proposed NE-GAN model was directly trained on the real dataset. Here the denoising scheme was firstly used to decompose the high-dose CT image with noise index of 10 to a clean CT image and a high-dose noise image. Then the NE-GAN was trained to map these decomposed images to the LDCT images with specific noise indices. The noise factors corresponding to noise indices of 20, 30, and 40 were set to 2.0, 3.0, and 4.0 respectively. The results on this real dataset set are shown in Fig. 7, where the first row shows the denoised clean image and the high-dose noise image decomposed from the real HDCT image with noise index of 10, the second row presents the real LDCT images with different noise indices, the third row gives the corresponding NE-GAN generated LDCT images with the same noise indices, and the last row shows the reference results by directly adding the scaled noise image with the same noise factors into the clean image. By comparing the second and the third row, we can see that the simulated images of different noise indices are similar to the corresponding real LDCT images. The results in the last row demonstrates that simply scaling the extracted noise image and adding it back to the clean image cannot generate images matching with the real LDCT images, while the proposed NE-GAN has the ability to simultaneously transfer and merge the high-dose noise image into the clean image to simulate more realistic LDCT images. More simulation results with NE-GAN with different noise factors beyond training are also shown in Fig. 7. Similarly, noise level increased continuously with the noise factors.

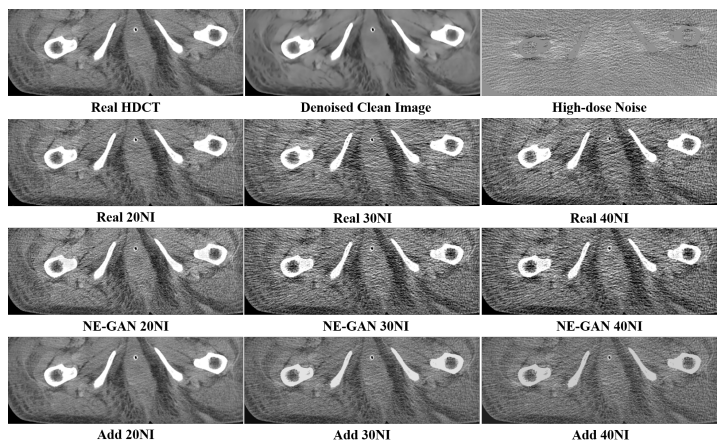

**Figure 7:** Results of NE-GAN on real dataset.

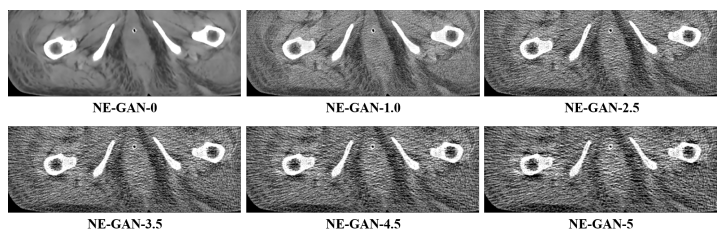

**Figure 8:** Results of NE-GAN on real dataset with additional noise factors beyond training.

## 4 Discussion and Conclusion

We presented a low-dose CT simulation method based on deep learning. Visual comparison and NPS-based noise property evaluation have demonstrated the effectiveness of the proposed method. One main advantage of the proposed NE-GAN is the high speed for simulation, which is extremely important when simulating a large number of LDCT images.

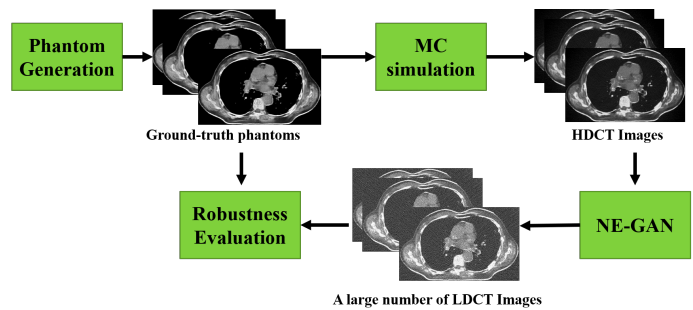

**Figure 9:** Virtual CT workflow for robustness evaluation of LDCT denoising algorithms.

For example, NE-GAN could be applied into a virtual CT workflow for robustness evaluation of LDCT denoising algorithms by generating a large number of LDCT images with the ground-truth, as shown in Fig. 9. Specifically, a CT generation method [9] is first used to generate many phantoms, which are forwarded to Monte Carlo (MC) simulation tools [10] to simulate the HDCT images. Then, NE-GAN can simulate a large number of LDCT images with a fast speed. Finally, the large number of LDCT images with the ground-truth can be used to test the robustness of the LDCT denoising algorithms. In the future, we will further improve the simulation quality by adding some statistical constrains in the NE-GAN loss function, such as those based on the noise variance map and the noise power spectra.

# Using Uncertainty in Deep Learning Reconstruction for Cone-Beam CT of the Brain

P. Wu,[1] A. Sisniega,[1] A. Uneri,[1] R. Han,[1] C. K. Jones,[1] P. Vagdragi,[1] X. Zhang,[1]
M. Luciano,[2] W. S. Anderson,[2] and J. H. Siewerdsen[1,2]

**[1]**Department of Biomedical Engineering, Johns Hopkins University, Baltimore MD USA 21205
**[2]**Department of Neurosurgery, Johns Hopkins Medical Institute, Baltimore MD USA 21287

**Purpose:** Contrast resolution beyond the limits of conventional cone-beam CT (CBCT) systems is essential to high-quality imaging of the brain for image-guided neurosurgery. We present a deep learning reconstruction method (dubbed DL-Recon) that integrates physically principled reconstruction models with DL-based image synthesis based on the statistical uncertainty in the synthesis image.

**Methods:** A synthesis network was developed to generate a synthesized CBCT image (DL-Synthesis) from an uncorrected filtered back-projection (FBP) image. To improve generalizability (including accurate representation of lesions not seen in training), voxel-wise epistemic uncertainty of DL-Synthesis was computed using a Bayesian inference technique (Monte-Carlo dropout). In regions of high uncertainty, the DL-Recon method incorporates information from a physics-based reconstruction model and artifact-corrected projection data. Two forms of the DL-Recon method are proposed: (i) image-domain fusion of DL-Synthesis and FBP (denoted DL-FBP) weighted by DL uncertainty; and (ii) a model-based iterative image reconstruction (MBIR) optimization using DL-Synthesis to compute a spatially varying regularization term based on DL uncertainty (denoted DL-MBIR). A high-fidelity forward simulator was developed to provide physically realistic simulated CBCT images over a broad range of exposure conditions as training and testing data for the synthesis network. The performance of DL-Recon was investigated using CBCT images with simulated and real low-contrast lesions in the brain.

**Results:** The error in DL-Synthesis images was correlated with the uncertainty in the synthesis estimate. Compared to FBP and PWLS, the DL-Recon methods (both DL-FBP and DL-MBIR) showed ~50% reduction in noise (at matched spatial resolution) and ~40-70% improvement in image uniformity. Conventional DL-Synthesis alone exhibited ~10-60% under-estimation of lesion contrast and ~5-40% reduction in lesion segmentation accuracy (Dice coefficient) in simulated and real brain lesions, suggesting a lack of reliability / generalizability for structures unseen in the training data. DL-FBP and DL-MBIR improved the accuracy of reconstruction by directly incorporating information from the measurements in regions of high uncertainty. Both maintained the advantages of DL-Synthesis. DL-FBP offered the runtime efficiency of FBP, and DL-MBIR offered a further ~10% improvement in contrast resolution compared to DL-FBP.

**Conclusion:** The image quality and robustness of CBCT of the brain were greatly improved with the proposed DL-Recon method incorporating uncertainty estimation with physically principled reconstruction models. Translation to clinical studies is underway.

## 1 Introduction

Cone-beam CT (CBCT) is increasingly prevalent in image-guided neurosurgery. Many implementations, however, are only suitable to visualization of high-contrast bone or surgical instrumentation. Challenges to imaging of low-contrast soft tissues are well established, including artifacts (e.g., x-ray scatter, beam-hardening)[1,2] and quantum and electronic noise that further limit contrast resolution.[2]

Recent advances in deep learning (DL) based reconstruction have opened the possibility for improved contrast resolution in CBCT.[3,4] A popular approach to DL-based reconstruction involves generation of a post-processed image from input given by conventional reconstruction [e.g., filtered backprojection (FBP) or model-based iterative reconstruction (MBIR)].[3,4] While DL methods provide a powerful tool for image synthesis, their accuracy and generalizability may not be guaranteed. Inaccuracy can arise especially when the input deviates strongly from the training cohort (e.g., pathology or imaging conditions not included in the training dataset). This is especially true for image-domain post-processing methods[3,4] that do not explicitly enforce fidelity to the projection data.

Important gains in the performance of DL reconstruction can be achieved by means of a principled approach that invokes understanding of mechanistic physical models underlying the data and/or the reconstruction method. Some researchers incorporate physical models by using the DL synthesized image as a prior (regularization) term in MBIR.[5,6] Such an approach permits deviations from the prior, as enforced by the data fidelity term, although conventional spatially invariant weighting of the regularization could underweight contributions of the prior in some regions and overweight the prior in regions in which the prior deviates from the image data due to inaccuracies in the DL synthesis image.

In this work, a DL reconstruction method (denoted DL-Recon) is presented. The method integrates physical models with image synthesis in a spatially varying manner. A Bayesian inference technique[7] is used to compute voxel-wise uncertainty in the DL synthesis image. In regions where the uncertainty is high, the DL-Recon leverages more contribution of the measured data, using either (i) FBP reconstruction (denoted DL-FBP); or (ii) a physics-based optimization model as in MBIR (denoted DL-MBIR). Thus, the contributions of both DL and physics model-based methods are leveraged in a physically principled manner for improved overall performance and reliability. The performance of the DL-Recon methods was validated in studies with CBCT images involving simulated and real low-contrast lesions in the brain.

## 2 Materials and Methods

As illustrated in Fig. 1, the proposed DL-Recon method involves three steps: (i) With an uncorrected FBP image ($\mu_{init}$) as input, the synthesis network generates a synthetic CBCT image (denoted DL-Synthesis, $\mu_s$) _and_ an uncertainty map ($\sigma$); (ii) Input projections $y$ are corrected for artifacts. In this work, $\mu_s$ is taken as the object model for correction of x-ray scatter and beam-hardening effects;[1,2] and (iii) Information from the corrected projection data and a physics-based reconstruction model (FBP or MBIR) is integrated with $\mu_s$ in relation to the uncertainty map ($\sigma$) to yield the final reconstruction output, $\mu$ (DL-Recon).

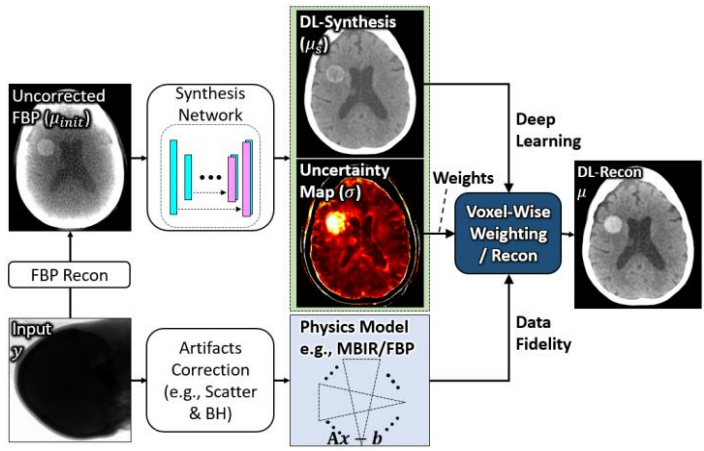

*Figure 1. Illustration of the proposed DL-Recon method incorporating DL-Synthesis and uncertainty information with a physics-based model. The DL-Recon result combines the performance of DL image synthesis with the reliability of physics-based models (FBP or MBIR).*

## 2.1 Uncertainty Estimation in DL Image Synthesis

Following the work of Gal et al.,[7] the variance of the network output is taken as a proxy for predictive uncertainty. Predictive uncertainty is interpreted in two forms that separately describe epistemic uncertainty due to noise in the network parameters (weights) and aleatoric uncertainty due to noise in the training data. The work reported below focuses on epistemic uncertainty which is associated with a lack of information available in the training data (e.g., previously unseen pathology).

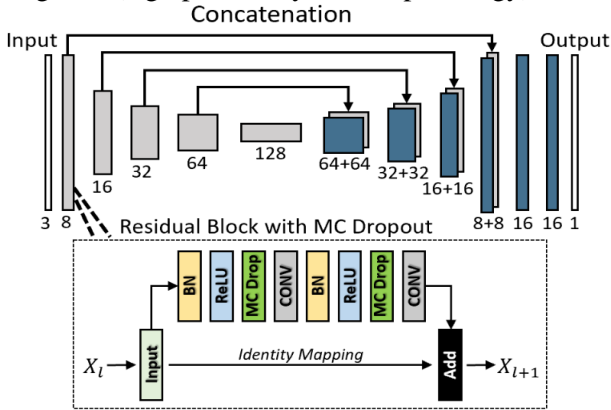

*Figure 2. CNN architecture for image-domain synthesis. Epistemic uncertainty is estimated via Monte Carlo (MC) dropout layers (dropout rate = 0.2) inserted in the downsampling and upsampling branch.*

The synthesis network used in this work is a residual U-Net[8] type of network as shown in Fig. 2. During network training, dropout is used to fit an approximate distribution over the weights of the CNN (Bayesian inference).[7] Then, during inference, dropout is applied in a MC manner to draw samples of the weights from the fitted approximate distribution. Inference is then performed multiple times, each with a different weights sample from the MC dropout. The variance of the network output (epistemic uncertainty) is then estimated from the sample variance:

$$\sigma^2 = \text{Var}(\mu_s) \approx \frac{1}{N} \sum_{n=1}^{N} f(\mu_{init}, P_n) f(\mu_{init}, P_n)^T -$$

$$\left( \frac{1}{N} \sum_{n=1}^{N} f(\mu_{init}, P_n) \right) \left( \frac{1}{N} \sum_{n=1}^{N} f(\mu_{init}, P_n) \right)^T \quad (1)$$

where $N$ is the number of weights samples (i.e., number of inferences, 16 in this work), and $\mu_s = f(\mu_{init}, P_n)$ is the network output for input $\mu_{init}$ and weights sample $P_n$.

## 2.2 The DL-Recon Methods

### 2.2.1 DL-Recon Method #1: DL-FBP

First, a straightforward approach is presented to integrate physics-based information with DL-Synthesis as a function of the (spatially varying) uncertainty via weighted fusion with an FBP reconstruction in the image domain:

$$\mu = \beta \odot \mu_s + (1 - \beta) \odot \mu_{fbp} \quad (2)$$

$$\beta = \left( \frac{[\sigma_m - D(\sigma)]_+}{\sigma_m} \right)^p \quad (3)$$

where $\mu_{fbp}$ is an (artifact-corrected) FBP reconstruction, and $\odot$ represents voxel-wise multiplication. The relative contribution of $\mu_s$ and $\mu_{fbp}$ to each voxel is controlled by the spatially varying map $\beta$ [Eq. (3)] ranging 0 to 1. The $\beta$ map is shaped by the scalar exponent $p$ [with $p = 2$ in this work]. The uncertainty ($\sigma$) of the synthesis ($\mu_s$) thus yields a normalized $\beta$ map, where $\sigma_m$ is the maximum allowed uncertainty (i.e., $\beta = 0$ for $\sigma > \sigma_m$). A dilation operator $D$ (5-voxel dilation in this work) was used to promote over-estimation of the uncertain region and ensure smooth-transition of the $\beta$ map. In this way, contributions from the physics-based / analytical reconstruction image ($\mu_{fbp}$) are greater where DL-Synthesis is less reliable (high uncertainty).

### 2.2.2 DL-Recon Method #2: DL-MBIR

Second, we propose integration of the DL synthesis result with a physically principled model via MBIR[9] – for example, iterative optimization of an objective combining the uncertainty-weighted DL-Synthesis based prior with a penalized weighted-least squares (PWLS)[9] estimate:

$$\hat{\mu} = \arg\min_{\mu} \frac{1}{2} \|A\mu - l\|_W^2 + \lambda \ \|\Psi(\mu)\|_1 + \lambda_{DL}\Psi_{DL} \quad (4)$$

$$\Psi_{DL} = \beta \odot \|\mu - \mu_s\|_1 \quad (5)$$

The first two terms in Eq. (4) are recognized simply as PWLS with an image roughness penalty (quadratic penalty in this work), where the system matrix $A$ denotes the linear forward projection operator, $l$ is the (artifact-corrected) line integral, W is the estimated variance for each measurement, and $\|\Psi(\mu)\|_1$ is the roughness penalty based on neighborhood differences with a scalar weighting $\lambda$ . The third term (i.e., the "deep learning" term) serves as an additional penalty on differences between the DL-Synthesis and the current estimate ($\mu$). The global contribution of this penalty is controlled by a constant scalar $\lambda_{DL}$, and – importantly – the spatially varying penalty strength is controlled by the uncertainty map via $\beta$ [Eq. (3)]. Thus, the penalty strength is inversely proportional to the uncertainty of DL-Synthesis, allowing greater contribution from the physics-based data fidelity term where DL-Synthesis is less certain. Compared to DL-FBP, DL-MBIR allows incorporation of an explicit projection domain data fidelity constraint and more accurate physics models.

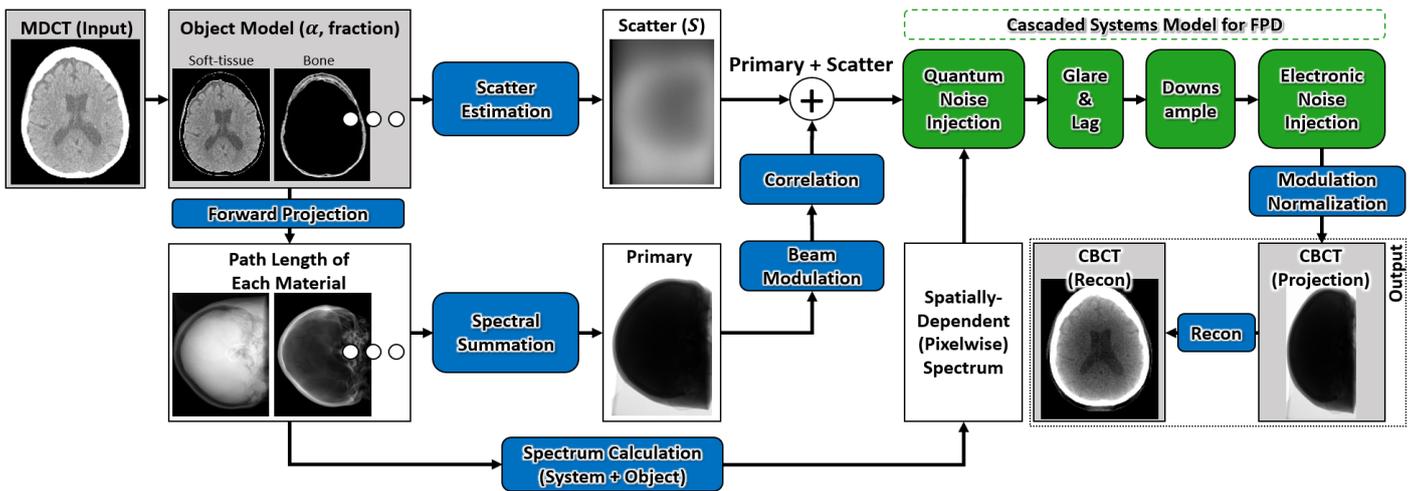

*Figure 3. Flowchart of the high-fidelity forward simulator. Input to the simulator is the high quality MDCT volumes. Output is the simulated CBCT.*

## 2.3 Training Data Generation

A high-fidelity forward simulator (Fig. 3) was developed to simulate realistic CBCT projection data from high-quality, helical multi-detector CT (MDCT) volumes. The simulator contains highly accurate, physics-based models of the full imaging chain and image formation process, comprising the following four main steps: (i) First, MDCT volumes were segmented to provide image-domain object models (e.g., soft-tissue, bone), which consist of voxel-wise fraction of each material. The object model was then used to estimate scatter and primary signal through MC simulation[1,2] and polyenergetic forward projection, respectively. These calculations included models of the system geometry, incident spectrum, focal spot blur, beam modulation, patient motion, antiscatter-grid, and detector response. (ii) An energy-dependent cascaded systems analysis model[10] was used to inject correlated quantum noise in the projection domain. The amount of injected noise was based on the number of photons and the incident spectrum (calculated based on the line integral for each material) at the surface of the detector for each pixel. (iii) Third, detector-domain artifacts including veiling glare and detector lag were added in the projection domain through spatial-temporal convolution with their associated kernel functions;[1,2] (iv)

Finally, realistic CBCT projection data were obtained by downsampling the projections to the specified readout binning mode and injecting uncorrelated electronic noise.

## 2.4 Experiments: CBCT of Low-Contrast Lesions

The synthesis network was trained using paired real MDCT and simulated CBCT images, which were the input and output of the simulator described in §2.3. Parameters for the simulation were adjusted to emulate the characteristics of a CBCT system common in image-guided surgery (the O2 O-arm, Medtronic; head scan protocol: 120 kV, 93 total mAs, 370 projections). A total of 22,000 slices were used for DL-Synthesis training (healthy and hydrocephalus subjects). Training was performed with the Adam optimizer (learning rate $5 \times 10^{-4}$, batch size = 8).

Two simulation studies involving low-contrast lesions (not present in the training set) were designed to investigate the performance of the proposed methods. Experiment #1 featured two types of simulated lesions added to a healthy patient: (i) simple circular lesions with varying contrast (-70 to +70 HU, pertinent to low-contrast features such as intracranial hemorrhage (ICH), ischemia, and abnormal fluid), size (diameter ranging from 10-40 mm), and location (random placement within the cranial vault); (ii) more complex star-polygon lesions with varying contrast [-70 to

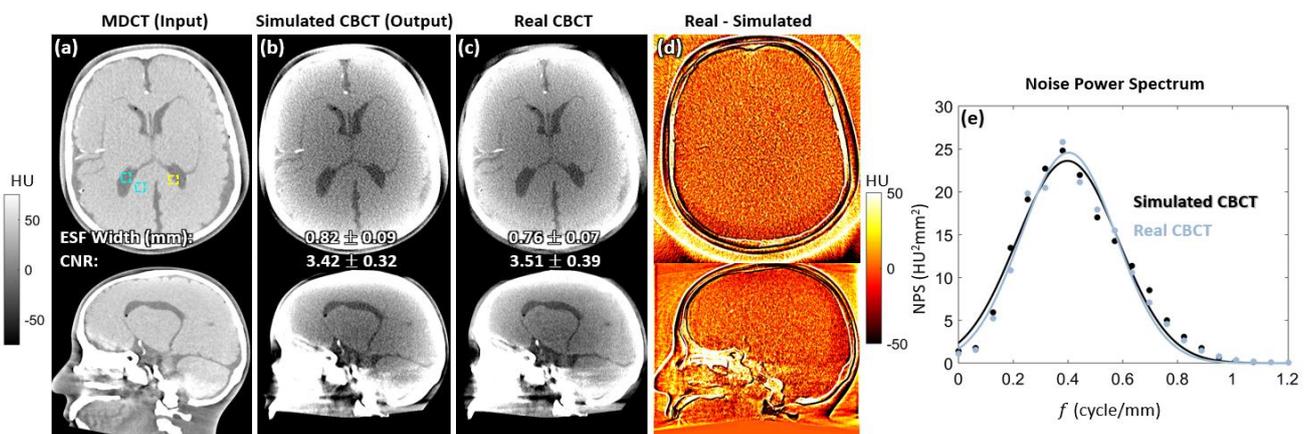

*Figure 4. Validation results for the high-fidelity forward simulator using an anthropomorphic head phantom. Top row: axial plane. Bottom row: sagittal plane. (a) MDCT images of the head phantom used as input to the simulator; (b) Simulated CBCT images. (c) Real CBCT images from the Medtronic O-arm system. (d) Difference map. (e) Axial plane noise-power spectrum (NPS) for the simulated and real CBCT image. Resolution [edge-spread function (ESF) width] and contrast-to-noise ratio (CNR) for the simulated and real CBCT image were also labeled in (b) and (c).*

-20 HU and +20 HU to +70 HU], size (inner diameter ranging from 8 – 25 mm; outer diameter ranging from 20-60 mm), shape (number of vertices ranging from 3-12), and location (random within the brain parenchyma). Experiment #2 used a dataset featuring real hypodense lesions (~ −30 HU contrast) of edema and ischemia.

### 3 Results

### 3.1. Validation of Training Data Generation
The performance of the high-fidelity forward simulator is summarized in Fig. 4. Side-by-side comparison shows that the simulated CBCT accurately reproduces the measured experimental data acquired with the O-arm system. The high level of agreement is also illustrated by the difference map in (d). Quantitative measurement shows <10% discrepancy in spatial resolution, CNR, and NPS between the simulated and real CBCT images.

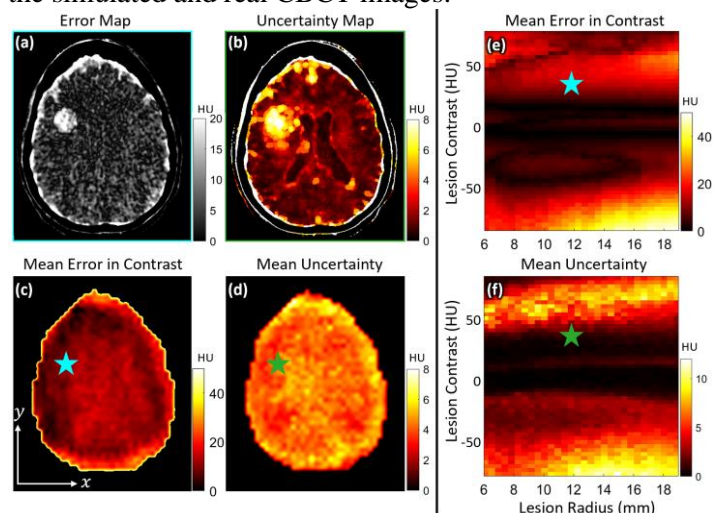

*Figure 5. Correlation between DL-Synthesis error and the statistical uncertainty in $\mu_s$ [Eq. (1)]. (a) Difference between the DL-Synthesis and ground truth for an example dataset in Exp #1 featuring a hyperdense lesion with a contrast of +40 HU. (b) Corresponding uncertainty map ($\sigma$) for the soft-tissue region. (c) Mean error (in lesion contrast) computed as a function of lesion location. (d) Mean uncertainty within the lesion computed as a function of lesion location. Lesion contrast and radius were fixed to +40 HU and 12 mm, respectively, in (c-d). (e) Mean error computed as a function of lesion contrast and size. (f) Mean uncertainty within the lesion computed as a function of lesion contrast and size. Lesion location was the same as (a) for calculations in (e) and (f). Values at each point-pair in (c-d) or (e-f) show results from one dataset [e.g., The point-pair marked with blue-green stars corresponds to the dataset in (a-b)].*

### 3.2. Validation of Uncertainty Estimation
Figure 5 shows the uncertainty of DL-Synthesis for datasets with circular lesions in Exp #1. Taking as an example the dataset featuring a simulated hyperdense lesion (+40 HU contrast) as shown in Fig. 6(a), we see that uncertainty is highest in the region of the simulated lesion, since such a lesion was not present in the training set, leading to errors in the conventional DL-Synthesis prediction [Fig. 5(a)]. Note the correlation between error and uncertainty [Fig. 5(a-b)]. Additionally, the correlation of DL-Synthesis error and uncertainty is shown: (i) as a function of the location of the lesion in Fig. 5(c-d), and (ii) as a function of the size and contrast of the lesion in Fig. 5(e-f). We also observe clear correlation between synthesis error and uncertainty for lesions of different location, size, and contrast – evident in the similar pattern between (c-d) and between (e-f). For example, lesions adjacent to or within the lateral ventricles resulted in greater synthesis error and uncertainty as seen in (c-d). Uncertainty was therefore taken as a reasonable surrogate for regions susceptible to error in the synthesis image.

### 3.3. Experiment #1: Simulated Lesion
Figure 6 shows images reconstructed with conventional methods (FBP, PWLS, and DL-Synthesis) and the proposed DL-Recon methods with uncertainty information (DL-FBP and DL-MBIR) for an example dataset in Exp #1 (simulated hyperdense ICH lesion of +40 HU contrast). Compared to (artifact-corrected) FBP and PWLS, the DL methods show ~50% reduction in noise (at matched spatial resolution measured at the wall of the lateral ventricle) and ~53% improvement in image uniformity. The improved noise-resolution tradeoff of DL-MBIR is shown in Fig. 6(i). Unfortunately, the conventional DL-Synthesis method alone exhibits ~52% reduction in contrast of the ICH lesion (compared to truth), showing a lack of reliability / generalizability for structures unseen in the training data. The DL-FBP and DL-MBIR methods, on the other hand, are significantly more robust against such contrast reduction by weighting the physical measurements in regions of high uncertainty. DL-MBIR shows the expected advantages in noise-resolution tradeoffs compared to DL-FBP.

Figure 6(j) illustrates the importance of the spatially varying uncertainty penalty in DL-MBIR, where the performance of two variants is shown – one in which the penalty varies according to the uncertainty map and one in which $\beta$ is held constant [i.e., no uncertainty information, denoted as DL-MBIR (constant $\beta$)]. With DL-Synthesis as a prior, the spatially varying, uncertainty-based model (DL-MBIR) maintains an accurate representation of lesion contrast via the data fidelity term. Incorporation of the physical model thus compensates for inaccuracies in DL-Synthesis in regions where uncertainty is high. Note that DL-MBIR with constant $\beta$ outperformed DL-Synthesis alone, showing the benefit of combining deep learning with physics models even when the uncertainty information is not available.

Figure 7 shows images reconstructed with conventional methods (FBP and DL-Synthesis) and the proposed DL-MBIR method for an example dataset in Exp #1 (hypodense star-polygon lesion with -40 HU contrast). Utilizing uncertainty information (highest in the hypodense lesion region), DL-Recon was able to mitigate biases introduced by the inaccuracy in DL-Synthesis (~45% improvement in lesion contrast and ~35% improvement in Dice coefficient) while maintaining the improved noise and uniformity characteristics of deep learning methods. The improvement in Dice coefficient for DL-MBIR reflects a higher degree of reliability in imaging pathologies unseen in the training data – an important aspect for many clinical scenarios.

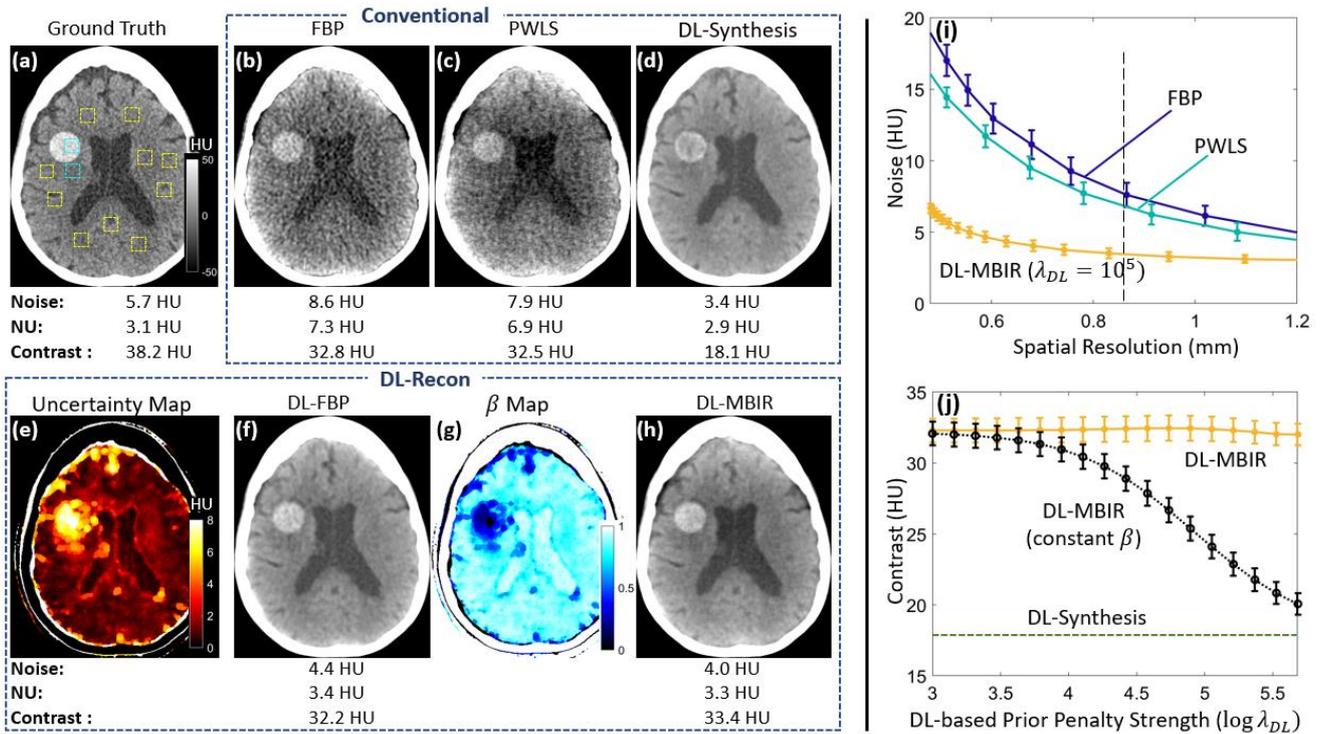

Figure 6. Image reconstructions and analysis for an example dataset in Exp #1. (a) Ground truth consisting of a clinical MDCT scan of the brain with the addition of a lesion of +40 HU contrast (ICH). (b) FBP. (c) PWLS. (d) DL-Synthesis. (e) Uncertainty map for DL-Synthesis. (f) DL-FBP. (g) Spatially varying penalty ("β map") computed by Eq. (3). (h) DL-MBIR. Note the more accurate ICH contrast in (f) [+32.2 HU] and (h) [+33.4 HU] compared to (d) [+18.1 HU] relative to truth (a) [+38.2 HU]. The spatial resolution in images (b, c, d, f, h) was matched at the boundary of the ventricle. The cyan ROIs in (a) were used to measure the lesion contrast, and the yellow ROIs were used to measure non-uniformity (NU) and noise in brain parenchyma. (i) Noise-resolution tradeoff for FBP, PWLS, and DL-MBIR. (j) ICH contrast for DL-MBIR with spatially varying beta and for DL-MBIR with constant β as a function of penalty strength $\lambda_{DL}$. Note that DL-MBIR maintains contrast despite the inaccurate DL-Synthesis prior.

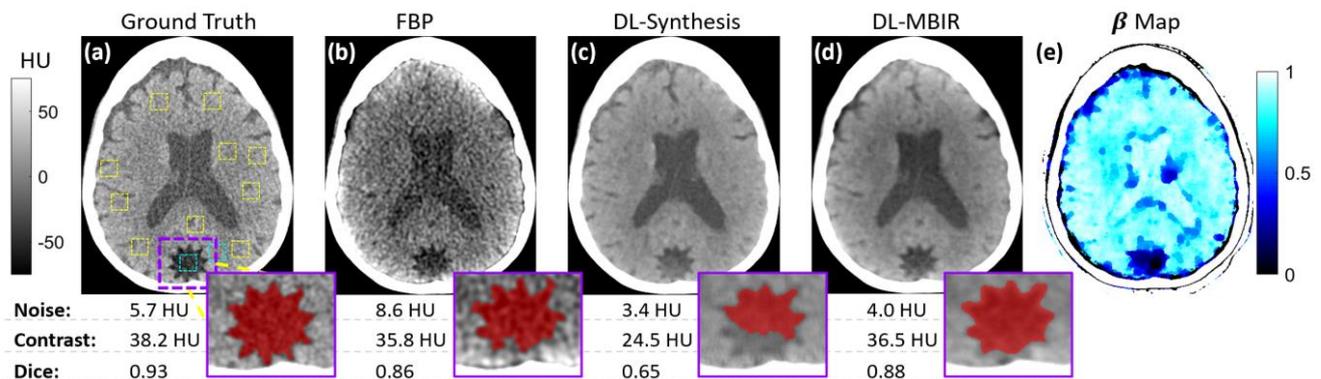

Figure 7. Reconstruction results for an example dataset in Exp #1, featuring a complex shaped (star-polygon) stimulus. (a) Ground truth consisting of a clinical MDCT scan of the brain with the addition of a hypodense lesion (ischemia) of -40HU contrast. (b) FBP. (c) DL-Synthesis. (d) DL-MBIR. (e) Spatially varying penalty ("β map") computed by Eq. (3). Dice coefficients from a threshold-based segmentation (threshold set to achieve the optimal segmentation in the FBP reconstruction) were measured within the purple ROI. Note the more accurate segmentation from DL-MBIR as compared to DL-Synthesis, allowing easier lesion analysis in a clinical workflow

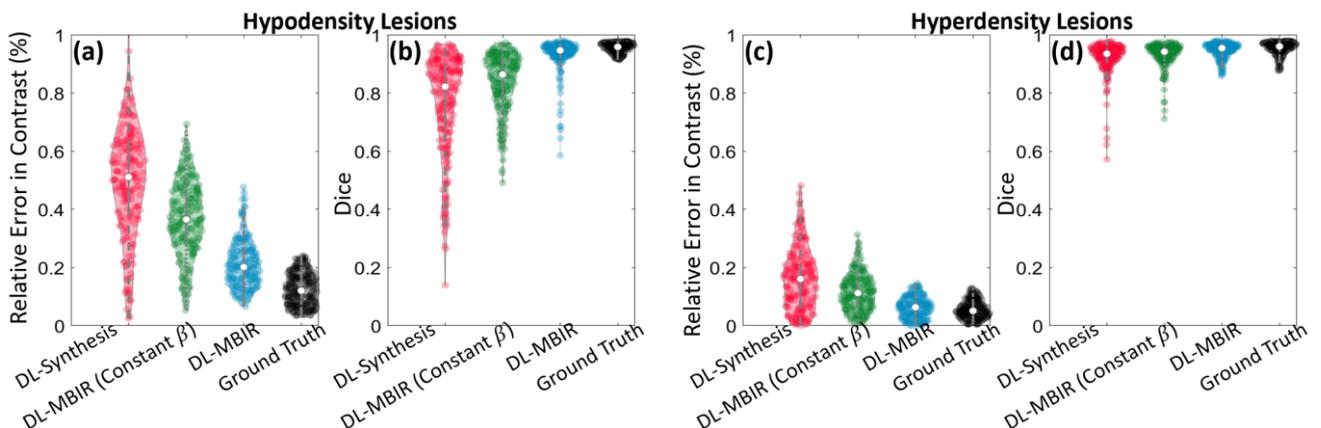

Figure 8. Quantitative analysis of reconstruction accuracy [DL-Synthesis, DL-MBIR (constant β), DL-MBIR, and ground truth] aggregated over all datasets in Exp #1 with star-polygon lesions. (a) Relative error in contrast for hyperdense lesions. (b) Dice coefficient for hyperdense lesions. (c) Relative error in contrast for hypodense lesions. (d) Dice coefficient for hypodense lesions.

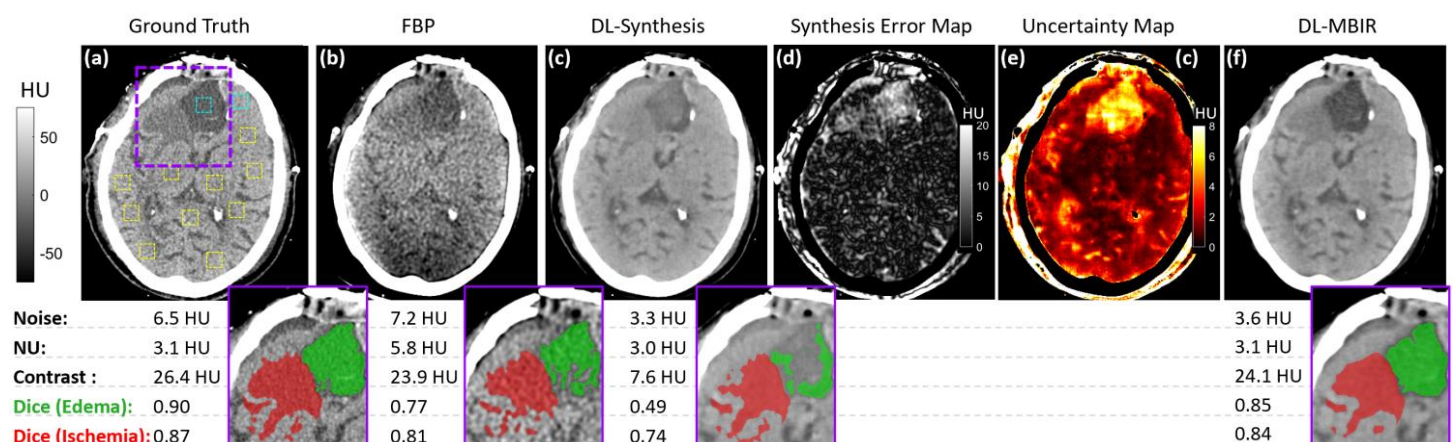

*Figure 9. Reconstruction results for Exp #2. (a) Ground truth consisting of a clinical MDCT scan of the brain with hypodense lesions. (b) FBP. (c) DL-Synthesis images. (d) Error map. (e) Uncertainty map for the DL-Synthesis image. (f) DL-MBIR. Note the improved lesion contrast in (e) and (f) compared to (c) [cyan ROIs]. The spatial resolution of (b, c, f) were matched at the boundary of the ventricle region (giving ESF width = 0.85 mm). Dice coefficients from a threshold-based segmentation were measured within the purple ROI for the edema (green) and ischemia lesion regions (red).*

Figure 8 shows the distribution of relative error in lesion contrast and Dice coefficient for all datasets in Exp #1 with star-polygon lesions. The results in Fig. 8 compare the performance of DL-Synthesis, DL-MBIR with spatially invariant (constant) $\beta$, and DL-MBIR methods. Compared with DL-Synthesis alone, DL-MBIR (constant $\beta$) improved the interquartile range (IQR) of the relative error in contrast and Dice coefficient for the hyperdense lesions by 31% and 10%, respectively, and DL-MBIR (spatially-varying beta) improved these characteristics by 51% and 23%, respectively. The performance of conventional DL-Synthesis was observed to be lower for reconstruction of hypodense lesions. Possible reasons for the decreased performance could lie in the similarity between hypodense lesions and most forms of CBCT artifacts, such as scatter and beam hardening, which tend to present as "hypodense" shading or streaks. With DL-MBIR, the IQR of the relative error in contrast and Dice coefficient for the hypodense lesions were improved by 58% and 68%, respectively, compared to DL-Synthesis.

### 3.4. Experiment #2: Real Pathology

Figure 9 shows images reconstructed with conventional methods (FBP and DL-Synthesis) and DL-MBIR. Compared to FBP, DL-MBIR shows ~50% reduction in noise and ~47% improvement in image uniformity throughout the brain parenchyma. The DL-Synthesis network exhibited highest uncertainty in the region of hypodense lesions, leading to inaccurate representation of the edematous lesion (~42% reduction in Dice; green overlay) and the ischemic lesion (~12% reduction in Dice; red overlay). By including uncertainty information and physics-based reconstruction models, DL-MBIR accurately depicted the contrast and shape of the lesions while maintaining the improved noise and uniformity characteristics of DL-Synthesis.

## 4 Discussion & Conclusion

This work presented a new type of DL-based image reconstruction method (termed DL-Recon) that integrates physics-based models with image synthesis based on epistemic uncertainty. To our knowledge, this represents a novel incorporation of Bayesian uncertainty in a neural network approach with physics-based and DL-based CBCT image reconstruction. Two variations of DL-Recon were proposed in this work, both maintaining the basic advantages of conventional DL-Synthesis: (i) the DL-FBP method improved the accuracy of reconstruction and offers practical advantages of runtime efficiency; and (ii) the DL-MBIR offered further image quality improvement due to the more accurate physical model and the explicit data-fidelity constraint. Compared with DL-Synthesis alone, both of the DL-Recon methods showed improved robustness to anatomical variations (e.g., pathologies) that were unseen in the training set. Besides image reconstruction, other applications of syntehsis uncertainty can be envisioned, for example: helping to identify abnornal anatomy and image features in image classifcaiton; providing a quantitative measurement of sufficiency in the size and/or variety of a training dataset; and helping to quantify improvements (or unexpected variations) in continuous learning. Ongoing work includes extension to fully 3D image reconstruction and investigation in clinical studies.

# Deep Positron Range Correction

Joaquin L. Herraiz[1,2], Alejandro López-Montes[1], Adrián Bembibre[1], Nerea Encina[1]

[1]Nuclear Physics Group, EMFTEL & IPARCOS, Complutense University of Madrid, Madrid, Spain
[2]Health Research Institute of the Hospital Clínico San Carlos (IdISSC)

**Abstract** Positron range is one of the main limiting factors to the spatial resolution achievable with Positron Emission Tomography (PET). Several PET radionuclides such as $^{68}$Ga and $^{82}$Rb, emit high-energy positrons, creating a significant blurring in the reconstructed images. In this work, we have trained a deep neural network (Deep-PRC) with a U-NET architecture to correct PET images for positron range effects. Deep-PRC has been trained with 3D input patches from reconstructed images from realistic Monte Carlo simulations that considers the positron energy distribution and the materials and tissues it propagates into, as well as acquisition effects. The quantification of the reconstructed PET images corrected with Deep-PRC shows that it may restore the images up to 95% without any significant noise increase. The proposed method can provide an accurate positron range correction in a few seconds for a typical PET acquisition.

## 1 Introduction

Positron range (PR) is one of the main limiting factors to the spatial resolution achievable with Positron Emission Tomography (PET) [1–6]. PR makes the spatial distribution of the annihilation points to be a somehow blurred version of the emission points one (Figure 1). Accurate PR modeling is complex, as it depends both on the kinetic energy of the emitted positrons and the electron density of the surrounding tissues.

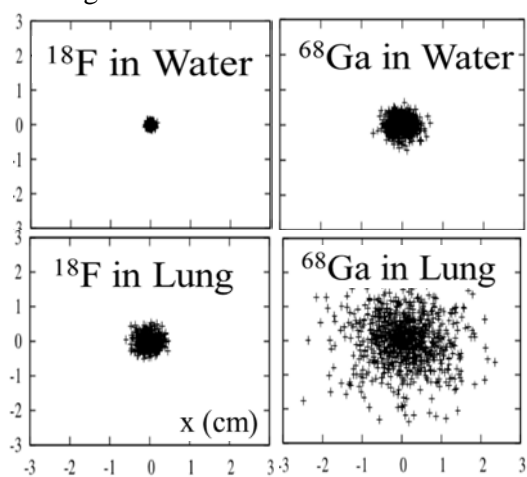

Fig. 1 – Distribution of annihilation points from a centered positron emitter for different radionuclides and tissues.

The quantitative accuracy of PET depends on an accurate positron range correction (PRC). As $^{18}$F, the most widely used PET radionuclide, has a relatively small PR in soft-tissue, accurate PRC for different radionuclides has been neglected to a large extent, and to the best of our knowledge is not explicitely performed yet in standard PET image reconstruction. Nevertheless, the improved resolution of current scanners and the use of radionuclides with large PR such as $^{68}$Ga requires to improve this correction.

Many PRC approaches have been proposed to remove the blurring caused by the PR on PET reconstructed images. They differ on the PR model they are based on and how they are applied in the image reconstruction process. On the one hand, PRC can be applied within the tomographic iterative image reconstruction. PR models can be included into a resolution kernel or Point-Spread-Function (PSF) [7-9], or within the System Response Matrix (SRM) used in the iterative reconstruction. This approach has some important limitations: on one hand, adapting an existing SRM to other radionuclides could be difficult, unless a SRM where PR is factored out is used [10]. Furthermore, fully realistic PR models would require evaluating the SRM in each acquisition. On the other hand, PRC can be applied as a post-processing step to the reconstructed images. In this approach, the reconstructed images before PRC are considered to represent the distribution of the positron annihilations, instead of the distribution of the positron emissions. Therefore, the goal of the post-processing PRC procedure is to convert the annihilation distribution into the positron emission distribution (the one expected to be seen with PET). Post-processing PRC has the advantage of being fast, simple, and to some extent independent of the procedures, algorithms and codes used in the image reconstruction.

Machine Learning (ML), and more specifically the area of ML known as deep learning are having a huge impact on many areas including medical imaging and PET [11]. Deep learning methods are able to create accurate mappings between inputs and outputs by means of an artificial neural network (NN) with a large (deep) number of layers. The fact that the same framework can connect many different inputs such as measurements, raw images, and outputs such as labels, and reference images allows its application in a large variety of problems and disciplines. A recent overview on ML and deep learning for PET imaging can be found in [12]. The components of the NN are learned from example training datasets. In the case of supervised learning, the example inputs are paired with their corresponding desired outputs. After the training, the NN can then be used on new input data to predict their output. Various studies based on convolutional neural networks (CNNs) have been proposed for medical images generation, especially for segmentation, many of them using the U-NET network structure [13]. However, to the best of our knowledge, it has not been applied yet for PRC.

In this work, we propose a deep-learning based PRC method (Deep-PRC) applied as a post-processing step to the reconstructed PET images. Our goal was to develop a fast PRC method for 3D PET imaging, that provides PET images for medium and large range radionuclides rivaling in spatial resolution to the ones reconstructed with the standard short-range $^{18}$F radionuclide. The NN is trained with realistic simulated cases of preclinical studies of reconstructed images of $^{68}$Ga and $^{18}$F corresponding to the same activity distribution, and it is able to produce accurate, and precise $^{68}$Ga PR-corrected images similar to the $^{18}$F ones. As we assume that the image reconstruction method has already incorporated the PRC for $^{18}$F, the goal of this work is to obtain a PRC for $^{68}$Ga that makes it look similar to the corresponding $^{18}$F counterpart, without the risk of double-correcting this effect. The source code of this work is available in Github [14].

## 2 Materials and Methods

Numerical mice models from a repository [17] were used to simulate the different cases of activity, material, and density distributions needed for training, testing and validating the NN. The material composition and density of each tissue in the models were directly obtained from the repository, while different activities were assigned to each tissue type such as heart, liver, kidneys, and tumors using a range of typical values found in $^{18}$F-FDG acquisitions. The numerical models consisted of $154 \times 154 \times 242$ cubic voxels of $0.28 \times 0.28 \times 0.28$ mm.

A total of 8 whole-body mouse models were used to generate with PenEasy (v2020) [15] the positron annihilation distributions from the initial positron emission, material and density distributions. PenEasy considers the path traveled by each positron until its annihilation taking into account their energy distribution and all the materials in the field-of-view. Each model was simulated twice, once for $^{18}$F and once for $^{68}$Ga. The energy distribution of the positrons emitted by $^{68}$Ga and $^{18}$F was obtained with PenNuc [16]. Each simulation consisted of around $3 \times 10^8$ positron emissions simulated at a rate of $3.4 \times 10^4$ histories per second for the $^{18}$F simulations and $2.2 \times 10^4$ histories per second for the $^{68}$Ga simulations in an Intel(R) Xeon(R) CPU @ 2.30GHz computer.

The positron annihilation distributions from PenEasy were used to simulate realistic PET acquisitions in the preclinical scanner Inveon PET/CT scanner [18] using the MC simulator MCGPU-PET [19], a PET-adapted version of the MC-GPU software [20]. MCGPU-PET allows simulating very fast and realistic PET acquisitions from voxelized activity, material, and density distribution. MCGPU-PET simulations contained around $1.2 \times 10^9$ coincidences including scatter and non-scattered true coincidences in a minute ($2 \times 10^7$ coincidences/second) in a computer with a GeForce GTX 1080 8Gb GPU. The output was stored into sinograms, with $147 \times 168 \times 1293$ bins, maximum ring difference of 79, axial compression of 11, and a radial bin size of 0.795 mm.

The sinograms were reconstructed with GFIRST [21], a GPU-accelerated version of FIRST [22], a 3D-OSEM algorithm which allows incorporating a physical model in the SRM. In this case, the SRM used was the standard one created based on $^{18}$F in water. We used 1 subset and 40 iterations. The final images consisted of $154 \times 154 \times 80$ voxels with a size of $0.28 \times 0.28 \times 0.795$ mm, as this is the typical size of the images reconstructed in the Inveon scanner [18]. The total reconstruction time was 50 seconds in a GTX 1080 8Gb GPU. The values of the reconstructed images were converted into standardized uptake value units (SUV) to make it easier to evaluate the performance of the method.

The CNN was implemented in Python within the Tensorflow framework (v 2.3.0) with Keras. It is based on the U-NET network [13] which has demonstrated to be useful in many medical imaging applications [23]. We directly used the U-NET model available in Keras with 4 levels, 64 filters and dropout factor of 0.2. The Swish activation function [24] was used instead of ReLU (except for the final output layer) as it performs better than ReLU with a similar level of computational efficiency. The loss function used was the L1-norm between the ground-truth of the $^{18}$F images and the output of the Deep-PRC network. The trained model was saved as Keras models in hdf5 format. The source code can be found in [14].

In a recent work [25], we trained a NN using several slices from the $^{68}$Ga and μ-map as channels in the input layer, while the output was the corresponding central slice from the $^{18}$F-PET image. In this work, in order to generate a more general NN, instead of slices we used 3D-patches from the $^{68}$Ga (PET) and μ-map (CT) volumes to train the neural network. Each input patch had $32 \times 32 \times 9$ voxels, and the output was the central $16 \times 16 \times 1$ voxels. This approach not only improves the results, and reduce the risk of overfitting, but it also makes it easier to fit the training data into the GPU memory. The input patches were normalized to make their values to be between 0 and 1. This normalization is restored in the output, so that the PRC preserves the appropriate units. We used an NVIDIA RTX 2080 Ti GPU with 11GB memory for the training. The model was trained for 200 epochs with 100 iterations each in around 1 hour.

One simulated case was set aside and not used in the training/validation process, to perform the final test. A quantitative analysis of the resulting image was performed obtaining the mean (μ) and standard deviation (σ) in different organs. The noise was defined as the ratio σ/μ in uniform regions away from any boundary and edges. The recovery coefficients were obtained defining regions over the whole organs, and their values were then normalized respect to the reference reconstruction with [18]F. The differences between the [68]Ga images before and after the proposed PRC can be easily evaluated from the obtained coefficients.

## 3 Results

Figure 2 shows a coronal view of a mouse with the u-map obtained from the CT, and the reconstructed images of [18]F, [68]Ga, and [68]Ga after the PRC. It can be easily seen that the proposed method is able to recover the resolution loss in [68]Ga images with respect to [18]F, and this increases the values in some areas with higher uptake. The time required for obtaining the PRC with the trained model on the whole volume was 2.14s in a RTX 2080 Ti GPU.

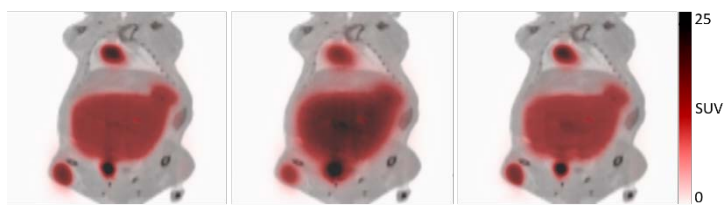

Fig. 2 Reconstructed images of [18]F (left), [68]Ga (center), and Deep-PRC [68]Ga (right), with the corresponding μ-map. The activity distribution in all cases was the same.

Profiles along some organs of interest in the [18]F, [68]Ga and [68]Ga with Deep-PRC images are shown in Figure 3. The significant impact of the PR in these cases is quite clear, as well as the capacity of the proposed method to correct for this effect.

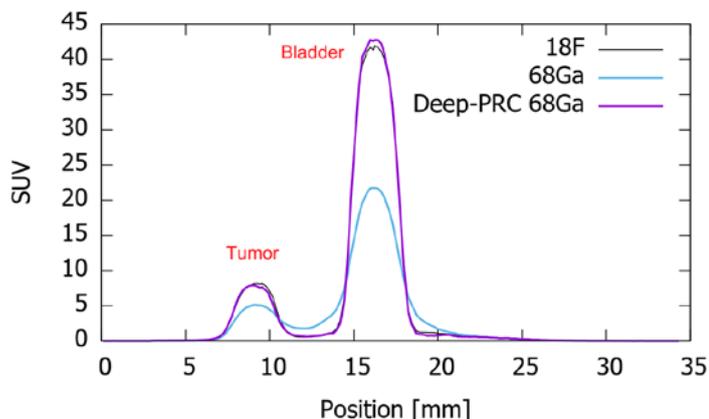

Fig. 3 Profiles along the tumor and the bladder in [18]F, [68]Ga and Deep-PRC [68]Ga images.

The quantitative analysis of the results is shown in Table 1. From the table, it is clear that the [68]Ga images corrected by PR images are very similar to the [18]F images (with recoveries greater than 95% of the reference values).

**Table 1.** Quantitative analysis of the regions.

|  | Recovery (%) | | Noise (%) |
|---|---|---|---|
|  | **Heart** | **Tumor** | **Heart** |
| **[18]F** | 100.00 | 100.00 | 6.0 |
| **[68]Ga** | 67 | 60 | 7.1 |
| **[68]Ga Deep-PRC** | 90 | 97 | 5.5 |

Additionally, the noise level of the estimated images is comparable to the reference one, which indicates that the proposed method does not trade noise for resolution, as it is the case in many deconvolution-based approaches for PRC.

## 4 Discussion

This paper presents the use of a deep convolutional neural network to provide an accurate PRC in PET. The method has been evaluated in simulations in preclinical studies and its performance characterized. To the best of our knowledge, this is the first work that successfully combines deep learning and PRC in a coherent framework.

Our results indicate that overall, the image quality produced by the learned model is comparable to that of the reference images, with recoveries going up from around 60% to more than 95%, while keeping low noise levels.
The training was based on the minimization of the L1-norm between the reference images and the estimated ones, but other loss functions could be explored in this context, including a loss term from an adversarial network (GAN).

We are working on a detailed comparison of the performance of the proposed method with previously proposed ones. In any case, the fact that the proposed deepPRC has no significant impact on the noise level of the images is a clear advantage respect to previous approaches [9]. It is important to note that although we proposed the method as a post-processing step, the same neural network architecture could be used to generate the PR model that can be applied in the forward projection within an image reconstruction (simply by inverting the input and outputs of the NN). We are currently working on this line of research.

In this work we have used the preclinical scanner Inveon and [68]Ga as a reference, but the proposed approach is flexible and suitable for any preclinical and clinical PET systems and with any other radionuclide.

## 5 Conclusion

We have developed and evaluated a deep convolutional neural network (Deep-PRC) that provides a fast and accurate PRC method to recover the resolution loss present in PET studies with radionuclides that emit positrons with large PR. We demonstrated its quantitative accuracy in realistic simulations of preclinical PET/CT studies with $^{68}$Ga. The correction of PR effects in PET image reconstruction is becoming mandatory due to the increasing use of high-energy positron emitters in preclinical and clinical PET imaging and their improved spatial resolution. CNN are very suitable for this type of correction.

## 6 Acknowledgements

We acknowledge support from the Spanish Government (RTI2018-095800-A-I00) and RTC2019-007112 (XPHASE-LASER), from Comunidad de Madrid (B2017/BMD-3888 PRONTO-CM), and NIH R01 CA215700-2 grant.

# Row Interpolation in Spiral CT with Deep Learning

Jan Magonov[1,2,3], Marc Kachelrieß[1], Eric Fournié[2], Karl Stierstorfer[2], Thorsten Buzug[3,4], and Maik Stille[3,4]

[1]Division of X-Ray Imaging and CT, German Cancer Research Center (DKFZ), Heidelberg, Germany
[2]Siemens Healthcare GmbH, Forchheim, Germany
[3]Institute of Medical Engineering, University of Lübeck, Lübeck, Germany
[4]Fraunhofer Research Institution for Individualized and Cell-Based Medical Engineering, Lübeck, Germany

**Abstract** Spiral computed tomography is a standard procedure in clinical diagnostics. Besides a faster measurement time, as compared to conventional CT, it can acquire 3D volumes. However, the limited sampling in the axial z-direction can lead to reconstruction artifacts. These spiral artifacts are referred to as windmill artifacts due to their characteristic appearance. Available methods to increase the sampling in z-direction, such as the *z-Flying Focal Spot* (zFFS), are technically intricate. This work aims to interpolate CT detector rows using a neural network trained with projection raw data from clinical patient images. The presented approach is abbreviated as the acronym RIDL (Row Interpolation with Deep Learning). In addition to analyzing the interpolation results with single projection data, the method was validated in the image domain. For this purpose, a reconstruction algorithm was applied to the output generated by the RIDL network and compared with data sets using zFFS and linear interpolation. Although the zFFS cannot be entirely replaced by the presented method, it was shown that the sampling in z-direction can be increased with RIDL while achieving better results than with linear interpolation.

## 1 Introduction

Since the introduction of spiral CT in 1989 [1], the method established itself as a standard procedure in modern clinical diagnostics. A characteristic feature is the spiral trajectory of the X-ray tube, caused by continuously advancing the patient table through the gantry. In comparison to conventional CT, faster measurement times and acquisition of three-dimensional volumes are enabled [2]. Further, spiral scans are less sensitive to motion artifacts. However, as with conventional CT, reconstruction artifacts may occur.

The windmill artifact is an image distortion in the axial plane that occurs in spiral multidetector CT scans. It is characterized by bright streaks diverging from a focal high-density structure, e.g. from a bone. The streaks appear to rotate while scrolling through the affected slices of a volume. Figure 1 shows an example of this type of artifact. Windmill artifacts are caused by inadequate data sampling in the z-plane since several detector rows cross the reconstruction plane with each rotation of the gantry [2]. Mathematically this arises from not satisfying the Nyquist-Shannon sampling criterion, which states that at least two samples per detector pixel should be recorded.

One method used to reduce the windmill artifact is the z-flying focal spot (zFFS). The zFFS enables the periodic movement of the X-ray focal spot in longitudinal direction to double the data sampling along the z-axis such that the above-mentioned Nyquist-Shannon sampling condition is met [2].

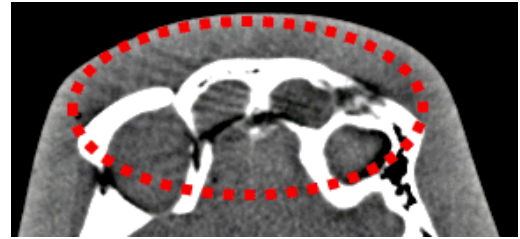

**Figure 1:** Representation of a slice from a CT scan of a skull phantom without applying artifact reduction methods. A windmill-shaped spiral artifact is clearly visible in the circled area.

The zFFS requires a specially manufactured X-ray tube that allows for alternating the focal spot between two positions on the anode surface by permanent electromagnetic deflection. The deflection amplitude is set so that the detector's two consecutive measured longitudinal sampling positions are shifted by half a detector row width in z-direction [2]. This doubles the effective sampling rate of the CT system while maintaining the same collimation. The resulting higher sampling rate in z-direction removes the aliasing in the acquired data and thus the windmill artifacts. However, the zFFS also has some disadvantages as it is technically intricate and expensive. Furthermore, it cannot be applied to CT systems that lack the technical requirements. Moreover, the zFFS requires to double the readout rate of the detector and thus may not be applicable for scan modes that run at maximum speed.

In recent years, deep neural networks have achieved remarkable results in the field of super-resolution, which aims at transforming input images into high-resolution versions [3]. The application of such neural networks has also been investigated to improve the resolution in CT images. However, previous works such as [4] mostly focus on the super-resolution of the reconstructed image. Only a few earlier works have explored the approach of super-resolution or super-sampling in the projection domain as well [5].

In this paper, we propose the RIDL network, which is similar to the zFFS, designed to double the effective number of acquired detector rows in projection domain. The network was trained on raw projection data from clinical CT scans, and the row interpolation was then compared with a linear interpolation in the projection and image domain.

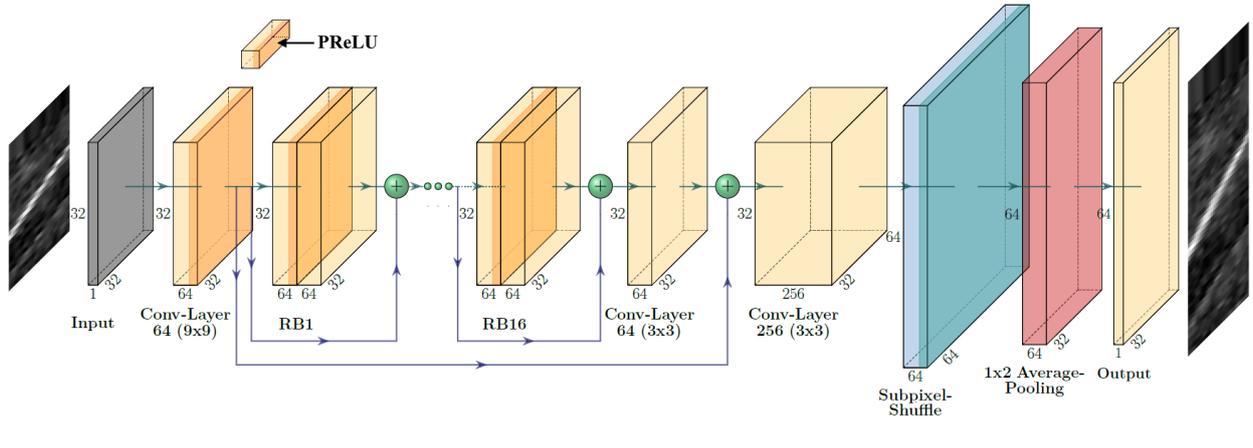

**Figure 2:** Structure of the RIDL network. It contains 16 residual blocks (RB) at its core, followed by a subpixel convolution (labeled blue). An $1 \times 2$ average pooling operation (labeled red) was added to halve the number of channels again.

## 2   Materials and Methods

### 2.1   Basic concept

The basic concept of the RIDL method is to train a neural network that receives input projection raw data and predicts the correct representation of rows in-between, i.e. to perform an upsampling of the input data to double the number of detector rows. The projection data used for the training process are divided into alternative rows, so the projections containing all rows (acquired with zFFS) are defined as the desired output $y$, which is the ground truth, and every second row out of the corresponding projections is used for the network input $x$. It is therefore trained in a supervised manner. The error between the network output and the ground truth is calculated by a loss function and the parameters of the network are adjusted via backpropagation to ideally calculate an absolute minimum of the loss function iteratively. The final trained network should receive a projection image $x$ with a number of $M \times L$ detector pixels as input, where $M$ describes the number of channels and $L$ the number of rows. Subsequently, the network should output a projection image $\hat{y}$ with $M \times (L \times 2)$ detector pixels according to the input.

### 2.2   Data preparation

To achieve the highest possible generalization of the network on unseen clinical data, raw projection data of clinical CT scans from different body regions were selected. The data set includes a total of 29 CT scans, which predominantly cover the body regions head, thorax, and abdomen. The individual scans were acquired with different Siemens CT systems, such as the SOMATOM Force, SOMATOM Definition Flash, and SOMATOM Definition AS/AS+. Projections were obtained after the rebinning, which is the rearrangement of the measured fan-beam data to parallel geometry. The data set was divided into two disjoint subsets for training the neural network: 24 of the CT scans were used as training data set and five scans as testing set. It was ensured that the acquired body regions were equally distributed in both data sets. Preprocess-ing steps were performed on the data sets prior to reading the data for training the RIDL network. Utilizing data augmentation, instead of the full projection data, randomized image patches were selected from the raw data. This procedure can significantly increase the effective number of training and test samples for the network. Furthermore, all values of the resulting patches were normalized globally to lie between 0 and 1.

### 2.3   Network architecture

The proposed architecture of the RIDL network is a modified version of the SRResNet presented by Ledig et al. in [6]. The network is designed to compute super-resolution images, i.e. very high-resolution images. For this purpose, the network is trained to receive low-resolution (LR) images as input, and the corresponding high-resolution (HR) images are defined as ground truth. In [6], HR images were transformed to LR images using bicubic downsampling to generate the training pairs. In the context of this work, however, the downsampling of the input patches should only take place in row-direction. For this purpose, image patches of size $64 \times 32 \times 1$ were generated for the network training from the training data set and defined as ground truth. As already mentioned, according to these patches, every second row was used as input data for the network so that the input had a size of $32 \times 32 \times 1$. The structure of the RIDL network is presented in Figure 2. The network is structured in such a way that an input layer is followed by a convolutional layer with 64 filters and a $9 \times 9$ kernel. The core of the network is formed by 16 residual blocks. Each of these blocks consists of two convolutional layers with 64 filters, $3 \times 3$ kernels and a parametric rectified linear unit (PReLU) as the non-linear activation between the convolutional layers. The residual blocks are followed by a convolutional layer with the same features of the previous layers, whose output is connected to the input of the first residual block by a final residual layer. In the following step, an upsampling of the input is performed based on the obtained feature maps, which is another spe-cial feature of the SRResNet. The network uses a so-called subpixel convolutional layer introduced by Shi et al. in [7]. This layer essentially uses the feature maps from previous layers followed by a specific type of image reshaping called phase shifting [7]. Instead of putting zeros between pixels and performing additional computations as in conventional upsampling, a subpixel convolution computes multiple convolutions at lower resolution and resizes the resulting feature map to an upscaled image. An upsampling ratio of $r = 2$ was used in the context of this work. Furthermore, the network architecture was extended with a $1 \times 2$ average pooling layer to halve the output of the subpixel convolution in the horizontal direction. This is necessary because otherwise both spatial dimensions would be increased by a factor of $r = 2$.

## 2.4   Implementation and training details

The proposed RIDL network was implemented using TensorFlow 2.2.0 in a Python 3.7.6 environment and training was performed using an NVIDIA RTX 2070 graphic unit with 32 GB RAM. A data set consisting of 500,000 examples from the training data set was used for the network training. Another 125,000 test examples from the test data set were used for network validation. The entire 1,377,921 trainable parameters of the network were initialized with random weights and the adaptive moment estimation algorithm (ADAM) was used to update the parameters during the training by minimizing the underlying loss function. In this work, a combined loss function was used, which takes into account the pixel-wise computed error between the network output $\hat{y}$ and ground truth $y$ by the mean absolute error (MAE) but also the structural similarity between the two images by the multi-scale structural similarity index (MS-SSIM) [8]. This combined loss function $L_{\text{comb}}$ was proposed in [9] and can be described by

$$L_{\text{comb}}(y, \hat{y}) = \alpha \cdot (1 - L_{\text{MS-SSIM}}(y, \hat{y})) + (1 - \alpha) \cdot L_{\text{MAE}}(y, \hat{y}),$$

where $\alpha$ is used to weight the terms of the function. We used the empirically determined value $\alpha = 0.84$ from [9]. The initial learning rate was set to $1 \times 10^{-5}$ and was halved during the training process once the validation error could not be minimized after 25 consecutive epochs. An early-stopping regularization was used to stop training the network if the validation error could not be reduced after 100 epochs. In total, the network was trained for 729 epochs with a batch size of 128, which corresponds to 3,907 update steps of the network parameters per epoch.

## 2.5   Evaluation and validation

The interpolation performance of the trained RIDL network was first evaluated on single projection data. Therefore an algorithm iterated over all projections from the test data, replacing every other row once with the RIDL network and once with a linear interpolation. The results were compared with the ground truth by calculating the error measures PSNR and structural similarity index measure (SSIM) in comparison to the ground truth. One problem that arises when directly comparing the interpolated images is that for an input image with dimension $m \times n$ and even number of rows $n$, only a maximum of $n - 1$ rows can be linearly interpolated. For this reason, the last row of each input image was removed for the comparison. The same procedure was used for the output projection, which was generated by the RIDL network.

| Method | $\overline{\text{PSNR}}(y, \hat{y})$ | $\overline{\text{SSIM}}(y, \hat{y})$ |
|--------|--------------------------------------|--------------------------------------|
| LI     | $62.894 \pm 5.376$                   | $0.9989 \pm 0.0009$                  |
| RIDL   | $63.463 \pm 5.498$                   | $0.9991 \pm 0.0008$                  |

**Table 1:** Quantitative results of row interpolation on all test projections with the RIDL network compared to a linear interpolation (LI). Error measures are given accordingly as mean values with standard deviation.

Finally, the row interpolation with the RIDL network was applied to reconstruct a CT scan for validation. For this purpose, a spiral CT of a skull phantom was acquired with a Siemens SOMATOM-Force CT system. The zFFS was enabled and the spiral pitch was set to $p = 1$, as it has been shown that more substantial spiral artifacts occur at this value compared to a lower pitch factor [2]. This acquisition was defined as ground truth. A plugin for the Siemens-specific reconstruction software was then implemented. Using the plugin, every second row of the zFFS-generated projection data was replaced by linearly interpolated rows or rows interpolated by the RIDL network. The results of all reconstructions were compared qualitatively and quantitatively with the error measures in the final step.

## 3   Results

### 3.1   Results in projection domain

Table 1 shows the quantitative results of the row interpolation on the total amount of test projections with the RIDL network compared to a linear interpolation. It can be observed that the RIDL network can increase the mean measures for PSNR and SSIM. Thus a better interpolation related to the ground truth was achieved compared to a linear interpolation of the rows. A relatively high standard deviation of the PSNR values is noticeable, which indicates that the individual projections have varying degrees of interpolation difficulty.

### 3.2   Results in image domain

Concerning this work's aim, the validation of the RIDL network with reconstructed CT images is most important. Only in the image domain it can be determined whether the method affects the occurrence of spiral artifacts. Figure 3 compares reconstructions of a specific slice, with differently modeled rows of the projection data, to the reconstruction with zFFS-generated projection data. While comparing the LI reconstruction to the RIDL reconstruction one can observe that the RIDL reconstruction provides the best result compared to the ground truth. This observation is confirmed by the quantitative results presented in Table 2. Furthermore, an influence of the RIDL method on the occurrence and distinctness of windmill artifacts could be noticed. A windmill artifact can be observed in the nasal bone region and the nasal cavity from the difference images of the LI reconstruction. In comparison, this artifact is less noticeable in the difference image for the RIDL reconstruction.

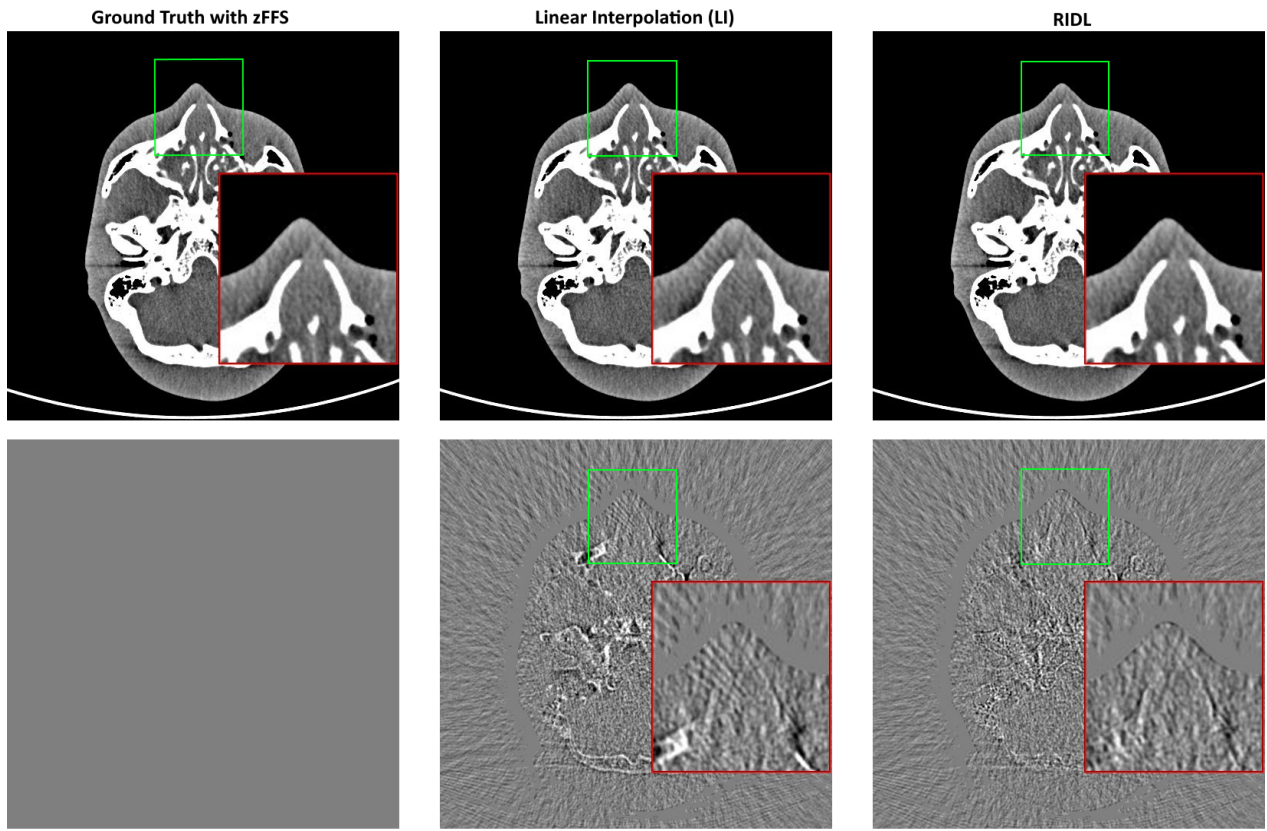

**Figure 3:** Representation of a reconstructed slice with the respective row adjustments LI and RIDL on the projection data and corresponding difference images to the ground truth. The reconstructions are windowed in $[60, 360]$ HU and the difference images to the ground truth in $[0, 150]$ HU.

| Method | RMSE [HU] | PSNR | SSIM |
|--------|-----------|--------|--------|
| LI | 14.036 | 49.303 | 0.9913 |
| RIDL | 12.920 | 50.022 | 0.9920 |

**Table 2:** Comparison of quantitative results for the reconstructed slice with different row adjustment of projection data.

## 4 Discussion and Conclusion

In this work, we proposed the RIDL network, which is designed to double the effective number of detector rows in acquired projections before image reconstruction to improve the sampling in z-direction and prevent windmill artifacts.

Our results can be considered a proof of concept that applying a neural network can meet the requirements of increasing the sampling of underlying projection data and has a positive impact on the prevention of windmill artifacts in spiral CT reconstruction. Future work will focus on the adaption of better network architecture, particularly the subpixel convolution, to the underlying task. The use of generative adversarial networks (GANs) is also conceivable. In general, however, the focus should be on simple, lightweight networks that are easy to integrate into a CT system. Another point is that the network results so far have only been compared with a simple linear interpolation. However, using a linear interpolation to simulate zFFS generated upsampling in z-direction is not a concrete approach to improve longitudinal sampling, since a linear interpolation is performed in each backprojection anyway. Replacing the zFFS-generated rows in the projection data with a linear interpolation should thus be equivalent to omitting the zFFS. Therefore, it is necessary to investigate more advanced interpolation methods and compare them with the network results. If a suitable method can be discovered that gives satisfactory results, the interpolated projections with this method could act as a better initialization of the input data for the network to further improve the results. Furthermore, it would be interesting to investigate whether the results could be improved with a larger amount of training data and the use of simulated data.

# Chapter 13

# Poster Session 3

**session chairs**
Klaus Mueller, *Stony Brook University (United States)*
Emil Sidky, *University of Chicago (United States)*

# Reconstruction of difference as a framework for spectral de-noising of photon-counting CT images

Thomas Wesley Holmes[1], Stefan Ulzheimer[2], and Amir Pourmorteza[1,3,4]

[1]Department of Radiology and Imaging Sciences, Emory University, Atlanta, GA, USA
[2]Siemens Healthcare, Forchheim, Germany
[3]Department of Biomedical Engineering, Georgia Institute of Technology - Emory University, Atlanta, GA, USA
[4]Winship Cancer Institute, Emory University, Atlanta, GA, USA

**Abstract**

Photon-counting CT is an emerging technology that provides many advantages over the conventional energy-integrating CT systems. One of these advantages is the availability of spectral information in the form of 4-6 energy bins. Since each energy bin includes a fraction of the entire detected photons, the bin images may suffer from noise and photon starvation. Many algorithms have been proposed that use prior information in order to reconstruct current low-fidelity CT projection data such as PICCS and PIRPLE. We have proposed a penalized likelihood approach that directly reconstructs the difference between prior and current data called Reconstruction of Difference (RoD). Direct regularization of the difference image may be advantageous in applications such as spectral material decomposition, where the difference between energy bins is important. Here we propose a spectral RoD framework to reconstruct each energy bin using the projection of entire photons as the prior.

## 1 Introduction

Many algorithms have been proposed that use prior images of a patient in order to reconstruct current CT projection data such as prior image constrained compressed sensing (PICCS)[1] and Prior Image Registration, Penalized-Likelihood Estimation (PIRPLE)[2].
Recently we have proposed and tested a penalized likelihood (PL) approach that directly reconstructs the difference between prior and current data called Reconstruction of Difference (RoD)[3], [4]. Unlike PIRPLE and PICCS where the prior information is used to regularize the current image, RoD incorporated the prior information in the data-fit term of the PL cost function. Direct regularization of the difference image may be advantageous in applications such as spectral material decomposition, where the difference between energy bins is important.
Major CT manufacturers have developed photon-counting CT prototypes, which can measure the energy of the detected photons into a certain number of energy bins (usually 4 to 6)[5]–[7].

Here we propose a spectral RoD framework which uses the projections made from all the detected photons as the prior, in order to reconstruct each energy bin of photon-counting CT data. Edge-preserving Huber norm of the difference image was used for regularization, and the penalty weight was chosen by exhaustive search.

## 2 Materials and Methods

**Spectral Reconstruction of difference**

We adopted the reconstruction of difference algorithm [8] reconstruct spectral bins. We consider the forward model as follows:

$$\overline{y}_i = b_i \cdot \exp(-[\mathbf{A}\mu]_i)$$

where $b_i$ is a gain term of unattenuated photons and detector gain of one energy bin, $\mu$ is vector of attenuation coefficients representing the energy bin image, $\mathbf{A}$ is the system matrix, and $[\mathbf{A}\mu]_i$ is the line integral associated with the $i^{th}$ measurement, and $y_i$ is independent and Poisson distributed measurements in the energy bin. The current energy bin image can be modeled as the sum of a prior image, $\mu_p$, and a difference image, $\mu_\Delta$ such that: $\mu = \mu_p + \mu_\Delta$.
We choose the projection data from all detected photons at all energy bins as the prior due to its higher fidelity compared to individual energy bins. The forward model can be rewritten as

$$\overline{y} = b \cdot \exp(-\mathbf{A}\mu_p) \cdot \exp(-\mathbf{A}\mu_\Delta),$$

Assuming the following term is independent of the current measurement,

$$g = b \cdot \exp(-\mathbf{A}\mu_p)$$

we have a new forward model that uses measurements of one energy bin in order to reconstruct the difference between that bin and the image reconstructed from all the detected photons.

$$\overline{y} = g \cdot \exp(-\mathbf{A}\mu_\Delta),$$

We chose the following cost function to solve for the difference image:

$$\hat{\mu}_\Delta = \arg\min_{\mu_\Delta}\{-L(\mu_\Delta; y, \mu_p) + \beta_R\|\Psi\mu_\Delta\|_1\}$$

where L represents log-likelihood, and the second term is a roughness penalty with edge-preserving Huber norm; $\Psi$ is a local pairwise voxel difference operator whose strength is controlled by $\beta_R$.

We used a separable quadratic surrogate method to optimize the cost function. The difference image was then added to the prior image to achieve de-noised energy bin images.

**Test subjects and photon-counting imaging protocol**

We compared the performance of spectral RoD to filtered backprojection (FBP) algorithm. We imaged a 20-cm water phantom with vials containing multiple concentrations of iodine-, gadolinium-, and bismuth-based contrast agents as well as samples of ferrous sulfate, hydroxyapatite, and fat. The phantom was imaged at: 140 kVp, 300 mAs, 1 s rotation time, with 4 energy thresholds: 25/50/75/90 keV on the Somatom CounT prototype PCCT scanner (Siemens, Germany). More information regarding the scanner can be found in [9]. In addition we reconstructed dual-energy PCCT brain scans a healthy human volunteer acquired at 120 kVp with thresholds set at 25/52 keV.

## 3 Results

Figure 1 and 2 summarize the findings. Joint histogram of bin1 and bin2 shows reduced image noise, while the slopes of the lines are preserved; indicating that the RoD did not introduce a bias in attenuation of images.

We also analyzed projection data from a brain scan of a human volunteer acquired at 120 kVp, 370 mAs, and two energy thresholds at 22/52 keV. Figure 3 shows the improvements in image quality with RoD.

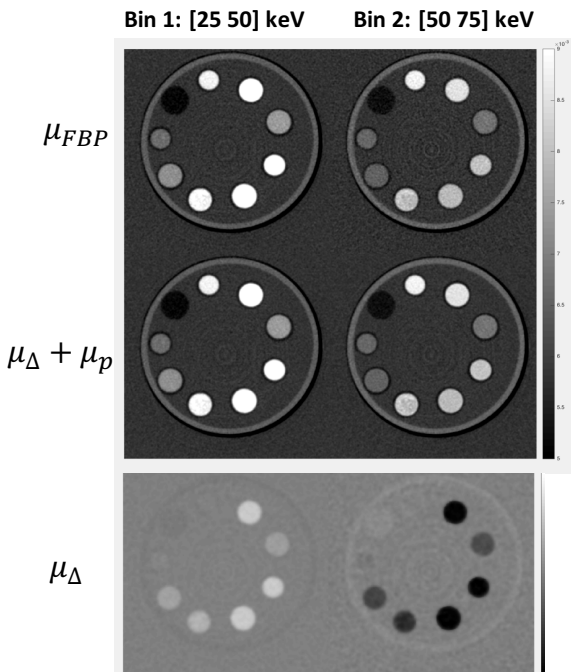

Figure 1- FBP and model-based RoD images reconstructed from photon-counting CT scan of a water phantom with vials filled with multiple contrast agents.

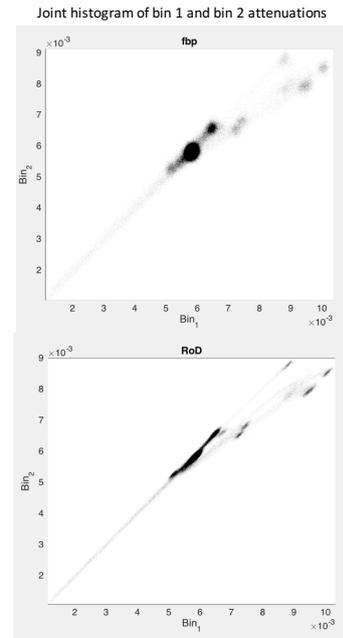

Figure 2- Joint-histograms of attenuation coefficients of bin 1 and bin 2 images reconstructed with FBP and RoD of the phantom in Figure 1.

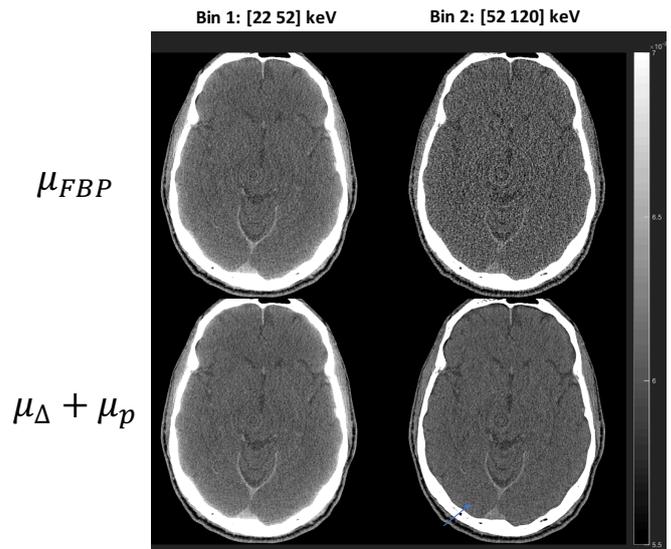

Figure 3- Energy bin images of a dual-energy photon-counting brain CT of a healthy volunteer, reconstructed with FBP (top row), and RoD (bottom row).

## 4 Discussion

We tested performance of RoD on data from phantoms and humans imaged on a prototype whole-body photon-counting CT scanner with 2 to 4 energy thresholds. The early results indicate significant noise reduction is achievable, while preserving the spectral fidelity of attenuation coefficients with RoD. Future work will include optimizing the penalty weight and joint reconstruction and noise matching of multiple energy bins.

# Two extensions of the separable footprint forward projector

Tim Pfeiffer[1,2], Robert Frysch[1,2], and Georg Rose[1,2]

[1]Institute for Medical Engineering, University of Magdeburg, Germany
[2]Research Campus *STIMULATE*, Magdeburg, Germany

**Abstract** Voxel-based forward projectors play an important role in iterative CT reconstruction, since they provide a suitable candidate for an adjoint pair of forward and backprojector with reasonable computational efficiency. A prominent example is the separable footprint technique, proposed by Long et al. However, some of the incorporated optimizations introduce constraints on the geometry the method can be applied to without loss in accuracy. Here, we present two extensions of the TT footprint approach that 1) allow generating accurate projections in a broader range of acquisition geometries, and 2) make use of sparse representations of an input volume to grant a huge increase in computation speed. Based on a series of simulation experiments, we demonstrate that accurate projections can be computed for geometries including tilted and twisted detectors with the proposed extension. Its sparse counterpart is shown to provide a speed-up of up to factor 65 in specific situations. The increased number of possible geometries can be of importance in many practical scenarios involving imperfect acquisition geometry, for example, when facing miscalibration, patient motion, or unconventional (non-circular) scan trajectories.

## 1 Introduction

Generating artificial X-ray images—typically called forward projections (FP)—is an important aspect of various topics in computed tomography, particularly in the field of iterative image reconstruction (IR). Usually, FP algorithms make a well-defined compromise between flexibility, accuracy, and computational effort. A variety of methods exists, all differing in their specific position within this "compromise phase space". In IR, a further aspect comes into play, as FP methods should ideally also have an adjoint backprojector (BP).

A well-known pair of adjoint FP/BP is provided by the separable footprint (SFP) technique proposed by Long et al [1]. SFP combines the efficiency of a voxel-driven BP with a FP that provides proper accuracy without excessive computational demand. To achieve that goal, an approximation is used that describes the shape of the footprint (i.e. the "shadow imprint" a single voxel leaves on the detector) as a multiplication of two separate functions in both detector dimensions. Part of its algorithmic efficiency comes from clever optimizations within the SFP routine that utilize properties found in typical acquisition geometries. However, these properties might not always be fulfilled under practical conditions, thus, limiting the usability of said approach in those situations—or accepting the unavoidable loss in accuracy.

Here we present an extension of the so-called TT footprint (trapezoidal shape in both dimensions) that allows for a broader range of geometries, while keeping the additional computational effort as limited as possible. Based on simulation experiments, we demonstrate that geometries containing a tilted and/or twisted detector can be properly handled by the proposed approach for angles up to 40 degrees. We investigate the difference to the unmodified version and discuss potential applications of the new generalized routine.

Another important advantage of voxel-based FPs is the possibility to apply them to an arbitrary subset of voxels from the original volume. Ray casting techniques can hardly make use of similar approaches without restricting themselves to highly specialized data representation such as surface models, which are typically poorly suited to describe medical datasets. In the second part, we show how the proposed footprint-based FP can be used efficiently to create projections from a sparsely-populated volume to substantially reduce computation times. We demonstrate the efficiency and accuracy of the routine on an artificially generated vessel tree phantom and investigate the achievable speed-up for different volume sparsity levels.

## 2 Materials and Methods

### 2.1 Generic TT footprint

With respect to space limitations, the original TR/TT footprint approaches will not be introduced here in detail. The interested reader is referred to the original publication [1] by Long et al. The two central properties of the algorithms that need to be considered here are: a) the assumption that the s-component of a footprint is independent of the voxel's z-coordinate (cf. Fig. 1), and—even more crucial— b) the approximation of full separability of the footprint in the two detector dimensions. A trivial solution for both issues would be computing everything on-the-fly for each detector pixel (and each voxel) individually. This, however, implies huge computational effort, which would render the method unsuitable for use in IR. Instead, the proposed extension aims at keeping as many of the simplifications/optimizations as possible, but still provide proper results when the originally-made assumptions are violated. Before going into detail of how a) and b) are addressed, one important concept of the footprint computation should be recapitulated here; it consists of two steps. Step 1: The profile of the footprint is described by a simple geometric shape in each direction (rectangle or trapezoid). This geometric shape can be characterized by supporting points (e.g. four points for a trapezoid: starting, plateau, and end points). Roughly speaking, these can be computed via forward projection of appropriate points on the voxel outline (i.e. edges or corners) and subsequent sorting of the results to map them to the corresponding supporting point of the shape. Step 2: To get the actual contribution to a specific detector pixel, one then needs to compute the integral of the shape function over the pixel extent.

Issue a) can be tackled rather simply, by computing the footprint in $s$-direction individually for each voxel in $z$-direction. Additionally, the points on the voxel that define the supporting points need to be refined. Instead of simply using the center positions on the four edges (parallel to $z$), one needs to choose points that are projected to the same $t$ coordinate on the detector (note that this was trivially fulfilled when $t$ was parallel to $z$). This is done by projecting the voxel center and using the resulting $t$ coordinate as reference $t_{\text{ref}}$. For projecting the four voxel edges, $t_{\text{ref}}$ is then used to determine which $z$ position on the edge needs to be chosen ($x$ and $y$ remain the same):

$$z(t_{\text{ref}}; x, y) = t_{\text{ref}}^{-1}(P_{21}x + P_{22}y + P_{24}) - \frac{P_{31}x + P_{32}y + P_{34}}{P_{33} - t_{\text{ref}}^{-1}P_{23}},$$

where $\mathbf{P} \in \mathbb{R}^{3 \times 4}$ denotes the projection matrix for the view to be computed and $\mathbf{r} = (x, y, z)^T$ is the world coordinate of the sought-for point on the voxel edge.

Dealing with issue b) requires introducing some kind of coupling between the $s$- and $t$-dimension. As mentioned earlier, computing the entire $s$-footprint for each $t$ would be far too expensive. Instead, we make use of the fact that—with some limitations at the lower and upper end of the voxel—the shape of the $s$-footprint still remains the same for the entire extent of the footprint. The only adjustment that needs to be done is a shift in position on the $s$-axis. This shift follows a simple linear relation: $s_{\text{shift}}(t) = (t - t_{\text{ref}}) \cdot \Delta s$, where $\Delta s$ denotes the required shift in $s$ direction per $t$, which corresponds to the slope of footprint's vertical edge. We compute $\Delta s$ by projecting the lower and upper end of a voxel edge and computing the slope: $\Delta s = \frac{s_2 - s_1}{t_2 - t_1}$. In principle, the full $s$-footprint—that means including integration of contributions to the detector pixels—could be precomputed in the step described in a); however, due to the required shift (typically a fraction of a pixel) in position, resampling of the contributions would become necessary (Step 2). This introduces potential inaccuracies and, interestingly, experiments (not provided here) showed that on-the-fly computation of the integrated values did even end up in faster computation times on the GPU than using precomputed values along with linear interpolation. Consequently, we decided to compute the integrated values on-the-fly for each $t$. Note that the supporting points for the $s$-direction trapezoid (Step 1), which are rather expensive due to four projection operations and one sorting step being involved, need to be computed only once for each voxel.

## 2.2 Sparse projector

As described earlier, voxel-based projectors can directly operate on any subset of voxels from the original volume and decrease computation times accordingly—because there are simply fewer voxels to process. Voxel data (i.e. coordinates and attenuation) can be stored efficiently within the RGBA channels of a GPU image object (`cl::Image1DBuffer`, where RGB components hold the voxel position in world coordinates (i.e. $x$, $y$, and $z$) and the alpha channel contains its actual attenuation value. This allows us to retrieve all four values with a single GPU read command. The remainder of the algorithm remains identical to non-sparse versions of the projector, granting free choice between TR, TT, and generic TT versions. However, most of the optimizations in the original TR/TT projection methods would not result in substantial benefit in case of an arbitrary sparse subset of voxels. Hence, we opt for the most flexible option and use the proposed generic version of the TT method as base.

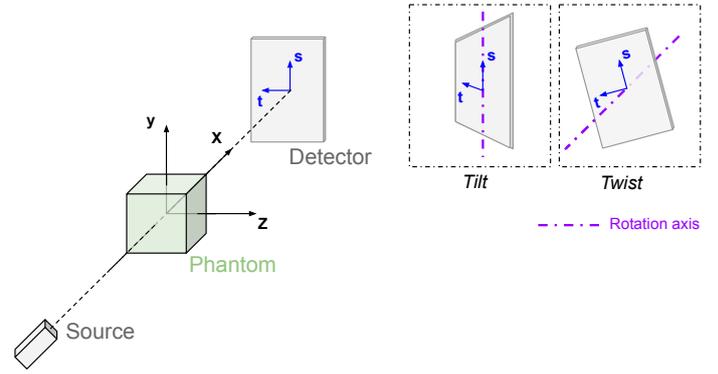

**Figure 1:** Schematic view on the undistorted geometry setting used in detector tilt and twist manipulation experiments. Projections are acquired along the $x$-axis of the world coordinate system (WCS); detector axes are aligned with the WCS $y$- and $z$-axis.

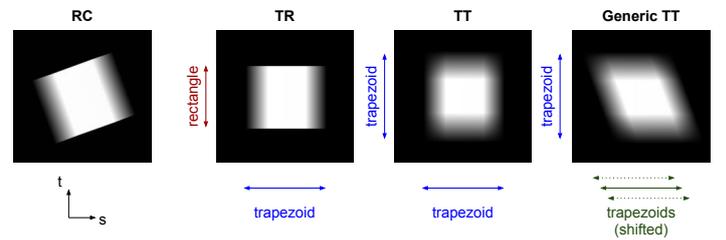

**Figure 2:** Projection of a single voxel (i.e. footprint) computed with a ray caster, the regular TR and TT projector, as well as the proposed generic extension in a setting with 20° twist angle.

## 2.3 Simulation experiments

To evaluate the performance of the proposed extensions (i.e. generic and sparse version) of the TT footprint algorithm, a series of simulation experiments has been carried out. All simulations feature a cone-beam projection setting with a flat-panel detector with 1280 x 960 pixels (0.25 mm pixel size), imaging the scene at a source-to-detector distance of 1000 mm and a source-to-object-distance of 750 mm. First, we analyze the accuracy of the proposed generic TT extension under two specific geometrical distortions. Figure 1 shows the baseline setting (i.e. undistorted geometry). The first manipulation, called *tilt*, is a rotation of the detector around the detector's $s$-axis; in our specific setting, this axis is parallel to the WCS $y$-axis. Tilting preserves orthogonality between the $s$-axis and WCS $z$-axis, whereas the $t$-axis will no longer be aligned with the WCS $z$-axis. The second manipulation, the *twist*, is a rotation of the detector around the principal ray; note that this corresponds to the WCS $x$-axis in our specific setting. This manipulation destroys parallelism between both detector axes with the corresponding WCS axes. By design, the "traditional" TR/TT method should be sensitive to such manipulations and lose accuracy rather quickly. Analysis is based on projections of a $256^3$ voxel phantom (0.5 mm voxel size) containing random values from [0,1] (see Figure 3).

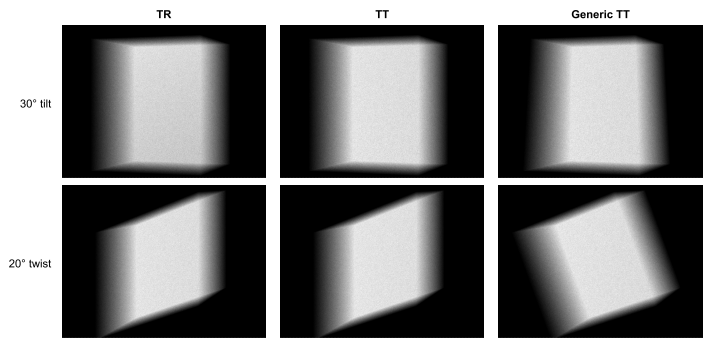

**Figure 3:** Comparison of projection images generated with the original footprint routines and the proposed generic extension.

The second experiment focuses on computation times of both extensions to assess: a) the additional cost of the generic TT extension and b) the potential speed-up achievable by using the sparse version of the projector. A software phantom representing an artificially "grown" (binary) vessel tree has been used to investigate the performances in a more or less realistic use case. The phantom is generated on a $512^3$ voxel grid (0.5 mm voxel size) and contains 126'663 non-zero voxels (approx. 0.14 % of total voxel count). To judge the accuracy of the results, projections are compared to reference projections (relative L2 differences) generated using a high-resolution ray casting approach (non-interpolating, constant step length of 5% of the voxel size, 10x10 subrays per detector pixel). An example of a projection image is shown in Figure 4.

In addition, the speed-up of the sparse version is evaluated in more detail using a simple randomized dummy volume. A cube of $M^3$ voxels ($M \in \{128, 256, 512\}$; voxel sizes: $\{2.0, 1.0, 0.5\}$ mm) is initialized with zero. Subsequently, a number of $N$ (randomly selected) voxels is set to 1, thus generating a phantom with sparsity level: $\sigma = N/M^3 \in [0,1]$. In both cases (i.e. vessel tree and dummy), sparse representations of the volumes are created with a simple thresholding approach, keeping all voxels with a value greater than zero.

## 2.4 Implementation

Open-source implementations of the proposed generic TT projector and its sparse counterpart, as well as the regular TR/TT methods, are available as part of our Computed Tomography Library (CTL) C++ toolkit [2], which is publicly available on GitLab (gitlab.com/tpfeiffe/ctl).

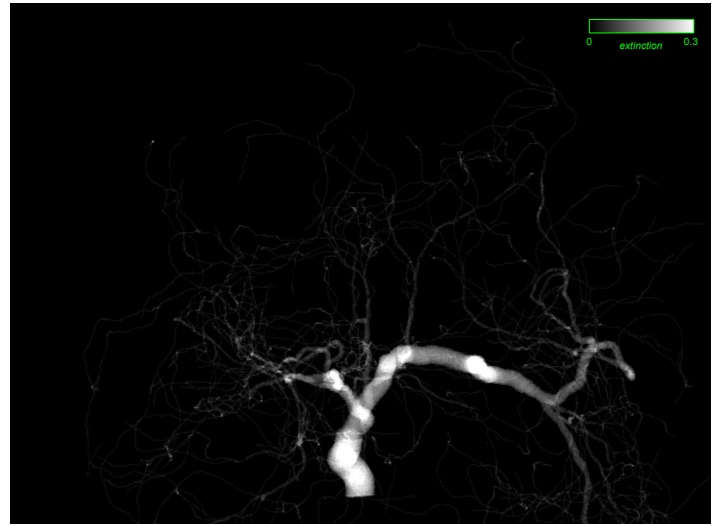

**Figure 4:** Example projection image of the vessel tree phantom (view 59). Displayed extinction range narrowed to [0, 0.3] to improve visibility (full value range: [0, 0.53]).

**Table 1:** Performance comparison of different projection techniques (sparsity: 0.14%). $t$ – computation time, Err. – relative L2 norm w.r.t. reference images, RC-LQ – low-quality ray caster.

|  | RC-LQ | TR | TT | Gen. TT | Sparse |
|---|---|---|---|---|---|
| $t$ [ms] | 6974 | 18895 | 46257 | 64289 | 1886 |
| Err. [%] | 4.560 | 0.162 | 0.155 | 0.155 | 0.155 |

## 3 Results and Discussion

**Detector tilt and twist**   Figure 3 shows example images of forward projections computed with the three different footprint approaches for 30° detector tilt and 20° twist. As expected, both TR and TT projector have issues with the introduced geometric distortions. For quantitative assessment of the effect, relative L2 errors (w.r.t. high-res. ray casting projections) have been computed for varying rotation angles of both manipulations (Figure 5). The results demonstrate that the proposed generic TT approach provides accurate projection results for arbitrary tilt angle (up to 45°) and twist angles up to about 25°(error <1%). For very small tilt angles, TR and TT provide reasonable accuracy; twists, however, are handled poorly even for smallest angles.

**Comparison and Runtime**   Projection images of the artificial vessel tree (cf. Section 2.3) are generated in a typical short scan geometry (angular range approx. 198°) for 100 equiangularly-spaced views. Runtimes are measured for the entire projection procedure, including data transfer to, and back from, the GPU device (OpenCL kernel compilation times are excluded). All benchmarks have been performed on a system equipped with an Intel i5-8400, 16 GB RAM, and a single NVIDIA GTX 1060 GPU. The results (Table 1) show that the generic TT extension incurs additional computation effort compared to regular TR/TT. Computation times in the given example are roughly 40% higher than for the TT method and about 3.4 times longer than those of TR. In the chosen example (artificial vessel tree, sparsity level 0.14%), the sparse version can lower the computation times substantially, leading to a speed-up of about factor 34 w.r.t. the non-sparse generic TT method. Even in comparision to a fast, low-quality ray casting technique, a speed-up factor of approximately 3.7 is achieved, while providing substantially more accurate results (0.15% vs. 4.5% error). Note that these results imply an average simulation time (incl. data transfers) of only 19 ms per projection.

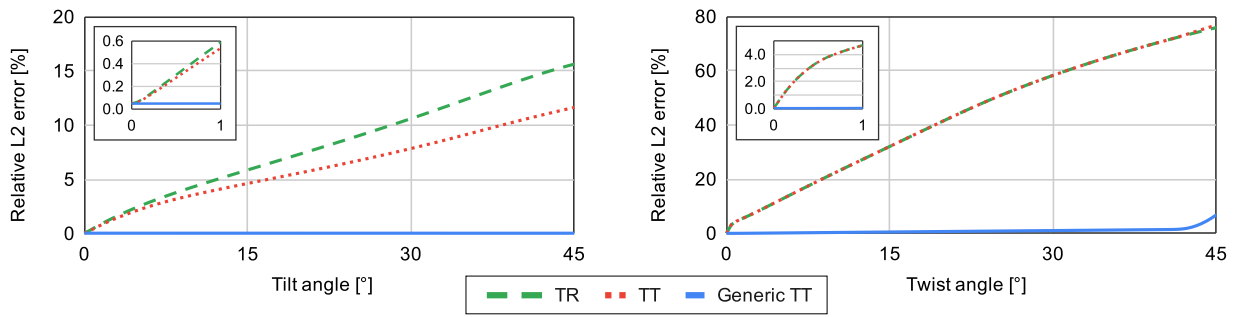

**Figure 5:** Relative L2 errors of projections generated with different footprint approaches as functions of detector tilt and twist angles.

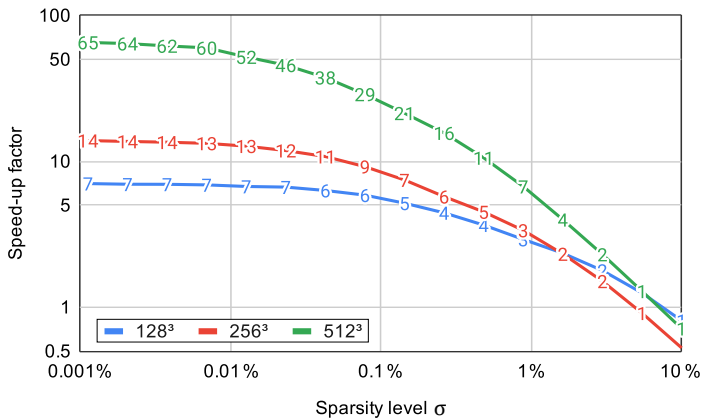

**Figure 6:** Speed-up factors of the sparse projector version compared to the non-sparse generic TT projector for varying sparsity level of three differently resolved dummy phantoms.

The speed-up over the non-sparse generic TT method in dependency of the sparsity level of the input volume is shown in Figure 6 for three different voxelizations. It becomes apparent that the speed-up runs into saturation when the volume becomes too sparse. This is due to data transfer overhead dominating the total computation time in these cases. Depending on the number of voxels in the full dataset, speed advantage starts just below a sparsity of 5 to 7%, which should constitute a reasonable size for many practical segmentations (e.g. vessel trees, bone structures, or implants).

## 4 Conclusion and Future Work

We presented an extension of the TT footprint approach that allows generating accurate projections in presence of acqui-sition geometries that deviate from perfectly aligned cases (t ∥ z); the additional computation effort could be kept at about 40% of the regular TT method. The gained flexibility can be of importance in all practical scenarios that involve imperfect acquisition geometry, for example, when facing miscalibration. Another important area of application might be motion compensated reconstruction. Patient motion can be interpreted as system miscalibration, and thus, leads to similar geometric deviation. Non-circular data acquisition schemes, e.g. ellipse-line-ellipse trajectories [3], constitute a further topic for meaningful application of the proposed approach. Due to the difficulties faced in analytical reconstruction of arbitrary trajectory data, IR is particularly interesting in these cases. We further showed that the method can easily be used with sparse representations of an input volume, granting huge speed advantages of up to factor 65 in specific situations. This provides a powerful tool for all tasks that require accurate, fast, and flexible FP of sparse objects. As a side effect, flexibility of use is increased even further, as off-the-grid positioning of individual voxels becomes possible at zero additional cost. Future work will focus primarily on the implementation of a corresponding backprojector for the proposed generic TT method, including evaluation of IR results produced with the resulting FP/BP pair in previously described target applications (esp. motion compensation).

**Acknowledgments**    This work was supported by the Federal Ministry of Education and Research within the Research Campus *STIMULATE* and KIDs-CT [grant numbers 13GW0095A, 13GW0229A].

# Sparse-View Joint Reconstruction and Material Decomposition for Dual-Energy Cone-Beam Computed Tomography

Suxer Alfonso Garcia[1], Alessandro Perelli[1], Alexandre Bousse[1], and Dimitris Visvikis[1]

[1]LaTIM, INSERM, UMR 1101, *Université de Bretagne Occidentale*, Brest, France.

**Abstract** In this work we address the challenge of exploiting the structural similarities of images acquired at different energy to improve the reconstruction and material estimation in dual-energy cone beam computed tomography (DE-CBCT). We focus on the sparse-view single-source fast KVp switching acquisition set-up to reduce scan time and the total dose delivered during a computed tomography (CT) acquisition. We propose to exploit the joint total variation (JTV) regularization between low- and high-energy images to reduce the artifacts due to the under-sampling of the angular views. We show through numerical experiments and patient data the benefit of the proposed method for material decomposition and estimation, both qualitatively and quantitatively, compared to individual total-variation (TV) regularization.

## 1 Introduction

Dual-energy cone beam computed tomography (DE-CBCT) allows quantitative imaging and improves tissue visualization [1]. The technique provides two sets of measurements from two distinct energy spectra (low- and high-energy) acquired over the same anatomical region which are reconstructed into material- and energy-selective images that enable enhanced tissue characterization. In addition, it has the ability for real tumor-guided radiation therapy in combination with a contrast agent.

The acquisition techniques in DE-CBCT can be classified into 4 different categories: single-source sequential acquisition; single source with dual-layer detector; dual source with two detectors positioned orthogonally; and single-source with rapid KVp switching. Rapid potential switching allows consecutive projection measurements with alternating tube potentials where both the low- and high-energy projection data are acquired throughout a whole gantry rotation [2, 3]. The tube voltage varies between low and high, and transmission data is acquired twice for adjacent projection angles. The major disadvantage of this method is the need of reducing the rotation speed of the system to acquire the extra projections and to account for the rise and fall times required for voltage modulation [4]. Due to fast switching it is not possible to modulate the tube current between low and high energy simultaneously. It remains constant during the acquisition. Thus, the tube current needs to be increased to reduce the noise on images obtained with lower peak voltage, which results in an increase of the radiation dose [5, 6].

A reduction in the acquisition time can be achieved by decreasing the number of projection angles. Moreover, sparse-view acquisitions can reduce the radiation dose. However, aliasing artifacts can appear in the reconstructed images if the number of projection angles does not follow the Shannon/Nyquist sampling theorem, which makes more challenging to reconstruct high-resolution, high-contrast and high-signal-to-noise ratio (SNR) images [7, 8].

The present work proposes a methodology for image reconstruction in sparse-view single-source rapid KVp switching DE-CBCT by exploiting structural similarity with JTV regularization. The hypothesis behind this approach is that the low- and high-energy images can inform each other, thus not only giving room for acquisition time and dose reduction but also enhancing the spatial resolution deficit due to the down-sampled projection data.

## 2 Materials and Methods

### 2.1 Dual Energy Image Reconstruction

Assuming a simplified single-source rapid KVp switching DE-CBCT setting, each sinogram $\boldsymbol{y}_\ell \in \mathbb{R}^m$, obtained from the energies $\ell \in \{L, H\}$ (low and high), is a modeled by random vector $\boldsymbol{y}_\ell = [y_{1,\ell}, \ldots, y_{n,\ell}]^\top$ with independent entries, where $n$ is the number of detectors. At each detector $i \in \{1, \ldots, n\}$, the number of detected photons $y_{i,\ell}$ follows a Poisson distribution:

$$y_{i,\ell} \sim \text{Poisson}(\bar{y}_{i,\ell}(\boldsymbol{\mu}_\ell)), \tag{1}$$

with

$$\bar{y}_{i,\ell}(\boldsymbol{\mu}_\ell) = b_i \exp(-[\boldsymbol{P}\boldsymbol{\mu}_\ell]_i) + s_{i,\ell} \tag{2}$$

where $\boldsymbol{\mu}_\ell \in \mathbb{R}^m$ is the attenuation image at energy $\ell$, $\boldsymbol{P}$ is a $n \times m$ matrix modeling the system, $s_{i,\ell}$ is a background term and $m$ is the number of voxels in the image.

In this work we propose to reconstruct the low- and high-energy attenuation images $(\boldsymbol{\mu}_L, \boldsymbol{\mu}_H)$ by penalized maximum-likelihood joint estimation from the sinograms $(\boldsymbol{y}_L, \boldsymbol{y}_H)$:

$$(\hat{\boldsymbol{\mu}}_L, \hat{\boldsymbol{\mu}}_H) = \underset{\boldsymbol{\mu}_L, \boldsymbol{\mu}_H \geq \boldsymbol{0}}{\arg\max} F(\boldsymbol{\mu}_L, \boldsymbol{y}_L) + F(\boldsymbol{\mu}_H, \boldsymbol{y}_H) - \beta R(\boldsymbol{\mu}_L, \boldsymbol{\mu}_H) \tag{3}$$

where $R(\boldsymbol{\mu}_L, \boldsymbol{\mu}_H)$ is a joint regularization term, $\beta$ is the regularization parameter and $F(\boldsymbol{\mu}_\ell, \boldsymbol{y}_\ell)$ is the log-likelihood defined as:

$$F(\boldsymbol{\mu}_\ell, \boldsymbol{y}_\ell) = \sum_{i=1}^n y_{i,\ell} \log \bar{y}_{i,\ell}(\boldsymbol{\mu}_{i,\ell}) - \bar{y}_{i,\ell}(\boldsymbol{\mu}_{i,\ell}). \tag{4}$$

In this work the maximization problem (3) is solved using a limited-memory Broyden-Fletcher-Goldfarb-Shanno (L-BFGS) algorithm [9].

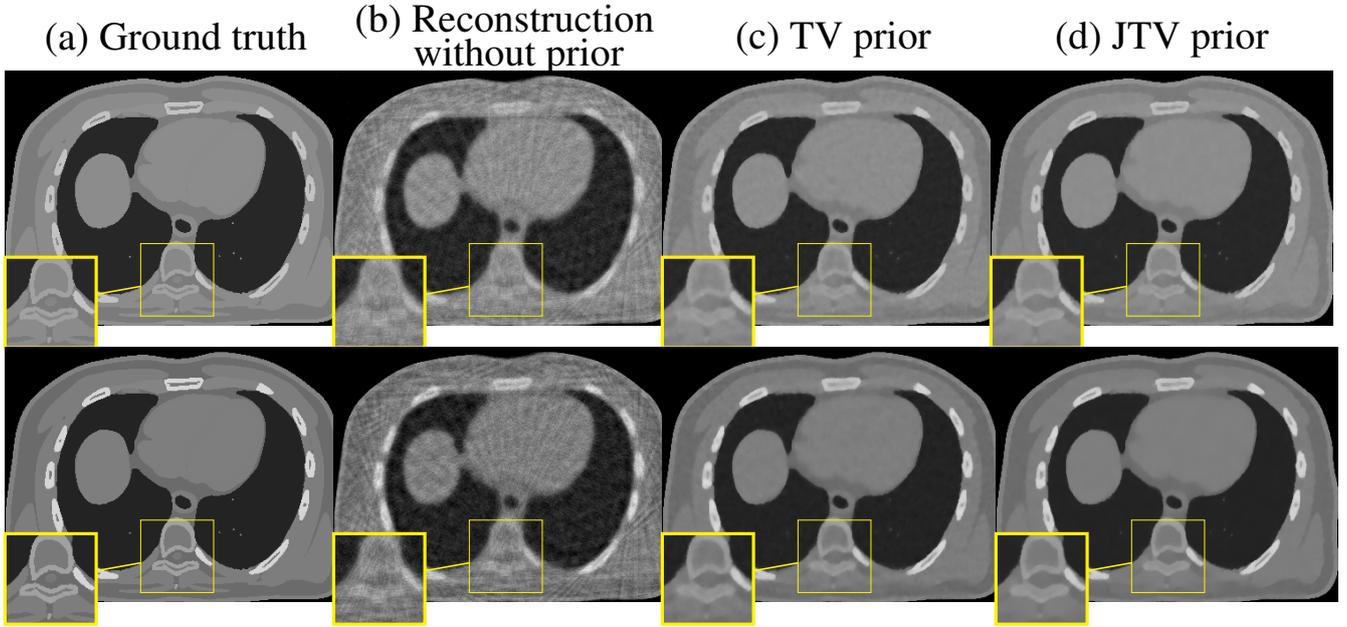

**Figure 1:** Comparison of reconstructed extended cardiac-torso (XCAT) phantoms using different reconstruction methods for sparse-view DE-CBCT with top row corresponding to high energy ($E = 140$ KeV) and bottom row to low energy ($E = 70$ KeV): (a) Ground truth, (b) reconstruction without prior, (c) TV reconstruction, (d) joint reconstruction using JTV prior.

## 2.2 Joint Total Variation Regularization

In this work, we used the JTV penalty term $R(\boldsymbol{\mu}_L, \boldsymbol{\mu}_H)$ inspired from [10]. The JTV regularization term can be written as:

$$R(\boldsymbol{\mu}_L, \boldsymbol{\mu}_H) = \sum_{j=1}^{m} \left( \|[\nabla \boldsymbol{\mu}_L]_j\|^2 + \|[\nabla \boldsymbol{\mu}_H]_j\|^2 + \gamma^2 \right)^{1/2} \quad (5)$$

where $\nabla \boldsymbol{\mu}_\ell \in \mathbb{R}^{m \times d}$ ($d = 2, 3$) is the gradient image of $\boldsymbol{\mu}_\ell$ and $[\nabla \boldsymbol{\mu}_\ell]_j \in \mathbb{R}^d$ is the gradient at voxel $j$, and $\gamma > 0$ tunes the smoothness of the prior (for differentiability). The role of this prior is to promote structural similarities by enforcing joint sparsity of the 2 gradient images. We compared the proposed approach of jointly reconstruct the images with JTV against reconstructing separately with TV as follows:

$$\hat{\boldsymbol{\mu}}_\ell = \arg\max_{\boldsymbol{\mu}_\ell \geq \mathbf{0}} F(\boldsymbol{\mu}_\ell, \boldsymbol{y}_\ell) - \delta S(\boldsymbol{\mu}_\ell) \quad (6)$$

with

$$S(\boldsymbol{\mu}_\ell) = \sum_{j=1}^{m} \left( \|[\nabla \boldsymbol{\mu}_\ell]_j\|^2 + \eta^2 \right)^{1/2} \quad (7)$$

where $\delta$ and $\eta$ play the same roles as $\beta$ and $\gamma$ respectively. With this approach each image is reconstructed independently without sharing structural information.

## 3 Experiments

We performed the dual-energy image reconstruction by iteratively alternating between (i) updating the low-energy image $\mu_L$ and (ii) updating the high-energy image $\mu_H$ using the L-BFGS algorithm. We initialized the images using an maximum-likelihood reconstruction for transmission tomography (MLTR) algorithm [11] without explicit prior.

## 3.1 Results on XCAT Phantom

The numerical down-sampled projection data was modeled by forward projection of a 0.85-mm pixel width $512 \times 512$ torso axial slice images generated from the extended cardiac-torso (XCAT) phantom at two energy levels [12]. We modeled the projector $\boldsymbol{P}$ with a 1-mm full width at half maximum (FWHM) fan beam system. We simulated sparse-view 60-angle sinograms. We distributed the projection angles such that, in a single gantry rotation, one projection angle corresponds to a low-energy projection and the consecutive angle corresponds to a high-energy projection. For each sinogram, we used a monochromatic source with $10^5$ incident photons and 100 background events. The values of the linear attenuation coefficients for each phantom were generated with X-ray energies of 70 KeV (low) and 140 KeV (high).

Figure 1 shows the reconstructed images using JTV regularization, TV and without prior. In absence of prior, the images suffer from under-sampling artifacts. The selected regions of interest (ROIs) in the images show the improved performance of JTV as compared with TV. Low-contrast features can be better identified with JTV than with TV. Furthermore, we quantitatively evaluated the performance of JTV using the peak signal-to-noise ratio (PSNR) defined as:

$$\text{PSNR(dB)} = 10 \cdot \log_{10} \left( \frac{\max_j \left\{ \mu_j^{\text{GT}} \right\}^2}{\sum_{j=1}^{m} \frac{1}{K} \left( \hat{\mu}_j - \mu_j^{\text{GT}} \right)^2} \right) \quad (8)$$

where $\hat{\mu}_j$ and $\mu_j^{\text{GT}}$ represent the intensity value at the pixel $j$ in the reconstructed image and the ground truth respectively. We also computed the structural similarity index measure (SSIM) using equation (7) in [13]. Table 1 shows the values of the metrics mentioned above. At both energy levels, the JTV approach results in higher PSNR and SSIM. For the low-energy image the gain was 4.58% in PSNR and 0.03% in SSIM while for the high-energy image the gain was 5.79% in PSNR and 0.06% in SSIM.

| 70 KeV | PSNR | SSIM | 140 KeV | PSNR | SSIM |
|---|---|---|---|---|---|
| **JTV** | 64.85 | 0.9996 | **JTV** | 66.66 | 0.9998 |
| **TV** | 62.01 | 0.9993 | **TV** | 63.01 | 0.9992 |
| **Gain(%)** | 4.58 | 0.030 | **Gain(%)** | 5.79 | 0.06 |

**Table 1:** PSNR in dB and PSNR for the JTV and TV reconstruction algorithms at low energy (70 KeV) and high energy (140 KeV). The gain is calculated as Gain(%) = $100 \cdot (\text{JTV} - \text{TV})/\text{TV}$ where the terms JTV and TV correspond to the values of the PSNR and SSIM for each regularization.

We analyzed the bias/variance trade-off of JTV and TV on the low- and high-energy images by plotting the absolute bias (AbsBias) against the variance (Var) of the total image, based on $K = 30$ realizations of $y_L$ and $y_H$, for different values of the regularization parameters, i.e.,

$$\text{AbsBias} = \frac{1}{K}\frac{1}{J}\sum_{k=1}^{K}\sum_{j=1}^{J}\left|\hat{\mu}_j^k - \mu_j^{\text{GT}}\right| \qquad (9)$$

$$\text{Var} = \frac{1}{K}\frac{1}{J}\sum_{k=1}^{K}\sum_{j=1}^{J}\left(\hat{\mu}_j^k - \bar{\mu}_j\right)^2$$

$$\text{with} \quad \bar{\mu}_j = \frac{1}{K}\sum_{k=1}^{K}\hat{\mu}_j^k$$

where $\hat{\mu}_j^k$ is the reconstructed image at pixel $j$ for the noise realization $k$ and $\mu_j^{\text{GT}}$ is the ground truth. Figure 2 and 3 show that JTV achieves lower absolute bias for any variance level in the two energy images.

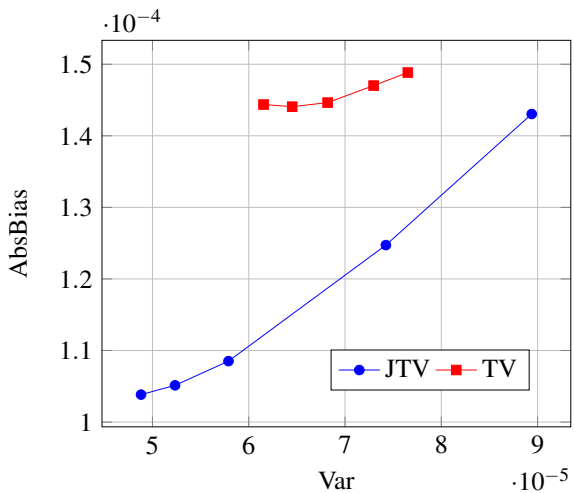

**Figure 2:** Plot of the absolute bias (AbsBias) versus the variance (Var) for the sparse-view reconstruction with XCAT data and high X-ray source energy, 140 keV.

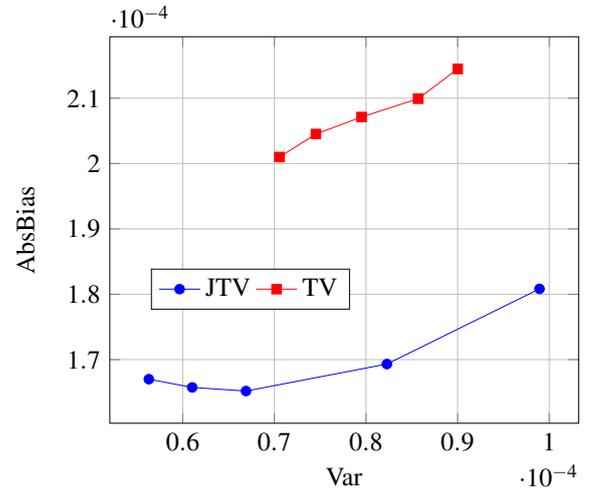

**Figure 3:** Plot of the absolute bias (AbsBias) versus the variance (Var) for the sparse-view reconstruction with XCAT data and low X-ray source energy, 70 keV.

## 4 Results on Clinical Data

The clinical dataset was acquired on the Philips IQon Spectral CT scanner from the Poitiers University Hospital. All patients used in the study provided signed permission for the use of their clinical data for scientific purposes and anonymous publication of data. We selected 2-dimensional (2-D) slices from a full body patient scan with 0.902-mm pixel width and $512 \times 512$ image size corresponding to the thorax area. The energies used in this study were 70 keV and 140 keV. To generate the sparse-view DE-CBCT measurements we used the same geometrical and noise settings as for the XCAT simulation.

In Figure 4 we observe that JTV outperforms TV for clinical data; TV-reconstructed images shows aliasing artifacts.

Figures 5 and 6 report on the AbsBias versus the Var for different values of the regularization parameters. We obtain a similar behavior as compared with the XCAT simulations; JTV outperforms TV.

### 4.1 Modulation Transfer Function

The spatial resolution of the DE-CBCT-reconstructed images was measured by computing the modulation transfer function (MTF) derived from an edge measurement. Initially, an edge spread function (ESF) was obtained at the slanted edge between the trachea and the lung. The ESF was resampled using linear interpolation and averaged across multiple ESF realizations to reduce variance. Then, a line spread function (LSF) was estimated by taking the derivative of the ESF. Finally, the MTF was obtained by applying the Fourier transform to the LSF [14, 15]. Figures 8 and 9 show the MTFs of the images reconstructed utilizing TV and JTV regularization for high- and low-energy images respectively. We observe that JTV produces higher spatial resolution than TV. The spatial resolution analysis reveals that JTV increases detectability and edge-preservation in comparison to TV.

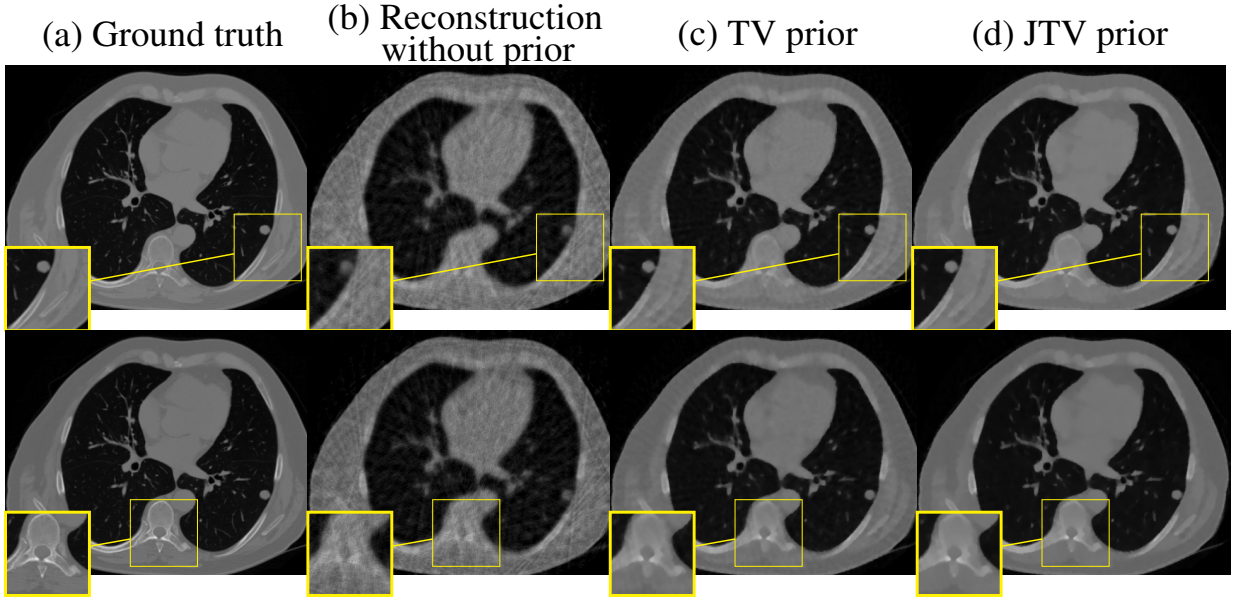

**Figure 4:** Comparison of reconstructed clinical data using different reconstruction methods for sparse-view with top row corresponding to high energy ($E = 140$ KeV) and bottom row to low energy ($E = 70$ KeV): (a) Ground truth, (b) reconstruction without prior, (c) TV reconstruction, (d) joint reconstruction using JTV prior.

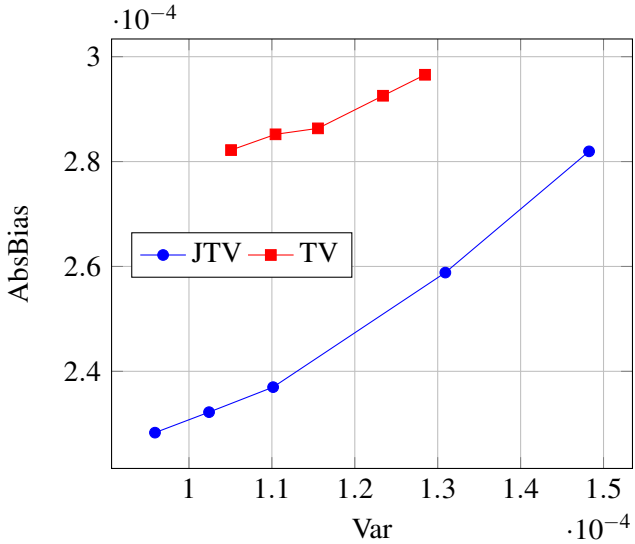

**Figure 5:** Plot of the absolute bias (AbsBias) versus the variance (Var) for the sparse-view reconstruction with clinical data and high X-ray source energy (140 keV).

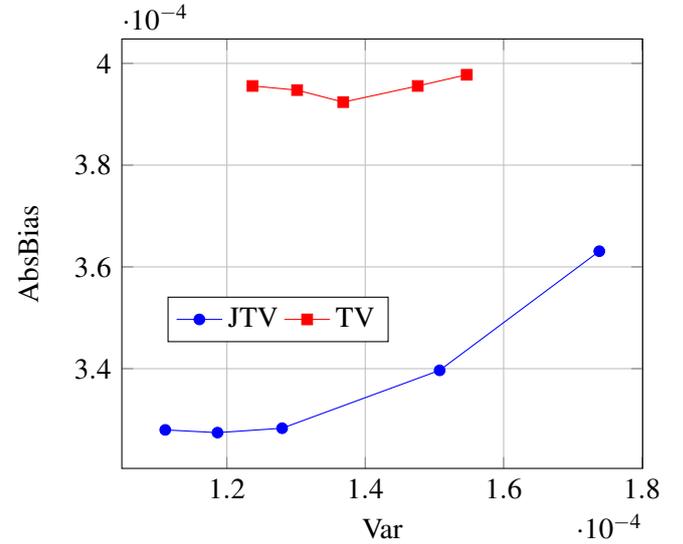

**Figure 6:** Plot of the absolute bias (AbsBias) versus the variance (Var) for the sparse-view reconstruction with clinical data and low X-ray source energy, (70 keV).

## 5    Results for Material Decomposition

An important application of DE-CBCT is material decomposition. It relies on the approximation of the linear attenuation coefficient at each pixel in the CT image by a linear combination of the attenuation values of basis materials. Thus, the material decomposition can be written as:

$$\begin{pmatrix} \mu_L \\ \mu_H \end{pmatrix} = \begin{pmatrix} \mu_{1,L} & \mu_{2,L} \\ \mu_{1,H} & \mu_{2,H} \end{pmatrix} \begin{pmatrix} x_1 \\ x_2 \end{pmatrix} \qquad (10)$$

where $\mu_{p,l}$ is the linear attenuation coefficient of material $p \in \{1, 2\}$ at energy $l \in (L, H)$, $x_1$ and $x_2$ are the volume fractions of the two basis materials at the same position of two basis material images and $\mu_L$ and $\mu_H$ are the low- and high-energy reconstructed attenuation values. The aim of material decomposition algorithms is to estimate the volume fractions knowing the linear attenuation coefficient of the basis materials. In the present study we utilized the methodology proposed in [16].

We decompose into soft tissue ($x_1$) and bone ($x_2$) with attenuation coefficient values (in $cm^{-1}$):

$$\begin{pmatrix} \mu_{1,L} = 0.1929 & \mu_{2,L} = 0.3432 \\ \mu_{1,H} = 0.1538 & \mu_{2,H} = 0.2237 \end{pmatrix} \qquad (11)$$

Figure 7 shows the image decomposition into bone and water (body) basis material from the reconstructed images using JTV, TV, and the ground truth. We observe that small bone structures can be better identified in the bone-decomposed image obtained from the JTV reconstruction.

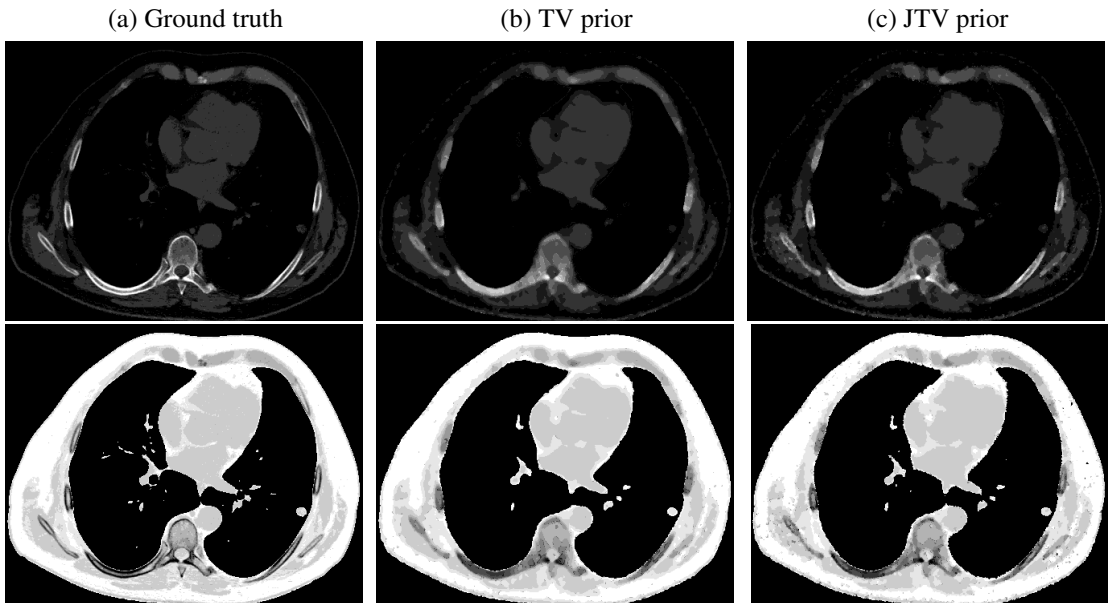

**Figure 7:** Decomposed images into Bone (top row) and Water (bottom row) basis materials utilizing the clinical images obtained form the (a) ground truth, (b) reconstruction with TV and (c) reconstruction using JTV prior

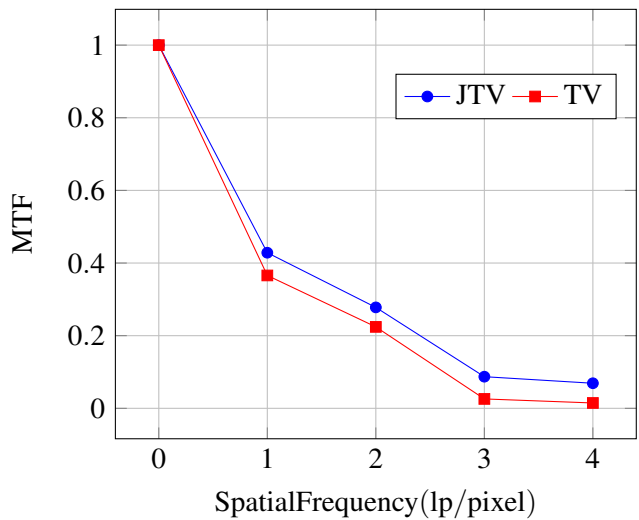

**Figure 8:** MTF obtained from the reconstructed images utilizing TV and JTV priors for high-energy clinical data , 140 keV.

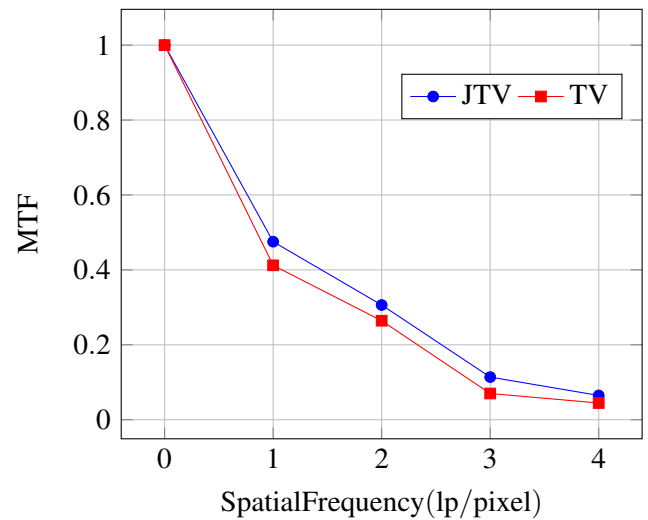

**Figure 9:** MTF obtained from the reconstructed images utilizing TV and JTV priors for low-energy clinical data, 70 keV.

## 6 Discussion

By using JTV and coupling the low- and high-energy images, is possible to incorporate joint structural information between the 2 energies. The results presented in this work show the ability of the JTV regularization to improve sparse-view reconstruction, even when the number projection angles are 6 times less than that of a full-view setting, which allows a significant decrease of the scanning time and the radiation dose to the patient. In comparison with TV regularization, JTV leads to improved accuracy both in reconstruction and material decomposition. The reconstruction with JTV results in better contrast and spatial resolution. The results obtained with patient data or more textured phantoms corroborate the high performance of JTV compared to TV.

Further analysis will involve using the proposed reconstruction framework in new CT scanner technologies, like photon-counting spectral CT, where the algorithm can leverage the joint structural similarities from an increased number of images at different energies, leading to an overall improved quantitative estimation even with a further reduction of the acquired projection angles.

## 7 Conclusion

The present work proposes an image reconstruction methodology for sparse-view DE-CBCT using a JTV regularization. The coupled regularizer exploits structural similarities between the two images acquired at low- and high-energy. We compared the performance of the proposed approach against the reconstruction of each image separately using TV regularization. Reconstruction with JTV resulted in improved contrast and spatial resolution as well as improved material decomposition.

## 8  Acknowledgment

This work has been supported by the French National Research Agency (ANR) under grant ANR-20-CE45-0020. The authors would like to thank Jean-Pierre Tasu and Nikolaos Efthimiadis from Department of Radiology, University Hospital Poitiers, Poitiers, France for providing the clinical data used in this project.

# GPU-Based Real-Time Software Coincidence Processing

Yu Shi, Fanzhen Meng, Chenfeng Li, Jianwei Zhou, Juntao Li, and Shouping Zhu*

Engineering Research Center of Molecular and Neuro Imaging of Ministry of Education, School of Life Science and Technology, Xidian University, Shaanxi 710126, China

**Abstract** The software coincidence processing is typically implemented on central processing units (CPUs) and then it is accelerated by CPU multithreading technology. However, the more detection modules a PET system has, the more CPU threads are used in acquisition and the fewer threads are available in coincidence processing when the number of threads is fixed, which results reduced processing performance of CPU-based software coincidence processing (CPU-SCP). In this paper, we proposed the GPU-based real-time software coincidence processing (GPU-SCP) method to solve the limited CPU thread problem. To evaluate the validity of the proposed GPU-SCP, we adapted it to our PET system. We did the speedup experiments and the image quality experiments. The speedup experiment results show that the proposed GPU-SCP achieve up to 14.5 times speedups on GTX1070 compared to the serial CPU algorithm, which is 7.6 times faster than parallel CPU-SCP with 10 CPU threads. Besides, the image quality experiments indicate the reconstruction images processed by GPU-SCPs are almost the same as the references in low activity nuclide imaging (differences 1% in image domain and projection domain). Owing to GPU-SCPs' faster processing speed, there are more coincidences extracted by it than that extracted by CPU-SCP in high activity nuclide imaging.

## 1 Introduction

The PET imaging is based on the gamma photon coincidence technology to find the annihilation photon pairs among many single photon events (singles) and finally realizes the quantification of the distribution of radionuclides in the body. In some digital PET systems, the photon coincidence processing is implemented on the host computers which provide strong computing power to increase the speed of coincidence processing and offer much more flexibility. They are called software coincidence processing (SCP) [1–3]. The online software coincidence processing is that the single events raw data are processed in real time on the host computer and the information of the coincidence events are acquired once the PET data acquisition is over. Based on the above characteristics, the processing speed of online SCP must be high enough to meet the real-time requirement. However, it is almost impossible to processing singles serially in real time with single central processing unit (CPU) thread. In order to realize lossless real-time acquisition and coincidence processing, the real-time SCP methods based on CPU multi-threading technology[2, 3] was proposed. However, the more basic detection modules (BDMs) a PET system has, the more threads are used in acquisition and the fewer threads are available in coincidence processing when the number of threads is fixed which results reduced processing performance of CPU multithreads-based software coincidence processing.

In the past decades, graphics processing unit (GPU), which has great potential in parallel computing, has been widely used in acceleration of medical image reconstruction algorithm [4, 5]. However, the implementation of GPU in the acceleration of SCP has been reported rarely so far. In this paper, we propose a novel GPU-based real-time software coincidence processing (GPU-SCP) approach, which solve the real-time processing challenges in PET systems with multiple modules. The proposed processing architecture simplifies the management of threads between acquisition and coincidence processing, accelerates the coincidence processing by GPU multiple threads and finally realizes the online coincidence processing in our PET system. Besides, the introduction of GPU relaxes the requirements on the number of CPU threads, thus greatly reducing the system cost.

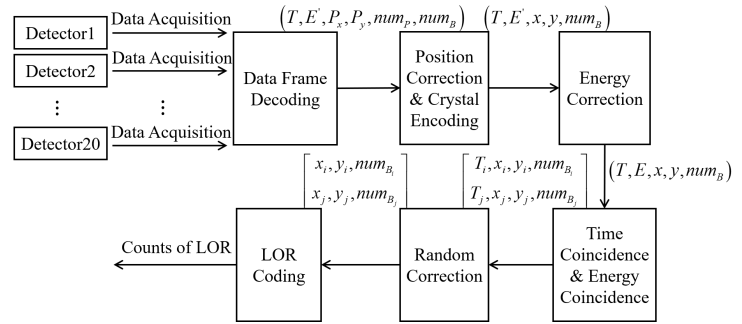

**Figure 1:** SCP flow of our PET system.

## 2 Materials and Methods

The SCP is applied to extract the pairs of singles from the same positron annihilation captured by the PET system on the host computer. As shown in Figure 1, SCP includes data frame decoding, crystal position encoding, energy correction, energy coincidence, time coincidence, random correction and line of response (LOR) coding in our PET system. After SCP, the PET system outputs the histogram of LORs. Considering the different levels of parallelization, the SCP is viewed as two parts, Before Coincidence Processing and Coincidence Processing. 'Before Coincidence Processing' consists data frame decoding, crystal position encoding and energy correction. 'Coincidence Processing' is composed of energy coincidence processing, time coincidence processing, random correction and LOR coding.

## 2.1 Before Coincidence Processing

Each single is encoded as a 16-byte data frame in the process of data acquisition. At the beginning of SCP, the data frame needs to be decoded into single data containing hit time, photon energy and location information. We first parallelize the process of data frame decoding. Equation (1) and (2) are applied to each 16-byte data frame, and communication is not required between different single photon data frames.

$$
\begin{bmatrix} T[tid] \\ E^{'}[tid] \\ P_x[tid] \\ P_y[tid] \\ N_B[tid] \\ N_P[tid] \end{bmatrix}_{6 \times 1} = A_{6 \times 15} \times \begin{pmatrix} d[tid][0] \\ d[tid][1] \\ \vdots \\ d[tid][j] \\ \vdots \\ d[tid][15] \end{pmatrix}_{15 \times 1} \quad (1)
$$

$$
A_{6 \times 15} = \begin{bmatrix} a_{1,1} & a_{1,2} & \cdots & a_{1,n} & 0 & \cdots & 0 & \vec{0} \\ 0 & 0 & \cdots & 0 & a_{2,n+1} & \cdots & a_{2,n+m} & \vec{0} \\ \vdots & \vdots & \vdots & \vdots & \vdots & \vdots & \vdots & \vdots \\ 0 & 0 & \cdots & 0 & 0 & \cdots & \vec{0} & a_{6,16} \end{bmatrix} \quad (2)
$$

where $d[i][j], j = 0, ..., 15$, represents the 16-byte $ith$ single data frame, $T$ is the hit time of the single photon, $E^{'}$ is the uncorrected photon energy, $P_x$ and $P_y$ are the cartesian coordinates of a hit on PMT, $N_P$ is the index of PMT and $N_B$ is the index of BDM. $A$ is the row echelon form representing the decoding rules. It is suitable for all single data frames. After data frame decoding, a single captured by PET system can be represented by $(T, E^{'}, P_x, P_y, N_P, N_B)$. This is a typical single instruction multiple data (SIMD) computing model, which is suitable for parallelization.

The goal of position encoding is to rearrange the position $(P_x, P_y)$, and then the position of each photon deposition was corresponding to the actual crystal position according to the crystal identification map ($CIM$) [6] established in advance. Besides, energy correction is to calibrate the energy value of a single captured by i-th crystal [7]. Similar to the data frame decoding, crystal position encoding and energy correction are also the SIMD models, except that it may read the energy correction factors or $CIM$ from the same address simultaneously. Through the above analysis, each GPU thread can perform data extraction, correction and energy coincidence processing for a 16-byte single data frame before coincidence processing. Therefore, $tid$ in Equation (1) represents the GPU thread index.

## 2.2 Coincidence Processing

In the energy coincidence processing, we record the index of singles that satisfied with the energy window by Equation (3).

$$
I[tid] = \varepsilon((E[tid] - downEnergy) \\ \times (upEnergy - E[tid])) \times tid \quad (3)
$$

where $\varepsilon$ is Heaviside function, $I$ is the index of singles. In the implementation, zero-padding is used at the end of each

BDM data stream so that the singles from the same BDM which are at the end of data stream will be extracted by the same warp.

Different from the 'Before Coincidence Processing' which are highly suitable for parallelization, the parallelism of time SCP is not obvious. In this section, the parallelism of time SCP is explored and then the GPU-SCP is proposed. There are many coincidence processing approaches at present, we only developed the sorting-based GPU-SCP in this paper.

In the sorting-based time SCP, the single data streams from all BDMs are merged into one data stream firstly. Then the total data stream is sorted according to the photon hit time, finally a global time-ordered data stream is obtained. A time window is applied to each single event in the global time-ordered data stream. If there are other singles in the time window, these singles are identified to be from the same annihilation event and constitute coincidences.

---

**Algorithm 1** Pseudocode implementation of GPU-SCP

// DEFINITIONS
// $\mathbf{T}^k$ is the hit time of singles which are accepted by energy window from $BDM_k$
// $\mathbf{T} = [\mathbf{T}^0, 0; \mathbf{T}^1, 0; ...; \mathbf{T}^{20}, 0]$,
// $value$ is the global index of a single before sorting
**Input:** T
**Output:** Time Coincidence Index
1: COIN_KERNEL<<< $blockSize, threadSize$ >>>( **T**, **value**)
2: **function** COIN_KERNEL(**T**,**value**)
3:     $tid \leftarrow blockDim.x \times (blockIdx.x + gridDim.x \times blockIdx.y) + threadIdx.x$
4:     i← **value**[tid]
5:     j← **value**[tid + 1]
6:     k← **value**[tid + 2]
7:     $\Delta time_0$ ←T[j]-T[i]
8:     $\Delta time_1$ ←T[k]-T[i]
9:     **if** $\Delta time_0 < TW$ **then**
10:       record i and j
11:     **end if**
12:     **if** $\Delta time_1 < TW$ **then**
13:       record i and k
14:     **end if**
15: **end function**

---

There are two main parts in the sorting-based time SCP, sorting and coincidence. In the process of parallelizing the sorting algorithm, according to the piecewise ordered data type of the sequence, the optimal sorting method should be merge sort. However, there have been developed parallel sorting libraries in CUDA. Therefore the function cub:: DeviceRadixSort::SortPairs, which is based on radix sort, is used to sort the continuous data stream to obtain the global time-ordered data stream. Besides, in order to avoid the migration of a large amount of data, including hit time, photon energy, position and so on, lexicographical sorting is used in this paper.

Specifically, the arrival time of the photon is taken as the key and the global index of the single is taken as the value. Next, we parallelize the process of time coincidence. The modified time coincidence processing is as follow. First, the hit time differences between the current single and the next two singles are obtained by making a first forward difference to the global time-ordered sequence in a GPU thread. And then by comparing the time differences and TW, it is determined whether these two pairs of singles constitute coincidence pairs. Finally, the subsequent LOR encoding is performed to encode photon positions as the LOR index. The specific implementation process is shown in Algorithm 1. Compared with the conventional time coincidence processing, the simplified time difference method could lose a certain accuracy. At the cost of certain processing precision, such modification greatly improves the parallelism of the algorithm.

## 3 Results

### 3.1 Experimental Setups

To evaluate the performance and illustrate the effectiveness of the proposed GPU-SCP, we did two phantom experiments on our PET system with 20 BDMs.
1) The Rat-like phantom was scanned in order to evaluate the speedups and real-time performance of GPU-SCP on high activity nuclide. The diameter of the phantom is 60mm and the length is 110mm. The line source with 44.4MBq (1.2mCi) fluorine-18 fluorodeoxyglucose (F-18-FDG) at the start of scanning was inserted into the hole of the phantom. The scan lasted 2 minutes.
2) Derenzo phantom was used to evaluate the speedups of GPU-SCP on low activity nuclide and evaluate the impact of GPU-SCP on the resolution of images. The diameter of the hot spots was 1mm, 1.5mm, 1.8mm, 2.4mm, 3.0mm, 3.6mm. The phantom was filled with F-18-FDG whose activity was 3.4MBq (92$\mu$Ci). The scan lasted 10 minutes.
The acquisition and SCP were run on a Dell Poweredge with two Intel Xeon Sliver 4108 (1.80 GHz, 16 (32 logical) cores) CPUs, 512 GByte DDR3-RAM and an NVIDIA GeForce GTX 1070 GPU. The acceleration strategies of the CPU-SCP are similar to that of GPU-SCP. In the acquisition, 20 logical cores were used to ensure the completeness of scan data of 20 BDMs. According to the above settings, there were 12 logical cores left for SCP. In the SCP, the energy window was set to 350-650 keV. The coincidences were extracted by 6ns TW. Images were reconstructed using an MLEM reconstruction algorithm with 50 iterations.

### 3.2 Speedup

The processing time and speedups of different algorithms in 'Before Coincidence Processing' are shown in Table 1. The parallel CPU-SCP with 10 threads can increase the processing speed by 3.4 times. Proposed GPU-SCP achieved the maximum speedups in data frame decoding and correction which are up to 80 times faster than serial CPU-SCP and 18.4 times the maximum speedups of parallel CPU-SCP with 10 threads. Even when the data transmission between RAM and video memory is considered, the advantages in speed of the GPU-SCP are still obvious.

| Phantom(Activity) | Rat-like(1.2mCi) | | Derenzo(92.0$\mu$ Ci) | |
|---|---|---|---|---|
| | Time | Ratio | Time | Ratio |
| CPU-SCP(serial) | 81.0 | 1.0 | 41.7 | 1.0 |
| CPU-SCP(10 threads) | 18.5 | 4.4 | 10.0 | 4.2 |
| GPU-SCP | 1.0 | 81.0 | 0.63 | 66.2 |
| GPU-SCP* | 4.4 | 18.4 | 3.1 | 13.5 |

**Table 1:** Processing time and speedups of data frame decoding, crystal positron encoding and energy correction.'Time' represents processing time in seconds and 'Ratio' represents the ratio of the processing time of serial CPU-SCP to the processing time of corresponding approach. '*' means the data transmission between RAM and video memory is considered.

| Phantom(Activity) | Rat-like(1.2mCi) | | Derenzo(92.0 Ci) | |
|---|---|---|---|---|
| | Time | Ratio | Time | Ratio |
| CPU-SCP(serial) | 177.8 | 1.0 | 81.5 | 1.0 |
| CPU-SCP(10 threads) | 114.4 | 1.6 | 72.9 | 1.1 |
| GPU-SCP | 16.8 | 10.6 | 11.6 | 7.0 |

**Table 2:** Processing time and speedups of energy coincidence processing, time coincidence processing and random correction.The abbreviations have the same meanings as Table1.

Table 2 shows the processing time and speedups of different algorithms in 'Coincidence Processing'. The sorting-based GPU-SCP achieved the maximum speedup which is 10.6 times faster than serial CPU-SCP at most. It is noted that the processing speed of the GPU-SCP is more than 6 times the speedups of the parallel GPU-SCP with 10 threads.

Table 3 shows the processing time and speedups of different algorithms in the overall SCP. The results are similar to Table 2 because the execution time of coincidence processing takes up the largest proportion in the whole process. The proposed GPU-SCP is up to 13.5 times faster than serial CPU-SCP and about 7.6 times the maximum speedups of the parallel CPU-SCP with 10 threads at most.

| Phantom(Activity) | Rat-like(1.2mCi) | | Derenzo(92.0 Ci) | |
|---|---|---|---|---|
| | Time | Ratio | Time | Ratio |
| CPU-SCP(serial) | 258.8 | 1.0 | 123.2 | 1.0 |
| CPU-SCP(10 threads) | 132.9 | 1.9 | 82.9 | 1.5 |
| GPU-SCP | 17.8 | 14.5 | 12.2 | 10.1 |
| GPU-SCP* | 21.2 | 12.2 | 15.7 | 7.8 |

**Table 3:** Processing time and speedups of entire process.The abbreviations have the same meanings as Table1.

| Methods | MRE on LORs | RE of Count of Coincidences |
|---------|-------------|----------------------------|
| CPU-SCP | 1.6% | 4.6% |
| GPU-SCP | 2.0% | 0.1% |

**Table 4:** Maximum relative error of LORs and relative error of total count of coincidences by different SCPs in the rat-like phantom experiment.

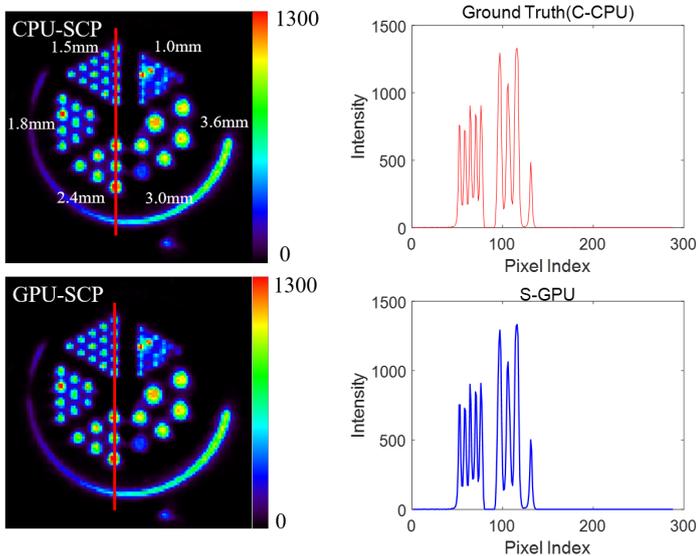

**Figure 2:** Reconstruction images of different SCPs for Derenzo phantom.

### 3.3 Image Quality

In the Rat-like phantom experiment, the output of offline serial CPU-SCP was used as the ground truth considering that the offline CPU-SCP does not cause the data loss. The ground truth was compared with the outputs of all online methods. We calculated the maximum relative error of LORs and the relative error between the total counts of coincidences of SCPs and the ground truth. As shown in the Table 4, the maximum relative error on LORs is less than 2% for all methods. However, the relative error of total count of coincidences by parallel CPU-SCP is 4.6%, indicating the data loss in high activity acquisition by parallel CPU-SCP.

Figure 2 shows reconstruction images of the Derenzo phantom for serial CPU-SCP and sorting-based GPU-SCP. The outputs of offline serial approach are used as ground truth. The line profiles across 1.5mm rods and 2.4mm rods were shown on the right side of Figure 2 There was no significant difference in image quality of GPU-SCPs and the ground truth. Besides, in order to compare performances of different methods in image domain and projection domain quantificationally, we extracted the max value of reconstruction images and the total count of coincidences to calculate the difference in image domain and projection domain respectively. The differences between the GPU-SCP and the ground truth are less than 1% in image domain and in projection domain.

### 4 Discussion and Conclusion

The speedups of GPU algorithm are stable and optimal in 'Before Coincidence Processing'. The reason for this phenomenon is the high parallelism of the process which is satisfied with SIMD model. In 'Coincidence Processing', the performance of GPU-SCP is not so surprising as in 'before coincidence processing'. However, the proposed GPU-SCP can still realize the real-time processing for nuclides of different activity without image degradation.

In this paper, we propose the GPU-SCP to solve the limited-thread problem. The proposed processing architecture simplifies the management of threads between acquisition and coincidence processing, accelerates the coincidence processing by GPU multiple threads and finally realizes the online coincidence processing.

### Acknowledgments

This work was partly supported by the National Key Research and Development Program of China under Grant No. 2016YFC0103800, the National Natural Science Foundation of China under Grant Nos. 62071362, 61901339, the Project funded by China Postdoctoral Science Foundation No. 2018M643589, the National Natural Science Foundation of Shaanxi Province under Grant No.2019JQ-655, and the Fundamental Research Funds for the Central Universities Nos. JB211209.

# A Deep Convolutional Framelet Network based on Tight Steerable Wavelet: application to sparse-view medical tomosynthesis

Luis F. Alves Pereira[1,2], Vincent Van Nieuwenhove[3], Jan De Beenhouwer[1], and Jan Sijbers[1]

[1]imec Vision Lab, University of Antwerp, Antwerp, Belgium
[2]Universidade Federal do Agreste de Pernambuco, Garanhuns, Brazil
[3]AGFA NV, Mortsel, Belgium

**Abstract** Deep Learning networks outperformed the state-of-the-art in many fields. However, the link between those methods and the classical signal processing theory is not yet fully understood. To address such gap, Deep Convolutional Framelet Networks (DCFNs) were proposed as a new scheme for signal representation composed of learning and non-learning components. DCFNs allow the representation of input signals using a fixed non-local basis $\Phi$ convolved with a data-driven, local basis $\Psi$. We propose a novel DCFN where $\Phi$ is a Steerable wavelet. The data representation within our network is translation- and rotation- invariant. We evaluated our method for recovering sparse-view tomosynthesis images from ripple artifacts. Both quantitative and qualitative evaluations show that our method performs better than U-net for this application.

## 1 Introduction

Despite the huge success of the Deep Learning methods in many fields, the effectiveness of those architectures cannot be demonstrated mathematically by classical signal processing approaches. Therefore, Deep Convolutional Framelet Networks (DCFNs) [1] were designed based on the theory of convolution framelets to fill this gap. DCFNs are regarded as a new scheme for signal representation that - in contrast to wavelets [2], SVD [3], and DCT [4] - is composed of learning and non-learning components.

Convolution framelets were first proposed by Yin *et al.* [5] for representing a signal using a fixed, non-local basis $\Phi$ convolved with a data-driven, local basis $\Psi$. Later, Ye *et al.* [1] showed that such convolution framelet representation can also be regarded as a DCFNs. In such new deep networks, local convolutions filters are learned after a given non-local basis $\Phi$ is fixed.

In this work, we propose a novel DCFN by introducing a Steerable wavelet as $\Phi$. Steerable wavelets are much more flexible than orthogonal separable Wavelets (such as Haar, Daubechies, and others) since no orthogonality constraints are applied to the filters. In a Steerable basis, the filters are only constrained to be *(i)* rotated copies of each other; and *(ii)* a linear combination of the basis filters [6]. Therefore, a large variety of filter sets can be chosen according to the application task.

We evaluate the proposed network for recovering digital X-ray tomosynthesis images in a sparse-view setup. Tomosynthesis is a medical imaging technology known for providing superior diagnostic information than 2D radiography at a lower cost and radiation than 3D X-ray Computed Tomogra-phy [7]. By acquiring sparse-view projections from a regular tomosynthesis setup, faster scans that operate with lower radiation levels could be obtained. However, the lack of projection data available generates clinically unacceptable images highly degraded by artifacts. To improve the image quality in X-ray tomosynthesis, the current baseline in the literature relies on the use of U-nets [8]. For this reason, we compare the accuracy of our method against an U-net.

## 2 Mathematical Background

A *frame* is a family of functions $\{\phi_k\}_{k\in\mathbb{N}}$ in a Hilbert space $H$ that decomposes any signal $x \in H$ into $K$ components $\langle x, \phi_k \rangle$ that satisfy the following:

$$\alpha||x||^2 \leq \sum_{k=1}^{K} |\langle x, \phi_k \rangle|^2 \leq \beta ||x||^2, \forall x \in H \qquad (1)$$

where $\alpha, \beta > 0$ are frame bounds. When $\alpha = \beta$, the frame $\Phi$ composed of $\{\phi_k\}$ is called tight. From the frame coefficients $c = \Phi^T x$, the original signal can be recovered using the dual frame $\tilde{\Phi}$ which satisfies the frame condition $\tilde{\Phi}\Phi^T = I$:

$$\hat{x} = \tilde{\Phi}c = \tilde{\Phi}\Phi^T x = x \qquad (2)$$

For any input signal $x \in \mathbb{R}^n$, let $\Phi = [\phi_1, \cdots, \phi_m] \in \mathbb{R}^{n \times m}$ (which interacts with all the $n-$elements of $x \in \mathbb{R}^n$) be a non-local basis, and $\Psi = [\psi_1, \cdots, \psi_q] \in \mathbb{R}^{d \times q}$ (which only interacts with $d-$neighborhood of $x \in \mathbb{R}^n$) be a local basis. Therefore, the deep convolutional framelet expansion states that [1]:

$$x = \frac{1}{d}\sum_{i=1}^{m}\sum_{j=1}^{q} \langle x, \phi_i \circledast \psi_j \rangle \tilde{\phi}_i \circledast \tilde{\psi}_j \qquad (3)$$

where $\circledast$ refers to the convolution operator. Furthermore, if $C$ is a framelet coefficient matrix, where [1]:

$$C = \Phi^T(x \circledast \overline{\Psi}), \qquad (4)$$

Equation (3) can be represented by:

$$x = (\tilde{\Phi}C) \circledast \nu(\tilde{\Psi}) \qquad (5)$$

where

$$v(\tilde{\Psi}) = \frac{1}{d} \begin{bmatrix} \tilde{\Psi}_1 \\ \cdots \\ \tilde{\Psi}_q \end{bmatrix} \in \mathbb{R}^{dq} \tag{6}$$

Based on Equations (4) and (5), Ye et al. derived the multi-layer implementation of convolution framelets composed of the following encoder stage [1]:

$$C_{low}^{(0)} = x, \forall x \in \mathbb{R}^n \tag{7}$$

$$C_{low}^{(l)} := \Phi_{low}^{(l)T} \left( C_{low}^{(l-1)} \circledast \overline{\Psi}^{(l)} \right) \tag{8}$$

$$C_{high}^{(l)} := \Phi_{high}^{(l)T} \left( C_{low}^{(l-1)} \circledast \overline{\Psi}^{(l)} \right) \tag{9}$$

where $\Phi_{low}$ and $\Phi_{high}$ are set of functions from the frame $\Phi$ that decomposes an input signal into low- and high-pass sub-bands, respectively. Then, the following decoder stage completes the multilayer convolution framelet:

$$\hat{C}_{low}^{(L-1)} := \left( \tilde{\Phi}^{(L)} \hat{C}^{(L)} \right) \circledast v \left( \tilde{\Psi}^{(L)T} \right) \tag{10}$$

$$\hat{x} = \hat{C}_{low}^{(0)} \tag{11}$$

where $\hat{C}^{(l)}$ is the composition of $\left\{ \hat{C}_{low}^{(l)}, \hat{C}_{high}^{(l)} \right\}$, so:

$$\Phi^{(l)} \hat{C}^{(l)} \tilde{\Psi}^{(l)T} = \Phi_{low}^{(l)} \hat{C}_{low}^{(l)} \tilde{\Psi}^{(l)T} + \Phi_{high}^{(l)} \hat{C}_{high}^{(l)} \tilde{\Psi}^{(l)T} \tag{12}$$

and $\hat{C}_{high}^{(l)}$ is given based on a high-frequency filter $H^{(l)}$:

$$\hat{C}_{high}^{(l)} = C_{high}^{(l)} \circledast H^{(l)} \tag{13}$$

## 3 Proposed Method

The Steerable Pyramid allows image decomposition into scale, and orientation sub-bands [6]. In this wavelet-like representation, the framelet basis are dilated and rotated versions of a single directional wavelet [9]. In this way, the steerable wavelet decomposition components capture orientations of local image features in any $k$ directions. Furthermore, they form a tight frame that satisfies the frame condition given in Equation (2).

The filter banks diagram illustrated in Figure 1 shows a complete stage of the steerable transform composed of decomposition ($\Phi$) and restoration ($\tilde{\Phi}$) for $k = 4$. In the diagram, $\{B_n(\omega_1, \omega_2), \ n = 0, 1, 2, 3\}$ are band-pass oriented filters, $H_0(\omega_1, \omega_2)$ and $L_1(\omega_1, \omega_2)$ are non-oriented high-pass and narrowband low-pass filters, respectively. Higher levels of decomposition are obtained recursively by inserting the subsystem enclosed in the dashed lines between the down-sampling and up-sampling blocks.

In this work, we propose a DCFN that employs a Steerable wavelet as a fixed non-local basis $\Phi^1$. This way, the network local basis $\Psi$ learns to suppress noise and artifacts into the low- and the high-pass sub-bands of the input image. The data representation in our method is rotation-invariant due to the use of steerable derivative operators and translation-invariant due to the nature of the band-pass decomposition in the Fourier space.

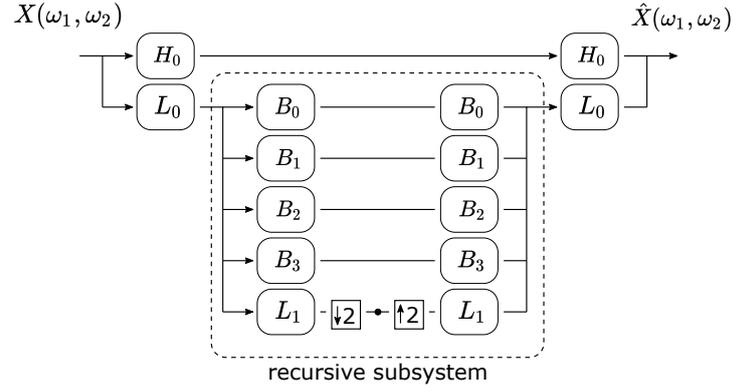

**Figure 1:** Diagram of filter banks presenting a complete stage of the steerable transform ($\Phi$) and its inverse ($\tilde{\Phi}$).

Furthermore, in contrast to the regular orthogonal separable Wavelets, the filter sets used at the Steerable wavelet are highly flexible since there are no orthogonality constraints. In this way, filters can be designed to focus on the characteristics of noise and artifacts to be suppressed in a particular denoising problem.

Finally, the proposed method also includes Mixed Scale Dense (MSD) [10] and Deconvolution [11] networks in its architecture. In contrast to architectures that enlarge the network receptive field via tensor scaling [12] and need to learn upscaling operations, the MSD networks exploit dilated convolutions. As a result, the models obtained are smaller and easier to train. To avoid image blurring, the Deconvolution networks apply large horizontal, vertical, and squared kernels along with its layers.

Figure 2 illustrates the architecture of the proposed DCFN. The gray tensors refer to the data input/output, the yellow tensors refer to the output of MSD networks, the red tensors are the high and low band of the $\Phi = \{\Phi_{low}, \Phi_{high}\}$ decomposition, the blue tensors are the result of $\tilde{\Phi}$ reconstructions, and the green tensors refer to the output of the Deconvolution networks.

## 4 Experiments

Our experiments to evaluate the proposed network were focused on recovering sparse-view X-ray tomosynthesis images from ripple artifacts. By doing so, we expect to allow faster scans that involve reduced levels of radiation dose.

Let $T$ be a training set composed of $I$ 3D reconstructions of full-sampled tomosynthesis with $J$ slices per volume, and the same amount of sparse-sampled data. Then, $T = \{(\mathbf{x}'_{ij}, \mathbf{x}_{ij}), \ \forall i \in [1, \cdots, I], \ \forall j \in [1, \cdots, J]\}$ where $\mathbf{x}'$, and $\mathbf{x}$ refer to 2D slices of sparsely and fully sampled tomosynthe-sis data, respectively. We trained our model $f_\theta(\cdot)$ to find the set of parameters $\theta$ that minimizes:

$$\arg\min_\theta \sum_{i=1}^{I} \sum_{j=1}^{J} ||f_\theta(\mathbf{x}'_{ij}) - \mathbf{x}_{ij}||_2^2 \qquad (14)$$

---

$^1$the code of the Steerable layers of our DCFN is available at: github.com/luisfilipeap/A-Deep-Convolutional-Framelet-Network-based-on- Tight-Steerable-Wavelet

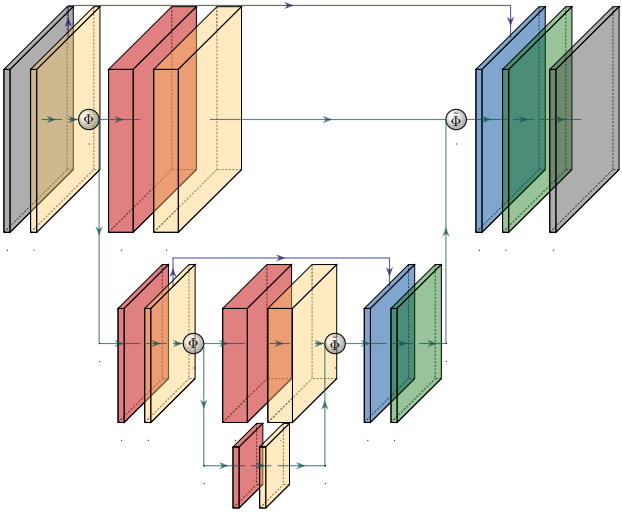

**Figure 2:** Architecture of the proposed DCFN: the gray tensors refer to the data input/output, the yellow tensors refer to the output of MSD networks, the red tensors are the high and low band of the $\Phi = \{\Phi_{low}, \Phi_{high}\}$ decomposition, the blue tensors are the result of $\bar{\Phi}$ reconstructions, and the green tensors refer to the output of the Deconvolution networks.

We designed our experimental setup in Python using the Tensorflow framework. During the training stage, the ADAM optimizer was used to update the network's weights with learning rate and batch size equal to $10^{-5}$ and 20, respectively. It took 9 days to run 300 epochs on an Nvidia DGX station using a single Tesla V100 GPU. Furthermore, the Steerable wavelet chosen for $\Phi$ was based on the $4^{th}$ order filters described by Karasaridis and Simoncelli [13].

### 4.1 Dataset

We used medical data from the Clinical Proteomic Tumor Analysis Consortium Pancreatic Ductal Adenocarcinoma (CPTAC-PDA) collection[2] as the training set. It contains Computed Tomography (CT), and Magnetic Resonance Imaging (MRI) data from 74 patients with pancreatic cancer. To design a solution with high generalization capacity, we used MRI and CT data. Reconstruction volumes with less than 120 slices were discarded. As a result, 152 volumes were used. For testing our model, we used the Visible Human Project dataset[3]. It contains 10 CT reconstructions of different human parts from male and female subjects, such as the ankle, pelvis, knee, and shoulder.

---

[2]https://wiki.cancerimagingarchive.net/display/Public/CPTAC-PDA
[3]https://mri.radiology.uiowa.edu/visible_human_datasets.html

### 4.2 Scanning setup

We simulated the scanned setup illustrated in Figure 3 using the Astra-Toolbox [14]. Such setup is composed of an X-ray source (a) that moves linearly to scan a patient lying down on a stationary detector (b). This is a particular challenging scanning geometry for 3D reconstructions due to its minimal angular view.

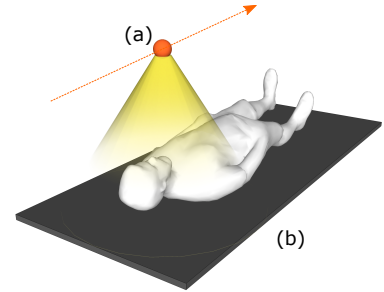

**Figure 3:** Tomosynthesis scanning setup simulated in this work.

Using $120 \times 512 \times 100$ volumes from our dataset as phantoms, fully sampled ($\mathbf{x}_{ij}$) and sparsely sampled ($\mathbf{x}'_{ij}$) acquisitions were reconstructed using SIRT from 100 and 20 X-ray projections, respectively. Furthermore, we used the source-to-object distance (SOD) as 1500 $mm$, the source-to-detector distance (SDD) as 1650 $mm$, the detector as a $1406 \times 1130$ grid of unit cells, and a linear track of 800 $mm$ for the movement of the radiation source.

### 5 Results and Discussion

Figure 4 shows central slices obtained from sparsely sampled tomosynthesis data reconstructed by SIRT without Deep Learning processing as well as slices obtained from datasets processed by U-net [15] and our method. It also shows error maps that stretch the difference between the reconstructions and the fully sampled tomosynthesis data. From Figure 4, it can be appreciated that the energy of the U-net error map is higher than that of our method.

Furthermore, Table 1 shows the SSIM, PSNR, and NRMSE metrics for a quantitative evaluation of the reconstruction images presented in Figure 4. Our method outperforms U-net in all metrics, notably in PSNR.

|          | SSIM   | PSNR   | NRMSE  |
|----------|--------|--------|--------|
| **SIRT** | 0.9978 | 54.063 | 0.0335 |
| **U-net**| 0.9983 | 55.412 | 0.0287 |
| **Proposed** | **0.9986** | **57.028** | **0.0238** |

**Table 1:** Quantitative evaluation of the proposed method in relation to SIRT and U-net.

### 6 Conclusion

We presented a Deep Convolutional Framelet Network (DCFN) that employs a Steerable wavelet as a fixed non-

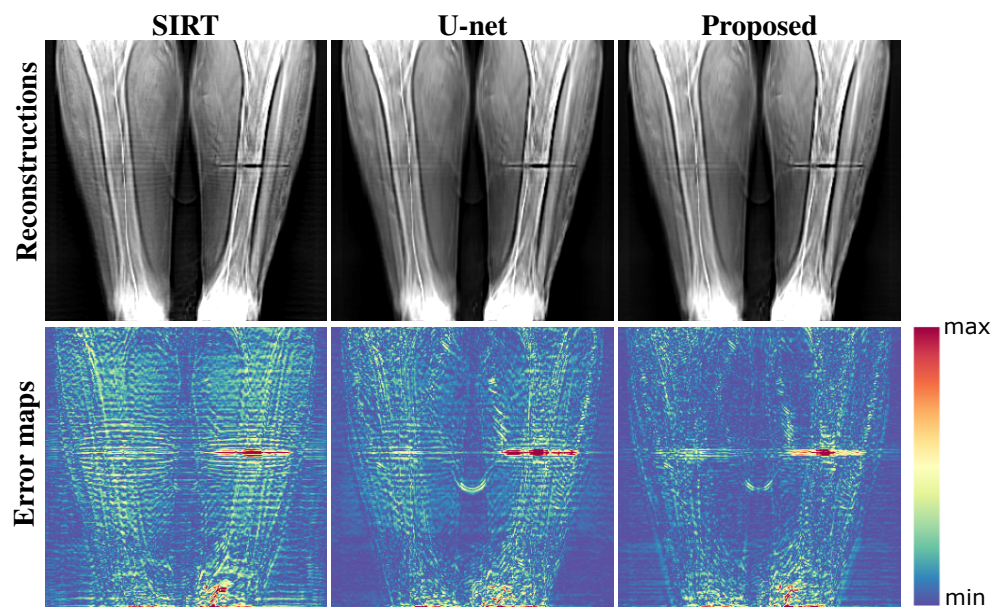

**Figure 4:** In the first line, reconstruction images obtained using SIRT, Unet, and the proposed method. In the second line, their respective error maps with contrast stretching.

local basis Φ. As a result, the data representation within the model is translation- and rotation-invariant. In contrast to regular orthogonal separable Wavelets, the filter set used here is flexible since there are no orthogonality constraints.

We applied our proposed data representation to the reduction of ripple artifacts in sparse-view tomosynthesis images. Quantitative and qualitative evaluations showed that our method outperforms the U-net, which is the current baseline in the literature for denoising X-ray tomosynthesis.

### Acknowledgement

This work was financially supported by VLAIO through the ANNTOM project HBC.2017.0595 and through the Flemish Government under the "Onderzoeksprogramma AI Vlaanderen" programme.

# Reducing Metal Artifacts by Restricting Negative Pixels

Gengsheng L. Zeng[1,2] and Megan Zeng[3]

[1]Department of Computer Science, Utah Valley University, Orem, USA
[2]Department of Radiology and Imaging Sciences, University of Utah, Salt Lake City, USA
[3]Department of Electrical Engineering and Computer Science, University of California at Berkeley, Berkeley, USA

**Abstract** When the object contains metals, its x-ray computed tomography images are normally affected by streaking artifacts. These artifacts are mainly caused by the x-ray beam hardening effects, which deviate the measurements from their true values. One interesting observation of the metal artifacts is that certain regions of the metal artifacts often appear as negative pixel values. Our novel idea in this paper is to set up an objective function that restricts the negative pixel values in the image. We must point out that the naïve idea of setting the negative pixel values in the reconstructed image to zero does not work. This paper proposes an iterative algorithm to optimize this objective function, and the unknowns are the metal affected projections. Once the metal affected projections are estimated, the filtered backprojection algorithm is used to reconstruct the final image. This paper also proposes a convex optimization formulation for this problem.

## 1 Introduction

Due to the wide energy spectrum of x-rays, beam hardening effects are severe when the object being imaged contains metals. The beam hardening effects introduce large errors in the x-ray computed tomography (CT) projection measurements. Those measurement errors in turn produce artifacts in the reconstructed CT images. Typical metal artifacts appear as dark and bright streakings. This metal artifact problem has been recognized for a long time and it is still an open problem.

Most methods to combat the metal artifacts are iterative algorithm based [1-8]. Among these iterative algorithms, projection data inpainting is popular. The basic principle of inpainting is first to remove the metal affected measurements and to assume that there is no metal in the object. Next, estimation methods such as interpolation, lowpass filtration, or some non-linear approaches are used to inpaint the measurements that are artifacially removed in the first step. Iterative algorithms are designed to optimize an objective function, which can contain Bayesian terms. For example, the total variation (TV) norm is effective in enforcing the peicewise constant prior [9-10]. Noise weighting is often incorporated in the objective function as well.

Our proposed method is inspired by the observation that the metal artifacts usually have regions with negative pixel values. The innovation of this paper is the establishment of an objective function that restricts the negative pixel values in the reconstructed images. The proposed methods will be presented in the next section. Results with real x-ray CT measurements are presented. The measurements are obtained from airport bags that contain metal objects inside.

## 2 Methods

A usual objective function in image reconstruction consists of two parts: the data fidelity part and the Bayesian part. The data fidelity part projects the image array to generate pseudo projections and then matches them to the measurements. Noise weighting can be applied in the data fidelity part. The main purpose of the Bayesian part is for regularization, because the image reconstruction problem may be ill posed. An $L_2$-nom of the reconstructed image can be used to regularize the image to enforce smoothness. The TV norm of the image can be used to denoise and maintain the sharp edges, by encouraging the piecewise constant constraint. Projection data inpainting is usually required before iterative image reconstruction. Unfortunately, inpainting methods are problematic and the pseudo projections are not the same as the projections when metals are not present.

### 2.1 Approach #1

It is observed that the metal artifacts often have regions with negative pixel values. This paper proposes a novel objective function, which is the total sum of the negative pixel values (or the total sum of the square of the negative pixel values).

Let the reconstructed image be $X$, represented as a vector. The proposed objective function is

$$F = \max_{p_M}\{\sum_i min(0, x_i)\}, \qquad (1)$$

subject to

$$p_M \geq p_M^{measure} \qquad (2)$$

where $x_i$ is the $i$th pixel of $X$ and $p_M$ is metal affected measurements. The set $p_M$ is a small proper subset of the total measurements. Therefore, the entire measurements consist two parts: $p_M$ and $p_0$, where $p_0$ represent the measurements not affected by the metals. The projections in $p_M$ are the optimization variables, but the projections in $p_0$ are fixed. Essentially, the optimization problem (1) is measurement inpainting. Unlike most other methods, which estimate the measurements as if the metals are not present, we estimate the measurements of metals. We adjust the metal projections until the metal artifacts are reduced. We choose the metal artifact indicator as the negative pixel values.

In the constraint (2), $p_M^{measure}$ is the set of metal affected projections in the original measurements, $p_M$ is the same set but is treated as variables. The beam hardening effects introduce large errors to $p_M^{measure}$. Our empirical evidence indicates that the beam hardening effects make the measurements smaller than they really are. The constraint (2) implies that the true values in $p_M$ are somewhat larger than the measured values.

Most optimization algorithms are gradient based. However, the proposed objective function (1) involves a non-differentiable function $min(0, x_i)$, which makes the optimization of (1) difficult. Currently, we use the following steps to optimize (1).

Step 1: Use the filtered backprojection (FBP) algorithm to generate a raw image $X_{raw}$ using $\{p_M, p_0\}$. The raw image may contain severe metal artifacts.

Step 2: Calculate the total sum of all the negative pixels $N = \sum_i min(0, x_i^{raw})$.

Step 3: Segment the raw image to obtain a metal-only image.

Step 4: Forward project the metal-only image to obtain the indices of $p_M$.

The above 4 steps are the preparation steps. The following Step 5 actually optimizes the objective function (1).

Step 5: This step is an iterative algorithm. At each iteration, loop through all indices of $p_M$.

   Step 5.1: Introduce small perturbations to $p_M$, obtaining $p_M^{temp}$.

   Step 5.2: Obtain the FBP reconstruction $X_{temp}$ using $\{p_M^{temp}, p_0\}$.

   Step 5.2: Calculate the total sum of all the negative pixels $N_{temp} = \sum_i min(0, x_i^{temp})$.

   Step 5.3: If $N < N_{temp}$, update $N$ as $N_{temp}$ and update $p_M$ as the perturbed $p_M^{temp}$.

Step 6: The final image is the FBP reconstruction with the latest $\{p_M^{temp}, p_0\}$, where $p_0$ is the measurements that are not affected by metals and is never changed in the algorithm.

The proposed algorithm was implemented in MATLAB and applied to some CT data of airport bags. The original projections of airport bags were acquired with an Imatron C300 clinical CT scanner. The original projections were rebinned and downsized in this paper. The number of views for the scaled-down version was 180 over 180°. The number of channels (i.e., the detection bin at each view) for the scaled-down version was 597. The reconstructed image size was 420 × 420.

## 2.2 Approach #2

The objective function formulated in (1) can be formulated in other ways. For example, we can use convex optimization to formulate the same strategy [11,12]. In doing so, we need to introduce a set of slack variables, $T = \{t_i\}$, as follows:

$$\text{Objective funtion} = \min_{p_M}\|T\| \qquad (3)$$

subject to:

$$t_i \leq 0 \qquad (4)$$
$$t_i \leq x_i \qquad (5)$$
$$p_M \geq p_M^{measure} \qquad (6)$$
$$\|A_M X - p_M\| \leq \varepsilon \qquad (7)$$
$$\|A_0 X - p_0\| \leq \varepsilon \qquad (8)$$

where $A_M$ and $A_0$ are the projection matrices corresponding to $p_M$ and $p_0$, respectively. Therefore, our proposed problem is a standard convex optimization problem. One has the freedom to choose the norm involved in (3), (7), and (8). If the $\ell_1$ norm (i.e., the Manhatton norm) is chosen, some outliers are forgiven. If the $\ell_\infty$ norm (the Chebyshev norm) is chosen, the maximum error is considered. If the $\ell_2$ norm (i.e., the Euclidean norm), the problem is more mathematically friendly with the ease of evaluation of gradients.

Over-the-shelf canned software is available to solve (2). For example, CVX is a Matlab software for disciplined convex programming, developed by Stanford University. The CVX software requires the matrices $A_M$ and $A_0$ completely formed and fed into the package. This is currently not practical for a CT application. One must develop their own practical algorithm to solve (2) for any real-world size CT applications.

## 3 Results

Some results from Approach 1 are shown in Figures 1-4. Two airport bags were used. Fig. 1 shows the raw FBP reconstruction of the first bag. The negative values are shown as the darkest color. The metals appear as the brightest color. Fig. 2 shows the result generated by the proposed method.

Fig. 2 shows the raw FBP reconstruction of the second bag. The negative values are shown as the darkest color. The metals appear as the brightest color. Fig. 3 shows the result generated by the proposed method.

It is osverved that the proposed iterative algorithm reduces the number of negative pixels in the image. The streaking artifacts are also reduced.

The display window for the raw image and the final image is the same.

## 4 Discussion and Conclusions

This paper uses a unique objective function for image reconstruction and metal artifact reduction. The traditional iterative algorithm's main goal is to iteratively reconstruct the image. On the other hand, we use the FBP to reconstruct the image in every step of the iteration in the algorithms.

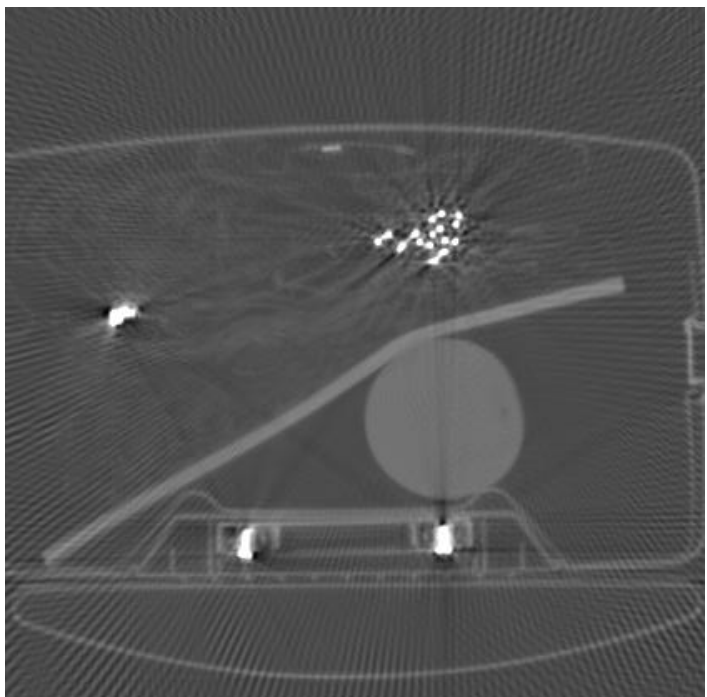

Fig. 1. Raw FBP reconstruction of bag #1.

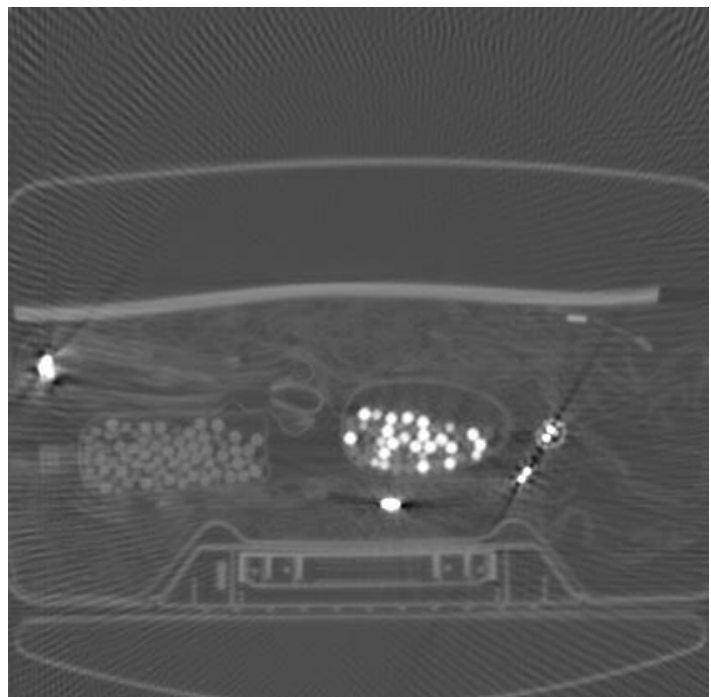

Fig. 3. Raw FBP reconstruction of bag #2.

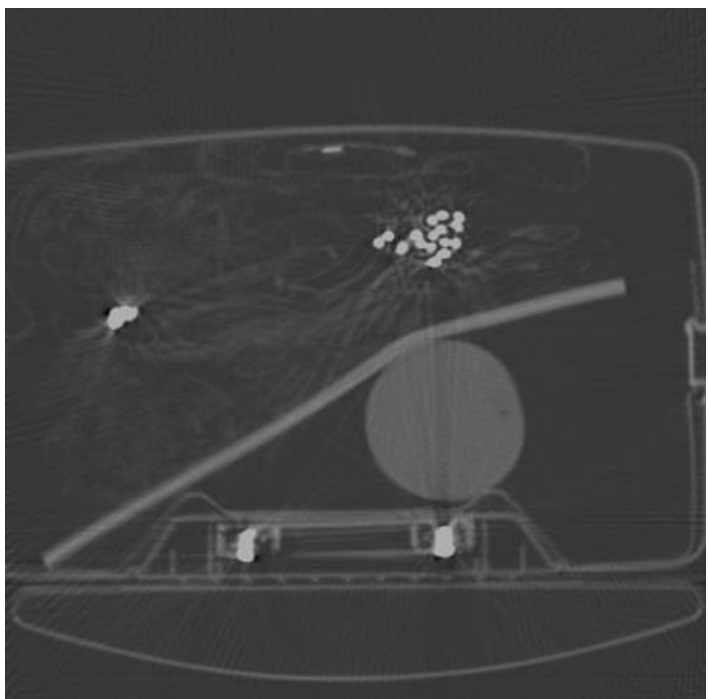

Fig. 2. FBP reconstruction of bag #1 using the processed sinogram.

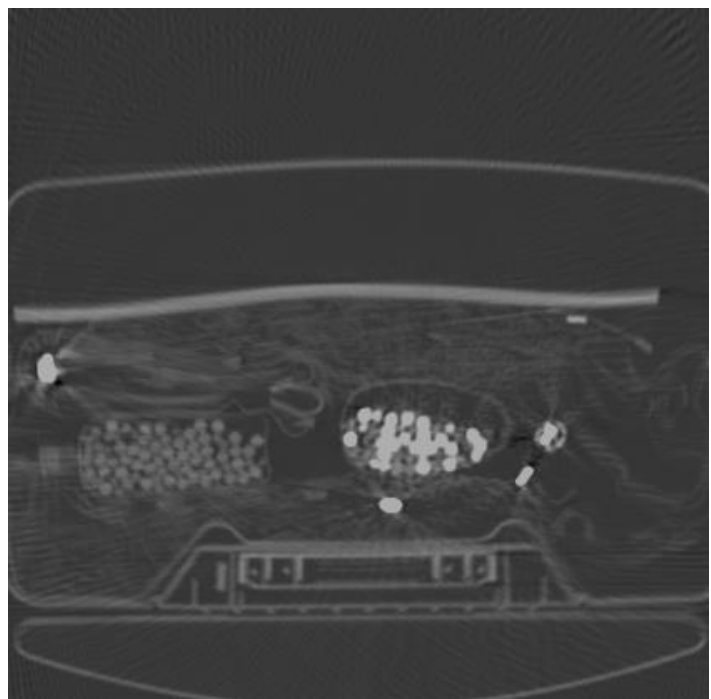

Fig. 4. FBP reconstruction of bag #2 using the processed sinogram.

From our knowledge, it is the first time in image reconstruction that the total sum of the negative pixels is used as the objective to be maximized. By reducing the total sum of the negtive pixels, the metal caused streaking artifacts are reduced accordingly.

The main difficulty of development of an algorithm to optimize the objective function is that we do not know the partial derivative of the objective function with respect to the varaibles, which are the metal affected projections. In traditional image reconstruction, the variables are the image pixels.

We do not treat the image pixels as variables, because the image can be readily reconstructed by the FBP algorithm once the projections are determined.

The motivation of the paper is to minimize the the features of the metal artifacts. The negative overshoots are not the only feature. There could be more features. Once we can express the features, we are able to minimize them. In our previous paper, the total variation (TV) was used as a feature for the metal artiacts [13]. The TV norm is useful and effective, but it may smooth the image too much.

ACKNOWLEDGMENT

The airport bag data was provided by the U.S. Department of Homeland Security, Science and Technology Directorate, under Task Order Number HSHQDC-12-J-00056. In this paper, we downgrade the spatial resolution of the original data on purpose. The views and conclusions are those of the author and should not be interpreted as necessarily representing the official policies, either expressed or implied, of the U.S. Department of Homeland Security.

# A Machine-learning Based Initialization for Joint Statistical Iterative Dual-energy CT with Application to Proton Therapy

Tao Ge[1], Maria Medrano[1], Rui Liao[1], David G. Politte[2], Jeffrey F. Williamson[3], and Joseph A. O'Sullivan[1]

[1]Department of Electrical & Systems Engineering, Washington University in St. Louis, St. Louis, United States
[2]Mallinckrodt Institute of Radiology, Washington University in St. Louis, St. Louis, United States
[3]Department of Radiation Oncology, Washington University in St. Louis, St. Louis, United States

**Abstract** Proton therapy is increasingly used in treating certain types of cancers because of decreased side effects compared to traditional radiation therapy. We previously proposed the dual-energy alternating minimization (DEAM) algorithm that has achieved sub-percentage uncertainty in estimating proton stopping power mappings from experimental phantom data. However, the high accuracy requirement leads to a large computational cost. To obtain accurate proton stopping power mappings in clinically acceptable time, a Convolutional Neural Network (CNN) based initialization method is introduced for DEAM. The CNN is trained on our former initialization and converged images. The simulation results show that our method generates denoised images with greatly improved estimation accuracy for adipose, tonsils, and muscle tissue. Also, it reduces elapsed time approximately 8-fold for DEAM to reach the same objective function value for both simulated and real data.

## 1 Introduction

Over the last 20 years, proton radiotherapy has been increasingly used to treat certain types of cancers because it has fewer side effects than traditional radiation. The absorbed dose reaches the maximum value after the proton beam traveling a certain depth in the object and then drops to near zero immediately. This peak is relatively narrow and is known as the Bragg peak. The estimated proton stopping power ratio (SPR) mapping allows radiation oncologists to align the Bragg peak to the tumor of the object, irradiating diseased tissues while sparing healthy cells. The current clinical practice estimates the SPR mappings from the single-energy CT (SECT) results, which leads to $2 - 3.5\%$ proton beam range uncertainty.

Dual-energy CT (DECT) SPR estimation methods were introduced to reduce the SPR uncertainty. Previous studies have shown that our iterative DECT algorithm, dual-energy alternating minimization (DEAM), has achieved sub-percentage uncertainty in estimating proton stopping-power mappings from experimental 3 mm collimated phantom data [1].

However, DEAM is quite time-consuming when reconstructing 3D image volumes from helical sinograms, due to the large system operator and its low convergence rate. Compared to SECT, DECT algorithms reconstruct two measured sinograms scanned at different peak energies, which at least doubles the required number of system operations per iteration. Moreover, because the objective function of DEAM is decoupled in two more domains than the objective function in single-energy monoenergetic CT alternating minimization (AM), DEAM converges much slower than the AM algorithm

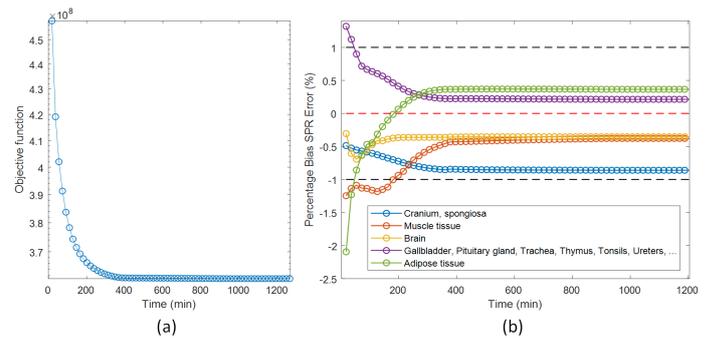

**Figure 1:** (a) plot objective function of DEAM versus time, (b) percentage biases of the SPR mapping derived from DEAM results with different elapsed time

with respect to the number of iterations. These factors make it difficult to get an accurate DECT result within a clinically acceptable time (20 minutes).

Several algorithm-based and implementation-based acceleration methods have been taken into account, including GPU computation and the ordered subsets method, but it still takes more than 4 hours for the DEAM algorithm to converge. Figure 1(a) shows the plotted objective function of DEAM versus time in minutes. It can be seen that the objective function converges after around 400 minutes. Figure 1(b) shows the percentage biases of the SPR mapping derived from DEAM results with different elapsed time. The SPR biases are calculated inside 5 regions of interest, corresponding to 5 different materials in a virtual human head phantom. It can be seen that the percentage biases reach the $(-1\%, +1\%)$ range after around 180 minutes and become steady after around 380 minutes, $9\times$ or $19\times$ greater than the target elapsed time.

Neural networks have been widely used in image processing and reconstruction because they are able to learn complicated potential image features which are difficult to capture with model-based methods. However, CNNs do not reliably generate accurate and critical information due to data susceptibility. As a result, they have been combined with the model-based optimization method to take advantage of the known physics knowledge. In [2, 3], researchers plug a pre-trained denoising CNN as a prior into a model-based optimization algorithm to solve different inverse imaging problems. In [4], an unrolled network of the model-based algorithm is constructed with trained hyper-parameters and a CNN regularizer for mask-based lensless imaging. In [5], photoacoustic tomography images are updated iteratively by a pre-trained CNN based on the previous image volume and the gradient computed by the model-based algorithm. In this paper, We introduce a CNN-based initialization method to better estimate the initial condition of DEAM, which takes advantage of CNN's speedup while sparing the data susceptibility.

## 2    Materials and Methods

### 2.1    Dual Energy Alternating Minimization (DEAM)

DEAM is a joint statistical iterative algorithm that minimizes the objective function given by the sum of I-divergence [6],

$$I(d||g) = \sum_j d_j(y) \ln \frac{d_j(y)}{g_j(y:c)} - d_j(y) + g_j(y:c), \quad (1)$$

and a penalty term,

$$R(c) = \lambda \sum_{i=1}^{2} \sum_x \sum_{\tilde{x} \in N_x} w(x,\tilde{x}) \Phi\left(c_i(x) - c_i(\tilde{x})\right), \quad (2)$$

$$\Phi(t) = \delta^2 \left( \left| \frac{t}{\delta} \right| + \log\left(1 - \left| \frac{t}{\delta} \right|\right) \right), \quad (3)$$

where $x, y$ denote the indices of the discretized image space and measurement space, respectively. $N_x$ denotes the set of the neighbouring voxels of the image index $x$, $w(x,\tilde{x})$ is the voxel weight calculated as the inverse physical distance between voxel $x$ and $\tilde{x}$ $\lambda$ and $\delta$ are two hyper-parameters that control the weight and sparsity of the regularization term, $i$ denotes image component index (specifically 1 for polystyrene and 2 for CaCl2 ), $j$ denotes measured data index (specifically 1 for 90 kVp and 2 for 140 kVp ), $d$ denotes measured data, $g(y:c)$ denotes the estimation of measured data based on image components $c_i$, which is the forward model, written as

$$g_j(y:c) = \sum_E I_{0,j}(y,E) \exp\left(-\sum_x h(x,y) \sum_{i=1}^{2} \mu_i(E)c_i(x)\right), \quad (4)$$

where $\mu_i(E)$ denotes the attenuation coeffcient of the $i^{th}$ material at energy $E$, $I_{0,j}$ denotes the photon counts of the $j^{th}$ peak energy in the absence of an object, which contains information of the spectrum and the bowtie filter, and $h(x,y)$ denotes the system operator that represents the helical fan beam CT system.

The original initial condition was estimated by a iterative filtered backprojection (iFBP) based algorithm [7] which requires less computational resources but has less in-practice accuracy than DEAM algorithm.

### 2.2    CNN-based initialization method

The architecture of the whole CNN-based estimation process is shown in figure 2. The main idea is to utilize CNN to estimate a better initial guess based on iFBP, our previous initialization method for DEAM. The proposed CNN has a widely used U-net structure originally from [8] with some modifications. Its encoding-decoding structure allows the neural network to learn global features and reduce noise. This CNN takes four inputs: iFBP $c_1$ image, iFBP $c_2$ image, the DEAM update direction of $c_1$, and the DEAM update direction of $c_2$. It has been observed that the CNN trained with update directions performs better than the CNN trained without update directions. All the input and output images share the same positioning and sampling information, so the entire process could take advantage of the image alignment. Due to the computational cost and their physical property, each slice of the 3D reconstructed image is trained or tested separately. The reconstructed image with the size $610 \times 610$ is zero-padded to $640 \times 640$ pixels to fit this U-net structure. The update directions for two basis vector model (BVM) components $c_1, c_2$ read

$$ud_i(x) = \log \frac{\sum_j p_{ij}^B(x)}{\sum_j q_{ij}^B(x)}, \quad (5)$$

where $p_{ij}^B$ denotes the backprojection of the $j^{th}$ measured data recalculated based on BVM and the spectrum for the $i^{th}$ basis, and $q_{ij}^B$ denotes the backprojection of the $j^{th}$ estimation data recalculated based on BVM and the spectrum for the $i^{th}$ basis. Therefore, the update direction gives the pixel-wise distance of the estimated components to the "truth."

In unregularized DEAM, two basis vector model components $c_1, c_2$ are updated by

$$c_i^{k+1}(x) = c_i^k(x) - \frac{1}{Z_i(x)} \log \frac{\sum_j p_{ij}^{B,k}(x)}{\sum_j q_{ij}^{B,k}(x)}, \quad (6)$$

where $Z_i(x)$ is the auxiliary variable to ensure convergence. Then, the CNN estimation

$$c_1, c_2 = CNN(c_1, c_2, ud_1, ud_2) \quad (7)$$

could be regarded as an update step of DEAM whose step size and regularization term are trained rather than embedded.

## 3    Results and Discussion

### 3.1    Simulation results

The measured data is simulated from the ICRP [9], a virtual female phantom. The simulation system operator has the same geometry as the Phillips Brilliance Big Bore CT scaner. The reconstruction and CNN training-testing process are done in a 20-threaded computer with 4 GTX 1080TI. The training process takes the iFBP result of ICRP chest and its corresponding update directions as the input and takes the ground truth of ICRP chest as the output. In the test process, the input of the trained CNN is the iFBP result of ICRP head. Figure 3 shows an example slice of the training and testing data. Chest iFBP is noisier than head iFBP since the larger area of the chest leads to the larger line integral of attenuation coefficients. Compared the CNN head result to the ground truth, the CNN eliminates most of the noise of iFBP images but introduces some artifacts.

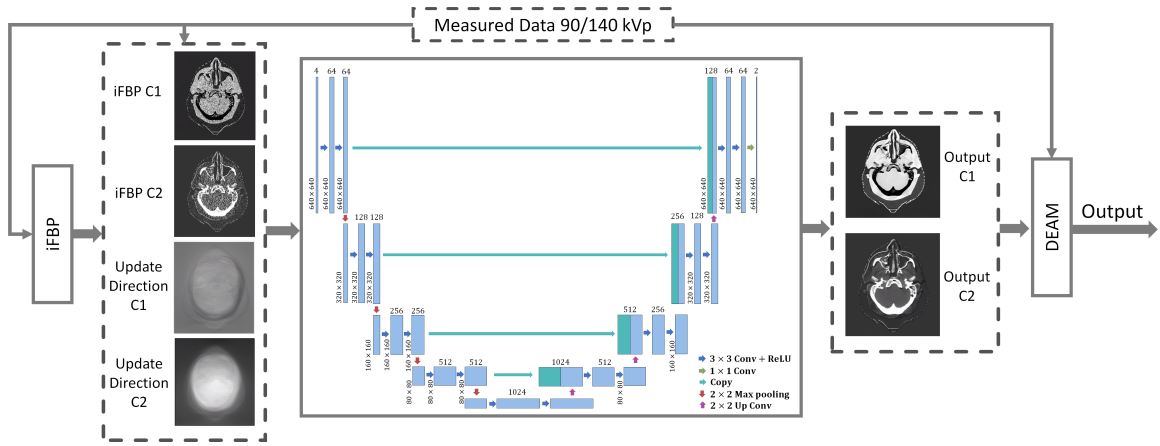

**Figure 2:** The overall process of DEAM initialization.

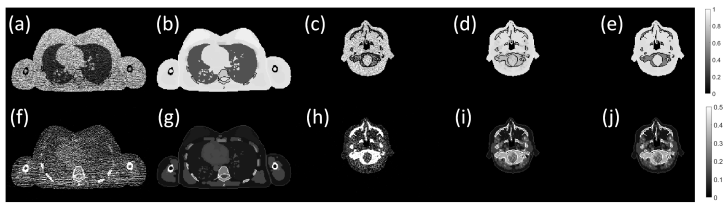

**Figure 3:** Examples of simulation images. From left column to right column: training input, training output, test input, test output, test ground truth (as reference). Top row: C1; Bottom row: C2.

Figure 4 shows the quantitative comparison. Five regions of interest are selected for different materials and indicated by red lines in 4(a). 4(b) shows the objective function of DEAM starting from different initial condition versus time. The objective function of iFBP is initially around 10 times larger than the objective function of CNN estimation. The objective function of CNN estimation after 3 minutes is close to the objective function of iFBP after 25 minutes, which means CNN introduces $\sim 8\times$ speedup in this case. 4(c) and (d) are the plots of percentage bias and percentage standard deviation of SPR mapping estimated by 20 minutes DEAM result starting from CNN estimation, the result of CNN estimation, iFBP, 20 minutes DEAM result starting from iFBP, the result of iFBP and converged DEAM.

Converged DEAM has the best overall performance. The uncertainty measures of converged DEAM for all ROIs are within 1%. iFBP has the worst overall performance with the largest uncertainty measures. CNN and CNN-DEAM-20min have less bias than iFBP-DEAM-20min in muscle tissue, tonsils, and adipose tissue, but the bias of CNN and CNN-DEAM-20min beyond $\pm 1\%$ range in spongiosa (cranium) and brain. Due to the absence of brain tissue and spongiosa (cranium) in training data, it is not surprising that CNN fails to estimate their BVM components. We hypothesize that it could be fixed by containing brain slices in the training data. iFBP got the highest standard deviation, which matches the noise level of iFBP images. Standard deviation of iFBP-DEAM-20min is the second highest, because the data fidelity term is much greater than the penalty term in early iterations, so DEAM tends to reduce the data fidelity term rather than the penalty term. CNN and CNN-DEAM-20min did a great job in the standard deviation analysis since the U-net structure is good at denoising.

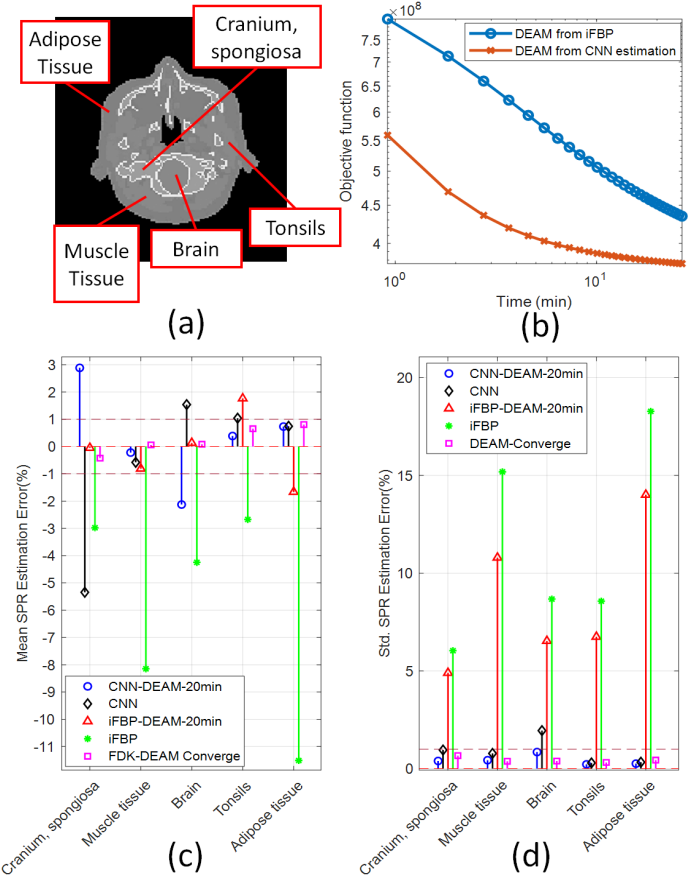

**Figure 4:** (a) Five ROIs in different materials, (b) plot of objective function of DEAM starting from iFBP result and DEAM starting from CNN estimation vs. time, (c) and (d) mean SPR estimation percentage error and SPR standard deviation of DEAM 20 minutes result with CNN estimation as initial condition, iFBP result, DEAM 20 minutes result with iFBP estimation as initial condition and DEAM converged result.

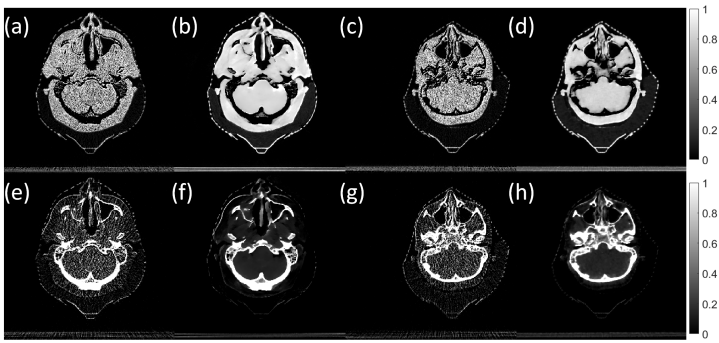

**Figure 5:** Examples of real patient images. From left column to right column: training input, training output, test input, test output. Top row: C1; Bottom row: C2

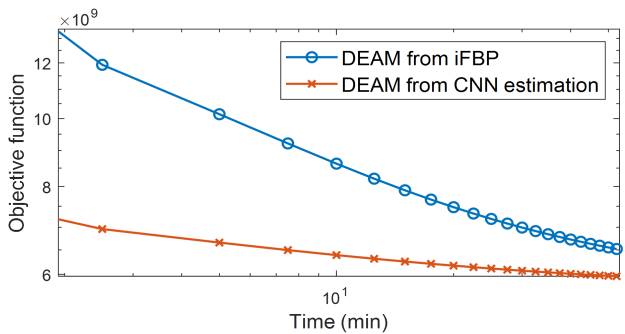

**Figure 6:** Real patient objective function of DEAM starting from iFDK and CNN estimation vs. time.

### 3.2 Clinic results

In this section, all 90 kVp and 140 kVp clinically measured data are acquired sequentially on the Phillips Brilliance Big Bore CT scanner, with 12 mm collimation. The CNN takes iFBP of the head scan as the training input and DEAM converged result of the same head scan as the training input. In the test process, iFBP of the head scan of a different patient is used as the input. The example images of the experimental training and testing data are shown in figure 5.

Figure 6 shows the plot of the real patient objective function of DEAM starting from iFDK and CNN estimation versus time. Similar to simulation results, the objective function of CNN estimation after 8 minutes is close to the objective function of iFBP after 70 minutes, which means CNN also introduces $\sim 8\times$ speedup in the real patient case.

## 4 Conclusion

We have proposed a CNN-based method that improves the initial guess for DEAM. We only show the CNN-based initialization of one iterative algorithm, but it is applicable for other iterative algorithms. This CNN initialization method takes iFBP result as the input, generating a denoised image with a great improvement of estimation uncertainty for adipose, tonsils, and muscle tissue in the simulation task. However, the method did not work well in estimating brain and spongiosa tissue, due to the absence of brain and spongiosa (cranium) in training data. We hypothesize that the misestimation of spongiosa and brain tissue could be fixed by containing brain slices in the training data. It is desirable to use CNN trained with head-neck images for head-neck data, and CNN trained with thorax images for thorax data. Also, in both the simulation and real patient tasks, the proposed method reduces elapsed time approximately 8-fold for DEAM to reach the same objective function value.

## 5 Acknowledgments

This study is supported by NIH R01 CA 212638.

# MBIR Training for a 2.5D DL network in X-ray CT

Obaidullah Rahman[1], Madhuri Nagare[2], Ken D. Sauer[1], Charles A. Bouman[2], Roman Melnyk[3], Brian Nett[3], and Jie Tang[3]

[1]Department of Electrical Engineering, University of Notre Dame, USA
[2]School of Electrical Electrical and Computer Engineering, Purdue University, West Lafayette, USA
[3]General Electric Healthcare, Waukesha, USA

**Abstract** In computed tomographic imaging, model based iterative reconstruction methods have generally shown better image quality than the more traditional, faster filtered backprojection technique. The cost we have to pay is that MBIR is computationally expensive. In this work we train a 2.5D deep learning (DL) network to mimic MBIR quality image. The network is realized by a modified Unet, and trained using clinical FBP and MBIR image pairs. We achieve the quality of MBIR images faster and with a much smaller computation cost. Visually and in terms of noise power spectrum (NPS), DL-MBIR images have texture similar to that of MBIR, with reduced noise power. Image profile plots, NPS plots, standard deviation, etc. suggest that the DL-MBIR images result from a successful emulation of an MBIR operator.

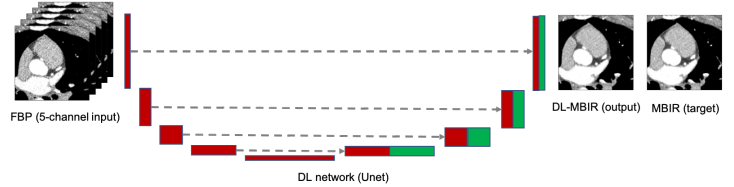

**Figure 1:** DL architecture and training setup. This is a modified form of Unet. Layers on the left denote the contracting path where features are compressed from image towards latent space but the number of features increases. Layers on the right denote the expanding path where feature are decompressed from latent space towards corrected image but the number of features decreases.

## 1 Introduction

X-ray computed tomography has become a important tool in applications such as healthcare diagnostics, security inspection, and non-destructive testing. The industry preferred method of reconstruction is filtered backprojection (FBP) and its popularity is owed to its speed and low computational cost. Iterative methods such as model-based iterative reconstruction (MBIR) generally have better image quality than FBP and do better in limiting image artifacts [1, 2].

MBIR is a computationally expensive and potentially slow reconstruction method since it entails repeated forward projection of the estimated image and back projection of the sinogram residual error. Even with fast GPUs becoming the norm, MBIR may take minutes compared to an FBP reconstruction that can be performed in seconds. The computational cost and reconstruction time have been deterents in wide adoption of MBIR.

In recent years, deep learning has made serious inroads in CT applications. It is applied in sinogram and image domains and sometimes in both. It has been applied in low signal correction [3], image denoising [4, 5], and metal artifact reduction [6].

Ziabari, et al [7] showed that a 2.5D deep neural network, with proper training, can effectively learn a mapping from an FBP image to MBIR. In this paper we expand on their work and study the characteristics of output from networks trained to simulate MBIR with a highly efficient neural network implementation.

## 2 Methods

We will first train a deep neural network, which we will, similarly to [7], entitle DL-MBIR. Our aim is to train the network to closely approximate MBIR images from FBP images. The training input is FBP images and the target is MBIR images from the same data. Let $X_{FBP}$ be the input to the network, $X_{MBIR}$ be the target, and $\sigma$ represents a hypothetical mapping such that $\sigma : X_{MBIR} \rightarrow X_{FBP}$. Let $f_{DL-MBIR_Z}$ be the DL neural network with $Z$ number of input channels. During the training phase:

$$\hat{f}_{DL-MBIR_Z} = \underset{f_{DL-MBIR_Z}}{\mathrm{argmin}} \left|\left| f_{DL-MBIR_Z}(X_{FBP}) - X_{MBIR} \right|\right|_2 \quad (1)$$

$f_{DL-MBIR_Z}$ can be thought of as the inverse of $\sigma$, i.e. $f_{DL-MBIR_Z} = \sigma^{-1}$. During the training phase, the weights of $f_{DL-MBIR_Z}$ are randomly initialized and then adjusted in several iterations using error backpropagation. Once the number of iterations is exhausted or the convergence criteria is met, the training stops.

For training, 4 pairs of clinical exams were selected. Each pair had one FBP image volume and the corresponding MBIR volume. Each image volume had about 200 slices, resulting in about 800 training image pairs. A modified version of Unet [8] was chosen as the network architecture. The learning rate was set to 0.0001 and 2 GPUs were used. Training and inferencing were done on Tensorflow/Keras. Training was run for 300 epochs. 3 versions of DL-MBIR were trained: $DL\text{-}MBIR_1$ was trained with inputs with 1 channel i.e. 1 axial slice, $DL\text{-}MBIR_3$ was trained with 3-channel inputs and $DL\text{-}MBIR_5$ was trained with 5-channel inputs. Having adjacent slices in the input provides additional information to the DL network [7] and helps train it better. Figure 1 shows the DL architecture and the training setup.

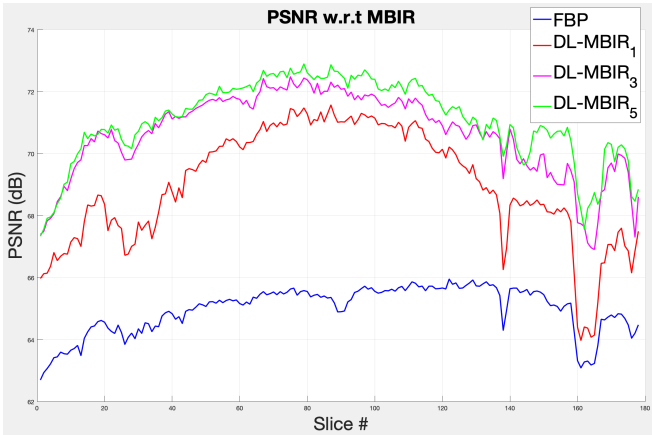

**Figure 2:** PSNR (with MBIR image as reference). PSNR values are in dB. X axis represents the axial slices in the image volume.

## 3    Results

A cardiac FBP image was inferenced on the trained DL-MBIR network. Inference time for every network was between 4 and 6 seconds, and it goes up with the increase in the number of input channels. The MBIR version of the same exam was also available. Figure 3 shows a comparison, for 4 slices – (a), (b), (c), and (d) in the image volume, among MBIR image, FBP image, and the outputs of *DL-MBIR$_Z$*, where $Z = 1$, 3, 5. Figure 4 shows a comparison, for the same slices in the image volume, among difference between images and the MBIR images. Figure 5 has a profile plot to show the comparison of *DL-MBIR$_Z$* and FBP images w.r.t the MBIR images.

Peak signal to noise ratio (PSNR) is another measure of similarity between images and is closely related to mean squared error. A higher value would mean that the image is closer to the reference image. Figure 2 is a plot of PSNR of all slices within the images with MBIR as the reference image. Table 1 has some other metrics of comparison among the images, such as averaged (across slices) PSNR, standard deviation (std) within regions of interest (RoIs), and average CT number within those RoIs.

The noise power spectrum (NPS) is a reliable tool for demonstrating similarity in the image texture. To measure NPS, uniform region patches from one of the cardiac chambers were extracted from each image. Then NPS was measured for all patches and averaged. Then 1D radial profile was measured from the 2D NPS. Figure 6 shows NPS in the uniform region within one of the cardiac chambers.

## 4    Discussion

Visually, all DL-MBIR images bear close resemblance to the MBIR images in figure 3. It is confirmed by the difference images in figure 4. In the profile plot of Figure 5, the DL-MBIR profiles closely follow that of MBIR.

All DL-MBIR images have higher PSNR than that of FBP, with *DL-MBIR$_5$* having the best. Ziabari, et al [7] achieved a PSNR gain over FBP of 3.4 dB for *DL-MBIR$_1$*, compared to 4.1 dB with the current implementation. For *DL-MBIR$_5$*, we have improved the result from 4.25 dB to 6.1 dB over FBP. DL-MBIR images have standard deviations nearly the same as MBIR, with *DL-MBIR$_5$* outperforming the rest. The average within the chosen RoI is more or less preserved in all images.

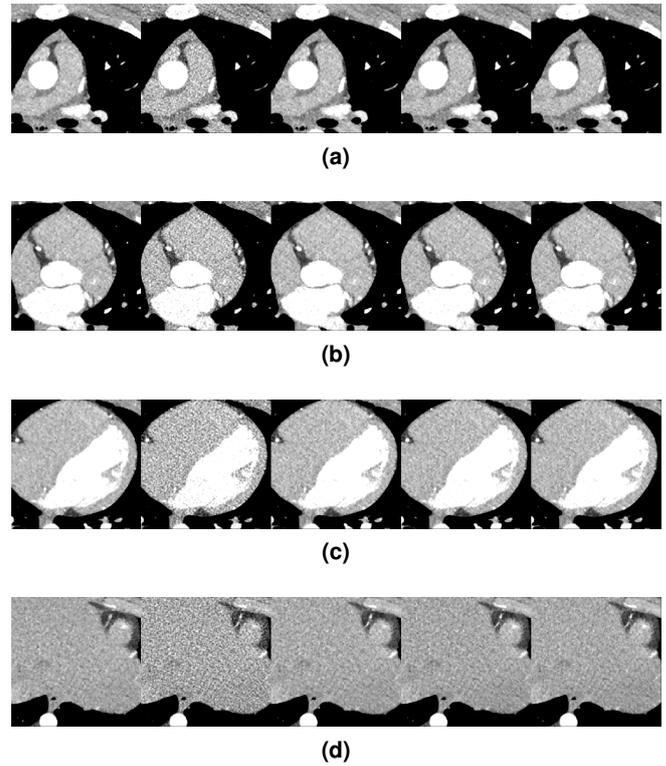

**Figure 3:** Reconstructed image. (left to right): *MBIR*, *FBP*, *DL-MBIR$_1$*, *DL-MBIR$_3$*, *DL-MBIR$_5$*. (a), (b), (c) and (d) represent different slices in the image volume. WW/WL 450/0 HU.

|  | *MBIR* | *FBP* | *DL-MBIR$_1$* | *DL-MBIR$_3$* | *DL-MBIR$_5$* |
|---|---|---|---|---|---|
| PSNR | - | 64.98 | 68.99 | 70.63 | 71.08 |
| std | 25.07 | 46.65 | 26.39 | 27.24 | 25.85 |
| average | 74.68 | 75.20 | 77.0 | 71.52 | 74.43 |

**Table 1:** Performance metrics. PSNR (dB) is calculated w.r.t the MBIR images and averaged for all axial slices. Standard deviation (std) and the average value are in HU, with water being 0 and air -1000.

In figure 6, noise power spectrum (NPS) plots of DL-MBIR images are quite close to that of MBIR, indicating that the DL-MBIR image texture is also similar to that of MBIR, and it appears this attribute is learned well by the network. Due to its type of adaptive regularization, MBIR may create distinctive texture in the surviving image noise. The attenuation of noise is a clear gain; however this texture and its effect on low-contrast detectability may be of concern to some users.

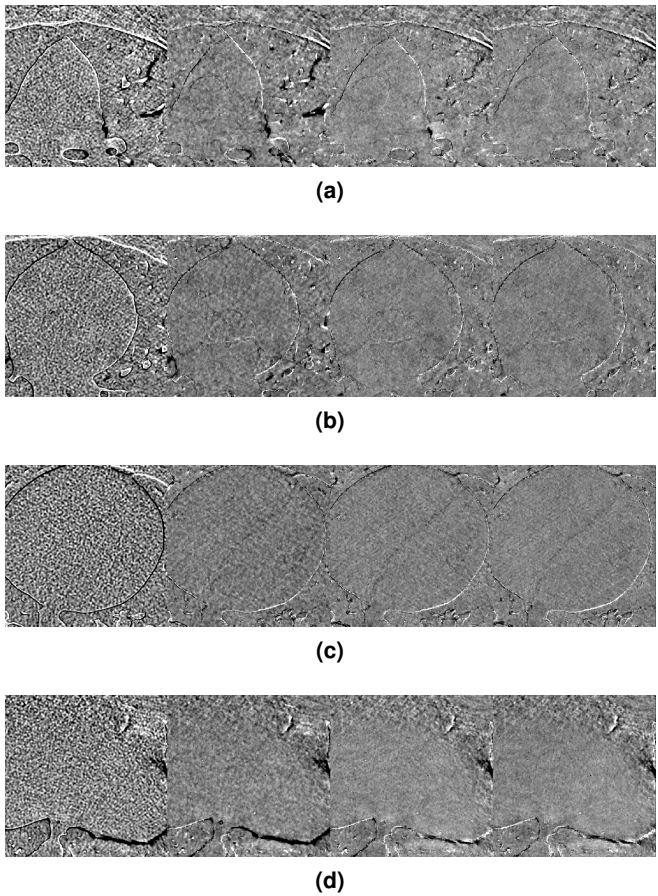

**Figure 4:** Difference image w.r.t. MBIR. (left to right): *FBP*, *DL-MBIR$_1$*, *DL-MBIR$_3$*, *DL-MBIR$_5$*. (a), (b), (c) and (d) represent different slices in the image volume. WW/WL 150/0 HU.

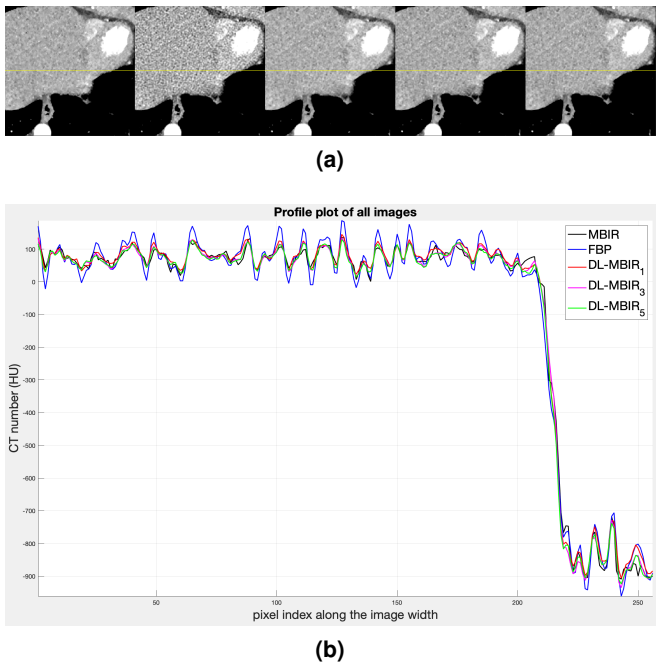

**Figure 5:** Image profile. (a) One axial slice, from left to right, of: *MBIR, FBP, DL-MBIR$_1$, DL-MBIR$_3$, DL-MBIR$_5$* (b) Profile plot of the images along the yellow line.

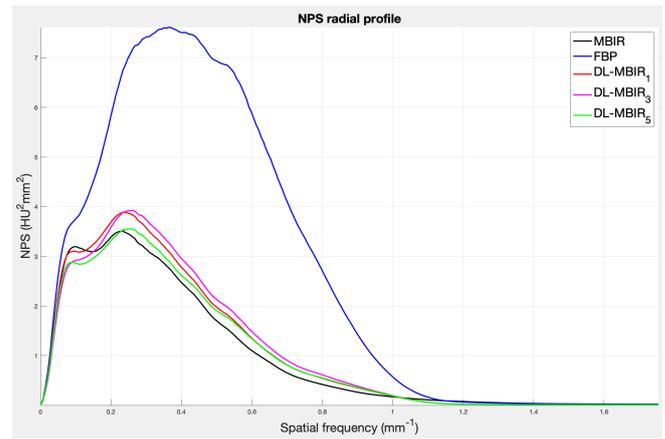

**Figure 6:** NPS plot demonstrating similar behavior of DL-MBIR with MBIR for noise power.

## 5 Conclusion

We trained a U-net, 2.5D DL network that effectively estimates MBIR results from FBP input images. The computation cost is also signifcantly less than that of MBIR. All metrics – NPS, PSNR, standard deviation, profile plots demonstrate that DL-MBIR images have all the features of MBIR including noise reduction and noise texture.

# Noise correlation in multi-material decomposition-based spectral imaging in photon-counting CT

Xiangyang Tang and Yan Ren

Department of Radiology and Imaging Sciences, Emory University School of Medicine, Atlanta, USA

**Abstract:** Aimed at further understanding the fundamentals of photon-counting spectral CT and providing guidelines on its design and implementation, we investigate the correlation of noise in the material-specific images obtained via multi-material decomposition (*m*-MD) and its impact on the performance of spectral imaging in photon-counting CT. By exercising the rule that governs the relation between the covariance matrix of random variables and that of their functions, we derive the equations that characterize the noise and noise correlation in the material specific images in *m*-MD. Via simulation studies using a specially designed phantom that mimics the soft and bony tissues in the head, we assess and verify those relations and their impact on the performance of spectral imaging. The simulation studies, in which the geometry of photon-counting CT is similar to a clinical CT, run over two-, three- and four-material decomposition based material specific imaging in energy range [18 150] *keV* under both ideal and realistic detector spectral responses. The results in 2-MD show that the noise correlation coefficient between the two basis materials always approaches -1 and is consistent with what has been published in the literature. The results in 3-MD are complicated and interesting, as the noise correlation coefficients among the three material specific images alternate between ±1, and so do the noise correlation coefficients among the four material specific images in 4-MD. Moreover, the distortion in detector's spectral response, which is inevitable in a realistic photon-counting detector due to Compton scattering, charge-sharing and fluorescent escaping, results in correlation in the acquired projection data and thus degrades the noise property. The finding of the alternation in noise correlation coefficient between ±1 in 3-MD, 4-MD and beyond (i.e., *m*-MD) is novel and thus of significance. The revealed relationship and the data obtained in this study may provide insightful information to help understand the fundamentals of material decomposition based spectral imaging guidelines on the implementation of spectral imaging in photon-counting CT and other x-ray related imaging modalities (e.g., radiography and tomosynthesis).

## 1 Introduction

We have recently been studying the conditioning of basis materials (functions) and spectral channelization (energy binning). With the singular value decomposition (SVD) based approaches[1-3], we demonstrated how the conditioning of basis materials (functions) and spectral channelization can impact the performance of multi-material decomposition (*m*-MD) based spectral imaging in photon-counting CT[2,3]. Here, we study another fundamental issue in photon-counting CT—the correlation of noise in the material-specific (basis) images and its impact on the performance of spectral imaging.

Initially, Alvarez and Macovski assumed no correlation in the noise between the projection data acquired at high and low tube peak voltages (namely correlation-vanished henceforth), but showed that there is a negative correlation in noise of the image corresponding to each basis material[4-6]. Focused on 2-MD based spectral imaging, Roessl *et al* carried out an analytic investigation to reveal the influence of noise correlation in the projection data on the noise and its correlation in the material-specific images[7]. By characterizing the noise propagation from projection space to the *A*-space[8], they derived a set of equations and arrived at the conclusions that the noise correlation in the projection data decreases (*i*) the noise in the image corresponding to each basis material and (*ii*) the correlation of noise in the basis images is always negative.

An immediate benefit of photon-counting detection for spectral CT is the facilitation of spectral channelization (energy binning[8-12]) by thresholding the energy of incident x-ray photons. With the engagement of multi-spectral channels and multi-basis materials, the dimension of the Jacobian increases accordingly, which in turn complicate the mechanism at which the correlation of noise in the projection data acquired in each spectral channel impacts the noise in each of the material-specific images. The research and development (R&D) in *m*-MD based spectral imaging in photon-counting CT is gaining the momentum. It is the time to investigate the relationship between the noise and noise correlation in the projection data and their counterparts in the material-specific images in *m*-MD based spectral imaging. Via analytical derivation, analysis and simulation studies using digital phantoms, we attempt to answer these fundamental questions in this work, by constraining our effort on *m*-MD that is implemented in projection domain, i.e., the so-called *A*-space approach[8,13].

## 2 Materials and Methods

By treating $I_k(L)$ ($k=1, …, K$) as random variables in the projection space and $A_p(L)$ ($p=1, …, P$) the random variables in the *A*-space, the transformation from projection space to *A*-space can be written as

$$I_k(L) = f_k(A_1(L), …, A_p(L)) \ (k = 1, …, K) \quad (1)$$

Then, the rule governing the relationship between the covariance matrix of random variables and that of their functions states[12,14]

$$V[A] = (F^{-1}) \cdot V[I] \cdot (F^{-1})^T \quad (2)$$

where each entry of the transformation matrix $F$ is defined as $F_{kp} = \partial I_k / \partial A_p$. Furthermore, following the way in reference[12], we define the effective attenuation coefficient, signal-to-noise ratio (SNR), the correlation coefficient in the projection space and A-space, respectively, as

$$\mu_{kp} = -\frac{1}{I_k}\frac{\partial I_k}{\partial A_p}, \ SNR_k = \frac{I_k}{\sigma_{I_k}}, \quad (3)$$

$$\rho_{I_{k1}I_{k2}} = \frac{\sigma_{I_{k1}I_{k2}}}{\sigma_{I_{k1}}\sigma_{I_{k2}}}, \rho_{A_{p1}A_{p2}} = \frac{\sigma_{A_{p1}A_{p2}}}{\sigma_{A_{p1}}\sigma_{A_{p2}}}, \quad (4)$$

In 3-MD, starting from $eq.$ (2), the covariance matrix $V[A]$ becomes

$$V_{3\times3}[A] = \begin{bmatrix} \sigma_{A_1}^2 & \sigma_{A_1 A_2} & \sigma_{A_1 A_3} \\ \sigma_{A_2 A_1} & \sigma_{A_2}^2 & \sigma_{A_2 A_3} \\ \sigma_{A_3 A_1} & \sigma_{A_2 A_2} & \sigma_{A_3}^2 \end{bmatrix}, \quad (5)$$

with each entry of $V_{3\times3}[A]$ being given in the Appendix. In $m$-MD while $m > 3$, the analytic expression of each entry of the covariance matrix becomes exhaustively complicated, though they can be readily determined using the definitions given in $eqs.$ (3)–(5).

The simulation study is carried out using a simulation software kit of photon-counting spectral CT[15] in which the geometric parameters are chosen to mimic a clinical CT for diagnostic imaging. The CT system is assumed working at $140 \, kVp$, $1000 \, mA$, and 1 rotation/sec gantry rotation speed. The criterion for spectral channelization is to assure that the photon counts in each spectral channel are roughly equal prior to their entering into the object to be imaged, though other criteria have been reported in the literature[16]. The spectral channels are implemented via energy thresholding as [1~58, 59~140] $keV$ in 2-MD, [1~51, 52~68, 69~140] $keV$ in 3-MD, [1~43, 44~58, 59~72, 73~140] $keV$ in 4-MD. Each detector element's spectral response is initially assumed ideal, followed by the cases with spectral distortion induced by scattering, charge-sharing and fluorescent escaping[17,18]. In data acquisition, the photon-counting detector is a curved array at dimension 864×16 and pitch $1.024 \times 1.092 \, mm^2$. The source-to-iso and source-to-detector distances are 541.0 $mm$ and 949.0 $mm$, respectively, leading to a nominal voxel size $0.5816 \times 0.5816 \times 0.625 \, mm^3$ in reconstructed image at 512×512 matrix.

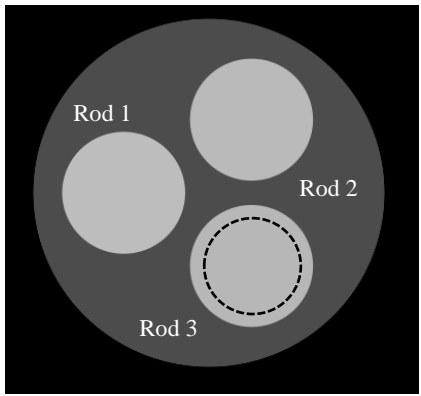

**Fig. 1**. A sectional view of the cylindrical phantom and the ROI for measurement of noise correlation.

A cylinder of water at 20 cm diameter, which consists of three rods at 7 cm diameter, is designed as the phantom to study and verify the correlation of noise in basis (material-specific) images in photon-counting CT. The mass attenuation coefficients of the materials in the phantom and their variation over energy are determined by consulting the authoritative publications[19,20] and the EDPL library[21]. A sectional view of the phantom is presented in Fig. 1, in which a region of interest (ROI) at 6 cm diameter within rod 2 is defined for noise measurement.

*Table 1. Material configuration (fraction in weight) of the three rods in the phantom used for verification of noise correlation in the material-specific (basis) images in 2-MD, 3-MD and 4-MD, respectively (I: iodine; Gd: Gadolinium).*

| Rod | 2-MD | 3-MD | 4-MD |
|---|---|---|---|
| 1 | 1/3 soft tissue<br>2/3 cortical bone | 2/7 soft tissue<br>4/7 (10 mg/ml) I<br>1/7 cortical bone | 2/7 soft tissue<br>2.5/7 (10 mg/ml) I<br>2/7 (10 mg/ml) Gd<br>0.5/7 cortical bone |
| 2 | 1/2 soft tissue<br>1/2 cortical bone | 3/7 soft tissue<br>3/7 (10 mg/ml) I<br>1/7 cortical bone | 3/7 soft tissue<br>2/7 (10 mg/ml) I<br>1.5/7 (10 mg/ml) Gd<br>0.5/7 cortical bone |
| 3 | 2/3 soft tissue<br>1/3 cortical bone | 4/7 soft tissue<br>2/7 (10 mg/ml) I<br>1/7 cortical bone | 3/7 soft tissue<br>2.5/7 (10 mg/ml) I<br>1/7 (10 mg/ml) Gd<br>0.5/7 cortical bone |

## 3 Results

In 3-MD, the correlation of noise in the images corresponding to basis materials soft tissue, iodine (10 mg/ml) and cortical bone under ideal detector spectral response are presented in Fig. 2 (a–c), while those under realistic detector response are in Fig. 2 (a′–c′). It is observed that the polarity of correlation in noise over the material specific images in 3-MD based spectral imaging alternates between ±1. The comparison between Fig. 2 (a–c) and (a′–c′) tells us that the overlapping in spectral channel caused by the spectral distortion in detector's response increase the noise and thus makes the noise correlation spread over a wider region.

In 4-MD, the correlation of noise in the images corresponding to basis materials soft tissue, gadolinium (10mg/ml), iodine (10 mg/ml) and cortical bone under ideal detector spectral response are presented in Fig. 3 (a–f), while those under realistic detector response are in Fig. 4 (a–f). Note that the polarity of correlation in noise over the material-specific images in 4-MD alternates between ±1, in a way that is even more complicated than that in the case of 3-MD. Again, a comparison between Fig. 3 and Fig. 4 tells us that the inter-channel overlapping caused by the spectral distortion in detector's response makes the image noisier and thus the noise correlation spread over a wider region.

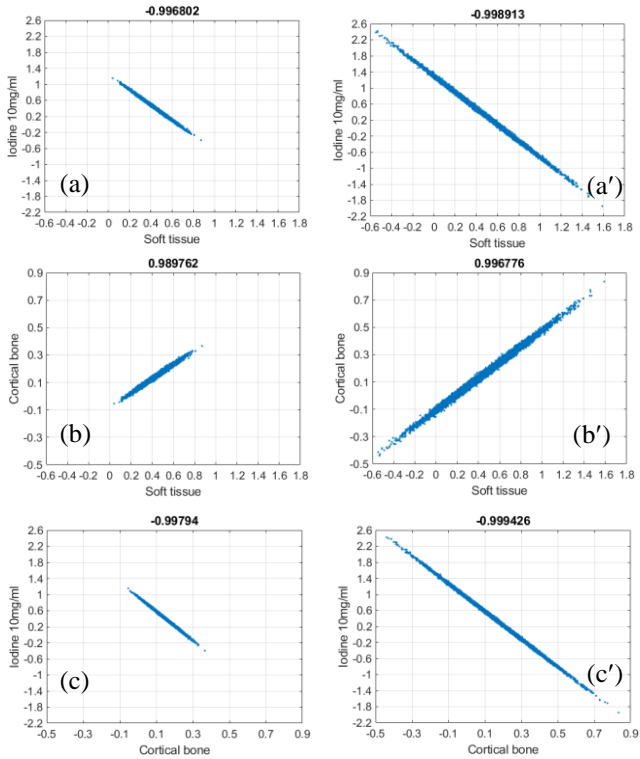

**Fig. 2.** Correlation of noise in 3-MD under both ideal (left) and realistic (right) detector spectral response: (a–a′) soft tissue vs. iodine, (b–b′) soft tissue vs. cortical bone and (c–c′) iodine vs. cortical bone.

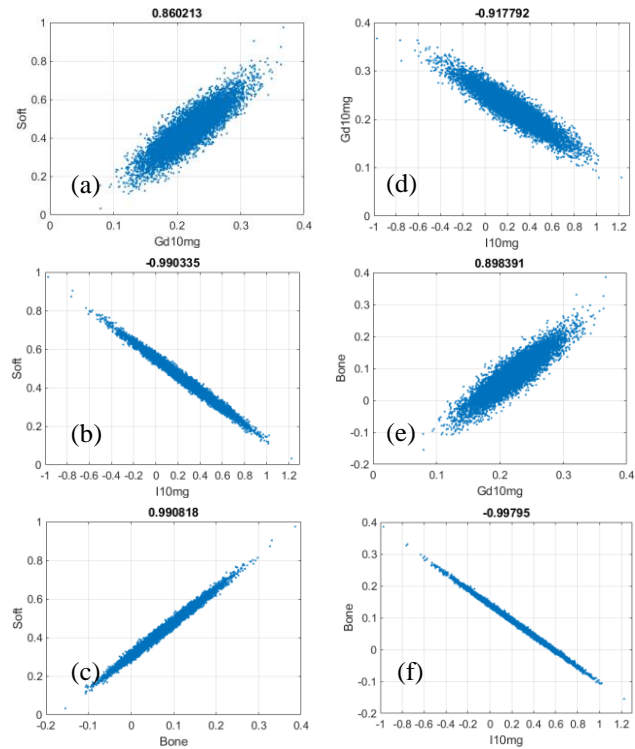

**Fig. 3.** Correlation of noise in 4-MD under ideal detector response: (a) soft tissue vs. Gd, (b) soft tissue vs. iodine, (c) soft tissue vs. bone, (d) Gd vs. iodine, (e) Gd vs. bone and (f) iodine vs. bone.

## 4 Discussion

Recognizing the increasing momentum in R&D of photon-counting CT and the potential of *m*-MD based spectral imaging, we derived the equations governing the behavior of noise and noise correlation in the material-specific images. Via phantom study supported by computer simulation and using 3-MD and 4-MD as the examples, we quantitatively evaluated and verified the equations derived by us to characterize the noise and correlation of noise in the material specific images in *m*-MD based spectral imaging in photon-counting CT. To the best of our knowledge, our work is novel and thus of innovative relevance. As observed in previous sections, the behavior of the noise correlation in the *m*-MD (3-MD, 4-MD and beyond) differs substantially from that in the 2-MD case. Below is a summary of the points that are believed to be of theoretical and practical prominence.

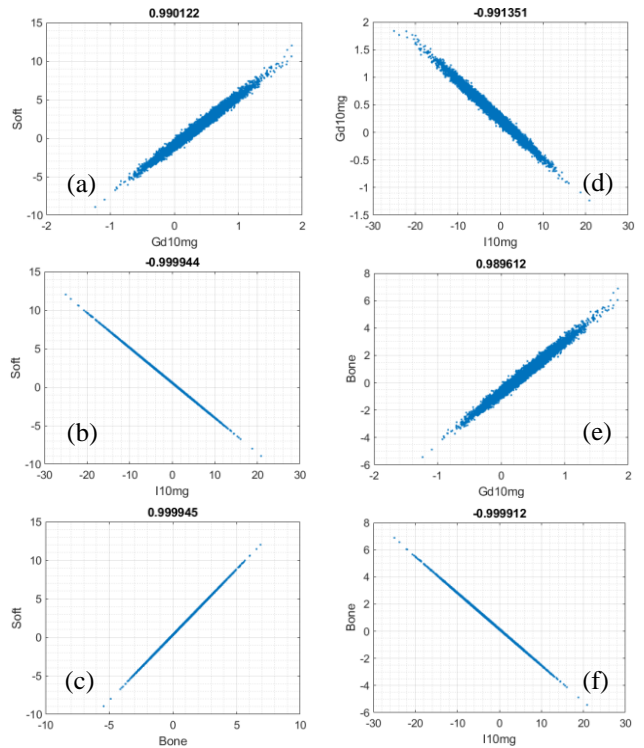

**Fig. 4.** Correlation of noise in 4-MD under realistic detector response: (a) soft tissue vs. gadolinium, (b) soft tissue vs. iodine, (c) soft tissue vs. bone, (d) gadolinium vs. iodine, (e) gadolinium vs. bone and (f) iodine vs. bone.

As we know, the correlation of noise in the material-specific images of 2-MD is always negative. However, as demonstrated in Figures 2, 3 and 4, the polarity of the correlation of noise in the basis (material specific) images of 3-MD and 4-MD varies between positive or negative. It is important to note that the magnitude of noise correlation between the material-specific images in 3-MD and 4-MD, no matter it is positive or negative, approaches one in magnitude, implying that the correlation is actually very strong. The negative correlation of noise in the material-specific images in 2-MD has been the ground on which sophisticated algorithms can be designed and implemented for de-noising in spectral (dual energy) CT[22]. As indicated above, the polarity in the correlation of noise in the material-specific images in 3-MD, 4-MD and beyond alternates between ±1, and such an alternation makes the noise property in $m$-MD based spectral imaging complicated in photon-counting CT, as has been demonstrated in our recent work[2,3]. It may bring about the opportunity for one to design and implement sophisticated algorithms for de-noising in $m$-MD based spectral imaging in photon-counting CT, similar to what has been reported in the literature[22]. We believe that this subject deserves further effort for an in-depth investigation.

Only 3-MD and 4-MD are considered in our study as the examples to verify the property of noise and noise correlation in $m$-MD. As indicated in our recent investigation on the conditioning of basis materials and spectral channelization and their impacts on the performance of spectral imaging in photon-counting CT[2,3], the chance for more than four basis materials (and accordingly four spectral channels) is really limited, in light of the fact that more basis materials (and thus more spectral channels) may radically impact the imaging performance, e.g., the occurrence of severe noise with increasing number of spectral channels. Finally, we would like to state that the data acquired in this work are informative and instructive on implementation of spectral imaging in either photon-counting or energy-integration CT.

## 5 Conclusion

In this study, by taking 3-MD and 4-MD as the examples, we investigated the correlation of noise in basis (material specific) images and its impact on the performance of $m$-MD based spectral imaging in photon-counting CT. It has been found by us that the noise correlation in 3-MD, 4-MD and beyond (i.e., $m$-MD) alternate between ±1, in addition to the existing observation that the noise correlation coefficient between the two basis materials in 2-MD always approaches -1. It is believed that the obtained results and conveyed information can help further understand the fundamentals of material decomposition based spectral imaging in either photon-counting or energy-integration CT and other x-ray related imaging modalities, such as radiography and tomosyn-thesis.

## Appendix

$\sigma^2_{A_1A_1} = ((\mu_{23}\mu_{32} - \mu_{33}\mu_{22})^2/SNR^2_1 + (\mu_{13}\mu_{32} - \mu_{33}\mu_{12})^2/SNR^2_2 + (\mu_{13}\mu_{22} - \mu_{12}\mu_{23})^2/SNR^2_3 - 2\rho_{112}(\mu_{13}\mu_{23}\mu^2_{32} + \mu^2_{33}\mu_{12}\mu_{22} - \mu_{33}\mu_{12}\mu_{23}\mu_{32} - \mu_{33}\mu_{13}\mu_{22}\mu_{32})/SNR_1SNR_2 - 2\rho_{123}(\mu^2_{13}\mu_{22}\mu_{32} + \mu_{33}\mu^2_{12}\mu_{23} - \mu_{12}\mu_{13}\mu_{23}\mu_{32} - \mu_{33}\mu_{12}\mu_{13}\mu_{22})/SNR_2SNR_3 - 2\rho_{113}(\mu_{12}\mu^2_{23}\mu_{32} + \mu_{33}\mu_{13}\mu^2_{22} - \mu_{13}\mu_{22}\mu_{23}\mu_{32} - $

$\mu_{33}\mu_{12}\mu_{22}\mu_{23})/SNR_1SNR_3)/\Delta^2_{3\times3}.$

(A-1)

$\sigma^2_{A_2A_2} = ((\mu_{23}\mu_{31} - \mu_{33}\mu_{21})^2/SNR^2_1 + (\mu_{13}\mu_{31} - \mu_{33}\mu_{11})^2/SNR^2_2 + (\mu_{11}\mu_{23} - \mu_{13}\mu_{21})^2/SNR^2_3 - 2\rho_{112}(\mu_{13}\mu_{23}\mu^2_{31} + \mu^2_{33}\mu_{11}\mu_{21} - \mu_{33}\mu_{11}\mu_{23}\mu_{31} - \mu_{33}\mu_{13}\mu_{21}\mu_{31})/SNR_1SNR_2 - 2\rho_{123}(\mu^2_{13}\mu_{21}\mu_{31} + \mu_{33}\mu^2_{11}\mu_{23} - \mu_{11}\mu_{13}\mu_{23}\mu_{31} - \mu_{33}\mu_{11}\mu_{13}\mu_{21})/SNR_2SNR_3 - 2\rho_{113}(\mu_{11}\mu^2_{23}\mu_{31} + \mu_{33}\mu_{13}\mu^2_{21} - \mu_{13}\mu_{21}\mu_{23}\mu_{31} - $

$\mu_{33}\mu_{11}\mu_{21}\mu_{23})/SNR_1SNR_3)/\Delta^2_{3\times3}.$

(A-2)

$\sigma^2_{A_3A_3} = ((\mu_{11}\mu_{32} - \mu_{12}\mu_{31})^2/SNR^2_2 + (\mu_{21}\mu_{32} - \mu_{22}\mu_{31})^2/SNR^2_1 + (\mu_{11}\mu_{22} - \mu_{12}\mu_{21})^2/SNR^2_3 - 2\rho_{112}(\mu_{11}\mu_{21}\mu^2_{32} + \mu_{12}\mu_{22}\mu^2_{31} - \mu_{11}\mu_{22}\mu_{31}\mu_{32} - \mu_{12}\mu_{21}\mu_{31}\mu_{32})/SNR_1SNR_2 - 2\rho_{123}(\mu^2_{12}\mu_{21}\mu_{31} + \mu^2_{11}\mu_{22}\mu_{32} - \mu_{11}\mu_{12}\mu_{21}\mu_{32} - \mu_{11}\mu_{12}\mu_{22}\mu_{31})/SNR_2SNR_3 - 2\rho_{113}(\mu_{12}\mu^2_{21}\mu_{32} + \mu_{11}\mu^2_{22}\mu_{31} - \mu_{11}\mu_{21}\mu_{22}\mu_{32} - $

$\mu_{12}\mu_{21}\mu_{22}\mu_{31})/SNR_1SNR_3)/\Delta^2_{3\times3}.$

(A-3)

$\sigma_{A_1A_2} = \sigma_{A_2A_1} = -((\mu^2_{23}\mu_{31}\mu_{32} + \mu^2_{33}\mu_{21}\mu_{22} - \mu_{33}\mu_{21}\mu_{23}\mu_{32} - \mu_{33}\mu_{22}\mu_{23}\mu_{31})/SNR^2_1 + (\mu^2_{13}\mu_{31}\mu_{32} - \mu_{33}\mu_{11}\mu_{13}\mu_{32} - \mu_{33}\mu_{12}\mu_{13}\mu_{31})/SNR^2_2 + (\mu_{11}\mu_{13}\mu^2_{23} + \mu^2_{13}\mu_{21}\mu_{22} - \mu_{11}\mu_{13}\mu_{22}\mu_{23} - \mu_{12}\mu_{13}\mu_{21}\mu_{23})/SNR^2_3 - \rho_{112}(\mu^2_{33}\mu_{11}\mu_{22} + \mu^2_{33}\mu_{12}\mu_{21} + 2\mu_{13}\mu_{23}\mu_{31}\mu_{32} - \mu_{33}\mu_{11}\mu_{23}\mu_{32} - \mu_{33}\mu_{13}\mu_{22}\mu_{31} - \mu_{33}\mu_{13}\mu_{21}\mu_{32} - \mu_{33}\mu_{12}\mu_{23}\mu_{31})/SNR_1SNR_2 - \rho_{123}(\mu^2_{13}\mu_{21}\mu_{32} + 2\mu_{33}\mu_{11}\mu_{12}\mu_{23} + \mu^2_{13}\mu_{22}\mu_{31} - \mu_{11}\mu_{13}\mu_{23}\mu_{32} - \mu_{12}\mu_{13}\mu_{23}\mu_{31} - \mu_{33}\mu_{11}\mu_{13}\mu_{22} - \mu_{33}\mu_{12}\mu_{13}\mu_{21})/SNR_2SNR_3 - \rho_{113}(\mu_{12}\mu^2_{23}\mu_{31} + 2\mu_{33}\mu_{13}\mu_{21}\mu_{22} + \mu_{11}\mu^2_{23}\mu_{32} - \mu_{13}\mu_{21}\mu_{23}\mu_{32} - \mu_{13}\mu_{22}\mu_{23}\mu_{31} - \mu_{33}\mu_{11}\mu_{22}\mu_{23} - $

$\mu_{33}\mu_{12}\mu_{21}\mu_{23})/SNR_1SNR_3)/\Delta^2_{3\times3}.$

(A-4)

$\sigma_{A_1A_3} = \sigma_{A_3A_1} = -((\mu_{21}\mu_{23}\mu^2_{32} + \mu_{33}\mu^2_{22}\mu_{31} - \mu_{22}\mu_{23}\mu_{31}\mu_{32} - \mu_{33}\mu_{21}\mu_{22}\mu_{32})/SNR^2_1 + (\mu_{11}\mu_{13}\mu^2_{32} + \mu_{33}\mu^2_{12}\mu_{31} - \mu_{12}\mu_{13}\mu_{31}\mu_{32} - \mu_{33}\mu_{11}\mu_{12}\mu_{32})/SNR^2_2 + (\mu^2_{12}\mu_{21}\mu_{23} + \mu_{11}\mu_{13}\mu^2_{22} - \mu_{12}\mu_{13}\mu_{21}\mu_{22} - \mu_{11}\mu_{12}\mu_{22}\mu_{23})/SNR^2_3 - \rho_{112}(\mu_{11}\mu_{23}\mu^2_{32} + 2\mu_{33}\mu_{12}\mu_{22}\mu_{31} + \mu_{13}\mu_{21}\mu^2_{32} - \mu_{12}\mu_{23}\mu_{31}\mu_{32} - \mu_{13}\mu_{22}\mu_{31}\mu_{32} - \mu_{33}\mu_{12}\mu_{21}\mu_{32} - \mu_{33}\mu_{11}\mu_{22}\mu_{32})/SNR_1SNR_2 - \rho_{123}(\mu^2_{12}\mu_{23}\mu_{31} + 2\mu_{11}\mu_{13}\mu_{22}\mu_{32} + \mu_{33}\mu^2_{12}\mu_{21} - \mu_{11}\mu_{12}\mu_{23}\mu_{32} - \mu_{12}\mu_{13}\mu_{21}\mu_{32} - \mu_{12}\mu_{13}\mu_{22}\mu_{31} - \mu_{33}\mu_{11}\mu_{12}\mu_{22})/SNR_2SNR_3 - \rho_{113}(\mu_{13}\mu^2_{22}\mu_{31} + 2\mu_{12}\mu_{21}\mu_{23}\mu_{32} + \mu_{33}\mu_{11}\mu^2_{22} - \mu_{11}\mu_{22}\mu_{23}\mu_{32} - \mu_{13}\mu_{21}\mu_{22}\mu_{32} - \mu_{12}\mu_{22}\mu_{23}\mu_{31} - $

$\mu_{33}\mu_{12}\mu_{21}\mu_{22})/SNR_1SNR_3)/\Delta^2_{3\times3}.$

(A-5)

$\sigma_{A_2A_3} = \sigma_{A_3A_2} = -((\mu_{33}\mu^2_{21}\mu_{32} + \mu_{22}\mu_{23}\mu^2_{31} - \mu_{21}\mu_{23}\mu_{31}\mu_{32} - \mu_{33}\mu_{21}\mu_{22}\mu_{31})/SNR^2_1 + (\mu_{33}\mu^2_{11}\mu_{32} + \mu_{12}\mu_{13}\mu^2_{31} - \mu_{11}\mu_{13}\mu_{31}\mu_{32} - \mu_{33}\mu_{11}\mu_{12}\mu_{31})/SNR^2_2 + (\mu^2_{11}\mu_{22}\mu_{23} + \mu_{12}\mu_{13}\mu^2_{21} - \mu_{11}\mu_{12}\mu_{21}\mu_{23} - \mu_{11}\mu_{13}\mu_{21}\mu_{22})/SNR^2_3 - \rho_{112}(\mu_{12}\mu_{23}\mu^2_{31} + 2\mu_{33}\mu_{11}\mu_{21}\mu_{32} + \mu_{13}\mu_{22}\mu^2_{31} - \mu_{11}\mu_{23}\mu_{31}\mu_{32} - \mu_{13}\mu_{21}\mu_{31}\mu_{32} - \mu_{33}\mu_{11}\mu_{22}\mu_{31} - $

$\mu_{33}\mu_{11}\mu_{21}\mu_{22})/SNR_1SNR_3)/\Delta^2_{3\times3}$.

$$(A-6)$$

where $\quad \Delta_{3\times3} = \mu_{11}\mu_{23}\mu_{32} - \mu_{13}\mu_{21}\mu_{32} - \mu_{12}\mu_{23}\mu_{31} + \mu_{13}\mu_{22}\mu_{31} - \mu_{33}\mu_{11}\mu_{22} + \mu_{33}\mu_{12}\mu_{21}$

$$(A-7)$$

If no correlation exists in projection data, ($\rho_{I_{ij}} = 0$, $i = 1, 2, 3, 4$; $j = 1, 2, 3, 4$), (A-1) - (A-6) degenerate into

$\sigma^2_{A_1A_1} = ((\mu_{23}\mu_{32} - \mu_{33}\mu_{22})^2/SNR^2_1 + (\mu_{13}\mu_{32} - \mu_{33}\mu_{12})^2/SNR^2_2 + (\mu_{13}\mu_{22} - \mu_{12}\mu_{23})^2/SNR^2_3)/\Delta^2_{3\times3}$.

$$(A-1')$$

$\sigma^2_{A_2A_2} = ((\mu_{23}\mu_{31} - \mu_{33}\mu_{21})^2/SNR^2_1 + (\mu_{13}\mu_{31} - \mu_{33}\mu_{11})^2/SNR^2_2 + (\mu_{11}\mu_{23} - \mu_{13}\mu_{21})^2/SNR^2_3)/\Delta^2_{3\times3}$.

$$(A-2')$$

$\sigma^2_{A_3A_3} = ((\mu_{11}\mu_{32} - \mu_{12}\mu_{31})^2/SNR^2_2 + (\mu_{21}\mu_{32} - \mu_{22}\mu_{31})^2/SNR^2_1 + (\mu_{11}\mu_{22} - \mu_{12}\mu_{21})^2/SNR^2_3)/\Delta^2_{3\times3}$.

$$(A-3')$$

$\sigma_{A_1A_2} = \sigma_{A_2A_1} = -((\mu^2_{23}\mu_{31}\mu_{32} + \mu^2_{33}\mu_{21}\mu_{22} - \mu_{33}\mu_{21}\mu_{23}\mu_{32} - \mu_{33}\mu_{22}\mu_{23}\mu_{31})/SNR^2_1 + (\mu^2_{33}\mu_{11}\mu_{12} + \mu^2_{13}\mu_{31}\mu_{32} - \mu_{33}\mu_{11}\mu_{13}\mu_{32} - \mu_{33}\mu_{12}\mu_{13}\mu_{31})/SNR^2_2 + (\mu_{11}\mu_{13}\mu_{22}\mu_{23} + \mu^2_{13}\mu_{21}\mu_{22} - \mu_{11}\mu_{13}\mu_{22}\mu_{23} - \mu_{12}\mu_{13}\mu_{21}\mu_{23})/SNR^2_3)/\Delta^2_{3\times3}$.

$$(A-4')$$

$\sigma_{A_1A_3} = \sigma_{A_3A_1} = -((\mu_{21}\mu_{23}\mu^2_{32} + \mu_{33}\mu^2_{22}\mu_{31} - \mu_{22}\mu_{23}\mu_{31}\mu_{32} - \mu_{33}\mu_{21}\mu_{22}\mu_{32})/SNR^2_1 + (\mu_{11}\mu_{13}\mu^2_{32} + \mu_{33}\mu^2_{12}\mu_{31} - \mu_{12}\mu_{13}\mu_{31}\mu_{32} - \mu_{33}\mu_{11}\mu_{12}\mu_{32})/SNR^2_2 + (\mu^2_{12}\mu_{21}\mu_{23} + \mu_{11}\mu_{13}\mu^2_{22} - \mu_{12}\mu_{13}\mu_{21}\mu_{22} - \mu_{11}\mu_{12}\mu_{22}\mu_{23})/SNR^2_3)/\Delta^2_{3\times3}$.

$$(A-5')$$

$\sigma_{A_2A_3} = \sigma_{A_3A_2} = -((\mu_{33}\mu^2_{21}\mu_{32} + \mu_{22}\mu_{23}\mu^2_{31} - \mu_{21}\mu_{23}\mu_{31}\mu_{32} - \mu_{33}\mu_{21}\mu_{22}\mu_{31})/SNR^2_1 + (\mu_{33}\mu^2_{11}\mu_{32} + \mu_{12}\mu_{13}\mu^2_{31} - \mu_{11}\mu_{13}\mu_{31}\mu_{32} - \mu_{33}\mu_{11}\mu_{12}\mu_{31})/SNR^2_2 + (\mu^2_{11}\mu_{22}\mu_{23} + \mu_{12}\mu_{13}\mu^2_{21} - \mu_{11}\mu_{12}\mu_{21}\mu_{23} - \mu_{11}\mu_{13}\mu_{21}\mu_{22})/SNR^2_3)/\Delta^2_{3\times3}$.

$$(A-6')$$

# An iterative image-based motion correction method for dynamic PET imaging

Tao Sun[1], Zhanli Hu[1], and Yongfeng Yang[1]

[1]Paul C. Lauterbur Research Center for Biomedical Imaging,
Shenzhen Institute of Advanced Technology, Chinese Academy of Sciences

**Abstract** As a non-invasive imaging tool, PET play an important role in brain science and disease research. Dynamic PET imaging is one way of brain PET data acquisition. However, its widely application in clinics research has often been hindered by practical challenges, such as patient involuntary movement, which could degrade both image quality and the accuracy of the quantification. This is even more obvious in scans of the patients with neurodegeneration and mental disorders. Traditional motion correction methods are either based on images or raw measured data, were shown to be able to compensate the motion to some extent. However, when the PET tracer kinetics are present, like in the early and middle scan phase, existing method may fail. In this work, we propose a motion compensation approach for dynamic PET imaging. Our method only requires reconstructed images, based on which the motion can be estimated and compensated. The simulations and patient study show that the proposed method can compensate the complex motion in scans with three different tracers. The recovered image quality and quantification was superior to the ones corrected with the conventional image-based method. The proposed method enables image quality control for dynamic PET imaging, hence facilitate its applications in clinics and research.

## 1 Introduction

Advanced imaging technology such as Positron Emission Tomography (PET), as a noninvasive tool, has led to remarkable improvement in our knowledge of brain. Dynamic PET is an acquisition method that allows clinicians and researchers to study the physiological or pathological processes of the human body, and in particular the brain via the use of specific tracers. The metabolism, or receptor binding, of the living body, can be quantitatively calculated from dynamic data based on proper kinetic modeling.

A factor affecting dynamic PET quality is voluntary and involuntary patient movement. For brain PET imaging, patient head movement during scanning presents a challenge for accurate PET image reconstruction and subsequent quantitative analysis [1]. In dynamic PET, these problems are often more sever since the scan times are commonly over one to two hours [2]. In such cases, subjects that have neurodegenerative disease or mental disorders cannot remain still. The ability to compensate head motion in dynamic imaging, would be of great value. Several types of retrospective correction methods already exist, from image-based methods such as multiple acquisition frame [3,4], to list-mode data-based methods [5,6]. However, these methods often cannot cope with dynamic PET imaging well, as their performance can be degraded by the changing tracer kinetics, e.g. in early-mid phase of a FDG scan or in a perfusion scan.

In this study, we propose a method which can compensate the motion in dynamic reconstructed PET images. The method has an iterative implementation, and in each iteration two subsequent steps are performed. First, abrupt inter-frame movement can be compensated by incorporating the tracer kinetics into a groupwise image registration process. Second, a multi-phase alignment can further compensate the residual slow movement across the entire scan. By iterating these steps, the reconstructed images are aligned, and subsequent kinetic modelling can be performed with confidence. In the following context, we will first illustrate the details of the methods, and then perform experiments to evaluate the proposed method.

## 2 Materials and Methods

Let us define the motion parameters $\mathbf{T}$ at each frame that representing the relative head position in scanner coordinate system:

$$\mathbf{T} = \{T_k\}_{k=1\ldots N\_frame} = \left\{r_x, r_y, r_z, t_x, t_y, t_z\right\}_{k \in N\_frame} \quad (1)$$

where k is the frame index, $r_x$, $r_y$ and $r_z$ are the rotations, $t_x$, $t_y$ and $t_z$ are the translations. For $\mathbf{T}$ we have a total number of N_frame×6 parameters to be estimated. Conventional image-based method registers all frames to a reference frame to obtain $\mathbf{T}$. However, the varying contrast in dynamic images could prevent the accurate registration, especially considering the relatively high noise in PET image. Therefore, we propose an iterative method to cope with the above problem. In each iteration, motion parameters were first estimated within a procedure called kinetics-driven estimation, and then refined within a procedure called multi-phase refinement (Figure 1). To perform the kinetics-driven estimation, we estimated the initial $\mathbf{T}$ by utilizing the fact that motion can affect the accuracy of the kinetic modelling. A cost function which reflects the degree that how much the dynamic images fit a kinetic model was introduced. We estimated the kinetic parameters p by minimizing the cost function L1 from the reconstructed images:

$$L1(p;T) = \sum_{j \in N}\left[\sum_{k \in F} w_k \left(I(\widehat{T}_k(x_j)) - I_S(\widehat{T}_k(x_j), p)\right)^2\right] \quad (2)$$

$$\hat{p} = \underset{p}{\operatorname{argmin}} L1(p;T)$$

which is essentially a modelling process, where $w_k$ is the weighting factor at each frame (proportional to the frame duration). $T_k$ is a zero vector in first iteration, $\hat{p}$ are the estimated kinetic parameters. We can generate a new set of images $I_S(x, \hat{p})$, which is less affected by motion artifacts compared with $I(x)$. For different tracer and the target organ, we can choose appropriate model to compute $I_S(x,p)$. For example, Patlak Graphical model can be applied for $^{18}$F-FDG; Simplified reference tissue model (SRTM) can be applied for $^{11}$C-raclopride. Performing the pairwise image registration for each frame in $I_S(x, \hat{p})$ and $I(x)$:

$$L2(T_k; p) = \sum_{j \in N} NCC\left( I(T_k(x_j)), \hat{I}_S(T_k(x_j), \hat{p}) \right)$$

$$\hat{T}_k = \arg\min_{T_k} L2(T_k; p) \qquad (3)$$

$$\hat{\mathbf{T}} = \{\hat{T}_k\}_{k=1...N\_frame}$$

where the similarity metric is chosen as summed normalized cross correlation (NCC) difference of all voxels. By optimizing L2, we can find the initially estimated motion parameters $\hat{\mathbf{T}}$.

It was shown that the most of the abrupt inter-frame movement can already be estimated by performing above steps. To further estimate the residual slow movement compenent, we introduced a multi-phase registration step. All frames were aligned to a reference position. Unlike conventional MAF method, it segemented the frames into three groups and for each group we registered the images to the middle frame in that group (Figure 1). Then these three middle reference frame were aligned, hence all frames were aligned.

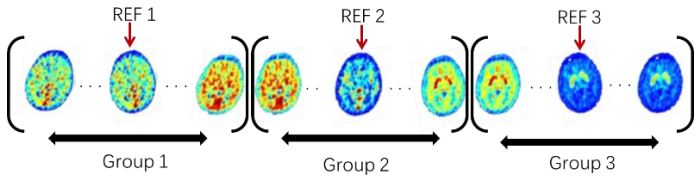

Figure 1 Multi-phase registration that refines the estimated motion parameters.

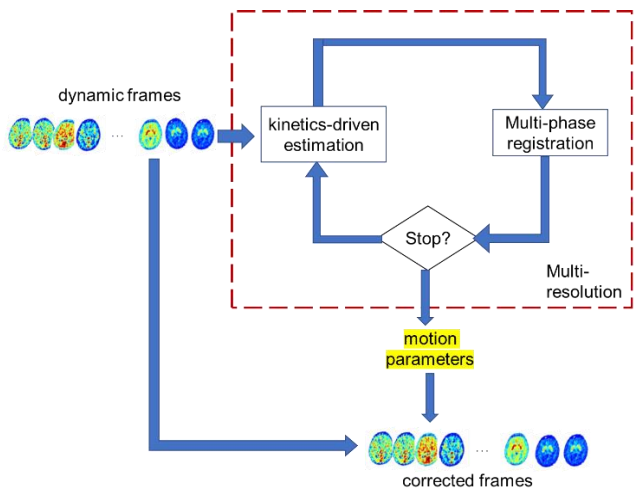

Figure 2 The iterative correction scheme proposed in this work.

We can alternate above estimation and refinement process, with the output from each as the input to another (Figure 2). Multi-resolution acceleration was applied to the scheme. At first iteration, we down-sampled all dynamic images in a designated resolution level and performed the estimation. Then the transformed images can be supplied into next iteration and repeated until next resolution level. Whether preceding to next resolution level or not was determined by checking if the change in motion estimates was less than predefined threshold. Upon obtaining final motion parameters, we can re-align the original reconstructed images to a reference position to finish correction. Subsequent kinetic modelling can be performed on the corrected images to derive either micro- or macro-parameters.

## 3 Results

We simulated 60-min dynamic acquisitions for three tracers: $^{18}$F-FDG with irreversible two-tissue compartment model; $^{11}$C-raclopride and $^{11}$C-WAY-100635 with reversible two-tissue compartment model. The simulated micro parameters were obtained from existing publications. Realistic Poisson noise was added into these images with normalization to the frame duration [7]. We applied rigid motion segments with slow and abrupt position change to the scan by transforming each individual image, hence only inter-frame movement was simulated. Motion correction was performed using the proposed and conventional MAF methods. The proposed method run for two iterations from coarse to fine resolution levels (downsample factor 4×4×2, 2×2×1). The proposed method recovered the dynamic PET image quality, as the time-activity-curve (TAC) plotted in Figure 3. For the simulation with $^{11}$C-Raclopride, proposed method can produce high quality image at static frame (50-60 mins summed, MSE=16.0, SSIM=0.99), while MAF correction produce inferior results (MSE=35.0, SSIM=0.92) when comparing to the truth. The derived parametric images from $^{11}$C-WAY-100635 simulation were shown in Figure 4, indicates the superiority of the proposed method in recovering the kinetic parameters. In a patient study, the proposed method was shown to be able to compensate the motion artifacts and improve the image quality in a dynamic FDG scan (Figure 5).

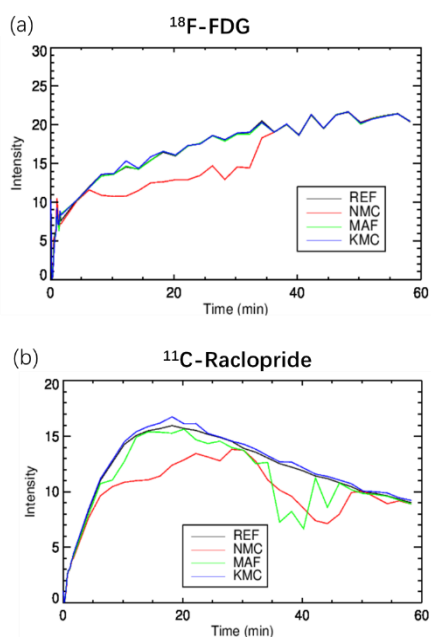

Figure 3 Time activity curves in left striatum from uncorrected, proposed correction method, MAF and truth images from (a) simulation with FDG, (b) simulation with $^{11}$C-raclopride. The meaning of the abbreviations is: NMC-uncorrected, MAF-multiple aligned frame, KMC-proposed method, REF-reference.

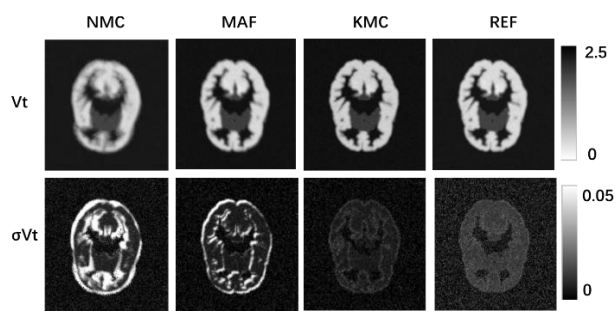

Figure 4. Results from noise-free $^{11}$C-WAY-100635 simulation. (a) Transaxial slice of voxelized volume of distribution image that is calculated with Logan graphical analysis. (b) The decrease in variance from Logan analysis indicates the effectiveness in recovery. The methods listed here are the same as in Figure 3.

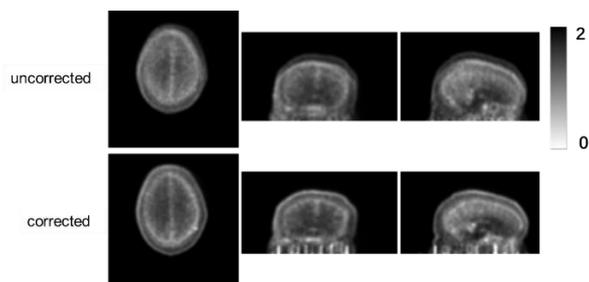

Figure 5 Motion correction was applied to a motion-contaminated 60-min $^{18}$F-FDG patient scan. After correction, the image quality is improved (summed over 5-10 min). The intensity value of a region in frontal lobe is 15.1 % higher for the corrected image.

## 4 Discussion and Conclusion

To conclude, in this preliminary study, we demonstrate a motion correction method that works for a dynamic brain PET scan. It can reduce the effect of the involuntary head movement on image quality and potential inaccurate quantification.

# Relaxed Equiangular Detector CT: Architecture and Image Reconstruction

Yingxian Xia[1,2], Zhiqiang Chen[1,2], Li Zhang[1,2], and Hewei Gao *[1,2]

[1]Department of Engineering Physics, Tsinghua University, Beijing 100084, China

[2]Key Laboratory of Particle and Radiation Imaging (Tsinghua University), Ministry of Education, Beijing 100084, China

**Abstract** Recently, a new concept of COplanar Transmission and Emission Guidance in Radiotherapy (Co-TeGRT) has been proposed to enable multi-modality imaging with great potential of better data registration and simplified workflow. In Co-TeGRT, a relaxed Equiangular Detector CT (RedCT) scan is formed naturally, where the source position is away from the focus of equiangularly distributed detector pixels, leading to a unique nonuniform data sampling on detector. In this work, based on an in-depth investigation of the novel geometry in RedCT, we present a direct filtered-backprojection (FBP) type reconstruction with no data rebinning needed, where an innovative weighting strategy before filtering is developed. Numerous simulation studies show that accurate reconstruction with better spatial resolution can be achieved by using our proposed FBP algorithm, validating our method to be effective and feasible in practical applications.

## 1 Introduction

Nowadays, multi-modality imaging such as positron emission tomography (PET) and computed tomography (CT) has been greatly advanced for image guidance in Radiotherapy[1]. Recently, a new concept of compact architecture of COplanar Transmission and Emission Guidance in RadioTherapy (Co-TeGRT) has been proposed[2], which has great potential of multi-modality imaging with benefits in better data registration and simplified workflow. The Co-TeGRT is enabled by leveraging a common detector developed to detect both X-ray and gamma ray photons. The system can be further optimized by combining the common detector with a small-sized KVCT detector to form a hybrid radiation sensor (as illustrated in Fig. 1), where PET and CT imaging and MV treatment can all be assembled in the same scan plane on a rotating gantry. Such a compact system design balances both PET and KVCT image performance during MV treatment. Regardless of the performance difference between the KVCT dedicated detector segment and the common detector segment, there exists an interesting problem on image reconstruction from such a novel geometry, specifically, from a relaxed equiangular detector CT (RedCT) scan.

## 2 Method

In Co-TeGRT, the arc-shape common/hybrid detector, in general, is formed symmetrically along a circle. Let the detect pixels be equiangularly distributed, whose focus is at the center of the rotating gantry. Apparently, the corresponding KV

*Author to whom correspondence should be addressed. Email address: hwgao@tsinghua.edu.cn.

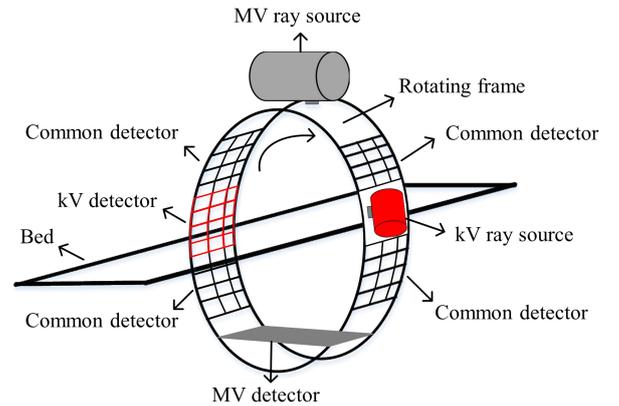

**Figure 1:** An illustration of a new concept of Compact architecture of coplanar Transmission and Emission Image Guided Radiotherapy (Co-TeGRT).

X-ray source should locate on the gantry as well and is therefore away from the focus of the detector, making it a unique imaging geometry. As a result, CT imaging in Co-TeGRT is quite different from a typical equiangular or equispaced detector CT scan.

### 2.1 The Architecture of RedCT

A simplified imaging geometry using the KV X-ray source and the common/hybrid detector in Co-TeGRT is shown in Fig. 2. Without loss of generality, the KV X-ray source may not be exactly positioned on the same orbiting circle as the detector due to the manufacturing limitations in practical applications. Let $R$ be the radius of the detector equiangularly distributed with a focus at location $O$ (namely the detector rotation center), the distance between the kV X-ray source and the detector rotation center can be denoted as $D = kR$. Here, $k$ is a ratio depending on the relative location of the source to the detector. Apparently, $k = 1$ when the source and detector are strictly co-circled.

Given the geometry in Fig. 2, the fan angle $\alpha$ and the central angle $\gamma$ always meet the following condition:

$$\alpha = g(\gamma) = atan(\frac{sin\gamma}{cos\gamma + k}). \tag{1}$$

From the Icnscribed Angle Theorem, it is seen that only when $k = 1$ will the fan angle $\alpha \equiv \gamma/2$ be equiangularly sampled (assuming $\gamma$ is uniformly sampled). Otherwise, it will be a nonuniform sampling on the detector (with respect to the projection rays). Therefore, a CT scan in Co-TeGRT, in essence, is a relaxed equiangular detector CT (RedCT).

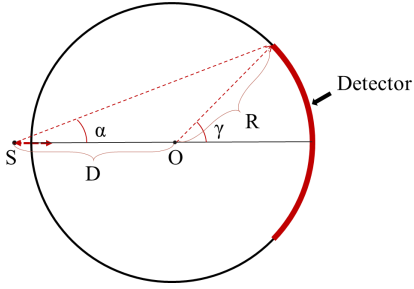

**Figure 2:** The imaging geometry of a RedCT scan in Co-TeGRT.

For a RedCT scan, due to the nonuniform sampling on detector, it is an interesting but open question if an analytic reconstruction can be done without data rebinning. In this work, we demonstrate that a practical filtered-backprojection (FBP) type algorithm is still available as long as the ratio $k$ is not too off from 1.

## 2.2  A Practical FBP-type Reconstruction

For a typical equiangular fan-beam CT with a uniformly sampled fan-angle $\alpha$, an FBP reconstruction formula can be written as:

$$f(x,y) = \frac{1}{2}\int_0^{2\pi} d\beta \int_{-\alpha_m}^{+\alpha_m} d\alpha \ p(\alpha,\beta) \\ \times \frac{Dcos\alpha}{L^2} h(sin(\alpha_0 - \alpha)),$$
(2)

where,

$$L^2 = (R + xcos\beta + ysin\beta)^2 + (-xsin\beta + ycos\beta)^2, \\ tan\alpha_0 = \frac{-xsin\beta + ycos\beta}{R + xcos\beta + ysin\beta},$$
(3)

and $\beta$ is the source-detector rotation angle.

For a RedCT scan, substituting $\alpha$ by $\gamma$ using Eq. (1), one gets

$$f(x,y) = \frac{1}{2}\int_0^{2\pi} d\beta \int_{-\gamma_m}^{+\gamma_m} d\gamma g(\gamma) \ p(\gamma,\beta) \\ \times \frac{kRcosg(\gamma)}{L^2} h(sin(g(\gamma_0) - g(\gamma))),$$
(4)

where,

$$\frac{sin\gamma_0}{cos\gamma_0 + k} = \frac{-xsin\beta + ycos\beta}{R + xcos\beta + ysin\beta}.$$
(5)

In Eq. (4), unfortunately, due to the $g(\cdot)$ function the filtering processing no long preserves a shift-invariant property (cannot be derived as a function of $(\gamma_0 - \gamma)$), making it difficult to achieve an exact filtered-backprojection algorithm. However, a good approximate formula can still be achieved as follows. First, Let us simplify $h(sin(g(\gamma_0) - g(\gamma)))$ as

$$h(sin(g(\gamma_0) - g(\gamma))) = h(sin(g'(\zeta)(\gamma_0 - \gamma))) \\ \approx h(sin(m(\gamma_0 - \gamma))),$$
(6)

where,

$$m = \frac{\int_{-\gamma_m}^{\gamma_m} g'(x)dx}{2\gamma_m}, \ g'(x) = \frac{1 + kcosx}{k^2 + 2kcosx + 1}.$$
(7)

Here, the scaling factor $m$ is the mean value of $g'(x)$ within a range of $[-\gamma_m, \gamma_m]$ that roughly scales a fan angle $\alpha$ in RedCT (nonuniformly sampled) back to a central angle $\gamma$ (uniformly sampled), with a maximum value being $\frac{1}{k+1}$. Comparing with using $\frac{1}{k+1}$ as the scaling factor directly($\alpha \approx \frac{\gamma}{k+1}$), $m$ can reduce overall reconstruction error across the entire fan angle ranges. After using such a scaling factor, the filtering processing becomes shift-invariant as desired.

Of course, using a simply scaling factor $m$ or $\frac{1}{k+1}$ will result in causing inconsistencies across CT projections. However, an empirical weighting strategy can be established to further reduce these inconsistencies and get quite accurate reconstruction. In this work, we apply a weighting factor $w$ into the original projection $p(\gamma,\beta)$ before filtering,

$$\widetilde{p}(\gamma,\beta) = w \cdot p(\gamma,\beta), \ w = \left(\frac{\gamma/\alpha}{mean(\gamma/\alpha)}\right)^4.$$
(8)

The $w$ can be treated as a compensation for the sampling variation from $\gamma$ to $\alpha$. Finally, a practical FBP-type reconstruction formula is obtained as

$$f(x,y) = \frac{1}{2}\int_0^{2\pi} d\beta \int_{-\gamma_m}^{+\gamma_m} d\gamma \frac{kRcosg(\gamma)}{L^2} g'(\gamma)\widetilde{p}(\gamma,\beta) \\ \times \frac{(\gamma_0 - \gamma)^2}{sin^2(m(\gamma_0 - \gamma))} h(\gamma_0 - \gamma),$$
(9)

## 3  Results

In order to evaluate the performances of RedCT reconstruction methods, fan-beam RedCT projections of various phantoms were generated with $k's$ ranging from 0.8 to 1.2. The radius $R$ (detector-to-focus distance) was set as 500 mm, with 480-mm-in-diameter as the nominal scan field-of-view of RedCT. An equiangular detector arc consisting of 1000 detector pixels with a pixel size of $1 \times 1$ mm$^2$ was used. The number of view angles was set as 1000 per one rotation. CT images were reconstructed using Eq. (9) and its variants. Rebinning-to-equiangular CT scan was also implemented for comparison of spatial resolution. Reconstruction from standard equiangular CT having the same detector-to-focus and source-to-focus distance(ie, $k = 1$) was taken as the ground truth for other $k$ values.

### 3.1  A water phantom

Here, a water clyinder with a radius of 240 mm was simulated. The source to center distance was 450 mm ($k = 0.9$). As shown in Fig. 3 and Table 1, CT images reconstructed using the scaling factor $m$ during the filtering step is biased, the scaling factor $\frac{1}{k+1}$ makes it better. Our proposed empirical weighting strategy before the filtering generates quite accurate results, with a CT number inaccuracy of about 1 Hounsfield Units (HU), largely acceptable for practical applications.

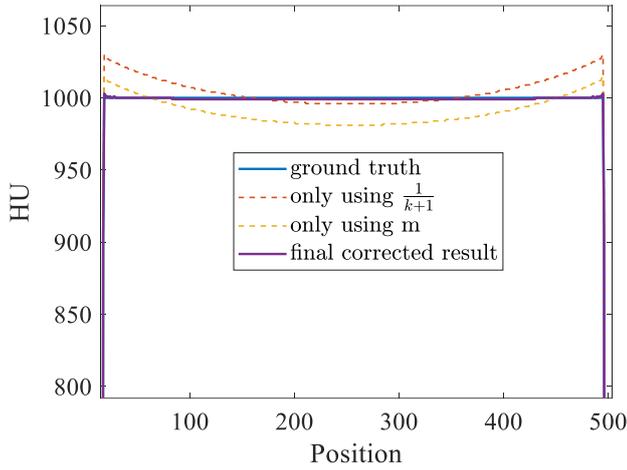

**Figure 3:** The central profiles of reconstructed images for a 480mm-in-diameter water phantom using different reconstruction approaches. It is clear that using scale factor $m$ or $\frac{1}{k+1}$ alone cannot decrease the CT nonuniformity as the peripheral region is higher than the central. With the weighting factor $w$, no significant nonuniformity is observed.

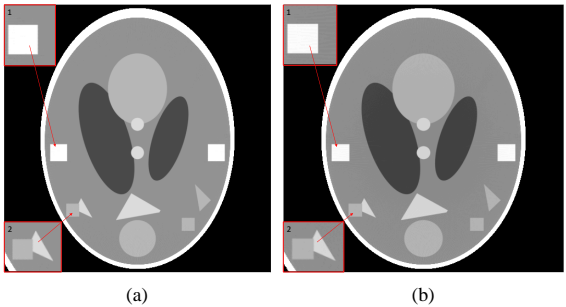

**Figure 4:** CT images of the modified Shepp-Logan phantom reconstructed from a fan beam equiangular CT (a), and from RedCT (b). Display window: [0.98, 1.05].

**Table 1:** Reconstruction PSNR and SSIM relative to ground truth for Fig. 3.

| Reconstructin method | PSNR(dB) | SSIM |
|---|---|---|
| Standard equiangular CT | 52.86 | 1 |
| Using $1/(k+1)$ only | 38.64 | 0.9998 |
| Using $m$ only | 41.18 | 0.9998 |
| Using both $m$ and $w$ | 51.84 | 0.9999 |

### 3.2 A modified Shepp-Logan phantom

To evaluate the performance of RedCT image reconstruction with respect to object shapes, we simulated a RedCT scan of a modified Shepp-Logan phantom with $k = 0.9$. As shown in Fig. 4, it is observed that our proposed weighted FBP method can generate high quality CT images with no obvious artifacts.

### 3.3 RedCT image from real CT data

To evaluate the performance of RedCT image reconstruction of object close to real patient, we simulated a RedCT scan using a dataset of CT images from a diagnostic CT scanner with $k = 0.9$. As shown in Fig. 5 and Table 2, our proposed weighted FBP method achieve reasonably accurate results as expected.

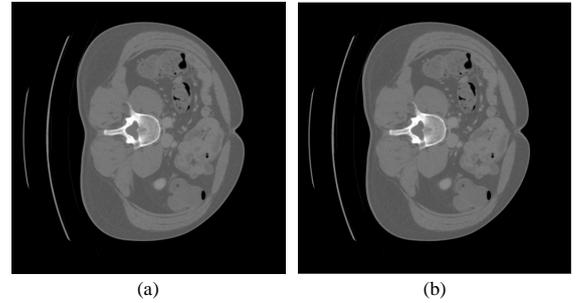

**Figure 5:** Abdomen CT images reconstructed from a fan beam equiangular CT (a), and from RedCT (b). Display window: [0.2, 0.8].

**Table 2:** PSNR and SSIM of CT images in Fig. 5.

| Architecture | PSNR(dB) | SSIM |
|---|---|---|
| The relaxed architecture($k = 0.9$) | 46.39 | 0.9924 |
| Standard euqiangular CT | 45.43 | 0.9912 |

### 3.4 A water phantom with iodine contrasts

It is expected that reconstruction inaccuracy increases as the ratio $k$ moves away from 1. As a result, an important question is under what range of $k$ our proposed FBP algorithm can work well. To quantitatively get the number, we simulated a water cylinder with four iodine contrasts shown in Fig. 6, with CT number average relative error (ARE) calculated as,

$$E_{ARE} = \frac{1}{N} \sum_N |\frac{CT_{result} - CT_{groundtruth}}{CT_{groundtruth}}| \times 100\%, \quad (10)$$

where, $N$ is the total number of image pixels in the selected region of interest (ROI).

The CT images reconstructed from RedCT scan at $k = 0.9$ is demonstrated in Fig. 6. Figure 7 shows the relationships between $k$ and $E_{ARE}$ error for the ROI being the entire water cylinder and for the four iodine contrasts, respectively. When $k = 1$, $E_{ARE} = 0$ as expected. It is seen that find $E_{ARE}$ errors increase slightly slower for $k < 1$ when compared with that of $k > 1$. Iodine contrasts shares a little bigger error when compared with pure water areas but the overall accuracy is acceptable for $k$ ranging from 0.89 to 1.05, with $E_{ARE} < 0.08\%$, indicating less than 1 HU of inaccuracy. These results validate our proposed weighting strategy in the practical FBP algorithm for RedCT.

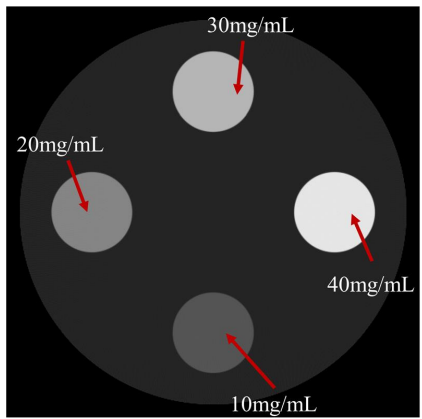

**Figure 6:** CT images of water cylinder with iodine contrasts used in RedCT scan at $k = 0.9$. Display window: [0.9, 1.6].

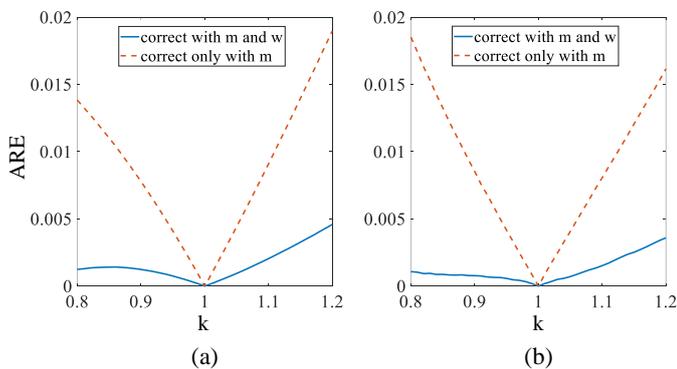

**Figure 7:** The $E_{ARE}$ error for ROI being (a) the entire water phantom and (b) the four iodine contrasts only.

### 3.5 Spatial Resolution Comparison

To evaluate the performance of RedCT reconstruction in terms of spatial resolution, we simulated a tungsten wire with a diameter of 40 $\mu m$ at $k = 0.9$. The modulation transfer function (MTF) curves were plotted along radial direction and azimuth direction in Fig. 8, for reconstructions by our proposed weighted FBP method with no rebinning and by the rebinning-to-equiangular CT method, respectively. As expected, interpolation during rebinning-to-equiangular step costs some loss in spatial resolution, while our weighted FBP method preserves better MTF values.

### 4 Discussion and Conclusions

A compact architecture of coplanar transmission and emission Guidance in Radiotherapy (Co-TeGRT) is a promising system design to allow PET and CT imaging and MV treatment all be assembled in the same scan plane on a rotating gantry. Due to the unique data sampling on detector, a relaxed equiangular detector CT (RedCT) is established in Co-TeGRT. For this novel geometry, a practical FBP-type reconstruction with an innovative weighting strategy before filtering is explored with higher CT number accuracy and better spatial resolution. Numerous phantom studies validate our method to be effective and feasible for practical applications. For the ratio $k$ ranging from 0.89 to 1.05, the $E_{ARE}$ error was $< 0.08\%$, indicating less than 1 HU of inaccuracy in reconstruction.

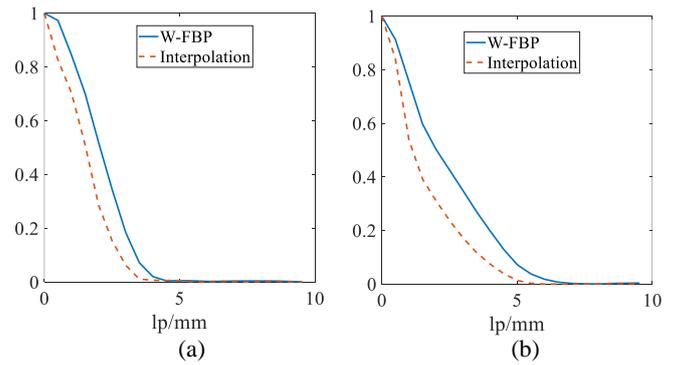

**Figure 8:** The MTF curve when k=0.9: (a) the results of radial direction, (b) the results of ring direction.

In theory, our proposed RedCT reconstruction method could be easily promoted to a regular equiangular geometry with source position away from the detector focus, where $k$ is close to 0 instead of 1 in RedCT. Further study of RedCT includes using neural network to further optimize the weighting strategy and extending our algorithm into 3D imaging geometry.

### 5 Acknowledgment

This project was supported in part by grants from the National Natural Science Foundation of China (No. U20A20169 and No. 12075130).

# Region-of-Interest CT Reconstruction using One-Endpoint Hilbert Inversion in Optimized Directions

Aurélien Coussat[1], Simon Rit[1], Michel Defrise[2], and Jean Michel Létang[1]

[1]Université de Lyon, INSA-Lyon, Université Claude Bernard Lyon 1, UJM-Saint Etienne, CNRS, Inserm, CREATIS UMR 5220, U1206, F-69373, Lyon, France

[2]Department of Nuclear Medicine, Vrije Universiteit Brussel, Brussels, Belgium

**Abstract** In computed tomography, scanning the entire object is sometimes impossible, causing truncated projection data. Reconstruction is however still possible: using differentiated backprojection, the Hilbert transform of the object can be calculated and inverted along line segments in the field-of-view. When two endpoints of the line segment are outside the object extent, a stable analytic reconstruction formula exists. When only one endpoint is outside the object extent, every pixel can be reconstructed, but no inversion formula is known yet. Uniqueness of the inverse Hilbert transform is nevertheless guaranteed along such segments, and a numerical inverse can be used. Most pixels of the field-of-view accept more than one "one-endpoint line segment" and one can choose the direction; we propose here to select the optimal direction based on an empirical criterion. Image quality improvement is assessed against a reconstruction that uses a single direction for every pixel.

## 1 Introduction

Conventional reconstruction procedures generally require the object to fit entirely within the scanner field-of-view (FOV), as they expect non-truncated tomographic data. However, in many imaging scenarios, such a condition cannot be met. Reconstruction must then use truncated projections. Recent theoretical results show that reconstruction is possible for some patterns of data truncation using, among other methods [1], the differentiated backprojection (DBP) followed by the inversion of the Hilbert transform. The reconstruction of some subset of the object, called region-of-interest (ROI), is then possible.

For simplicity, we focus on two-dimensional (2D) parallel beam tomography, although a generalization of the proposed method to three-dimensional cone-beam CT is possible. The source rotates along a $180°$ arc around the object. The scanner FOV is defined as the circular region where all points are illuminated by every source location. Knowledge of an approximate object extent $\Omega$ is assumed. If the FOV is entirely contained in the interior of $\Omega$, then the problem is interior and will not be addressed in this paper. We rather focus on the problem occurring when the FOV partly overlaps the object extent. The ROI is defined as this overlap: ROI = FOV $\cap \Omega$. Consider any line segment that overlaps $\Omega$ and whose two endpoints lie on the FOV border. The reconstructibility of this line segment varies depending on the number of its endpoints located inside $\Omega$. When the two endpoints are located outside $\Omega$, an analytic reconstruction formula can be applied [2]. When only one endpoint is located outside $\Omega$, the line segment admits a unique and mathematically stable reconstruction [3], but no analytic formula is known yet. Unlike the "two-endpoint line segments", the "one-endpoint line segments" can reconstruct any pixel of the FOV. We refer to the problem of reconstructing such line segments as the "one-endpoint Hilbert inversion". This problem can be solved numerically by several techniques, such as projection onto convex sets [3] or singular value decomposition (SVD) [4, 5].

To our knowledge, the choice of a particular direction for the one-endpoint Hilbert line segment is a question that has never been addressed. In this work, each pixel of the image is reconstructed using a direction that seems optimal with respect to an empirical criterion. The reconstruction itself is achieved using extended SVD (XSVD) inversion [4]. A comparison with XSVD that uses a single Hilbert direction [4] is discussed.

## 2 Materials and Methods

Let $f : \mathbb{R}^2 \to \mathbb{R}$ be the sought object such that $\forall \boldsymbol{x} \notin \Omega, f(\boldsymbol{x}) = 0$. Let $p : [0; \pi[ \times \mathbb{R} \to \mathbb{R}$ be the projections of $f$, corresponding to the scanner measurements, defined as

$$p(\phi, r) = p_\phi(r) = \int_{-\infty}^{+\infty} f\left(r\boldsymbol{\alpha}_\phi + s\boldsymbol{\beta}_\phi\right) \mathrm{d}s \qquad (1)$$

where $\boldsymbol{\alpha}_\phi = (\cos\phi, \sin\phi)$ and $\boldsymbol{\beta}_\phi = (-\sin\phi, \cos\phi)$. Since projections are truncated, $p(\phi, r)$ is unavailable for $r$ below or above a certain threshold in some directions $\phi$.

DBP links the projections to the sought object with

$$g_\theta(\boldsymbol{x}) = \frac{1}{2\pi} \int_\theta^{\theta+\pi} \frac{\partial p(\phi, r)}{\partial r}\bigg|_{r=\boldsymbol{x}\cdot\boldsymbol{\alpha}_\phi} \mathrm{d}\phi = H_\theta f(\boldsymbol{x}) \qquad (2)$$

where $H_\theta$ is the one-dimensional Hilbert transform of a line of $f$ in the direction $\theta \in [-\pi; \pi[$, defined as

$$H_\theta f(\boldsymbol{x}) = \fint_{-\infty}^{+\infty} \frac{f(\boldsymbol{x} - t\boldsymbol{\beta}_\theta)}{\pi t} \mathrm{d}t, \qquad (3)$$

$\boldsymbol{\beta}_\theta = (-\sin\theta, \cos\theta)$ and $\fint$ represents the Cauchy principal value of the integral [2]. Since the derivative is a local operation, the DBP is not impacted by projection truncation and yields the correct Hilbert transform of $f$ for points measured by all projections, i.e., within the FOV. Therefore, $f$ can be reconstructed inside the FOV by applying an inverse truncated Hilbert transform to the DBP over enough segments to cover the FOV.

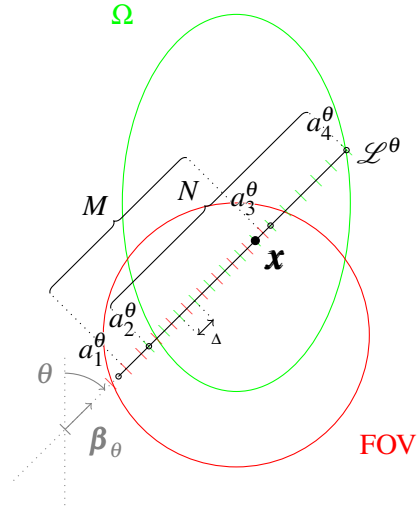

**Figure 1:** Notations employed in this work. The line segment $\mathscr{L}^\theta$ passes through the point $\boldsymbol{x}$ with an angle $\theta$. The points $a_{1,\dots,4}^\theta$ are defined such that the data are measured on the segment $[a_1^\theta, a_3^\theta]$, and the object extent $\Omega$ is included in segment $[a_2^\theta, a_4^\theta]$. The samples of $\boldsymbol{g}$ are illustrated as the red ticks, and the samples of $\boldsymbol{f}$ as the green ticks.

We now focus on the reconstruction of a single pixel $\boldsymbol{x} \in$ ROI, and we omit the $\boldsymbol{x}$ dependency of all symbols defined below for clarity. Let $\mathscr{L}^\theta$ be a one-endpoint line segment that passes through $\boldsymbol{x}$ at some angle $\theta$, and let $\Delta > 0$ be the sampling step along $\mathscr{L}^\theta$. We define $\boldsymbol{g} \in \mathbb{R}^M$ as the vector whose components are defined as

$$g_{i-a_1^\theta+1} = g_\theta\left(\boldsymbol{x} + \left(i - \frac{1}{2}\right)\Delta\boldsymbol{\beta}_\theta\right) \quad \text{for} \quad a_1^\theta \leq i \leq a_3^\theta \quad (4)$$

with $M = a_3^\theta - a_1^\theta + 1$. The integers $(a_1^\theta, a_3^\theta) \in \mathbb{Z}^2$ are chosen such that $\boldsymbol{g}$ spans the entire FOV along $\mathscr{L}^\theta$.

Similarly, the vector $\boldsymbol{f} \in \mathbb{R}^N$ has its components defined as

$$f_{j-a_2^\theta+1} = f\left(\boldsymbol{x} + j\Delta\boldsymbol{\beta}_\theta\right) \quad \text{for} \quad a_2^\theta \leq j \leq a_4^\theta \quad (5)$$

with $N = a_4^\theta - a_2^\theta + 1$. The integers $(a_2^\theta, a_4^\theta) \in \mathbb{Z}^2$ are chosen such that $\boldsymbol{f}$ entirely bounds $\Omega$ along $\mathscr{L}^\theta$. Note that the integers $a_{1,\dots,4}^\theta$ depend on $\boldsymbol{x}$ and $\theta$ since the distances between the FOV borders, $\Omega$ borders and $\boldsymbol{x}$ vary with $\theta$. The sought value at $\boldsymbol{x}$ corresponds to Equation (5) with $j = 0$; we thus have $a_1^\theta < a_2^\theta < 0 < a_3^\theta < a_4^\theta$. If these inequalities do not hold, then $\mathscr{L}^\theta$ is not one-endpoint. Figure 1 summarizes these notations.

As shown in Equation (2), the link between $\boldsymbol{f}$ and $\boldsymbol{g}$ is the Hilbert transform, which, in its discrete form, can be expressed as $\mathbf{H}$, an $M \times N$ matrix whose values are defined as

$$H_{i-a_1^\theta+1, j-a_2^\theta+1} = \frac{1}{\pi}\frac{1}{i-j-\frac{1}{2}} \quad \text{for} \quad \begin{cases} a_1^\theta \leq i \leq a_3^\theta \\ a_2^\theta \leq j \leq a_4^\theta \end{cases} \quad (6)$$

such that $\mathbf{H}\boldsymbol{f} = \boldsymbol{g}$. The $\theta$ dependency of $\boldsymbol{f}$, $\boldsymbol{g}$ and $\mathbf{H}$ has been omitted for clarity. Note the term $-\frac{1}{2}$ appearing in Equations (4) and (6): this half-pixel shift improves the reconstruction resolution [6]. We used here the XSVD procedure, based on truncated SVD, to reconstruct $\boldsymbol{f}^{\text{XSVD}}$ [4]. The value at $\boldsymbol{x}$ is then $f_{-a_2^\theta+1}^{\text{XSVD}}$, i.e. Equation (5) for $j = 0$.

Any angle $\theta$ such that $\mathscr{L}^\theta$ is a one-endpoint line segment theoretically works to apply the reconstruction procedure. Note that points having a two-endpoint line segment should be reconstructed using the analytic formula [2], but we force here the usage of one-endpoint line segments for simplicity and to better evaluate the proposed method. We have previously observed a residual artifact with XSVD and that some Hilbert directions give better results than others. We therefore propose to select the direction that minimizes some criterion $\mathscr{C}$ for a given point $\boldsymbol{x}$ depending on the problem parameters $a_{1,\dots,4}^\theta$, that is

$$\hat{\theta} = \arg\min_\theta \mathscr{C}\left(a_1^\theta, a_2^\theta, a_3^\theta, a_4^\theta\right). \quad (7)$$

Two different criteria were considered in this study. We first empirically decided to minimize the distance between $a_3^\theta$ and $a_4^\theta$, as it is the part of the segment that cannot be reconstructed, relatively to the number of values $a_3^\theta - a_2^\theta$ that can be reconstructed. This criterion is modeled as

$$\mathscr{C}_1\left(a_1^\theta, a_2^\theta, a_3^\theta, a_4^\theta\right) = \frac{a_4^\theta - a_3^\theta}{a_3^\theta - a_2^\theta}. \quad (8)$$

A known artifact of XSVD reconstructions is an offset whose intensity decreases with the distance to the inner FOV border [4]. We designed another criterion aiming at maximizing the distance between $\boldsymbol{x}$ and the FOV boundary in the direction $\theta$. This second criterion is expressed as

$$\mathscr{C}_2\left(a_1^\theta, a_2^\theta, a_3^\theta, a_4^\theta\right) = -a_3^\theta. \quad (9)$$

The proposed method was evaluated using computer simulations of the 2D Shepp-Logan phantom scaled up 96 times. A set of 720 parallel projections of 800 rays each were analytically computed over an arc of $180°$. The pixel spacing of the projections was 0.25 mm. Reconstructions were computed on a $256 \times 256$ pixel grid, in which only the pixels within the FOV were reconstructed. A FOV was simulated by using the DBP data only from inside a circular region located near the bottom of the phantom. The estimated object extent is an ellipse that encompasses the phantom with a 1 % margin. The sampling step $\Delta$ was set to 0.25 mm. The resulting setup is as in Figure 1. The pixel spacing of the reconstruction was 1 mm. For comparison, the same reconstruction was performed with the Hilbert direction $\theta$ set to 0 (vertical) for all pixels.

For each pixel of the ROI and for a given angle $\theta$, our implementation analytically computes the locations of the intersections between the segment and the FOV or $\Omega$. The

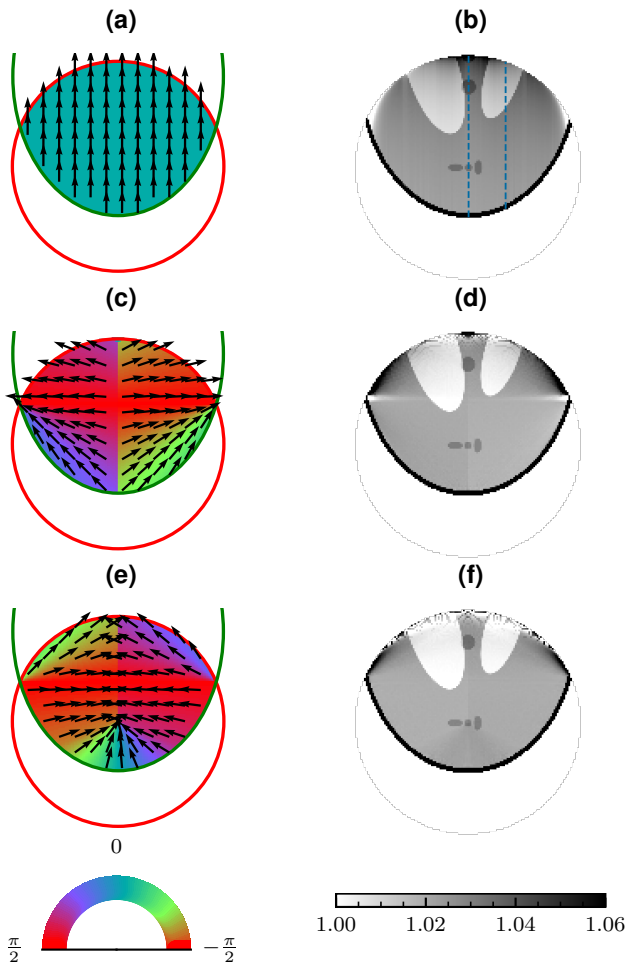

0

$\frac{\pi}{2}$                    $-\frac{\pi}{2}$

1.00   1.02   1.04   1.06

**Figure 2:** (a) and (b): directions and reconstruction when all directions are vertical. (c) and (d): directions and reconstruction for criterion $\mathscr{C}_1$. (e) and (f): directions and reconstruction for criterion $\mathscr{C}_2$.

optimization problem of Equation (7) is solved numerically by evaluating its right-hand side for every value of $\theta$ with a $0.25°$ step. Then, the DBP is computed on every sample of $\boldsymbol{g}$ to apply the XSVD. All simulations were implemented using Python 3.8.5, NumPy 1.19.4 and RTK 2.1.0 [7].

## 3   Results

Figure 2 shows directions and reconstructed images using either a vertical direction for every pixel (Figures 2a and 2b), the criterion $\mathscr{C}_1$ (Figures 2c and 2d) or the criterion $\mathscr{C}_2$ (Figures 2e and 2f).

Figure 3 shows profiles through two columns of the reconstructed images displayed in Figures 2b, 2d and 2f. Figure 3a shows the central column; Figure 3b shows the column located at 22 mm on the right hand side of the FOV center. The profile locations are displayed in Figure 2b.

## 4   Discussion

Using a single Hilbert direction aligned to the pixel grid to reconstruct the entire FOV has some advantages: a single inversion is sufficient to reconstruct a whole line of pixels, and the procedure is computationally inexpensive. However, it also has major drawbacks: a single direction is not enough for a full-FOV reconstruction in many configurations, and the chosen direction can be sub-optimal for some pixels, deteriorating the reconstruction. This procedure was reproduced and the result is shown in Figure 2b. This reconstruction also displays slight vertical streaks in the Hilbert direction because each line is treated as an independent problem. The method proposed here avoids these limitations, because each pixel is treated as an independent problem.

Figures 2d and 2f show that the reconstructions are less accurate than that of Figure 2b for pixels lying close to the FOV inner boundary. This poor quality is probably due to the interpolation used to compute the DBP in Figures 2d and 2f, whereas Figure 2b uses exact values, as it is less computationally demanding to do so in a single Hilbert direction. However, in the rest of the ROI, the quality is higher by the use of multiple Hilbert directions; this improvement is visible on Figure 3. The discretization of the angle $\theta$, when solving Equation (7), may explain the small streak artifacts visible in the reconstructions.

The pixels close to the interior endpoint are known to be difficult to reconstruct. We attribute this difficulty to the decreasing numerical stability of the one-endpoint Hilbert inversion model [3]. The same instability was observed in previous works performing one-endpoint reconstructions using XSVD [4, 5], where it was shown that this instability creates an offset whose intensity increases towards the interior endpoint. On Figure 3, this effect is clearly visible, especially looking at the profile of the reconstruction shown in Figure 2b; the same offset, although not as intense, is observed when each pixel uses its own direction. Interestingly, the offset is also observed in other works not using SVD inversion [3, 8–10], suggesting that this effect is not directly caused by the one-endpoint Hilbert inversion but is instead intrinsic to the ROI tomography problem.

As an attempt to reduce this offset, the criterion $\mathscr{C}_2$ (Equation (9)) was designed to favor directions that maximize the distance between the FOV border and the sought pixel. Figure 3 illustrates the improvements yielded by $\mathscr{C}_2$: the offset is considerably reduced compared to that of a reconstruction performed in a fixed direction. Figure 2e shows that the bottom of $\Omega$ is reconstructed using almost vertical directions using this criterion; this observation explains the similar quality to that of the vertical-only reconstruction, as shown on the left side of Figure 3a. Criterion $\mathscr{C}_1$ also diminishes the offset, but two areas near the intersections between the FOV and $\Omega$, visible in Figure 2d, are reconstructed with a poor quality. Neither $\mathscr{C}_1$ nor $\mathscr{C}_2$ stands out and the optimal criterion remains to be found.

The line that connects the two intersection points between $\Omega$ and the FOV boundaries forms a border between two regions. Pixels located above this border are only crossed by line segments that are at most one-endpoint, which fits

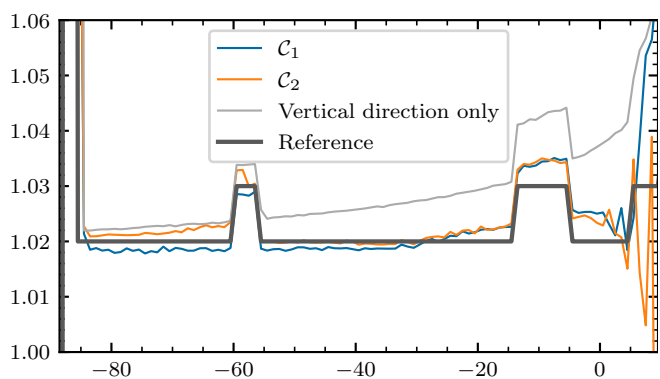

**(a)** Profile through the central column.

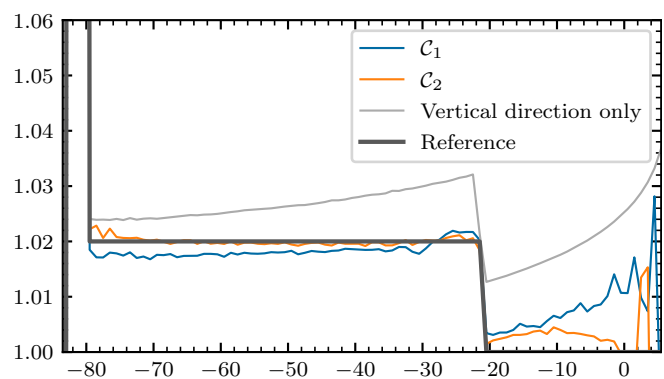

**(b)** Profile through the column located at 22 mm on the right hand side of the FOV center.

**Figure 3:** Two vertical profiles, shown in Figure 2b, of the reconstructions displayed in Figures 2b, 2d and 2f. The x-axis is limited to the ROI; the y-axis is limited to the grayscale limit $[1; 1.06]$.

the proposed approach. On the other hand, pixels that lie below the line accept at least one two-endpoint segment; along these segments, an analytic inversion formula should instead be applied [2] since this two-endpoint inversion is more accurate than the one-endpoint inversion. However, full-FOV reconstruction is impossible using two-endpoint line segments only. A previous work explored the possibility of combining two- and one-endpoint reconstructions in a single image [5]; here, for simplicity and to better evaluate the proposed method, the full FOV is reconstructed using one-endpoint line segments only, although it is not optimal.

## 5 Conclusion

We have proposed a method for ROI reconstruction from truncated projections using one-endpoint Hilbert inversion techniques. This method selects a different Hilbert direction for each pixel, chosen to optimize some criterion; here, we empirically designed two criteria. The proposed method is able to reconstruct any non-interior ROI tomography problem, while maintaining a satisfactory image quality. Noteworthy improvements were observed compared to a similar method that uses a fixed Hilbert direction for all pixels.

### Acknowledgments

This work was supported by grant ANR-17-CE19-0006 (ROIdoré project) from the Agence Nationale de la Recherche (France). This work was performed within the framework of the SIRIC LYriCAN Grant INCa_INSERM_DGOS_12563 and of the LABEX PRIMES (ANR-11-LABX-0063) of Université de Lyon, within the program 'Investissements d'Avenir' (ANR-11-IDEX-0007) operated by the French National Research Agency (ANR).

# Joint reconstruction with a correlative regularisation technique for multi-channel neutron tomography

Evelina Ametova[1,2,*], Genoveva Burca[3,4], Gemma Fardell[5], Jakob S. Jørgensen[4,6], Evangelos Papoutsellis[1,5], Edoardo Pasca[5], Ryan Warr[1], Martin Turner[7], William R. B. Lionheart[4], and Philip J. Withers[1]

[1]Henry Royce Institute, Department of Materials, The University of Manchester, United Kingdom
[2]Laboratory for Application of Synchrotron Radiation, Karlsruhe Institute of Technology, Germany
[3]ISIS Pulsed Neutron and Muon Source, STFC, UKRI, Rutherford Appleton Laboratory, United Kingdom
[4]Department of Mathematics, The University of Manchester, Manchester, United Kingdom
[5]Scientific Computing Department, STFC, UKRI, Rutherford Appleton Laboratory, United Kingdom
[6]Department of Applied Mathematics and Computer Science, Technical University of Denmark, Denmark
[7]Research IT Services, The University of Manchester, United Kingdom

**Abstract** Time-of-flight (ToF) energy-dispersive neutron tomography is complimentary to X-ray tomographic imaging method aimed at the reconstruction of wavelength-dependent material response in every three-dimensional volume element. ToF neutron tomography has already demonstrated great potential for both mapping of crystallographic properties and elemental composition in micrometer scale. However, available neutron beams have inherently low fluxes and high ToF resolution comes at the cost of prohibitively long exposure times. In this paper we investigate application of advanced iterative reconstruction algorithms with both spatial and spectral regularisation to reduce exposure time. The capability of advanced reconstruction algorithms is demonstrated on a specifically designed multi-material sample.

## 2 Introduction

Time-of-flight (ToF) energy-dispersive neutron computed tomography (CT) provides a complimentary technique to X-ray CT. As neutrons interact with atomic nuclei rather than an atom's electron cloud and can penetrate materials at wavelengths comparable to lattice spacings, they can be used to investigate the chrystallographic structure of materials. Governed by Bragg's law, coherent elastic scattering produces characteristic jumps in the transmitted neutron spectrum at wavelength equal twice the spacing between lattice planes. As neutrons are uncharged particles and can penetrate much deeper into material than X-rays, they allow to probe atomic structures in bulk samples. In fig. 1, left, we show the wavelength-dependent macroscopic total neutron cross-section $\Sigma_{tot}(\lambda)$, $[cm^{-1}]$ for materials employed in the present study. The neutron cross-section defines probability of interactions to occur, *i.e.* decrease in transmitted intensity. Such acquisition is also commonly called Bragg-edge neutron CT due to characteristic shape of the transmitted spectrum.

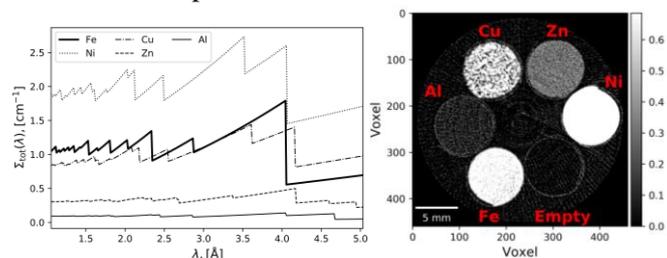

Fig. 1. Left: Theoretical neutron spectra for materials employed in this study. Right: White beam (sum of all channels) reconstruction of the sample cross-section with the conventional FBP method. Colour range was adjusted to highlight low intensity features.

ToF imaging utilises a pulsed neutron source and measures arrival time of each neutron with respect to the pulse. Pixellated counting ToF detector discretises recorded information into pixel elements, each counting the individual incident neutrons and registers each of them in one of multiple ToF channels depending on arrival time. ToF values are subsequently converted into wavelength values. Data acquisition in ToF neutron CT is very time consuming since neutron fluxes are typically low (compared to synchrotron X-ray sources) and because the detected neutrons are shared among multiple ToF channels. Thus, prohibitively long exposure time (in order of several hours per projection) is needed to acquire sufficiently high counts in each ToF channel. In practice, shorter exposure is used and acquired multi-channel projections are binned, in spatial and/ or spectral dimension, to improve signal-to-noise ratio [1, 2].

Regularised iterative reconstruction has already proven to be a viable approach to improve reconstruction quality from noisy and/or undersampled data in X-ray CT. Iterative reconstruction formulates reconstruction as an optimisation problem and allows to incorporate available prior knowledge to produce satisfactory results for otherwise unsolvable tomographic problems. In spectral tomography, priors exploiting structural similarities across energy channels are particularly promising. Here, we investigate application of two regularisation techniques: Total Nuclear Variation (TNV) and a dedicated tailored regularisation technique. The former method is a recent regulariser for reconstructing spectral CT images which enforces common edges across all channels [3-5]. The second technique is specifically tailored for Bragg edge neutron CT and combines Total Variation (TV) regularisation [6, 7] in the spatial dimension and Total Generalised Variation (TGV) regularisation [8] in the spectral dimension. TV preserves edges and suppresses noise by encouraging sparsity in the finite difference domain, while TGV regularisation promotes characteristic piece-wise smooth behaviour in the spectral dimension. Implementation of the advanced reconstruction methods has been made possible by the CCPi Core Imaging Library (CIL) reconstruction framework [9, 10]. The capability of advanced reconstruction algorithms is demonstrated on a specifically designed multi-material sample consisting of aluminium cylinders filled with metallic powder of high purity (fig. 1, right). Although the case study has been performed on a single two-dimensional slice, the results can be generalised to the third dimension.

## 2 Materials and methods

### 2.1. Data acquisiton

The dataset in the present study was imaged at the Imaging and Materials Science & Engineering (IMAT) beamline operating at the ISIS spallation neutron source (Rutherford Appleton Laboratory, U.K.) [11, 12]. The ToF detector [13] has 512×512 active pixels, 0.055 mm pixel size. The detector was configured to measure 2843 energy channels between 1 Å and 5 Å with wavelength resolutions between $0.7184 \cdot 10^{-3}$ Å and $2.8737 \cdot 10^{-3}$ Å. A set of spectral projections was acquired at 120 equally-spaced angular positions over 180° rotation. Each projection was acquired with 15 min exposure time. Additionally, 8 flat field images (4 before and 4 after the sample acquisition) were recorded with the same settings. Detector related corrections [14] and flat field correction were applied to all projections. Finally, spectral images were cropped spatially to 460 pixels and binned in the spectral dimension to 339 channels with a uniform bin width of $11.5 \cdot 10^{-3}$ Å.

In fig. 2 (upper row) we demonstrate sinograms for selected individual wavelength channels. Spanning the most valuable spectral range for the present sample, they show differences in noise levels and intensity of features depending on wavelength. The incident spectrum on IMAT has a crude "bell shape" with a peak around 2.6 Å [11, 12]. Therefore the elevated noise level is noticeable in the 4.5 Å wavelength channel. Only three most attenuative materials (Cu, Fe and Ni) are visible in sinograms; Zn and Al are obscured in higher levels of noise.

For comparison purposes, the images were reconstructed using filtered backprojection implemented as FBP in CIL with a Hann filter (fig. 2, bottom row). As expected, FBP produces extremely noisy reconstructions. Only Ni is clearly visible in the reconstructed cross-section. Reconstruction of 4.5 Å wavelength channel is substantially noisier due to both lower incident flux and lower neutron attenuation of the selected materials.

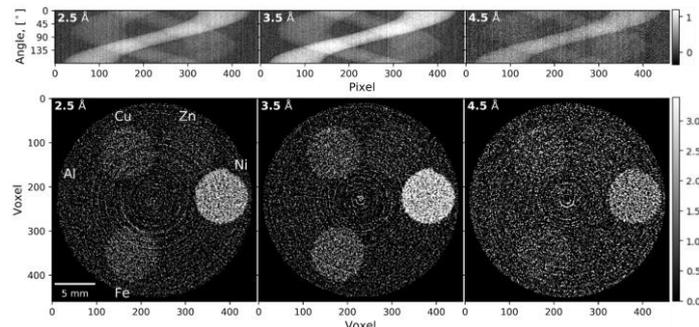

Fig 2. Top row: Sinograms for selected wavelength channels. Bottom row: FBP reconstruction of the selected sinograms.

### 2.2. Reconstruction

In every detector channel $k, k = 1, \dots, K$, where $K$ is the total number of channels, the measurement model in neutron CT can be well approximated by the Beer-Lambert law. Let us consider an incident beam of neutrons of given intensity $I_0$ at specific wavelength $\lambda_k$, the intensity $I$ which reach the detector element, will be reduced according to:

$$I(\lambda_k) = I_0(\lambda_k) \exp(-\int_L \mu(x, \lambda_k) \mathrm{d}x),$$

where $L$ is a linear path through the object and $\mu$ is the wavelength-dependent attenuation coefficient at the physical position $x$ in the object. Given an appropriate discretisation of Radon transform $A$ (projection operator), tomographic reconstruction in every wavelength channel $k$ can be modelled as a system of linear equations:

$$\bar{b}(\lambda_k) = -\ln\left(\frac{I(\lambda_k)}{I_0(\lambda_k)}\right) \approx \bar{A}\bar{u}(\lambda_k), \qquad (1)$$

where $\bar{b}$ is the discrete measured data, $\bar{u}$ is the to-be-reconstructed attenuation map discretised onto a Cartesian grid. The attenuation map $\bar{u}$ is typically represented as a column vector with $N = D^2$ elements (voxels), where $D$ is the number of elements in a detector row. The discrete measured data $\bar{b}$ is vectorised as a column vector with $M = PD$ elements (pixels), with $P$ being the total number of projections. The projection operator $\bar{A}$ contains $M \times N$ elements. If $i, i = \{0, 1, \dots M - 1\}$ and $j, j = \{0, 1, \dots, N - 1\}$, then $\bar{a}_{i,j}$ is the length of intersection of the $i$.th ray with the $j$.th voxel.

Similar to low-dose medical CT, the problem (1) is ill-posed in mathematical sense. Therefore, we seek for a way to compensate for the ill-posedeness of the problem by incorporating some prior knowledge about the solution. Unlike channel-by-channel methods which reconstruct each channel individually, we explore methods to jointly reconstruct all channels and exploit interchannel correlations. Then, the spectral CT data is modeled as

$$b = Au,$$

where $b$ and $u$ are obtained by stacking $K$ column vectors $\bar{b}(\lambda_k)$ and $\bar{u}(\lambda_k)$, respectively, and $A = I_{K \times K} \otimes \bar{A}$, $\otimes$ is the Kronecker product, and $I_{K \times K}$ is the identity matrix of order $K$. The reconstruction problem is formulated as

$$\arg\min\{F(u) = f(b, Au) + \alpha g(u)\},$$

where $f(b, Au)$ is a data fidelity metric which measures the discrepancy between the projection of solution $u$ and the acquired data $b$. The regularisation term $g(u)$ penalises undesired solutions and "guides" the optimization algorithm towards a solution with expected properties, which are commonly formulated in terms of image smoothness and sharp boundaries. The parameter $\alpha$ balances two terms and has to be tuned for each specific regulariser and dataset.

TNV is a recent extension of TV for multichannnel images [3-5]. TNV encourages the rank-sparsity by penalising the singular values of the Jacobian matrix. TNV has similar properties to TV regularisation, *i.e.* it also promotes a sparse image gradient in the spatial dimension, but also favours reconstructions with common edges across all channels. Consequently, TNV correlates channels and improves reconstruction quality by promoting common structures in multichannel images. The reconstruction problem is then formulated as,

$$F(u) = \|Au - b\|_2^2 + \alpha \text{TNV}(u). \quad (2)$$

Similar to TV, TNV suffers from a loss of contrast. Secondly, TNV does not allow the decoupling of regularisation parameters for the spatial and spectral dimensions which makes impossible to balance the level of regularisation between dimensions.

Here, we propose a novel tailored regulariser which combines TV [6,7] in the spatial dimension with TGV [8] the spectral (channel) dimension to jointly reconstruct low-count multi-channel neutron CT data. The proposed approach allows enforcing different image properties in respective dimensions. As the TV model captures piece-wise constant image properties in the spatial dimension, we rely on another regulariser to support reconstruction in the spectral dimension. Here, we rely on TGV to recover piece-wise smooth features in the spectral dimension because TGV allows balancing the first and the higher-order derivatives of images and consequently alleviates the staircasing effect inherent to TV. In this case the reconstruction problem is formulated as,

$$F(u) = \|Au - b\|_2^2 + \beta \text{TV}_{x,y,z}(u) + \gamma \text{TGV}_c(u). \quad (3)$$

Here we use $\text{TV}_{x,y,z}$ to designate a TV operator over three spatial dimensions $x$, $y$ and $z$, whereas $\text{TGV}_c$ operates over the spectral (channel) dimension.

Both methods (2) and (3) were implemented based on the CCPi Core Imaging Library (CIL) [9, 10]. CIL wraps the ASTRA toolbox [15] to perform forward- and back-projection operations and provides a set of various regularisers through the CCPi Regularisation Toolkit [16]. FISTA [17] and PDHG [18] were used to solve (2) and (3), respectively. Regularisation parameters were chosen manually to achieve both noise suppression and feature preservation in both spatial and spectral dimensions ($\alpha$=0.01, $\beta$=0.0075 and $\gamma$=0.3).

## 3 Results

Fig. 3 shows two-dimensional slices for selected (individual) wavelength channels reconstructed using the regularised iterative methods discussed in this paper. Both TNV and TV+TGV demonstrate drastic improvement in reconstruction quality and noise suppression. TNV produces "patchy" images and smearing of features is visible especially between the Cu and Zn cylinders (fig. 3, top row). Overall, features appear sharper in the TV+TGV reconstruction (fig. 3, bottom row).

TNV uses a small pixel neighbourhood information to correlate structural information along the spectral dimension. This acts similarly to a low-pass filter. Thus, TNV suppresses ring artifacts visible in FBP reconstruction

(fig. 2) but causes blurred and enlarged rings especially prominent in 4.5 Å channel, where counts are much lower. Al has very low neutron attenuation and is invisible in the TNV reconstruction due to contrast loss; a known drawback of both TV and TNV regularisation methods.

The TV+TGV reconstruction does not suffer from the ring artefacts and the faint Al cylinder is distinguishable in the reconstructed slices and profile lines (fig. 4, bottom row). Fine features inside the Cu cylinder are also partially preserved in the TV+TGV reconstruction (fig. 4, top row).

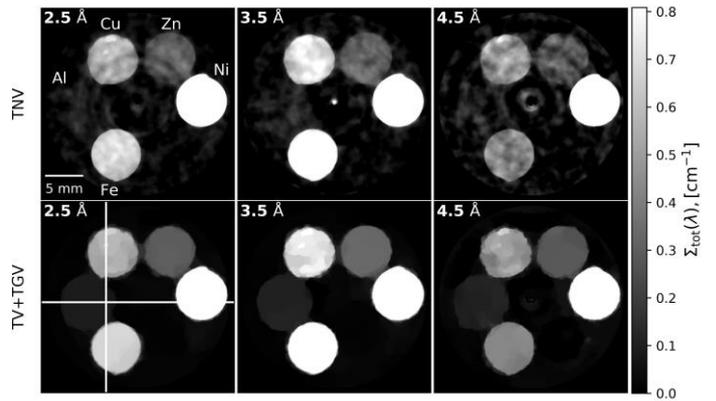

Figure 3: Two dimensional reconstructions of selected (individual) wavelength channels reconstructed with TNV regularisation (top row) and with TV+TGV regularisation (bottom row). White lines mark profile lines chosen for examination in the next figure. All slices are visualised with a common colour range. Colour range was adjusted to highlight low intensity features (maximum value of the display range was set to 30% of maximum intensity value in the reconstructed volume).

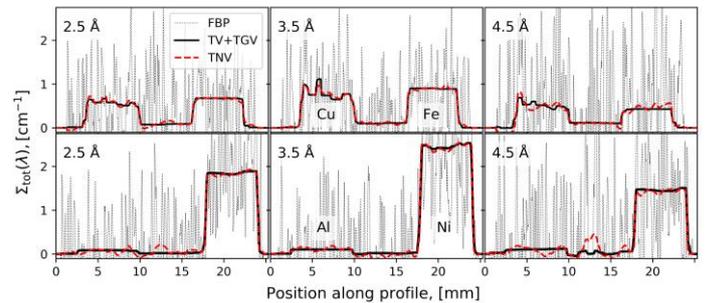

Figure 4: Profiles corresponding to the vertical (top row) and horizontal (bottom row) white lines in figure 3 (bottom left) passing through the Cu and Fe cylinders and the Al and Ni cylinders, respectively.

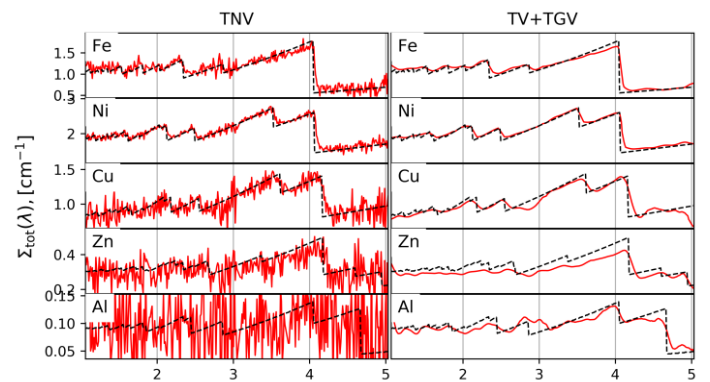

Figure 5: Individual spectra (solid line) reconstructed with TNV and TV+TGV for one representative 0.055³ mm³ voxel located within each material alongside the theoretical predictions (dotted black line).

Individual spectra reconstructed for one $0.055^3$ mm$^3$ voxel in each of the 5 materials are plotted in fig. 5 alongside the theoretical predictions. The voxel locations inside the cylinders were chosen arbitrarily. The reconstructed spectra for Fe, Ni and Cu closely follow the predicted spectra for both TNV and TV+TGV reconstructions. The amplified noise visible in the TNV reconstructions between 2.5 Å and 3 Å is caused by the increased noise levels in input data due to the detector dead time [14]. While the TV+TGV produces a much smoother spectra the Bragg edges appear to be less prominent due to smearing (for instance, small edges around 2 Å in Fe). In the case of the TNV regularisation, the noise dominates over smaller Bragg edges. For the materials with lower neutron attenuation (Al and Zn), TV+TGV clearly outperforms TNV as some Bragg edges are visible in the reconstructed spectrum, but are completely lost in the noise in the case of TNV.

Reconstruction is an intermediate step in Bragg edge imaging. The Bragg edge positions (in terms of $d$-spacing) allows compositional mapping, as each crystalline structure has a unique set of lattice spacings and hence a fingerprint in the neutron transmitted spectrum. The shape of the detected Bragg edges, *i.e.* deviation from abrupt predicted edge, supports characterisation of crystallographic properties. Quantitative comparison of the reconstruction results in terms of Bragg edge detection and characterisation, as well as material decomposition is a topic of future work.

## 5 Conclusion

We have demonstrated the capabilities of advanced reconstruction methods to improve reconstruction quality of low-count ToF neutron CT data. We investigated application of two regularisation techniques: the dedicated regulariser for multi-channel CT data and the tailored regularisation term encoding the prior information about neutron spectra. Both regularisation techniques showed drastic improvement of reconstruction quality compared to the standard FBP method. The tailored regularisation provided better reconstruction quality. Our study serves as an initial demonstration of a dedicated reconstruction technique that facilitates significant reduction of required exposure time – a major bottleneck in low flux ToF neutron CT studies.

## 5 Funding

This work was funded by EPSRC grants "A Reconstruction Toolkit for Multichannel CT" (EP/P02226X/1), "CCPi: Collaborative Computational Project in Tomographic Imaging" (EP/M022498/1 and EP/T026677/1). We gratefully acknowledge beamtime RB1820541 at the IMAT Beamline of the ISIS Neutron and Muon Source, Harwell, UK. EA was partially funded by BMBF and the Baden-Württemberg Ministry of Science as part of the Excellence Strategy of the German Federal and State Governments. JSJ was partially supported by The Villum Foundation (grant No. 25893). WRBL acknowledges support from a Royal Society Wolfson Research Merit Award. PJW acknowledges support from the European Research Council grant No. 695638 CORREL-CT.

# Comparison of MR acquisition strategies for super-resolution reconstruction using the Bayesian mean squared error

M. Nicastro[1], B. Jeurissen[1,2], Q. Beirinckx[1], C. Smekens[3], D. H. J. Poot[4], J. Sijbers[1,2], and A. J. den Dekker[1,2]

[1]imec - Vision Lab, Dept of Physics, University of Antwerp, Antwerp, Belgium.

[2]μNEURO Research Centre of Excellence, University of Antwerp, Antwerp, Belgium.

[3]Siemens Healthcare NV/SA, Beersel, Belgium.

[4]Biomedical Imaging Group Rotterdam, Dept of Radiology and Medical Informatics, Erasmus Medical Center, Rotterdam, The Netherlands.

**Abstract** In multi-slice super-resolution reconstruction (MS-SRR), a high resolution image, referred to as the SRR image, is estimated from a series of multi-slice images with a low through-plane resolution. This work proposes a framework based on the Bayesian mean squared error of the Maximum A Posteriori estimator of an SRR image to compare the accuracy and precision of two widely adopted magnetic resonance acquisition strategies in MS-SRR. The first strategy consists of acquiring a set of multi-slice images in which each image is shifted in the through-plane direction by a different, sub-pixel distance. The second consists of acquiring a set of multi-slice images in which each image is rotated around the frequency or phase-encoding axis by a different rotation angle. Results show that MS-SRR based on rotated multi-slice images outperforms MS-SRR based on shifted multi-slice images in terms of accuracy, precision and mean squared error of the reconstructed image.

## 1  Introduction

In conventional magnetic resonance imaging, a direct high resolution (HR) acquisition with a high signal-to-noise ratio (SNR) is often impractical due to the long scan time required. Previous studies have demonstrated the potential of multi-slice super-resolution reconstruction (MS-SRR) to address this issue by improving the inherent trade-off between resolution, SNR, and scan time [1]. The MS-SRR method consists of estimating an HR image, named SRR image, from a series of multi-slice images with a low through-plane resolution, hereafter referred to as the low resolution (LR) images [2]. Two acquisition strategies are more commonly adopted. The first consists of acquiring a set of LR images in which each image is shifted in the through-plane direction by a different, sub-pixel distance [3]. The second acquisition strategy consists of acquiring a set of LR images in which each image is rotated around the frequency or phase-encoding axis by a different rotation angle [4, 5]. The rotated scheme allows for a better sampling of the $k$-space compared to the shifted scheme since each LR image samples a different part of the $k$-space. Conversely, in the shifted scheme, all the LR images sample the same part of the $k$-space, causing the MS-SRR to rely exclusively on recovering the aliased frequencies in the slice-encoding direction.

The first comparison among MS-SRR acquisition protocols based on the two strategies was proposed in [4, 5]. A second comparison was proposed more recently in the context of fetal imaging, in which the segmentation quality of the SRR image was adopted as a performance criterion [6]. In both cases, the performance analysis focused on the non-regularized MS-SRR problem. However, MS-SRR estimation consists of solving an inverse problem, and regularization is required to find a stable solution [7]. Therefore, in this work, we extend the analysis to the regularized case, developing a framework in which the Bayesian mean squared error (BMSE) of the Maximum A Posteriori (MAP) estimator is proposed as a performance criterion [8]. The MAP estimator is built incorporating prior knowledge about the reconstruction target. The BMSE is chosen as a performance criterion to compare the acquisition strategies in terms of accuracy and precision for the class of reconstruction targets described by the prior distribution. The BMSE results are verified with Monte Carlo simulation experiments.

## 2  Materials and Methods

### 2.1  Super-resolution model

Let $\mathbf{r} \in \mathbb{R}^{N_r \times 1}$ be the vector containing the intensities of the noiseless HR target magnitude image. Furthermore, let $\mathbf{s}_m \in \mathbb{R}^{N_s \times 1}$, with $m = 1, .., M$, be the vector containing the intensities of the $m$-th noiseless LR multi-slice magnitude image. Then, $\mathbf{s}_m$ can be modeled as:

$$\mathbf{s}_m(\mathbf{r}) = \mathbf{D}(\text{AF})\mathbf{B}\mathbf{G}(\Phi_m)\mathbf{r}, \tag{1}$$

with $\mathbf{G} \in \mathbb{R}^{N_r \times N_r}$, $\mathbf{B} \in \mathbb{R}^{N_r \times N_r}$, $\mathbf{D} \in \mathbb{R}^{N_s \times N_r}$ linear operators that describe a geometric transformation, blurring and down-sampling, respectively. $\mathbf{G}$ is a function of the geometric transformation parameter $\Phi_m$, which represents the rotation angle or the shift of the $m$-th multi-slice image according to the acquisition strategy. $\mathbf{B}$ models the sampling function of the magnetic resonance imaging (MRI) acquisition method. For multi-slice acquisitions, the sampling function can be separated into three functions applied in orthogonal directions aligned with the MR image coordinates. The in-plane directions (frequency and phase-encoding) are modelled by a periodic sinc and the through-plane direction (slice-encoding) by a smoothed box. $\mathbf{D}$ is a function of the anisotropy factor AF, representing the ratio of the slice thickness to the in-plane resolution.

The sampling of the LR images can be expressed here as a matrix-vector multiplication $\mathbf{s}(\mathbf{r}) = \mathbf{A}\mathbf{r}$ where $\mathbf{s} =$ $[\mathbf{s}_1^T, ..., \mathbf{s}_M^T]^T \in \mathbb{R}^{MN_s \times 1}$ and $\mathbf{A} = [\mathbf{A}_1^T, ..., \mathbf{A}_M^T]^T \in \mathbb{R}^{MN_s \times N_r}$ with $\mathbf{A}_m = \mathbf{D}(\text{AF})\mathbf{B}\mathbf{G}(\Phi_m) \in \mathbb{R}^{N_s \times N_r}$. For a detailed description of the implementation, we refer to [9].

## 2.2 Conditional data distribution

Let $\tilde{\mathbf{s}} \in \mathbb{R}^{MN_s \times 1}$ be the vector containing the intensities of the M acquired magnitude LR images, subject to noise. Because of the relatively high SNR of the thick slices composing the LR images, the noise distribution can be well approximated by a zero-mean Gaussian distribution [10]. If all voxels are assumed to be statistically independent and the standard deviation of the noise $\sigma$ to be temporally and spatially invariant, the conditional probability density function (PDF) of the data points $p(\tilde{\mathbf{s}}|\mathbf{r})$ is equal to the product of the marginal PDFs of the individual data points and can be expressed as follows:

$$p(\tilde{\mathbf{s}}|\mathbf{r}) \propto \exp\left(-\frac{1}{2\sigma^2}\|\tilde{\mathbf{s}} - \mathbf{s}(\mathbf{r})\|_2^2\right). \qquad (2)$$

## 2.3 Prior distribution

The prior distribution is modelled as a stationary Gaussian Markov Random Field [11]. This corresponds with the assumption of a multivariate Gaussian prior of the form:

$$p(\mathbf{r}) \propto \exp\left(-\frac{1}{2}(\mathbf{r} - \bar{\mathbf{r}})^T \mathbf{K}^{-1}(\mathbf{r} - \bar{\mathbf{r}})\right), \qquad (3)$$

parametrised in terms of its mean $\bar{\mathbf{r}}$ and precision (inverse-covariance) matrix $\mathbf{K}^{-1}$, which is sparse, positive definite, and encodes statistical assumptions regarding the value of each HR image voxel based on the values of its neighboring voxels. Let $r_i$ be the $i$-th HR voxel and $\mathbf{r}_{\partial_i} \in \mathbb{R}^{N_n \times 1}$ the voxels from the neighborhood surrounding $r_i$, where $N_n$ is the number of neighborhood voxels, and $\partial_i$ represents the neighborhood voxels indices. We assume the conditional probability of the $i$-th HR voxel given the neighborhood voxels $p(r_i|\mathbf{r}_{\partial_i})$ to be Gaussian and of the form:

$$p(r_i|\mathbf{r}_{\partial_i}) \propto \exp\left(-\frac{\lambda^2}{2}\left(r_i - \sum_{j \in \partial_i} \alpha_j r_j\right)^2\right), \qquad (4)$$

where $\boldsymbol{\alpha} = \{\alpha_j\}_{j=1}^{N_n}$ is the vector of the so-called field potentials. It can be demonstrated [12] that Eq. (4) holds if and only if the joint PDF $p(\mathbf{r})$ assumes the form in Eq. (3) with:

$$\mathbf{K}_{i,j}^{-1} = \lambda^2 \begin{cases} 1, & i = j, \\ -\alpha_j & j \in \partial_i. \end{cases} \qquad (5)$$

Therefore, the hyperparameters $\boldsymbol{\alpha}$, $\bar{\mathbf{r}}$, and $\lambda$ characterize the prior distribution.

## 2.4 MAP estimator

The MAP estimator of $\mathbf{r}$ maximizes the posterior PDF $p(\mathbf{r}|\tilde{\mathbf{s}})$ with respect to $\mathbf{r}$, where $p(\mathbf{r}|\tilde{\mathbf{s}})$ is defined according to Bayes'

theorem [8] as:

$$p(\mathbf{r}|\tilde{\mathbf{s}}) \propto p(\tilde{\mathbf{s}}|\mathbf{r})p(\mathbf{r}). \qquad (6)$$

Therefore, the MAP estimator assumes the form:

$$\begin{aligned} \hat{\mathbf{r}} &= \arg\max_{\mathbf{r}} \ln p(\mathbf{r}|\tilde{\mathbf{s}}) \\ &= \arg\min_{\mathbf{r}} \frac{1}{\sigma^2}\|\tilde{\mathbf{s}} - \mathbf{A}\mathbf{r}\|_2^2 + (\mathbf{r} - \bar{\mathbf{r}})^T \mathbf{K}^{-1}(\mathbf{r} - \bar{\mathbf{r}}), \end{aligned} \qquad (7)$$

which admits the closed-form solution:

$$\hat{\mathbf{r}} = \left(\sigma^{-2}\mathbf{A}^T\mathbf{A} + \mathbf{K}^{-1}\right)^{-1}\left(\sigma^{-2}\mathbf{A}^T\tilde{\mathbf{s}} + \mathbf{K}^{-1}\bar{\mathbf{r}}\right). \qquad (8)$$

## 2.5 Bayesian MSE

The BMSE is proposed as a performance criterion to compare the two MS-SRR acquisition protocols described in the introduction section. Let us first define the component-wise MSE of $\hat{\mathbf{r}}$ as:

$$\text{MSE}(\mathbf{r})_j = \mathbb{E}_{\tilde{\mathbf{s}}}\left[(\hat{\mathbf{r}} - \mathbf{r})(\hat{\mathbf{r}} - \mathbf{r})^T\right]_{j,j}, \qquad (9)$$

where $\mathbb{E}_{\tilde{\mathbf{s}}}[.]$ is the expectation operator over $\tilde{\mathbf{s}}$. The MSE can be decomposed as the sum of a variance term and a squared bias term:

$$\text{MSE}(\mathbf{r})_j = \Sigma_{j,j} + \left[\boldsymbol{\beta}(\mathbf{r})\boldsymbol{\beta}^T(\mathbf{r})\right]_{j,j}, \qquad (10)$$

where $\boldsymbol{\Sigma} \in \mathbb{R}^{N_r \times N_r}$ and $\boldsymbol{\beta} \in \mathbb{R}^{N_r \times 1}$ are the covariance matrix and the bias vector of $\hat{\mathbf{r}}$, respectively. For the MAP estimator defined in the previous subsection, we have:

$$\boldsymbol{\Sigma} = \sigma^{-2}\mathbf{Q}\mathbf{A}^T\mathbf{A}\mathbf{Q}, \qquad (11)$$

$$\boldsymbol{\beta}(\mathbf{r}) = \mathbf{Q}\mathbf{K}^{-1}(\mathbf{r} - \bar{\mathbf{r}}), \qquad (12)$$

with

$$\mathbf{Q} = \left(\sigma^{-2}\mathbf{A}^T\mathbf{A} + \mathbf{K}^{-1}\right)^{-1}. \qquad (13)$$

The BMSE of the estimator of $\mathbf{r}$ can now be defined from the MSE as [8]:

$$\text{BMSE}(\mathbf{r})_j = \mathbb{E}_{\mathbf{r}}[\text{MSE}(\mathbf{r})]_j, \qquad (14)$$

where $\mathbb{E}_{\mathbf{r}}[.]$ is the expectation operator over $\mathbf{r}$. The BMSE can also be decomposed as the sum of a variance and a squared bias term, which can be linked to the MSE components as follows:

$$\begin{aligned} \text{BMSE}(\mathbf{r})_j &= \mathbb{E}_{\mathbf{r}}[\boldsymbol{\Sigma}]_{j,j} + \mathbb{E}_{\mathbf{r}}\left[\boldsymbol{\beta}(\mathbf{r})\boldsymbol{\beta}^T(\mathbf{r})\right]_{j,j} \\ &= \Sigma_{j,j} + \mathbb{E}_{\mathbf{r}}\left[\boldsymbol{\beta}(\mathbf{r})\boldsymbol{\beta}^T(\mathbf{r})\right]_{j,j}, \end{aligned} \qquad (15)$$

where $\mathbb{E}_{\mathbf{r}}[\boldsymbol{\Sigma}] = \boldsymbol{\Sigma}$, since $\boldsymbol{\Sigma}$ does not depend on $\mathbf{r}$, and the squared bias term of the BMSE can be calculated as the expectation over $\mathbf{r}$ of the MSE squared bias component in Eq. (12):

$$\mathbb{E}_{\mathbf{r}}\left[\boldsymbol{\beta}(\mathbf{r})\boldsymbol{\beta}^T(\mathbf{r})\right] = \mathbf{Q}\mathbf{K}^{-1}\mathbf{Q}. \qquad (16)$$

To simplify the notation, we define:

$$\mathbf{RBMSE} = \left\{ \sqrt{\mathrm{BMSE}\,(\mathbf{r})_j} \right\}_{j=1}^{N_r}, \qquad (17)$$

$$\mathbf{v} = \left\{ \sqrt{\Sigma_{j,j}} \right\}_{j=1}^{N_r}, \qquad (18)$$

$$\mathbf{b} = \left\{ \sqrt{\mathbb{E}_{\mathbf{r}} \left[ \boldsymbol{\beta}\,(\mathbf{r})\,\boldsymbol{\beta}^{\mathrm{T}}\,(\mathbf{r}) \right]_{j,j}} \right\}_{j=1}^{N_r}, \qquad (19)$$

where the dependency of **RBMSE** and **b** on **r** was omitted. In the following sections, we will refer to **v** and **b** as the standard deviation component and absolute bias component of the BMSE, respectively.

## 2.6 Acquisition protocols

The acquisition protocols are shown in Table 1. For all the protocols we fixed M/AF = 2. This choice ensures that the MS-SRR estimation problem is not under-determined $(M/AF \geq 1)$ [5] and that the $k$-space is efficiently sampled when the LR images are acquired with the rotated scheme $\left(M > \frac{\pi}{2}AF\right)$ [13]. Furthermore, it ensures that all the acquisition protocols require the same scan time. The HR protocol is included as a reference and represents a conventional multi-slice acquisition with AF = 1, repeated twice. In the SRrot protocols, the acquired images are simulated rotated around the phase-encoding axis. The rotation angles are uniformly distributed in the open interval [0, 180), with steps of 180/M°. In the SRsh protocols, the acquired images are simulated shifted in the through-plane direction. The shifts, expressed in HR voxel indices, are uniformly distributed in the closed interval $[-AF(M-1)/(2M), AF(M-1)/(2M)]$, with steps of AF/M.

| Protocols | AF | M | $\boldsymbol{\Phi} = \{\Phi_m\}_{m=1}^M$ |
|---|---|---|---|
| HR | 1 | 2 | [0, 0]° |
| SRrot$_1$ | 1 | 2 | [0, 90]° |
| SRrot$_2$ | 2 | 4 | [0, 45, 90, 135]° |
| SRrot$_3$ | 3 | 6 | [0, 30, 60, 90, 120, 150]° |
| SRrot$_4$ | 4 | 8 | [0, 22.5, 45, 67.5, 90, 112.5, 135, 157.5]° |
| SRsh$_1$ | 1 | 2 | [-0.25, 0.25] |
| SRsh$_2$ | 2 | 4 | [-0.75, -0.25, 0.25, 0.75] |
| SRsh$_3$ | 3 | 6 | [-1.25, -0.75, -0.25, 0.25, 0.75, 1.25] |
| SRsh$_4$ | 4 | 8 | [-1.75, -1.25, -0.75, -0.25, 0.25, 0.75, 1.25, 1.75] |

**Table 1:** MS-SRR acquisition protocols.

## 2.7 Prior hyperparameters estimation

The translational symmetry of the acquisition strategies along the phase-encoding axis was exploited to evaluate the framework in 2D, thereby reducing computational complexity and memory consumption. In order to estimate the prior hyperparameters, a training dataset was generated. The dataset, composed of 500 synthetic noiseless HR 2D T1-weighted (T1-w) magnitude brain images of size 120 × 120, was simulated starting from 10 anatomical brain models available in the Brainweb database [14]. The images, each representing an independent realization of **r**, were simulated with different acquisition planes (sagittal, transverse, and coronal) and T1 contrast. Additionally, each image was slightly rotated to simulate different head orientations, where the rotation angles were independently sampled from a Gaussian distribution with mean 0 and standard deviation 1. The hyperparameters $\boldsymbol{\alpha}$ and $\lambda$ were estimated from the images within the training dataset using the kernel-regression approach proposed in [15] from the non-zero voxels within the training dataset and their respective 3 × 3 neighborhoods. All the elements of the prior mean $\bar{\mathbf{r}}$ were set equal to the mean intensity of the non-zero voxels within the training dataset. The choice of setting the prior mean of all voxels equal to the same constant ensures the prior to be invariant to the positioning (translation) of the head within the field of view.

## 2.8 Protocols comparison

We assumed the images acquired with the HR protocol to have an SNR = 20, where the SNR was defined as the ratio of the mean intensity of the brain voxels within the training dataset to the standard deviation of the noise. The thus obtained standard deviation of the noise was fixed for all protocols, resulting in an SNR that increases with AF, as more signal is received from thicker slices. The BMSE as well as its separate squared bias and variance components were computed for each acquisition protocol using the closed-form expressions derived in the subsection 2.5.

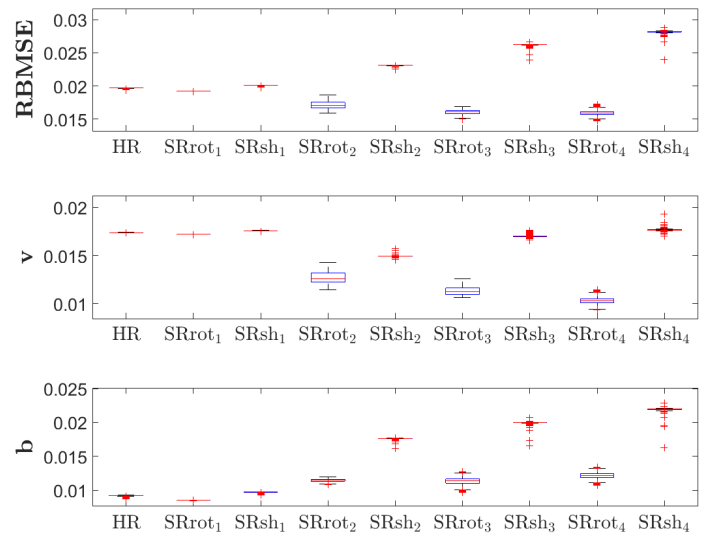

**Figure 1:** Boxplots of the **RBMSE** and of the BMSE standard deviation component **v** and absolute bias component **b** computed inside a ROI.

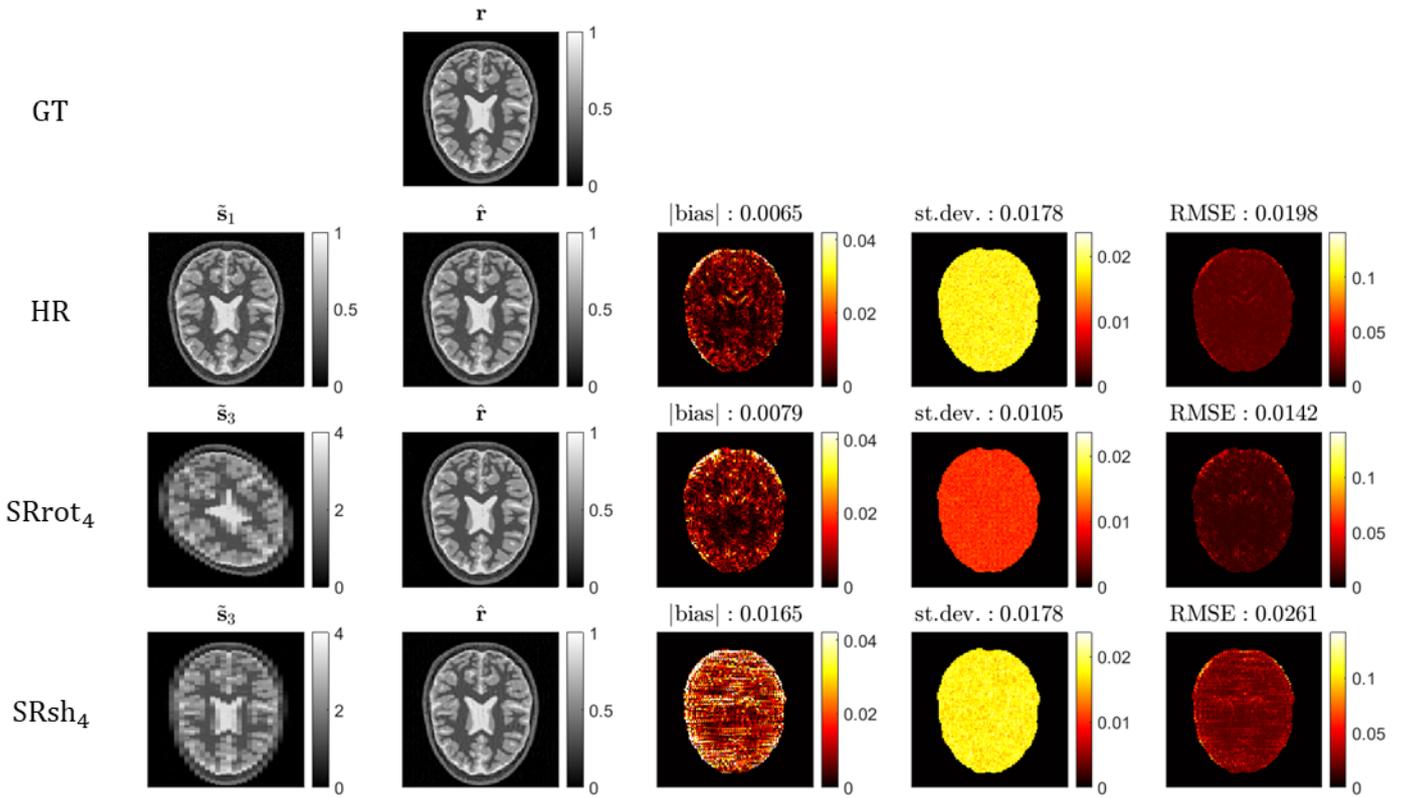

**Figure 2:** Monte Carlo simulation results for the protocols HR, SRrot4 and SRsh4. $\tilde{\mathbf{s}}_n$, $\mathbf{r}$, $\hat{\mathbf{r}}$ represent the $n$-th acquired LR image, ground truth image and estimated SRR image, respectively. The mean absolute bias, the mean standard deviation, and the mean RMSE values computed inside the brain mask are reported.

## 2.9   Monte Carlo simulation

A 2D T1-w axial brain slice, initially excluded from the training dataset, was used as ground truth (GT). The acquisition process was simulated for the protocols HR, SRrot$_4$ and SRsh$_4$ using the MS-SRR forward model in Eq. (1), and the simulated images were corrupted with noise, as described in the previous subsection. The Conjugate Gradient method [16] was used to solve the minimization problem in Eq. (7). Absolute bias, standard deviation, and RMSE maps were computed from 100 noise realizations for each protocol.

## 3   Results and Discussion

The distributions of the RBMSE maps inside a region of interest (ROI) for all the acquisition protocols are reported in Fig. 1, where the ROI was defined as the part of the field of view common to all the acquired images of all the acquisition protocols. The SR protocols based on the rotated acquisition scheme SRrot showed lower RBMSE values compared to the HR protocol and the SR protocols based on the shifted acquisition scheme SRsh. Increasing the AF led to an RBMSE improvement for the SRrot protocols, while the SRsh protocols showed the opposite trend. This difference is caused by both the absolute bias component $\mathbf{b}$, which increases severely with AF for the SRsh protocols, and the standard deviation component $\mathbf{v}$, which reduces significantly with AF for the SRrot protocols. The results of the Monte Carlo simulation for the HR, SRrot$_4$ and SRsh$_4$ protocols are reported in Fig. 2. The close agreement between the Monte Carlo results and the BMSE results demonstrates that the prior distribution was able to describe the statistics of the target image. The observed difference in terms of RBMSE between the SRrot and SRsh protocols (up to a factor 2, approximately) suggests that adopting the rotated acquisition scheme over the shifted scheme in a MS-SRR experiment can lead to a substantially reduced scan time while preserving the same MSE of the estimated SRR image. The main limitations of this work are the assumptions that the image registration parameters and the point spread function of the MRI acquisition process are perfectly known. The effect on the current analysis of nonidealities, such as motion artifacts and inconsistent modelling of the slice profile, will be the subject of future work. Additionally, real data experiments will be included to validate the proposed framework. Furthermore, we plan to extend the current study by applying the optimal experimental design theory principles to find the optimal acquisition settings for an MS-SRR experiment in terms of BMSE.

## 4   Conclusion

The potential of the BMSE framework for optimal experiment design was demonstrated by comparing two conventionally adopted MS-SRR acquisition protocols. The MS-acquisition strategy based on rotated multi-slice images out-performed the strategy based on shifted images in terms of estimation accuracy and precision, evaluated by the squared bias and variance terms of the BMSE of the MAP estimator, respectively. The results confirmed and extended the conclusion of [4, 5] and [6] to regularized MS-SRR.

## 5 Acknowledgments

The project B-Q MINDED has received funding from the European Union's Horizon 2020 research and innovation programme under the Marie Sklodowska-Curie grant agreement No 764513. B. J. is a postdoctoral fellow of FWO Vlaanderen.

# Status update on the Synergistic Image Reconstruction Framework: version 3.0

Richard Brown[1], Christoph Kolbitsch[2,3], Evgueni Ovtchinnikov[4], Johannes Mayer[2], Ashley G. Gillman[5], Edoardo Pasca[4], Claire Delplancke[6], Evangelos Papoutsellis[4,7], Gemma Fardell[4], Radhouene Neji[3,8], Casper da Costa-Luis[3,9], Jamie McClelland[9], Bjoern Eiben[9], Matthias J. Ehrhardt[6,10], and Kris Thielemans[1]

[1]Institute of Nuclear Medicine, University College London, London, UK
[2]Physikalisch-Technische Bundesanstalt, Braunschweig and Berlin, Germany
[3]School of Biomedical Engineering and Imaging Sciences, King's College London, London, UK
[4]Scientific Computing Department, STFC, UKRI, Rutherford Appleton Laboratory, Harwell Campus, Didcot, UK
[5]Australian e-Health Research Centre, Commonwealth Scientific and Industrial Research Organisation, Townsville, Australia
[6]Department of Mathematical Sciences, University of Bath, UK
[7]Henry Royce Institute, Department of Materials, The University of Manchester, Manchester, UK
[8]MR Research Collaborations, Siemens Healthcare, Frimley, UK
[9]Centre for Medical Image Computing, Radiotherapy Image Computing Group, Department of Medical Physics and Biomedical Engineering, University College London, UK
[10]Institute for Mathematical Innovation, University of Bath, UK

**Abstract** The Synergistic Image Reconstruction Framework (SIRF) is a research tool for reconstructing data from multiple imaging modalities, currently most prominently PET and MR. Included are acquisition models, reconstruction algorithms, registration tools, and regularisation models. In this work, we demonstrate the capabilities added since SIRF 2.0. PET/MR cardiac imaging results are presented with estimation of respiratory motion from the MR data, and motion compensation combined with various regularisation strategies used for both MR and PET reconstruction. The use of SIRF to facilitate this work enabled a range of techniques to be compared quickly and efficiently.

## 1 Introduction

Current trends in medical imaging continue to focus on the increased use of multiple modalities for imaging. Different properties of each modality can be combined together to complement each other and increase diagnostic power. One prominent example is simultaneous positron emission tomography (PET) and magnetic resonance (MR), where the speed and resolution of MR is able to improve upon the limitations of PET imaging and provide quantitative functional imaging with reduced imaging times and improved resolution.

As such, there is considerable interest in the development and refining of algorithms to share information between the previously independent images. This can be done subsequent to image reconstruction [1], or preferably by combining the modalities during the reconstruction process itself [2]. However, this is only feasible when used in combination with motion estimation and correction strategies to prevent misalignment. Research into such techniques requires considerable software infrastructure for reading and converting data, modelling acquisitions, reconstructing images, registration, etc. Medical imaging hardware vendors do often provide such infrastructure, however, it is often cumbersome or impossible to modify the internal components of these software required for such research. The purpose of the Synergistic Image Reconstruction Framework (SIRF) is to provide an open source software (OSS) tool to facilitate investigation into such algorithms.

Other OSS packages for image reconstruction are available and include: Gadgetron [3, 4] and the Berkeley Advanced Reconstruction Toolbox (BART) [5], which reconstruct MR data; the Software for Tomographic Image Reconstruction (STIR) [6], NiftyPET [7] and Customizable and Advanced Software for Tomographic Reconstruction (CASToR) [8] which have varying support for PET, SPECT and CT; and the Reconstruction Toolkit (RTK) [9], with CBCT, CT and in the future SPECT support. However, none of these packages support a diverse range of modalities, specifically combining MR and tomographic imaging. We are therefore developing SIRF [10–12] to address this gap.

SIRF development is led by the Collaborative Computational Platform on Synergistic Reconstruction for Biomedical Imaging CCP SyneRBI www.ccpsynerbi.ac.uk. SIRF uses several of the above mentioned packages as "engines" and integrates them into a consistent framework. The software includes documentation on exporting scanner data; functionality for converting and reading the data from supported hardware; modules for reading and writing acquisition data and images; acquisition models and reconstruction algorithms able to reconstruct images from acquisition data; models for regularising image reconstructions, some of which are able to model synergism between the modalities; and data processing tools for registering images to account for gantry shifts and patient motion. SIRF integrates with another OSS called the Core Imaging Library (CIL) [13, 14], which provides advanced optimisation and regularisation methods.

In this work, we demonstrate the currently implemented motion estimation and compensation strategies, together with examples of regularisation models in a cardiac PET/MR application.

Please note that since the submission of the conference abstract, SIRF 3.1 has been released [15]. In addition, some of the results in these proceedings were published recently [11] as part of a Special Issue on Synergistic Image Reconstruction [16, 17].

## 2 Methods and results

To be able to do motion correction, the data are split into several motion states, usually called "gates". There are numerous techniques for performing the motion correction, see a recent review on strategies for PET-MR [18]. The most common methods are the reconstruct-transform-add (RTA) scheme [19, 20], in which correction is performed after reconstruction, and the motion-compensated image reconstruction (MCIR) scheme [21, 22], in which the motion is incorporated into the acquisition model, one for each gate. Both of schemes need the motion to be known. One common way to determine the required motion information is to reconstruct motion resolved images (i.e., one for each gate) and then estimate the spatial transformation between the gates using image registration [19, 23, 24].

In the following, we present an example of the above-described framework using an *in vivo* cardiac scan. A simultaneous PET/MR scan was performed on a patient 182 min after the injection of 341 MBq $^{18}$F-FDG. Data was acquired for 3:18 min during free-breathing.

### 2.1 Respiratory motion estimation and correction for cardiac MR

In this section, a demonstration is given of the estimation of respiratory motion from a 3D non-Cartesian MR scan. The motion information is then used in an MCIR to improve the MR image quality. A new acquisition model was combined with the iterative reconstruction schemes available in CIL to ensure high image quality, even for highly undersampled data. 3D non-rigid motion fields are obtained using spline-based image registration and then applied during image reconstruction to minimise respiratory motion artifacts.

#### 2.1.1 Golden Radial Phase Encoding

Non-Cartesian MR sampling schemes are of great interest for motion-estimation and motion-correction. Even if the data are separated retrospectively into different motion gates (e.g., different phases of the breathing cycle), the $k$-space data are still well distributed in $k$-space covering both high and low spatial frequencies. In addition, high image quality can be achieved even from very few acquired $k$-space points (i.e., high undersampling) utilising iterative image reconstruction schemes. Here, a golden radial phase encoding (GRPE) sampling scheme was used [25, 26]. This is a 3D acquisition scheme which combines Cartesian frequency encoding (i.e. along $k_x$) with non-Cartesian sampling in the 2D phase-encoding plane $k_y - k_z$. The MR acquisition used here was a three-point Dixon scan (echo times: 1.2, 2.7 and 4.2 ms) with a field-of-view of $400 \times 400 \times 400$ mm and a spatial resolution of 1.9 mm along foot-head and $3.2 \times 3.2$ mm in the transverse plane. In the following, only the first echo was used.

SIRF was extended to use the non-uniform fast Fourier transform (NUFFT) which allowed for the transformation between Cartesian image data and non-Cartesian $k$-space data.

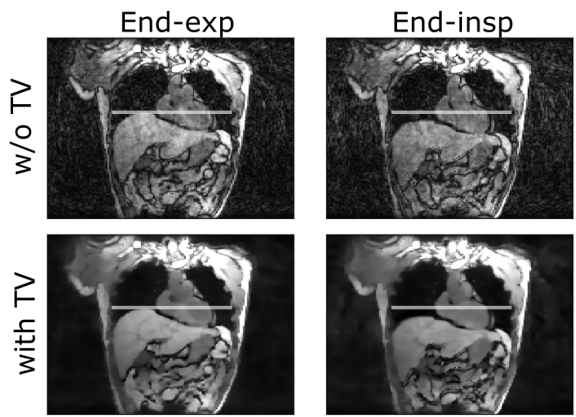

**Figure 1:** End-expiratory (end-exp) and end-inspiratory (end-insp) gate reconstructed without and with total variation (TV) regularisation. The horizontal line represents the superior-most diaphragm position in the reference gate, end-expiration.

#### 2.1.2 Self-gating and Reconstruction of respiratory gates

For the GRPE sampling scheme, the central ($k_y = k_z = 0$) $k_x$-line is acquired repeatedly. This allows for the extraction of a self-navigator signal [27, 28]. Each gate was then reconstructed using the implementation of fast iterative shrinkage-thresholding algorithm (FISTA) [29] in CIL with spatial TV regularisation [30].

Fig. 1 shows the end-expiration (which was later used as reference for the MCIR) and the end-inspiration gates, comparing both reconstruction algorithms. Changes in the anatomy during the breathing cycle mainly along the foot-head direction are clearly visible. The TV regularisation leads to suppression of undersampling artifacts and an improved depiction of the anatomy, which is beneficial for the next step.

#### 2.1.3 Estimation of respiratory motion fields

A non-rigid image registration scheme was then used to calculate the 3D respiratory motion fields from the respiratory gates. Motion deformation fields were estimated using a pairwise image registration, using the SIRF wrapper to the NiftyReg spline-based registration algorithm [31].

#### 2.1.4 Motion-corrected MR image reconstruction

The MCIR optimisation problem was solved with FISTA. Fig. 2 shows the final MCIR images reconstructed with FISTA with regularisation. MCIR leads to a clear reduction of respiratory motion artifacts (e.g., blurring of anatomical structures such as the liver and the heart. TV further improves image quality by minimising residual undersampling artifacts while ensuring a clear depiction of the anatomy.

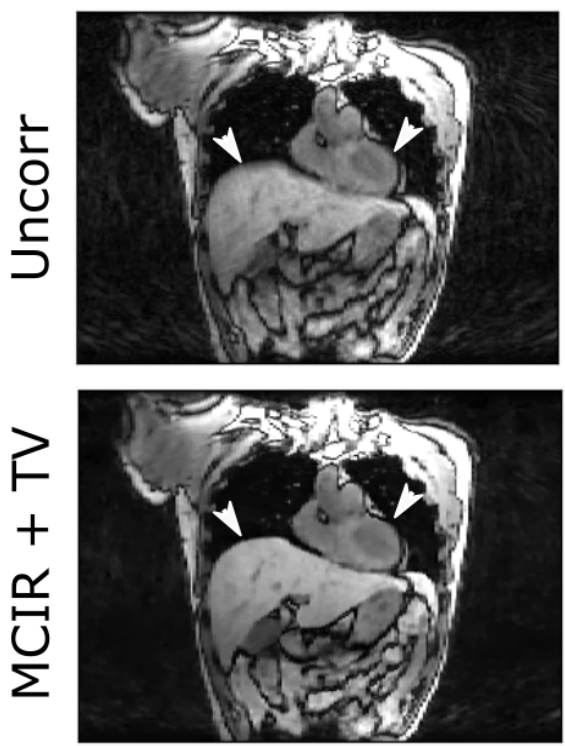

**Figure 2:** Uncorr: image reconstruction without motion correction with blurring due to respiratory motion clearly visible (white arrow heads). MCIR+TV: MCIR with TV regularisation. MCIR leads to a clear reduction of motion blurring and improves the visualisation of the anatomy. TV reduces undersampling artifacts and further improves image quality.

### 2.1.5 Motion-corrected PET image reconstruction

The motion fields from the previous section were used to reconstruct a motion-corrected PET image. We first estimated a coordinate transformation between the PET and MR images to cope with, for instance, gantry misalignment by performing a rigid registration between simultaneous MR and PET images reconstructed without attenuation correction (AC).

The GRPE acquisition was used for the separation of fat and water tissue and the calculation of a segmentation-based AC map [32] in the reference position. The construction of the MR-based AC map was not carried out in SIRF as it required segmentation tools not yet implemented in SIRF. The AC map was then deformed to each of the gates. An average AC map was computed for the ungated data. Randoms and scatter were computed from the ungated data and evenly divided over the gates.

Data were then reconstructed as follows: a single iteration of OSEM (24 subsets) [33] was used for initialisation of relaxed OSSPS (90 iterations, 7 subsets) [34] with resolution modelling and a quadratic Gibbs prior. Local weights were used

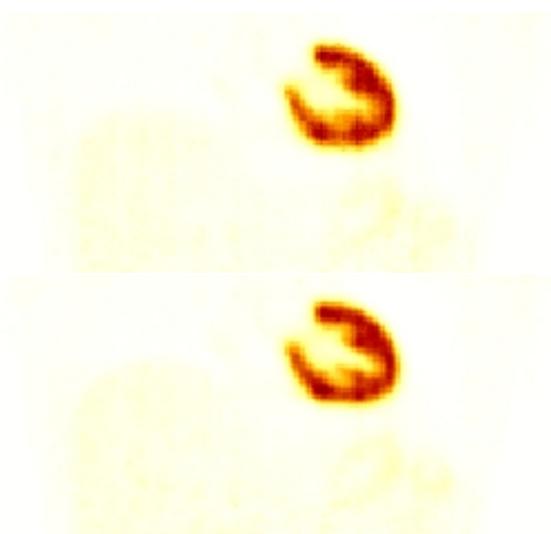

**Figure 3:** Comparison of (relaxed) OSSPS reconstructions without motion correction (top) and with gating and RTA (bottom). Both reconstructions after 420 updates with regularisation strength $\alpha = 0.0005$.

in the prior to obtain approximately uniform resolution [35]. Two example reconstructions are shown:

- no motion correction, i.e., using the ungated data

- RTA, where each gate was reconstructed separately, and resulting images were warped back to the reference position using the MR-derived deformation fields and then averaged.

### 3 Discussion and Outlook

We have presented recent improvements of SIRF, concentrating on motion correction and its integration with CIL for regularised reconstruction. Respiratory gates were reconstructed from a non-Cartesian 3D MR, and non-rigid respiratory motion fields were obtained using the NiftyReg integration in SIRF. These motion fields were then used for motion-compensation of both MR and PET.

We used MCIR for the MR reconstruction, while the presented example for PET reconstruction used RTA. However, RTA is known to have limitations due to count statistics of the gated data [36]. Please refer to [11] for an example of MCIR for PET with SIRF.

We intend to continue to develop SIRF for researchers to be able to exploit synergy in multi-modal, multi-contrast, multi-time point information for a greater range of applications. We welcome contributions via `https://github.com/SyneRBI/SIRF`.

### Acknowledgements

This work was funded by the UK EPSRC grants "Computational Collaborative Project in Synergistic PET/MR Reconstruction" (CCP PETMR) EP/M022587/1 and its associated Software Flagship project EP/P022200/1; the "Computational Collaborative Project in Synergistic Reconstruction for Biomedical Imaging" (CCP SyneRBI) EP/T026693/1; "A Reconstruction Toolkit for Multichannel CT" EP/P02226X/1 and "Collaborative Computational Project in tomographic imaging" (CCPi) EP/M022498/1 and EP/T026677/1; "PET++: Improving Localisation, Diagnosis and Quantification in Clinical and Medical PET Imaging with Randomised Optimisation" EP/S026045/1. This work made use of computational support by CoSeC, the Computational Science Centre for Research Communities, through CCP SyneRBI and CCPi.

We would like to thank Simon Arridge, David Atkinson, Julian Matthews, Andrew Reader, Steven Sourbron, Charalampos Tsoumpas and Martyn Winn for co-organising the CCP PETMR/SyneRBI network, its community for feedback, and Jakob S. Jørgensen and other members of CCPi for interactions on the software and algorithms. Computing resources were provided by STFC Scientific Computing Department's SCARF cluster and the STFC Cloud.

# DIRA-3D-wFBP - a Model-based Dual-Energy Iterative Algorithm for accurate dual-energy dual-source 3D helical CT

Maria Magnusson[1,2,3], Markus Tuvesson[1], Åsa Carlsson Tedgren[2,3,4], and Alexandr Malusek[2,3]

[1]Department of Electrical Engineering, Linköping University, Linköping, Sweden
[2]Department of Health, Medicine and Caring Sciences, Linköping University, Linköping, Sweden
[3]Center for Medical Image Science and Visualization (CMIV), Linköping University, Linköping, Sweden
[4]Department of Medical Radiation Physics and Nuclear Medicine, Karolinska University Hospital, Stockholm, Sweden.

**Abstract** Quantitative dual-energy computed tomography may improve the accuracy of treatment planning in radiation therapy. Of special interest are algorithms that can estimate material composition of the imaged object. One example of such an algorithm is the 2D algorithm DIRA. The aim of this work is to extend this algorithm to 3D so that it can be used with cone-beam helical scanning in multi-slice spiral CT. In the new algorithm, the parallel FBP method was replaced with the weighted FBP method from Siemens. We used the implementation available from the FreeCT project. Its performance was tested using a mathematical phantom consisting of six ellipsoids. The algorithm substantially reduced the beam-hardening artefacts, and simultaneously artefacts due to approximate reconstruction, after ten iterations for bone (and even earlier for soft tissues). Compared to Alvarez-Macovski's base material decomposition, DIRA-3D-wFBP does not require geometrically consistent projections and hence can be used in dual-source CT scanners. Also, it can use several tissue-specific material bases at the same time to represent the imaged object.

## 1 Introduction

Dual-energy computed tomography (DECT) may improve the accuracy of radiation treatment planning [1]. The field may especially benefit from algorithms providing quantitative information about CT numbers or material composition of the imaged object. The latter is provided by projection-based basis material decomposition (PBBMD) and image-based basis material decomposition (IBBMD) methods [2]. PBBMD methods, such as the well-known Alvarez-Macovski's base material decomposition [3], require geometrically consistent projections and only two base materials can be assigned for the whole object. IBBMD methods can use any number of tissue-specific material doublets or triplets, nevertheless the methods suffer from beam hardening artefacts caused by the polyenergetic projection data. Better results are achieved by applying Alvarez-Macovski's base material decomposition before IBBMD [4] or by using model-based iterative reconstruction algorithms, like MDIR [5] or DIRA [6] developed by the authors. The original DIRA algorithm is based in 2D projections and 2D FBP. It uses two- and three-material decomposition to tissue-specific, user-defined base material doublets and triplets for the characterization of the imaged object. The output from DIRA are several base material images as well as two monoenergetic images. If desired, it is possible to generate any monoenergy image from the base material images.

Clinical scanners shorten acquisition times by using helical

(spiral) scanning and multi-row detectors. Reconstruction is performed by FBP or iterative algorithms, where backprojection is part of the inner loop algorithm [7]. Examples of FBP algorithms for helical scanning are Siemens' weighted FBP (wFBP) [8] and the PI-method [9], which are approximate only, and Katsevich's FBP [10], which is exact. In [11] and [12], a 3D version of DIRA was presented, where the 2D FBP was replaced with the PI-method. Both the PI and Katsevich's methods discard projection data outside the Tam window, which makes them less favorable in clinical environment compared to the wFBP method.

The aim of this paper is to develop a DIRA algorithm for 3D helical geometry that uses wFBP. More detailed information about DIRA-3D-wFBP is in [13].

## 2 Theory

### 2.1 Material Decomposition

Two-material decomposition assumes that a mixture is composed of two materials. This decomposition determines mass fractions, $w_1$ and $w_2$, of the two base materials and the mass density, $\rho$, of the mixture. Three-material decomposition assumes that a mixture consists of three base materials. This decomposition determines mass fractions, $w_1$, $w_2$ and $w_3$, and the mass density of the mixture, which is calculated as $\rho^{-1} = \sum_{k=1}^{3} w_k/\rho_k$, where $\rho_k$ and $w_k$ are the mass density and mass fraction, respectively, of the $k$th material. In both methods, the mass fractions are normalized so that $\sum_k w_k = 1$. More information on the resulting systems of linear equations is in [6].

### 2.2 Forward projection generation

The logarithm of attenuation, here referred to as the polyenergetic projection $P$, is calculated as

$$P = \ln \frac{I_0}{I}, \quad (1)$$

where $I$ and $I_0$ are the detector responses with and without, respectively, the imaged object. The intensity $I_0$, is calculated for an ideal energy integrating detector as

$$I_0 = \int_0^{E_{max}} EN(E)\,dE, \quad (2)$$

where $E$ is the photon energy and $N(E)$ is the energy spectrum of photons emitted from the x-ray tube. The intensity $I$ is calculated as

$$I = \int_0^{E_{max}} EN(E) \exp\left[-\int_L \mu(x,y,z,E)dl\right] dE, \quad (3)$$

where $\mu(x,y,z,E)$ is the linear attenuation coefficient (LAC) of pixel $(x,y,z)$ at energy $E$ and $\int_L dl$ is a line integral through the object. This calculation is time consuming since the line integrals must be calculated for all energies in the energy spectrum. The calculation of projections change slightly when material decomposition is introduced. The line integrals are calculated through volume fractions of the different base materials. The intensity $I$ is then

$$I = \int_0^{E_{max}} EN(E) \exp\left[-\sum_k \mu_k(E)l_k\right] dE, \quad (4)$$

where $\mu_k$ is the LAC of the $k$th base material and $l_k$ is computed as

$$l_k = \rho_k^{-1} \int_L \rho(x,y,z)w_k(x,y,z)dl, \quad (5)$$

where $\rho_k$ is the tabulated density of the $k$th base material, $\rho(x,y,z)$ is the calculated density in voxel $(x,y,z)$ and $w_k(x,y,z)$ is the mass fraction of the $k$th material in voxel $(x,y,z)$. The density $\rho$ and the mass fractions $w_k$ are obtained from the two-material or three-material decomposition.

A monoenergetic projection $P_{E_i}$, where $E_i$, $i = 1,2$ is a specific energy, is calculated as

$$P_{E_i} = \sum_k \mu_k(E_i)l_k. \quad (6)$$

The line integral $\int_L dl$ can be calculated using a 3D version of Joseph's method [14].

## 2.3 The Dual-source helical CT geometry

Fig. 1 represents the geometry of a dual-source helical CT-scanner. In the figure, $\kappa$ is the cone-angle, $\gamma$ is the fan-angle and $P$ represents the pitch of the helix. $\kappa_{max}$ is the maximum value of the cone angle and is called cone-beam angle. $\gamma_{max}$ is the maximum value of the fan angle and is called fan-beam angle. The $s$-axis is aligned with the $z$-axis and starts at the cross-section $C$ between the central ray and the $z$-axis. The blue and red spherical markers represents the X-ray sources. The detectors are cylindrical, with axes parallel to the $z$-axis, and they travel in unison with their corresponding X-ray source in a helical trajectory around the $z$-axis. Projection data from this geometry can be reconstructed with weighted filtered backprojection (wFBP) [8]. The wFBP-method is not an exact method and for large cone-beam angles, small artefacts become visible in the reconstructed images.

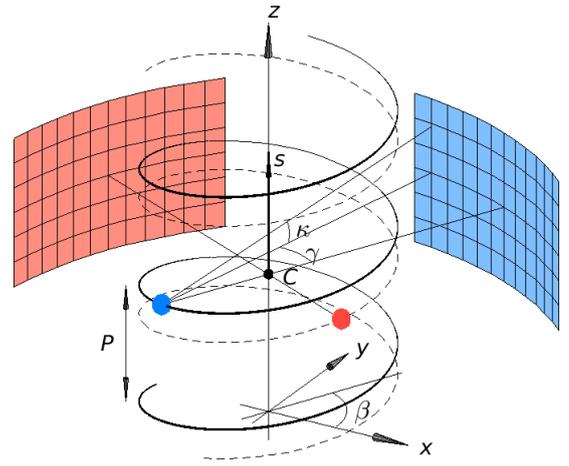

**Figure 1:** Dual-source helical CT geometry. Image source: [12]

## 2.4 DIRA-3D-wFBP

DIRA-3D-wFBP is an extension of the 2D algorithm DIRA presented in [6] and a modification of DIRA-3D presented in [11] and [12]. The algorithm is illustrated in Fig. 2 and performs the following steps:

1. Obtain helical cone-beam measured polyenergetic projections, $P_{M,U_i}$, for two different tube voltages, $U_i$, $i = 1,2$, giving $P_{M,U_1}$ and $P_{M,U_2}$, see Fig. 1.

2. Use wFBP to reconstruct from these projections so that the reconstructed LAC at energy $E_i$ is $\mu_i = \Delta\mu_i + \mu_{P_i}$, where $\mu_{P_i}$ is the reconstructed LAC from monoenergetic calculated parallel projections at energy $E_i$ and $\Delta\mu_i$ is the result of reconstructing from $P_{M,U_i} - P_{U_i}$, where $P_{U_i}$ are the calculated cone-beam projections. For the first iteration, $P_{U_i} = P_{E_i} = 0$ and thus $\mu_i$ is the reconstruction from the $P_{M,U_i}$ only.

3. Perform automatic threshold segmentation on $\mu_1$ and $\mu_2$.

4. Classify tissues using the material decomposition methods (section 2.1).

5. Calculate polyenergetic projections $P_{U_i}$ for cone-beam geometry, see equations (1), (2) and (4).

6. Calculate monoenergetic projections $P_{E_i}$ for parallel geometry, see equation (6).

Points 2-6 are repeated a predefined number of times. To formulate DIRA in mathematical terms, set

$$\boldsymbol{\mu} = \begin{pmatrix} \mu_1 \\ \mu_2 \end{pmatrix}, \quad \mathbf{P_{M,U}} = \begin{pmatrix} P_{M,U_1} \\ P_{M,U_2} \end{pmatrix}, \quad (7)$$

for the reconstructed images and the measured projections, respectively. Furthermore, denote the filtered backprojection operators $\mathscr{B}_C$ and $\mathscr{B}_{\text{II}}$; the former represents wFBP working

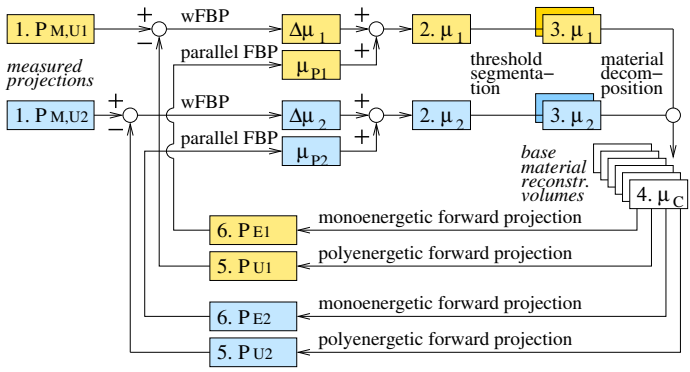

**Figure 2:** A flowchart of the DIRA-3D-wFBP algorithm.

with cone-beam projections, the latter represents ordinary parallel FBP. Set

$$\mathscr{P}_U = \begin{pmatrix} \mathscr{P}_{U_1} \\ \mathscr{P}_{U_2} \end{pmatrix}, \quad \mathscr{P}_E = \begin{pmatrix} \mathscr{P}_{E_1} \\ \mathscr{P}_{E_2} \end{pmatrix}, \tag{8}$$

for the projection operator for polyenergetic projections and monoenergetic projections, respectively. These projection operators include the automatic tissue segmentation and classification. The linear attenuation coefficient $\boldsymbol{\mu}^{(n+1)}$ obtained at the $(n+1)$th iteration is

$$\boldsymbol{\mu}^{(n+1)} = \mathscr{B}_C(\mathbf{P_{M,U}}) - \mathscr{B}_C \mathscr{P}_U(\boldsymbol{\mu}^{(n)}) + \mathscr{B}_{II} \mathscr{P}_E(\boldsymbol{\mu}^{(n)}). \tag{9}$$

Ideally, the calculated polyenergetic projections $\mathscr{P}_U(\boldsymbol{\mu}^{(n)})$ converge towards the measured projection $\mathbf{P_{M,U}}$. The $(n+1)$th iteration then gives $\boldsymbol{\mu}^{(n+1)} \approx \mathscr{B}_{II} \mathscr{P}_E(\boldsymbol{\mu}^{(n)})$, which is the filtered backprojection result of the monoenergetic projections. The generation of monoenergetic projections followed by backprojection in DIRA serves as a regularization [6].

## 3    Methods

The DIRA-3D-wFBP algorithm was implemented according to the description in sections 2.4 and 3.2. It was tested in the geometry described in section 3.1.

### 3.1    Mathematical phantom and projection geometry

The phantom consisted of six ellipsoids, see Fig. 3. Two ellipsoids consisting of protein and water had their centers located at slice 26. Four ellipsoids consisting of adipose tissue, lipid, protein and compact bone had their centers located at slice 10. The scanning geometry for the cone-beam projection generation is shown in figure 3. The parameters were: number of projection angles = 800 $(0-799)$, number of helix turns = 2, helical pitch = 32 voxels, fan-beam angle $\gamma_{max} = 25.13°$, cone-beam angle $\kappa_{max} = 6.26°$, detector size = $192 \times 64$ pixels, voxelsize $\Delta x = \Delta y = \Delta z = 2.76$ mm, reconstructed volume size $128 \times 128 \times 48$ voxels.

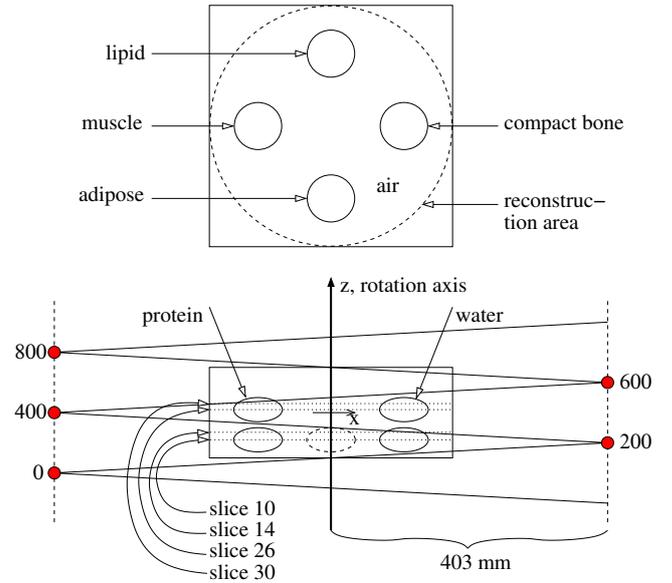

**Figure 3:** Top: Phantom slice at $z = 10$. Bottom: The scanning geometry of the semi-parallel projection generation through the phantom showing projection number 0, 200, 400, 600 and 800 and highlighting slice positions at $z = 10, 14, 26, 30$.

### 3.2    Implementation details

The program *take* [15] was used for the cone-beam projection generation. One of its functionalities is to calculate line integrals through voxel volumes using an extended version of Joseph's method [14]. The "measured" cone-beam projections were simulated using the line integrals in (4) with $l_k = \int_L m_k(x,y,z) dl$, where $m_k$ are the masks for the different ellipsoids of different materials. Energy spectra for the x-ray tube voltages of 80 and 140 kV with Sn filtration were used. The chosen mono-energies where $E_1 = 50.0$ keV and $E_2 = 88.5$ keV.

At each iteration, reconstructed volumes $\mu_1$ and $\mu_2$ were threshold segmentated into air, soft tissue and bone regions. Air was then decomposed into a (lipid, water) doublet, soft tissue was decomposed into a (lipid, protein, water) triplet and bone was decomposed into a (compact bone, bone marrow) doublet. As a consequence, the ellipsoid containing bone was decomposed into the (compact bone, bone marrow) doublet and the other ellipsoids were decomposed into the (lipid, protein, water) triplet.

The calculated cone-beam polyenergetic projections were computed using equations (1), (2), (4) and the calculated parallel monoenergetic projections were computed using (6). The generation of parallel projections with subsequent reconstruction was done slice by slice. The number of parallel detector elements was 128 and the number of projection angles was 400, with the angular interval $[0°, 180°)$.

A GPU-focused implementation of wFBP from the FreeCT Project was used [16].

### 3.3  Error calculation

The relative error of reconstructed LAC was estimated as $\delta(\bar{\mu}) = (\bar{\mu} - \mu_t)/\mu_t$, where $\mu_t$ is the tabulated value and $\bar{\mu}$ is the average of the calculated LAC in a spherical region of interest (ROI). The ROI was defined as a sphere with a radius of one third of the evaluated ellipsoid radius (in the $x, y$ plane) and positioned in the center of the evaluated ellipsoid.

## 4  Results

Fig. 4 shows how the reconstructed LAC for the different materials in slice 10 changes with iteration number for the energies $E_1 = 50.0$ keV and $E_2 = 88.5$ keV. The LAC for compact bone at low energy is the only LAC that shows a clearly notable improvement, changing from approximately $59.80 \, \text{m}^{-1}$ in iteration 1 to $79.14 \, \text{m}^{-1}$ in iteration 10, and to $79.30 \, \text{m}^{-1}$ in iteration 25. Notwithstanding, Fig. 5 shows that the relative error of the LAC in all the six different ellipsoids converged to a value close to zero already after 10 iterations (and even earlier for soft tissues).

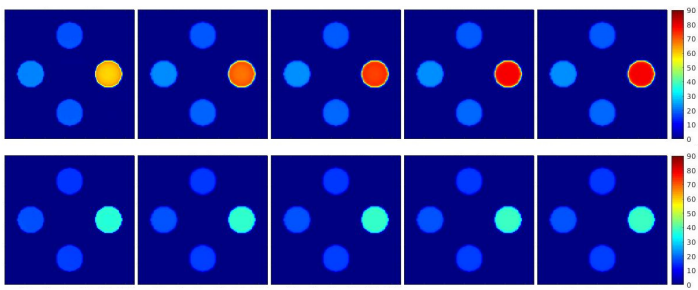

**Figure 4:** Reconstructed LAC images of the phantom for iterations 1, 2, 3, 10 and 25, slice 10, photon energies $E_1 = 50.0$ keV (top row) and $E_2 = 88.5$ keV (bottom row).

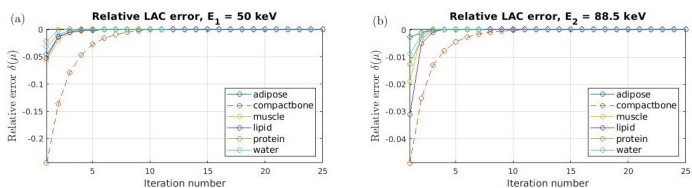

**Figure 5:** The relative errors of the LAC in the six different ellipsoids as a function of the number of iterations and photon energies (a) $E_1 = 50.0$ keV and (b) $E_2 = 88.5$ keV.

Fig. 6 shows the suppression of approximate-reconstruction artefacts for slice 14. The LAC was restricted to the interval $[-1, 1] \, \text{m}^{-1}$, since the approximate-reconstruction artefacts are mainly visible in the air region, where the LAC is close to zero. Note that the shape of the artefacts for low and high energies is different; the reason being that the X-ray sources are positioned $90°$ with respect to each other.

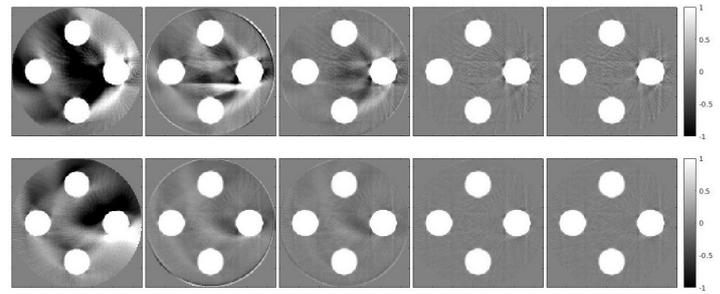

**Figure 6:** Suppression of approximate-reconstruction artefacts. Images of reconstructed LAC (in $\text{m}^{-1}$) in slice 14 for iterations 1, 2, 3, 10 and 25 and photon energies $E_1 = 50.0$ keV (top row) and $E_2 = 88.5$ keV (bottom row). The range of LAC values was restricted to the interval $[-1, 1] \, \text{m}^{-1}$.

## 5  Discussion and Conclusion

We have presented DIRA-3D-wFBP, a model-based iterative reconstruction algorithm for helical cone-beam scanning to be used in multislice spiral CT. The two dual energy X-ray sources were placed orthogonally and on different helices. DIRA-3D-wFBP performs material decomposition of the imaged object within the iterative loop, Weighted FBP (wFBP) is used for reconstruction within the iterative loop.

DIRA-3D-wFBP generated two monoenergetic image volumes, free from beam-hardening caused by the polyenergetic projection data. Also artefacts caused by the approximate reconstruction in wFBP were eliminated.

As mentioned in the introduction, DIRA also produces several base material images. They were not shown here, however. If desired, it is possible to generate any monoenergy image from the base material images.

The algorithm was evaluated using computer simulations with a simple phantom consisting of ellipsoids of different materials. More work is needed to test the stability of the algorithm in the presence of quantum noise. Also more complicated phantoms, as well as real data should be evaluated.

# Convolutional Encoder-Decoder Networks for Volumetric CT Scout Reconstructions from Single and Dual-View Radiographs

Siddharth Bharthulwar[1], Peter Noël[1], and Nadav Shapira[1]

[1]Department of Radiology, University of Pennsylvania School of Medicine, Philadelphia, PA, USA

**Abstract** Computed tomography (CT) is the most extensively used imaging modality capable of generating detailed images of a patient's anatomy for diagnostic and interventional procedures. Within CT research, radiation dosage reduction is a topic of importance, with recent advances in low-dose or limited-angle imaging. However, few studies have considered possibilities of 2D-to-3D stereo image reconstruction methods applied to CT reconstruction. In this study, we develop modified 2D-to-3D encoder-decoder neural network architectures to reconstruct CT volumes from single and dual-view radiographs for patient-specific planning and dose reductions through optimized tube voltage and current modulations. We then validate the developed neural networks on synthesized chest radiographs from a large-scale publicly available thoracic CT dataset. Finally, we assess the viability of the proposed transformational encoder-decoder architecture on both common image similarity metrics and quantitative clinical use case metrics, a first for 2D-to-3D CT reconstruction research. Results indicate the dual-input neural network shows promise improved scan planning, dose modulation, and other advanced scout/surview radiographic applications.

## 1 Introduction

X-ray computed tomography (CT) is an extensively used imaging modality that captures three-dimensional anatomical structures, primarily for clinical and research applications. Conventional CT scanners involve measuring individual radiographic projections in a circular or helical pattern, rotating around the entire body to produce detailed, high-resolution volumetric data which provide significant diagnostic advantages over two-dimensional radiographic modalities. Through state-of-the-art reconstruction techniques such as filtered back-projection (FBP) and iterative reconstruction (IR), modern CT scanners are able to produce high-resolution volumetric images of a patient's anatomy [1]. However, by measuring hundreds of radiographic projections at varying angles, full-dose CT scans entail considerable radiation exposure to patients, raising numerous health concerns. Prior research efforts have focused on reducing the radiation dosage required for volumetric CT scans [2]. Such efforts include reconstructing full-dose CT scans from low-dose CT scans or limited-angle scans [3]. Additionally, in recent years the appeal of deep learning and artificial intelligence in the medical imaging domain has inspired several studies investigating deep-learning-facilitated CT reconstruction. Notable studies involve transforming and denoising low-dose CT scans with convolutional autoencoders and convolutional neural network (CNN) facilitated limited-angle reconstruction [4, 5]. Within the specific subfield of 2D-to-3D CT reconstruction, Shen *et al* investigate deep learning methods for single-view radiographic projections [6]. Few studies investigate stereo 2D-to-3D CT reconstruction problems. Notably, Katsen *et al* develop a CNN to reconstruct three-dimensional knee bone segmentations from pairs of two-dimensional knee radiographs [7]. However, the challenge of reconstructing thoracic CT scans from two opposing projections has not been investigated before. In this study, we develop and implement single-view and dual-view encoder-decoder neural networks for few-view radiographic CT reconstructions. Then, we train both neural networks on synthetic x-ray/CT data pairs and determine clinical viability through both quantitative image similarity and use-case-specific metrics for exam planning and dose modulation.

## 2 Methods

### 2.1 Encoder-Decoder Architecture

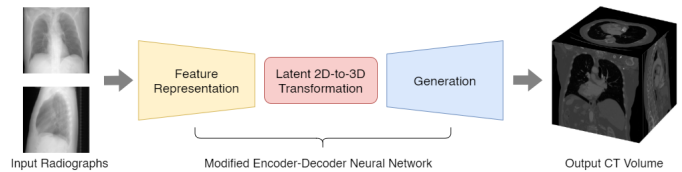

**Figure 1:** Schematic diagram of proposed deep learning framework for volumetric CT scout (surview) scan reconstruction

We consider the problem of volumetric CT reconstruction from single or dual 2D projection(s) as an image translation task with an additional transformation component to increase image dimensionality. Our formulated solution is defined as a modified encoder-decoder neural network architecture with convolutional layers. The larger neural networks consist of three subnetworks: one or two input representation networks, a feature transformation network, and a feature generation network. Considering coronal and sagittal radiographic projections $X_1$ and $X_2$, our goal is to generate an estimated output CT volume $Y_{\text{pred}}$. We define two deep learning mapping functions to encompass the task of volumetric reconstruction, $F_1$ and $F_2$, such that $F_1(X_1) = Y_{\text{pred}}$ and $F_2(X_1, X_2) = Y_{\text{pred}}$.

### 2.2 Representation Subnetwork

The representation subnetwork(s) are tasked with reducing the input 2D radiograph dimensionalities into smaller latent tensor(s). Considering input radiograph $X_1$ and output latent tensor $L_1$, the single-view representation subnetwork, abstracted as $F$, is applied such that $F(X_1) = L_1$. In the case of the dual-view neural network, two representation subnetworks with identical architectures process input radiographs in opposing views. Considering input radiographs $X_1$ and $X_2$ in the coronal and sagittal planes, respectively, and respective output tensors $L_1$ and $L_2$, the representation subnetworks $F_1$ and $F_2$ are applied such that $F_1(X_2) = L_2$, $F_1(X_2) = L_2$. Each arrow in the representation subnetworks (a) in Fig. 2, and Fig. 3 represent convolutional layers.

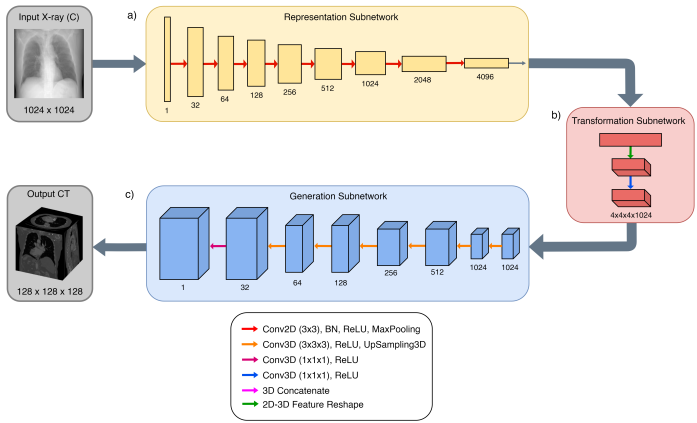

**Figure 2:** Architecture of single-input neural network based on Shen *et al*, featuring a representation subnetwork, a transformation subnetwork, and a generation subnetwork. The number of channels is indicated by the number below each hidden layer.

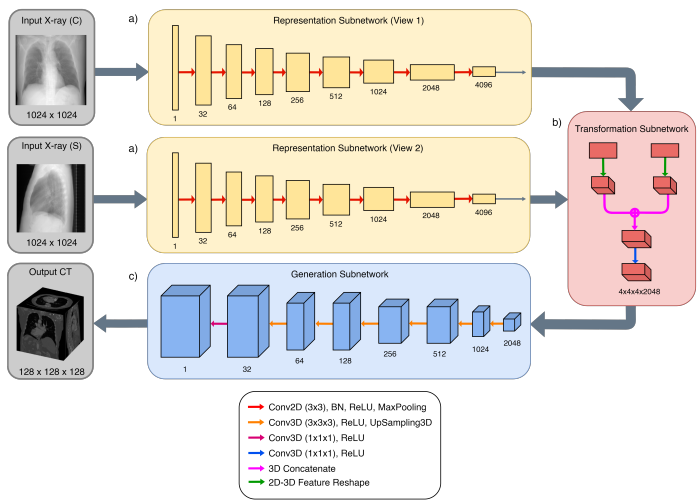

**Figure 3:** Architecture of novel dual-input neural network, featuring four distinct modules: two representation subnetworks, a bifurcated transformation subnetwork, and a generation subnetwork. The number of channels is indicated by the number below each hidden layer.

### 2.3    Transformation Subnetwork

The transformation subnetwork, denoted as (b) of Fig. 2 and Fig. 3, is tasked with combining the latent tensor representations of each radiographic projection and increasing dimensionality. Considering transformation subnetwork $T$, latent tensor $L$, and reshaped latent tensor $R$, the single-view subnetwork is invoked such that $T(L) = Z$. In the dual-view architecture, both previously generated latent tensors $L_1$ and $L_2$ are concatenated to form a higher-dimensionality latent tensor $Z$. Hence, we give the operation as $T(L_1, L_2) = Z$. In both variants of the transformation subnetwork, a single convolution with kernel size $1x1x1$ is invoked to learn the new reshaped spatial hierarchies.

### 2.4    Generation Subnetwork

The generation subnetwork is the final component of the developed set of encoder-decoder neural networks, and is tasked with enlarging the reshaped latent tensor into a final output volume. Considering reshaped latent tensor $Z$ and final output CT $Y_{\text{pred}}$, the generation subnetwork, abstracted as $G$, is invoked such that $G(Z) = Y_{\text{pred}}$. The data flow of the hidden convolutional layers in the generation subnetwork is given in (c) of Fig. 2 and Fig. 3.

### 2.5    Synthetic Paired Dataset

To train both neural networks, we synthesize a paired x-ray/CT training dataset from existing CT datasets. The Lung Image Data Consortium (LIDC) dataset totals 1050 helical thoracic CT scans in the DICOM format, compiled from seven academic centers and eight medical imaging companies. Each exam contains both volumetric pixel information as well as relevant scan parameters (slice thickness, tube current, etc) [8]. The synthetic radiograph creation process involves simulating beams passing through CT volumes from an x-ray point source. To calculate the intensities of each radiograph, the number and distance of intersections between traversed voxels of each ray must be calculated. Once the traversed voxels and their respective distances are calculated, the radiographs can be synthesized with the Beer-Lambert Law for the attenuation of light [9].

### 2.6    Training Details

To determine the accuracy of both single-view and dual-view architectures, both neural networks are trained on the synthesized x-ray/CT paired dataset. For both models, an initial learning rate of 0.0002 on the Adam optimizer is used to minimize the mean-squared-error (MSE) loss function via stochastic gradient descent and backpropagation. All training is conducted on two SLI-connected NVIDIA Tesla P100 GPUs, each with 16GB of VRAM. Model weights are saved locally every 10 epochs, and the training process automatically terminates after convergence. The weights with the lowest average loss are preserved and serialized.

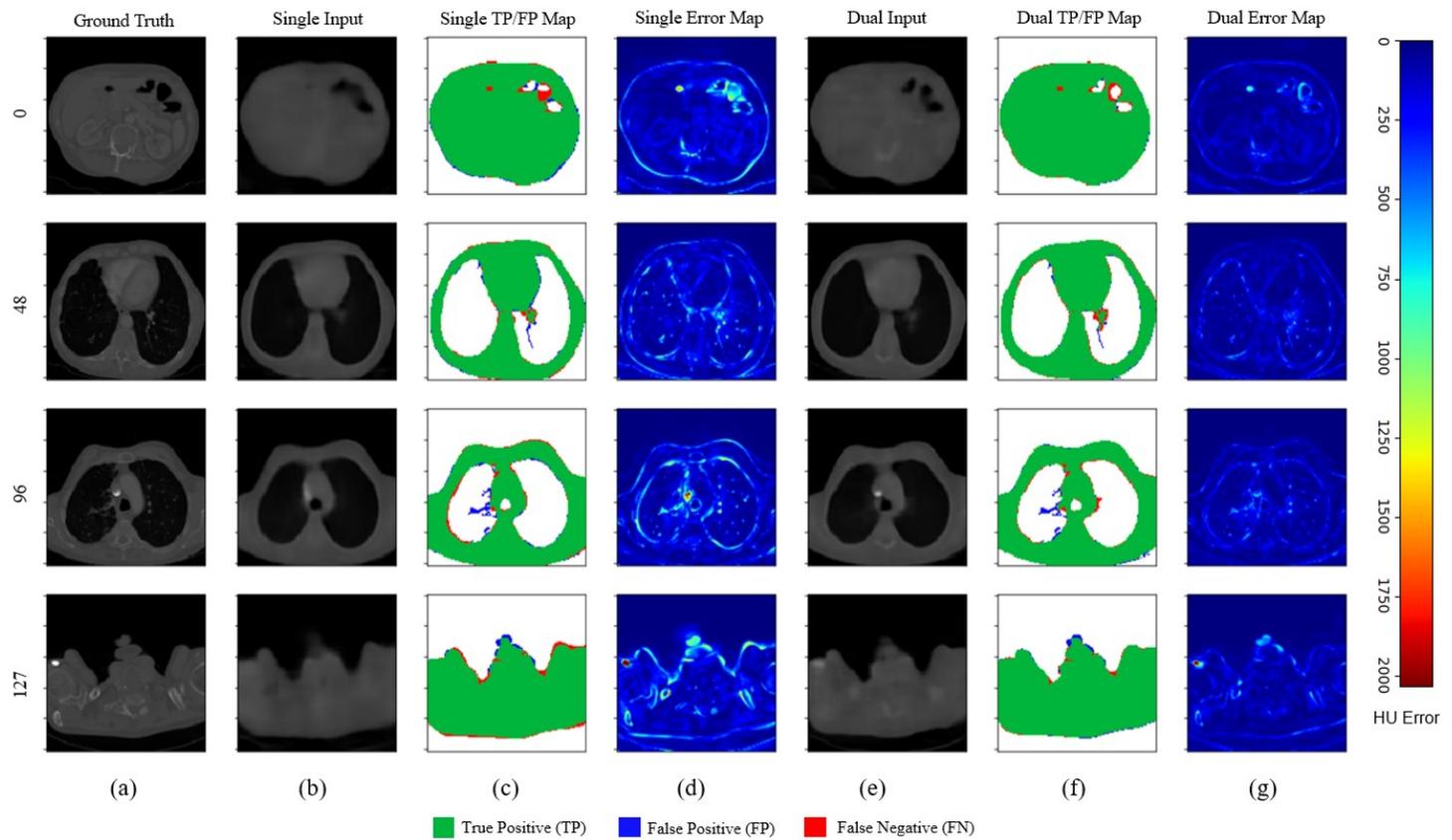

**Figure 4:** Reconstruction results viewed in axial, coronal, and sagittal planes. Displayed are (a) ground truth slices, (b) slices reconstructed from the single-input network, (c) true positive/false positive/false negative map for single-input reconstruction, (d) absolute error map for single-input reconstruction, (e) slices reconstructed from the dual-input network, (f) true positive/false positive/false negative map for dual-input reconstruction, and (g) absolute error map for dual-input reconstruction. The key for the true-positive/false-positive/false-negative maps (c, f) are shown at the bottom, along with a HU colormap for subfigures (d, g).

## 3    Results and Discussion

### 3.1    Image Similarity Metrics

For quantitative evaluation, four common image similarity metrics are calculated: mean-squared error (MSE), mean-average error (MAE), structural similarity (SSIM), peak signal-to-noise ratio (PSNR). Additionally, a new metric is developed to quantify anatomical similarity of lung and tissue segmentations in the resultant three-dimensional scout (3D-scouts) outputs. The DICE score is used to quantify the accuracy of the segmentations produced by the neural networks.

| Model | MAE | MSE | SSIM | PSNR | DICE |
|-------|-----|-----|------|------|------|
| Single | 0.0013 | 0.0012 | 0.8893 | 29.4531 | 0.9401 |
| Dual | 0.0004 | 0.0009 | 0.9396 | 31.5240 | 0.9676 |

**Table 1:** Image similarity metrics for single and dual view reconstructions

Reconstruction results from both neural networks are displayed in Fig. 4 and Table 1. The single-view neural network is capable of accurately reconstructing scout CTs from a single radiograph, as indicated by the error maps (d), true-positive (TP) /false-positive (FP) /false-negative (FN) maps (c), reconstructed slices (b), and MSE/MAE/SSIM/PSNR scores. The dual-view model demonstrates higher overall performance, as shown by the higher MSE, MAE, and SSIM scores and error maps with lower overall HU error intensities (g). However, as given by the nearly identical TP/FP/FN maps and DICE scores, both architectures have similar structural/spatial segmentation capabilities, with the dual-view model still outperforming the single-view model. Higher PSNR scores yielded by the dual-view neural network indicate increased effectiveness in denoising and artifact reduction. These observations indicate that the addition of an extra sagittal radiograph input improves structural/spatial accuracies as well as reducing artifacts.

### 3.2    Use-Case-Specific Metrics

Both neural networks are also validated on case-specific metrics for clinical applications. The applicability of both architectures to attenuation correction applications is assessed through radiodensity distribution similarity. Reconstructed volumes from both neural networks are normalized to a Hounsfield Unit (HU) scale and partitioned into histogram bins, allowing for a qualitative assessment of radiodensity similarity. The histograms of the two CNNs indicate con-siderable similarity but fail to model peaks/valleys of the ground truth distribution, as shown in Fig. 5. Moreover, smoother distributions generated by the neural networks, as compared to distinct ground truth distributions, indicate considerable uncertainty in estimating radiodensity distributions for diagnostic purposes.

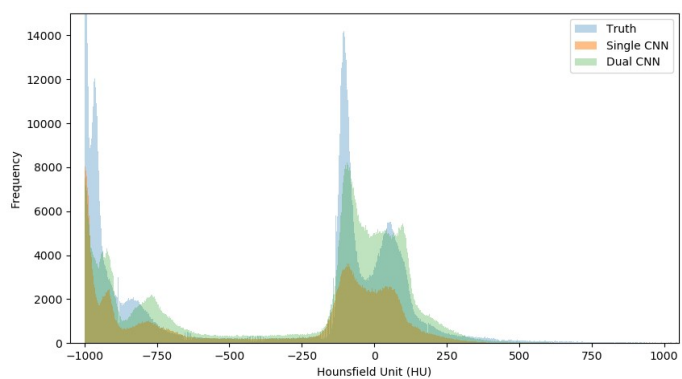

**Figure 5:** Histograms of Hounsfield Units (HU) of ground truth CT scan (blue), single-view reconstruction (red), and dual-view reconstruction (green).

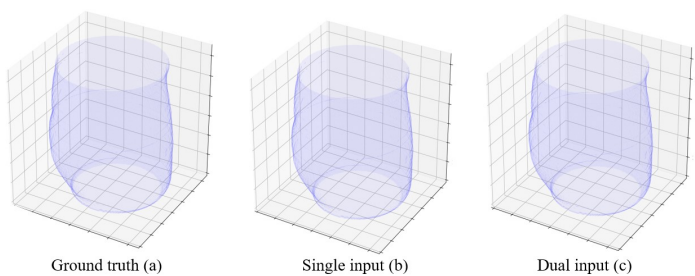

Ground truth (a)      Single input (b)      Dual input (c)

**Figure 6:** Volumetric external bodies and volumetric water equivalent areas ($A_w$) for ground truths and generated volumes. Displayed are meshes created from the water-equivalent-areas ($A_w$) of a ground truth volume (a), a volume generated by the single-view neural network (b), and a volume generated by the dual-view neural network (c).

Additionally, radiation dose modulation is often necessary in volumetric imaging applications to minimize radiation exposure while maintaining image quality [10]. We develop a methodology to determine dose modulation parameter calculation accuracy from reconstructions of both neural networks. From the methodology described in [11], we calculate water-equivalent area ($A_w$) and water-equivalent diameter ($D_w$). We calculate average water-equivalent-diameter accuracies for both neural networks, with the single-view architecture scoring 0.911 and the dual-view architecture scoring 0.925. Additionally, estimated patient shapes derived from $D_w$ (a, b, c) are displayed in Fig. 6. Note significant similarities in the mesh geometries between ground-truth (a), single-view reconstructions (b), and dual-view reconstructions (c). Based on these data, the two neural networks, especially the dual-view variant, demonstrate considerable promise in serving as a replacement or alternative to dose modulation facilitated by conventionally reconstructed CT volumes.

## 4 Conclusion

In this investigation, we refine and develop two encoder-decoder neural networks for CT reconstruction from ultra-sparse radiographic projections. The developed models are trained and validated on synthetic x-ray/CT paired data. Both the single and dual-view networks perform well on image similarity and use-case-specific metrics, with results indicating the dual-view CNN outperforming the single-view CNN in most cases. These data indicate that the developed deep neural networks are a promising method for CT reconstruction from only one or two radiographic projections. Potential applications include automatic kVp selection, automatic protocol or scan parameter selection, improved scan planning by the operator, and dose modulation procedures.

# Chapter 14

# Oral Session - CT imaging 2

**session chairs**

Marc Kachelrieß, *DKFZ (Germany)*

Xuanqin Mou, *Xi'an Jiaotong University (China)*

# A preliminary study of image reconstruction with single directional-TV constraint from limited-angular-range data

Zheng Zhang[1], Buxin Chen[1], Dan Xia[1], Emil Y. Sidky[1], and Xiaochuan Pan[1,2]

[1]Department of Radiology, The University of Chicago, Chicago, IL, USA
[2]Department of Radiation & Cellular Oncology, The University of Chicago, Chicago, IL, USA

**Abstract** In computed tomography (CT) imaging, an optimization-based method with directional total-variation (DTV) constraints has demonstrated its advantage for image reconstruction from limited-angular-range data. The DTV optimization problem contains two DTV constraints, which are upper-bounded $\ell_1$ norms of an image's partial derivatives along $x$ and $y$ axes, while the scan angular range is symmetric to the $y$ axis in the coordinator system. DTV constraints along different directions, however, may not have equal impact on image reconstruction from limited-angular-range data. In this work, we investigate the impact of each single DTV constraint on image reconstructions. We first design two convex optimization problems, each of which contains a single image DTV constraint, and develop single-DTV algorithms for solving the problems. We then carry out a simulation study with a numerical bar phantom, collect data over a variety of limited-angular ranges, and reconstruct images by using the single-DTV algorithms. Results show that, for image reconstruction from limited-angular-range data, the algorithm employing a single DTV constraint along $x$ axis is more effective than that using a single DTV constraint along $y$ axis in terms of eliminating artifacts.

## 1 Introduction

Recent advances in iterative reconstruction method with directional image constraints [1, 2] have gained attention for computed tomography (CT) image reconstruction from limited-angular-range data. In our previous work [3, 4], we have designed a convex optimization problem with directional total-variation (DTV) image constraints and developed a DTV algorithm based upon a general convex primal-dual algorithm [5–8] for solving the problem. Results of our previous study have demonstrated that the DTV algorithm can accurately reconstruct cross-section images, and that it can considerably diminish artifacts observed otherwise in reconstructions by use of existing algorithms.

The DTV constraints are defined as upper-bounded $l_1$-norms of image's partial derivatives along two orthogonal directions [3, 4]. As the limited-angular range is symmetric to one direction (e.g., $y$ axis) but not to the other ($x$ axis), the DTV constraints along different directions may not have equal impact on image reconstruction performance. It is thus of interest and significance to investigate image reconstruction with each single DTV constraint from data collected over limited-angular ranges, and to evaluate the corresponding impact on reconstruction performance.

In the work, we first design two convex optimization problems, each of which contains a single image DTV constraint, along $y$ or $x$ axis. We then achieve image reconstruction by developing convex primal-dual optimization algorithms [5–8] to solve the single-DTV optimization problems. For comparison, we also consider several existing algorithms including FBP and our previously developed DTV [3, 4] algorithms. A numerical bar phantom is designed for mimicking the cross-section of a phantom in industrial-CT imaging application. Using the phantom, we demonstrate that the DTV constraint along the $x$ axis is more effective than that along the $y$ axis in terms of eliminating artifacts due to limited-angular range.

## 2 Materials and Methods

### 2.1 Data generation

In this work, we carry out a simulation study by using a numerical bar phantom which is discretized on a $150 \times 210$ image grid of size $0.15 \times 0.15$ cm$^2$, as shown in panel (a) of Fig. 2. We generate data from the numerical bar phantom with a fan-beam CT configuration, with a pair of source and detector rotating over a limited-angular range, as illustrated in Fig. 1. The scan angular range is $\alpha$ which is symmetric to the $y$-axis. The source-to-rotation-axis and source-to-detector distances are 100 cm and 150 cm, respectively. The linear detector consists of 512 detector bins of size 0.11 cm. Using the configuration, we generated noiseless data from the numerical bar phantom over a variety of angular ranges with an angular interval of 1° per view. In particular, we in this work focus on image reconstruction from data collected over the limited-angular range of 30°.

### 2.2 Single-DTV optimization program

In this study, we consider two optimization programs with single DTV constraint along $y$ or $x$ axis, which are defined in Eqs. (1) and (2) below:

$$\mathbf{f}^\star = \operatorname*{argmin}_{\mathbf{f}} \left\{ \frac{1}{2} \parallel \mathcal{H}\mathbf{f} - \mathbf{g}^{[\mathcal{M}]} \parallel_2^2 \right\}$$
$$\text{s.t.} \quad ||\mathscr{D}_y\mathbf{f}||_1 \leq t_y, \text{ and } f_i \geq 0, \tag{1}$$

and

$$\mathbf{f}^\star = \operatorname*{argmin}_{\mathbf{f}} \left\{ \frac{1}{2} \parallel \mathcal{H}\mathbf{f} - \mathbf{g}^{[\mathcal{M}]} \parallel_2^2 \right\}$$
$$\text{s.t.} \quad ||\mathscr{D}_x\mathbf{f}||_1 \leq t_x, \text{ and } f_i \geq 0, \tag{2}$$

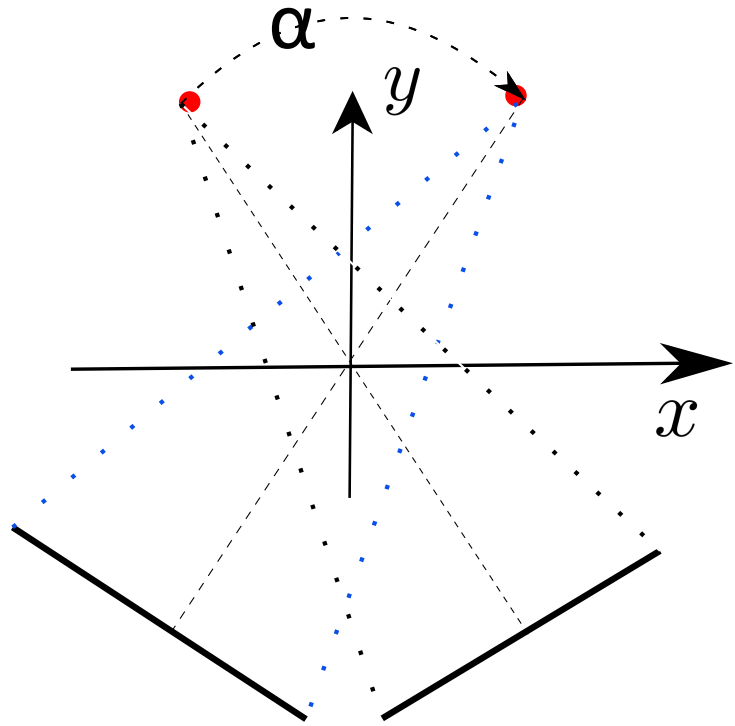

**Figure 1:** Illustration of a fan-beam CT scan configuration over limited-angular range $\alpha$.

where vector $\mathbf{g}^{[\cdot\mathcal{M}]}$ of size $M$ denotes discrete measured data; vector $\mathbf{f}$ of size $N$ is a 2D discrete image; $f_i$ is the entry $i$ of $\mathbf{f}$; $\mathcal{H}$ the system matrix of size $M \times N$, with element $h_{ji}$ representing the intersection length of ray $j$ within pixel $i$; matrices $\mathcal{D}_y$ and $\mathcal{D}_x$ of size $N \times N$ denote an image's partial derivative along $y$ and $x$ axes, respectively; vectors $\mathcal{D}_y\mathbf{f}$ and $\mathcal{D}_x\mathbf{f}$ are of size $N$; $||\cdot||_1$ indicates the $\ell_1$-norm of the input vector; and parameters $t_y$ and $t_x$ depict the upper bounds on the DTV constraints along $y$ and $x$ axes. For simplicity, we refer to the optimization program in Eq. (1) with a single DTV constraint along $y$ axis as the YDTV program, and that in Eq. (2) with a single DTV constraint along $x$ axis as the XDTV program.

### 2.2.1 Reconstruction algorithm

With the designed YDTV and XDTV optimization programs in Eqs. (1) and (2), we derive algorithms as new instances of the convex primal-dual (CPD) algorithm [5, 6] by tailoring the proximal problems in the CPD algorithm to each of the programs, and refer to the new algorithms as YDTV and XDTV algorithms, respectively.

The developed image reconstructions involve several reconstruction parameters that can either affect the final solution or have impact on the convergence rate. In particular, constraint parameters $t_y$ in the YDTV program and $t_x$ in the XDTV program are the most important parameters that specify the optimization programs and confine the feasible solution set. In this study, we compute the DTV values from the truth phantom image, referred to as $t_{y0}$ and $t_{x0}$, and use them as the DTV constraint parameters for studies in Sec. 3.

It is worth noting that the YDTV and XDTV programs are convex, which should be mathematically exactly and numerically accurately solved by the corresponding algorithms. Therefore, we also define algorithm's convergence conditions [4, 6–9] that specify the stopping criteria and determine the final solution to the designed optimization program. Details of the convergence conditions can be found in Ref. [4].

### 2.3 Reconstruction algorithms for comparison

For comparison, we also reconstruct images by using the FBP algorithm and CPD algorithms for solving a data-$\ell_2$-minimization problem in Eq. (3),

$$\mathbf{f}^\star = \underset{\mathbf{f}}{\arg\min} \left\{ \frac{1}{2} \parallel \mathcal{H}\mathbf{f} - \mathbf{g}^{[\cdot\mathcal{M}]} \parallel_2^2 \right\} \quad \text{s.t. } f_i \geq 0, \quad (3)$$

and our recently developed image DTV constrained data-$\ell_2$-minimization problem in Eq. (4),

$$\mathbf{f}^\star = \underset{\mathbf{f}}{\arg\min} \left\{ \frac{1}{2} \parallel \mathcal{H}\mathbf{f} - \mathbf{g}^{[\cdot\mathcal{M}]} \parallel_2^2 \right\}$$
$$\text{s.t. } ||\mathcal{D}_x\mathbf{f}||_1 \leq t_x, \ ||\mathcal{D}_y\mathbf{f}||_1 \leq t_y, \text{ and } f_i \geq 0, \quad (4)$$

where terms and symbols have been specified in Sec. 2.2. We refer to the CPD algorithms for solving optimization programs in Eqs. (3) and (4) as L2 and DTV algorithms, respectively.

## 3 Results

### 3.1 Algorithm verification study

Although the YDTV and XDTV algorithms can theoretically solve the convex optimization programs in Eqs. (1) and (2), we must also establish in numerical studies that these algorithms can indeed accurately solve the optimization programs in terms of yielding the convergence conditions up to the computer precision. In particular, we have carried out a simulation study using data generated from the numerical bar phantom over $2\pi$ angular range, and the result confirms that the convergence conditions of the each algorithm are satisfied up to computer precision.

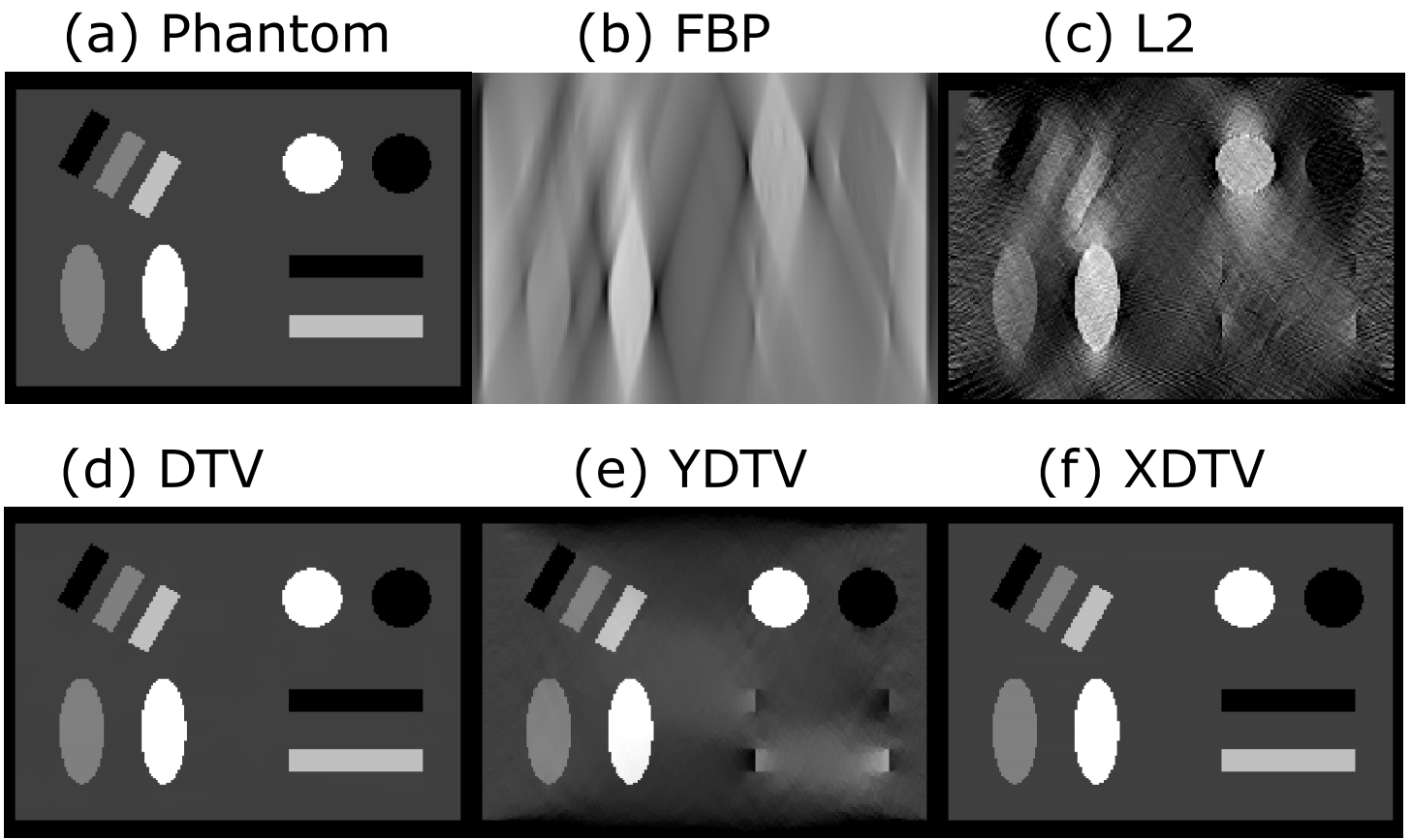

**Figure 2:** Numerical bar phantom (a) and images reconstructed by use of FBP (b), L2 (c), DTV (d), YDTV (e), and XDTV (f) algorithms from data collected over angular range of $30°$. Display window $[0.1, 0.5]$ cm$^{-1}$.

With the algorithms numerically verified, we present below some of the study results exploring the performance each algorithm for image reconstruction from data over the limited-angular range.

### 3.2 Reconstruction from $30°$-data

Using the YDTV and XDTV algorithms developed, we reconstruct images from the noiseless bar-phantom data collected over the limited-angular range of $30°$, as described in Sec. 2.1. We show the YDTV and XDTV reconstructions in panels (e) and (f) of Fig. 2. For references, we also show FBP, L2, and DTV results in panels (b), (c), and (d) of Fig. 2, respectively. It can be observed that severe artifacts and structure distortion exist in the FBP and L2 reconstructions, and most structures cannot be identified. The DTV reconstruction, however, is free of artifacts and visually identical to the truth bar phantom. By inspecting the YDTV reconstruction, we observe that although it generally outperforms FBP and L2 results in terms of artifact reduction, artifacts and distortion are still visible. The XDTV reconstruction, on the other hand, shows significant improvement comparing to its YDTV counterpart, and is visually identical to the truth bar phantom.

### 3.3 Reconstruction as a function of iterations

Reconstructions of the bar-phantom images above were obtained when the DTV algorithm's convergence conditions are satisfied. We also investigated how the reconstructions of the bar phantom evolve as functions of the iteration number. In particular, we focus on reconstructions by use of the DTV, XDTV, and YDTV algorithms. We show in Fig. 3 reconstructions of the bar phantom at iteration 10, 500, 1000, 5000, and 20000, along with the final convergent reconstruction. It can be observed that the DTV reconstruction at about iteration 5000 and the the XDTV reconstruction at about iteration 20000, visually resemble their final convergent reconstructions, respectively, which are also visually identical to the truth image. The YDTV reconstruction becomes visually similar to its final convergent reconstruction at iteration 5000, which suffers from considerable limited-angular-range artifacts such as distortion and streaks.

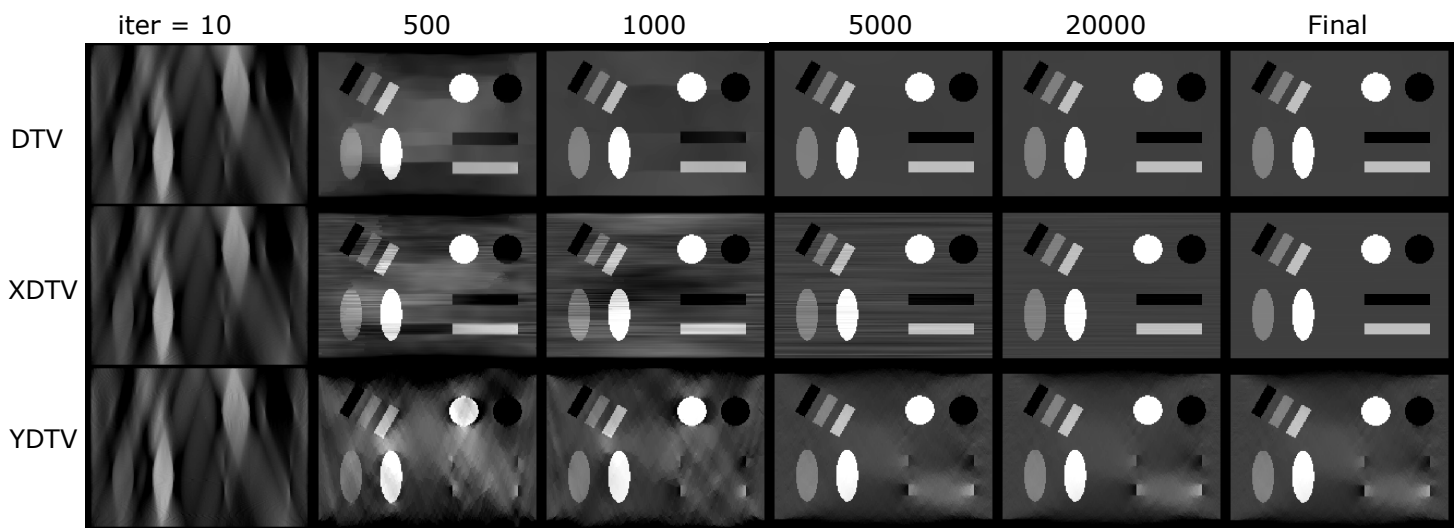

**Figure 3:** Image reconstructions by use of the DTV (row 1), XDTV (row 2), and YDTV (row 3) algorithms at iteration 10 (column 1), 500 (column 2), 1000 (column 3), 5000 (column 4), and 20000 (column 5), along with final convergent reconstruction (column 6).

## 4 Discussion & conclusion

In the work, we have developed and investigated algorithms for image reconstruction with a single DTV constraint from data collected over limited-angular ranges. For references, we also use the existing FBP, L2, and DTV algorithms to reconstruct images. Results show that the XDTV algorithm is more effective than the YDTV algorithm in terms of eliminating artifacts for reconstruction from limited-angular-range data. It is worth noting that XDTV reconstruction is equivalent to the DTV reconstruction by setting $t_x$ to $t_{x0}$ and $t_y$ to infinity. It is therefore of interest in the future to investigate DTV reconstruction with $t_x = t_{x0}$ and varying $t_y$, and study how different $t_y$ parameters impacts the performance of DTV reconstruction. In addition, for future studies, we will investigate the XDTV and YDTV algorithms, as well as the DTV algorithm, for image reconstruction from limited-angular-range data containing inconsistencies, such as noise, beam hardening, and scatter.

## Acknowledgments

This work was supported in part by NIH R01 Grant Nos. EB026282, EB023968, and Grayson-Jockey Club Research. The computation of the work was performed in part on the computer cluster funded by NIH S10-OD025081, S10-RR021039, and P30-CA14599 awards. The contents of this paper are solely the responsibility of the authors and do not necessarily represent the official views of NIH.

# Clinical Indication of the Potential of Fully 3D Reconstruction from Ultra-low-dose Pre-log CT Data for Lung Nodule Screening

Lin Fu[1], Yongfeng Gao[2], Bruno De Man[1], Mike Reiter[2], Haifang Li[2],
Charles Mazzarese[2], Amit Gupta[2], and Zhengrong Liang[2,3]

[1]Radiation Imaging, GE Research, Niskayuna, NY, USA
Departments of Radiology[2] and Biomedical Engineering[3], Stony Brook University, NY, USA

**Abstract**: Lung cancer causes the highest cancer-related death rate among all cancers. The most effective approach to reduce the death rate is to detect the precursor, i.e. the pulmonary nodules, at as early stage as possible. Current clinical-low-dose computed tomography (i.e. 1.05 mSv LDCT) screening has shown great value for detecting nodules of 5 mm and larger. This exploratory study aims to clinically demonstrate the potential of ultra-low-dose CT (i.e. 0.12 mSv ULDCT) to accomplish the same task by fully 3D reconstruction from pre-log data, instead of the current CT reconstruction from post-log data. A patient with a 4 mm nodule was recruited undergoing a clinical LDCT, followed by a research ULDCT scan. Reconstructions of these scans were assessed by a thoracic radiologist in terms of both image quality and nodule detectability. Our fully 3D reconstruction from the research pre-log data performed comparably with the current clinical-low-dose CT from the post-log data.

**Keywords**: Lung cancer screening, Ultra-low-dose CT, Pre-log reconstruction, Pulmonary nodule detection.

## 1 Introduction

Current low-dose computed tomography (LDCT) screening for detection of the lung cancer precursor, i.e. the pulmonary nodules, has demonstrated a reduction of 24% of the death rate [1]. As a massive screening modality, LDCT operating at ~1 mSv dose level remains a concern due to the potential risk of radiation-induced cancer [2]. Ultra-low-dose computed tomography (ULDCT) operates in a dose regime that is a factor of 10 lower than clinical-low-dose CT techniques [3], reaching a level of patient dose similar to a chest x-ray exam (~0.1 mSv), which makes it an attractive modality for the screening of lung nodules [4]. Other potential applications of ULDCT include 4D dynamic imaging, attenuation correction for PET/CT imaging, and virtual CT colonoscopy screening, etc.

Model-based iterative reconstruction (MBIR) has been a powerful algorithmic technique for CT dose reduction and image quality improvement [5], [6]. Reduction of patient dose of up to 80% relative to filtered backprojection (FBP) reconstruction has been reported [4]. Most MBIR algorithms developed for clinical applications operate in the post-log data domain, i.e., x-ray transmission data are first logarithmically transformed to post-log line integral values before being fed to the reconstruction algorithm. Post-log MBIR can leverage existing pre-correction steps for the FBP algorithm, but transforming data to post-log domain can cause loss of information, especially in the ultra-low-dose regime [7]. In ULDCT, the mean number of x-ray photons received per detector cell exposure is commonly in the range of tens [8], [9], and photon starvation is commonplace. The detector measurements may be zero due to the quantum nature of x-ray and even become negative due to the detector electronic noise [9]–[12], in which case the logarithm simply cannot be taken. These non-positive values must be either discarded [13], replaced by some artificial positive values [8], corrected by some recursive mean-preserving operations [14], or interpolated by sinogram smoothing or denoising methods [15], [16]. When the extent of such pre-correction is aggressive, such as in ULDCT, handling and logarithmically transforming such photon-starved measurements can cause strong bias and artifacts in post-log reconstructed images.

Compared with post-log MBIR, pre-log MBIR incorporates Beer's law in the forward model and use appropriate statistical model to directly reconstruct images from pre-log x-ray transmission measurements. Pre-log MBIR has the potential to achieve further dose reduction capability beyond post-log MBIR by fully incorporating the information in photon-starved detector measurements into image reconstruction. Various pre-log MBIR algorithms have been proposed in the literature [17]–[21], and has recently been shown to be feasible to be applied to clinical CT data [22], achieving promising improvement of image quality in ultra-low-dose scans.

The present study is an initial effort to evaluate the potential of pre-log MBIR in the context of ultra-low-dose lung nodule screening CT and its impact to nodule detection capability. A patient with a 4 mm nodule was recruited undergoing a clinical-low-dose scan, followed by a research ultra-low-dose scan. Reconstructions from the pre-log and post-log datasets of the clinical-low-dose and research ultra-low-dose scans were performed. We compared images reconstructed by pre-log MBIR, post-log MBIR, and FBP at both the clinical-low-dose and the ultra-low-dose levels. A thoracic radiologist assessed these reconstructions in terms of image quality and nodule detectability.

## 2 Theory

We use a unified MBIR framework to reconstruct CT images from either pre-log or post-log CT data. Major components of the MBIR framework include a CT forward model, a maximum *a posterior* (MAP) statistical model, and an iterative solver.

*CT forward model*

We use a discrete-discrete forward model of CT imaging system:

$$\hat{y}_i \triangleq E[y_i] = I_i e^{-f_i([\mathbf{A}x]_i)}, \qquad i = 1, \dots, M$$

where $\hat{y}_i$ is the ensemble mean of the $i$th detector measurement $y_i$; $x \in \mathbb{R}^N$ is a vector denoting the image to be reconstructed; $I_i > 0$ is an air measurement; $\mathbf{A} = \{a_{ij}\}$ with $a_{ij} \geq 0$ is an $M \times N$ system matrix representing the Radon or x-ray transform; and $[\mathbf{A}x]_i \triangleq \hat{p}_i = \sum_{j=1}^{N} a_{ij}x_j$ is the line integral value along ray $i$; $f_i(\cdot)$ is a function to model the beam-hardening effect due to a polyenergetic x-ray beam passing through the object and detector. In this study $f_i(\cdot)$ is approximated by a polynomial obtained from detector calibration procedures.

*MAP statistical model*

The reconstructed image $\hat{x} \in \mathbb{R}^N$ is the one that maximizes *a posterior* probability given the measurement vector $y \in \mathbb{R}^M$:

$$\hat{x} = \underset{x}{\mathrm{argmin}}\, \Phi(x|y).$$

The cost function $\Phi(x|y)$ consists of a data likelihood term and an image-space regularization term:

$$\begin{aligned}
\Phi(x) &\triangleq -\phi(x) + U(x) \\
&= -\sum_{i=1}^{M} h(y_i|[\mathbf{A}x]_i) + \sum_{j}^{N}\sum_{k>j}^{N} b_{jk}\rho(x_j - x_k),
\end{aligned}$$

where the data likelihood term $\phi(x)$ is the sum of the log-likelihood $h_i(y_i|\hat{p}_i)$ of individual measurements (assuming the noise in the $y_i$'s is statistically independent), and the regularization term $U(x)$ is a Markov random field (MRF) with $b_{jk}$ representing the penalty strength between pixel $j$ and $k$, and $\rho(\cdot)$ being a prior potential function.

In pre-log reconstruction, we use shifted Poisson (SP) log-likelihood:

$$\begin{aligned}
h_{\text{prelog}}(y_i|\hat{p}_i) &= (y_i + \sigma^2)\log\!\big(I_i e^{-f_i(\hat{p}_i)} + \sigma^2\big) \\
&\quad - \big(I_i e^{-f_i(\hat{p}_i)} + \sigma^2\big),
\end{aligned}$$

where $\sigma^2$ is the variance of the detector electronic noise.

In post-log reconstruction, we use a Gaussian model, bypassing Beer's law and leading to a linear weighted-least-squares formulation with respect to $\hat{p}_i$:

$$h_{\text{postlog}}(y_i|\hat{p}_i) = -\frac{1}{2}W_i(\hat{p}_i - p_i)^2$$

where $W_i$ is the estimated inverse variance of $p_i$ and

$$p_i \triangleq f_i^{-1}\left(\log\frac{I_i}{\max(y_i, \delta)}\right)$$

is the post-log data obtained by taking logarithm of the ratio between the pre-log data and the air scan, followed by the beam-hardening correction $f_i^{-1}$. Small or non-positive values in $y_i$ are clipped by a threshold $\delta$ [8][23]. The weight factor is determined by [6]:

$$W_i = f_i'(p_i)^2 \frac{y_i^2}{y_i + \sigma^2},$$

where small or non-positive values in $y_i$ are also clipped by the threshold $\delta$.

*Image update equations*

We use a preconditioned gradient descent algorithm for both pre-log and post-log reconstructions. The image update equation for pre-log reconstruction is:

$$\begin{aligned}
\hat{x}_j^{(n+1)} = \hat{x}_j^{(n)} + \frac{1}{m_i} \\
\left\{ \left[\sum_i^M a_{ij}\left(f_i'(\hat{p}_i^{(n)})\frac{\hat{y}_i^{(n)}}{\hat{y}_i^{(n)} + \sigma^2}\right)(\hat{y}_i^{(n)} - y_i)\right] - \sum_{k>j}^N b_{jk}\rho'(\hat{x}_j^{(n)} - \hat{x}_k^{(n)}) \right\}
\end{aligned}$$

where the preconditioner $m_i \triangleq \sum_i^M \big[a_{ij}y_i\big(\sum_h^N a_{ih}\big)\big]$. The error sinogram $\hat{y}_i^{(n)} - y_i$ is evaluated in the pre-log domain thus can accommodate non-positive values in $y_i$.

The image update equation for post-log reconstruction is:

$$\begin{aligned}
\hat{x}_j^{(n+1)} = \hat{x}_j^{(n)} + \frac{1}{m_i} \\
\left\{ \left[\sum_i^M a_{ij}\left(f_i'(p_i)^2 \frac{y_i^2}{y_i + \sigma^2}\right)(p_i - \hat{p}_i^{(n)})\right] - \left[\sum_{k>j}^N b_{jk}\rho'(\hat{x}_j^{(n)} - \hat{x}_k^{(n)})\right] \right\}
\end{aligned}$$

where the error sinogram $p_i - \hat{p}_i^{(n)}$ is evaluated in the post-log domain, and the WLS weights $W_i = f_i'(p_i)^2 \frac{y_i^2}{y_i + \sigma^2}$ are fully pre-determined (after clipping non-positive values).

# 3 Experiments

We compared different reconstruction algorithms with patient data acquired as part of clinical work-up for lung nodule screening at Stony Brook University Hospital, approved by institutional review board approval and with written informed consent. Both a clinical-low-dose and an ultra-low-dose dataset were acquired for comparison. The scans were acquired on a GE Lightspeed VCT scanner (GE Healthcare, Waukesha, WI), with 64-row collimation, a helical pitch of 0.516, and 1.0 s rotation. The clinical-low-dose dataset was acquired with the standard lung nodule screening CT protocol at 120 kVp and automatic mA modulation (1.05 mSv). The ultra-low-dose dataset was acquired with a research protocol at 80 kVp and 10 mA (0.12 mSv). For each dataset, three reconstruction algorithms were compared: FBP with standard kernel, Post-log MBIR, and Pre-log MBIR.

All images were reconstructed on a 512×512 grid with a field-of-view of 50 cm, an in-plane pixel size of 0.98 mm, and a slice thickness of 0.625 mm. In post-log MBIR, the threshold $\delta$ was set to $10^{-5}$ relative to the air scan. The threshold is roughly equivalent to the attenuation of about 57 cm of water (assuming $\mu_{\text{WATER}} = 0.2 \text{ cm}^{-1}$ at 60 keV). In this initial study $\sigma^2$ was set to 0. All MBIR reconstructions used the distance-driven forward and back projectors [24] and the $q$-GGMRF regularization [6]. Regularization strengths were empirically selected. All MBIR reconstructions were initialized with the corresponding standard FBP images and run 2000 iterations for practical convergence.

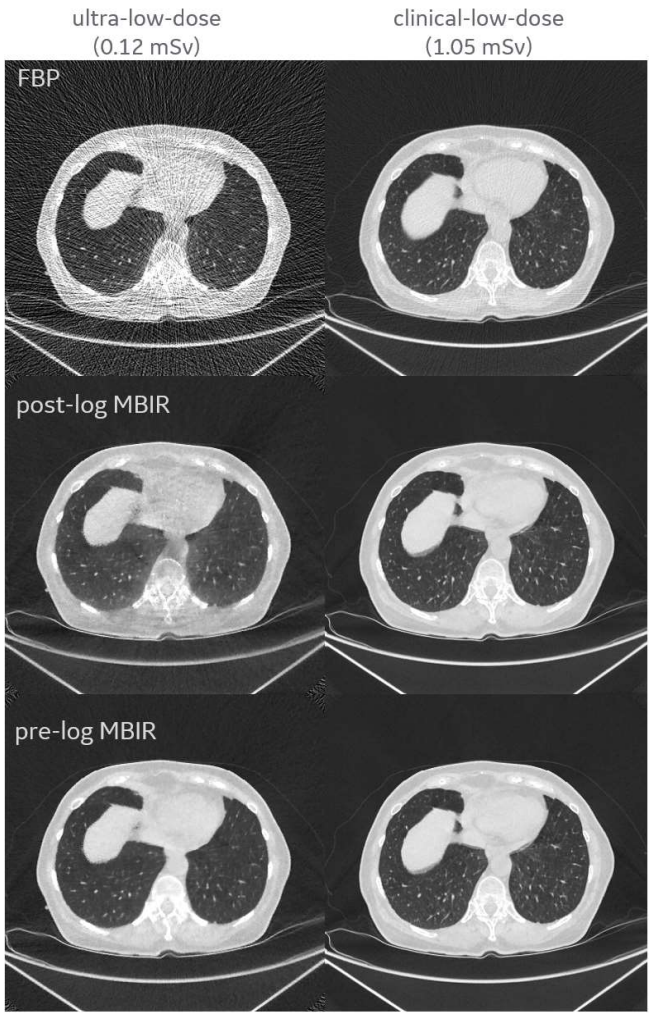

Fig. 1. Examples of images reconstructed from clinical-low-dose and ultra-low-dose data with different algorithms. Window width/level = 1500/-500 HU.

These six reconstructed volumetric images were de-identified by their reconstruction algorithms and randomly displayed to the thoracic radiologist, who had all the visualization tools in current clinical setting to exam each volumetric image. The radiologist firstly scored the image diagnostic quality by a range from 1 (worst quality) to 10 (highest quality) and then scored the confidence on detection of the 4mm nodule in the range from 1 to 10.

## 4 Results

Fig. 1. shows reconstructed CT images produced by different reconstruction algorithms at the two dose levels. The ultra-low-dose FBP reconstruction is extremely noisy. The ultra-low-dose post-log MBIR image contains much less noise but still shows strong artifacts and negative bias. The pre-log MBIR algorithm shows remarkable improvement of HU accuracy relative to post-log MBIR, almost fully removing the dark shading in the image, and the appearance of lung and chest wall are comparable to the clinical-low-dose reference image. At clinical-low-dose, the post-log and pre-log MBIR images appear very comparable. Figs. 2 and 3 show the horizontal profiles through the center portion of the images for more quantitative comparison.

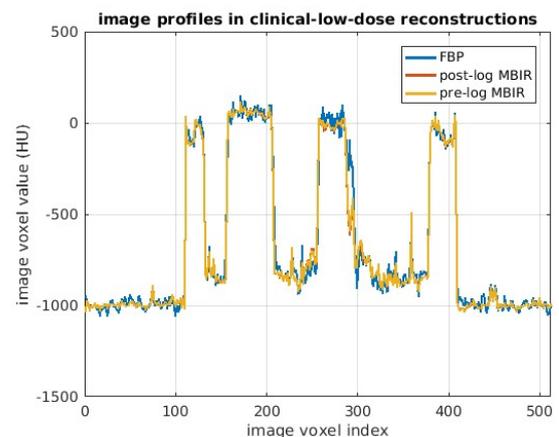

Fig. 2. Image profiles through the center of reconstructed images from the clinical-low-dose data. The profiles of post-log and pre-log MBIR images closely overlap.

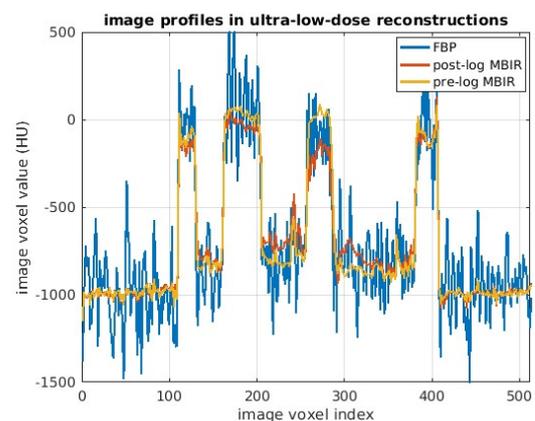

Fig. 3. Image profiles through the center of reconstructed images from the ultra-low-dose data. The FBP image is extremely noisy. The post-log MBIR image contains much less noise but shows strong bias on the order of 100 HU in the central area. The pre-log MBIR image gives remarkable improvement of HU accuracy when compared with post-log MBIR.

Table 1 shows the scores of the thoracic radiologist on the six reconstructions in terms of image quality and nodule detectability. It should be noted that in this initial study the regularization parameters used in pre-log and post-log

MBIR algorithms have not been optimized for the specific clinical task. Nevertheless, at either dose levels, the pre-log MBIR received the highes score among all reconstruction options. At clinical-low-dose level, post-log and pre-log MBIR images revceived different scores, although they are visually very similar, as indicated by the image profiles in Fig. 2. The lower score of post-log MBIR image at clinical-low-dose might be attributed to slight over-regularization when compared with the pre-log MBIR image, which will be fine-tuned in future studies, after which post-log MBIR is expected to match pre-log MBIR scores at clinical-low-dose. At ultra-low-dose level,  post-log MBIR may also be poentially improved by more advanced post-log low-signal correction techniques [16].

Table 1. Scores of Image Quality and Nodule Detectability

| Reconstruction Algorithms | Image Quality | Nodule Detectability |
|---|---|---|
| Post-log MBIR (ultra-low-dose) | 1 | 1 |
| FBP (ultra-low-dose) | 2 | 2 |
| Pre-log MBIR(ultra-low-dose) | 4 | 4 |
| Post-log MBIR (clinical-low-dose) | 6 | 6 |
| FBP (clinical-low-dose) | 7 | 7 |
| Pre-log MBIR(clinical-low-dose) | 10 | 10 |

## 5 Conclusion and Discussion

We applied pre-log MBIR to lung nodule screening CT with an ultra-low-dose research protocol that has not been used in routine clinical practice.  Our results show that at the ultra-low-dose level, pre-log MBIR can remarkably improve the overall visual impression and quantitative accuracy when compared with post-log MBIR and FBP, thanks to its ability to better incorporate the information from photon-starved measurements. More importantly, the performance of our fully 3D pre-log reconstruction at the ultra-low-dose (0.12mSv) level approaches that of the FBP at the clinical-low-dose (1.05mSv) level, indicating a great potential of fully 3D reconstruction of ultralow-dose CT screening for the pulmonary nodules or the precursor of lung cancer.  This clinical observation concurs with our simulation study of characterizing the pre-log Bayesian CT reconstruction by the relationship between radiation dose level and tissue texture measure [21].   Pre-log MBIR may play an important role in enabling ultra-low-dose lung screening CT applications. Future work will entail further adjustment of algorithm parameters based on the clinical feedback.

## Acknowledgements
This work was partially supported by the NIH/NCI grant #CA206171.  The authors would appreciate the assistance from Ms. Danielle Giulietti and Ms. April Ida Plank for data acquisition.

# Sparse View Deep Differentiated Backprojection for Circular Trajectories in CBCT

Philipp Ernst[1,2], Georg Rose[2], and Andreas Nürnberger[1]

[1]Faculty of Computer Science, University of Magdeburg, Germany
[2]Institute for Medical Engineering and Research Campus *STIMULATE*, University of Magdeburg, Germany

**Abstract** In this paper, we present a method for removing streak artifacts from reconstructions of sparse cone beam CT (CBCT) projections along circular trajectories. The differentiated backprojection on 2-D planes is combined with convolutional neural networks for both artifact reduction and the ill-posed inversion of the Hilbert transform. Undersampling errors occur at different stages of the algorithm, so the influence of applying the neural networks at these stages is investigated. Spectral blending is used to combine coronal and sagittal planes to a full 3-D reconstruction. Experimental results show that using a neural network to reconstruct a plane-of-interest from the differentiated backprojection of few projections works best by additionally providing FDK reconstructed planes to the network. This approach reduces streaking and cone beam artifacts compared to the direct FDK reconstruction and is also superior to post-processing CNNs.

## 1 Introduction

For several years, convolutional neural networks (CNNs) and deep learning have found their way into medical imaging and CT image reconstruction. CNNs are especially beneficial for closely approximating complex tasks since they are trained with domain-specific data and in most cases need less computation time during inference than their exact analytical counterparts. However, training CNNs on 3-D data sets is usually not feasible due to the large memory requirements and is often stripped down to 2-D problems or patch-based 3-D approaches.

In medical interventions, surgeons usually use fluoroscopic images to guide them during the operation. This keeps the dose on both the surgeons and patients low but does not provide correct spatial information due to distortions of the cone beam projections. This can be overcome using CBCT reconstructions but increases the dose. To keep the radiation as low as possible, the number of projections or the radiation per projection must be minimal, both of which introduces different kinds of artifacts in the final reconstructions. The approach proposed in this work will focus on the first case of a reduced number of projections.

There are algorithms that are able to reduce the streaking artifacts caused by a low number of projections, but many of them are not applicable in an interventional setting due to their computation time, especially for iterative methods.

Han et al. [1] combine a factorization approach of the 3-D problem onto 2-D planes using the differentiated backprojection (DBP) of [2] with a CNN-learned inversion of the Hilbert transform to remove cone beam artifacts. As shown by [3], the DBP is less prone to artifacts caused by truncated projections that are also usually acquired during interventions, e.g. to reconstruct volumes of interest. Keeping the radiation dose low by reducing the number of projections also creates artifacts in the DBP domain. However, the following sections will show that CNNs can be used to reduce these artifacts and outperform post-processing FDK reconstructions with CNNs of the same architecture.

## 2 Mathematical Preliminaries

### 2.1 Differentiated Backprojection

Following the notation of [1], let $f(\mathbf{x})$ denote the scanned object. Acquired cone beam projections are treated as X-ray transforms $D_f(\mathbf{a}, \theta)$ from source locations $\mathbf{a}(\lambda)$ along a circular trajectory of radius $R$

$$\mathbf{a}(\lambda) = R \cdot [\cos\lambda, \sin\lambda, 0]^T, \quad \lambda \in \mathbb{R}, \quad (1)$$

along lines of direction $\theta \in S^2 \subset \mathbb{R}^3$ such that

$$D_f(\mathbf{a}, \theta) = \int_{\mathbb{R}} f(\mathbf{a} + t\theta) \, dt. \quad (2)$$

Applying the partial derivative along the source trajectory and backprojecting between the source locations $\mathbf{a}(\lambda), \lambda \in [\lambda_-, \lambda_+]$ results in the differentiated backprojection (DBP)

$$g(\mathbf{x}) = \int_{\lambda_-}^{\lambda_+} \frac{1}{\|\mathbf{x} - \mathbf{a}(\lambda)\|} \left. \frac{\partial}{\partial \mu} D_f(\mathbf{a}(\mu), \theta) \right|_{\mu=\lambda} d\lambda, \quad (3)$$

which is related to the object function $f(\mathbf{x})$ by the Hilbert transform.

### 2.2 Factorization onto 2-D Planes

Eq. 3 can be evaluated on 2-D planes perpendicular to the circular source trajectory, i.e. parallel to the $z$-axis, which allows this 3-D problem to be converted to successive 2-D problems [2]. If the chosen plane-of-interest $\mathscr{P}$ contains the source locations $\mathbf{a}(\lambda_-)$ and $\mathbf{a}(\lambda_+)$, then $g(\mathbf{x})$ is the convolution of the Hilbert kernel over lines of the object $f$ from $\mathbf{x}$ to $\mathbf{a}(\lambda_-)$ and from $\mathbf{x}$ to $\mathbf{a}(\lambda_+)$. This suggests a deconvolution algorithm to retrieve the object function $f(\mathbf{x})$ on $\mathscr{P}$. CNNs can be trained to approximate this ill-posed deconvolution problem without much computational effort (during inference), in contrast to analytical algorithms [1, 2].

## 3  Method

### 3.1  Discretization Errors and Sparse Views

Applying the DBP algorithm to real world data necessarily introduces discretization errors: (1) Cone beam projections are inherently discrete, so $D_f(\mathbf{a}(\lambda), \theta)$ can only be evaluated for discrete subsets of $\lambda$ and $\theta$, (2) the partial derivative $\partial D_f / \partial \lambda$ needs to be approximated as well as (3) the integral in Eq. 3. Theoretically exact reconstructions can only be achieved for planes parallel to the $z$-axis containing $\mathbf{a}(\lambda_\pm)$. (4) This limits the resolution of the reconstructed volume (out-of-plane) to the distance of neighboring source locations.

A low number of projections makes these errors even more prominent. Sticking to the approach of [1], the partial derivative $\partial D_f / \partial \lambda$ is approximated using the view-dependent differentiation of [4], which only involves choosing a resolution parameter that is set to $\varepsilon = 2 \times 10^{-3}$ empirically for a good trade-off between accuracy and resolution.

### 3.2  Approach

The errors caused by discretization occur at different stages in the reconstruction algorithm. For this reason, it is necessary to dedicate different networks to these stages and evaluate if combined networks can approximate the Hilbert inversion with errors from different stages more accurately than others. In total, six networks are trained. (1) For comparison, a post-processing network is trained that enhances the FDK reconstruction of 36 projections for sagittal or coronal slices. (2) A network that enhances the DBP (Eq. 3) of 36 projections to approximate the DBP of fully sampled projections. (3) A Hilbert inversion network that inverts fully sampled DBP planes. (4) Like (3) but with an additional FDK reconstructed (360 projections) plane as input. (5) A Hilbert inversion network that inverts DBP planes from 36 projections and enhances them to approximate reconstructions of fully sampled projections. (6) Like (5) but with an additional FDK reconstructed (36 projections) plane as input.

All networks share the same U-Net-like architecture except for the number of input/output channels and are trained on both coronal and sagittal planes-of-interest.

For the final reconstructions, the following combinations of networks are investigated: Network (1) for comparison (`fdkconv`). Network (2) + Network (3) (`s2f_inv`). Network (2) + Network (4) (`s2f_inv3`). Network (5) (`inv_sp`). Network (6) (`inv_sp3`).

### 3.3  Spectral Blending

As described in [1], the reconstructed planes of the different Hilbert directions can be combined using spectral blending in order to minimize the missing frequency information. A bow-tie mask is multiplied with the Fourier transforms of the reconstructed planes and added. The masks are chosen such that the frequency information from both planes complement each other. By angular blurring of the mask, frequency information that is contained in both planes can be combined, as well.

| Method | NMSE [%] | PSNR [dB] | SSIM [%] |
|---|---|---|---|
| fdkconv | 2.05 | 81.02 | 99.67 |
| s2f_inv | 3.02 | 80.17 | 99.45 |
| s2f_inv3 | 6.95 | 75.73 | 98.67 |
| inv_sp | 2.95 | 80.70 | 99.42 |
| inv_sp3 | **1.05** | **84.06** | **99.83** |

Table 1: Errors w.r.t. ground truth of reconstructions from coronal planes-of-interest averaged over axial planes.

### 3.4  Datasets and Training

The data of eleven subjects from the CT Lymph Nodes collection [5] of The Cancer Imaging Archive [6] is used, consisting of reconstructed volumes of the abdomen that serve as ground truth. Cone beam projections were generated using the CTL toolkit [7] equiangularly along a circular trajectory with a source to detector distance (SDD) of 1000 mm and a source to isocenter distance (SID) of 750 mm. The flat panel detector consists of $1024 \times 1024$ elements with a pixel size of 1 mm$^2$ (cone angle of 54.2°).

A slightly modified U-Net [8] with a depth of 5 is used. The encoder doubles the number of layers after each average pooling, whereas the decoder halves the number of layers after each bilinear upsampling. SGD is used as the optimizer with a weight decay of $1 \times 10^{-4}$ and a learning rate of $5 \times 10^{-2}$ that gradually drops to $1 \times 10^{-2}$ by a factor of 0.8 after every 10 epochs of no improvement in validation loss. Every network was trained for 300 epochs using mean squared error (MSE). Eight subjects were used for training, two for validation and the remaining one for testing. For faster convergence, the reconstructed planes are normalized between 0 and roughly 1 by dividing by the 99th percentile of all axial planes of all datasets. Similarly, the Hilbert planes are normalized by dividing by the standard deviation of all Hilbert planes of all datasets. Random horizontal flips were used as augmentation during training.

## 4  Results

Tab. 1 shows the mean errors of axial slices using coronal planes-of-interest for the different combinations of networks as described in Sec. 3.2, which include normalized mean squared error (NMSE), peak signal-to-noise ratio (PSNR) and structural similarity index measure (SSIM). The lowest errors are achieved using `inv_sp3`, followed by the simple post-processing network `fdkconv`. All other combinations result in worse errors, the worst being `s2f_inv3` with an NMSE which is almost seven times higher than the best

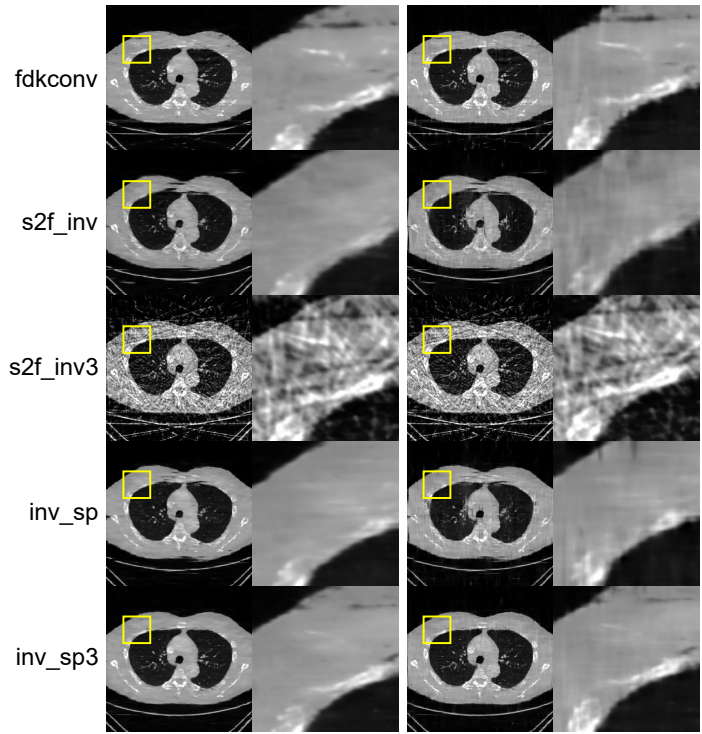

**Figure 1:** Exemplary reconstruction of different methods. Left: using coronal planes-of-interest. Right: after spectral blending.

| Method | NMSE [%] | PSNR [dB] | SSIM [%] |
|---|---|---|---|
| fdkconv | 1.69 | 81.99 | 99.71 |
| s2f_inv | 3.29 | 79.33 | 99.42 |
| s2f_inv3 | 8.06 | 75.06 | 98.38 |
| inv_sp | 2.99 | 79.97 | 99.47 |
| inv_sp3 | **1.19** | **83.50** | **99.81** |

**Table 2:** Errors w.r.t. ground truth of reconstructions from sagittal planes-of-interest averaged over axial planes.

`inv_sp3`. An important thing to note here is that the additional FDK plane of Network (4) was reconstructed using 360 projections while training, whereas during the inference for the combination with Network (2), only 36 projections were available for the FDK reconstruction and necessarily introduced streaking artifacts. The other combinations `s2f_inv` and `inv_sp` have only slightly worse errors than `fdkconv`.

The left column of Fig. 1 shows an axial slice reconstructed from coronal planes using the different methods. Except for `s2f_inv3`, all combinations result in less discontinuous reconstructions than `fdkconv`. `s2f_inv` seems to smooth out highly absorbing tissues. The best visual appearance for this slice is achieved using `inv_sp` with the least discontinuities and the highest edge preservation. As described earlier, `s2f_inv3` necessarily performs worse because of the way it was trained. However, since the streaking artifacts are very prominent, it can be assumed that Network (4) mainly focuses on the FDK input rather than the DBP plane.

The same behavior as in Tab. 1 can be seen in Tab. 2, but here for sagittal planes-of-interest. Interestingly, all errors are slightly worse than their counterpart on coronal planes-of-interest except for `fdkconv`. For brevity, the qualitative results are not shown.

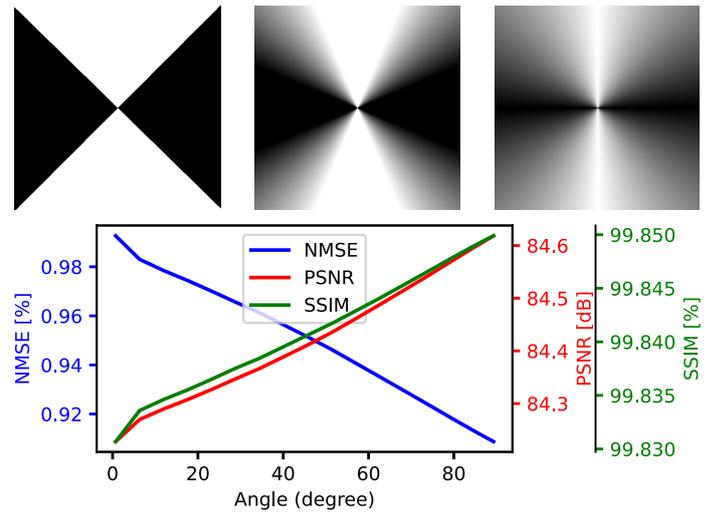

**Figure 2:** Top: Masks without (left), 45° (center) and 90° blurring (right). Bottom: Error metrics for `inv_sp3` reconstructions after spectral blending depending on blurring radius of masks.

| Method | NMSE [%] | PSNR [dB] | SSIM [%] |
|---|---|---|---|
| sparse_fdk | 15.80 | 72.09 | 96.95 |
| fdkconv | 1.39 | 82.86 | 99.76 |
| s2f_inv | 1.98 | 81.53 | 99.64 |
| s2f_inv3 | 7.87 | 75.17 | 98.44 |
| inv_sp | 1.80 | 82.14 | 99.67 |
| inv_sp3 | **0.90** | **84.61** | **99.84** |

**Table 3:** Errors after spectral blending. `sparse_fdk` shows the errors of an FDK reconstruction from 36 projections for reference.

## 4.1 Spectral Blending

The blurring radius of the bow-tie mask for the spectral blending of reconstructions from coronal and sagittal planes-of-interest seems to be an essential parameter for the quality of the final reconstructions, as prior tests have shown. The influence of different blurring radii is shown in Fig. 2 for `inv_sp3`. There seems to be an almost linear dependency between the radius and the different error metrics: the higher the radius, the closer the reconstruction to the ground truth. This is reasonable because more and more frequencies from both planes are accounted for when increasing the radius. For this reason, the blurring radius is set to 90°.

Tab. 3 shows the errors of the different methods after spectral blending. Compared to the reconstructions without spectral blending, the errors are even lower. For the best method `inv_sp3`, the NMSE is reduced by 0.15 % and 0.19 % compared to coronal and sagittal plane-of-interest reconstruc-tions.

The right column of Fig. 1 shows the reconstructions after spectral blurring. Almost all methods benefit from the additional sagittal information. Visual differences cannot be observed for `s2f_inv3` because of the focus of Network (4) on the FDK input which does not incorporate different information for sagittal or coronal planes. The reconstruction using `inv_sp` introduces some additional discontinuities, probably caused by the worse quality of the network on sagittal planes.

## 5    Discussion

In general, only one of the proposed combinations in fact improves the simple post-processing baseline `fdkconv`, which is `inv_sp3`. A possible explanation for this is that, compared to the other combinations, `inv_sp3` can directly learn to extract the most useful information from both the sparse view FDK (which already contains correct frequency information) and DBP (which is able to incorporate information of truncated projections as well as different information from sparse views compared to FDK). The other combinations do not include the FDK reconstruction (`inv_sp` and `s2f_inv`) or are not trained end-to-end (`s2f_inv` and `s2f_inv3`), which does not allow the gradients to flow back completely.

As described earlier, the additional input of Network (4) was the FDK reconstruction of 360 projections while training and of 36 projections while testing `s2f_inv3`. For further tests, the output of `fdkconv` could be used as this additional input to be closer to what the network was trained on.

Moreover, there is a significant difference in accuracy of all networks between coronal and sagittal planes, which might be caused by less variance in the sagittal planes. Additional data or different augmentation techniques could resolve this.

The spectral blending results in even lower errors but depends on the masks that are used. The almost linear dependency of the blurring radius of the mask on the final error metrics (Fig. 2) suggests increasing the radius even further or using masks of non-bow-tie shape.

We discovered that all networks including some kind of Hilbert inversion need high learning rates $\geq 10^{-2}$. Setting them lower resulted in both higher loss values and less robust trainings, which seems rather counter-intuitive. Trainings with learning rates between $10^{-4}$ and $10^{-5}$ (cf. [1]) did not converge at all, which might be related to the data set or normalization of the data.

To further increase the reconstruction quality, it is conceivable to additionally input neighboring planes-of-interest to the networks to gain more spatial information. In addition, the effect of choosing different values for $\varepsilon$ for creating the partial derivatives was not investigated and needs further tests. The code is available on Github[1].

## 6    Conclusion

In this paper, an approach for enhancing CT reconstructions from sparse cone beam projections combining convolutional neural networks and differentiated backprojection was presented. Since errors caused by undersampling of the trajectory occur at different stages of the algorithm, analytical steps and CNNs were joined in different combinations to find the optimal strategy. We discovered that inverting the sparse view backprojection with a CNN from 36 projections works best when additionally providing the FDK reconstruction to the network, increasing the PSNR by almost 3 dB and 1.5 dB compared to a simple post-processing CNN for coronal and sagittal planes-of-interest, respectively. Spectral blending further increases the PSNR to 84.61 dB when using a radially blurred bow-tie mask.

## Acknowledgements

This work was conducted within the International Graduate School MEMoRIAL at OVGU Magdeburg, supported by the ESF (project no. ZS/2016/08/80646).

---

[1] https://github.com/phernst/sparse_dbp

# Learning Overcomplete or Undercomplete Models in Clustering-based Low-dose CT Reconstruction

Ling Chen[1], Yong Long[1], and Saiprasad Ravishankar[2]

[1]University of Michigan-Shanghai Jiao Tong University Joint Institute, Shanghai Jiao Tong University, Shanghai 200240, China

[2]Department of Computational Mathematics, Science and Engineering and Department of Biomedical Engineering, Michigan State University, East Lansing, MI 48824, USA

**Abstract** The development of computed tomography (CT) image reconstruction methods that significantly reduce patient radiation exposure, while maintaining high image quality is an important area of research in low-dose CT imaging. We propose a new method for CT image reconstruction that combines penalized weighted-least squares reconstruction (PWLS) with regularization based on a union of undercomplete (and alternatively overcomplete) sparsifying transforms learned from datasets. The proposed cost function can be efficiently solved by alternating between an image update step and a sparse coding and clustering step. Simulations on the XCAT phantom show that for low-dose levels the proposed method significantly improves the quality of reconstructed images compared to PWLS reconstruction with a nonadaptive edge-preserving regularizer and is comparable with the previous union of learned.

## 1 Introduction

Model-based image reconstruction (MBIR) methods also known as statistical image reconstruction methods [1] improve computed tomography (CT) image quality while greatly reducing patient exposure to potentially harmful levels of radiation. MBIR methods can be divided into two categories: nonadaptive methods and learning-based methods. Edge-preserving (EP) regularization is widely used for non-adaptive image reconstruction. For learning-based methods, dictionary learning [2] has shown promising performance in CT image reconstruction, but the efficacy is limited by expensive sparse coding, e.g., with the orthogonal matching pursuit (OMP) algorithm. Ravishankar and Bresler [3] proposed efficiently learning square sparsifying transforms (ST) for images. In contrast to the often highly non-convex and NP-hard dictionary learning problems, the transform model can be learned efficiently [4] due to the simple thresholding-based sparse coding solution in the transform domain. The square transform learning approach was later extended to handle overcomplete transforms [5], and shown to outperform synthesis dictionary learning (K-SVD) for image denoising, while being faster.

Zheng et al. [6] proposed a PWLS-ST method which has demonstrated the superior performance of learned ST for low-dose CT reconstruction. Zheng et al. [6] then extended PWLS-ST to learning a more general union of square sparsifying transforms (ULTRA). Deep neural networks (NNs) have been applied to low-dose CT image reconstruction problems. Early works use the powerful mapping capacity of deep learning to transform noise contaminated CT images into the desired high quality images [7]. However, the global mapping capability in deep learning also leads to the fundamental challenge of generalizability issue that causes some artificial features when test images are not similar to training images.

In this work, we propose a PWLS-Union-UST (PWLS-Union-OST) method (UST and OST denote undercomplete sparsifying transforms and overcomplete sparsifying transforms, respectively) which combines the penalized weighted-least squares (PWLS) estimation with regularization based on a union of pre-learned undercomplete (overcomplete) sparsifying transforms. We propose efficient algorithms for learning and reconstruction and show that the proposed scheme outperforms conventional PWLS-EP and is comparable to PWLS-ULTRA with the image quality improving somewhat with increasing transform size in our schemes.

## 2 Problem Formulations

### 2.1 Learning a Union of Undercomplete Sparsifying Transforms

To learn a Union of UST $\{\boldsymbol{\Omega}_k\}_{k=1}^K$ ($\boldsymbol{\Omega}_k \in \mathbb{R}^{m \times l}$, $m < l$) from $N'$ (vectorized) image patches, we optimize:

$$\min_{\{\boldsymbol{\Omega}_k, \mathbf{Z}_i, \mathbf{C}_k\}} \sum_{k=1}^K \left\{ \sum_{i \in \mathbf{C}_k} \left\{ \|\boldsymbol{\Omega}_k \mathbf{X}_i - \mathbf{Z}_i\|_2^2 + \gamma^2 \|\mathbf{Z}_i\|_0 \right\} + \mu \sum_{\substack{l,m,l \neq m}} \langle \boldsymbol{\Omega}_{kl}, \boldsymbol{\Omega}_{km} \rangle^2 + \eta \sum_m (\|\boldsymbol{\Omega}_{km}\|_2^2 - 1)^2 \right\} \quad \text{(P0)}$$

Problem (P0) groups the training signals $\mathbf{X}_i \in \mathbb{R}^l$ into $K$ classes according to the transforms that best sparsify them, and $\mathbf{C}_k$ denotes the set of indices of patches matched to the $k$th cluster. $\{\mathbf{Z}_i\}_{i=1}^{N'}$ denotes the sparse codes of the training signals (vectoried patches) $\{\mathbf{X}_i\}_{i=1}^{N'}$. $\boldsymbol{\Omega}_{km}$ means the $m$th row of the $k$th transform matrix $\boldsymbol{\Omega}_k$. The $l_0$ "norm" counts the number of non-zeros in a vector. The term $\|\boldsymbol{\Omega}_k \mathbf{X}_i - \mathbf{Z}_i\|_2^2$ is called the sparsification error and measures the deviation of the signals in the transform domain from their sparse approximations. The penalties $\mu \sum_{l,m,l \neq m} \langle \boldsymbol{\Omega}_{kl}, \boldsymbol{\Omega}_{km} \rangle^2$ and $\eta \sum_m (\|\boldsymbol{\Omega}_{km}\|_2^2 - 1)^2$ together help control the condition number of $\boldsymbol{\Omega}_k \boldsymbol{\Omega}_k^T$. The parameters $\mu = \mu_0 \|\mathbf{X}_{\mathbf{C}_k}\|_F^2$ and $\eta = \eta_0 \|\mathbf{X}_{\mathbf{C}_k}\|_F^2$, where $\mu_0 > 0$ and $\eta_0 > 0$ are constants and $\mathbf{X}_{\mathbf{C}_k}$ is a matrix whose columns are the training signals in the $k$th cluster. The weight $\|\mathbf{X}_{\mathbf{C}_k}\|_F^2$ for the penalties allows them to scale similarly as the sparsification error term. The param-eter $\gamma = \gamma_0 \|\mathbf{X}\|_F$, and the weight $\|\mathbf{X}\|_F$ for the sparse codes scale appropriately with the data.

## 2.2 Learning a Union of Overcomplete Sparsifying Transforms

To learn a Union of OST $\{\mathbf{\Omega}_k\}_{k=1}^K$ ($\mathbf{\Omega}_k \in \mathbb{R}^{m \times l}$, $m > l$) from $N'$ (vectorized) patches, we solve:

$$\min_{\{\mathbf{\Omega}_k, \mathbf{Z}_i, \mathbf{C}_k\}} \sum_{k=1}^K \left\{ \sum_{i \in \mathbf{C}_k} \left\{ \|\mathbf{\Omega}_k \mathbf{X}_i - \mathbf{Z}_i\|_2^2 + \gamma^2 \|\mathbf{Z}_i\|_0 \right\} - \lambda \log \det \left( \mathbf{\Omega}_k^T \mathbf{\Omega}_k \right) \right.$$
$$\left. + \mu \sum_{l \neq m} \langle \mathbf{\Omega}_{kl}, \mathbf{\Omega}_{km} \rangle^2 + \eta \sum_m (\|\mathbf{\Omega}_{km}\|_2^2 - 1)^2 \right\} \quad \text{(P1)}$$

The terms $\mu \sum_{l,m,l \neq m} |\langle \mathbf{\Omega}_{kl}, \mathbf{\Omega}_{km} \rangle|^2$, $\eta \sum_m (\|\mathbf{\Omega}_{km}\|_2^2 - 1)^2$ and $\lambda \log \det \left( \mathbf{\Omega}_k^T \mathbf{\Omega}_k \right)$ together help control the conditioning and incoherence of $\mathbf{\Omega}_k$. The parameter $\lambda = \lambda_0 \|\mathbf{X}_{C_k}\|_F^2$. Note that unlike (P0), where $\mathbf{\Omega}_k$ has more columns than rows, it has fewer columns than rows in (P1), and the log determinant penalty allows controlling the condition number of $\mathbf{\Omega}_k^T \mathbf{\Omega}_k$ [5].

## 2.3 LDCT Reconsturction

We solve the following optimization problem to reconstruct an image $\mathbf{x} \in \mathbb{R}^{N_p}$ from noisy sinogram data $\mathbf{y} \in \mathbb{R}^{N_d}$ using a pre-learned Union of transforms [6] $\{\mathbf{\Omega}_k\}_{k=1}^K$:

$$\min_{\mathbf{x} \geq 0} \frac{1}{2} \|\mathbf{y} - \mathbf{A}\mathbf{x}\|_{\mathbf{W}}^2 + \beta \mathbf{R}(\mathbf{x}) \quad \text{(P2)}$$

where $\mathbf{W} = diag\{w_i\} \in \mathbb{R}^{N_d \times N_d}$ is a diagonal weighting matrix with elements being the estimated inverse variance of $y_i$, $\mathbf{A} \in \mathbb{R}^{N_d \times N_p}$ is the system matrix of a CT scan, the parameter $\beta > 0$ controls the noise and resolution trade-off, and the regularizer $\mathbf{R}(\mathbf{x})$ based on a union of UST $\{\mathbf{\Omega}_k\}_{k=1}^K$ is defined as:

$$\mathbf{R}(\mathbf{x}) \triangleq \min_{\{\mathbf{z}_j, \mathbf{C}_k\}} \sum_{k=1}^K \sum_{j \in \mathbf{C}_k} \left\{ \|\mathbf{\Omega}_k \mathbf{P}_j \mathbf{x} - \mathbf{z}_j\|_2^2 + \gamma^2 \|\mathbf{z}_j\|_0 \right\}. \quad (1)$$

The operator $\mathbf{P}_j \in \mathbb{R}^{l \times N_p}$ extracts the $j$th patch of $l$ voxels of $\mathbf{x}$ as $\mathbf{P}_j \mathbf{x}$. The regularizer includes a sparsification error term and a $l_0$ "norm"-based sparsity penalty with weight $\gamma^2$.

## 3 Algorithms

### 3.1 Algorithm for Learning a Union of Sparisifying Transforms

We use an alternating algorithm to solve (P0) that alternates between updating $\{\mathbf{\Omega}_k\}$ (transform update step) and $\{\mathbf{Z}_i, \mathbf{C}_k\}$ (sparse coding and clustering step). The algorithm for (P1) is similar and so we describe for (P0) below.
In the first step called transform update step, we solve problem (P0) with $\{\mathbf{Z}_i, \mathbf{C}_k\}$ fixed as follows:

$$\min_{\{\mathbf{\Omega}_k\}} \sum_{k=1}^K \sum_{i \in \mathbf{C}_k} \|\mathbf{\Omega}_k \mathbf{X}_i - \mathbf{Z}_i\|_2^2 + \sum_{k=1}^K \left\{ \mu \sum_{l \neq m} \langle \mathbf{\Omega}_{kl}, \mathbf{\Omega}_{km} \rangle^2 \right.$$
$$\left. + \eta \sum_m (\|\mathbf{\Omega}_{km}\|_2^2 - 1)^2 \right\}. \quad (2)$$

We use the nonlinear conjugate gradient (CG) method to update each $\mathbf{\Omega}_k$ in (2). The details are shown in Algorithm 1.

---

**Algorithm 1** Learning a Union of Undercomplete Sparsifying Transforms

**Inputs:** initial transform: $\{\mathbf{\Omega}_k^0\}_{k=1}^K$, the step size $\alpha$ (could be also obtained by line search), the parameters $\mu_0$ and $\eta_0$, the training matrix $\mathbf{X}$, number of outer iterations P, number of inner iterations N.

**Outputs:** learned transform $\{\tilde{\mathbf{\Omega}}_k^P\}_{k=1}^K$

**for** $p = 0, 1, 2, \ldots, P-1$ **do**
  **for** $k = 0, 1, 2, \ldots, K-1$ **do**
    **(1) Transform Update:** with $\{\tilde{\mathbf{Z}}_i^p, \tilde{\mathbf{C}}_k^p\}$ fixed,
    **Initialization:**
    **for** $n = 0, 1, 2, \ldots, N-1$ **do**
      $\nabla_{\mathbf{\Omega}_k^{(n)}} \|\mathbf{\Omega}_k^{(n)} \mathbf{X}_i - \tilde{\mathbf{Z}}_i^p\|_F^2 = 2\mathbf{\Omega}_k^{(n)} \mathbf{X}_i \mathbf{X}_i^T - 2\tilde{\mathbf{Z}}_i^p \mathbf{X}_i^T = \mathbf{G}_1^{(n)}$
      $\nabla_{\mathbf{\Omega}_k^{(n)}} \sum_{l \neq m} \left\langle \mathbf{\Omega}_{kl}^{(n)}, \mathbf{\Omega}_{km}^{(n)} \right\rangle^2 = 2 \left( \mathbf{C}^{(n)} \mathbf{\Omega}_k^{(n)} - \mathbf{B}^{(n)} \right) = \mathbf{G}_2^{(n)}$
      $\nabla_{\mathbf{\Omega}_k^{(n)}} \sum_m (\|\mathbf{\Omega}_{km}^{(n)}\|_2^2 - 1)^2 = \mathbf{D}^{(n)}$

      The matrices $\mathbf{C}^{(n)}$ and $\mathbf{B}^{(n)}$ above have entries $c_{ij}^{(n)} = \left\langle \mathbf{\Omega}_{ki}^{(n)}, \mathbf{\Omega}_{kj}^{(n)} \right\rangle$ and $b_{ij}^{(n)} = c_{ii}^{(n)} (\mathbf{\Omega}_k^{(n)})_{ij}$. The matrix $\mathbf{D}^{(n)} \in \mathbb{R}^{m \times l}$ has entries $d_{ij}^{(n)} = 4(\mathbf{\Omega}_k^{(n)})_{ij} (\sum_{q=1}^l (\mathbf{\Omega}_k^{(n)})_{iq}^2 - 1)$.
      $\mathbf{G}^{(n)} = \mathbf{G}_1^{(n)} + \mu_0 \|\mathbf{X}_{\tilde{\mathbf{C}}_k^p}\|_F^2 \mathbf{G}_2^{(n)} + \eta_0 \|\mathbf{X}_{\tilde{\mathbf{C}}_k^p}\|_F^2 \mathbf{D}^{(n)}$
      **if** $n = 0$ **then**
        $\mathbf{d}^{(n)} = -\mathbf{G}^{(n)}$
      **else**
        $\mathbf{d}^{(n)} = -\mathbf{G}^{(n)} + \frac{\|\mathbf{G}^{(n)}\|_2^2}{\|\mathbf{G}^{(n-1)}\|_F^2} \mathbf{d}^{(n-1)}$
      **end if**
      $\mathbf{\Omega}_k^{(n+1)} = \mathbf{\Omega}_k^{(n)} + \alpha \mathbf{d}^{(n)}$
    **end for**
  **end for**
  $\tilde{\mathbf{\Omega}}_k^{p+1} = \mathbf{\Omega}_k^{(N)}$
  **(2) Sparse Coding and Clustering:** with $\tilde{\mathbf{\Omega}}_k^p$ fixed, the optimal cluster membership for each $\mathbf{X}_i$ is: $\hat{k}_i = \arg\min_{1 \leq k \leq K} \|\tilde{\mathbf{\Omega}}_k^p \mathbf{X}_i - H_\gamma(\tilde{\mathbf{\Omega}}_k^p \mathbf{X}_i)\|_2^2 + \gamma^2 \|H_\gamma(\tilde{\mathbf{\Omega}}_k^p \mathbf{X}_i)\|_0 + \mu_0 \|\mathbf{X}_i\|_2^2 \sum_{l \neq m} \left\langle \tilde{\mathbf{\Omega}}_{kl}^p, \tilde{\mathbf{\Omega}}_{km}^p \right\rangle^2 + \eta_0 \|\mathbf{X}_i\|_2^2 \sum_m (\|\tilde{\mathbf{\Omega}}_{km}^p\|_2^2 - 1)^2$. Then the optimal sparse codes are $\tilde{\mathbf{Z}}_i^{(p+1)} = H_\gamma(\tilde{\mathbf{\Omega}}_{\hat{k}_i}^{p+1} \mathbf{X}_i)$.
**end for**

---

In the sparse coding and clustering step of (P0) (or (P1)), we solve for $\{\mathbf{Z}_i, \mathbf{C}_k\}$ with fixed $\{\mathbf{\Omega}_k\}$ as follows:

$$\min_{\{\mathbf{Z}_i, \mathbf{C}_k\}} \sum_{k=1}^K \sum_{i \in \mathbf{C}_k} \left\{ \|\mathbf{\Omega}_k \mathbf{X}_i - \mathbf{Z}_i\|_2^2 + \gamma^2 \|\mathbf{Z}_i\|_0 \right.$$
$$\left. + \mu_0 \|\mathbf{X}_i\|_2^2 \sum_{l \neq m} \langle \mathbf{\Omega}_{kl}, \mathbf{\Omega}_{km} \rangle^2 + \eta_0 \|\mathbf{X}_i\|_2^2 \sum_m (\|\mathbf{\Omega}_{km}\|_2^2 - 1)^2 \right\}. \quad (3)$$

For each $k$, we can replace $\mathbf{Z}_i$ in (3) with the optimal sparse codes, $\mathbf{Z}_i = H_\gamma(\mathbf{\Omega}_k \mathbf{X}_i)$, where the hard-thresholding operator $H_\gamma(\cdot)$ sets entries with magnitude less than $\gamma$ to zero, leaving other entries unchanged. Then, the optimal cluster membership for each $\mathbf{X}_i$ can be obtained as:

$$\hat{k}_i = \arg\min_{1 \leq k \leq K} \|\mathbf{\Omega}_k \mathbf{X}_i - H_\gamma(\mathbf{\Omega}_k \mathbf{X}_i)\|_2^2 + \gamma^2 \|H_\gamma(\mathbf{\Omega}_k \mathbf{X}_i)\|_0$$
$$+ \mu_0 \|\mathbf{X}_i\|_2^2 \sum_{l \neq m} \langle \mathbf{\Omega}_{kl}, \mathbf{\Omega}_{km} \rangle^2 + \eta_0 \|\mathbf{X}_i\|_2^2 \sum_m (\|\mathbf{\Omega}_{km}\|_2^2 - 1)^2. \quad (4)$$

Then the optimal $\hat{\mathbf{Z}}_i = H_\gamma(\mathbf{\Omega}_{\hat{k}_i} \mathbf{X}_i)$.

### 3.2 LDCT Reconstruction Algorithm

We use an alternating algorithm to solve (P2) that alternates between upating $\mathbf{x}$ (image update step), and $\{\mathbf{z}_j, \mathbf{C}_k\}$ (sparse coding and clustering step).
In the image update step, With $\{\mathbf{z}_j, \mathbf{C}_k\}$ fixed, (P2) reduces

---

**Algorithm 2** Image Reconstruction Algorithm

**Inputs:** initial image $\bar{\mathbf{x}}^{(0)}$, pre-learned $\{\mathbf{\Omega}_k\}$, threshold $\gamma$, $\alpha = 1.999$, $\mathbf{D}_{\mathbf{A}} \succeq \mathbf{A}^T \mathbf{W} \mathbf{A}$, $\mathbf{D}_{\mathbf{R}} \triangleq 2\beta \sum_{j=1}^{N} \mathbf{P}_j^T \mathbf{P}_j \lambda_{max}(\mathbf{\Omega}_k^T \mathbf{\Omega}_k)$, number of outer iterations T , number of inner iterations N, and number of subsets M.

**Outputs:** reconstructed image $\bar{\mathbf{x}}^{(T)}$, cluster indices $\bar{\mathbf{C}}_k^{(T)}$ reconstructed image: $\bar{\mathbf{x}}$

    **for** $t = 0, 1, 2, \ldots, T - 1$ **do**
        **(1) Image Update:** with $\{\bar{\mathbf{z}}_j^{(t)}, \bar{\mathbf{C}}_k^{(t)}\}$ fixed,

        **Initialization:** $\rho = 1$, $\mathbf{x}^{(0)} = \bar{\mathbf{x}}^{(t)}$, $\zeta^{(0)} = \mathbf{g}^{(0)} = M\mathbf{A}_M^T \mathbf{W}(\mathbf{A}_M \bar{\mathbf{x}}^{(t)} - \mathbf{y}_M)$, $\mathbf{h}^{(0)} = \mathbf{D}_{\mathbf{A}} \bar{\mathbf{x}}^{(t)} - \zeta^{(0)}$
        iteratively update x using OS-LALM [6] with N iterations and M ordered subsets
        $\bar{\mathbf{x}}^{(t+1)} = \mathbf{x}^{(NM)}$

        **(2) Sparse Coding and Clustering:** with $\bar{\mathbf{x}}^{(t+1)}$ fixed, the optimal cluster membership for each $\mathbf{X}_i$ is: $\hat{k}_j = \arg\min_{1 \leq k \leq K} \|\mathbf{\Omega}_k \mathbf{P}_j \bar{\mathbf{x}}^{(t+1)} - H_\gamma(\mathbf{\Omega}_k \mathbf{P}_j \bar{\mathbf{x}}^{(t+1)})\|_2^2 + \gamma^2 \|H_\gamma(\mathbf{\Omega}_k \mathbf{P}_j \bar{\mathbf{x}}^{(t+1)})\|_0$. Then the optimal sparse codes are $\bar{\mathbf{z}}_j^{(t+1)} = H_\gamma(\mathbf{\Omega}_{\hat{k}_j} \mathbf{P}_j \bar{\mathbf{x}}^{(t+1)})$.

    **end for**

---

to the following weighted least squares problem:

$$\min_{\mathbf{x} \geq 0} \frac{1}{2} \|\mathbf{y} - \mathbf{A}\mathbf{x}\|_{\mathbf{W}}^2 + \beta \sum_{k=1}^{K} \sum_{j \in \mathbf{C}_k} \left\{ \|\mathbf{\Omega}_k \mathbf{P}_j \mathbf{x} - \mathbf{z}_j\|_2^2 + \gamma^2 \|\mathbf{z}_j\|_0 \right\}. \quad (5)$$

We solve the problem using the relaxed linearized augmented lagrangian method with ordered-subsets (relaxed OS-LALM) [8]. The algorithmic details are shown in Algorithm 2. In the sparse coding and clustering step for (P2), with x fixed, we update $\{\mathbf{z}_j, \mathbf{C}_k\}$ by solving:

$$\min_{\{\mathbf{z}_j, \mathbf{C}_k\}} \sum_{k=1}^{K} \sum_{j \in \mathbf{C}_k} \left\{ \|\mathbf{\Omega}_k \mathbf{P}_j \mathbf{x} - \mathbf{z}_j\|_2^2 + \gamma^2 \|\mathbf{z}_j\|_0 \right\}. \quad (6)$$

The optimal cluster membership for each $\mathbf{X}_i$ is:

$$\hat{k}_j = \arg\min_{1 \leq k \leq K} \|\mathbf{\Omega}_k \mathbf{P}_j \mathbf{x} - H_\gamma(\mathbf{\Omega}_k \mathbf{P}_j \mathbf{x})\|_2^2 + \gamma^2 \|H_\gamma(\mathbf{\Omega}_k \mathbf{P}_j \mathbf{x})\|_0. \quad (7)$$

Then the optimal sparse codes are $\mathbf{z}_j = H_\gamma(\mathbf{\Omega}_{\hat{k}_j} \mathbf{P}_j \mathbf{x})$.

## 4 Experiment Results

### 4.1 Experiment Setup

We trained both a UST and an OST (1 cluster) along with a union of UST and OST (5 clusters) from 5 different slices of an XCAT phantom [9] using (P0). We extracted $8 \times 8$ overlapping image patches with a patch stride of 1 pixel from the five $512 \times 512$ XCAT slices.

We simulated a 2D fan-beam CT scan using a $1024 \times 1024$ XCAT phantom slice, which is different from the learning slices, with pixel dimensions $\Delta_x = \Delta_y = 0.4883mm$. Noisy (Poisson noise) sinograms of size $888 \times 984$ were numerically generated with GE LightSpeed fan-beam geometry corresponding to a monoenergetic source with $10^4$ incident photons per ray and no scatter. We reconstructed a $512 \times 512$ image with a coarser grid, where $\Delta_x = \Delta_y = 0.9766mm$.

To compare the performance quantitatively, we computed the root mean square error (RMSE) in Hounsfield units (HU) for the reconstructed images. For a reconstructed image $\hat{\mathbf{x}}$, RMSE is defined as $\sqrt{\sum_{j=1}^{N_p} (\hat{x}_j - x_j^\star)^2 / N_p}$, where $x_j^\star$ denotes the down-sampled true image intensity at the $j$th pixel location and $N_p$ is the number of pixels in the phantom support (a circle removes all the background area outside the image that is not interesting).

### 4.2 Transform Learning and LDCT Reconstruction Results

We evaluate PWLS-UST (OST) and compare its performance to PWLS-EP [8] and PWLS-ST [6]. We ran 2000 iterations of the CG algorithm to make sure the learned UST and OST completely converged. For training the UST (transform size: $56 \times 64$), we chose parameters, $\gamma = 100$, $\alpha = 3 \times 10^{-14}$ and $\lambda_0 = \mu_0 = 10^{-1}$. We also tuned parameters to train an OST ($128 \times 64$) and other different sizes of UST. Figure 1 shows the pre-learned UST and the OST. Each row of these transforms is displayed as an $8 \times 8$ patch. During image reconstruction, we used the image obtained after a few iterations of the PWLS-EP method as initialization for the adaptive schemes to realize faster convergence. We set $\gamma$ as 20 and $\beta$ as $1.5 \times 10^5$ for PWLS-UST ($56 \times 64$) to achieve a good trade-off between reconstructed image quality and convergence speed. In each iteration, we ran 2 inner iterations of the image update step. We ran 500 outer iterations to ensure convergence.

We tuned the parameters of all transform learning-based PWLS methods to obtain the best performances. Figure 2 shows the reconstructed images with PWLS-EP, PWLS-OST ($128 \times 64$), PWLS-ST, and PWLS-UST ($56 \times 64$ and $48 \times 64$) along with the ground truth. We can observe that PWLS-OST reduces the severe noise and artifacts observed in the PWLS-EP image, and reconstructs more details of the image such as the zoom-in areas. Table 1 lists the RMSE and SSIM values for ST, EP, UST of different sizes and OST. The metrics improve with increasing transform size.

**Table 1:** RMSE and SSIM values for the testing image with EP, OST, ST and UST with different transform sizes.

|  | EP | OST($128 \times 64$) | ST | UST($56 \times 64$) |
|---|---|---|---|---|
| RMSE | 39.4 | 35.5 | 36.5 | 36.5 |
| SSIM | 0.892 | 0.965 | 0.963 | 0.960 |
|  | UST($48 \times 64$) | UST($40 \times 64$) | UST($32 \times 64$) | UST($24 \times 64$) |
| RMSE | 36.6 | 36.6 | 36.7 | 37.0 |
| SSIM | 0.956 | 0.956 | 0.955 | 0.948 |

### 4.3 LDCT Results with Unions of Learned Transforms

Next, we evaluate PWLS-Union-UST (and OST) and compare its performance to PWLS-EP and PWLS-ULTRA [6]. We ran 2000 iterations of the CG algorithm to make sure the learned unions of transforms completely converged. For a Union of UST (transform size: $56 \times 64$), we chose parameters, $\gamma = 50$, $\alpha = 10^{-14}$ and $\lambda_0 = \mu_0 = 10^{-2}$. We also tuned

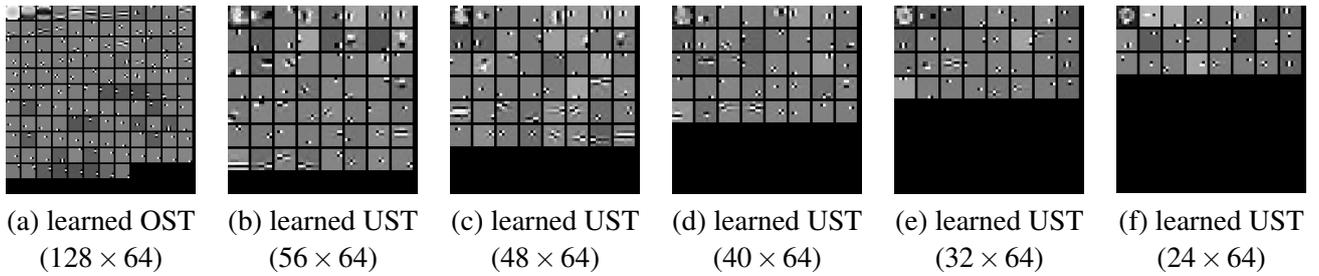

(a) learned OST          (b) learned UST          (c) learned UST          (d) learned UST          (e) learned UST          (f) learned UST
$(128 \times 64)$          $(56 \times 64)$          $(48 \times 64)$          $(40 \times 64)$          $(32 \times 64)$          $(24 \times 64)$

**Figure 1:** Rows of the learned sparsifying transform (only 1 cluster) shown as $8 \times 8$ patches for (a) learned OST $(128 \times 64)$, (b) learned UST $(56 \times 64)$, (c) learned UST $(48 \times 64)$, (d) learned UST $(40 \times 64)$, (e) learned UST $(32 \times 64)$ and (f) learned UST $(24 \times 64)$.

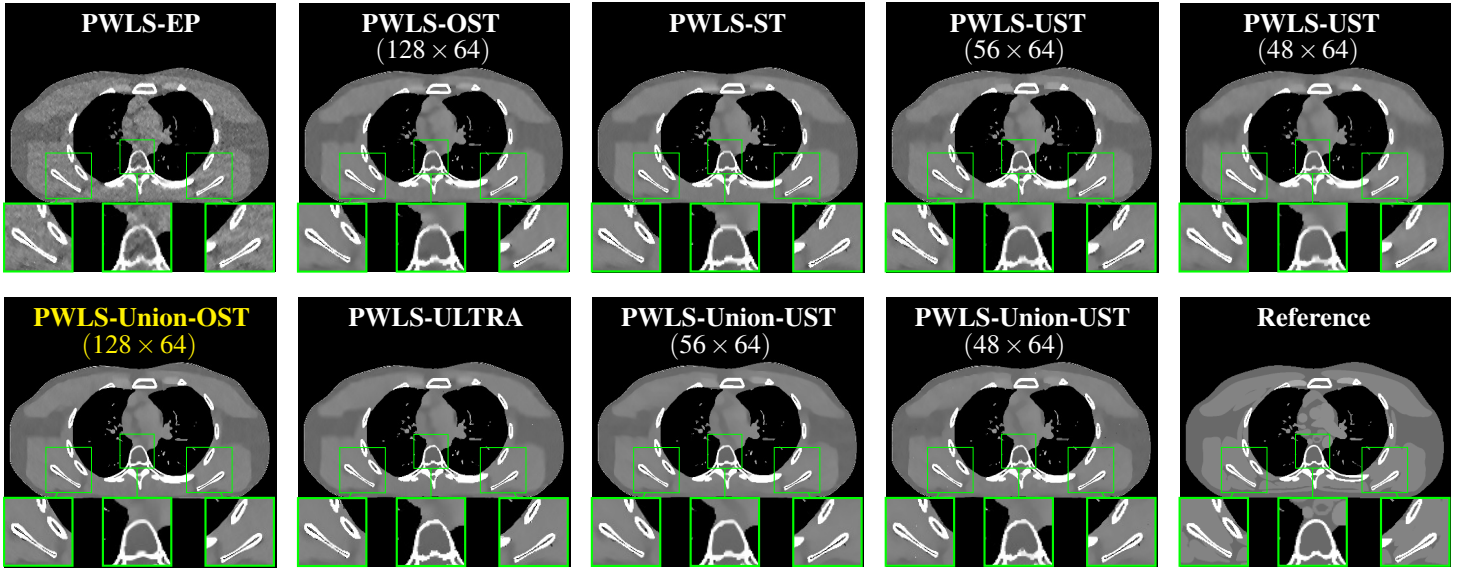

**Figure 2:** Comparison of reconstructions of XCAT phamtom with PWLS-EP, PWLS-OST $(128 \times 64)$, PWLS-ST, PWLS-UST $(56 \times 64)$, PWLS-UST $(48 \times 64)$, PWLS-Union-OST $(128 \times 64)$, PWLS-ULTRA, PWLS-Union-UST $(56 \times 64)$, PWLS-Union-UST $(48 \times 64)$ and Reference. The display window is [800, 1200] HU. The incident photon intensities $I_0 = 10^4$.

parameters to train a Union of OST $(128 \times 64)$ and Unions of UST with different sizes.

**Table 2:** RMSE and SSIM values for the testing image with Union-OST, ULTRA, and Union-UST with different transform sizes. We use 5 clusters.

| | Union-OST $(128 \times 64)$ | ULTRA | Union-UST $(56 \times 64)$ | Union-UST $(48 \times 64)$ |
|---|---|---|---|---|
| RMSE | 34.0 | 34.4 | 34.7 | 35 |
| SSIM | 0.967 | 0.967 | 0.967 | 0.966 |
| | Union-UST $(40 \times 64)$ | Union-UST $(32 \times 64)$ | Union-UST $(24 \times 64)$ | |
| RMSE | 35.2 | 35.6 | 35.8 | |
| SSIM | 0.965 | 0.959 | 0.957 | |

We used the image obtained after a few iterations of the PWLS-EP method as initialization to obtain faster convergence for the adaptive methods. We set $\gamma$ as 15 and $\beta$ as $5 \times 10^5$ for PWLS-Union-UST $(56 \times 64)$ to achieve a good trade-off between reconstructed image quality and convergence speed. We ran 500 outer iterations of PWLS-Union-UST (OST) for convergence and ran 2 inner iterations of the image update step. We also tuned the parameters of all reconstruction algorithms for best RMSE and SSIM.

Figure 2 shows the reconstructed images with PWLS-Union-OST, PWLS-ULTRA and PWLS-Union-UST $(56 \times 64$ and $48 \times 64)$ along with the ground truth. The zoom-in areas show that PWLS-Union-OST can reconstruct image details better. Moreover, the union of transforms approaches outperform the corresponding single transform (1 cluster) schemes. Table 2 lists the corresponding RMSE and SSIM values with Union-OST, ULTRA, and different sized transforms in Union-UST. The Union-OST scheme achieves the best RMSE and SSIM.

## 5 Conclusion

We propose a new method for CT image reconstruction that combines the conventional PWLS estimation with a regularizer involving a pre-learned union of undercomplete (overcomplete) sparsifying transforms. We demonstrated that the proposed PWLS-Union-UST (PWLS-Union-OST) method outperformed the PWLS-EP method (which uses a fixed finite differencing type sparsifying model) in terms of image quality in low-dose simulations. The proposed method is also comparable with PWLS-ULTRA (that uses a union of pre-learned square sparsifying transforms), with RMSE improving somewhat with increasing transform size. In future work, we will apply the proposed methods to clinical data.

# A Simulation Study of a Novel High-Resolution CT Imaging Technique: Zoom-In Partial Scans (ZIPS)

Lin Fu, Eri Haneda, Bernhard Claus, Uwe Wiedmann, and Bruno De Man

Radiation Imaging, GE Research, Niskayuna, NY, USA

**Abstract** The spatial resolution of clinical multi-slice CT is mainly limited by the size of CT detector cells. Upgrading the entire CT detector array to one with finer-pitched cells is expensive for widespread application. Here we propose a novel high-resolution CT scheme, named Zoom-In Partial Scans (ZIPS), to improve the resolution of clinical CT scanners without a costly upgrade of the detector array. Unlike a conventional CT scan where a region of interest (ROI) is positioned at the center of the scanner's field of view, in ZIPS CT the ROI is positioned off-center to increase the geometric magnification of the ROI and overcome the resolution limit of the detector cell size. A ZIPS reconstruction algorithm is developed to merge two off-center limited-angle partial scans to form a single high-resolution image of the ROI. The proposed ZIPS technique was evaluated in a CatSim simulation environment. Improvement in modulation transfer functions from 30% to 100% relative to a conventional centered scan was observed.

**Keywords:** high-resolution CT, magnification, limited angle, image reconstruction

## 1 Introduction

Revealing finer anatomical details in CT imaging has long been sought in clinical areas such as inner ear imaging and assessment of bone microstructure. Various efforts have been devoted to the improvement of spatial resolution of clinical multi-slice CT. Some high-resolution scanners are based on a redesign of the entire CT detector array to one with half-pitched cells [1]–[3], which are costly for widespread clinical application as the number of detector cells would quadruple. Some experimental scanners combine flat panel detector technology with a standard clinical CT gantry to achieve high resolution [4]–[6], but the slow read out speed, marked hysteresis effects, and high scatter-to-primary ratio preclude their use in many standard clinical CT applications. Specialty CT scanners, such as high-resolution extremity or dental CT [7]–[9] are less ubiquitous and accessible, and their small bore size and limited X-ray tube power preclude imaging of larger anatomies such as the spine and pelvis.

Here we propose a novel CT scanning scheme, named Zoom-In Partial Scans (ZIPS), to improve the intrinsic spatial resolution of CT imaging of a local region of interest (ROI) without upgrading the CT detector array. Unlike conventional CT imaging where the ROI is placed at the center of the scanner's field of view (FOV), ZIPS CT introduces a novel off-center scanning scheme to increase the geometric magnification of the ROI when projected onto the detector, combined with a small focal spot size. The high magnification can "zoom in" the ROI to overcome the resolution limit imposed by the detector cell pitch. For rotating-gantry-based CT, because the high magnification of an off-center ROI can only be achieved over a limited angular range, i.e., when the X-ray source rotates on the same side as the ROI relative to the rotation center, we call such an off-center scan a Zoom-In Partial Scan (ZIPS). We utilize *two* partial scans to acquire the complete high-resolution projection data of the ROI. A ZIPS image reconstruction algorithm is developed to merge the two partial scans and reconstruct a single high-resolution image of the ROI.

In this paper we introduce the ZIPS CT scheme, show the feasibility of an initial ZIPS reconstruction algorithm, and evaluate improvement of the spatial resolution in a CatSim [10] simulation environment. Modulation transfer functions (MTF) and examples of reconstructed images are presented as results.

## 2. Theory

### Magnification of an off-center ROI

On a clinical CT scanner, the fan-beam magnification ratio at the isocenter is typically ~1.7X (source-to-detector distance ≈ 1 m, source-to-isocenter distance ≈ 0.6 m). However, an off-center ROI can have higher magncation (Fig. 1). The magnification can reach up to ~2.5X for an ROI at 20 cm off center when the X-ray source rotates on the same side as the ROI. The magnification can further reach up to ~3.3X for an ROI at 30 cm off center. The diameter of the bore opening of a modern clinical CT scanner is typically about 80 cm. Depending on the patient size and the anatomy of interest, a patient may be positioned off center when possible to increase the magnification of an ROI. We intend to utilize the high magnification of the off-center ROI to acquire high-resolution projection data and overcome the resolution limit of the detector cell size.

As the CT gantry rotates, however, the magnification ratio of an off-center ROI varies. Fig. 2 plots the magnification ratio as a function of projection angles. Relatively high magnification is achieved over a limited angular range (about 90°) when the source rotates on the same side as the ROI relative to the isocenter. However, the magnification is reduced when the X-ray source is on the opposite side of the ROI.

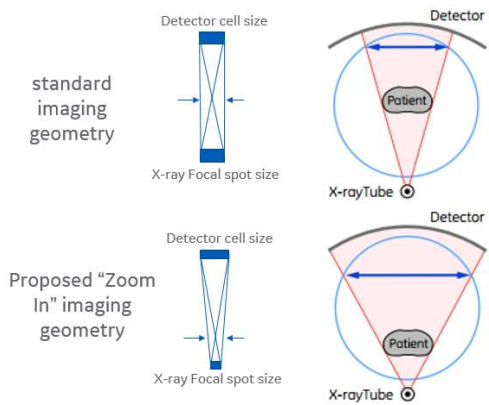

Fig 1. Illustration of the higher magnification of an off-center scan. The high magnification combined with a small focal spot can improve the spatial resolution of imaging.

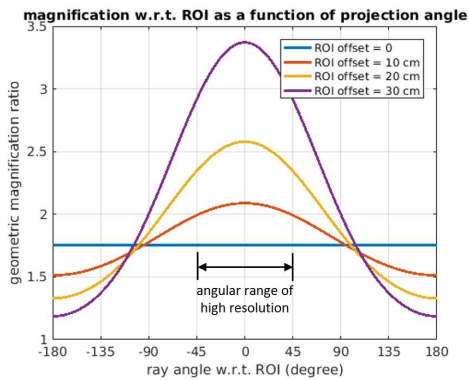

Fig. 2. The magnification ratio of an off-center ROI plotted as a function of projection ray angle. Zero ray angle is defined as when the X-ray source is closest to the ROI. (based on source-to-detector distance ≈ 1.1 m, source-to-isocenter distance ≈ 0.63 m.)

## Dual partial scans

In order to acquire high-resolution data over the complete Radon projection space of the ROI, we propose a dual partial scan scheme. As illustrated in Fig. 3, the patient is scanned at two bed positions. The scanner bed remains still during each of the two scans. After the first scan is completed, the patient is translated to the second bed position, where the second scan is performed. Each of the two scans acquires high-resolution projection data of the ROI over a limited angular range (about 90°). The two partial scans together acquire high-resolution projection of the complete Radon space (180°). More generally, let $(d_1\cos\varphi_1,\ d_1\sin\varphi_1)$ and $(d_2\cos\varphi_2, d_2\sin\varphi_2)$ denote positions of the ROI (relative to the iso-center) in the two scans, respectively. For the configuration shown in Fig. 3, $d_1 = d_2$, $\varphi_1 = 45°$, $\varphi_2 = 135°$ (angle zero is defined at 12 o'clock position), and the scanner bed only moves up or down between the two partial scans. In general, $|\varphi_1 - \varphi_2| \approx \frac{\pi}{2}$, so that the two partial scans cover complimentary angular ranges of high-resolution data, and the two partial scans together cover the complete Radon space of the ROI with high resolution. The exact displacement between the two scans may not be known exactly and will be estimated by a registration algorithm.

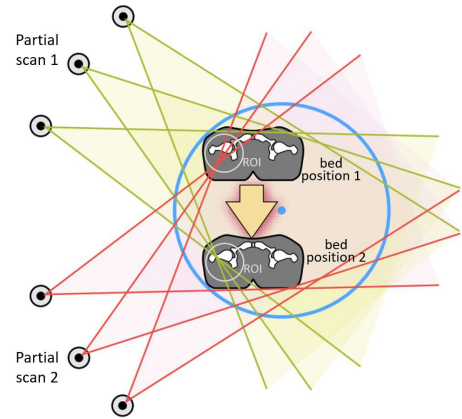

Fig. 3. Illustration of the concept of ZIPS CT. Two limited-angle partial scans are utilized to acquire high-resolution projection data of the ROI in complementary angular ranges.

Fig. 4 represents the acquired Radon samples in each partial scan in the ROI's local coordinate system, i.e., with respect to the center of the ROI, to visualize how the two partial scans acquire high-resolution projection data in complementary angular ranges. In the $i$th partial scan, a line integral sample $(r_i, \theta_i)$ relative to the iso-center can be mapped to a line integral sample $(r', \theta')$ in the ROI's local coordinate system through $r' = r_i - d_i \sin(\theta - \varphi_i)$ and $\theta' = \theta_i$. As shown in Fig. 4, in the first partial scan, high-resolution data (indicated by larger markers) are acquired over the angular range [0, 90°]. In the second partial scan, high-resolution data are acquired over the angular range [90°, 180°].

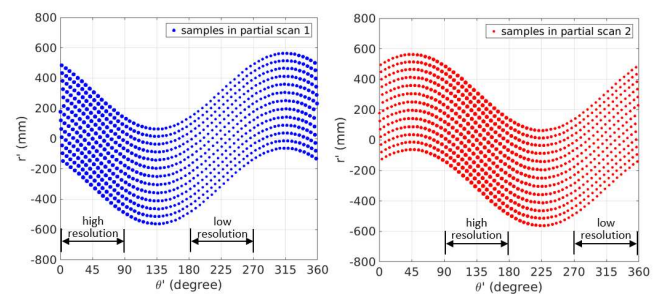

Fig. 4. Illustration of the sampling of the Radon space in the two partial scans. The samples are represented in the ROI's local coordinate system. Larger markers indicate higher-resolution samples due to higher magnification of the ROI.

## ZIPS CT Reconstruction

Because each partial scan covers only half of the Radon projection space with high resolution, these spatial frequencies from the two partial scans need to be properly merged by an image reconstruction algorithm to form a regular reconstructed image. To show the feasibility, we developed an initial approximate algorithm for joint reconstruction from the two partial datasets. In this method, we perform a Parker-weighted filtered backprojection (FBP) reconstruction from each of the partial datasets, then we merge the resulting images in Fourier space

$$x_{\text{zips}} = \mathcal{F}^{-1}(m\mathcal{F}x_1 + (1-m)\mathcal{F}Tx_2),$$

where the output $x_{\text{zips}}$ is a high-resolution ZIPS reconstructed image, $\mathcal{F}$ and $\mathcal{F}^{-1}$ denote the Fourier transform and its inverse, $x_1$ and $x_2$ are the Parker weighted half-scan reconstruction from the two partial scans respectively, $T$ is a registration operator that compensates for the displacement of the ROI between the two partial scans, $m$ is a Fourier-domain mask to select a 90° angular range of spatial frequencies from $x_1$ or $x_2$

$$m(\rho\cos\zeta, \rho\sin\zeta) = \begin{cases} 1, & \text{when } |\zeta - \varphi_1 + k\pi| \leq \dfrac{\pi}{4} \\ 0, & \text{otherwise} \end{cases}$$

Fig. 5 shows an example of the Fourier-domain mask $m$.

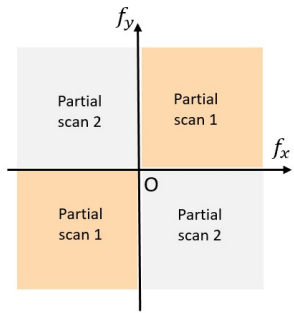

Fig. 5. An example of a Fourier-domain mask that extracts a 90° angular range of spatial frequencies from a partial reconstruction ($\varphi_1 = 45°$, $\varphi_2 = 135°$).

It should be noted that because of the shorter distance between the X-ray source and the off-center ROI, a relatively short rotation of the scanner gantry can cover a relatively large angular range of spatial frequencies of the ROI. We calculated that to acquire a 90° range of projection rays with respect to an off-center position, the X-ray source only needs to rotate by an angle $\alpha = 90° - 2\sin^{-1}(d/\sqrt{2}R)$, where $d$ is the offset relative to the isocenter, and $R$ is the source-to-isocenter distance. For example, $\alpha \approx 63°$ for $R = 600$ mm and $d = 200$ mm, and $\alpha \approx 49°$ for $R = 600$ mm and $d = 300$ mm. Therefore, the 90° limited-angle high-resolution projection data in each partial scan can be acquired with less than 90° of gantry rotation. The X-ray flux can be modulated down outside the small angular range to reduce patient dose.

## 3 Simulation Results

The proposed ZIPS technique was evaluated in a CatSim [10] simulation environment with a detector cell pitch of 1.1 mm, source-to-isocenter distance of 626 mm, and source-to-detector distance of 1098 mm. A 50 μm diameter tungsten wire phantom contained in a 5 cm diameter water cylinder and a line pair gauge phantom contained in a 22 cm diameter water cylinder were scanned. The phantoms were placed at the isocenter, 20 cm off-center, 30 cm off-center, and 35 cm off-center.

For high-resolution imaging with ZIPS CT, the X-ray focal spot size needs to be sufficiently small. We evaluated four focal spot sizes in this study: 0.8 mm, 0.4 mm, 0.3 mm, and 0.1 mm. A 10° anode angle was modeled in the simulation. High-resolution imaging also requires sufficient CT angular sampling rate to reduce azimuthal blur. Three angular sampling rates were evaluated in this study: 1000 views/rot, 4000 views/rot, and 8000 views/rot.

The line pair gauge was scanned with a 70 keV monochromatic X-ray and a tube current of 800 mA. Both X-ray quantum noise and detector electronic noise were modeled. The rotation period of the CT gantry was 1 s, 2 s, and 4 s, for conventional CT, ZIPS CT with 20 cm offset, and ZIPS CT with 30 cm offset, respectively. The slower gantry speeds in the ZIPS CT experiments were intended to approximately compensate for the very short source trajectory over which the high-resolution data can be acquired (approximately 63° with 20 cm offset, and 49° with 30 cm offset, compared with the full 360° for the conventional CT). We expect that mA modulation can be used in ZIPS CT so that patient dose will not significantly increase relative to a conventional CT scan, which will be a future topic.

All CT images were reconstructed on a 1024x1024 grid over a ROI of 10 cm diameter. A standard fan-beam FBP algorithm with Shepp-Logan ramp filter was used for image reconstruction. In ZIPS CT reconstruction, Parker weighting was applied to the CT sinogram to exclude the low-resolution portion of sinogram from the reconstruction. The Parker-weighted reconstructions from the two partial scans were merged in Fourier domain to form the ZIPS reconstructed image.

Fig. 6 shows the azimuthally averaged MTF measured on the reconstructed image from a noiseless simulation of the wire phantom. MTF@10% measured with the standard clinical CT (wire at the isocenter) was 10 lp/cm. Reducing the focal spot size and increasing the angular sampling rate only marginally increased the MTF@10% to 11 lp/cm in standard CT, because the bottleneck of spatial resolution in this case is the detector cell size. In ZIPS CT, MTF@10% measured with 20 cm offset, 30 cm offset, and 35 cm offset was 13 lp/cm, 17 lp/cm, and 20 lp/cm, respectively, which correspond to 30% to 100% (two-fold) improvement of spatial resolution relative to the conventional centered scan.

Fig. 7 shows reconstructed images of the line pair gauge phantom from noisy simulations. The 400 μm features are not distinguishable in the standard CT images but they are clearly revealed by the ZIPS CT images. The visual sharpness and contrast of 500 μm and 600 μm features are also improved by the ZIPS CT compared with the standard CT.

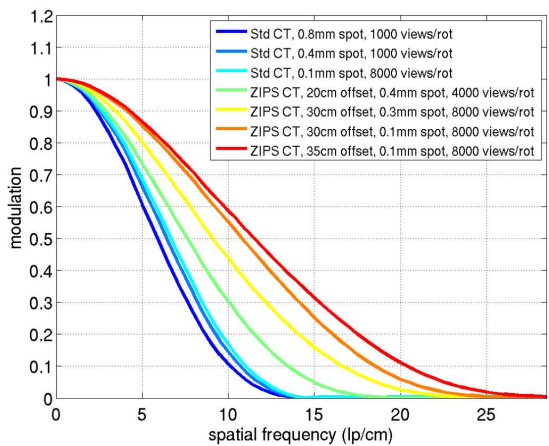

Fig. 6. MTFs measured with a 50 μm diameter wire. ZIPS CT provides 30% to 100% improvement of spatial resolution relative to the standard CT.

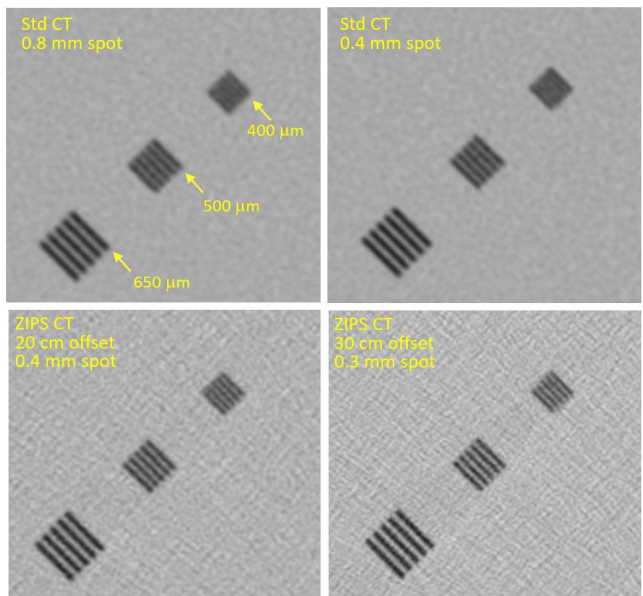

Fig. 7. Reconstructed images of a line pair gauge phantom with different CT techniques.

## 4 Conclusion and discussion

We proposed a ZIPS CT technique to improve the spatial resolution of rotating-gantry-based CT scanners without upgrading the CT detector array to one with finer-pitched cells. ZIPS CT introduces a novel dual partial scan scheme and leverages the high magnification of off-center ROIs to improve spatial resolution. Promising simulation results showed that ZIPS CT can achieve up to two-fold (100%) improvement of the MTF when compared with standard CT. Clear improvement of visual detectability of fine features was observed with a line pair gauge phantom. ZIPS CT does not require an upgrade of the CT detector array and thus has the potential to be applied to existing clinical CT systems. ZIPS CT is also orthogonal to pure algorithmic resolution-boosting methods and a combination may give further improvement. Future work includes more advanced ZIPS reconstruction algorithms, image registration between the two partial scans, and evaluation of patient dose.

### Acknowledgement

Research reported in this publication was supported by the NIH/NIBIB grant# 1R01EB028270-01A1. The content is solely the responsibility of the authors and does not necessarily represent the official views of the NIH. The authors would like to thank Drs. Ryan Breighner, Fernando Quevedo Gonzalez, Hollis Potter, and Marjolein van der Meulen for valuable discussions.

# Chapter 15

# Oral Session - DL in SPECT/PET

**session chairs**

Joaquin L. Herraiz, *Complutense University of Madrid (Spain)*

Rongping Zeng, *US Food and Drug Administration (United States)*

# A GAN-based Approach to Attenuation Map Generation for AdaptiSPECT-C Imaging

Akshay Iyer[1], Clifford Lindsay[1], Benjamin Auer[1], Kesava Kalluri[1], Jan De Beenhouwer[2], Navid Zeraatkar[1,3], Phillip H. Kuo[4], Lars R. Furenlid[4], and Michael A. King[1]

[1]Department of Radiology, University of Massachusetts Medical School, Worcester, USA
[2]imec-VisionLab, University of Antwerp, Antwerp, Belgium
[3]Department of Biomedical Engineering University of California Davis, Davis, USA
[4]Department of Medical Imaging & Wyant's College of Optical Sciences, University of Arizona, Tucson, AZ, USA

**Abstract** During Single-photon emission computed tomography, attenuation correction has been shown to enhance image quality and increase the quantitative accuracy of the images. A common method of acquiring attenuation maps is from attached CT-based scanners, but these systems can have significantly higher cost than SPECT-only systems. Herein, we report preliminary results which shows that we can generate realistic synthetic attenuation maps from SPECT emission data without the need for CT-based attenuation data.

Our method utilizes a deep learning network called a GAN, which were trained on a collection of 100 different realistic digital brains phantoms. Prior to training, phantoms acquisitions were simulated and reconstructed using the configuration of the AdaptiSPECT-C dedicated brain imaging system. The resulting synthetic attenuations maps achieved a high-level of accuracy when compared to ground truth, with a low average Normalized Root Mean Square Error of 0.1454 and Structural Similarity Index Measure of .99 across all test sets.

## 1 Introduction

Single-photon emission computed tomography (SPECT) is a widely used imaging technique that provides 3D images of the targeted radiotracer in-vivo distribution. A contributing factor to its widespread clinical use is the reduced cost of owning and operating SPECT systems, which can further be reduced by the omission of an anatomical acquisitions such as MRI or CT for generating/obtaining attenuation maps to be used in attenuation/scatter compensation. With such systems, it is often still desirable to perform attenuation/scatter correction during the reconstruction as it has been shown to enhance image quality [1] and quantitative accuracy [2]. Multiple alternatives for generating the linear attenuation correction factors at the SPECT tracer energy without a combined CT/MRI system are available, such as using additional stand-alone CT or MRI imaging systems. However, the latter approaches have the downside of needing to register, resize, and transform these images into the SPECT space, as well as to increasing acquisition time and cost, and patient radiation exposure (*in case of CT*). For the AdaptiSPECT-C project [3], we have taken a different approach by which mitigates the limitation of the previous methods by attempting to estimate the patient specific attenuation factors from the emission data itself using deep learning. Our approach uses a Generative Adversarial Network (GAN) approach, like the one Shi et al [4] used for cardiac perfusion imaging, but our approach differs in that we apply our method for a dedicated brain SPECT system.

In that sense, our work is more similar to that of Chen et al. [5] where our focus is on modelling the soft-tissue and cortical bone of the head. Unlike Chen et al., we instead of utilizing a conventional convolutional neural network (CNN), our technique focuses on employing a GAN optimized for Image-to-Image translation.

## 2 Materials and Methods

### 2.1    Dataset

Herein, we employed a dataset consisting of a family of 100 realistic digital head and brain phantoms [6] generated from the morphometric analysis of brain shape and volume [7]. Each phantom used for training our GAN, was derived from the anatomy of a unique volunteer patient which has wider distribution of shape, size, brain folds than other phantoms and provided up to two hundred different labeled tissue regions. Each phantom was simulated for $^{99m}$Tc-HMPAO perfusion imaging with the 5-pinhole configuration of the dedicated brain imaging system AdaptiSPECT-C [8]. A total number of 24 million detected counts million consistant with imaging a clinically appropriate activity level for an imaging period of 20 minutes, as per clinically at our site. Additionally, the counts correspond to an equal imaging time based on the simulated sensitivity compared to clinical imaging with a dual-head parallel system commonly used in brain SPECT procedures [8,9].

The ground truth attenuation maps were composed of four attenuation regions (i.e. air, soft tissue, bone, brain). The linear attenuation coefficients at 140.5 keV (i.e. $^{99m}$Tc principal energy) of these regions were exported from the GATE software [10]. In-house software derived from GATE [11], was used to compute forward projections with attenuation modeling and reconstructing the projections with Pixel-based Ordered Subset Expectation Maximization [12] without attenuation compensation for 7 iterations with a matrix size of $120^3$ and 2mm voxels.

### 2.2    Data Pre-processing

The training dataset was organized by pairing each of 100 reconstructed images with the corresponding ground truth attenuation maps. A sample slice of SPECT image and its corresponding attenuation map can be seen in Fig. 2. Both images were edge-padded to a size of $128^3$ to match the input size of the network. Augmentations in the form of random rotation and horizontal flips were performed for each image-map pair. essentially tripling the dataset size. For rotation, the image-map pair is rotated for the same angle randomly sampled between ±10 degrees. This would make the network robust against slight perturbations in head orientation which could be induced during acquisition. The horizontal flip results in a new pair of image-map which closely resembles a real-life pair due to the symmetry in the original pair but slight variations. The entire dataset, including augmented images, was split into 5-folds, stratified by subject group to ensure proportionate distribution into train and validation.

### 2.3    Network Architecture

We used a deep learning framework called a Generative Adversarial Network (GAN) to synthesize synthetic attenuation maps from emission data only. GANs can be considered as two separate deep learning networks, consisting of a Generator network and Discriminator network, which are in competition to outperform each other in a specific task. For this project, the task is to synthesize plausible attenuation data from emission data only. Our specific network architecture was inspired by *Vox2Vox* [13], which tries to solve the Image-to-Image translation problem [14](in our case volume-to-volume), by conditionally training a generator in the form of an encoder-decoder network (see Fig. 1a), to learn a translation process from one image type (emission image) to another (attenuation map). For this work, the generator is conditioned on the 100 unique emission image sets, for which each exhibit a wider variety in shape of the anatomical structures of real humans (see [6] for more details) in order to produce attenuation maps consistent with that person's physiology. Additional structural details regarding the generator sub-network can be found in [15] (U-Net). Input to the generator consists only of reconstructed SPECT images without attenuation compensation. The output is a synthetic attenuation map of the same size and is matched against a similar-sized ground truth attenuation map.

The second half of the GAN network (i.e., the Discriminator, see Fig. 1b) consists of a convolutional neural network CNN, which is trained to distinguish between real and synthetic attenuation maps. The discriminator was implemented as a CNN in the style of *Patch-GAN* [14], which takes input pairs of reconstructed SPECT images and attenuation maps (alternating synthetic and real) and produces an output indicating its prediction for identifying the synthetic attenuation map. Typical discriminator outputs would provide a binary decision (one bit, indicating a prediction) on whether the input image was synthetic. Instead our discriminator produces an output of size $8^3$, which can be interpreted as the confidence level for the prediction at patched regions of the input voxels (i.e., inverse of the quality of the synthetic attenuation map). In other words, the discriminator tries to model high-frequencies features in the attenuation map by considering correctness of structure in $N^3$ patches from the original and provides a corresponding prediction level for each patch. This allows a flexible size input, which can be adapted to the image sizes that are multiples of 8.

### 2.4    GAN Image-to-Image Translation Training

The loss function $L_G$ (eq. 1) used for training the generator $G$ was a weighted sum of two different loss functions, the first being the $L_2$ norm between a tensor of ones and the output of the discriminator $D$ on the synthetic attenuation map produced by the generator and the second being the $L_1$ norm between the ground truth attenuation map $A_{GT}$ and synthetic attenuation map $A_S$. The $L_2$ norm part is useful for incorporating feedback, indirectly from the discriminator, in order to learn how to generate images which the discriminator finds indistinguishable from the ground truth. Whereas the $L_1$ part of the equation ensures the generated attenuation maps are similar to the original ground truth maps.

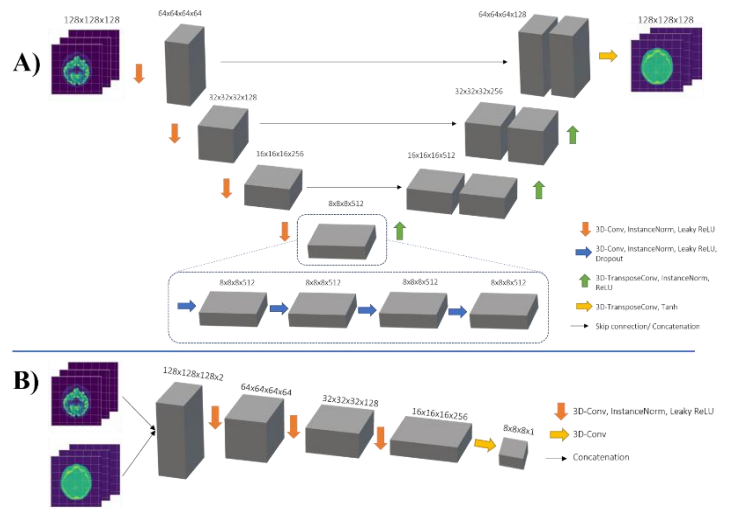

**Fig.1: The GAN network architecture diagram. Part A on top shows the overall structure of the generator, which is used to create the attenuation maps. Notice the U-Net type shape. Part B on the bottom shows the Discriminator network, whose is trained to differentiate between synthetic and ground truth attenuation maps. The structure of this network follows a typical CNN type structure.**

$$L_G = L_2[1, D(A_S)] + \alpha L_1[A_{GT}, A_S] \quad \text{(eq. 1)}$$

The loss function $L_D$ (eq. 2) used for training the discriminator was the average of two different loss functions, the $L_2$ norm between a tensor of ones and the output of the discriminator $D$ for input SPECT image and ground truth attenuation map $A_{GT}$ and the $L_2$ norm a tensor of zeros and of the output of the discriminator $D$ for input SPECT image and synthetic attenuation map $A_S$ produced by the generator. In other words, it is the average of the discriminator's predictions for both the ground truth and synthetic attenuation maps.

$$L_D = L_2[D(A_{GT}), 1] + L_2[D(A_S), 0] \text{ (eq. 2)}$$

Training was performed using the Adam optimizer with a learning rate of 2e-4 for both the generator and discriminator and $\beta_1 = 0.5$ and $\beta_2 = 0.999$ as was done in [14]. Dropouts with a rate of 0.2 was performed after every 3D convolution in the bottleneck section of the Generator. This provides regularization in the network training which has been shown to improve training in conditional-GANs and avoid mode collapse [14]. Training was implemented using the PyTorch framework [16] and run on a Nvidia GeForce GTX 1080 Ti GPU. Each fold was trained for 200 epochs, which required approximately 5 hours per fold.

## 3 Results

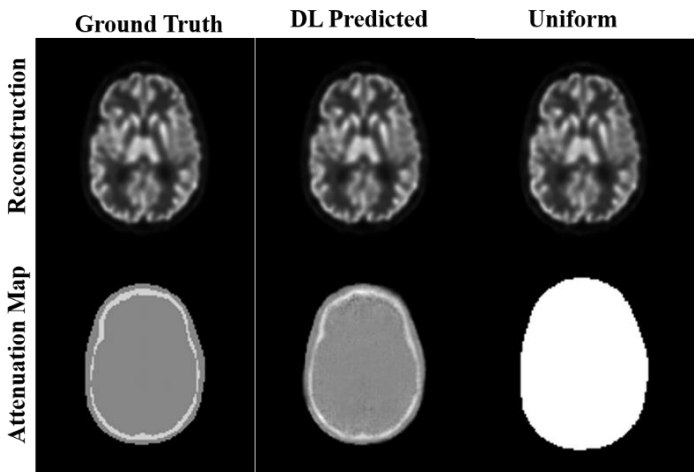

*Fig 2. Results comparing ground truth, predicted by DL, and uniform attenuation maps in reconstruction. Top row, from left to right shows the reconstructed activity phantoms with corresponding attenuation maps (middle row). Qualitatively, the reconstructions show the ground truth and DL predicted are very similar in appearance with slightly more differences visible in the uniform attenuation map.*

The accuracy of attenuation map generation process was determined from the difference between the ground truth attenuation map and the predicted attenuation map. The difference was quantified as the normalized root mean square error (NRMSE) between the two maps. For each fold, we calculated the mean and standard deviation of the normalized root mean square error between the synthetic and ground truth attenuation maps for all data/phantom in the validation set (i.e., not part of training for that fold). The results of these calculations are listed in Table 1. Qualitatively, as illustrated in Fig. 2, we show slice 67 of a set of reconstructions (top row), attenuation maps (middle row) from a validation subject which highlights our results.

*Table 1: Average Normalized Root Mean Squared Error (NRMSE) and Peak Signal to Noise Ratio (PSNR) across the overall phantom dataset after a 5-fold cross-validation.*

| Fold # | 1 | 2 | 3 | 4 | 5 |
|---|---|---|---|---|---|
| Avg. NRMSE | .265±.014 | .121±.020 | .115±.012 | .117±.015 | .109±.007 |
| Avg. PSNR | 39.9± .683 | 47.4± 1.64 | 47.9± 1.28 | 47.7± 1.31 | 48.5± .717 |

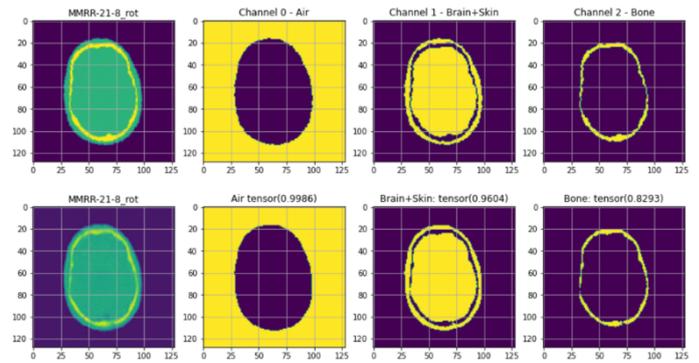

*Fig. 3: Dice/overlap images illustrating the process of thresholding and discretizing the continuous output values.*

The left column presents the ground truth and the middle and right column are the generated and uniform attenuation maps, respectively. The uniform attenuation map was created by setting all the tissue values from the ground truth to a uniform value consistent with a soft tissue linear attenuation coefficient at 140.5 keV.

During the training, we also calculated the Structural Similarity Index Measure (SSIM) and the Dice scores to assess the performance and accuracy of the GAN network in estimating attenuation maps. Again, each metric was calculated only for the output of the validation sets of the network for each fold then averaged across all folds. SSIM was calculated between the ground truth and synthetic attenuation maps after converting the attenuation maps to discrete representation (i.e., 4 distinct values representing the 4 tissue types) through a thresholding of the continuous values from the networks output. This representation is necessary for calculating the Dice scores and more accurate for calculating the SSIM metrics. All linear attenuation coefficient values less than 0.1 cm$^{-1}$ were set to 0, values between 0.1 and 0.19 were set to 1, and values greater than 0.19 were set to 2. The threshold parameters were determined after creating the confusion matrices for the outputs and balancing the voxels for the 3 classes.

| Air Dice Score | Brain & Soft tissue Dice Score | Bone Dice Score | SSIM |
|---|---|---|---|
| 0.99 | 0.97 | 0.83 | 0.99 |

*Table 2: Average Dice scores for the 3 classes and average SSIM for the overall output after a 5-fold cross-validation.*

## 4 Discussion

### 4.1    *Analysis*

From the results, it can be seen that our method produces a plausible set of attenuation maps from the out of the deep learning network. In particular, we are encouraged by the faithful and accurate recreation of the different shapes of the tissue values, reaching a Dice score of *0.99* for the outline (air) and 0.97 for the soft tissues (brain, soft tissue, muscle, etc.). For the cortical bone, a Dice score of 0.83 is also quite encouraging and not unexpected as the bone voxels represent the least frequent tissue type and therefore the network has less opportunity to learn these shapes (similar to a class imbalance problem). Additionally, the NRMSE and PSNR values are also quite encouraging, and with the exception for Fold 1, the values hover between *0.109-0.121*. This might be explained by a slight screen-door effect that occurs on some of the images which can be attributed to our method of deconvolution.

### 4.2    Work To be done

Going forward, we plan to explore the approach of training the network to output discrete attenuation values and compare this type of training to the current method of using a continuous output and thresholding the values to discretize them. This will provide us the opportunity to utilize a variety of different techniques that can be borrowed from the Semantic Segmentation research, which will might help mitigating the need to optimize the threshold values by training the networks to learn this from the data.

Currently, our bone tissue categorization does not consider or differentiate different types of bone structures within the head. We would like to expand our categorization to include both compact and spongy bone regions to more accurately account for their attenuation. Additionally, we plan to include noise levels within our training scheme and plan to accurately account for variations in uptake and shorter acquisition times than we are currently modelling with the noise-free images. We hope this will more realistically model clinical acquisitions for use in future studies. Our plan is to perform both fine-tuning with noisy images as well as retraining from scratch with the noisy images only using the same family of digital phantoms.

Finally, we expect that photons/counts coming from higher attenuation and absorptions areas are mostly scattered/attenuated and therefore might not be detected in the photopeak window. We would like to incorporate scatter windows to improve bone recovery by including scatter in the simulation, reconstruction, and training.

## 5 Conclusion

In this work, we have utilized a deep learning network called a GAN which is able to generate realistic-looking, plausible attenuation maps with high-level of accuracy (SSIM of 0.99) from the emission data alone. The training of this network was performed using a family of realistic digital phantoms for which acquisitions were simulated and reconstructed using the 5-pinhole configuration of the brain dedicated AdaptiSPECT-C system.

## 6 Acknowledgements

Research reported in this publication was supported by the National Institute of Biomedical Imaging and Bioengineering (NIBIB) of the National Institutes of Health under Award Number R01 EB022521. The content is solely the responsibility of the authors and does not necessarily represent the official views of the National Institutes of Health.

# Deep Learning based Model Observers for Multi - Modal Imaging

Hamidreza Naderi Boldaji[1], Mayank Patwari[1,2], Maximilian P. Reymann[1,2,3], Ralf Gutjahr[2], Rainer Raupach[2], and Andreas Maier[1]

[1]Friedrich-Alexander University Erlangen-Nürnberg, Pattern Recognition Lab, Erlangen, Germany
[2]Siemens Healthineers AG, Forchheim, Germany
[3]Clinic for Nuclear Medicine, University Hospital Erlangen, Germany

**Abstract** Model Observers (MO) are useful tools for assessing task-based image quality in medical imaging systems. Detection and characterization of anatomical structures in clinical imaging are some of the most significant tasks of MOs. Various linear model observers have been widely applied for these tasks with great success. However, we are unable to deploy such observers without a high prior knowledge of the task, the signal, and the background. Additionally, most model observers currently used in medical imaging research are used to optimize single modality systems such as standalone CT or PET scanners. In this paper, we aim to solve these problems by means of a Convolutional Neural Network (CNN). We employ supervised learning with CNNs to approximate the Human Observer (HO) for a signal-known-exactly and background-known-statistically (SKE/BKS) signal detection task. We use this observer in a novel multi - modality imaging setting. First, we propose a CNN-based anthropomorphic model observer to predict human observer detection performance for PET images. Then, we suggest another CNN-based MO for multi - modal PET and CT images. We trained both networks using 160 sets of joint PET/CT images of a dataset from the MICCAI2020 HECKTOR challenge, and tested on 41 patients from the same dataset. Our CNN - MOs have high accuracy in detecting the lesions present in the image (AUC = 0.95). Furthermore, our MOs degrade monotonically with increase in noise levels, similar to how a HO would. Including both modalities into an MO results in improved lesion detection performance, especially in the presence of image noise, compared to single modality CNN - MOs.

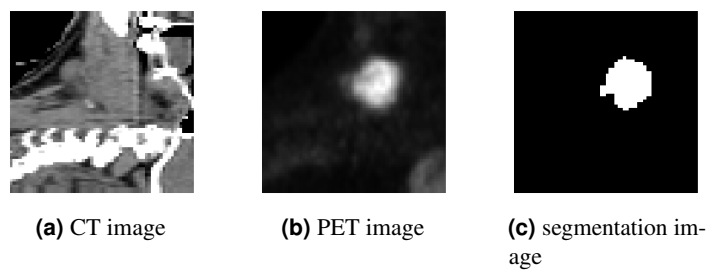

**(a)** CT image    **(b)** PET image    **(c)** segmentation image

**Figure 1:** Sample images of MICCAI2020 HECKTOR challenge dataset. This dataset contains (a) aligned CT (window [-100, 200]) and (b) PET images, along with (c) segmentation masks

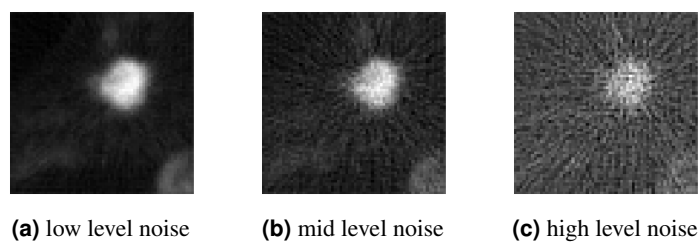

**(a)** low level noise    **(b)** mid level noise    **(c)** high level noise

**Figure 2:** Sample of noisy image in at 3 different levels of (a) low level noise, (b) medium level noise, (c) high level noise

## 1 Introduction

Image quality is assessed by the performance of some "observer" on a specific pre - defined task. The observer (either a human or a mathematical model) measures how well the desired information for a given task can be extracted. Model Observers (MOs) have been widely used in medical domain in classification tasks, such as a task that requires classifying patients into healthy vs. diseased, or estimation tasks, such as a task that requires estimating the volume of a tumor[1, 2]. Image evaluation for a particular trial with Human Observers (HOs) can be challenging due to limitations in time, money and availability of trained and certified observers. Observer 'jitter' is also a problem hindering reproducibility of results, even with the same HO on the same data over time [3]. Therefore to address such problems, researchers have focused on the development of a reliable MO that assists rapid, accurate, efficient, tireless and repeatable assessment of a particular task or parameter. While the mathematical and linear MOs follows the performance of the HO in different tasks (e.g. Channelized Hotelling Observer (CHO) observer model to assess image quality of X-ray angiography system[4]), the performances are still suboptimal in many cases. Furthermore, the signal and background need to be very specifically designed for mathematical observers to be useful.

In recent years, with the growth in popularity of Artificial Intelligence (AI) and Deep Learning (DL), there have been numerous studies applying convolutional neural network (CNN) as MOs in various tasks e.g. detection, characterization and localization of lesion in the image. DL based MOs perfectly reflect the non-linearity of the observer within the network, producing more accurate performance.

The first work that uses DL for MO was proposed by Massanes et al. [5] in 2017, where they trained a CNN to perform detection and localization of a small Gaussian object in the presence of a correlated noise. De Man et al.[6] compared DL based and human observer performance for detection of simulated lesions. Kim et al. [7] implemented an ideal observer model using CNNs for breast CT images. Lorente et al. [8] introduced a DL based MO based on U-Net for a localization task. Fan et al. [9] suggested a DL based MO that replicated human observerd for PET imaging.

However, DL based MOs have received little study in multi - modal imaging systems where there are different sources of information for a possible lesion detection task. In this study, we propose two DL based MOs. The first proposed MO introduces a CNN to detect the presence of abnormalities in small patches of PET images. The second proposed MO introduces a network which analyses PET and CT image patches for same task simultaneously. We also analyze MO performance in presence of different levels of random noise.

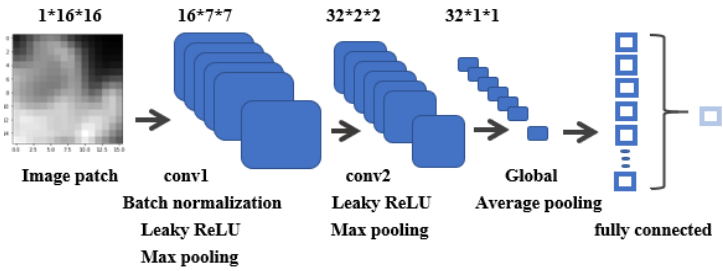

**Figure 3:** Structure of first proposed network

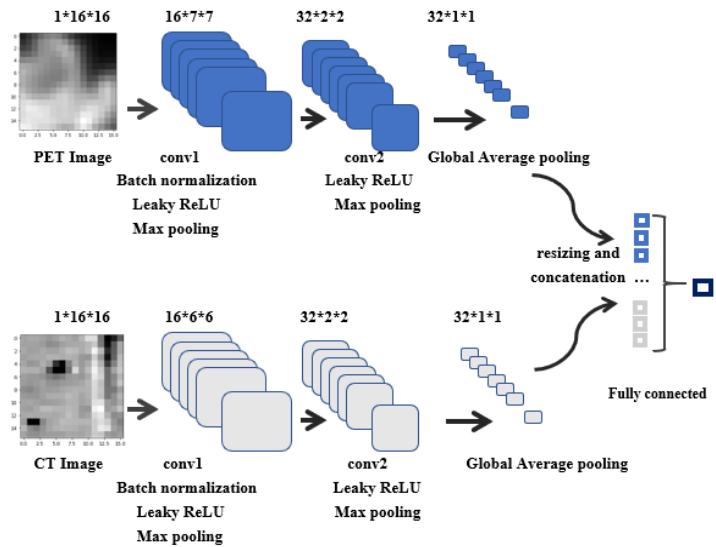

**Figure 4:** Structure of second proposed network

## 2   Methods

### 2.1   Data preparation

In this study we use the dataset from the MICCAI 2020 HECKTOR challenge, which includes multi - modal PET/CT scans from 201 different patients, collected at 4 different sites in Switzerland. Each set includes FDG-PET, CT and a ground truth volume labeled by experts. These scans were head and neck scans containing oropharyngeal tumors. This dataset was available at https://www.aicrowd.com/challenges/miccai-2020-hecktor. We perform patchwise tumor detection, so that our CNN - MOs could later be used for both detection and localization. First, we resampled all images into a fixed size, determined by a bounding box of the lesion provided by the challenge organizers, and subsequently divide them into small non-overlapping patches. Therefore, the dataset contained 204800 image patches of size $16 \times 16$. To test the performance of our CNN - MOs in a noisy situation, we added Poisson noise to the Radon transform of the images. This resembles the noise present in clinical PET/CT images.

We created 3 datasets in which 3 different levels of random Poisson noise (Poisson $\lambda = 10$, 1, 0.1) were applied to the original images. Figure 1 displays the original images in the dataset, while Figure 2 displays artificially created noisy images.

In previous studies[7, 9–11] evaluation is based on 2 or 4 Alternative-Forced Choice (2AFC or 4AFC) to detect lesion presence or absence. Since we are using CNN - MOs, we instead classify each patch independently, as a signal present/signal absent binary task.

For training labels, we use the ratio of labeled tumor tissue to the whole patch size. Let $V_{li}$ be the sum of lesion pixels in the $i$-th patch or lesion volume and $V_{pi}$ be the sum of $i$-th patch pixels or total patch volume. The $i$-th patch label $L_i$ is calculated according to Eq. 1.

$$L_i = \frac{V_{li}}{V_{pi}} \tag{1}$$

### 2.2   Training the Networks

We divided the dataset into a training set of 160 patients (163840 patches) and a test set containing 41 patients (40960 patches) and employed data augmentation by means of horizontal and vertical flipping. To focus on soft tissue, the CT data was preprocessed by windowing all values outside of the range $[-100, 200]$ HU.

The first network consisted of two $3 \times 3$ convolutional layers with 16 and 32 feature maps respectively, with Leaky Rectified Linear Unit (Leaky ReLU) [12] activation functions, two $2 \times 2$ max-pooling layers, one global average pooling, and one fully connected layer as shown in Figure 3. This network was trained for single modality lesion detection task in PET images.

The second proposed network which was aimed at multi-modality image assessment has a Y-shape structure. It contains two branches of convolutional blocks to analyze PET and CT images separately. Apart from the difference in the first convolutional layer kernel size of $5 \times 5$ in the CT analyzing branch, both branches are identical to the previously presented network. The outputs of these two branches are reshaped and passed through a fully connected layer as shown in Figure 4. The network architectures were selected empirically.

In this study we focused on image patches of size $16 \times 16$, however, the use of a global average pooling layer in both networks enables us to process image patches of arbitrary sizes. Patch-wise assessment not only enables tumor detection, but also enables tumor localization.

Since most of the labels are zero (no lesion in the patch), the dataset is highly unbalanced. To avoid gradients vanishing, we modified the MSE loss function for our network. Additionally, we desired the network to focus more on the patches that contain lesions. Therefore, the errors of non-zero labels (with lesion patches) are weighted 4 times (selected empiri-cally) more than errors of patches with zero labels (see Eq. 2). $L_i$ is i-th patch label and $P_i$ is i-th patch network prediction.

**Table 1:** RMSE and AUC results of proposed networks for test set

| | First Network | | | | Second Network | |
| | PET images | | CT images | | joint PET/CT | |
| | RMSE | AUC | RMSE | AUC | RMSE | AUC |
|---|---|---|---|---|---|---|
| **No additional noise** | $5.3 \times 10^{-4}$ | 0.95 | $22.6 \times 10^{-4}$ | 0.86 | $4.2 \times 10^{-4}$ | 0.95 |
| **low-level noise** | $5.8 \times 10^{-4}$ | 0.92 | $86.2 \times 10^{-4}$ | 0.58 | $5.0 \times 10^{-4}$ | 0.94 |
| **mid-level noise** | $10.4 \times 10^{-4}$ | 0.85 | $89.7 \times 10^{-4}$ | 0.57 | $7.0 \times 10^{-4}$ | 0.90 |
| **high-level noise** | $15.1 \times 10^{-4}$ | 0.65 | $90.3 \times 10^{-4}$ | 0.54 | $12.4 \times 10^{-4}$ | 0.77 |
| **AUC across all noise levels** | **0.79** | | **0.64** | | **0.86** | |

**Table 2:** Specificity and Sensitivity results

| | First Network | | | | Second Network | |
| | PET images | | CT images | | joint PET/CT | |
| | sensitivity | specificity | sensitivity | specificity | sensitivity | specificity |
|---|---|---|---|---|---|---|
| **No additional noise** | 0.877 | 0.877 | 0.771 | 0.772 | 0.881 | 0.882 |
| **low-level noise** | 0.841 | 0.841 | 0.574 | 0.573 | 0.874 | 0.875 |
| **mid-level noise** | 0.769 | 0.768 | 0.573 | 0.573 | 0.827 | 0.828 |
| **high-level noise** | 0.610 | 0.610 | 0.557 | 0.558 | 0.700 | 0.700 |

$$\mathcal{L} = \frac{1}{N} \sum_i^N \varepsilon_i \times (P_i - L_i)^2 \text{ with } \varepsilon_i = \begin{cases} 0.25, & \text{if } L_i = 0 \\ 1, & \text{else.} \end{cases} \quad (2)$$

We used the Adam optimizer[13] for training with a learning rate of $5e^{-4}$ and a batch size of 64. We trained the network for 30 epochs with a modified Mean Squared Error (MSE) loss function (Eq. 2).

## 3 Experiments and results

The evaluation of the CNN-MO was done by measuring the performance of the trained CNN-MO on a simple tumor detection/localization task. The images were divided into overlapping patches of fixed size, and the CNN-MO was independently tasked with delivering a single signal present (positive) and signal absent (negative) response for each patch. Patches where any amount of tumor was present were defined as positive, while patches with no tumor whatsoever were defined as negative.

To examine performance under noisy conditions, we report the performance of all networks with and without noise by computing the Area Under the Curve (AUC) of the patchwise detection task, as well as the sensitivity and specificity. The Root Mean Squared Error (RMSE) of the results is reported as well. Table 1 shows the RMSE and AUC results of these tests, while Table 2 shows the sensitivity and specificity of the detection tasks at different noise levels.

We tested all networks in presence of 3 different levels of noise to analyse performance. Network uncertainty increases with increase in noise level, as shown by the decrease in AUC in Table 1 (Spearman Coefficient = 1) and Figure 5. This resembles the decline in detection performance with increase in noise levels as shown by HOs [1].

Results indicate that CT images contain less significant information regarding lesions (AUC = 0.86) compared to PET images (AUC = 0.95). Integrating both modalities does not improve detection accuracy in the no noise case (AUC = 0.95), however, the integrated modalities improve the detection performance with the increase in noise levels. In the case of high noise, joint PET/CT network is more robust to tumor detection (AUC = 0.77) compared to other single modality networks (AUC = 0.65 for PET images and AUC= 0.54 for CT images). The increase in the severity of the image noise decreases the detection performance.

## 4 Conclusion

In this work, we proposed two CNN - MOs for SKE/BKS lesion detection tasks. We trained the first CNN - MO for single modality detection task with just PET images and repeated for only CT images. We trained a Y-shaped CNN - MO for multi-modality cases. We also applied 3 different levels of Poisson distributed random noise to the dataset to investigate the performance of the network.

Overall, the presented results show the second proposed MO leads to better lesion detection in higher noise levels. This implies that, combination of information in CT and PET images help the network to find the lesion better when the noise levels increase. The magnitude of improvement increases with the increase in noise levels. In both proposed CNN / MOs, Spearman's coefficient of the AUC with respect to the level of noise equals to 1 which means by increasing the noise level, the MO uncertainty increased monotonically as expected.

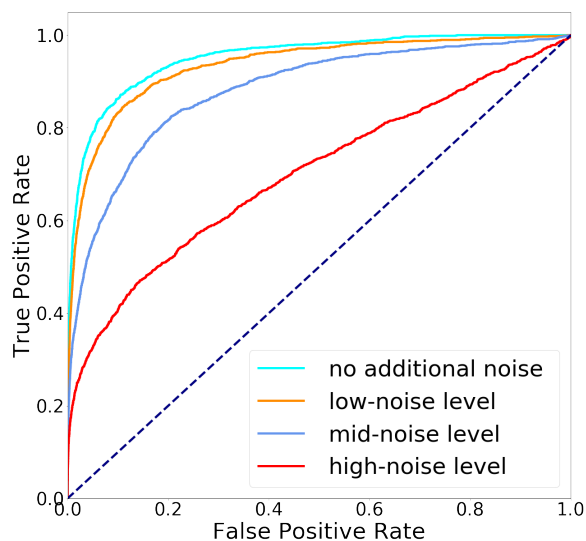

**Figure 5:** ROC curves for the different noise levels evaluated on the joint PET/CT network

In future work, we plan to compare these results to other anthropomorphic model observers. We plan to improve the proposed network and explore various network architectures to obtain better performance in replicating human observers. Finally, we will attempt to use the developed CNN-MOs to optimize both uni-modal and multi-modal imaging systems.